\documentclass[11pt,paper=a4]{report}

\usepackage[a4paper,landscape]{geometry}
\usepackage{fancyhdr}
\usepackage{authblk}
\usepackage{amsmath}
\usepackage{color}
\usepackage{graphicx}
\usepackage{cancel}
\usepackage[usenames,dvipsnames]{xcolor}
\usepackage{chngcntr}
\usepackage{natbib}

\usepackage{hyperref}
\hypersetup{
        colorlinks = true,
        linkcolor = blue,
        anchorcolor = red,
        citecolor = blue,
        filecolor = red,
        urlcolor = blue
}

\newcommand{\Msun}{\mbox{M$_\odot$\,}}         

\newcommand{\eht}{\overline}    
\newcommand{\fht}{\widetilde}    
\newcommand{\dr}{\frac{\partial}{\partial r}}

\newcommand{\fav}{\widetilde}    
\newcommand{\av}{\overline}  

\def\ef#1{#1'}
\def\ff#1{#1''}

\def\erho{\eht{\rho}}

\newcommand{\dgr}{\mbox{$^\circ$}}           

\counterwithout{section}{chapter}

\title{Compressible Hydrodynamic Mean-Field Equations in Spherical Geometry and their Application to Turbulent Stellar Convection Data}

\author[1]{Miroslav Moc\'ak}
\author[1,2,3]{Casey A. Meakin}
\author[4]{Maxime Viallet}
\author[2]{David Arnett}
\affil[1]{Theoretical Division, Los Alamos National Laboratory, Los Alamos, NM 87545, USA}
\affil[2]{Steward Observatory, University of Arizona, Tucson, AZ 85721, USA}
\affil[3]{New Mexico Consortium, Los Alamos, NM 87545, USA}
\affil[4]{Max-Planck-Institut f\"ur Astrophysik, Karl Schwarzschild Strasse 1, Garching, D-85741, Germany}
\affil[ ]{}
\affil[ ]{\textit {miroslav.mocak@gmail.com, casey.meakin@gmail.com, mviallet@mpa-garching.mpg.de, wdarnett@gmail.com}}

\date{\today}

\begin{document}
\bibliographystyle{plainnat}
\maketitle

\tableofcontents

\newpage

\section{Introduction}
\label{sect:intro}

\par We present a statistical analysis of turbulent convection in stars within our Reynolds-Averaged Navier Stokes (RANS) framework in spherical geometry which we derived from first principles (see \S\ref{sect:mean-field-derivation} and further sections for details). The primary results reported in this document include (1) an extensive set of mean-field equations for compressible, multi-species hydrodynamics and (2) corresponding mean-field data computed from various simulation models. Some supplementary scale analysis data is also presented.

\par The simulation data which is presented includes (1) shell convection during oxygen burning in a 23 $\Msun$ supernova progenitor, (2) envelope convection in a 5 $\Msun$ red giant, (3) shell convection during the helium flash, and (4) a hydrogen injection flash in a 1.25 $\Msun$ star. These simulations have been partially described previously in \citet{MeakinPhD2006,MeakinArnett2007,meakin2007a,meakin2010,arnett2009,ArnettMeakin2010,viallet2011,viallet2013,VialletMeakin2013} and \citet{Mocak2009,Mocak2011}. New data is also included in this document with several new domain and resolution configurations as well as some variations in the physical model such as convection zone depth and driving source term.

\vspace{0.25cm}

\par {\bf The long term goal of this work is to aid in the development of more sophisticated models for treating hydrodynamic phenomena (e.g., turbulent convection) in the field of stellar evolution  by providing a direct link between 3D simulation data and the mean fields which are modeled by 1D stellar evolution codes.} As such, this data can be used to test previously proposed turbulence models found in the literature and sometimes used in stellar modeling (see e.g. \S\ref{sec:model-comparison}). This data can also serve to test basic physical principles for model building and  inspire new prescriptions for use in 1D evolution codes.

\subsection{Derivation of RANS Equations}

\par We present  an extensive set of mean-field evolution equations in the same spirit as those studied in classical turbulence research \citep[e.g.,][]{hinze1975} but extended to the fully compressible, multi-species treatment in spherical geometry relevant to stellar interiors.  (A complimentary  set of mean-field equations in the ideal magnetohydrodynamics (MHD) approximation will be released in the future.) Although the equations presented here are similar to those recently published in \citet{canuto2011part1,canuto2011part2,canuto2011part3,canuto2011part4,canuto2011part5} we have adopted a different approach.  In particular, we present our equation set completely unmodeled and unclosed and without approximation.  Our intention is to provide exact evolution equations together with what is in essence raw data to be used to check approximations in models such as those presented in Canuto's and similar works (an example of this is presented in \S\ref{sec:model-comparison}). Therefore, this work is largely pedagogical and includes lengthy derivations in later sections.

\par The equation set presented describes the time evolution of a variety of mean quantities.  Examples include the mean specific internal energy, the mean specific entropy, the mean convective flux (enthalpy flux), the mean entropy flux, the mean specific kinetic energy (velocity variance), the mean composition, and the Reynolds stresses, to name just a few.  The mean fields are all one-dimensional (a function of only radial coordinate or enclosed mass) and time-dependent such that they correspond to the quantities modeled in traditional 1D stellar evolution codes (see e.g. \cite{kippenhahn1990,paxton2011}). While we believe that this document is relatively complete and self-contained, we encourage readers that are not that familiar with RANS equation sets to peruse \citet{VialletMeakin2013} for additional background and discussion on the general approach as well as basic textbooks on turbulence modeling such as that by \citet{Pope2000}.

\vspace{0.25cm}
\par We obtain our 1D RANS equations by introducing two types of averaging:  statistical averaging and  horizontal averaging \citep{Besnard1992,VialletMeakin2013}. In practice, statistical averages are computed by performing a time average (the ergodic hypothesis). Therefore, the combined average of a quantity $q$ is defined as

\begin{align}\label{eq:eht}
\eht{q}(r,t) = \frac{1}{T\Delta\Omega}\int_{t- T/2}^{t+T/2} q(r,\theta,\phi,t')~d\Omega~dt'
\end{align}

\noindent where $d \Omega = \sin \theta d \theta d \phi$ is the solid angle in spherical coordinates, $T$ is the averaging time period, and $\Delta\Omega$ is total solid angle being averaged over. This type of average is referred to as a {\em Reynolds average}.

\par The flow variables are then decomposed into mean and fluctuation $q = \eht{q} + q'$, noting that $\eht{q'} = 0$ by construction. Similarly, we introduce Favre (or density weighted) averaged quantities by 

\begin{align}\label{eq:fhb}
\fht{q} = \frac{\eht{\rho q}}{\eht{\rho}}
\end{align}

\noindent which defines a complimentary decomposition of the flow into mean and fluctuations according to $q = \fht{q} + \ff{q}$. Here, $\ff{q}$ is the Favrian fluctuation and its mean is zero when Favre averaged $\fht{\ff{q}} = 0$. For a more complete elaboration on the algebra of these averaging procedures we refer the reader to \S\ref{sec:reynolds-decomposition} and \ref{sec:favre-decomposition} as well as \citet{Chassaing2010}.

\vspace{0.25cm}

\par {\bf A note on domain geometry:} All of the simulation data examined in this document has been modeled using a spherical coordinate system and all of the mean-fields are presented as one-dimensional mean fields in spherical coordinates.  The choice of a spherical coordinate system results in several terms due solely to curvilinear coordinates such as the fictitious Coriolis and centrifugal forces. For readability and concision we denote these so-called "geometric terms" with the super- and sub-scripted symbols $G$ and $\mathcal G$ (see definitions in Tables~\ref{tab:rans-def} to \ref{tab:rans-def-cont-3}).

\par On a practical note, all of the simulation data studied in this document use computational domains that are restricted to sectors  (or wedges) in the spherical coordinate system and have periodic boundary conditions in both angular directions. Therefore the angular components of divergence terms vanish upon averaging (see \S\ref{sec:periodic-bc}).

\par Finally, it is worth mentioning that while we are able to present  the current data in a 1D mean-field format, flows that involve rotation and/or large-scale magnetic field must be extended to at least a 2D mean-field description in order to retain self-consistency since these phenomena break the spherical symmetry of the system. This issue is discussed a bit further in \citet{VialletMeakin2013} (see their \S\S3 and 5).

\vspace{0.25cm}

\par {\bf A note on nomenclature:} Mean fields are often referred to as {\em moment equations} and the order of the equation is simply the number of items in the product which defines the quantity. For example, the evolution equation for the mean velocity is a first-order moment equation since the mean velocity, e.g.  $\fav{u_i}$, is not a product.  The entropy flux is a second-order moment since it is defined by the covariance of two quantities, $ f_s = \eht{\rho} \fht{s'' u''_r}$, that of the Favrian entropy and velocity fluctuation (the mean density which multiplies this quantity is not counted by convention). 

\par Another convention which we adopt in some places in this document is to refer to the evolution equations for  mean-field quantities as {\em budget equations}.

\subsection{Data Analysis and Presentation}

\par We have computed the various terms appearing in a large number of the mean-field evolution equations that we have derived  as well as the mean-fields themselves in several cases (e.g., mean turbulent kinetic energy and mean entropy profiles), all for a diverse set of stellar convection simulation data. In general, the mean fields are calculated from the simulation data according to the averaging operators defined in eqs.~\ref{eq:eht} and ~\ref{eq:fhb} above.  Two averaging methods have been employed, one more accurate than the other.  The most accurate method is to calculate averages during the course of the simulation so that every time-scale in the problem is sampled.  This turns out to be an important issue because of the presence of acoustic phenomena that is often characterized by timescales smaller than the increment in simulation time used to store output data when using a  compressible hydrodynamics simulation code.

\par For cases in which we do not have run-time averaged data stored, we find that we can filter troublesome acoustic phenomena (which is primarily radial in nature, including pressure waves trapped by the inner and outer boundaries) because it is generally of such a low amplitude that its impact on the other fields is negligible.  In particular, as described previously in the appendix of \citet{arnett2009}, we are able to successfully identify the instantaneous fluctuations in any given snapshot by using a horizontally averaged mean rather than the full time-and-horizontally averaged mean.  

\par A comparison with run-time averaged data shows that this procedure is very accurate for the mean-field data presented below.  However, it should be kept in mind that if one were interested in studying the interaction of high frequency waves with turbulence or other physics (e.g., with nuclear burning as in an epsilon mechanism) then a more accurate run-time averaged data set would be needed if a fully sampled set of flow snapshot data were not available.  In general, data sets which sample the acoustic time-scale are prohibitively large at present, even for modest resolution calculations and can easily exceed a petabyte.

\vspace{0.25cm}
\par {\bf A note on numerics:}  The data presented in this document was computed using three separate, but similar codes: MUSIC \citep{viallet2011,viallet2013}, PROMPI \citep{MeakinArnett2007}, and Herakles \citep{Mocak2009}, all built around conservative, finite-volume solvers.   PROMPI and Herakles both use piecewise parabolic method (PPM; see \citet{colella1984,colella1985}) solvers and have been used to simulate the oxygen-burning shell models (PROMPI; Tables~\ref{tab:ob-models}), and the helium burning models (Herakles; Tables~\ref{tab:chf-hif-models}). The MUSIC code, which uses a finite-volume scheme on a staggered grid was used to simulate red giant envelope convection (Table~\ref{tab:rg-models}).

\par  The extremely large Reynolds numbers expected for stellar turbulence (larger than $10^{10}$) makes it impossible to model all scales of the flow, from the stellar scale (or even the scale of a pressure or density scale height) down to the dissipation scale, on the current generation of computers.  Therefore, as discussed in \citet{VialletMeakin2013} and \citet{MeakinArnett2007}, we do not model viscosity explicitly and instead opt to solve the inviscid Euler equations and relegate the action of viscosity to the numerics, an approach referred to as implicit large eddy simulation (ILES;  \citet{grinstein2007}).  This approach is to be distinguished from both direct numerical simulation (DNS), wherein the viscosity (and other relevant molecular properties of the gas) are directly simulated down to the scales on which the flow is smooth within the continuum approximation; as well as large eddy simulation (LES), which employs a model to account for the impact that sub-grid scales (SGS) have on resolved scales. LES methods are strongly dependent on the choice of SGS model and, despite interesting developments, remains an active area of research with difficult outstanding issues regarding flows under circumstances expected in stellar interiors (e.g., \citet{sullivan2011}). In fact, one of the main conclusions of this last reference was that LES was found to converged with resolution only once an inertial range was captured by the simulation, a condition which is in fact suited for ILES and thus obviating the need for an SGS model. Another motivation for choosing an ILES approach over LES is that, in some sense, ILES allows one to achieve the highest effective Reynolds number (or equivalently, the smallest effective Kolmogorov length scale) for a given number of grid zones since LES models generally result in additional mixing at the grid scale.  This last point also applies to DNS.

\par While ILES provides arguably the highest effective Reynolds number for a given computational cost (see e.g. \citet{benzi2008}) the role played by dissipative processes (such as viscosity and molecular diffusion) must be quantified in a manner indirect compared to DNS. Furthermore, these processes will depend on the specific flow and computational grid choice in a way that is not readily predicted prior to simulation.  Therefore, ILES simulations are best suited to circumstances in which the flow is not dependent on the details at the dissipation scale, a circumstance typified by the Kolmogorov cascade found in high Reynolds number turbulent flow \citep{kolmogorov1941,aspden2008}.

\par Despite the indirect treatment of dissipative processes in ILES simulations one can use mean-field equations to quantify their effect to the extent that the hydrodynamics algorithm is formulated in a conservative manner.  As an example, consider the turbulent kinetic energy (TKE)  evolution equation  (eq.~\ref{eq:rans_tke}).  This budget equation will not balance for a dissipative momentum-conserving hydrodynamics scheme  but will instead result in a residual term which we denote by $\mathcal N_k$ that can be recovered by summing all of the other explicitly modeled terms.  In this example $\mathcal N_k$ quantifies the rate at which TKE is dissipated by the numerical scheme and provides a measure of the degree to which the solution deviates from that of truly inviscid hydrodynamics.  (As discussed in \citet{meakin2008,MeakinArnett2007,arnett2009,VialletMeakin2013}, the TKE dissipation found by this method is consistent with fully developed turbulence and the presence of an inertial range). In our analysis, we have defined an analogous residual term for each of the mean-field budget equations that can be directly calculated from the simulation data and which provides insight into and quantitative information about the dissipation and transport phenomena taking place near the grid scale in ILES calculations. This technique  provides a powerful tool for analyzing the turbulent phenomena being modeled (see super- and sub-scripted ${\mathcal N}$ terms in  \S\ref{sec:rans-summary} and Tables~\ref{tab:rans-def} to \ref{tab:rans-def-cont-3} where they are referred to as "numerical effects").

\vspace{0.25cm}
\par {\bf A note on units:} cgs units are used throughout. In addition, it should be noted that all calculated mean fields presented in this document are multiplied by the area factor $4 \pi r^2$ to reflect the volume increment in spherical geometry  which helps visualize the volume integral budgets of the individual mean fields which are then equivalent to the area below the corresponding mean field profiles.  This area factor is {\em not} reflected in the legends for each figure but it {\em is} reflected by the units in each y-axis label.


\subsection{Availability of Raw Data}

\par We will continue to make the data presented in this document available for download.  At present  the subset of data discussed in \citet{VialletMeakin2013} is available at \url{http://www.stellarmodels.org}.

\subsection{Structure of this Document}

\par While this document is primarily meant as a  reference for our research group, we are posting it publicly because we believe it has pedagogical significance for the field of stellar hydrodynamics and contains unique data and formulae that do not have a natural outlet for publishing and which can not be found elsewhere.  We suggest that an interested reader simply browse the table of contents for orientation and then scroll through the remainder of the document in order to familiarize themselves with its content which we feel is fairly self explanatory.  The authors are happy to address any questions that you might have or provide you with raw data if you are interested. Casey Meakin (casey.meakin@gmail.com) is a good first point of contact.

\subsection{Acknowledgments}

\par At Los Alamos we would like to thank Rob Gore, Daniel Livescu, Ray Ristorcelli, Fernando Grinstein, and Len Margolin for sharing some of their insights into the art and science of turbulence modeling. 
C.M. and W.D.A. acknowledge support from NSF grant 1107445 at the University of Arizona and through a subcontract to the New Mexico Consortium. This work used the Extreme Science and Engineering Discovery Environment (XSEDE), which is supported by National Science Foundation grant number OCI-1053575.

\newpage

\section{Summary of Reynolds-averaged Navier Stokes equations in spherical geometry}\label{sec:rans-summary}

\subsection{Various mean fields equations (first-order moments)}

\begin{table}[!h]
\label{tab:rans}
\begin{align}
\fav{D}_t \av{\rho} =& -\av{\rho} \fav{d} + {\mathcal N_\rho}  \label{eq:rans_density}\\
\av{\rho}\fav{D}_t\fav{u}_r = & -\nabla_r \fav{R}_{rr} -\av{G^{M}_r} - \partial_r \av{P} + \av{\rho}\fav{g_r} + {\mathcal N_{ur}} \\ 
\av{\rho}\fav{D}_t\fav{u}_\theta = & -\nabla_r \fav{R}_{\theta r} -\av{G^{M}_\theta} - (1/r)\av{\partial_\theta P} + {\mathcal N_{u \theta}}  \\
\av{\rho}\fav{D}_t\fav{u}_\phi = & -\nabla_r \fav{R}_{\phi r} -\av{G^{M}_\phi} + {\mathcal N_{u \phi}} \\
\av{\rho} \fav{D}_t \fav{\epsilon}_I = & -\nabla_r  ( f_I + f_T ) - \av{P} \ \av{d} - W_P  + {\mathcal S} + {\mathcal N_{\epsilon I}} \\
\av{\rho} \fav{D}_t \fav{\epsilon}_k = & -\nabla_r  ( f_k +  f_P ) - \fht{R}_{ir}\partial_r \fht{u}_i + W_b + W_P +\av{\rho}\fav{D}_t (\fav{u}_i \fav{u}_i / 2) + {\mathcal N_{\epsilon k}} \label{eq:rans_mke} \\
\av{\rho} \fav{D}_t \fav{\epsilon}_t = &  -\nabla_r ( f_I + f_T + f_k + f_P ) - \fht{R}_{ir}\partial_r \fht{u}_i - \av{P} \ \av{d} + W_b + {\mathcal S} + \av{\rho}\fav{D}_t (\fav{u}_i \fav{u}_i / 2) + {\mathcal N_{\epsilon t}}  \label{eq:rans_etot} \\
\erho\fav{D}_t \fav{h} = & -\nabla_r f_h - \Gamma_1\eht{P} \ \eht{d} - \Gamma_1 W_P + \Gamma_3 {\mathcal S} + \Gamma_3 \nabla_r f_T + {\mathcal N_h} \label{eq:rans_h} \\
\av{\rho} \fav{D}_t \fav{s} = &  -\nabla_r  f_s  - \av{(\nabla \cdot F_T)/T}+ \av{{\mathcal S}/T} + {\mathcal N_s}  \label{eq:rans_entropy} \\
\av{D}_t \av{P} = &  -\nabla_r f_P - \Gamma_1 \eht{P} \ \eht{d} + (1 -\Gamma_1) W_P + (\Gamma_3 -1){\mathcal S} + (\Gamma_3 - 1)\nabla_r f_T + {\mathcal N_P} \\
\av{D}_t \av{T} = & -\nabla_r f_T + (1-\Gamma_3)\eht{T}\ \eht{d} + (2-\Gamma_3)\eht{T'd'} + \eht{(\nabla \cdot F_T) / \rho c_v} + \eht{(\tau_{ij}\partial_i u_j)/\rho c_v} + \eht{\epsilon_{\rm nuc} / c_v} + {\mathcal N_T} \\
\erho\fav{D}_t \fav{X}_\alpha = & -\nabla_r f_\alpha + \av{\rho}\fav{\dot{X}}_\alpha^{\rm nuc} + {\mathcal N_\alpha} \label{eq:rans_comp} \\
\erho\fav{D}_t \fav{A} = & -\nabla_r f_A - \av{\rho A^2\Sigma_\alpha (\dot{X}_\alpha^{\rm nuc} / A_\alpha)} + {\mathcal N_A}  \label{eq:rans_abar}\\
\erho\fav{D}_t \fav{Z} = & -\nabla_r f_Z  - \av{\rho Z A \Sigma_\alpha (\dot{X}_\alpha^{\rm nuc} / A_\alpha)}  + \overline{\rho A \Sigma_\alpha (Z_\alpha \dot{X}_\alpha^{\rm nuc} / A_\alpha)} + {\mathcal N_Z} \label{eq:rans_zbar} \\
\erho\fav{D}_t \fav{j}_z = & -\nabla_r f_{jz}  + {\mathcal N_{jz}} \label{eq:rans_jz} 
\end{align}
\end{table}

\newpage

\subsection{Mean Reynolds stress equations (second-order moments)}

\begin{table}[!h]
\begin{align}
\eht{\rho}\fht{D}_t \left( \fht{R}_{rr} / \eht{\rho} \right) = & -\nabla_r ( 2 f_k^r + 2 f_P ) + 2 W_b - 2\fht{R}_{rr}\partial_r \fht{u}_r + 2\eht{P'\nabla_r u''_r} + 2 {\mathcal G_k^r} + {\mathcal N_{Rrr}} \\
\eht{\rho}\fht{D}_t \left( \fht{R}_{\theta \theta} / \eht{\rho} \right) = & -\nabla_r ( 2 f_k^\theta )- 2\fht{R}_{\theta r}\partial_r \fht{u}_\theta +2\eht{P' \nabla_\theta u''_\theta}  + 2 {\mathcal G_k^\theta} + {\mathcal N_{R \theta \theta}}  \\
\eht{\rho}\fht{D}_t \left( \fht{R}_{\phi \phi} / \eht{\rho} \right) = & - \nabla_r ( 2 f_k^\phi ) - 2\fht{R}_{\phi r}\partial_r \fht{u}_\phi + 2\eht{P' \nabla_\phi u''_\phi}   + 2 {\mathcal G_k^\phi} + {\mathcal N_{R \phi \phi}}
\end{align}
\end{table}

\subsection{Mean turbulent kinetic energy equations (second-order moments)}

\begin{table}[!h]
\label{tab:rans}
\begin{align}
\av{\rho} \fav{D}_t \fav{k}^{ } = &  -\nabla_r ( f_k +  f_P ) - \fht{R}_{ir}\partial_r \fht{u}_i + W_b + W_P + {\mathcal N_k}  \label{eq:rans_tke} \\
\av{\rho} \fav{D}_t \fav{k}^r =  & -\nabla_r  ( f_k^r + f_P )  - \fht{R}_{rr}\partial_r \fht{u}_r + W_b  + \eht{P'\nabla_r u''_r} + {\mathcal G_k^r} +  {\mathcal N_{kr}} \label{eq:rans_ekin_r} \\
\av{\rho} \fav{D}_t \fav{k}^h =  &  -\nabla_r f_k^h - (\fht{R}_{\theta r}\partial_r \fht{u}_\theta + \fht{R}_{\phi r}\partial_r \fht{u}_\phi) + (\eht{P' \nabla_\theta u''_\theta} + \eht{P' \nabla_\phi u''_\phi}) + {\mathcal G_k^h} + {\mathcal N_{kh}} \label{eq:rans_ekin_h} \\
\end{align}
\end{table}

\subsection{Mean turbulent mass flux and mean density-specific volume covariance equation (second-order moments)}

\begin{table}[!h]
\label{tab:rans}
\begin{align}
\eht{\rho}\fht{D}_t \eht{u''_r} =&  -(\eht{\rho'u'_ru'_r}/\eht{\rho})\partial_r\eht{\rho} + (\fht{R}_{rr}/\eht{\rho})/\partial_r \eht{\rho} - \eht{\rho} \nabla_r (\eht{u''_r} \ \eht{u''_r}) + \nabla_r \overline{\rho' u'_r u'_r} - \eht{\rho}\eht{u''_r} \nabla_r \eht{u}_r + \eht{\rho} \eht{u'_r d''} - b\partial_r \eht{P} + \eht{\rho' v \partial_r P'} +{\mathcal G_a} + {\mathcal N_a} \label{eq:rans_a}\\
\eht{D}_t b = &  +\eht{v} \nabla_r \eht{\rho} \eht{u''_r} -\eht{\rho}\nabla_r (\eht{u'_r v'}) + 2\eht{\rho}\eht{v'd'} +  {\mathcal N_b} \label{eq:rans_b}
\end{align}
\end{table}

\newpage

\subsection{Mean flux equations (second-order moments)}

\begin{table}[!h]
\begin{align}
\erho \fav{D}_t (f_I / \eht{\rho}) = &  -\nabla_r f_I^r  - f_I \partial_r \fht{u}_r  - \fht{R}_{rr} \partial_r \fht{\epsilon_I} - \eht{\epsilon''_I} \partial_r \eht{P} - \eht{\epsilon''_I \partial_r P'}  - \eht{u''_r \left( P d \right)}  + \overline{u''_r ({\mathcal S} + \nabla \cdot F_T)} + {\mathcal G_I} +  {\mathcal N_{fI}} \label{eq:rans_fi} \\
\erho \fav{D}_t (f_h / \eht{\rho}) = &  -\nabla_r f_h^r - f_h \partial_r \fht{u}_r - \fht{R}_{rr} \partial_r \fht{h} -\eht{h''}\partial_r \eht{P} - \eht{h''\partial_r P'} - \Gamma_1\eht{u''_r \left( P d \right) } + \Gamma_3 \overline{u''_r ({\mathcal S} + \nabla \cdot F_T)} + {\mathcal G_h} + {\mathcal N_{h \ }} \label{eq:rans_fh} \\
\erho \fav{D}_t (f_s / \eht{\rho}) = & -\nabla_r f_s^r - f_s \partial_r \fht{u}_r - \fht{R}_{rr} \partial_r \fht{s} -\eht{s''}\partial_r \eht{P} - \eht{s''\partial_r P'} + \eht{u''_r ( {\mathcal S} + \nabla \cdot F_T)  / T} + {\mathcal G_s} + {\mathcal N_{fs}}  \label{eq:rans_fs} \\
\erho \fav{D}_t (f_\alpha / \eht{\rho}) = &  -\nabla_r f_\alpha^r  - f_\alpha \partial_r \fht{u}_r - \fht{R}_{rr} \partial_r \fht{X}_\alpha -\eht{X''_\alpha} \partial_r \eht{P} - \eht{X''_\alpha \partial_r P'} + \overline{u''_r \rho \dot{X}_\alpha^{\rm nuc}} + {\mathcal G_\alpha} + {\mathcal N_{f\alpha}}  \label{eq:rans_falpha} \\
\erho \fav{D}_t (f_A / \eht{\rho}) = &  -\nabla_r f_A^r - f_A \partial_r \fht{u}_r - \fht{R}_{rr} \partial_r \fht{A} -\eht{A''} \partial_r \eht{P} - \eht{A'' \partial_r P'} - \overline{u''_r \rho A^2\Sigma_\alpha \dot{X}_\alpha^{\rm nuc} / A_\alpha} + {\mathcal G_A} + {\mathcal N_{fA}}                \label{eq:rans_fabar} \\
\erho \fav{D}_t (f_Z / \eht{\rho}) = &  -\nabla_r f_Z^r  - f_Z \partial_r \fht{u}_r - \fht{R}_{rr} \partial_r \fht{Z} -\eht{Z''} \partial_r \eht{P} - \eht{Z'' \partial_r P'} - \overline{u''_r \rho Z A \Sigma_\alpha (\dot{X}_\alpha^{\rm nuc}/ A_\alpha)} - \overline{u''_r \rho A \Sigma_\alpha (Z_\alpha \dot{X}_\alpha^{\rm nuc} / A_\alpha)}  + \nonumber \\
&  + {\mathcal G_Z} + {\mathcal N_{fZ}}   \label{eq:rans_fzbar}  \\
\erho \fav{D}_t (f_{jz} / \rho) = & -\nabla_r f_{jz}^r  - f_{jz} \partial_r \fht{u}_r - \fht{R}_{rr} \partial_r \fht{j_z} -\eht{j''_z} \partial_r \eht{P} - \eht{j''_z \partial_r P'} + {\mathcal G_{jz}} + {\mathcal N_{jz}} \label{eq:rans_fjz} \\
\fht{D}_t f_T = & -\nabla_r f_T^r - f_T \partial_r \eht{u}_r - \eht{u'_r u''_r} \partial_r \eht{T} - \eht{T'\partial_r P / \rho} - (\Gamma_3 -1)(\eht{T} \ \eht{u'_r d''}+\fht{d} \ \eht{u'_r T'} + \eht{u'_r T'd''}) +\eht{T'u'_rd''} + \eht{u'_r \epsilon_{\rm nuc}/c_v} + \nonumber \\
& +  \eht{u'_r \nabla \cdot F_T / \rho c_v} + {\mathcal G_{T}} +  {\mathcal N_{fT}} 
\end{align} 
\end{table}

\subsection{Mean Reynolds variance equations (second-order moments)}

\begin{table}[!h]
\label{tab:rans-variances}
\begin{align}
\fht{D}_t \sigma_\rho =  &  - \nabla_r \eht{(\rho' \rho ' u''_r)}  - 2\eht{\rho} \ \eht{\rho'd''} - 2 \eht{\rho'u''_r} \partial_r \eht{\rho} - 2 \fht{d} \ \sigma_\rho - \eht{\rho'\rho'd''} + {\mathcal N_{\sigma_\rho}} \\
\fht{D}_t \sigma_P = & - \nabla_r \eht{(P' P' u''_r)} - 2\Gamma_1 \eht{P} \ W_P - 2 f_P \partial_r \eht{P} - 2\Gamma_1 \widetilde{d} \ \sigma_P - (2 \Gamma_1 -1) \eht{P'P'd''} + 2(\Gamma_3 - 1)\eht{P' {\mathcal S}} + {\mathcal N_{\sigma_P}} \\
\fht{D}_t \sigma_T = & -\nabla_r \eht{(T' T' u''_r)} - 2(\Gamma_3 -1)\eht{T} \eht{T'd''} - 2\eht{T'u''_r} \partial_r \eht{T} - 2(\Gamma_3-1)\fht{d}\sigma_T + (3-2\Gamma_3)\eht{T'T'd''} + \eht{2 T' \nabla \cdot F_T/ \rho c_v} + \nonumber \\ 
& + \eht{2 T' \epsilon_{\rm nuc} / c_v} + {\mathcal N_{\sigma T}}  
\end{align}
\end{table}

\subsection{Mean Favrian variance equations (second-order moments)}

\begin{table}[!h]
\label{tab:rans-variances}
\begin{align}
\eht{\rho} \fht{D}_t \sigma_{ur} = & -\nabla_r ( \eht{\rho u''_r u''_r u''_r} ) + 2 \nabla_r f_P + 2 W_b - 2\fht{R}_{rr}\partial_r \fht{u}_r + 2 \overline{P'\nabla_r u''_r} + {\mathcal G_{\sigma_{ur}}} + {\mathcal N_{\sigma_{ur}}} \\
\eht{\rho} \fht{D}_t \sigma_{\epsilon I} = &  -\nabla_r (\eht{\rho \epsilon''_I \epsilon''_I u''_r} ) - 2 f_I \partial_r \fht{\epsilon_I} - 2\overline{\epsilon''_I}\ \eht{P} \ \fht{d} - 2\eht{P} \ \eht{\epsilon''_I d''} - 2\fht{d} \ \eht{\epsilon''_I P'} - 2\overline{\epsilon''_I P' d''} + 2\eht{\epsilon''_I {\mathcal S}} + {\mathcal N_{\sigma_{\epsilon I}}} \\
\eht{\rho} \fht{D}_t \sigma_s = & -\nabla_r ( \eht{\rho s'' s'' u''_r} ) - 2 f_s \partial_r \fht{s} - 2\eht{s'' \nabla \cdot F_T / T} + 2 \eht{s'' {\mathcal S} / T} + {\mathcal N_{\sigma_{s}}}\\
\eht{\rho} \fht{D}_t \sigma_\alpha = & -\nabla_r (\eht{\rho X''_\alpha X''_\alpha u''_r} ) - 2 f_\alpha \partial_r \fht{X}_\alpha + 2 \eht{X''_\alpha \rho \dot{X}_\alpha^{\rm nuc}} + {\mathcal N_{\sigma_\alpha}}
\end{align}
\end{table}

\newpage

\begin{table*}
\caption{Definitions\label{tab:rans-def}}
\begin{align}                                                      
& \rho \ \ \mbox{density}                                           & & g_r  \ \ \mbox{radial gravitational acceleration} \nonumber \\
& T \ \ \mbox{temperature}                                          & & {\mathcal S} = \rho \epsilon_\mathrm{nuc} (q) \ \ \mbox{nuclear energy production (cooling function)} \nonumber \\
& P \ \ \mbox{pressure}                                             & & \tau_{ij} = 2\mu S_{ij} \ \ \mbox{viscous stress tensor}  \ \ (\mu \ \ \mbox{kinematic viscosity}) \nonumber \\ 
& u_r, u_\theta, u_\phi \ \ \mbox{velocity components}                 & & S_{ij} = (1/2)(\partial_i u_j + \partial_j u_i) \ \ \mbox{strain rate} \nonumber \\
& {\bf u} = u (u_r, u_\theta, u_\phi) \ \ \mbox{velocity}               & & \fht{R}_{ij} = \eht{\rho}\fht{u''_i u''_j} \ \ \mbox{Reynolds stress tensor} \nonumber \\              
& j_z = r \sin{\theta} \ u_\phi \ \ \mbox{specific angular momentum} & & F_T = \chi \partial_r T \ \ \mbox{heat flux}   \nonumber \\
& d = \nabla \cdot {\bf u} \ \ \mbox{dilatation}                     & & \Gamma_1 = (d \ ln \ P/ d \ ln \ \rho)|_s   \nonumber \\ 
& \epsilon_I \ \ \mbox{specific internal energy}                     & & \Gamma_2 / (\Gamma_2 -1) =  (d \ ln \ P/ d \ ln \ T)|_s \nonumber \\
& h \ \ \mbox{specific enthalpy}                                    & &  \Gamma_3 -1 =  (d \ ln \ T/ d \ ln \ \rho)|_s \nonumber    \\  
& k = (1/2) \fht{u''_iu''_i} \ \ \mbox{turbulent kinetic energy}    & & \fht{k}^r = (1/2) \fht{u''_ru''_r} = (1/2) \fht{R}_{rr}/\eht{\rho} \ \ \mbox{radial turbulent kinetic energy}  \nonumber \\  
& \epsilon_k \ \ \mbox{specific kinetic energy}                      & & \fht{k}^\theta = (1/2)\fht{u''_\theta u''_\theta} = (1/2)\fht{R}_{\theta \theta}/\eht{\rho} \ \ \mbox{angular turbulent kinetic energy}  \nonumber \\
& \epsilon_t \ \ \mbox{specific total energy}                        & & \fht{k}^\phi = (1/2)\fht{u''_\phi u''_\phi} = (1/2) \fht{R}_{\phi \phi}/\eht{\rho} \ \ \mbox{angular turbulent kinetic energy} \nonumber \\ 
& s \ \ \mbox{specific entropy}                                      & & \fht{k}^h = \fht{k}^\theta + \fht{k}^\phi \ \ \mbox{horizontal turbulent kinetic energy}                                    \nonumber \\
& v = 1/\rho \ \ \mbox{specific volume}                               & & f_k = (1/2)\eht{\rho} \fht{u''_i u''_i u''_r} \ \ \mbox{turbulent kinetic energy flux}                                    \nonumber \\    
& X_\alpha \ \ \mbox{mass fraction of isotope $\alpha$}               & & f_k^r = (1/2)\eht{\rho} \fht{u''_r u''_r u''_r} \ \ \mbox{radial turbulent kinetic energy flux}                          \nonumber \\
& \dot{X}_\alpha^{\mathrm nuc} \ \ \mbox{rate of change of $X_\alpha$}     & & f_k^\theta = (1/2)\eht{\rho} \fht{u''_\theta u''_\theta u''_r} \ \ \mbox{angular turbulent kinetic energy flux}              \nonumber \\    
& A_\alpha \ \ \mbox{number of nucleons in isotope $\alpha$}           & & f_k^\phi = (1/2)\eht{\rho} \fht{u''_\phi u''_\phi u''_r} \ \ \mbox{angular turbulent kinetic energy flux}                   \nonumber \\ 
& Z_\alpha \ \ \mbox{charge of isotope $\alpha$}                     & & f_k^h = f_k^\theta + f_k^\phi \ \ \mbox{horizontal turbulent kinetic energy flux}                                           \nonumber \\   
& A \ \ \mbox{mean number of nucleons per isotope}          & & W_p = \eht{P'd''} \ \ \mbox{turbulent pressure dilatation}      \nonumber \\                                            
& Z \ \ \mbox{mean charge per isotope}                     & &   W_b = \eht{\rho} \eht{u''_r} \fht{g}_r \ \ \mbox{buoyancy}          \nonumber  \\
& f_P = \eht{P' u'_r} \ \ \mbox{acoustic flux}                                    & &  f_T = -\eht{\chi \partial_r T} \ \ \mbox{heat flux ($\chi$ thermal conductivity})             \nonumber 
\end{align} 
\end{table*}

\newpage

\begin{table*}
\caption{Definitions (continued)\label{tab:rans-def-cont-1}}
\begin{align}
& f_I = \eht{\rho} \fht{\epsilon''_I u''_r} \ \ \mbox{internal energy flux}        & & f_\alpha = \eht{\rho} \fht{X''_\alpha u''_r} \ \ \mbox{$X_\alpha$ flux}                   \nonumber \\        
& f_s = \eht{\rho} \fht{s'' u''_r} \ \ \mbox{entropy flux}                & & f_{jz} = \eht{\rho}\fht{j''_z u''_r} \ \ \mbox{angular momentum flux}    \nonumber \\ 
& f_T = \eht{u'_r T'} \ \ \mbox{turbulent heat flux}                      & & f_A = \eht{\rho}\fht{A''u''_r} \ \ \mbox{A (mean number of nucleons per isotope) flux}  \nonumber \\
& f_h = \eht{\rho}\fht{h''u''_r} \ \ \mbox{enthalpy flux}        & &  f_Z = \eht{\rho}\fht{Z''u''_r} \ \ \mbox{Z (mean charge per isotope) flux}  \nonumber \\         
& b = \overline{v'\rho'} \ \ \mbox{density-specific volume covariance}                & & \mathcal N_\rho, \mathcal N_{ur}, \mathcal N_{u\theta}, \mathcal N_{u\phi}, \mathcal N_{jz}, \mathcal N_{\alpha}, \mathcal N_{A}, \mathcal N_{Z} \ \ \mbox{numerical effect} \nonumber \\              
& f_\tau = f_\tau^r + f_\tau^\theta + f_\tau^\phi \ \ \mbox{viscous flux}                    & &  \mathcal N_{\epsilon I} = -\nabla_r f_\tau +\varepsilon_k \ \ \mbox{numerical effect} \nonumber \\ 
& f_\tau^r = -\eht{\tau'_{rr}u'_r}  \ \ \mbox{viscous flux}                               & &  \mathcal N_{\epsilon k} = -\varepsilon_k \ \ \mbox{numerical effect} \nonumber \\              
& f_\tau^\theta = -\eht{\tau'_{\theta r} u'_\theta }  \ \ \mbox{viscous flux}                 & &  \mathcal N_{\epsilon t} = -\nabla_r f_\tau \ \ \mbox{numerical effect} \nonumber \\  
& f_\tau^\phi = -\eht{\tau'_{\phi r} u'_\phi}  \ \ \mbox{viscous flux}                       & &   \mathcal N_{s} = \eht{-\varepsilon_k/T} \ \ \mbox{numerical effect} \nonumber \\          
& f_\tau^h = f_\tau^\theta + f_\tau^\phi   \ \ \mbox{viscous flux}                           & & \mathcal N_{h} = -\nabla_r f_\tau + (\Gamma_3 -1)\varepsilon_k \ \ \mbox{numerical effect} \nonumber \\              
& f_I^r = \eht{\rho}\fht{\epsilon''_I u''_r u''_r} \ \ \mbox{radial flux of $f_I$}      & &  \mathcal N_{P} = +(\Gamma_3 -1)\varepsilon_k \ \ \mbox{numerical effect} \nonumber \\
& f_s^r = \eht{\rho}\fht{s'' u''_r u''_r} \ \ \mbox{radial flux of $f_s$}               & &  \mathcal N_{T} = +\eht{(\tau_{ij} \partial_j u_i)/(c_v \rho)}  \ \ \mbox{numerical effect} \nonumber \\ 
& f_h^r = \eht{\rho}\fht{h'' u''_r u''_r} \ \ \mbox{radial flux of $f_h$}               & & \mathcal N_{Rrr} = -2\nabla_r f_\tau^r - 2\varepsilon_k^r \ \ \mbox{numerical effect} \nonumber \\
& f_T^r = \eht{T' u'_r u'_r} \ \ \mbox{radial flux of $f_T$}                            & &  \mathcal N_{R\theta \theta} = -2\nabla_r f_\tau^\theta - 2\varepsilon_k^\theta  \ \ \mbox{numerical effect} \nonumber \\
& f_{jz}^r = \eht{\rho}\fht{j''_z u''_r u''_r} \ \ \mbox{radial flux of $f_{jz}$}                           & &  \mathcal N_{R\phi \phi} = -2\nabla_r f_\tau^\phi - 2\varepsilon_k^\phi \ \ \mbox{numerical effect} \nonumber \\
& f_\alpha^r =  \eht{\rho}\fht{X''_\alpha u''_r u''_r} \ \ \mbox{radial flux of $f_\alpha$} & &  \mathcal N_{k \ } = -\nabla_r f_\tau - \varepsilon_k  \ \ \mbox{numerical effect} \nonumber \\                          
& f_A^r = \eht{\rho}\fht{A'' u''_r u''_r} \ \ \mbox{radial flux of $f_A$}               & &  \mathcal N_{kr} = -\nabla_r f_\tau^r - \varepsilon_k^r \ \ \mbox{numerical effect} \nonumber \\                                 
& f_Z^r = \eht{\rho}\fht{Z'' u''_r u''_r} \ \ \mbox{radial flux of $f_Z$}               & &  \mathcal N_{kh} = -\nabla_r f_\tau^h - \varepsilon_k^h \ \ \mbox{numerical effect} \nonumber \\                                        
& \mathcal G_k^r = -(1/2)\eht{G_{rr}^R} - \eht{u''_rG_r^M}                 & &  \mathcal N_a = -\varepsilon_a \ \ \mbox{numerical effect} \nonumber                          
\end{align}
\end{table*}

\newpage

\begin{table*}
\caption{Definitions (continued)\label{tab:rans-def-cont-2}}
\begin{align} 
& \mathcal G_k^\theta = -(1/2)\eht{G_{\theta \theta}^R} - \eht{u''_\theta G_\theta^M}  & &  \mathcal N_b \ \ \mbox{numerical effect} \nonumber \\               
& \mathcal G_k^\phi  = -(1/2)\eht{G_{\phi \phi}^R} - \eht{u''_\phi G_\phi^M}   & &  \mathcal N_{fI} = -\nabla_r (\eht{\epsilon''_I \tau'_{rr}}) + \eht{u''_r \tau_{ij} \partial_i u_j} -\varepsilon_I  \ \ \mbox{numerical effect} \nonumber \\               
& \mathcal G_k^h = +\mathcal G_k^\theta + \mathcal G_k^\phi                  & &  \mathcal N_{fh} = -\nabla_r (\eht{h'' \tau'_{rr}}) + \eht{u''_r (\Gamma_3 - 1) \tau_{ij} \partial_i u_j} - \eht{ u''_r \nabla_i u_i \tau_{ji}} - \varepsilon_h   \ \ \mbox{numerical effect} \nonumber \\                                   
& \mathcal G_a = +\eht{\rho' v G_r^M}                                     & &  \mathcal N_{fs} = -\nabla_r (\eht{s''\tau'_{rr}}) + \eht{u''_r \tau_{ij} \partial_i u_j/T} -\varepsilon_s  \ \ \mbox{numerical effect} \nonumber \\     
& \mathcal G_I = -\eht{G_{r}^I} - \eht{\epsilon''_I G_r^M}  & &  \mathcal N_{fA} = -\nabla_r (\eht{A''\tau'_{rr}}) - \varepsilon_A \ \ \mbox{numerical effect} \nonumber \\                                       
& \mathcal G_\alpha =  -\eht{G_{r}^\alpha} - \eht{X''_\alpha G_r^M} & &  \mathcal N_{fZ} = -\nabla_r (\eht{Z''\tau'_{rr}}) - \varepsilon_Z \ \ \mbox{numerical effect} \nonumber \\                             
& \mathcal G_A =  -\eht{G_{r}^A} - \eht{A'' G_r^M}  & & \mathcal N_{f\alpha} = -\nabla_r (\eht{\alpha''\tau'_{rr}}) - \varepsilon_\alpha  \ \ \mbox{numerical effect} \nonumber \\
& \mathcal G_Z =  -\eht{G_{r}^Z} - \eht{Z'' G_r^M}  & & \mathcal N_{fjz} =  -\nabla_r (\eht{j''_z \tau'_{rr}}) - \varepsilon_{jz}  \ \ \mbox{numerical effect} \nonumber \\ 
& \mathcal G_h =  -\eht{G_{r}^h} - \eht{h'' G_r^M}  & & \mathcal N_{fT} =  + \eht{T'\partial_i \tau_{ri} / \rho} + \eht{u'_r \tau_{ij} \partial_i u_j / \rho c_v} \ \ \mbox{numerical effect} \nonumber \\ 
& \mathcal G_T =  -\eht{G_{r}^T} - \eht{T' G_r^M}  & & \mathcal N_\alpha \ \ \mbox{numerical effect}  \nonumber  \\
& \mathcal G_{s} = -\eht{G_{r}^s} - \eht{s'' G_r^M} & & \mathcal N_A \ \ \mbox{numerical effect}  \nonumber \\
& \mathcal G_{jz} = -\eht{G_{r}^{jz}} - \eht{j''_z G_r^M} &  & \mathcal N_Z \ \ \mbox{numerical effect}   \nonumber \\
& \sigma_\rho = \eht{\rho'\rho'} & & \mathcal N_{\sigma_\rho} \ \ \mbox{numerical effect} \nonumber \\
& \sigma_P = \eht{P'P'} & & \mathcal N_{\sigma_P} = +2 (\Gamma_3 -1)\eht{P'\tau_{ij}\partial_i u_j} \ \ \mbox{numerical effect} \nonumber \\
& \sigma_T = \eht{T'T'} & & \mathcal N_{\sigma_T} = +\eht{2 T' \tau_{ij} \partial_i u_j / \rho c_v} \ \ \mbox{numerical effect} \nonumber \\
& \sigma_{ur} = \fht{u''_r u''_r} & & \mathcal N_{\sigma_{ur}} = +2\nabla_r f_{\tau}^r - 2\varepsilon_{k}^r \ \ \mbox{numerical effect} \nonumber \\
& \sigma_{s} =  \fht{s''s''} & & \mathcal N_{\sigma_s} = +  2 \eht{s'' \tau_{ij} \partial_j u_i / T} \ \ \mbox{numerical effect} \nonumber \\
& \sigma_\alpha = \fht{X''_\alpha X''_\alpha} & & \mathcal N_{\sigma_\alpha} \ \ \mbox{numerical effect}  \nonumber \\
& \sigma_{\epsilon I} = \fht{\epsilon''_I \epsilon''_I} & &  N_{\sigma \epsilon_I} =  + 2\eht{\epsilon''_I \tau_{ij} \partial_j u_i} \ \ \mbox{numerical effect} \nonumber
\end{align}
\end{table*}

\newpage

\begin{table*}
\caption{Definitions (continued)\label{tab:rans-def-cont-3}}
\begin{align} 
& \varepsilon_k^r = \eht{\tau'_{rr}\partial_r u''_r} + \eht{\tau'_{r\theta}(1/r)\partial_\theta u''_r} + \eht{\tau'_{r\phi}(1/r\sin{\theta})\partial_\phi u''_r} & & \eht{G^{M}_r}     = -\eht{\rho u_\theta u_\theta/r} - \eht{\rho u_\phi u_\phi/r} \nonumber \\
& \varepsilon_k^\theta = \eht{\tau'_{\theta r}\partial_r u''_\theta} + \eht{\tau'_{\theta \theta}(1/r)\partial_\theta u''_\theta} + \eht{\tau'_{\theta \phi}(1/r\sin{\theta})\partial_\phi u''_\theta} & & \eht{G_\theta^M} = +\eht{\rho u_\theta u_r/r} - \eht{\rho u_\phi u_\phi/(r\tan{\theta})} \nonumber \\                                       
&  \varepsilon_k^\phi = \eht{\tau'_{\phi r}\partial_r u''_\phi} + \eht{\tau'_{\phi \theta}(1/r)\partial_\theta u''_\phi} + \eht{\tau'_{\phi \phi}(1/r\sin{\theta})\partial_\phi u''_\phi} & & \eht{G_\phi^M} = +\eht{\rho u_\phi u_r/r} + \eht{\rho u_{\phi} u_{\theta}/(r \tan{\theta})}  \nonumber \\
& \varepsilon_k = \varepsilon_k^r + \varepsilon_k^\theta + \varepsilon_k^\phi  & & \eht{G^{R}_{rr}}  = -\eht{\rho u''_\theta u''_\theta u''_r/r} - \eht{\rho u''_\theta u''_r u''_\theta/r} - \eht{\rho u''_\phi u''_\phi u''_r/r} - \eht{\rho u''_\phi u''_r u''_\phi /r}  \nonumber \\
& \varepsilon_k^h = \varepsilon_k^\theta + \varepsilon_k^\phi & & \eht{G^{R}_{\theta \theta}} = +\eht{\rho u''_\theta u''_r u''_\theta/r} + \eht{\rho u''_\theta u''_\theta u''_r/r} - \eht{\rho u''_\phi u''_\phi u''_\theta/(r\tan{\theta})} - \eht{u''_\phi u''_\theta u''_\phi/(r\tan{\theta})}  \nonumber \\
& \varepsilon_a = \eht{\rho' v \nabla_r  \tau'_{rr}} & & \eht{G^{R}_{\phi \phi}} = +\eht{\rho u''_\phi u''_r r_\phi /r} + \eht{\rho u''_\phi u''_\theta u''_\phi /(r\tan{\theta})} + \eht{\rho u''_\phi u''_\phi u''_r/r} + \eht{\rho u''_\phi u''_\phi u''_\theta / (r\tan{\theta})} \nonumber \\
& \varepsilon_I = \eht{\tau'_{rr}\partial_r \epsilon''_I} + \eht{\tau'_{r\theta}(1/r)\partial_\theta \epsilon''_I} + \eht{\tau'_{r\phi}(1/r\sin{\theta})\partial_\phi \epsilon''_I} & & \eht{G^{I}_r} = -\eht{\rho \epsilon''_I u''_\theta u''_\theta/r} - \eht{\rho \epsilon''_I u''_\phi u''_\phi/r} \nonumber \\
& \varepsilon_s = \eht{\tau'_{rr}\partial_r s''} + \eht{\tau'_{r\theta}(1/r)\partial_\theta s''} + \eht{\tau'_{r\phi}(1/r\sin{\theta})\partial_\phi s''} & & \eht{G^{s}_r} = -\eht{\rho s'' u''_\theta u''_\theta/r} - \eht{\rho s'' u''_\phi u''_\phi/r} \nonumber \\
& \varepsilon_\alpha = \eht{\tau'_{rr}\partial_r X''_\alpha} + \eht{\tau'_{r\theta}(1/r)\partial_\theta X''_\alpha} + \eht{\tau'_{r\phi}(1/r\sin{\theta})\partial_\phi X''_\alpha} & & \eht{G^{\alpha}_r} =  -\eht{\rho X''_\alpha u''_\theta u''_\theta/r} - \eht{\rho X''_\alpha u''_\phi u''_\phi/r} \nonumber \\  
& \varepsilon_A = \eht{\tau'_{rr}\partial_r A''} + \eht{\tau'_{r\theta}(1/r)\partial_\theta A''} + \eht{\tau'_{r\phi}(1/r\sin{\theta})\partial_\phi A''} & & \eht{G^A_r} =  -\eht{\rho A'' u''_\theta u''_\theta/r} - \eht{\rho A'' u''_\phi u''_\phi/r} \nonumber \\
& \varepsilon_Z = \eht{\tau'_{rr}\partial_r Z''} + \eht{\tau'_{r\theta}(1/r)\partial_\theta Z''} + \eht{\tau'_{r\phi}(1/r\sin{\theta})\partial_\phi Z''}& & \eht{G^Z_r} =  -\eht{\rho Z'' u''_\theta u''_\theta/r} - \eht{\rho Z'' u''_\phi u''_\phi/r} \nonumber \\
& \varepsilon_{h} = \eht{\tau'_{rr}\partial_r h''} + \eht{\tau'_{r\theta}(1/r)\partial_\theta h''} + \eht{\tau'_{r\phi}(1/r\sin{\theta})\partial_\phi h''} & & \eht{G^h_r} = -\eht{\rho h'' u''_\theta u''_\theta/r} - \eht{\rho h'' u''_\phi u''_\phi/r} \nonumber \\
& \varepsilon_{jz} = \eht{\tau'_{rr}\partial_r j''_z} + \eht{\tau'_{r\theta}(1/r)\partial_\theta j''_z} + \eht{\tau'_{r\phi}(1/r\sin{\theta})\partial_\phi j''_z} & & \eht{G^T_r} = -\eht{\rho T' u'_\theta u'_\theta/r} - \eht{\rho T' u'_\phi u'_\phi/r} \nonumber \\
& & & \eht{G^{jz}_r} = -\eht{\rho j''_z u''_\theta u''_\theta/r} - \eht{\rho j''_z u''_\phi u''_\phi/r} \nonumber 
\end{align}
\begin{align}
\nabla (.) = \nabla_r (.) + \nabla_\theta (.) + \nabla_\phi (.) = \frac{1}{r^2} \partial_r (r^2 . ) + \frac{1}{r\sin{\theta}} \partial_\theta (\sin{\theta} . ) + \frac{1}{r\sin{\theta}} \partial_\phi (.) \nonumber
\end{align}
\end{table*}

\clearpage

\section{Properties of oxygen shell burning and red giant simulation data}

\subsection{Summary of the Oxygen Burning Simulations}


\begin{table}[!h]
\centerline{
\begin{tabular}{|l c c c c c c c c|}
\hline
{\bf Parameter} & {\sf ob.3D.lr} & {\sf ob.3D.mr} & {\sf ob.3D.hr} & {\sf ob.3D.1hp} & {\sf ob.3D.2hp} & {\sf ob.3D.4hp} & {\sf ob.3d.1hp.vc} & {\sf ob.3d.1hp.vh} \\
& & & & & & & & \\
Grid zoning &   192$\times 128^2$ & 384$\times 256^2$ & 786$\times 512^2$ & 200$\times 50^2$ & 400$\times 100^2$ & 320$\times 50^2$ & 200$\times 50^2$ & 400$\times 100^2$ \\
$r_\mathrm{in}$, $r_\mathrm{out}$ ($10^{9}$ cm)  & 0.3, 1.0 & 0.3, 1.0 & 0.3, 1.0 & 0.3, 0.9 & 0.3, 1.0 & 0.3, 1.6 & 0.3, 0.9 & 0.3, 0.9 \\
$r_\mathrm{b}^c$, $r_\mathrm{t}^c$ ($10^{9}$ cm)  &  0.43, 0.85 &  0.43, 0.85 &  0.43, 0.84 & 0.43, 0.68 & 0.43, 0.84 & 0.42, 1.4 & 0.43, 0.65 & 0.43, 0.68 \\
$\Delta \theta$, $\Delta \phi$  &  45$^\circ$ & 45$^\circ$ &45$^\circ$ & 27.5$^\circ$ & 27.5$^\circ$ & 27.5$^\circ$ & 27.5$^\circ$ & 27.5$^\circ$ \\
CZ stratification ($H_P$) & 1.9 & 1.9 & 1.9 & 1.2 & 1.9 & 4.1 & 1.1 & 1.2  \\
$\Delta t_\mathrm{av}$ (s) & $230$ & $230$ & $165$ & $900$ & $230$ & $500$ & 300 & 300    \\
$v_\mathrm{rms}$  ($10^6$ cm/s)       & 10.7 & 10.9 & 10.9 & 5.28  & 9.15 & 4.88 & 4.66 & 4.97 \\
$\tau_\mathrm{conv}$  (s)             & 78.2 & 77.1 & 75.6 & 94.7  & 89.2 & 403. & 94.6 & 95.6  \\
$P_{turb}/P_{gas} (10^{-4})$            & 3.88 &  4.05 &  4.03 &  0.96  &  3.01 &  1.73 & 0.79 & 0.96   \\
$L$ ($10^{46}$ erg/s)                & 2.74 & 2.63 & 2.58 & 0.44 & 2.86 & 0.26 & 0.40 & -0.42  \\
$L_\mathrm{d}$ ($10^{46}$  erg/s)      & 0.29 & 0.28 & 0.26 & 0.04 & 0.31  & 0.08 & 0.03 & 0.03 \\
$l_\mathrm{d}$  ($10^{8}$ cm)          & 7.39 & 7.92 & 8.73 & 3.87 & 4.15 & 5.1 & 2.85 & 4.1  \\
$\tau_\mathrm{d}$  (s)                & 34.48 & 36.47 & 39.98 & 36.72 & 22.64 & 52.04 & 30.56 & 41.1 \\
$\tau_\mathrm{dr}$ (s)                & 38.06 & 39.27 & - & 75.65 & 46.59 & 130.73 & 90.20 & 90.8  \\
$\tau_\mathrm{dh}$ (s)                & 30.77 & 32.14 & - & 22.91 & 14.21 & 28.42 & 18.4 & 25.24  \\
\hline
\end{tabular}}
\caption{boundaries of computational domain $r_\mathrm{in}$, $r_\mathrm{out}$; boundaries of convection zone at bottom and top $r_\mathrm{b}^c$, $r_\mathrm{t}^c$; angular size of computational domain $\Delta \theta$, $\Delta \phi$ ; depth of convection zone ``CZ stratification'' in pressure scale height $H_P$; averaging timescale of mean fields analysis $\Delta t_\mathrm{av}$; global rms velocity $v_\mathrm{rms}$; convective turnover timescale $\tau_\mathrm{conv}$; average ratio of turbulent ram pressure and gas pressure $p_{turb}/p_{gas}$; total luminosity of the hydrodynamic model $L$; total rate of kinetic energy dissipation $L_d$; dissipation length-scale  $l_d$; turbulent kinetic energy dissipation time-scale $\tau_d$; radial turbulent kinetic energy dissipation time-scale $\tau_{dr}$; horizontal turbulent kinetic energy dissipation time-scale $\tau_{dh}$. The numerical values may vary in time up to 20$\%$ due to limited amount of data for averaging out the time dependence.\label{tab:ob-models} }
\end{table}

\newpage

\subsection{Summary of Red Giant Simulations}

\vspace{1.cm}

\begin{table}[!h]
\centerline{
\begin{tabular}{|l c c c|}
\hline
{\bf Parameter}  & {\sf rg.3D.lr} & {\sf rg.3D.mr} & {\sf rg.3D.4hp} \\
& & & \\
Grid zoning & 216$\times 128^2$ & 432$\times 256^2$ & 176$\times 128^2$ \\
$r_\mathrm{in}$, $r_\mathrm{out}$ ($10^{12}$ cm) & 0.82, 4.09 & 0.82, 4.09 & 0.82, 0.34\\
$r^c_\mathrm{in}$, $r^c_\mathrm{out}$ ($10^{12}$ cm) & 2.05, 3.86 & 2.07, 3.88 & 2.16, 3.33\\
$\Delta \theta$, $\Delta \phi$ & 45$^\circ$ & 45$^\circ$ & 45$^\circ$\\
CZ stratification ($H_p$)         & 7.0 & 7.2 & 3.5\\
$\Delta t_\mathrm{av}$ (days)       & $800$ & $800$ & $800$ \\
$v_\mathrm{rms}$ ($10^5$ cm/s)      & 2.59 & 2.66  & 2.01 \\
$\tau_\mathrm{conv}$ (days)         & 161. & 158.  & 134. \\
$P_{turb}/P_{gas} (10^{-3})$          & 4.68 &  4.98   & 0.98   \\ 
$L_\mathrm{cool}$ ($10^{36}$ erg/s)  & -8.57 & -7.13  & -9.2\\
$L_\mathrm{d}$ ($10^{36}$ erg/s)     & 7.26 & 7.24  & 2.27 \\
$l_\mathrm{d}$ ($10^{11}$ cm)        & 9.95 & 10.4  & 11.6\\
$\tau_\mathrm{d}$ (days)            & 22.2 & 22.7  & 33.3\\
$\tau_\mathrm{dr}$ (days)           & 36.7 & 44.7  & 53.0 \\
$\tau_\mathrm{dh}$ (days)           & 18.3 & 17.9  & 28.2 \\
\hline
\end{tabular}}
\caption{boundaries of computational domain $r_\mathrm{in}$, $r_\mathrm{out}$; boundaries of convection zone at bottom and top $r_\mathrm{b}^c$, $r_\mathrm{t}^c$; angular size of computational domain $\Delta \theta$, $\Delta \phi$ ; depth of convection zone ``CZ stratification'' in pressure scale height $H_P$; averaging timescale of mean fields analysis $\Delta t_\mathrm{av}$; global rms velocity $v_\mathrm{rms}$; convective turnover timescale $\tau_\mathrm{conv}$; average ratio of turbulent ram pressure and gas pressure $p_{turb}/p_{gas}$; total luminosity of the hydrodynamic model $L$; total rate of kinetic energy dissipation $L_d$; dissipation length-scale  $l_d$; turbulent kinetic energy dissipation time-scale $\tau_d$; radial turbulent kinetic energy dissipation time-scale $\tau_{dr}$; horizontal turbulent kinetic energy dissipation time-scale $\tau_{dh}$. The numerical values may vary in time up to 20$\%$ due to limited amount of data for averaging out the time dependence.\label{tab:rg-models} }
\end{table}

\newpage

\subsection{Snapshots of TKE in a meridional plane of the oxygen burning and red giant models}

\vspace{1.cm}

\begin{figure}[!h]
\centerline{
\includegraphics[width=11.8cm]{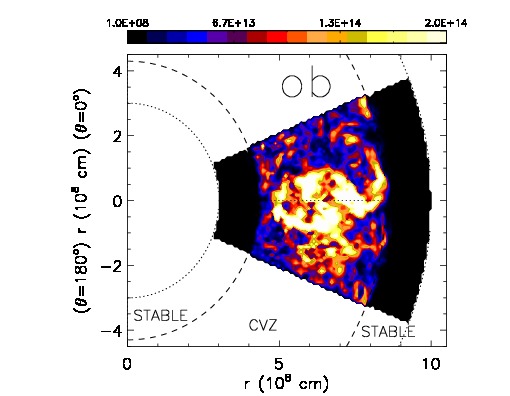}
\includegraphics[width=11.8cm]{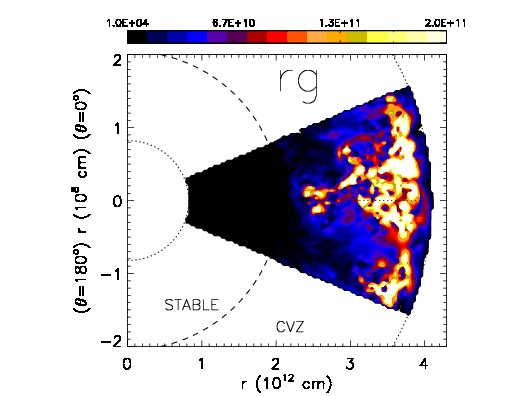}}
\caption{Snapshots of turbulent kinetic energy (in erg g$^{-1}$) in a meridional plane of 3D oxygen burning shell model ob.3D.mr (left) and red giant envelope convection model rg.3D.mr (right). Convectively unstable (CVZ) and stable layers (STABLE) are separated by dashed lines.}
\label{fig:ob-rg-tke-cuts}
\end{figure}

\newpage

\subsection{Background structure of oxygen burning and red giant models}

\begin{figure}[!h]
\centerline{
\includegraphics[width=7.cm]{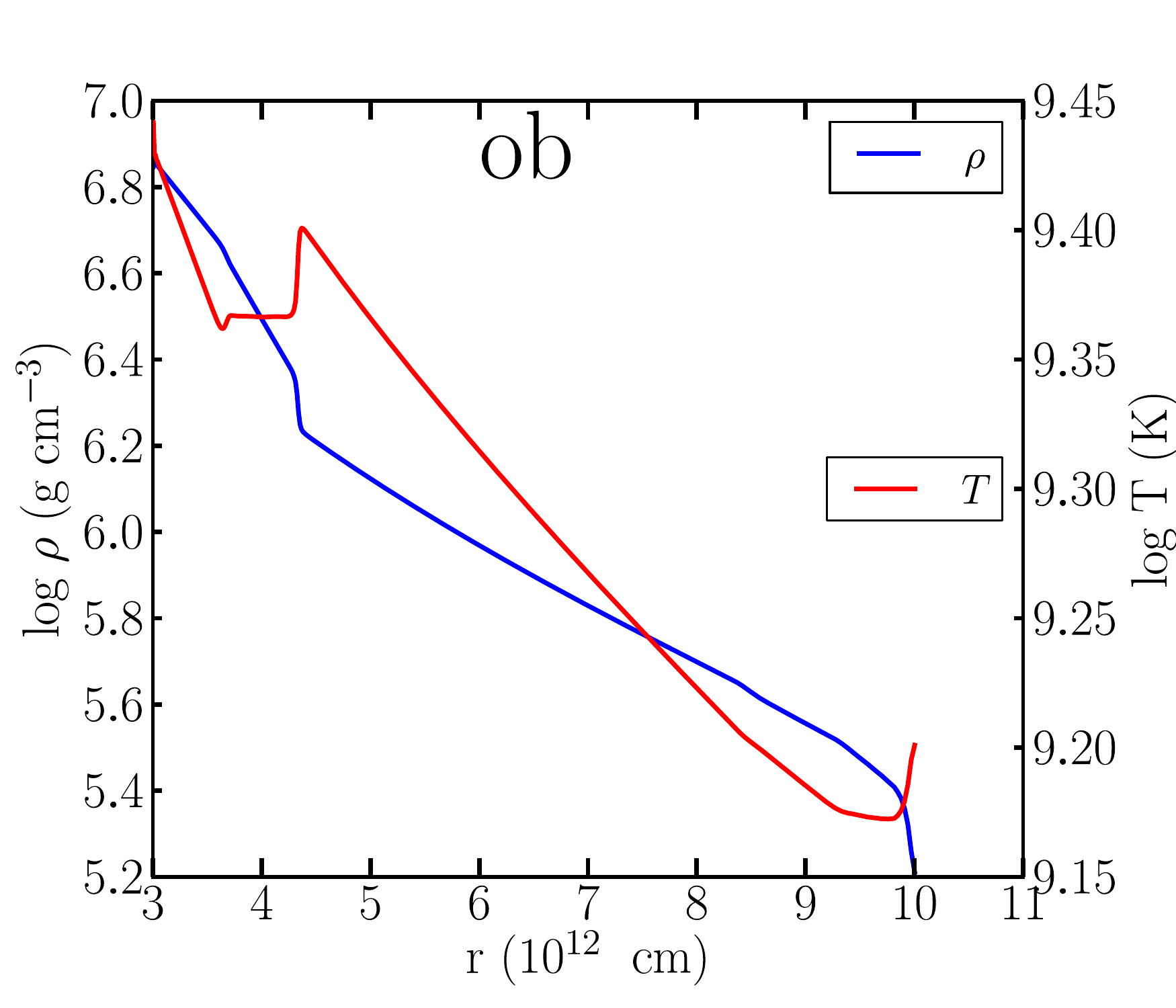}
\includegraphics[width=7.cm]{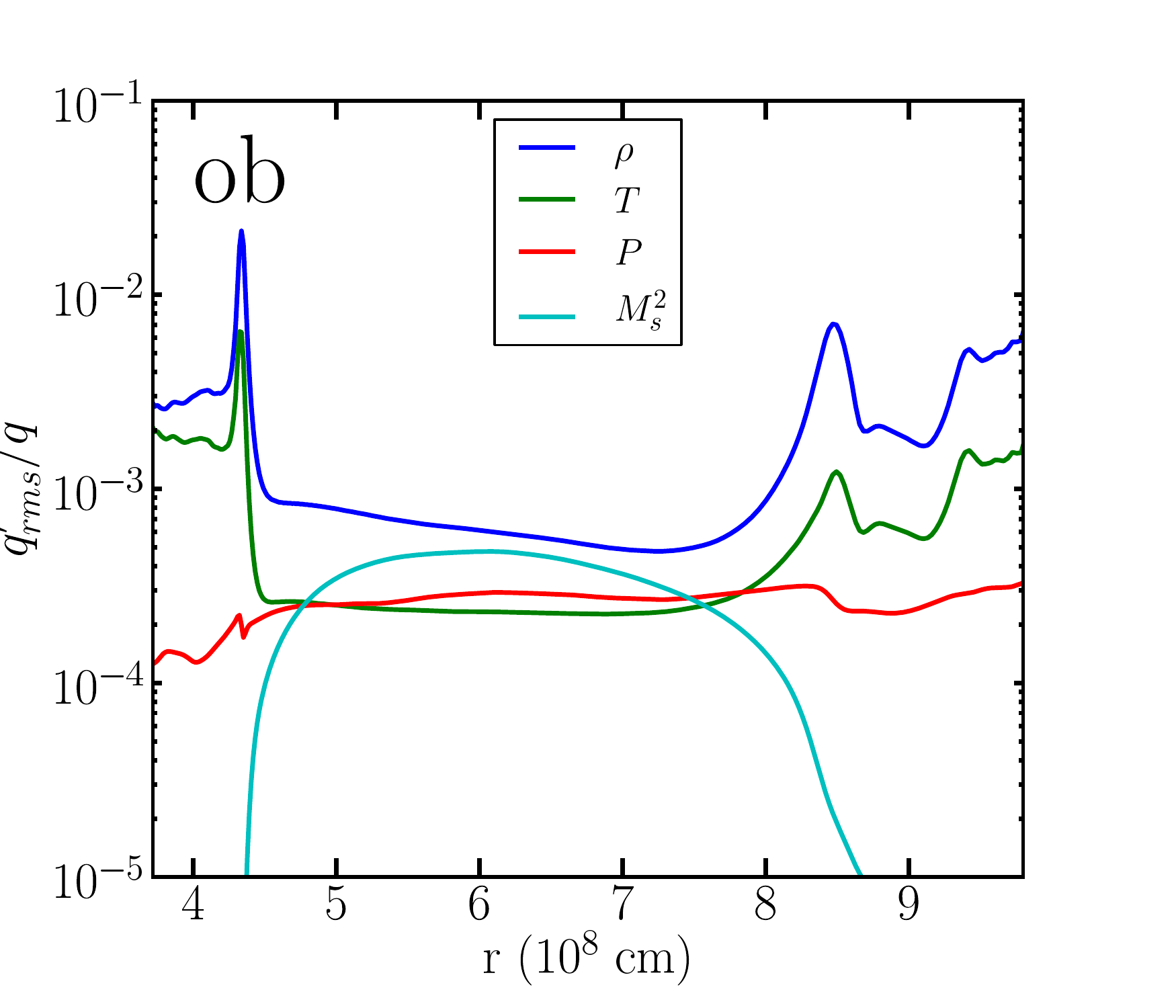}
\includegraphics[width=7.cm]{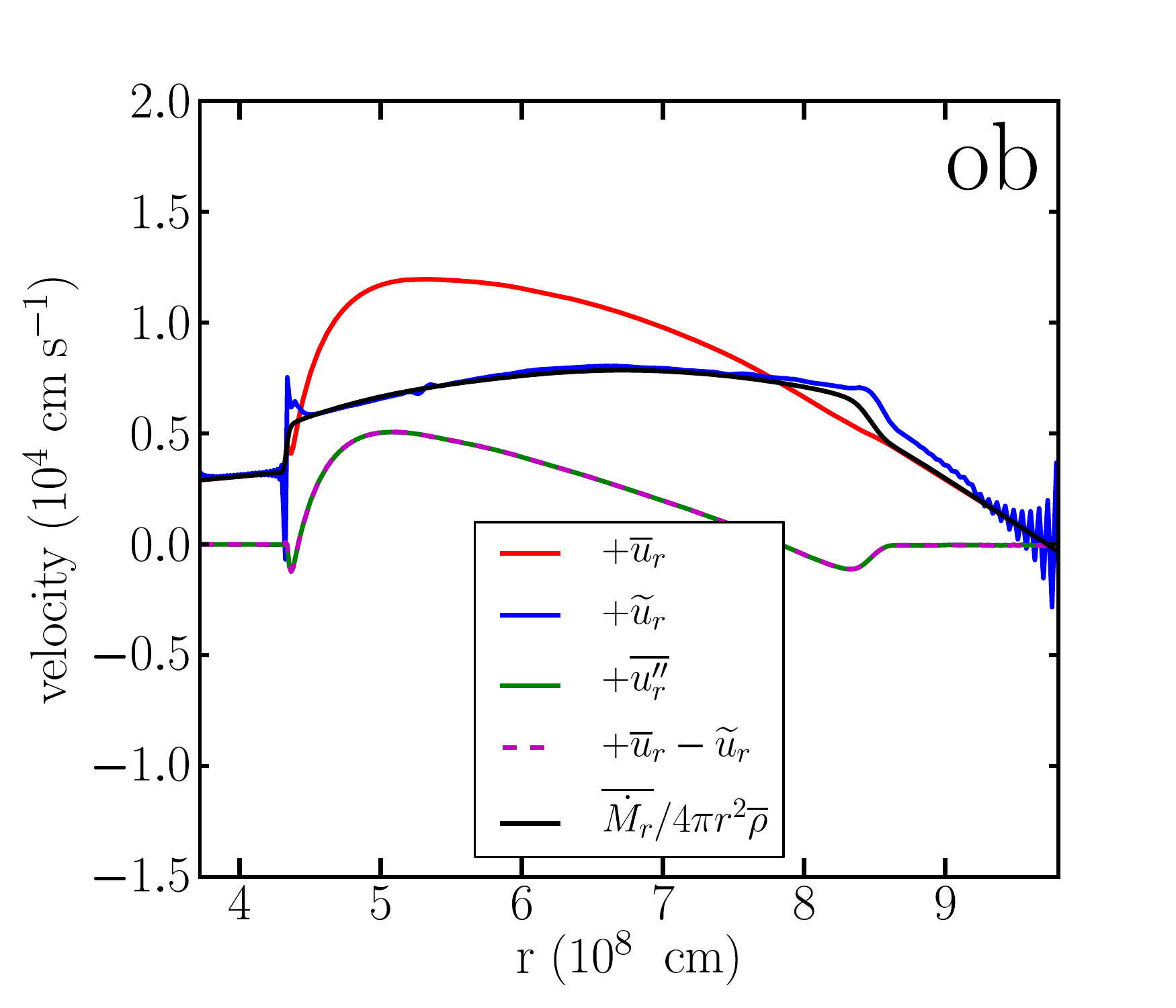}}

\centerline{
\includegraphics[width=7.cm]{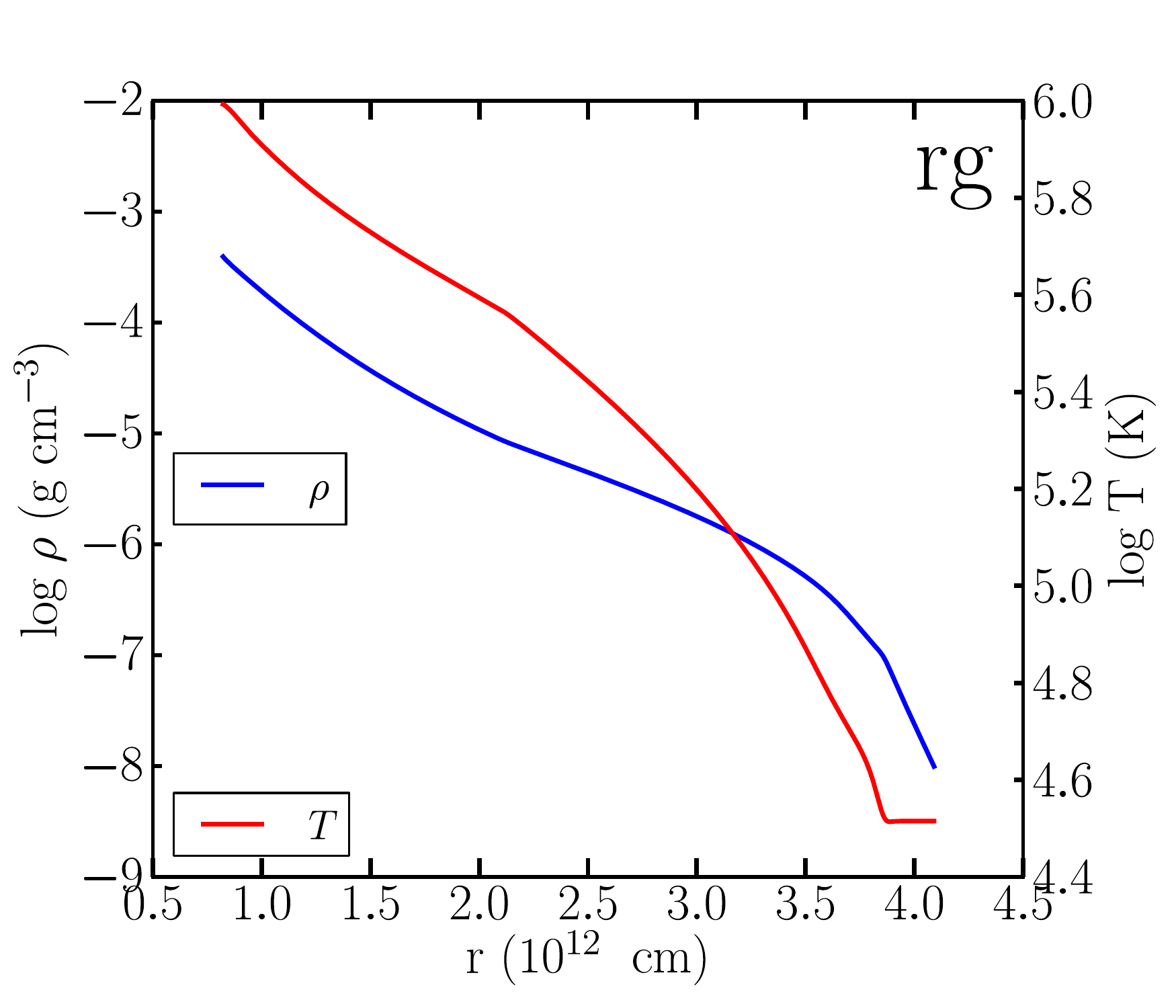}
\includegraphics[width=7.cm]{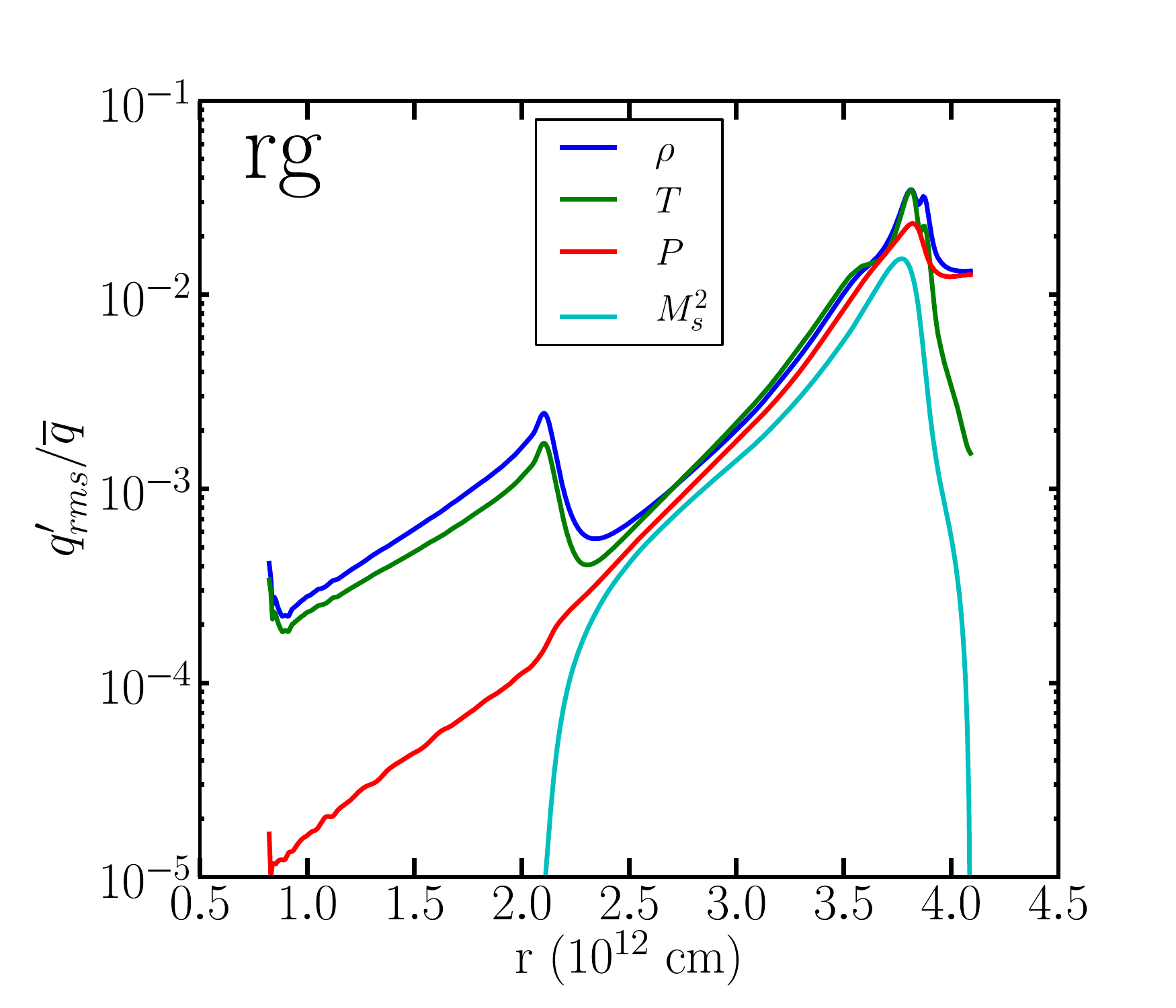}
\includegraphics[width=7.cm]{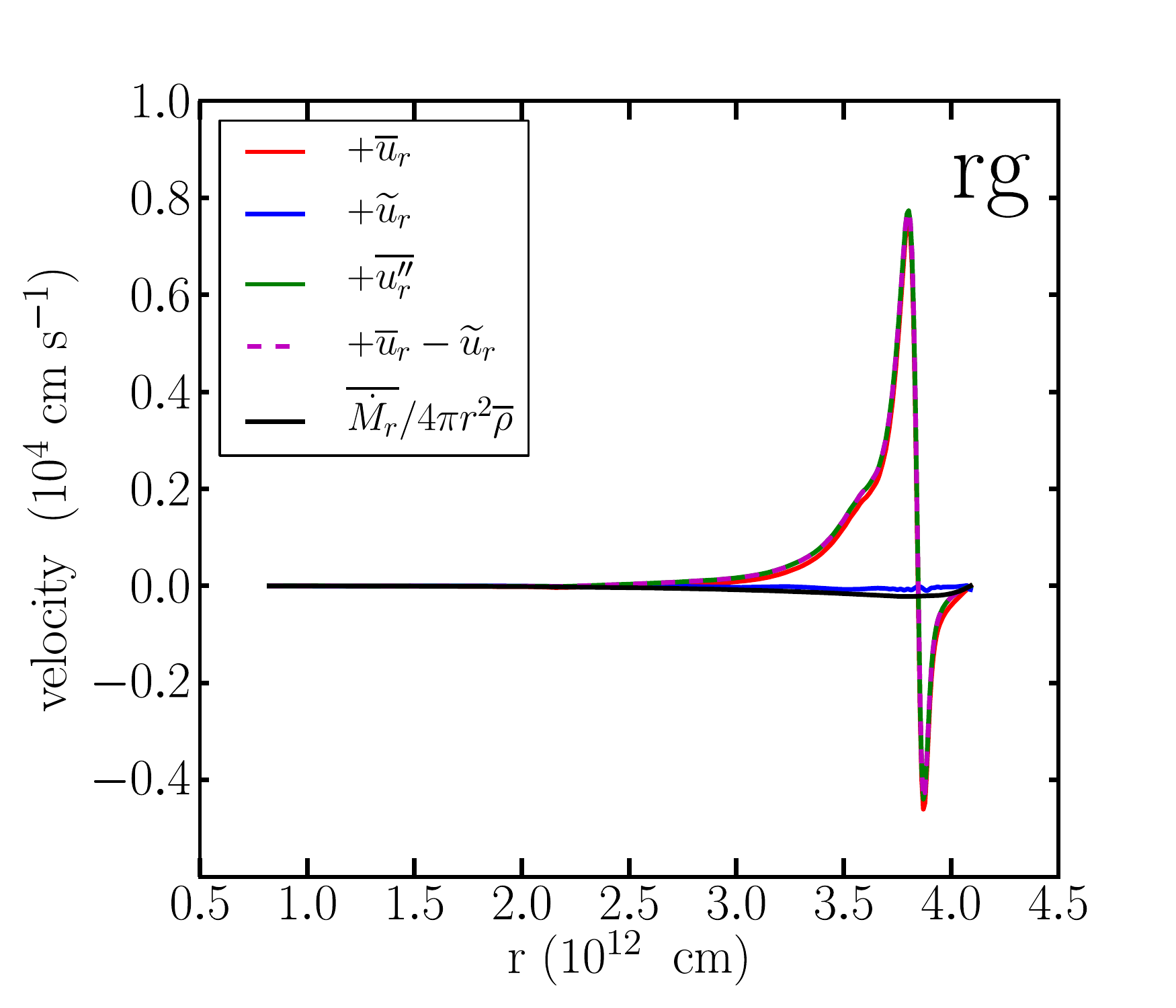}}
\caption{Properties of our data. Model {\sf ob.3D.mr} (upper panels) and model {\sf rg.3D.mr} (lower panels). \label{fig:data}}
\end{figure}
  
\clearpage

\section{Profiles and integral budgets of mean fields}

\subsection{Mean continuity equation}

\begin{align}
\fav{D}_t \av{\rho} =& -\av{\rho} \fav{d} + {\mathcal N_\rho}  \label{eq:rans_density}
\end{align}

\begin{figure}[!h]
\centerline{
\includegraphics[width=6.cm]{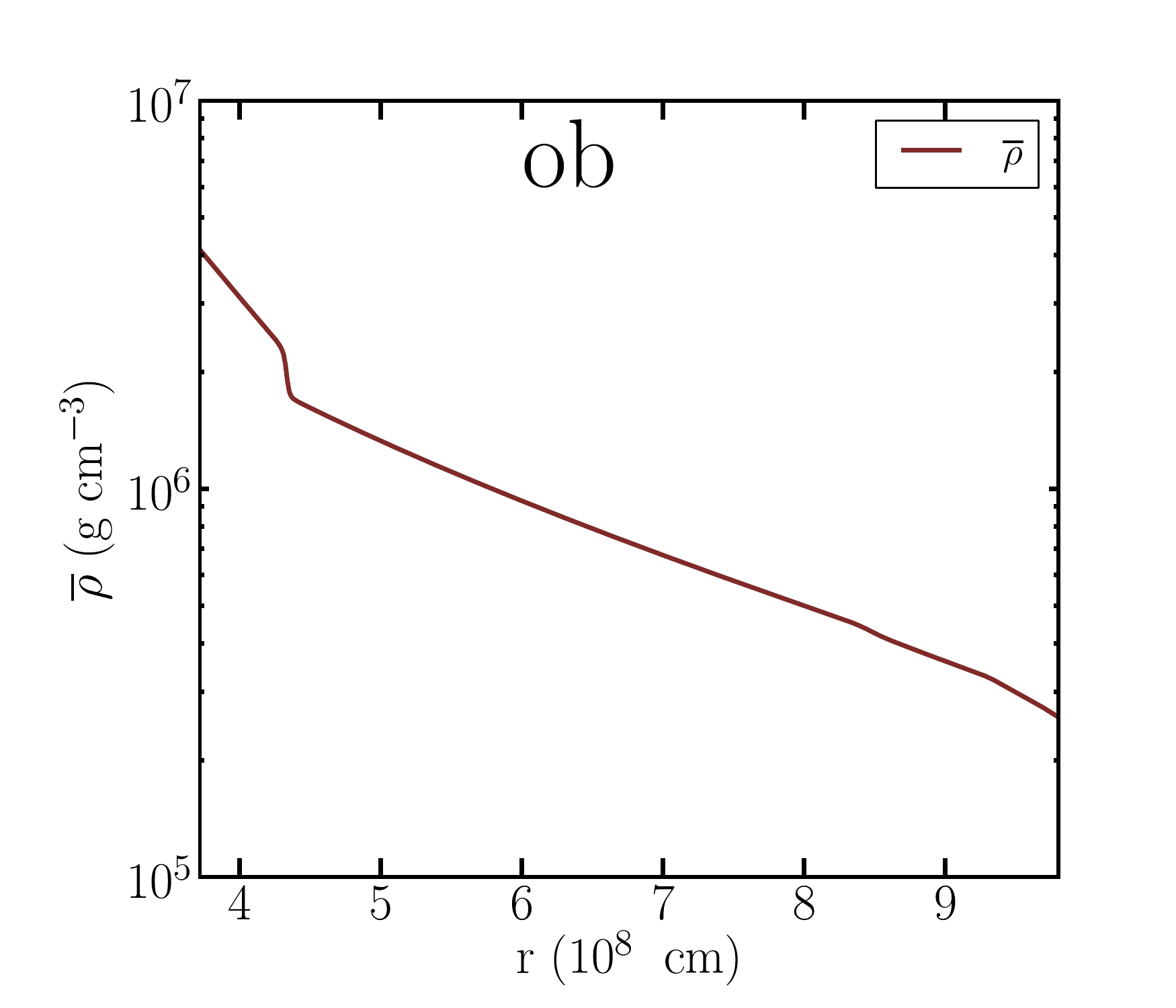}
\includegraphics[width=6.cm]{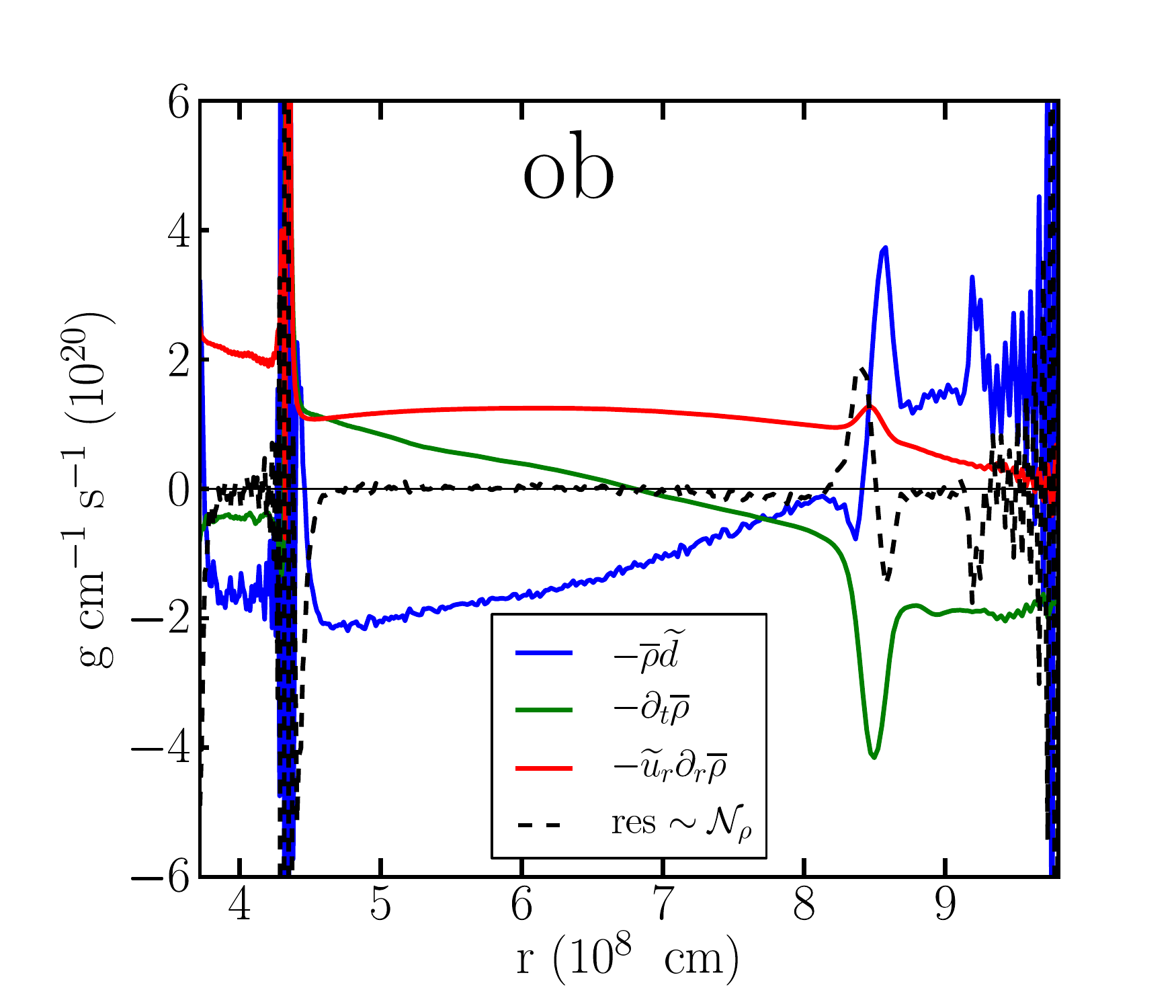}
\includegraphics[width=6.cm]{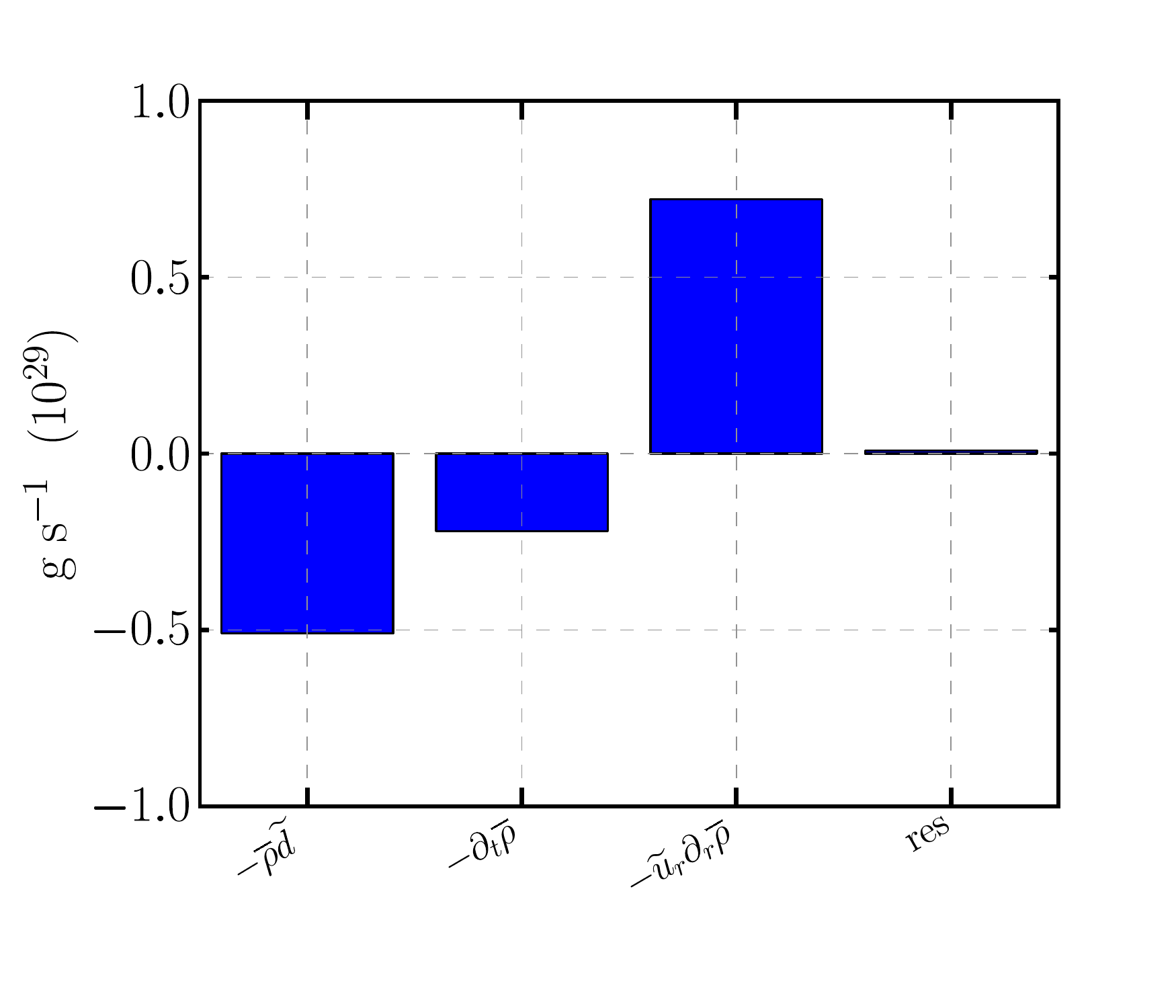}}

\centerline{
\includegraphics[width=6.cm]{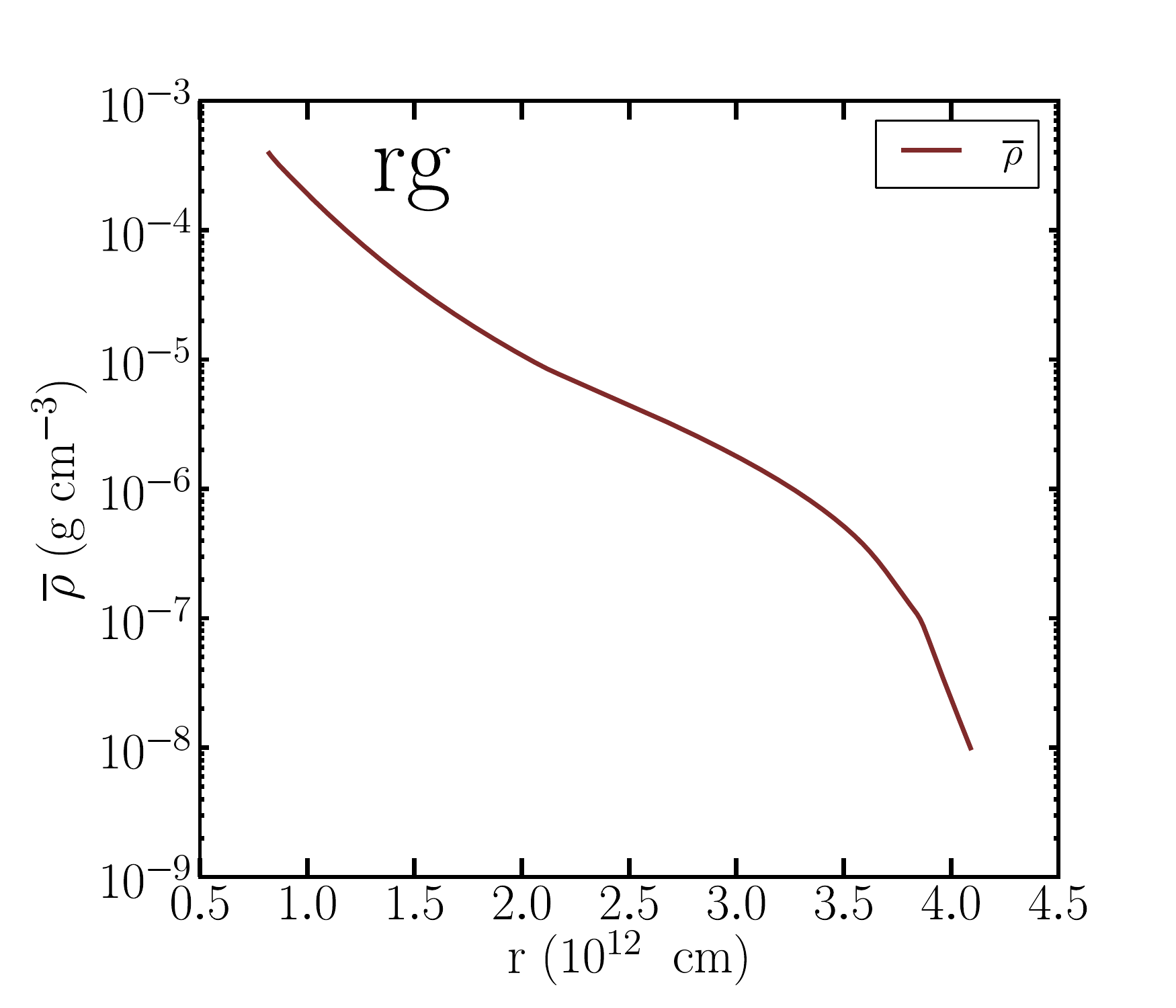}
\includegraphics[width=6.cm]{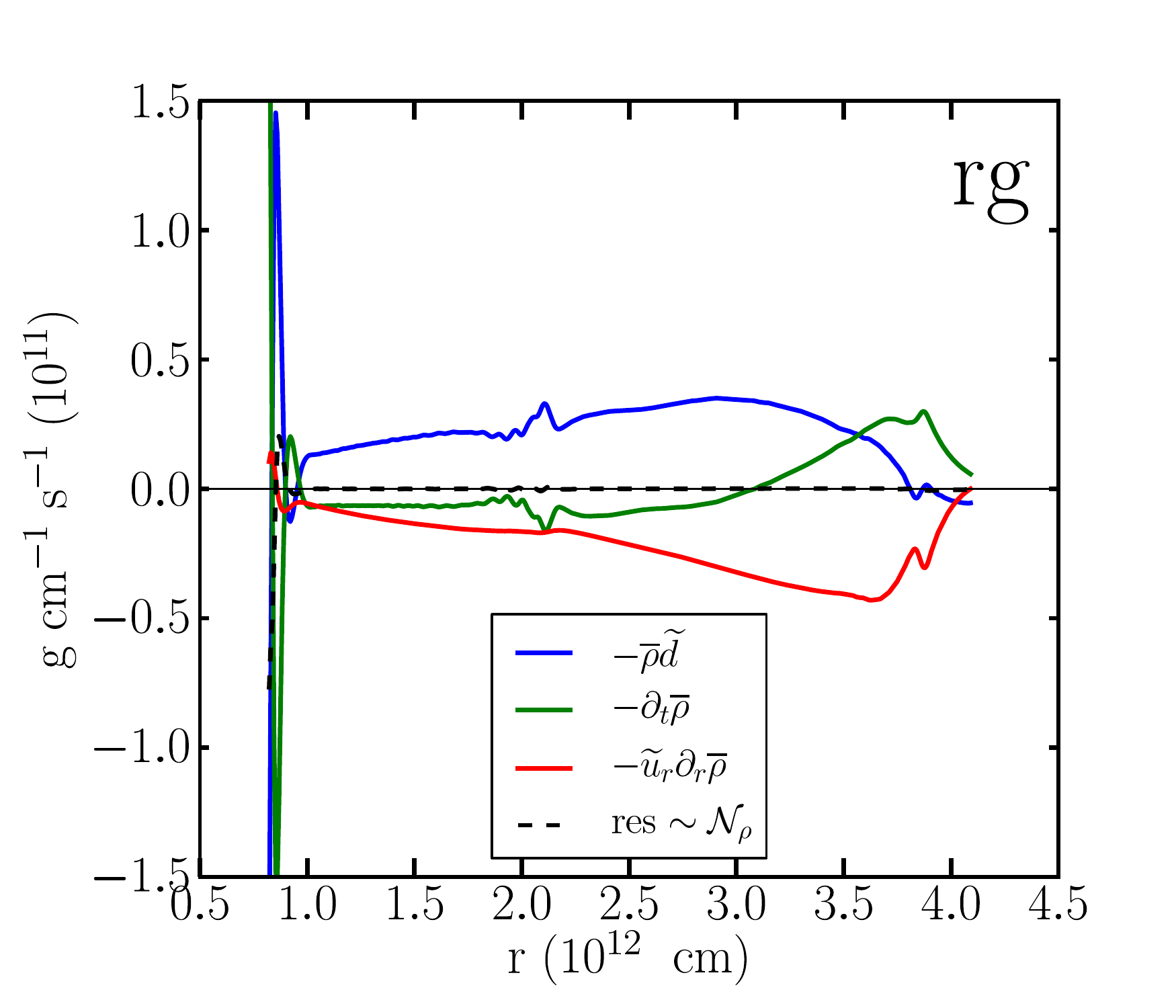}
\includegraphics[width=6.cm]{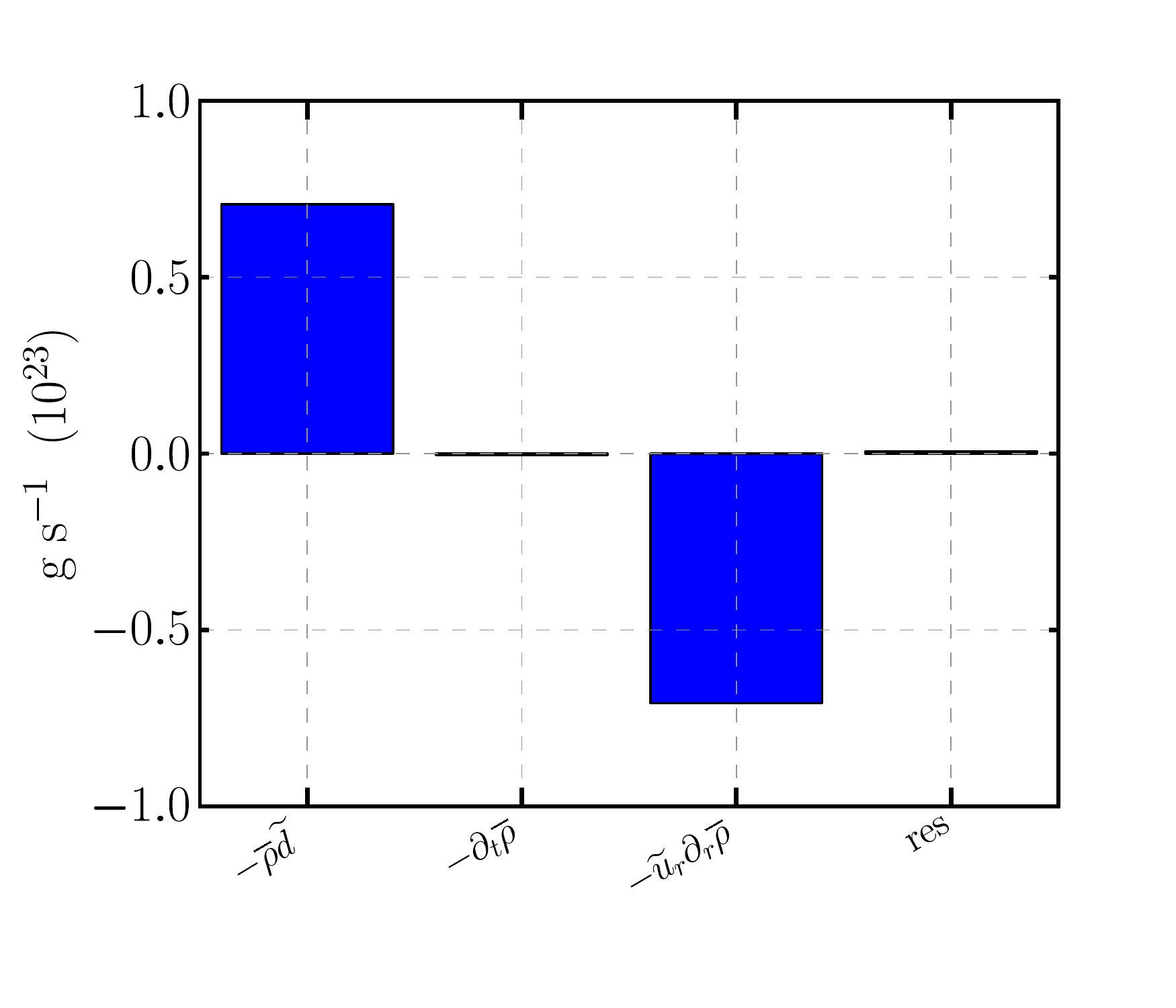}}
\caption{Mean continuity equation. Model {\sf ob.3D.mr} (upper panels) and model {\sf rg.3D.mr} (lower panels). \label{fig:continuity-equation}}
\end{figure}

\newpage

\subsection{Mean radial momentum equation}

\begin{align}
\av{\rho}\fav{D}_t\fav{u}_r = & -\nabla_r \fav{R}_{rr} -\av{G^{M}_r} - \partial_r \av{P} + \av{\rho}\fav{g_r} + {\mathcal N_{ur}}
\end{align}

\begin{figure}[!h]
\centerline{
\includegraphics[width=6.5cm]{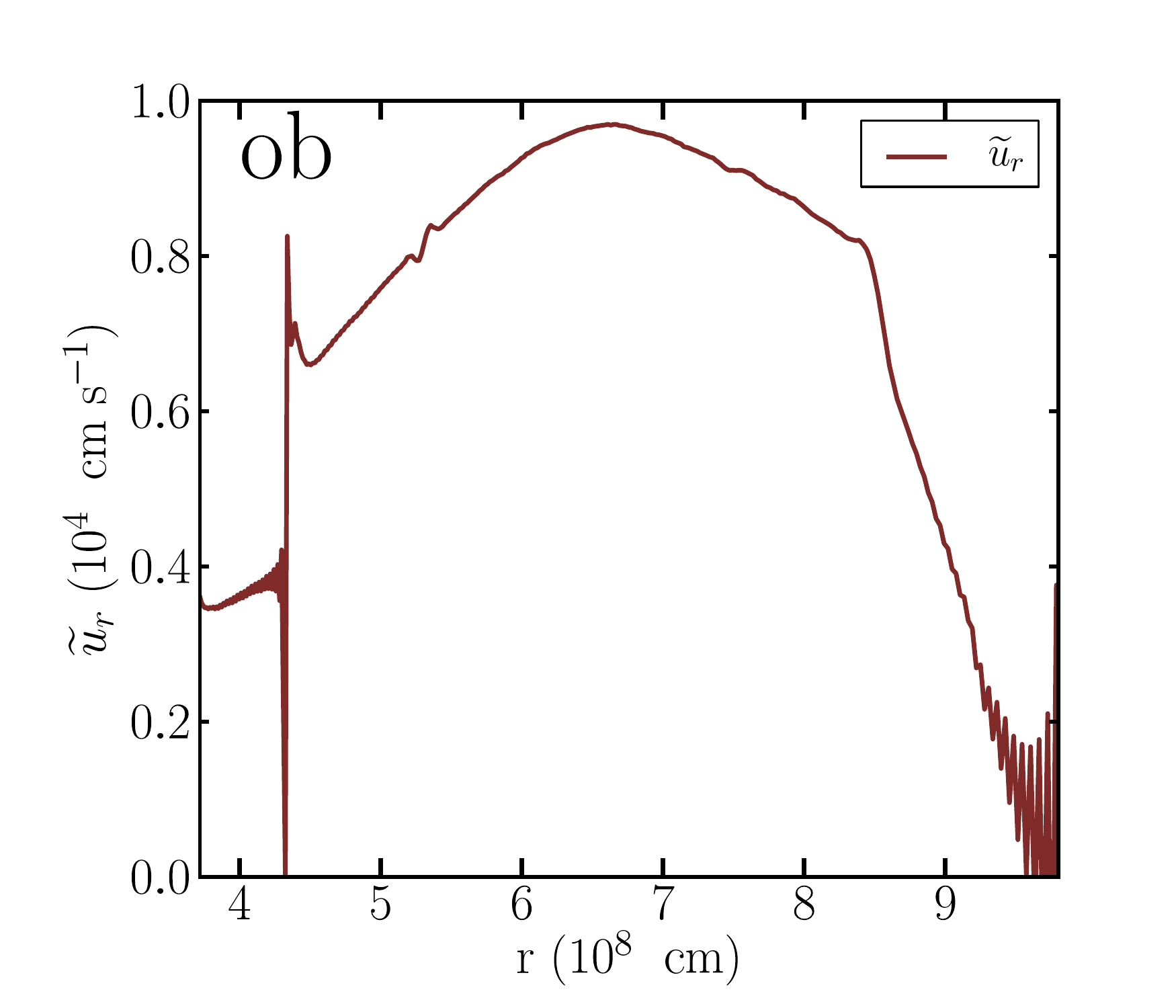}
\includegraphics[width=6.5cm]{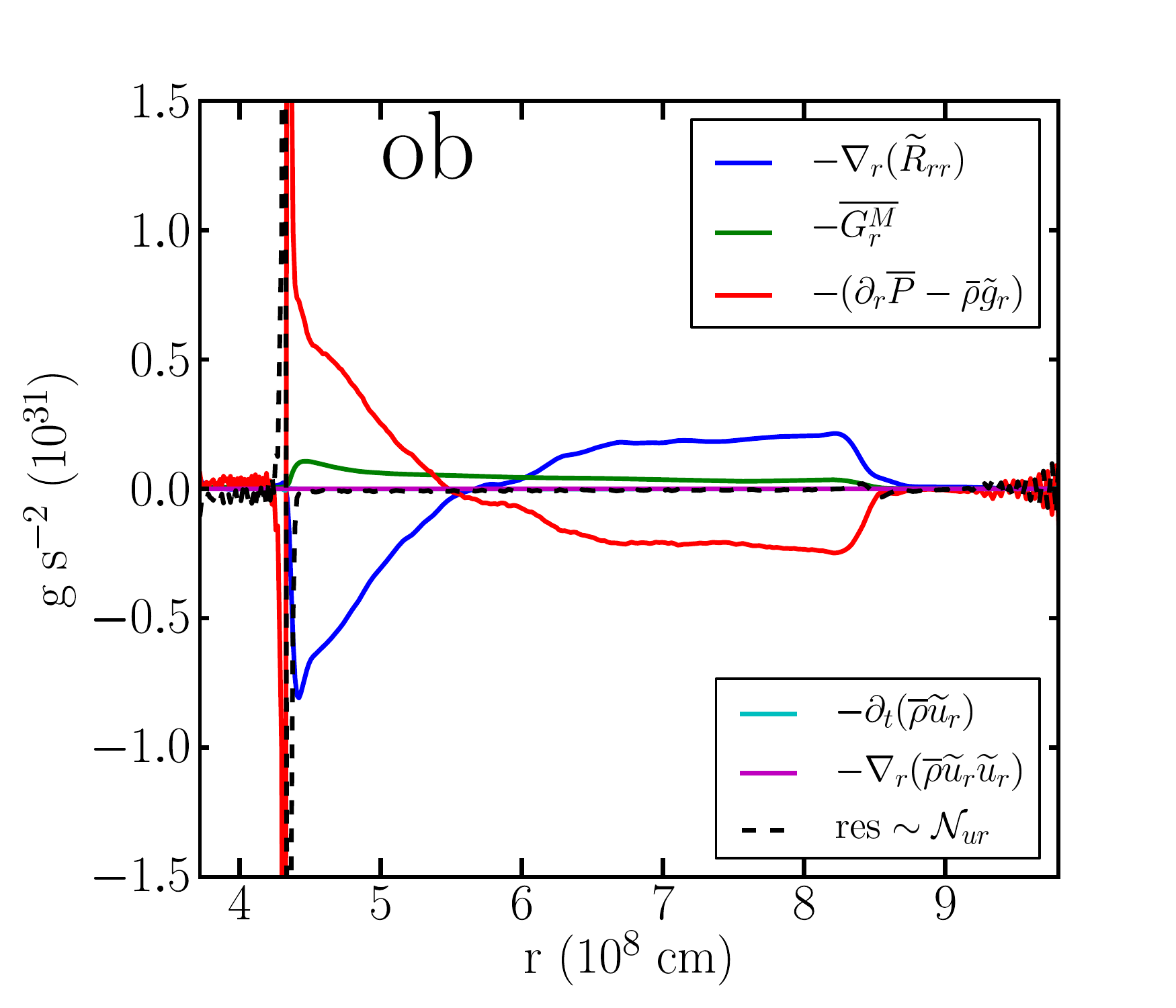}
\includegraphics[width=6.5cm]{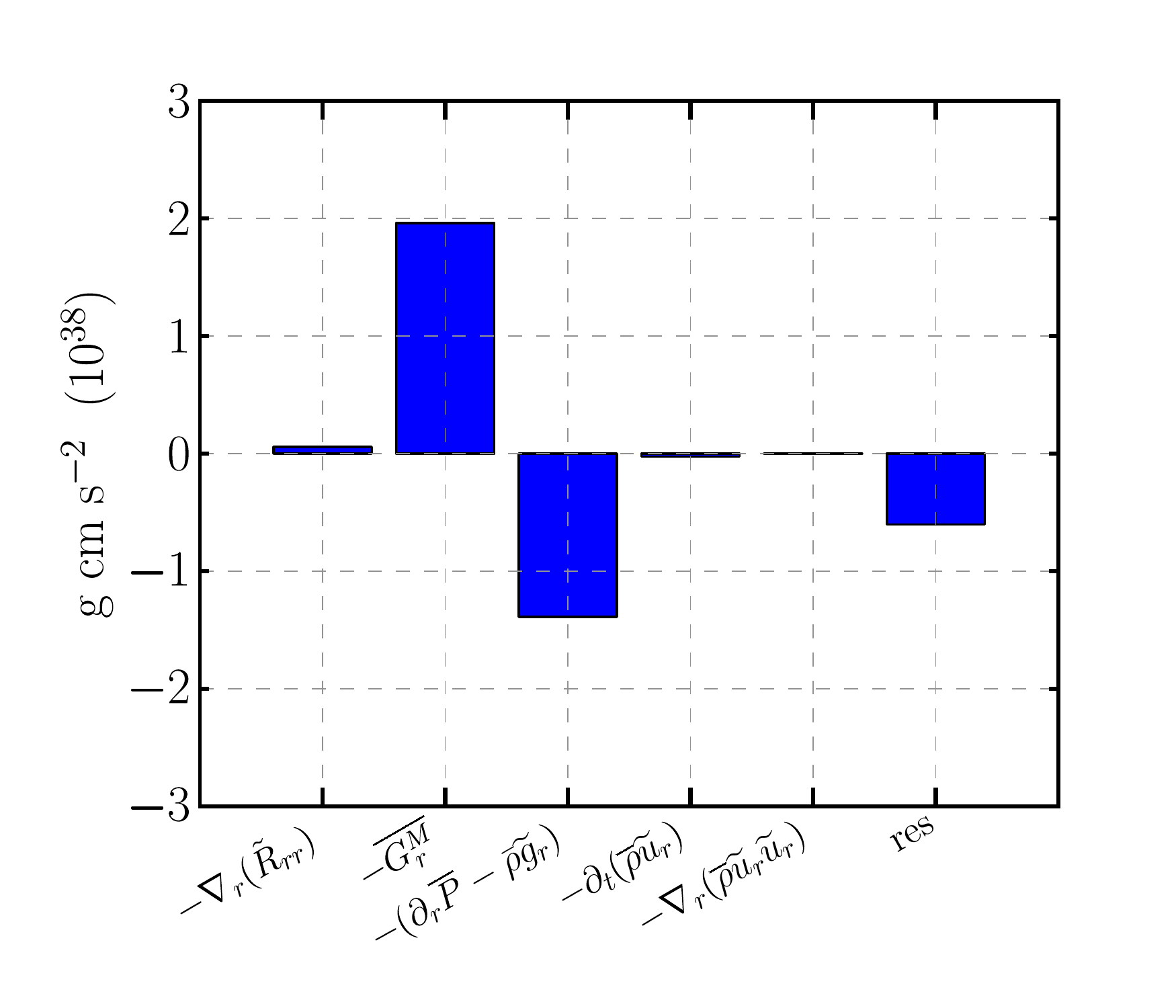}}

\centerline{
\includegraphics[width=6.5cm]{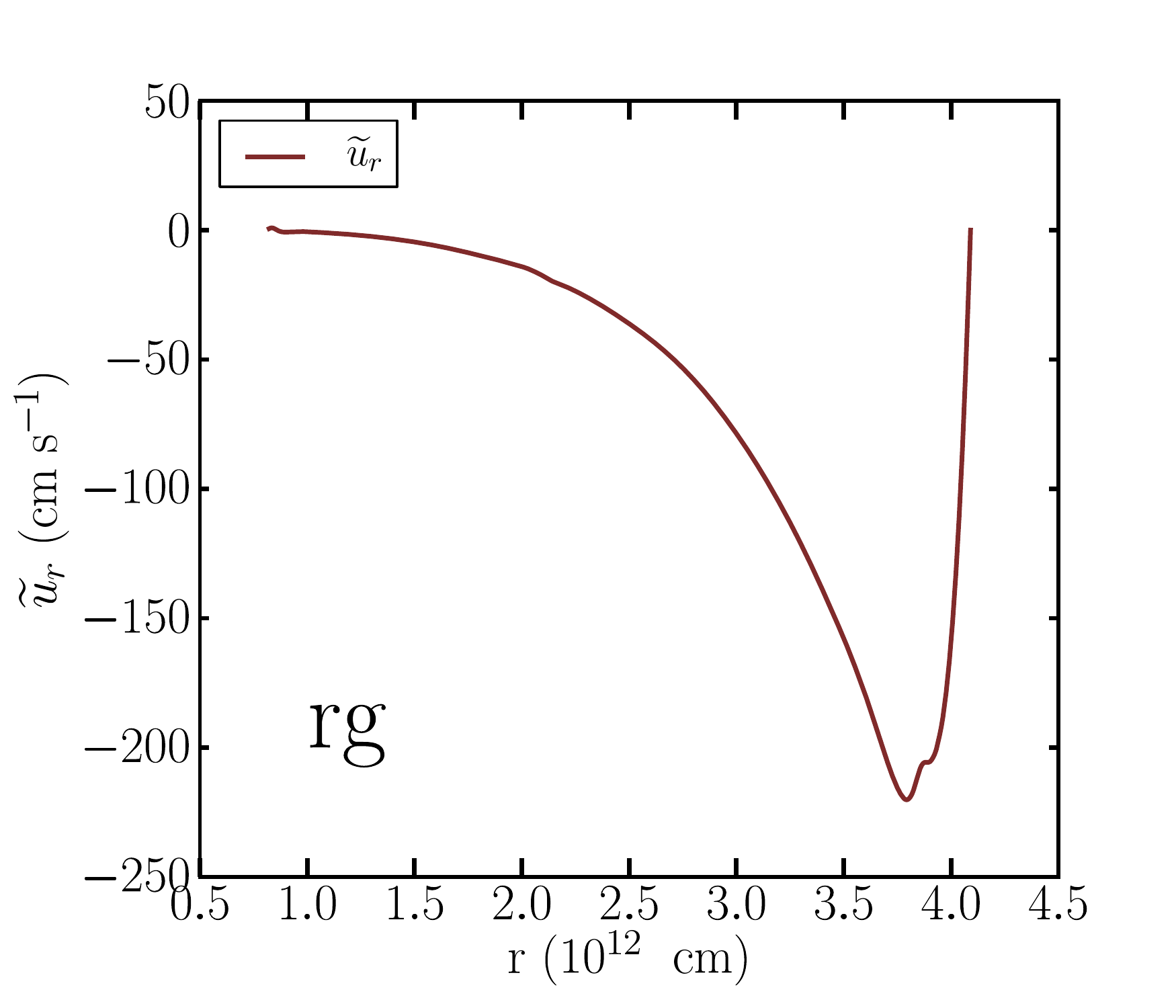}
\includegraphics[width=6.5cm]{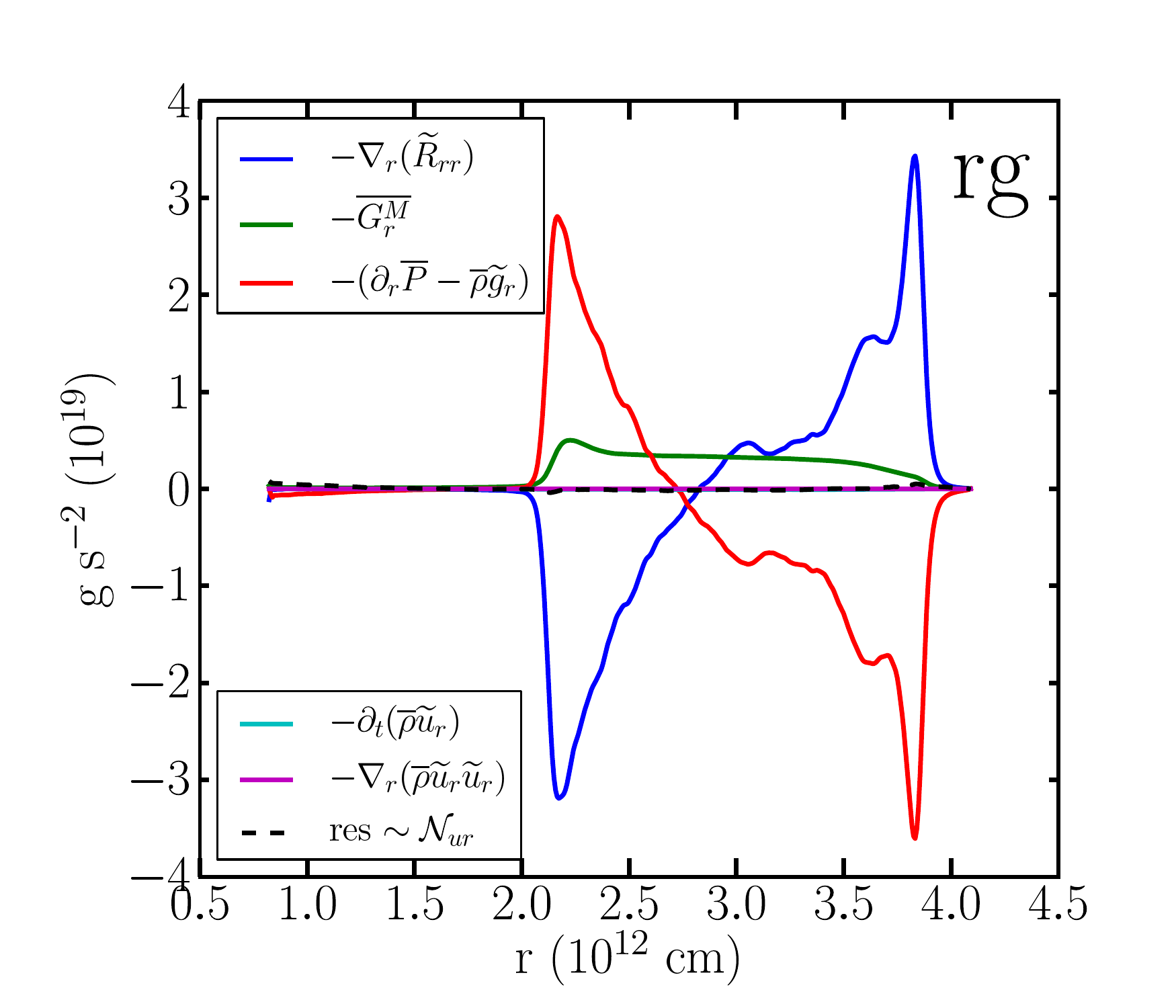}
\includegraphics[width=6.5cm]{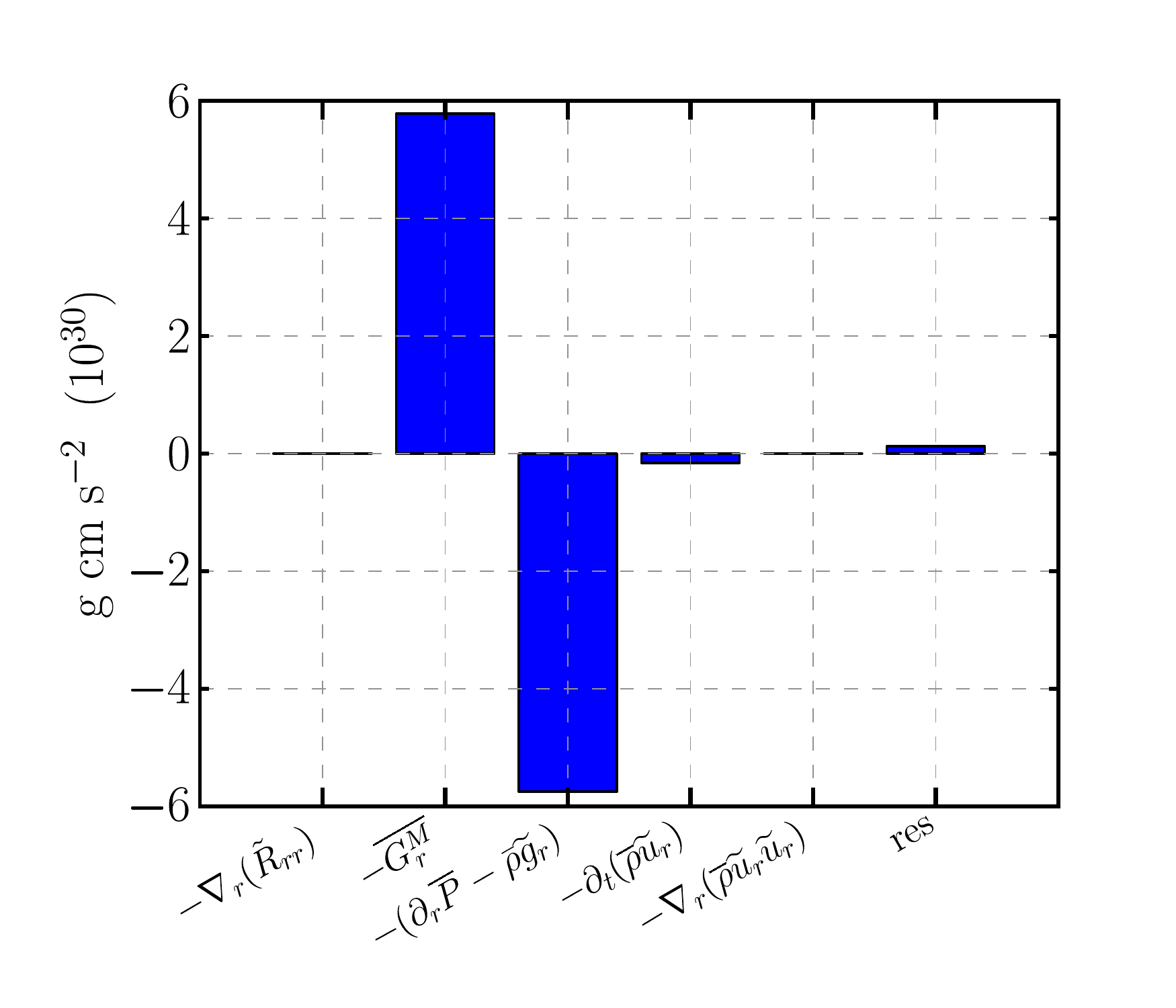}}
\caption{Mean radial momentum equation. Model {\sf ob.3D.mr} (upper panels) and model {\sf rg.3D.mr} (lower panels). \label{fig:rm-equation}}
\end{figure}

\newpage

\subsection{Mean azimutal momentum equation}

\begin{align}
\av{\rho}\fav{D}_t\fav{u}_\theta = &  -\nabla_r \fav{R}_{\theta r} -\av{G^{M}_\theta} - (1/r)\av{\partial_\theta P} + {\mathcal N_{u \theta}}
\end{align}

\begin{figure}[!h]
\centerline{
\includegraphics[width=6.5cm]{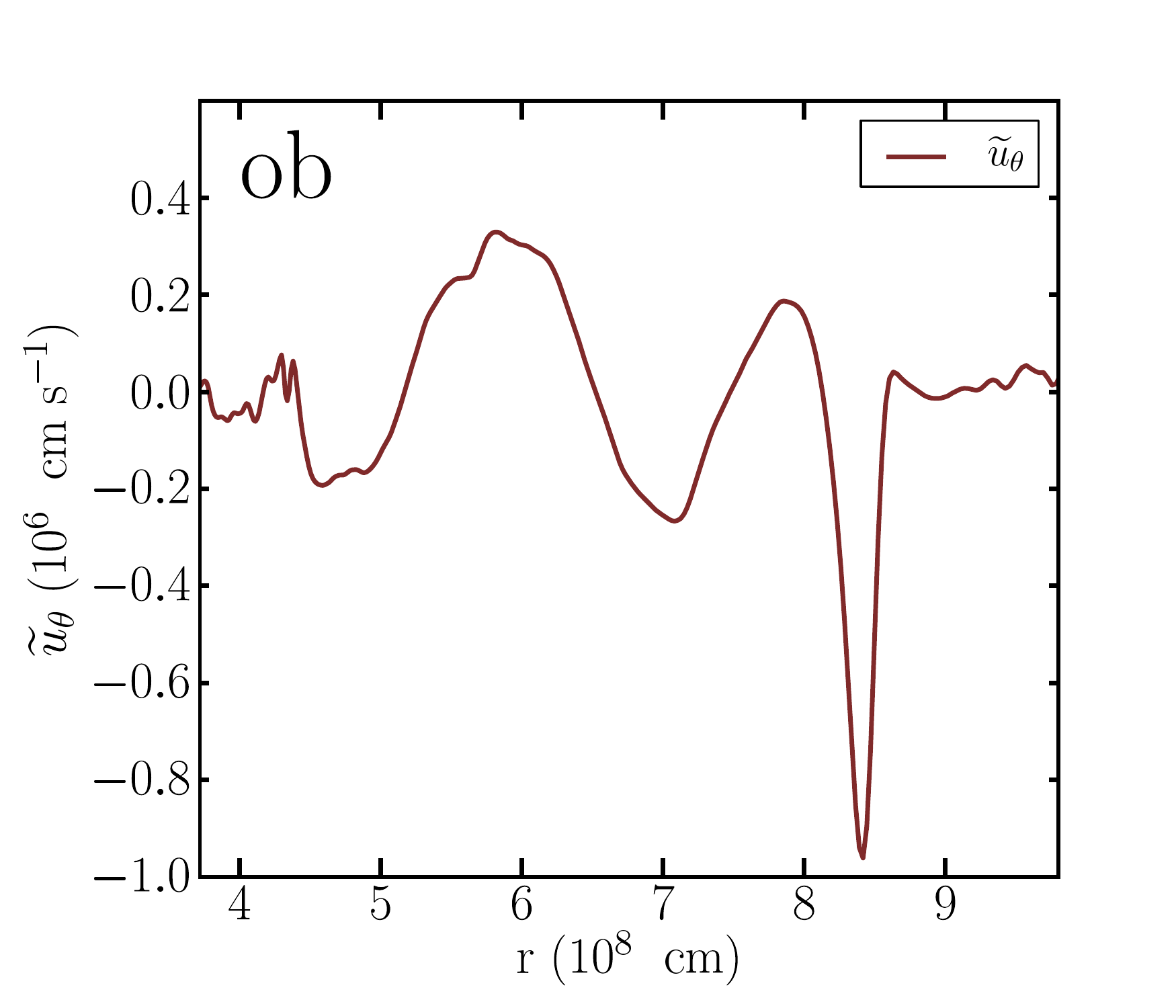}
\includegraphics[width=6.5cm]{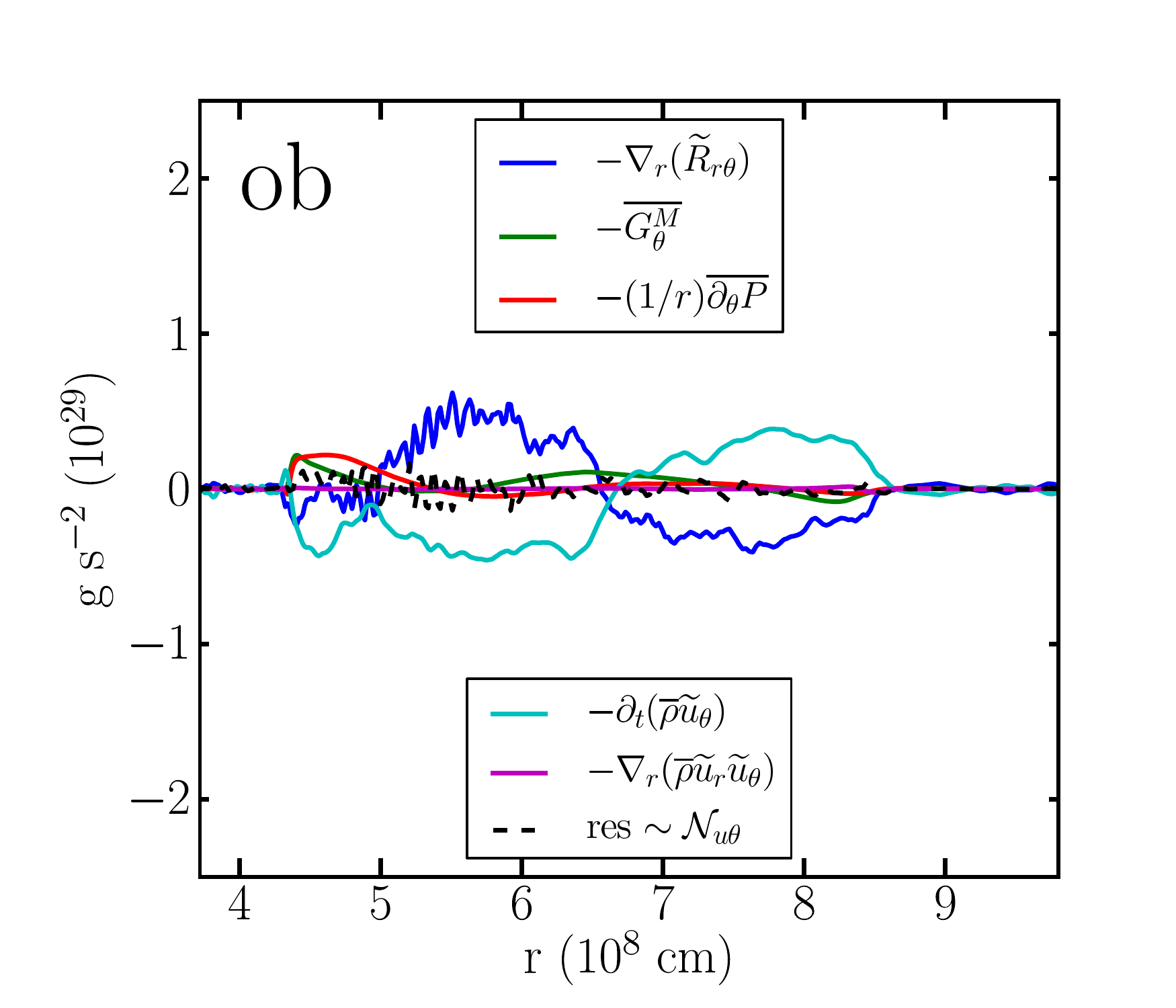}
\includegraphics[width=6.5cm]{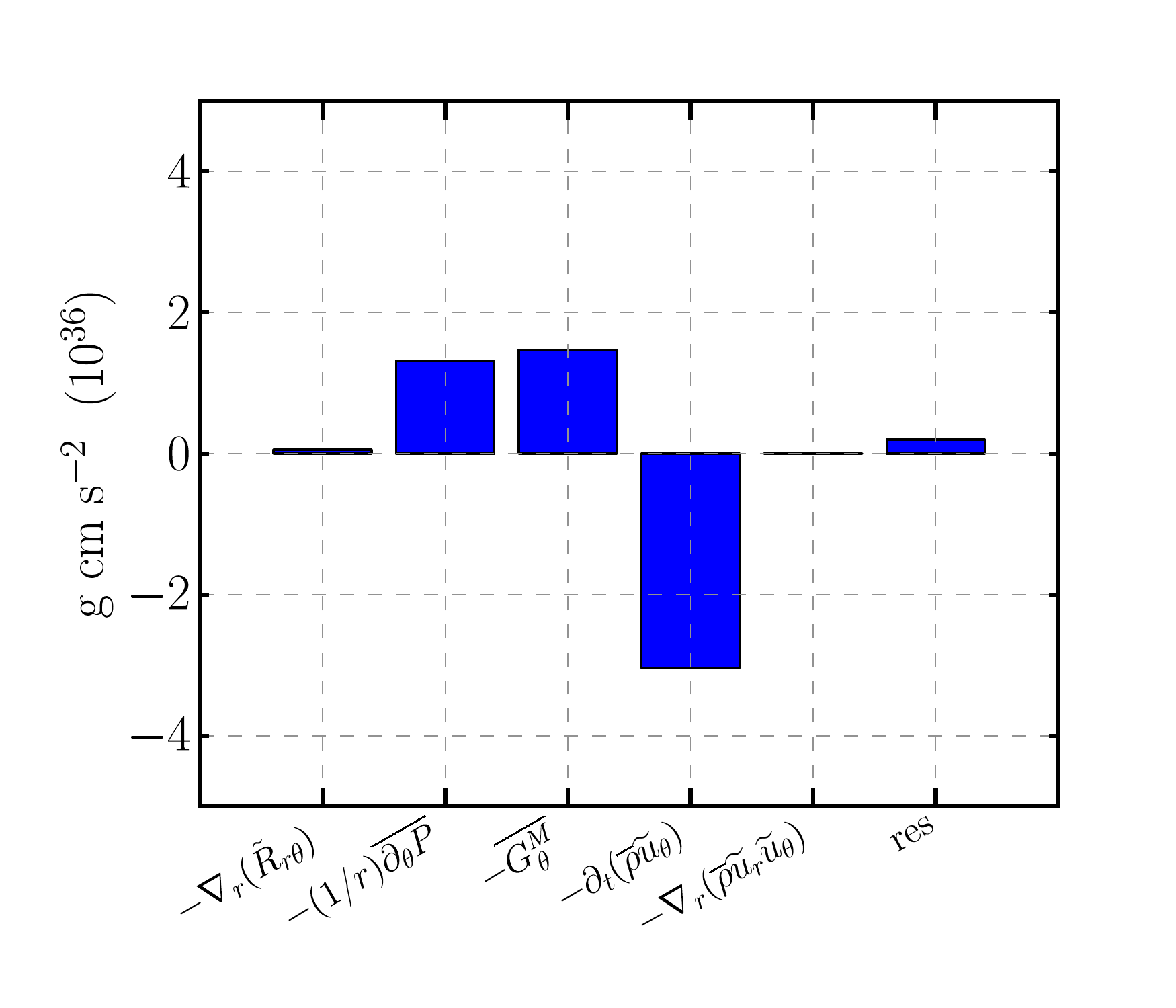}}

\centerline{
\includegraphics[width=6.5cm]{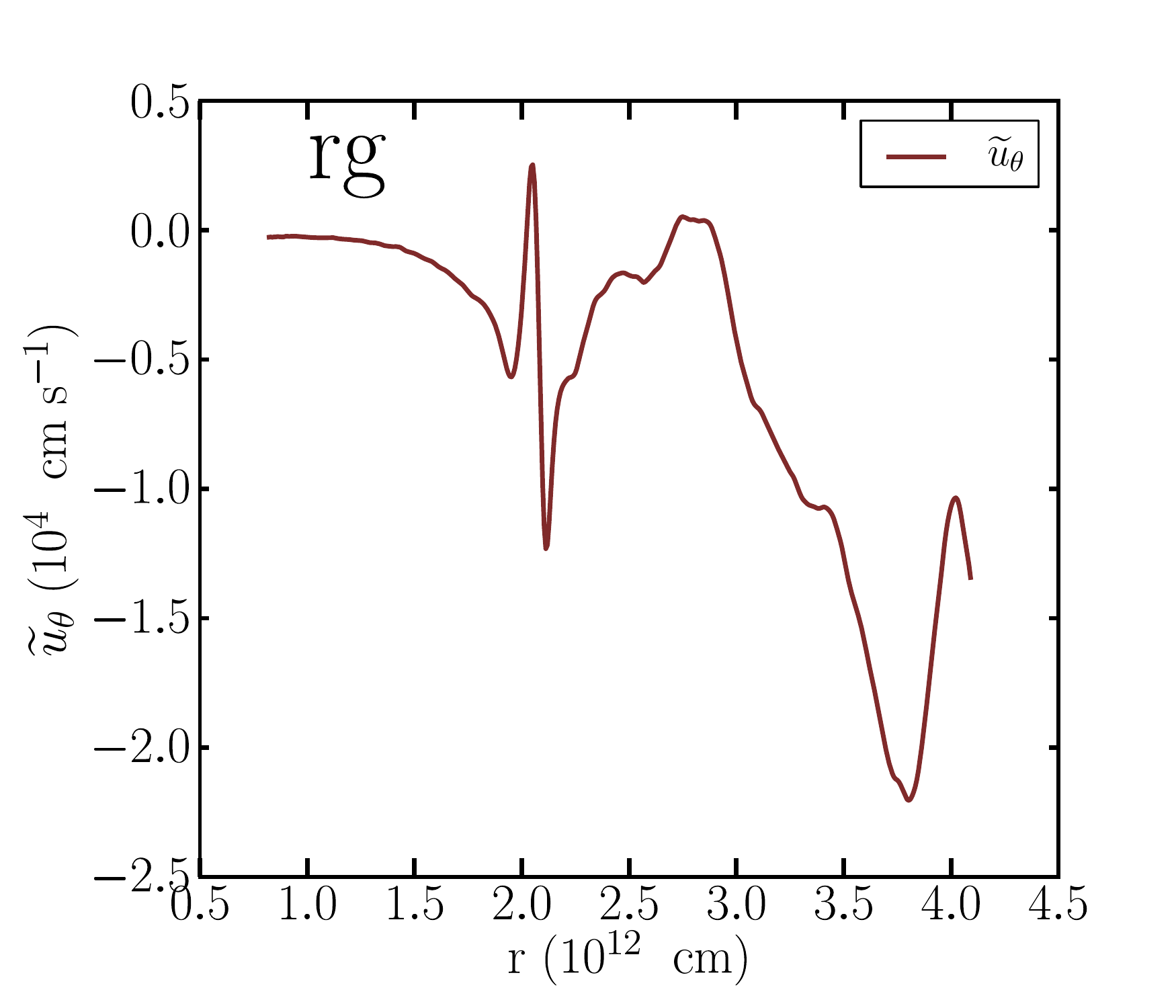}
\includegraphics[width=6.5cm]{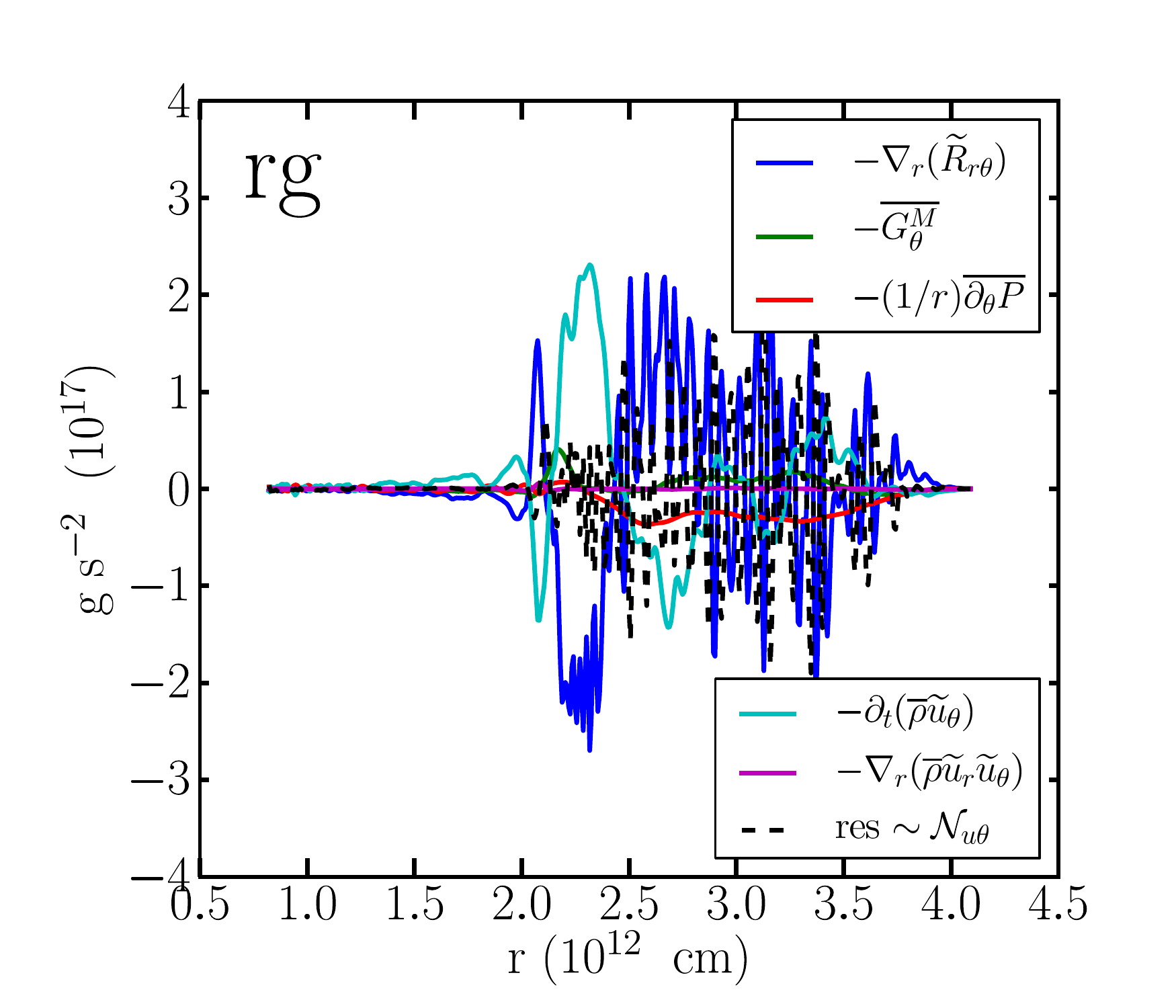}
\includegraphics[width=6.5cm]{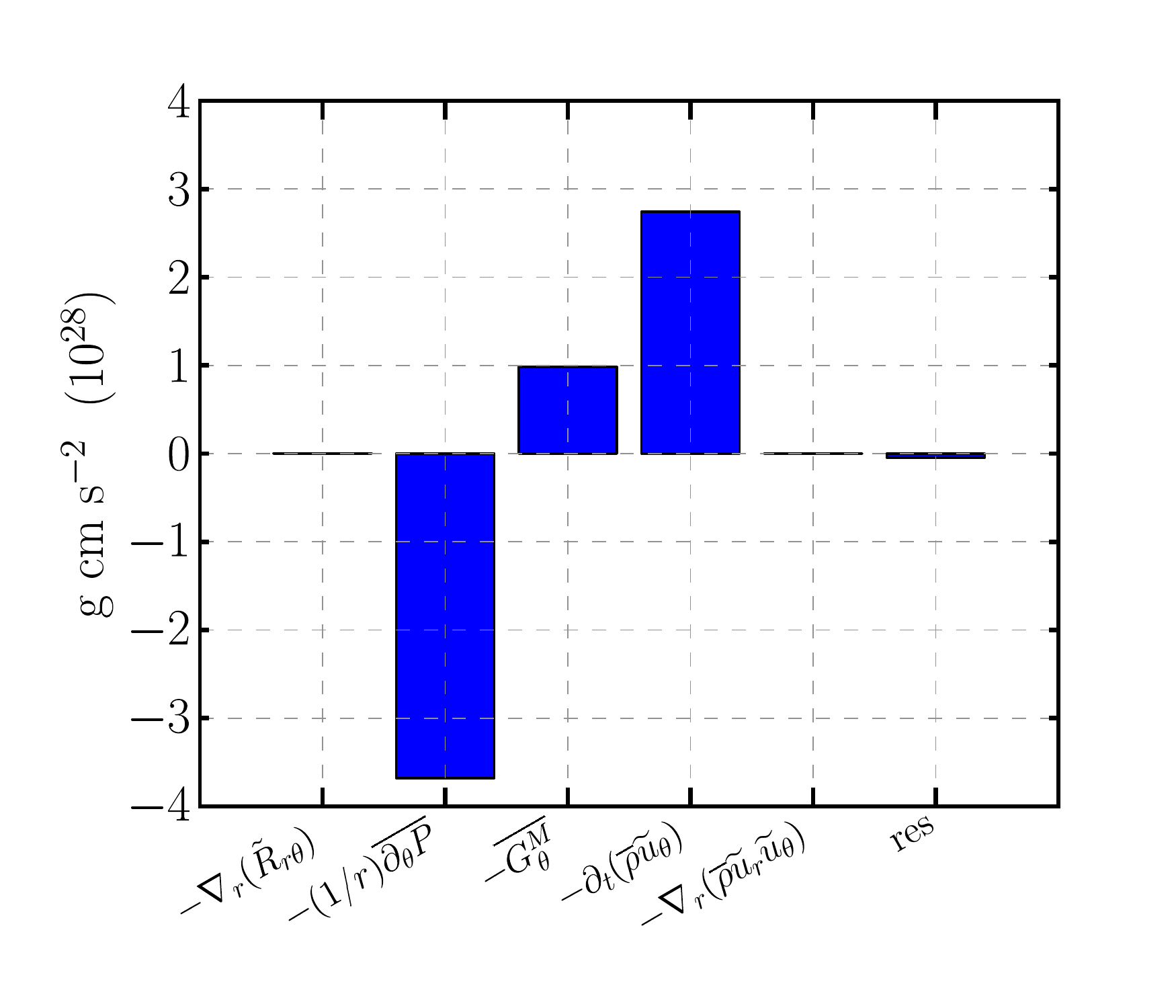}}
\caption{Mean azimuthal momentum equation. Model {\sf ob.3D.mr} (upper panels) and model {\sf rg.3D.mr} (lower panels). \label{fig:tm-equation}}
\end{figure}

\newpage

\subsection{Mean polar momentum equation}

\begin{align}
\av{\rho}\fav{D}_t\fav{u}_\phi = & -\nabla_r \fav{R}_{\phi r} -\av{G^{M}_\phi} + {\mathcal N_{u \phi}}
\end{align}

\begin{figure}[!h]
\centerline{
\includegraphics[width=6.5cm]{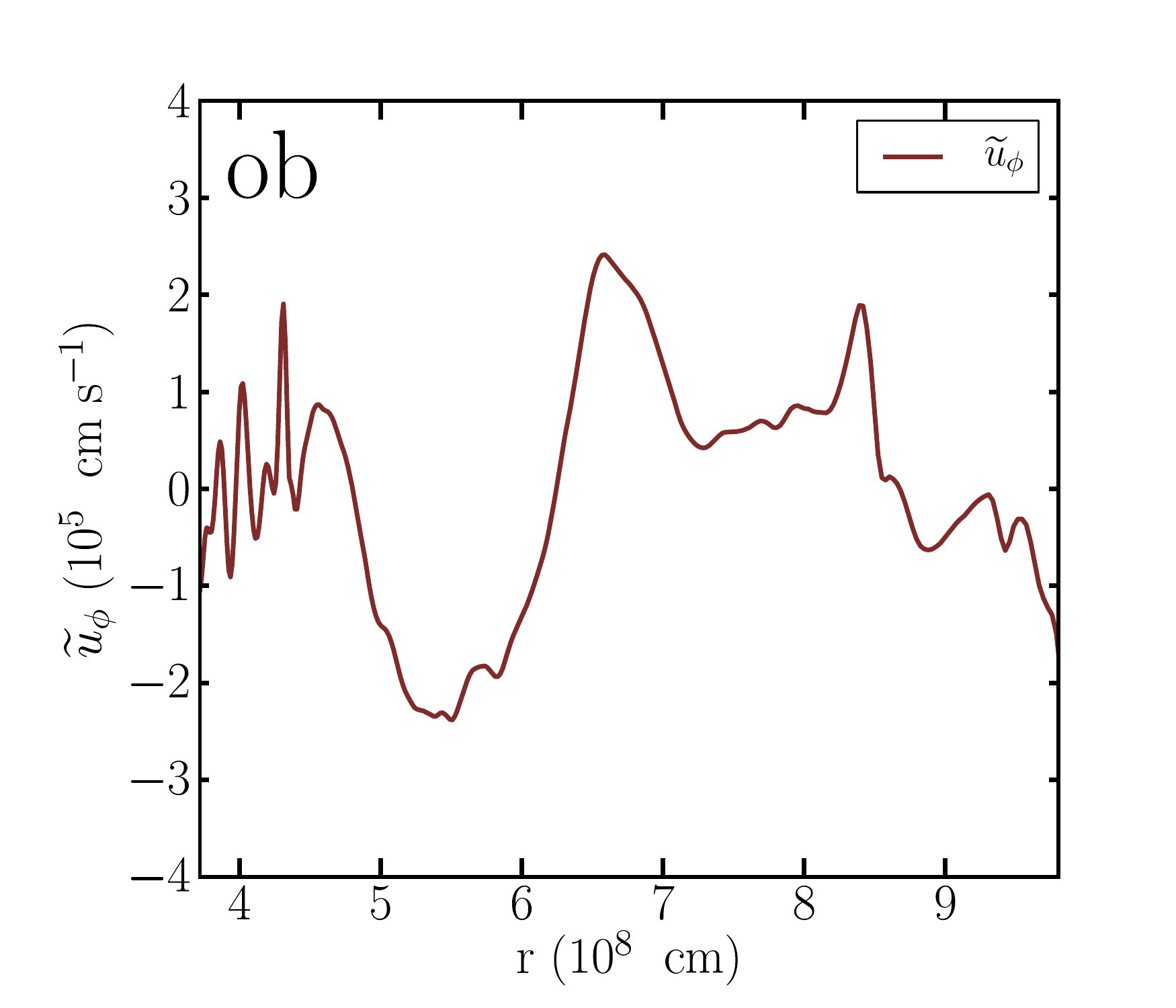}
\includegraphics[width=6.5cm]{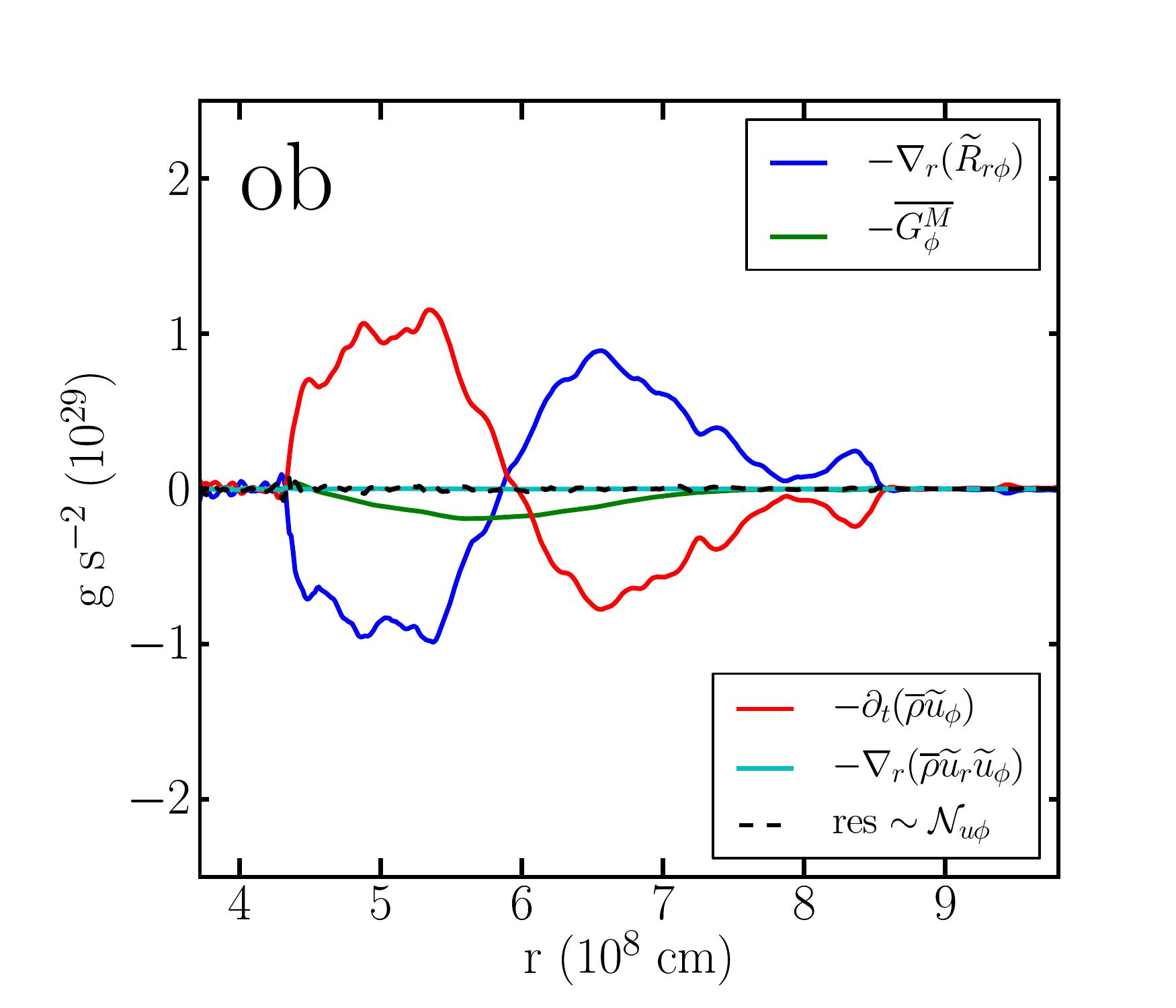}
\includegraphics[width=6.5cm]{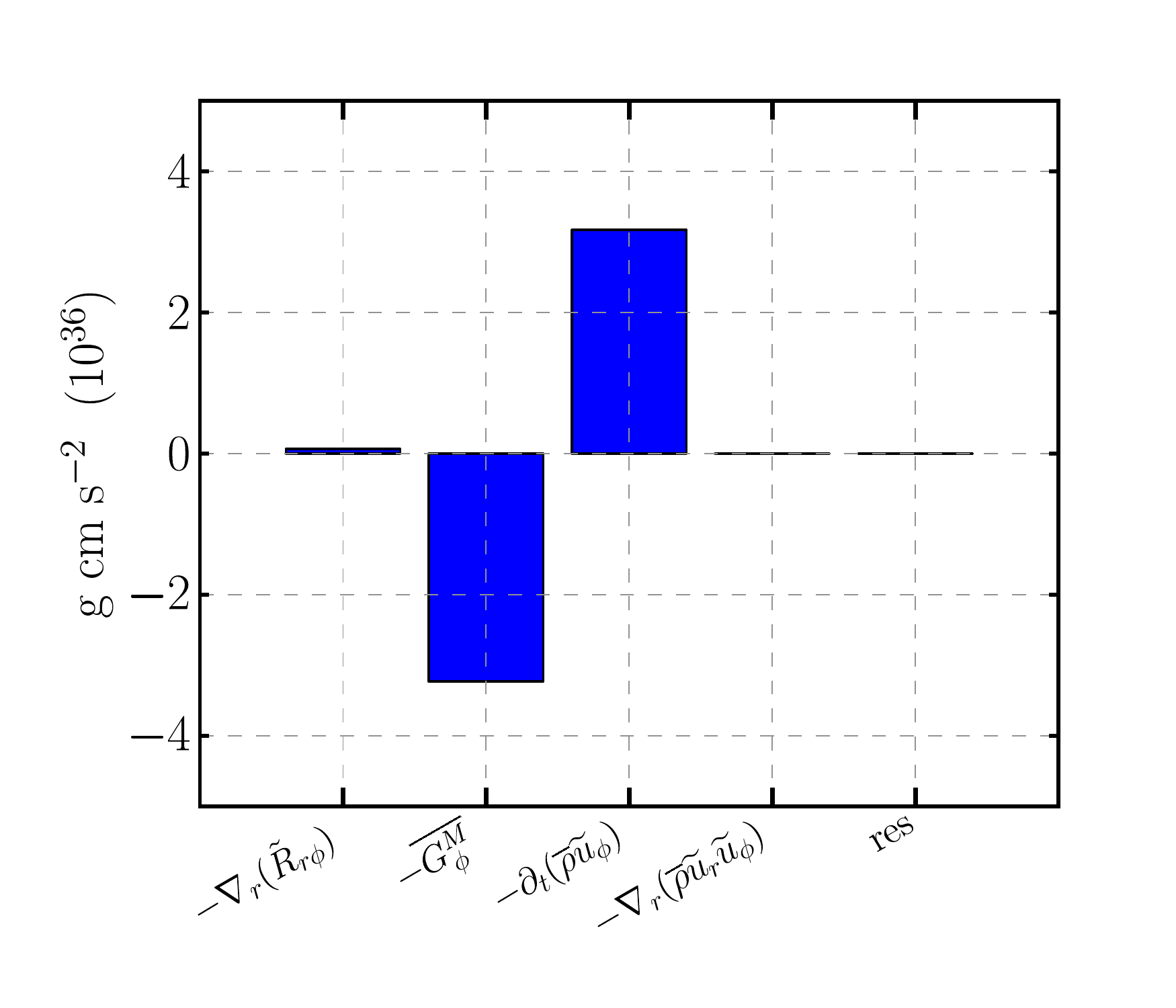}}

\centerline{
\includegraphics[width=6.5cm]{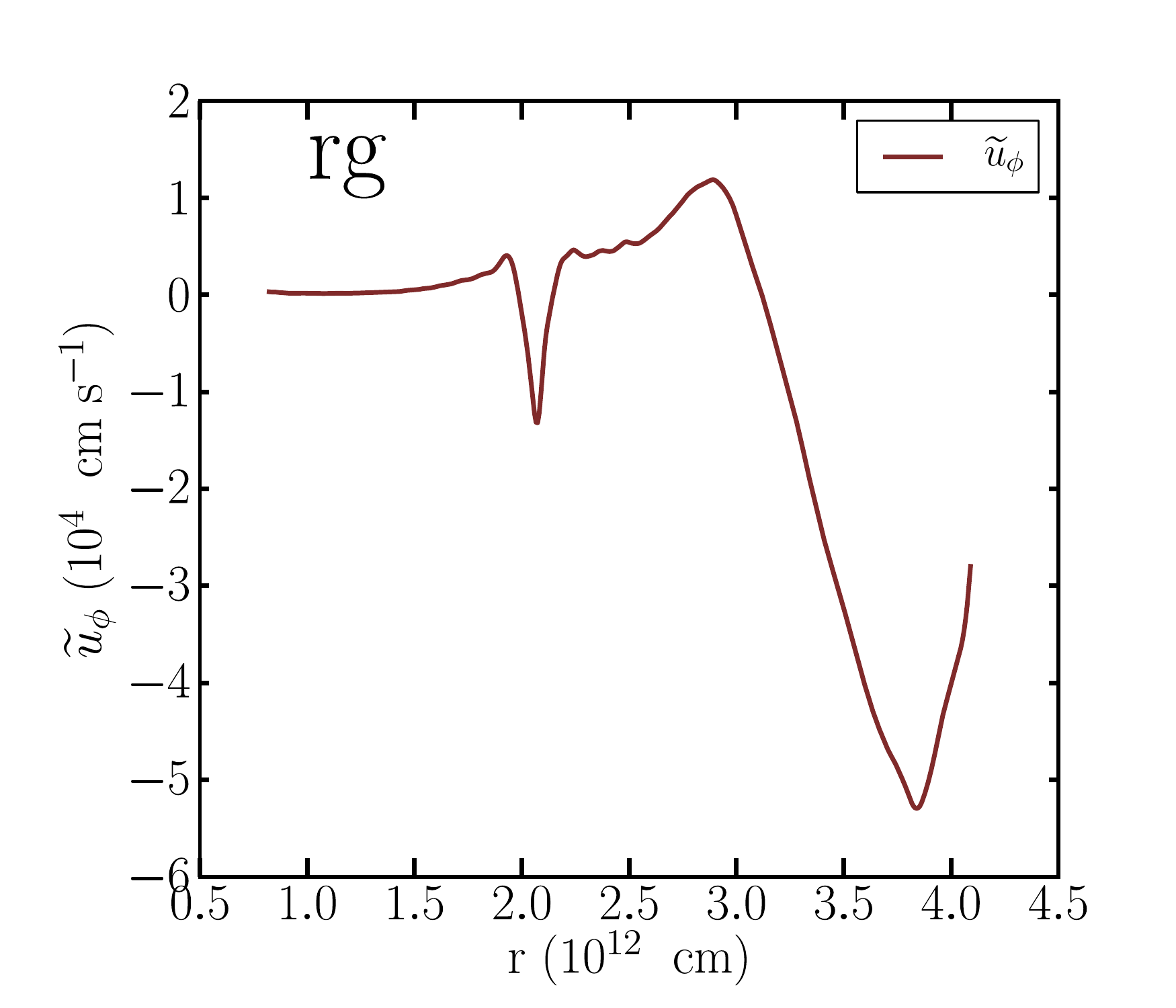}
\includegraphics[width=6.5cm]{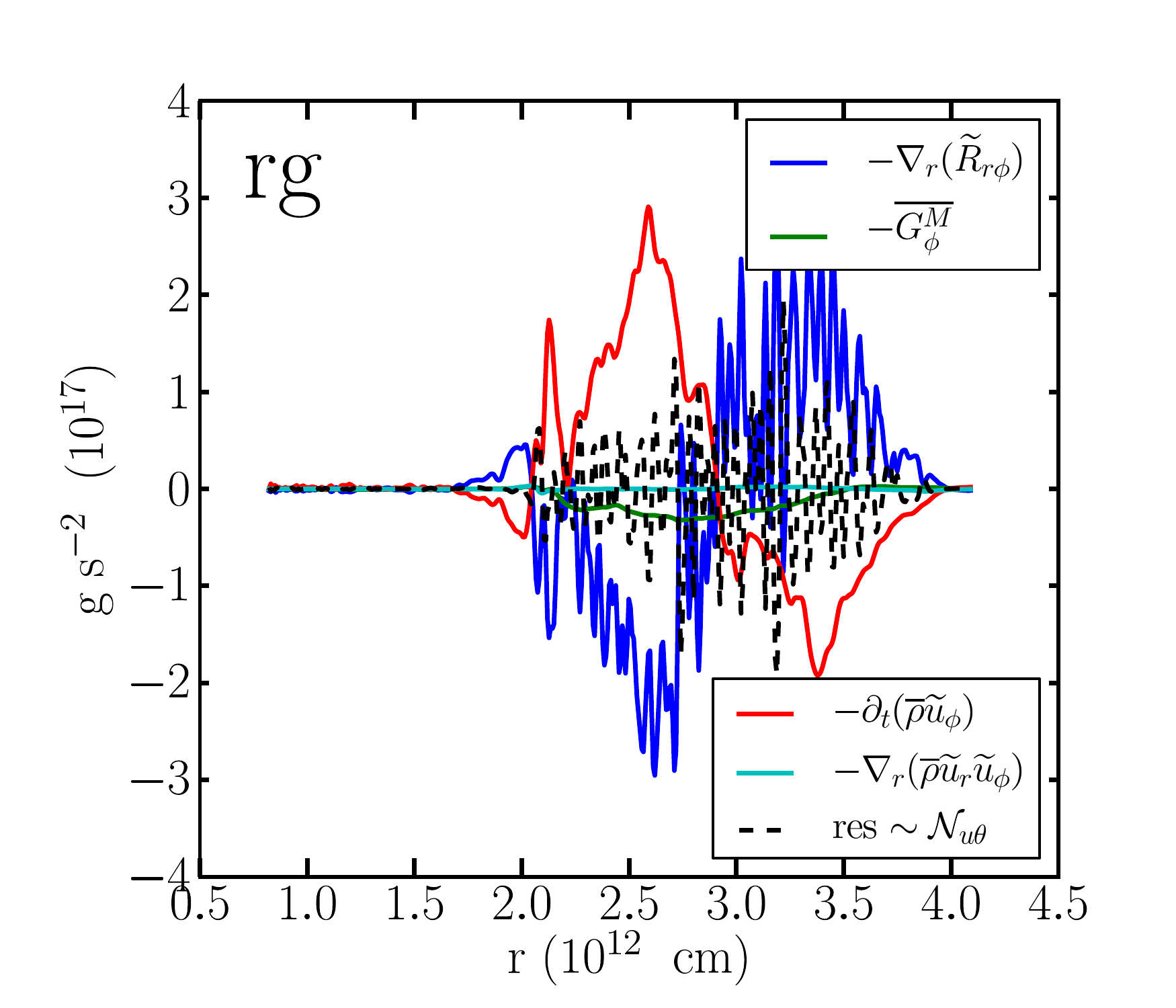}
\includegraphics[width=6.5cm]{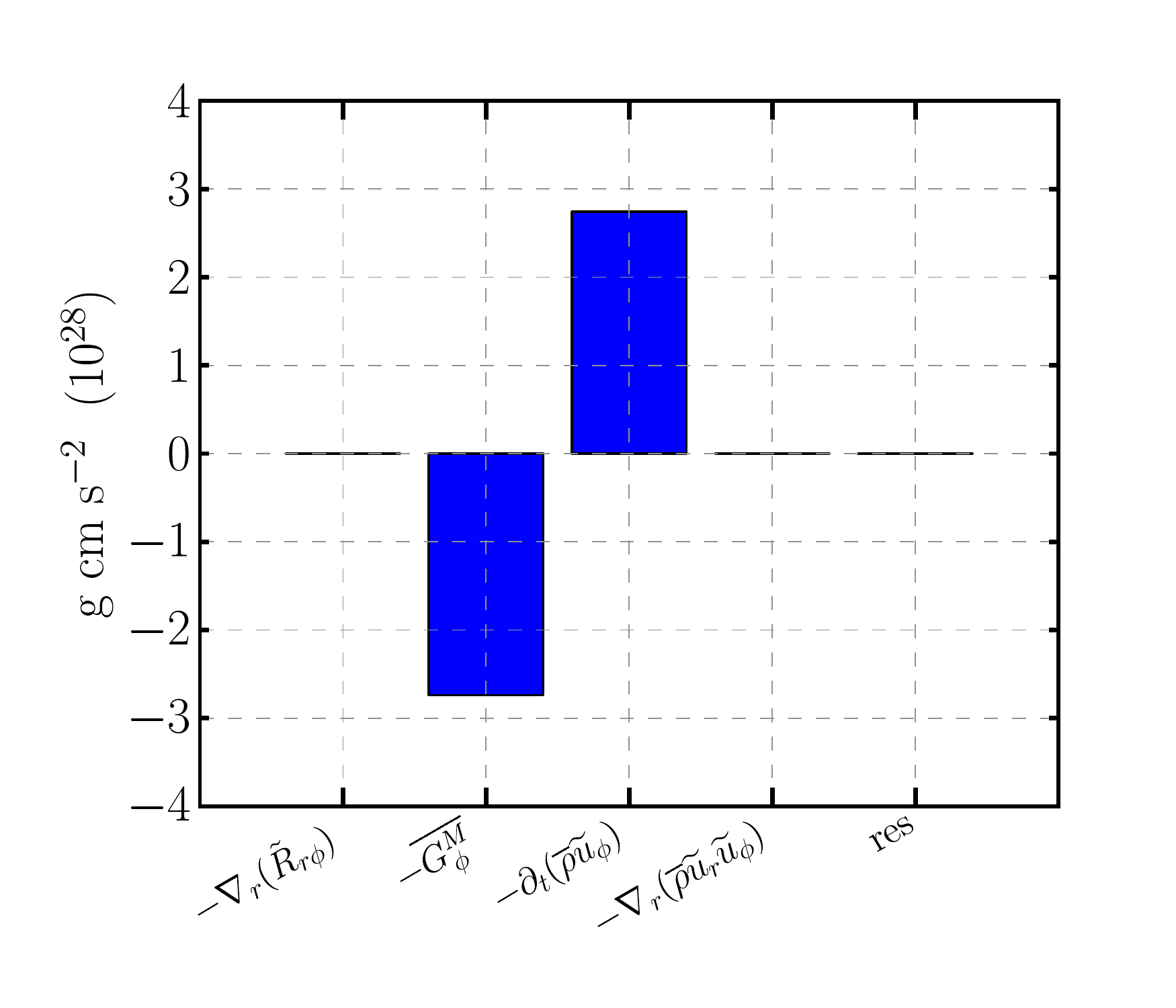}}
\caption{Mean polar omentum equation. Model {\sf ob.3D.mr} (upper panels) and model {\sf rg.3D.mr} (lower panels). \label{fig:pm-equation}}
\end{figure}

\newpage

\subsection{Mean internal energy equation}

\begin{align}
\av{\rho} \fav{D}_t \fav{\epsilon}_I = & - \nabla_r  ( f_I + f_T ) - \av{P} \ \av{d} - W_P  + {\mathcal S} + {\mathcal N_{\epsilon I}}
\end{align}

\begin{figure}[!h]
\centerline{
\includegraphics[width=6.5cm]{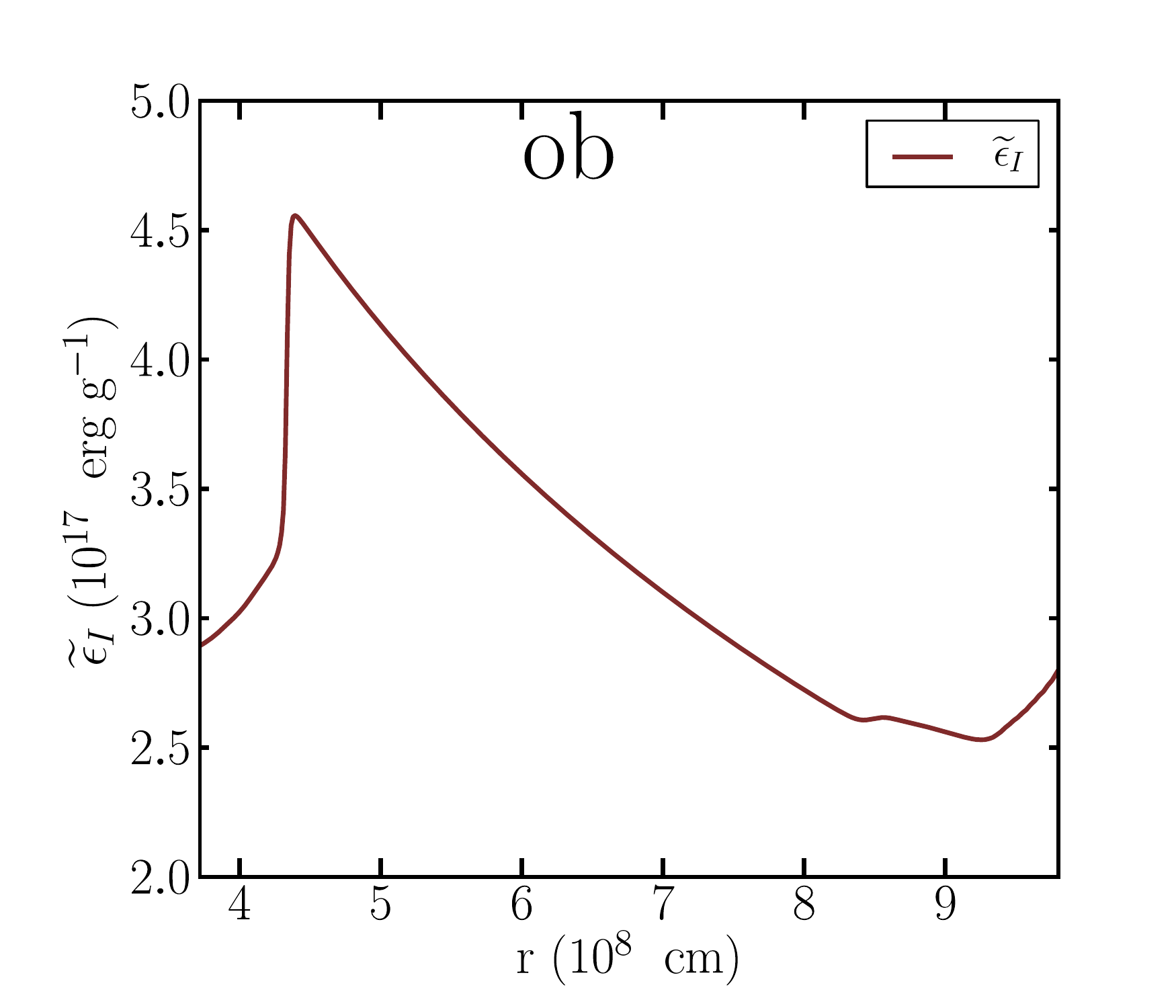}
\includegraphics[width=6.5cm]{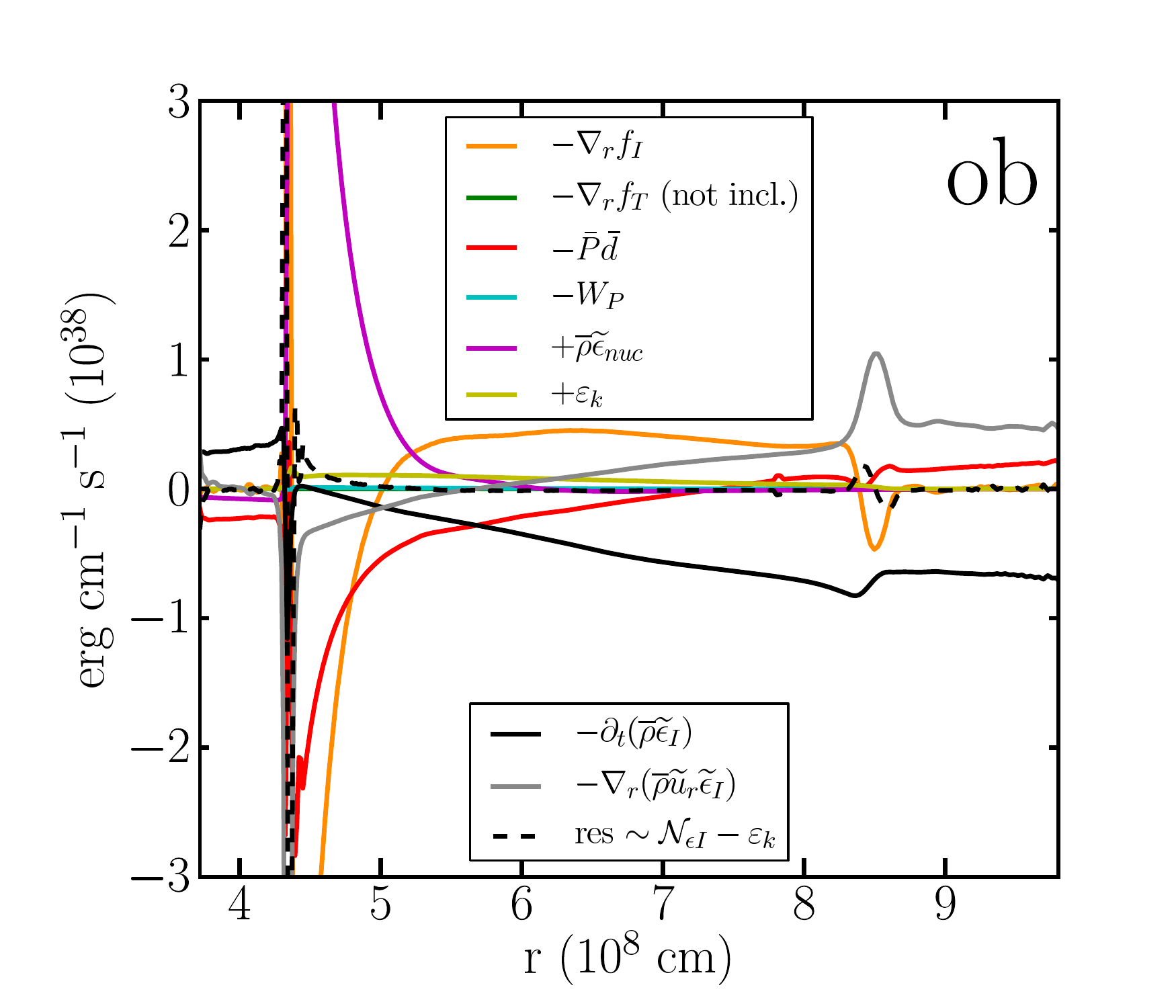}
\includegraphics[width=6.5cm]{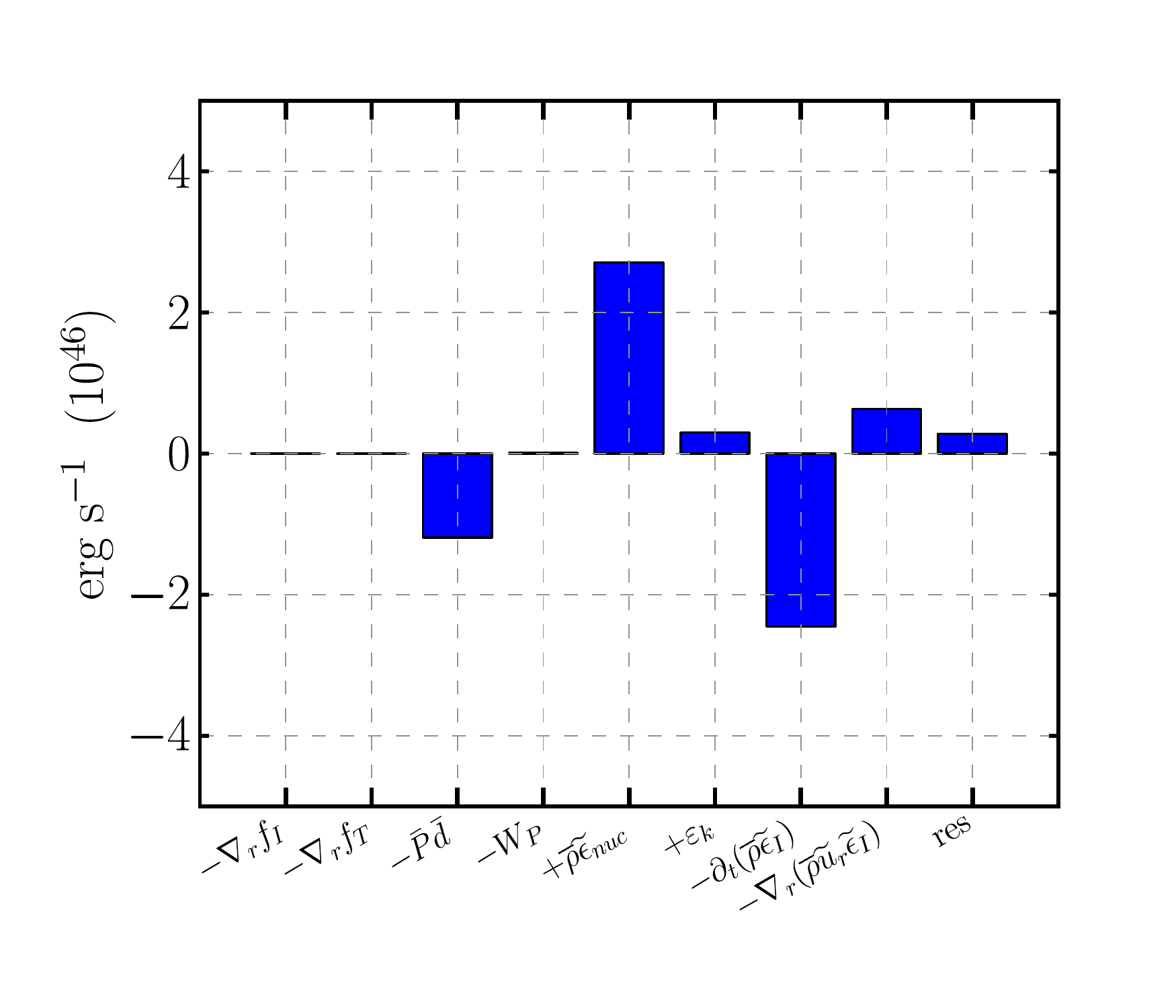}}

\centerline{
\includegraphics[width=6.5cm]{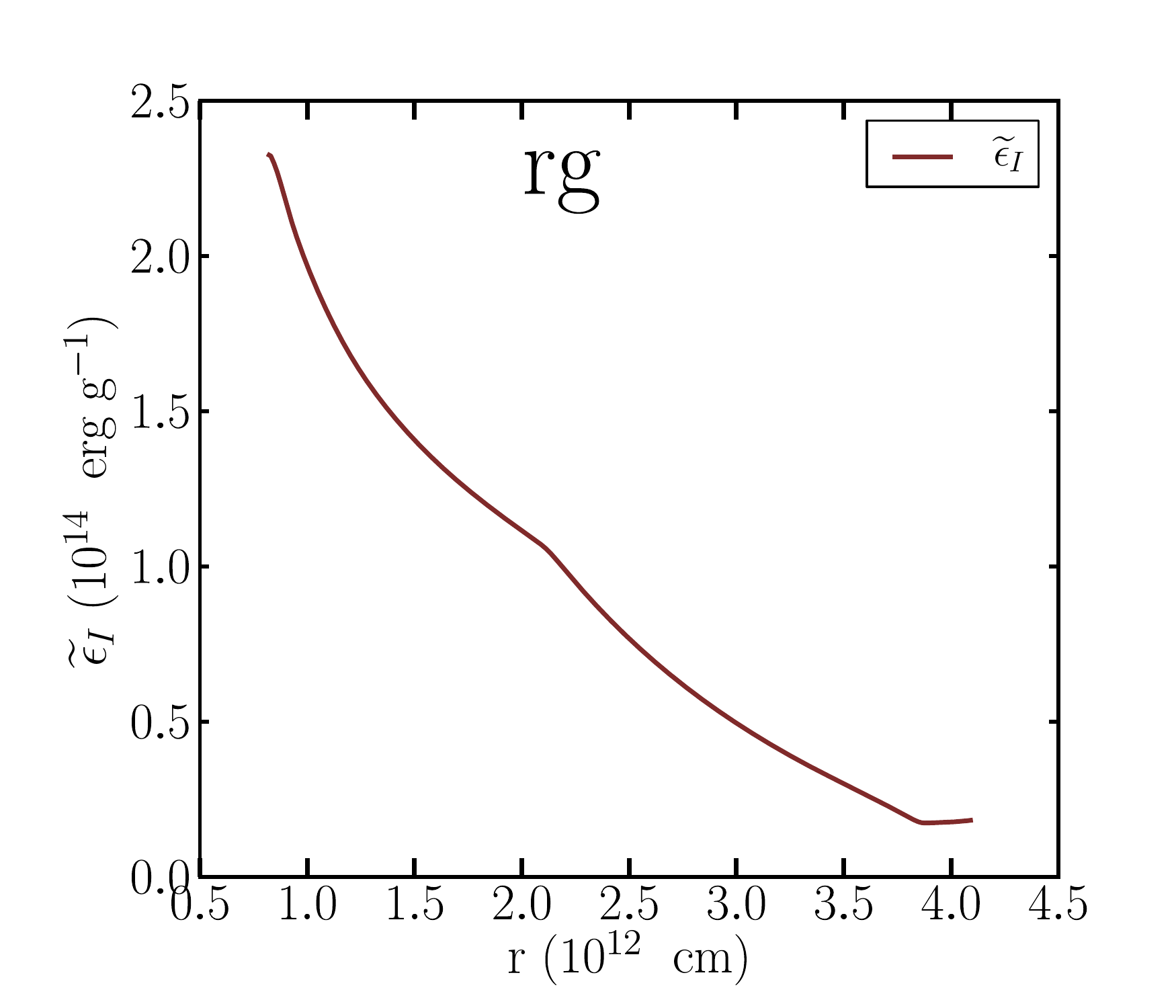}                      
\includegraphics[width=6.5cm]{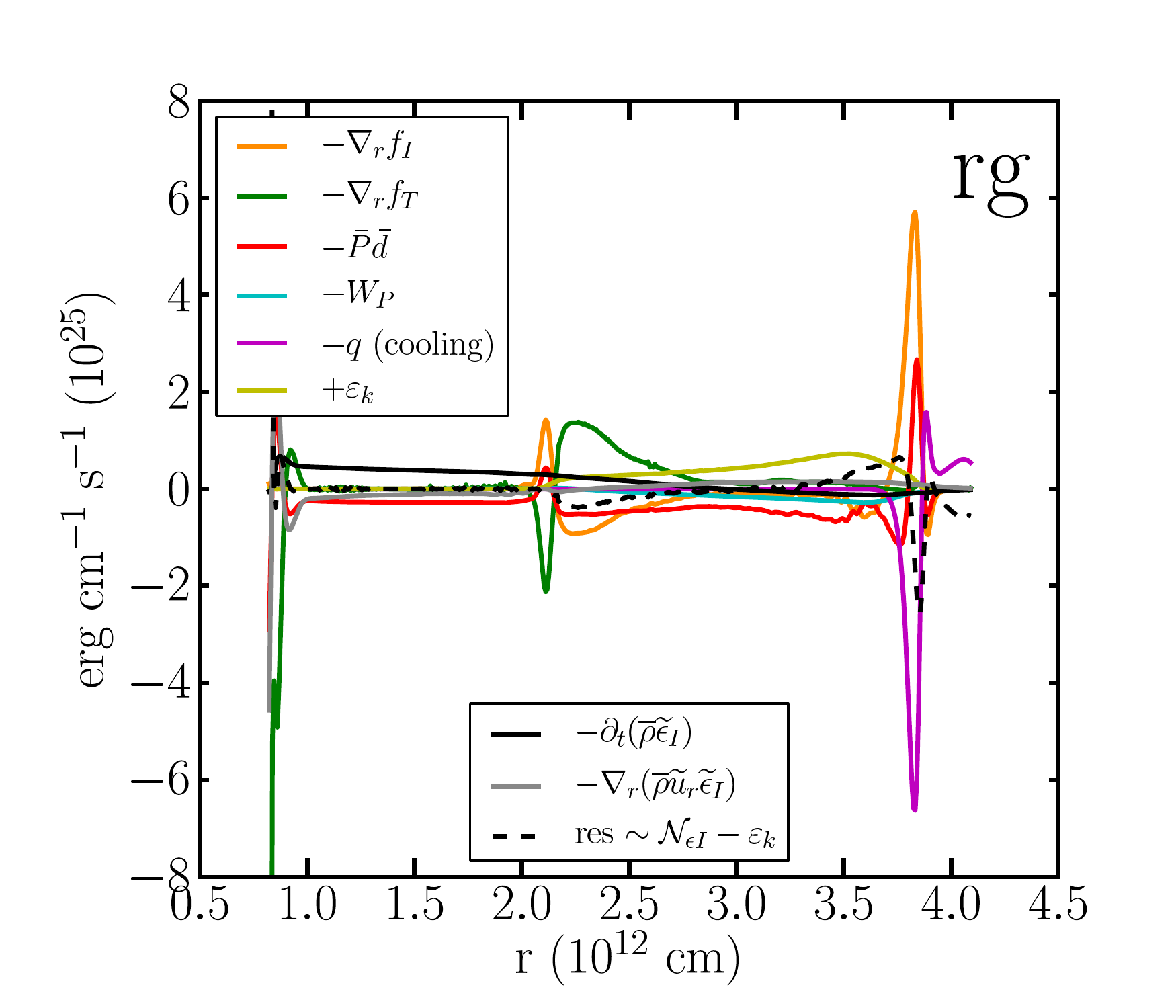}     
\includegraphics[width=6.5cm]{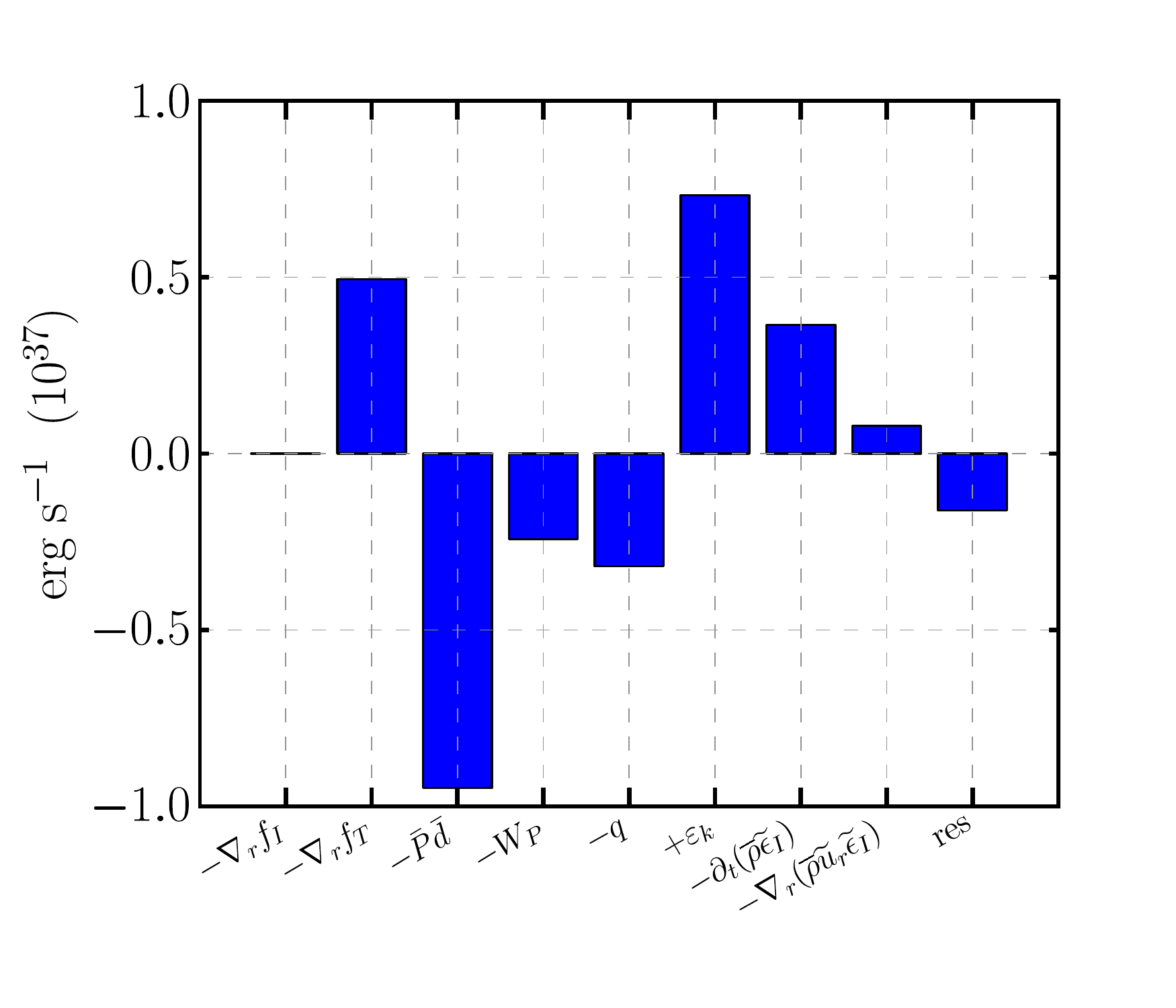}}
\caption{Mean internal energy equation. Model {\sf ob.3D.mr} (upper panels) and model {\sf rg.3D.mr} (lower panels). \label{fig:ei-equation}}
\end{figure}

\newpage

\subsection{Mean kinetic energy equation}

\begin{align}
\av{\rho} \fav{D}_t \fav{\epsilon}_k = &  -\nabla_r  ( f_k +  f_P ) - \fht{R}_{ir}\partial_r \fht{u}_i + W_b + W_P +\av{\rho}\fav{D}_t (\fav{u}_i \fav{u}_i / 2) + {\mathcal N_{\epsilon k}} \label{eq:rans_mke} 
\end{align}

\begin{figure}[!h]
\centerline{
\includegraphics[width=6.5cm]{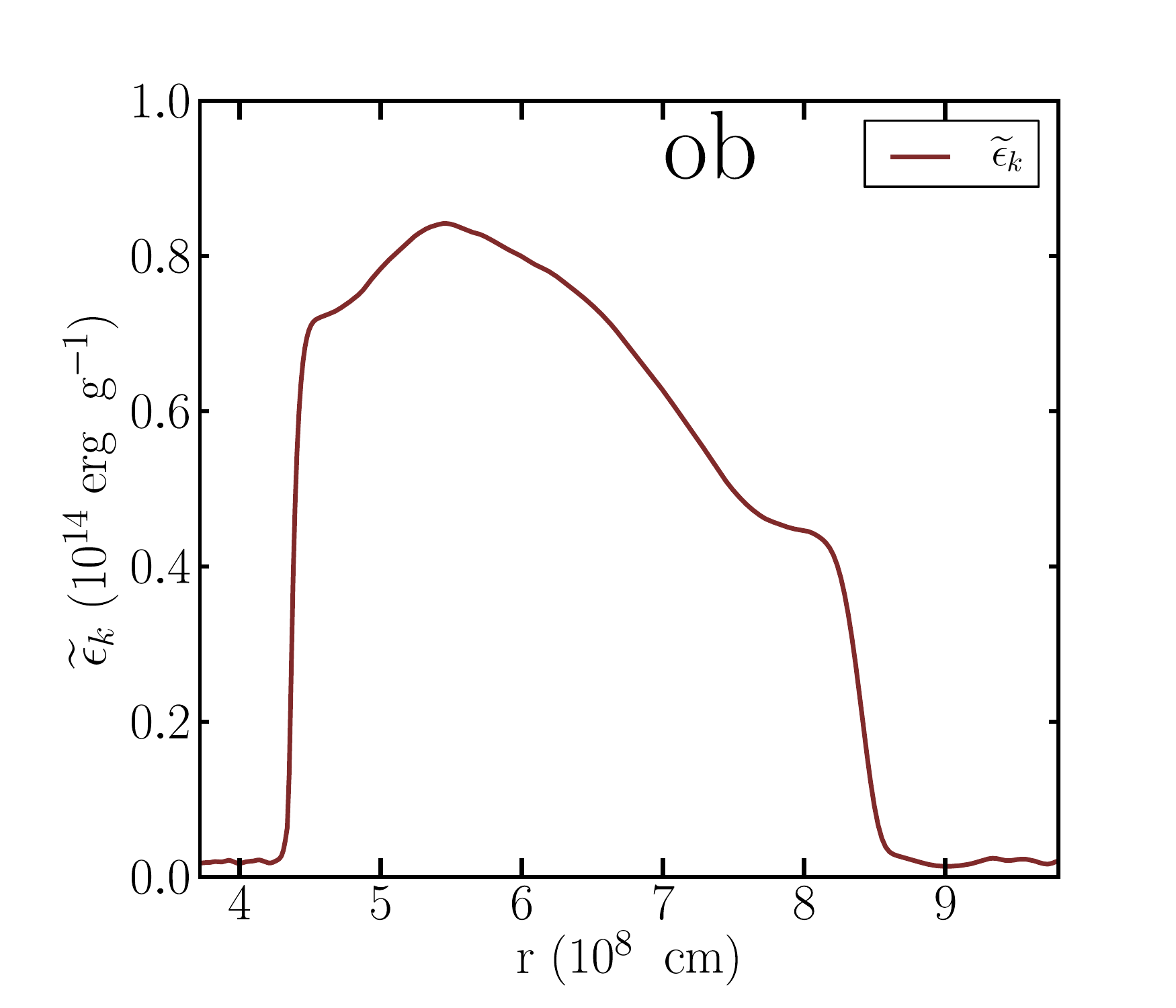}
\includegraphics[width=6.5cm]{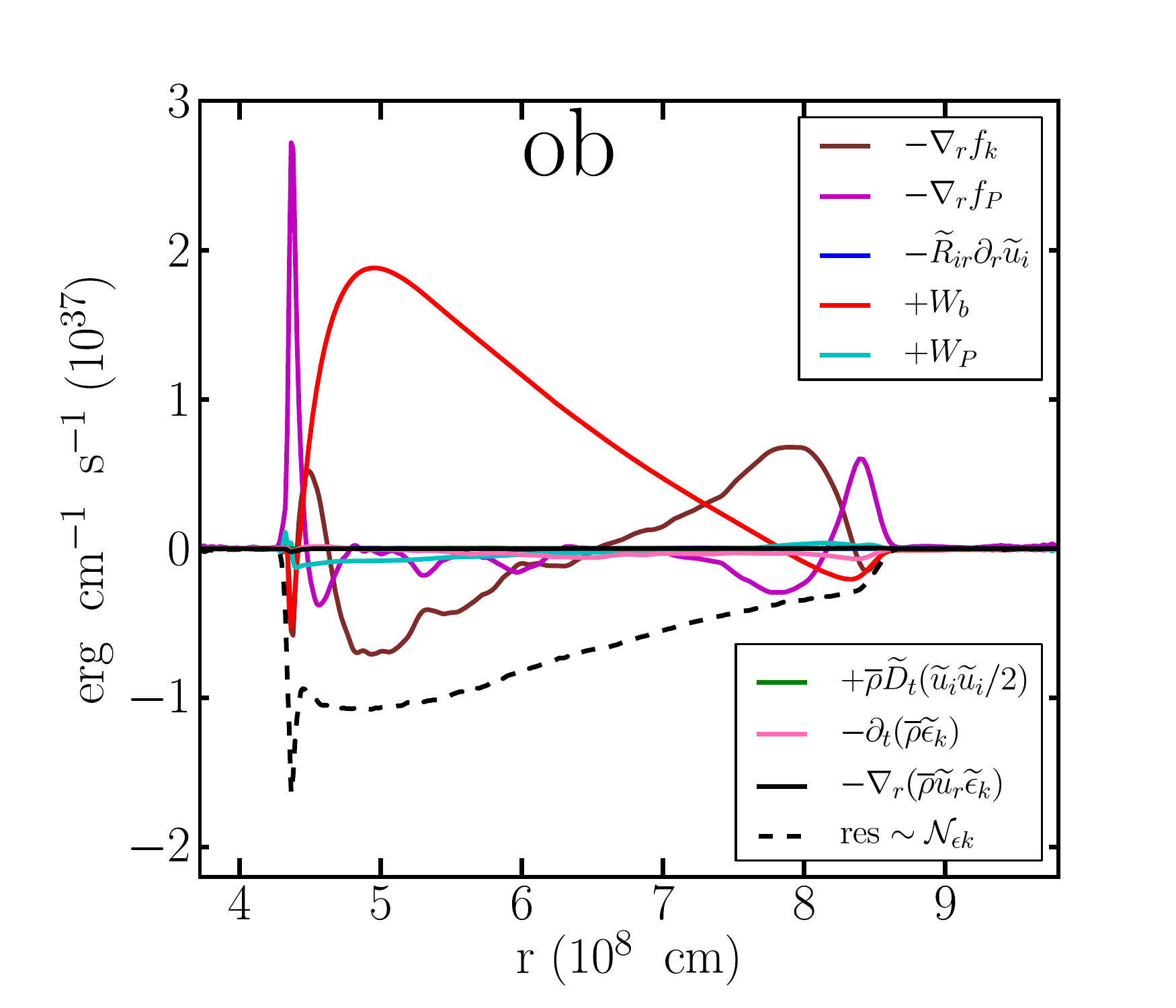}
\includegraphics[width=6.5cm]{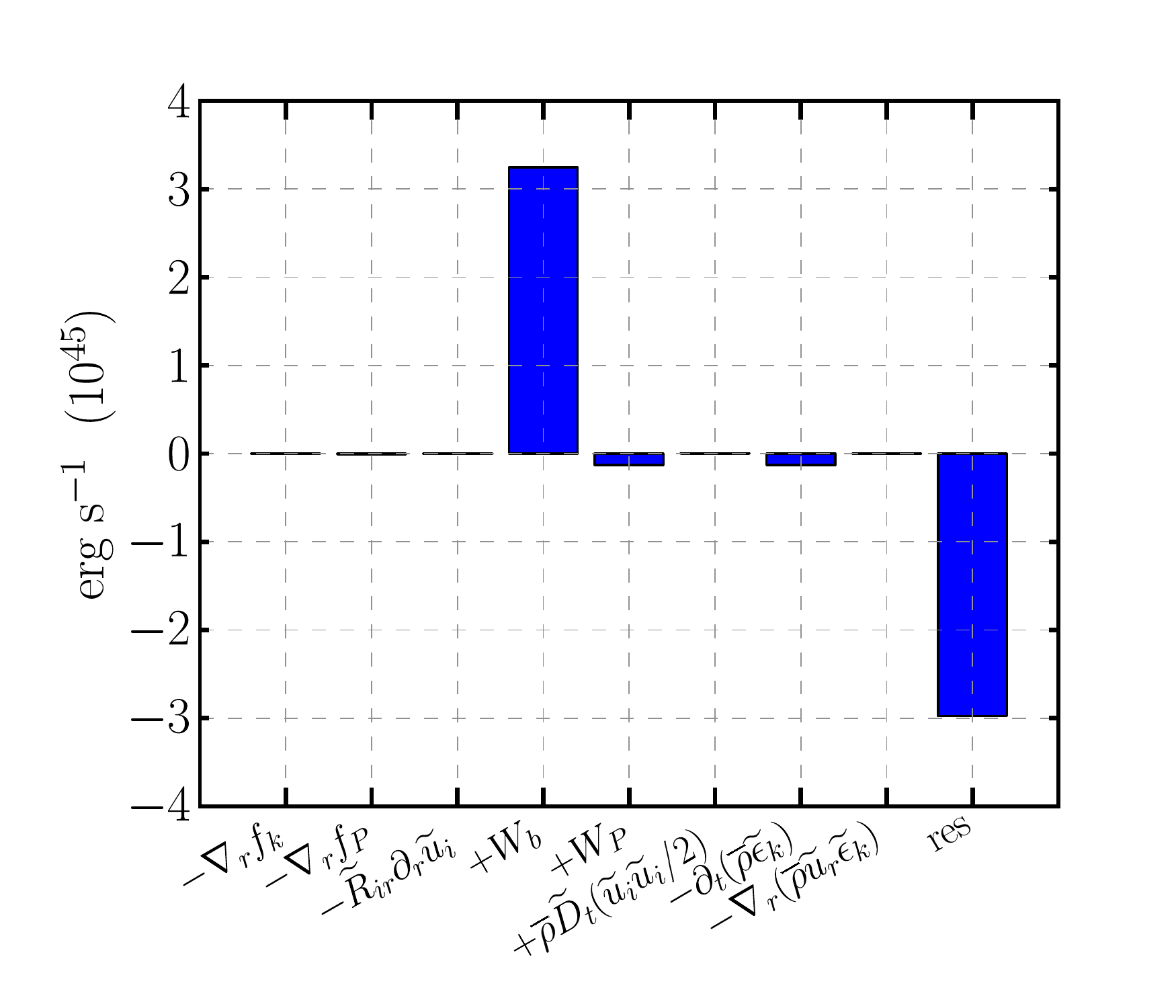}}

\centerline{
\includegraphics[width=6.5cm]{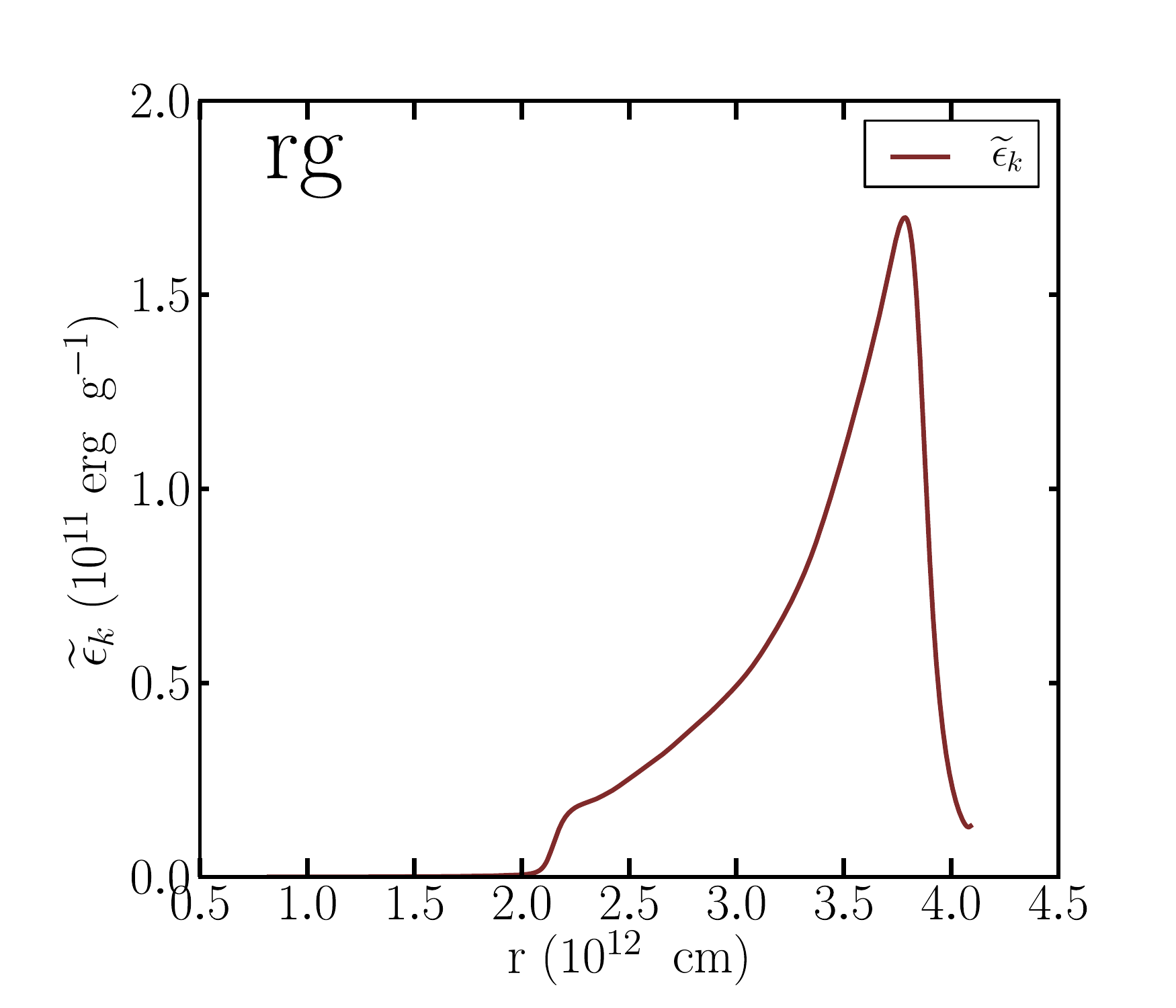}                      
\includegraphics[width=6.5cm]{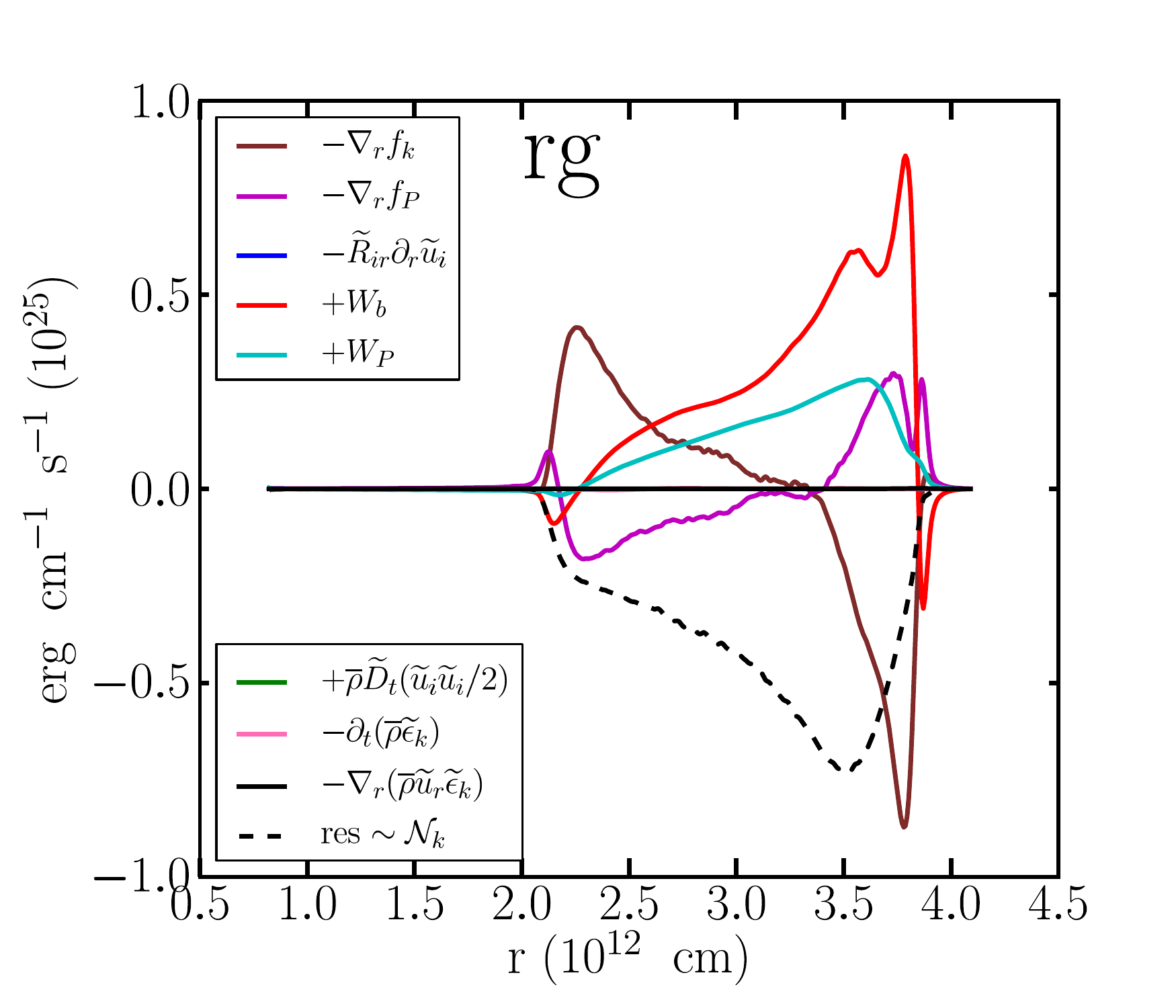}     
\includegraphics[width=6.5cm]{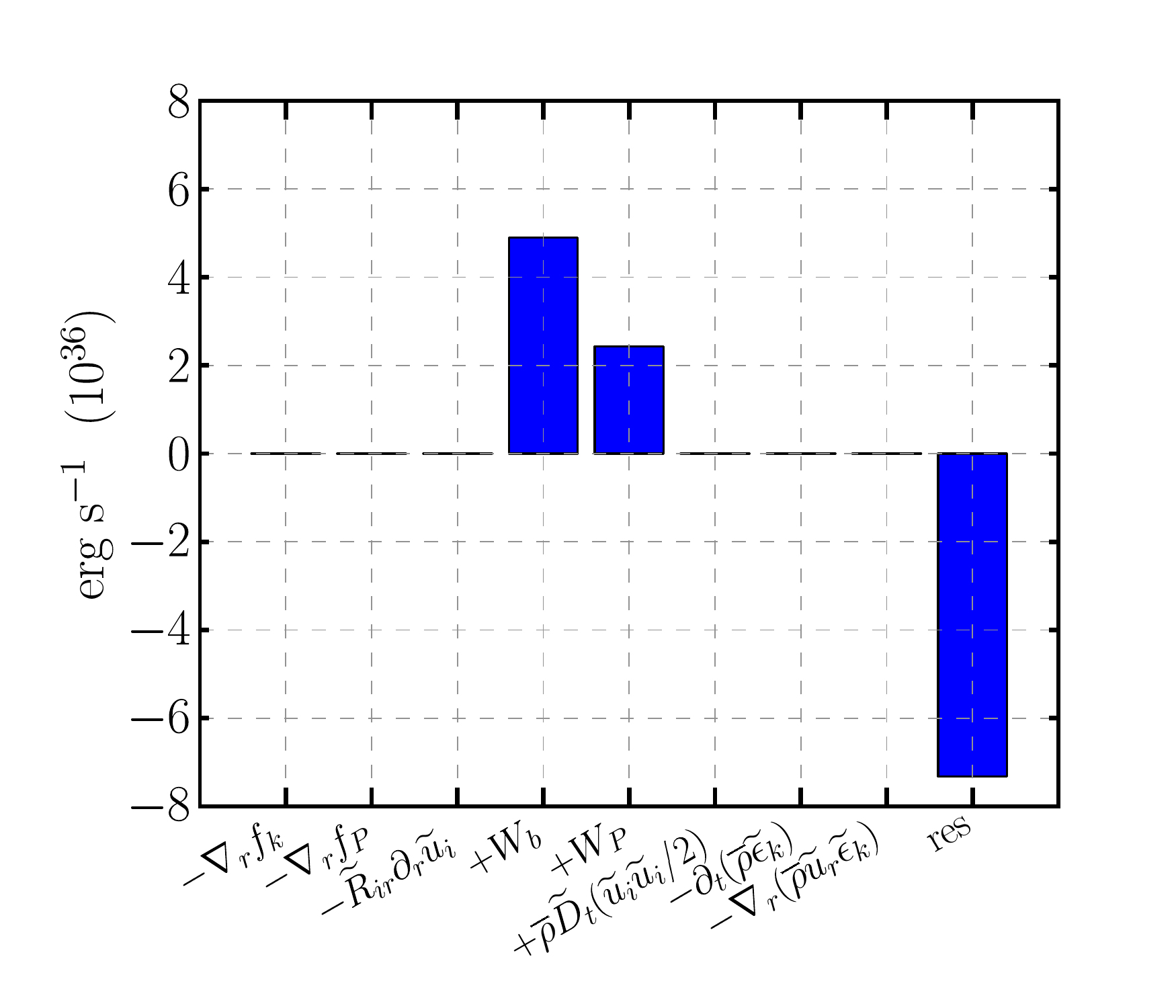}}
\caption{Mean kinetic energy equation. Model {\sf ob.3D.mr} (upper panels) and model {\sf rg.3D.mr} (lower panels). \label{fig:ek-equation}}
\end{figure}

\newpage

\subsection{Mean total energy equation}

\begin{align}
\av{\rho} \fav{D}_t \fav{\epsilon}_t = &  - \nabla_r ( f_I + f_T + f_k + f_P ) - \fht{R}_{ir}\partial_r \fht{u}_i - \av{P} \ \av{d} + W_b + {\mathcal S} + \av{\rho}\fav{D}_t (\fav{u}_i \fav{u}_i / 2) + {\mathcal N_{\epsilon t}} \label{eq:rans_etot}
\end{align}

\begin{figure}[!h]
\centerline{
\includegraphics[width=6.5cm]{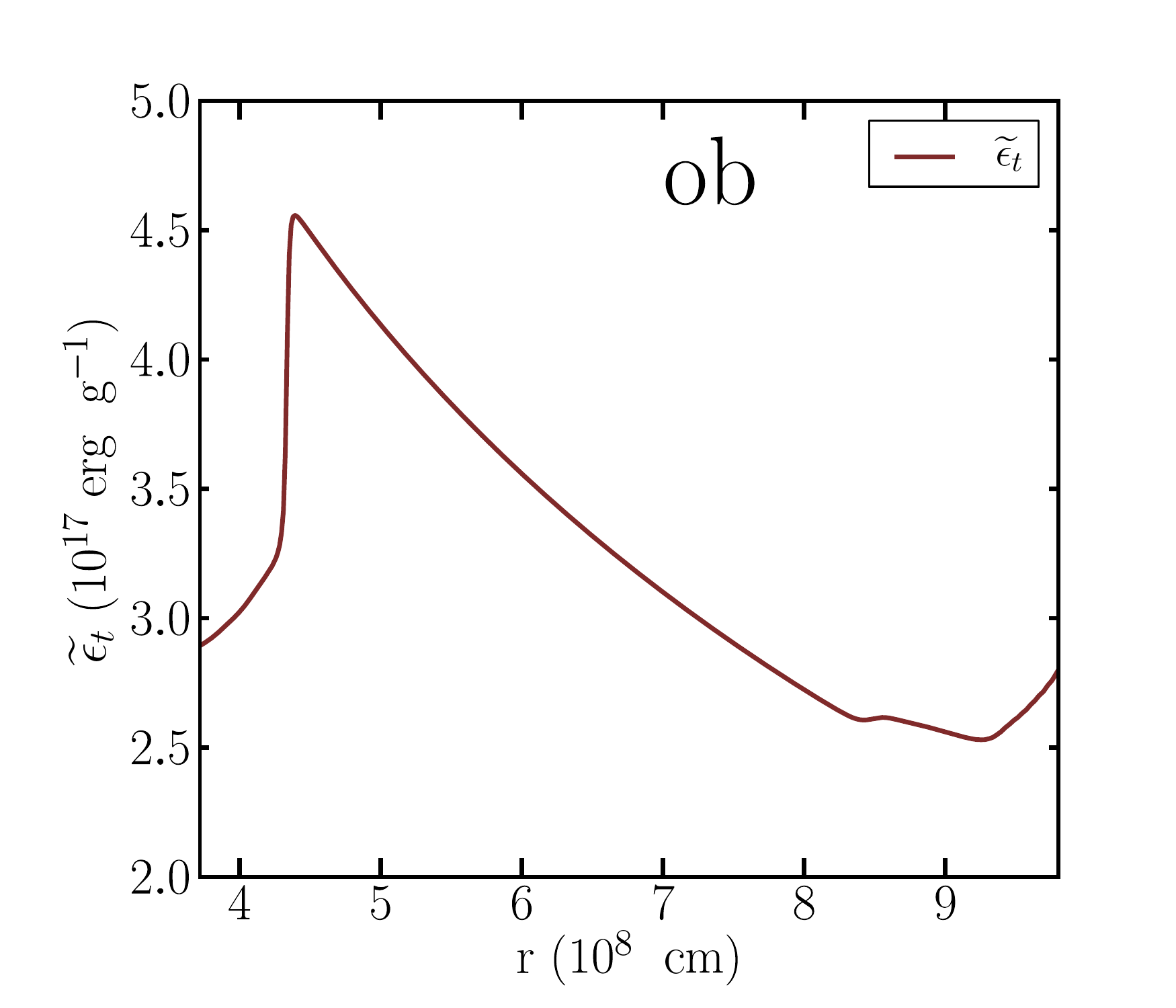}
\includegraphics[width=6.5cm]{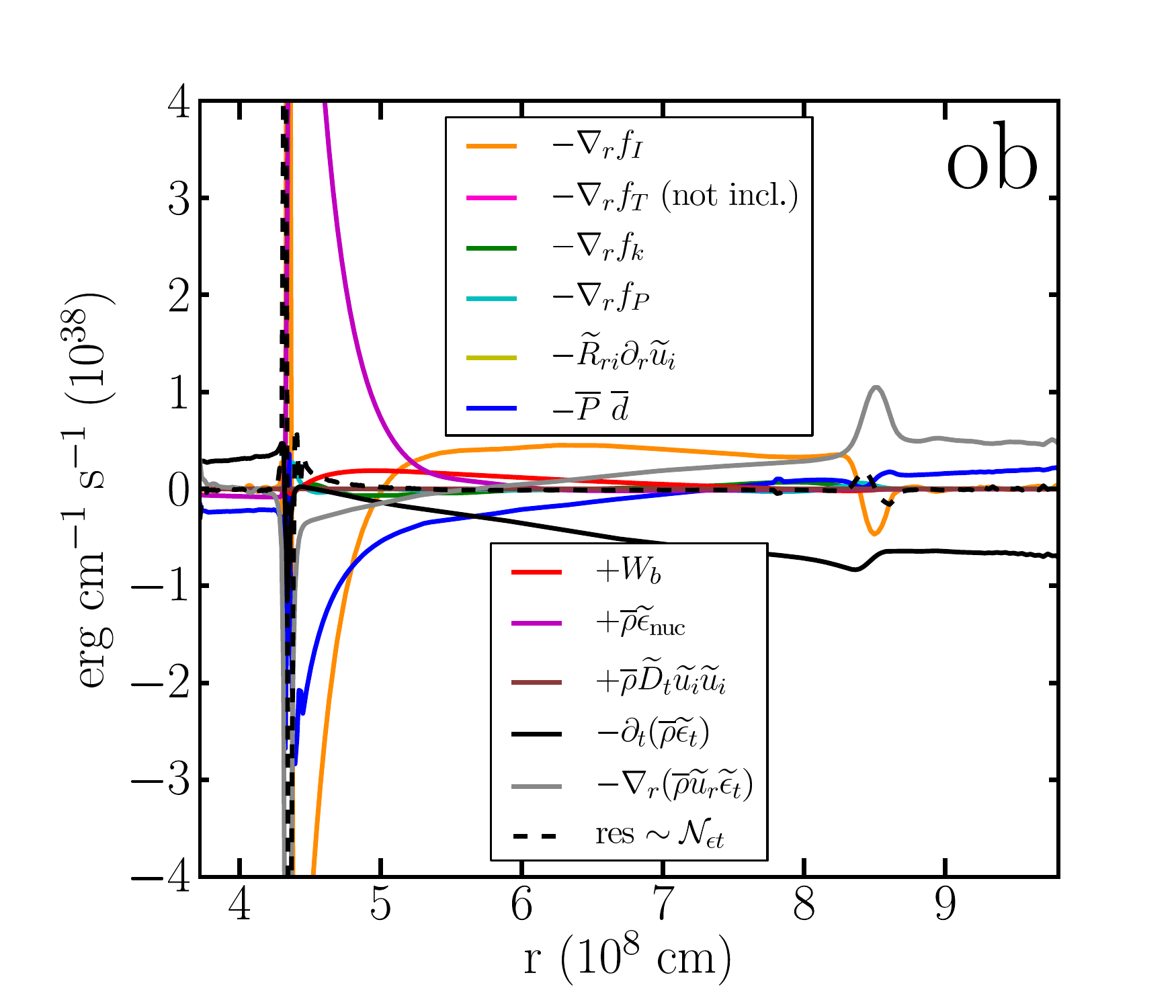}
\includegraphics[width=6.5cm]{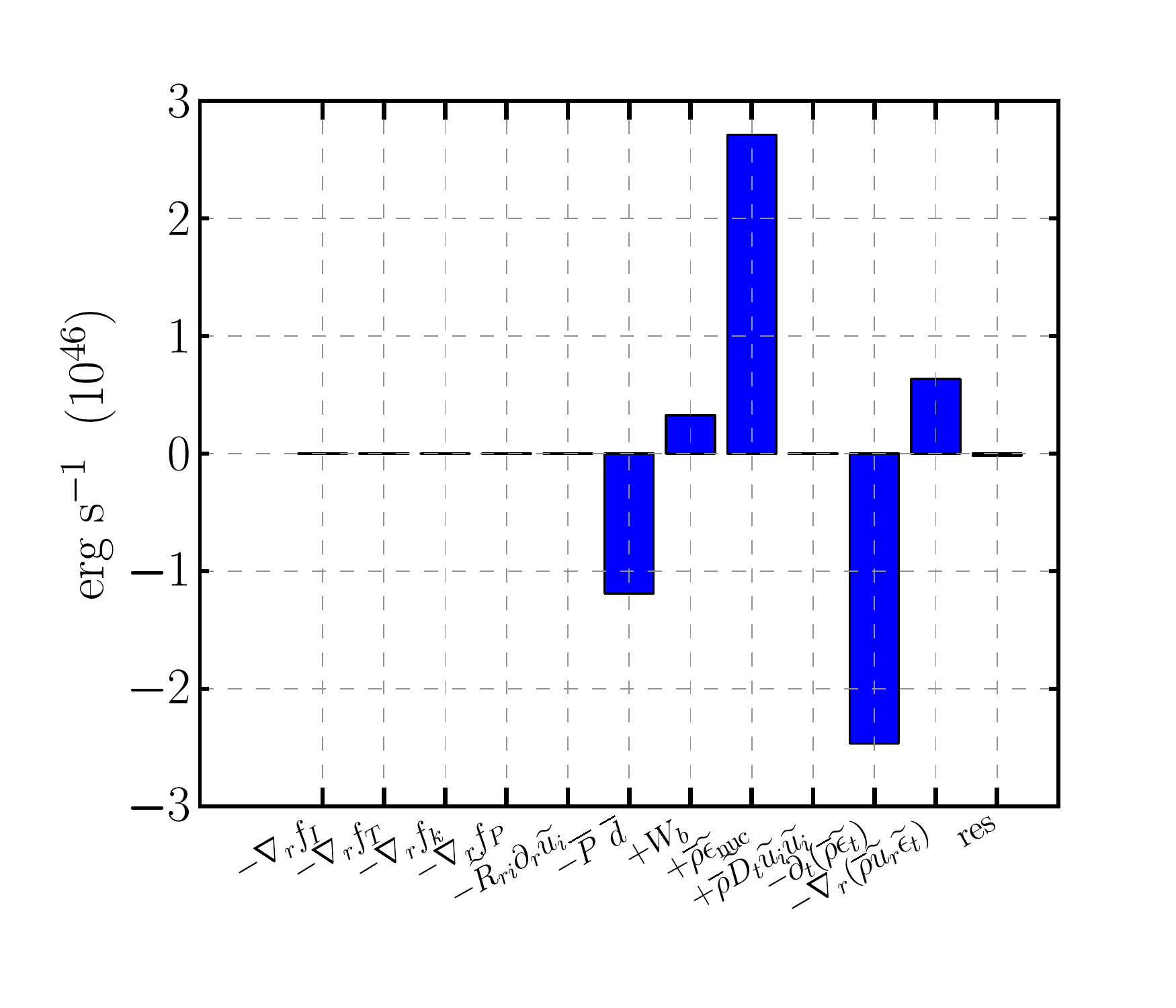}}

\centerline{
\includegraphics[width=6.5cm]{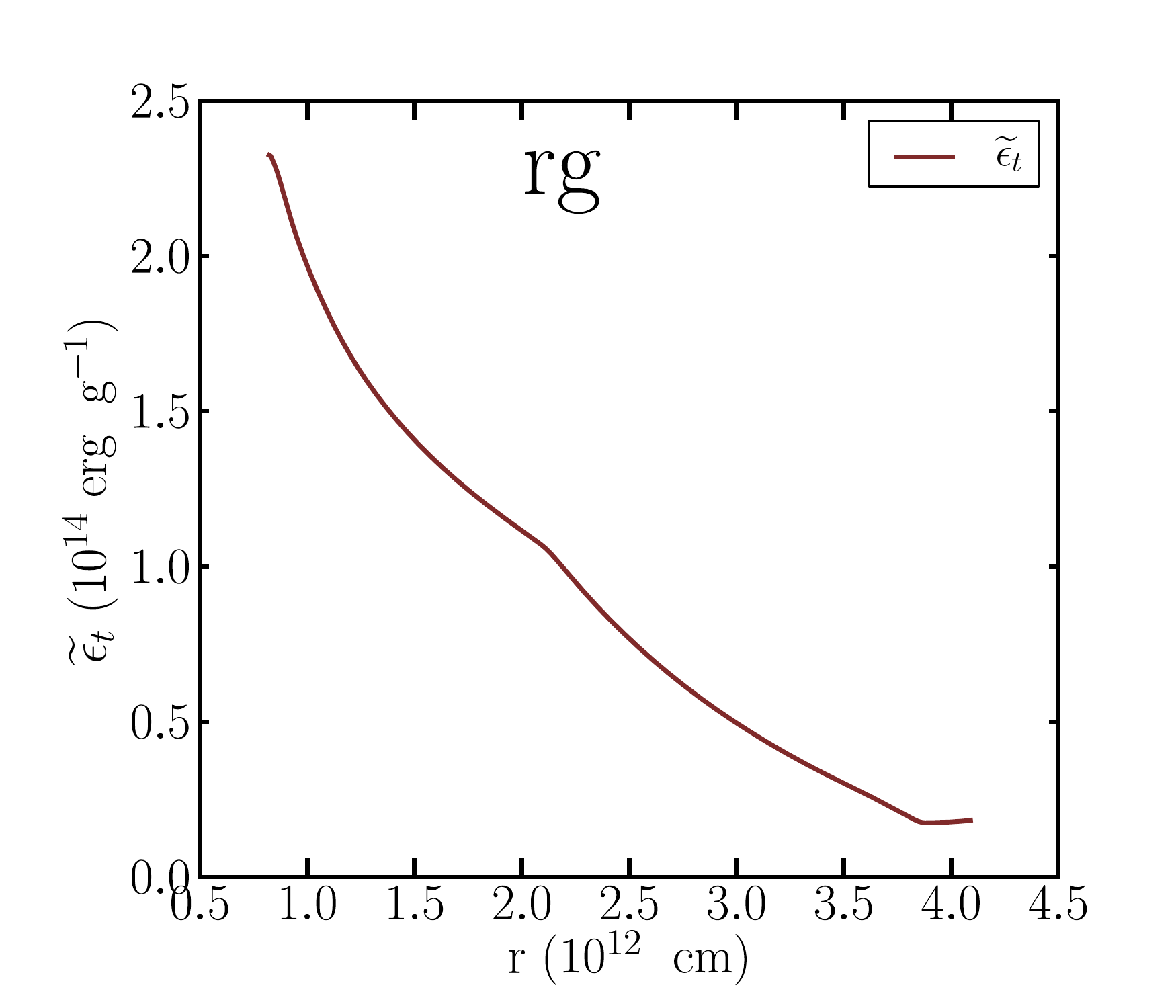}                      
\includegraphics[width=6.5cm]{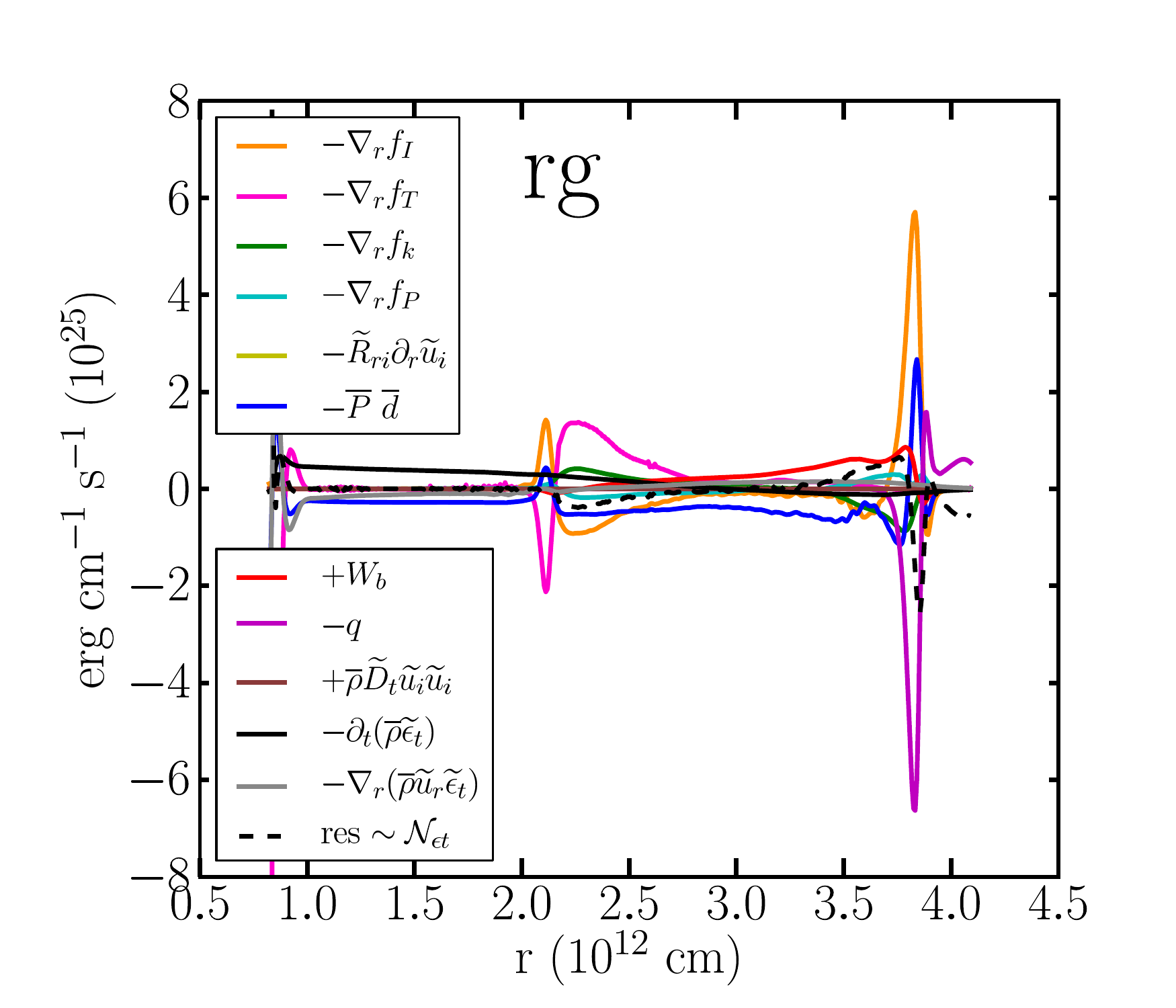}     
\includegraphics[width=6.5cm]{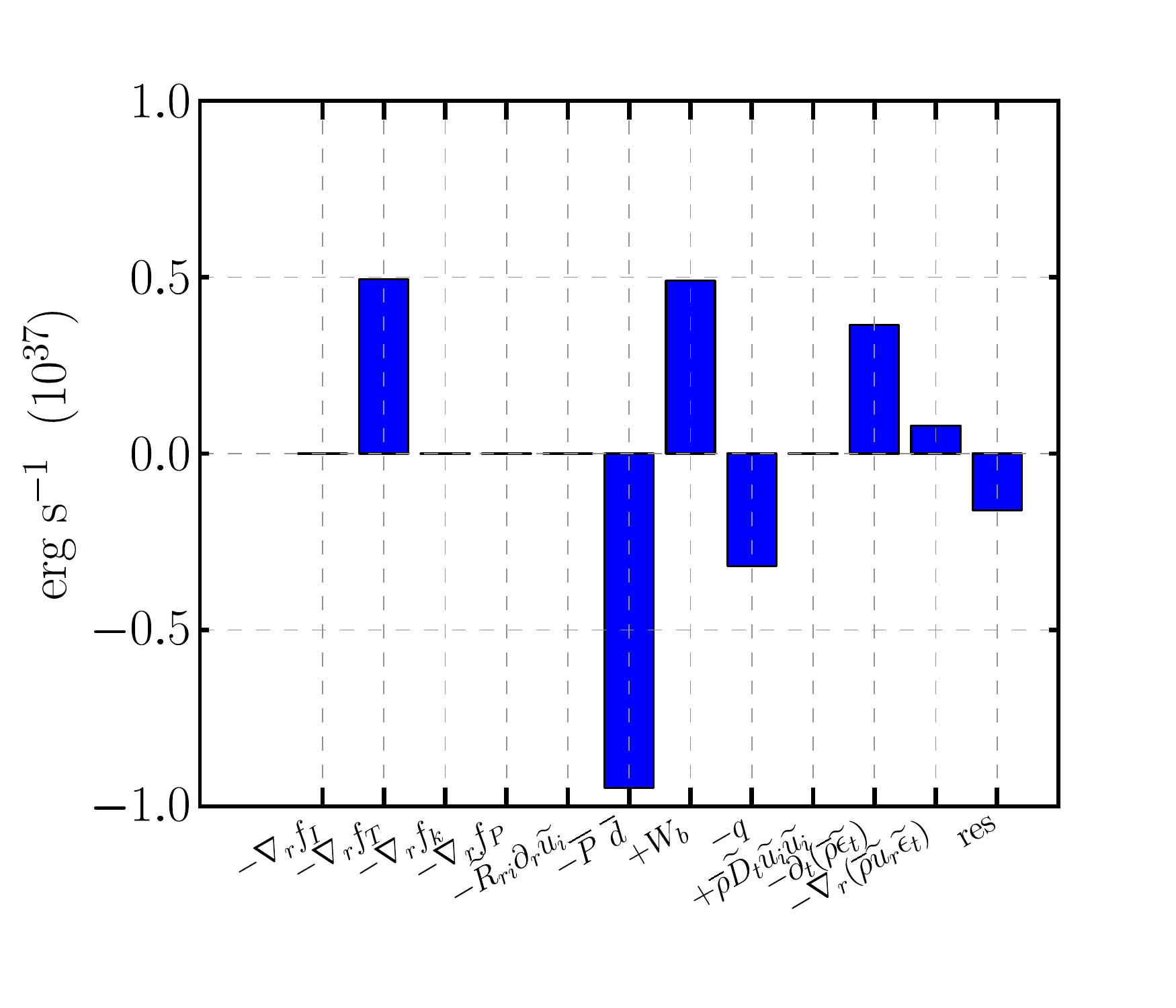}}
\caption{Mean total energy equation. Model {\sf ob.3D.mr} (upper panels) and model {\sf rg.3D.mr} (lower panels). \label{fig:et-equation}}
\end{figure}

\newpage

\subsection{Mean entropy equation}

\begin{align}
\av{\rho} \fav{D}_t \fav{s} = & - \nabla_r  f_s    - \av{(\nabla \cdot F_T)/T}+ \av{{\mathcal S}/T} + {\mathcal N_s}  \label{eq:rans_entropy} 
\end{align}

\begin{figure}[!h]
\centerline{
\includegraphics[width=6.5cm]{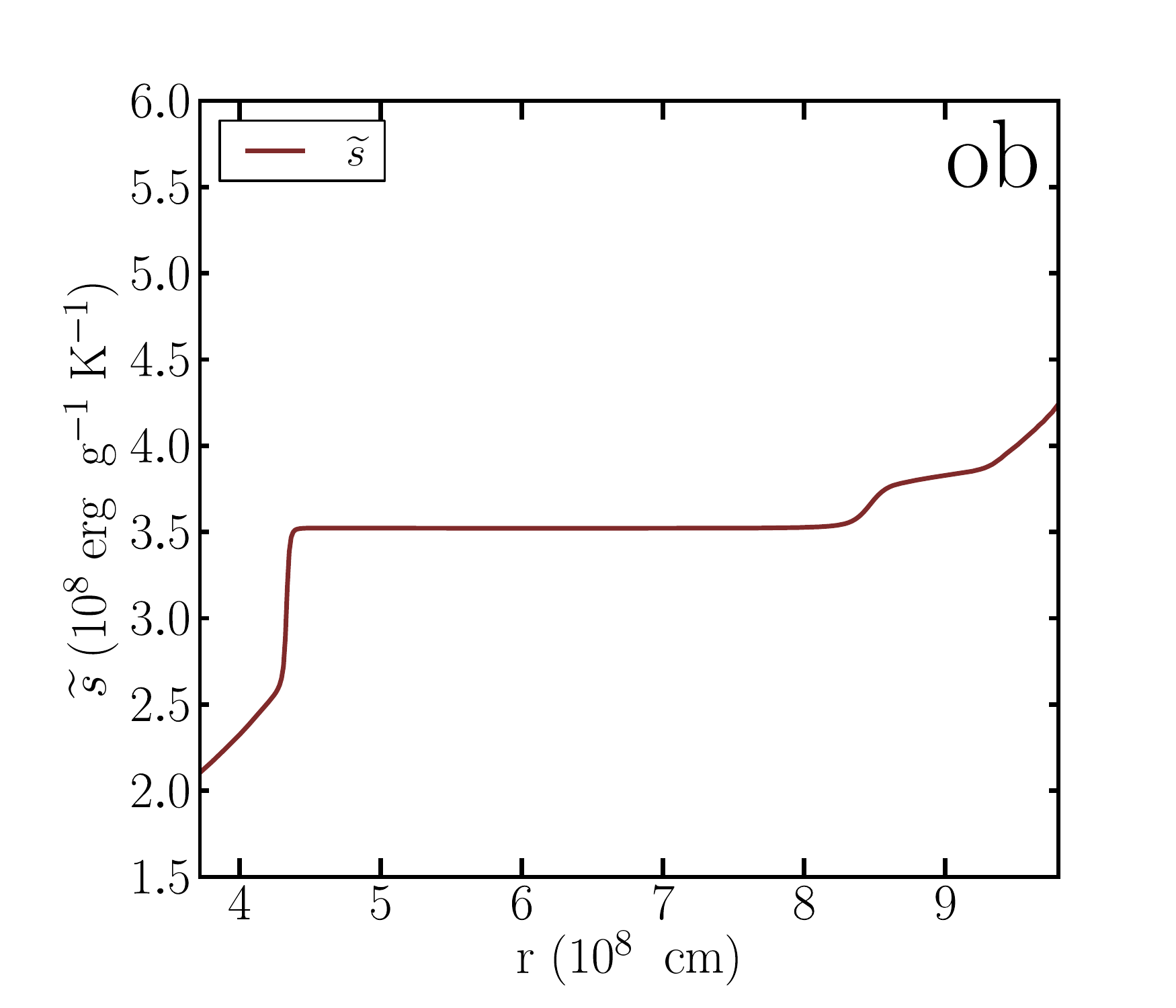}
\includegraphics[width=6.5cm]{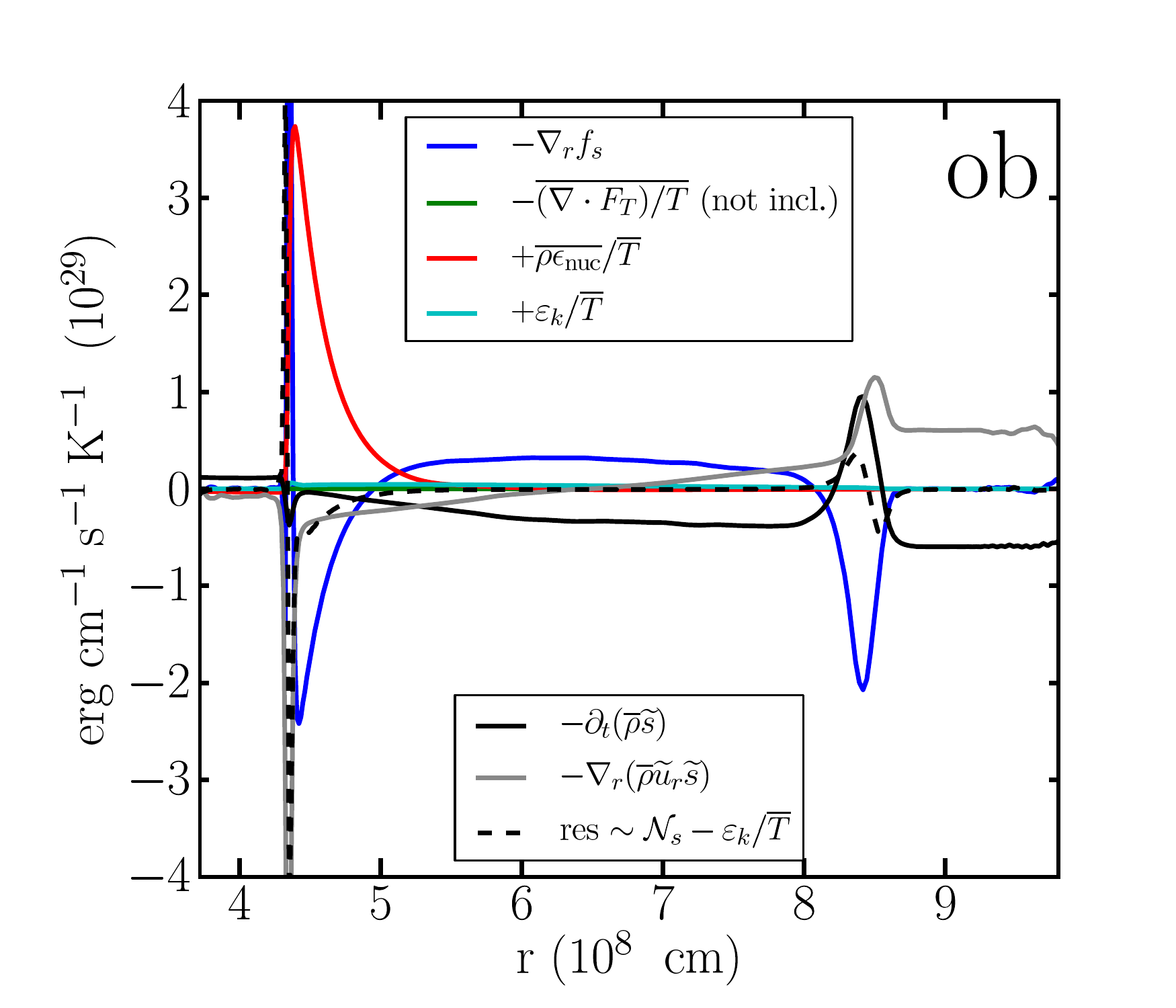}
\includegraphics[width=6.5cm]{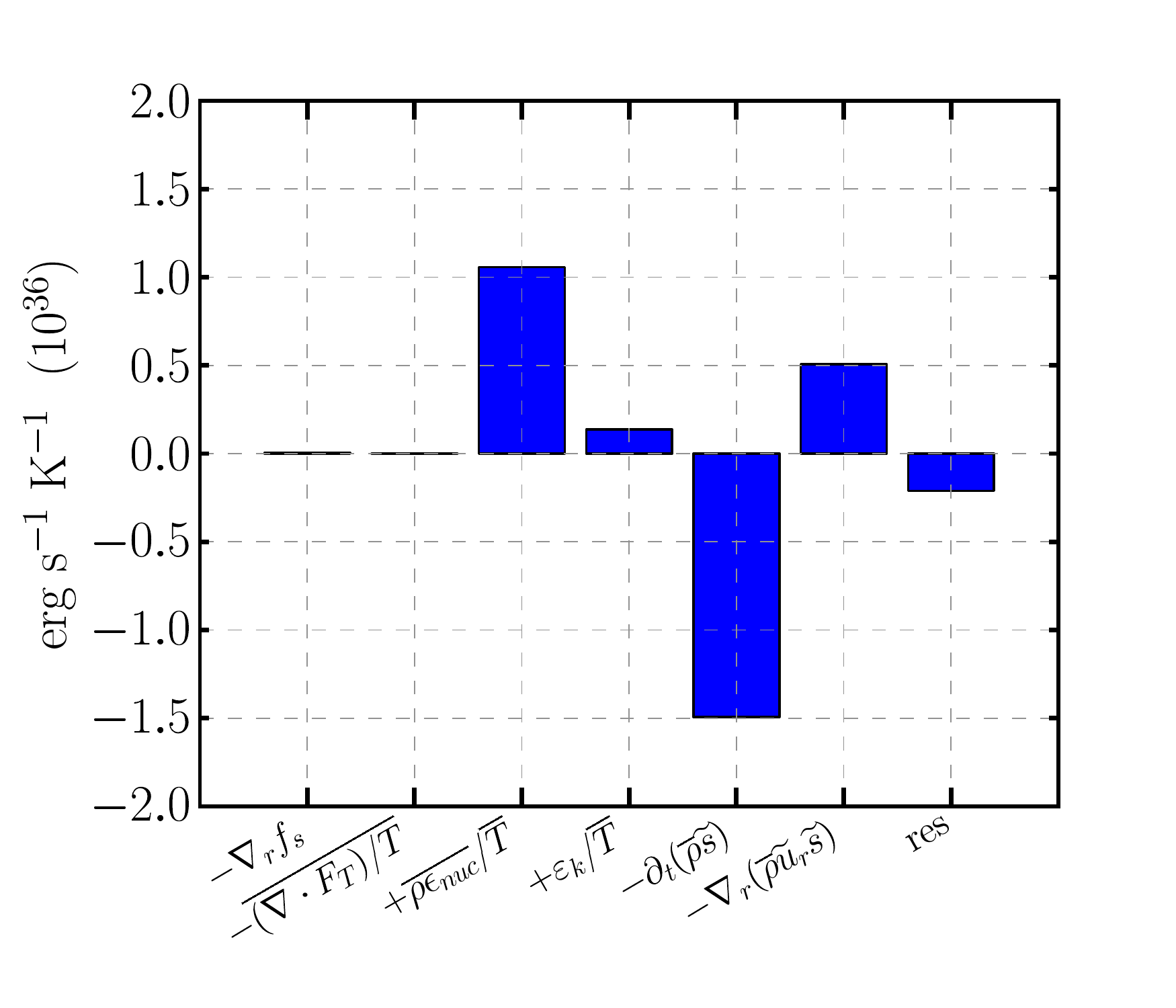}}

\centerline{
\includegraphics[width=6.5cm]{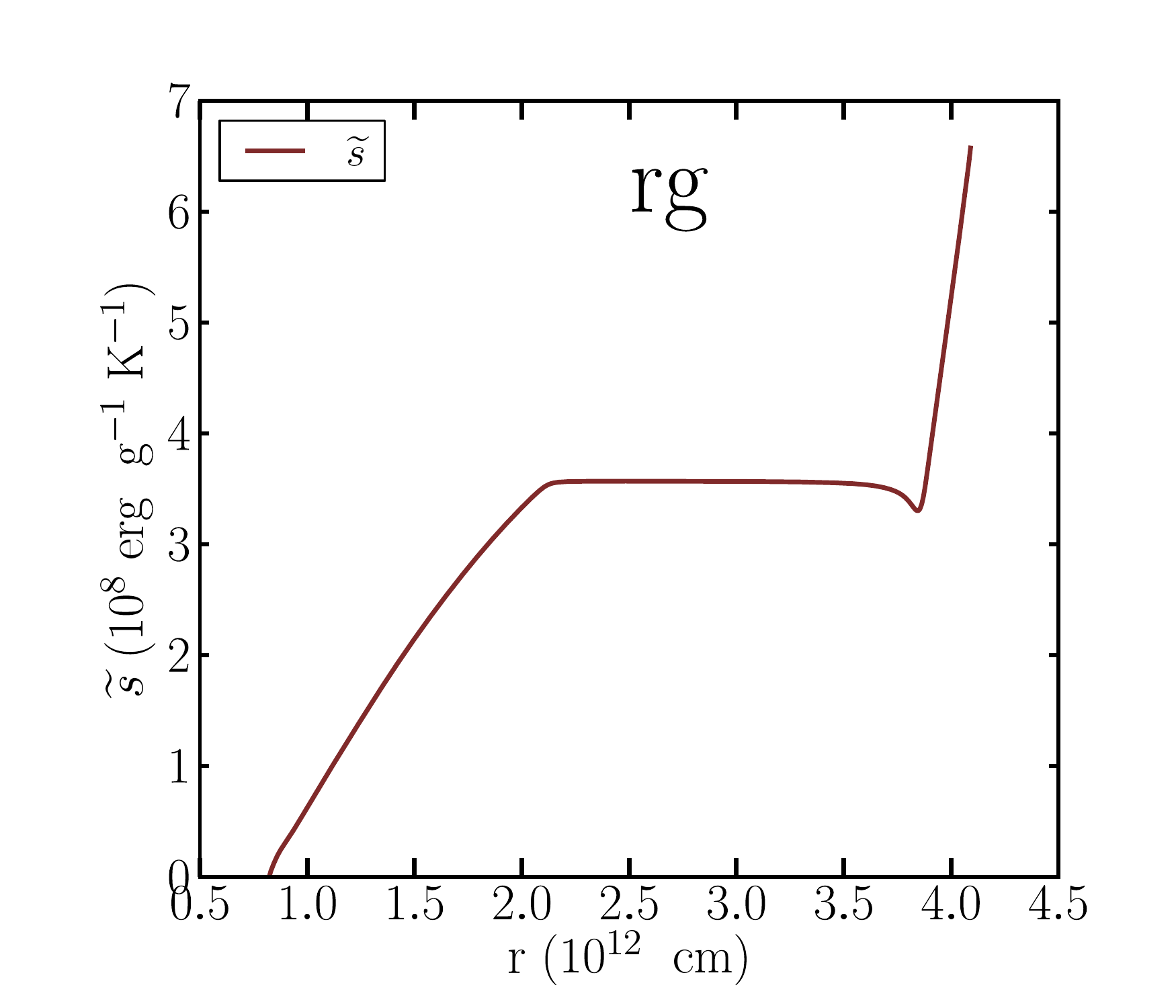}                      
\includegraphics[width=6.5cm]{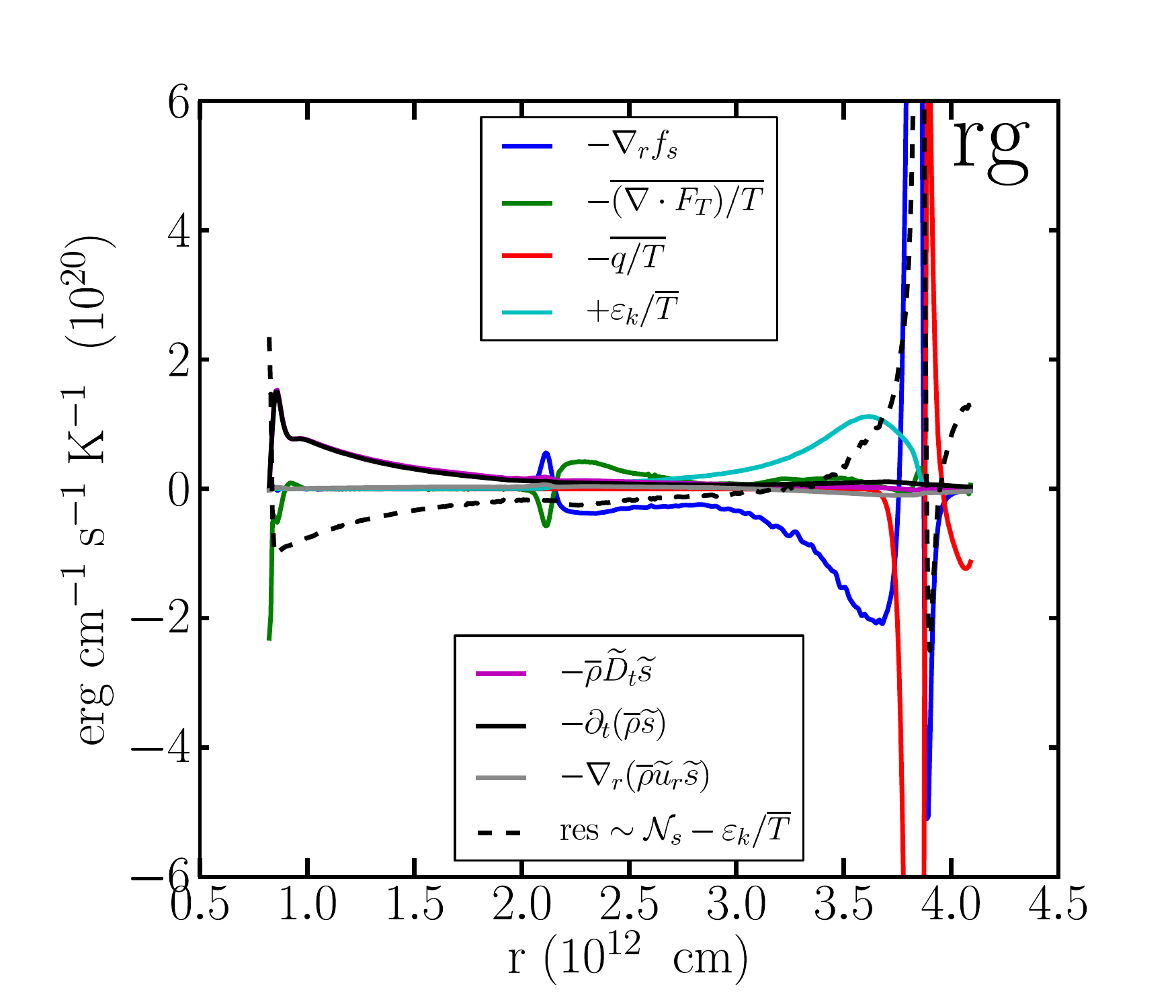}     
\includegraphics[width=6.5cm]{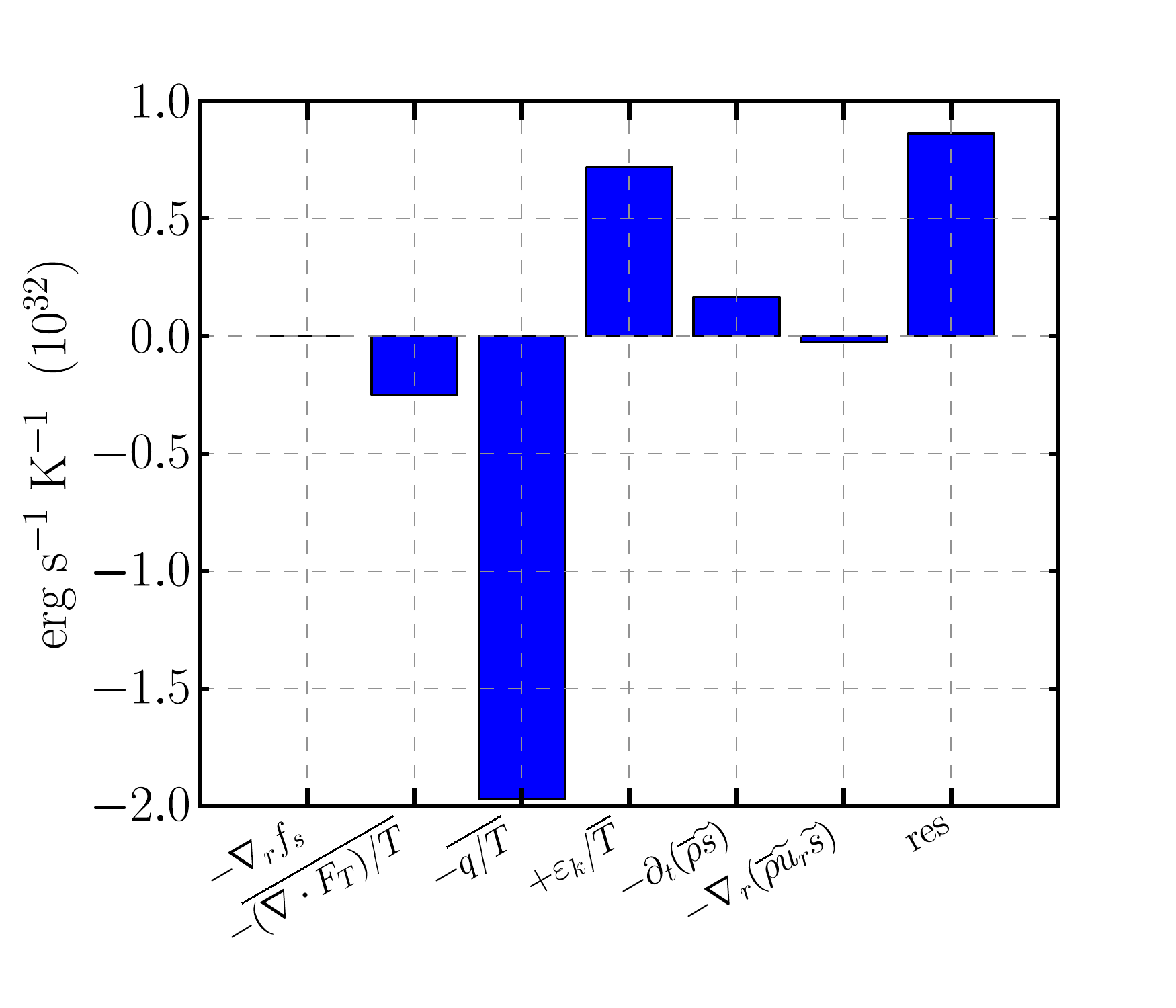}}
\caption{Mean entropy equation. Model {\sf ob.3D.2hp} (upper panels) and model {\sf rg.3D.mr} (lower panels). \label{fig:ss-equation}}
\end{figure}

\newpage

\subsection{Mean pressure equation}

\begin{align}
\av{D}_t \av{P} = & -\nabla_r f_P - \Gamma_1 \eht{P} \ \eht{d} + (1 -\Gamma_1) W_P + (\Gamma_3 -1){\mathcal S} + (\Gamma_3 - 1)\nabla_r f_T + {\mathcal N_P}
\end{align}

\begin{figure}[!h]
\centerline{
\includegraphics[width=6.5cm]{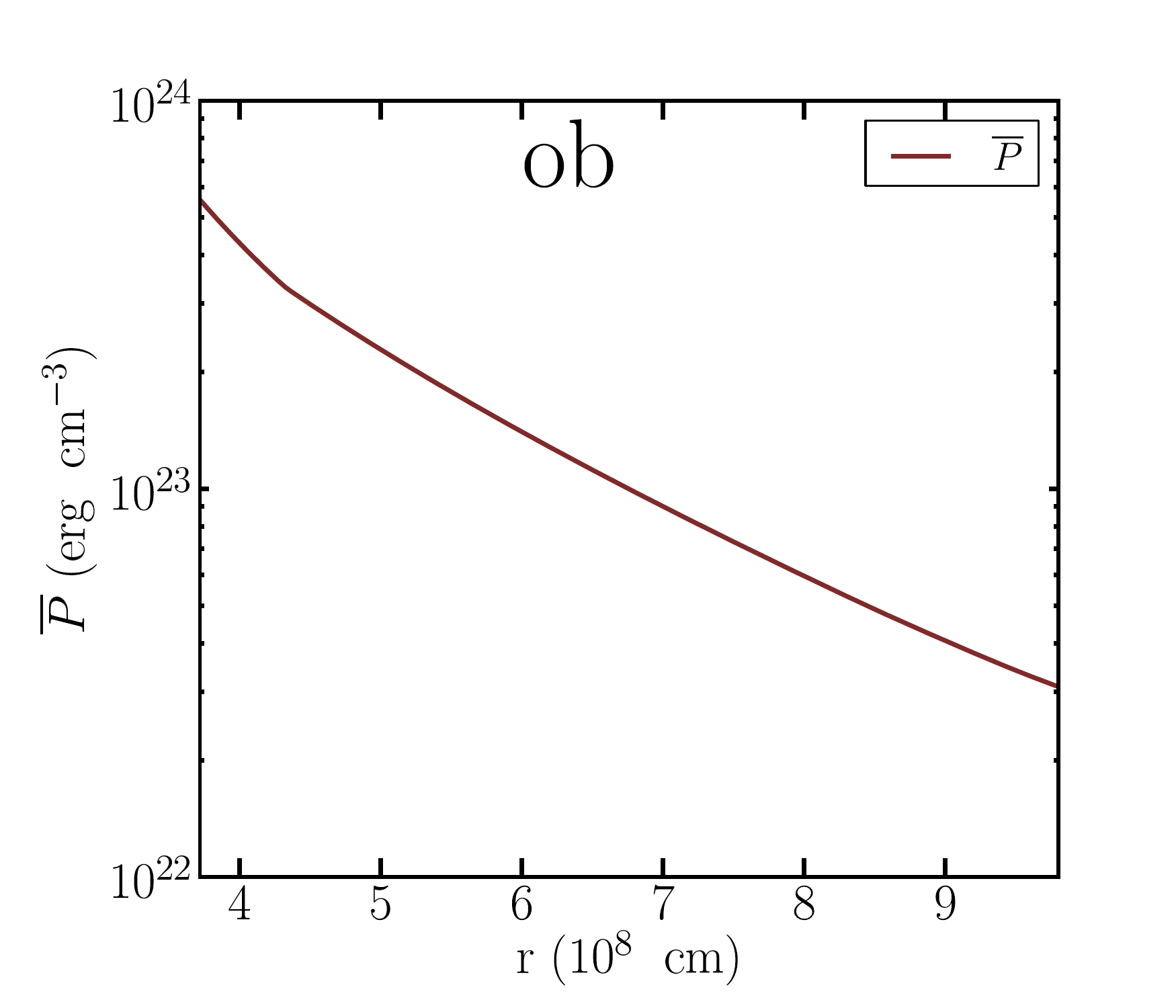}
\includegraphics[width=6.5cm]{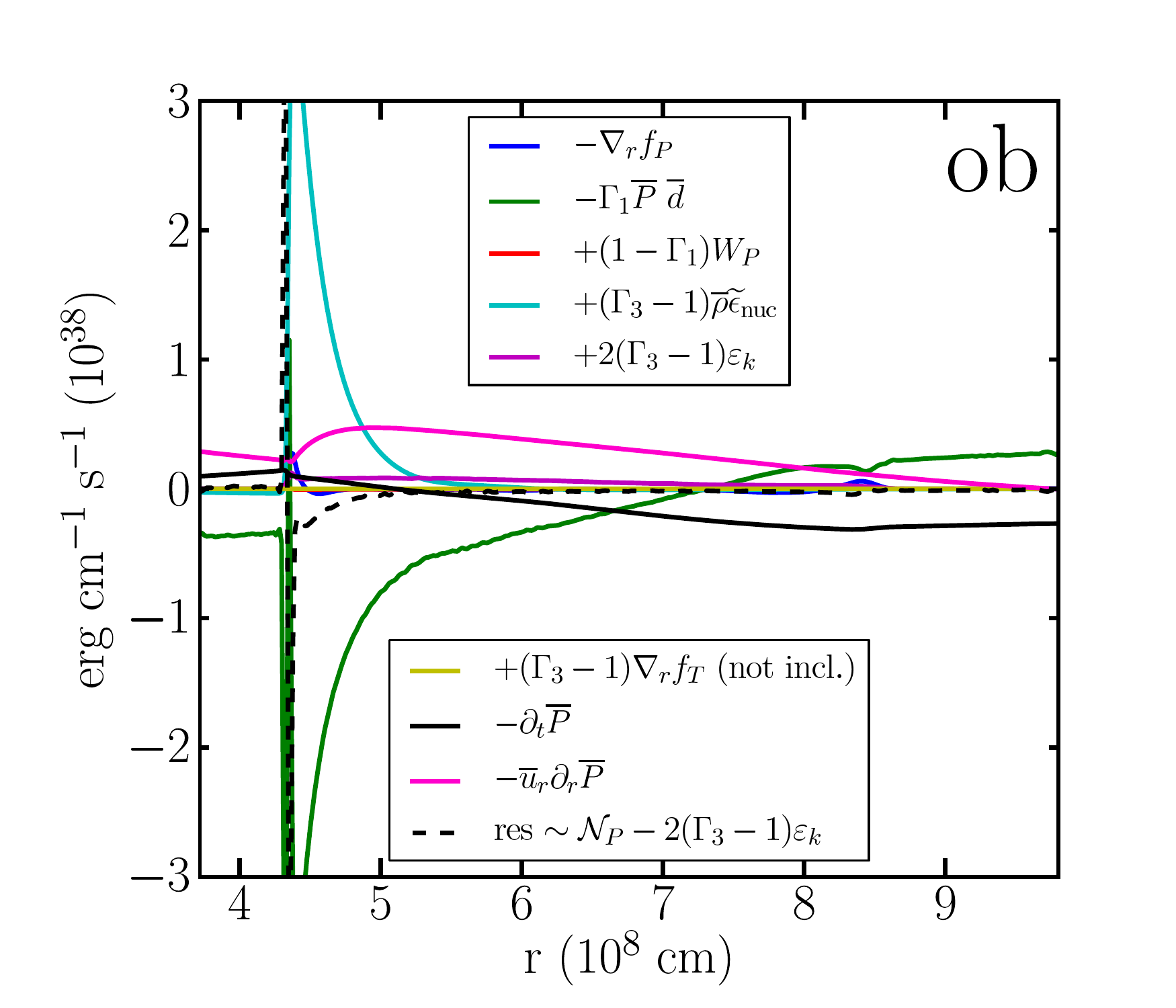}
\includegraphics[width=6.5cm]{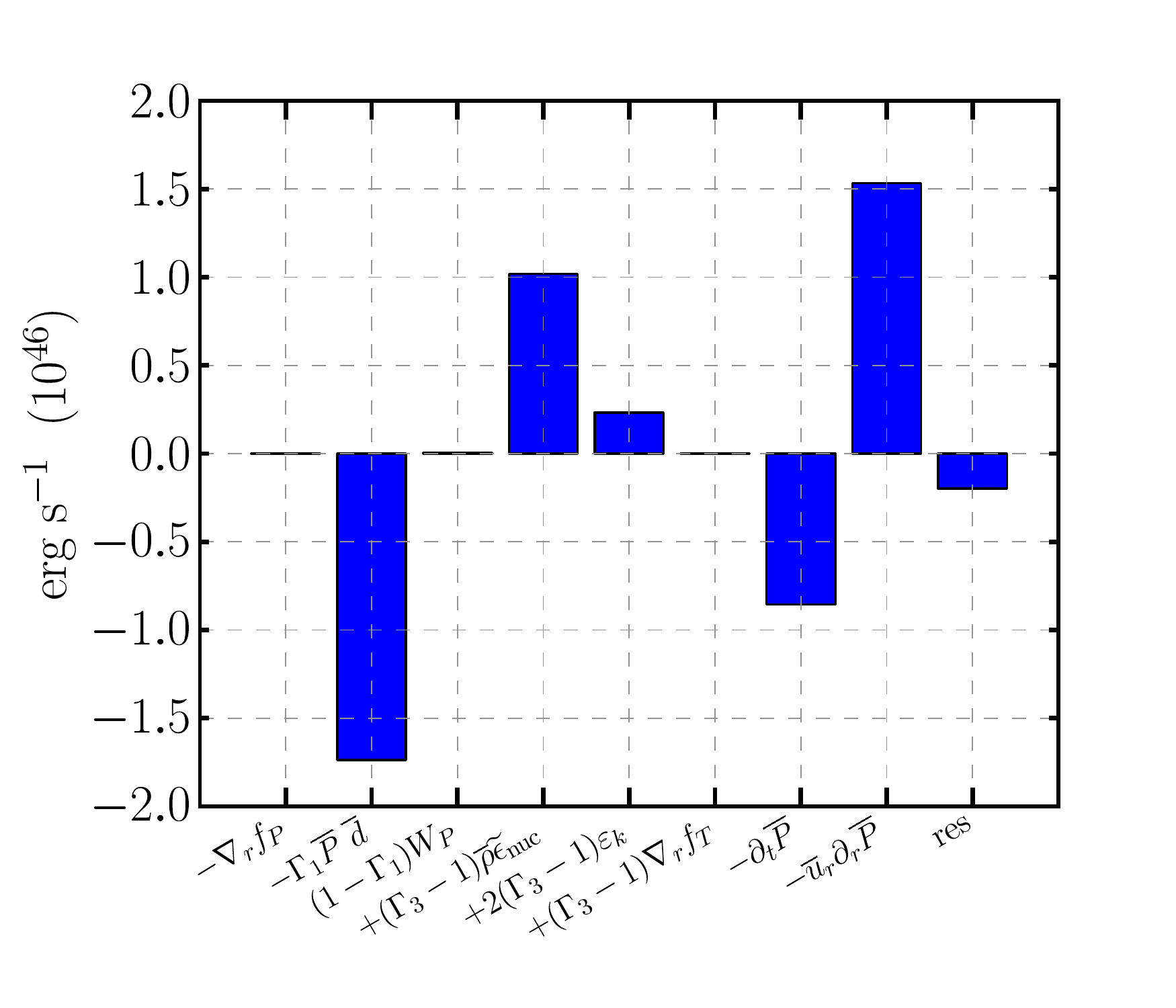}}

\centerline{
\includegraphics[width=6.5cm]{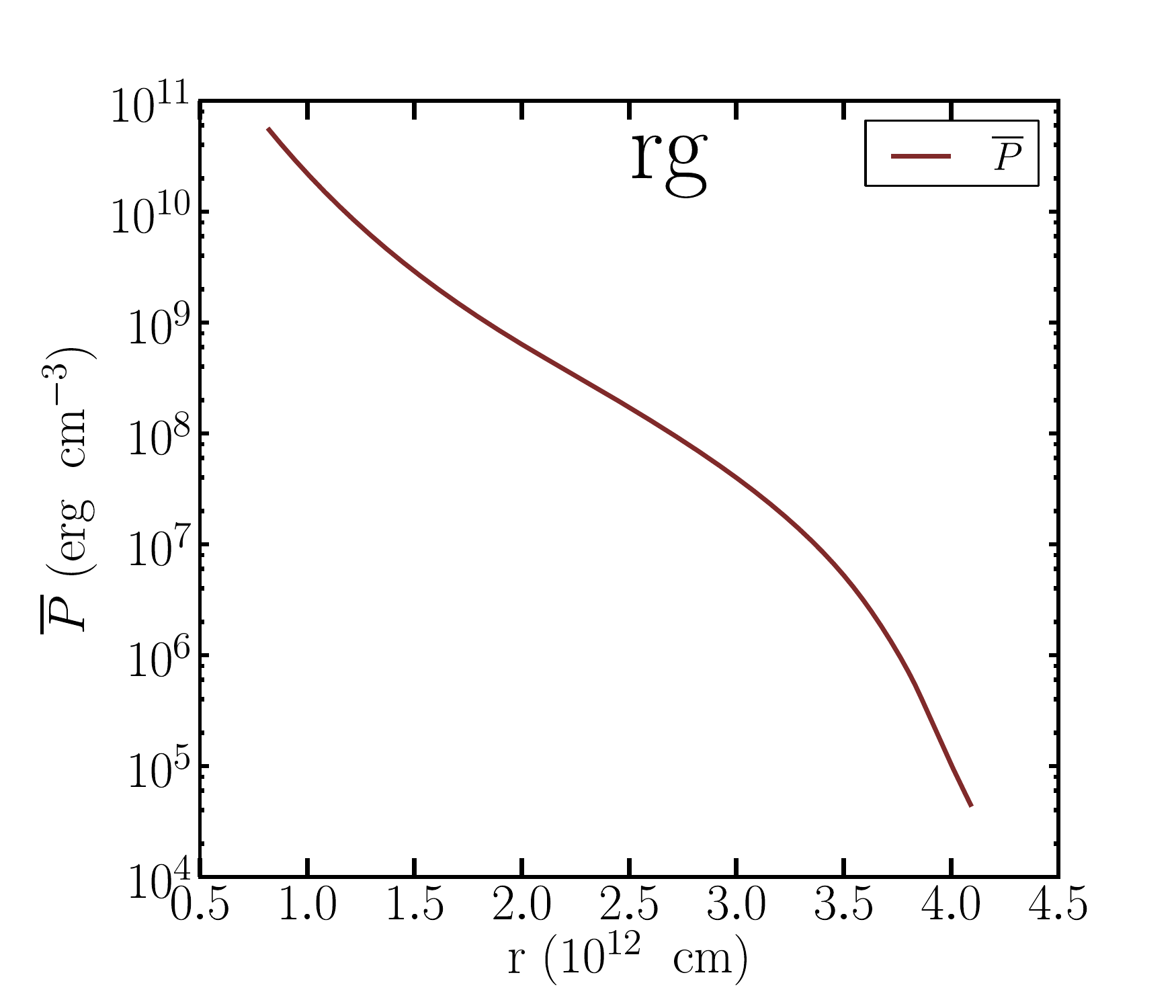}                      
\includegraphics[width=6.5cm]{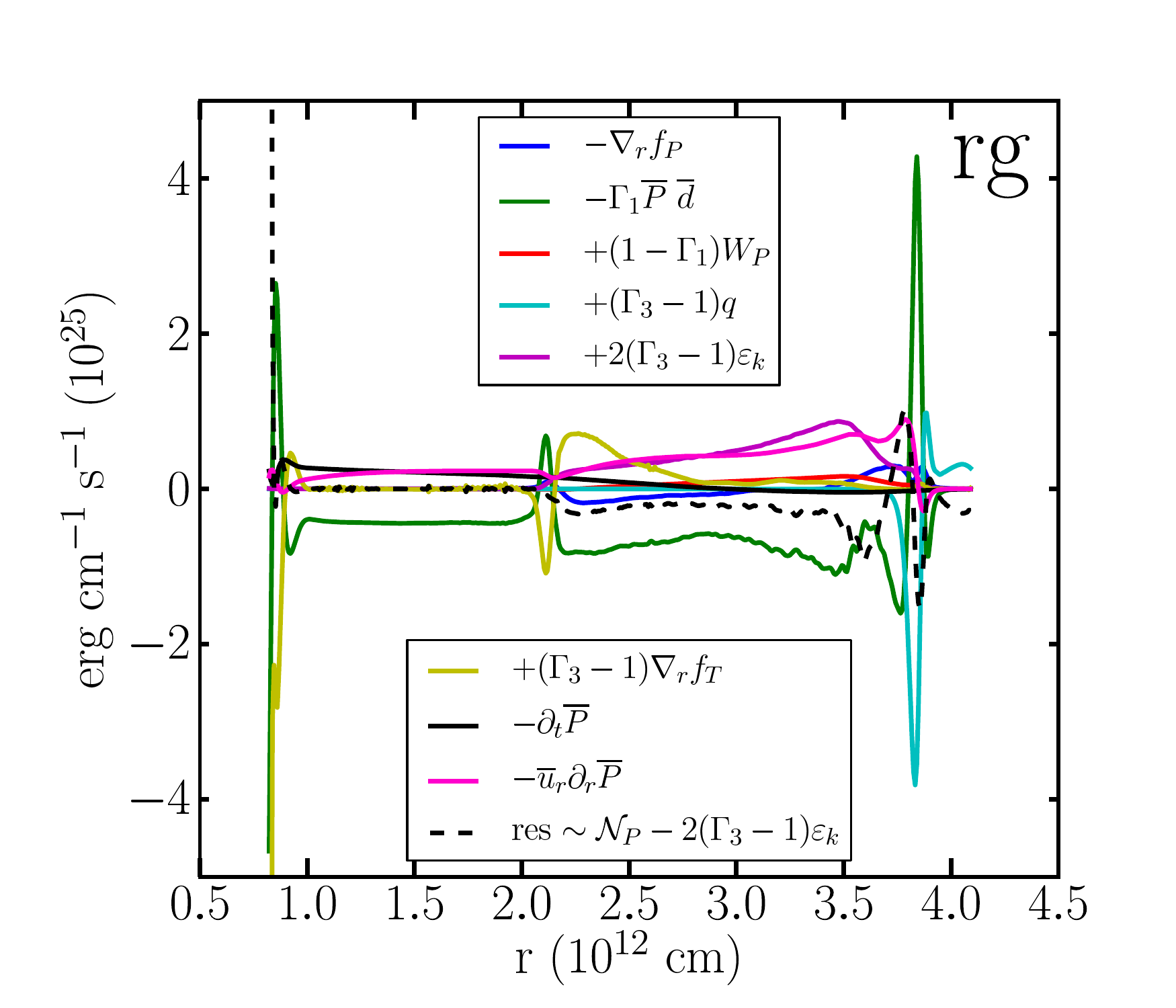}     
\includegraphics[width=6.5cm]{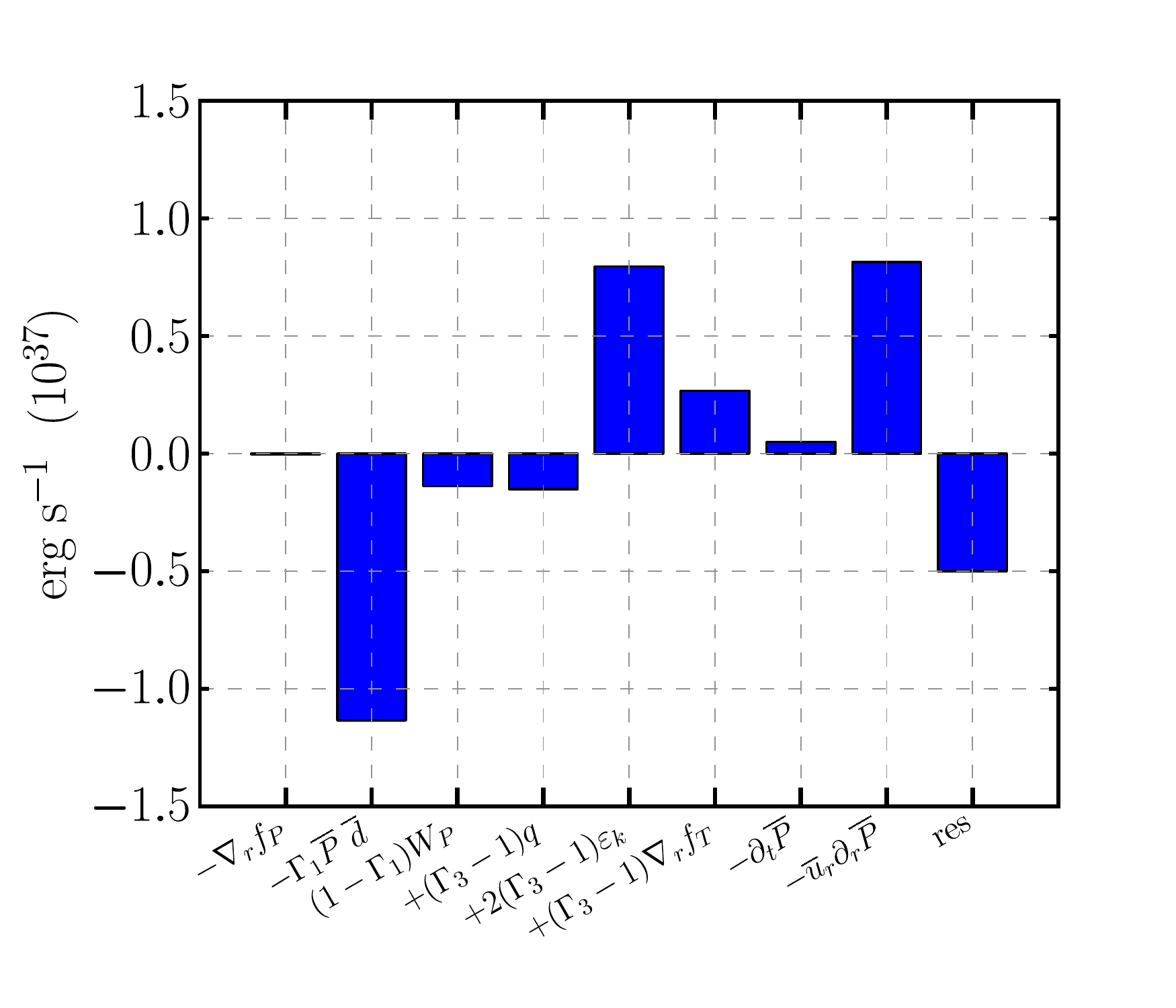}}
\caption{Mean pressure equation. Model {\sf ob.3D.mr} (upper panels) and model {\sf rg.3D.mr} (lower panels). \label{fig:pp-equation}}
\end{figure}

\newpage

\subsection{Mean enthalpy equation}

\begin{align}
\erho\fav{D}_t \fav{h} = & -\nabla_r f_h - \Gamma_1\eht{P} \ \eht{d} - \Gamma_1 W_P + \Gamma_3 {\mathcal S} + \Gamma_3 \nabla_r f_T +  {\mathcal N_h} \label{eq:rans_h}
\end{align}

\begin{figure}[!h]
\centerline{
\includegraphics[width=6.5cm]{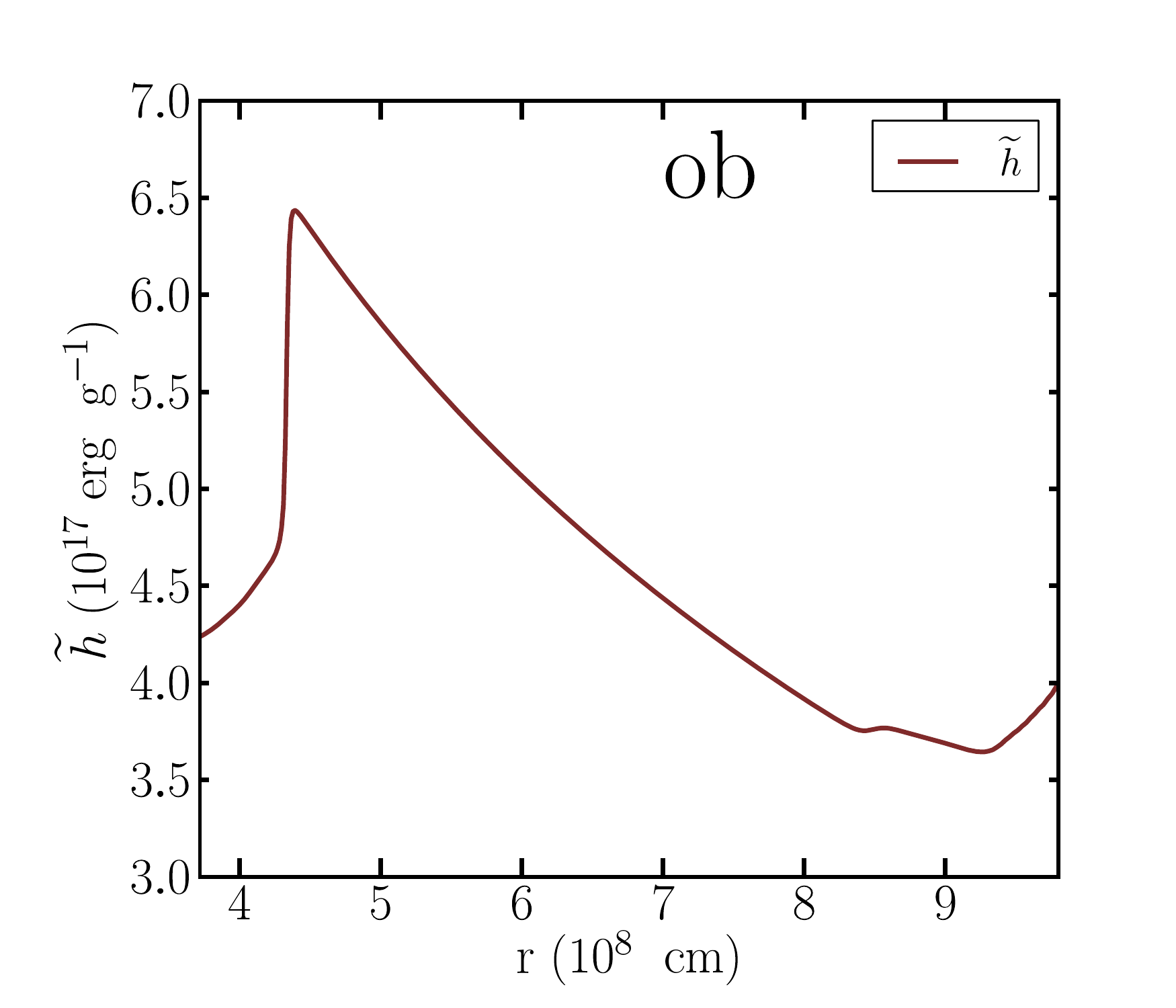}
\includegraphics[width=6.5cm]{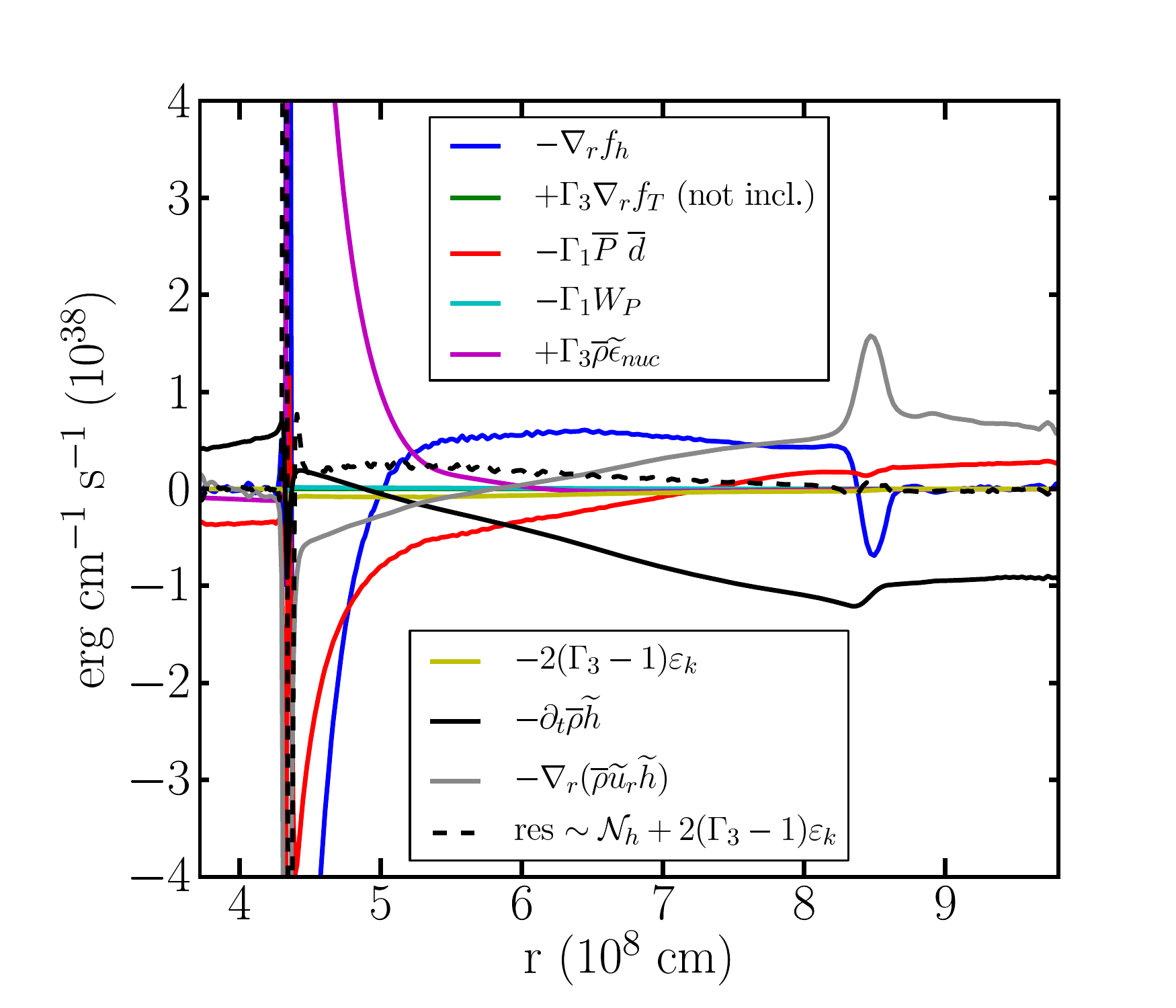}
\includegraphics[width=6.5cm]{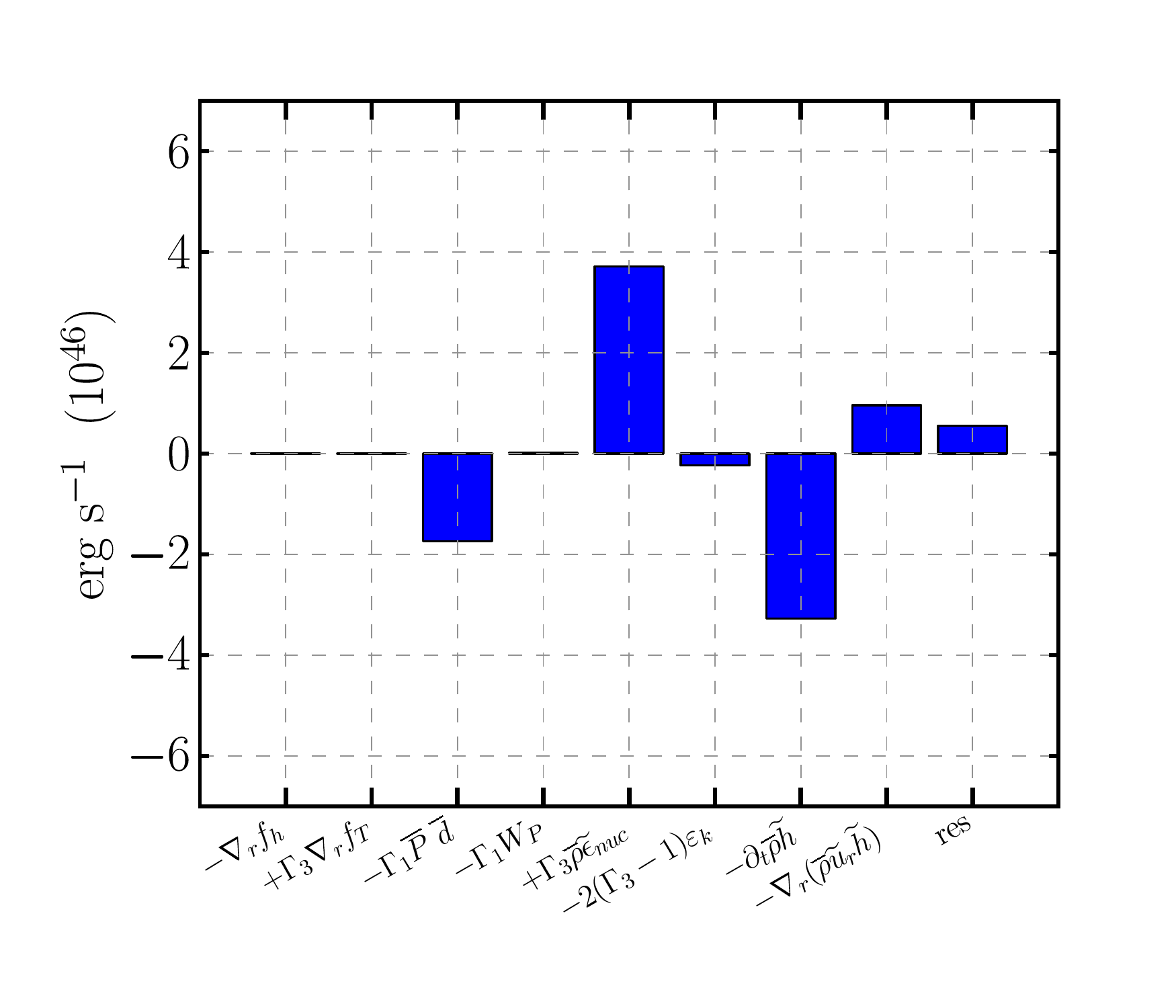}}

\centerline{
\includegraphics[width=6.5cm]{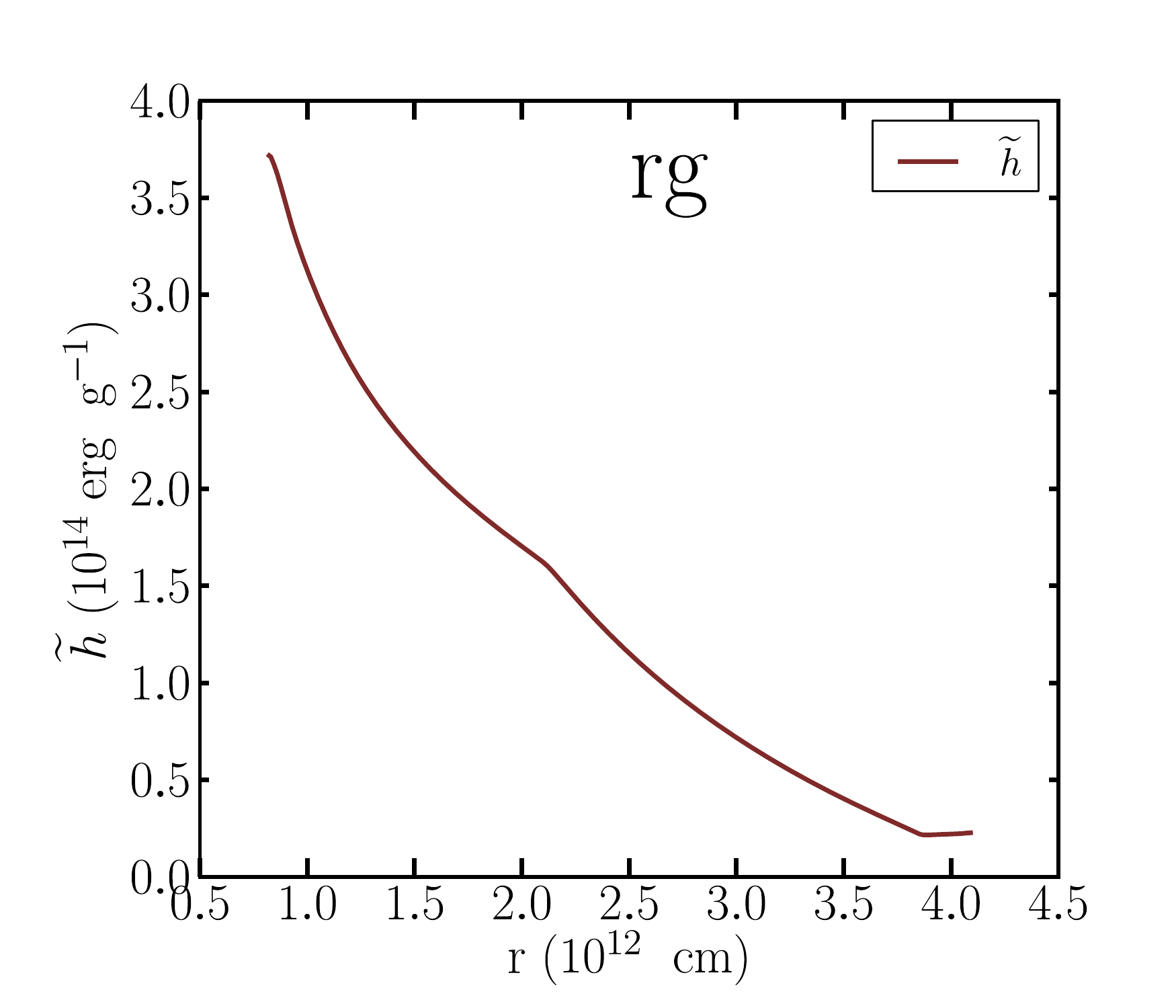}                      
\includegraphics[width=6.5cm]{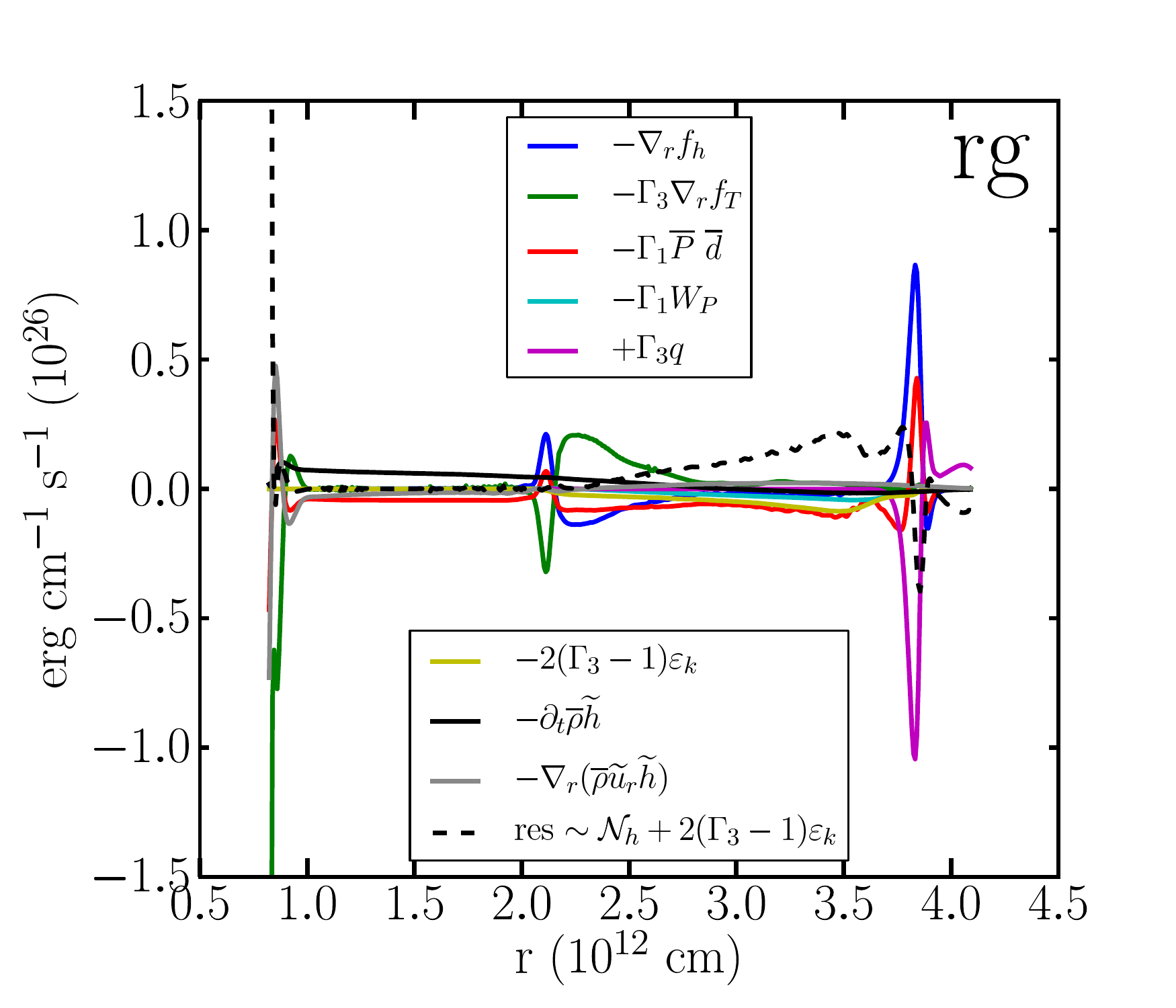}     
\includegraphics[width=6.5cm]{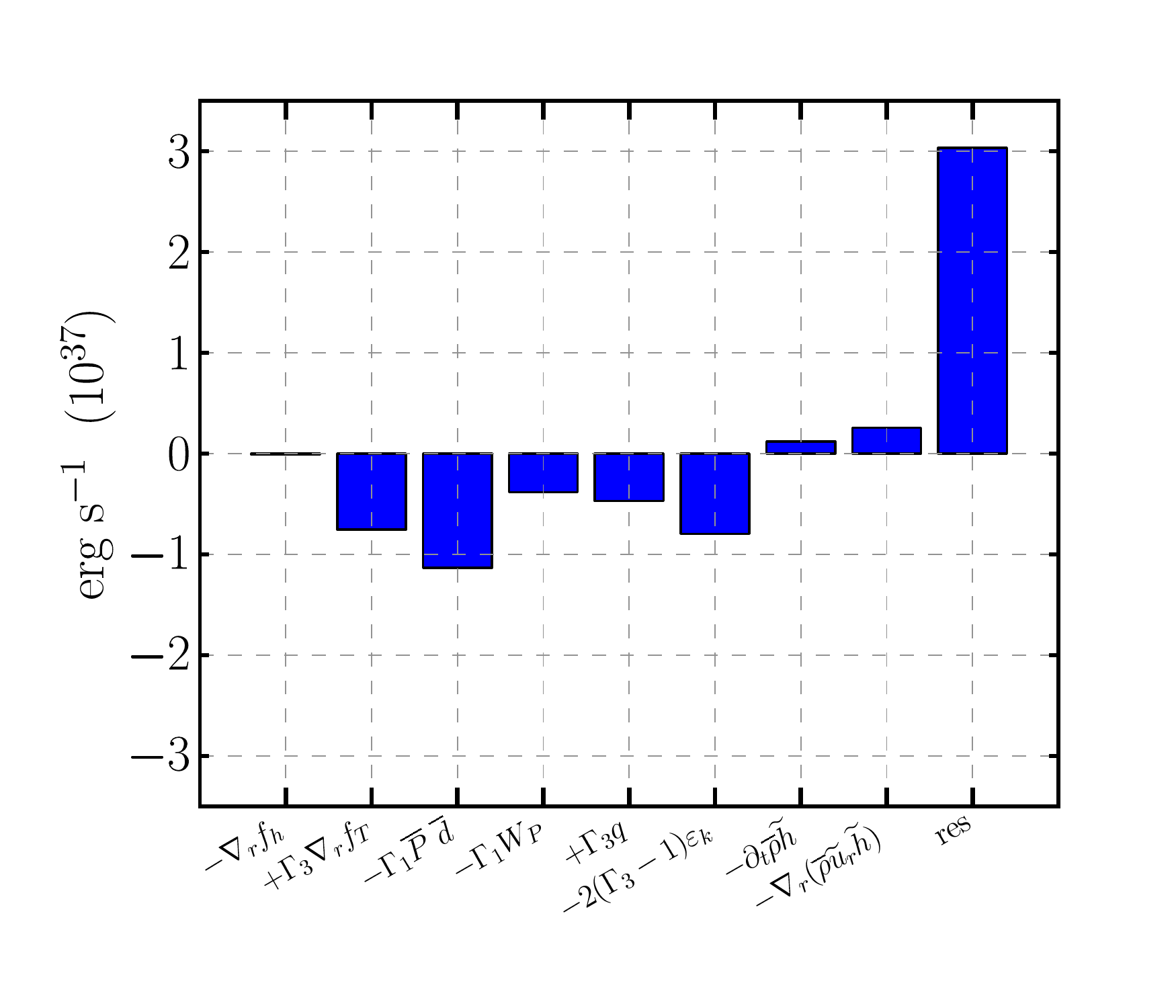}}
\caption{Mean enthalpy equation. Model {\sf ob.3D.mr} (upper panels) and model {\sf rg.3D.mr} (lower panels). \label{fig:hh-equation}}
\end{figure}

\newpage

\subsection{Mean angular momentum equation (z-component)}

\begin{align}
\erho\fav{D}_t \fav{j}_z = & -\nabla_r f_{jz} + {\mathcal N_{jz}}
\end{align}

\begin{figure}[!h]
\centerline{
\includegraphics[width=6.5cm]{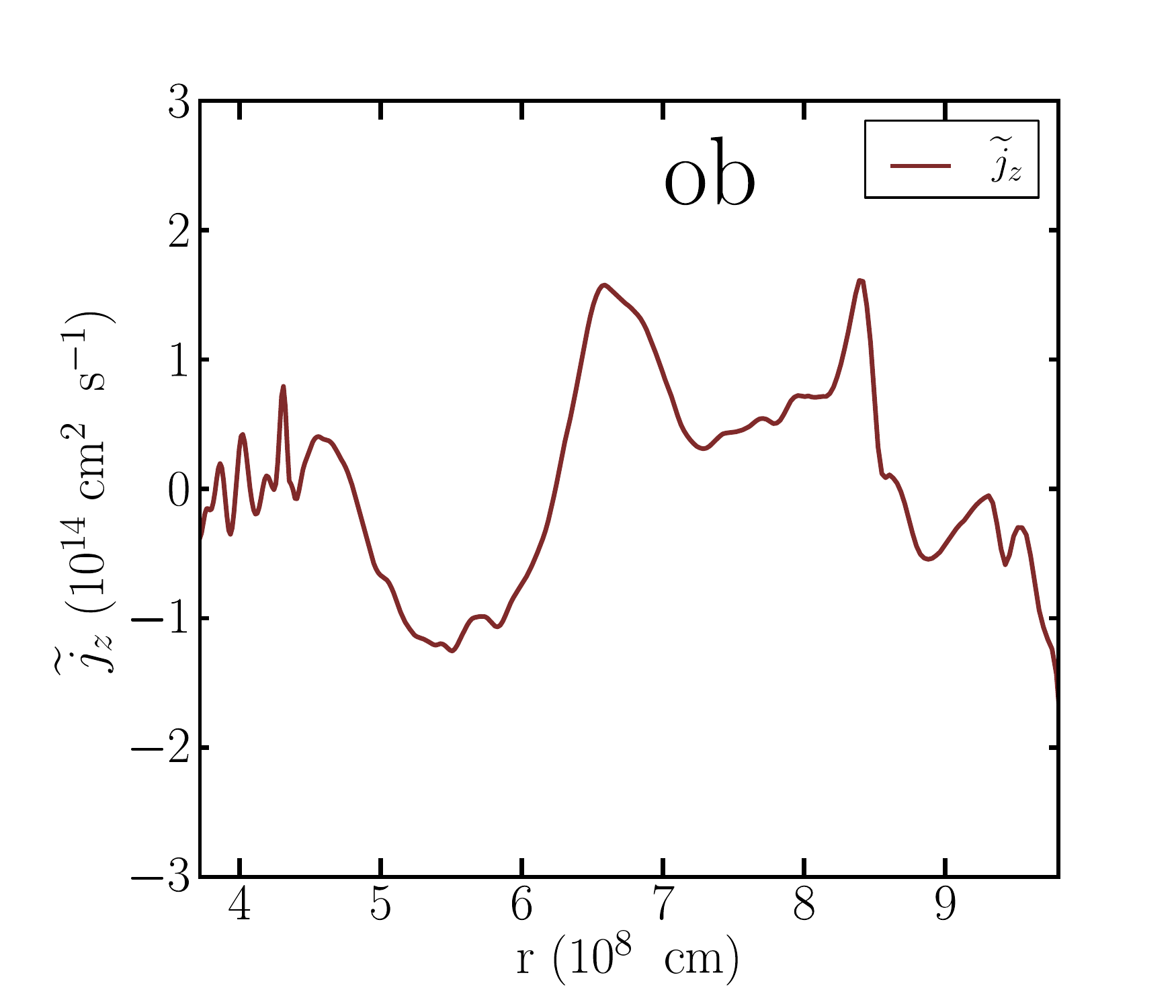}
\includegraphics[width=6.5cm]{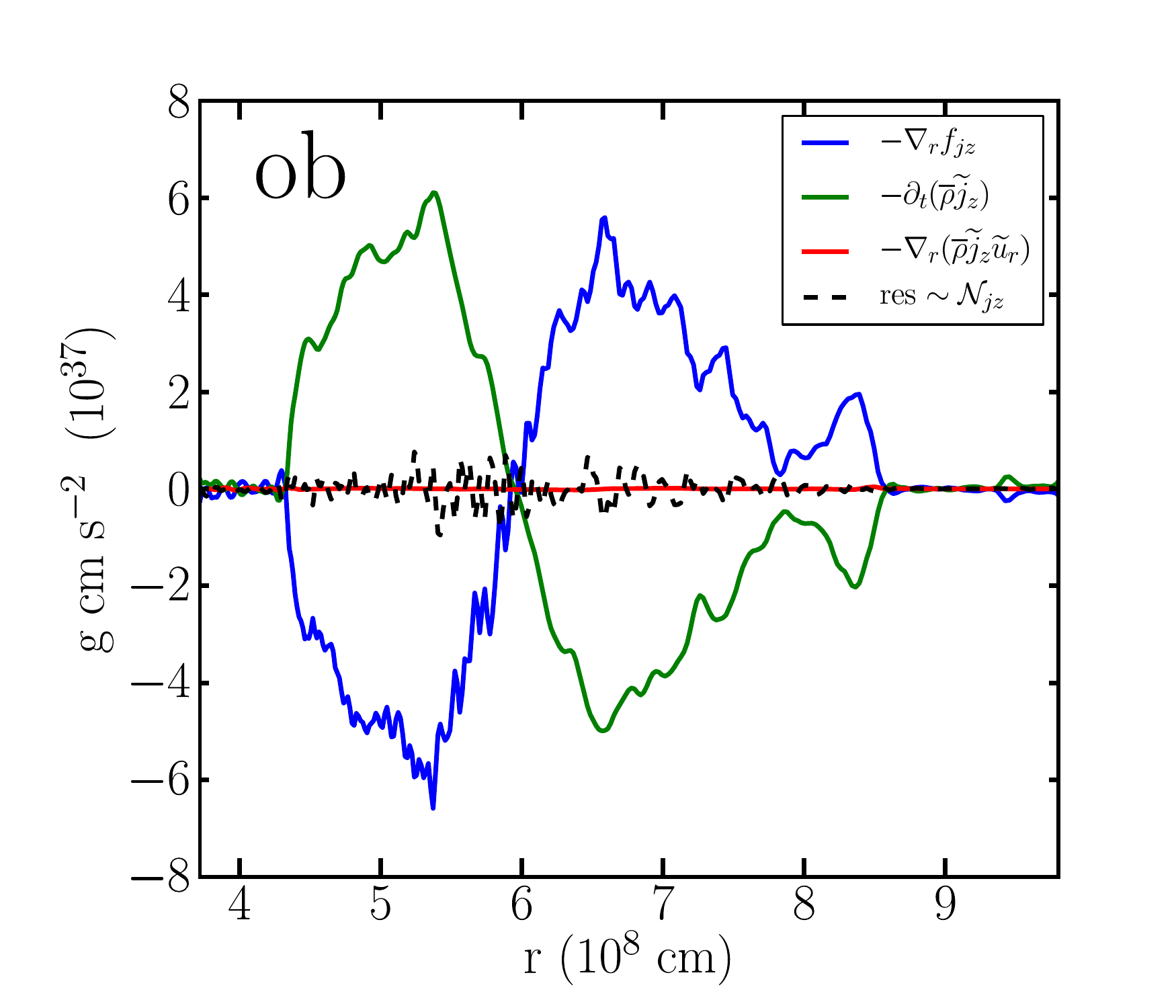}
\includegraphics[width=6.5cm]{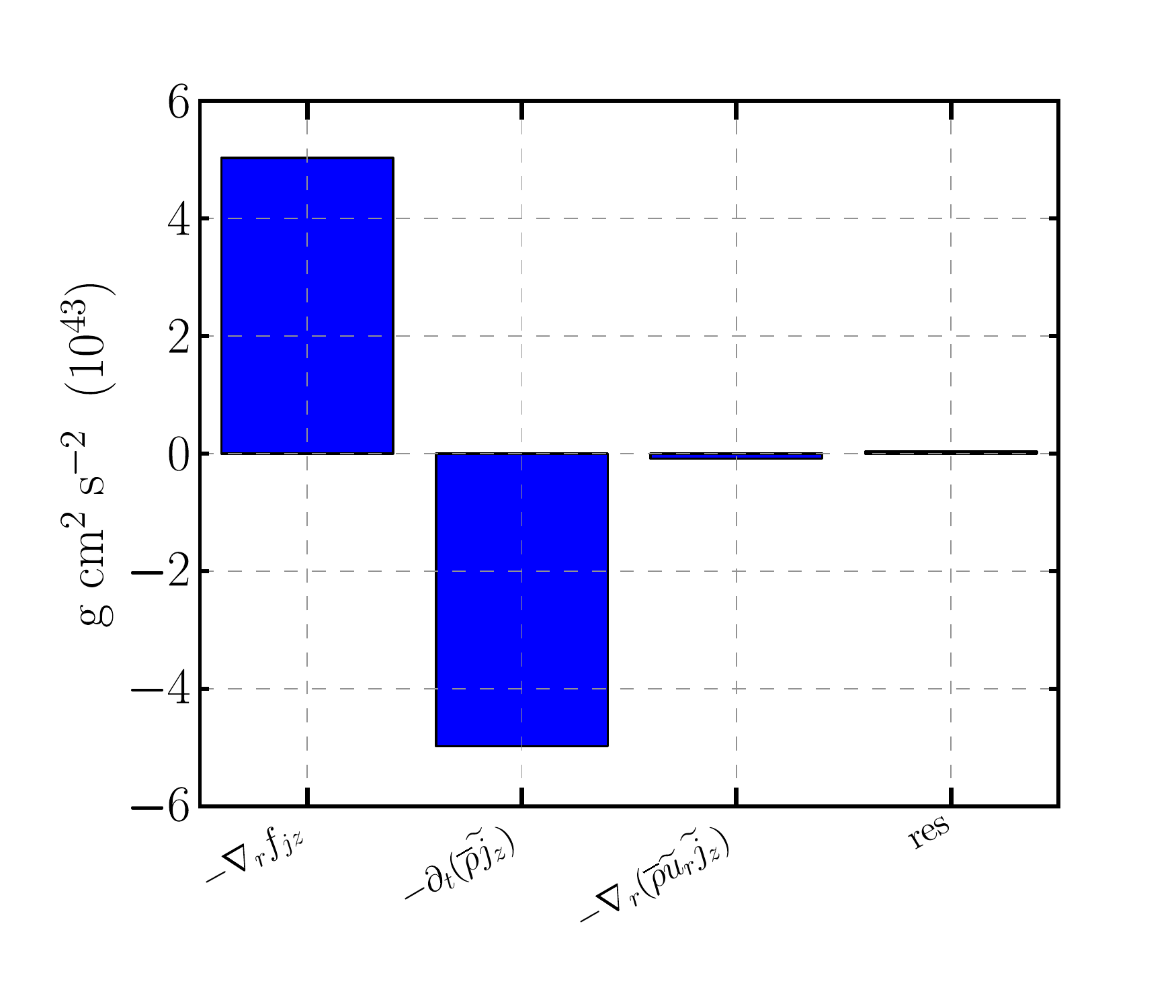}}

\centerline{
\includegraphics[width=6.5cm]{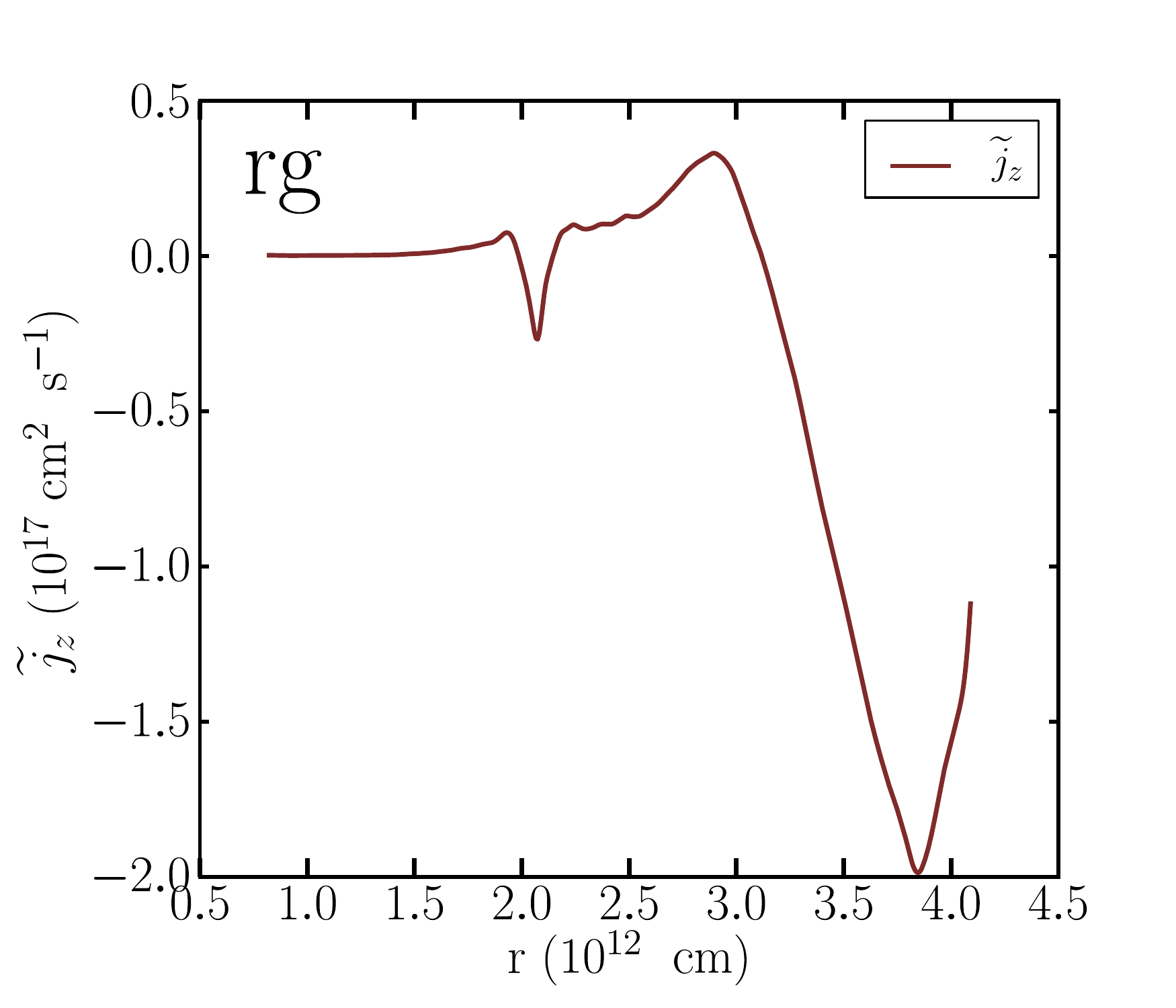}                      
\includegraphics[width=6.5cm]{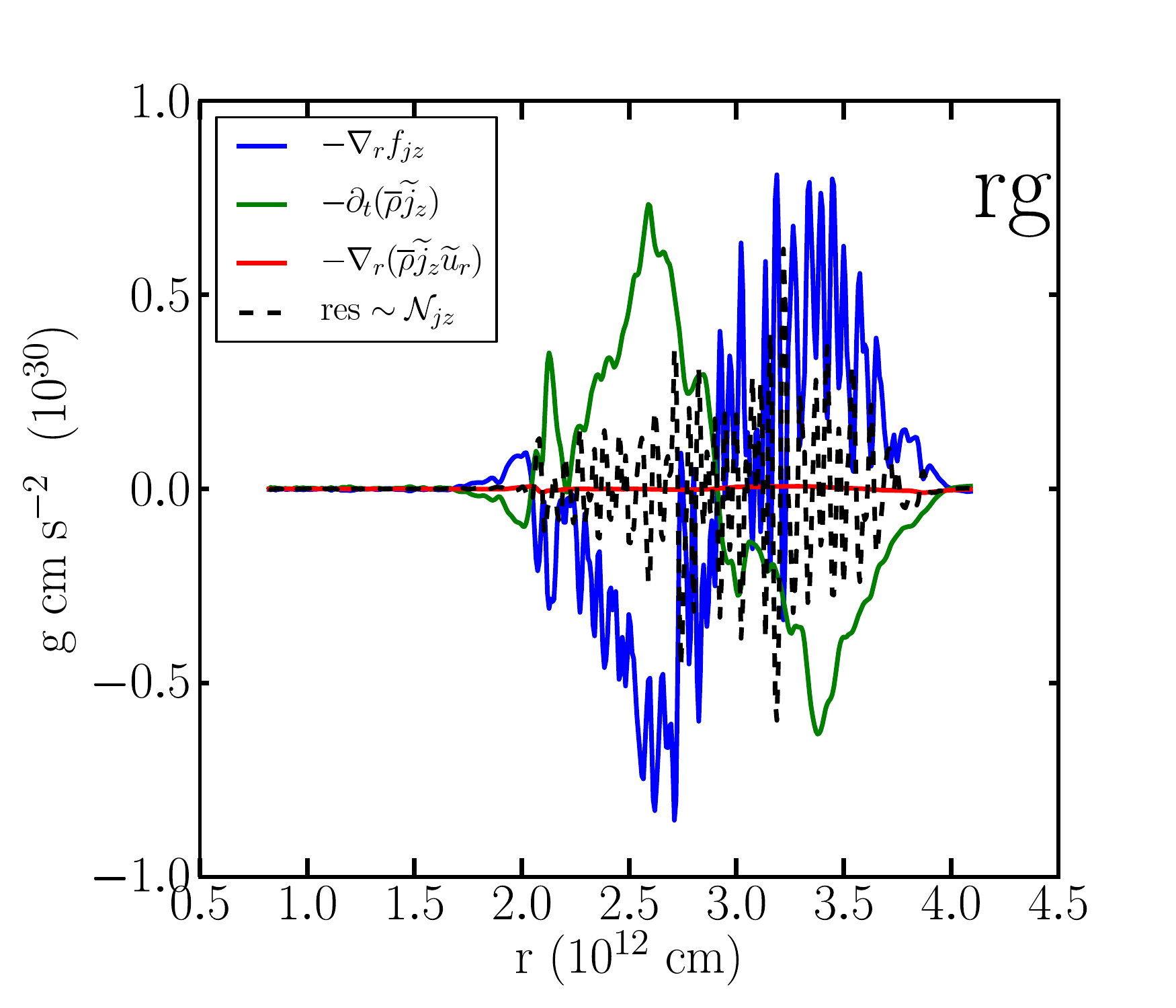}     
\includegraphics[width=6.5cm]{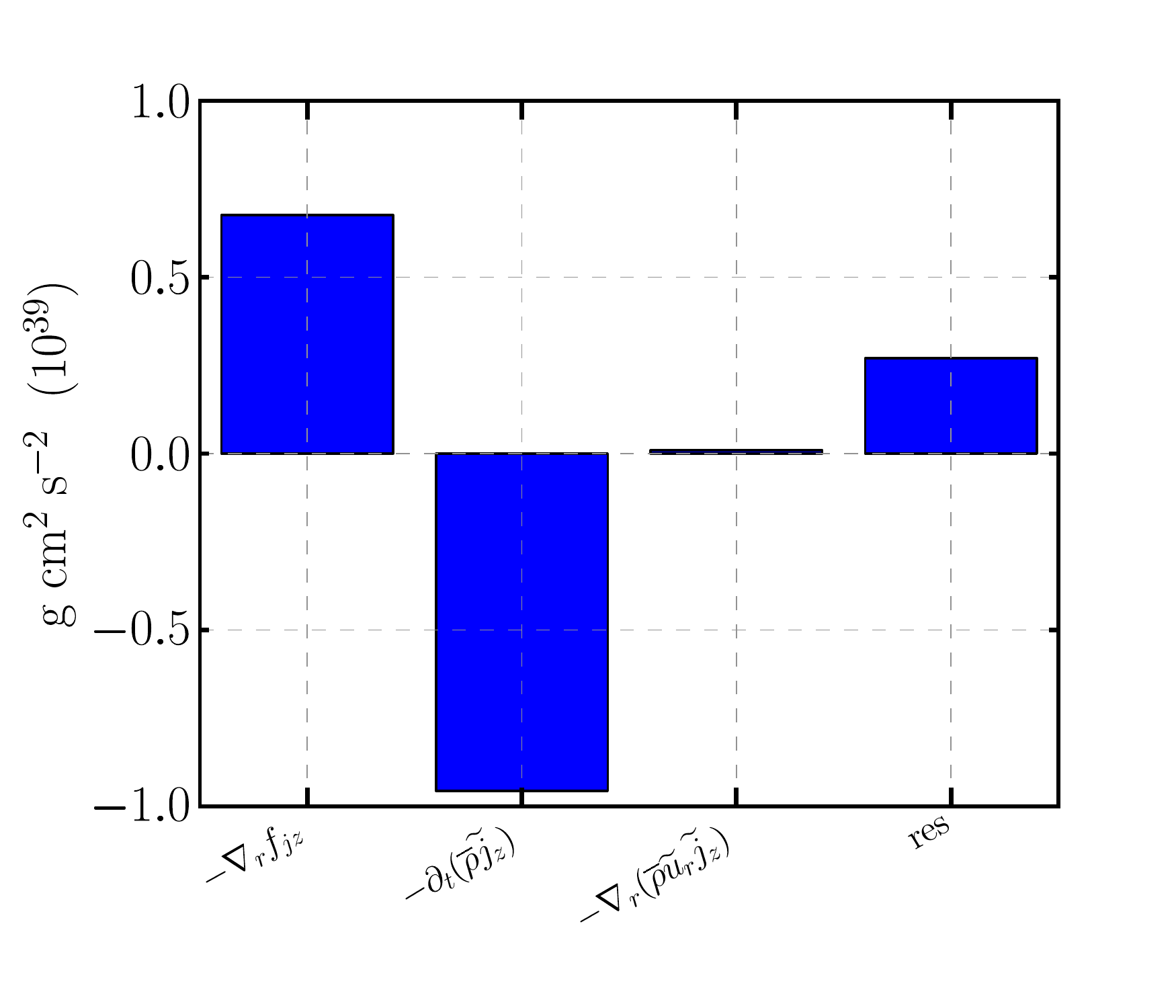}}
\caption{Mean angular momentum equation. Model {\sf ob.3D.mr} (upper panels) and model {\sf rg.3D.mr} (lower panels). \label{fig:jz-equation}}
\end{figure}

\newpage






\subsection{Mean turbulent kinetic energy equation}

\begin{align}
\av{\rho} \fav{D}_t \fav{k}^{ } = & -\nabla_r ( f_k +  f_P ) - \fht{R}_{ir}\partial_r \fht{u}_i + W_b + W_P + {\mathcal N_k}  \label{eq:rans_tke_with_fig}
\end{align}

\begin{figure}[!h]
\centerline{
\includegraphics[width=6.5cm]{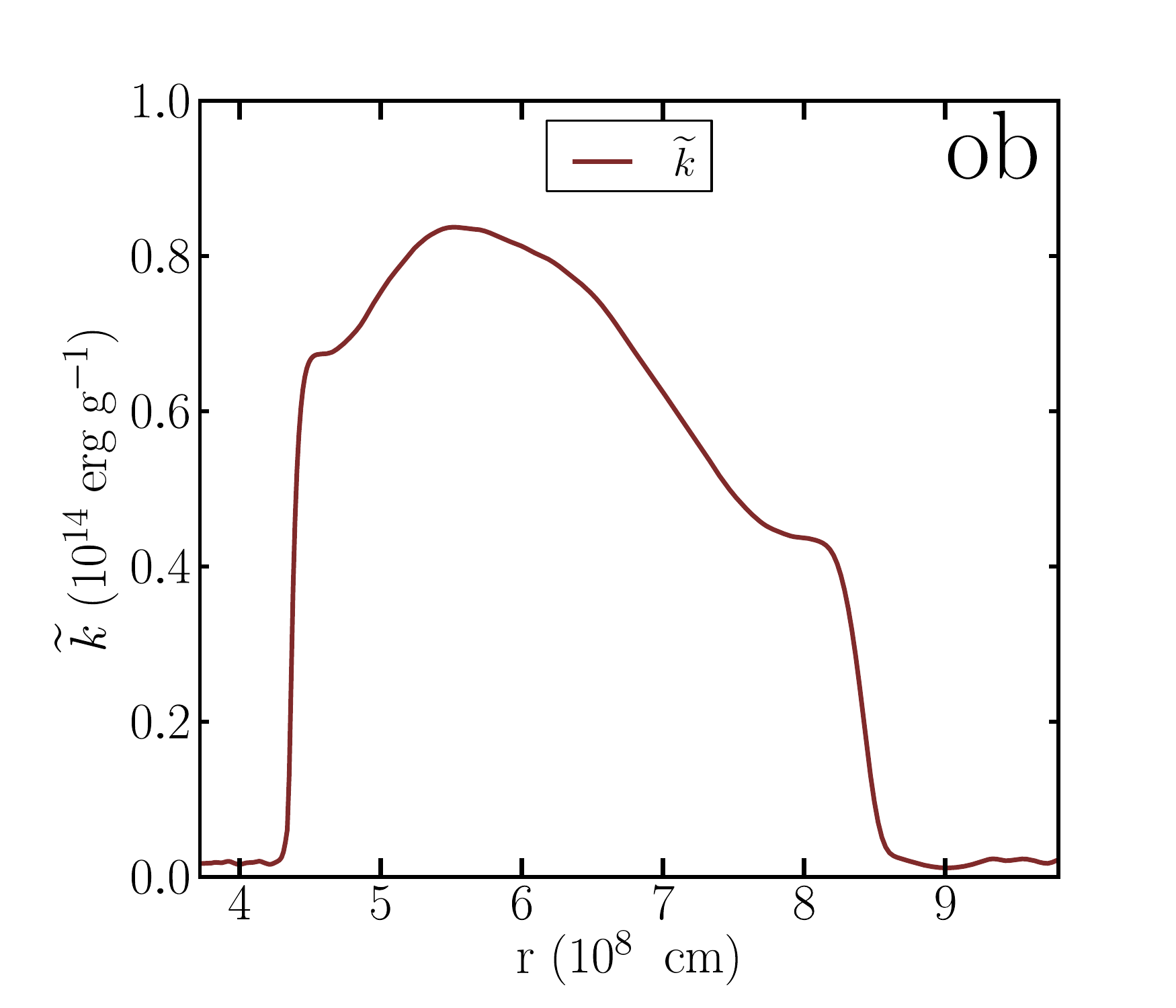}
\includegraphics[width=6.5cm]{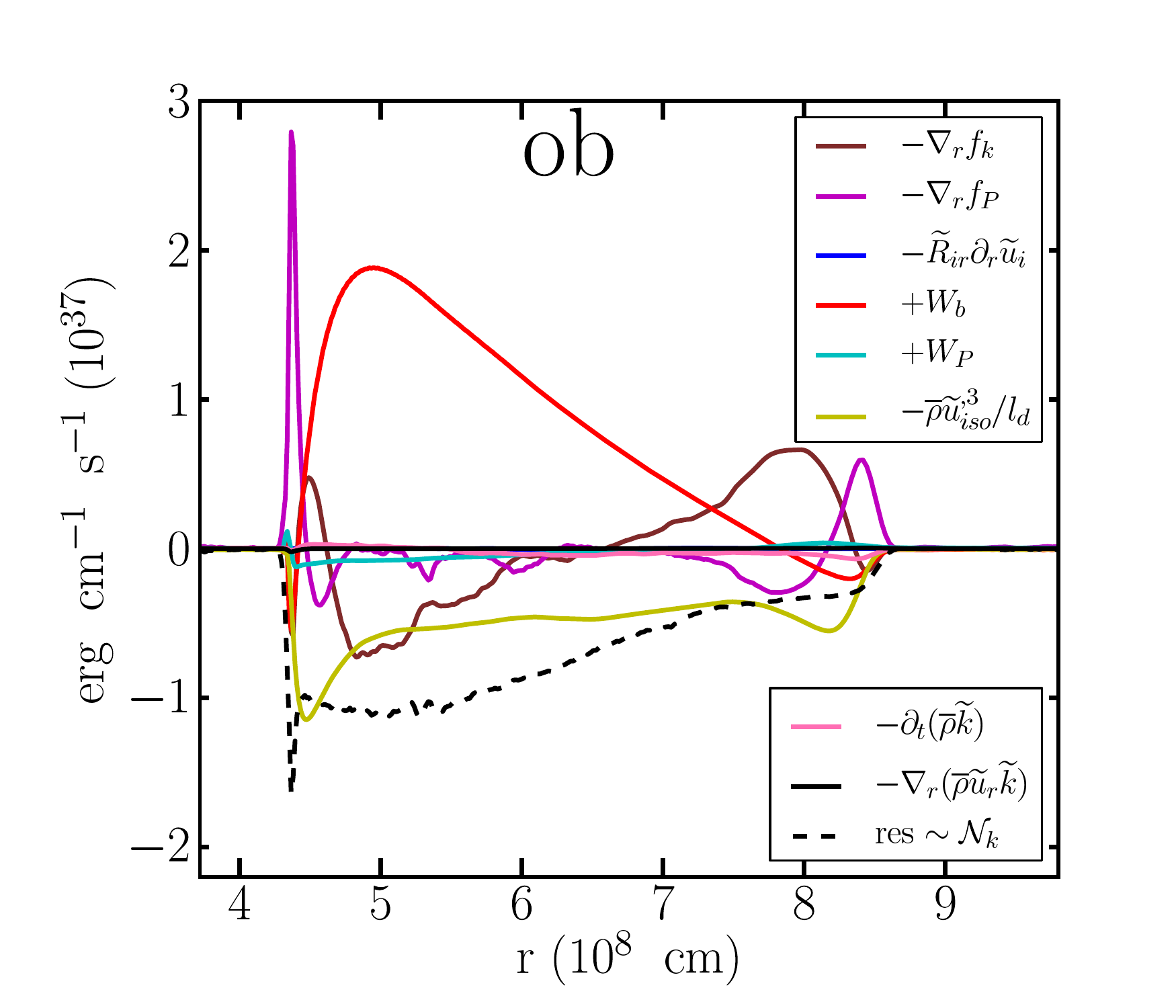}
\includegraphics[width=6.5cm]{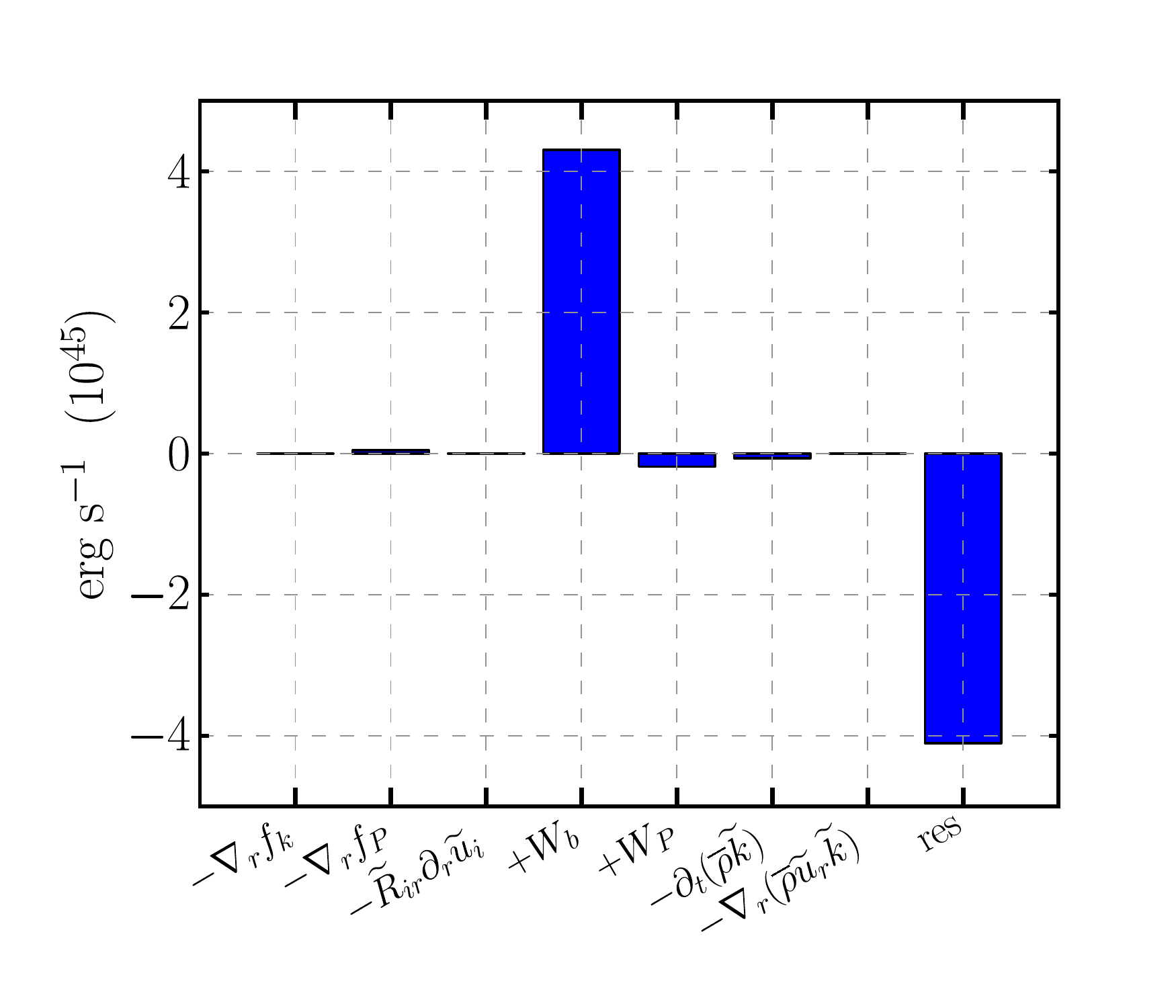}}

\centerline{
\includegraphics[width=6.5cm]{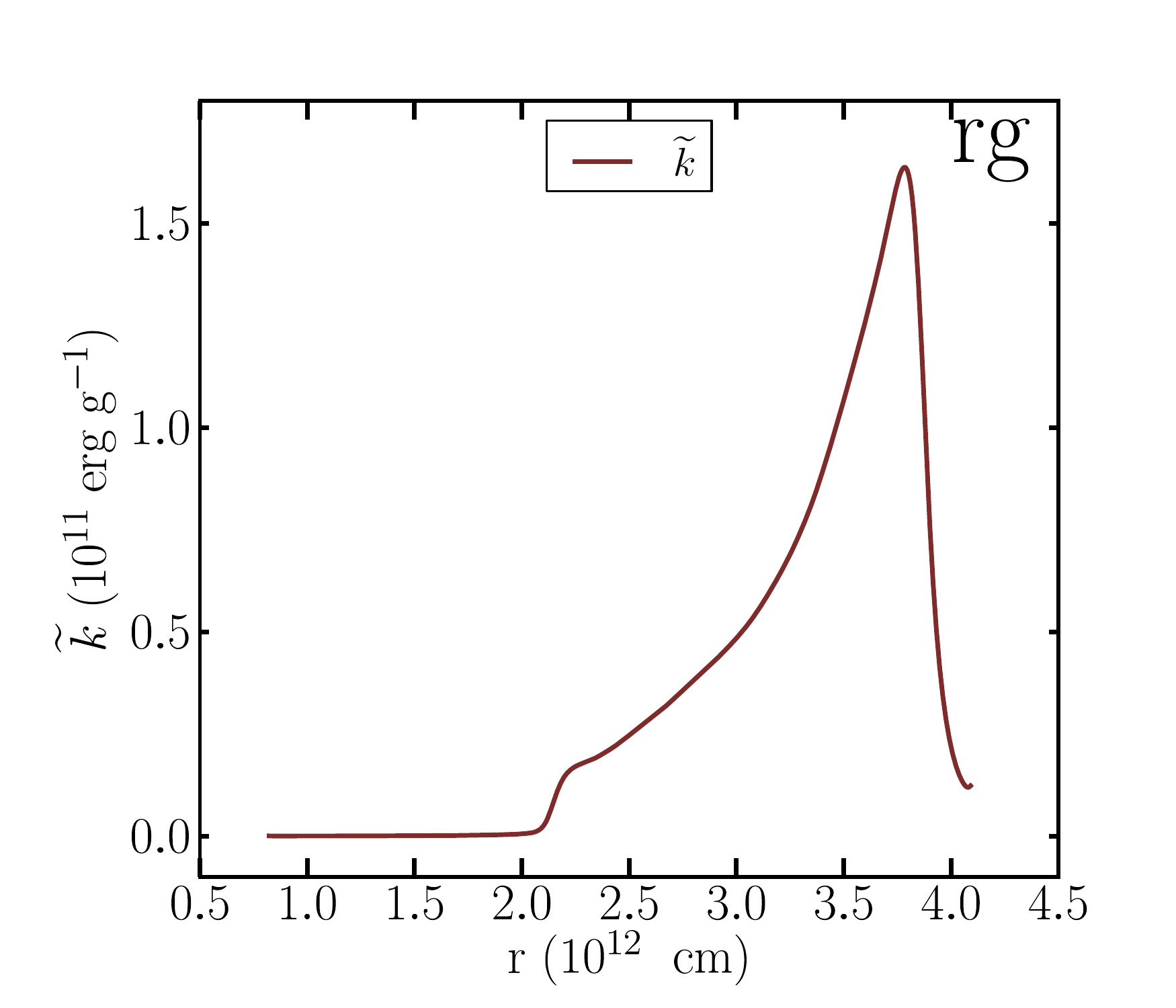}
\includegraphics[width=6.5cm]{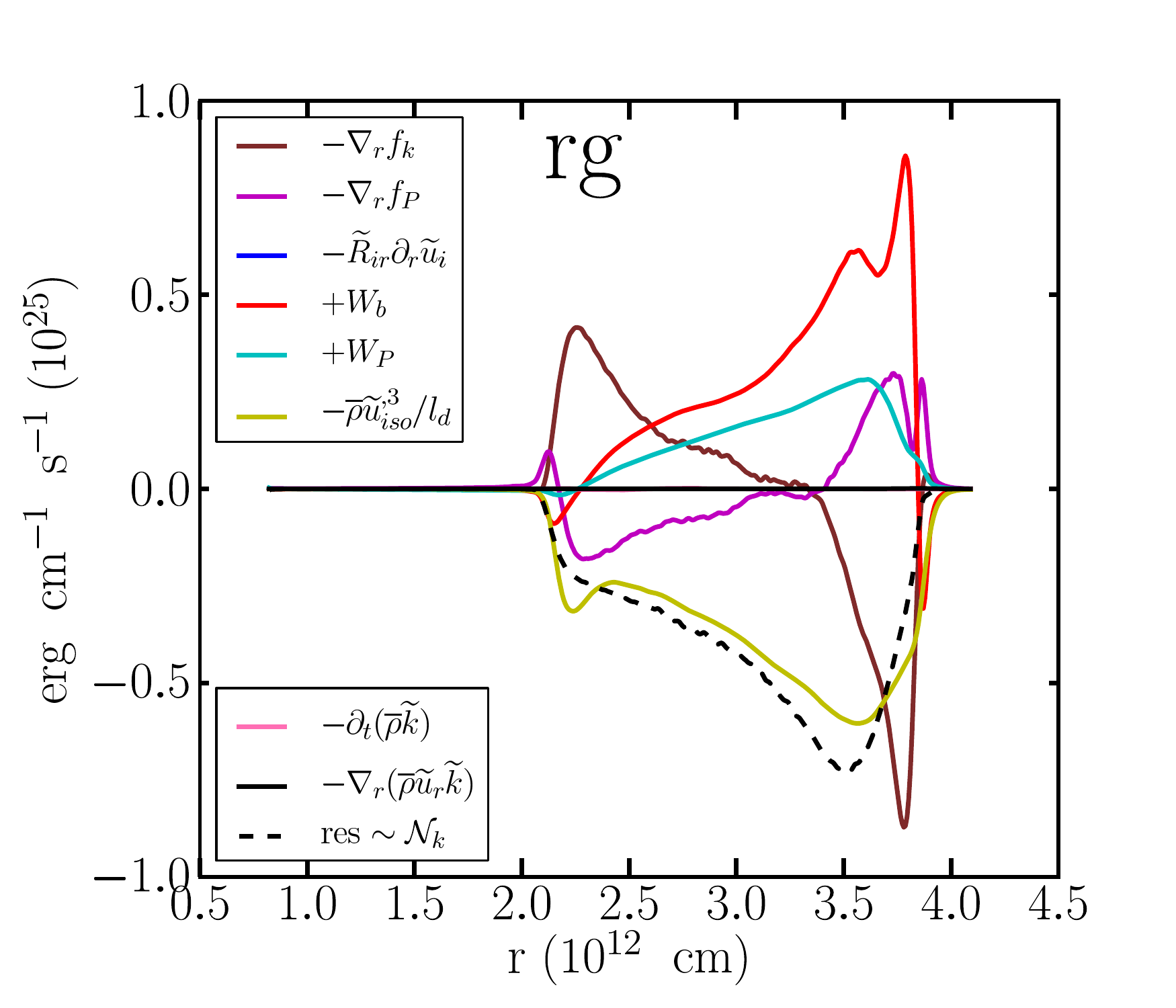}
\includegraphics[width=6.5cm]{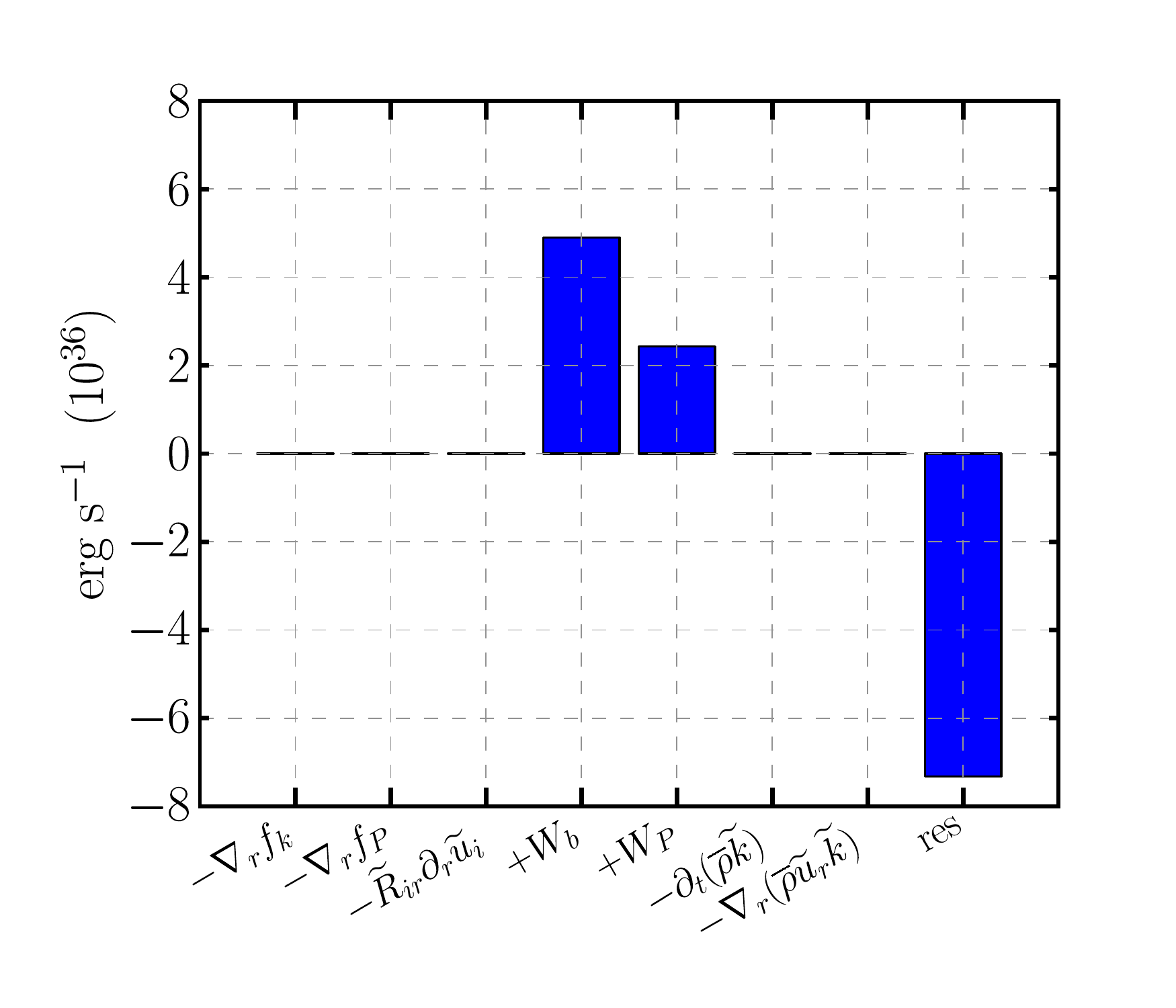}}
\caption{Turbulent kinetic energy equation. Model {\sf ob.3D.mr} (upper panels) and model {\sf rg.3D.mr} (lower panels). \label{fig:k-equation}}
\end{figure}

\newpage

\subsection{Mean turbulent kinetic energy equation (radial part)}

\begin{align}
\av{\rho} \fav{D}_t \fav{k}^r =  &  -\nabla_r  ( f_k^r + f_P )  - \fht{R}_{rr}\partial_r \fht{u}_r + W_b  + \eht{P'\nabla_r u''_r} + {\mathcal G_k^r} + {\mathcal N_{kr}} \label{eq:rans_ekin_r}
\end{align}

\begin{figure}[!h]
\centerline{
\includegraphics[width=6.5cm]{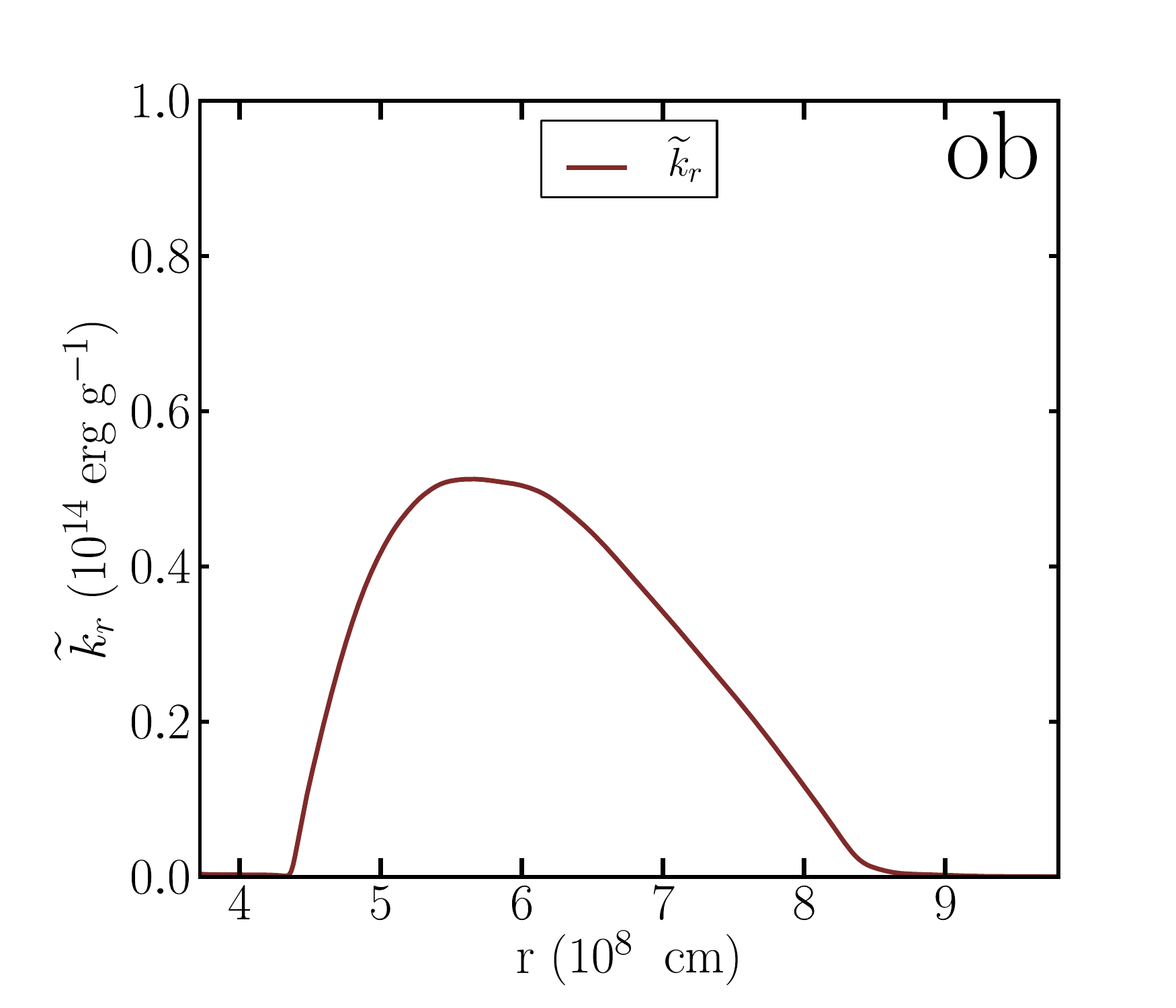}
\includegraphics[width=6.5cm]{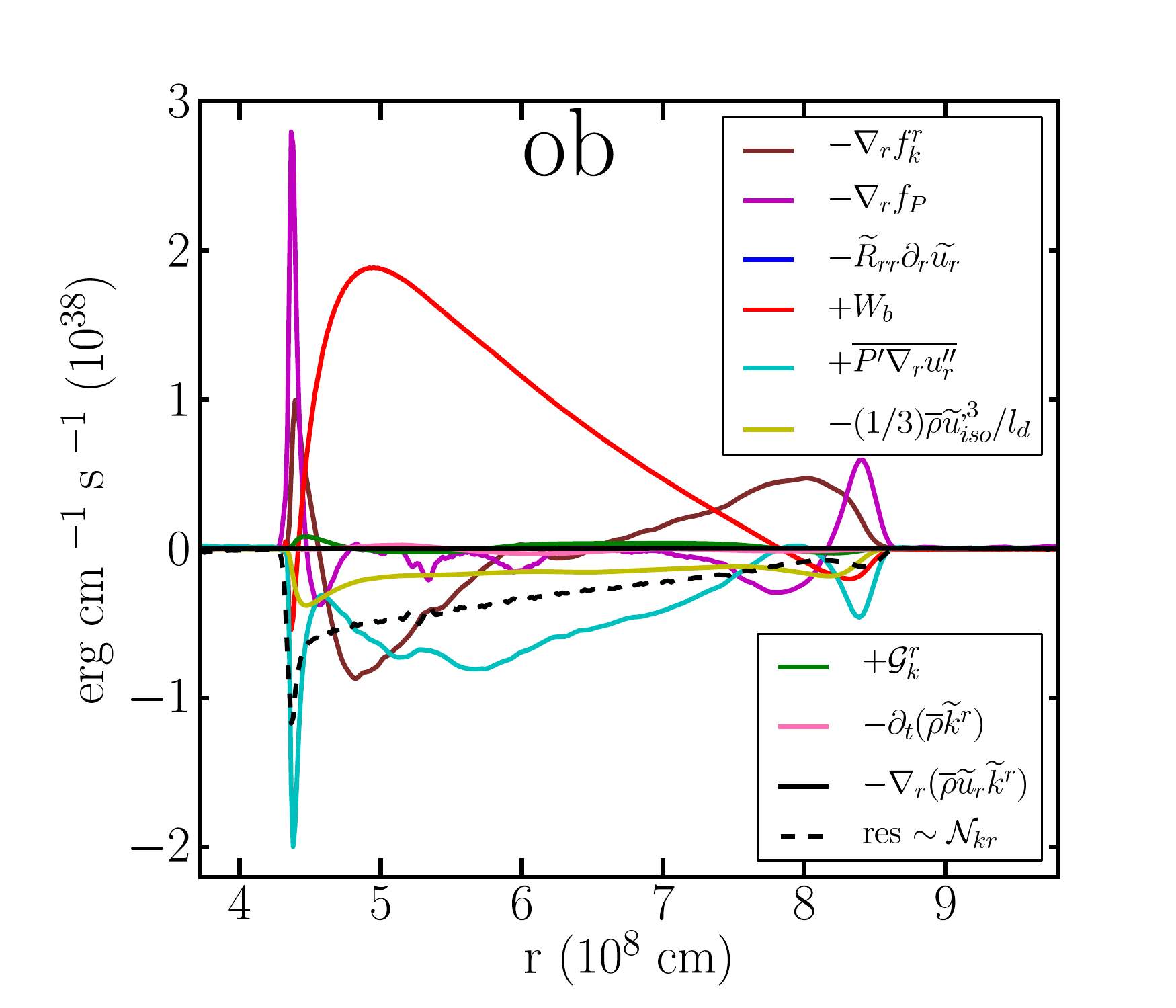}
\includegraphics[width=6.5cm]{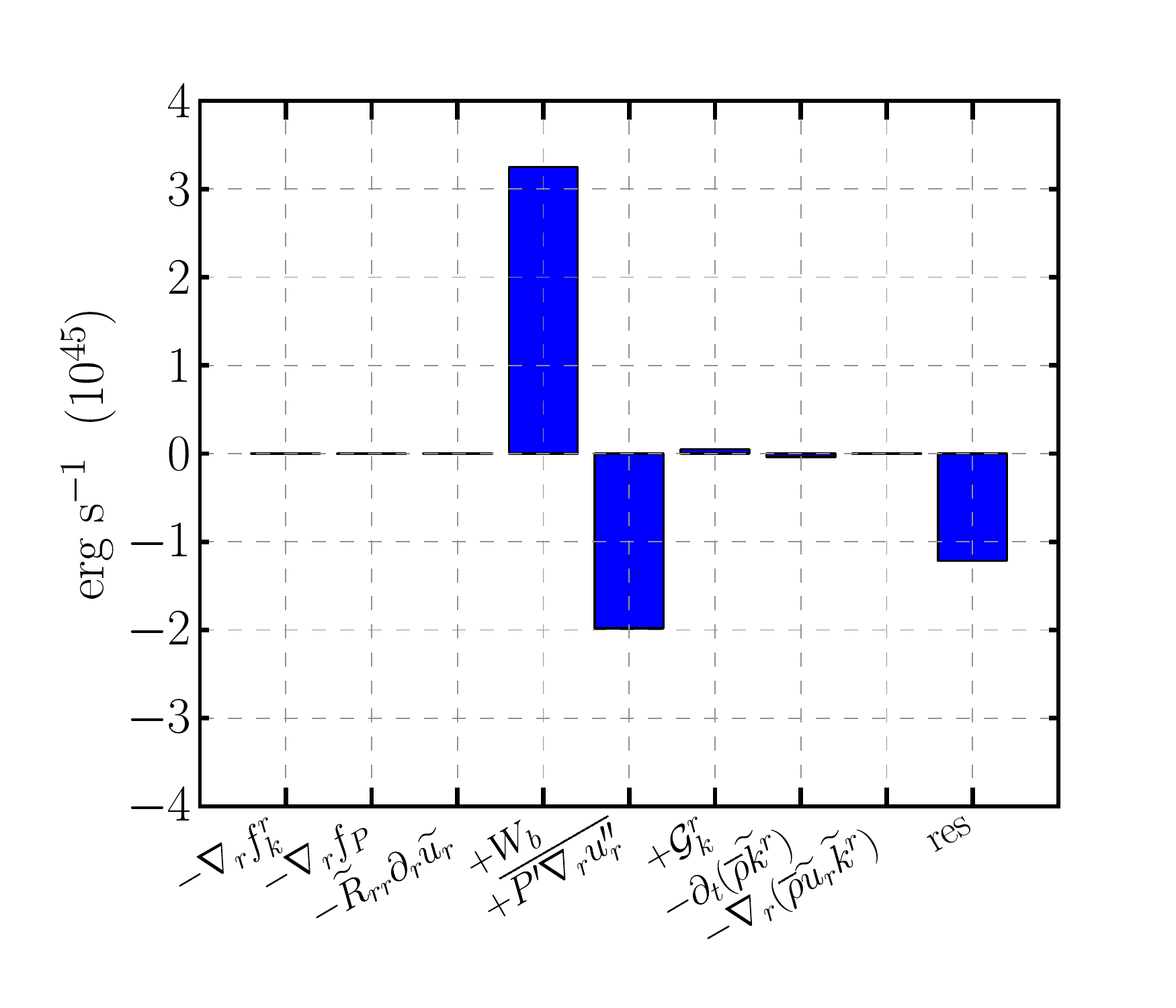}}

\centerline{
\includegraphics[width=6.5cm]{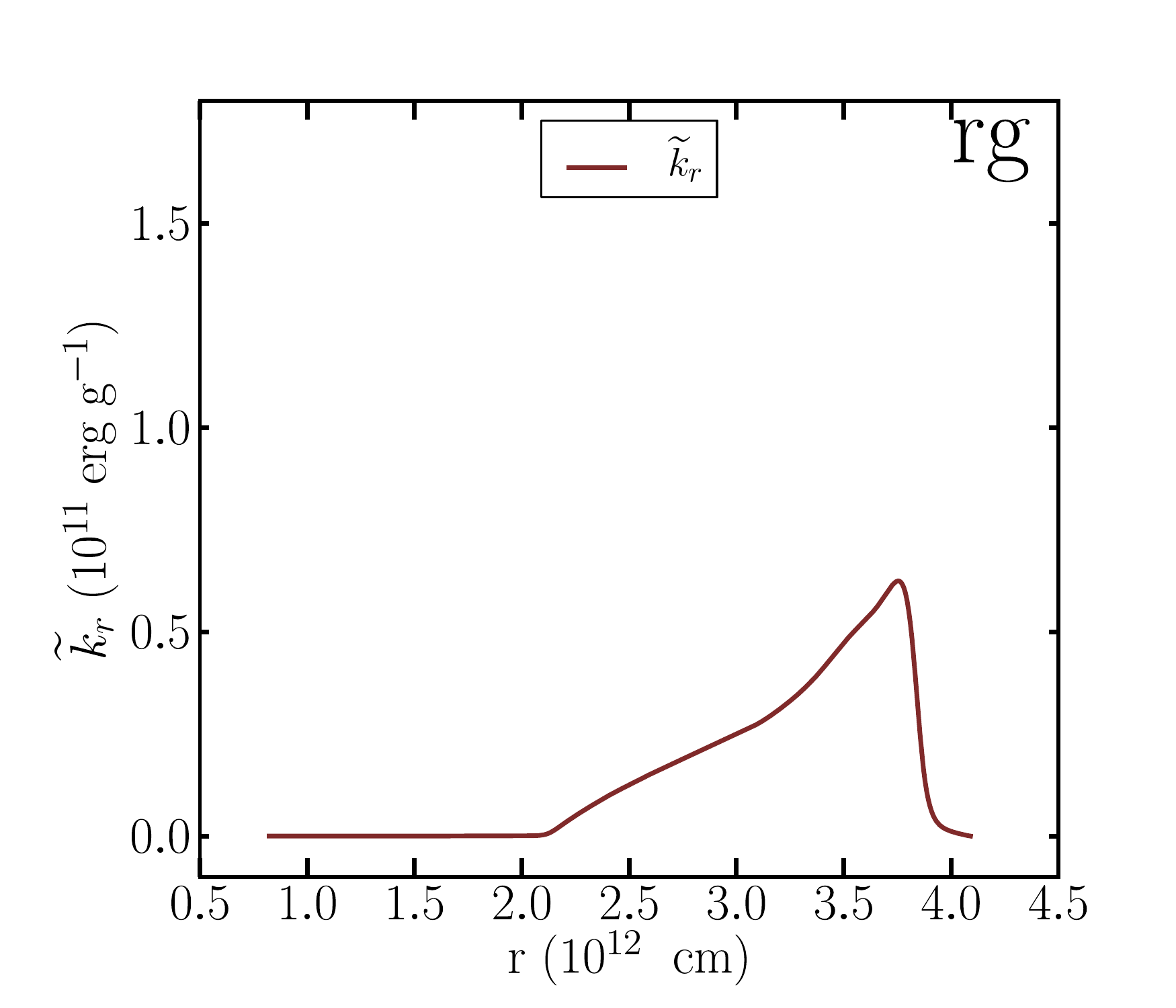}
\includegraphics[width=6.5cm]{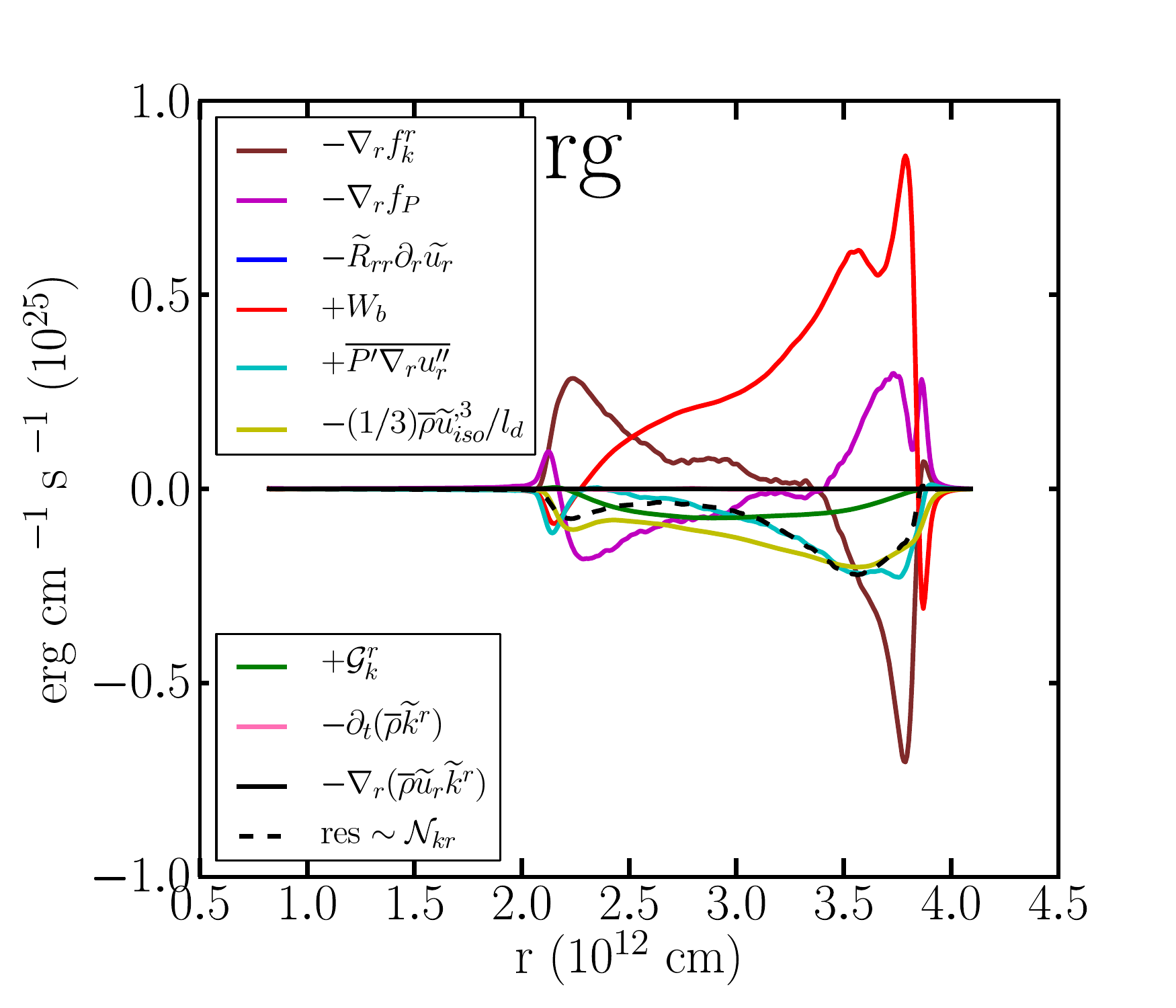}
\includegraphics[width=6.5cm]{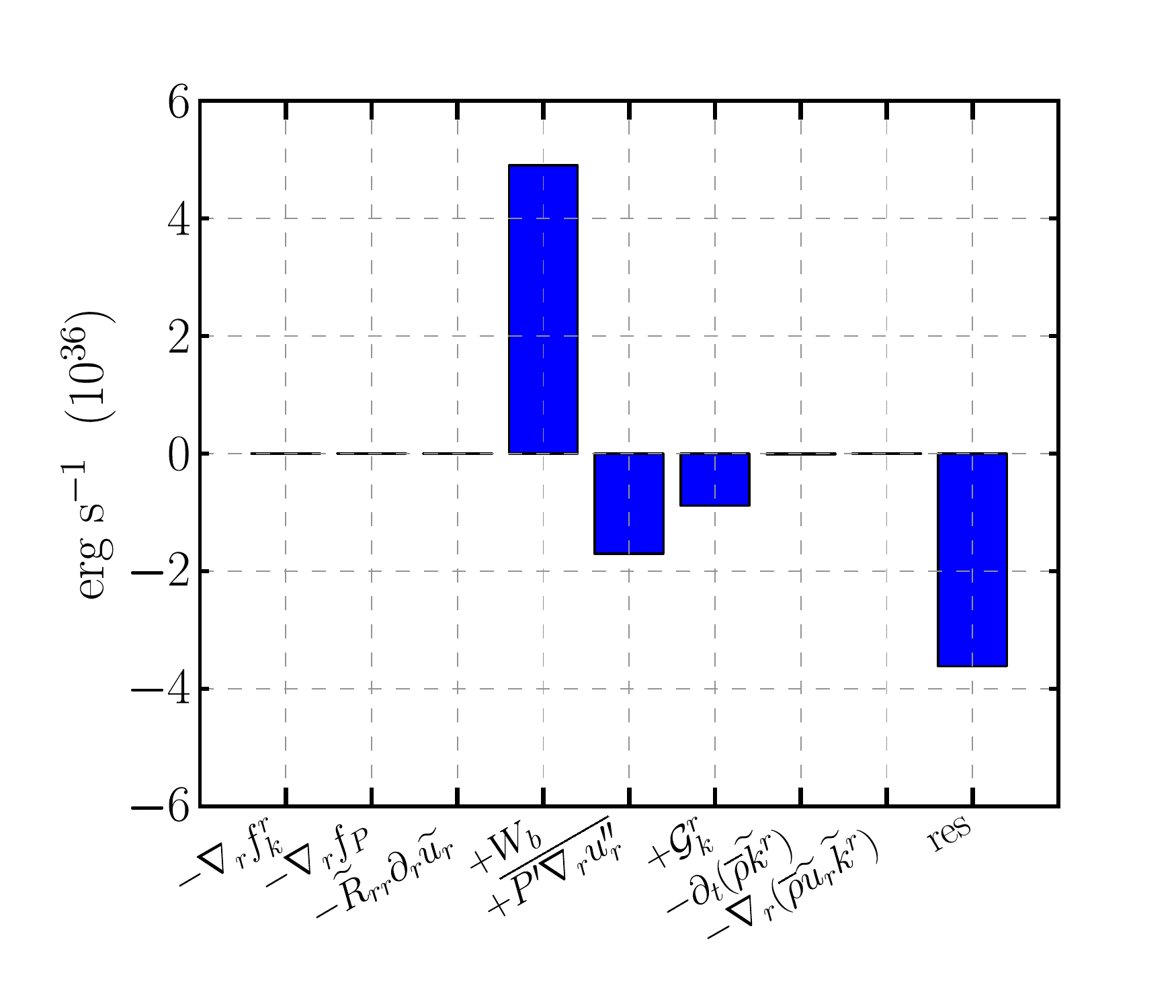}}
\caption{Turbulent radial kinetic energy equation. Model {\sf ob.3D.mr} (upper panels) and model {\sf rg.3D.mr} (lower panels). \label{fig:kr-equation}}
\end{figure}

\newpage

\subsection{Mean turbulent kinetic energy equation (horizontal part)}

\begin{align}
\av{\rho} \fav{D}_t \fav{k}^h =  &  -\nabla_r f_k^h - (\fht{R}_{\theta r}\partial_r \fht{u}_\theta + \fht{R}_{\phi r}\partial_r \fht{u}_\phi) + (\eht{P' \nabla_\theta u''_\theta} + \eht{P' \nabla_\phi u''_\phi}) + {\mathcal G_k^h} + {\mathcal N_{kh}} \label{eq:rans_ekin_h} 
\end{align}

\begin{figure}[!h]
\centerline{
\includegraphics[width=6.3cm]{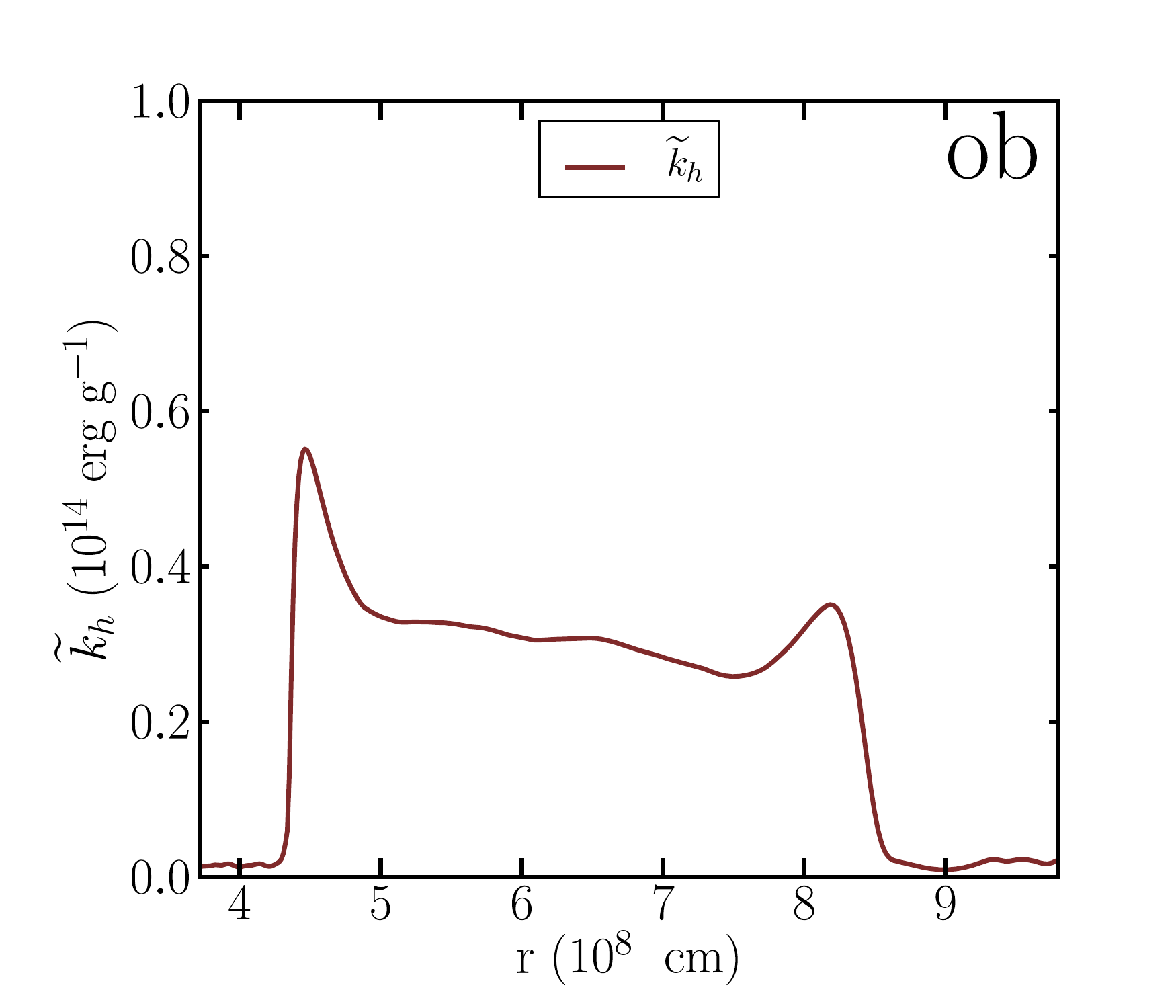}
\includegraphics[width=6.3cm]{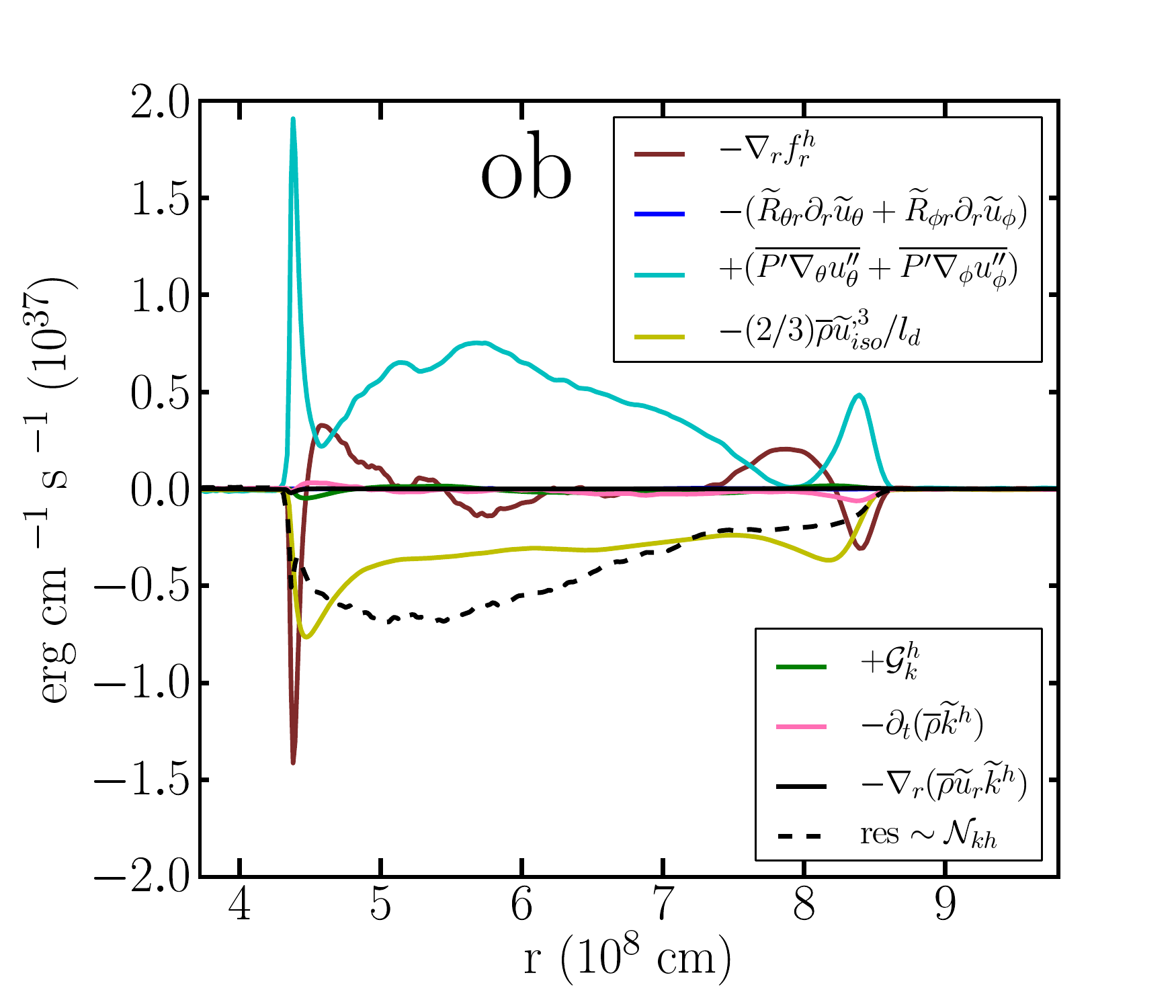}
\includegraphics[width=6.3cm]{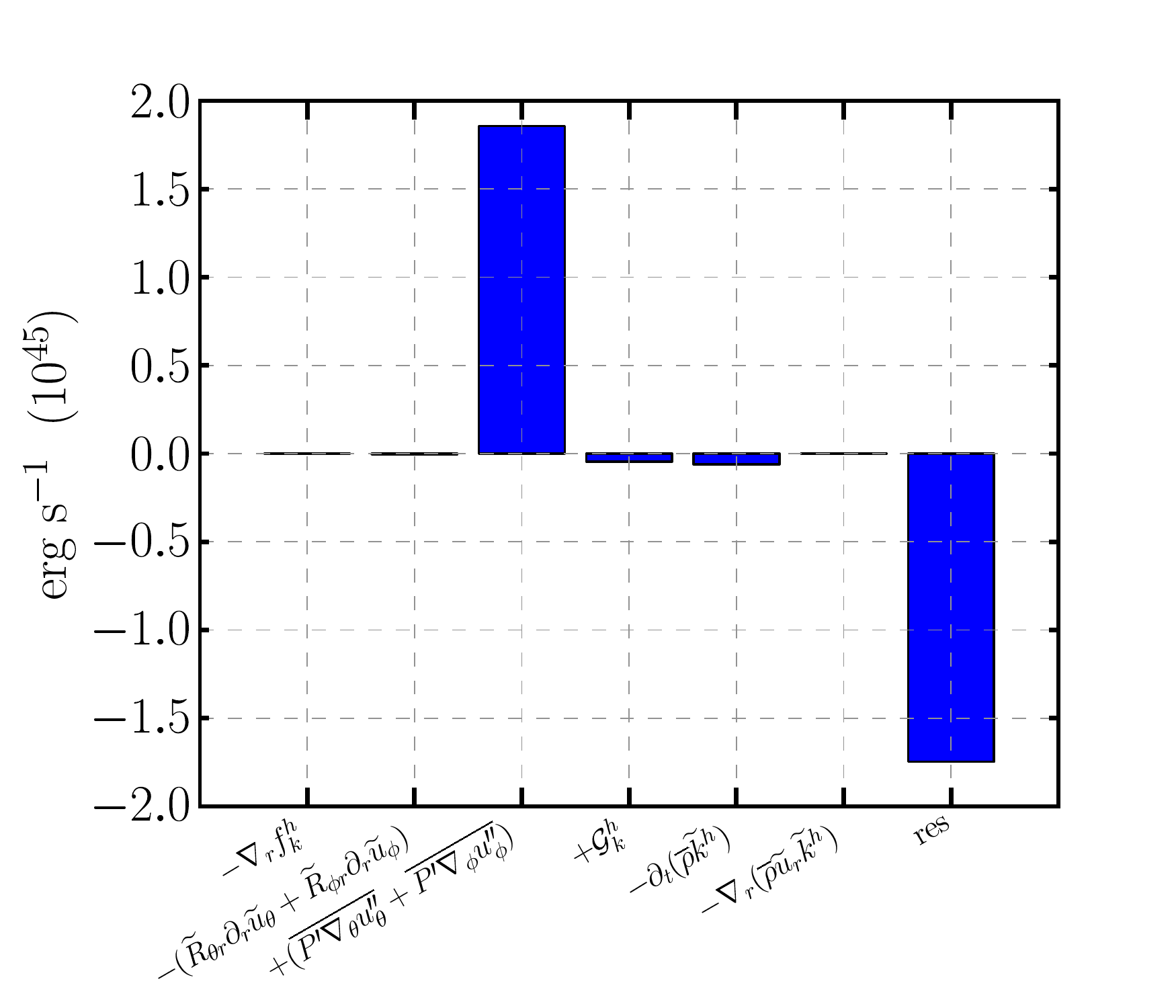}}

\centerline{
\includegraphics[width=6.3cm]{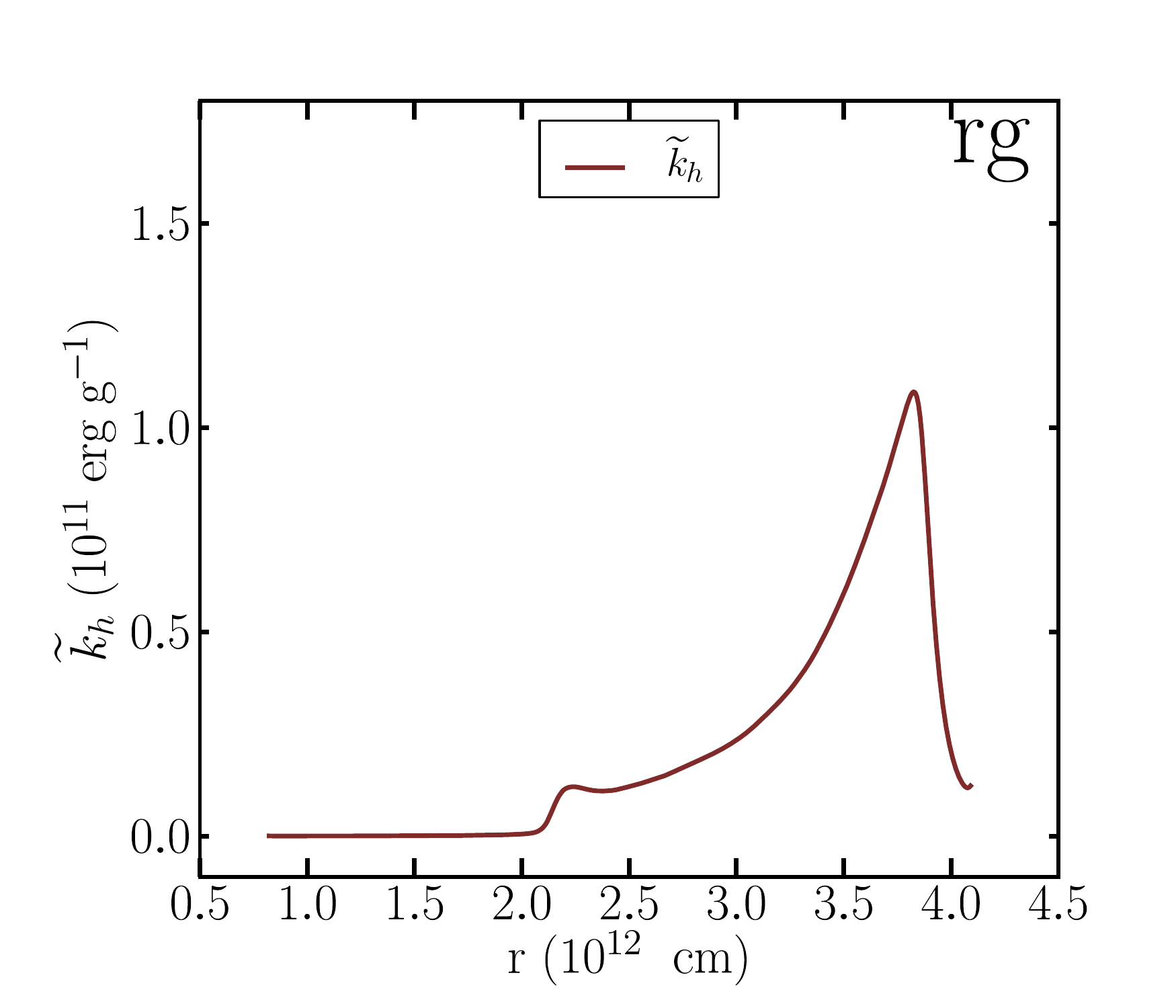}
\includegraphics[width=6.3cm]{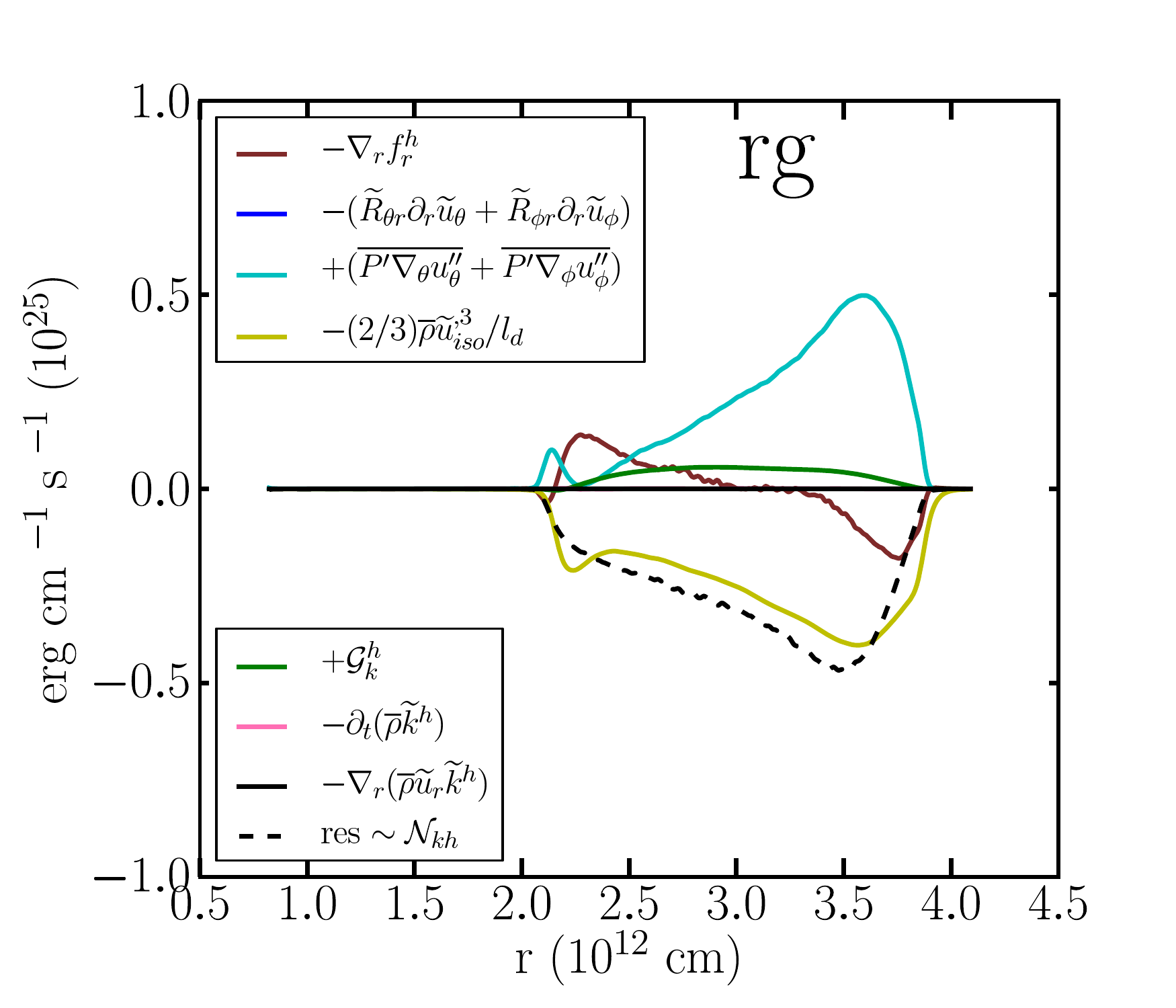}
\includegraphics[width=6.3cm]{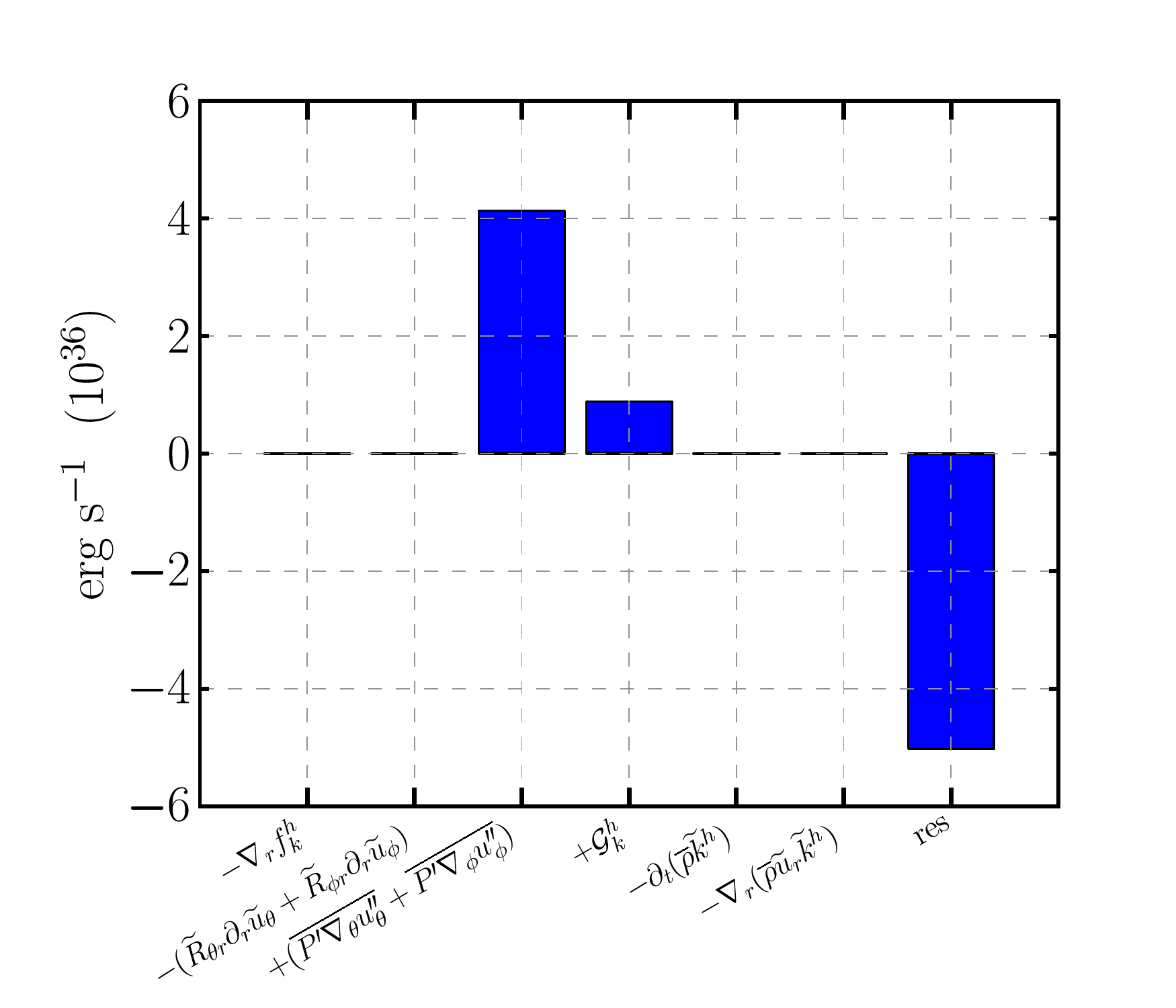}}
\caption{Turbulent horizontal kinetic energy equation. Model {\sf ob.3D.mr} (upper panels) and model {\sf rg.3D.mr} (lower panels). \label{fig:kh-equation}}
\end{figure}

\newpage

\subsection{Mean turbulent mass flux equation}

\begin{align}
\eht{\rho}\fht{D}_t \eht{u''_r} = & -(\eht{\rho'u'_ru'_r}/\eht{\rho})\partial_r\eht{\rho} + (\fht{R}_{rr}/\eht{\rho})/\partial_r \eht{\rho} - \eht{\rho} \nabla_r (\eht{u''_r} \ \eht{u''_r}) + \nabla_r \overline{\rho' u'_r u'_r} - \eht{\rho}\eht{u''_r} \nabla_r \eht{u}_r + \eht{\rho} \eht{u'_r d''} - b\partial_r \eht{P} + \eht{\rho' v \partial_r P'} +{\mathcal G_a} + {\mathcal N_a}
\end{align}

\begin{figure}[!h]
\centerline{
\includegraphics[width=6.5cm]{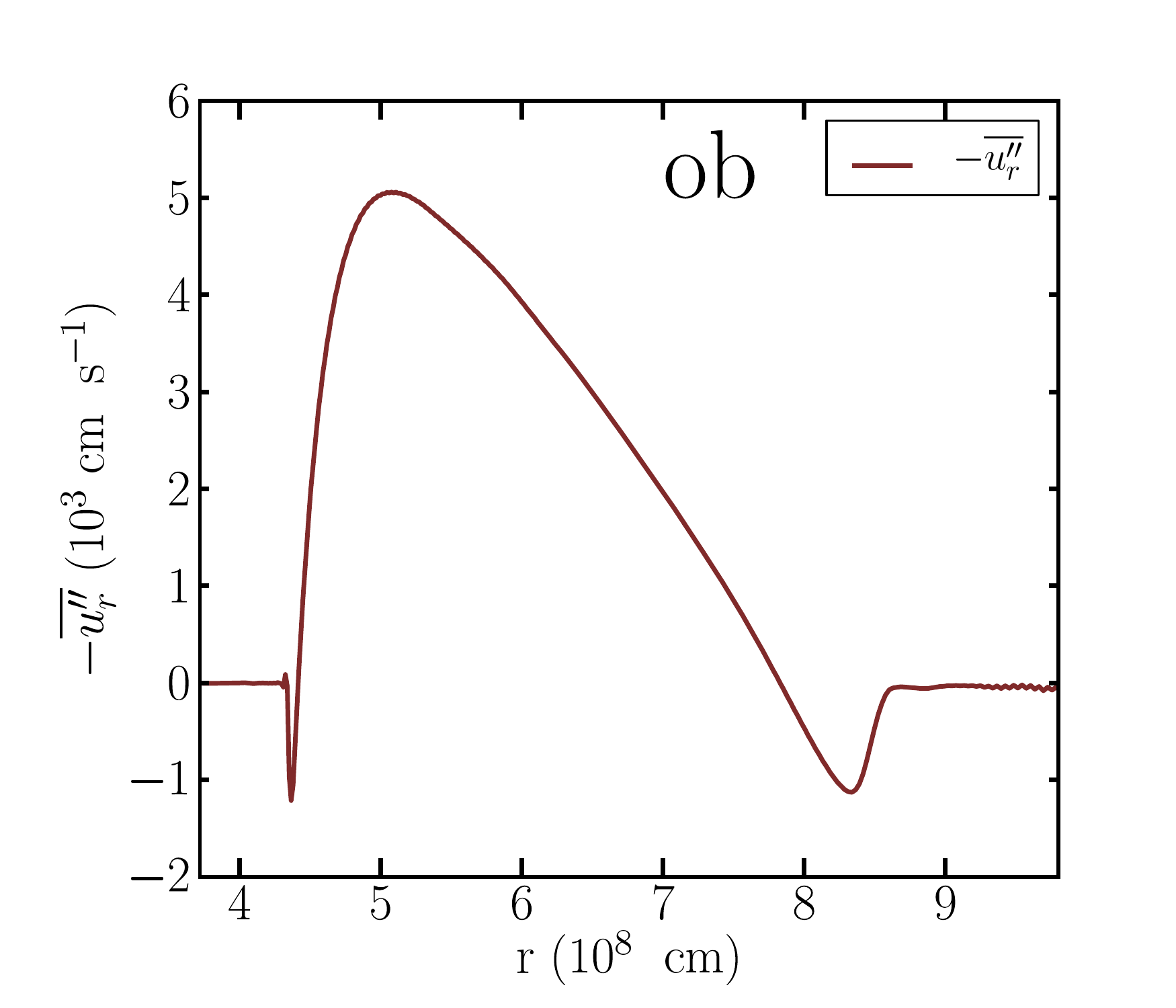}
\includegraphics[width=6.5cm]{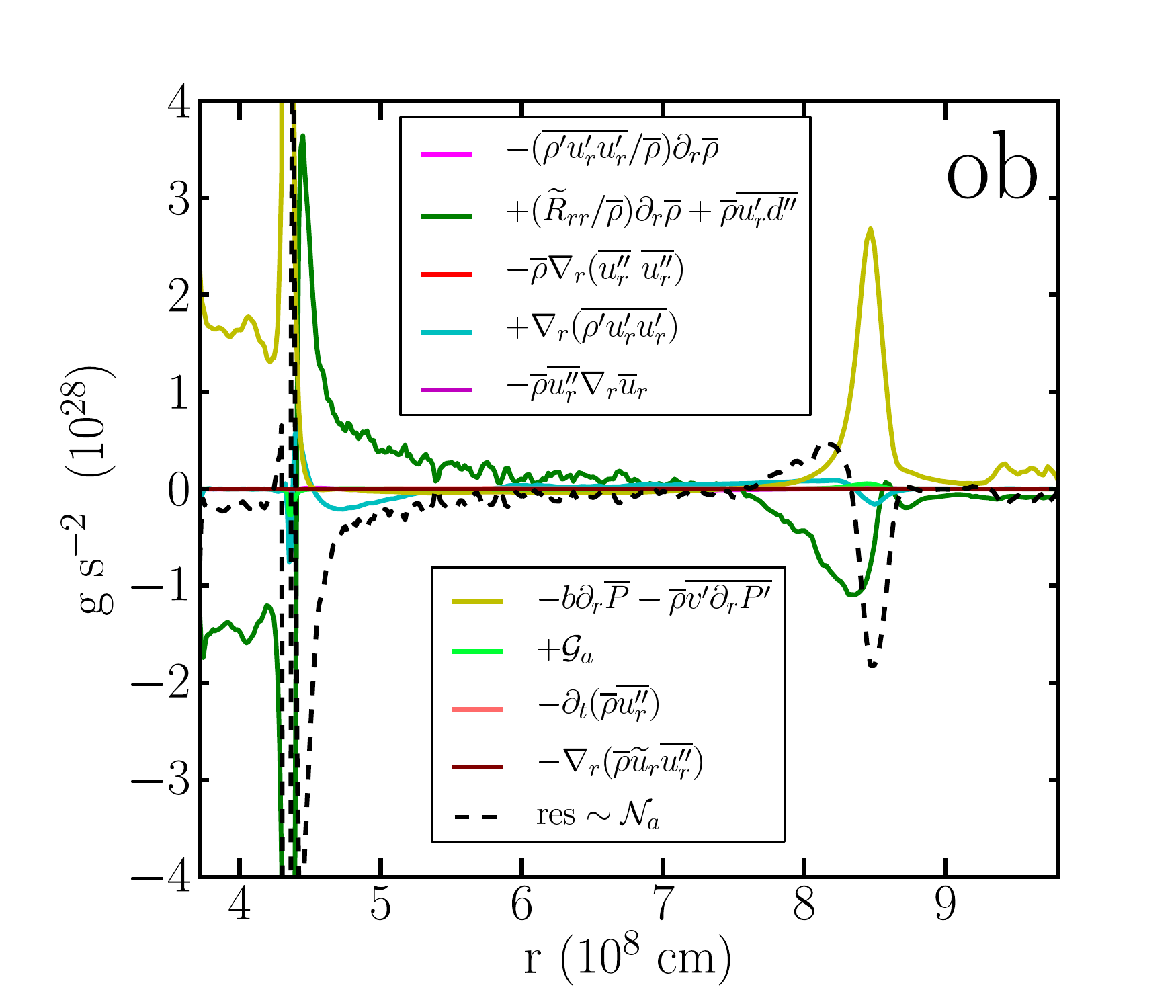}
\includegraphics[width=6.5cm]{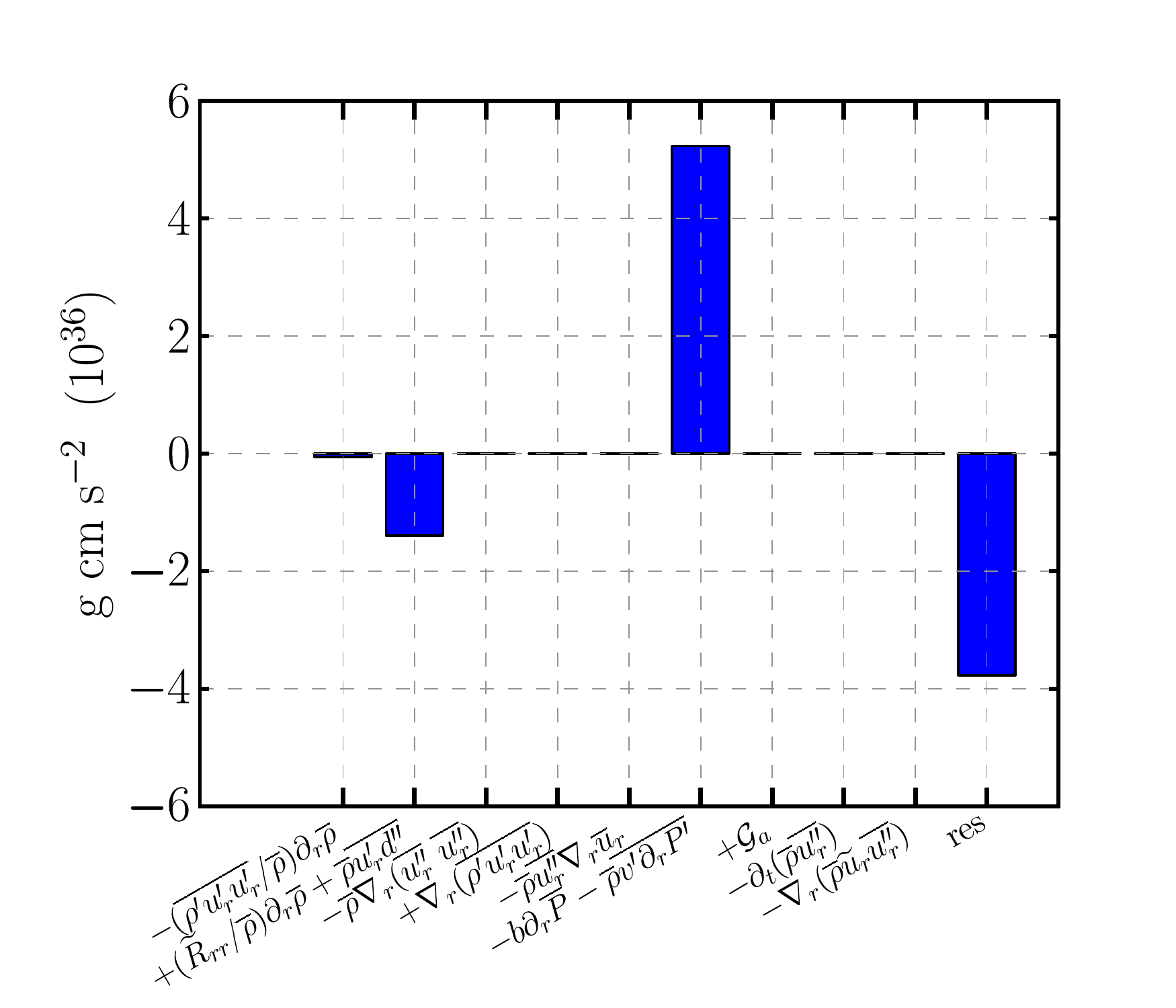}}

\centerline{
\includegraphics[width=6.5cm]{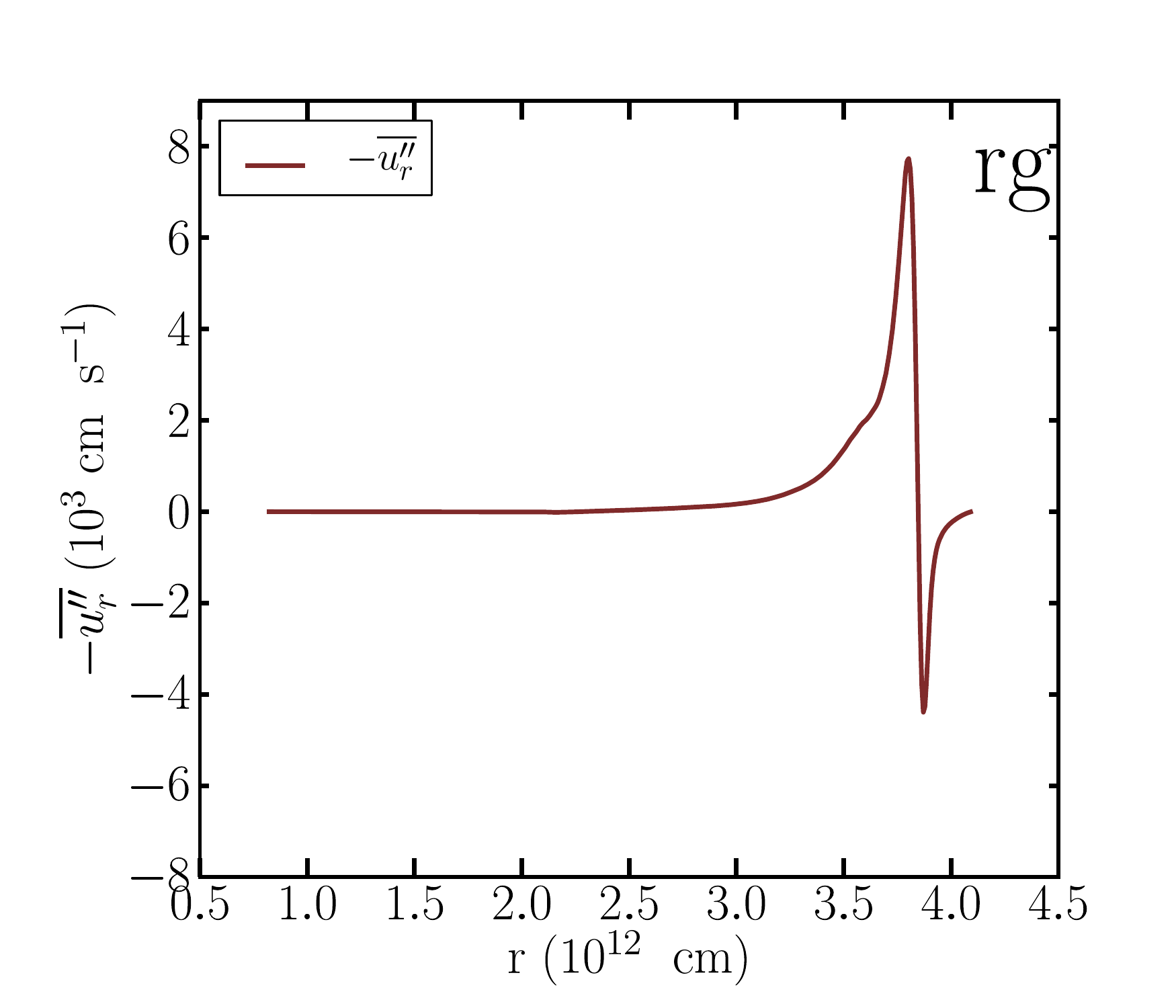}
\includegraphics[width=6.5cm]{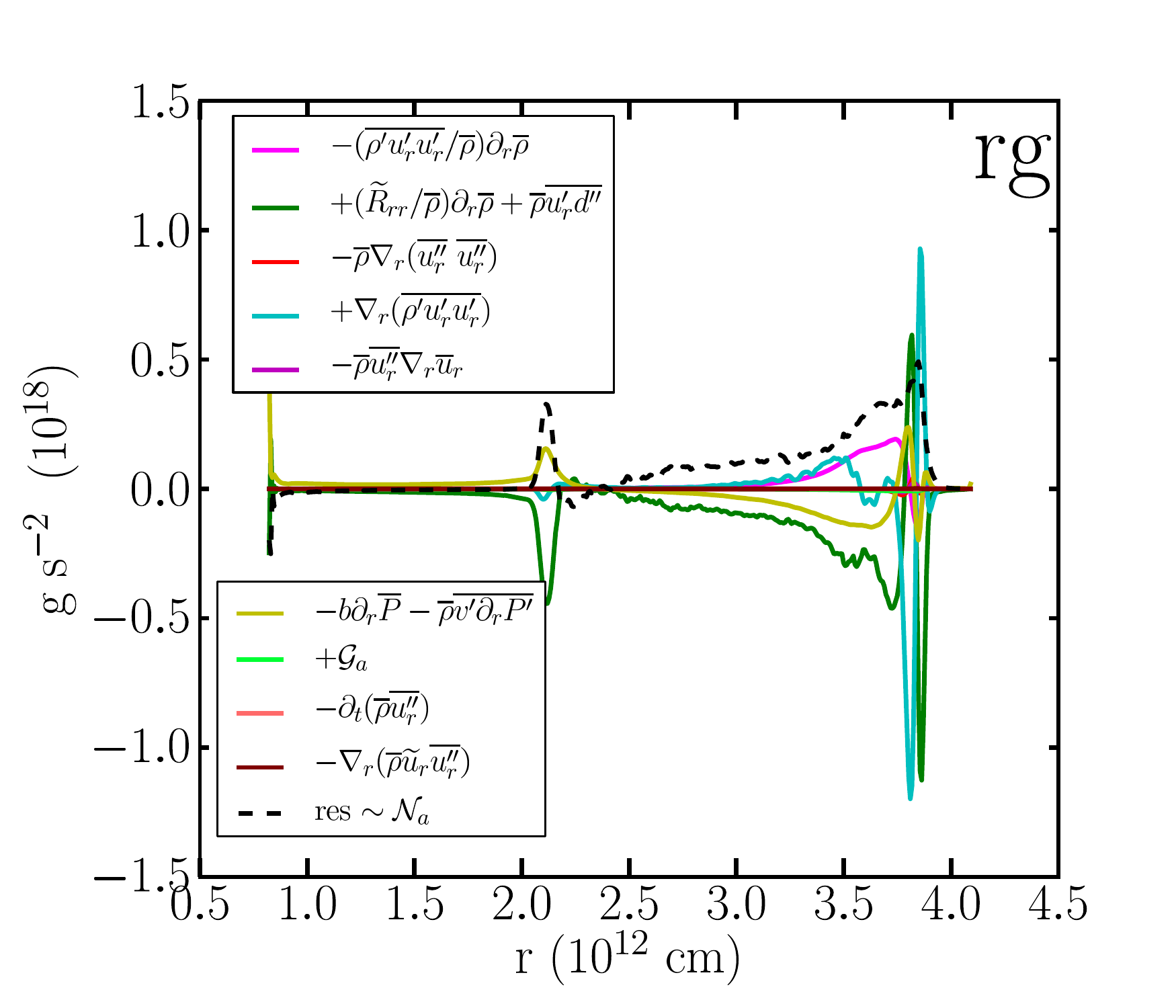}
\includegraphics[width=6.5cm]{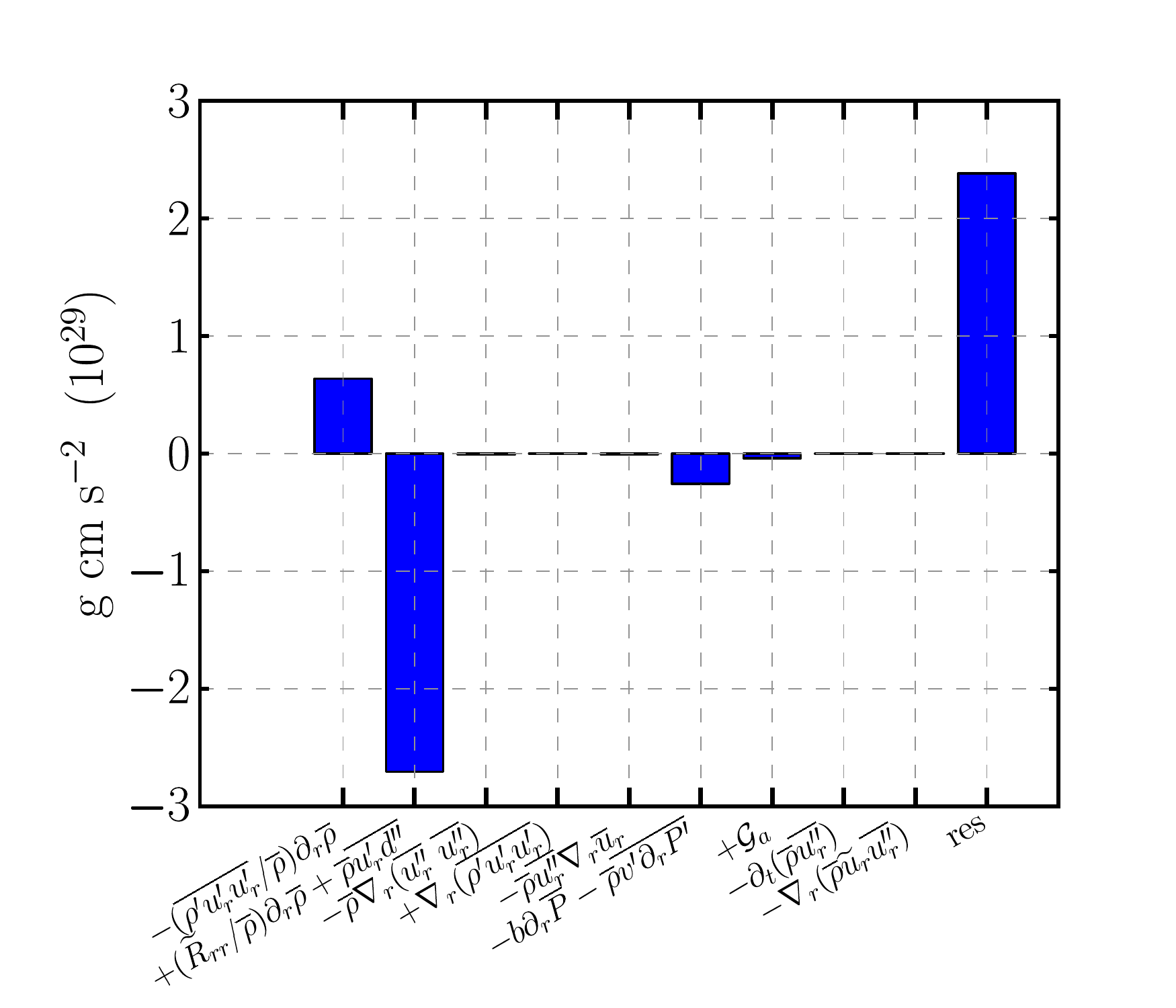}}
\caption{Turbulent mass flux equation. Model {\sf ob.3D.mr} (upper panels) and model {\sf rg.3D.mr} (lower panels). \label{fig:a-equation}}
\end{figure}

\newpage

\subsection{Mean density-specific volume covariance equation}

\begin{align}
\eht{D}_t b = &  +\eht{v} \nabla_r \eht{\rho} \eht{u''_r} -\eht{\rho}\nabla_r (\eht{u'_r v'}) + 2\eht{\rho}\eht{v'd'}+ {\mathcal N_b} \label{eq:rans_b}
\end{align}

\begin{figure}[!h]
\centerline{
\includegraphics[width=6.3cm]{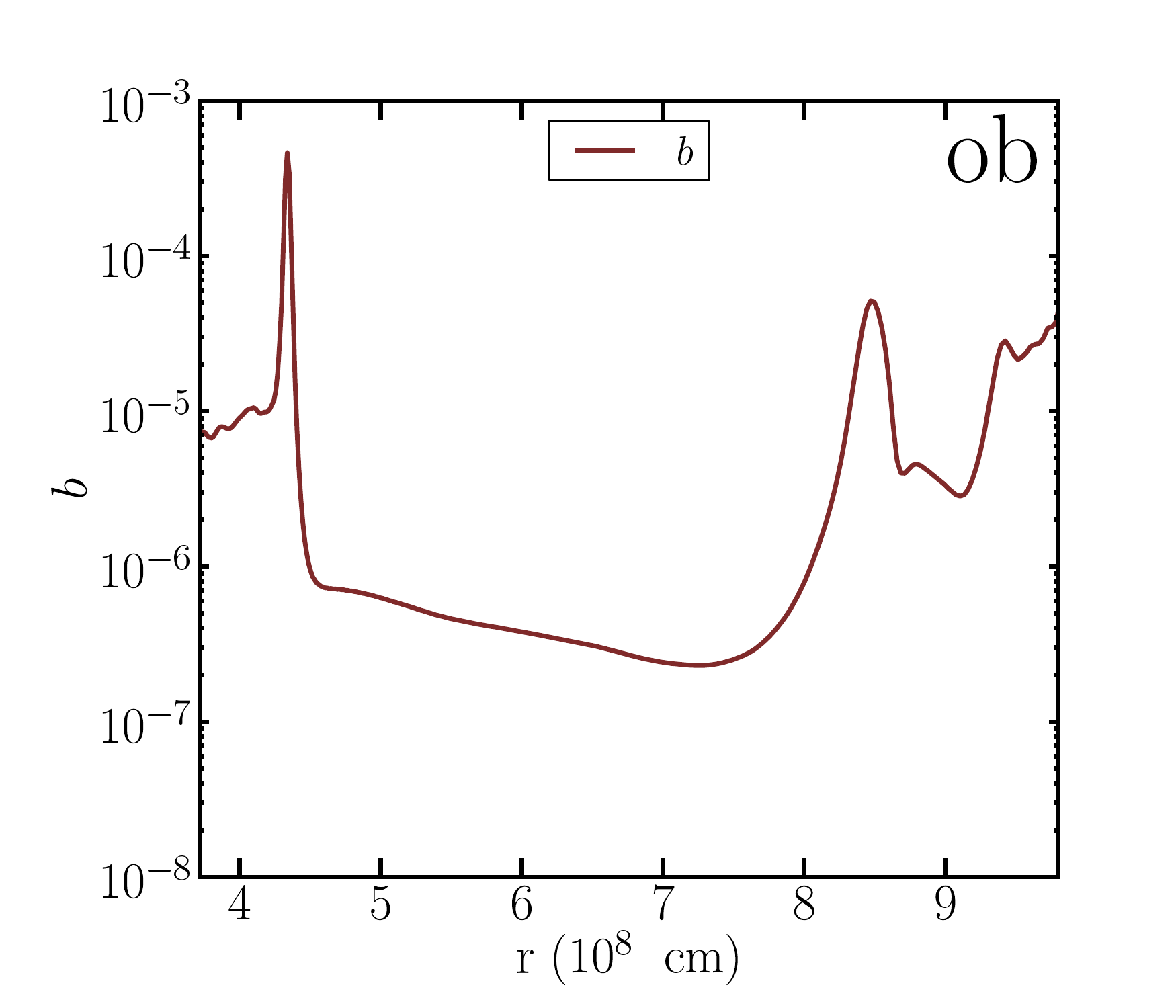}
\includegraphics[width=6.3cm]{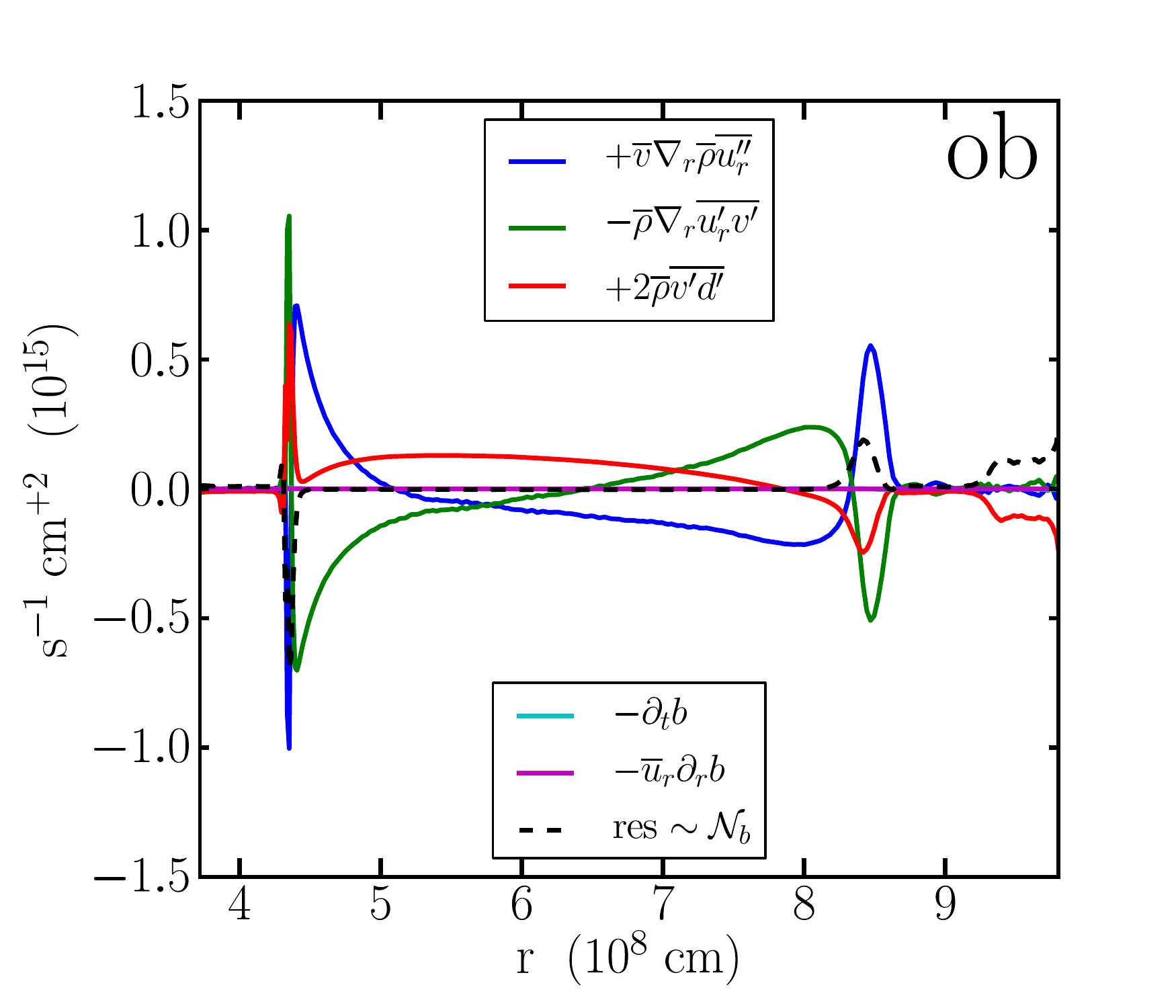}
\includegraphics[width=6.3cm]{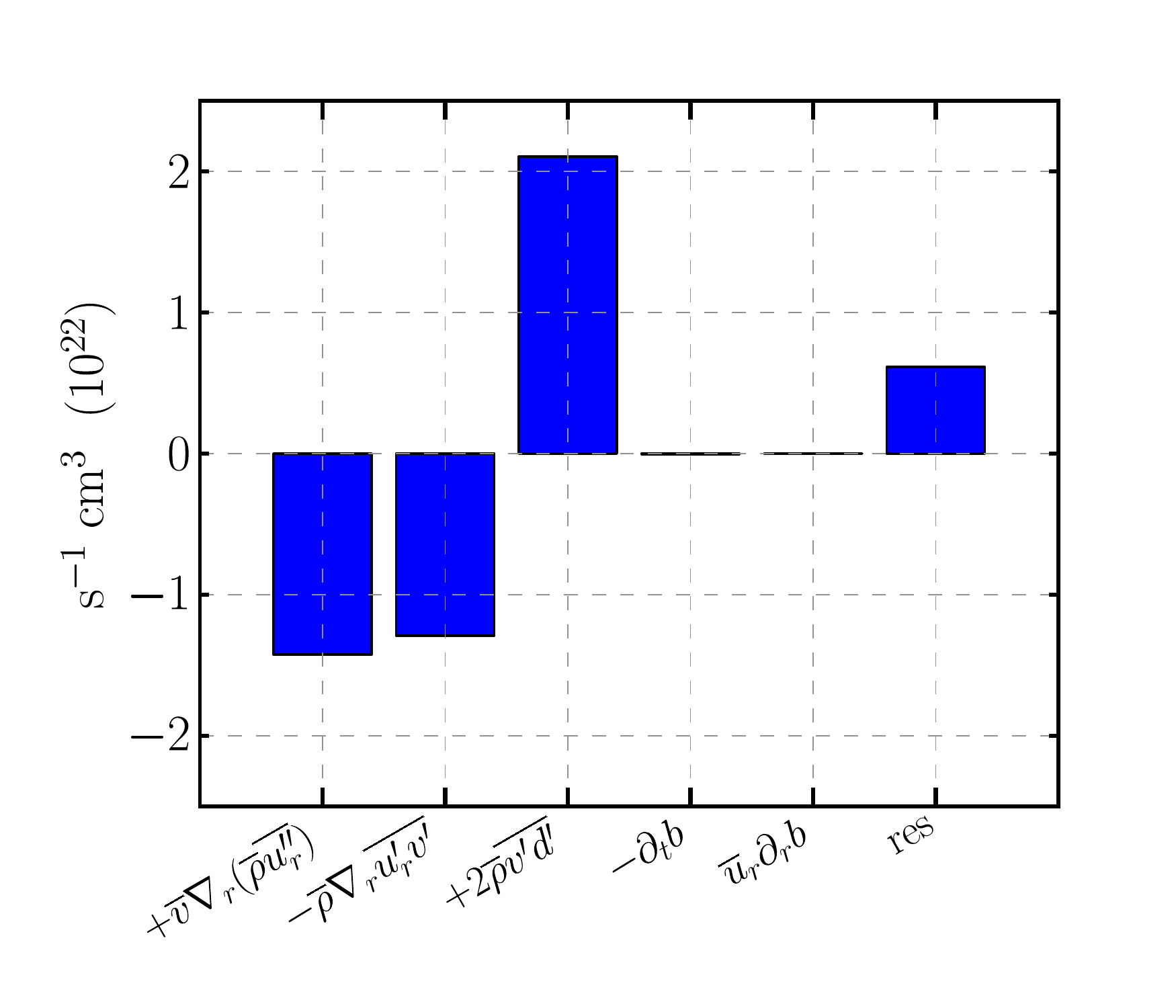}}

\centerline{
\includegraphics[width=6.3cm]{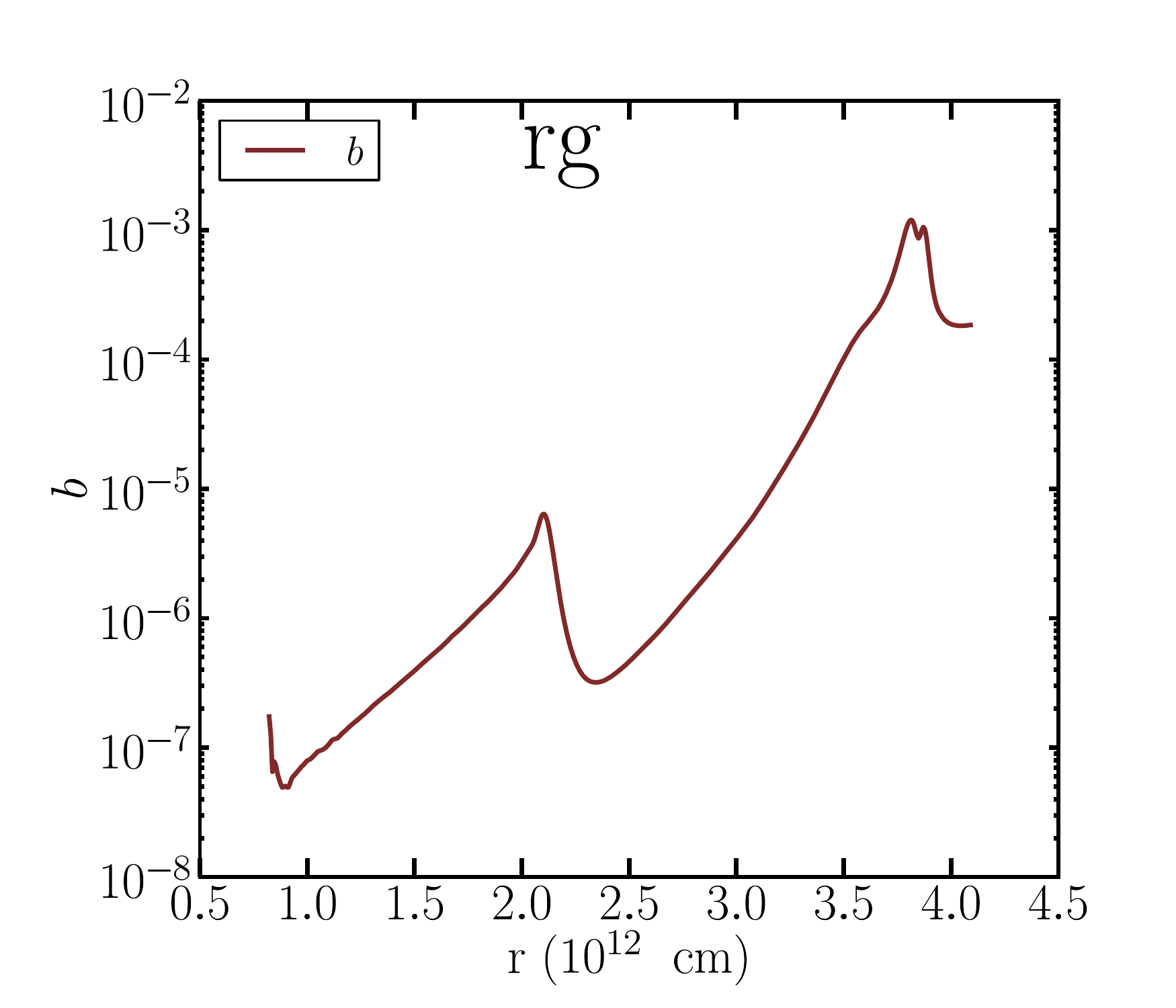}
\includegraphics[width=6.3cm]{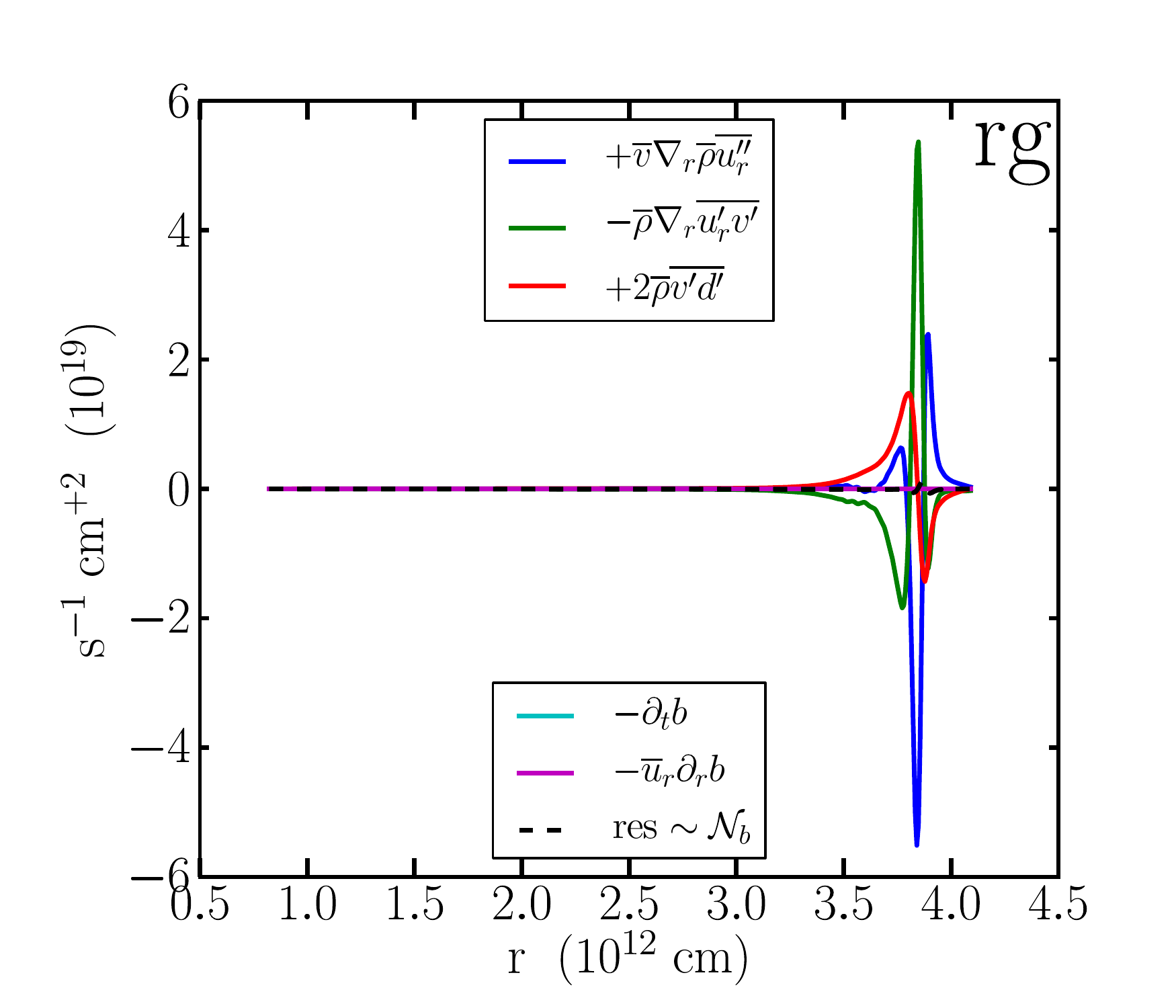}
\includegraphics[width=6.3cm]{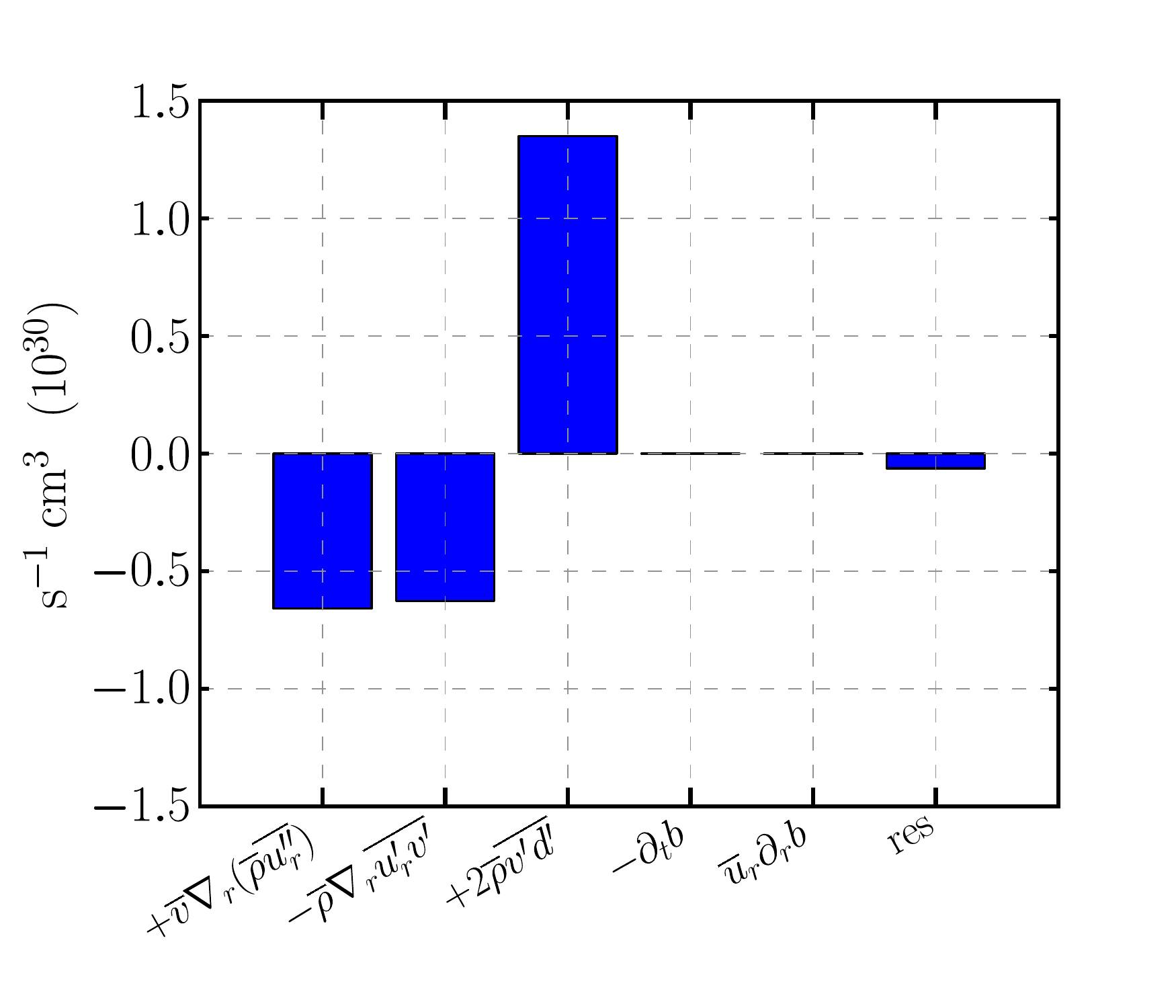}}
\caption{Density-specific volume covariance equation. Model {\sf ob.3D.mr} (upper panels) and model {\sf rg.3D.mr} (lower panels). \label{fig:b-equation}}
\end{figure}

\newpage

\subsection{Mean internal energy flux equation}

\begin{align}
\erho \fav{D}_t (f_I / \eht{\rho}) = &  {\mathcal N_{fI}} -\nabla_r f_I^r  - f_I \partial_r \fht{u}_r  - \fht{R}_{rr} \partial_r \fht{\epsilon_I} - \eht{\epsilon''_I} \partial_r \eht{P} - \eht{\epsilon''_I \partial_r P'}  - \eht{u''_r \left( P d \right)}  + \overline{u''_r ({\mathcal S} + \nabla \cdot F_T)} + {\mathcal G_I} + {\mathcal N_{fI}}\label{eq:rans_fi}
\end{align}

\begin{figure}[!h]
\centerline{
\includegraphics[width=6.5cm]{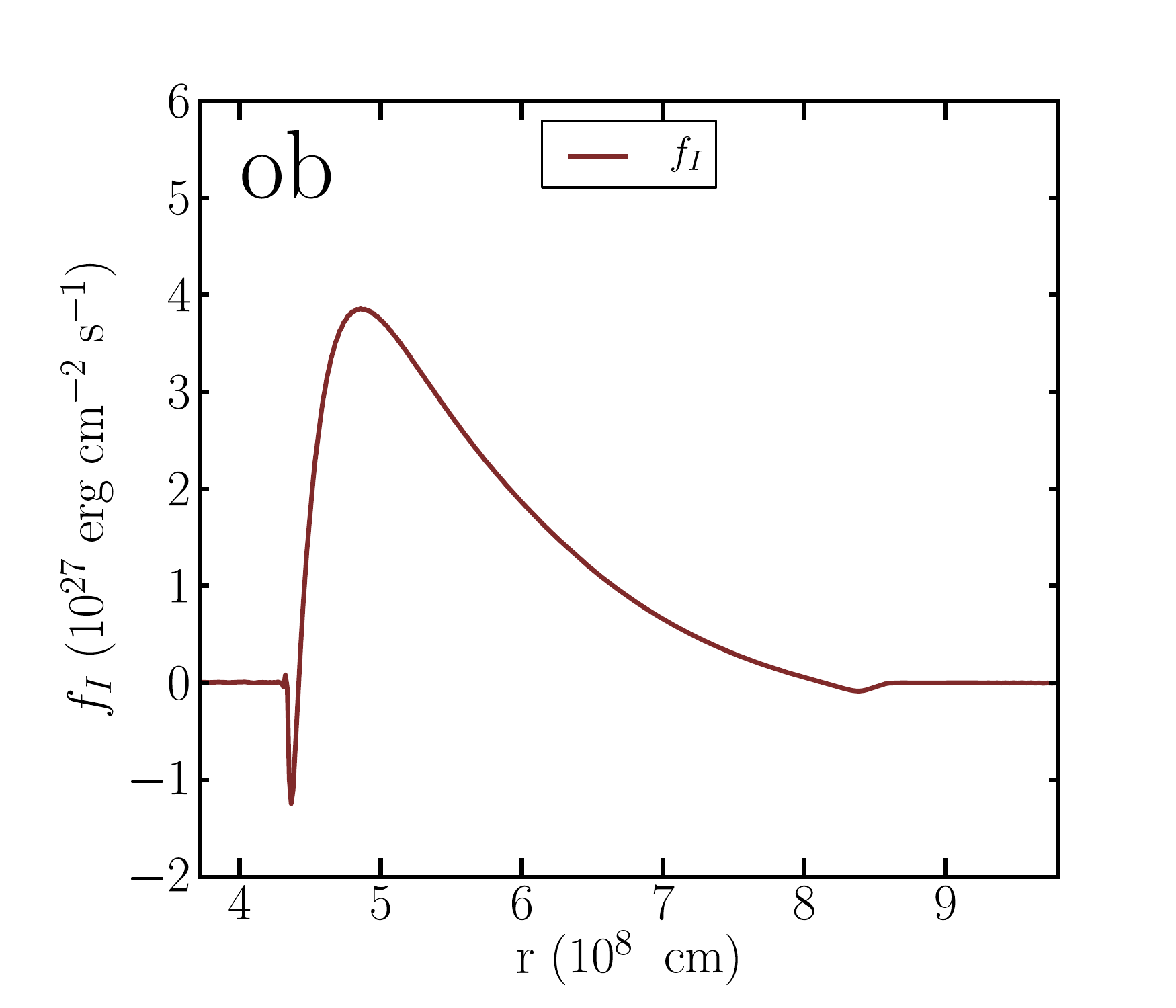}
\includegraphics[width=6.5cm]{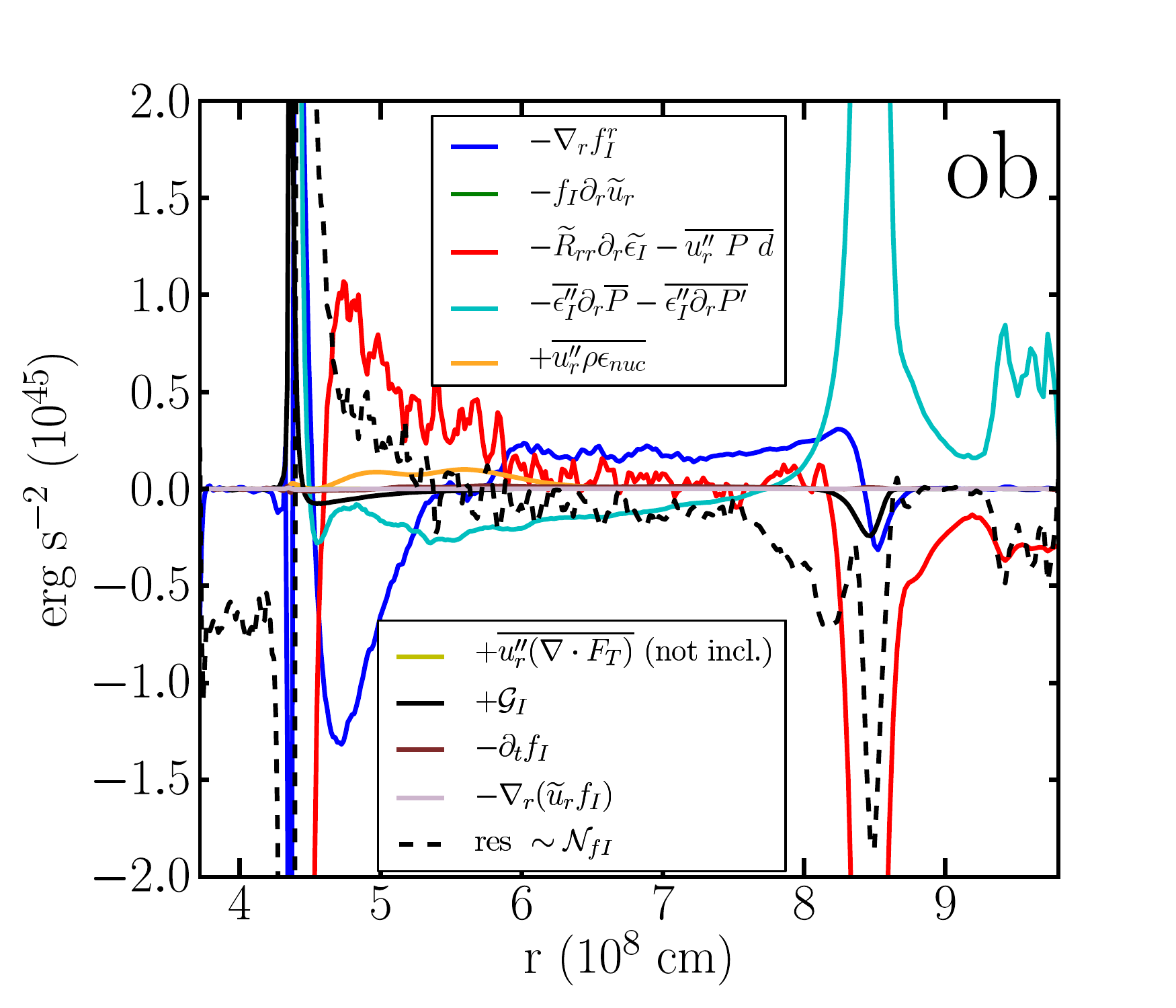}
\includegraphics[width=6.5cm]{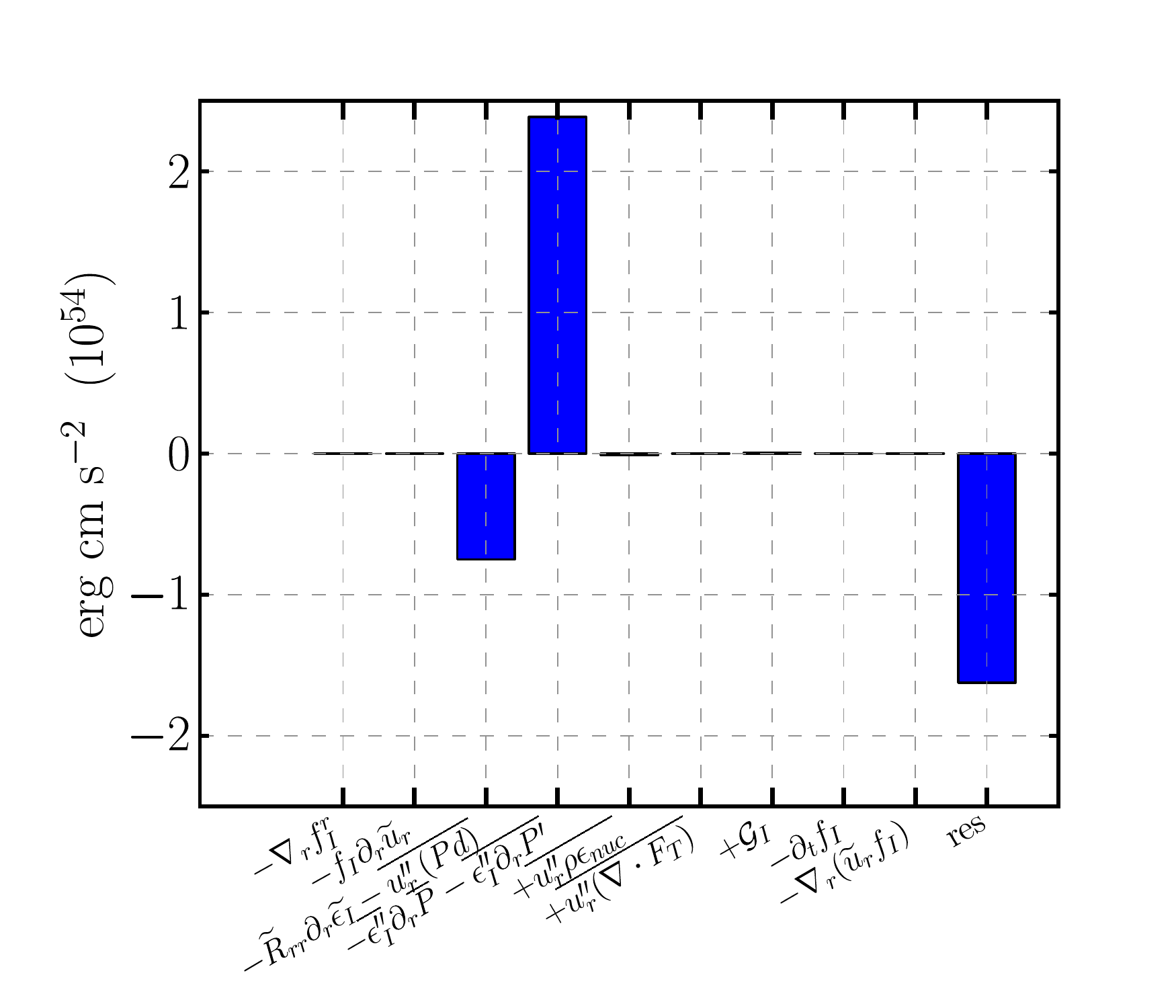}}

\centerline{
\includegraphics[width=6.5cm]{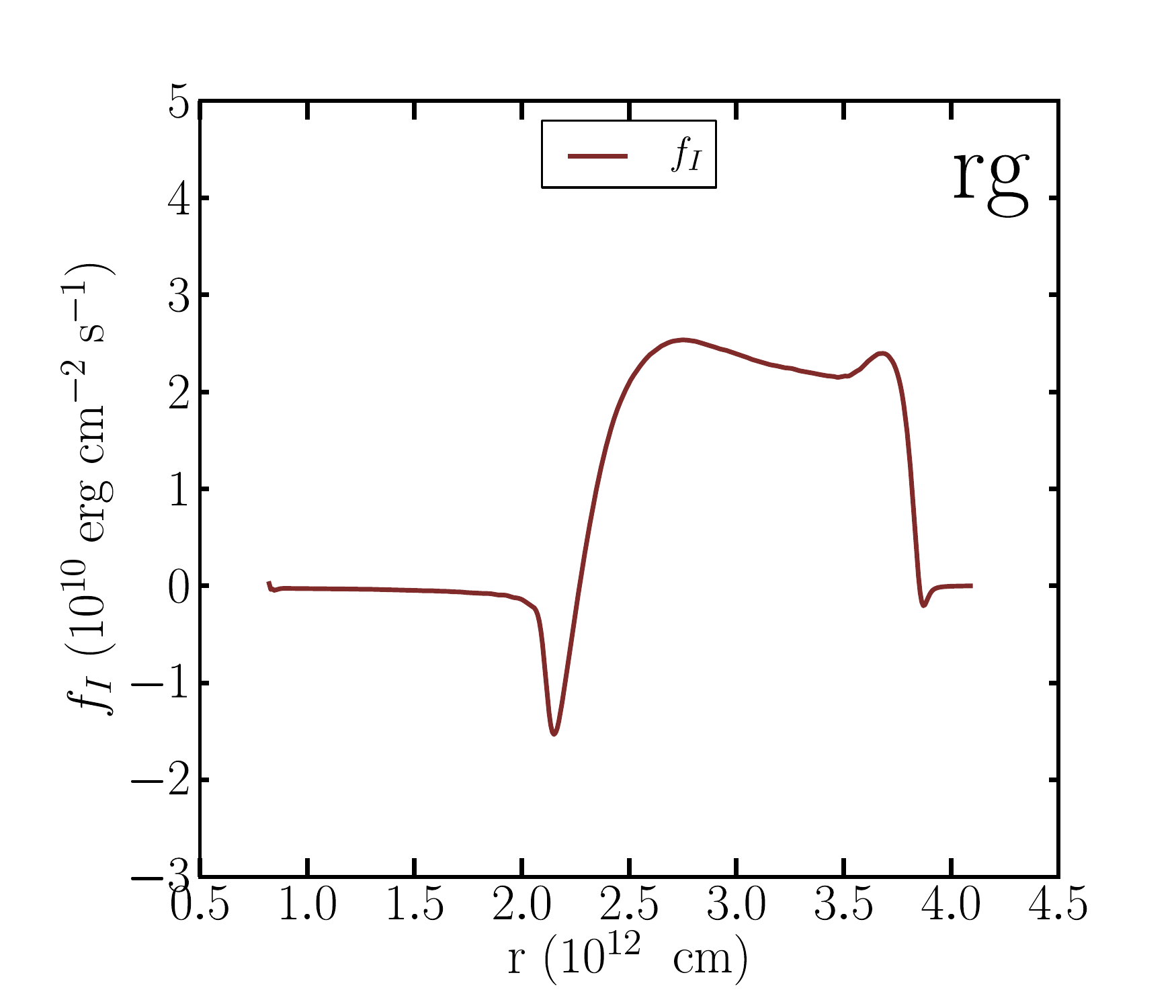}
\includegraphics[width=6.5cm]{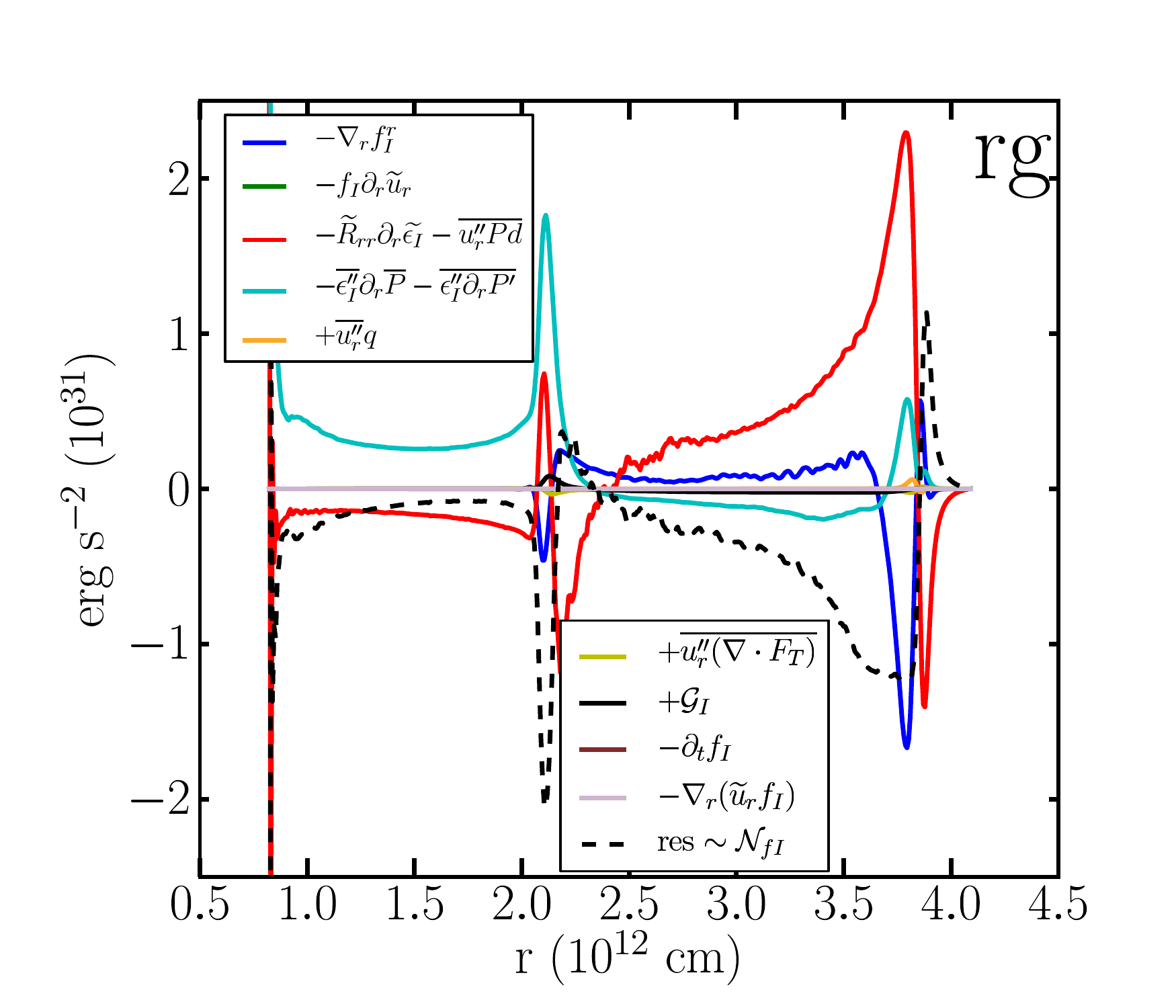}
\includegraphics[width=6.5cm]{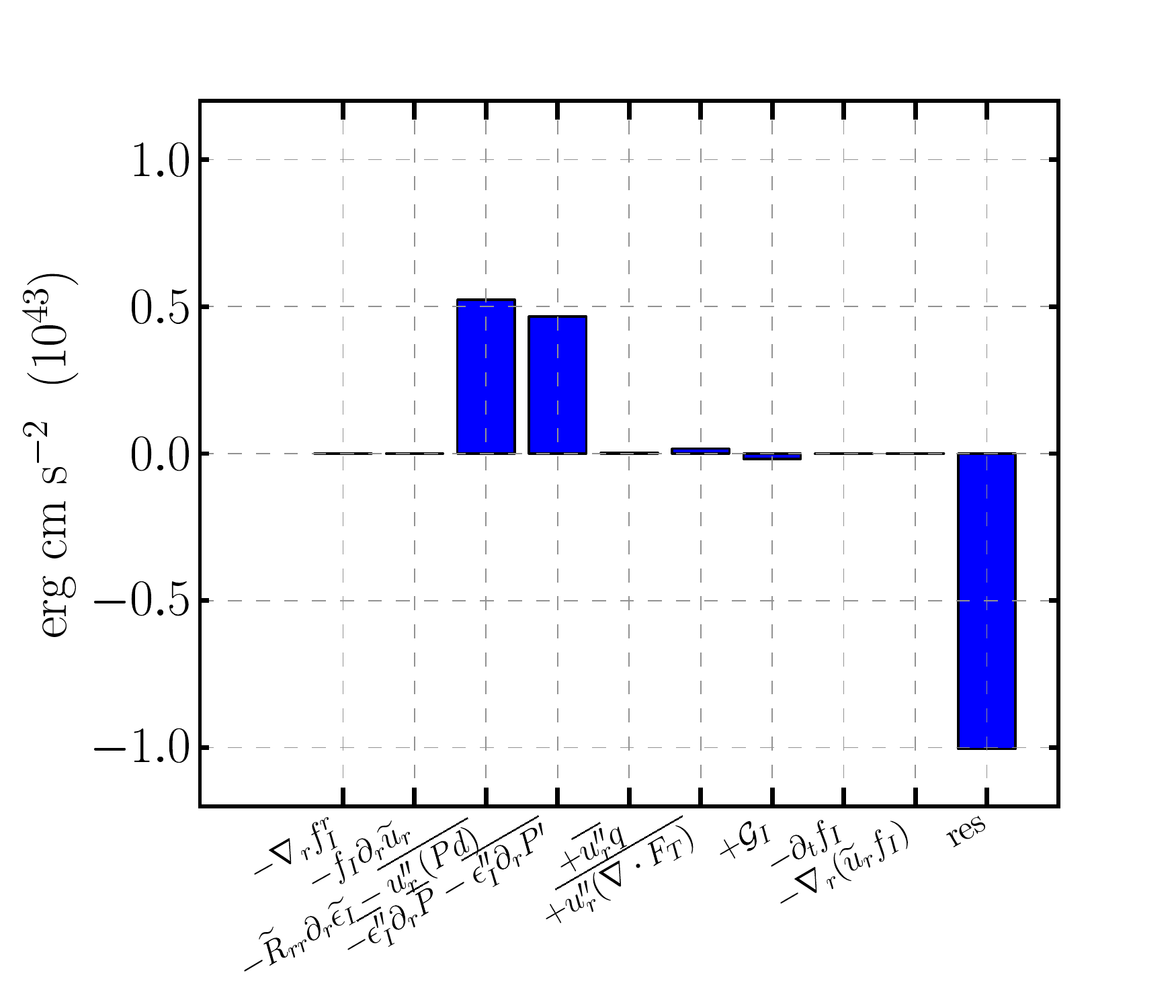}}
\caption{Mean internal energy flux equation. Model {\sf ob.3D.mr} (upper panels) and model {\sf rg.3D.mr} (lower panels). \label{fig:fi-equation}}
\end{figure}

\newpage

\subsection{Mean entropy flux equation}

\begin{align}
\erho \fav{D}_t (f_s / \eht{\rho}) = &  -\nabla_r f_s^r - f_s \partial_r \fht{u}_r - \fht{R}_{rr} \partial_r \fht{s} -\eht{s''}\partial_r \eht{P} - \eht{s''\partial_r P'} + \eht{u''_r ( {\mathcal S} + \nabla \cdot F_T)  / T} + {\mathcal G_s} + {\mathcal N_{fs}}  \label{eq:rans_fs}
\end{align}

\begin{figure}[!h]
\centerline{
\includegraphics[width=6.5cm]{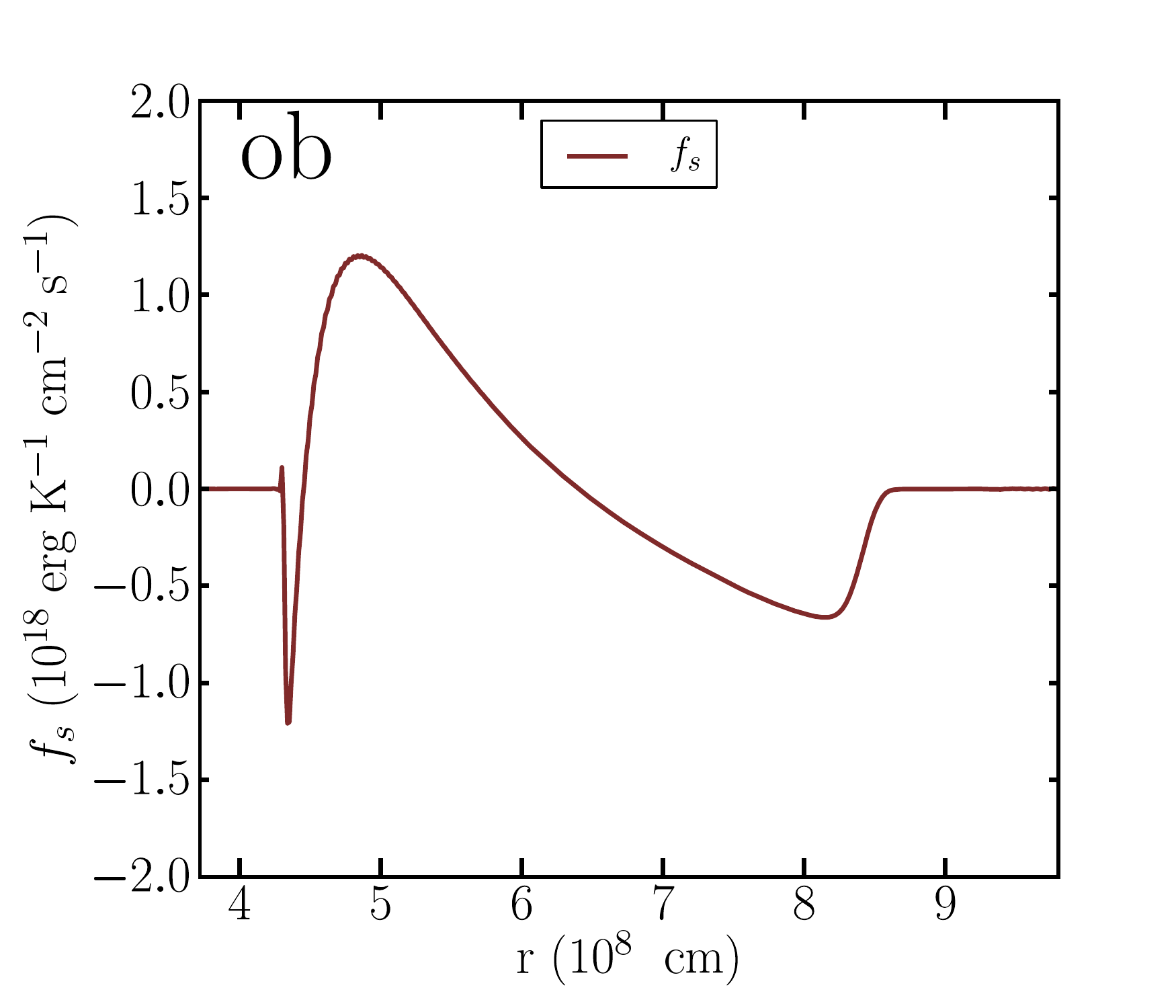}
\includegraphics[width=6.5cm]{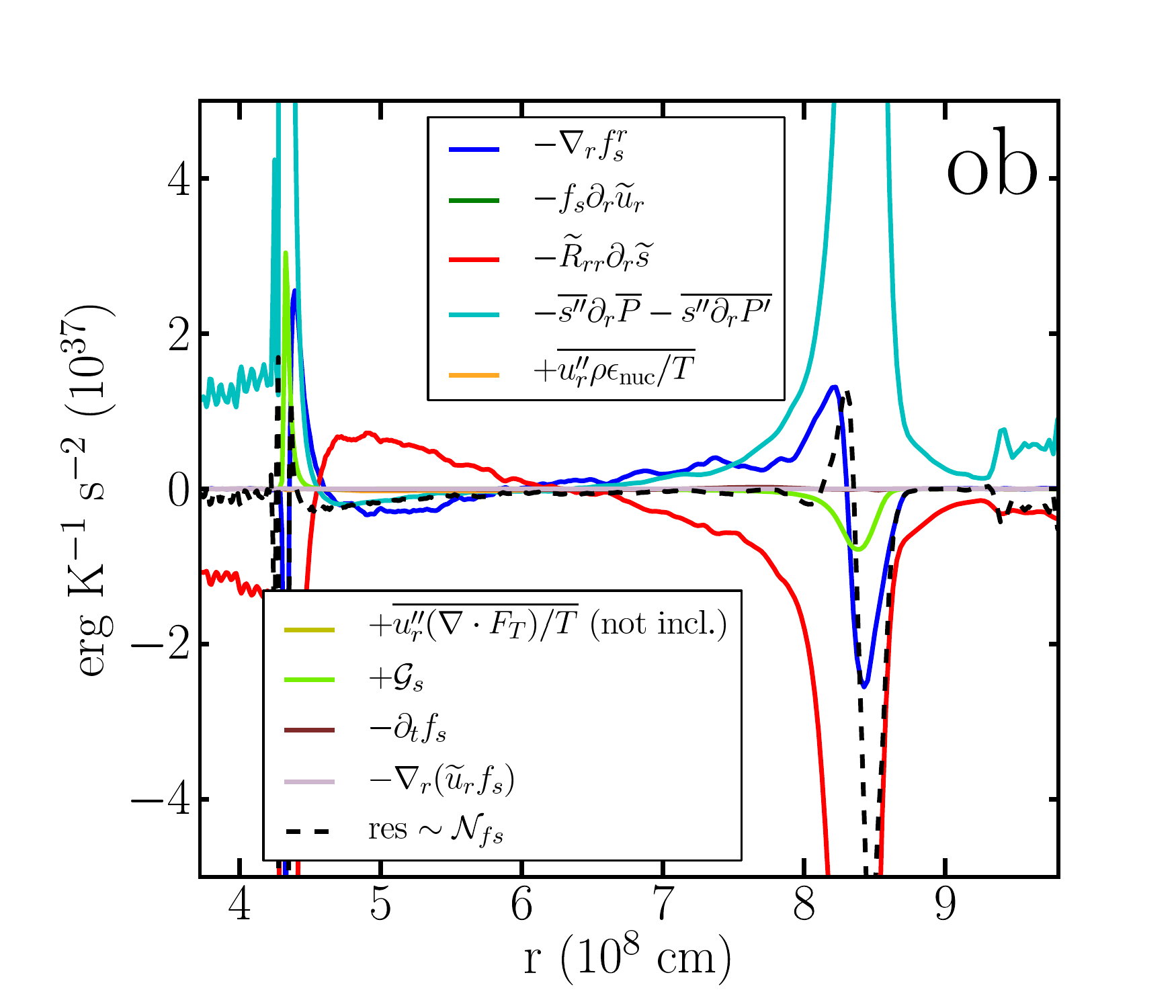}
\includegraphics[width=6.5cm]{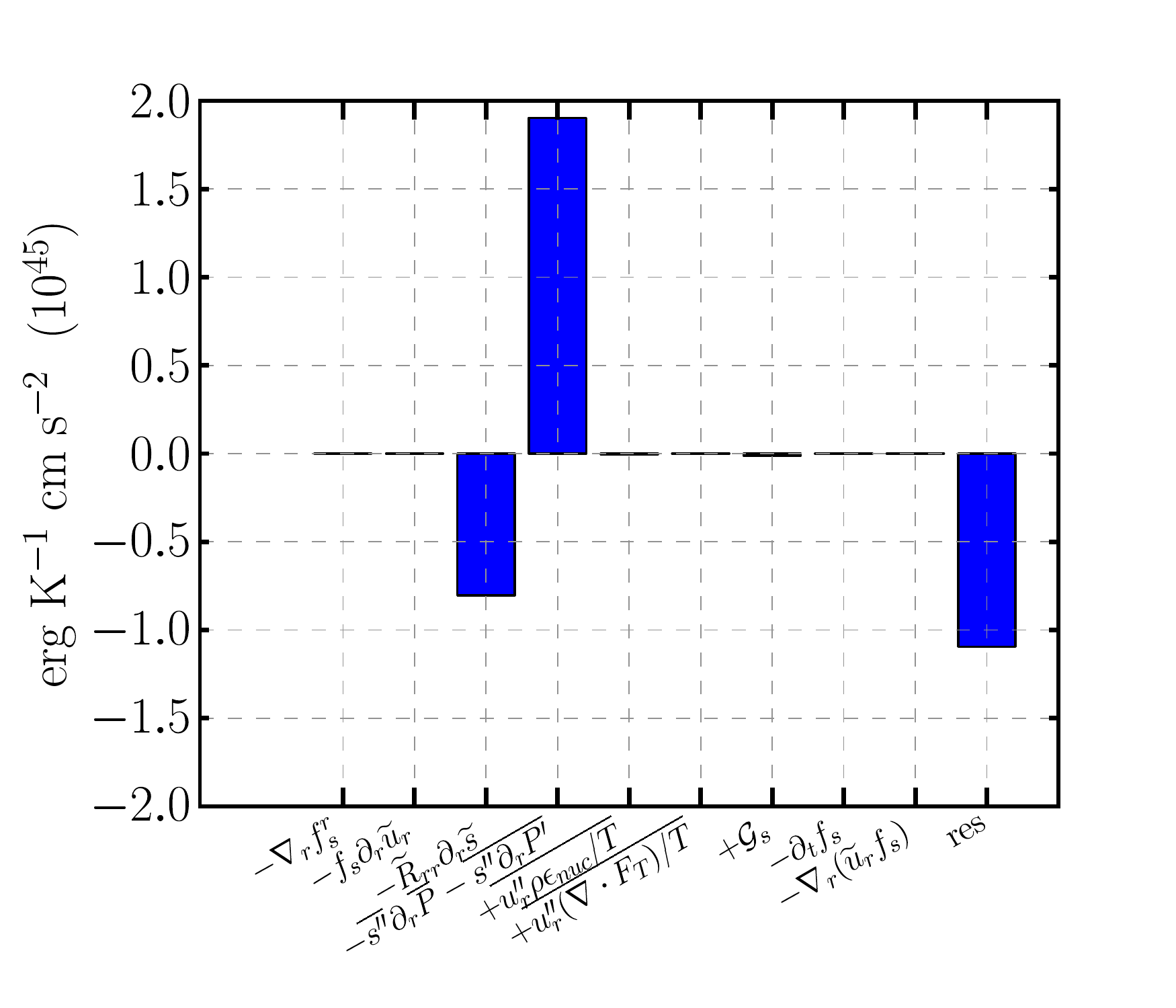}}

\centerline{
\includegraphics[width=6.5cm]{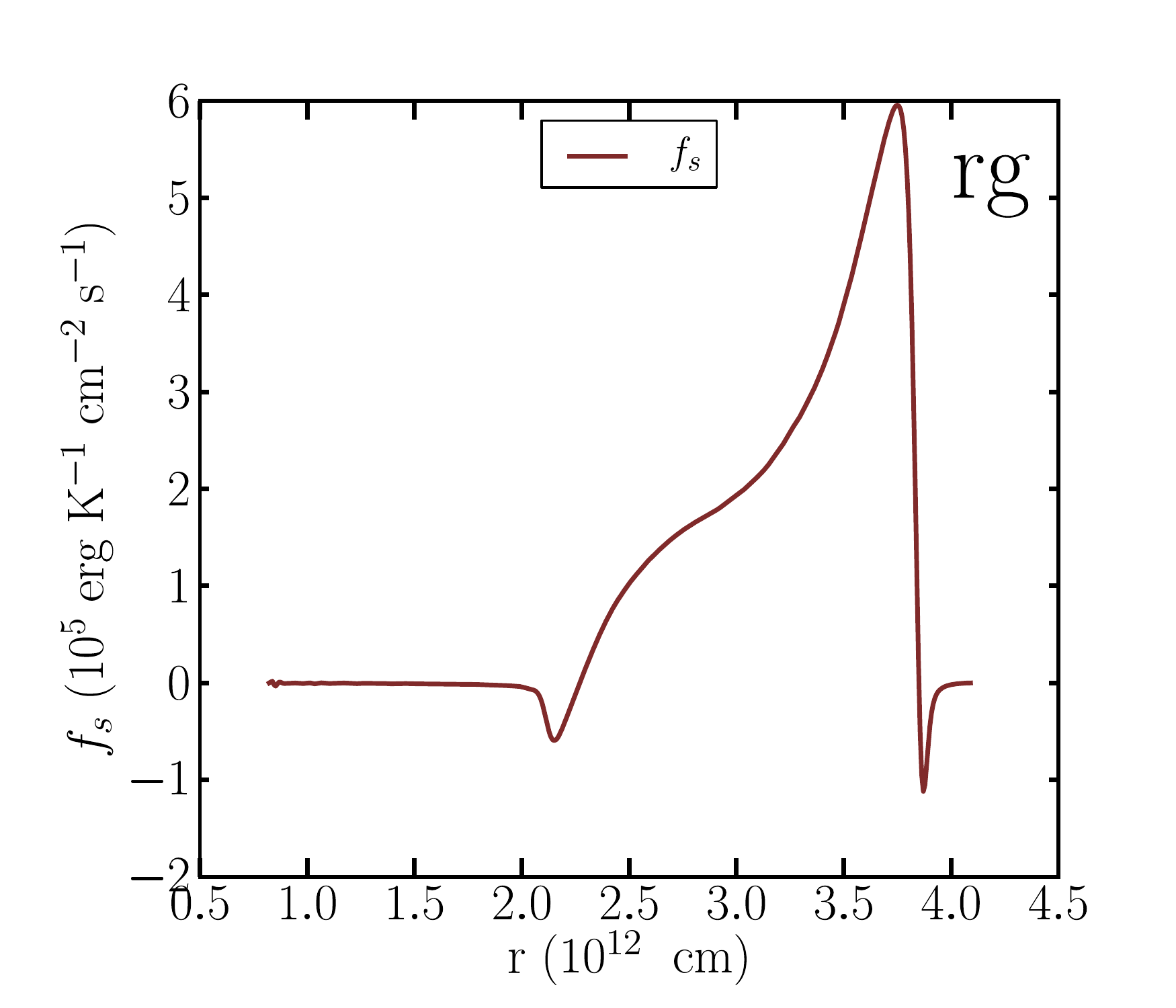}
\includegraphics[width=6.5cm]{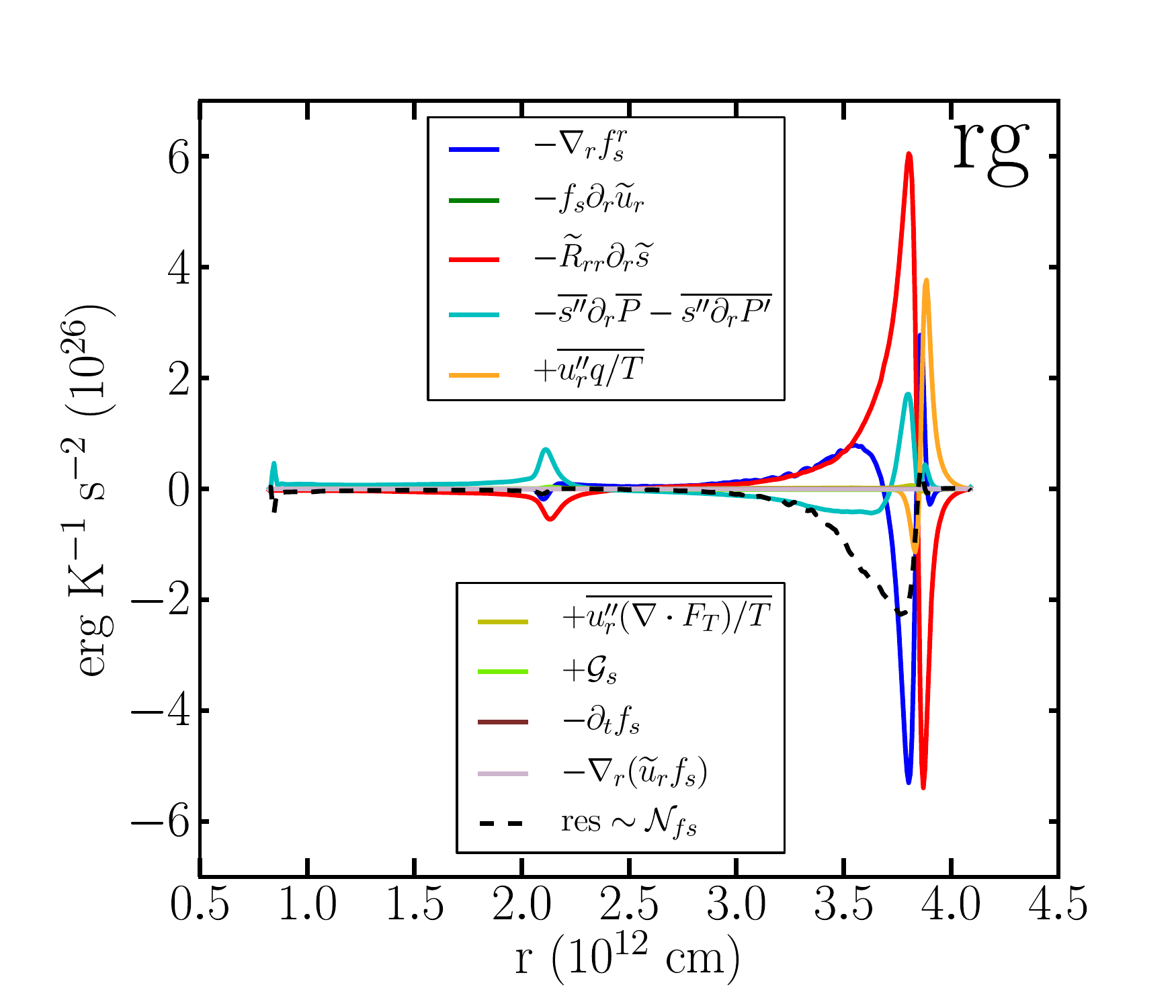}
\includegraphics[width=6.5cm]{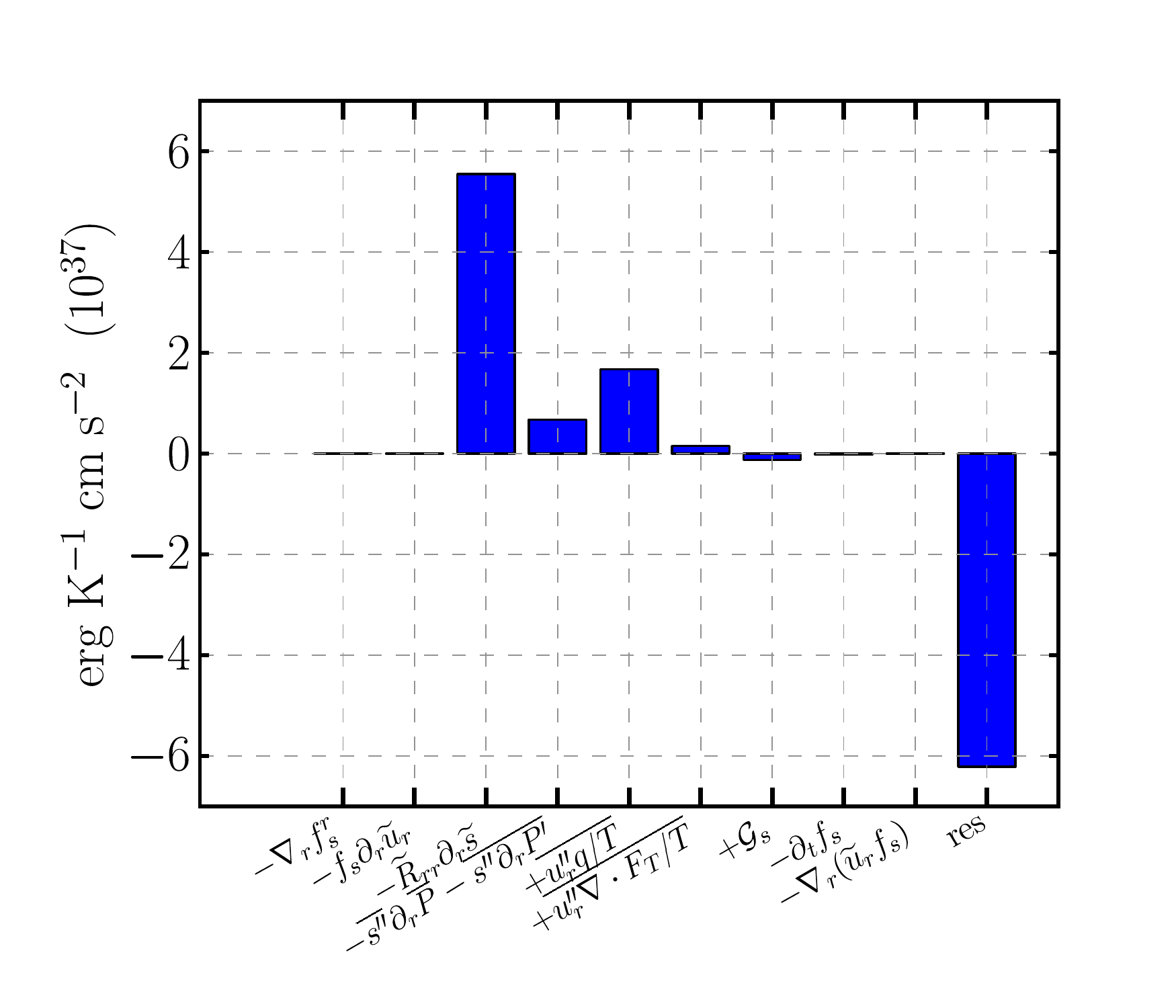}}
\caption{Mean entropy flux equation. Model {\sf ob.3D.2hp} (upper panels) and model {\sf rg.3D.mr} (lower panels). \label{fig:fs-equation}}
\end{figure}

\newpage

\subsection{Mean composition ($^{16}$O) flux equation} 

\begin{align}
\erho \fav{D}_t (f_\alpha / \eht{\rho}) = &  -\nabla_r f_\alpha^r  - f_\alpha \partial_r \fht{u}_r - \fht{R}_{rr} \partial_r \fht{X}_\alpha -\eht{X''_\alpha} \partial_r \eht{P} - \eht{X''_\alpha \partial_r P'} + \overline{u''_r \rho \dot{X}_\alpha^{\rm nuc}} + {\mathcal G_\alpha} + {\mathcal N_{f\alpha}} \label{eq:rans_falpha} \\
\end{align}

\vspace{1.cm}

\begin{figure}[!h]
\centerline{
\includegraphics[width=6.5cm]{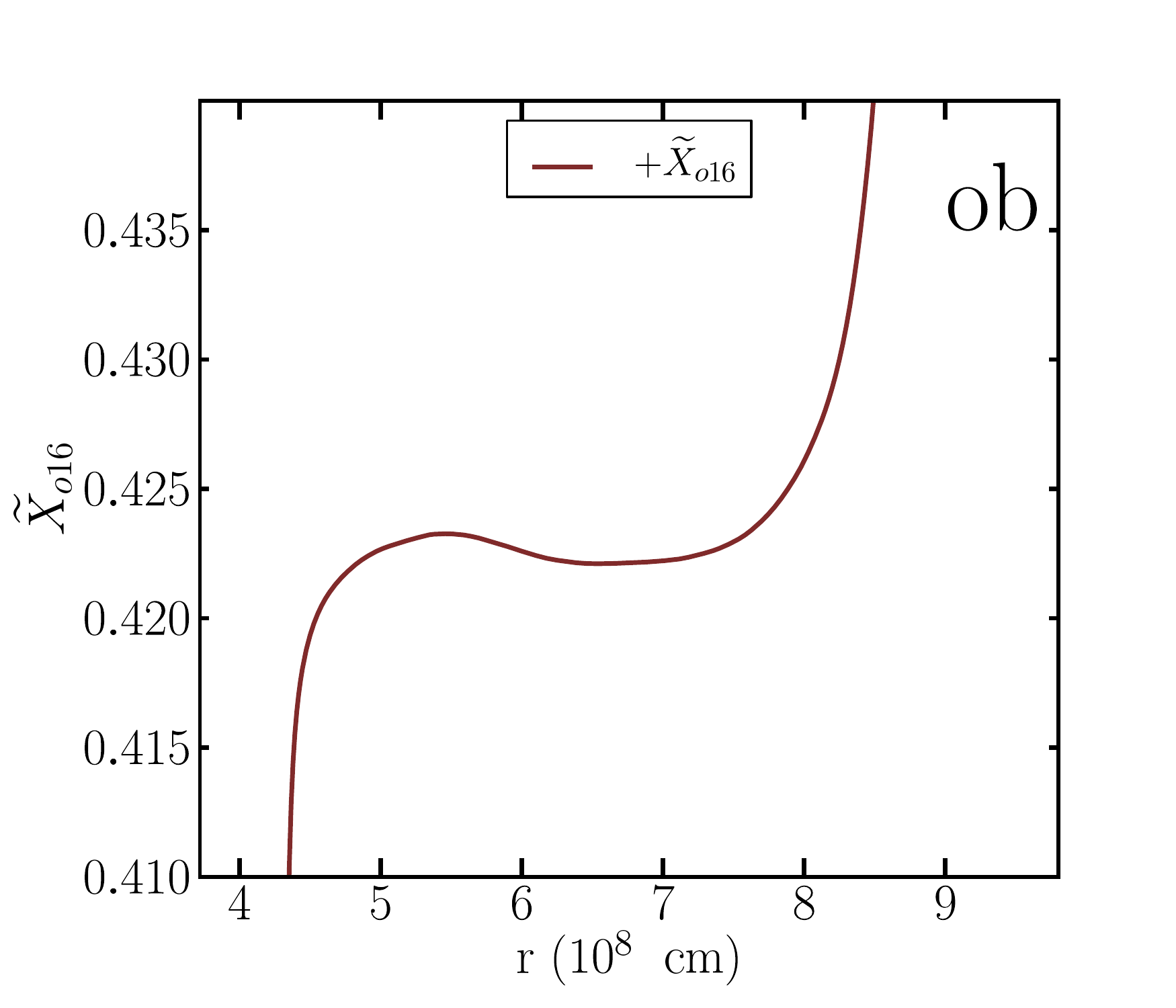}
\includegraphics[width=6.5cm]{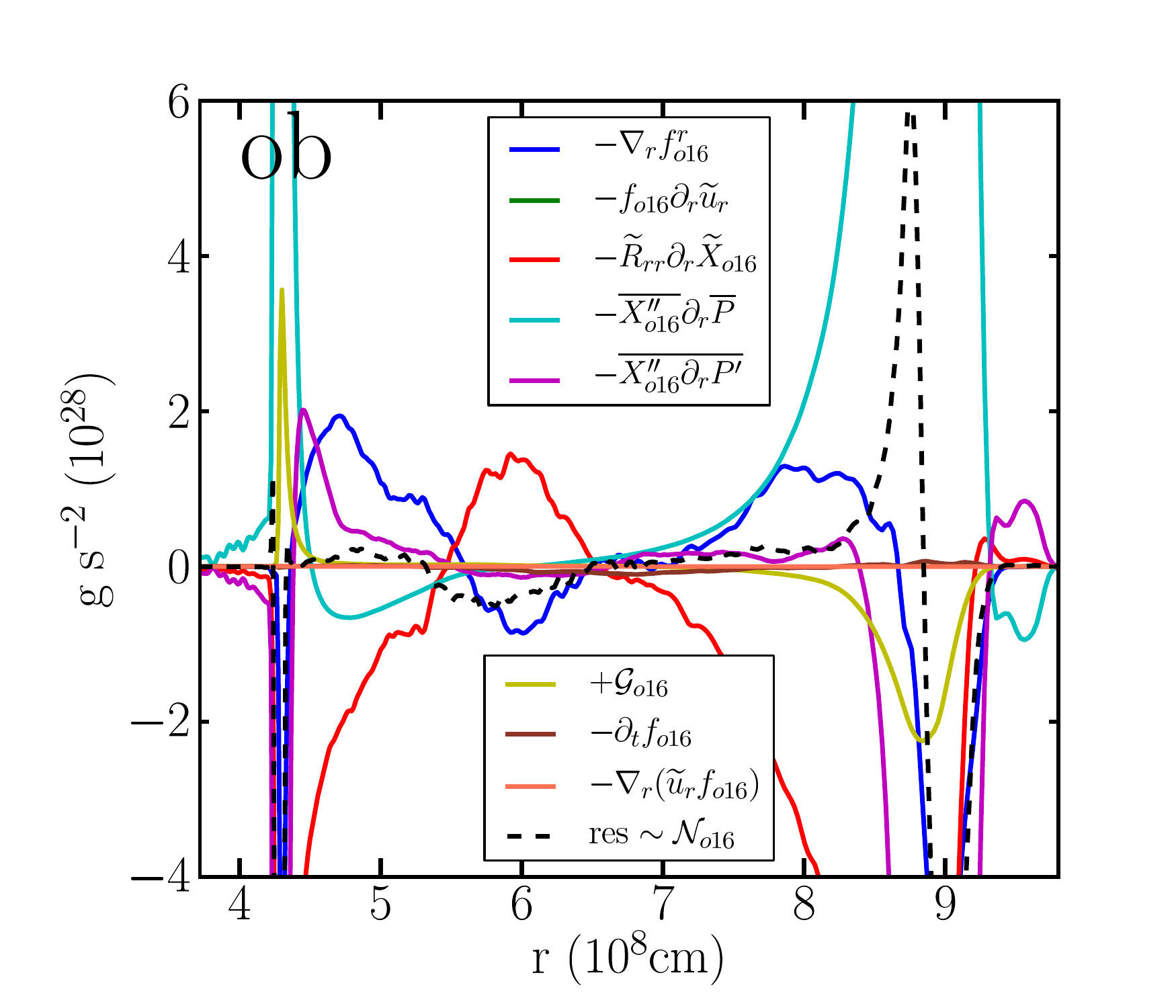}
\includegraphics[width=6.5cm]{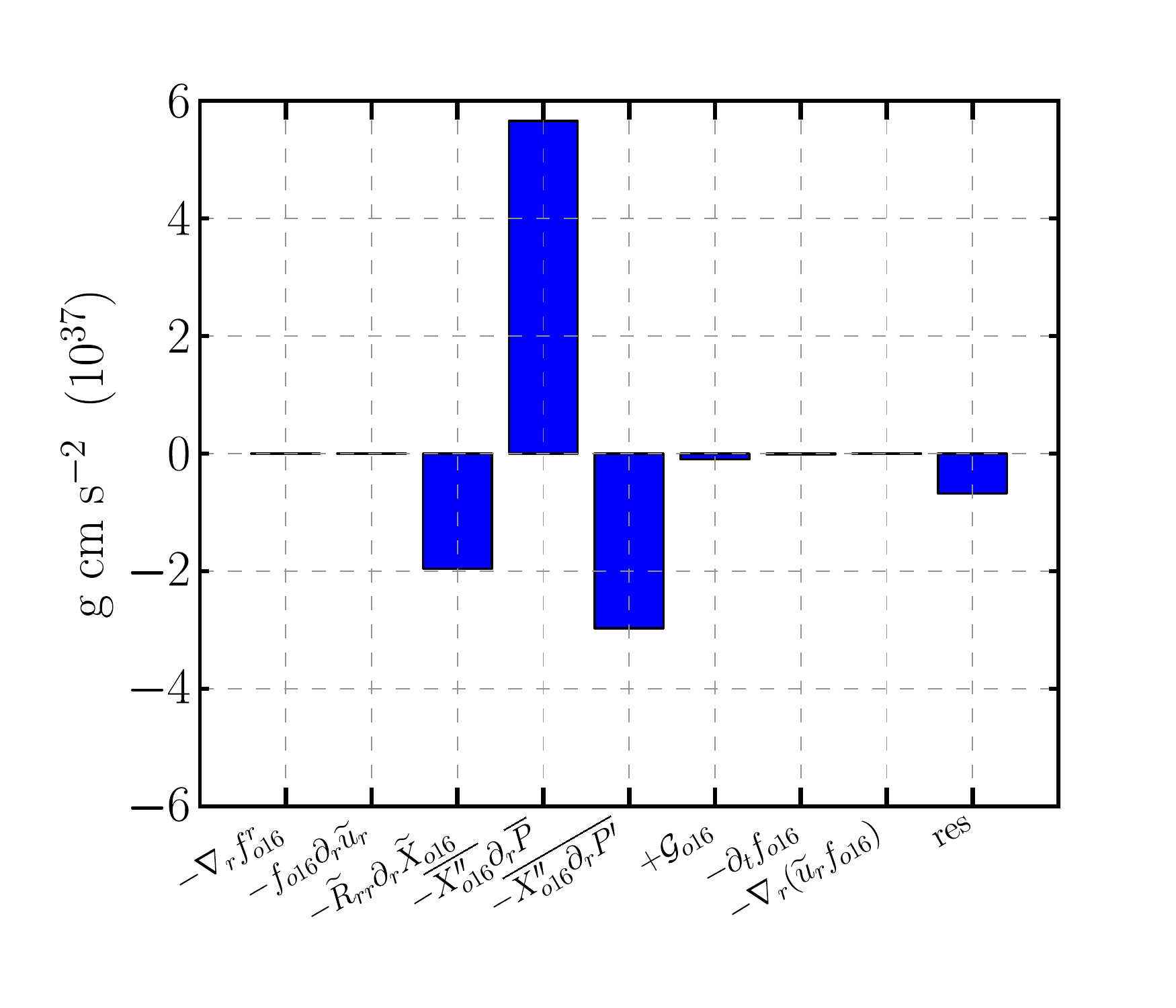}}
\caption{Mean composition ($^{16}$O) flux  equation. Model {\sf ob.3D.2hp}.\label{fig:fcomp-equation}}
\end{figure}

\newpage

\subsection{Mean A flux equation}

\begin{align}
\erho \fav{D}_t (f_A / \eht{\rho}) = &  \ {\mathcal N_{fA}} -\nabla_r f_A^r - f_A \partial_r \fht{u}_r - \fht{R}_{rr} \partial_r \fht{A} -\eht{A''} \partial_r \eht{P} - \eht{A'' \partial_r P'} - \overline{u''_r \rho A^2\Sigma_\alpha \dot{X}_\alpha^{\rm nuc} / A_\alpha} + {\mathcal G_A}                 \label{eq:rans_fabar} \\
\end{align}

\vspace{1.cm}

\begin{figure}[!h]
\centerline{
\includegraphics[width=6.5cm]{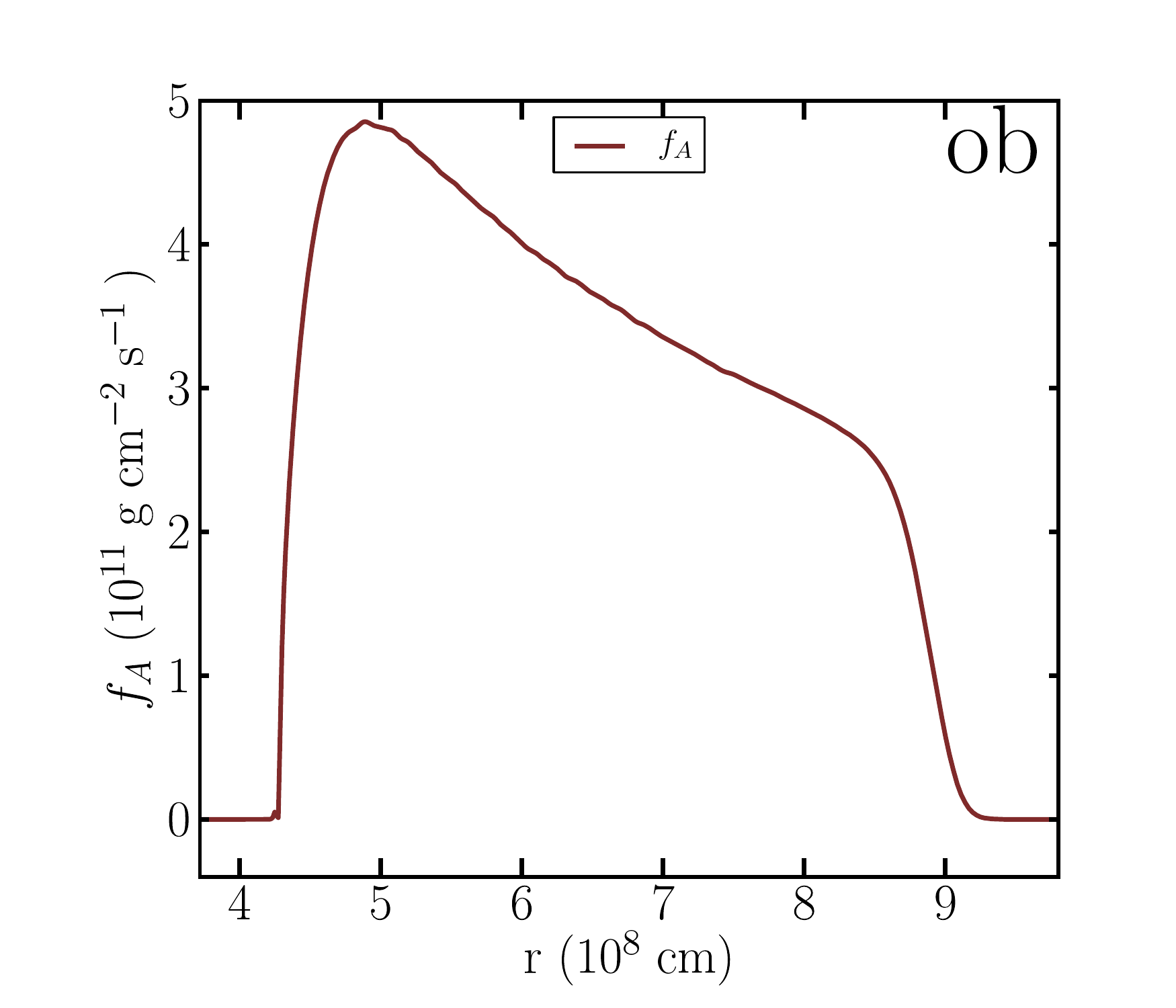}
\includegraphics[width=6.5cm]{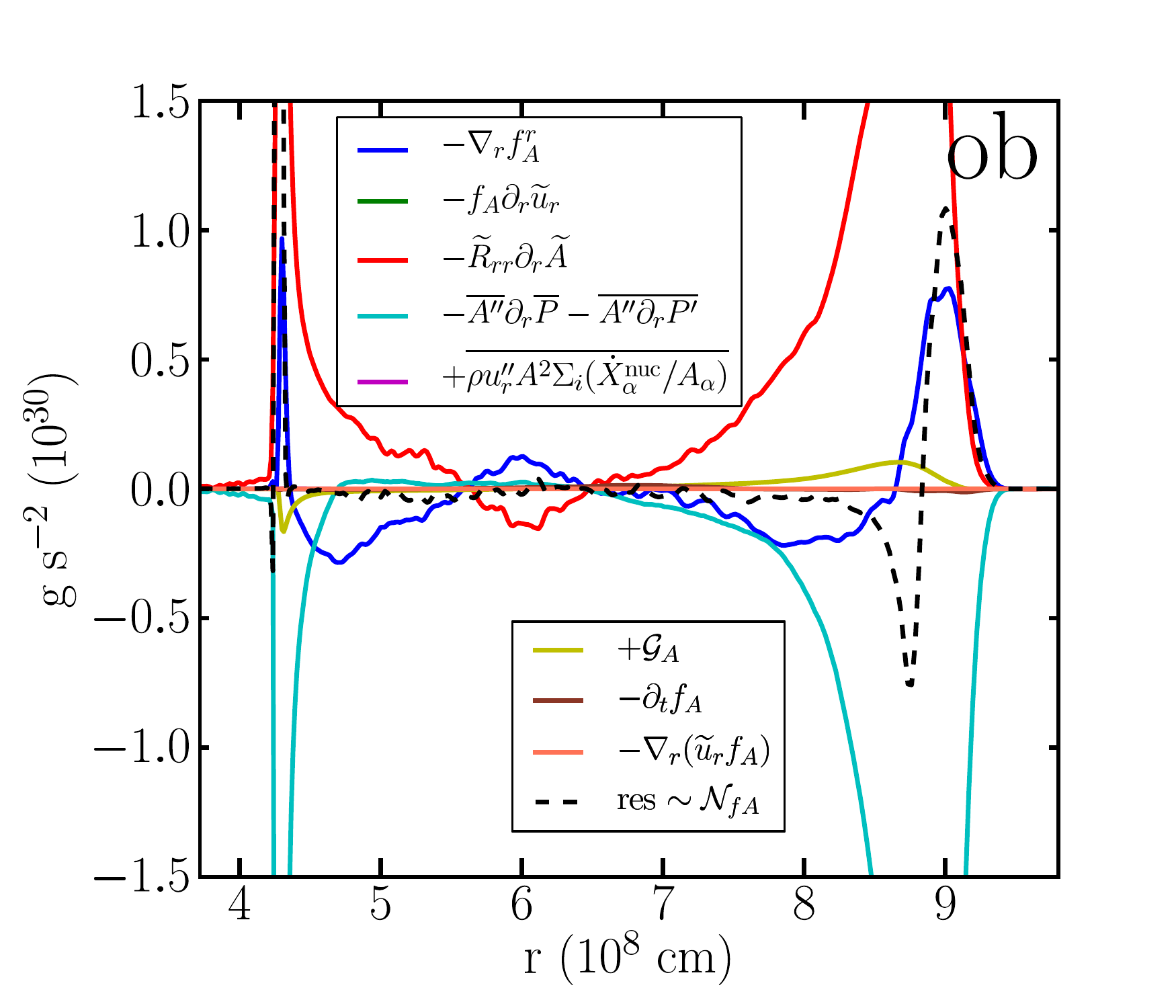}
\includegraphics[width=6.5cm]{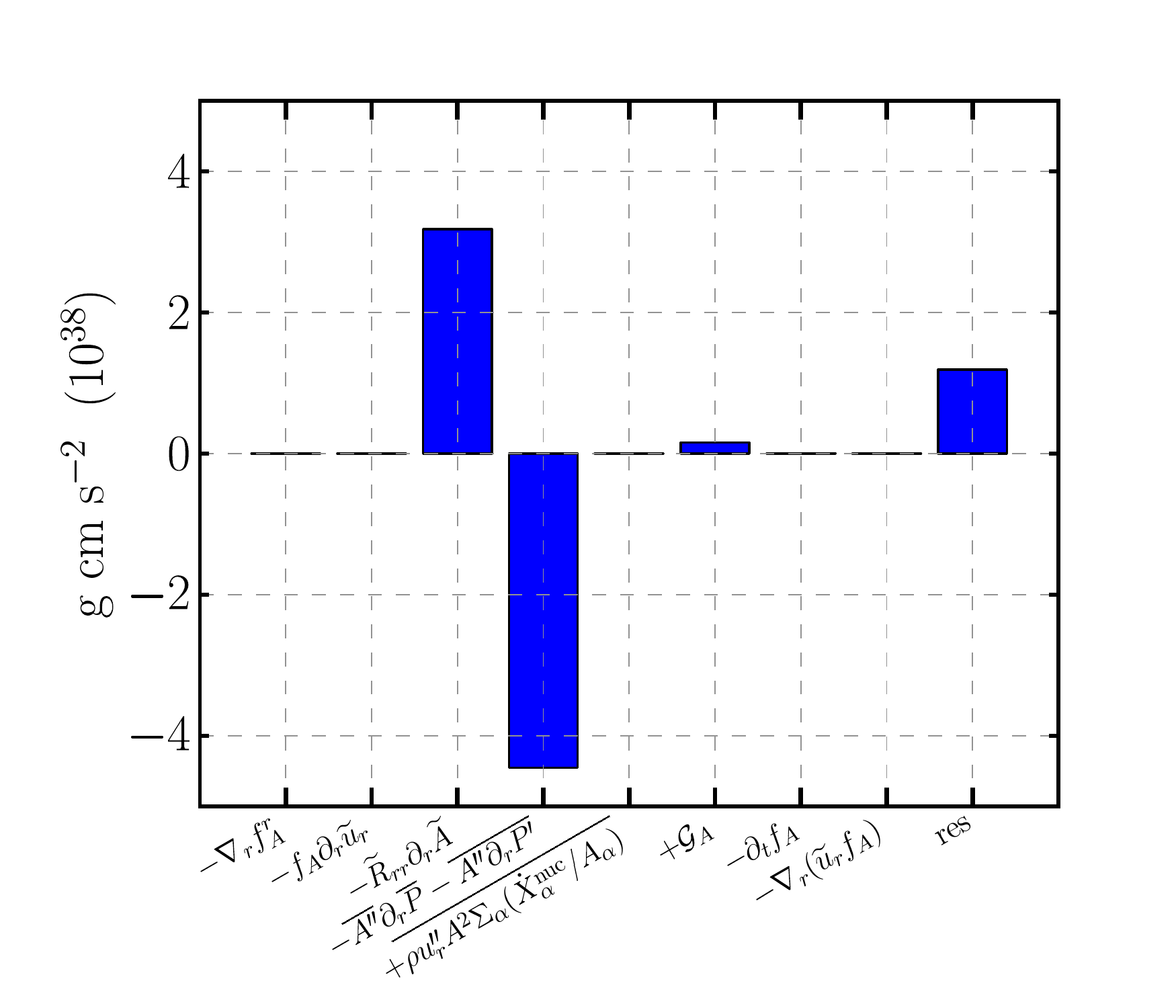}}
\caption{Mean A flux equation. Model {\sf ob.3D.2hp}. \label{fig:fcomp-A-equation}}
\end{figure}

\newpage

\subsection{Mean Z flux equation}

\begin{align}
\erho \fav{D}_t (f_Z / \eht{\rho}) = &  \ {\mathcal N_{fZ}} -\nabla_r f_Z^r  - f_Z \partial_r \fht{u}_r - \fht{R}_{rr} \partial_r \fht{Z} -\eht{Z''} \partial_r \eht{P} - \eht{Z'' \partial_r P'} - \overline{u''_r \rho Z A \Sigma_\alpha (\dot{X}_\alpha^{\rm nuc}/ A_\alpha)} - \nonumber \\ 
& - \overline{u''_r \rho A \Sigma_\alpha (Z_\alpha \dot{X}_\alpha^{\rm nuc} / A_\alpha)}  + {\mathcal G_Z}   \label{eq:rans_fzbar} 
\end{align}

\vspace{1.cm}

\begin{figure}[!h]
\centerline{
\includegraphics[width=6.5cm]{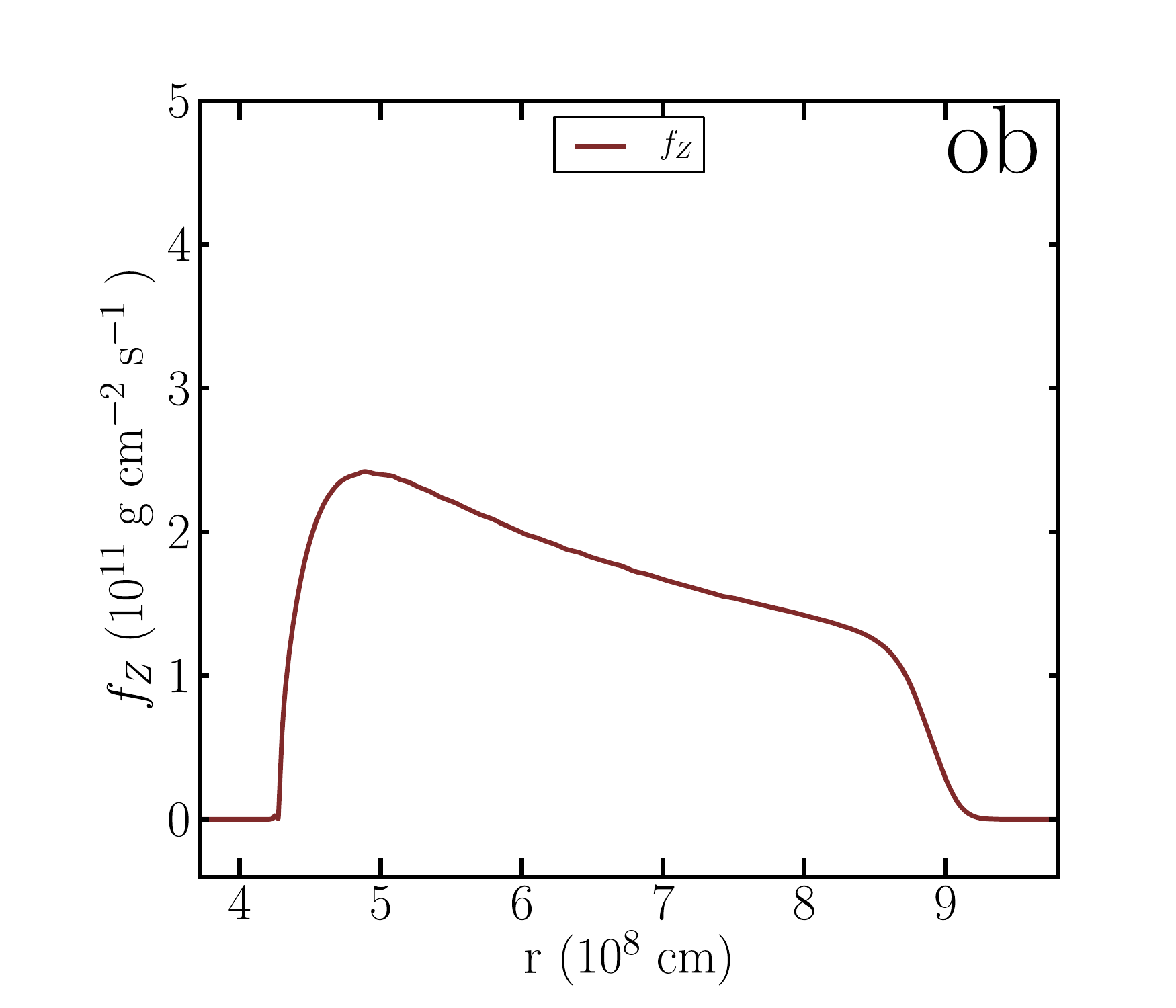}
\includegraphics[width=6.5cm]{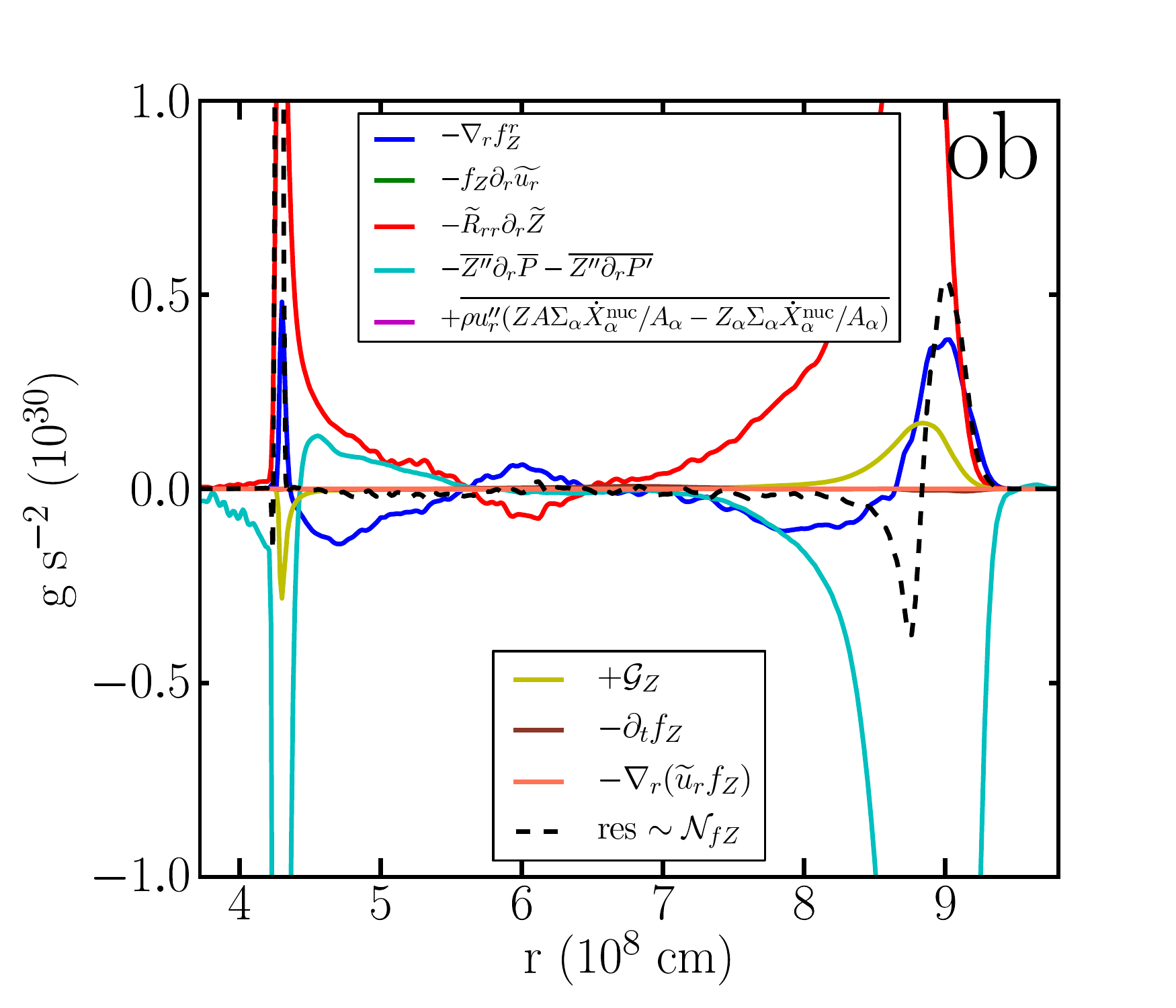}
\includegraphics[width=6.5cm]{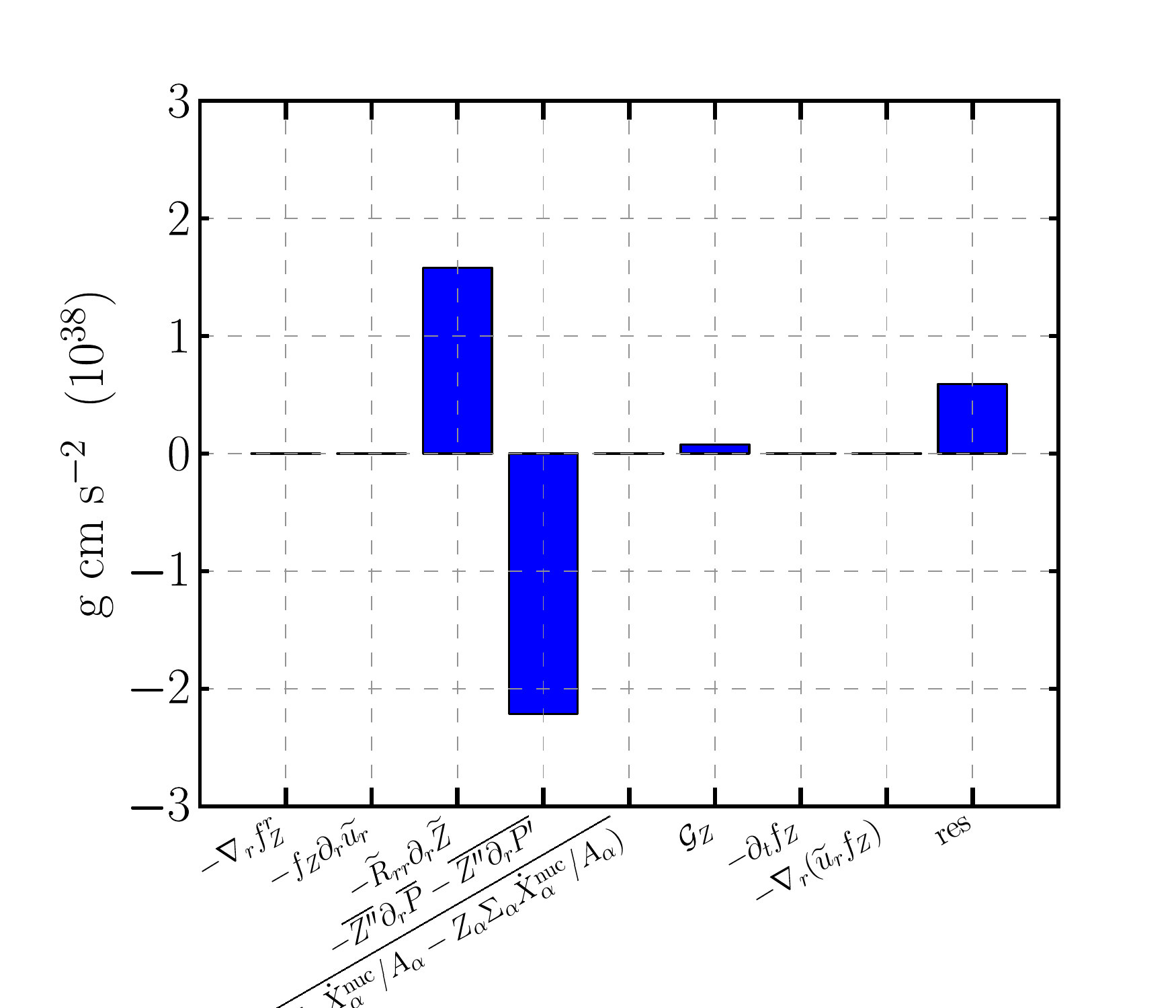}}
\caption{Mean Z flux equation. Model {\sf ob.3D.2hp}. \label{fig:fcomp-Z-equation}}
\end{figure}

\newpage

\section{Mean field composition data for the  oxygen shell burning model ob.3D.2hp}

\subsection{Mean C$^{12}$ and O$^{16}$ equation}

\begin{figure}[!h]
\centerline{
\includegraphics[width=6.8cm]{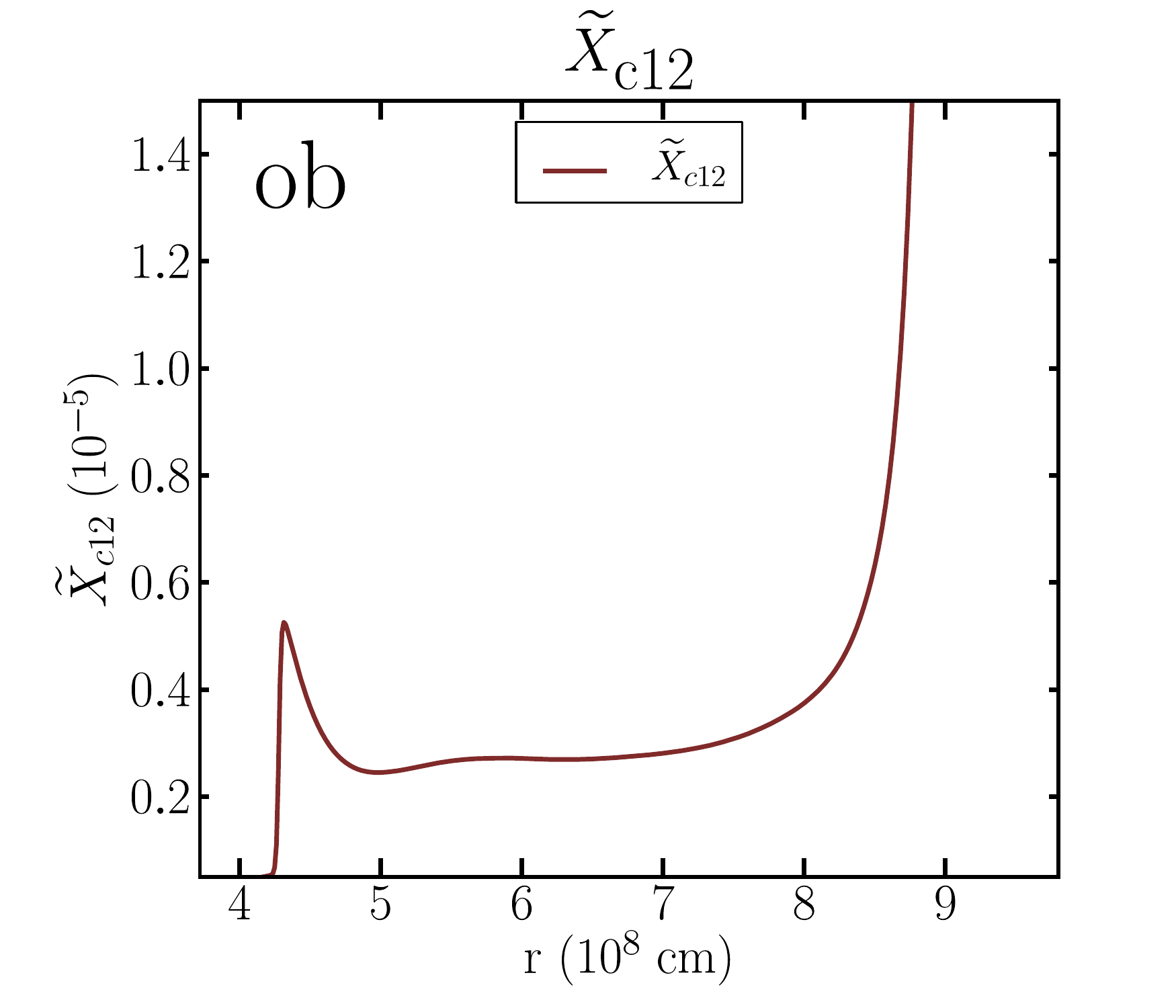}
\includegraphics[width=6.8cm]{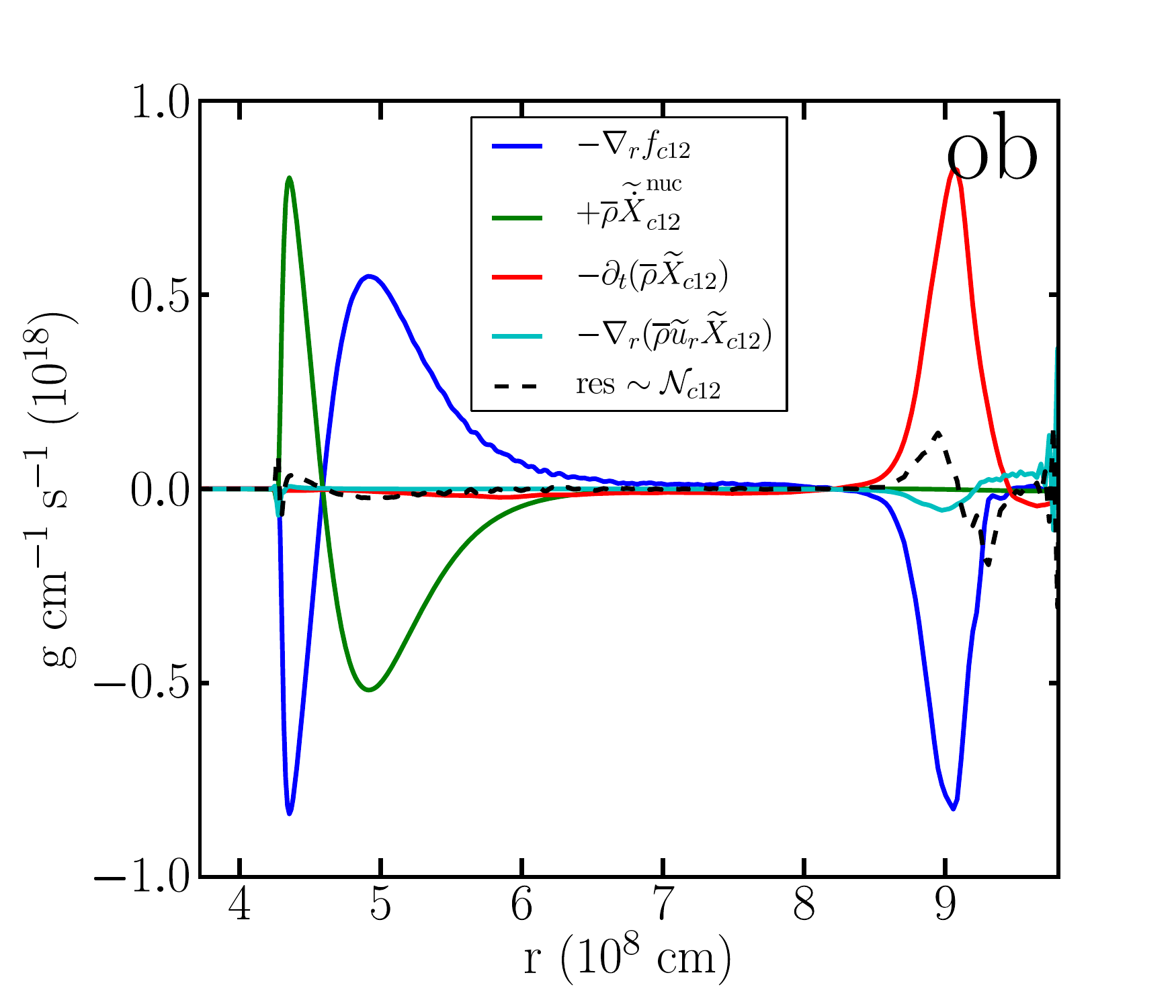}
\includegraphics[width=6.8cm]{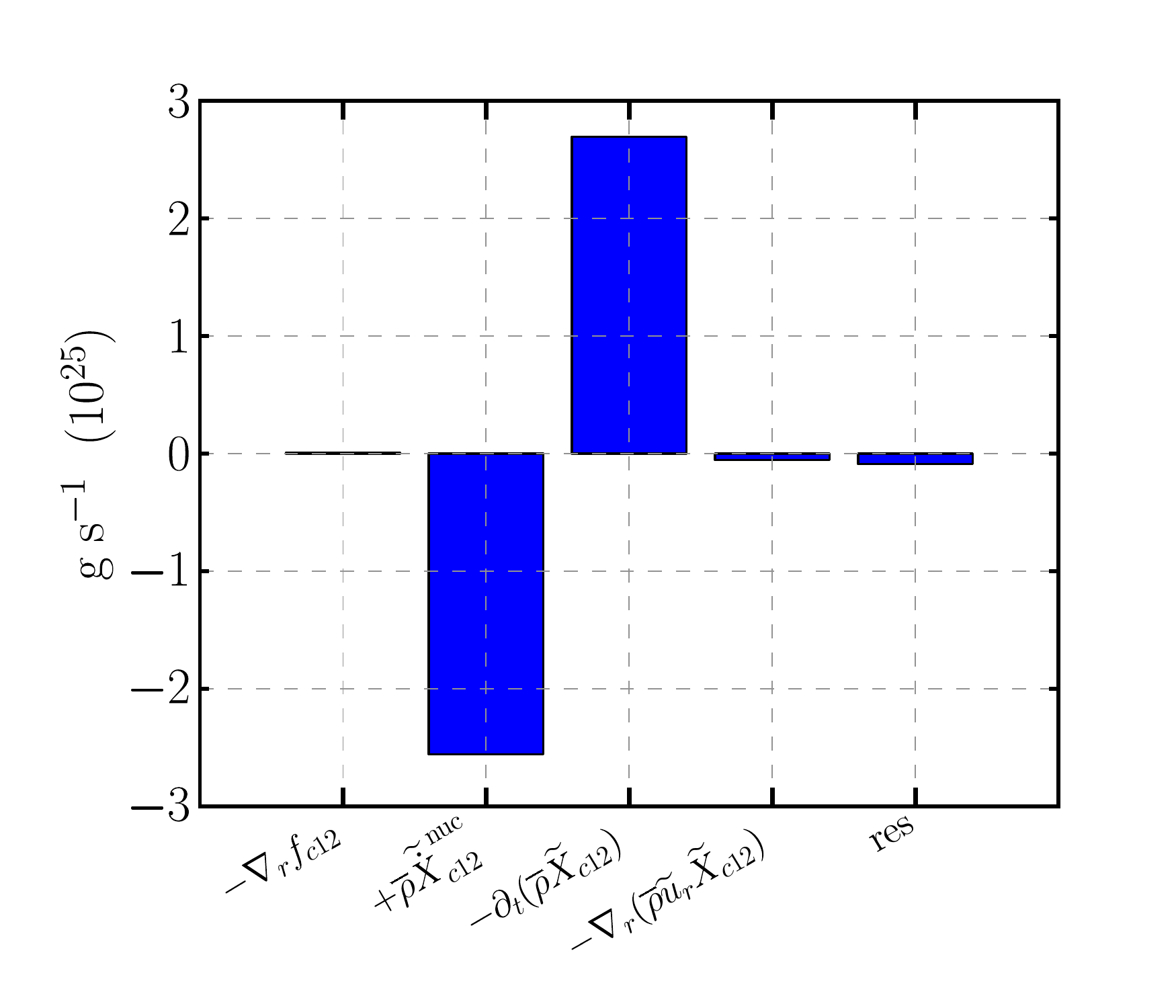}}

\centerline{
\includegraphics[width=6.8cm]{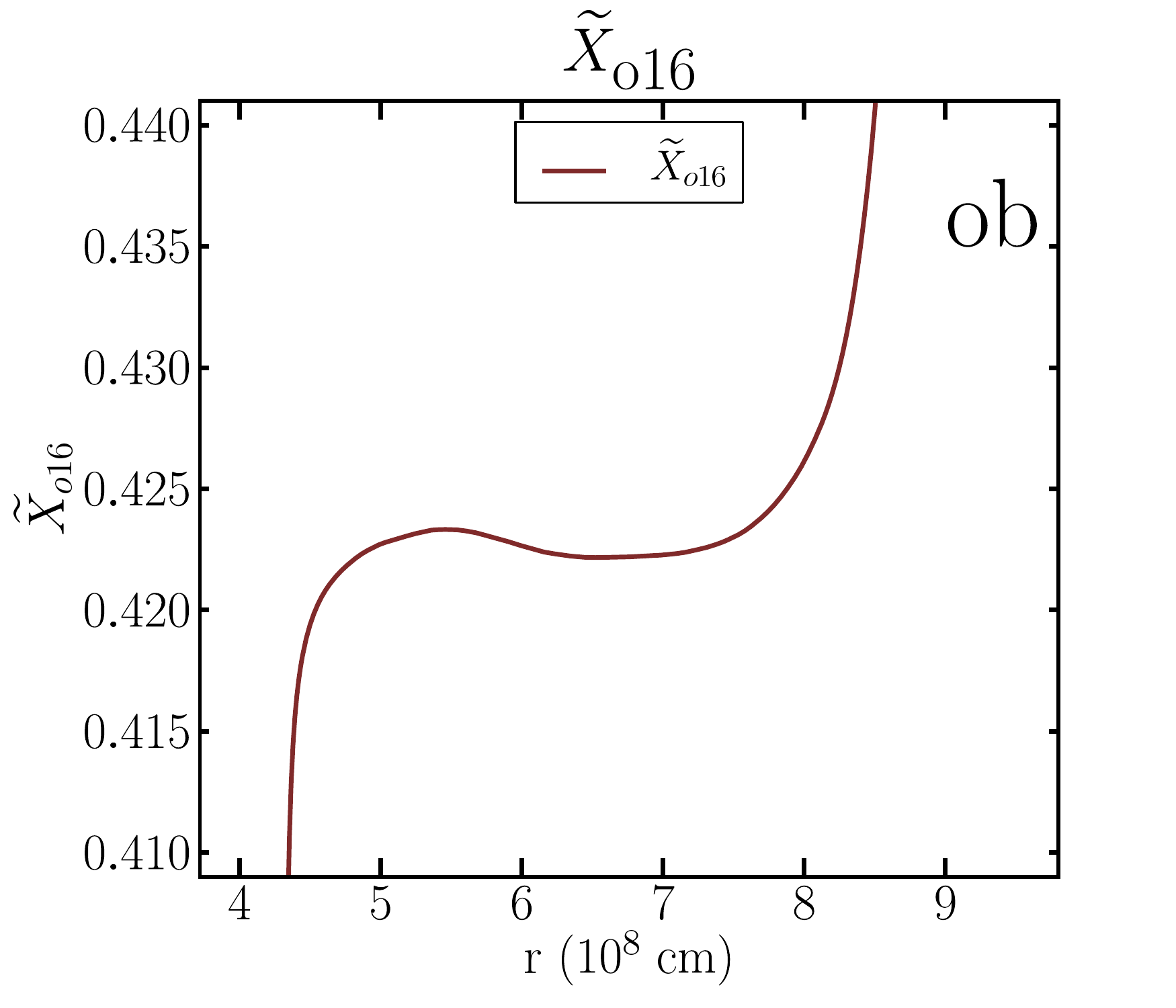}
\includegraphics[width=6.8cm]{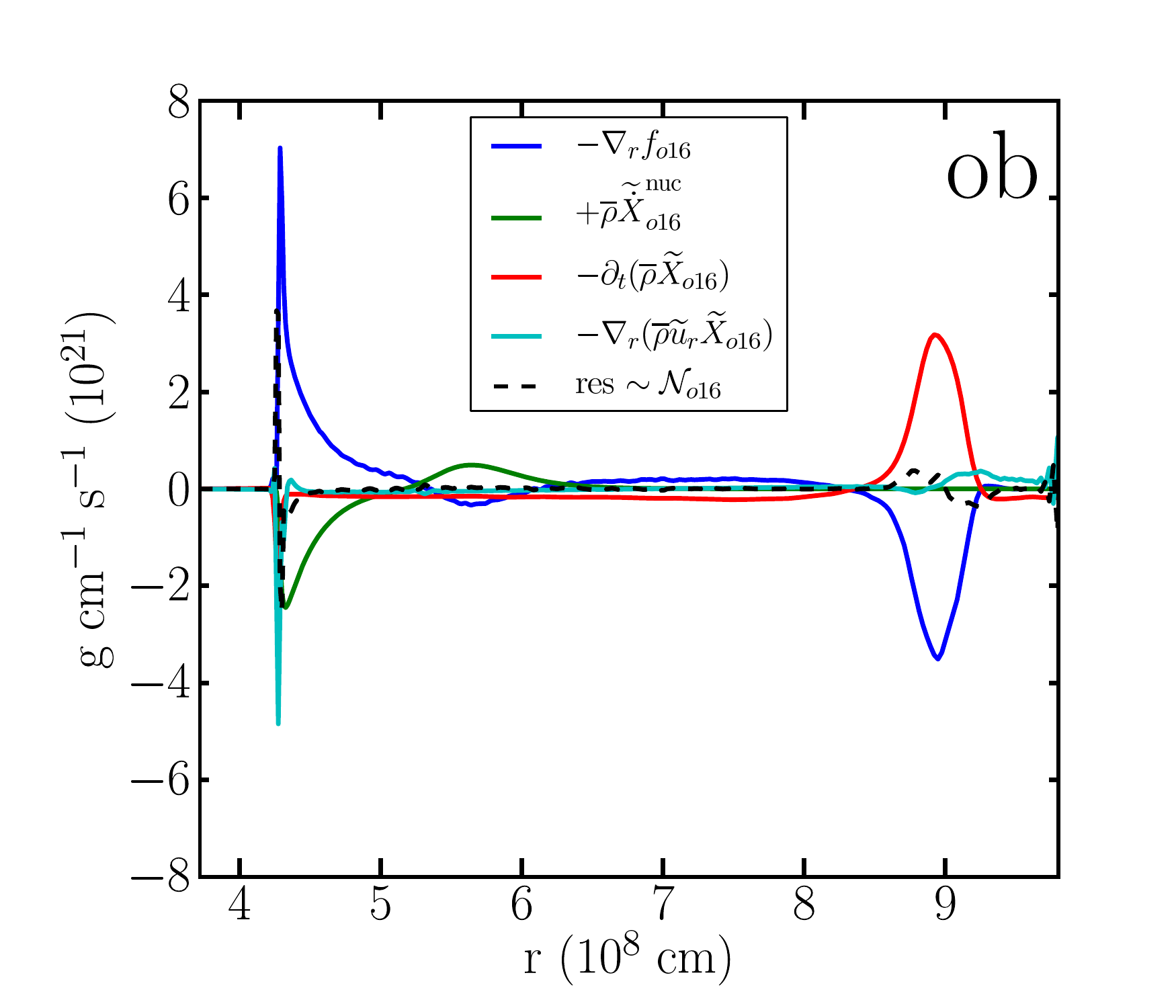}
\includegraphics[width=6.8cm]{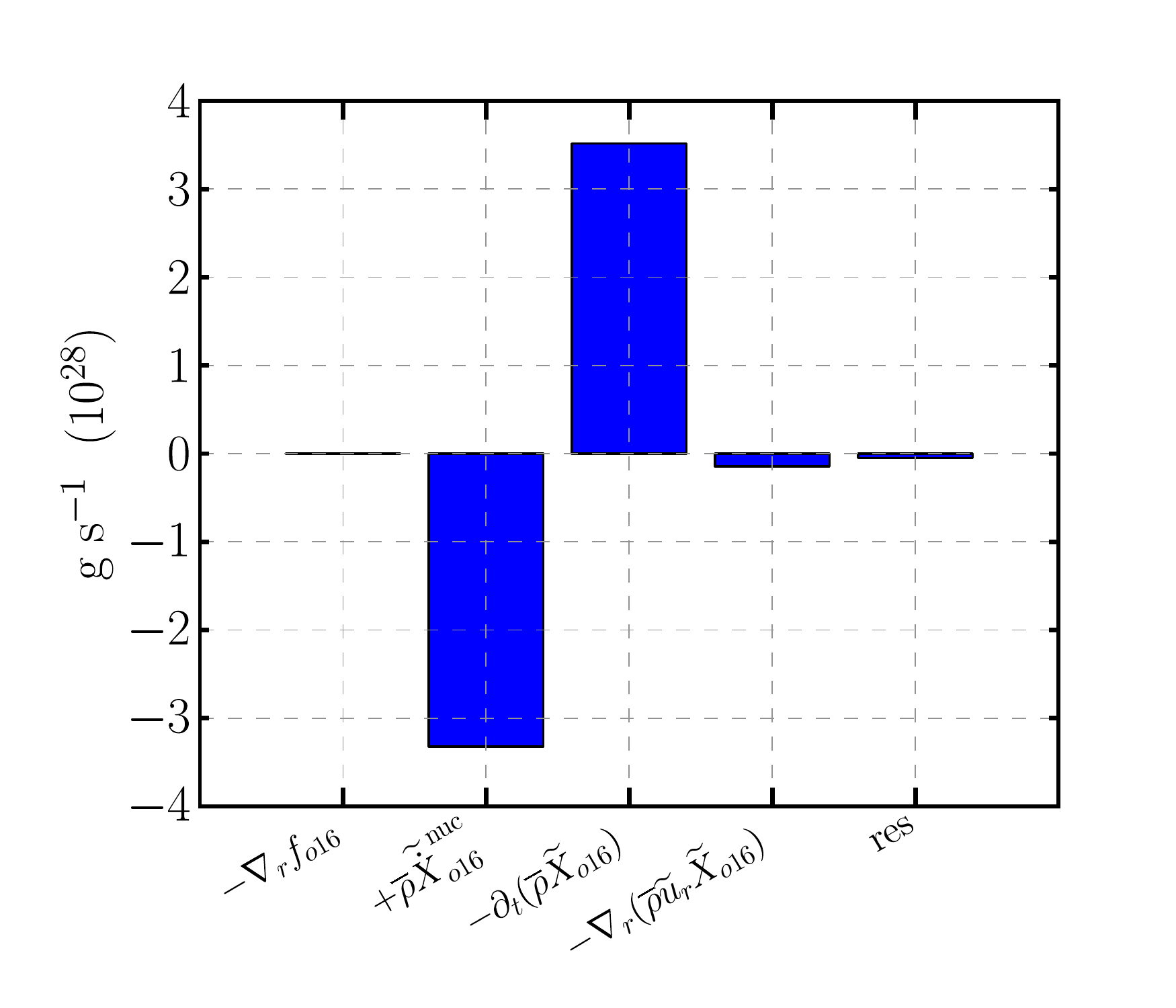}}
\caption{Mean composition equations. Model {\sf ob.3D.2hp}. \label{fig:xhe4-xc12-equations}}
\end{figure}

\newpage

\subsection{Mean Ne$^{20}$ and Na$^{23}$ equation}

\begin{figure}[!h]
\centerline{
\includegraphics[width=6.8cm]{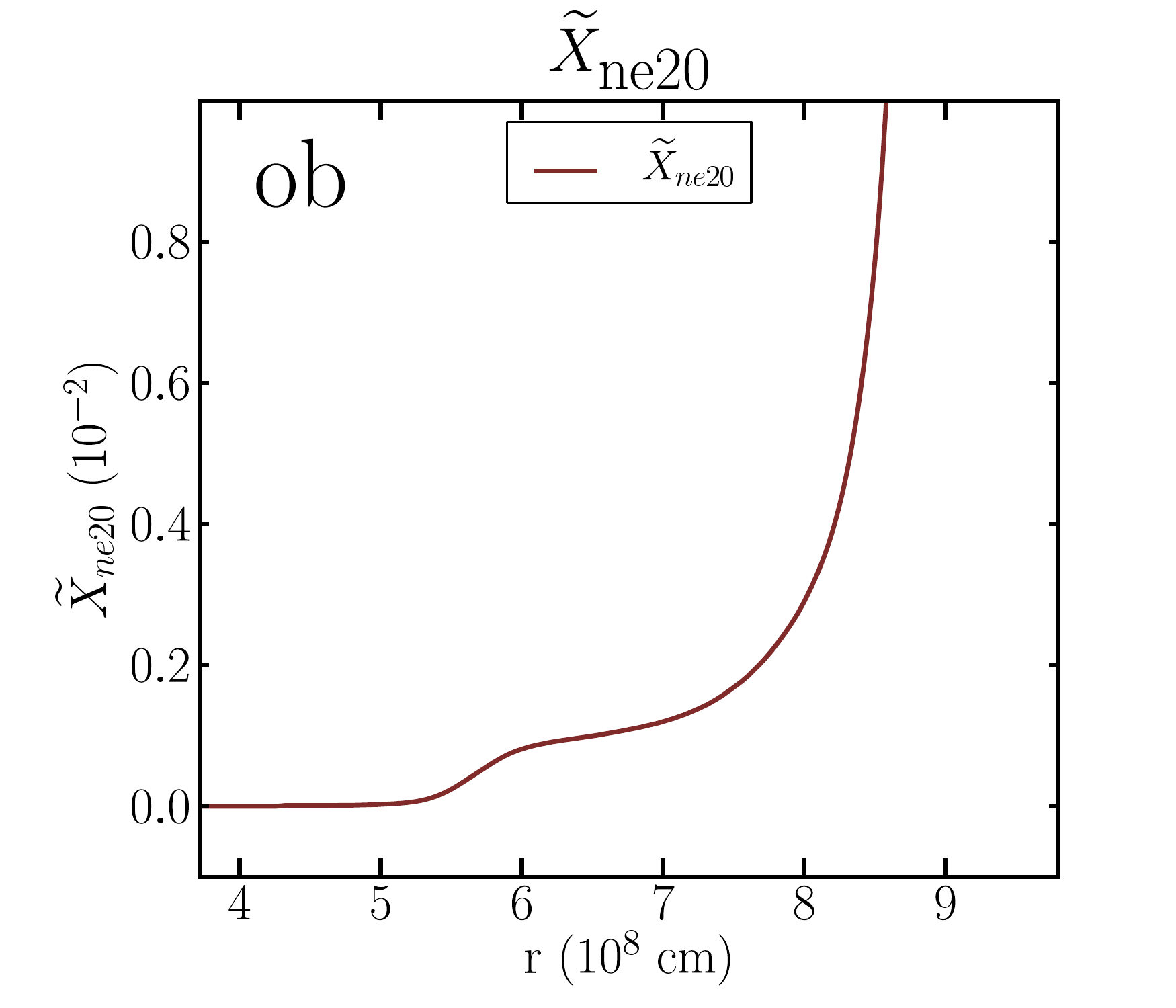}
\includegraphics[width=6.8cm]{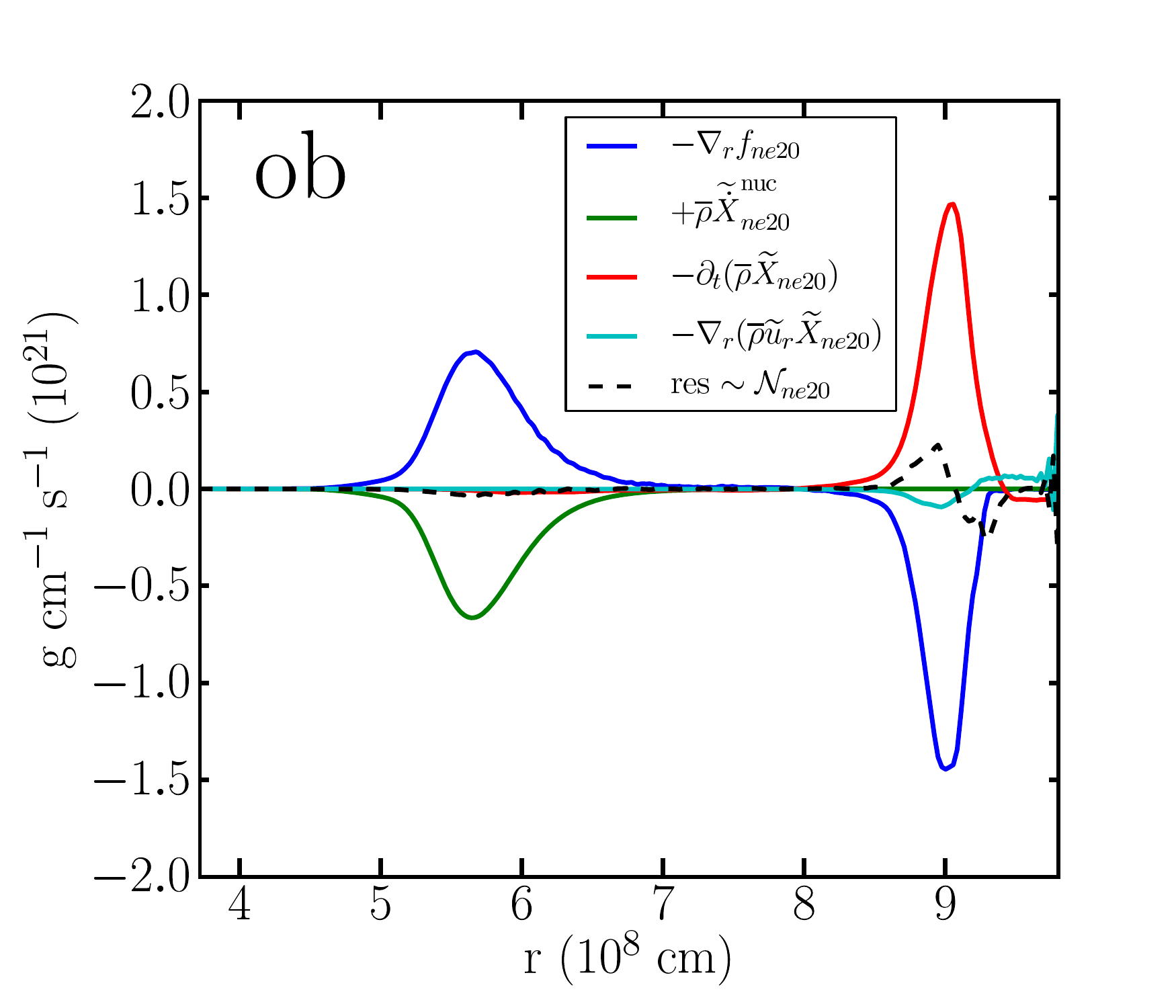}
\includegraphics[width=6.8cm]{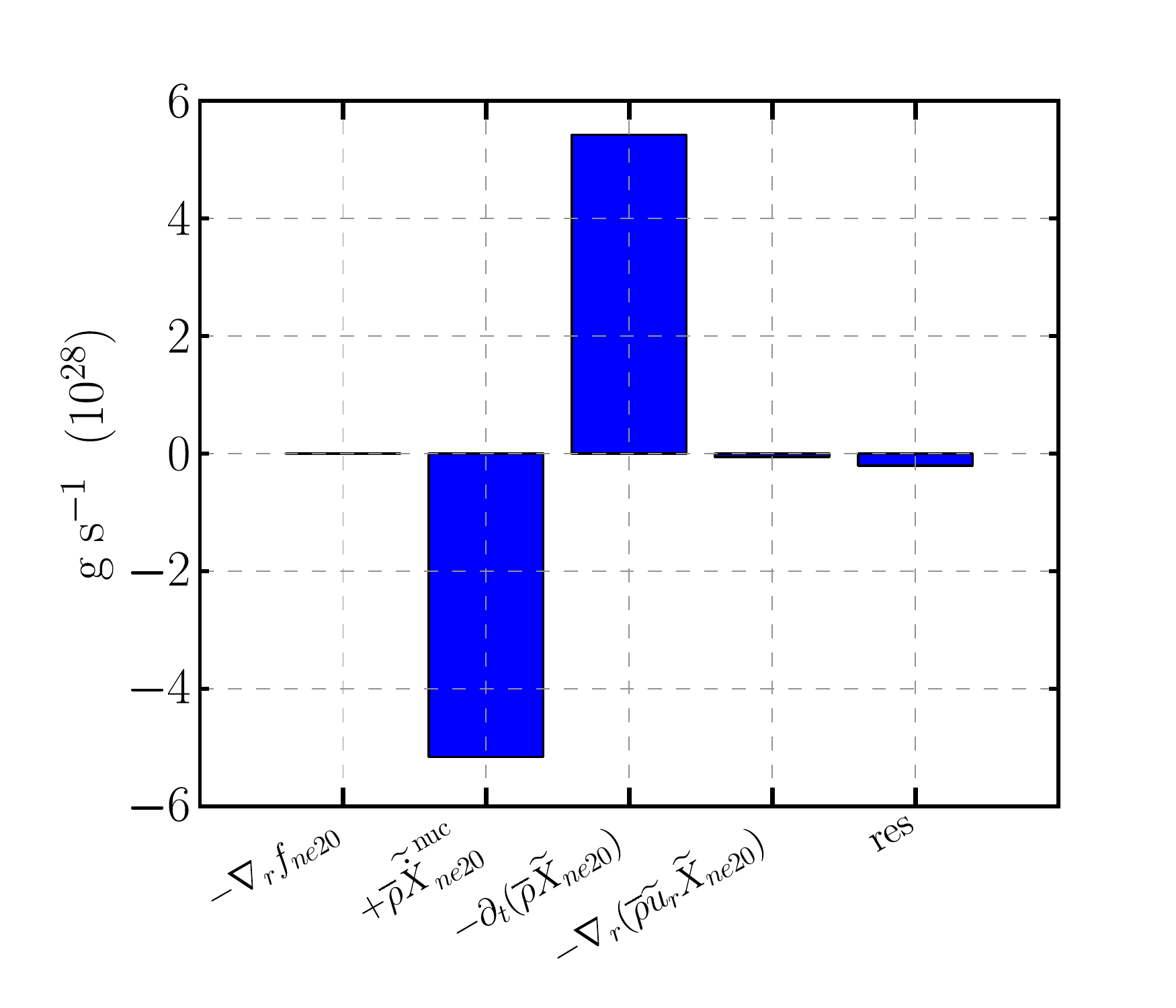}}

\centerline{
\includegraphics[width=6.8cm]{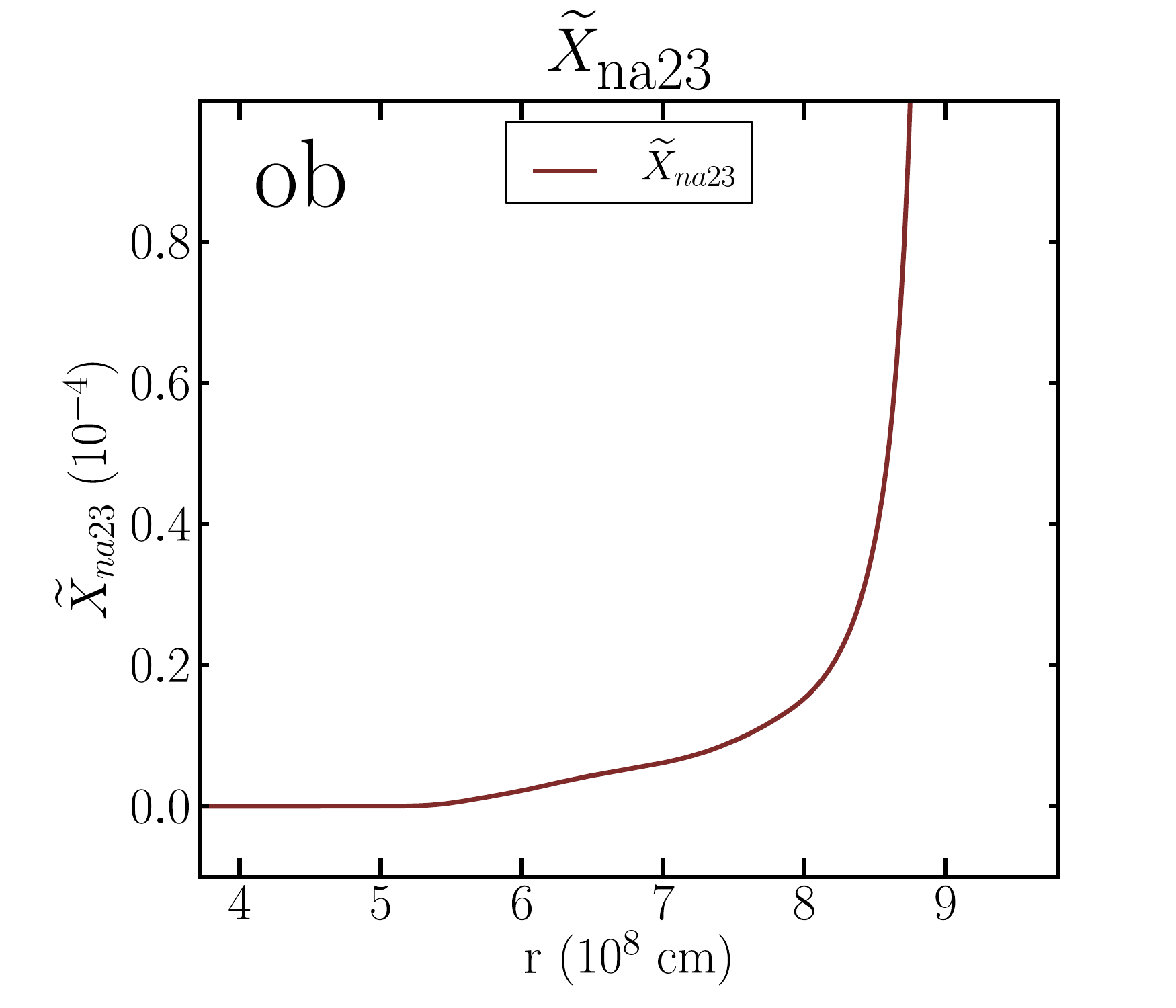}
\includegraphics[width=6.8cm]{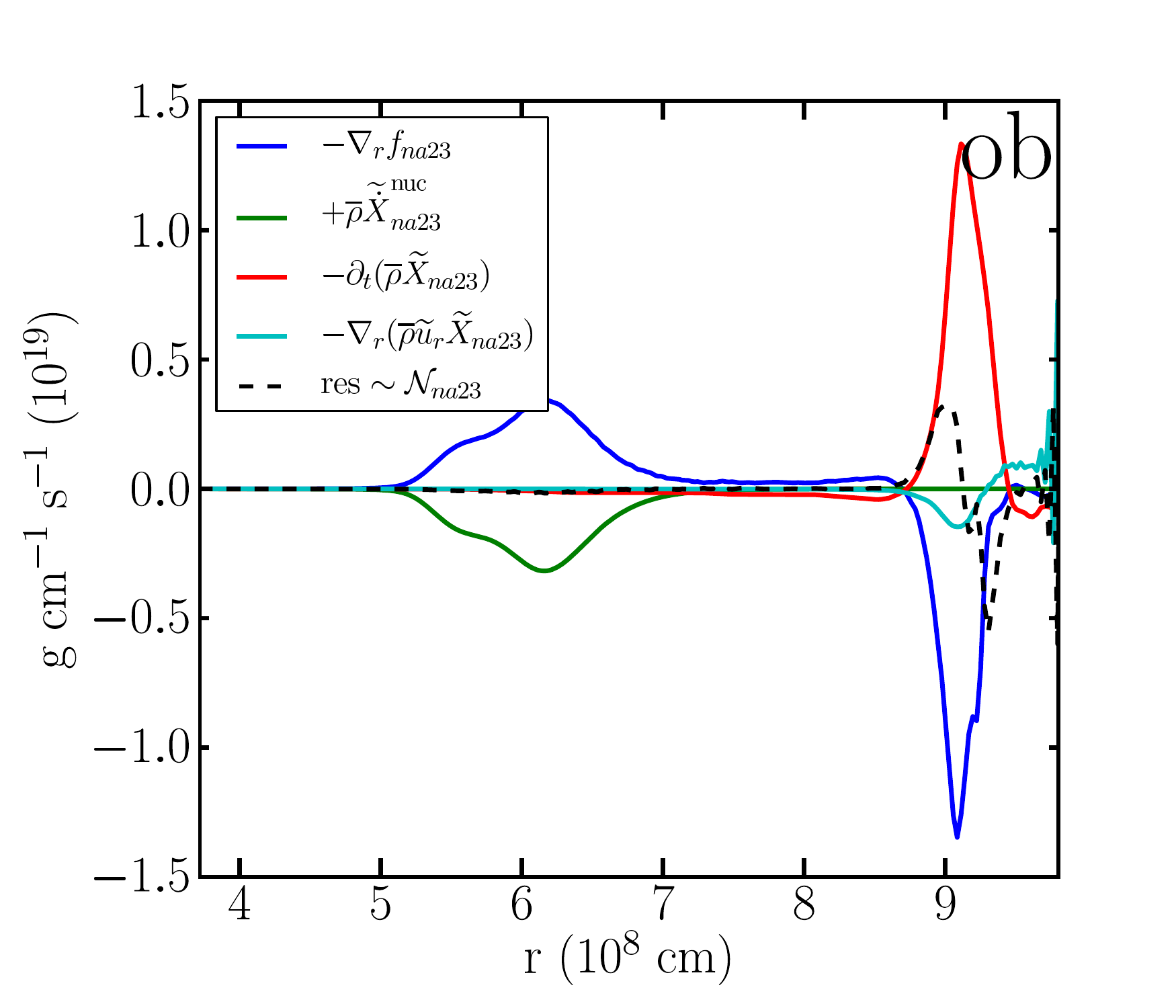}
\includegraphics[width=6.8cm]{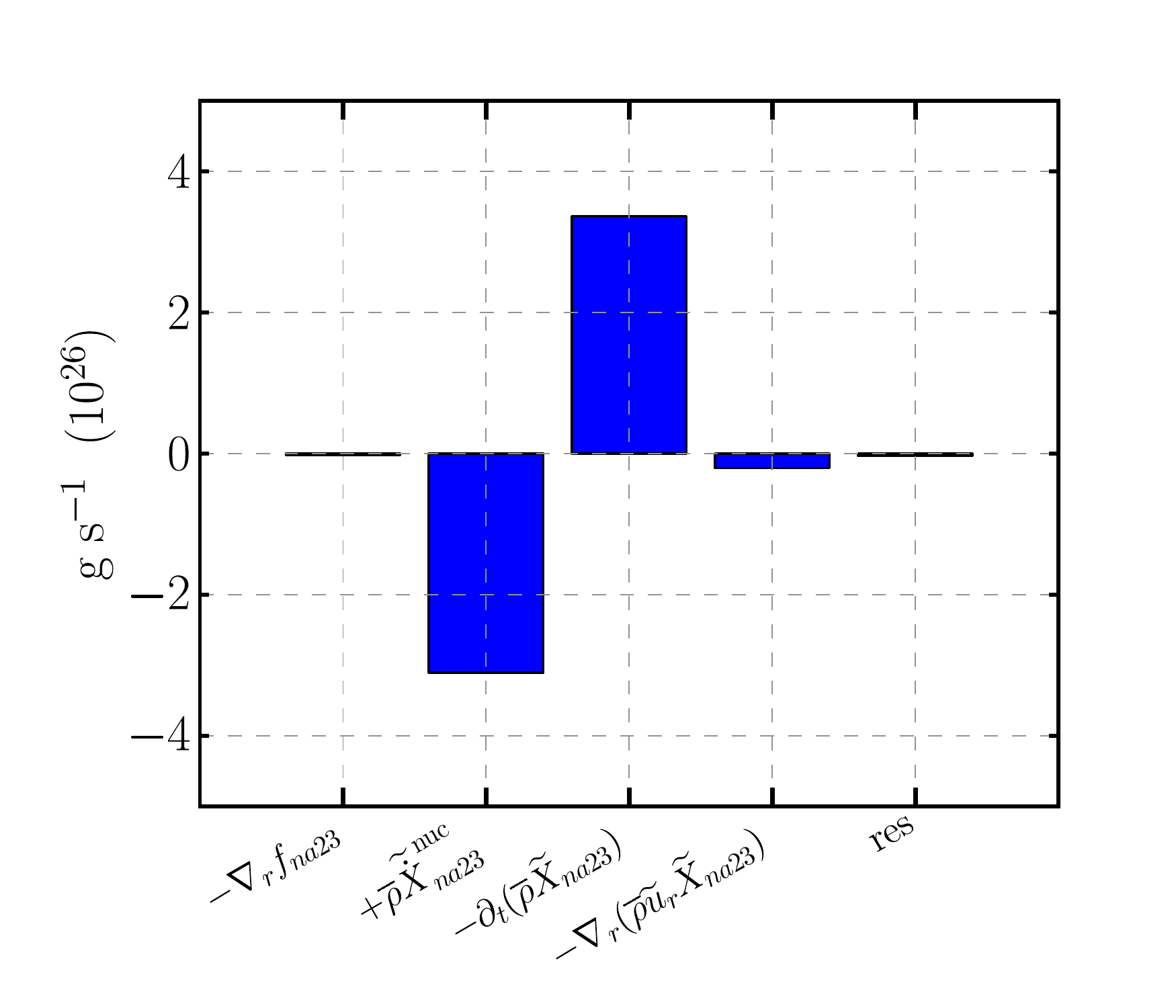}}
\caption{Mean composition equations. Model {\sf ob.3D.2hp}. \label{fig:xo16-xne20-equations}}
\end{figure}

\newpage

\subsection{Mean Mg$^{24}$ and Si$^{28}$ equation}

\begin{figure}[!h]
\centerline{
\includegraphics[width=6.8cm]{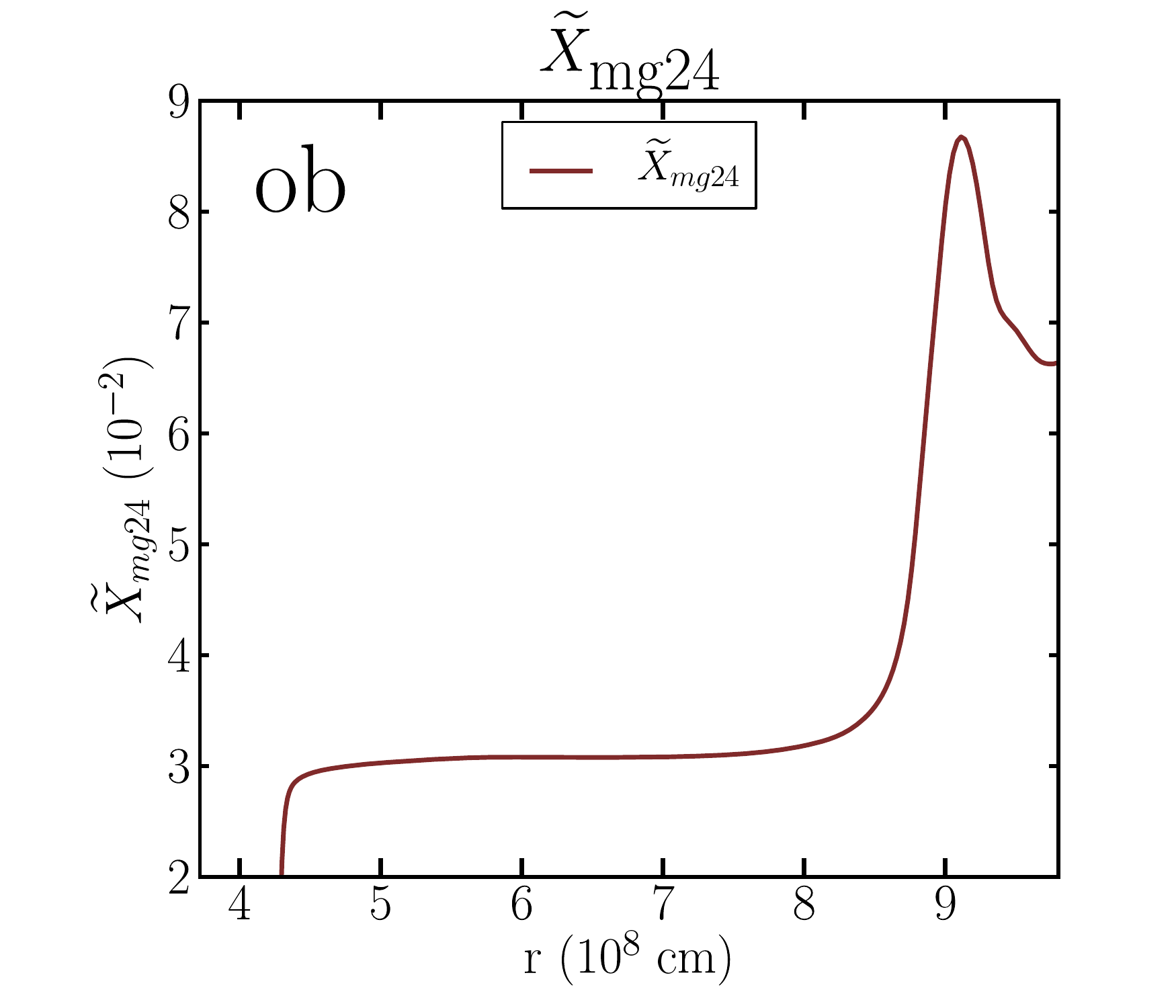}
\includegraphics[width=6.8cm]{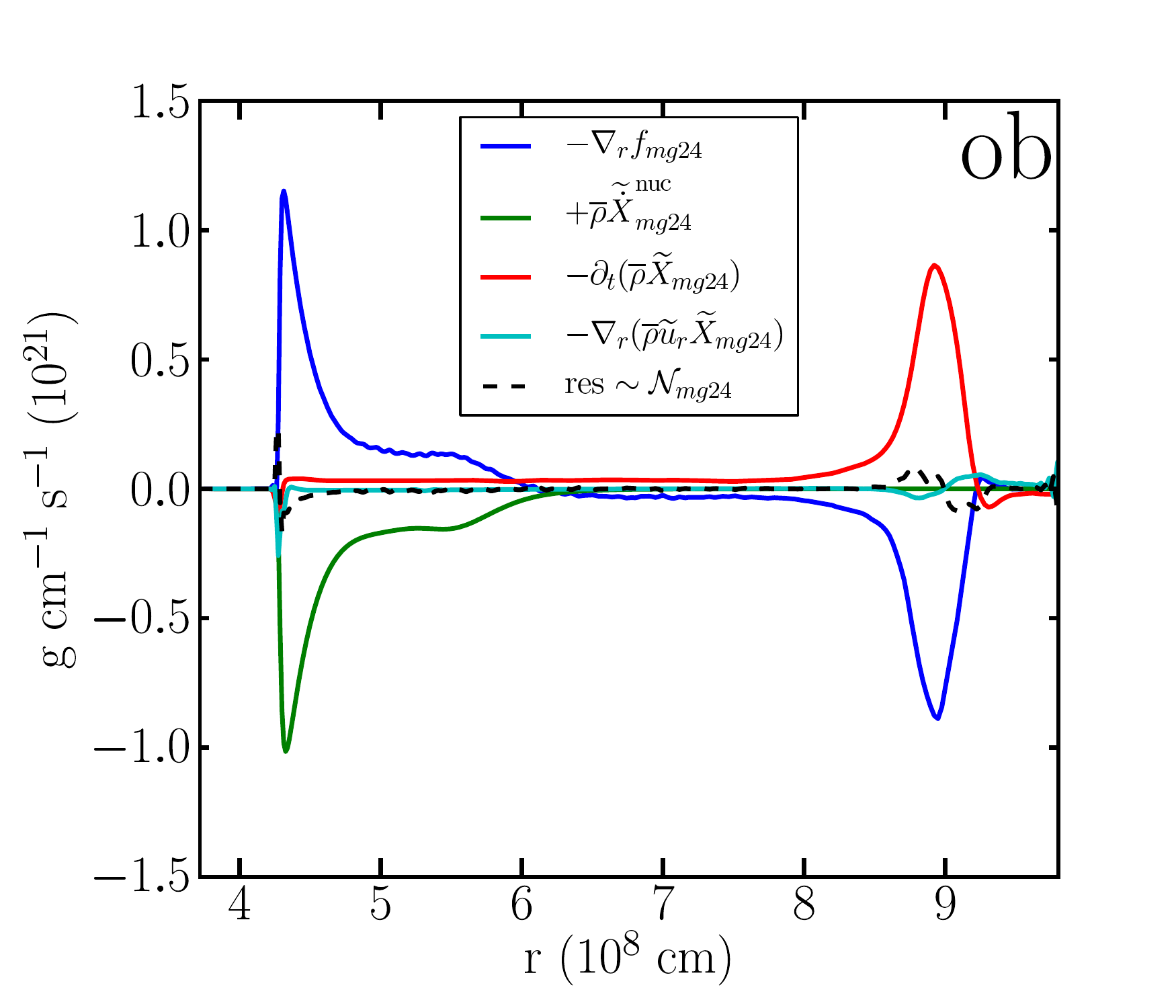}
\includegraphics[width=6.8cm]{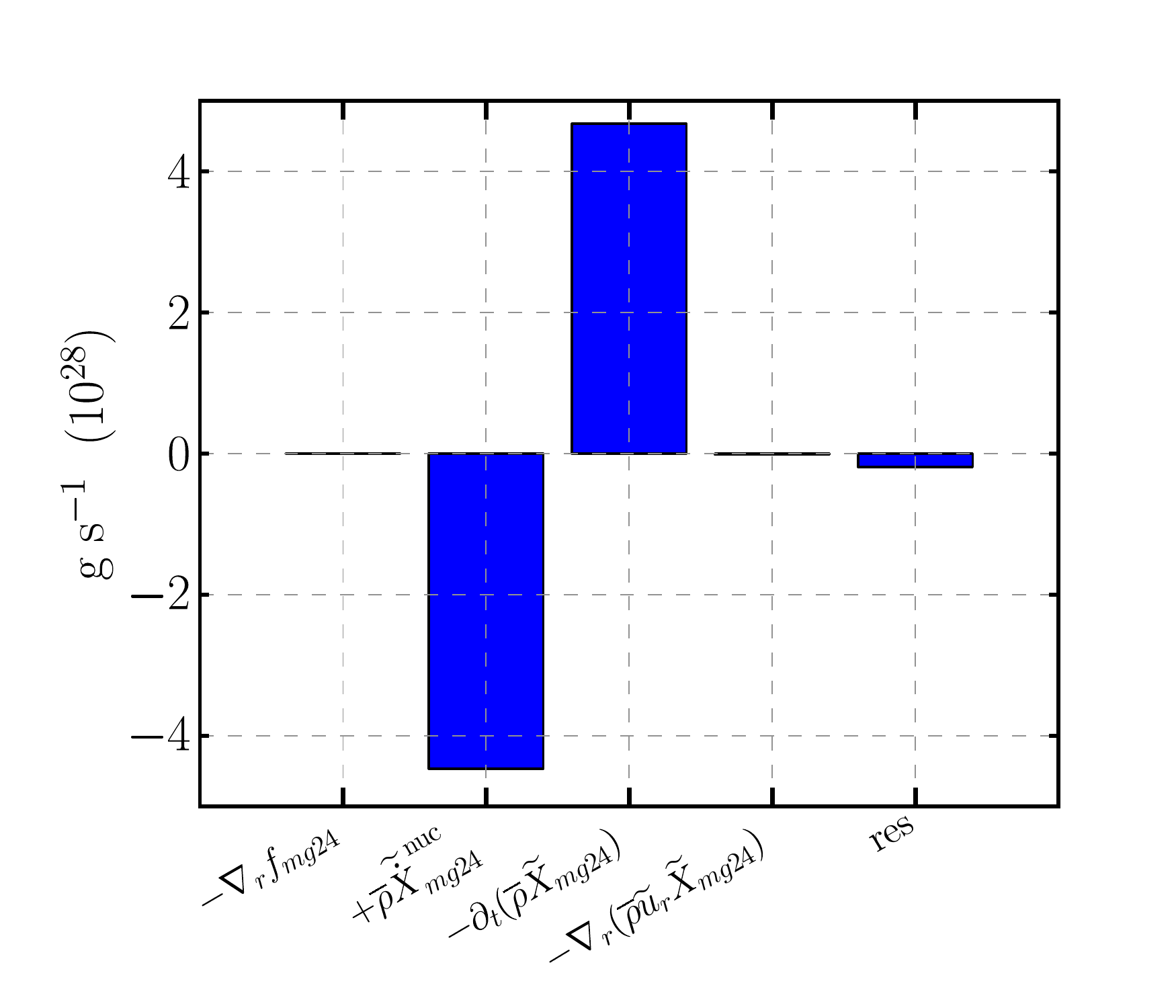}}

\centerline{
\includegraphics[width=6.8cm]{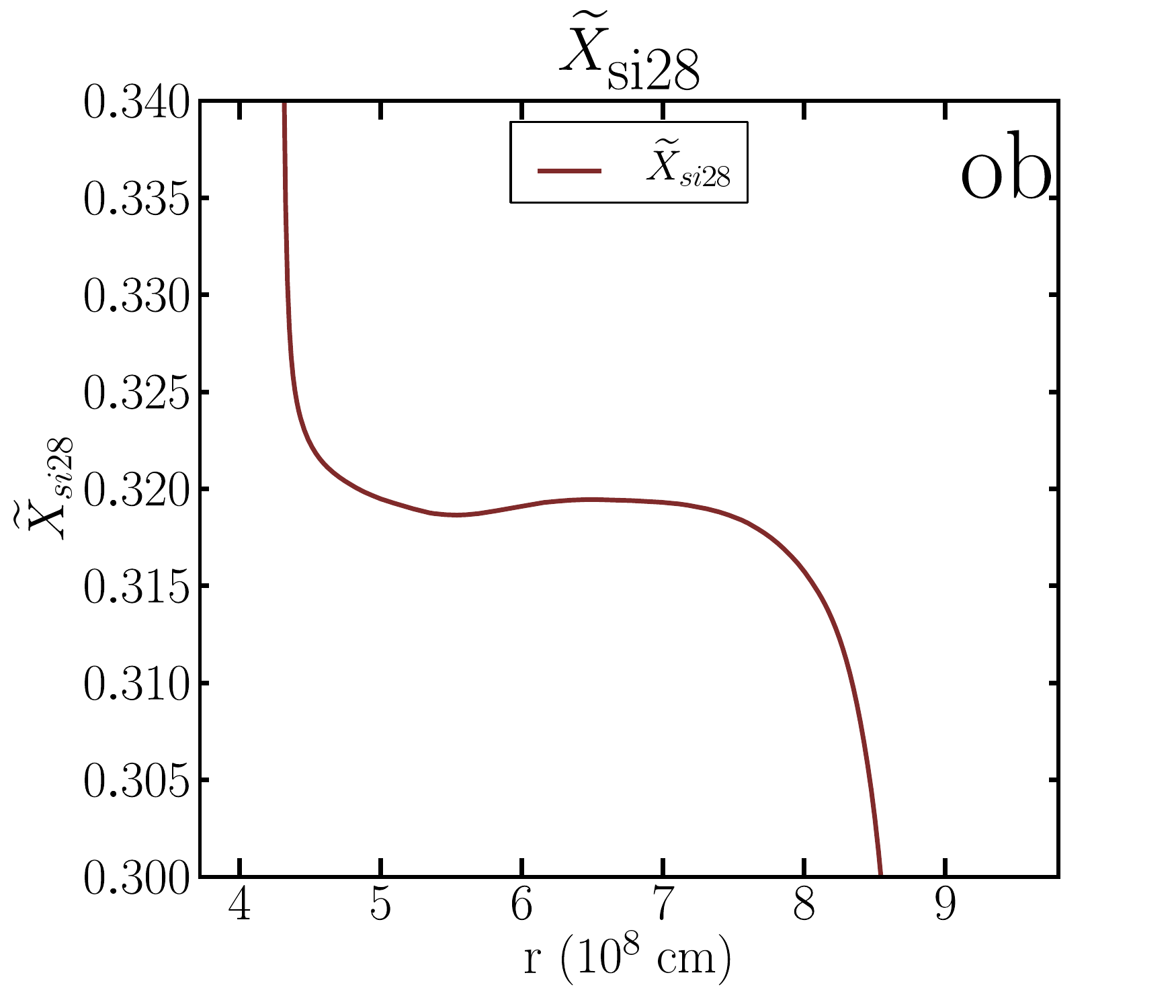}
\includegraphics[width=6.8cm]{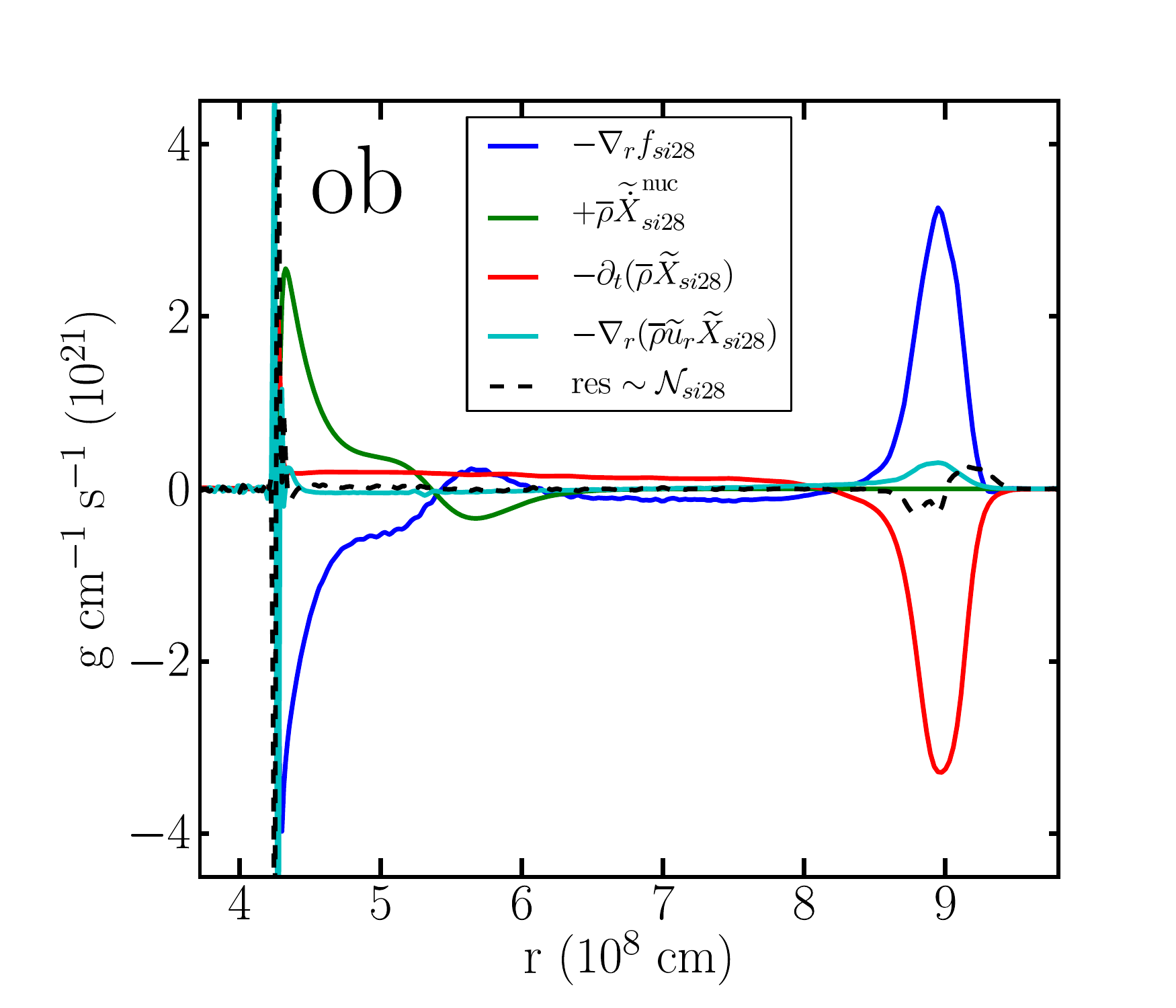}
\includegraphics[width=6.8cm]{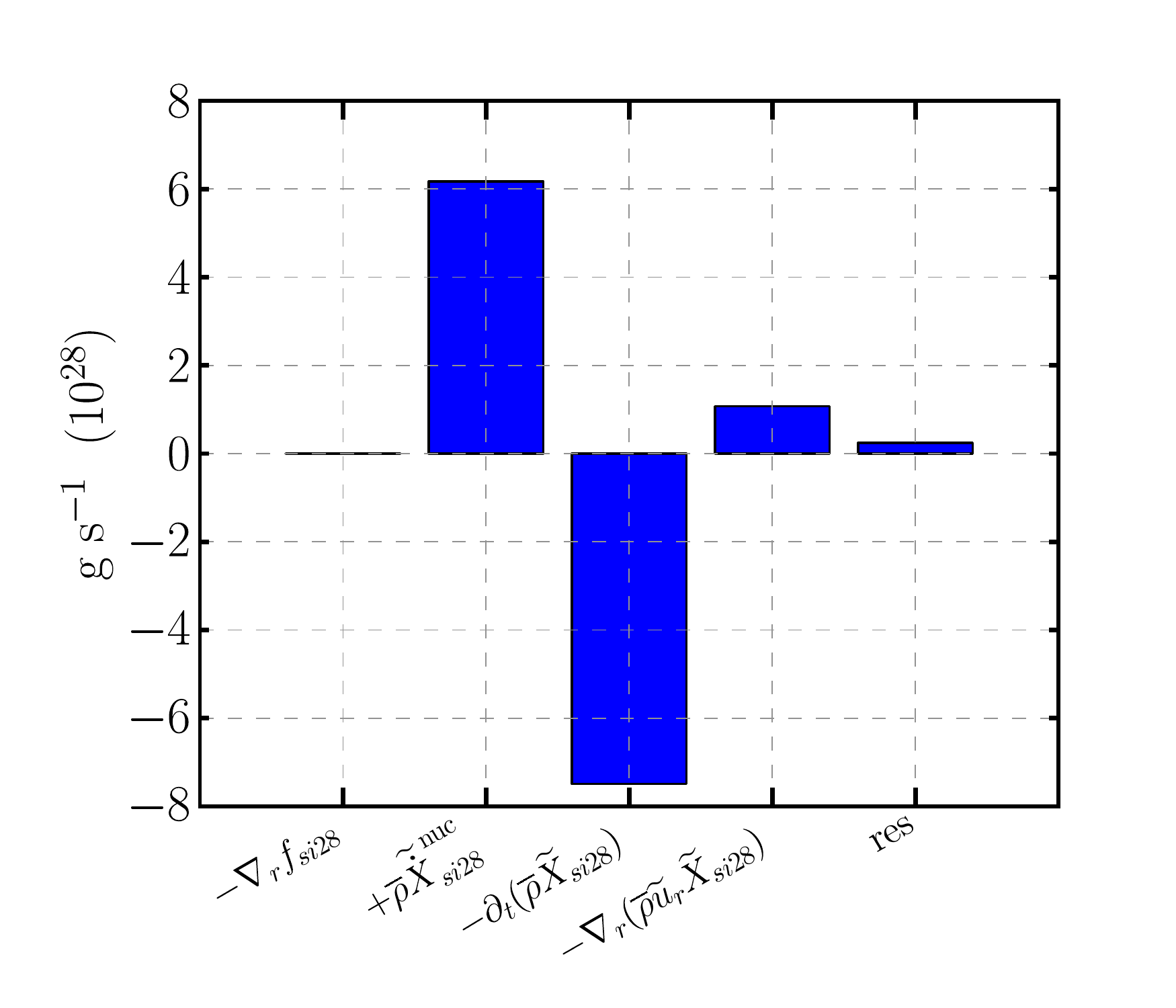}}
\caption{Mean composition equations. Model {\sf ob.3D.2hp}. \label{fig:xna23-xmg24-equations}}
\end{figure}

\newpage

\subsection{Mean P$^{31}$ and S$^{32}$ equation}

\begin{figure}[!h]
\centerline{
\includegraphics[width=6.8cm]{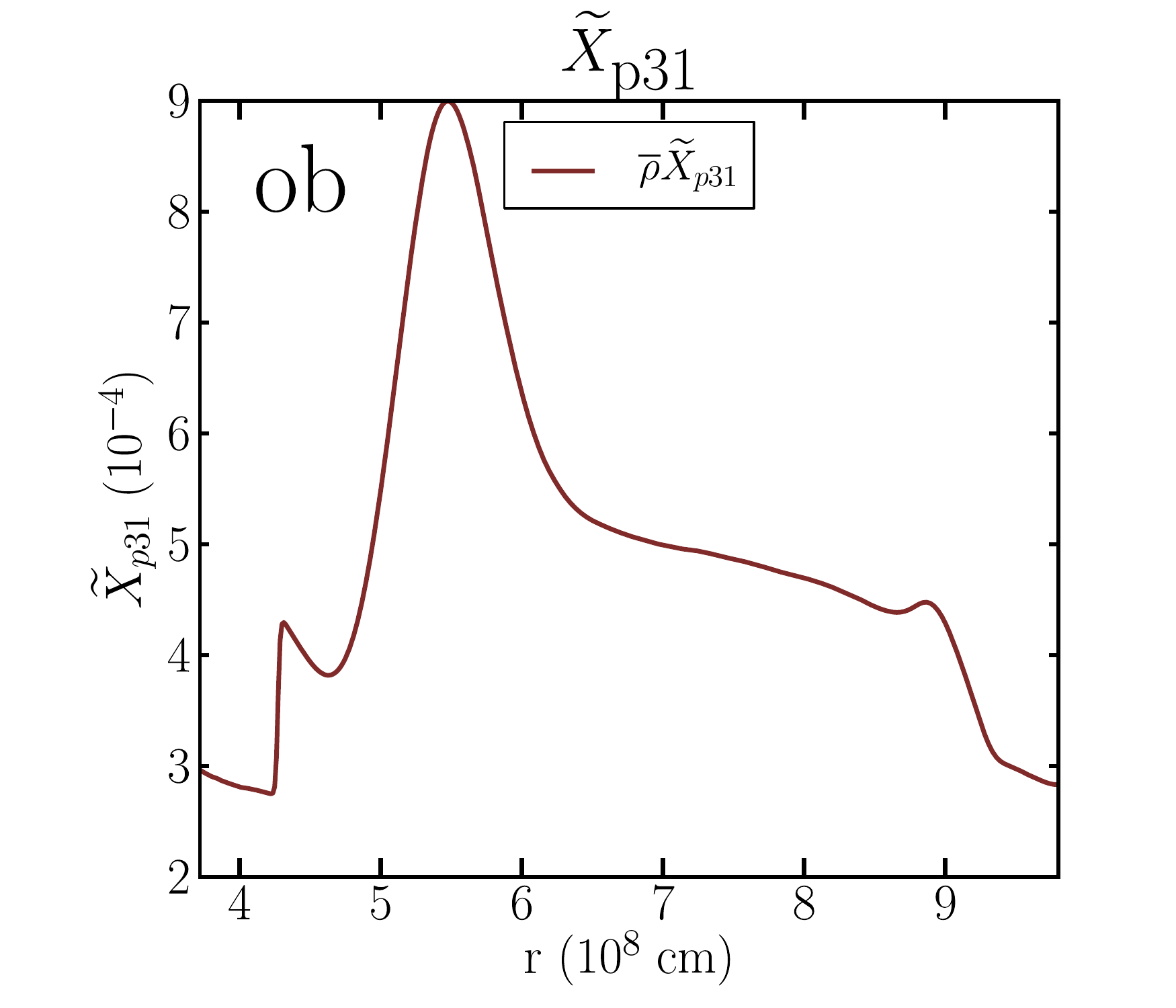}
\includegraphics[width=6.8cm]{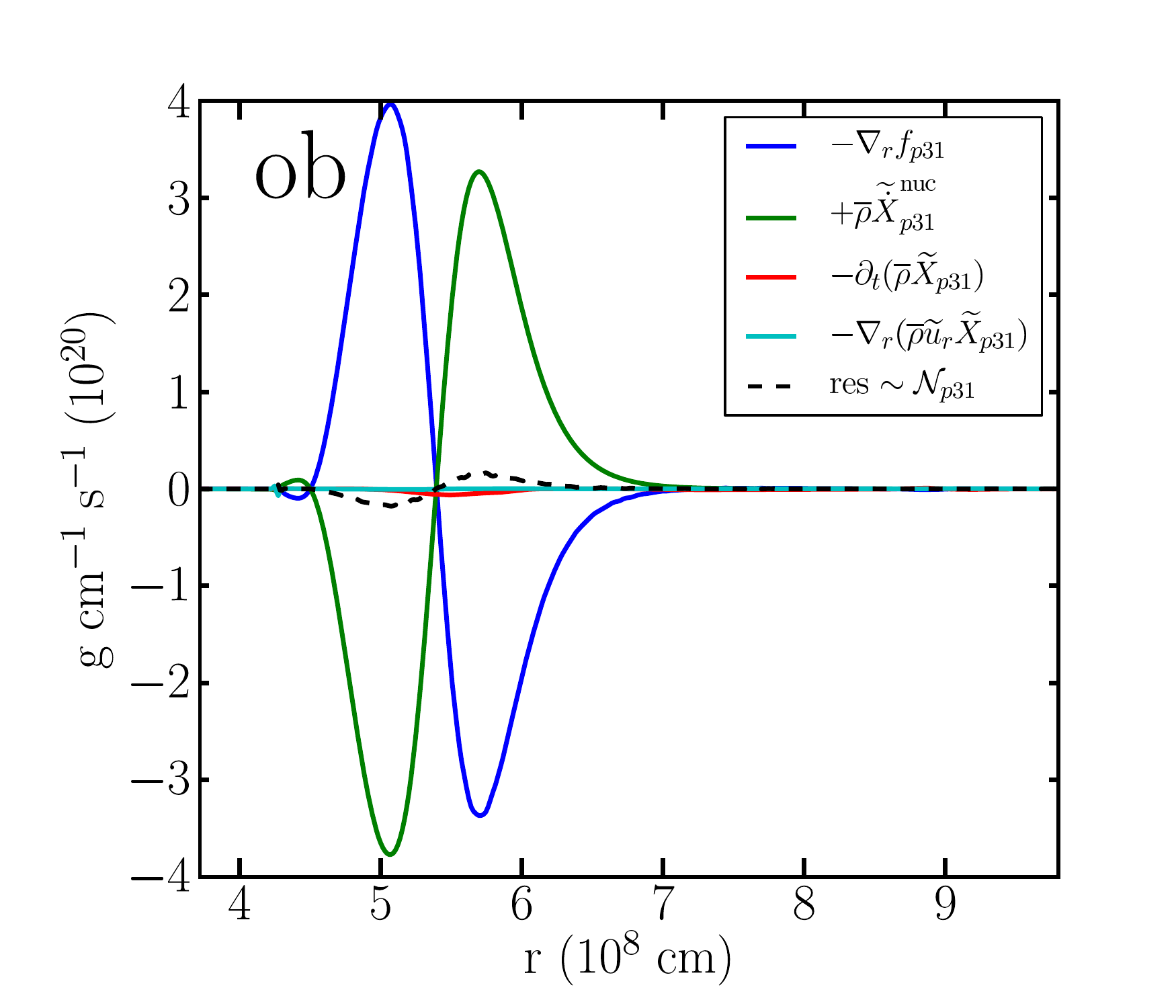}
\includegraphics[width=6.8cm]{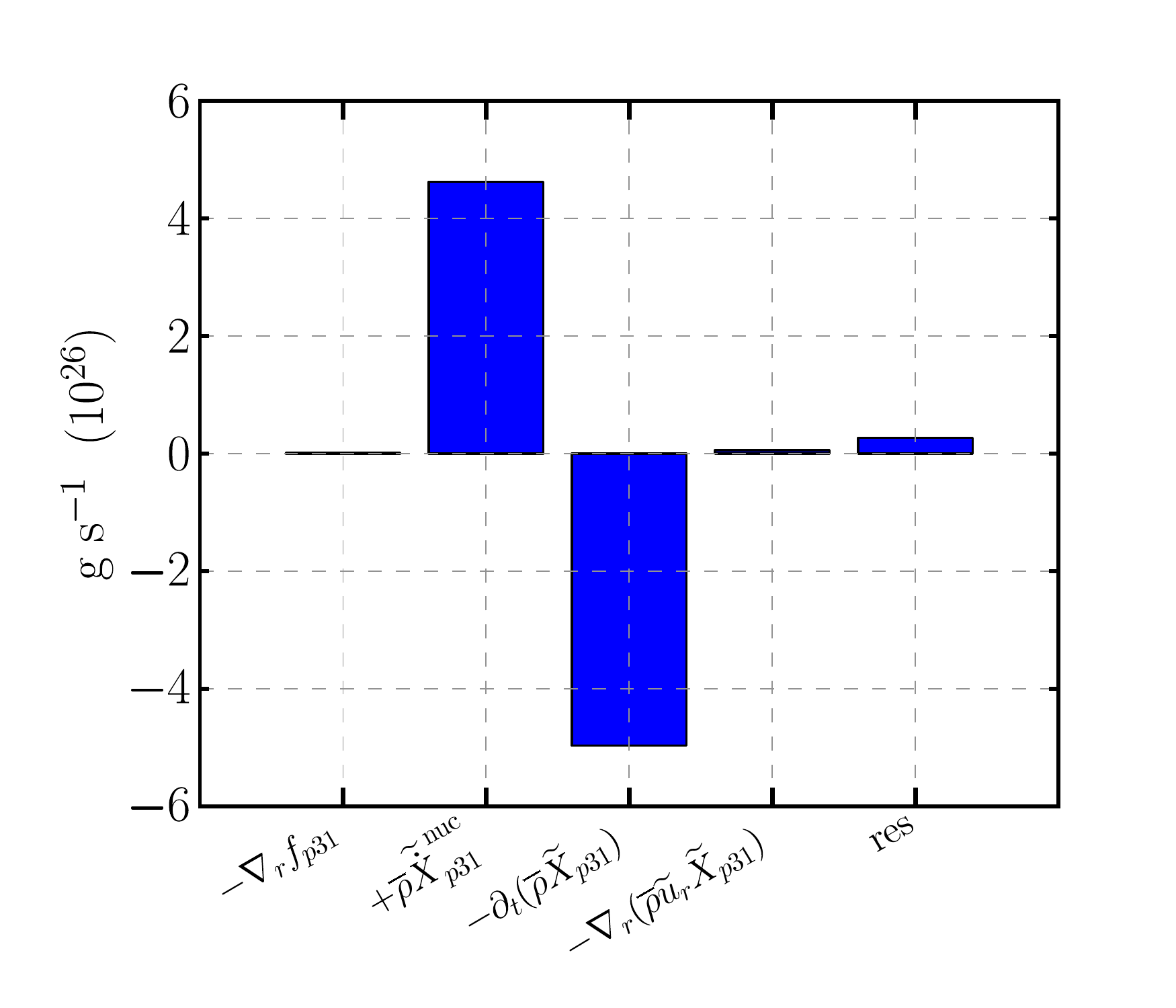}}

\centerline{
\includegraphics[width=6.8cm]{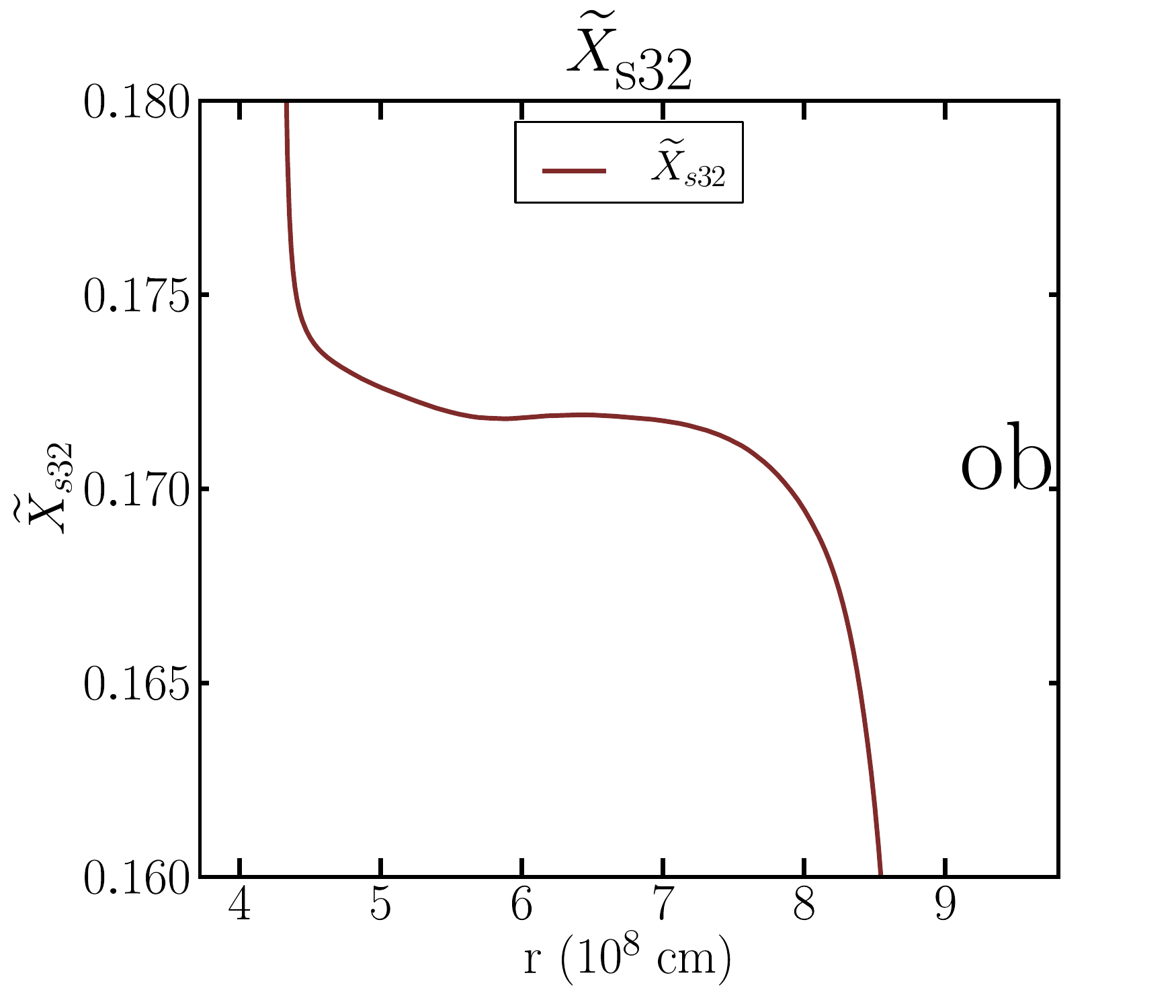}
\includegraphics[width=6.8cm]{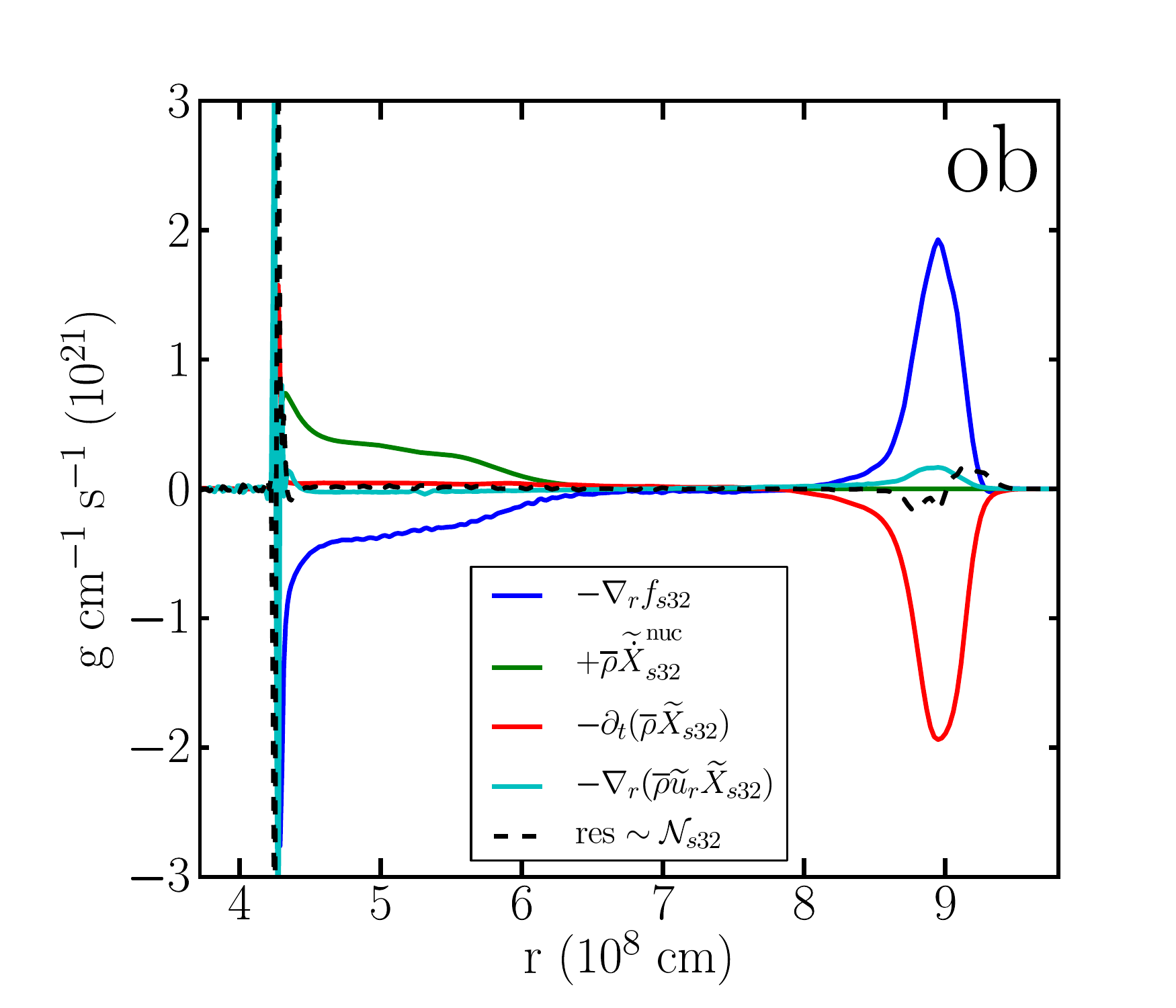}
\includegraphics[width=6.8cm]{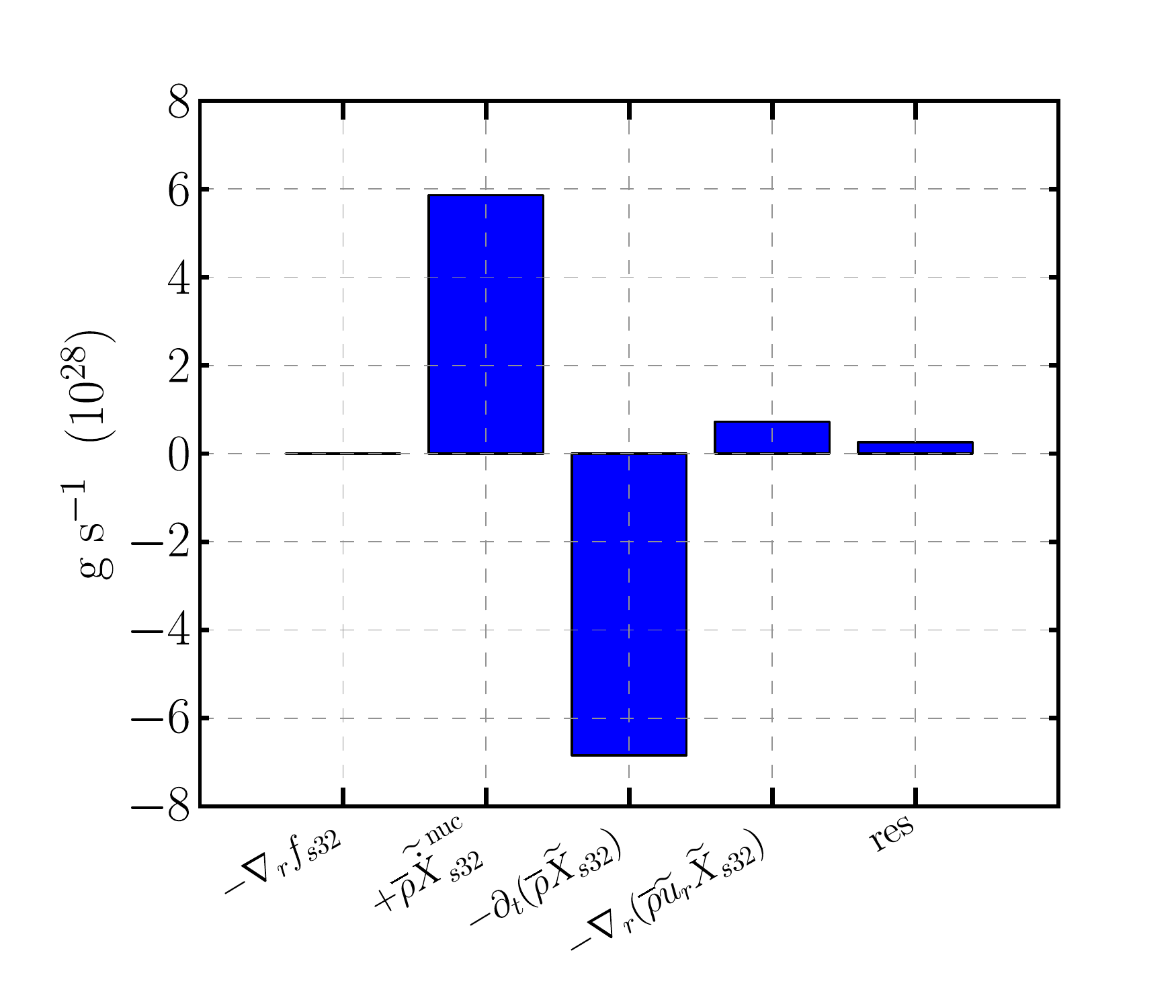}}
\caption{Mean composition equations. Model {\sf ob.3D.2hp}. \label{fig:xsi28-xp31-equations}}
\end{figure}

\newpage

\subsection{Mean S$^{34}$ and Cl$^{35}$ equation}

\begin{figure}[!h]
\centerline{
\includegraphics[width=6.8cm]{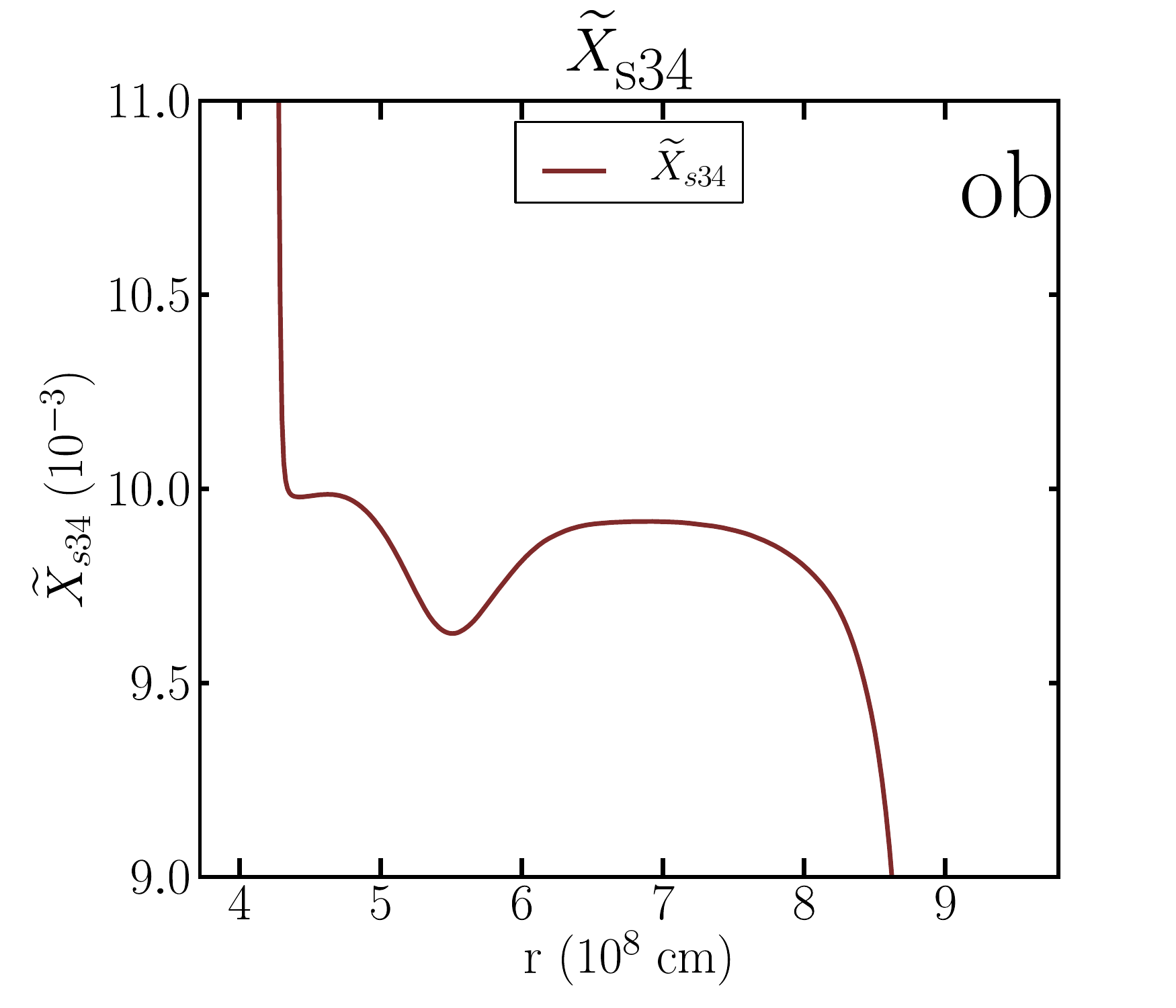}
\includegraphics[width=6.8cm]{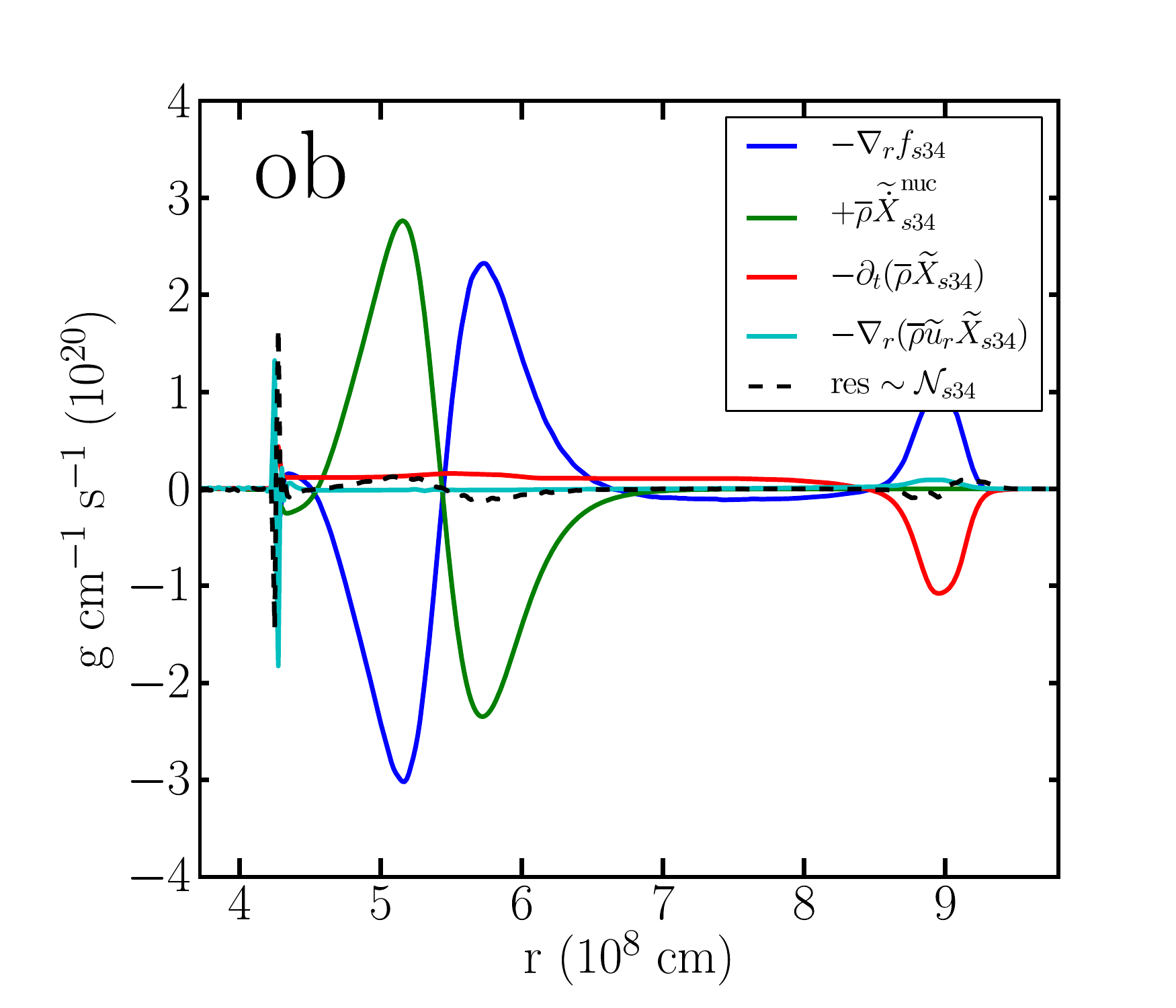}
\includegraphics[width=6.8cm]{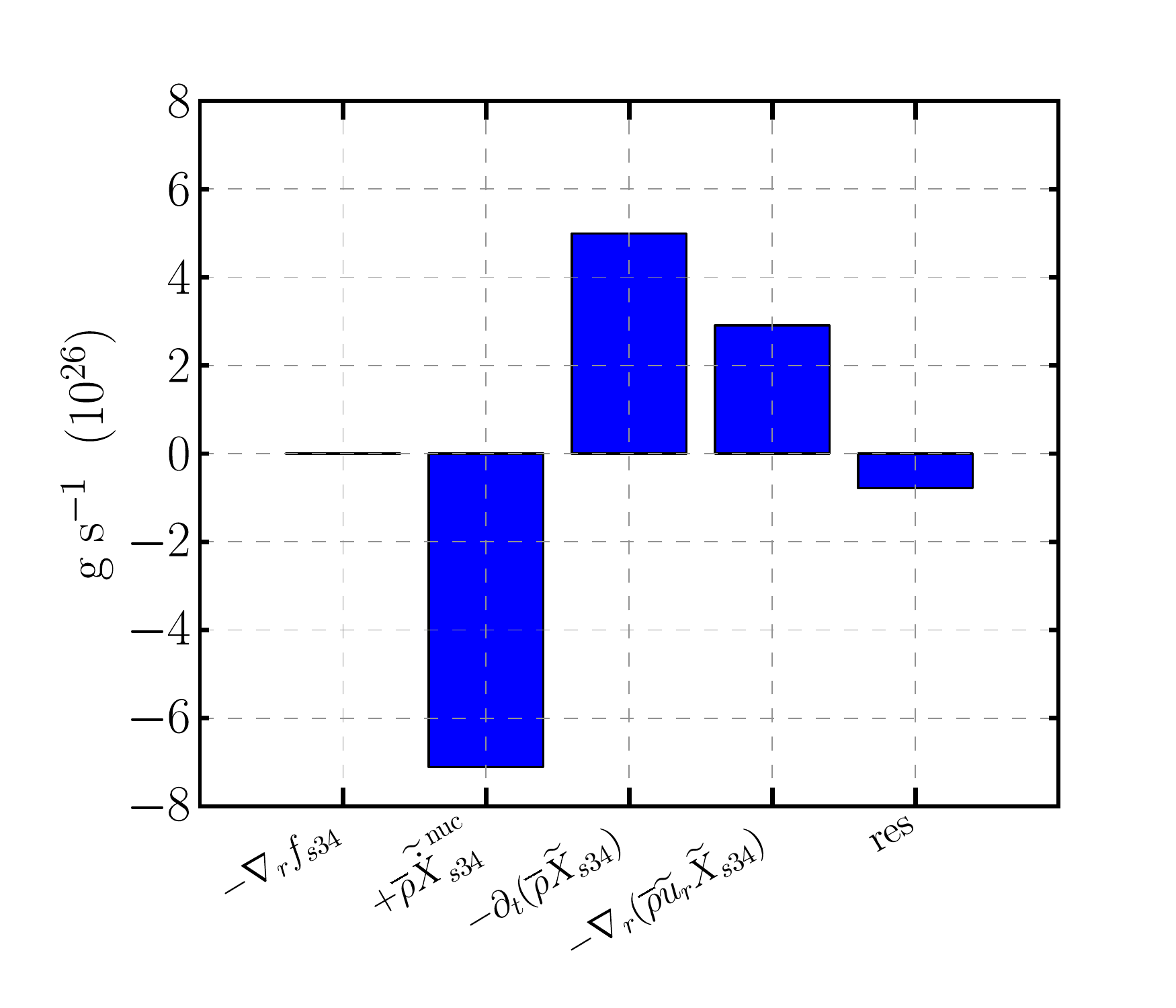}}

\centerline{
\includegraphics[width=6.8cm]{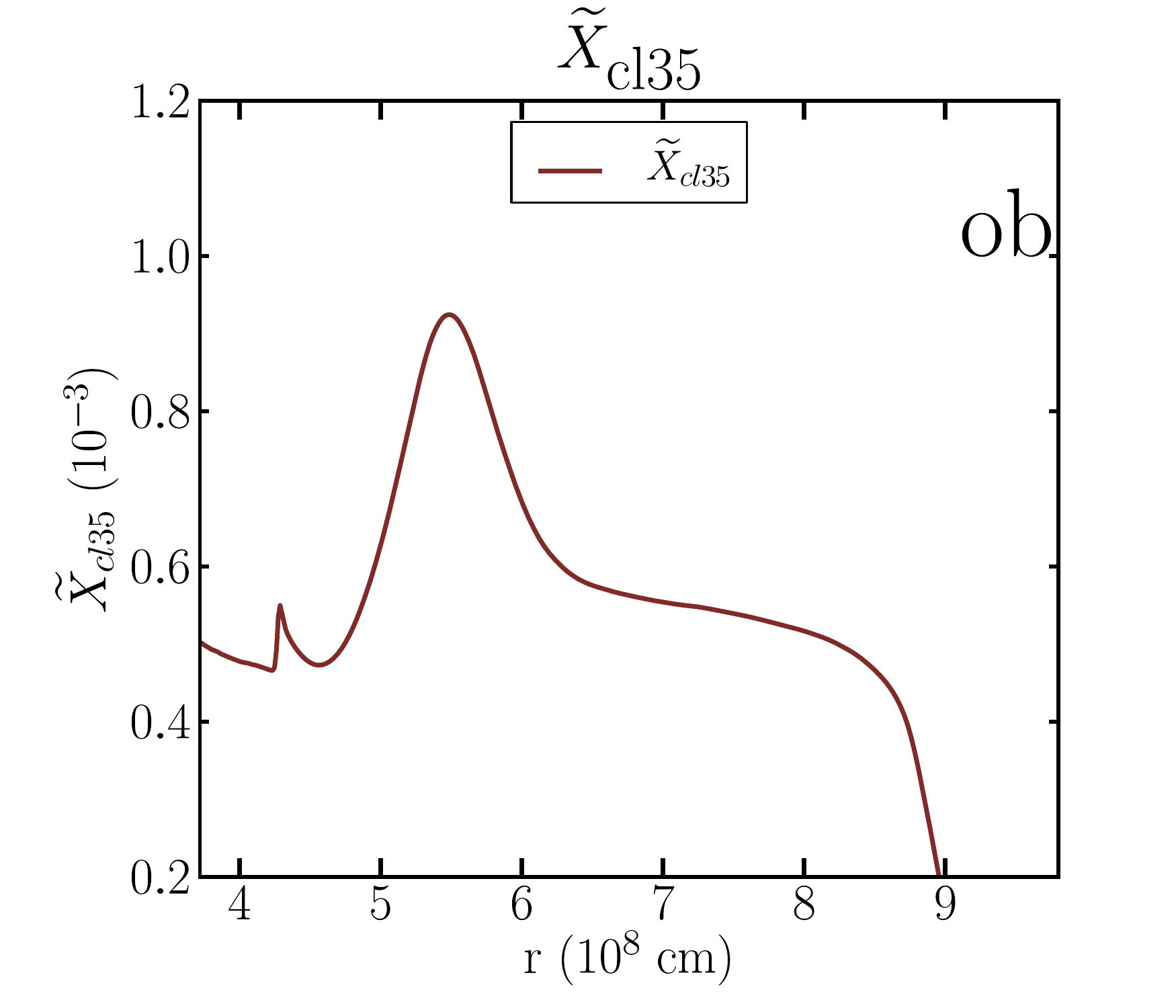}
\includegraphics[width=6.8cm]{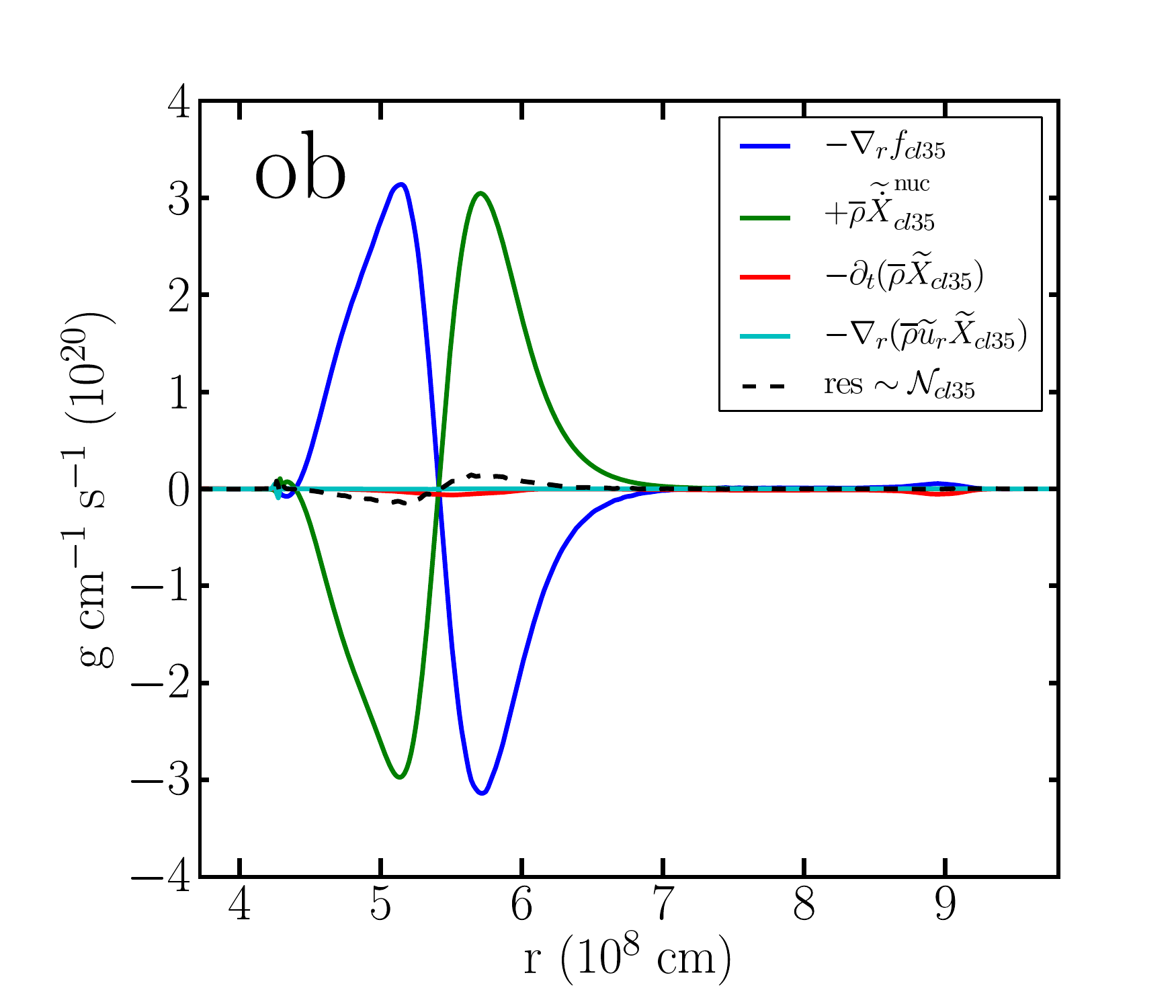}
\includegraphics[width=6.8cm]{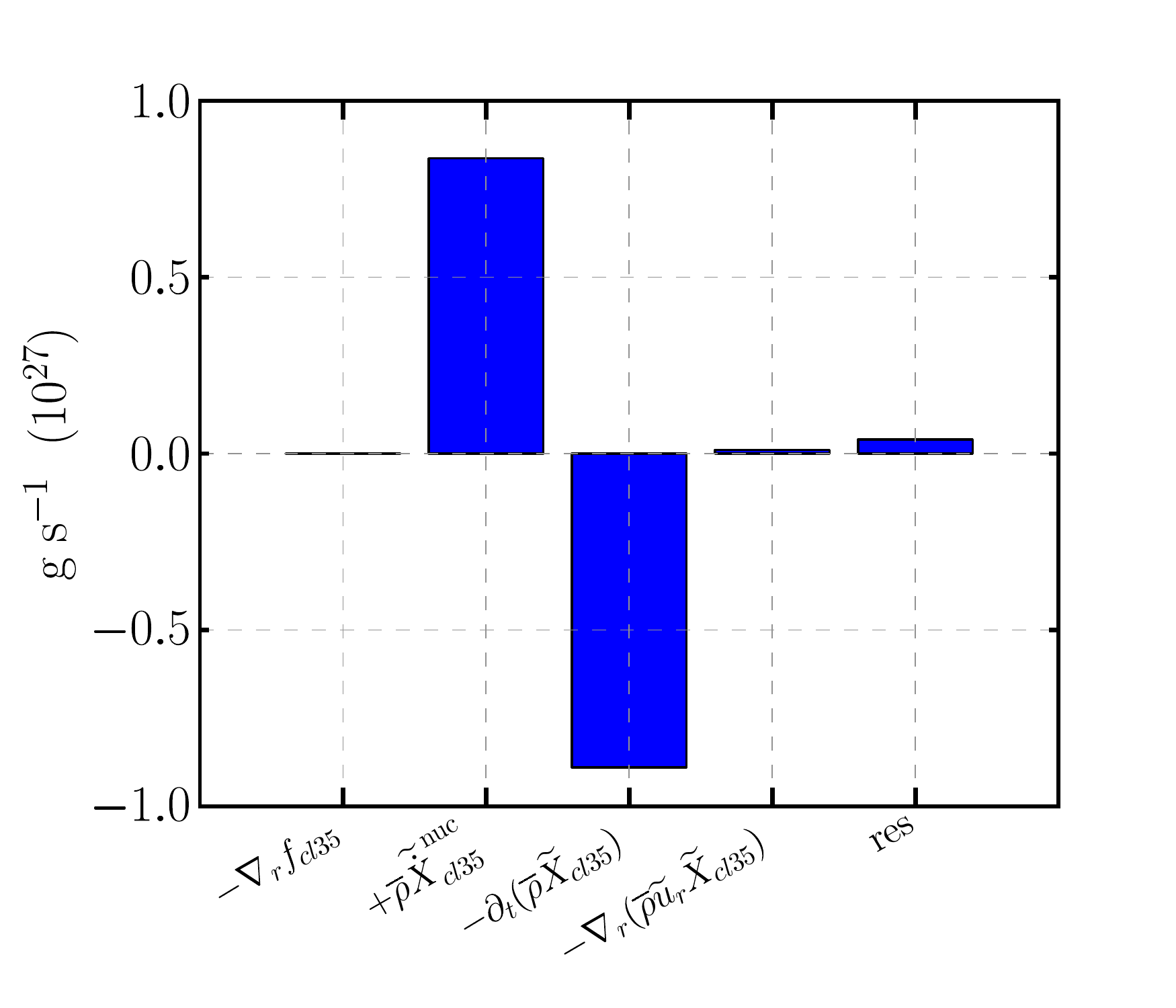}}
\caption{Mean composition equations. Model {\sf ob.3D.2hp}. \label{fig:xs32-xs34-equations}}
\end{figure}

\newpage

\subsection{Mean Ar$^{36}$ and Ar$^{38}$ equation}

\begin{figure}[!h]
\centerline{
\includegraphics[width=6.8cm]{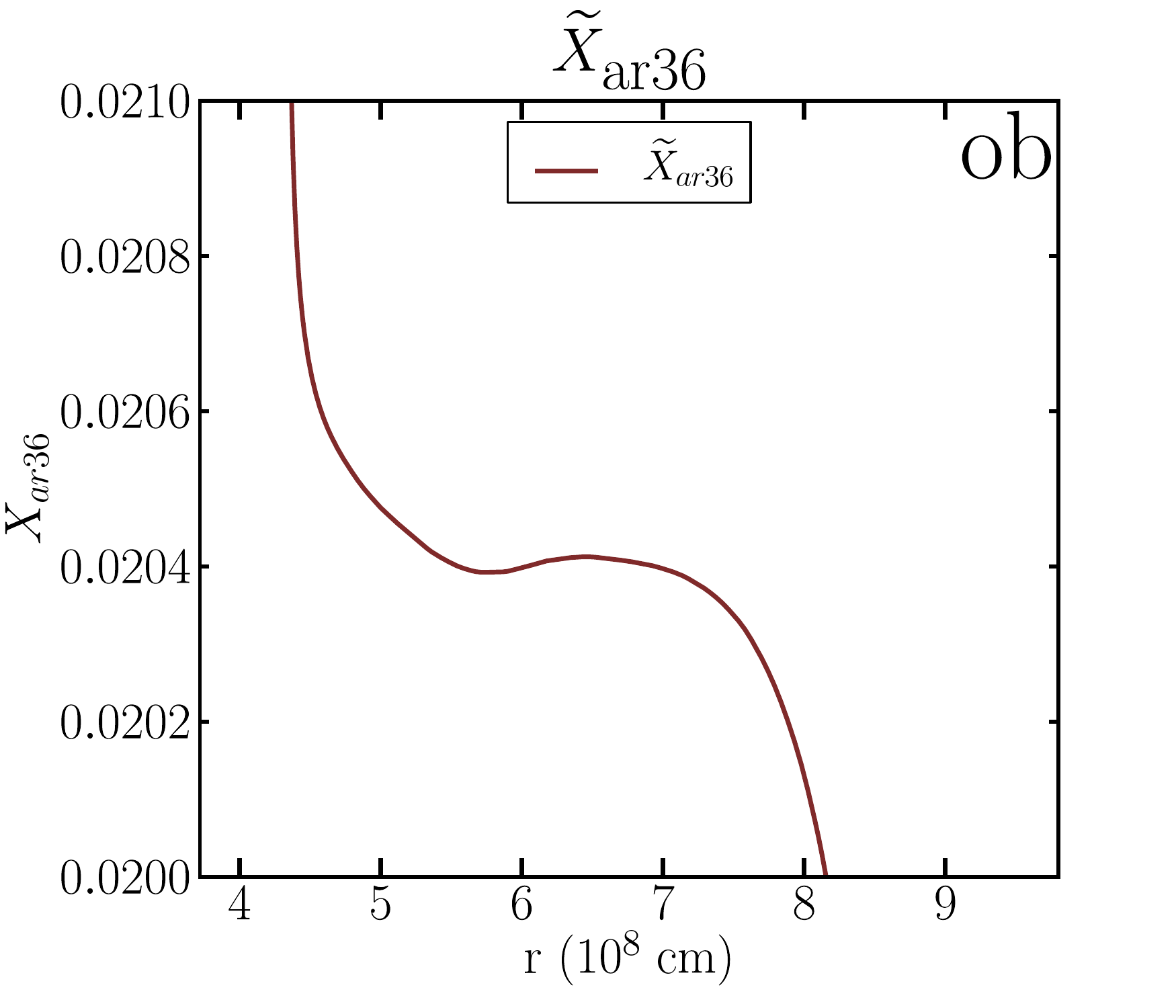}
\includegraphics[width=6.8cm]{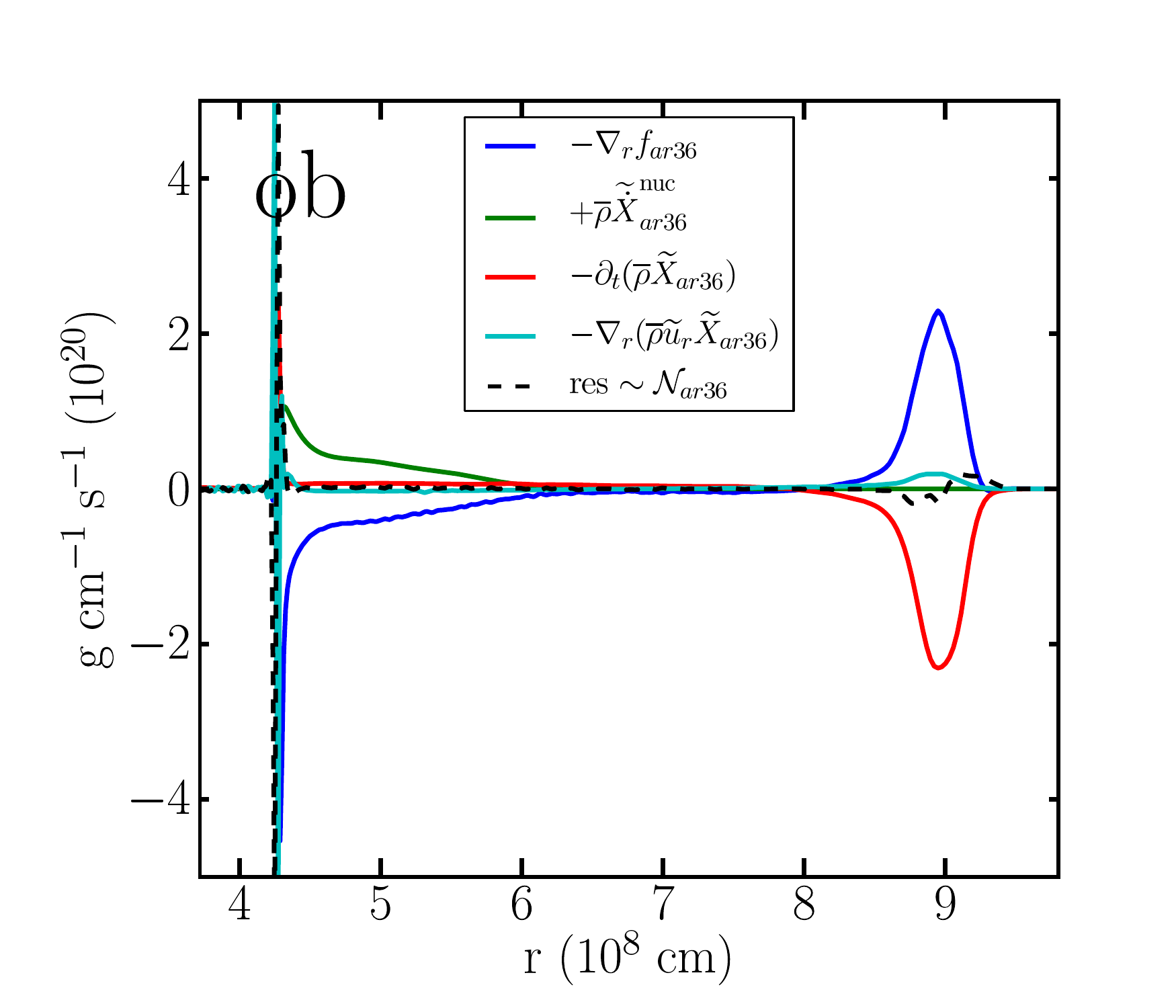}
\includegraphics[width=6.8cm]{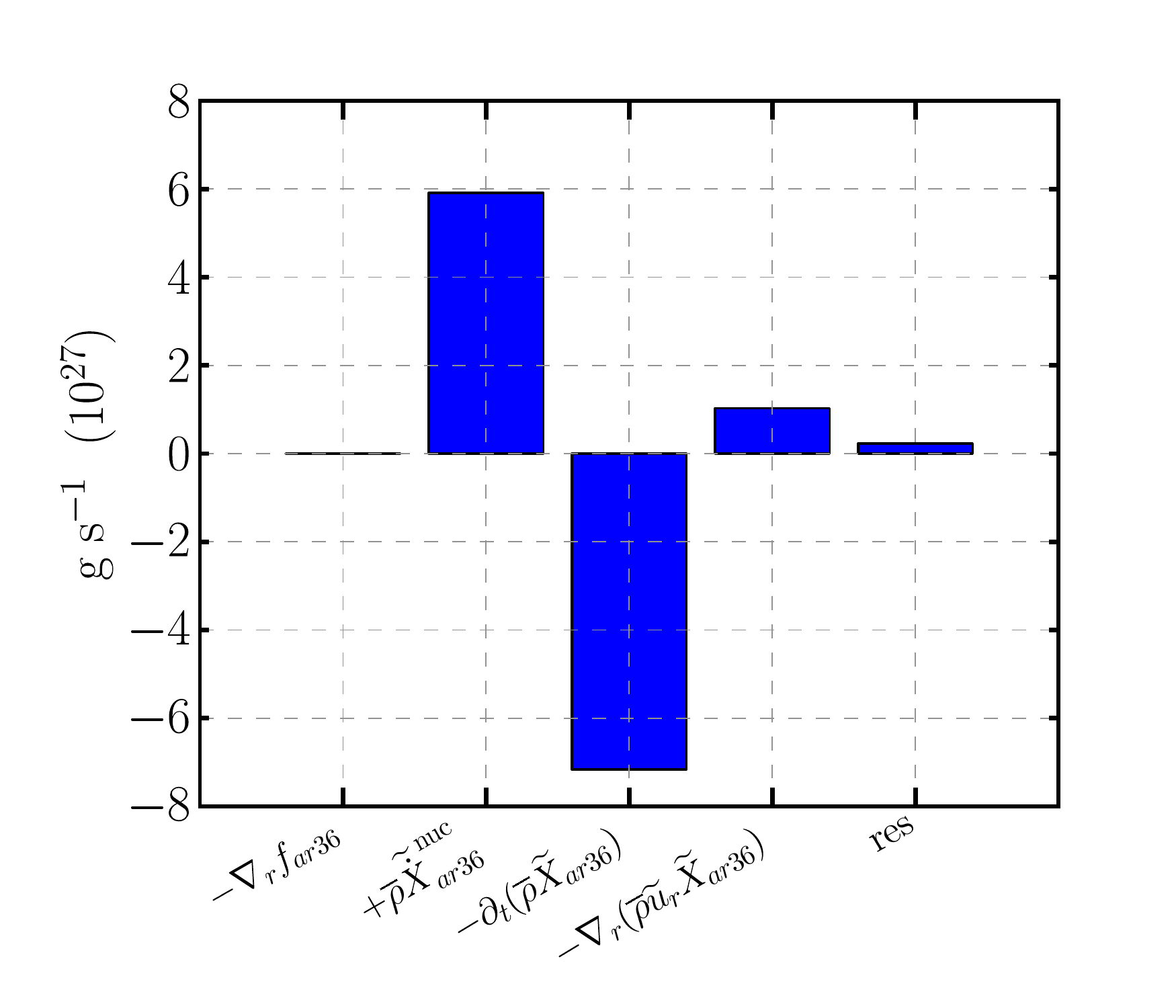}}

\centerline{
\includegraphics[width=6.8cm]{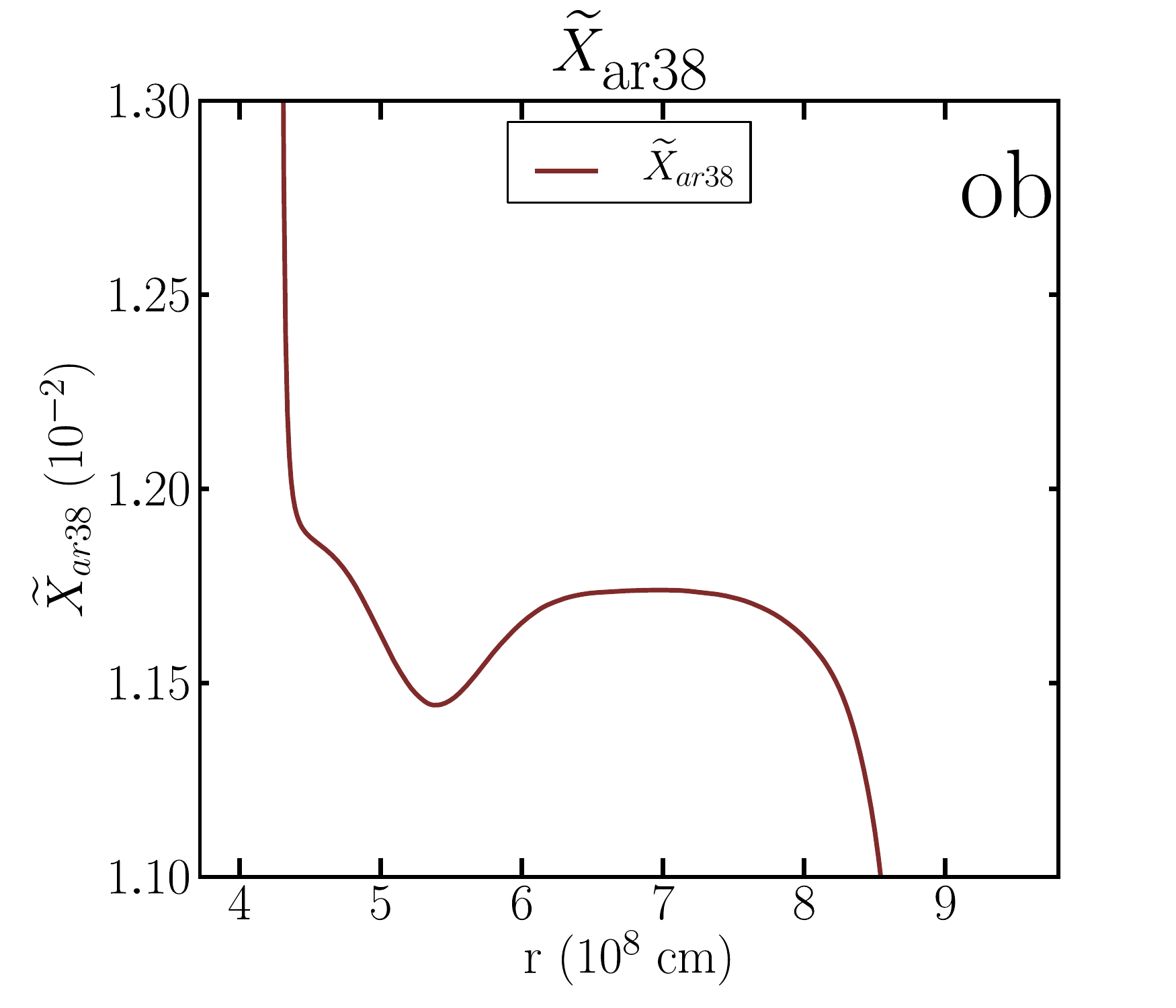}
\includegraphics[width=6.8cm]{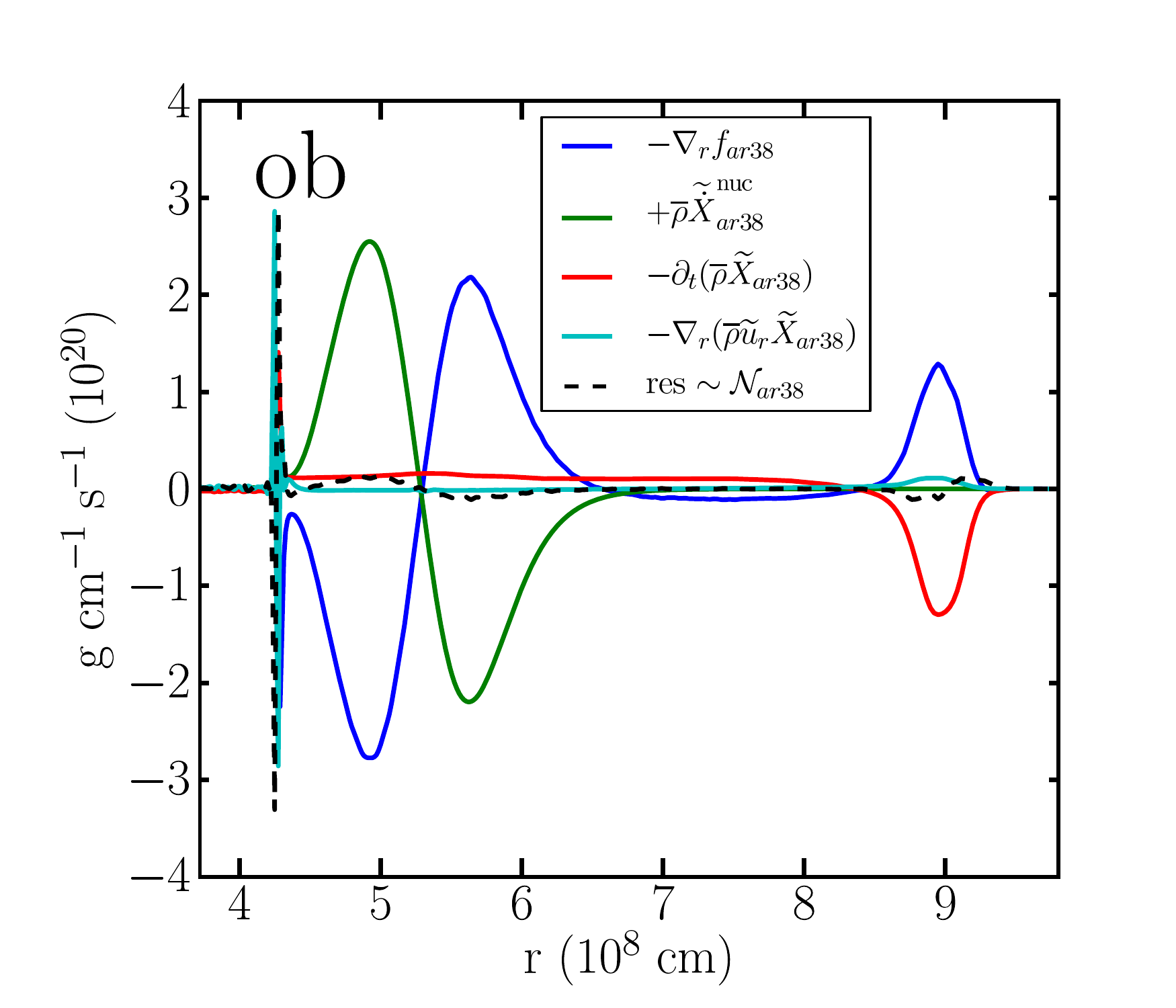}
\includegraphics[width=6.8cm]{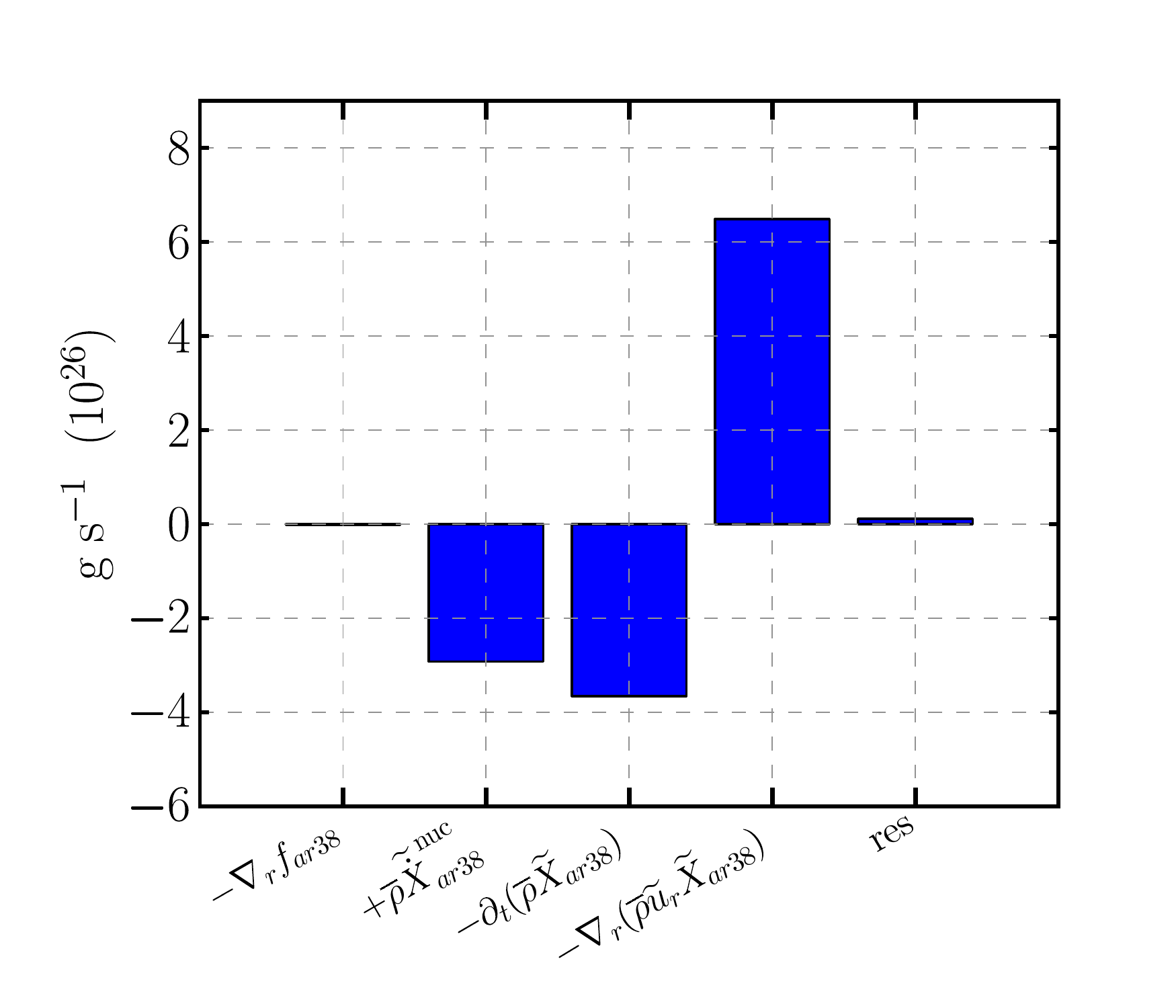}}
\caption{Mean composition equations. Model {\sf ob.3D.2hp}. \label{fig:xcl35-xar36-equations}}
\end{figure}

\newpage

\subsection{Mean K$^{39}$ and Ca$^{40}$ equation}

\begin{figure}[!h]
\centerline{
\includegraphics[width=6.8cm]{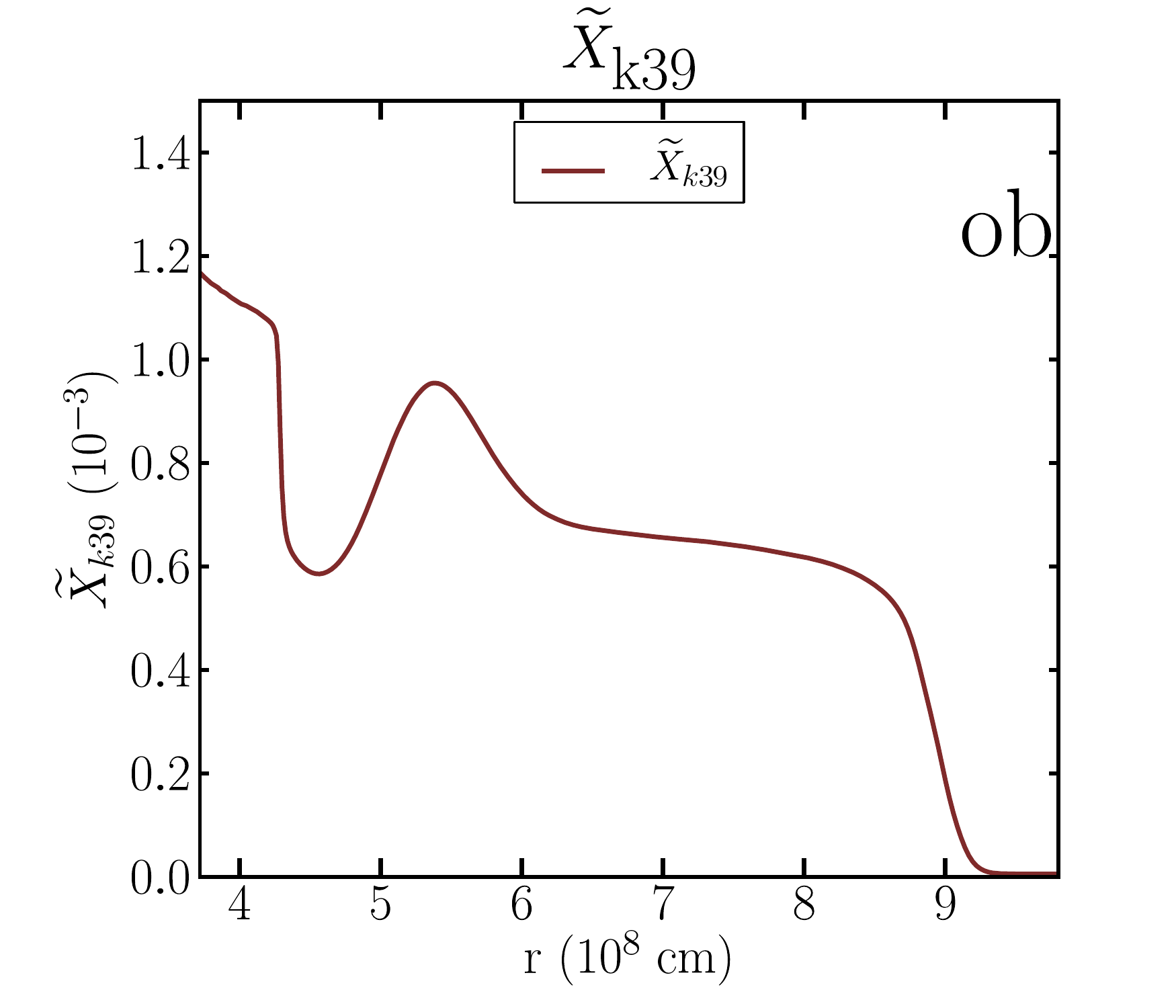}
\includegraphics[width=6.8cm]{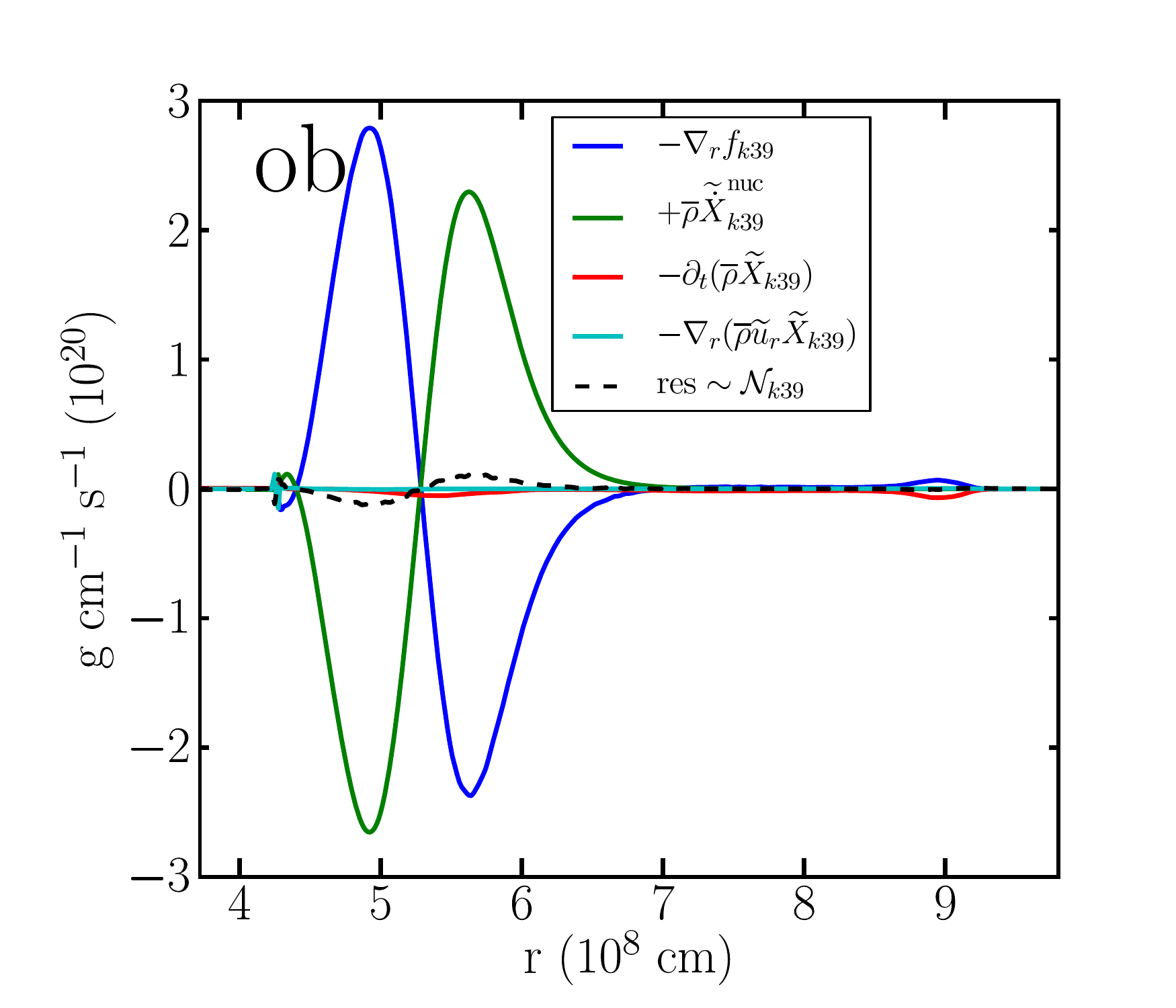}
\includegraphics[width=6.8cm]{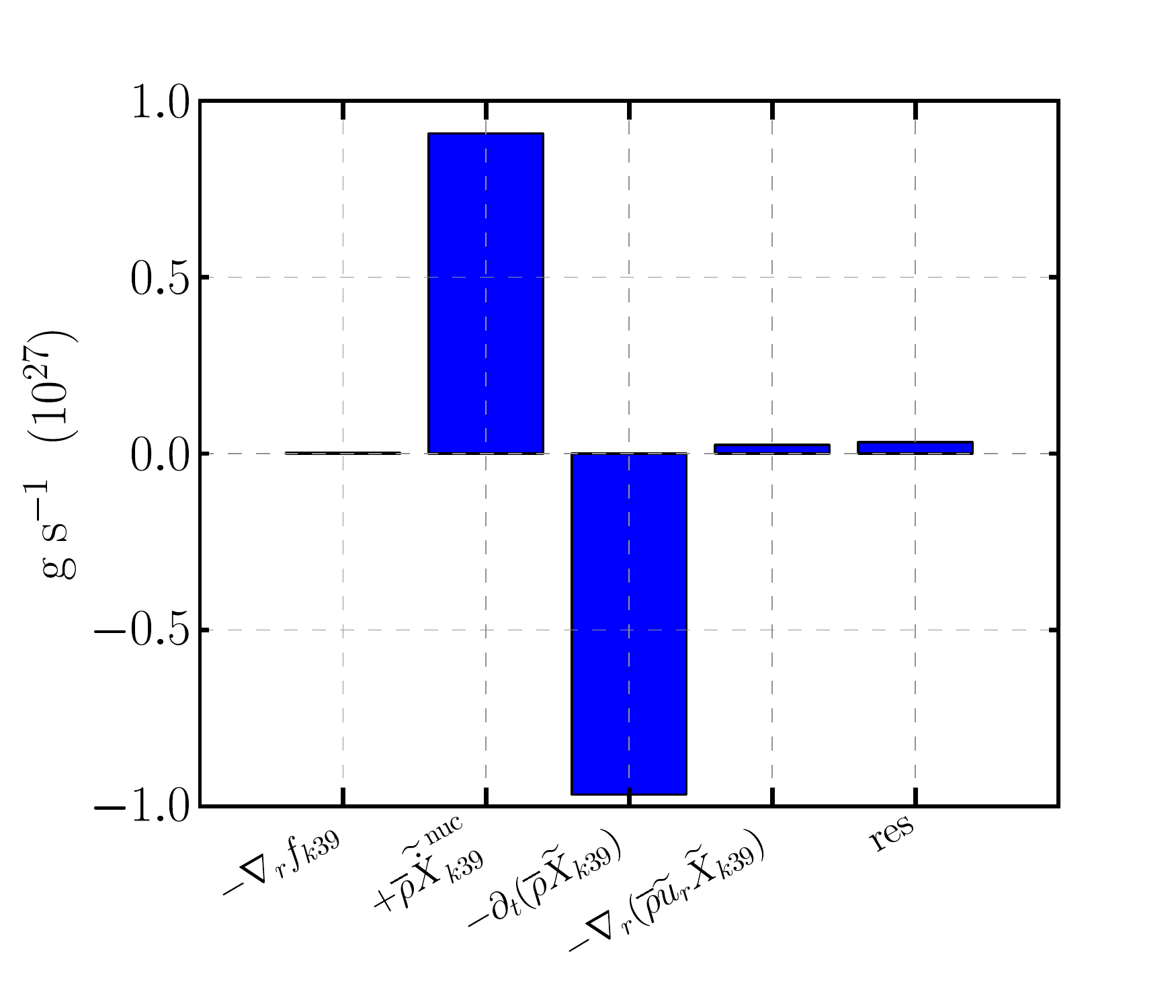}}

\centerline{
\includegraphics[width=6.8cm]{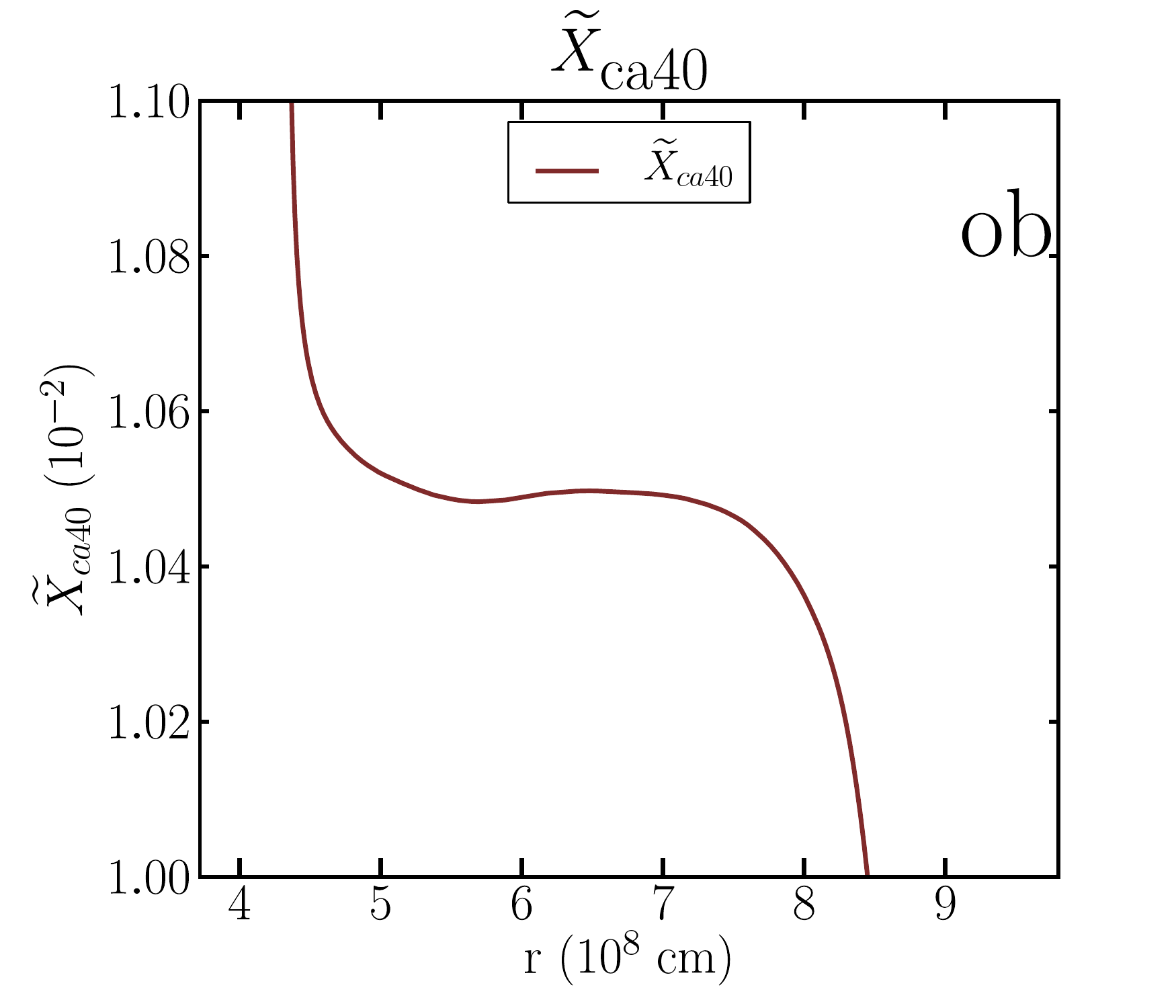}
\includegraphics[width=6.8cm]{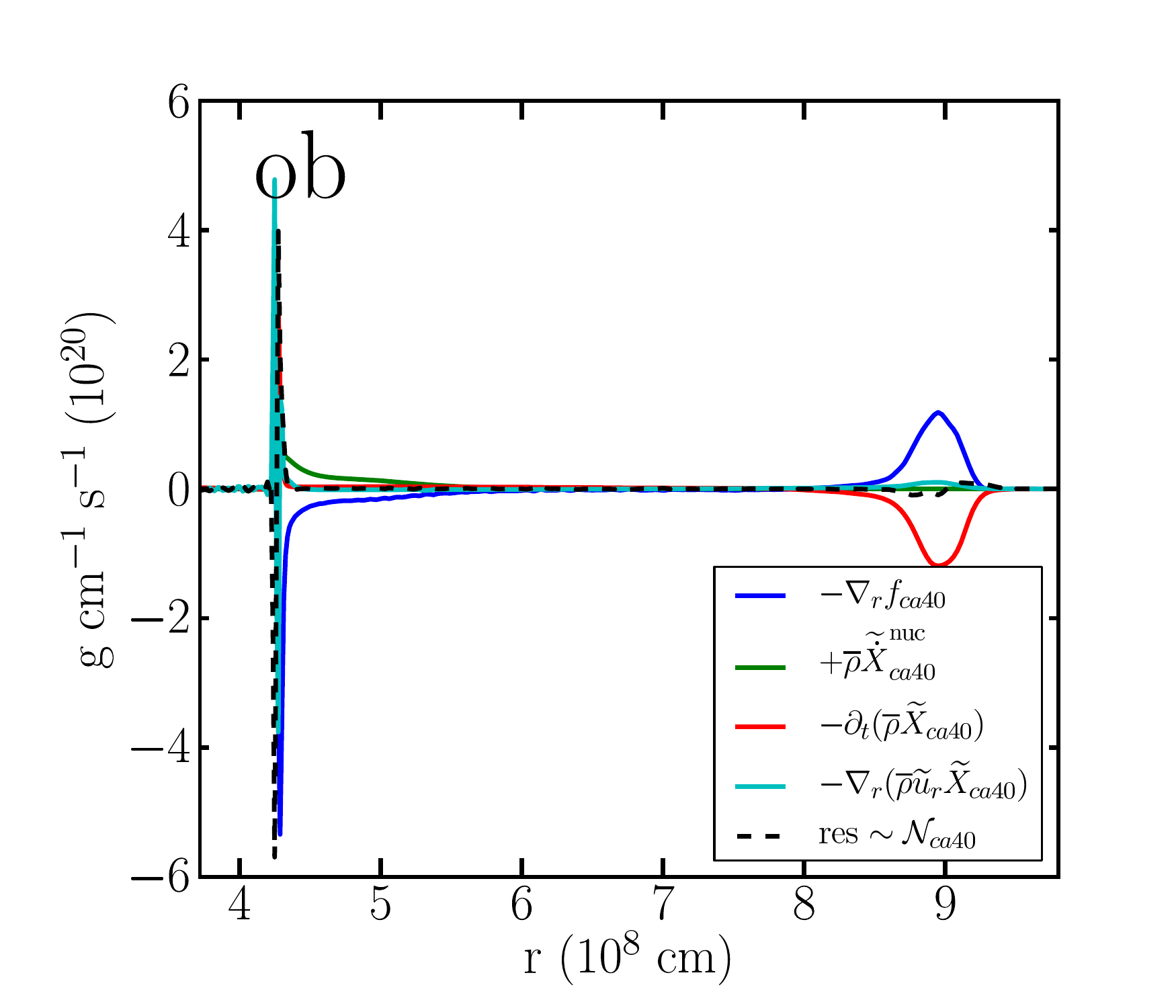}
\includegraphics[width=6.8cm]{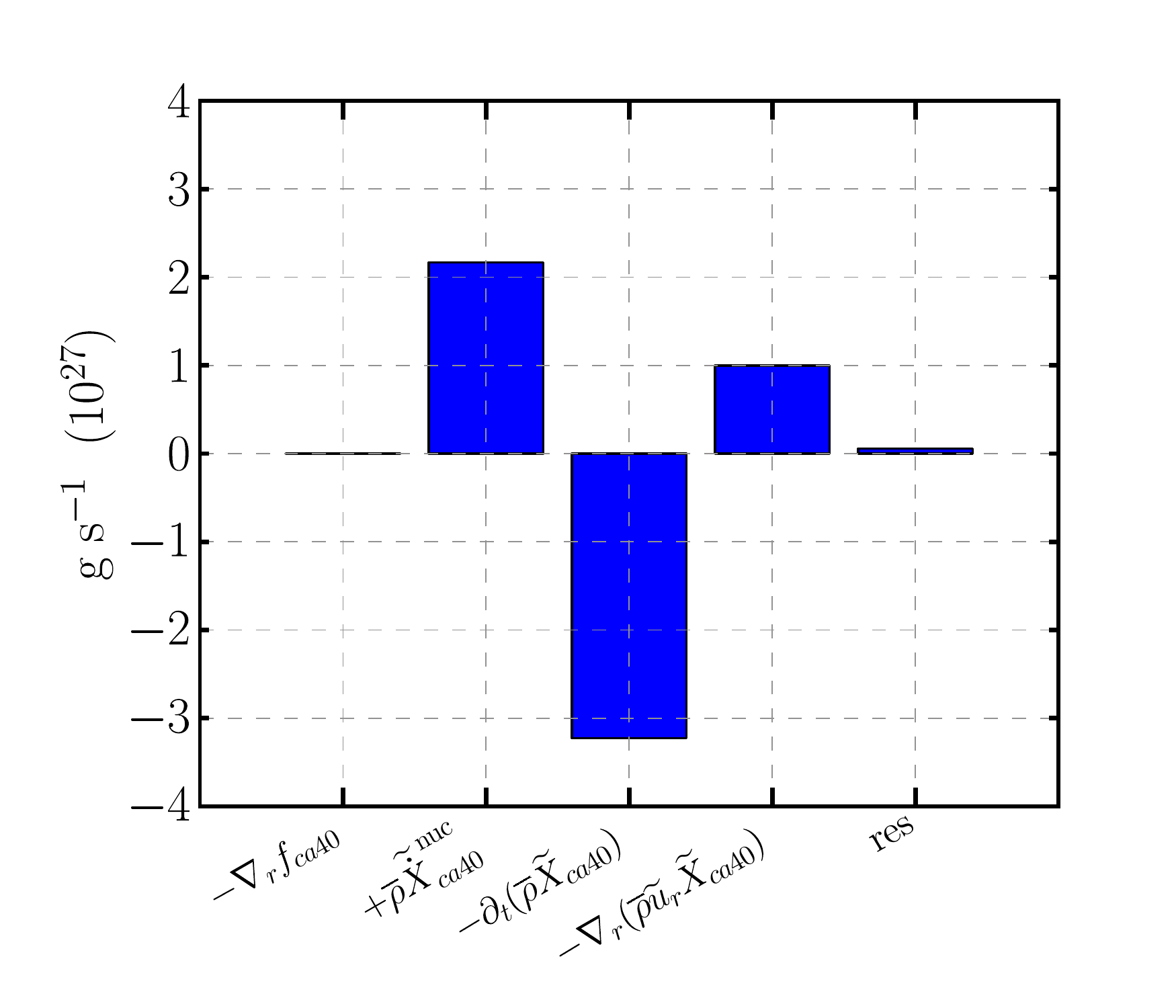}}
\caption{Mean composition equations. Model {\sf ob.3D.2hp}. \label{fig:xar38-xk39-equations}}
\end{figure}

\newpage

\subsection{Mean Ca$^{42}$ and Ti$^{44}$ equation}

\begin{figure}[!h]
\centerline{
\includegraphics[width=6.8cm]{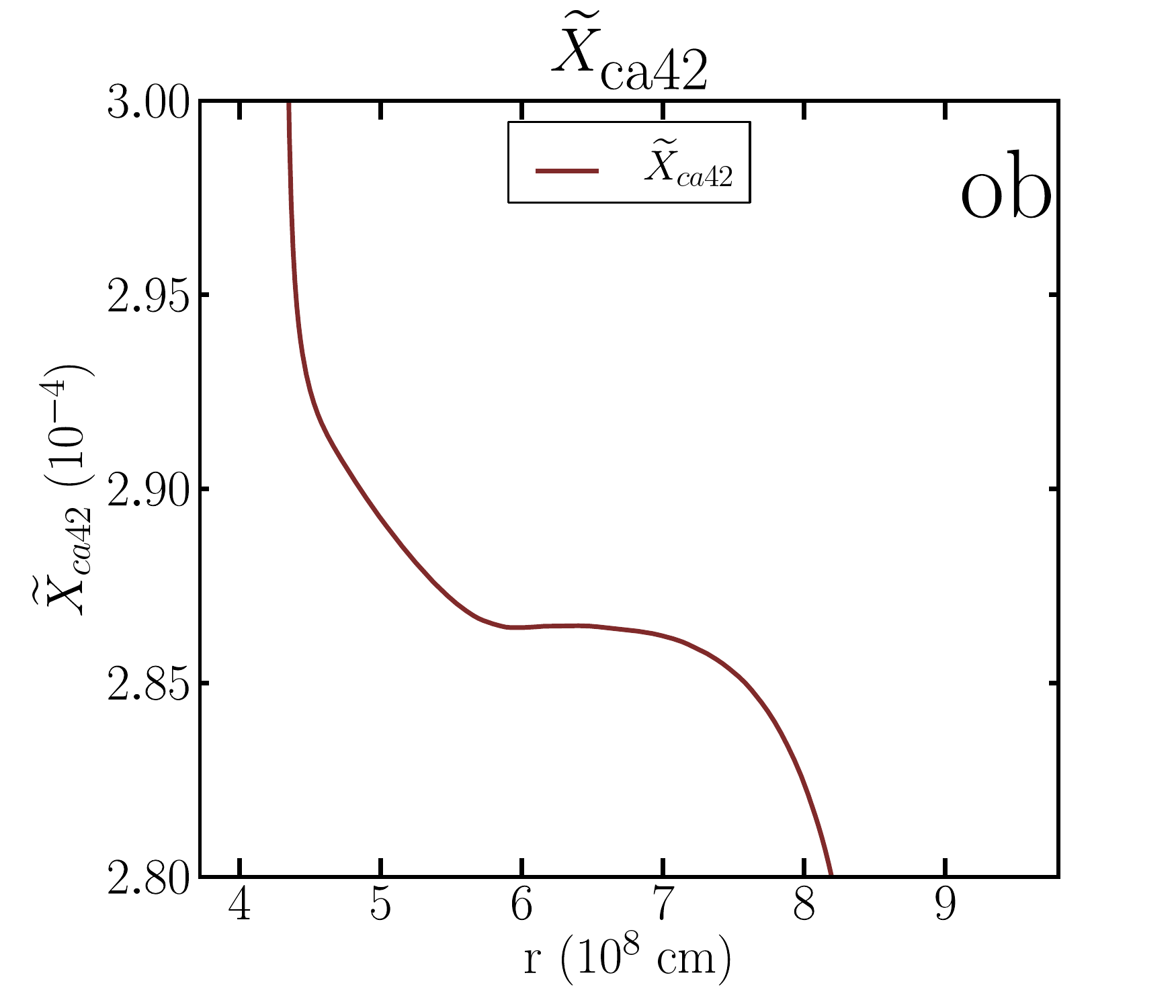}
\includegraphics[width=6.8cm]{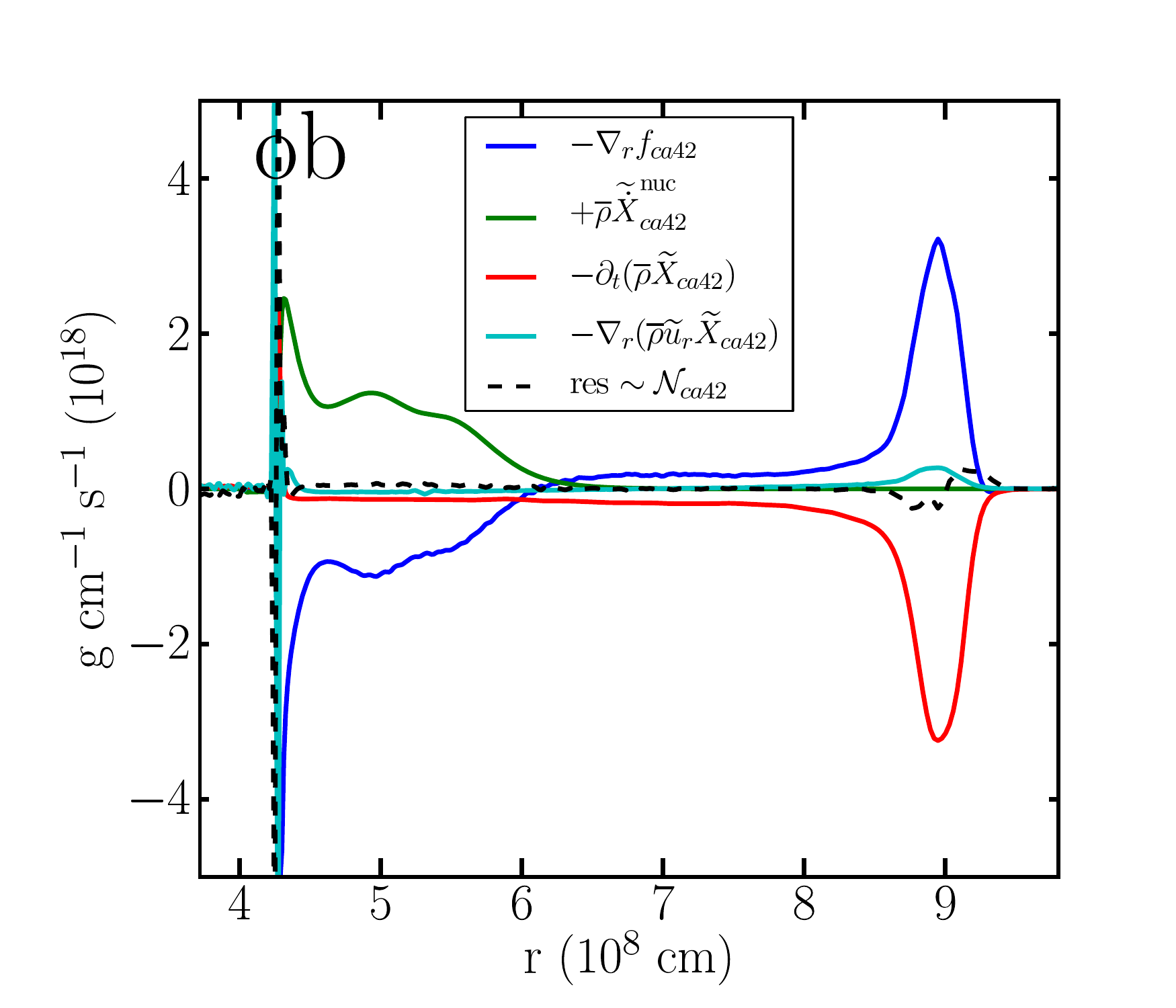}
\includegraphics[width=6.8cm]{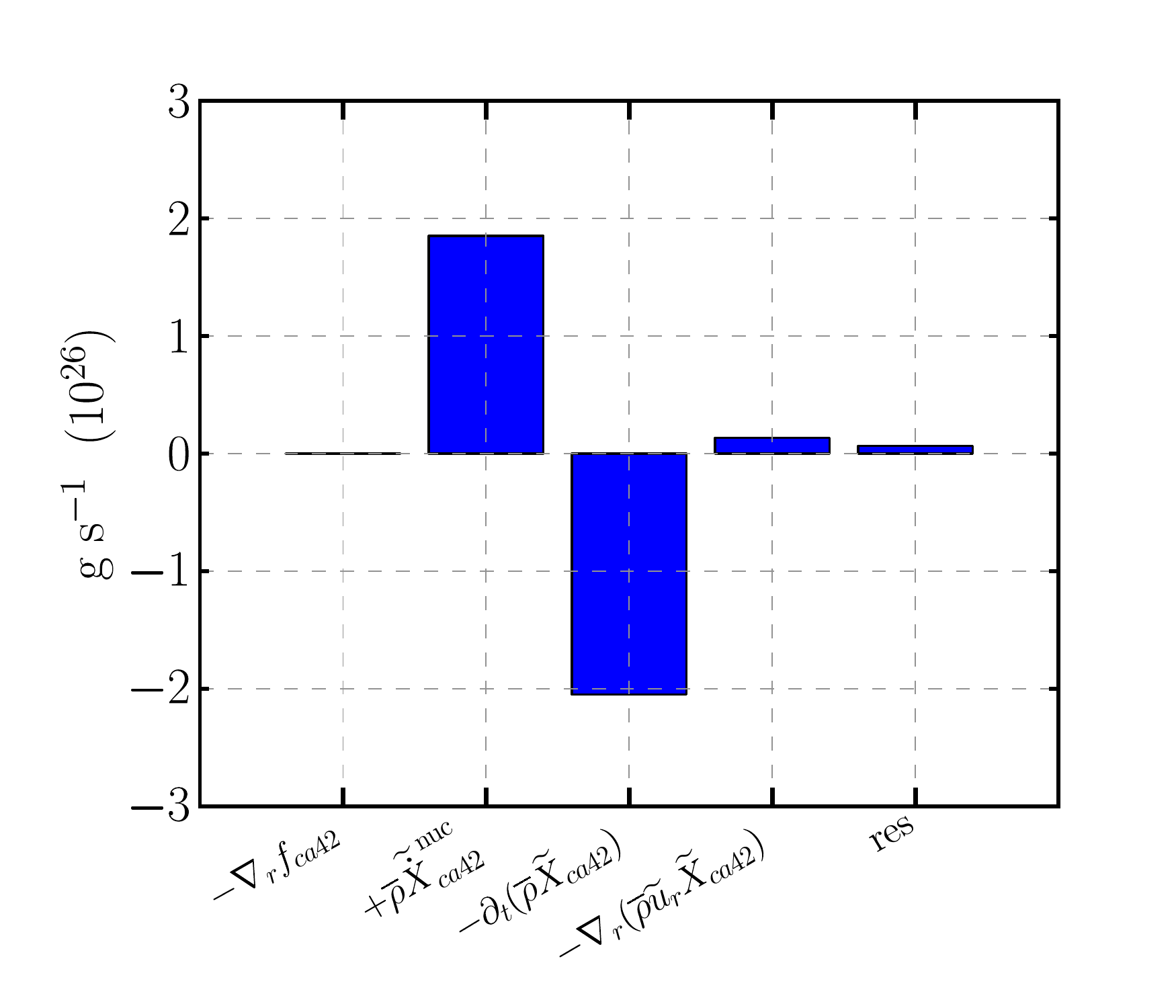}}

\centerline{
\includegraphics[width=6.8cm]{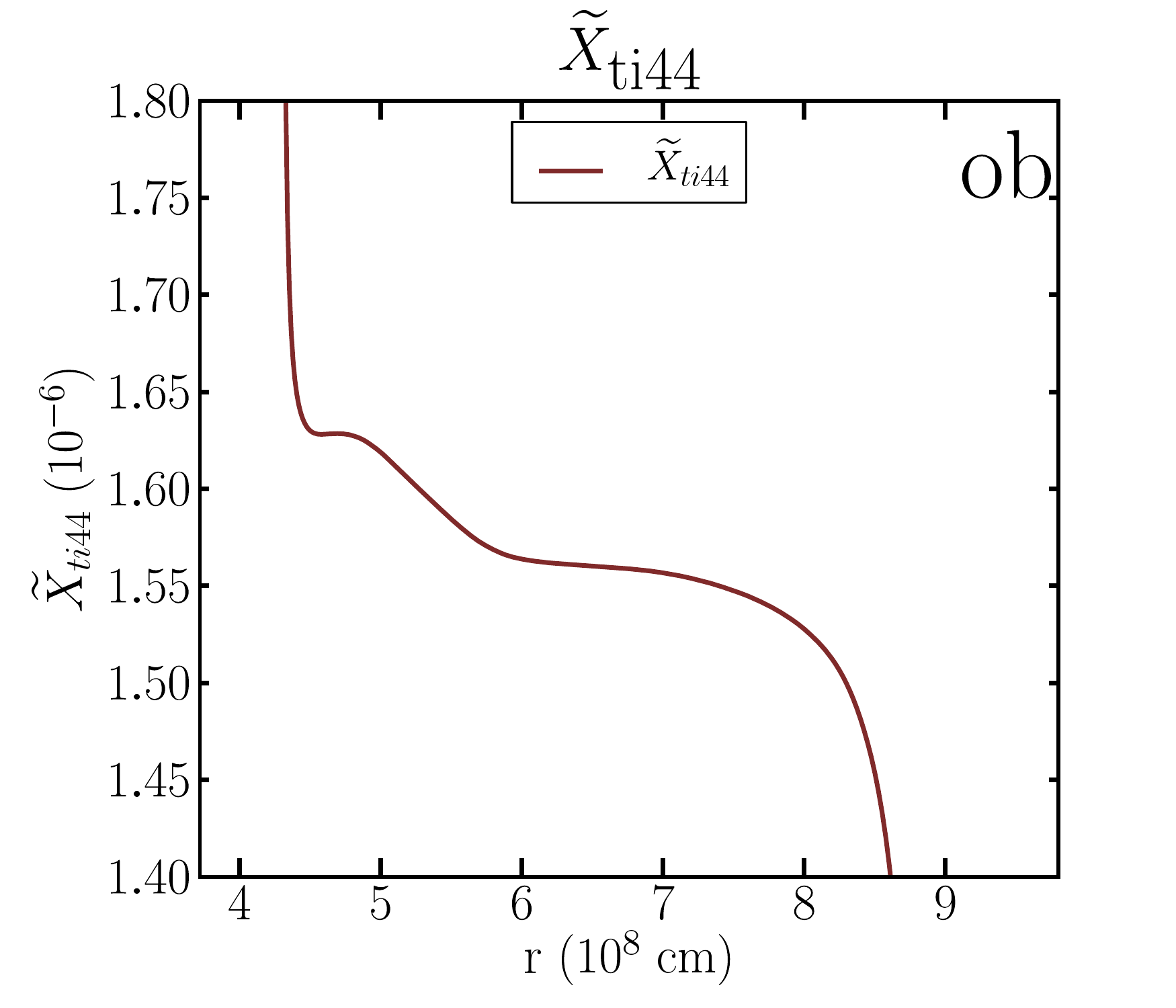}
\includegraphics[width=6.8cm]{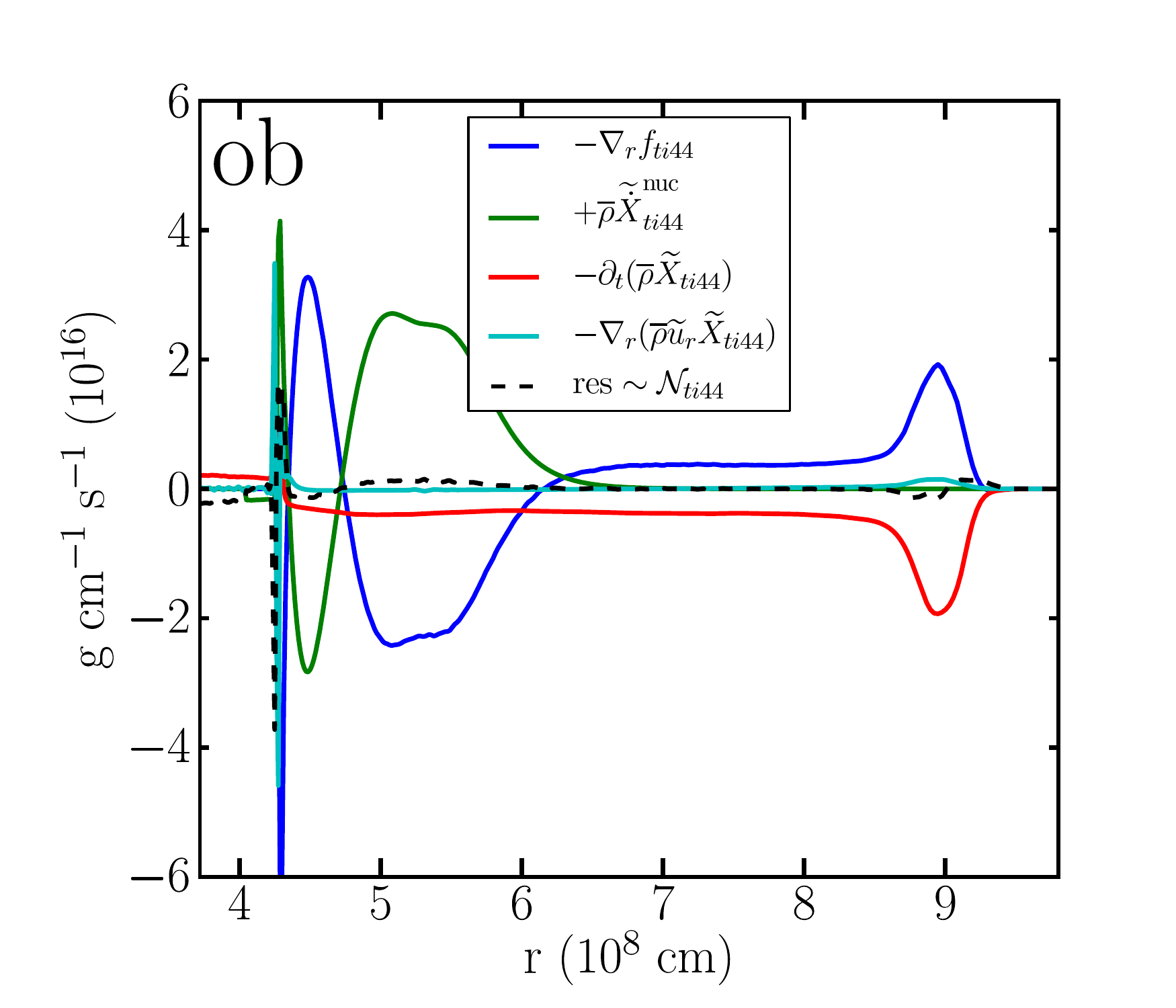}
\includegraphics[width=6.8cm]{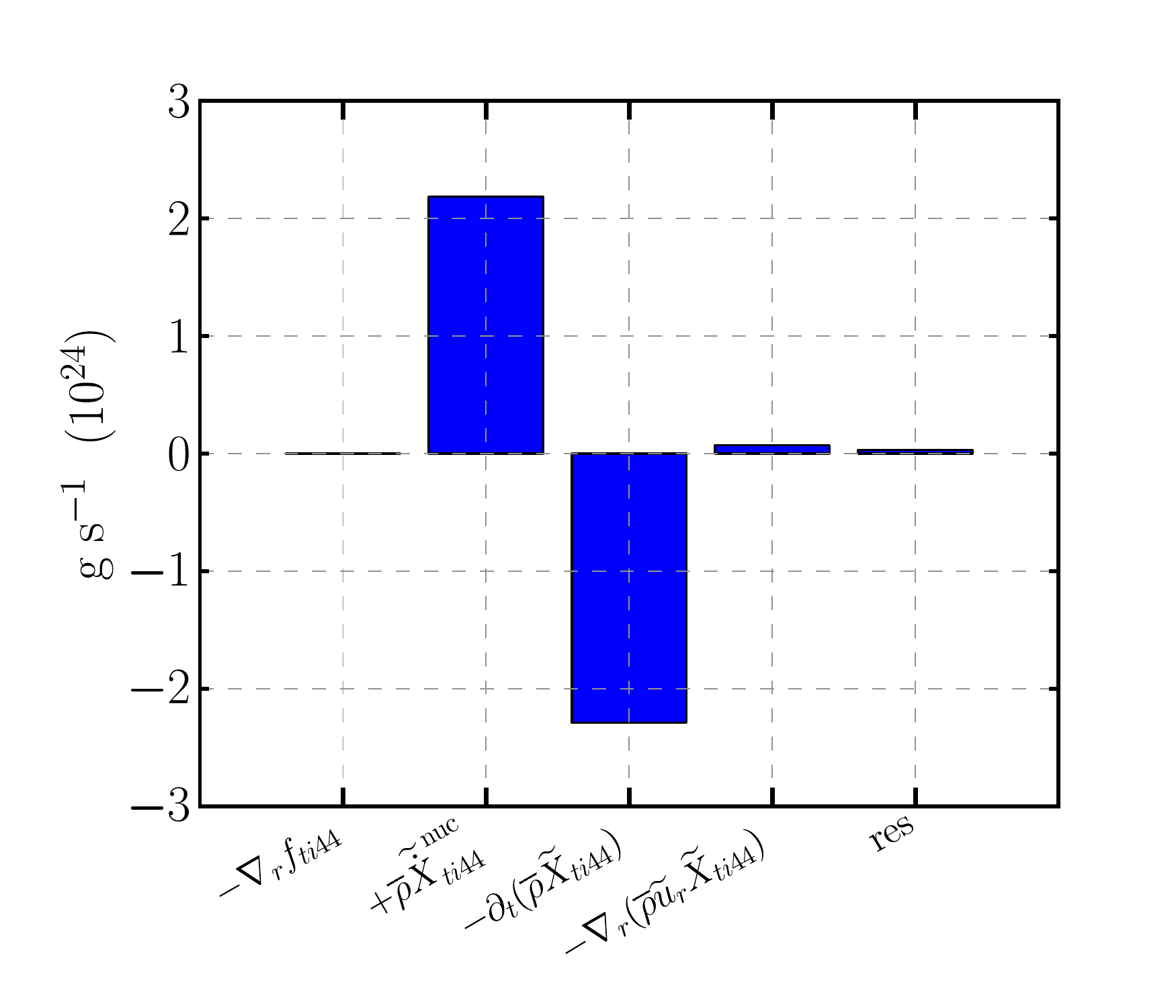}}
\caption{Mean composition equations. Model {\sf ob.3D.2hp}. \label{fig:xca40-ca42-equations}}
\end{figure}

\newpage

\subsection{Mean Ti$^{46}$ and Cr$^{48}$ equation}

\begin{figure}[!h]
\centerline{
\includegraphics[width=6.8cm]{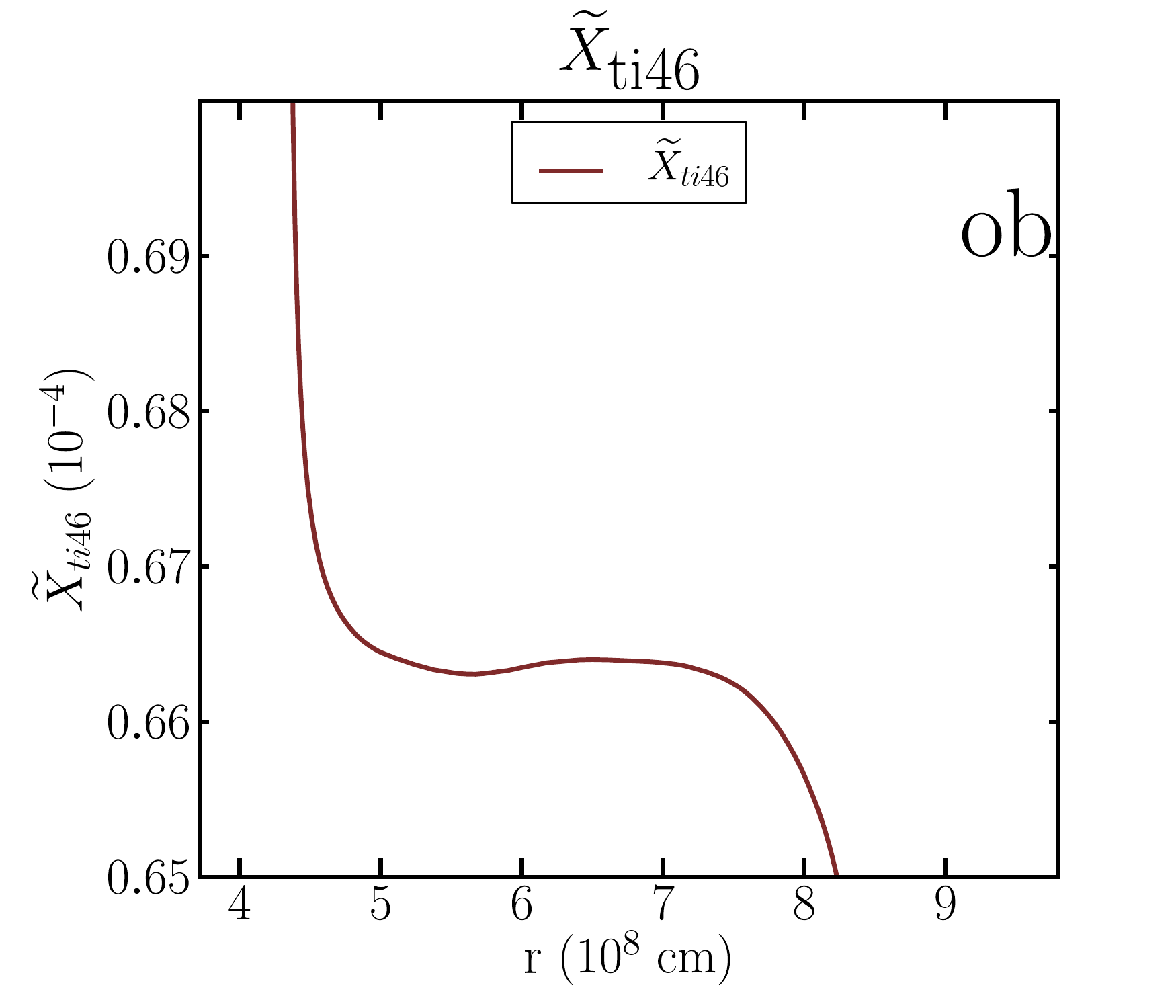}
\includegraphics[width=6.8cm]{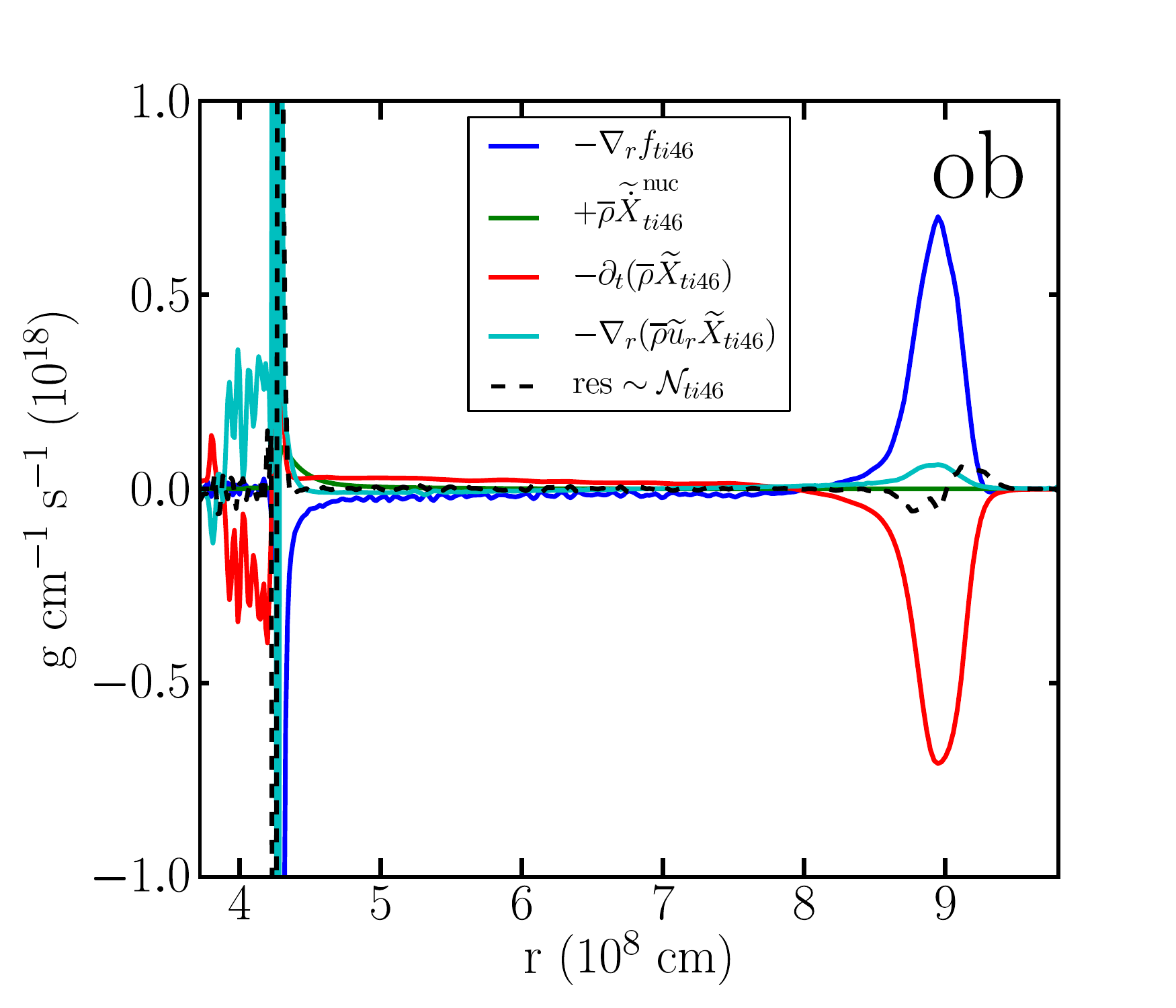}
\includegraphics[width=6.8cm]{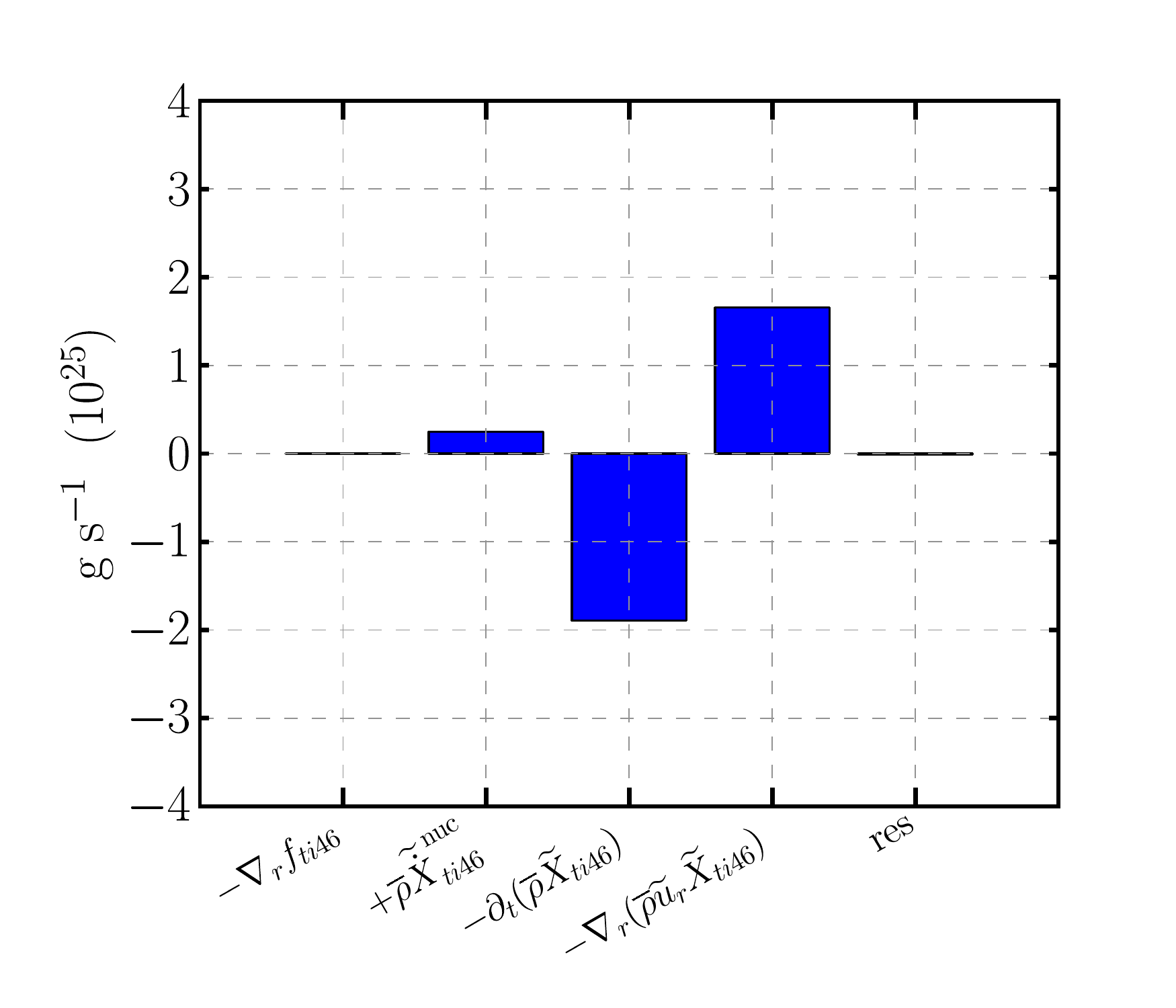}}

\centerline{
\includegraphics[width=6.8cm]{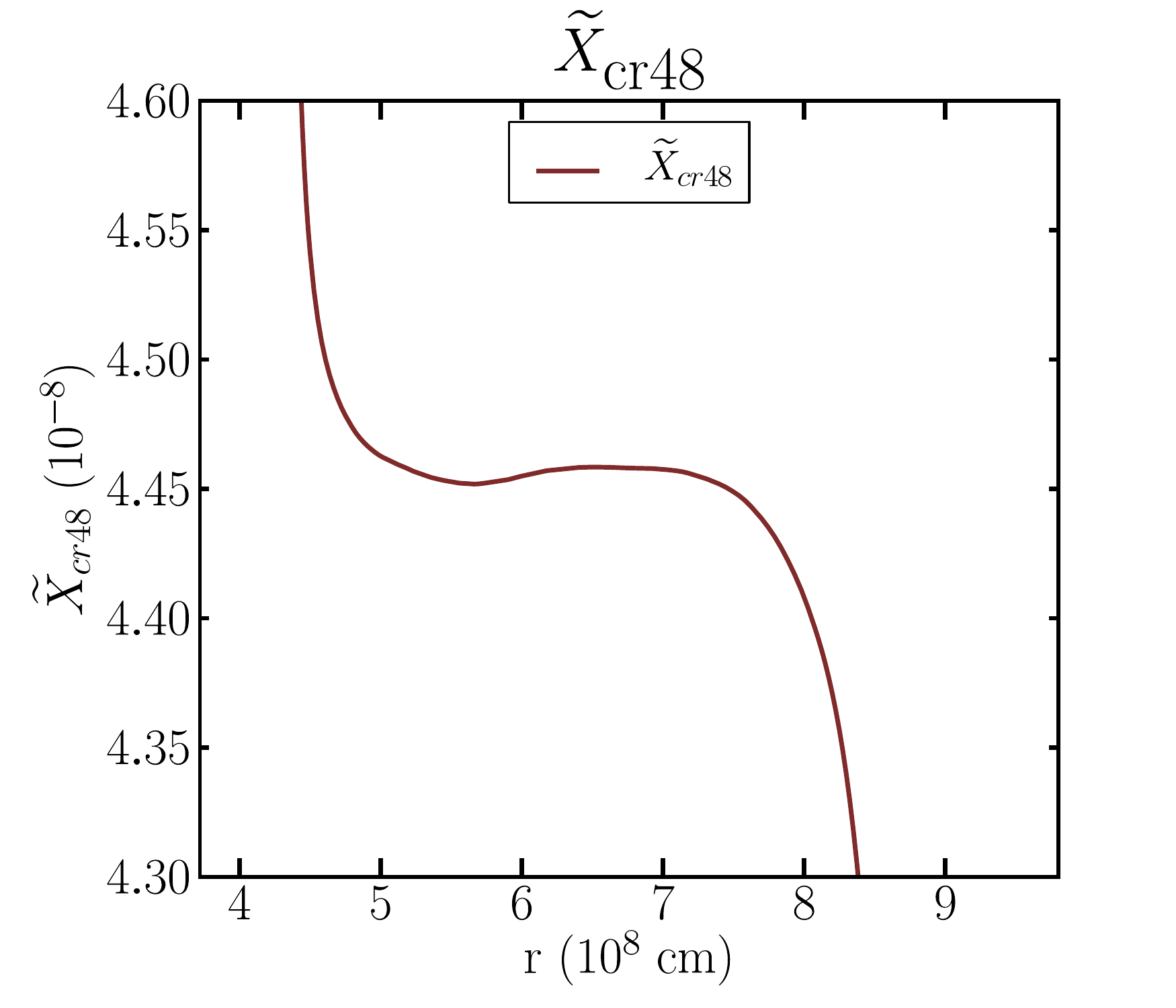}
\includegraphics[width=6.8cm]{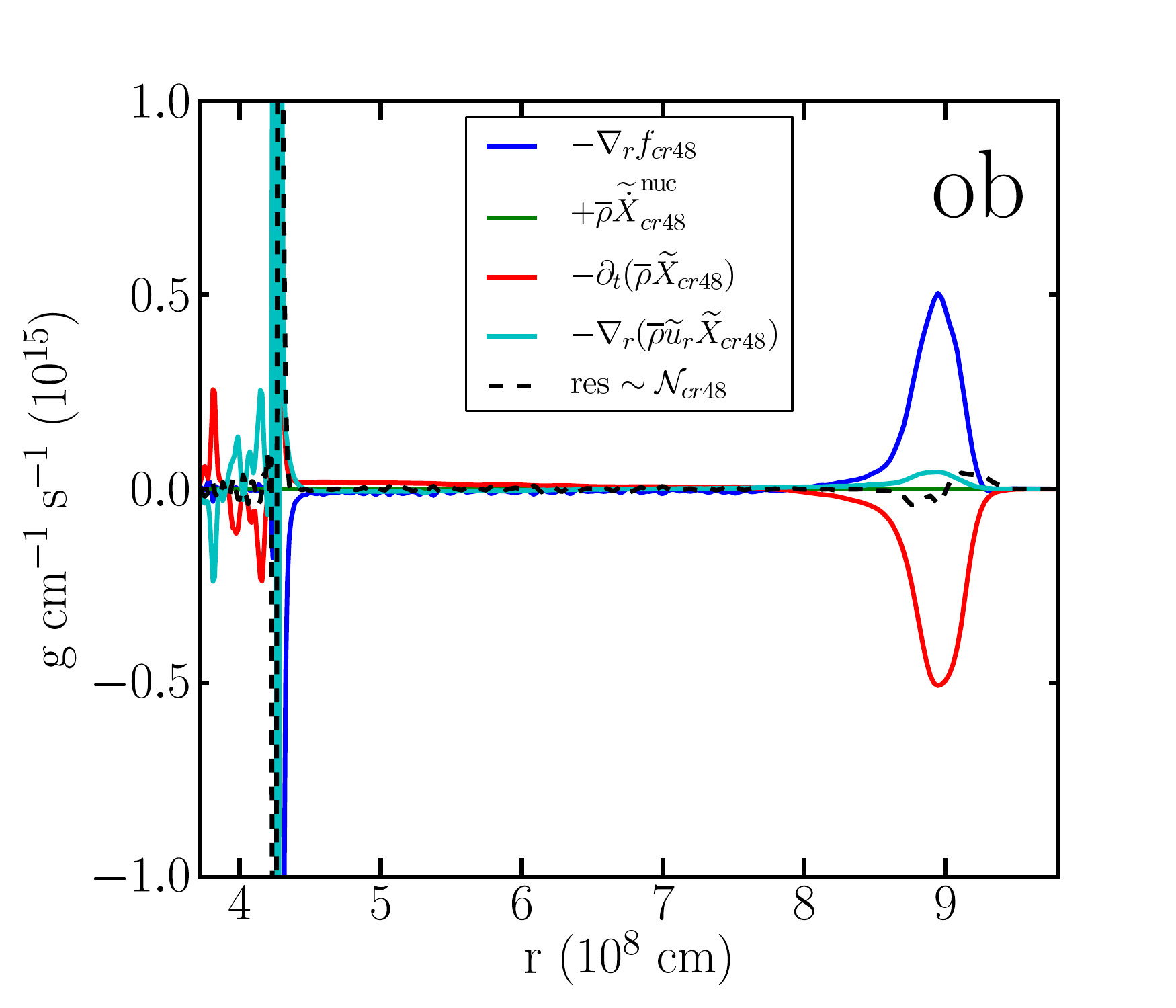}
\includegraphics[width=6.8cm]{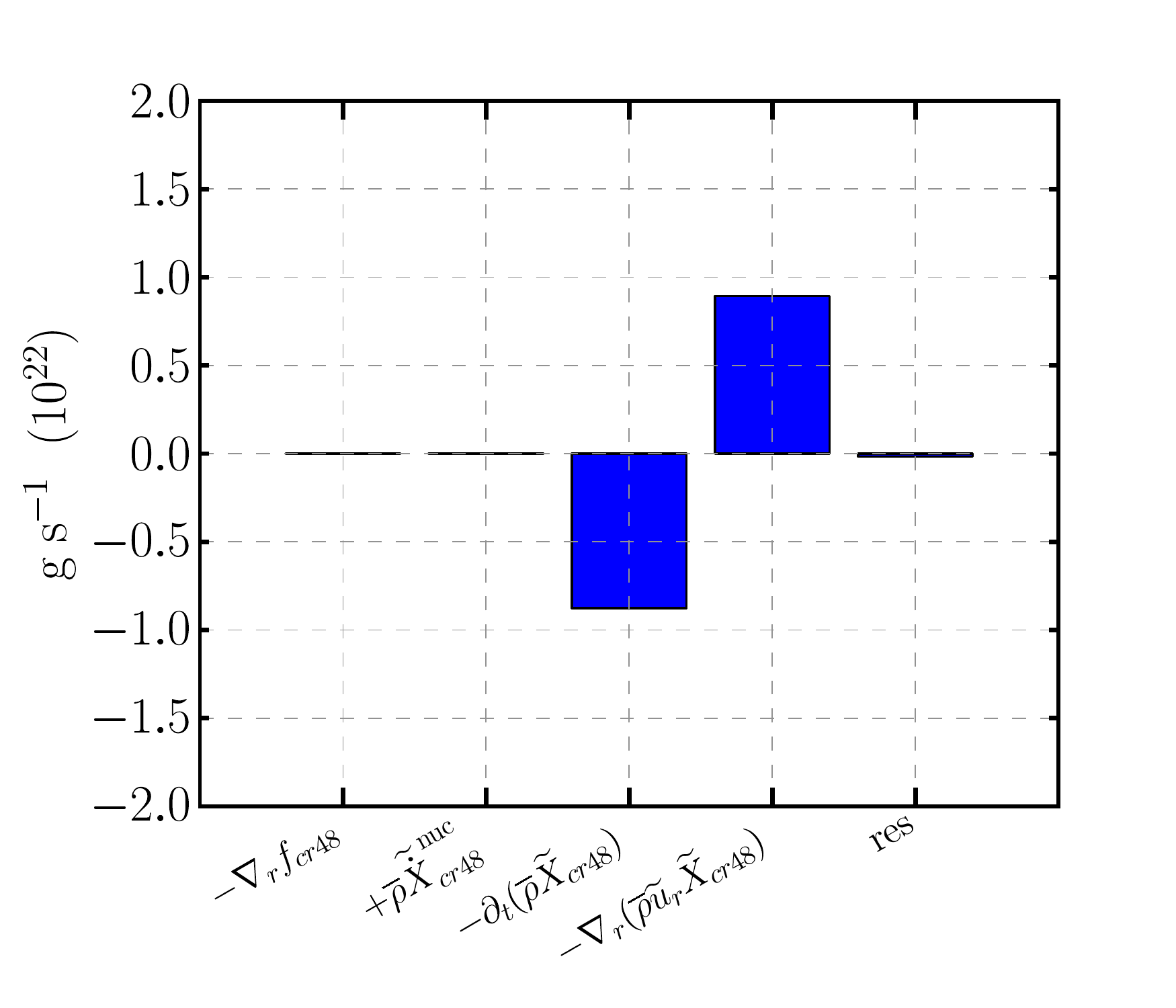}}
\caption{Mean composition equations. Model {\sf ob.3D.2hp}. \label{fig:xti44-xti46-equations}}
\end{figure}

\newpage

\subsection{Mean Cr$^{50}$ and Fe$^{52}$ equation}

\begin{figure}[!h]
\centerline{
\includegraphics[width=6.8cm]{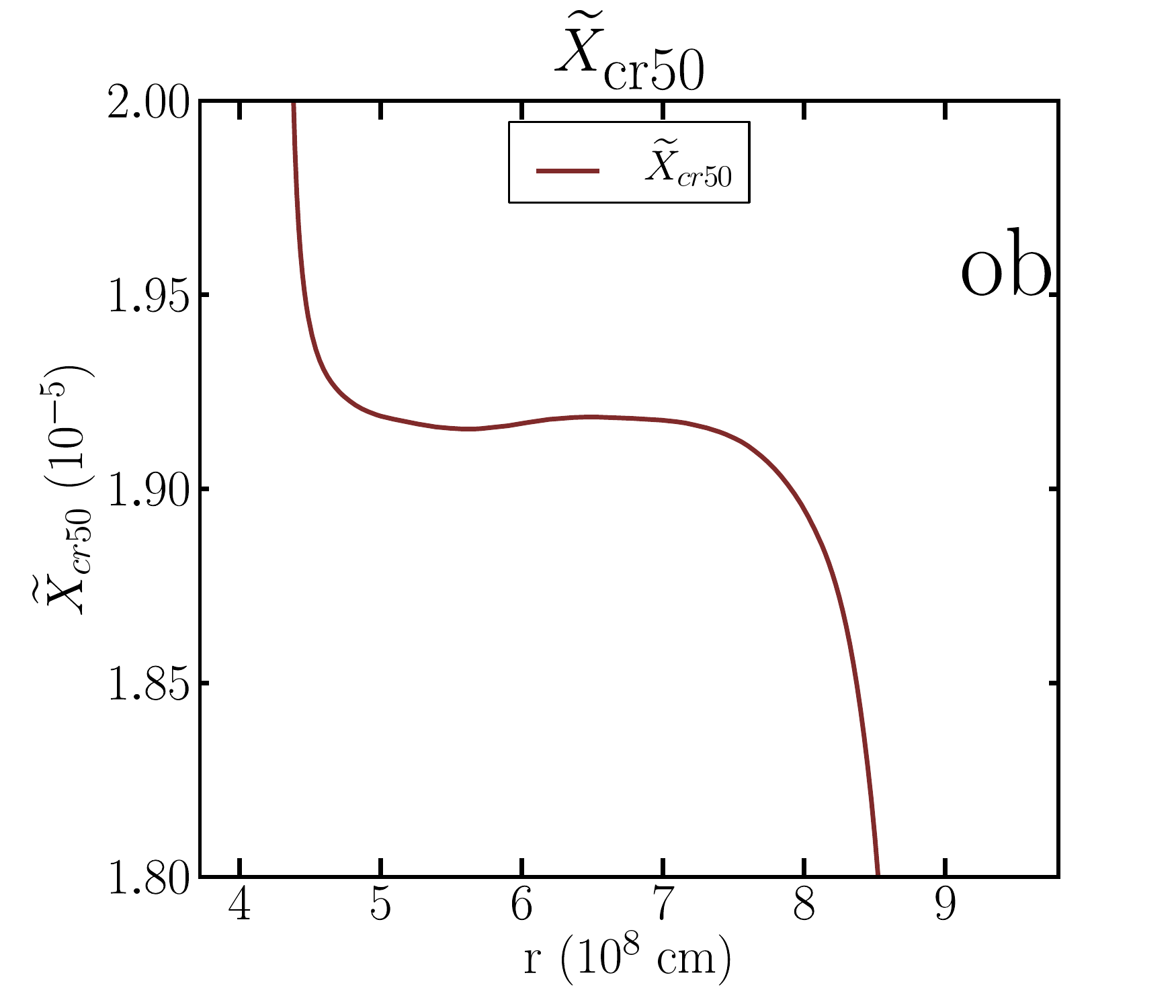}
\includegraphics[width=6.8cm]{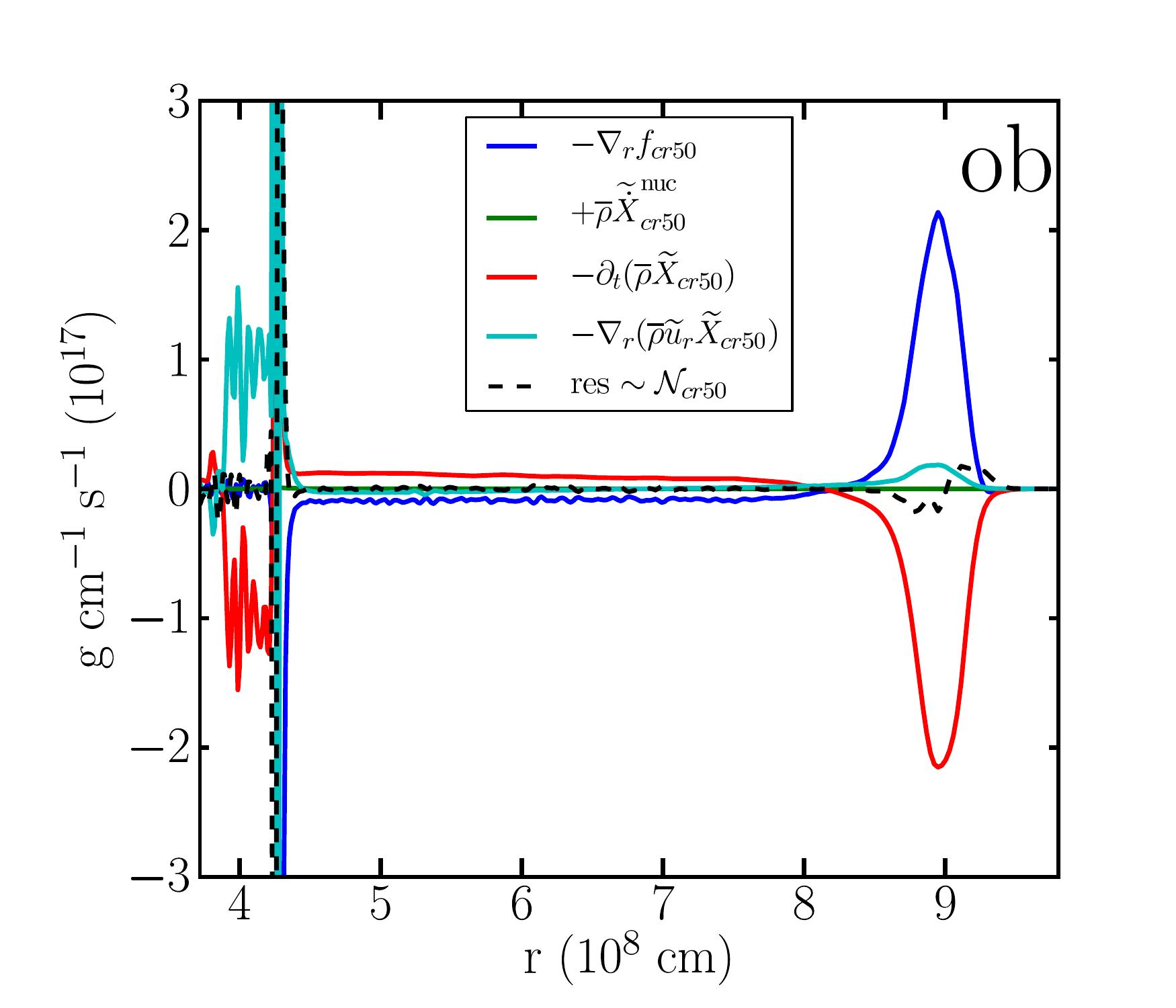}
\includegraphics[width=6.8cm]{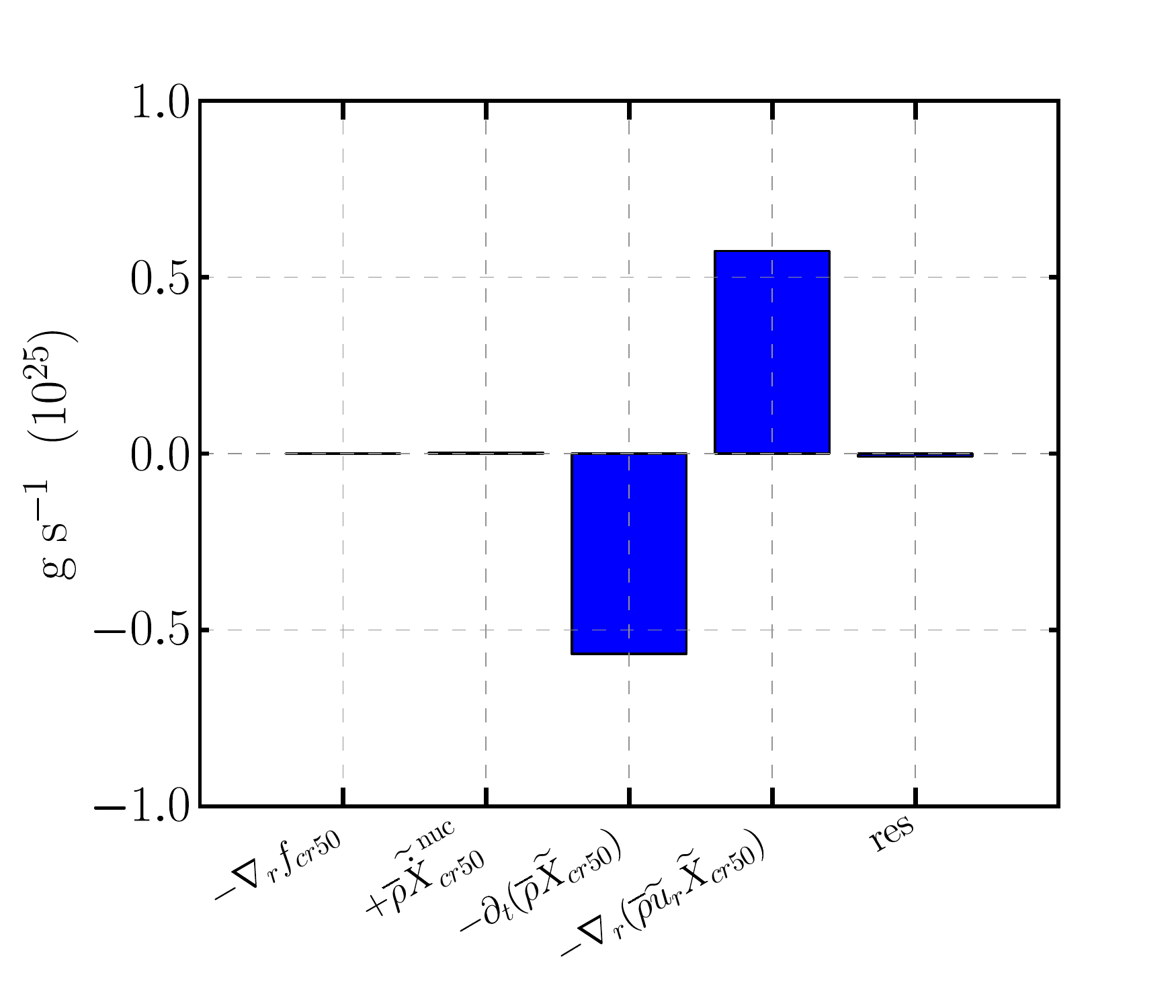}}

\centerline{
\includegraphics[width=6.8cm]{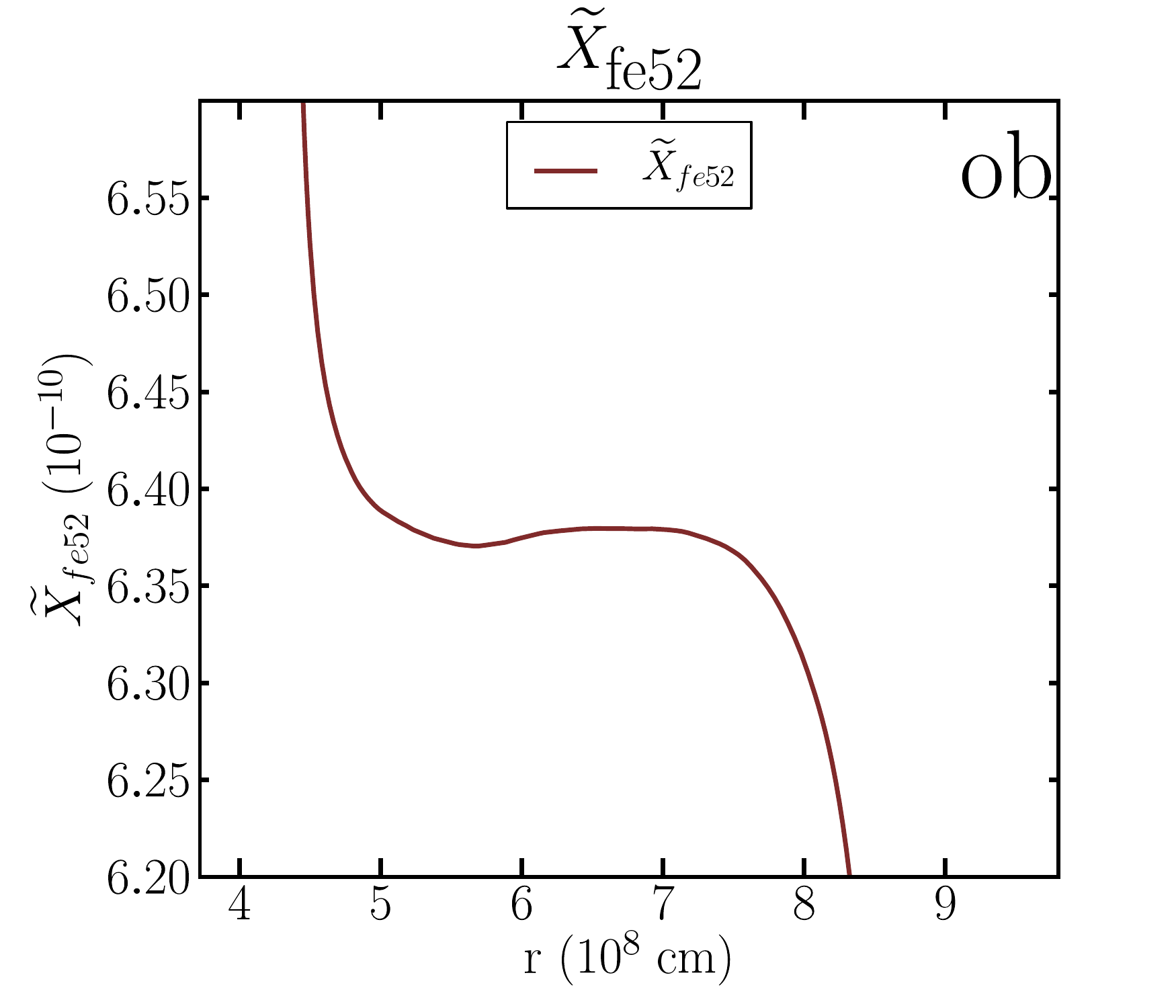}
\includegraphics[width=6.8cm]{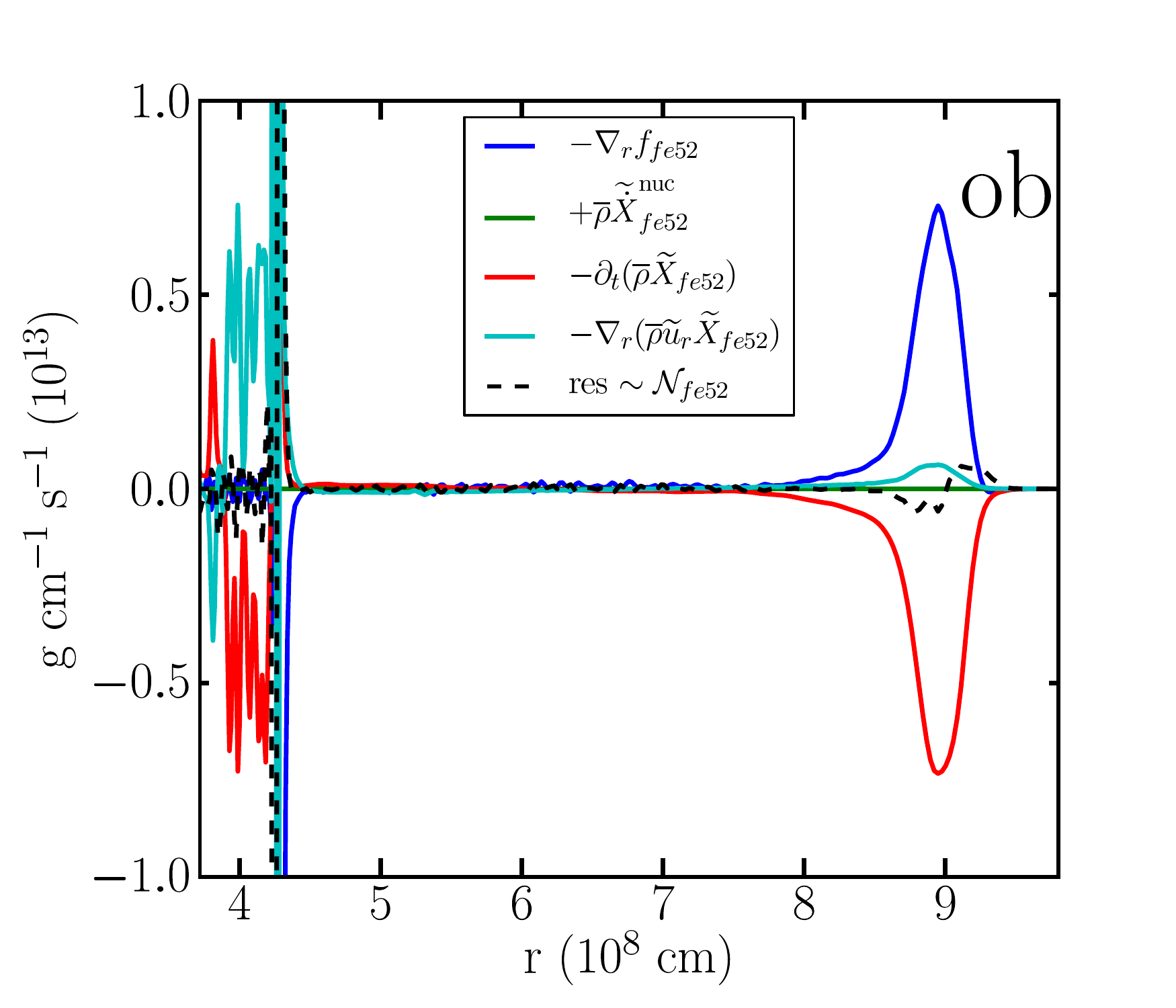}
\includegraphics[width=6.8cm]{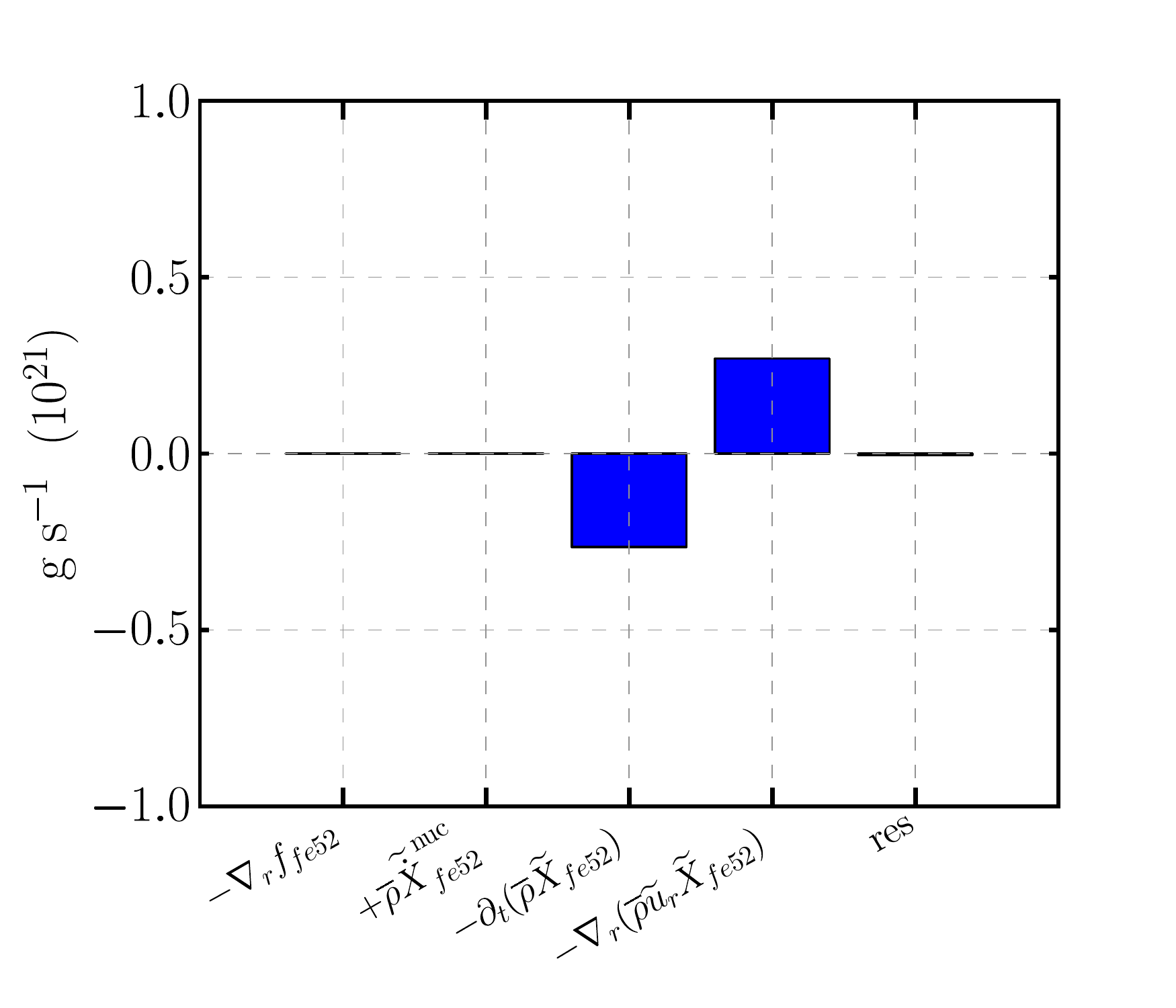}}
\caption{Mean composition equations. Model {\sf ob.3D.2hp}. \label{fig:xcr48-xcr50-equations}}
\end{figure}

\newpage

\subsection{Mean Fe$^{54}$ and Ni$^{56}$ equation}

\begin{figure}[!h]
\centerline{
\includegraphics[width=6.8cm]{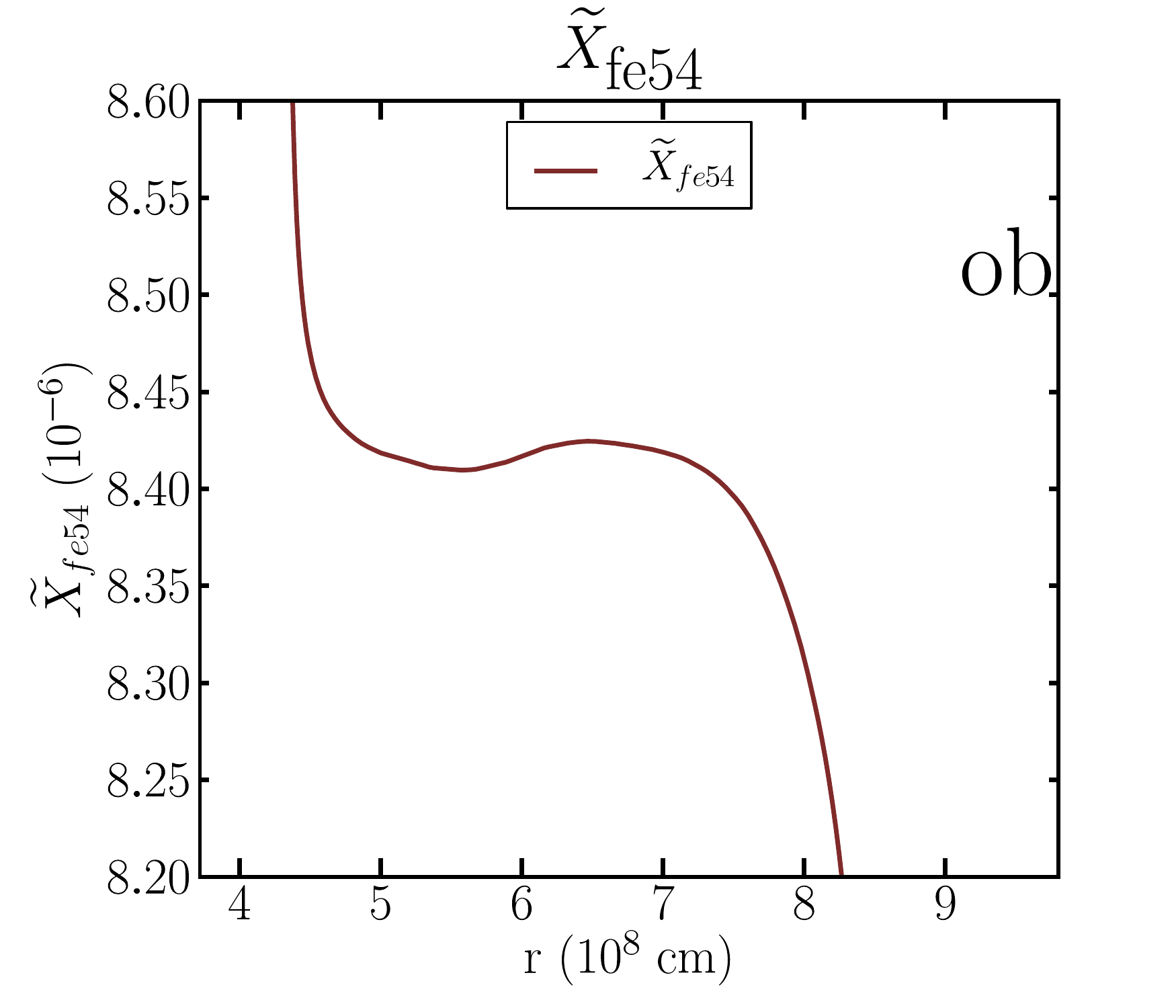}
\includegraphics[width=6.8cm]{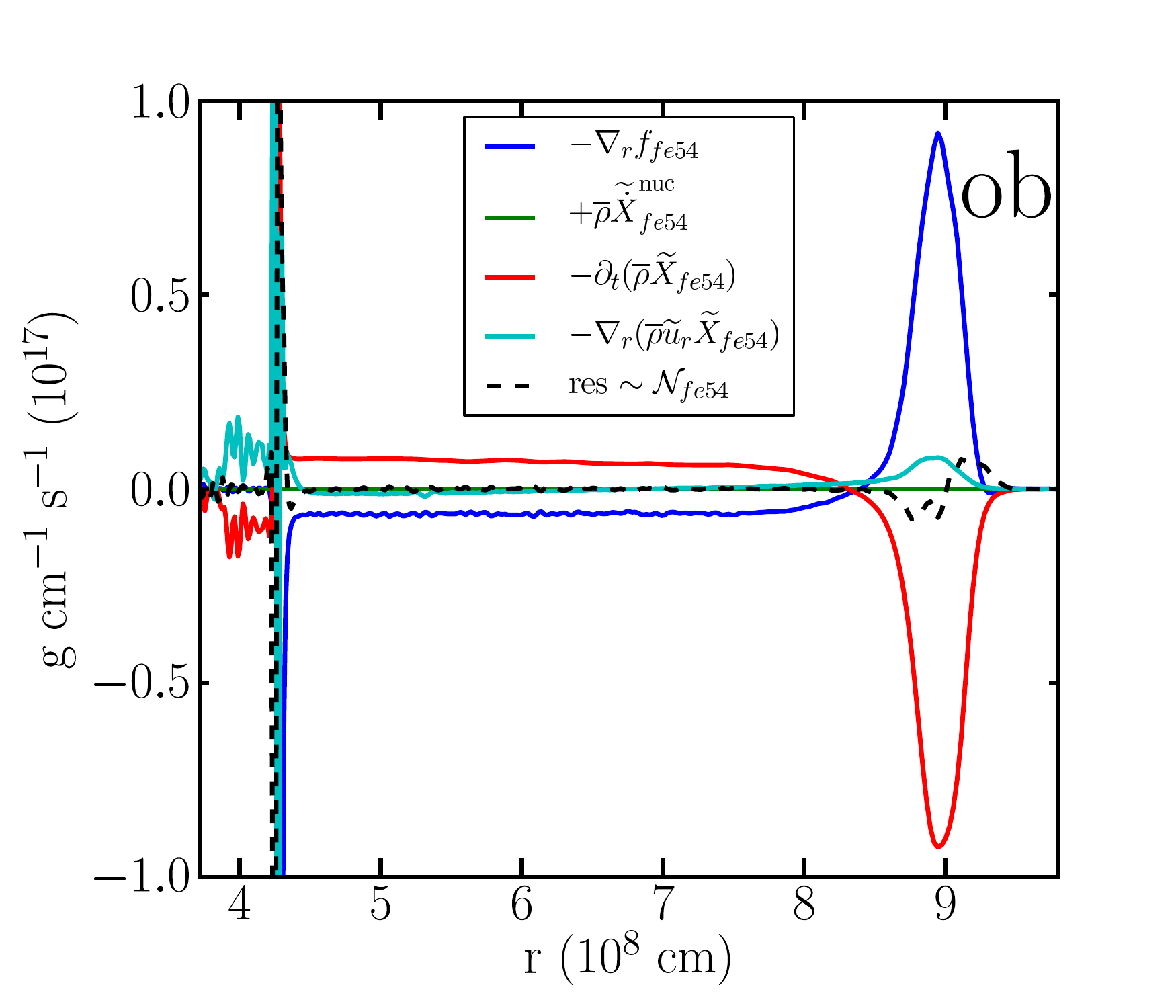}
\includegraphics[width=6.8cm]{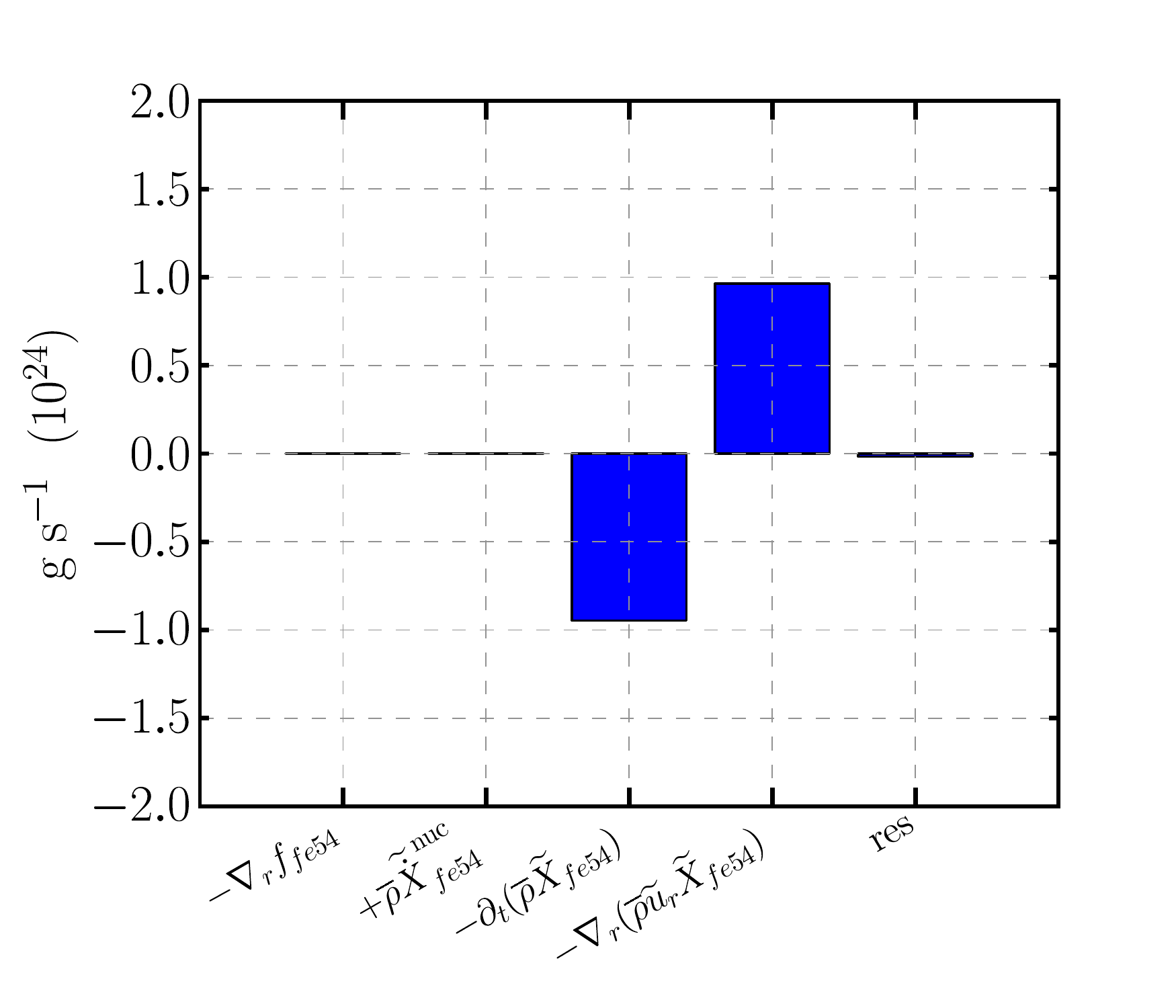}}

\centerline{
\includegraphics[width=6.8cm]{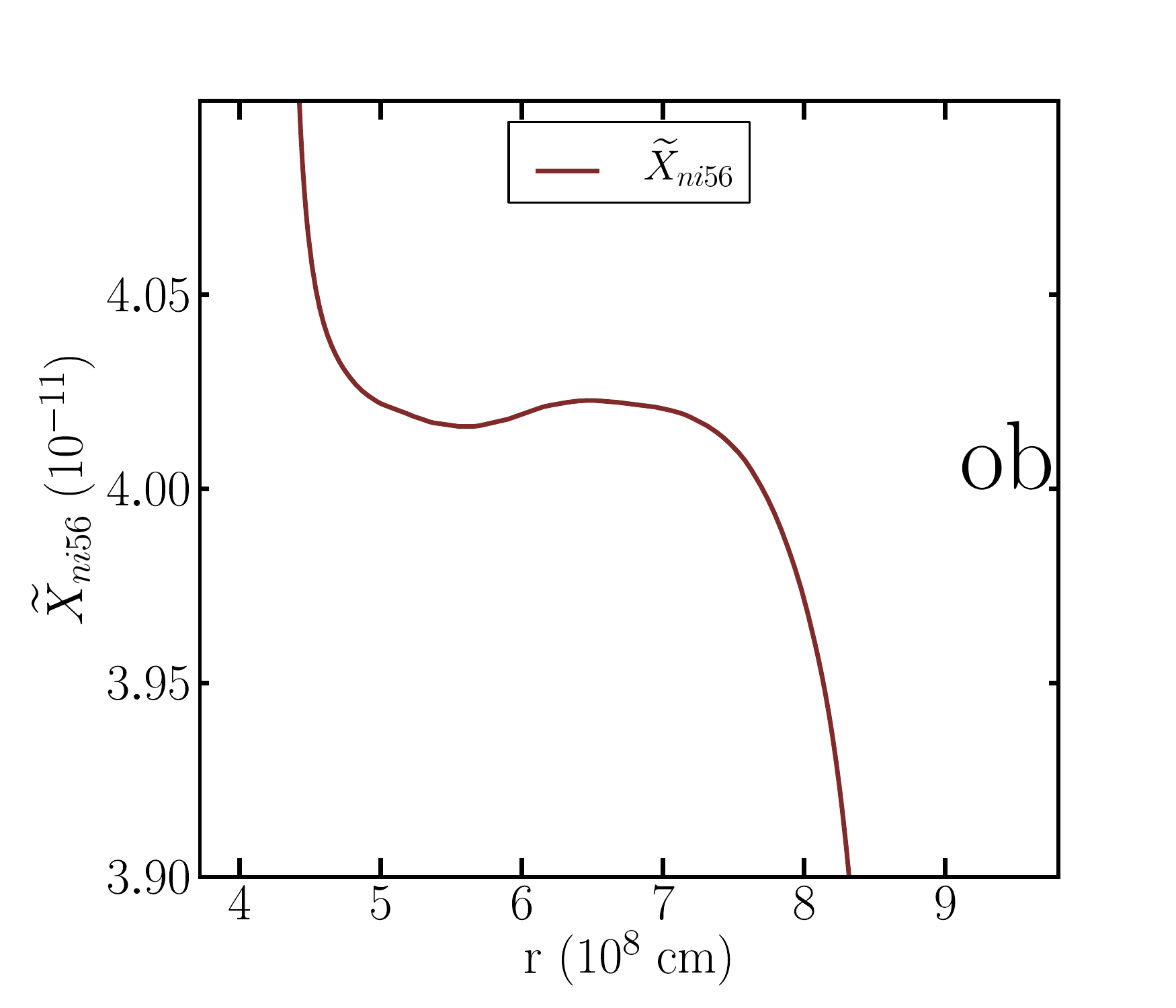}
\includegraphics[width=6.8cm]{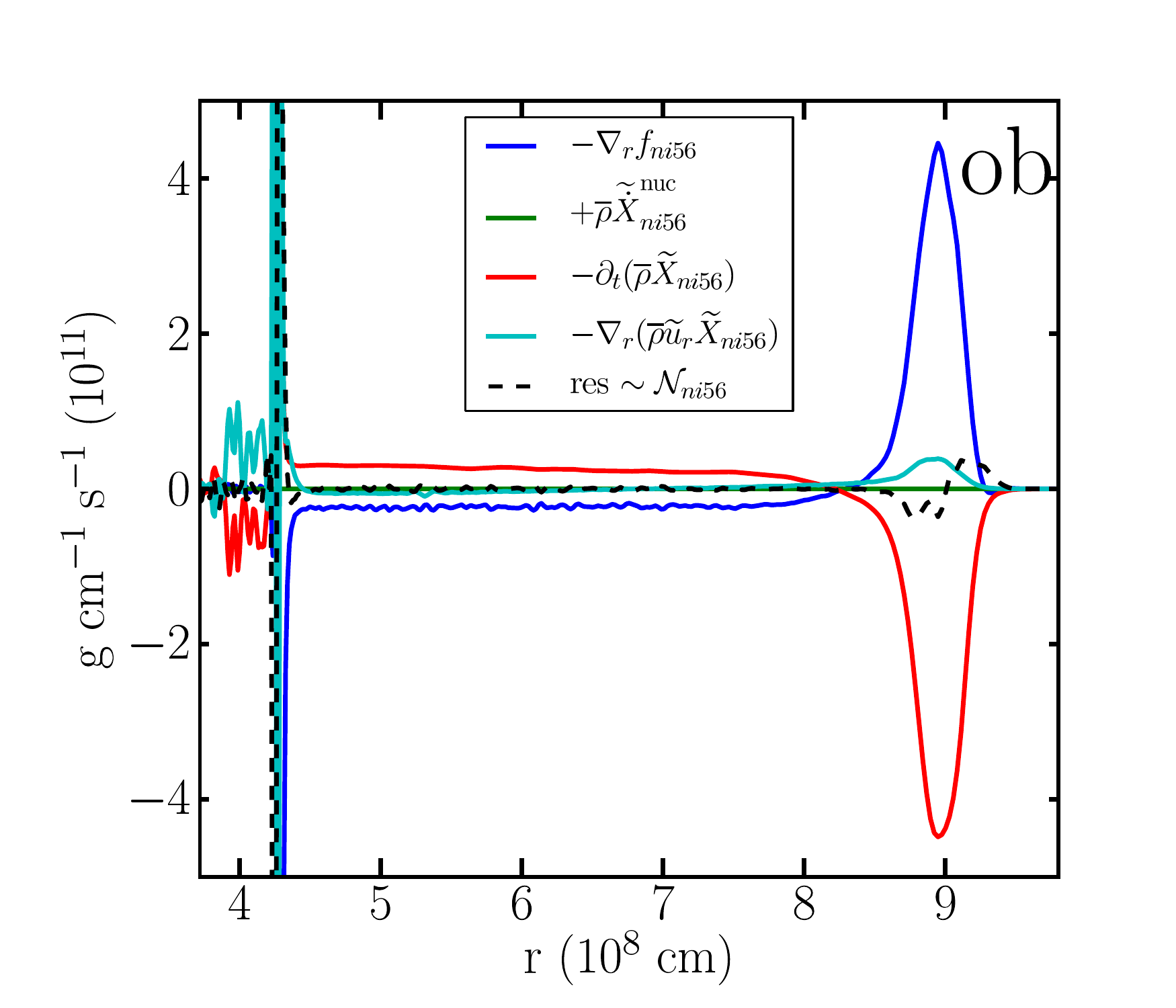}
\includegraphics[width=6.8cm]{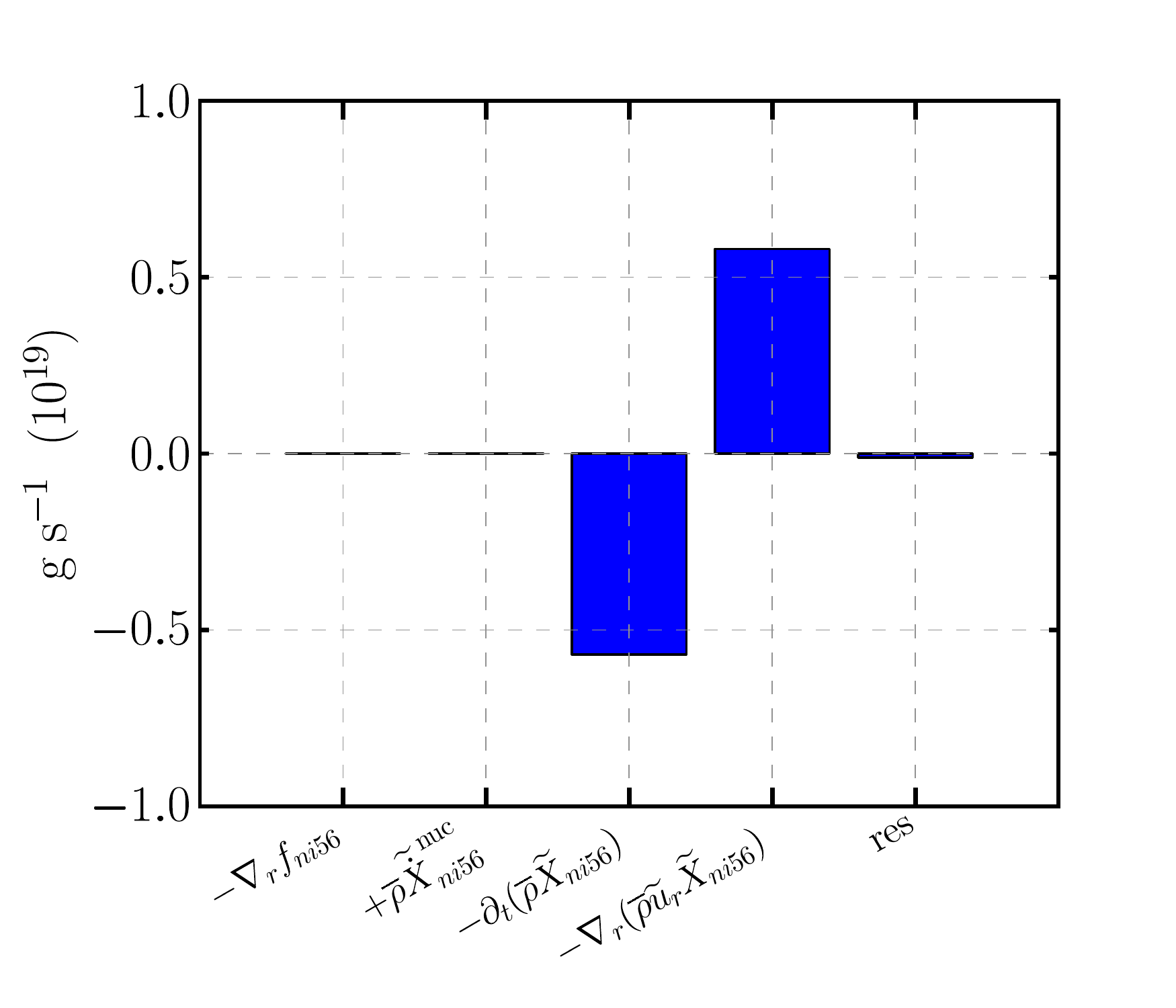}}
\caption{Mean composition equations. Model {\sf ob.3D.2hp}. \label{fig:xfe52-xfe54-equations}}
\end{figure}

\newpage

\section{Dependence on Numerical Resolution}

\subsection{Oxygen burning shell models}

\subsubsection{Mean continuity equation and mean radial momentum equation}

\begin{figure}[!h]
\centerline{
\includegraphics[width=6.1cm]{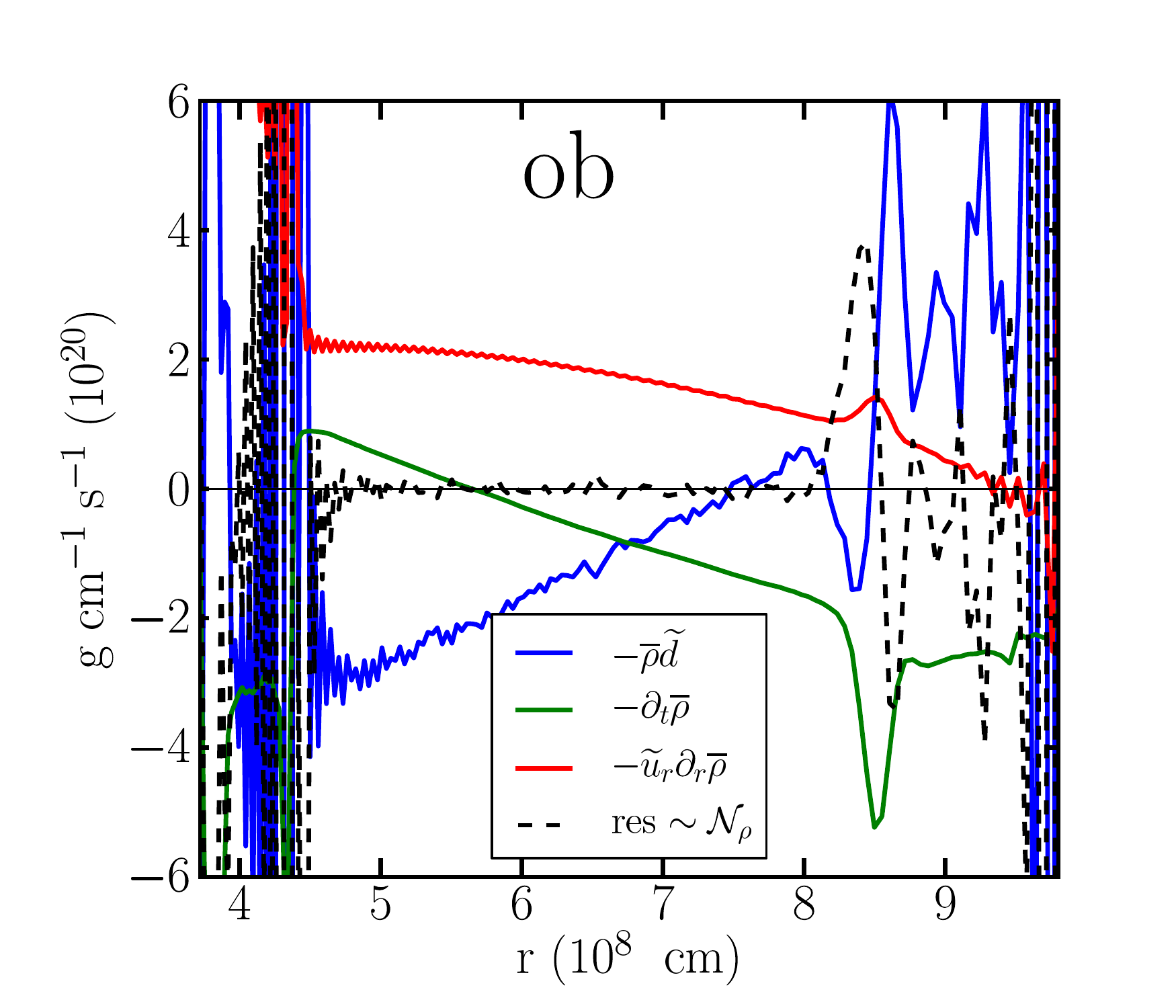}
\includegraphics[width=6.1cm]{obmrez_tavg230_continuity_equation_ransdat-eps-converted-to.pdf}
\includegraphics[width=6.1cm]{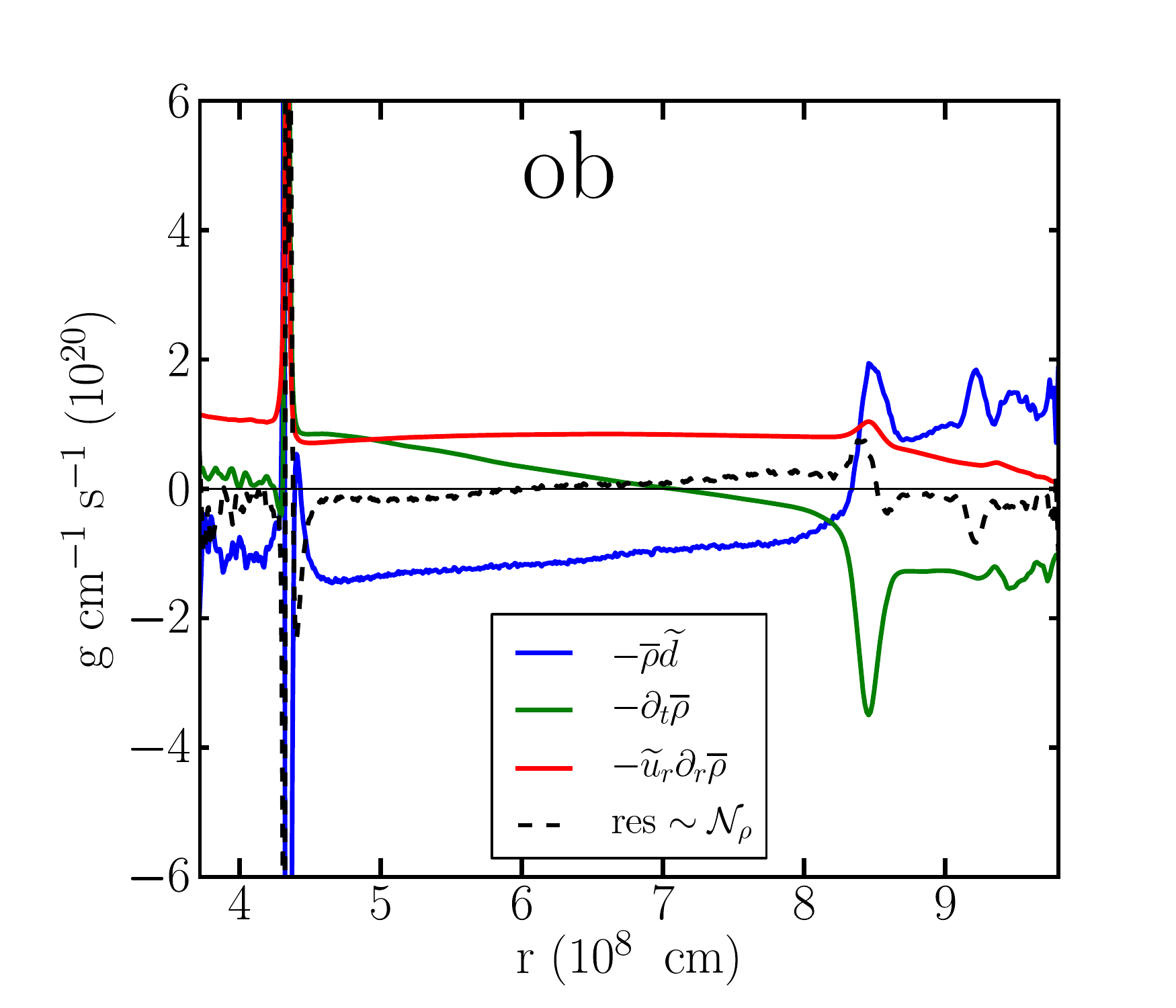}}

\centerline{
\includegraphics[width=6.1cm]{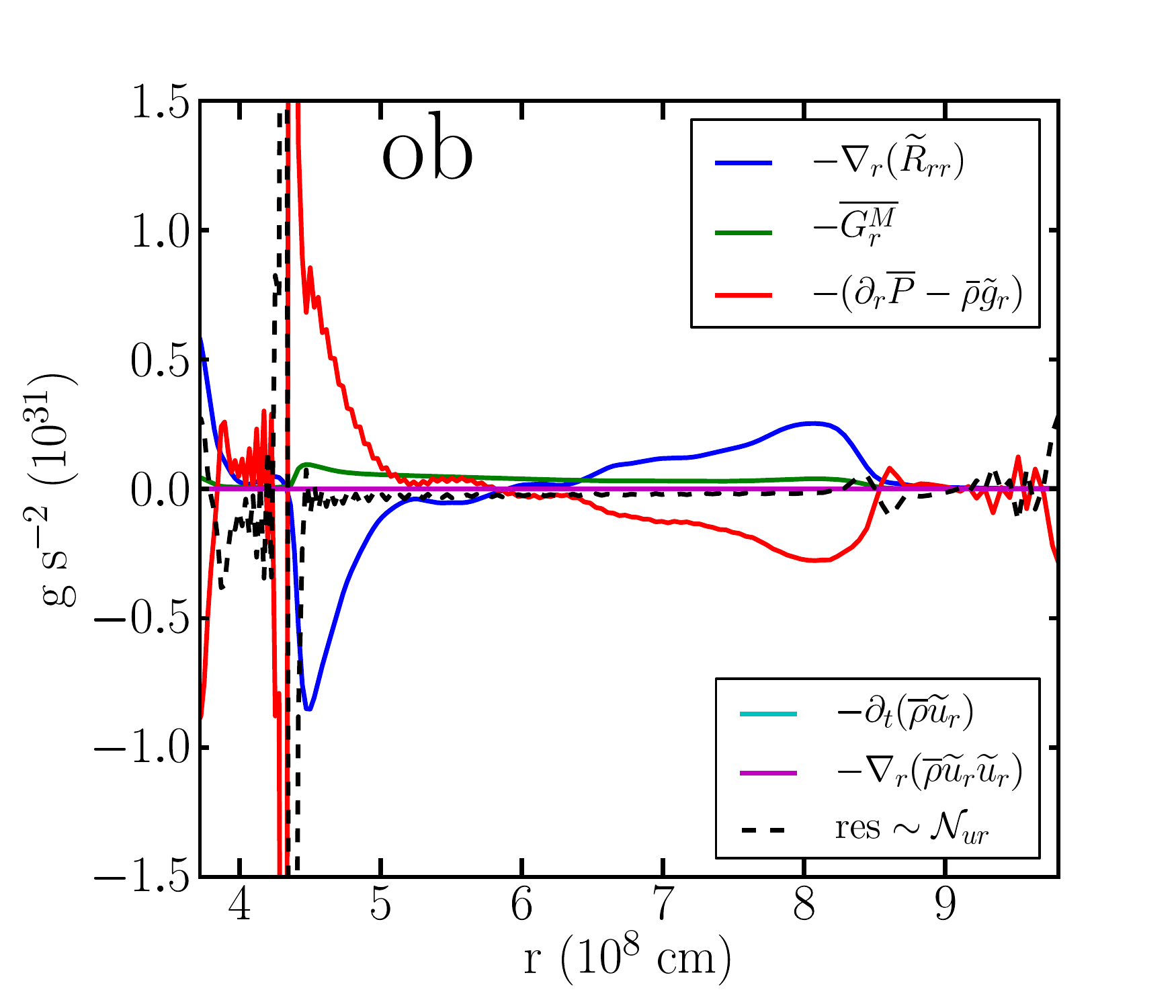}
\includegraphics[width=6.1cm]{obmrez_tavg230_rmomentum_equation_ransdat-eps-converted-to.pdf}
\includegraphics[width=6.1cm]{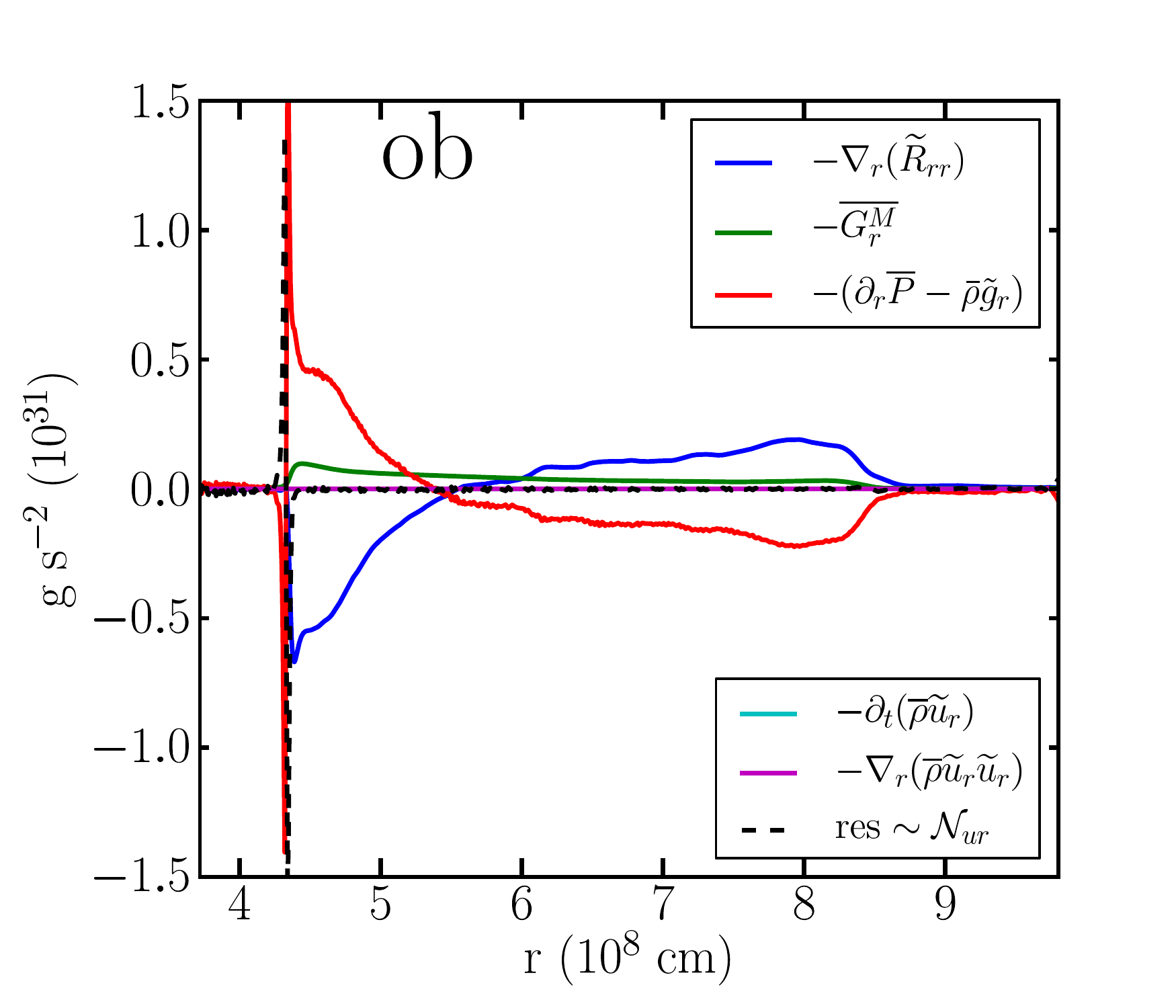}}
\caption{Mean continuity equation (upper panels) and radial momentum equation (lower panels). Model {\sf ob.3D.lr} (left), {\sf ob.3D.mr} (middle), {\sf ob.3D.hr} (right) \label{fig:ob-res-cont-rmomentum-equation}}
\end{figure}

\newpage

\subsubsection{Mean azimuthal and polar momentum equations}

\begin{figure}[!h]
\centerline{
\includegraphics[width=6.8cm]{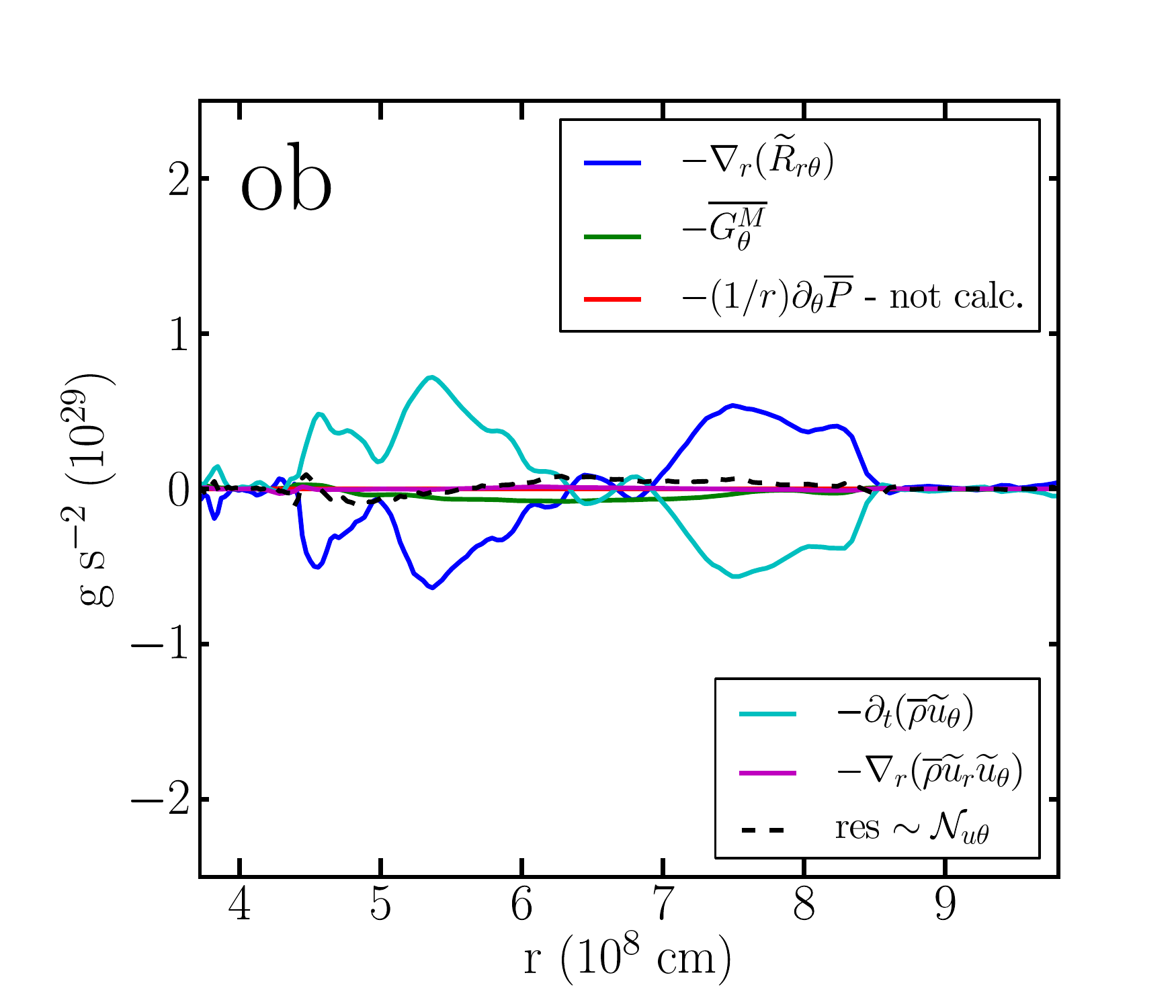}
\includegraphics[width=6.8cm]{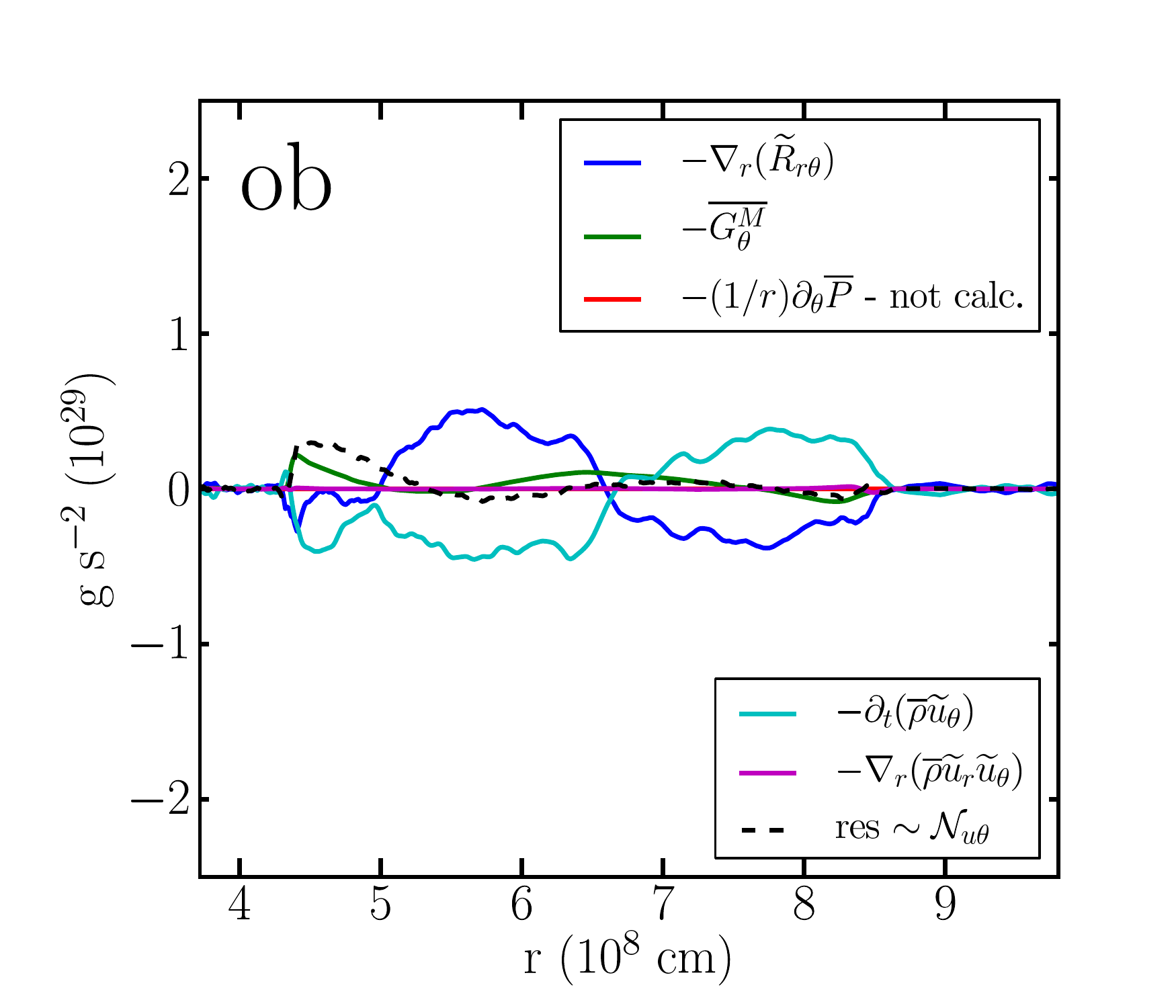}
\includegraphics[width=6.8cm]{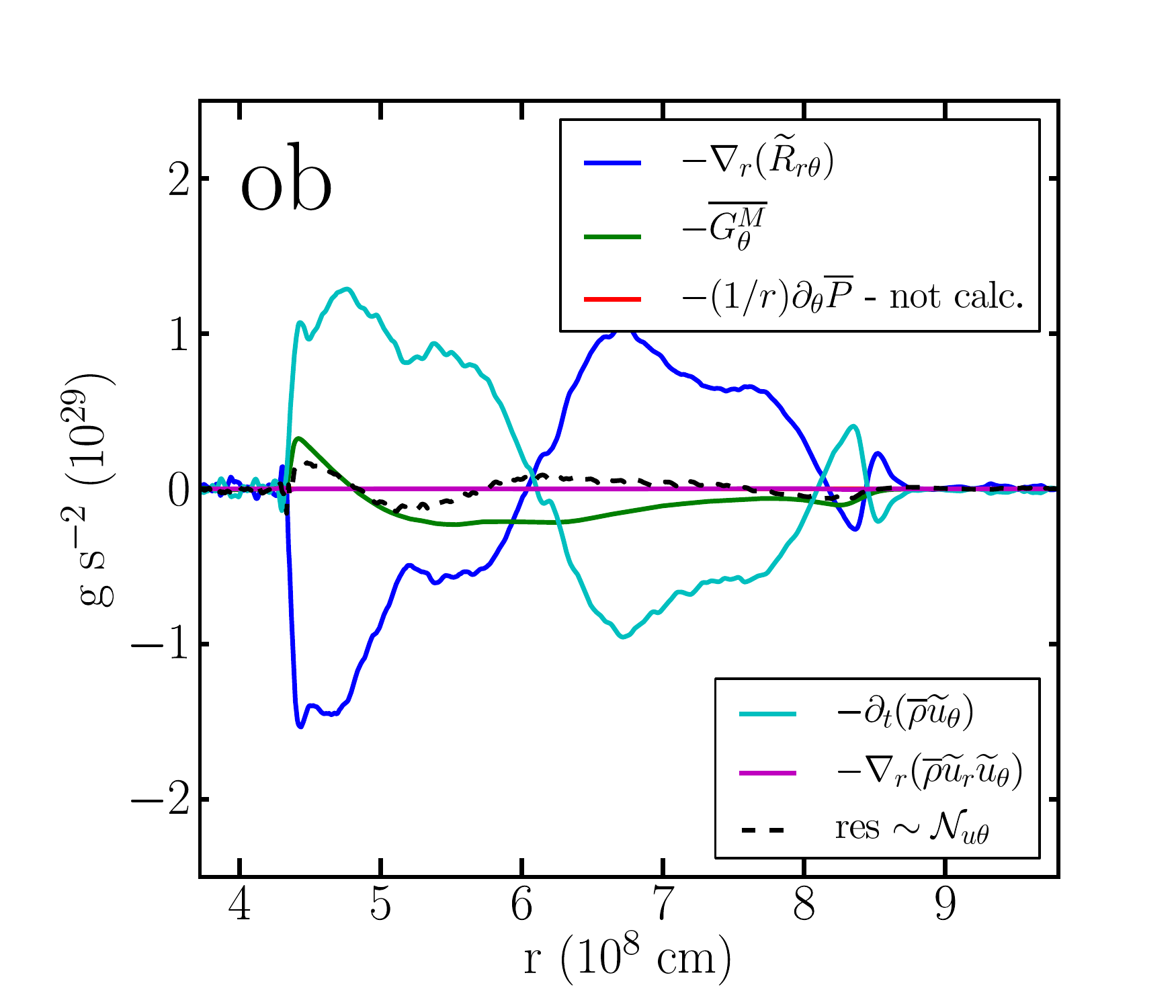}}

\centerline{
\includegraphics[width=6.8cm]{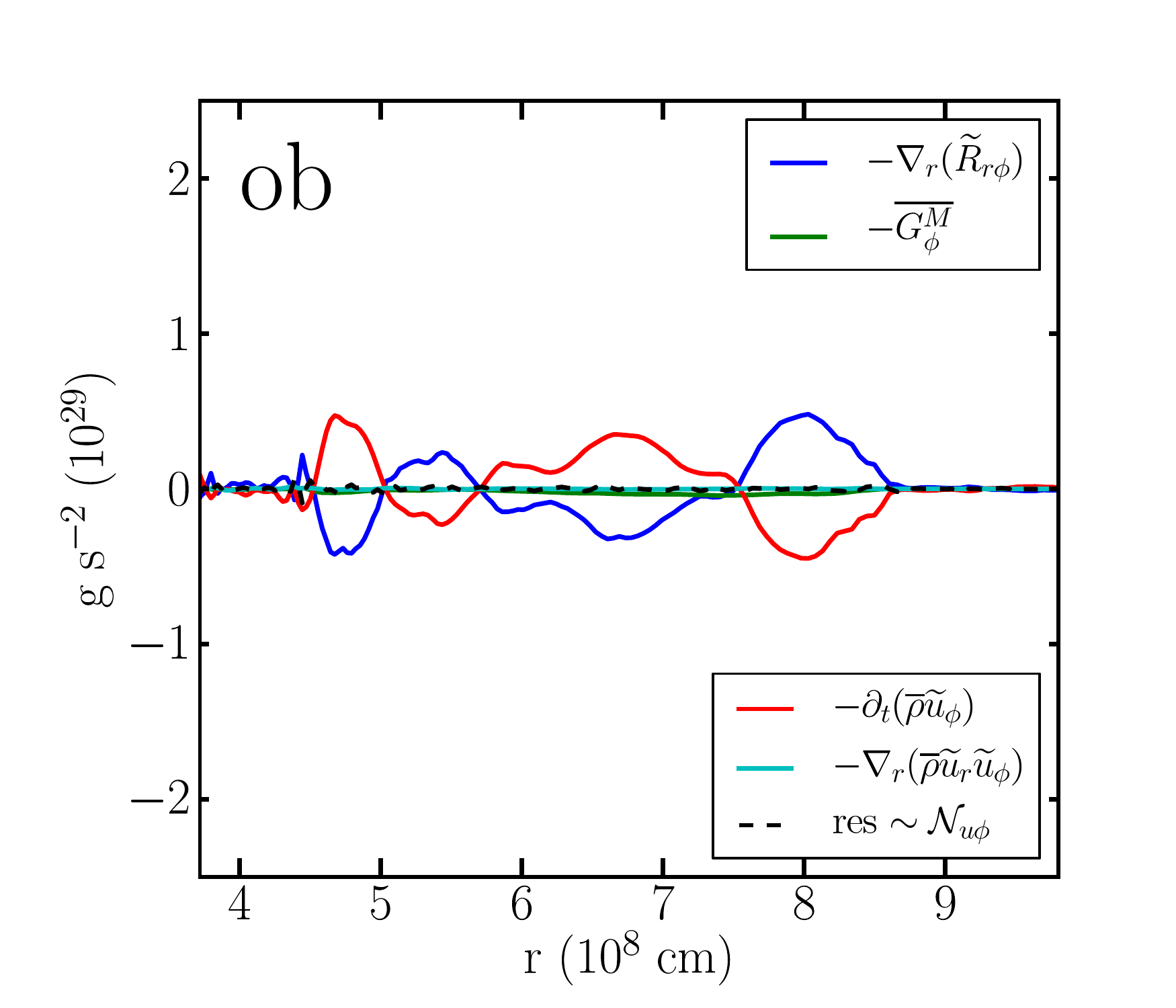}
\includegraphics[width=6.8cm]{obmrez_tavg230_pmomentum_equation_ransdat-eps-converted-to.pdf}
\includegraphics[width=6.8cm]{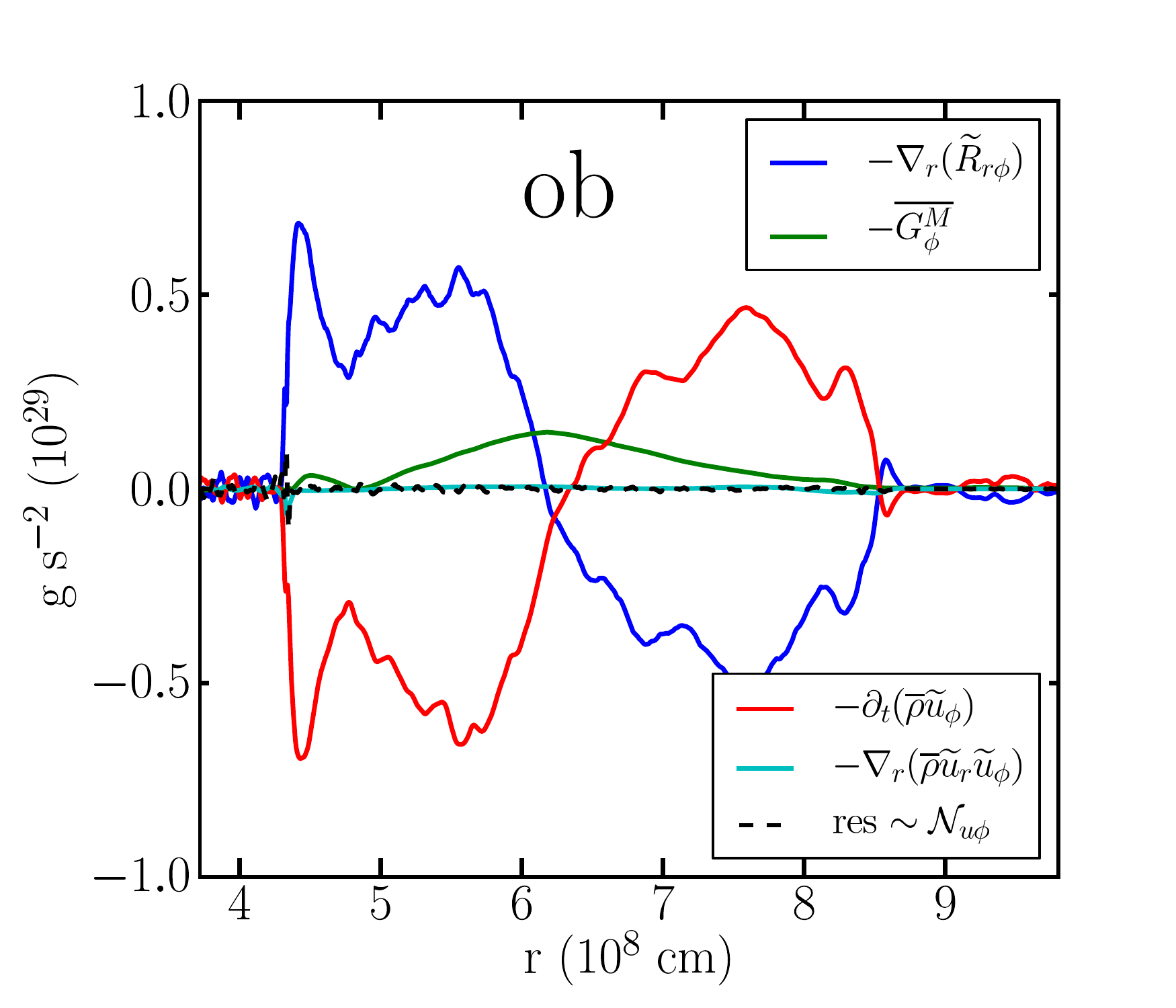}}
\caption{Mean azimuthal momentum equation (upper panels) and polar momentum equation (lower panels). Model {\sf ob.3D.lr} (left), {\sf ob.3D.mr} (middle), {\sf ob.3D.hr} (right) \label{fig:ob-res-tmomentum-pmomentum-equation}}
\end{figure}

\newpage

\subsubsection{Mean internal and kinetic energy equation}

\begin{figure}[!h]
\centerline{
\includegraphics[width=6.8cm]{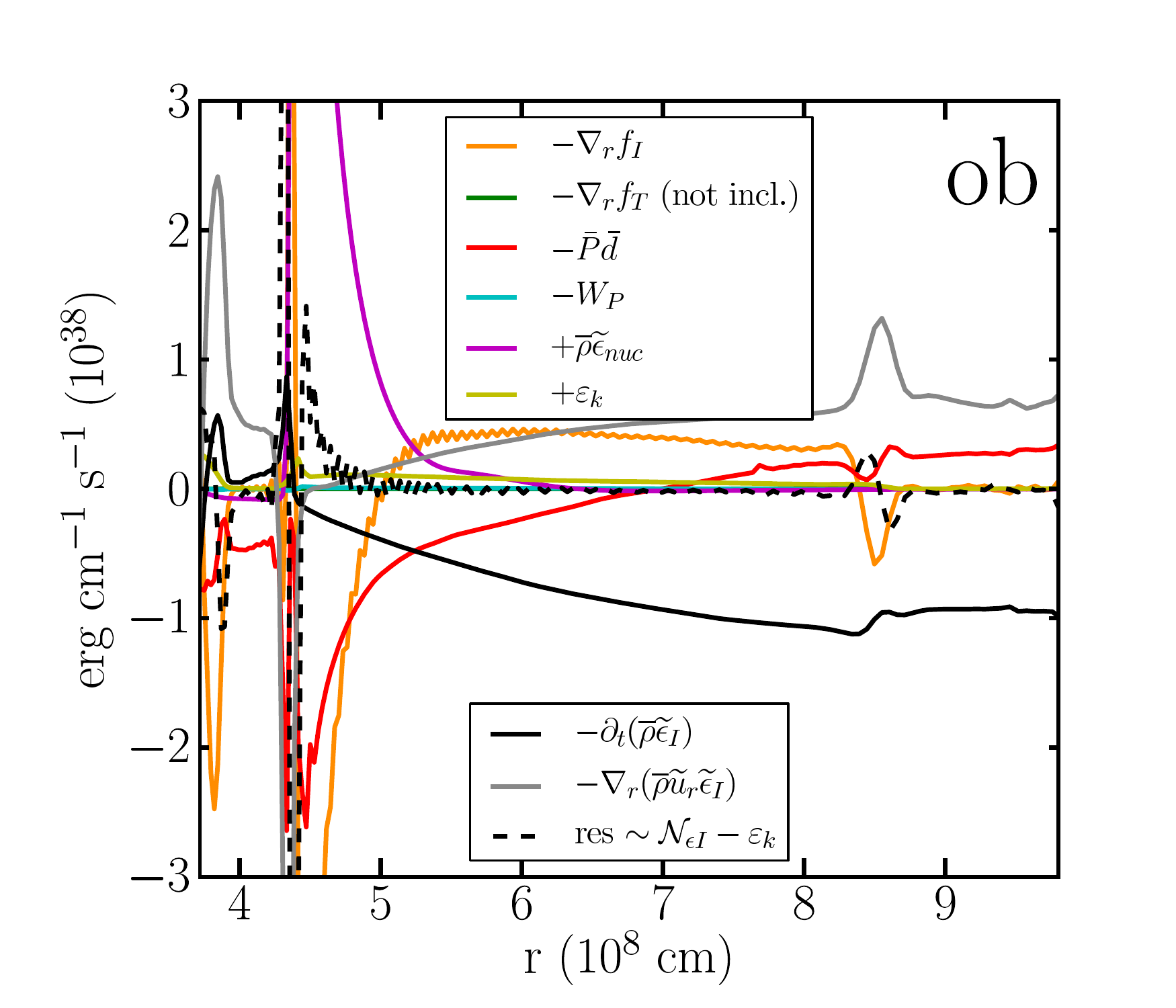}
\includegraphics[width=6.8cm]{obmrez_tavg230_internal_energy_equation_ransdat-eps-converted-to.pdf}
\includegraphics[width=6.8cm]{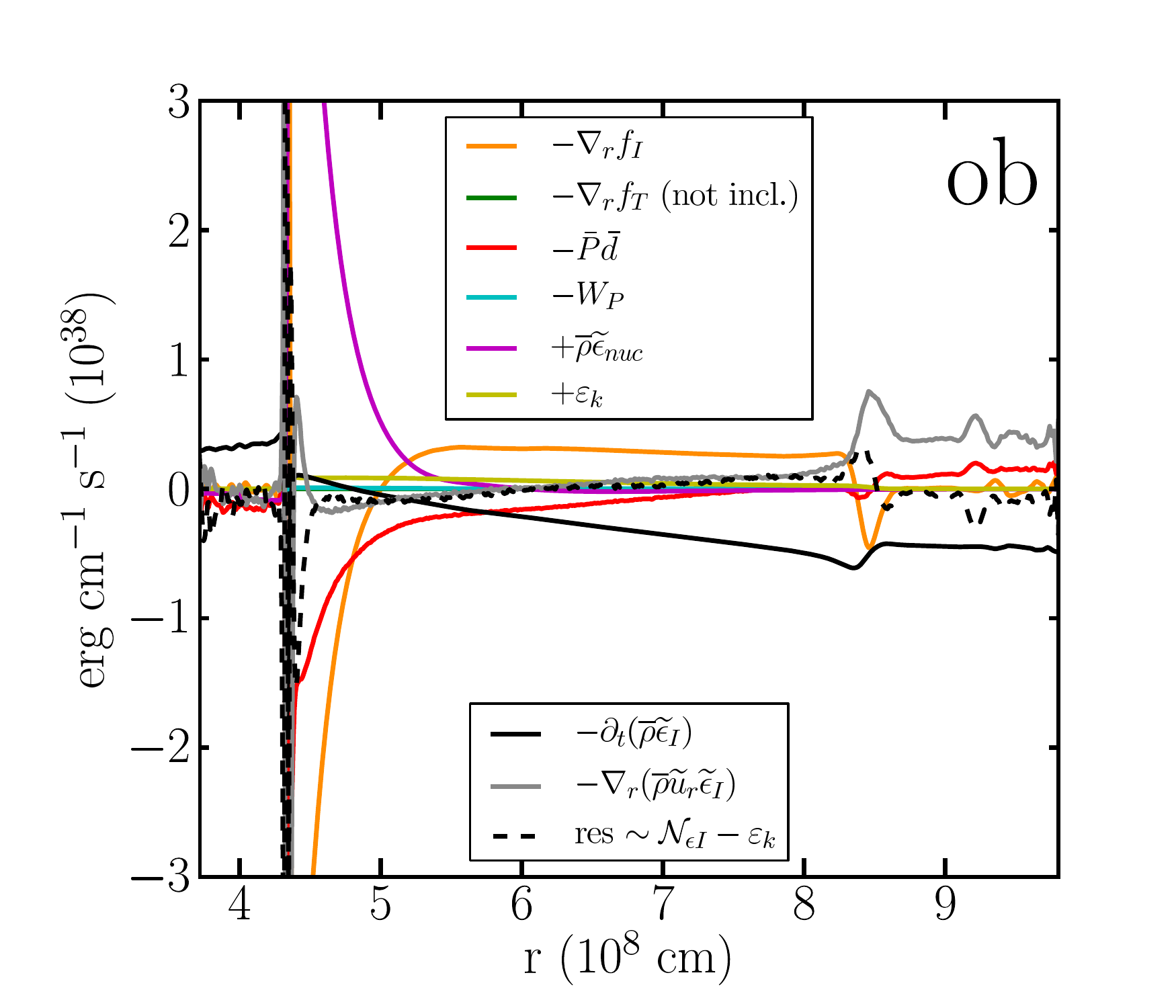}}

\centerline{
\includegraphics[width=6.8cm]{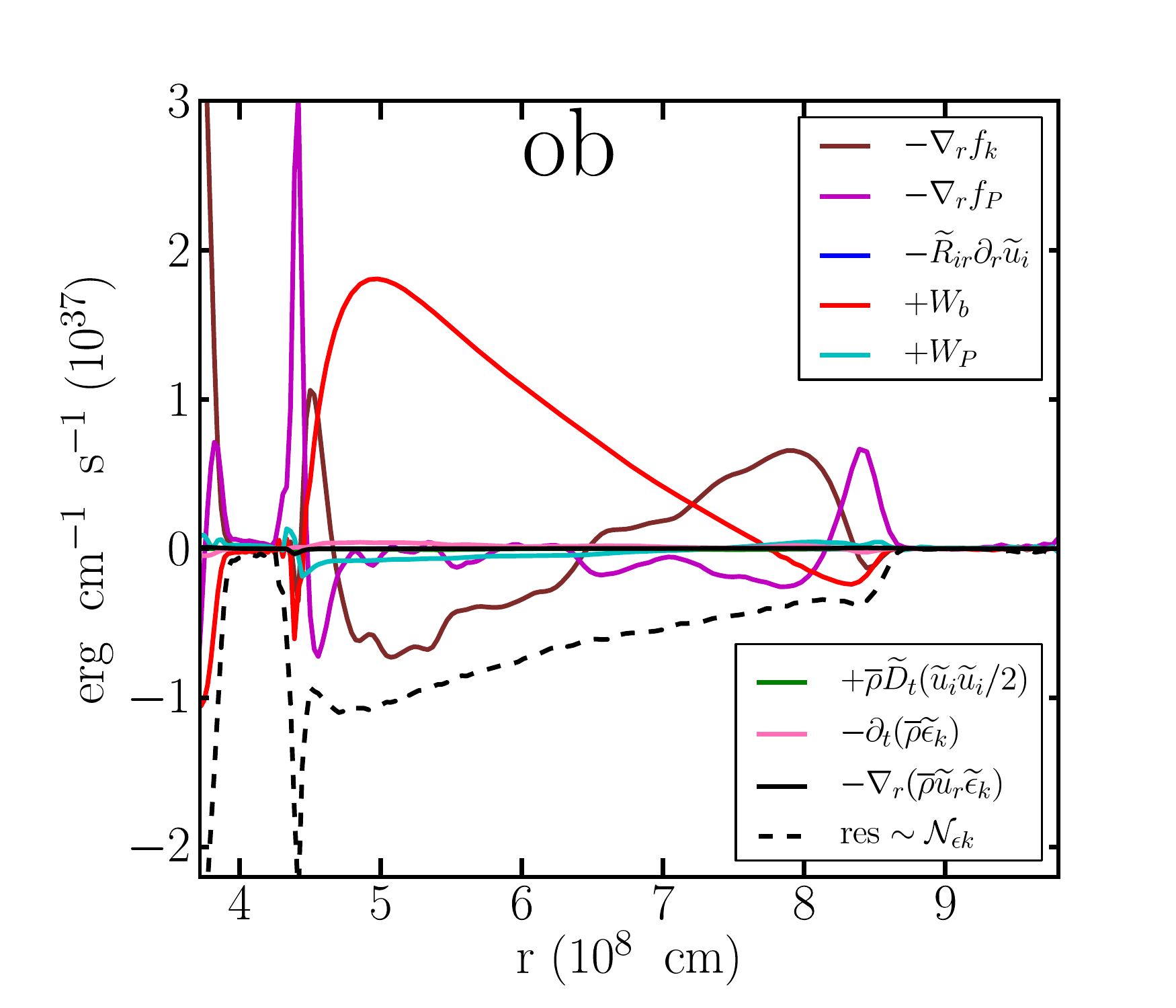}
\includegraphics[width=6.8cm]{obmrez_tavg230_kinetic_energy_equation_ransdat-eps-converted-to.pdf}
\includegraphics[width=6.8cm]{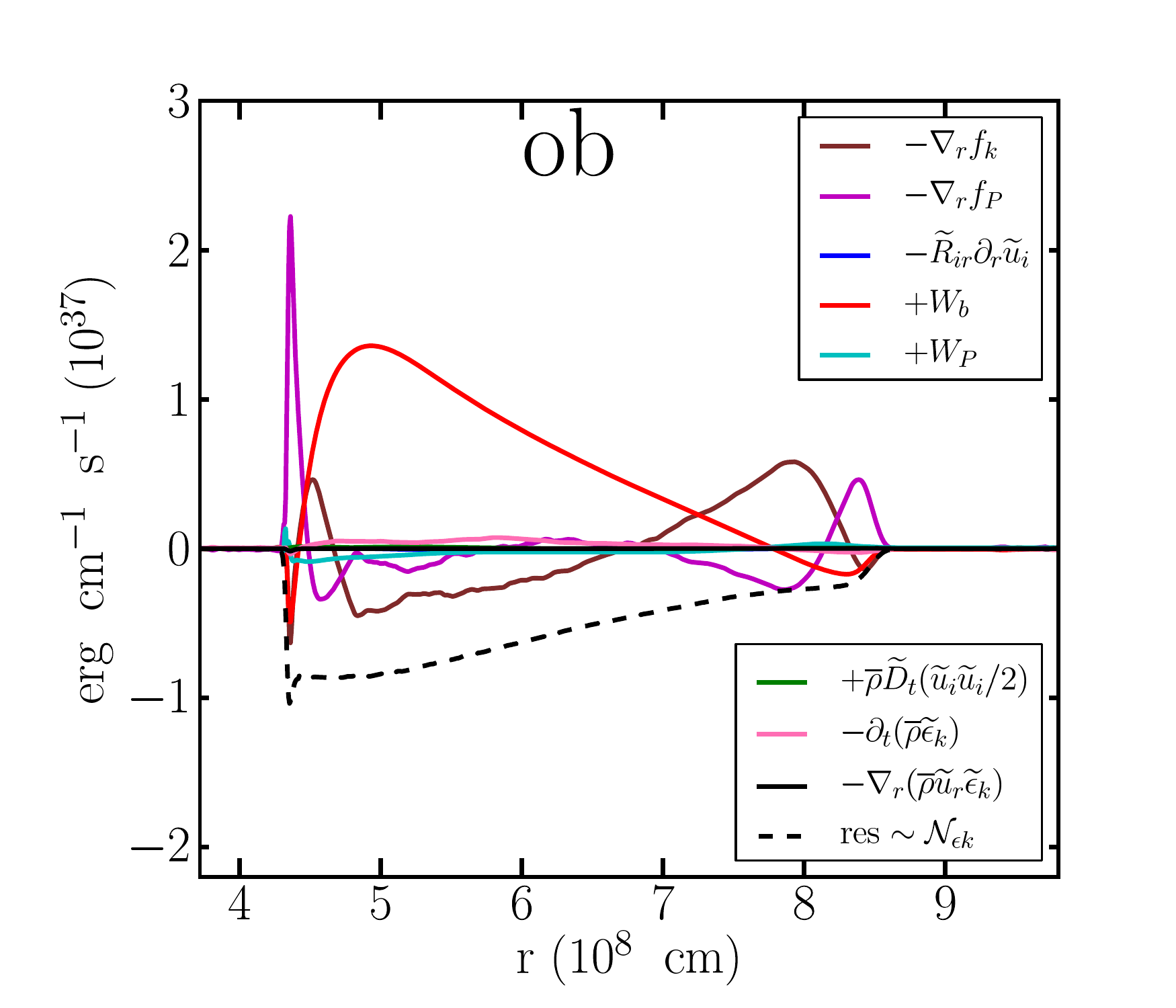}}
\caption{Mean internal energy equation (upper panels) and kinetic energy equation (lower panels). Model {\sf ob.3D.lr} (left), {\sf ob.3D.mr} (middle), {\sf ob.3D.hr} (right) \label{fig:ob-res-ei-ek-equation}}
\end{figure}

\newpage

\subsubsection{Mean total energy and entropy equation}

\begin{figure}[!h]
\centerline{
\includegraphics[width=6.8cm]{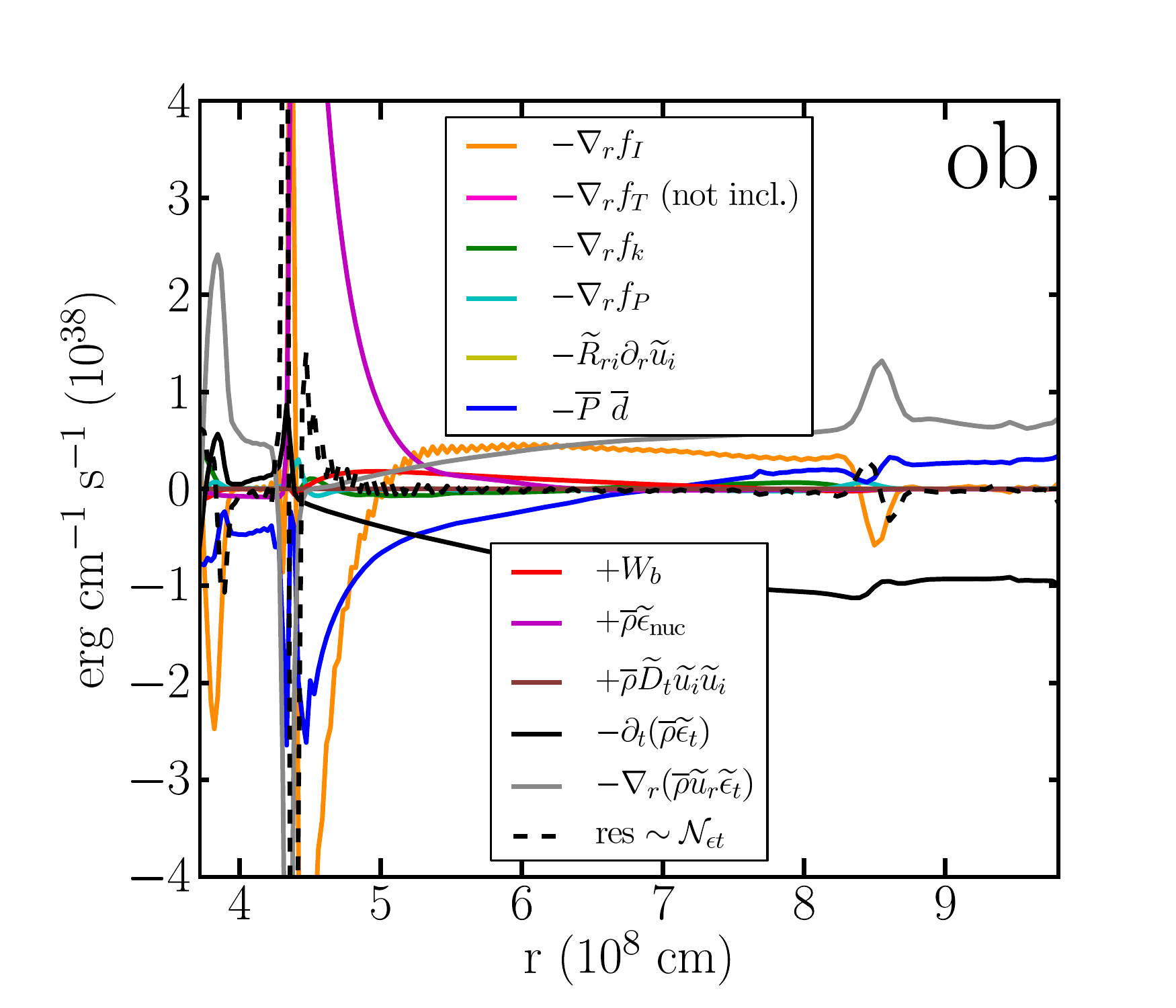}
\includegraphics[width=6.8cm]{obmrez_tavg230_total_energy_equation_ransdat-eps-converted-to.pdf}
\includegraphics[width=6.8cm]{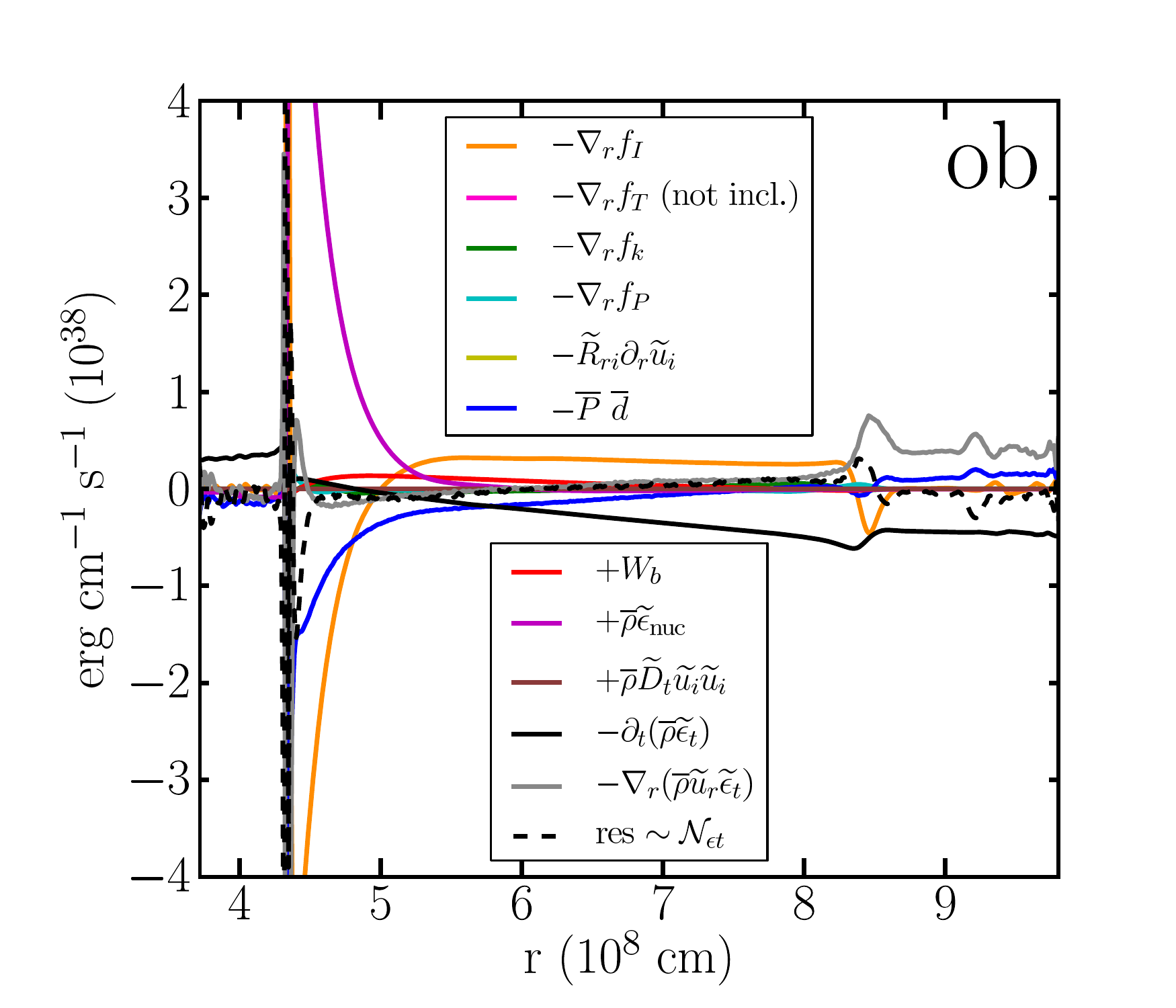}}

\centerline{
\includegraphics[width=6.8cm]{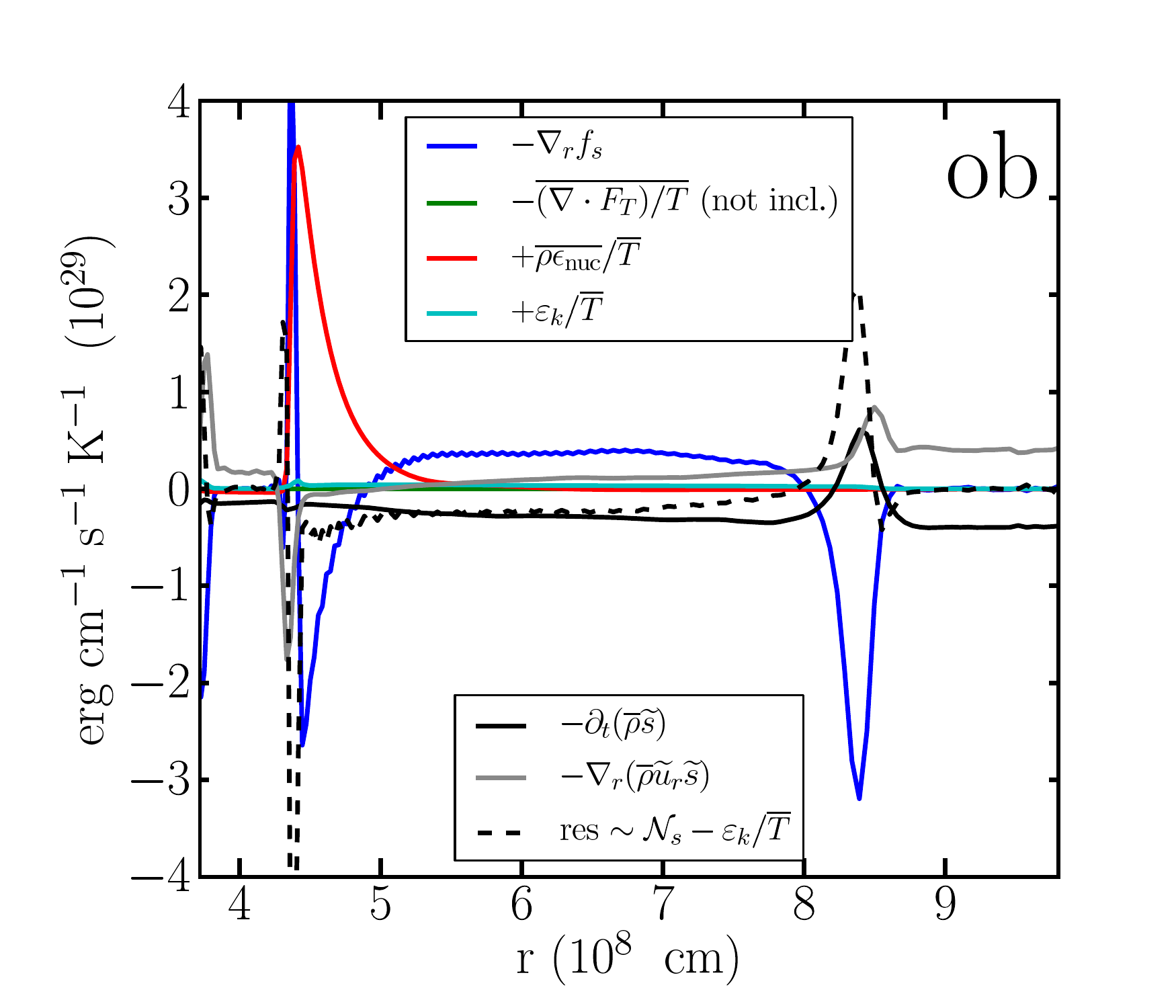}
\includegraphics[width=6.8cm]{obmrez_tavg230_entropy_equation_ransdat-eps-converted-to.pdf}
\includegraphics[width=6.8cm]{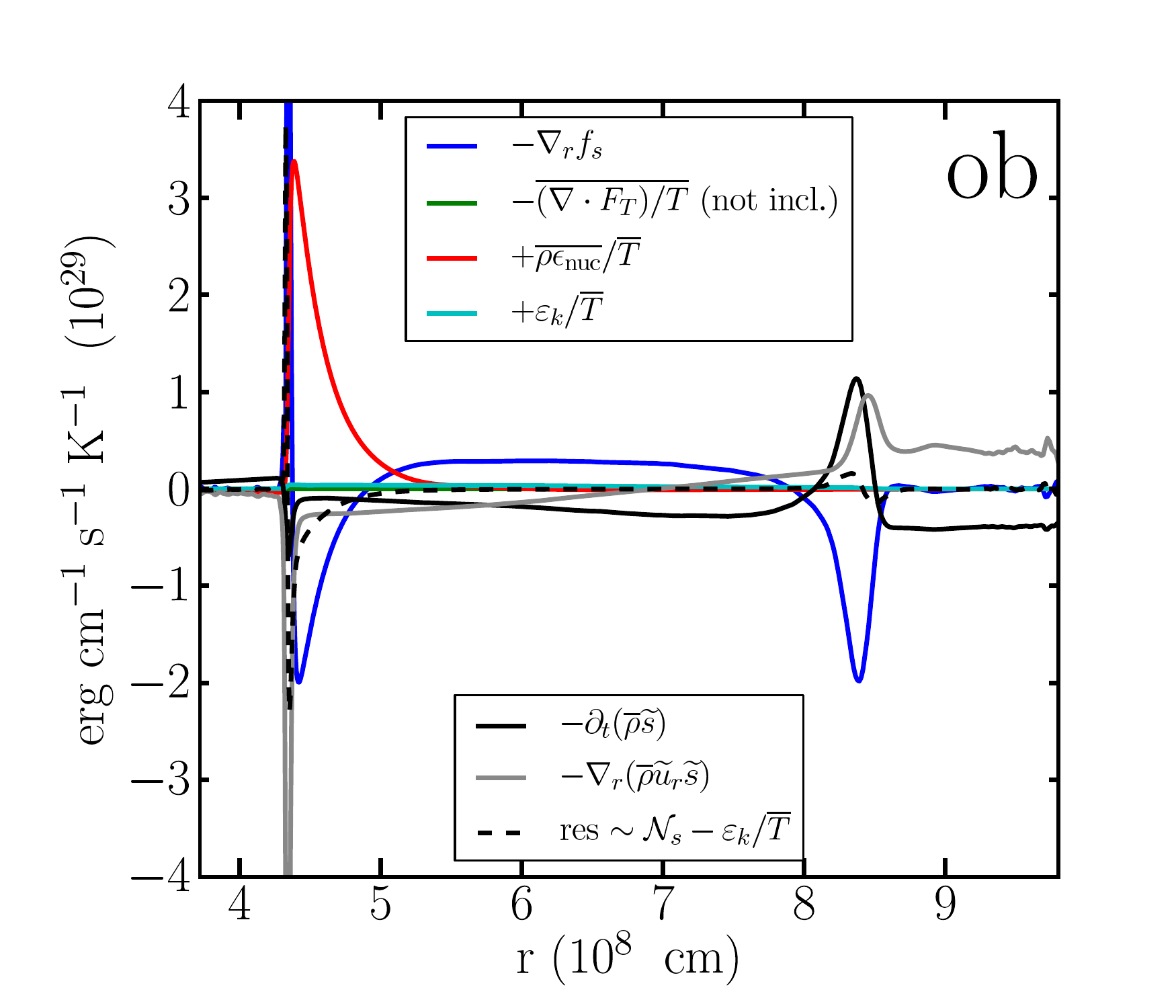}}
\caption{Mean total energy equation (upper panels) and mean entropy equation (lower panels). Model {\sf ob.3D.lr} (left), {\sf ob.3D.mr} (middle), {\sf ob.3D.hr} (right) \label{fig:ob-res-et-ss-equation}}
\end{figure}

\newpage

\subsubsection{Mean density-specific volume covariance equation and mean number of nucleons per isotope equation}

\begin{figure}[!h]
\centerline{
\includegraphics[width=6.8cm]{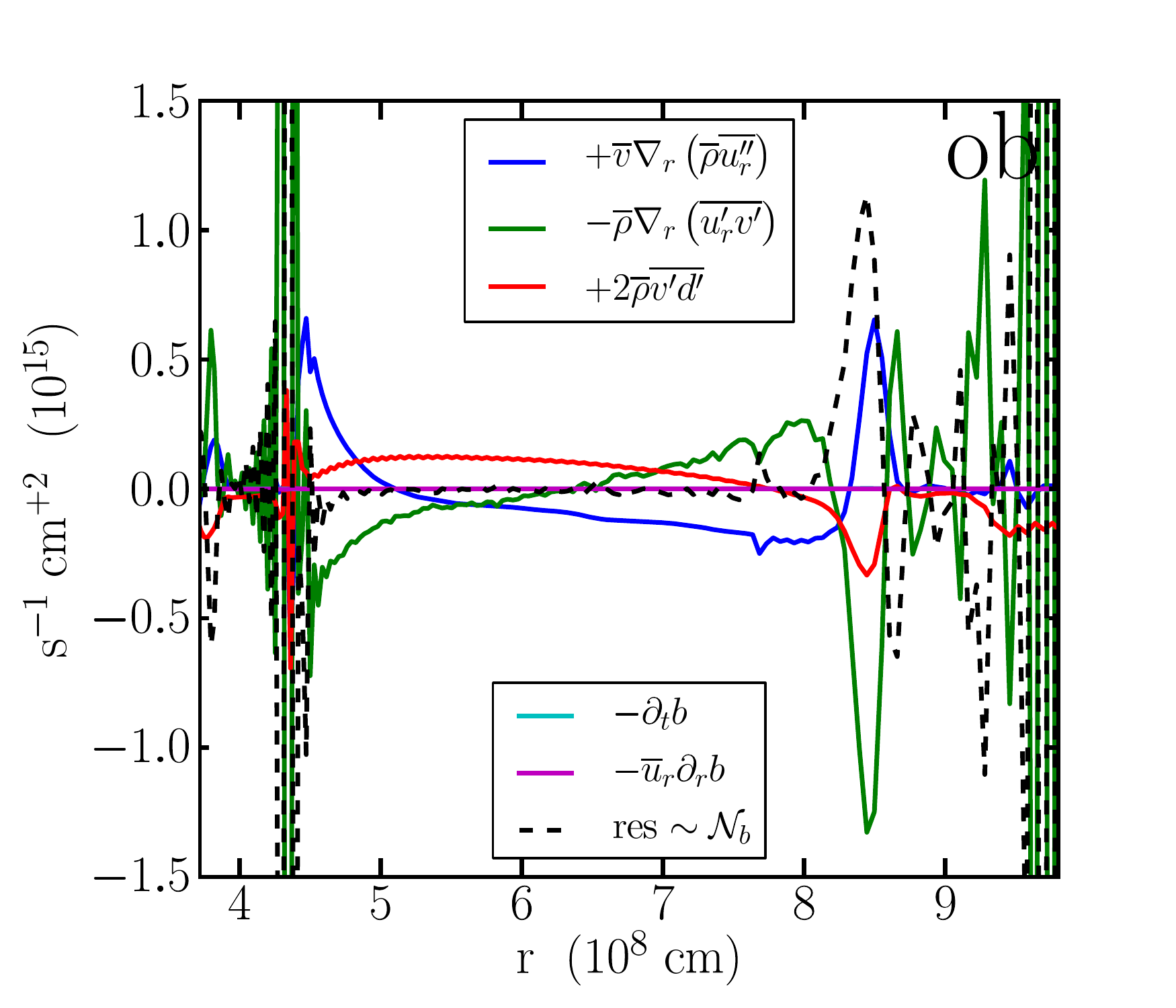}
\includegraphics[width=6.8cm]{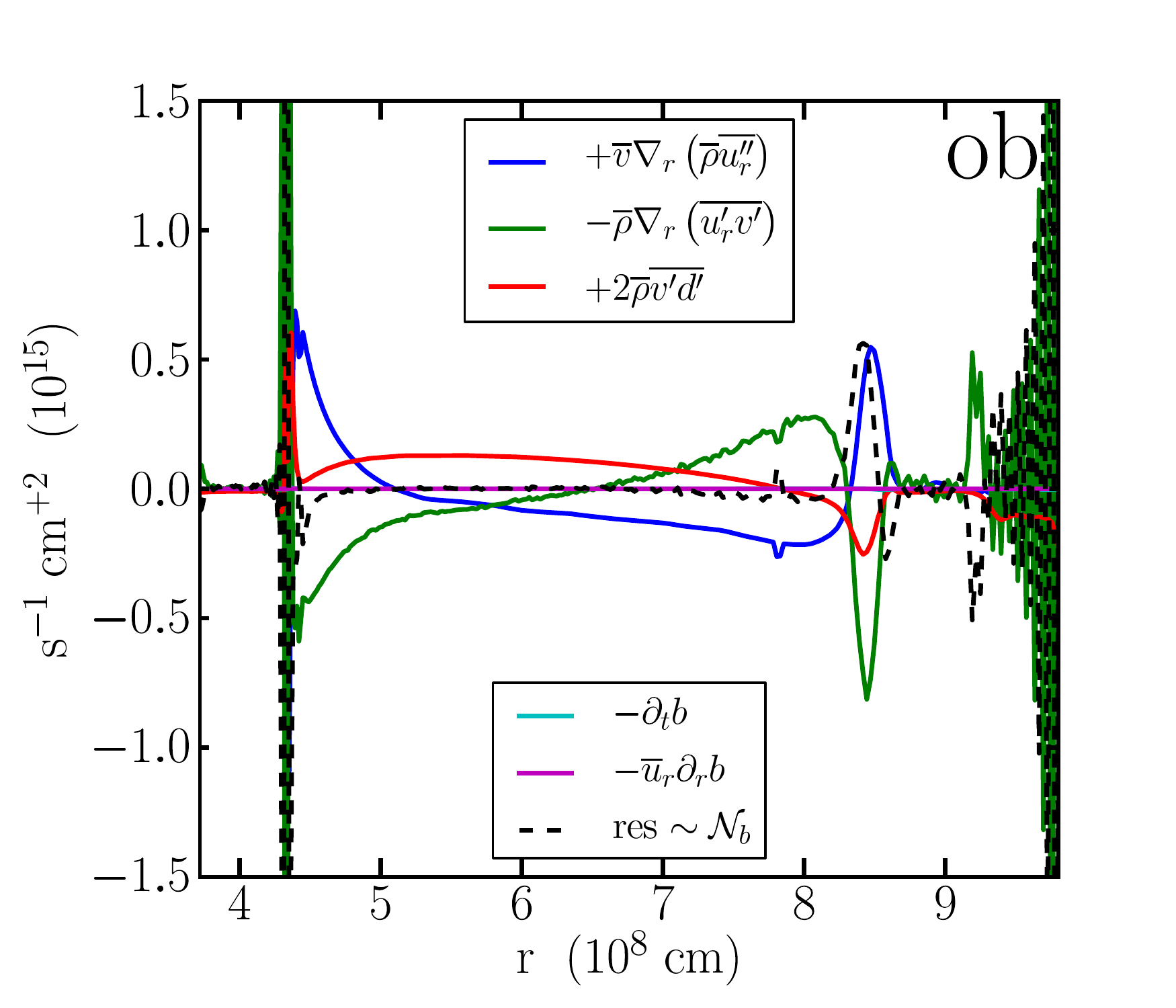}
\includegraphics[width=6.8cm]{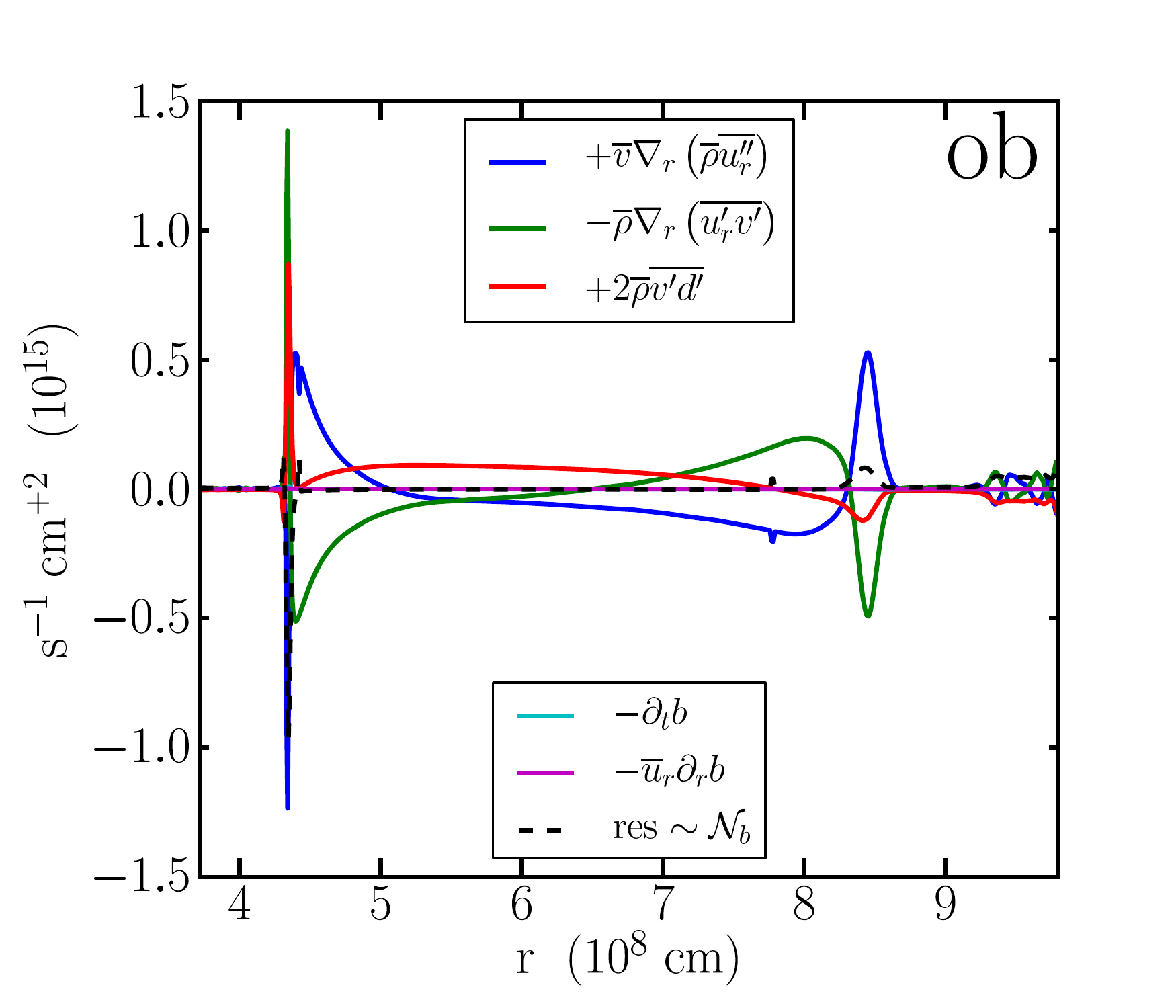}}

\centerline{
\includegraphics[width=6.8cm]{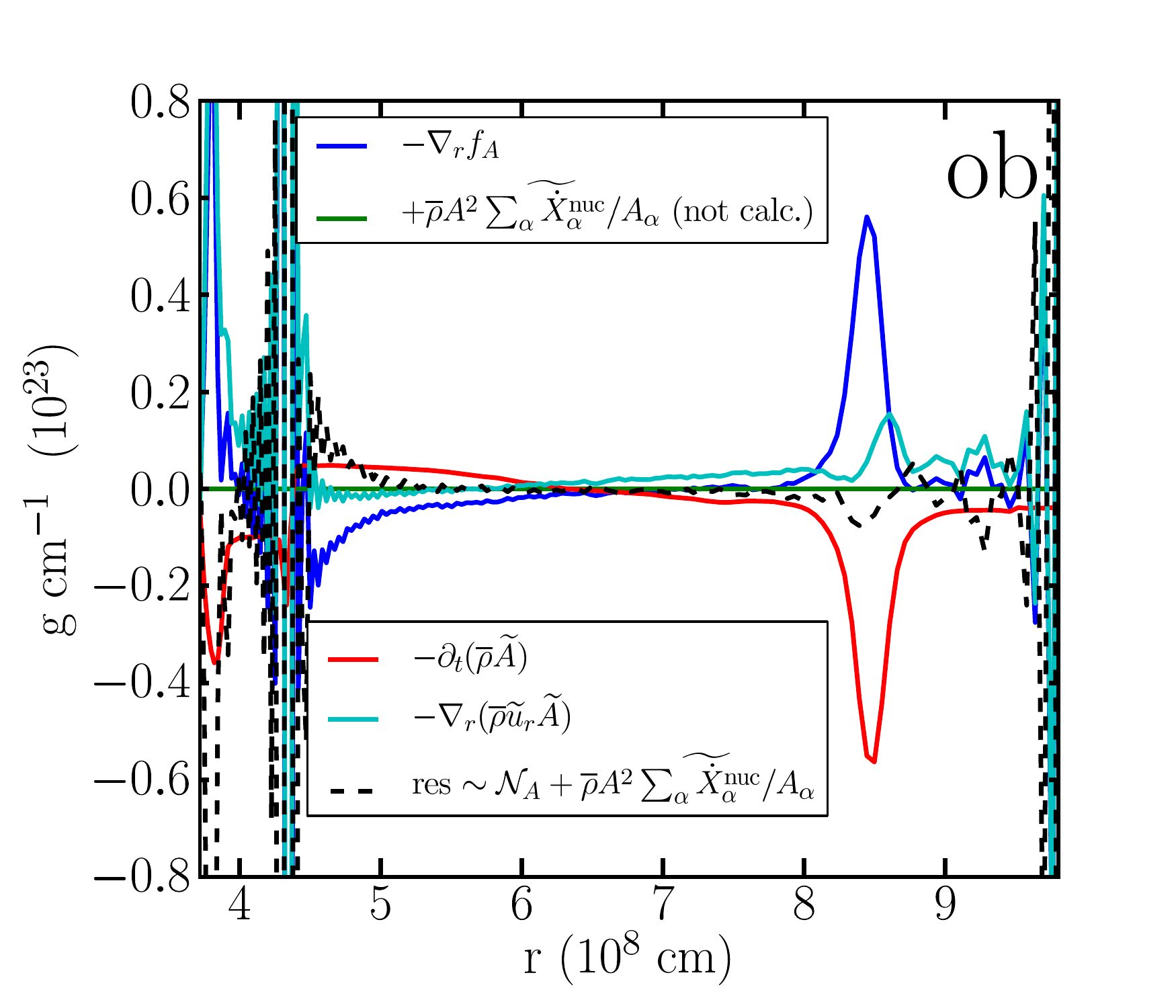}
\includegraphics[width=6.8cm]{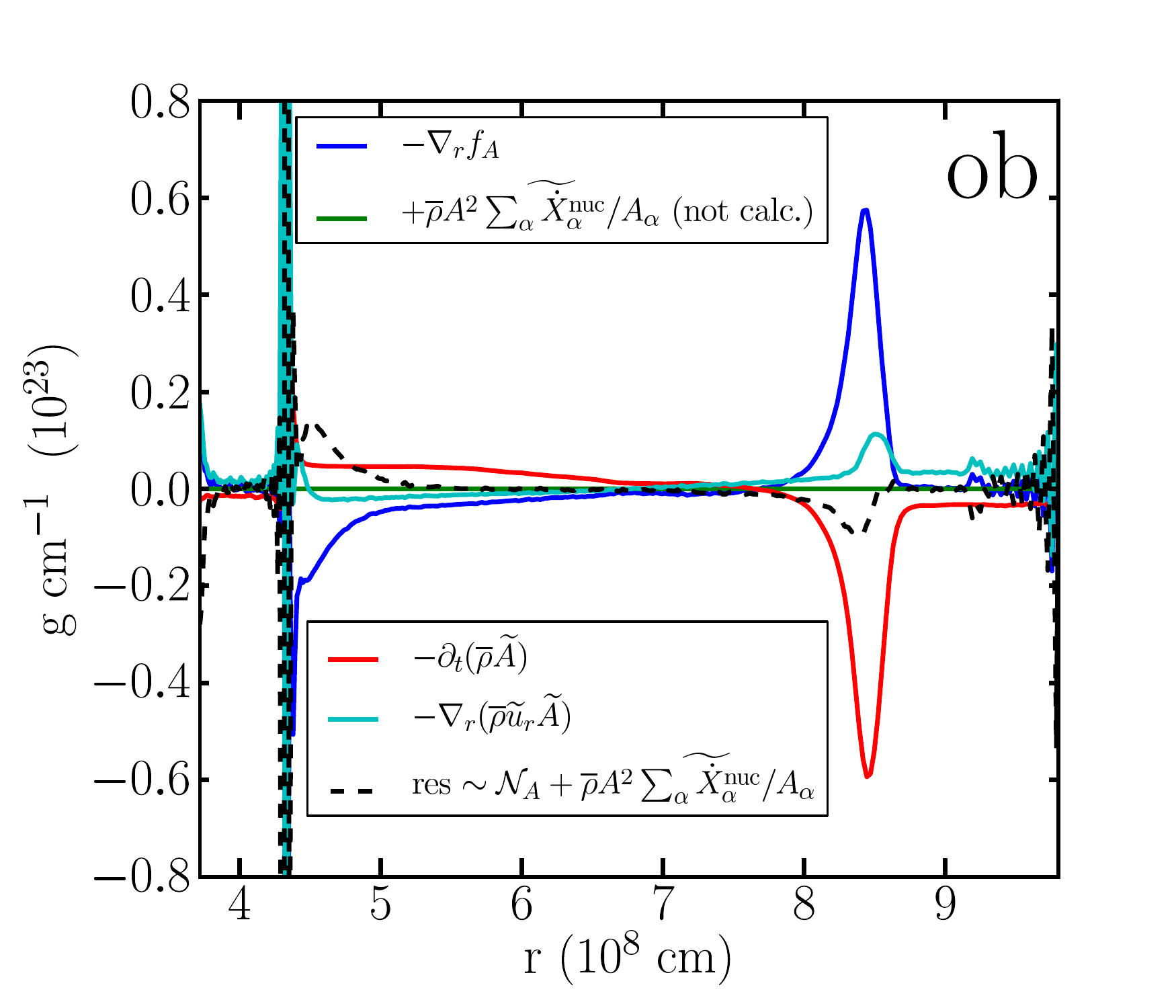}
\includegraphics[width=6.8cm]{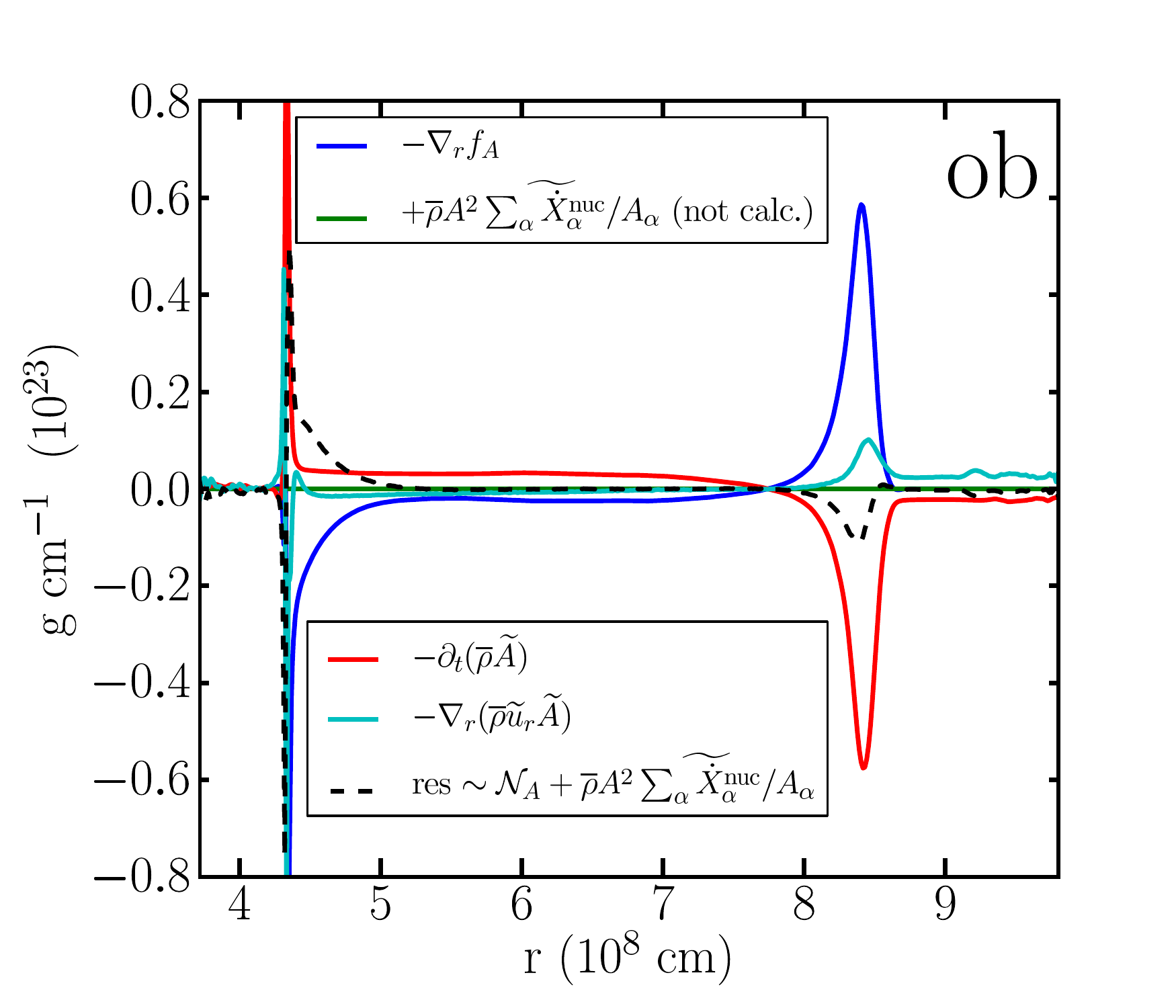}}
\caption{Mean density-specific volume covariance equation (upper panels) and mean number of nucleons per isotope equation (lower panels). Model {\sf ob.3D.lr} (left), {\sf ob.3D.mr} (middle), {\sf ob.3D.hr} (right) \label{fig:ob-res-b-A-equation}}
\end{figure}

\newpage

\subsubsection{Mean turbulent kinetic energy equation and mean velocities}

\begin{figure}[!h]
\centerline{
\includegraphics[width=6.8cm]{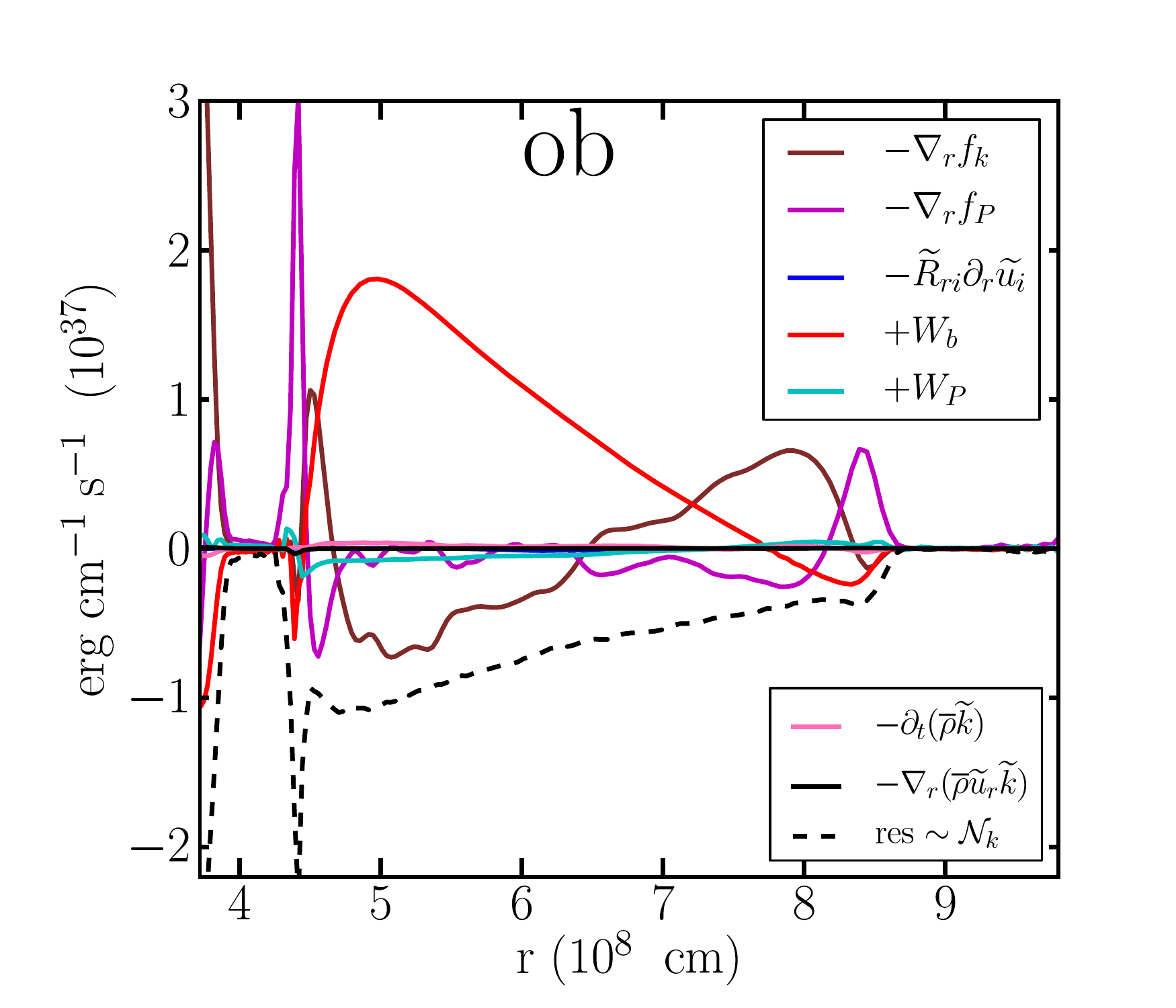}
\includegraphics[width=6.8cm]{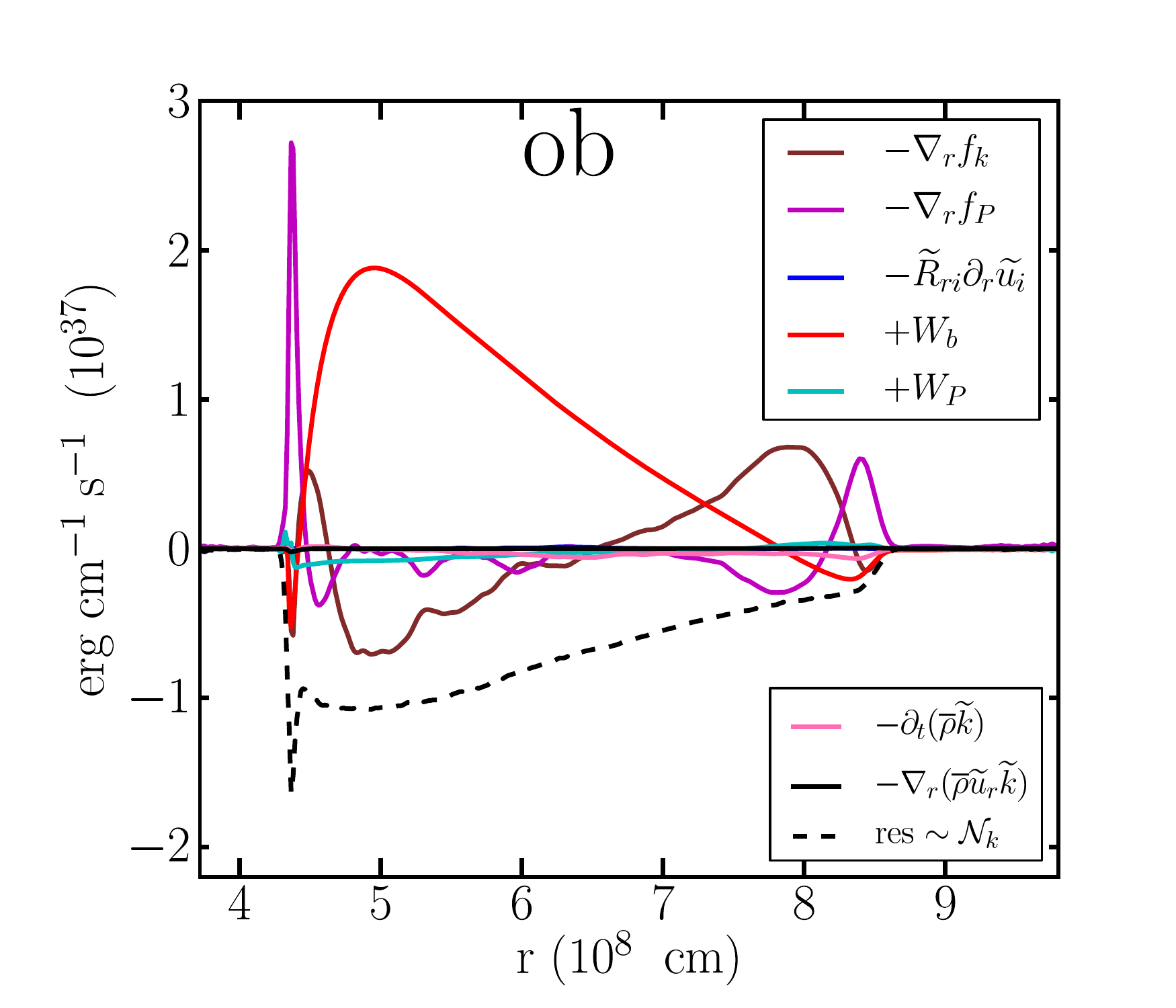}
\includegraphics[width=6.8cm]{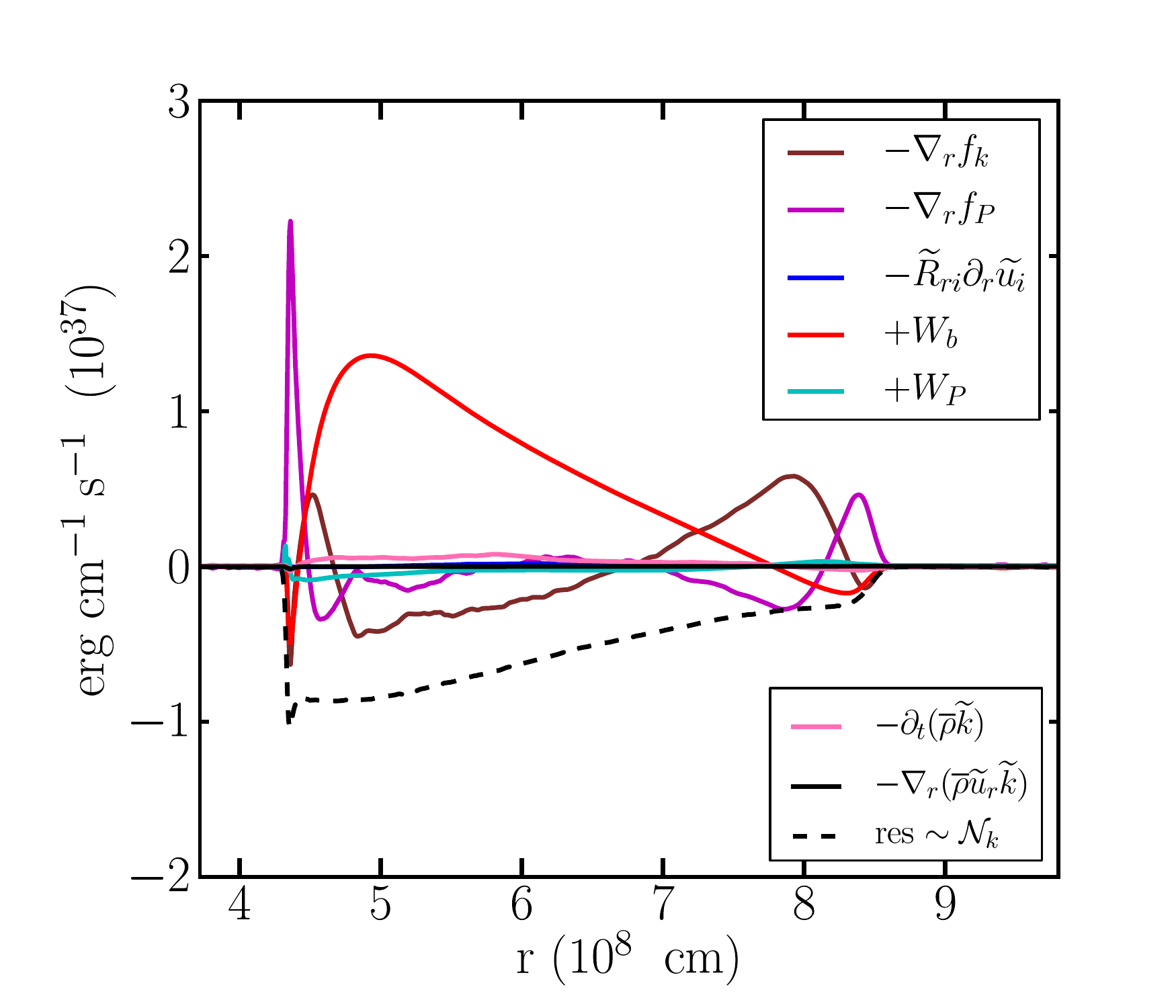}}

\centerline{
\includegraphics[width=6.8cm]{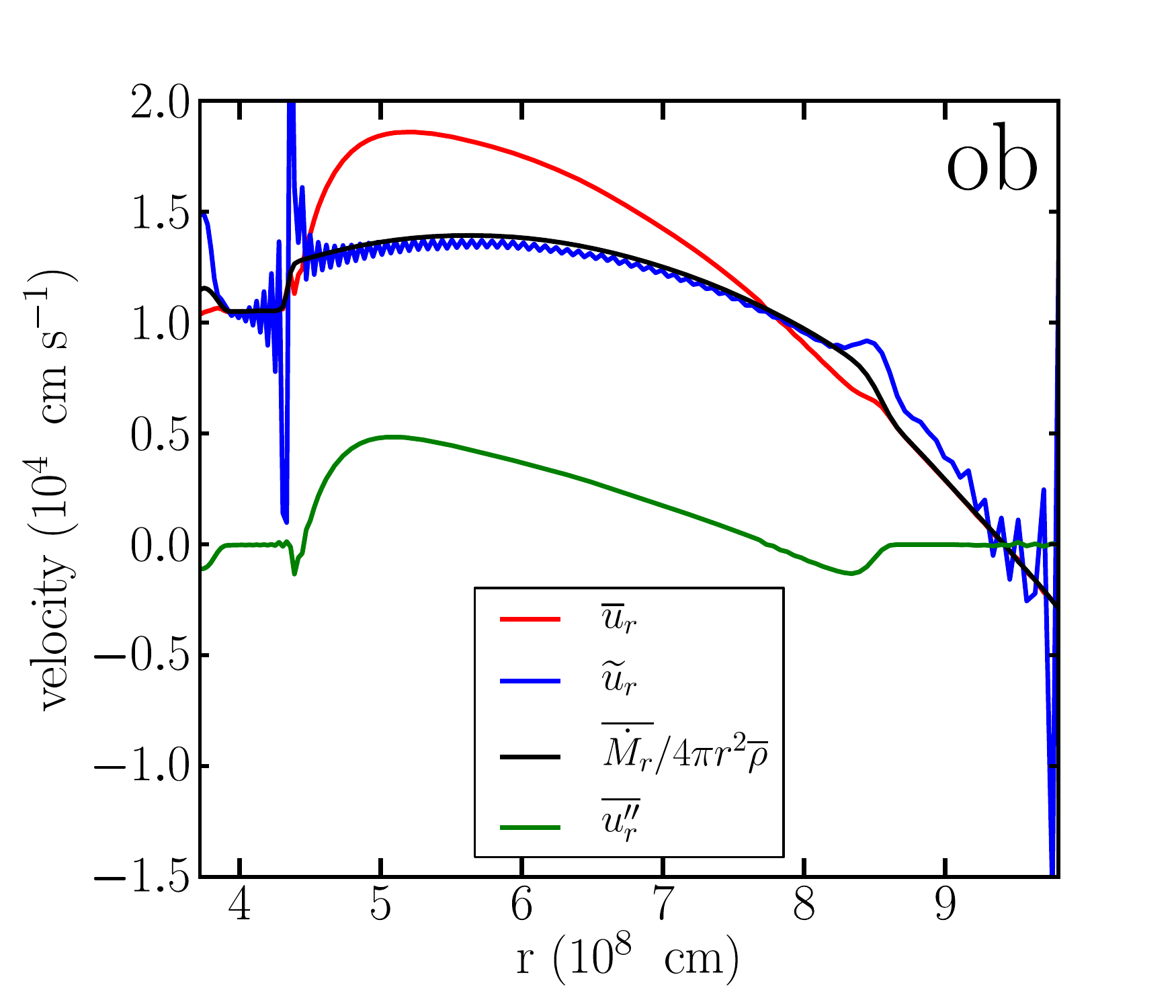}
\includegraphics[width=6.8cm]{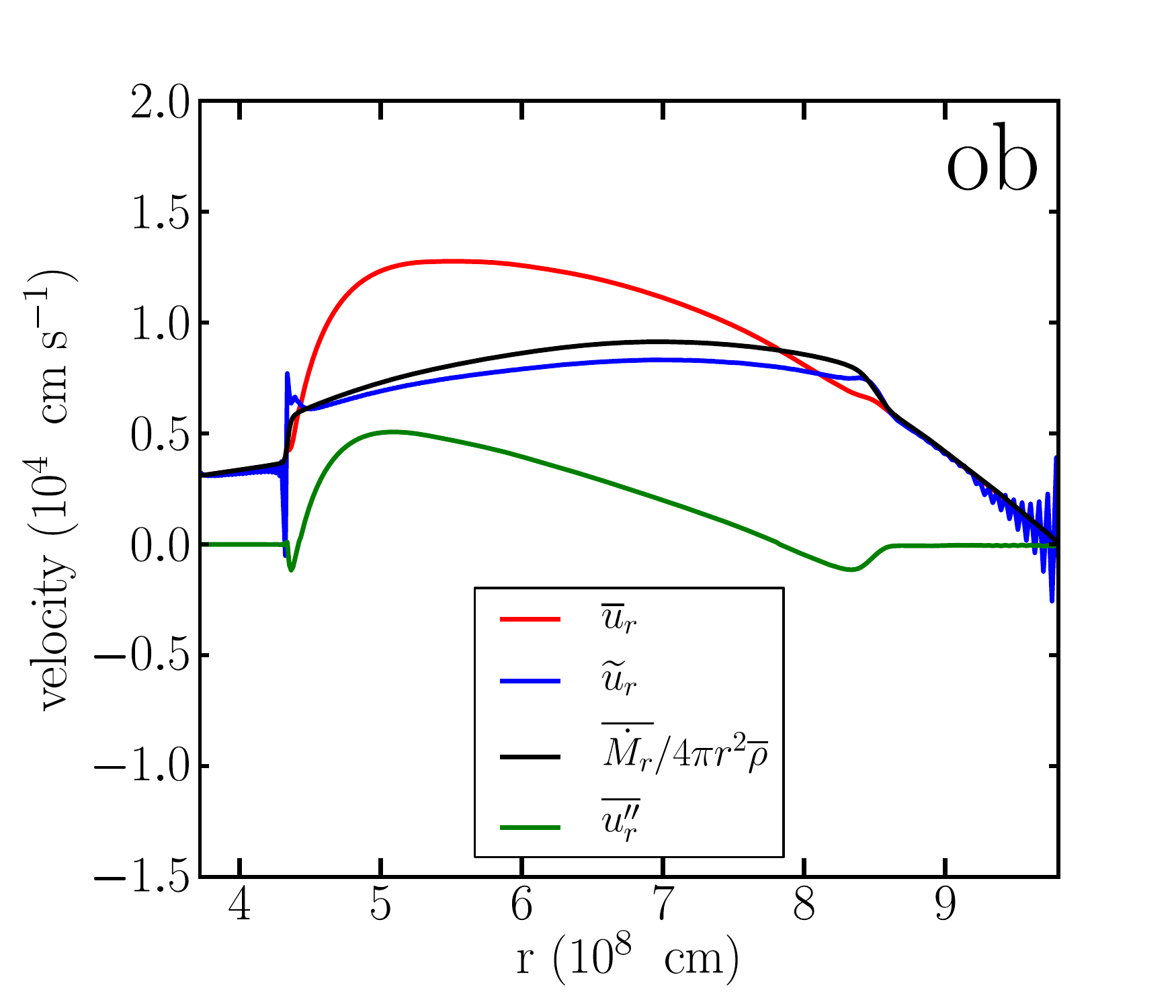}
\includegraphics[width=6.8cm]{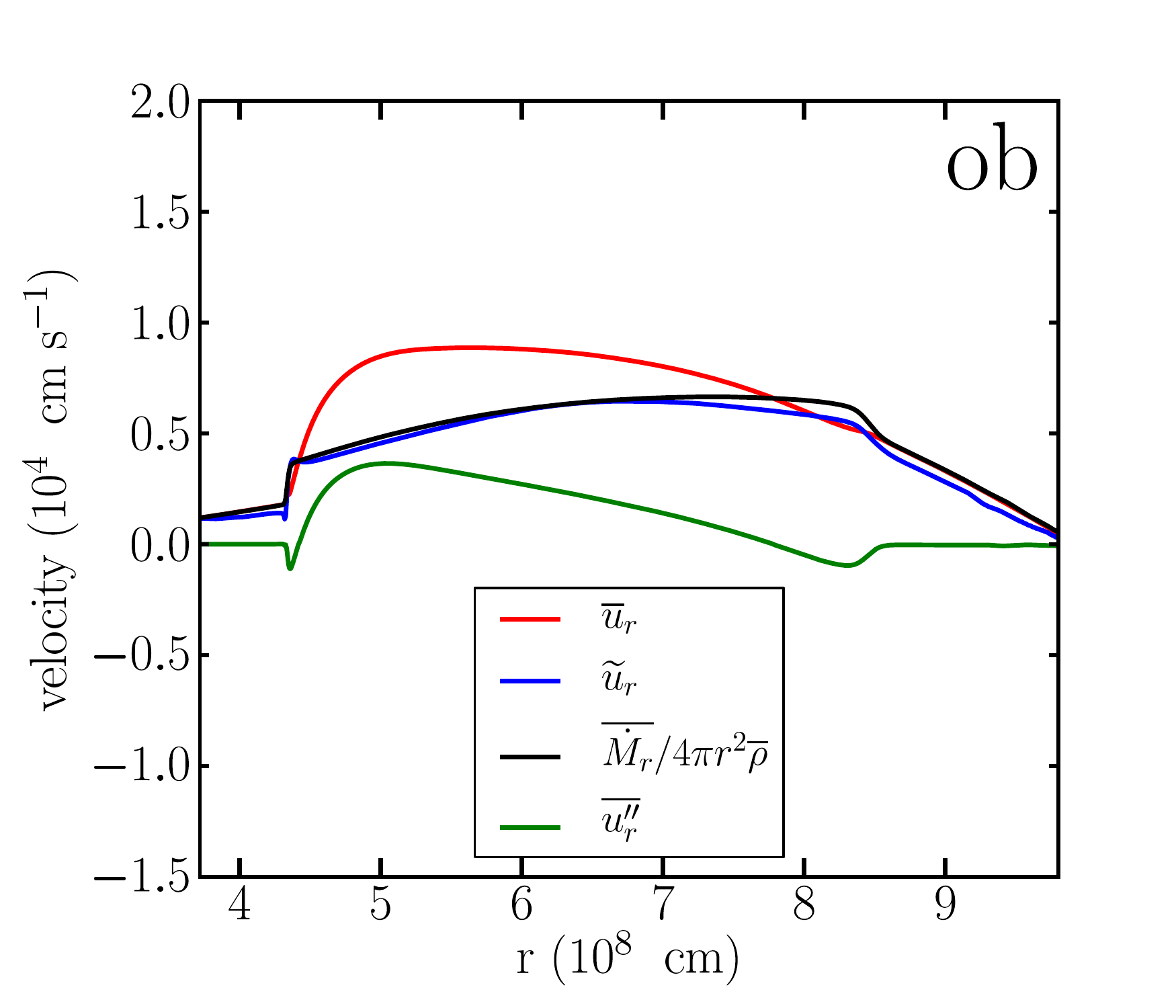}}
\caption{Mean turbulent kinetic energy equation (upper panels) and mean velocities (lower panels). Model {\sf ob.3D.lr} (left), {\sf ob.3D.mr} (middle), {\sf ob.3D.hr} (right) \label{fig:ob-res-k-vel-equation}}
\end{figure}

\newpage

\subsection{Red giant envelope convection}

\subsubsection{Mean continuity equation and mean radial momentum equation}

\begin{figure}[!h]
\centerline{
\includegraphics[width=6.6cm]{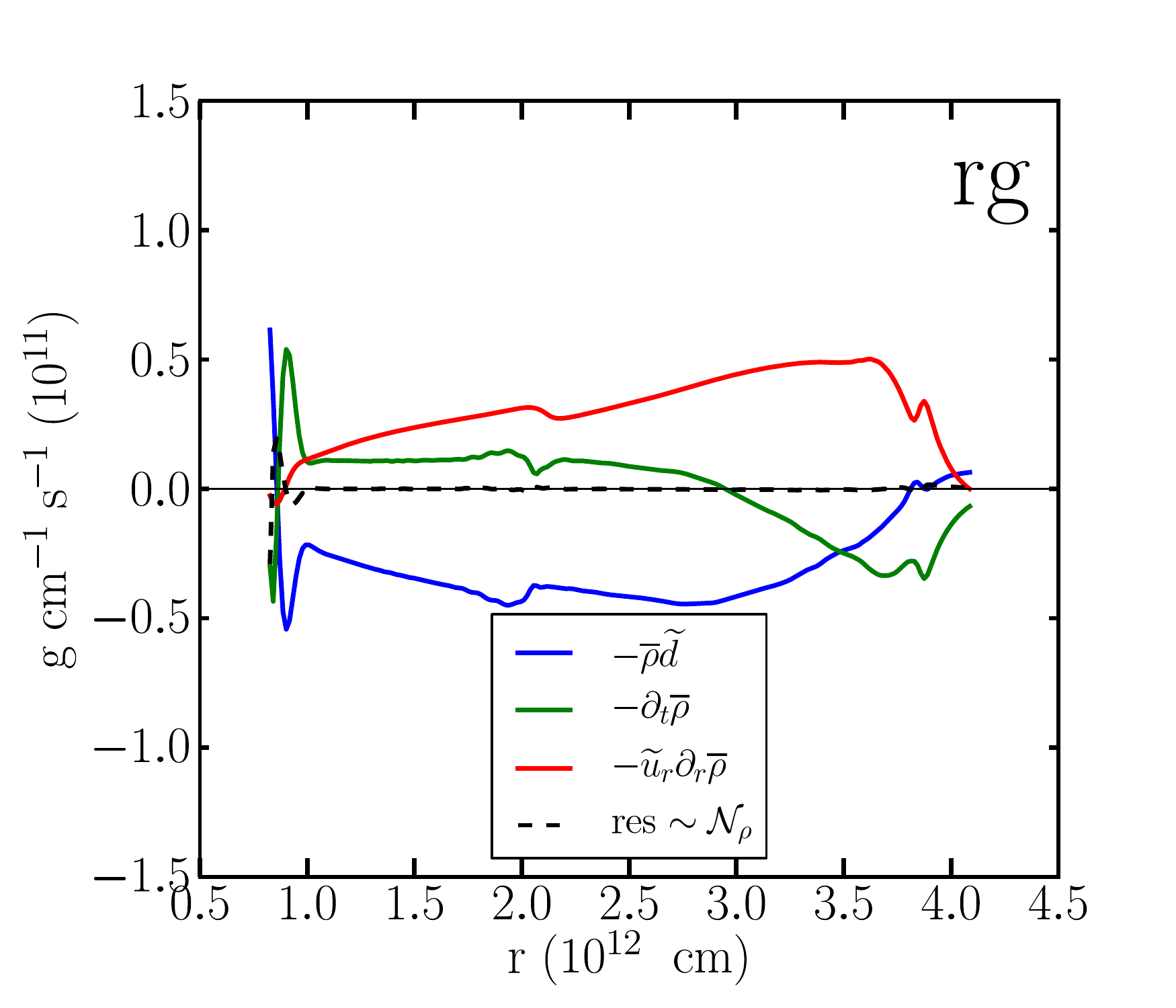}
\includegraphics[width=6.6cm]{rgmrez_tavg800_continuity_equation_insf-eps-converted-to.pdf}}

\centerline{
\includegraphics[width=6.6cm]{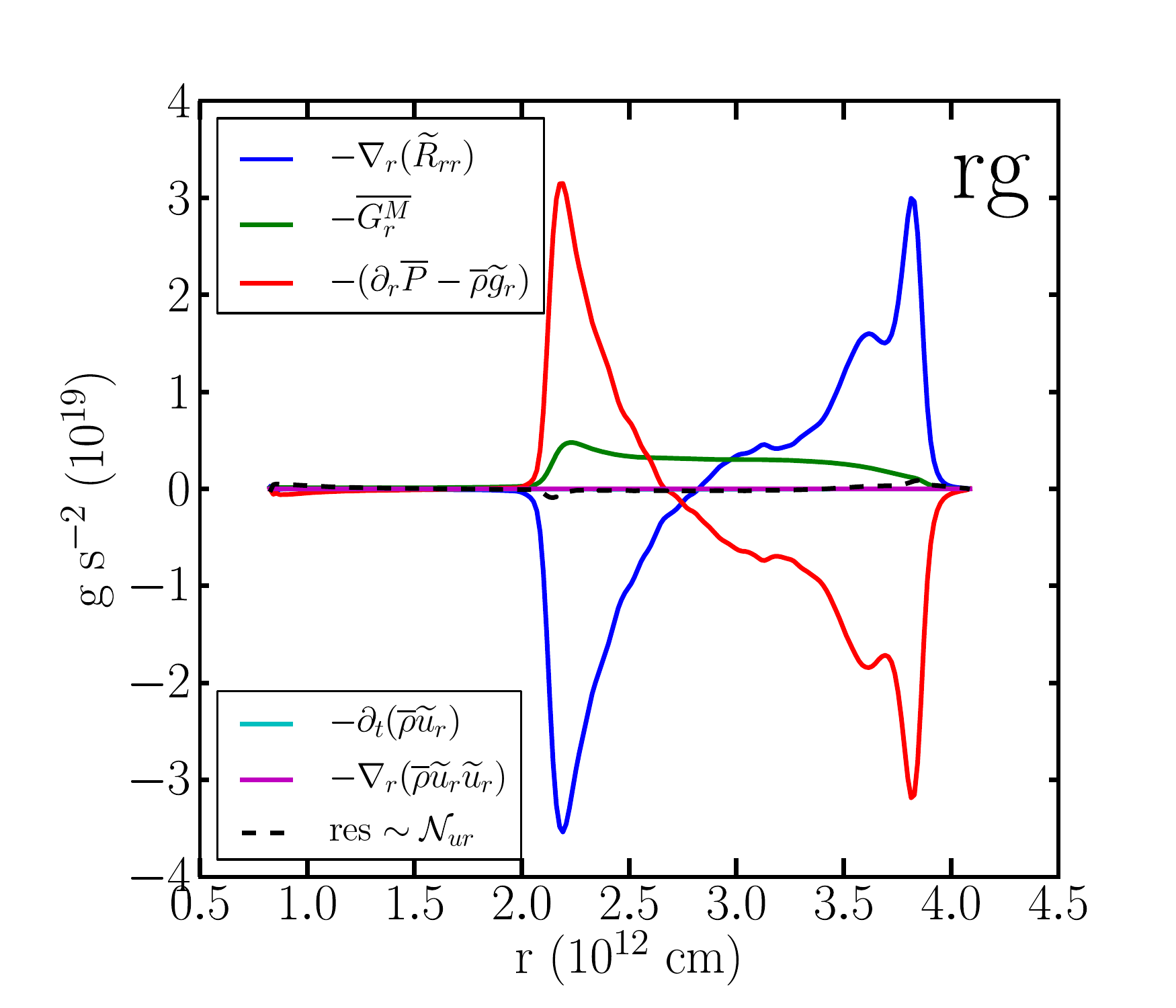}
\includegraphics[width=6.6cm]{rgmrez_tavg800_rmomentum_equation_insf-eps-converted-to.pdf}}
\caption{Mean continuity equation (upper panels) and radial momentum equation (lower panels). Model {\sf rg.3D.lr} (left) and {\sf rg.3D.mr} (right) \label{fig:rg-res-cont-rmomentum-equation}}
\end{figure}

\newpage

\subsubsection{Mean azimuthal and polar momentum equation}

\begin{figure}[!h]
\centerline{
\includegraphics[width=7.cm]{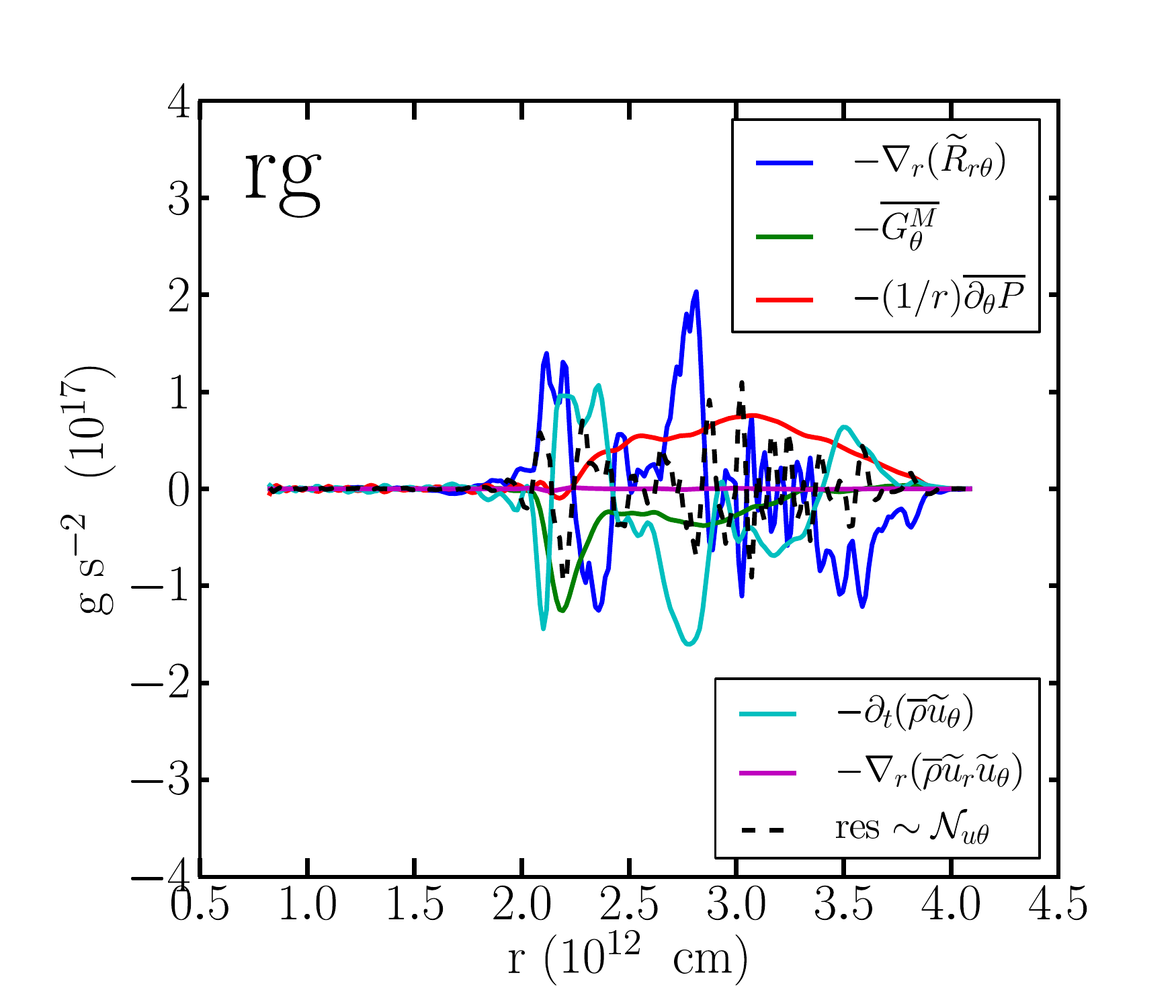}
\includegraphics[width=7.cm]{rgmrez_tavg800_tmomentum_equation_insf-eps-converted-to.pdf}}

\centerline{
\includegraphics[width=7.cm]{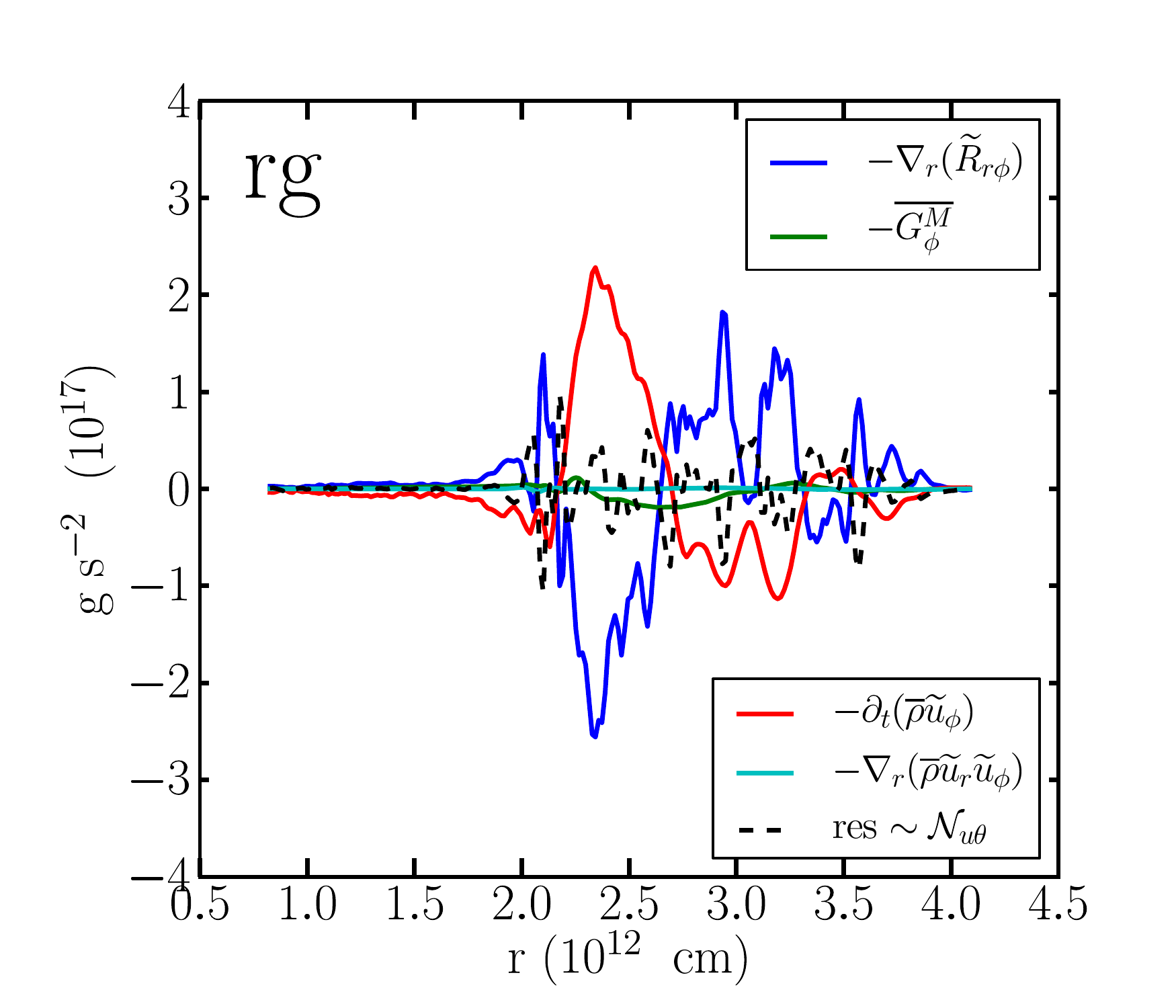}
\includegraphics[width=7.cm]{rgmrez_tavg800_pmomentum_equation_insf-eps-converted-to.pdf}}
\caption{Mean azimuthal momentum (upper panels) and polar momentum equation (lower panels). Model {\sf rg.3D.lr} (left) and {\sf rg.3D.mr} (right) \label{fig:rg-res-rmomentum-tmomentum-equation}}
\end{figure}

\newpage

\subsubsection{Mean internal and kinetic energy equation}

\begin{figure}[!h]
\centerline{
\includegraphics[width=7.cm]{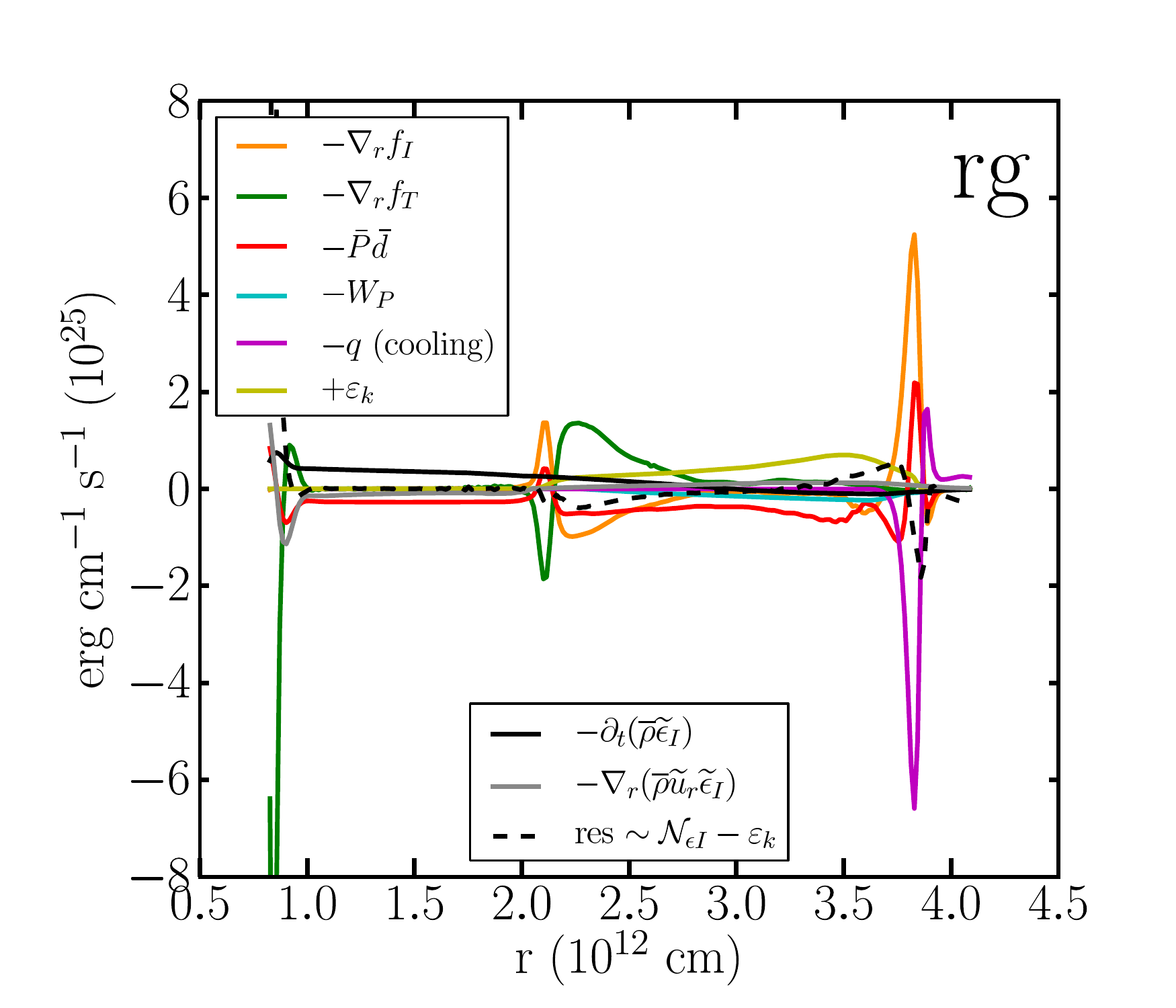}
\includegraphics[width=7.cm]{rgmrez_tavg800_internal_energy_equation_insf-eps-converted-to.pdf}}

\centerline{
\includegraphics[width=7.cm]{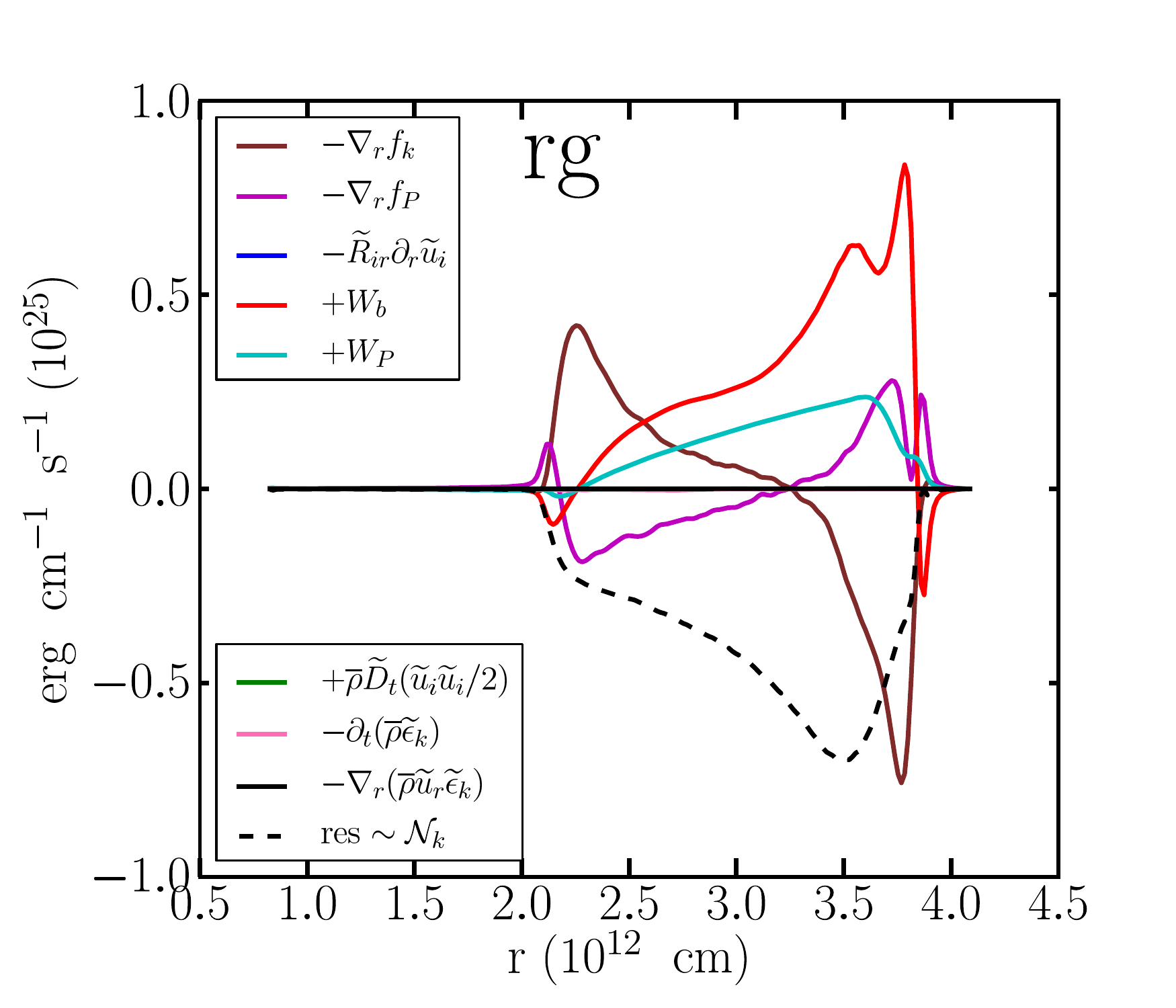}
\includegraphics[width=7.cm]{rgmrez_tavg800_kinetic_energy_equation_insf-eps-converted-to.pdf}}
\caption{Mean internal energy equation (upper panels) and kinetic energy equation (lower panels). Model {\sf rg.3D.lr} (left) and {\sf rg.3D.mr} (right) \label{fig:rg-res-ei-ek-equation}}
\end{figure}

\newpage

\subsubsection{Mean total energy equation and mean entropy equation}

\begin{figure}[!h]
\centerline{
\includegraphics[width=7.cm]{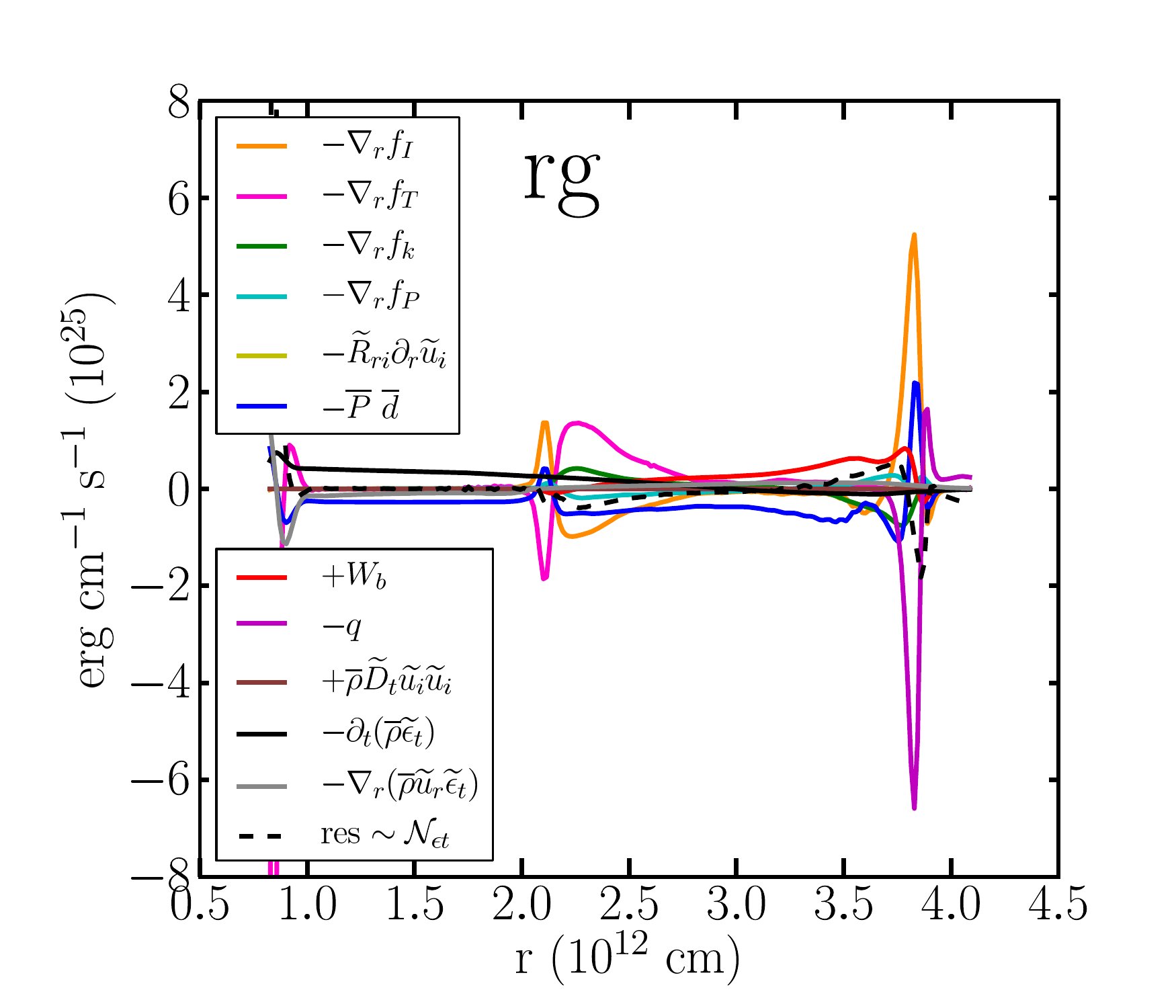}
\includegraphics[width=7.cm]{rgmrez_tavg800_total_energy_equation_insf-eps-converted-to.pdf}}

\centerline{
\includegraphics[width=7.cm]{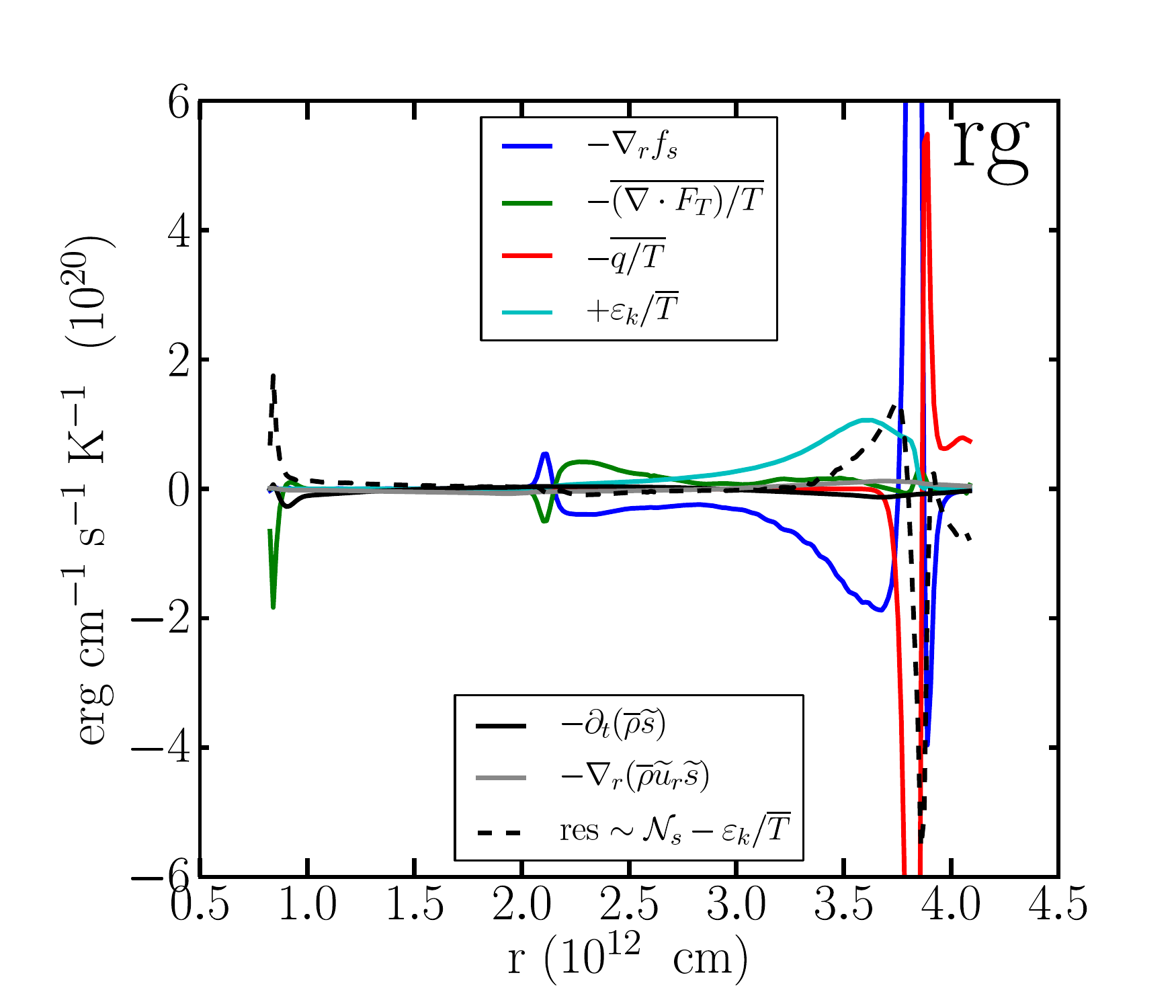}
\includegraphics[width=7.cm]{rgmrez_tavg800_entropy_equation_insf-eps-converted-to.pdf}}
\caption{Mean total energy equation (upper panels) and mean entropy equation (lower panels). Model {\sf rg.3D.lr} (left) and {\sf rg.3D.mr} (right) \label{fig:rg-res-et-ss-equation}}
\end{figure}

\newpage

\subsubsection{Mean density-specific volume covariance and entropy flux equation}

\begin{figure}[!h]
\centerline{
\includegraphics[width=7.cm]{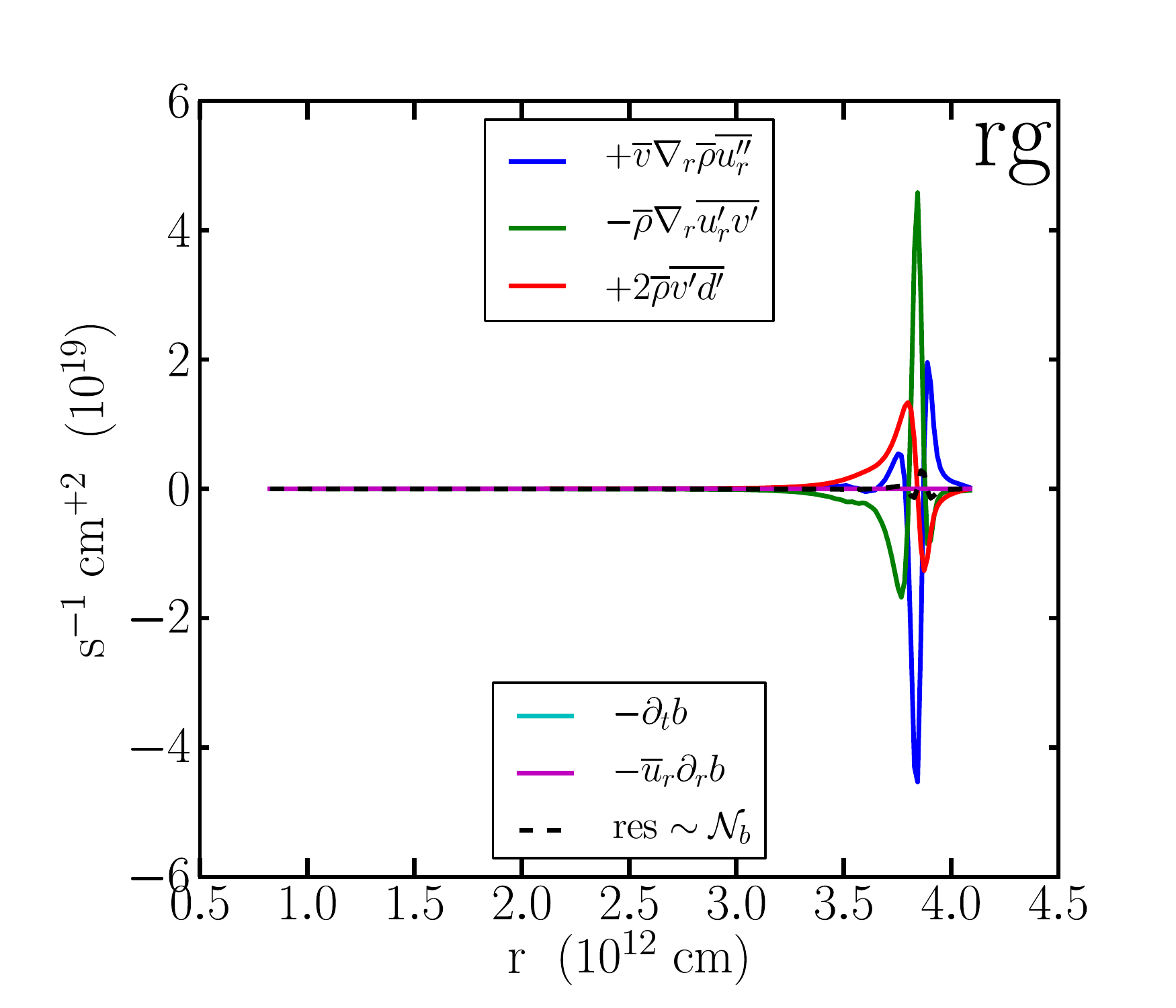}
\includegraphics[width=7.cm]{rgmrez_tavg800_mfields_b_equation_insf-eps-converted-to.pdf}}

\centerline{
\includegraphics[width=7.cm]{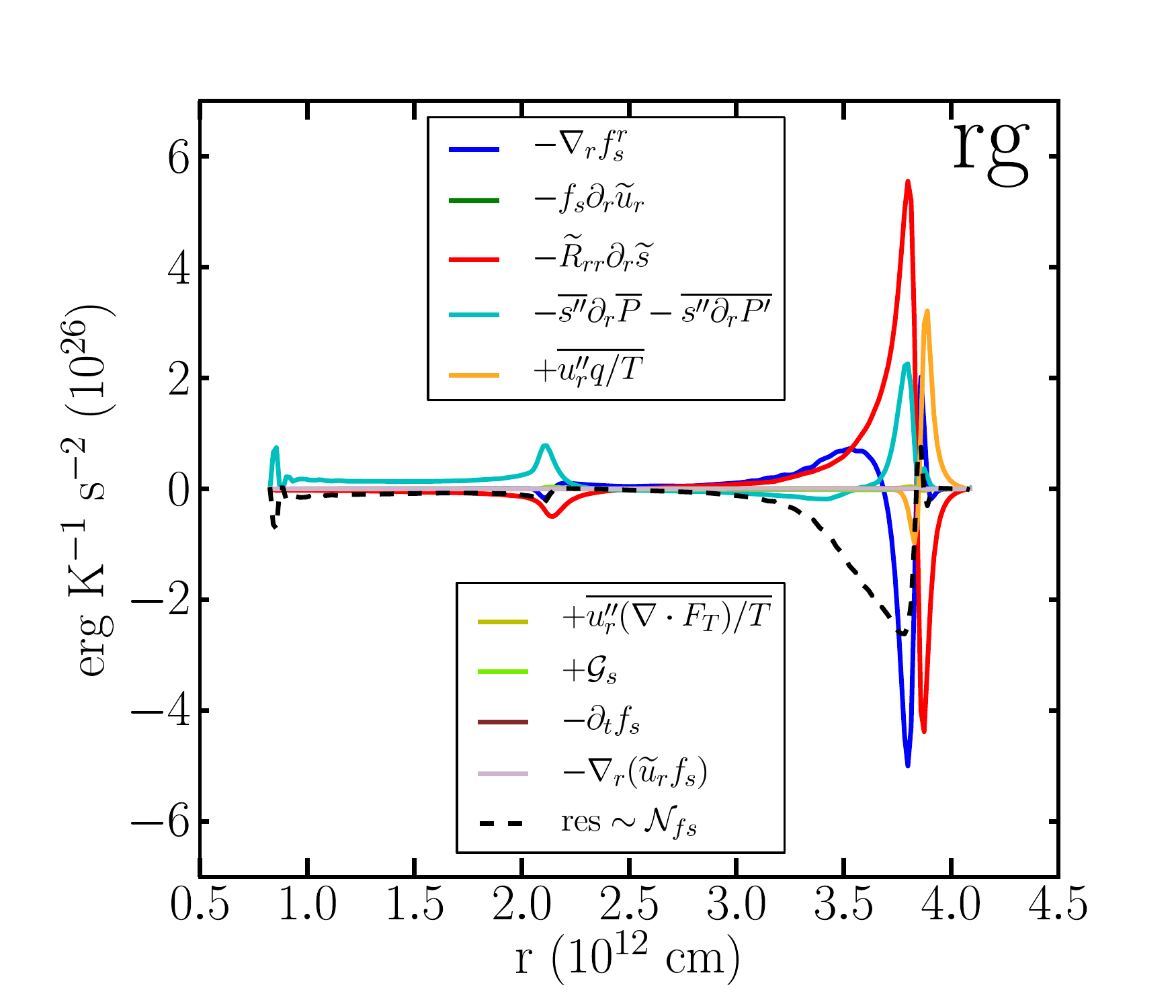}
\includegraphics[width=7.cm]{rgmrez_tavg800_mfields_s_equation_insf-eps-converted-to.pdf}}
\caption{Mean density-specific volume covariance equation (upper panels) and entropy flux equation (lower panels). Model {\sf rg.3D.lr} (left) and {\sf rg.3D.mr} (right) \label{fig:rg-res-ss-fssx-equation}}
\end{figure}

\newpage

\subsubsection{Mean turbulent kinetic energy equation and mean turbulent mass flux equation}

\begin{figure}[!h]
\centerline{
\includegraphics[width=7.cm]{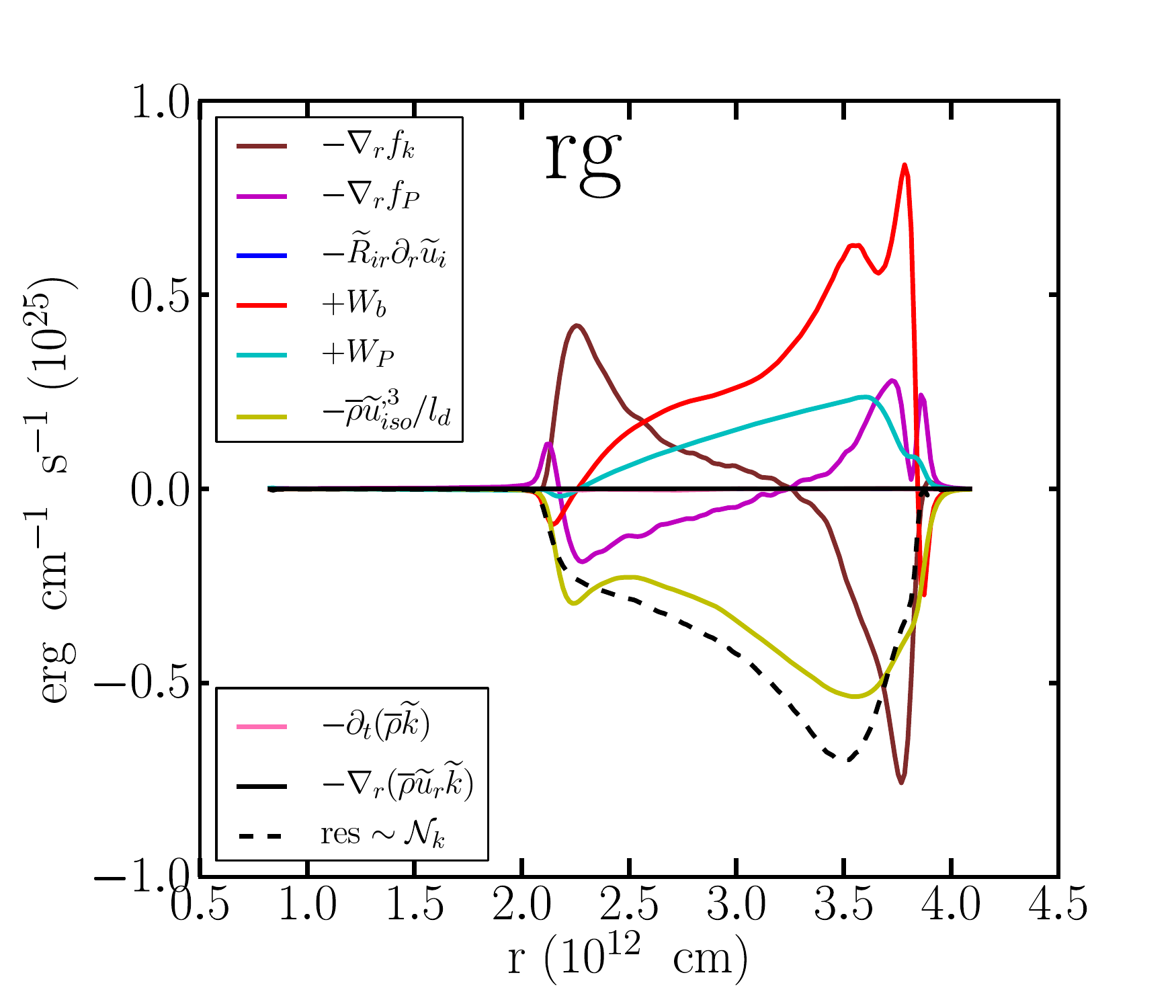}
\includegraphics[width=7.cm]{rgmrez_tavg800_mfields_k_equation_insf-eps-converted-to.pdf}}

\centerline{
\includegraphics[width=7.cm]{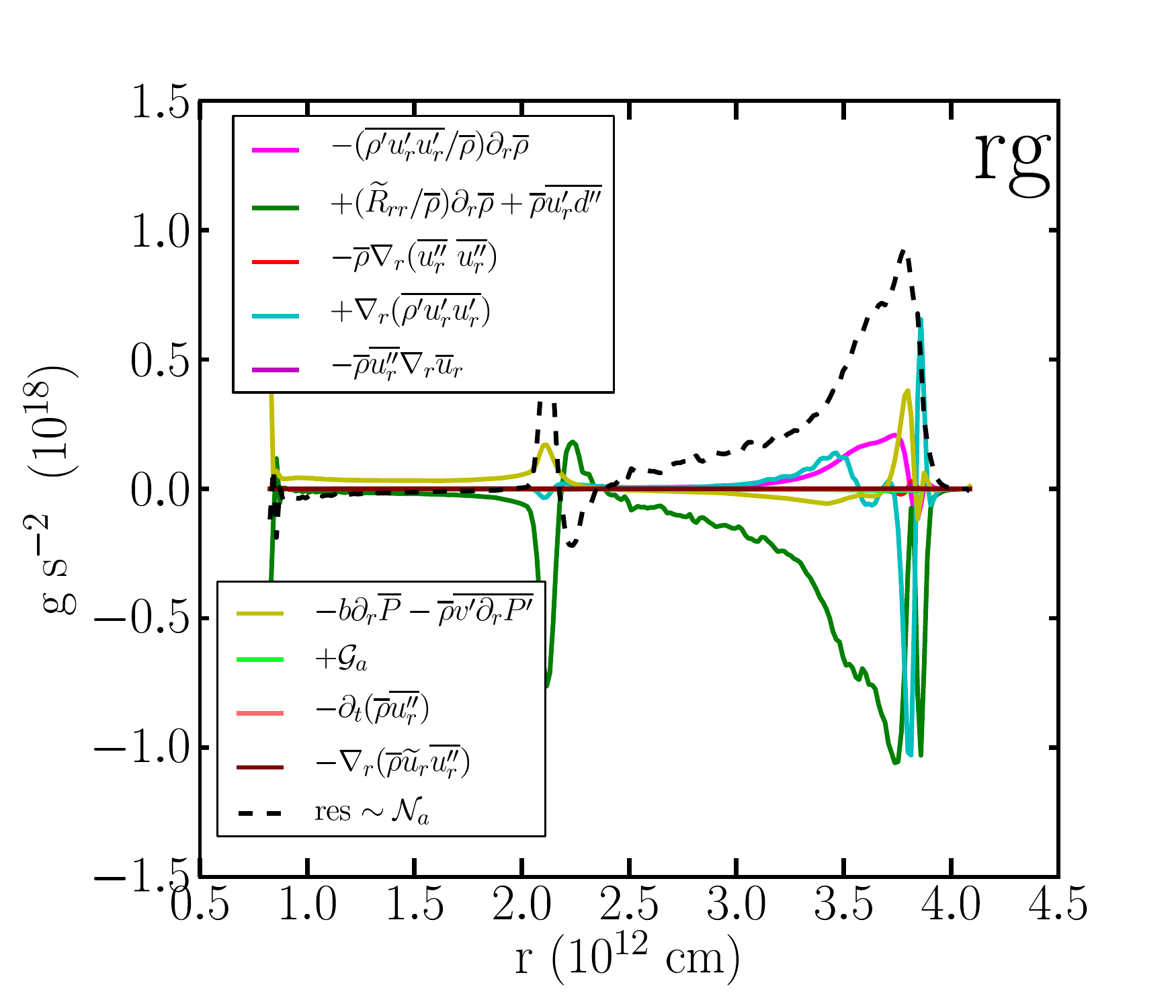}
\includegraphics[width=7.cm]{rgmrez_tavg800_mfields_a_equation_insf-eps-converted-to.pdf}}

\caption{Mean turbulent kinetic energy equation (upper panels) and mean turbulent mass flux equation (lower panels). Model {\sf rg.3D.lr} (left) and {\sf rg.3D.mr} (right) \label{fig:rg-res-k-vel-equation}}
\end{figure}

\newpage

\section{Dependence on Computational Domain Size}

\subsection{Oxygen burning shell models}

\subsubsection{Mean continuity equation and mean radial momentum equation}

\begin{figure}[!h]
\centerline{
\includegraphics[width=6.cm]{obmrez_tavg230_continuity_equation_ransdat-eps-converted-to.pdf}
\includegraphics[width=6.cm]{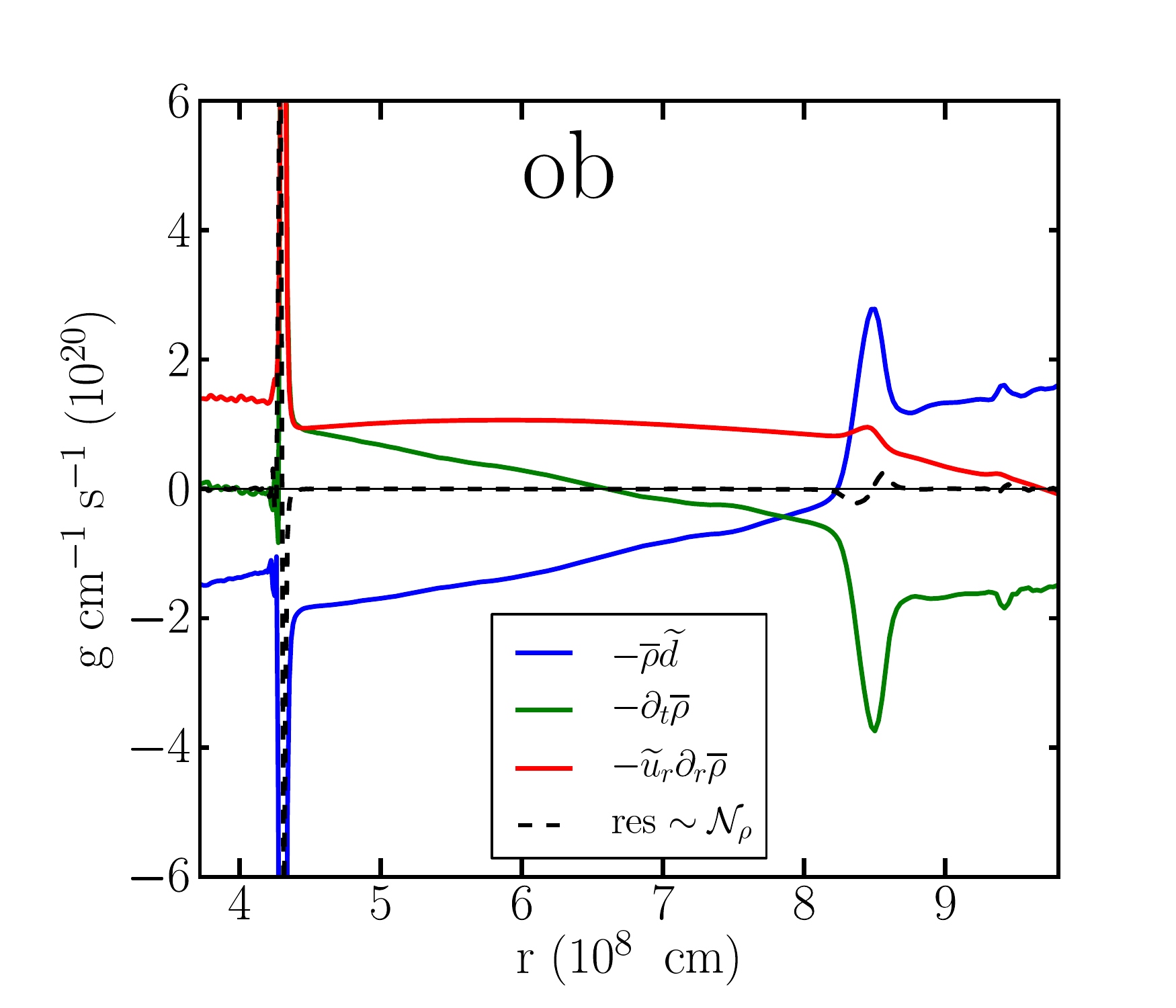}}

\centerline{
\includegraphics[width=6.cm]{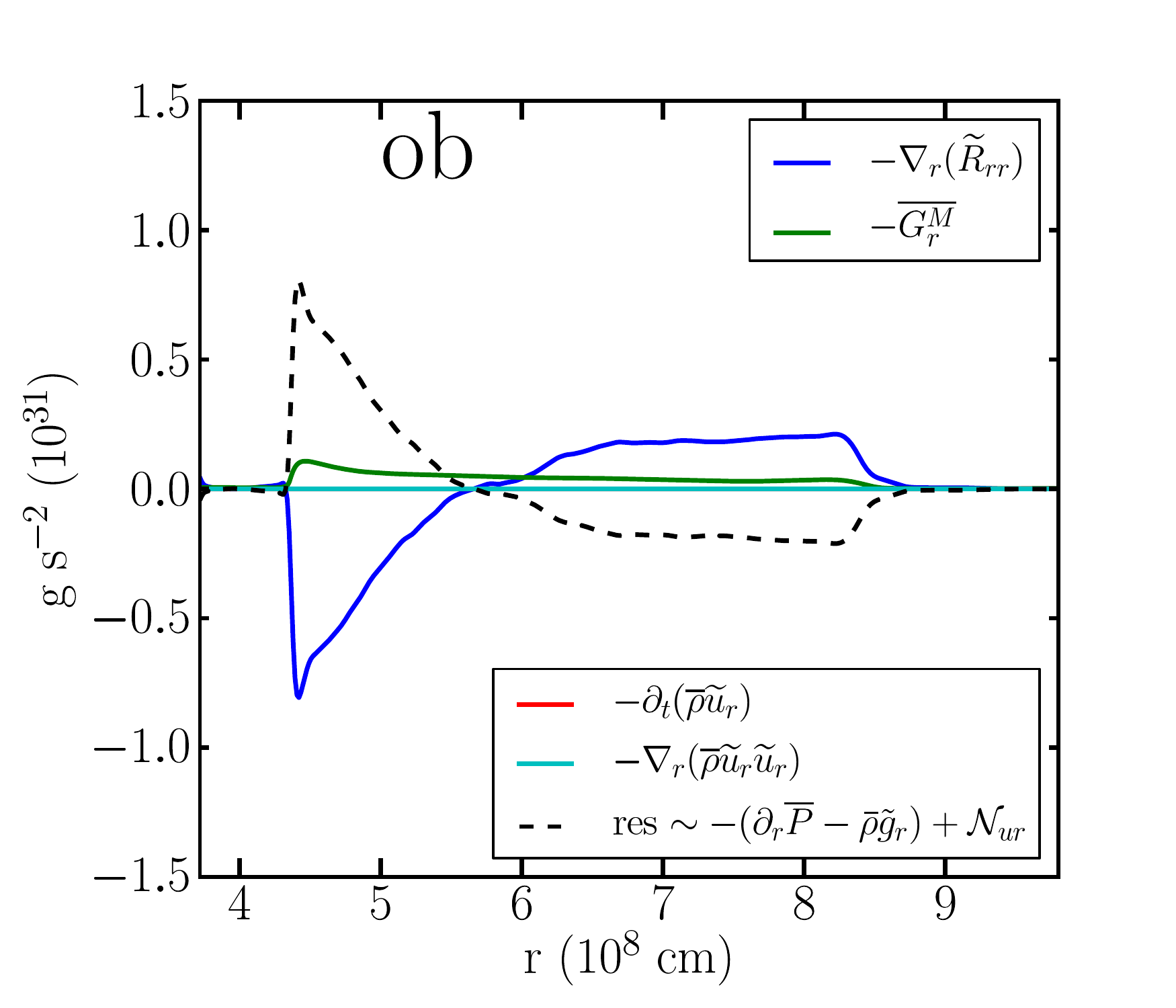}
\includegraphics[width=6.cm]{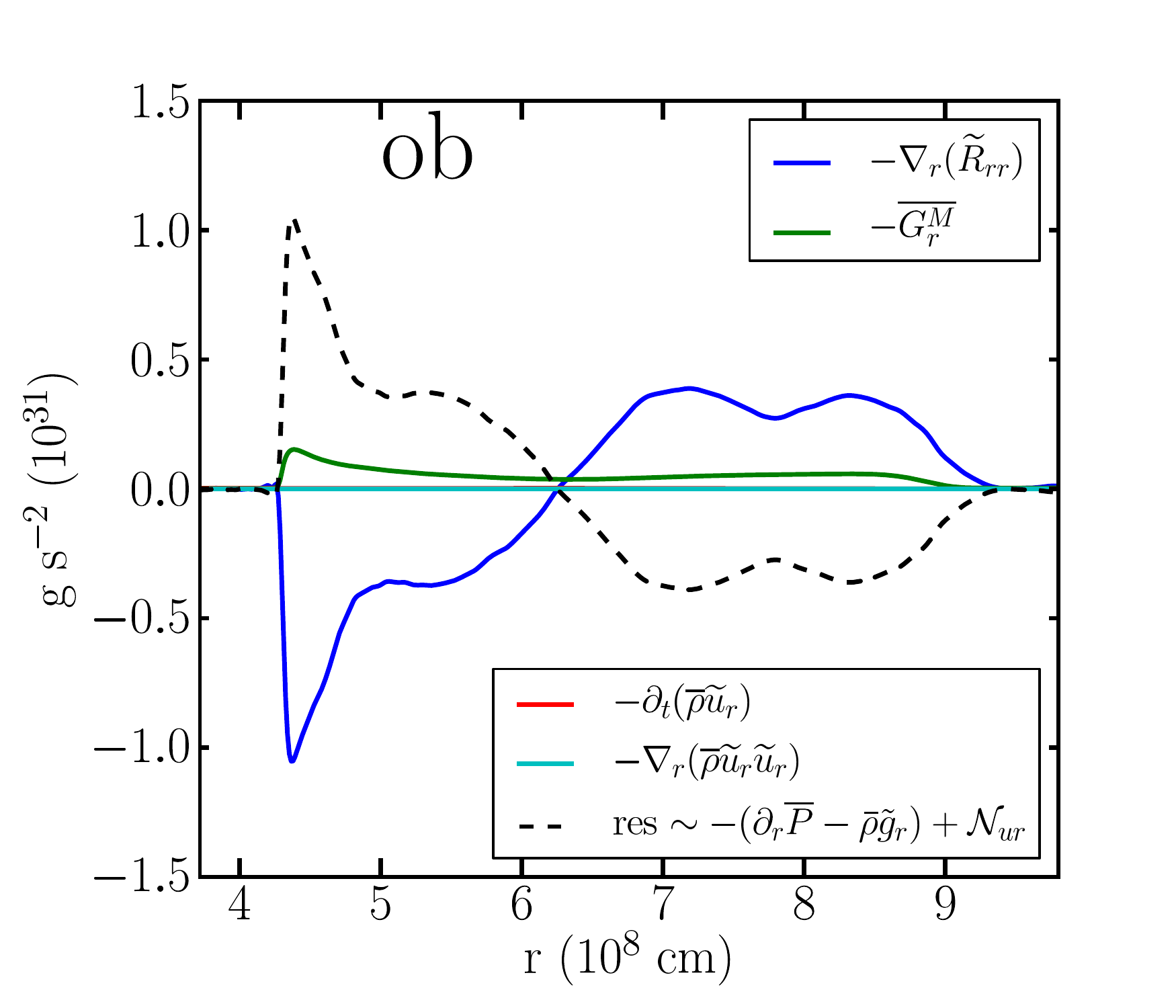}}
\caption{Continuity equation (upper panels) and radial momentum equation (lower panels). Model {\sf ob.3D.mr} (45\dgr wedge - left) and {\sf ob.3D.2hp} (27.5\dgr wedge - right). \label{fig:ob-wedge-effects-cont-rmomentum-eq}}
\end{figure}

\newpage

\subsubsection{Mean azimuthal momentum and polar momentum equation}

\begin{figure}[!h]
\centerline{
\includegraphics[width=7.cm]{obmrez_tavg230_tmomentum_equation_insf-eps-converted-to.pdf}
\includegraphics[width=7.cm]{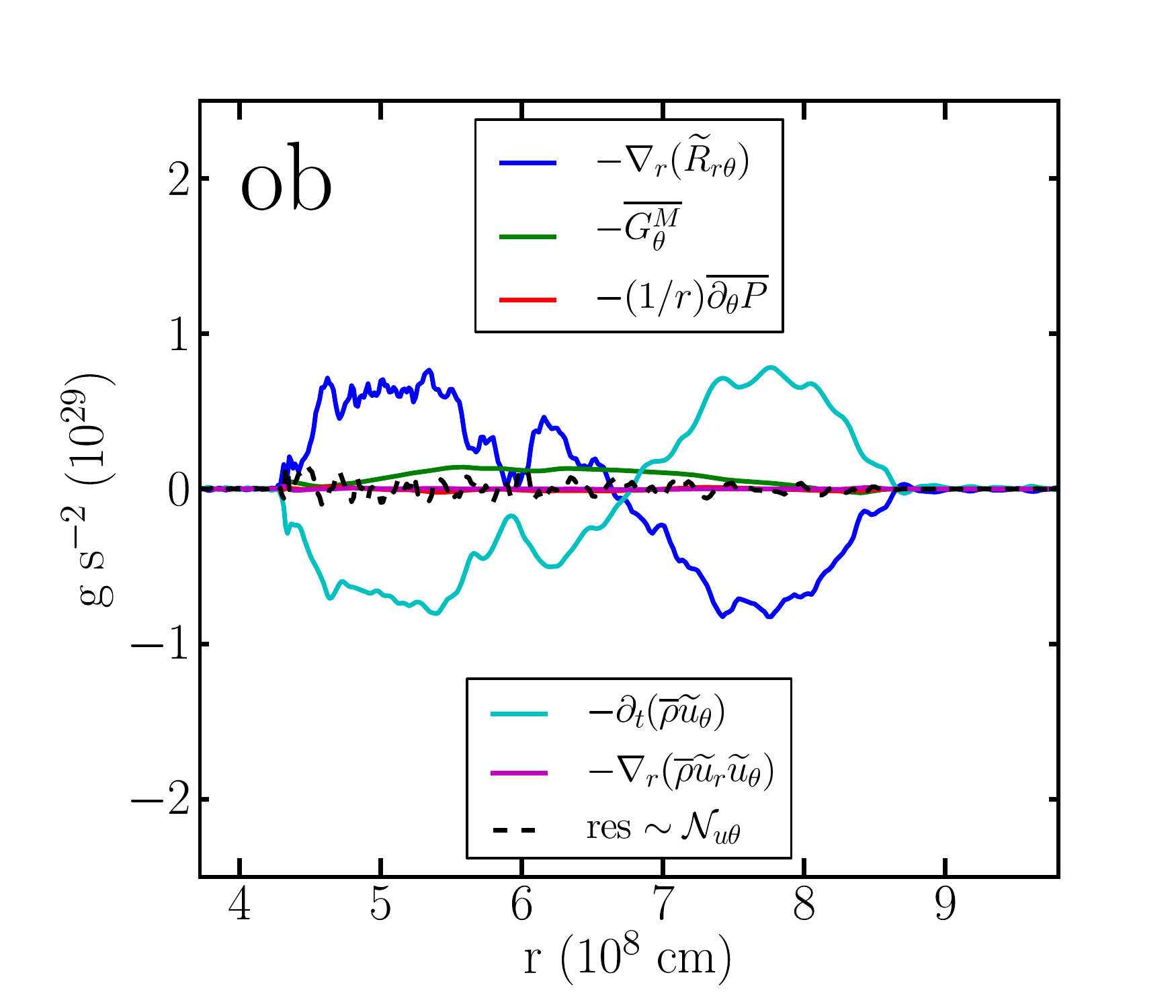}}

\centerline{
\includegraphics[width=7.cm]{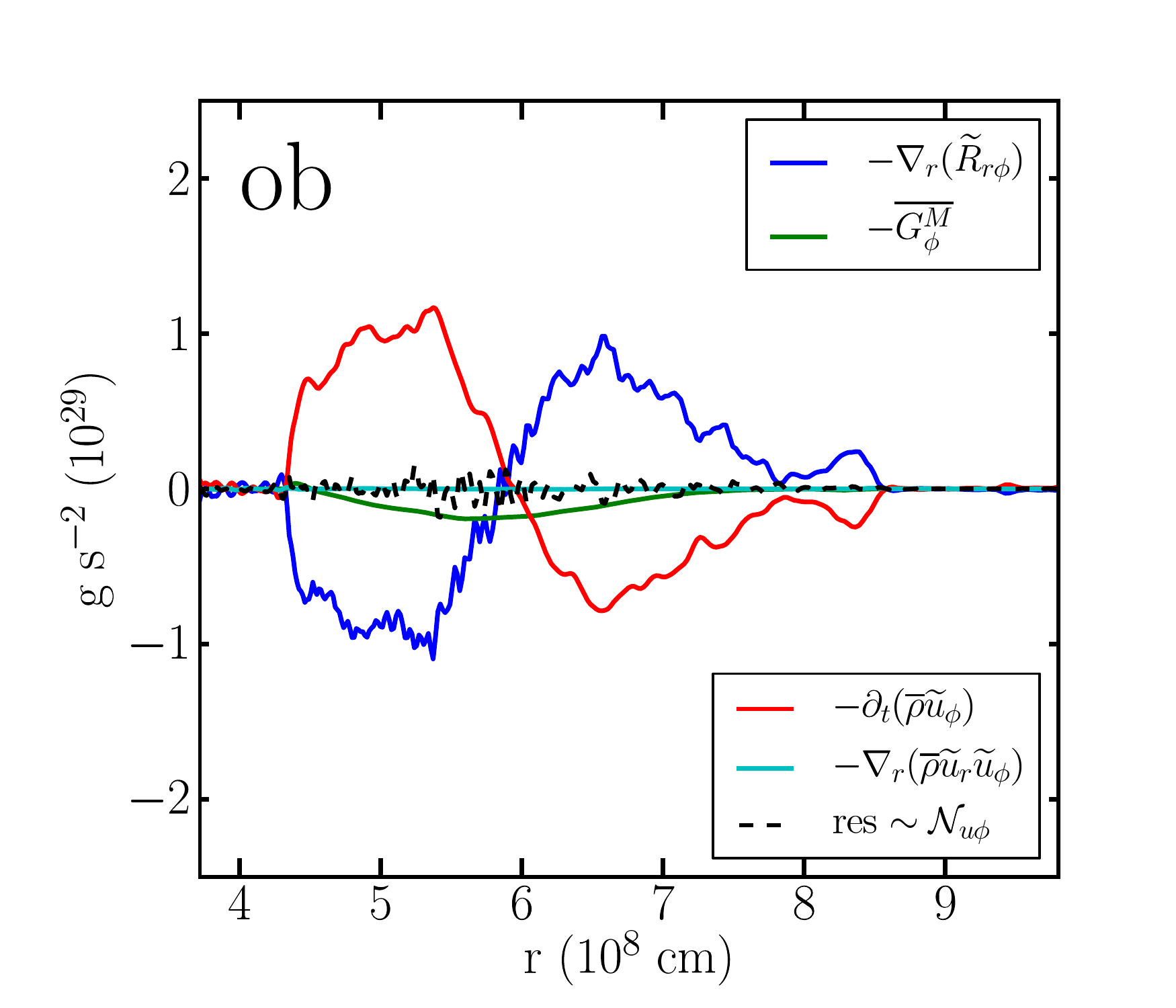}
\includegraphics[width=7.cm]{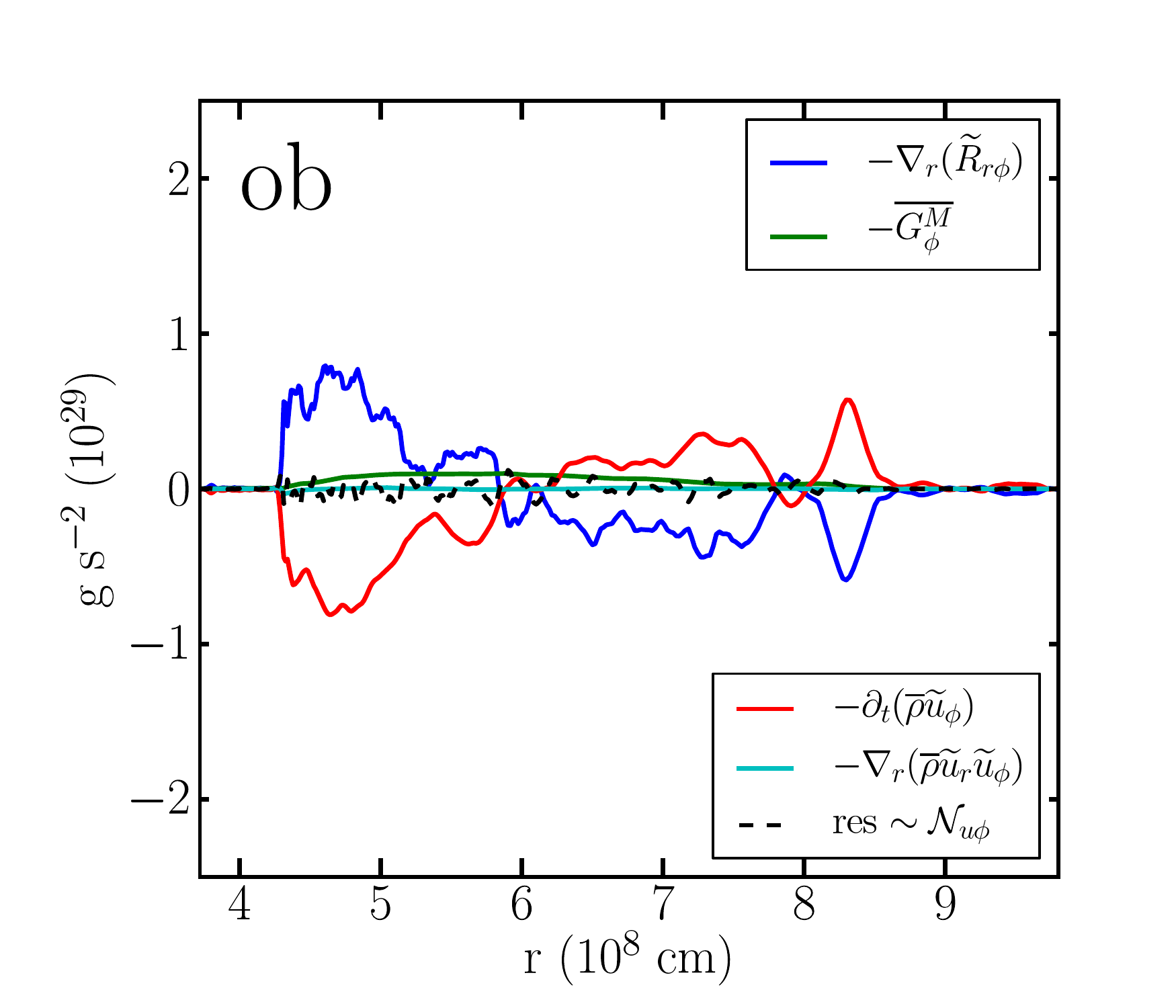}}
\caption{Mean azimuthal momentum (upper panels) and polar momentum equation (lower panels). Model {\sf ob.3D.mr} (45\dgr wedge - left) and {\sf ob.3D.2hp} (27.5\dgr wedge - right). \label{fig:ob-wedge-pmomentum-tmomentum-equation-eq}}
\end{figure}

\newpage

\subsubsection{Mean internal energy equation and total energy equation}

\begin{figure}[!h]
\centerline{
\includegraphics[width=7.cm]{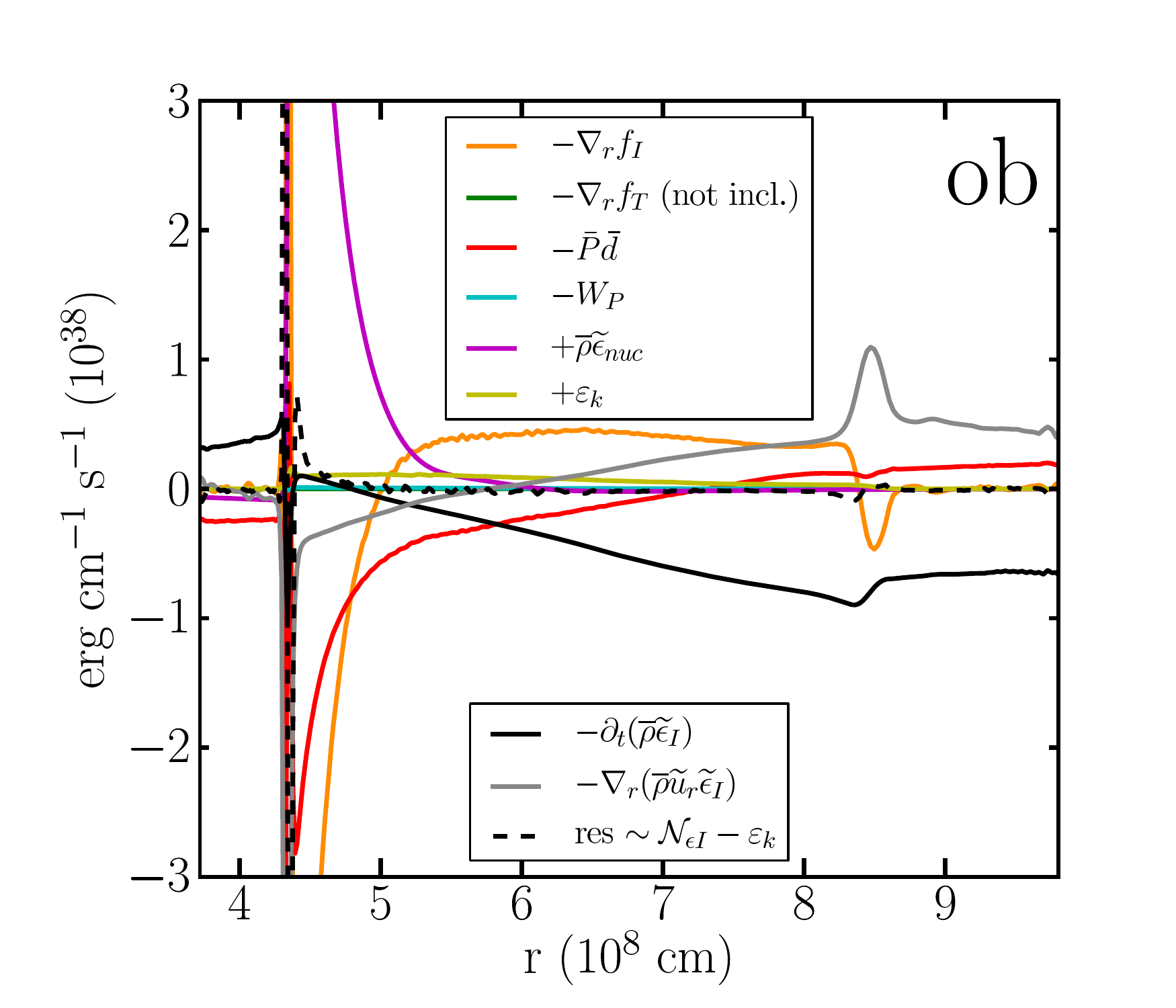}
\includegraphics[width=7.cm]{obmrez_tavg230_internal_energy_equation_insf-eps-converted-to.pdf}}

\centerline{
\includegraphics[width=7.cm]{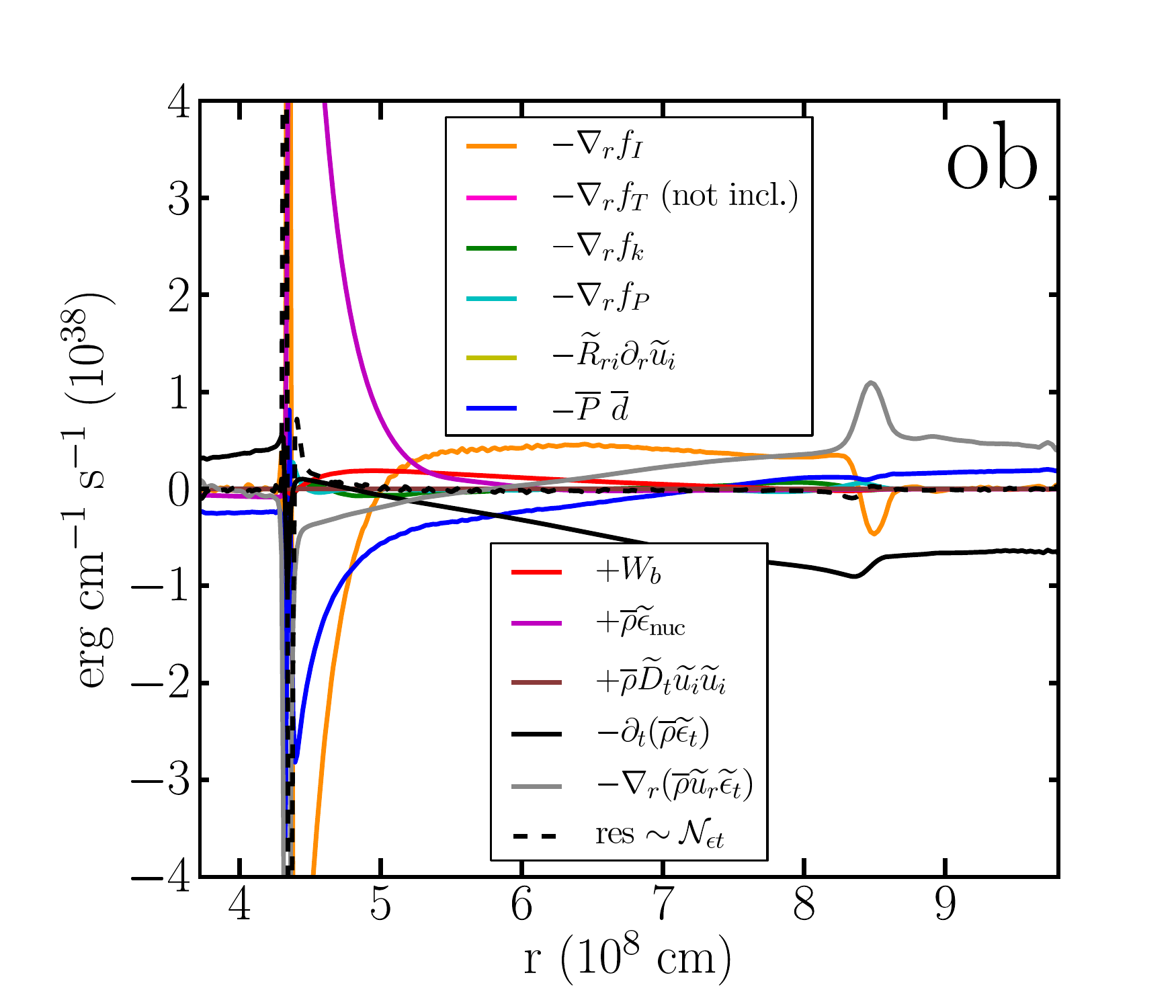}
\includegraphics[width=7.cm]{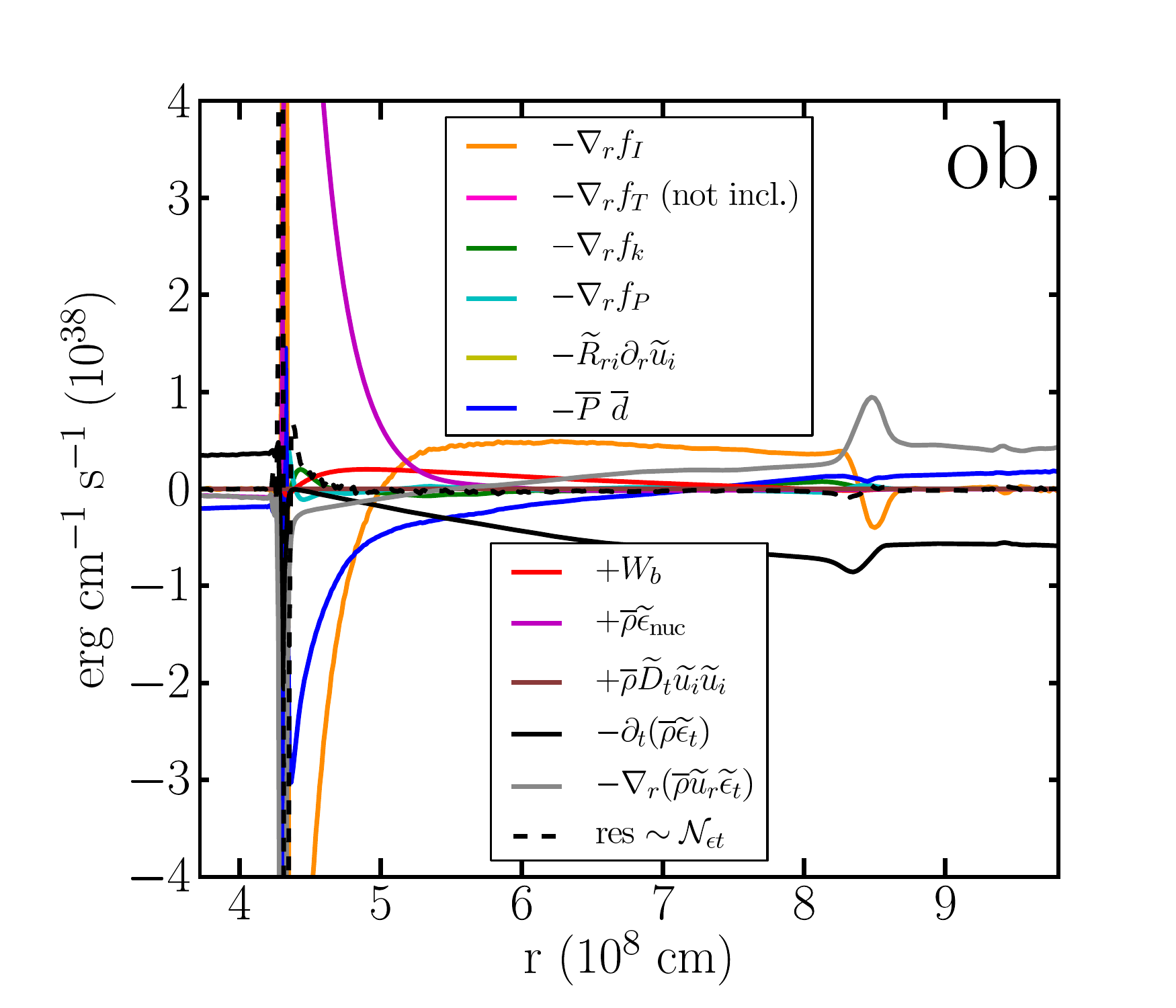}}
\caption{Mean internal energy (upper panels) equations and mean total energy equation (lower panels). Model {\sf ob.3D.mr} (45\dgr wedge - left) and {\sf ob.3D.2hp} (27.5\dgr wedge - right). \label{fig:obwedge-ei-et-eq}}
\end{figure}

\newpage

\subsubsection{Mean turbulent kinetic energy equation and turbulent mass flux equation}

\begin{figure}[!h]
\centerline{
\includegraphics[width=7.cm]{obmrez_tavg230_mfields_k_equation_insf-eps-converted-to.pdf}
\includegraphics[width=7.cm]{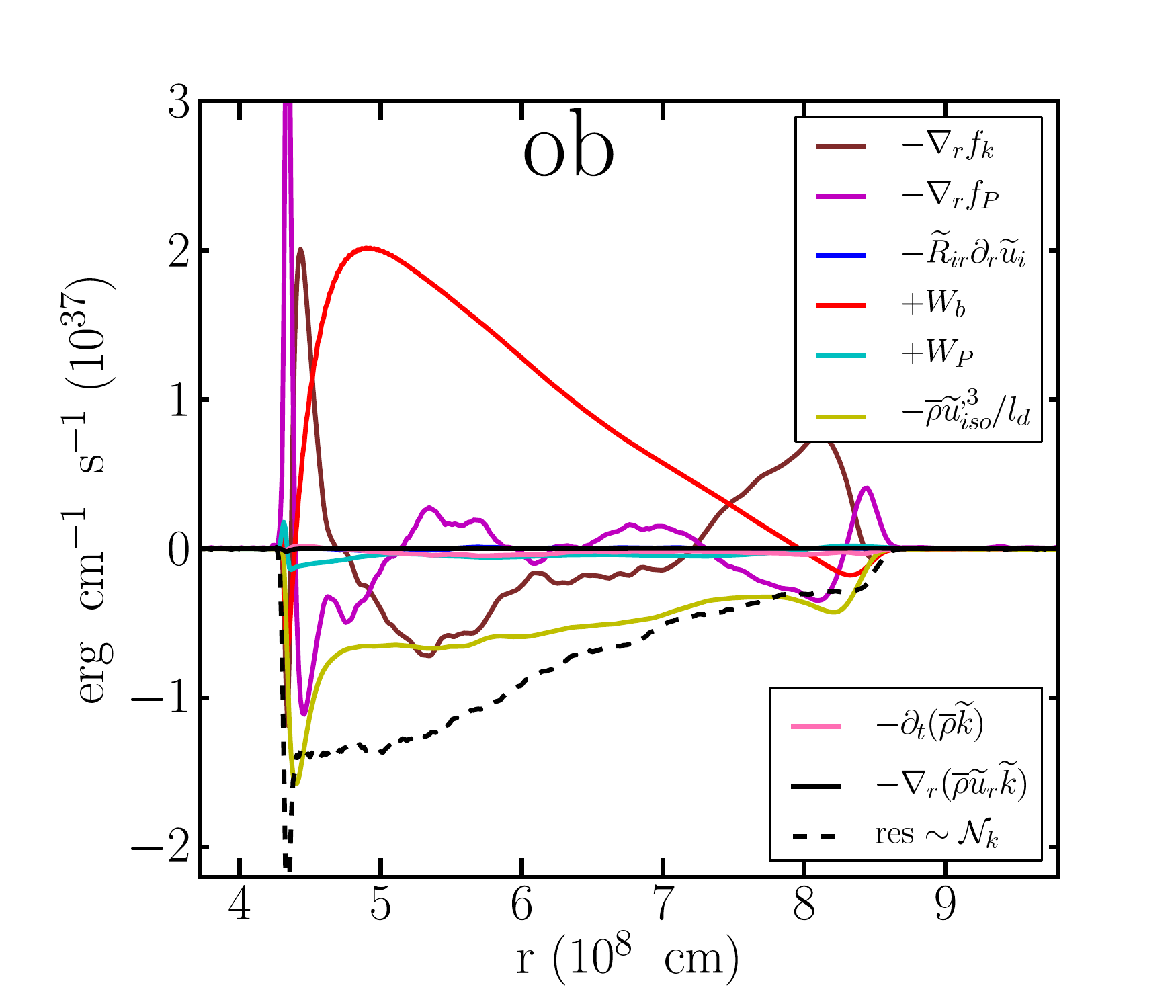}}

\centerline{
\includegraphics[width=7.cm]{obmrez_tavg230_mfields_a_equation_insf-eps-converted-to.pdf}
\includegraphics[width=7.cm]{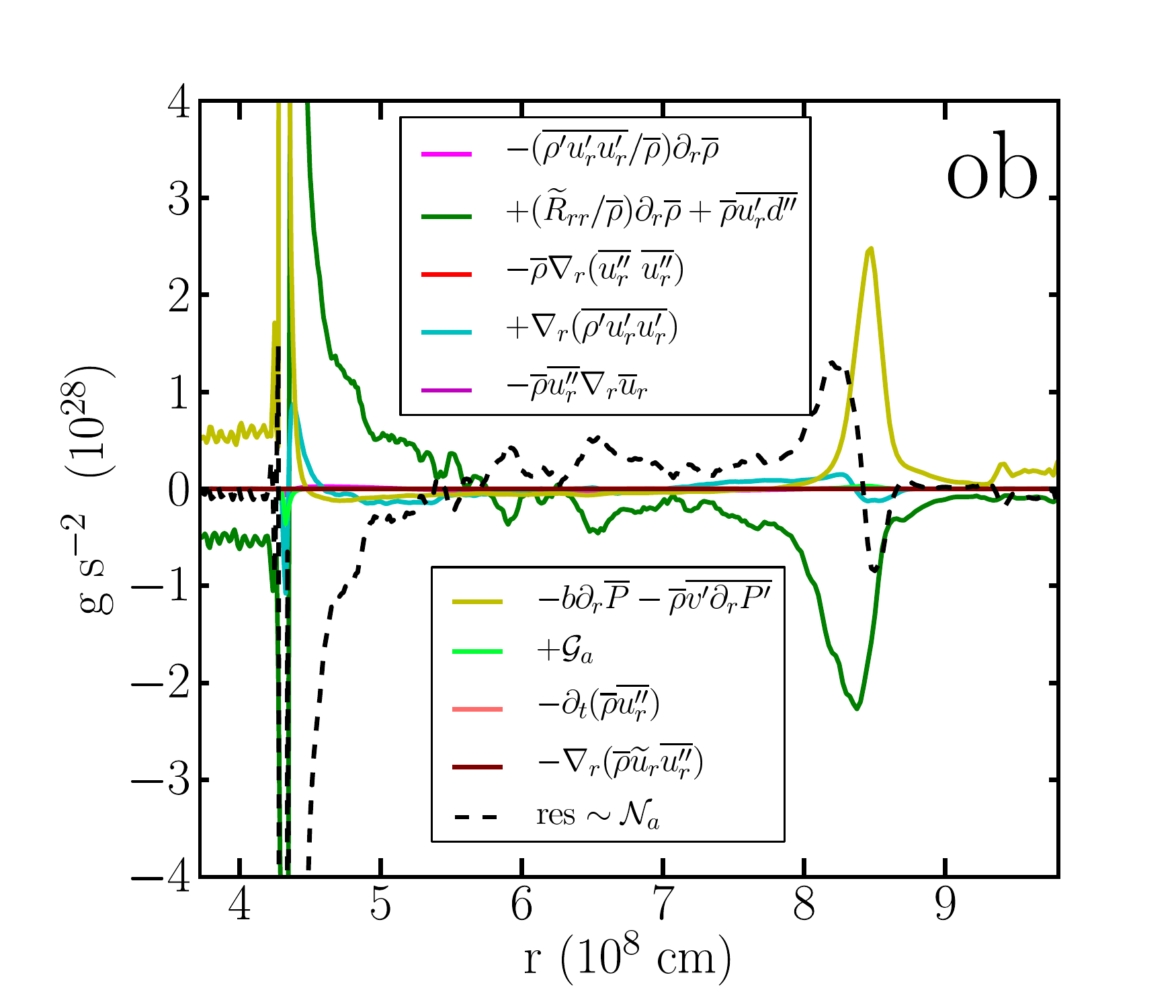}}
\caption{Mean turbulent kinetic energy equation and turbulent mass flux equation. Model {\sf ob.3D.mr} (45\dgr wedge - left) and {\sf ob.3D.2hp} (27.5\dgr wedge - right). \label{fig:ob-wedge-k-a-eq}}
\end{figure}

\newpage

\subsubsection{Mean density-specific volume covariance and internal energy flux equation}

\begin{figure}[!h]
\centerline{
\includegraphics[width=7.cm]{obmrez_tavg230_mfields_b_equation_insf-eps-converted-to.pdf}
\includegraphics[width=7.cm]{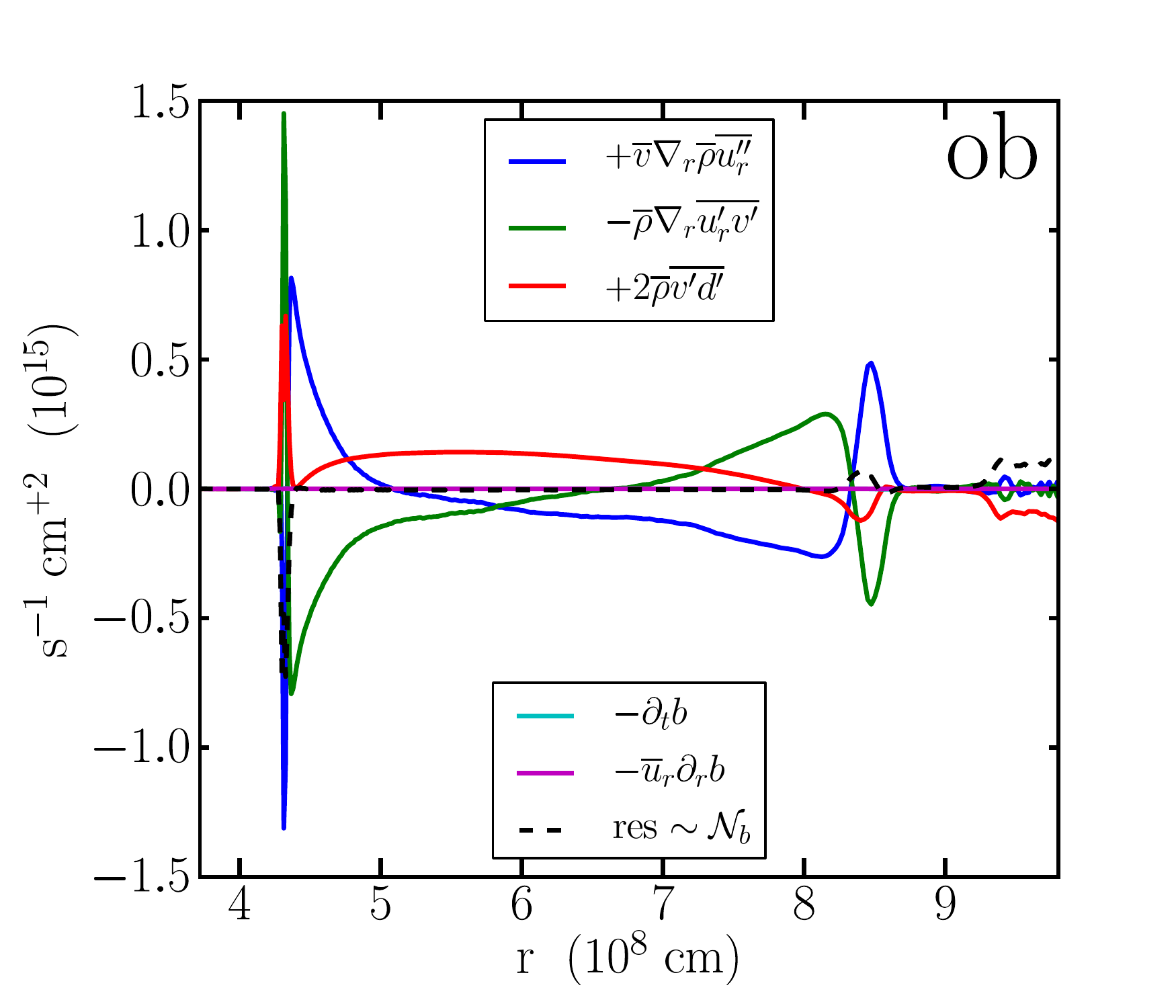}}

\centerline{
\includegraphics[width=7.cm]{obmrez_tavg230_mfields_i_equation_insf-eps-converted-to.pdf}
\includegraphics[width=7.cm]{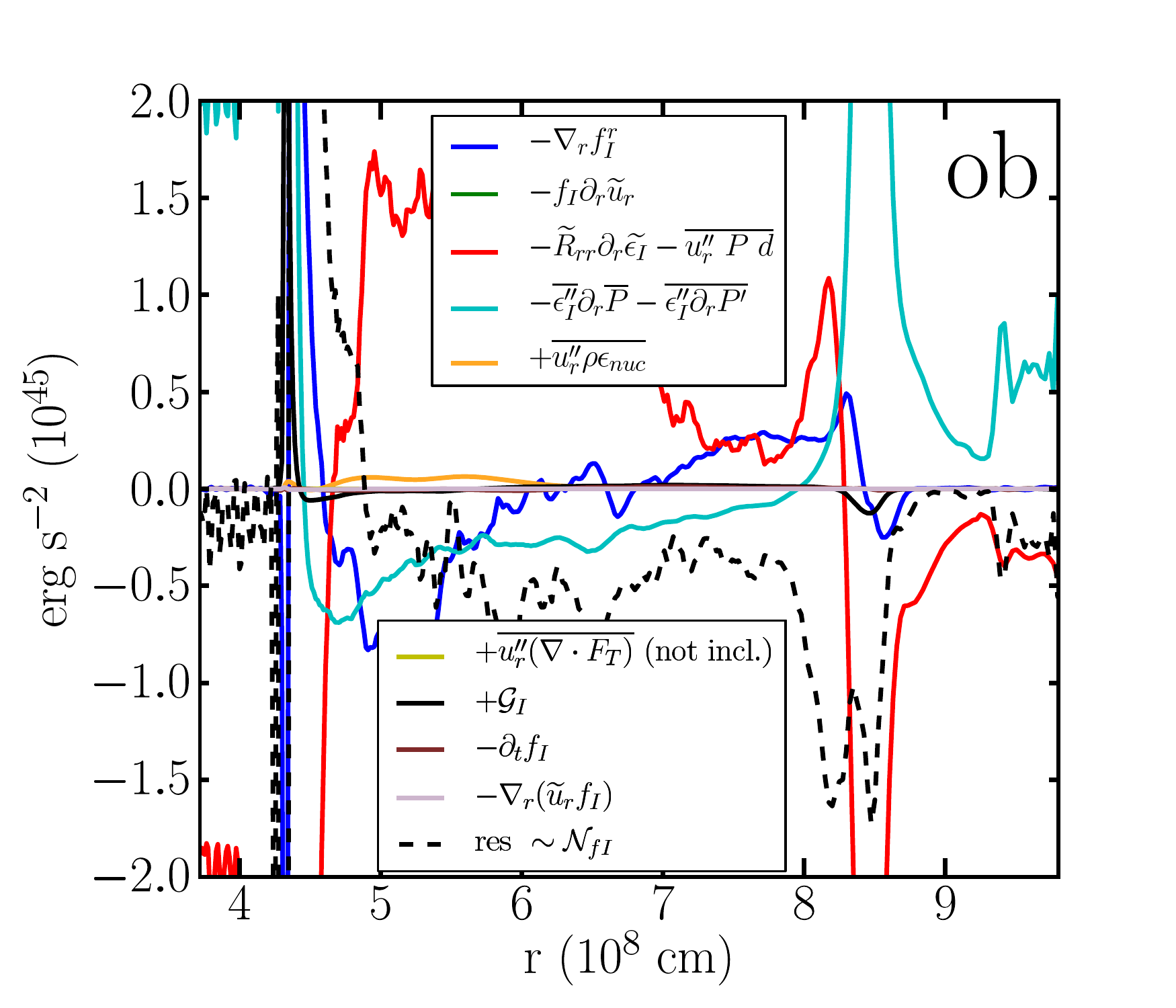}}
\caption{Mean density-specific volume covariance equation (upper panels) and mean internal energy flux equation (lower panels). Model {\sf ob.3D.mr} (45\dgr wedge - left) and {\sf ob.3D.2hp} (27.5\dgr wedge - right). \label{fig:ob-wedge-b-i-eq}}
\end{figure}

\newpage

\subsubsection{Mean kinetic energy equation and mean velocities}

\begin{figure}[!h]
\centerline{
\includegraphics[width=7.cm]{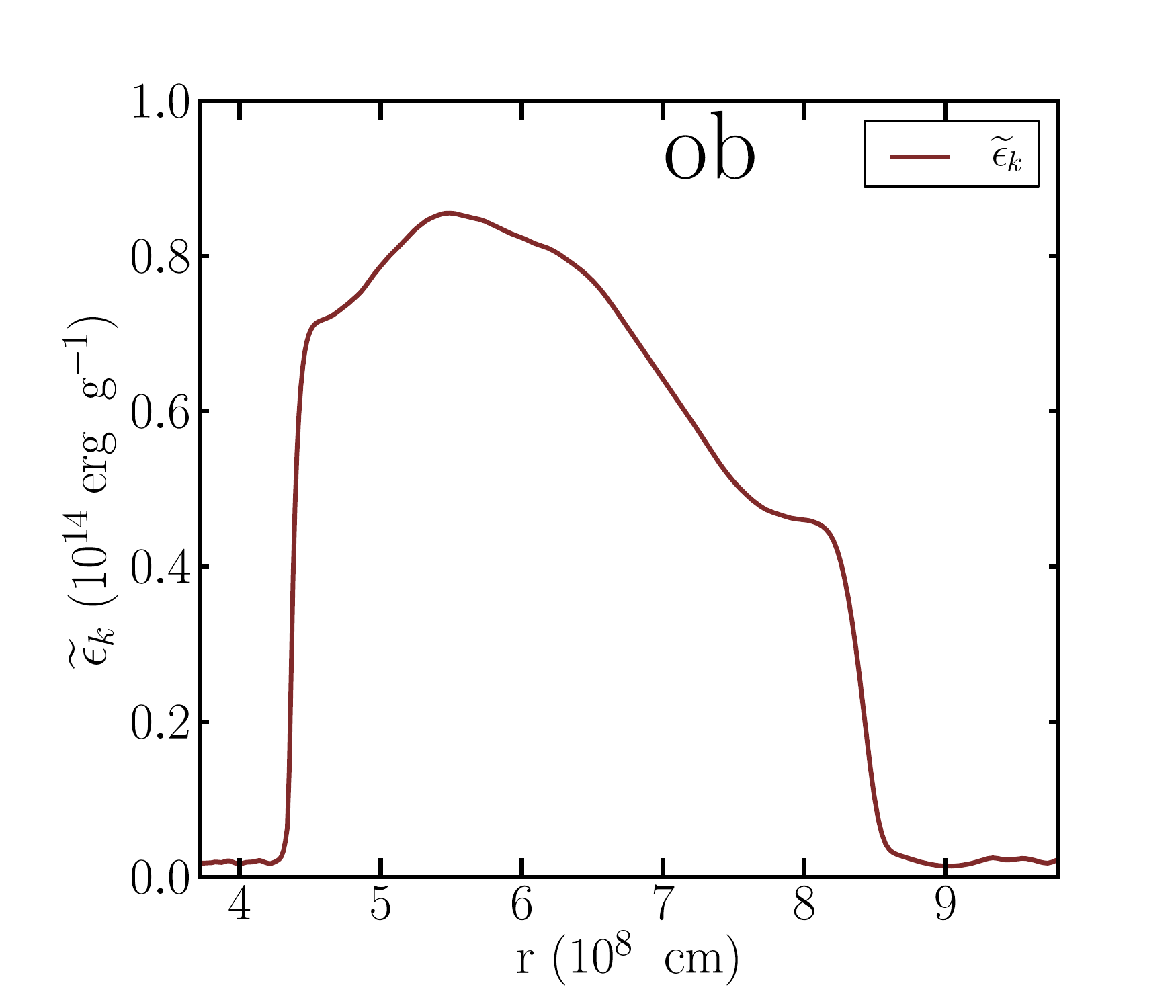}
\includegraphics[width=7.cm]{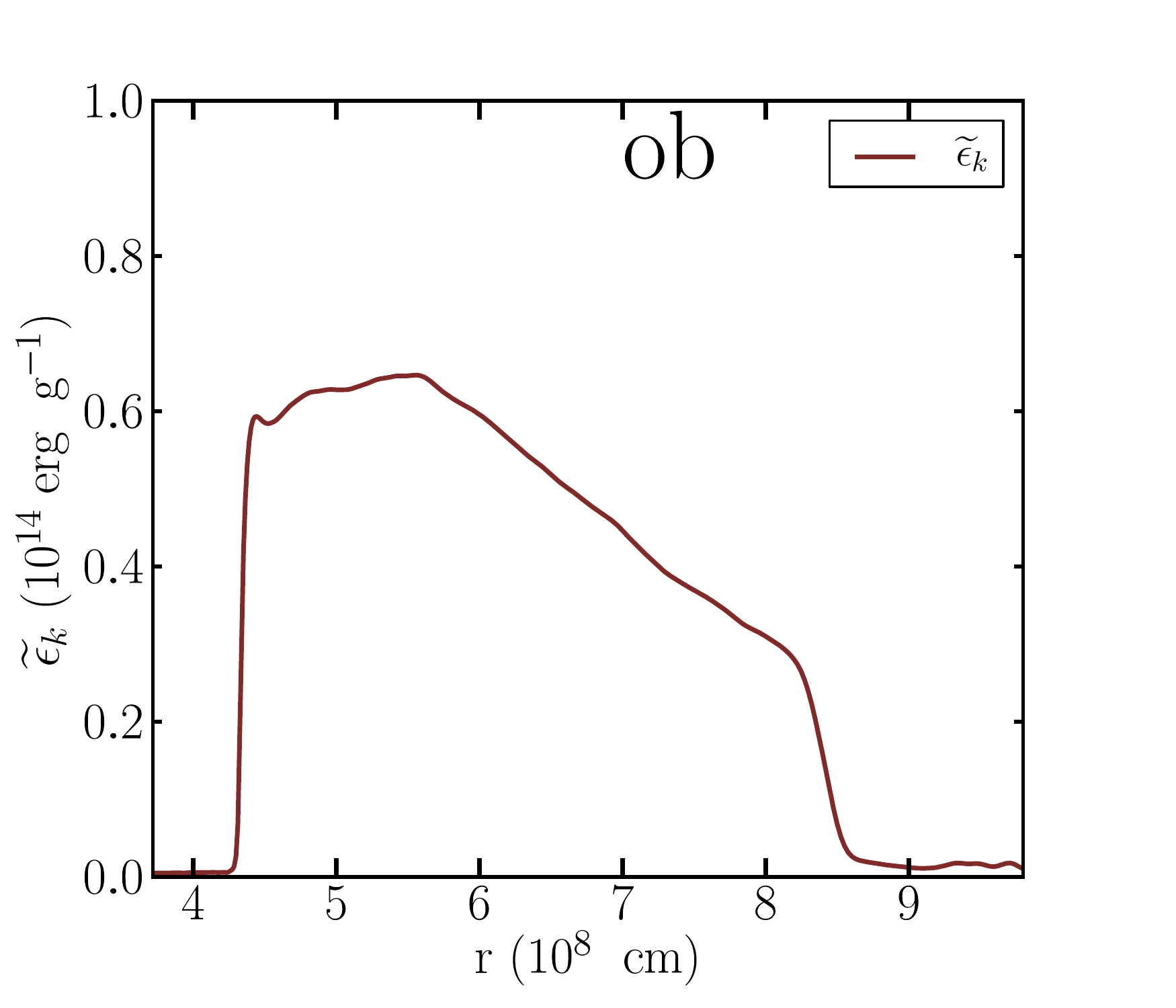}}

\centerline{
\includegraphics[width=7.cm]{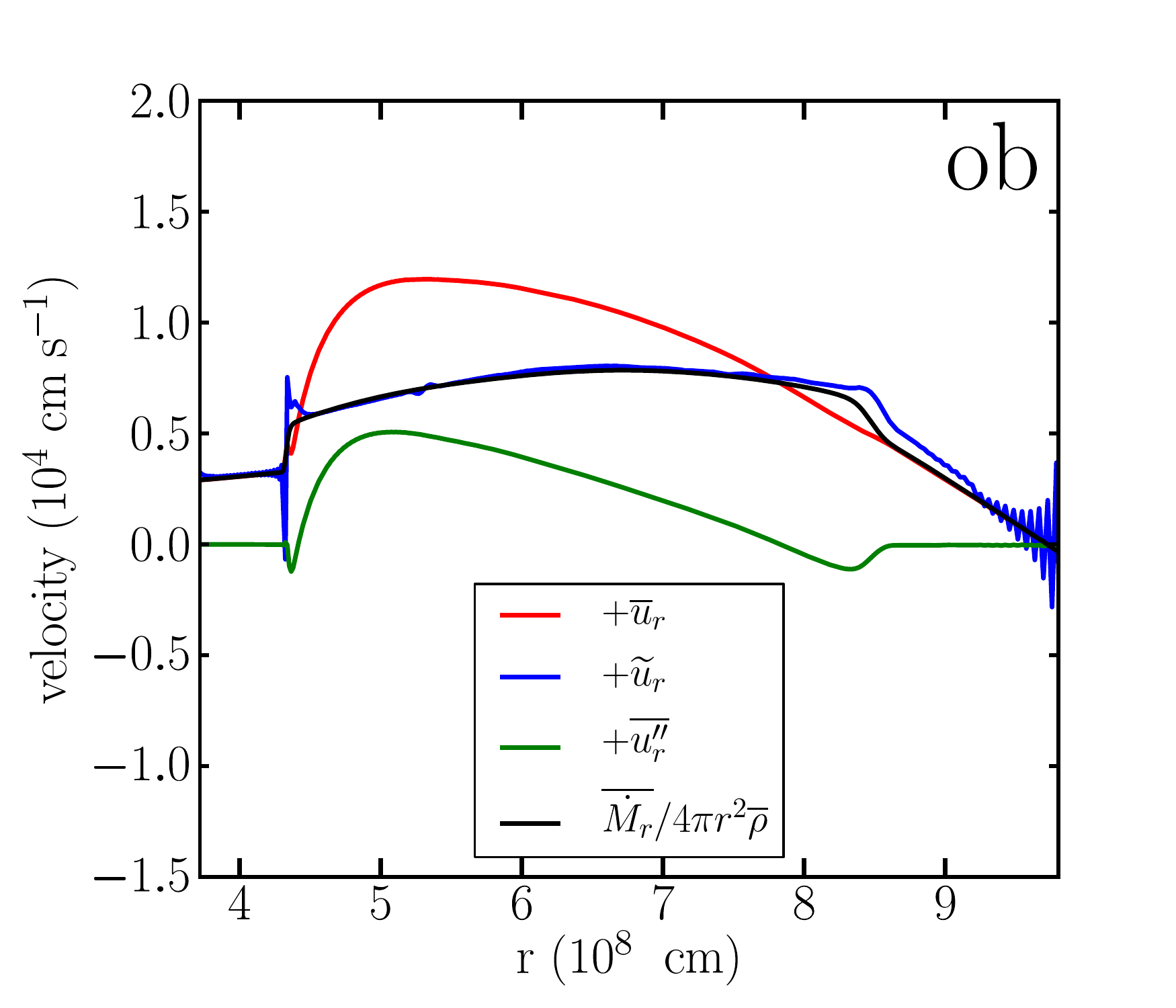}
\includegraphics[width=7.cm]{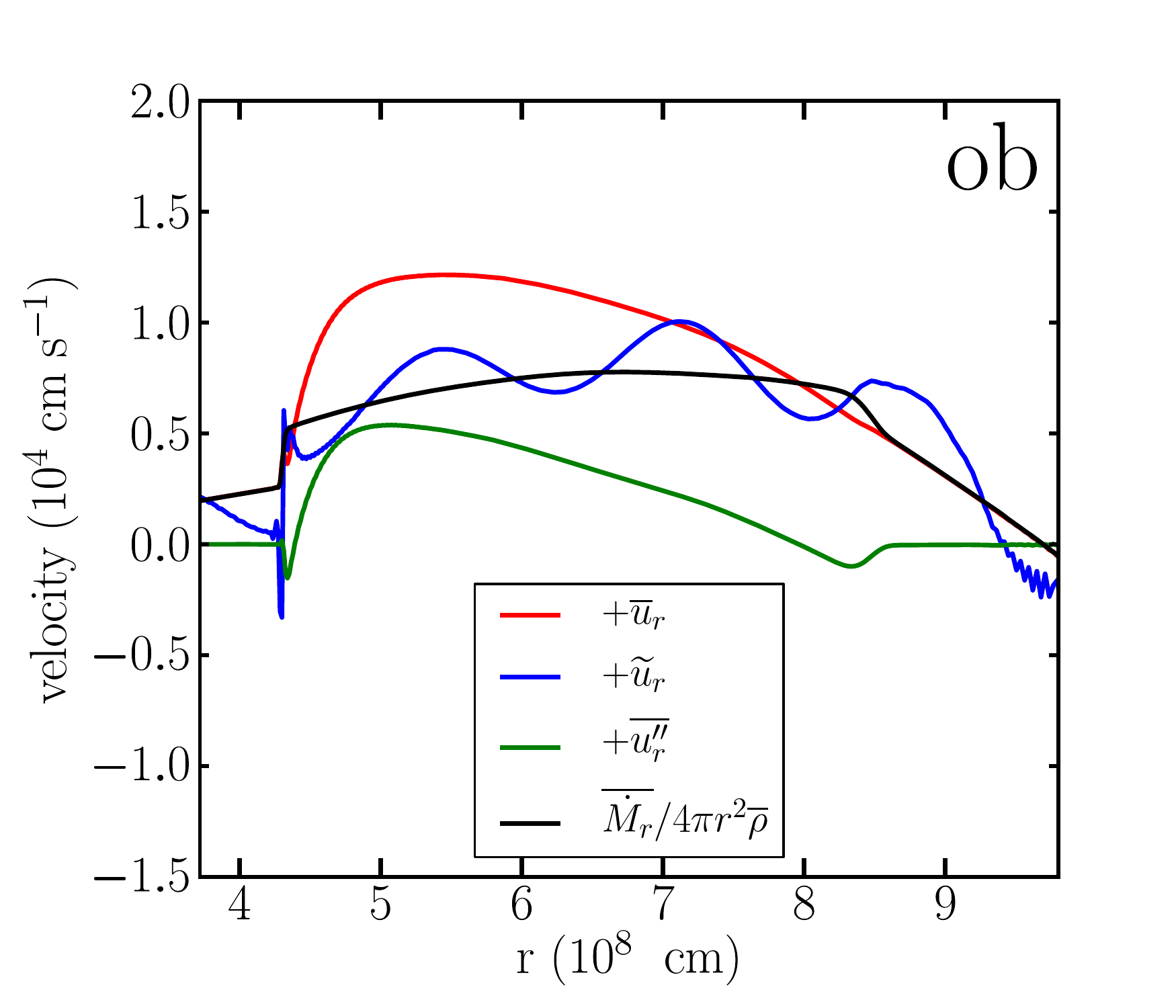}}
\caption{Mean kinetic energy equation (upper panels) mean velocities (lower panels). Model {\sf ob.3D.mr} (45\dgr wedge - left) and {\sf ob.3D.2hp} (27.5\dgr wedge - right).  \label{fig:ob-wedge-ek-vel}}
\end{figure}

\newpage

\section{Dependence on Time Averaging Window}

\par 

\subsection{Oxygen burning shell model}

\subsubsection{Mean continuity equation and mean radial momentum equation}

\begin{figure}[!h]
\centerline{
\includegraphics[width=5.6cm]{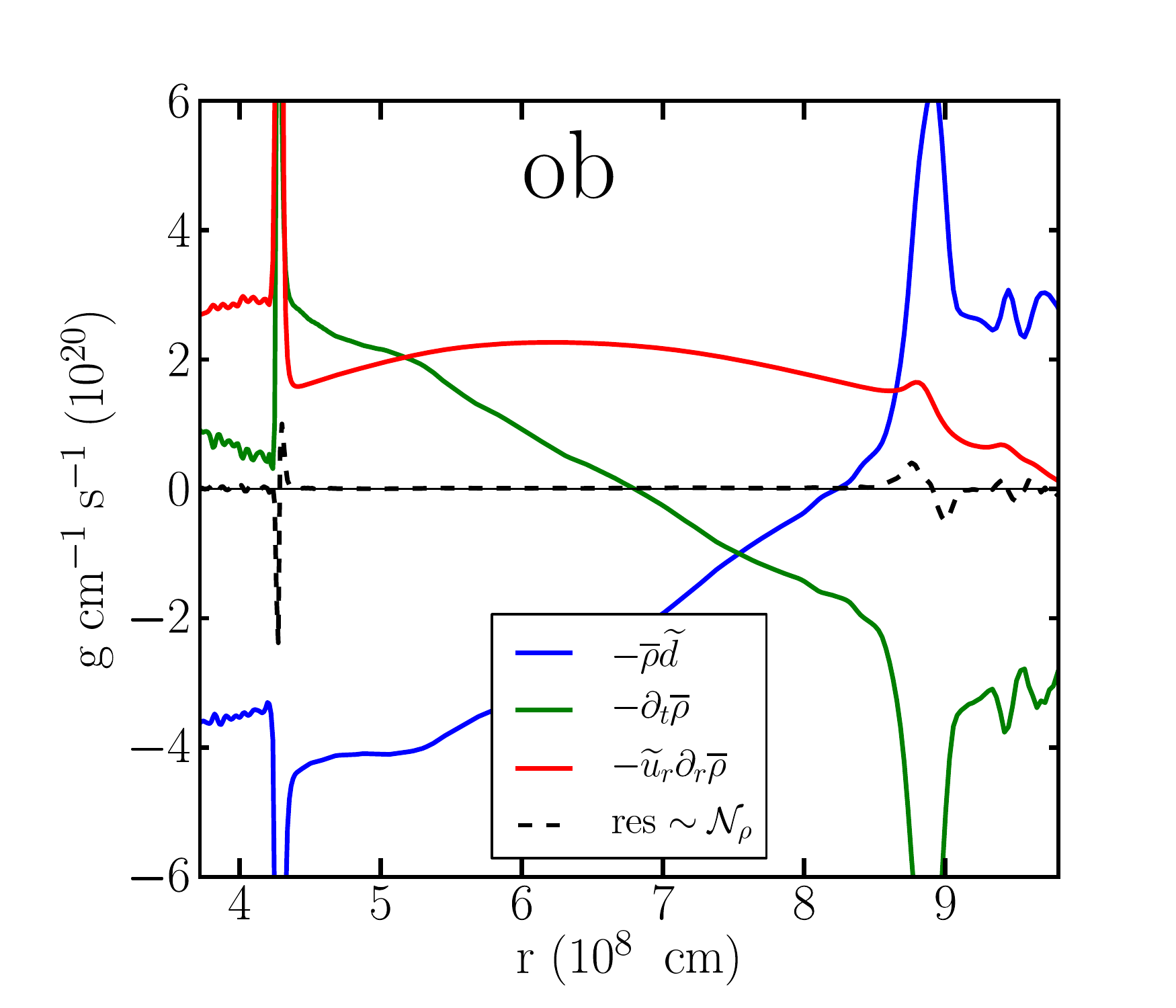}
\includegraphics[width=5.6cm]{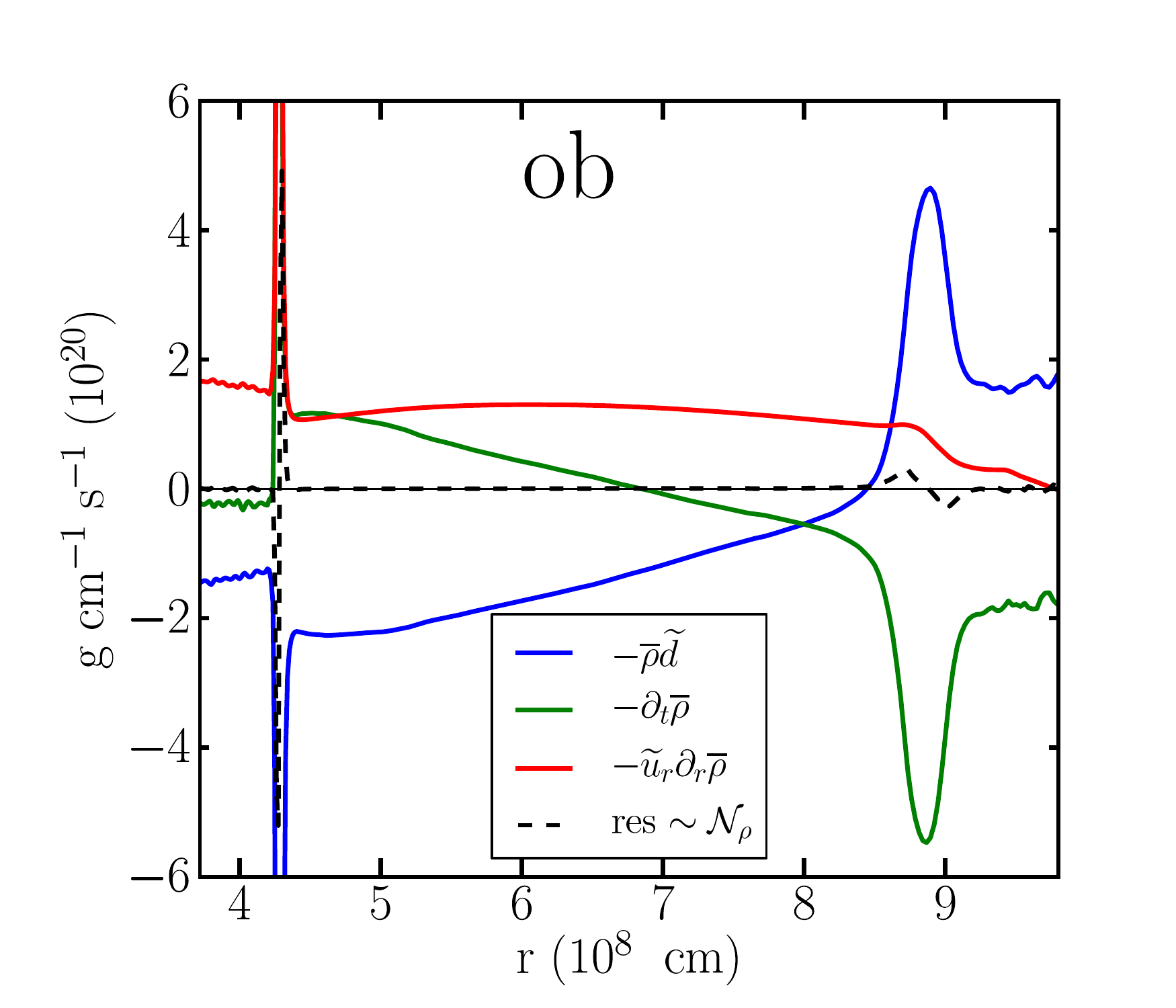}
\includegraphics[width=5.6cm]{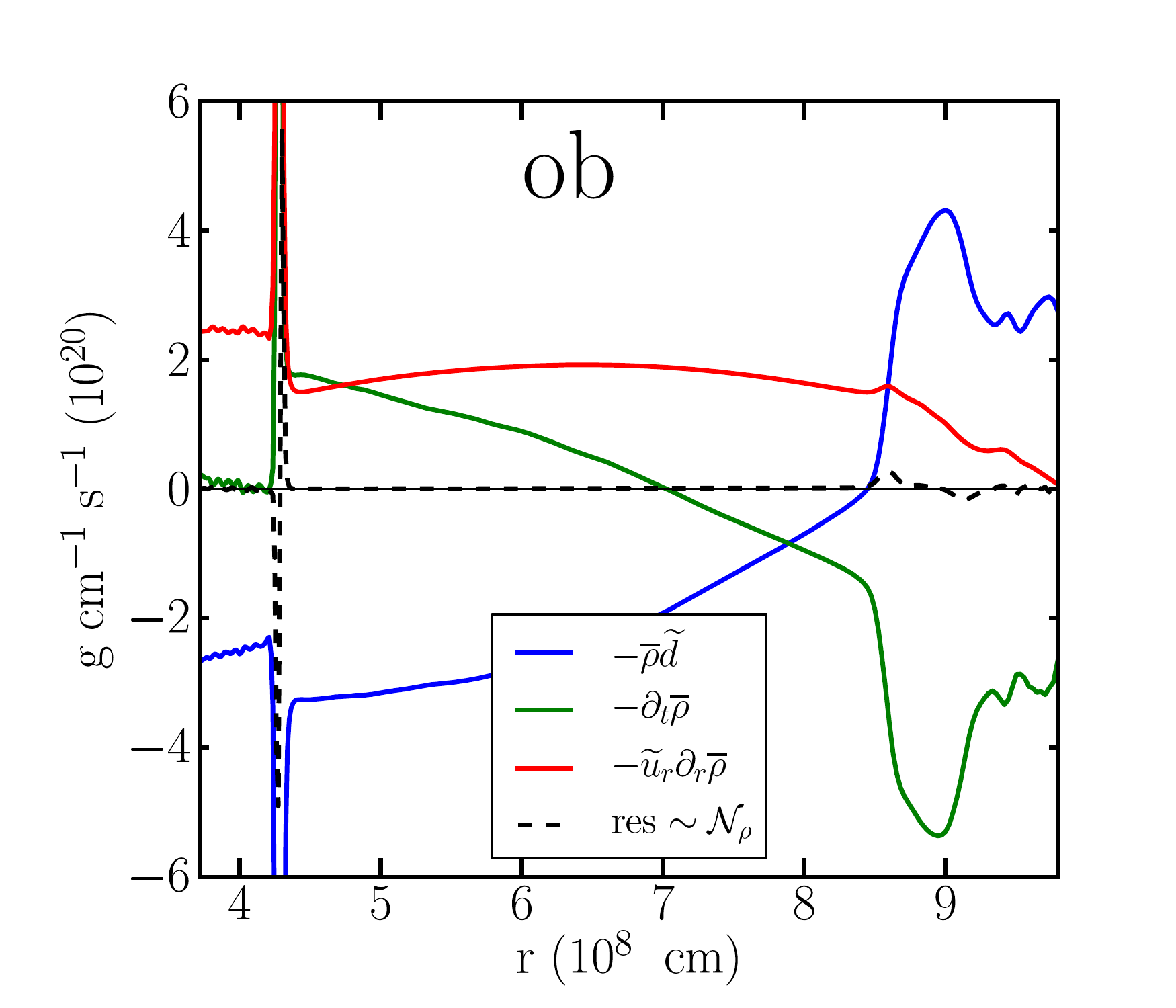}}

\centerline{
\includegraphics[width=5.6cm]{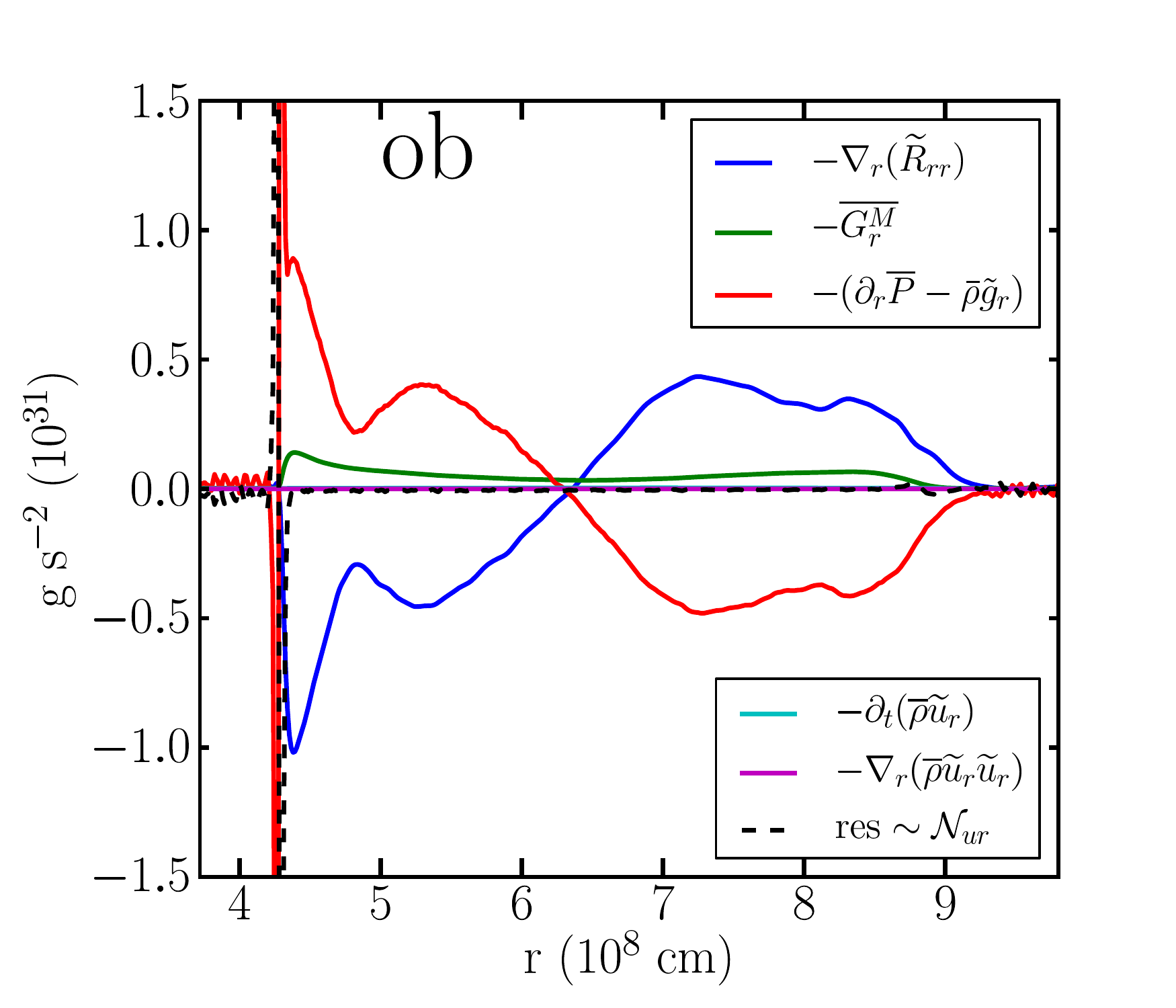}
\includegraphics[width=5.6cm]{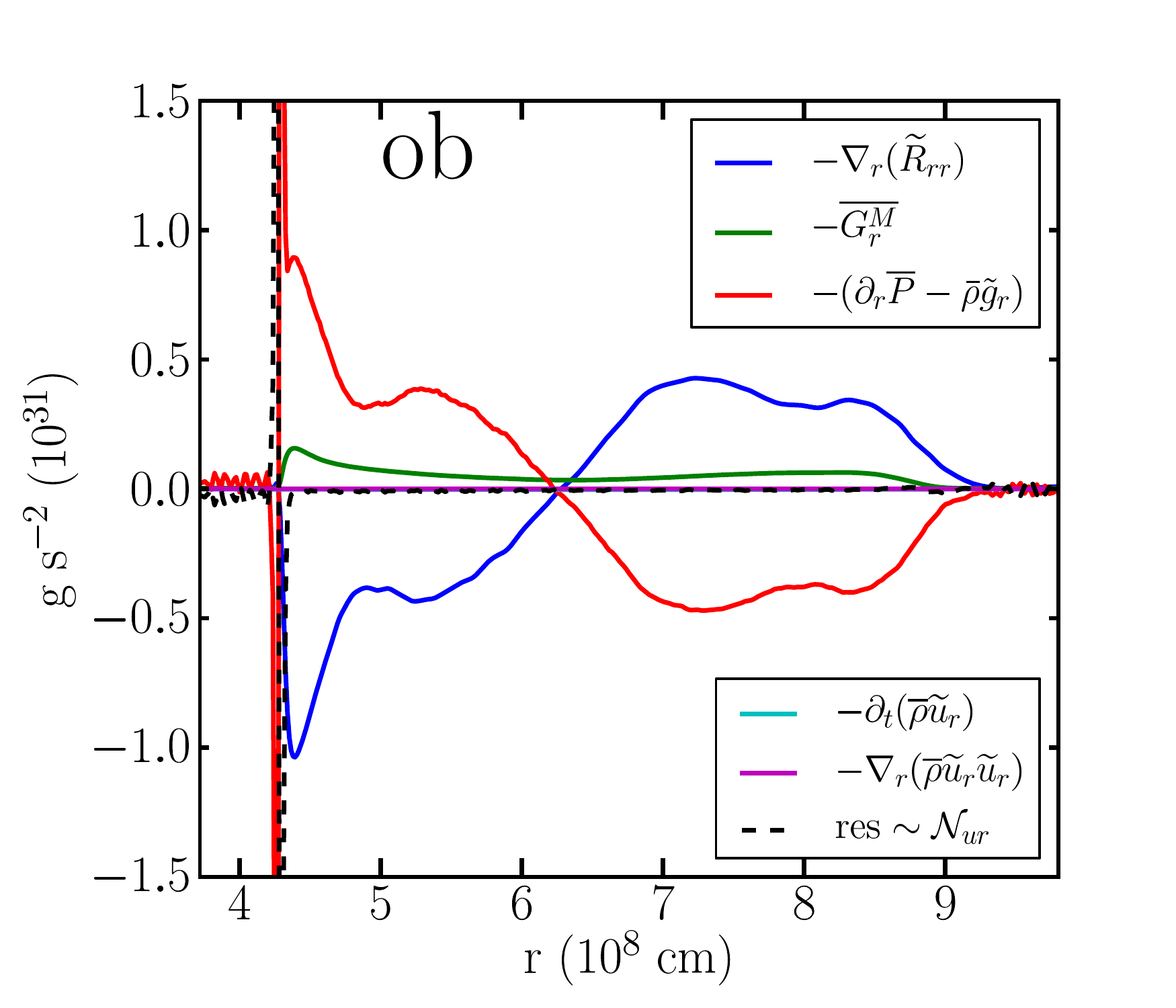}
\includegraphics[width=5.6cm]{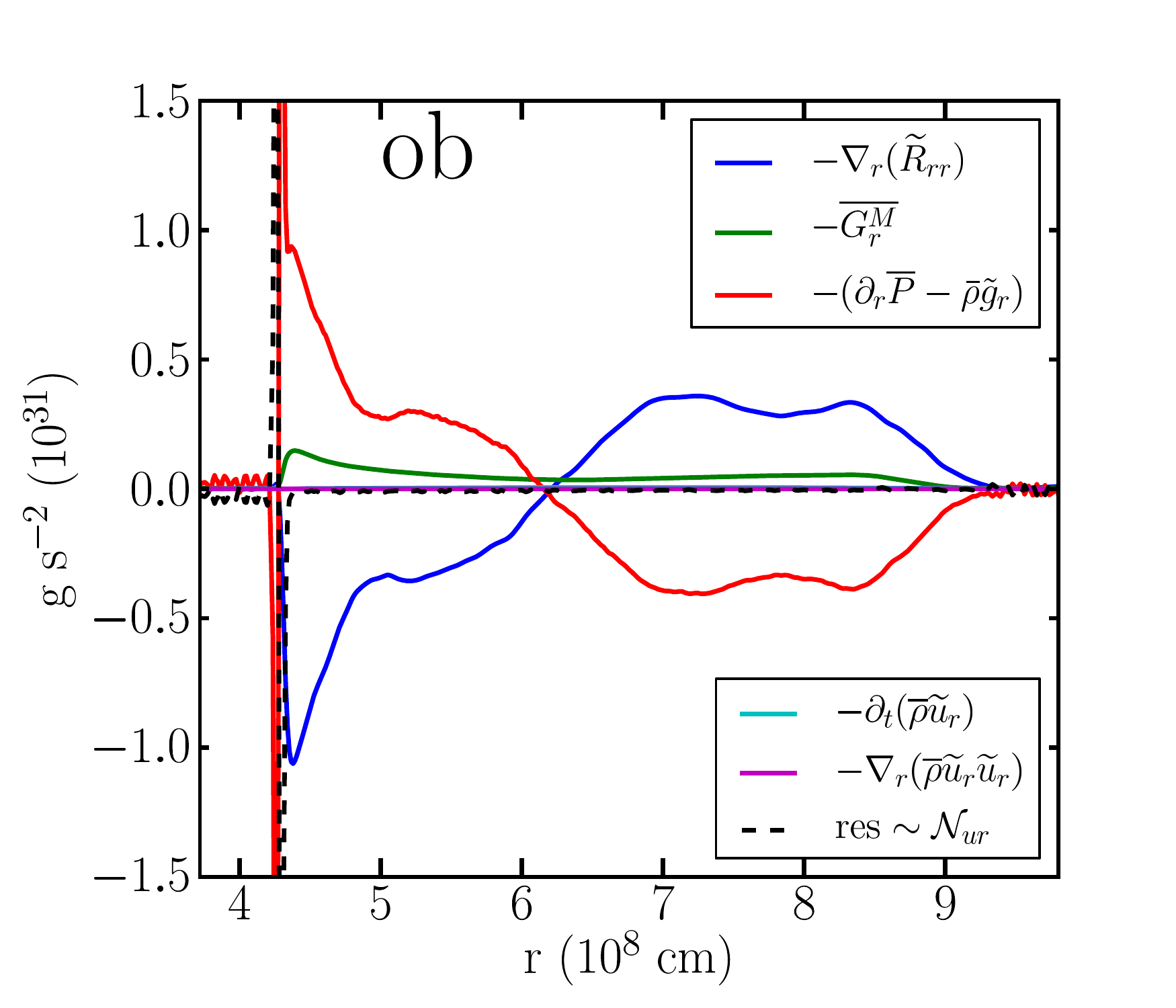}}
\caption{Mean continuity equation (upper panels) and radial momentum equation (lower panels) from model {\sf ob.3D.2hp}. Averaging window over roughly 2 convective turnover timescales 150 s (left), 3 convective turnover timescales 230 s (middle) and 4 convective turnover timescales 460 s (right).}
\end{figure}

\newpage

\subsubsection{Mean azimuthal and polar momentum equations}

\begin{figure}[!h]
\centerline{
\includegraphics[width=6.5cm]{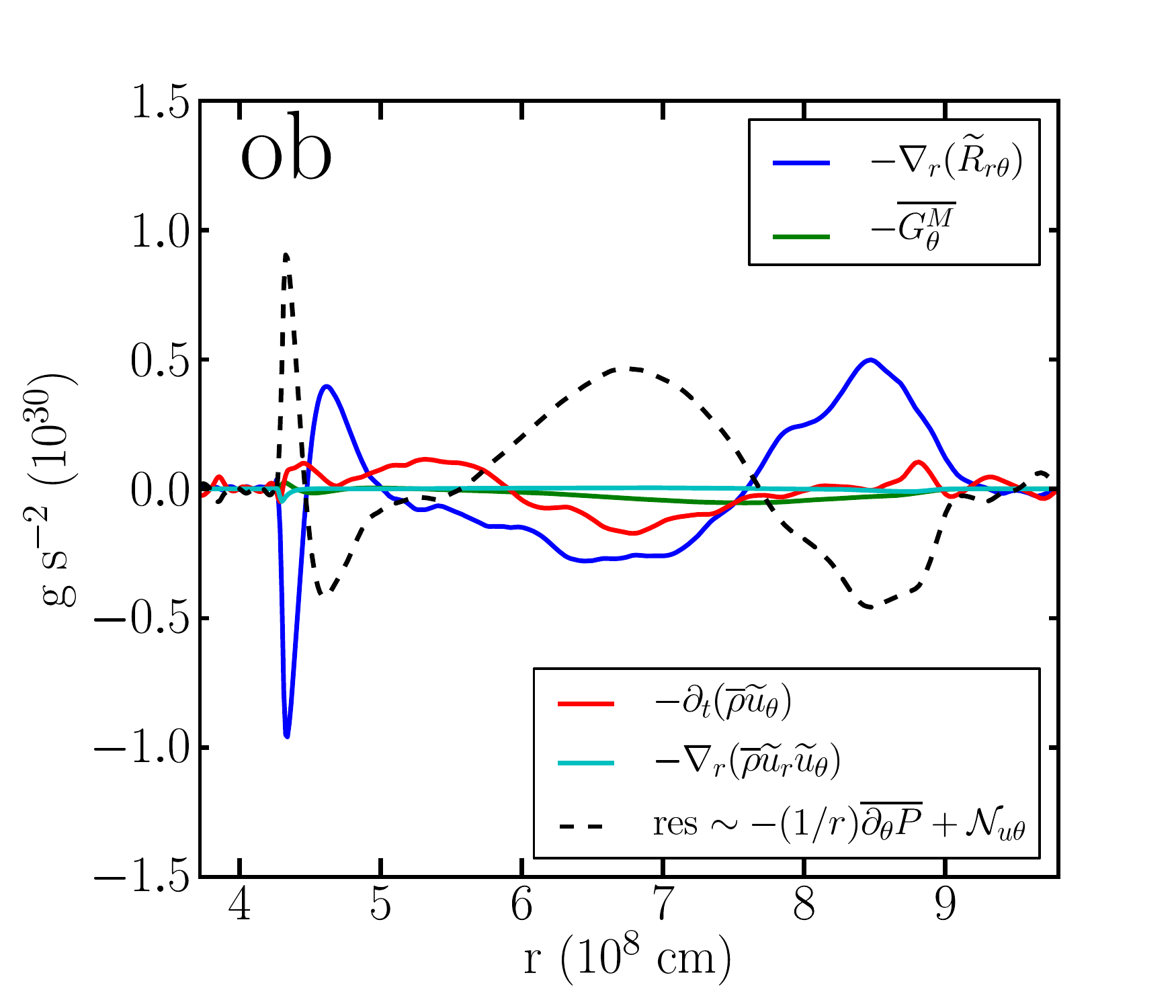}
\includegraphics[width=6.5cm]{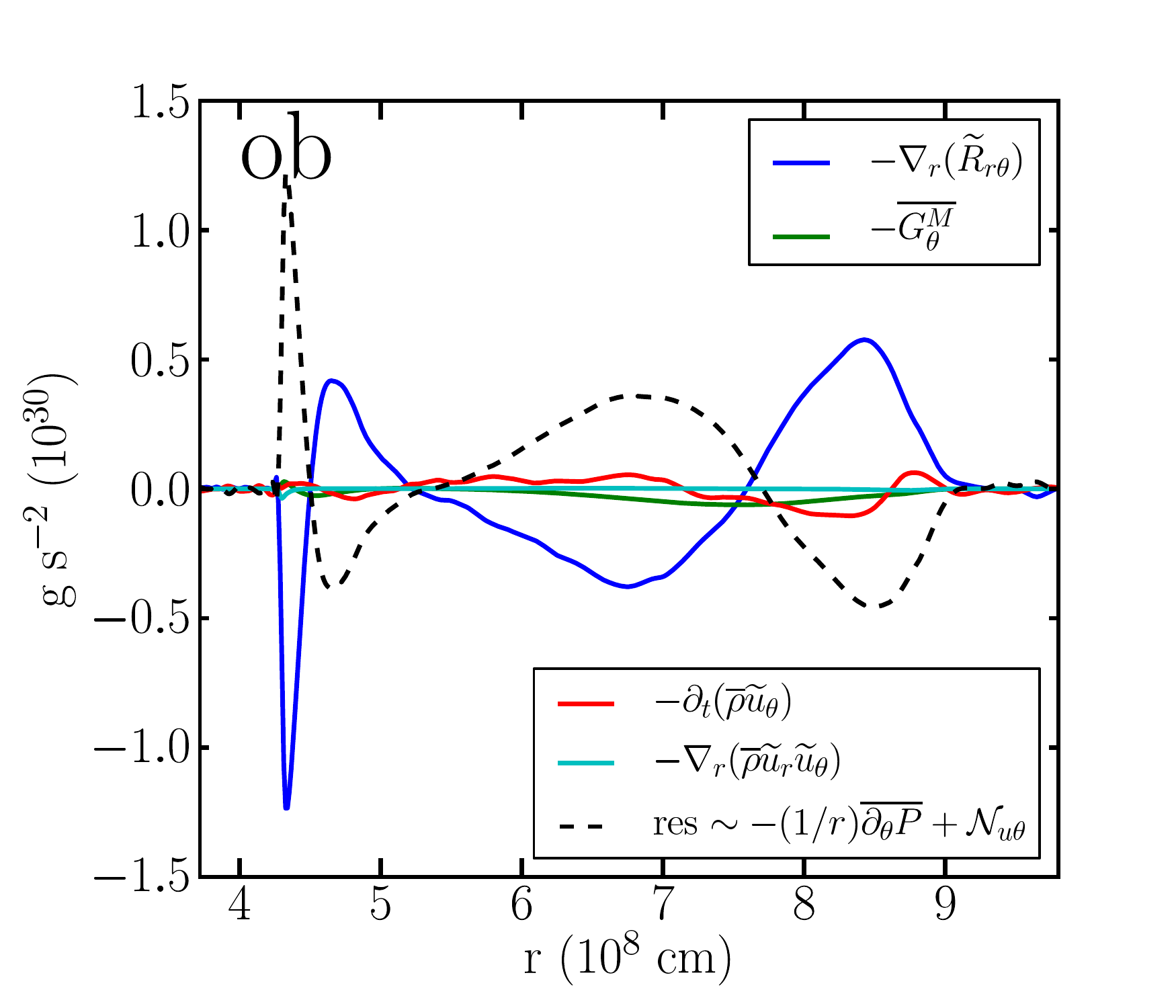}
\includegraphics[width=6.5cm]{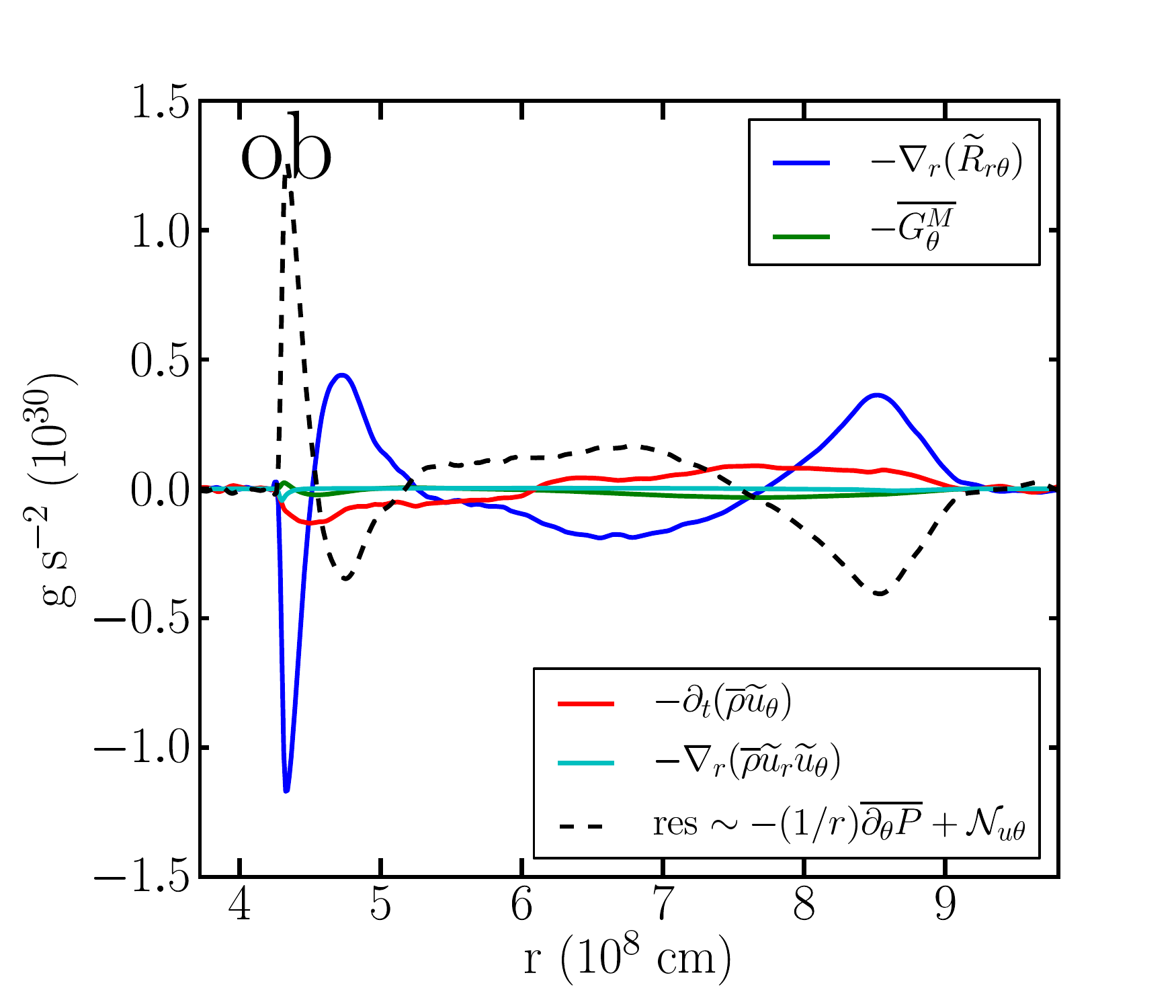}}

\centerline{
\includegraphics[width=6.5cm]{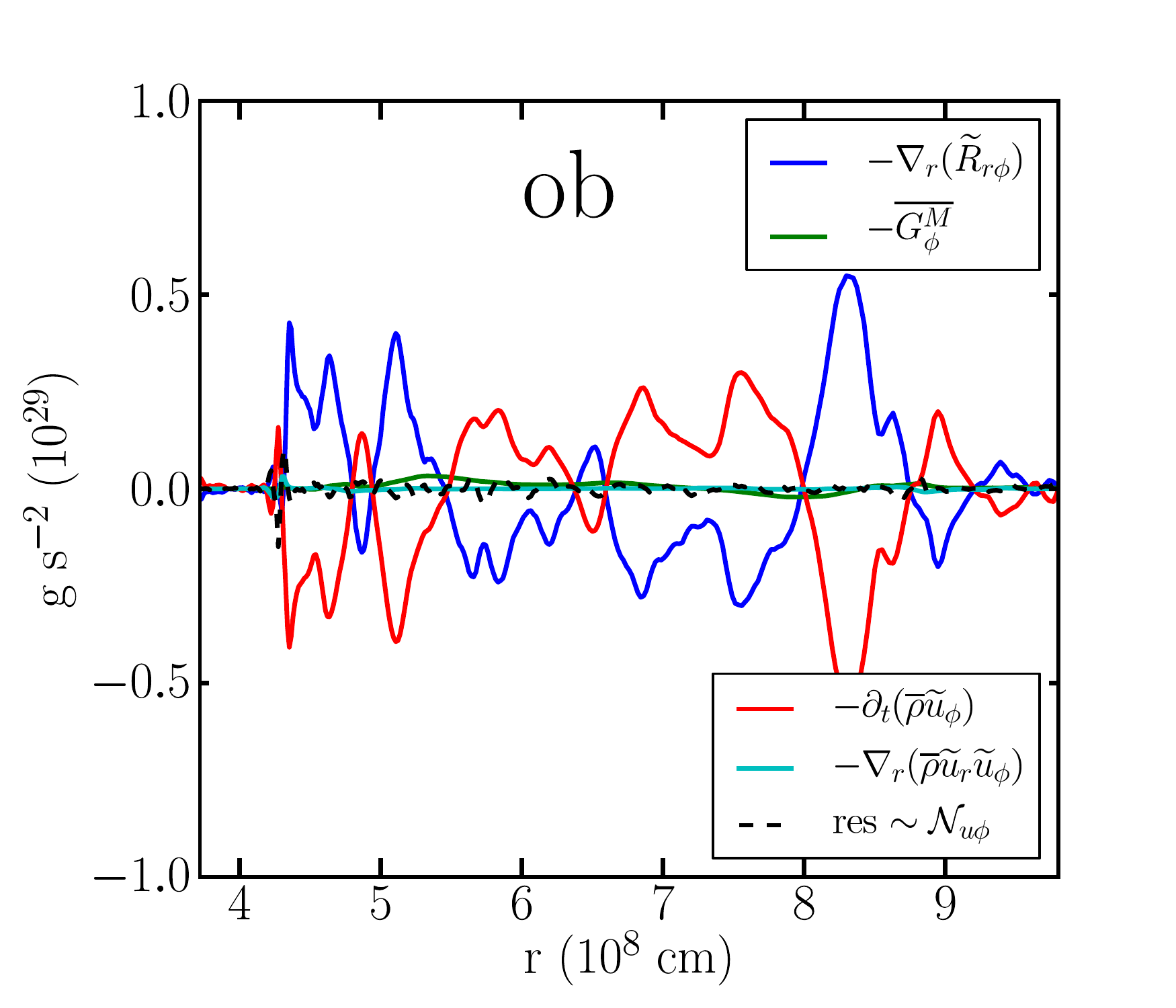}
\includegraphics[width=6.5cm]{obmrez275_tavg230_pmomentum_equation_ransdat-eps-converted-to.pdf}
\includegraphics[width=6.5cm]{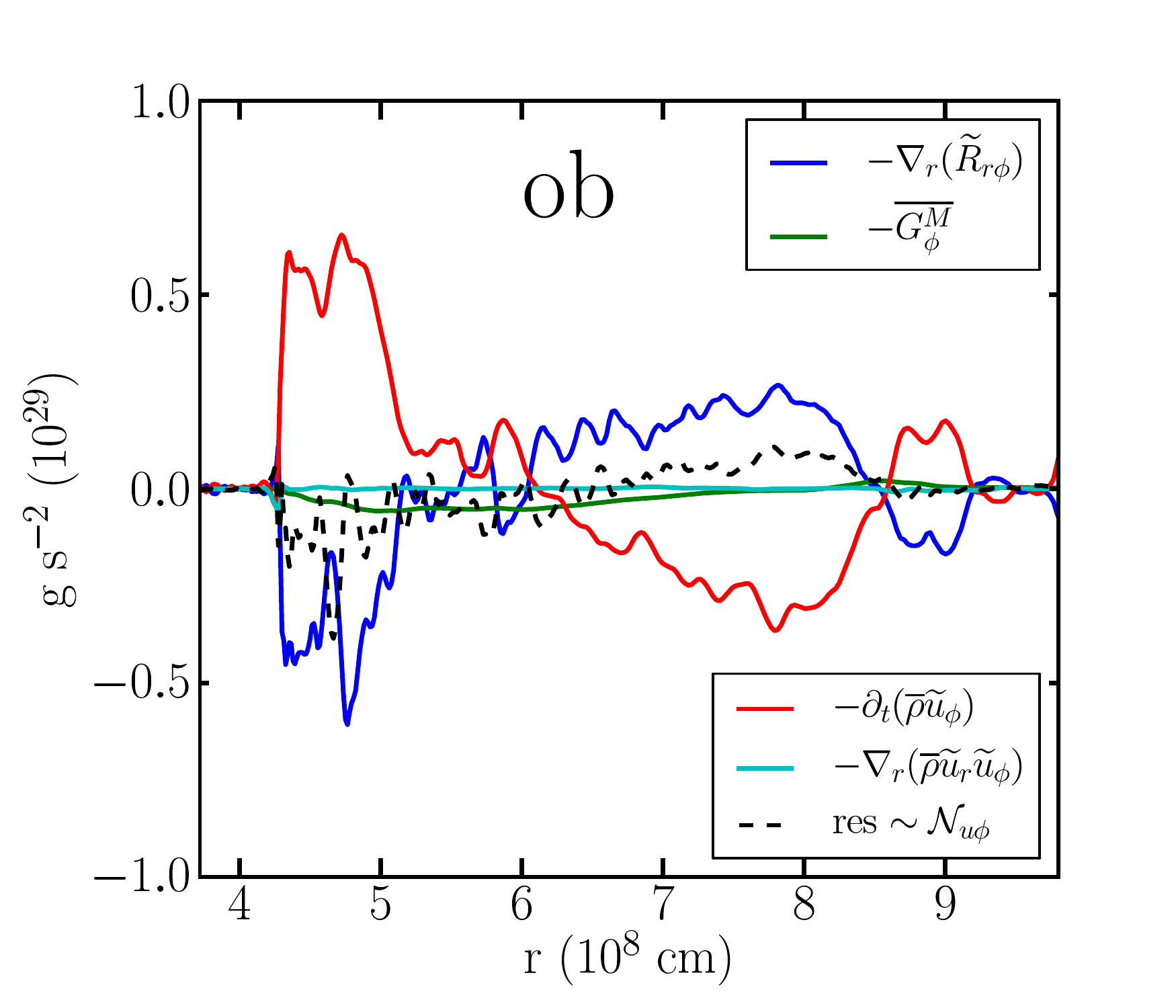}}
\caption{Mean azimuthal equation (upper panels) and mean polar momentum equation (lower panels) from model {\sf ob.3D.2hp}. Averaging window over roughly 2 convective turnover timescales 150 s (left), 3 convective turnover timescales 230 s (middle) and 4 convective turnover timescales 460 s (right). }
\end{figure}

\newpage

\subsubsection{Mean total energy equation and mean turbulent kinetic energy equation}

\begin{figure}[!h]
\centerline{
\includegraphics[width=6.5cm]{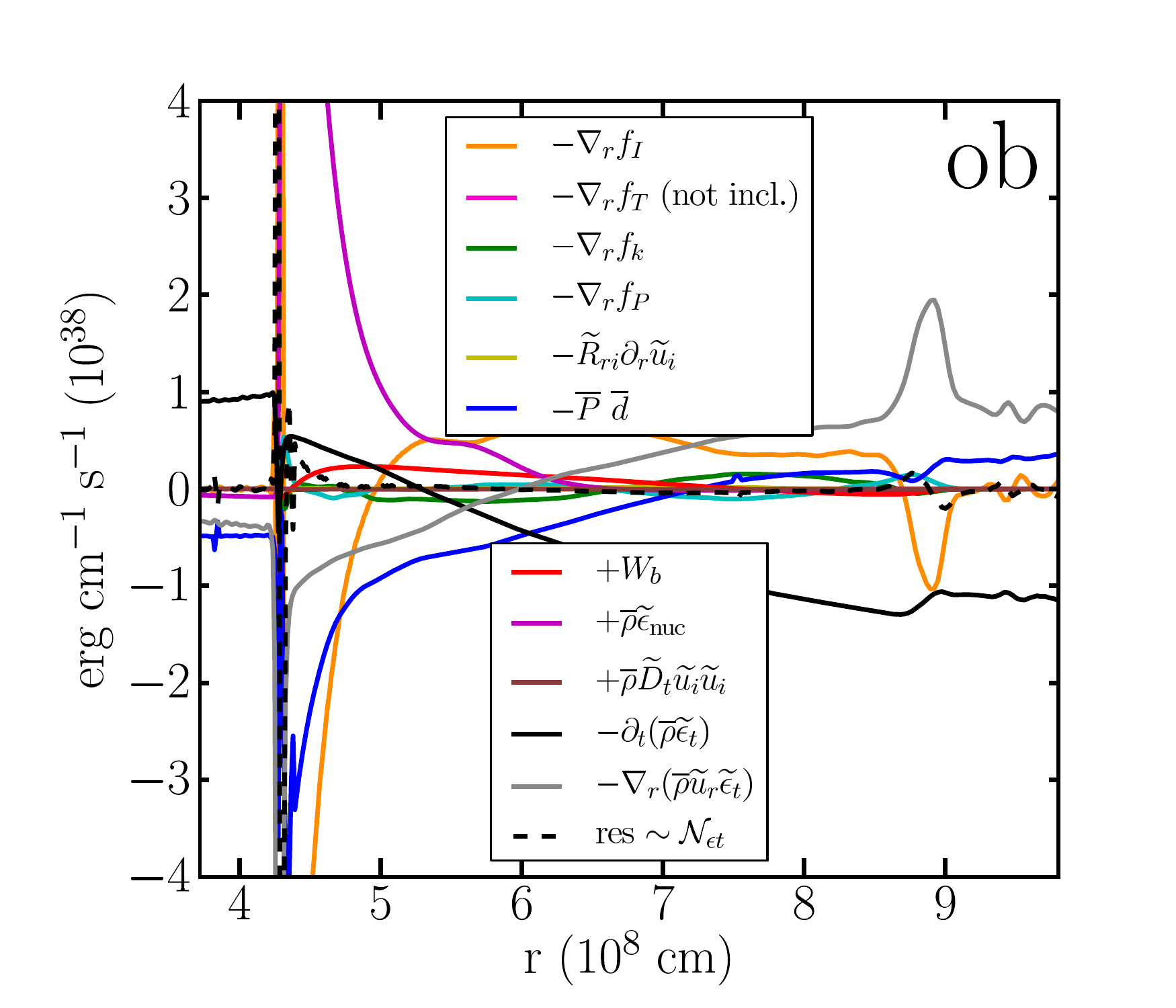}
\includegraphics[width=6.5cm]{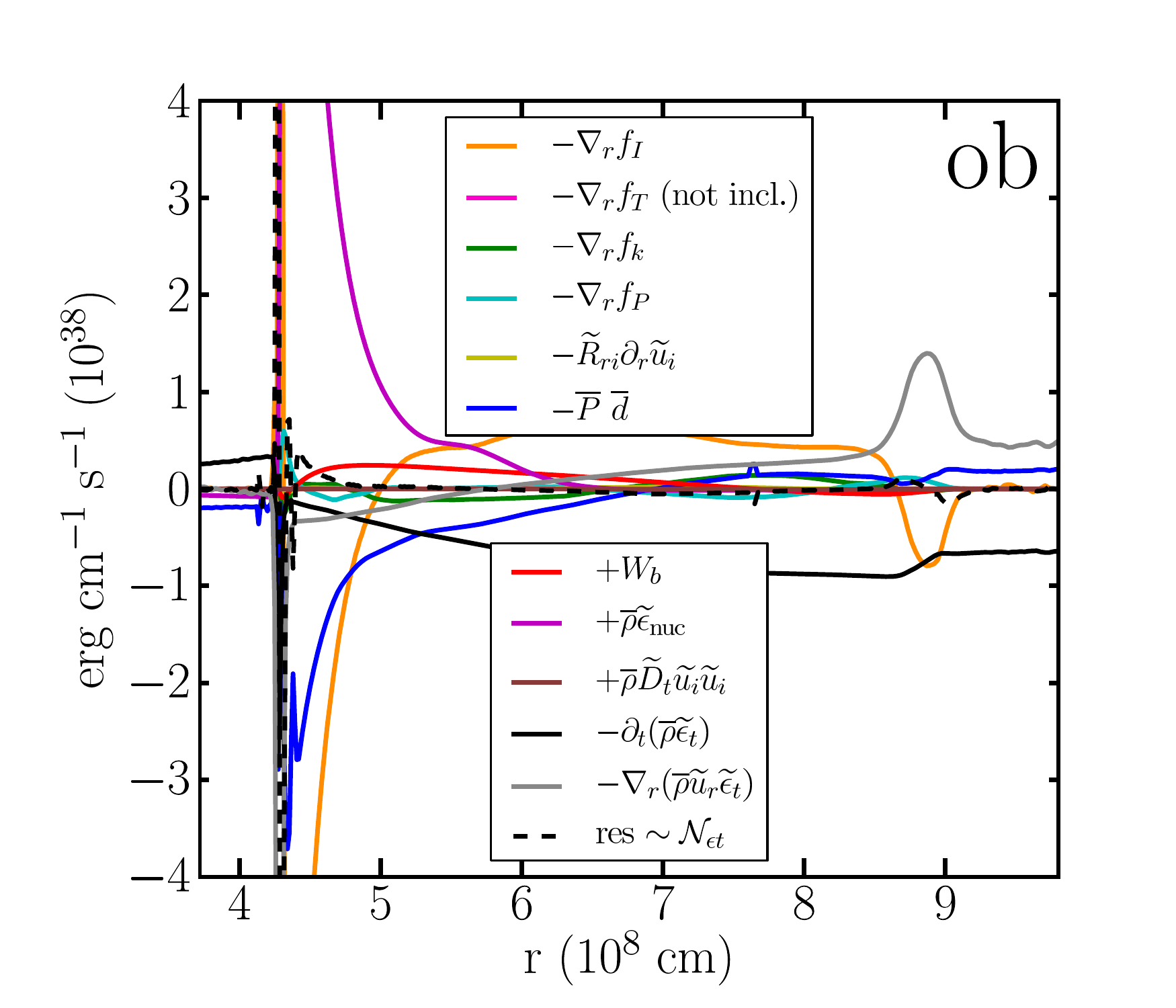}
\includegraphics[width=6.5cm]{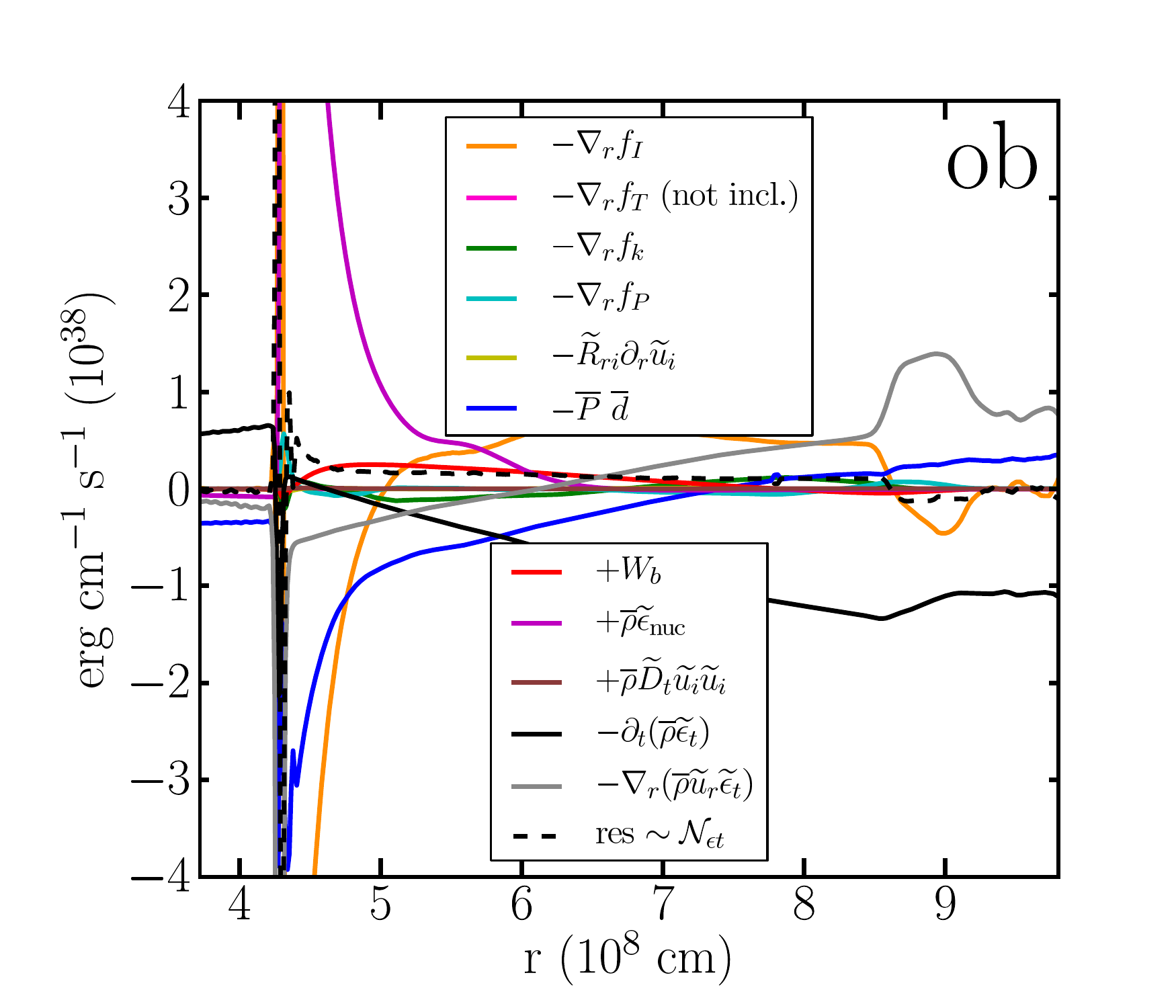}}

\centerline{
\includegraphics[width=6.5cm]{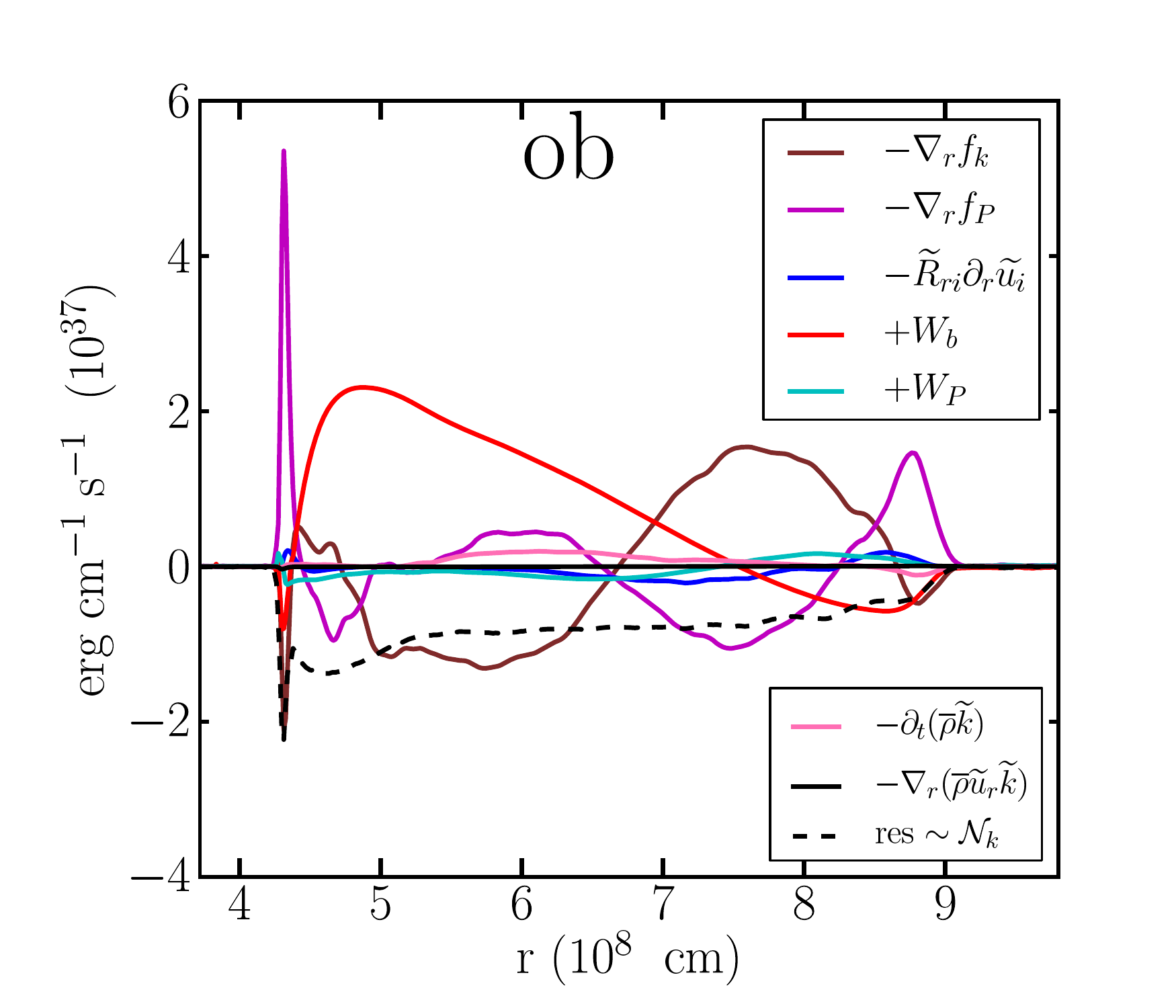}
\includegraphics[width=6.5cm]{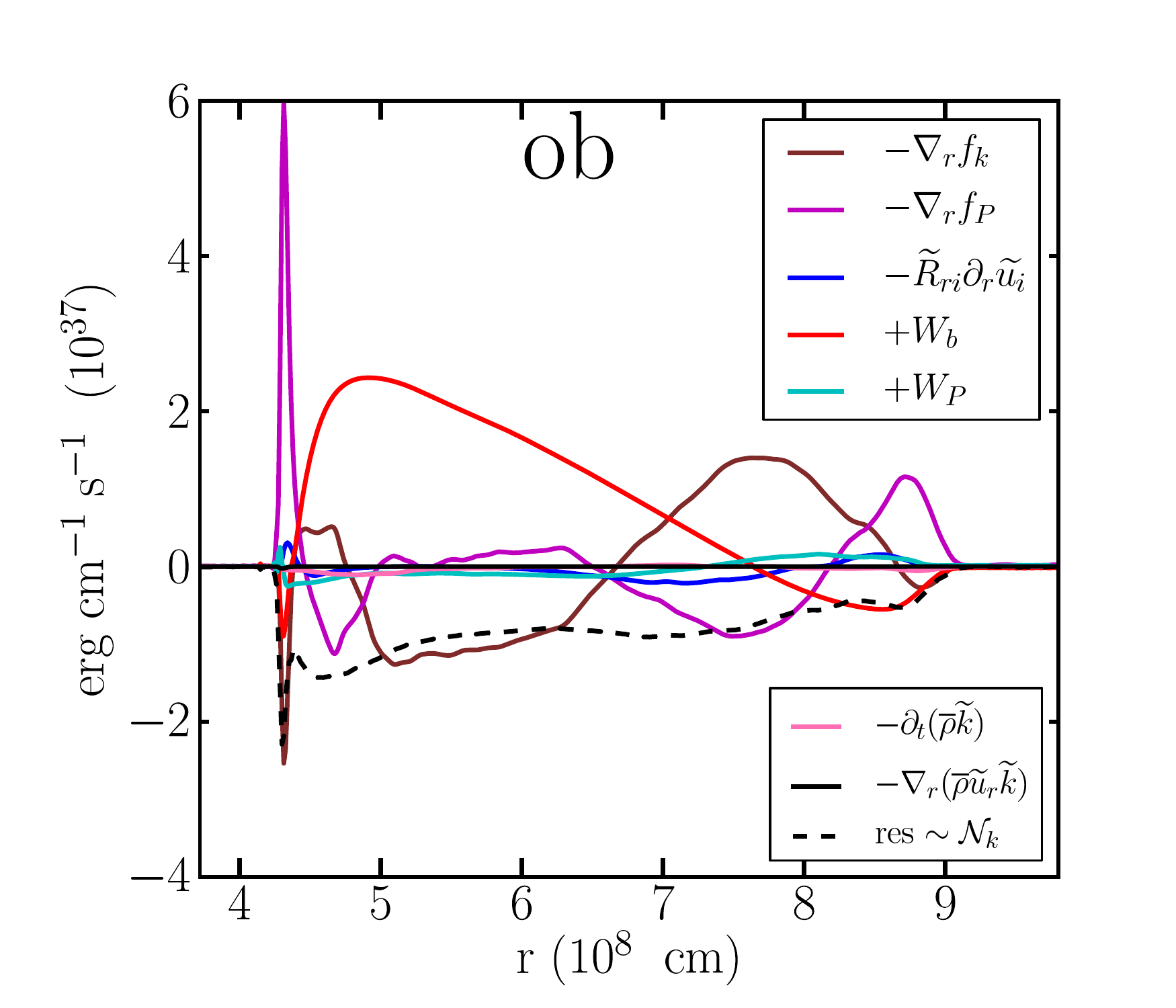}
\includegraphics[width=6.5cm]{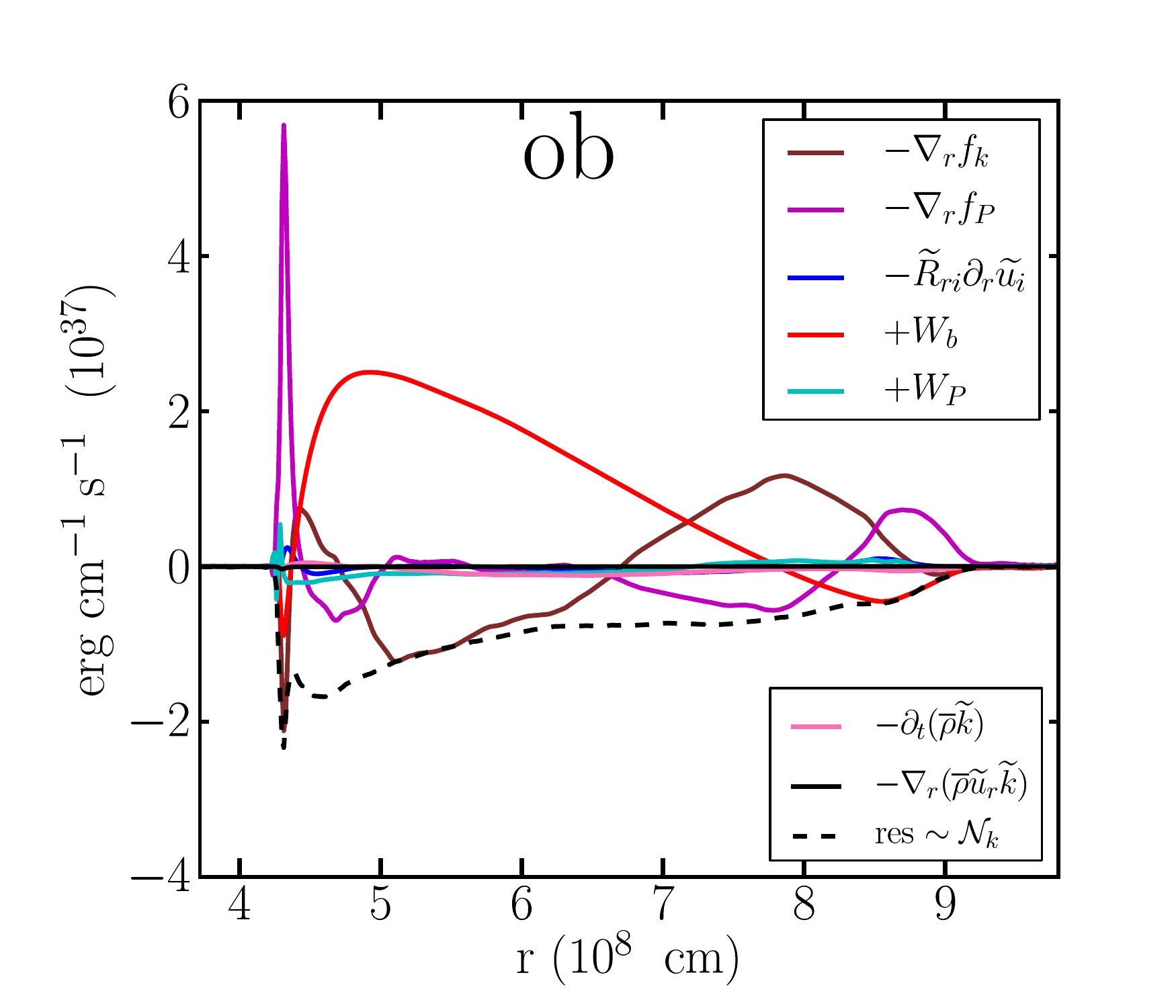}}
\caption{Mean total energy equation (upper panels) and mean turbulent kinetic energy equation (lower panels) from model {ob.3D.2hp}. Averaging window over roughly 2 convective turnover timescales 150 s (left), 3 convective turnover timescales 230 s (middle) and 4 convective turnover timescales 460 s (right).}
\end{figure}

\newpage

\subsubsection{Mean entropy equation and mean number of nucleons per isotope equation}

\begin{figure}[!h]
\centerline{
\includegraphics[width=6.5cm]{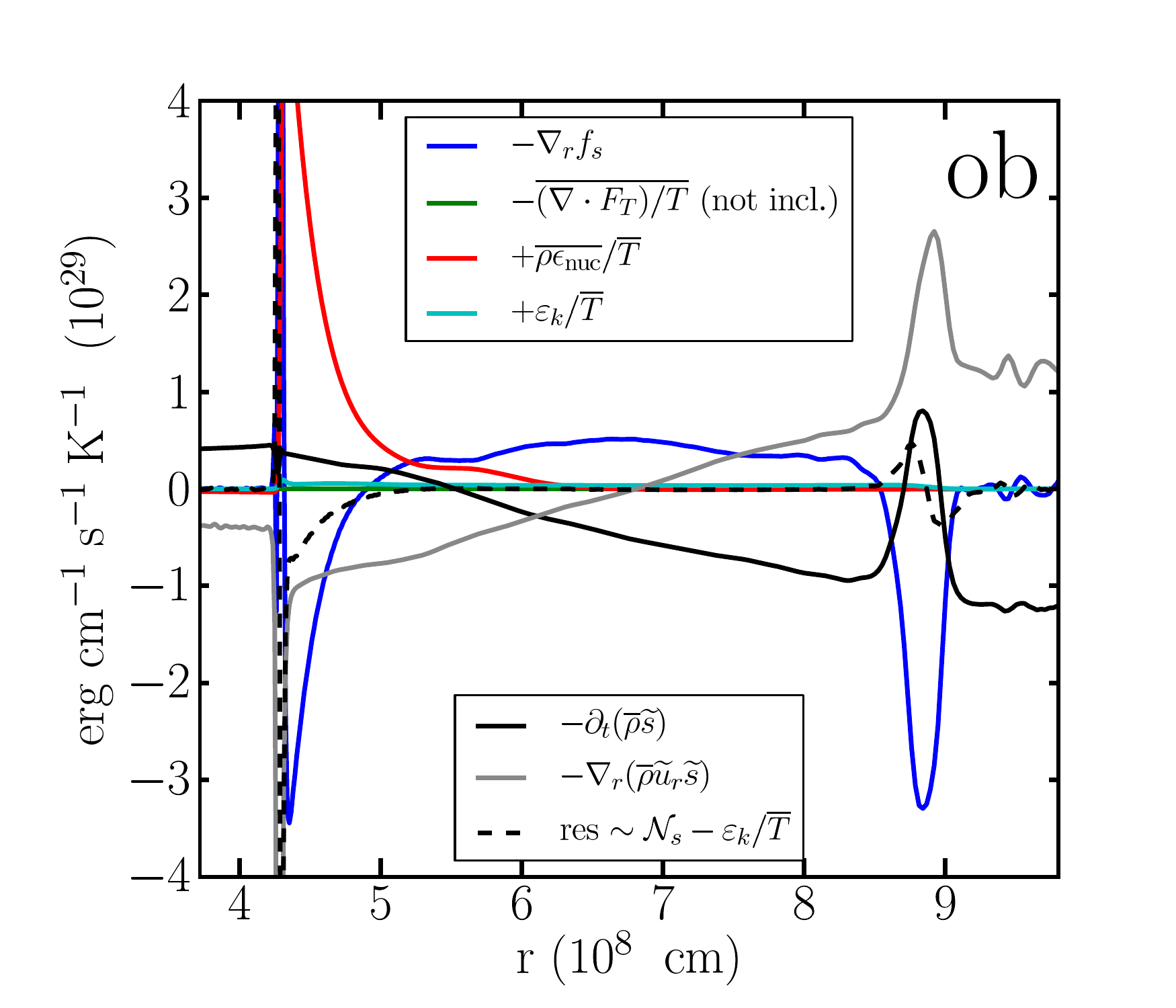}
\includegraphics[width=6.5cm]{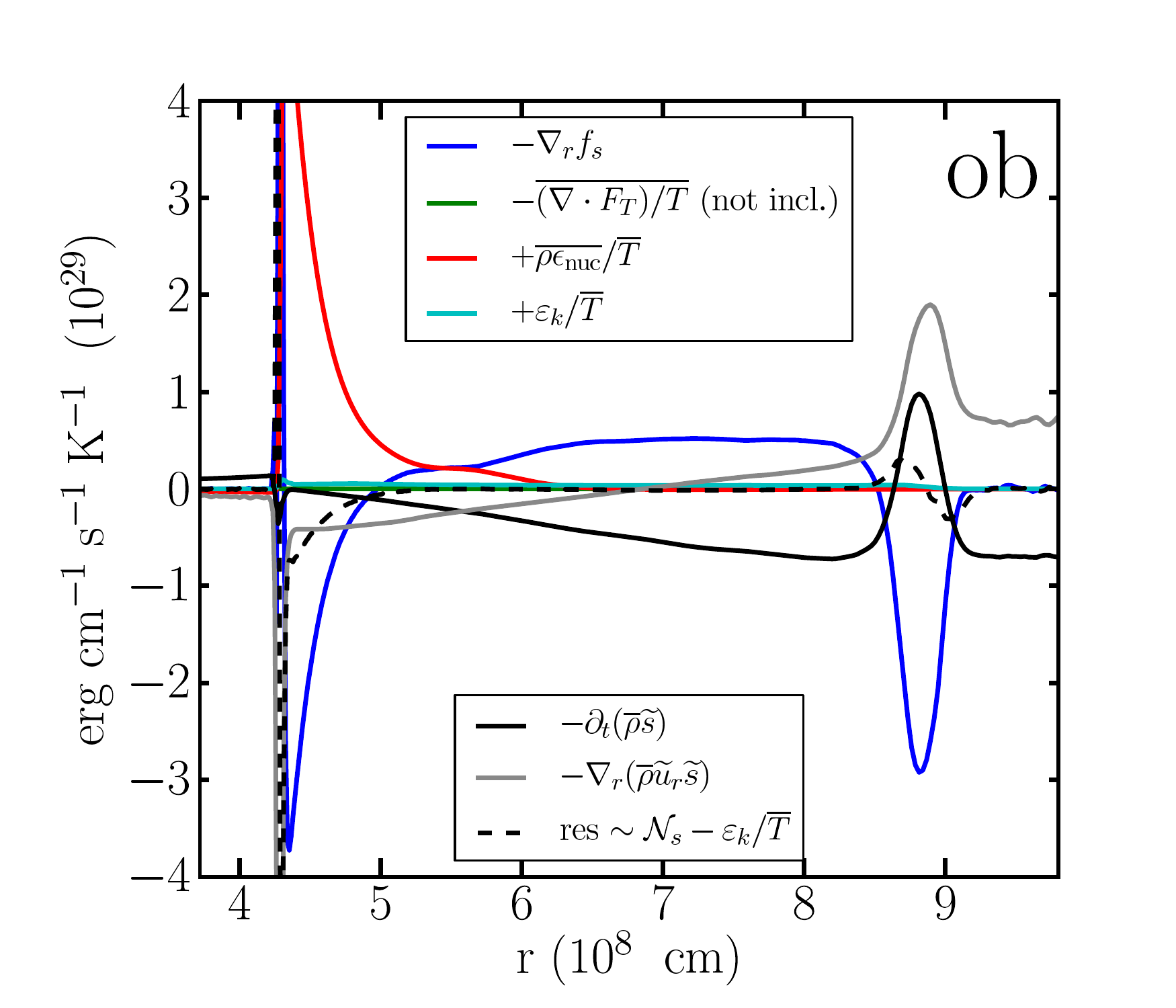}
\includegraphics[width=6.5cm]{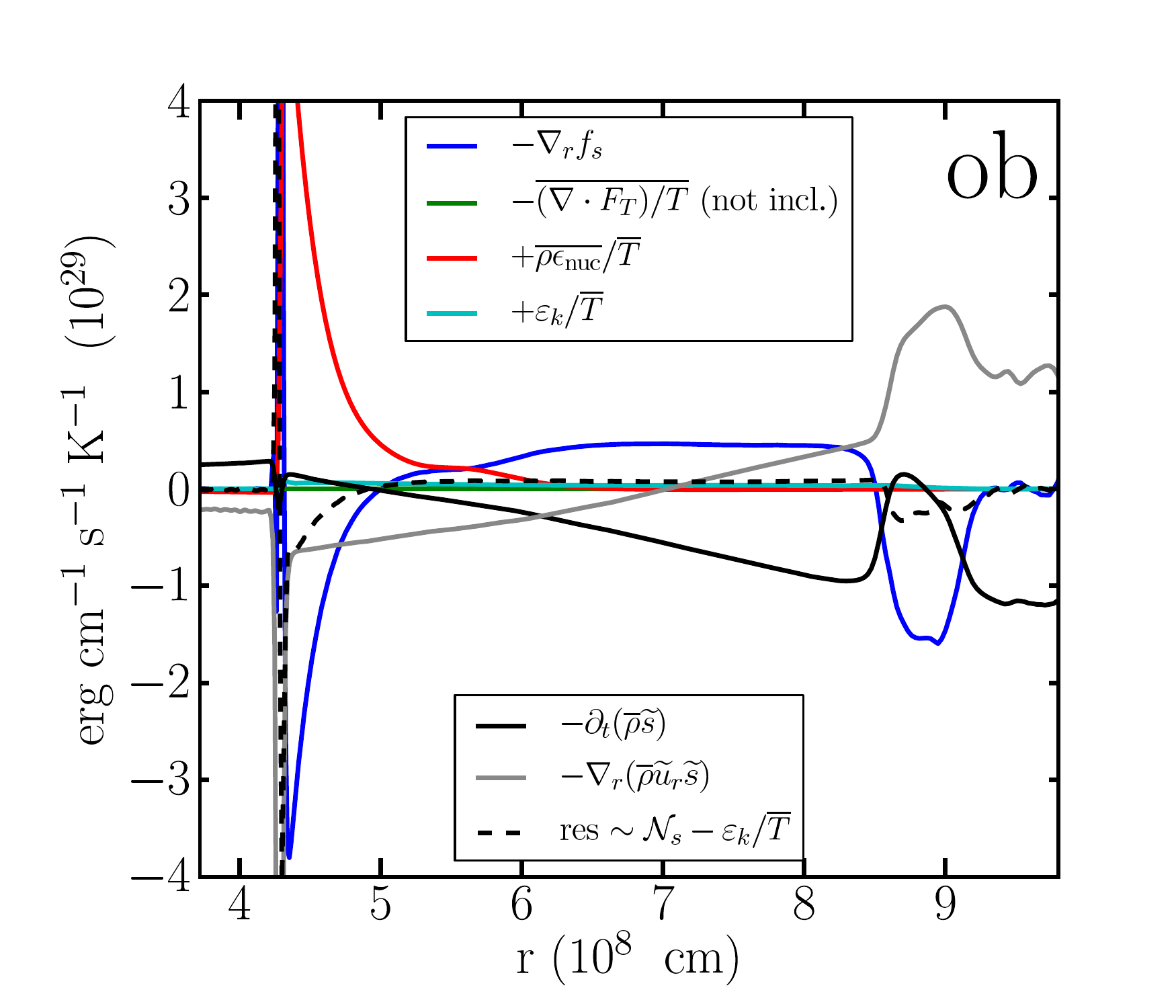}}

\centerline{
\includegraphics[width=6.5cm]{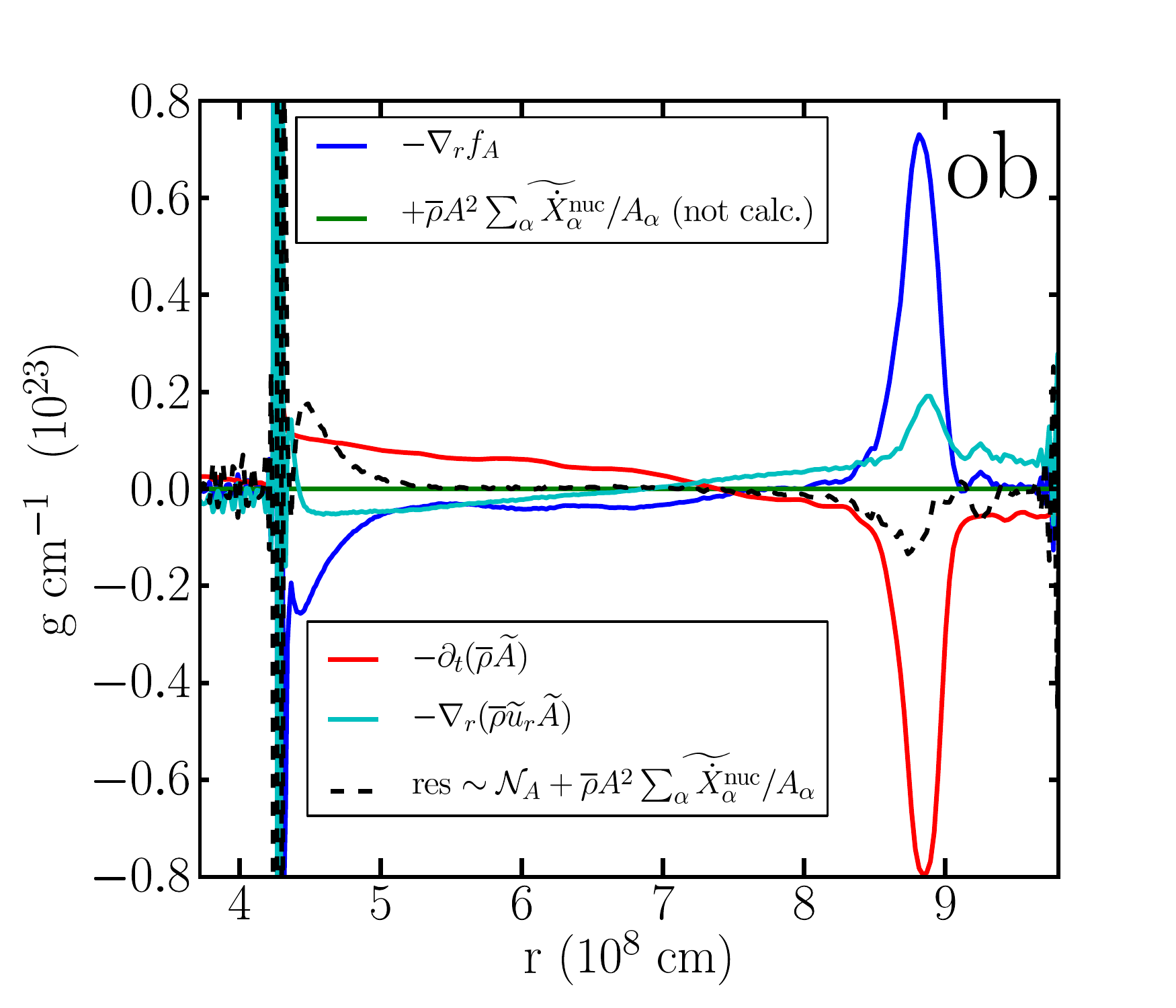}
\includegraphics[width=6.5cm]{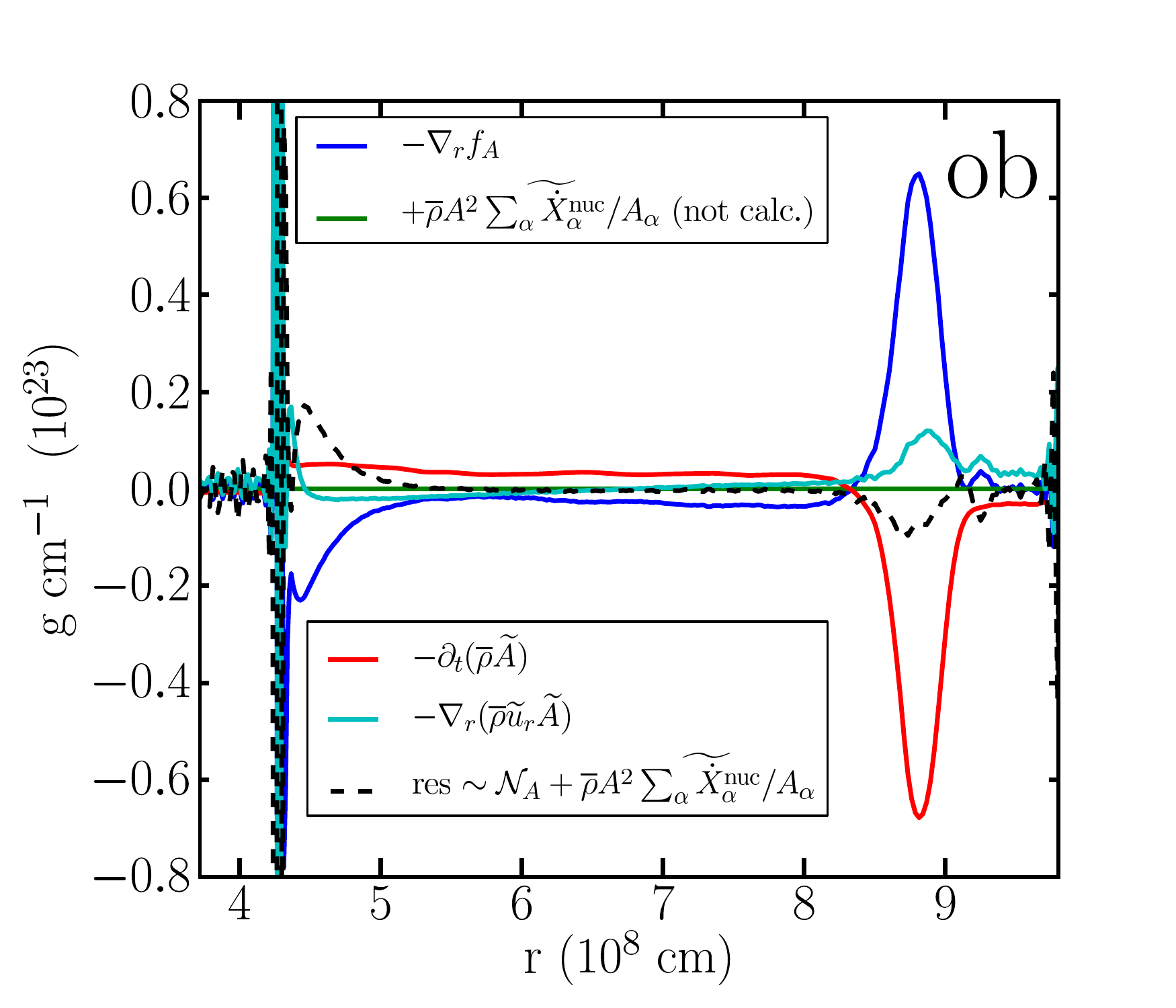}
\includegraphics[width=6.5cm]{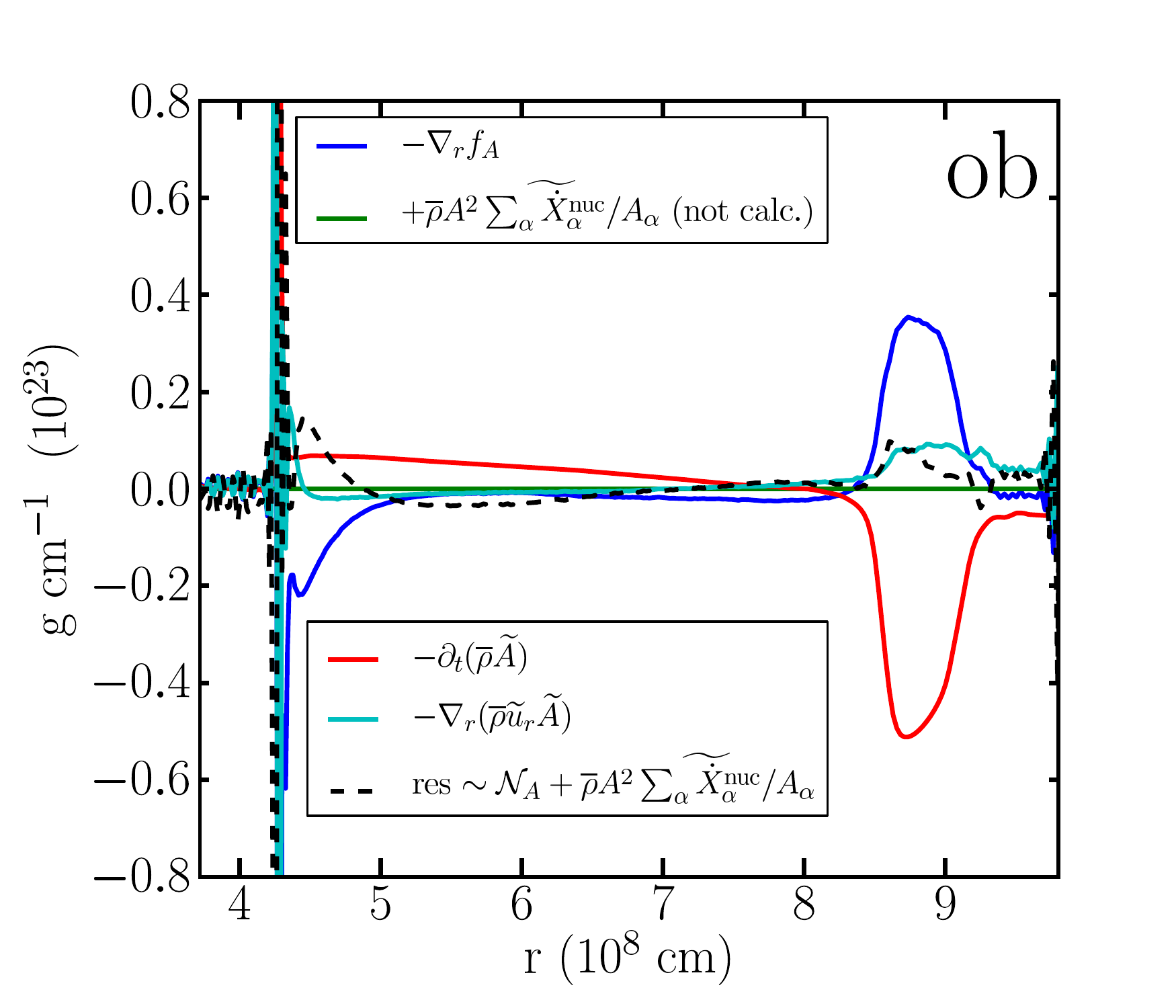}}
\caption{Mean entropy equation (upper panels) and mean number of nucleons per isotope (upper panels) from model {\sf ob.3D.2hp}. Averaging window over roughly 2 convective turnover timescales 150 s (left), 3 convective turnover timescales 230 s (middle) and 4 convective turnover timescales 460 s (right).}
\end{figure}

\newpage

\subsubsection{Mean turbulent kinetic energy and mean velocities}

\begin{figure}[!h]
\centerline{
\includegraphics[width=6.5cm]{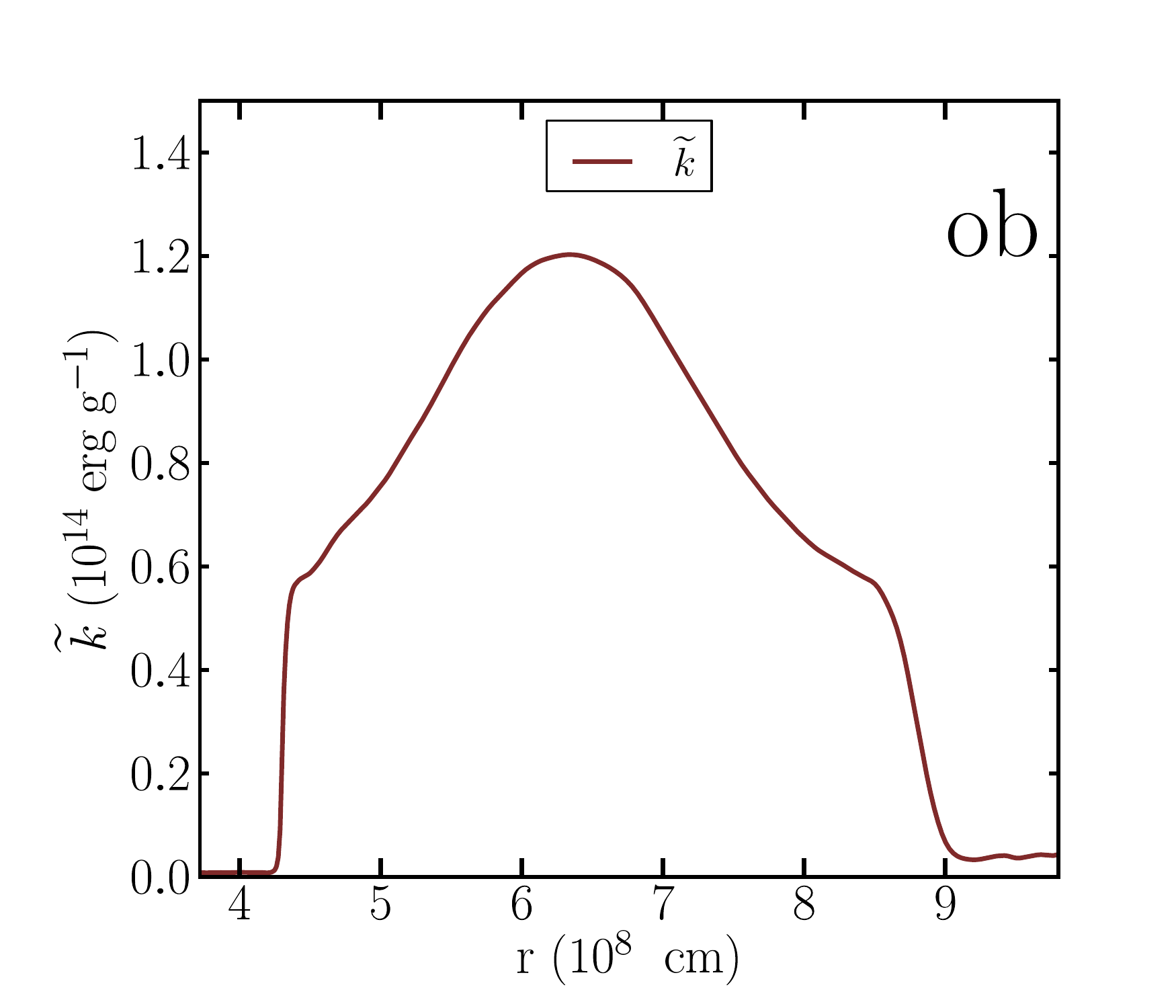}
\includegraphics[width=6.5cm]{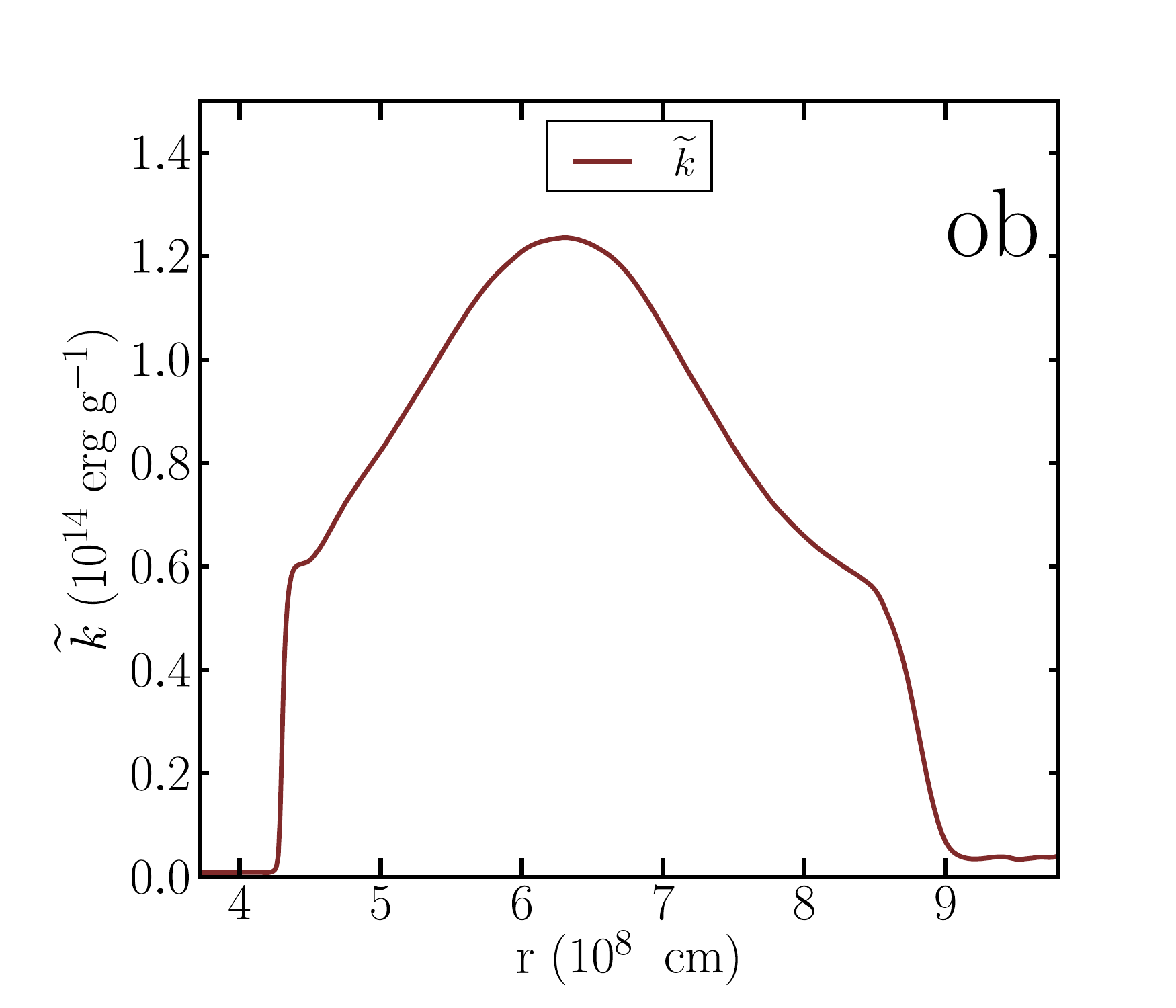}
\includegraphics[width=6.5cm]{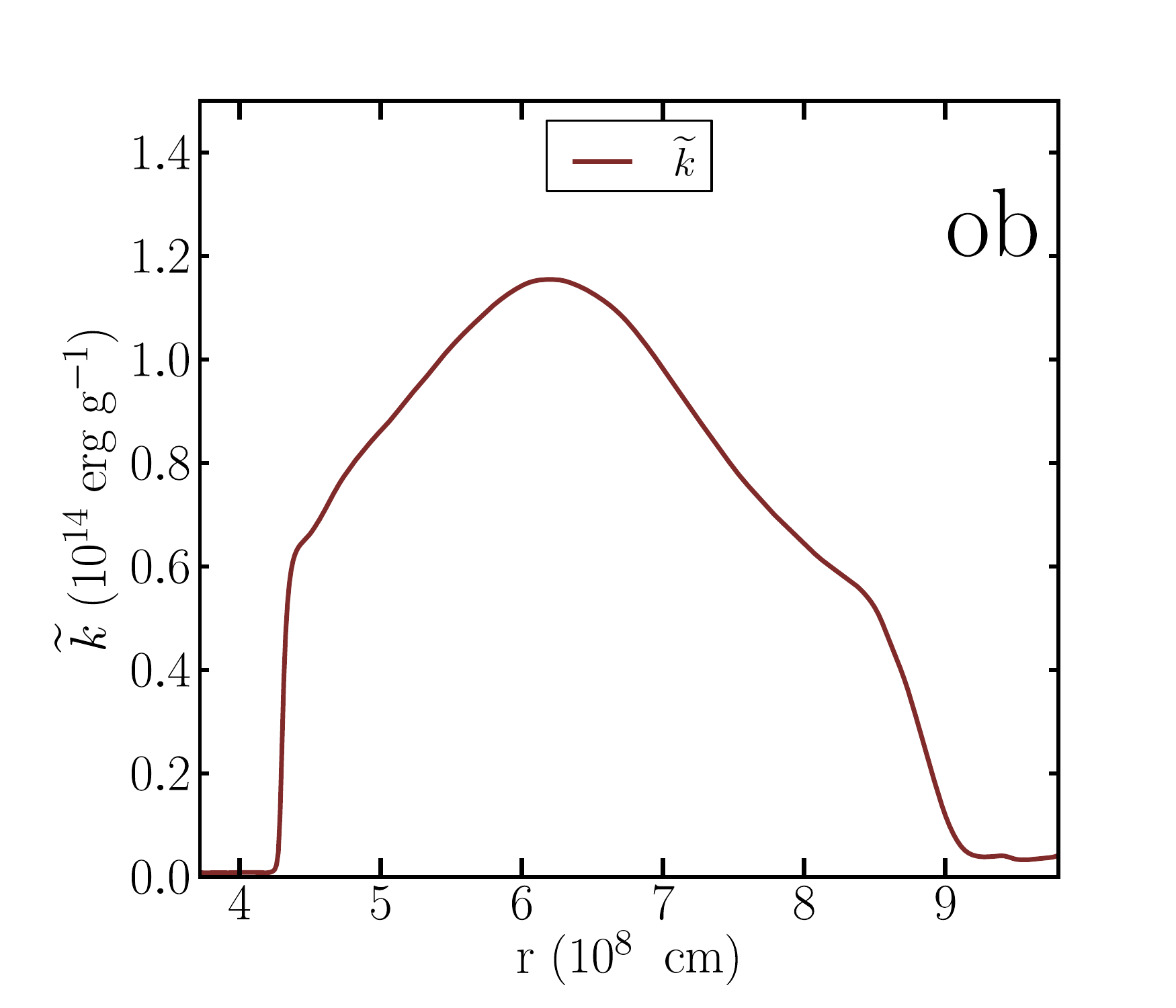}}

\centerline{
\includegraphics[width=6.5cm]{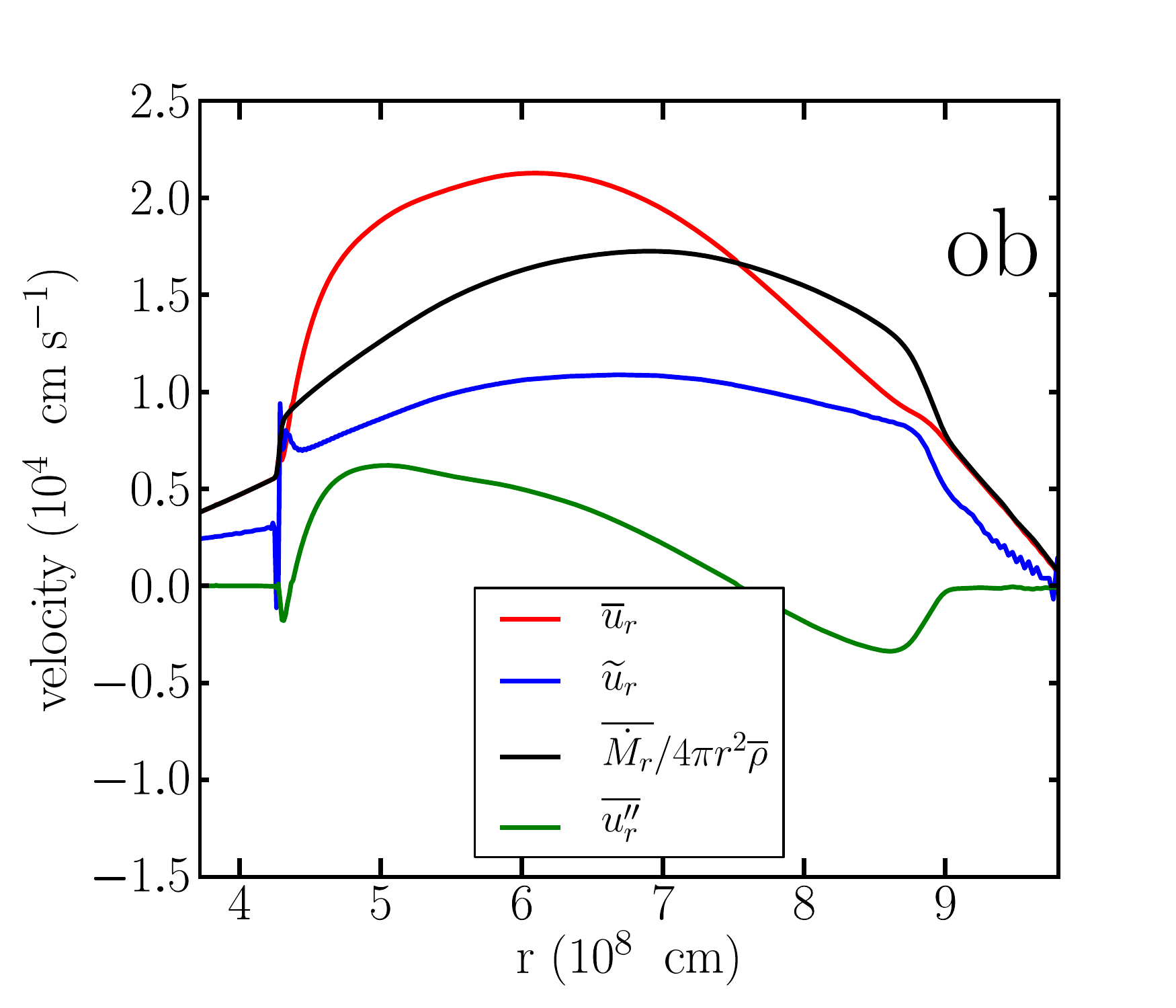}
\includegraphics[width=6.5cm]{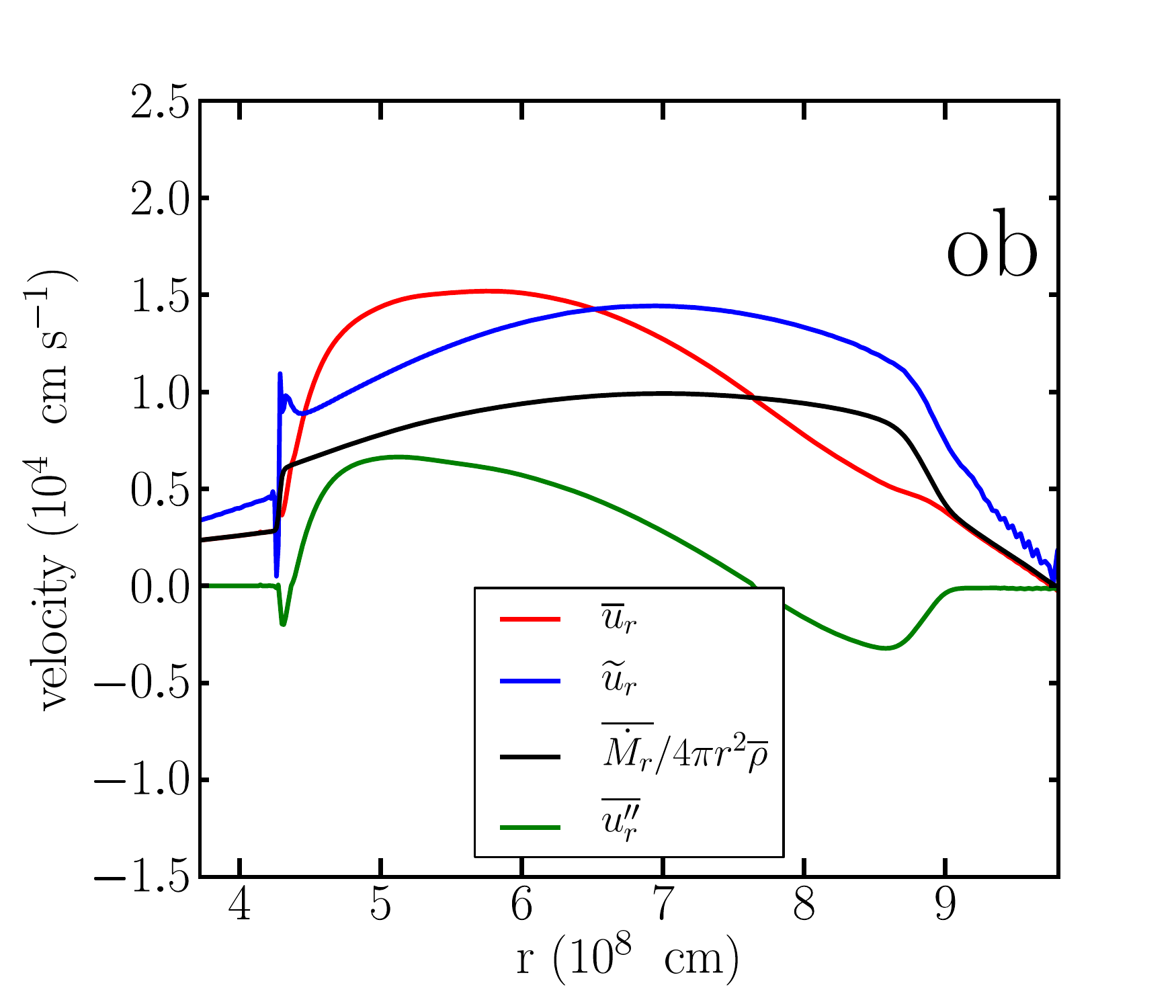}
\includegraphics[width=6.5cm]{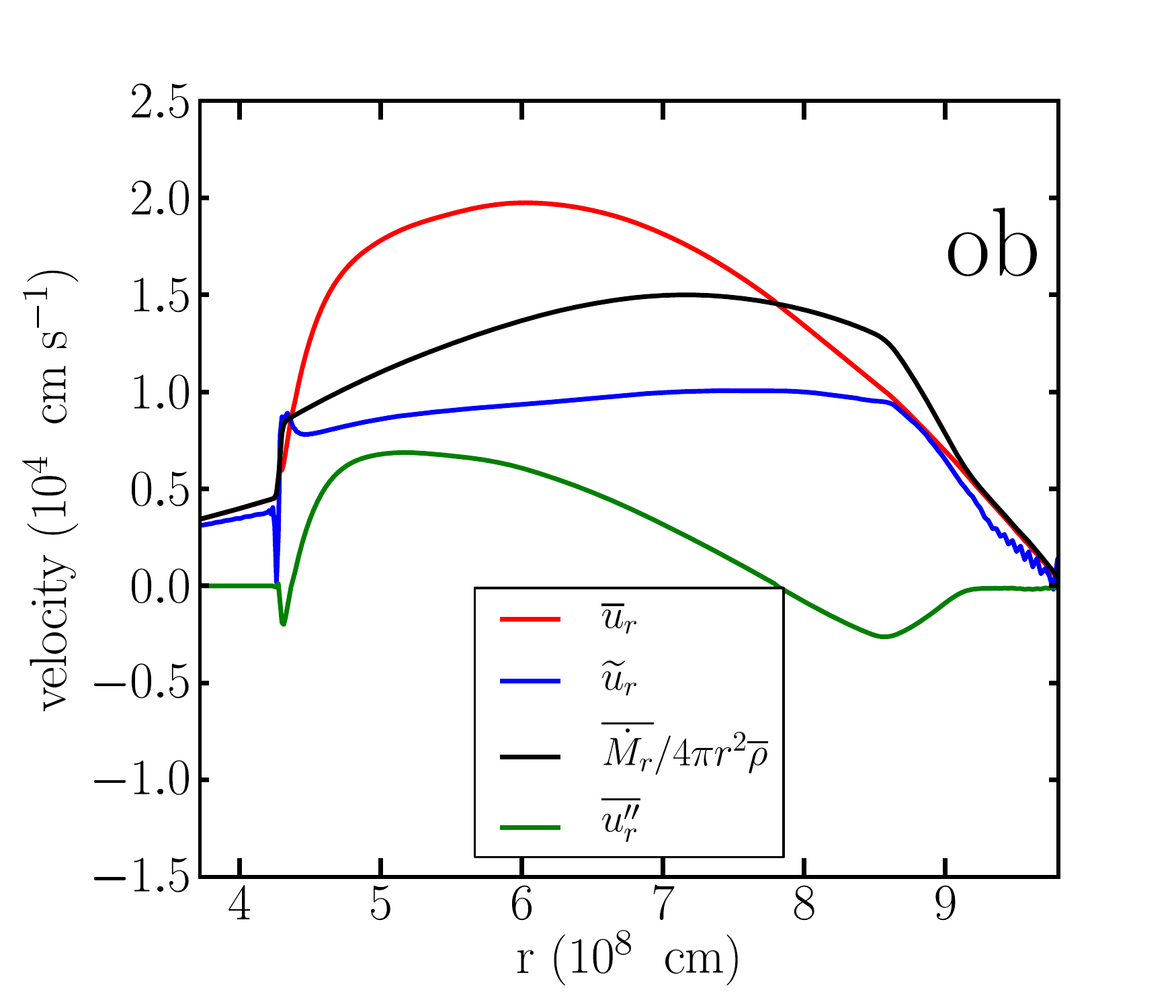}}
\caption{Mean turbulent kinetic energy equation (upper panels) and mean velocities from model {\sf ob.3D.2hp}. Averaging window over roughly 2 convective turnover timescales 150 s (left), 3 convective turnover timescales 230 s (middle) and 4 convective turnover timescales 460 s (right).}
\end{figure}

\newpage

\section{Dependence on Convection Zone Depth}

\subsection{Oxygen burning shell model}

\subsubsection{Additional information about background structure of compared models}

\begin{figure}[!h]
\centerline{
\includegraphics[width=5.8cm]{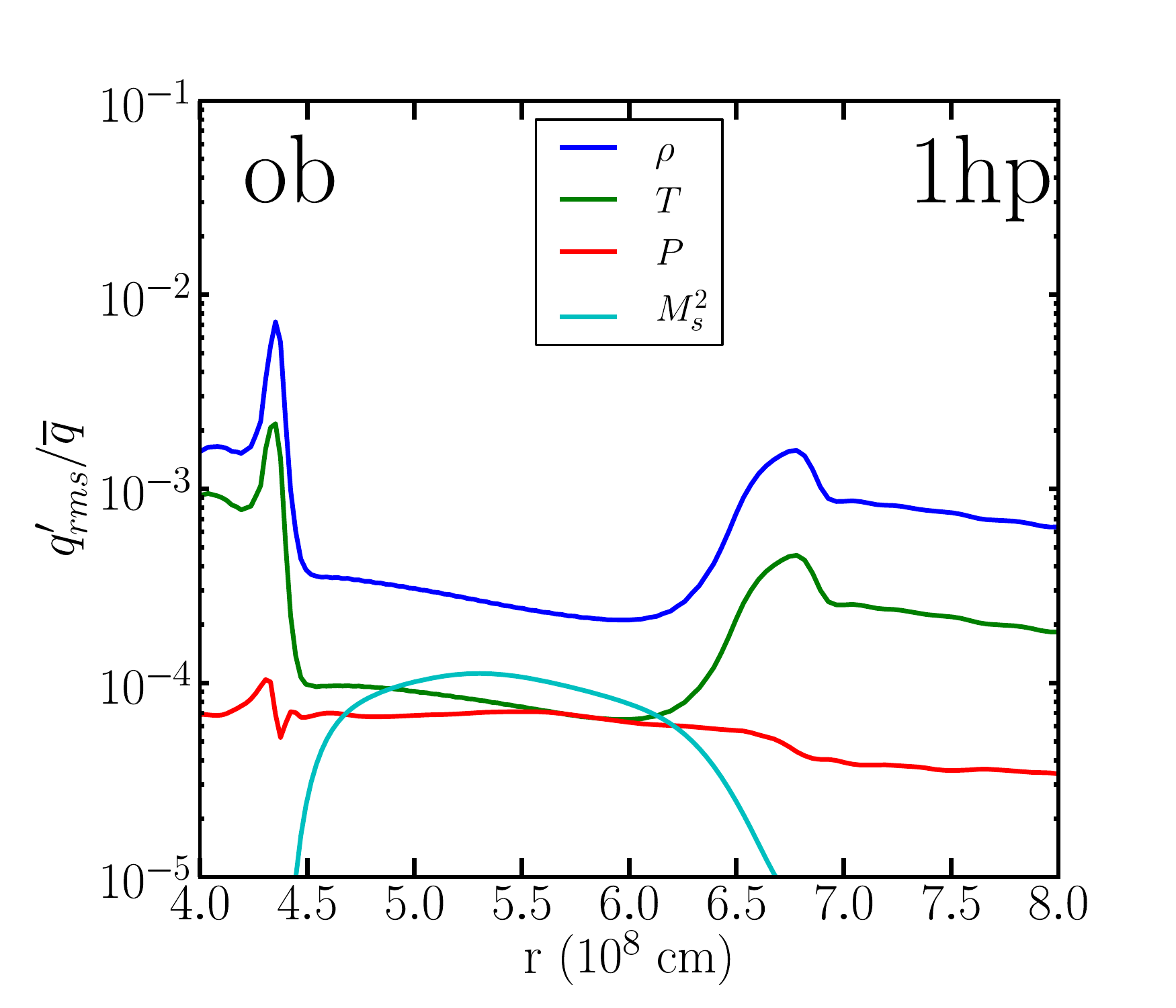}
\includegraphics[width=5.8cm]{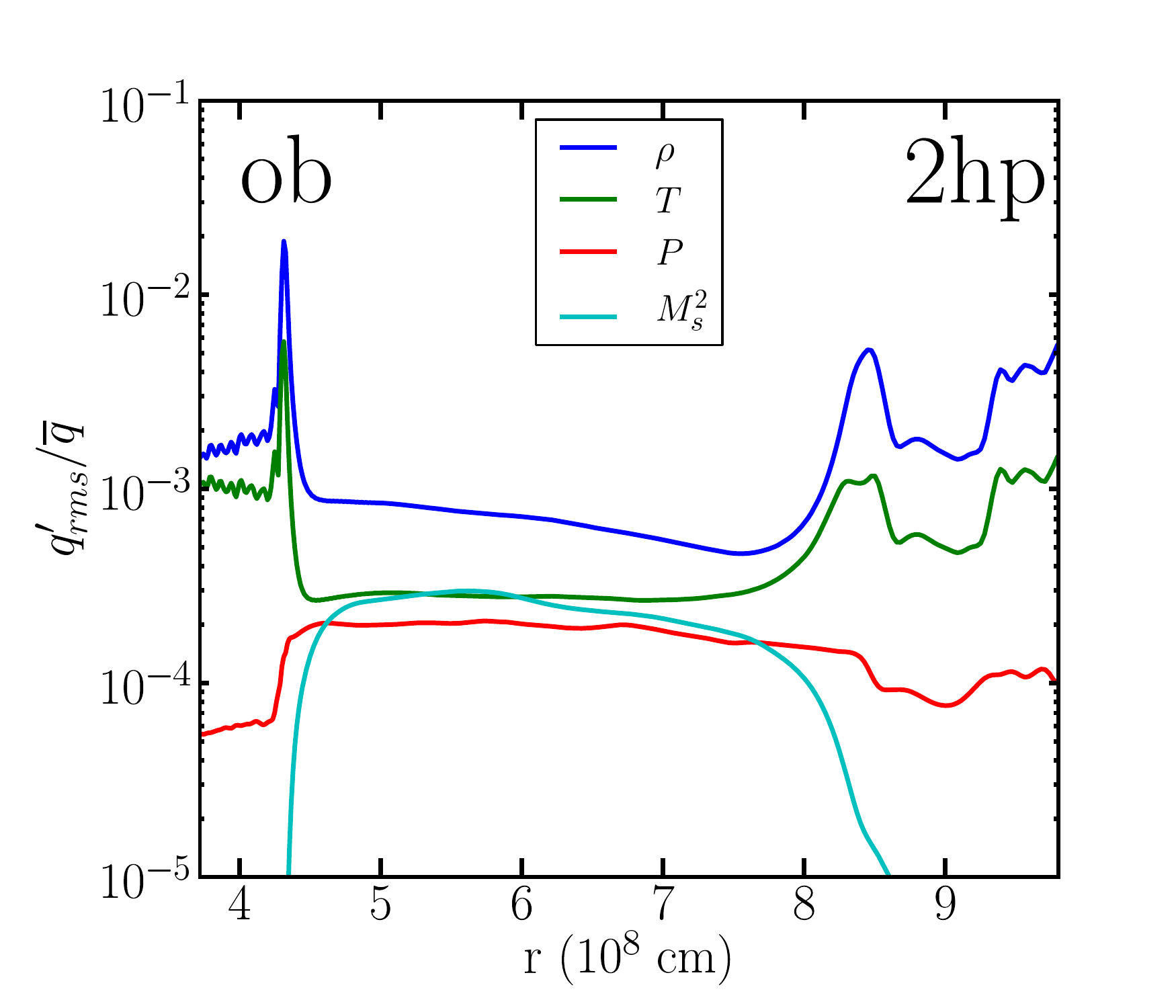}
\includegraphics[width=5.8cm]{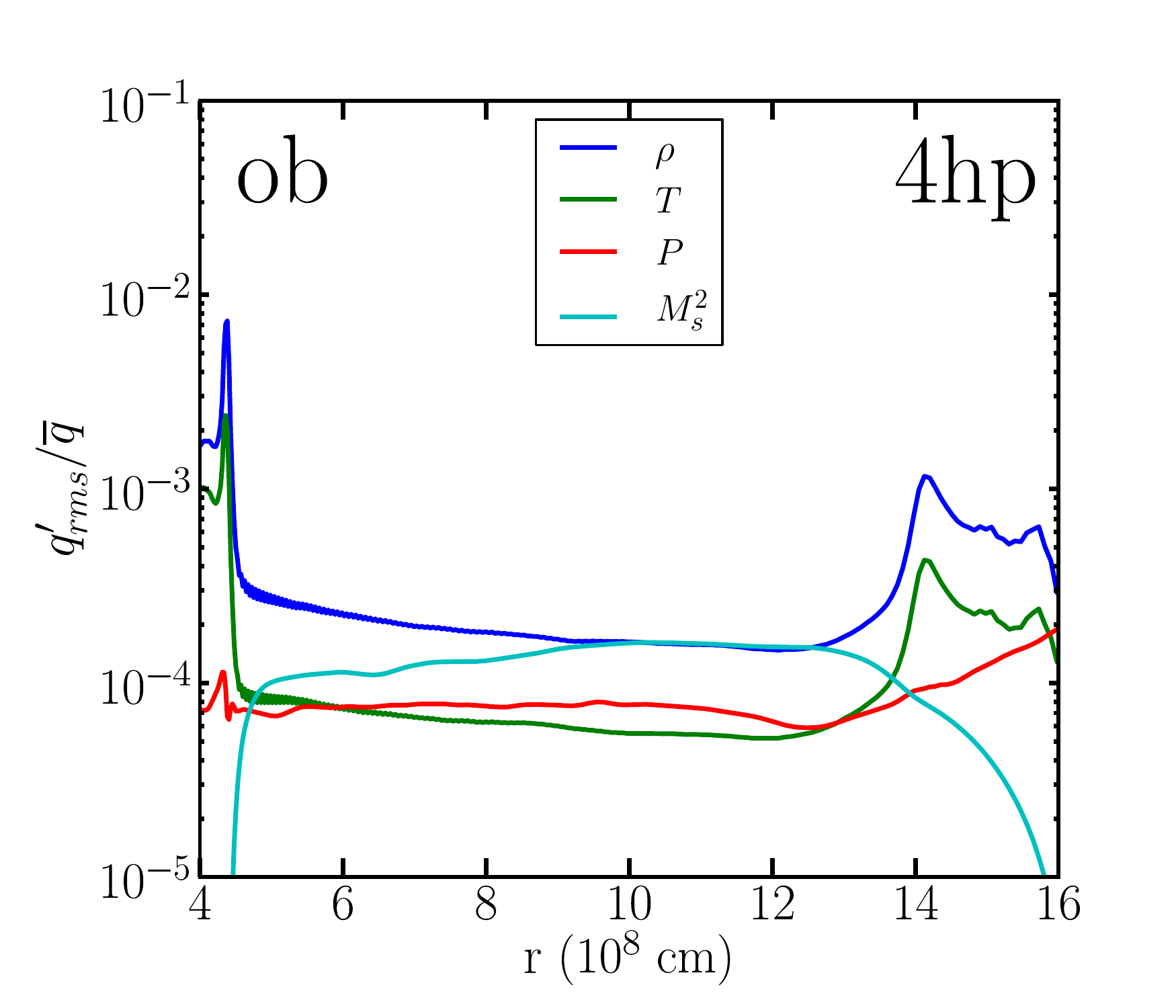}}

\centerline{
\includegraphics[width=5.8cm]{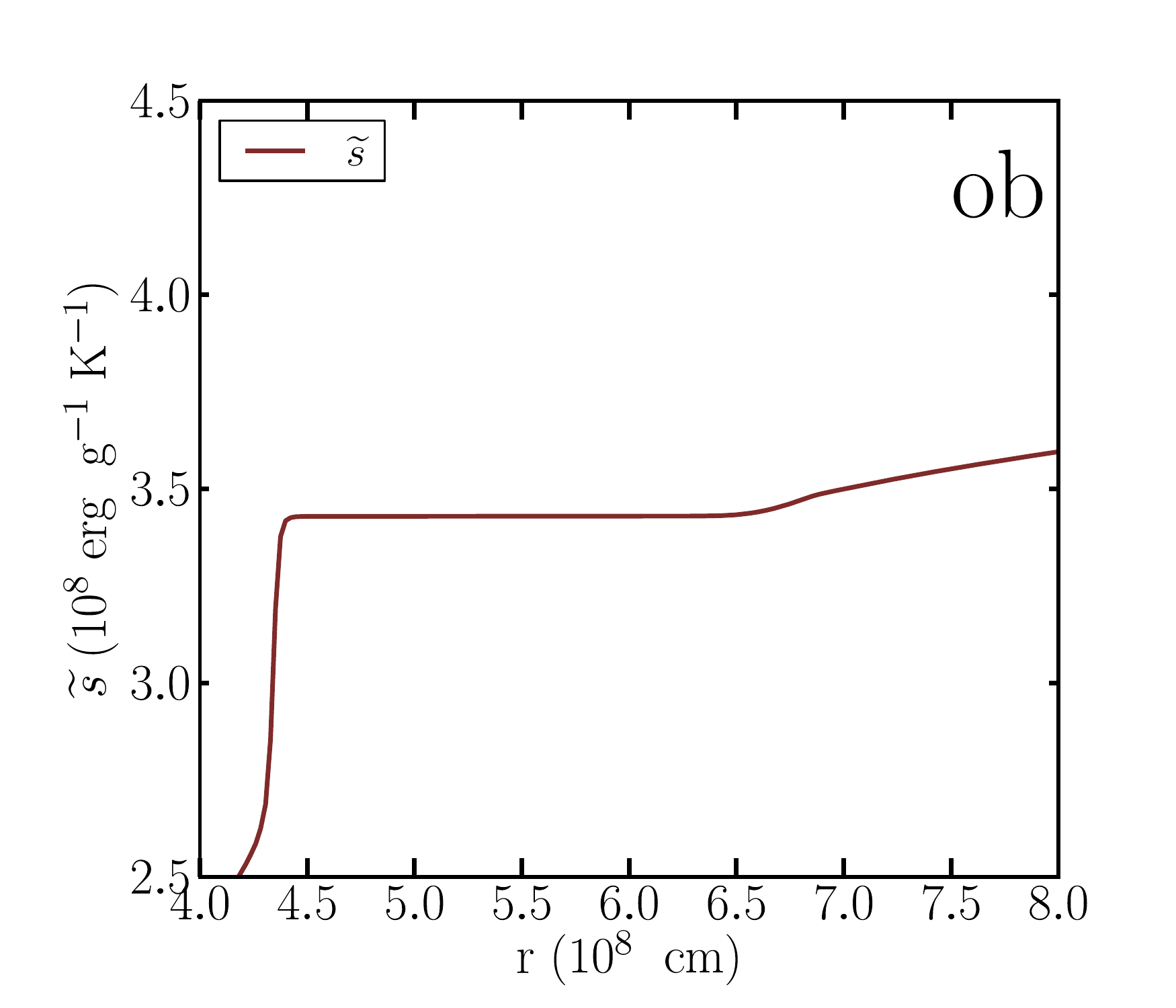}
\includegraphics[width=5.8cm]{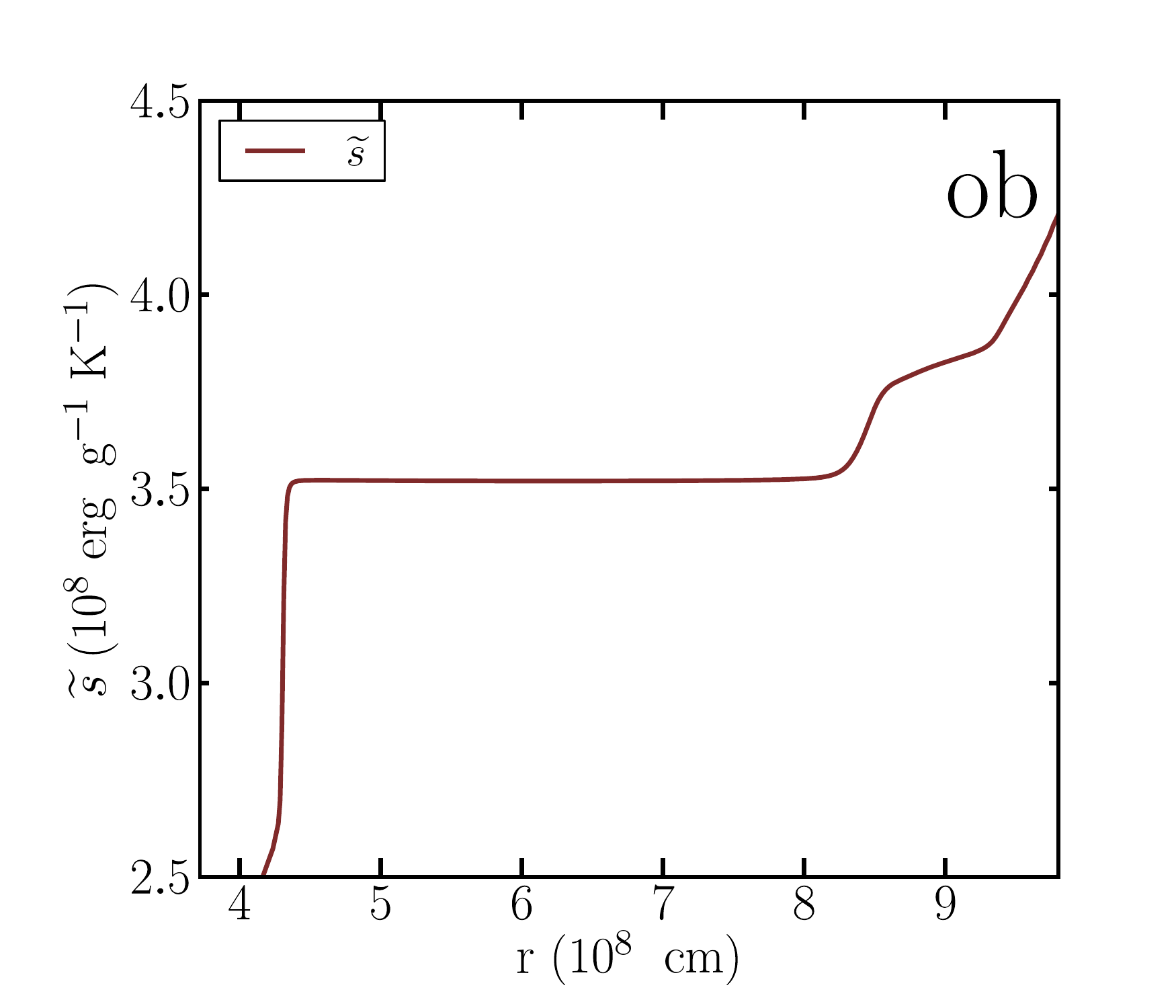}
\includegraphics[width=5.8cm]{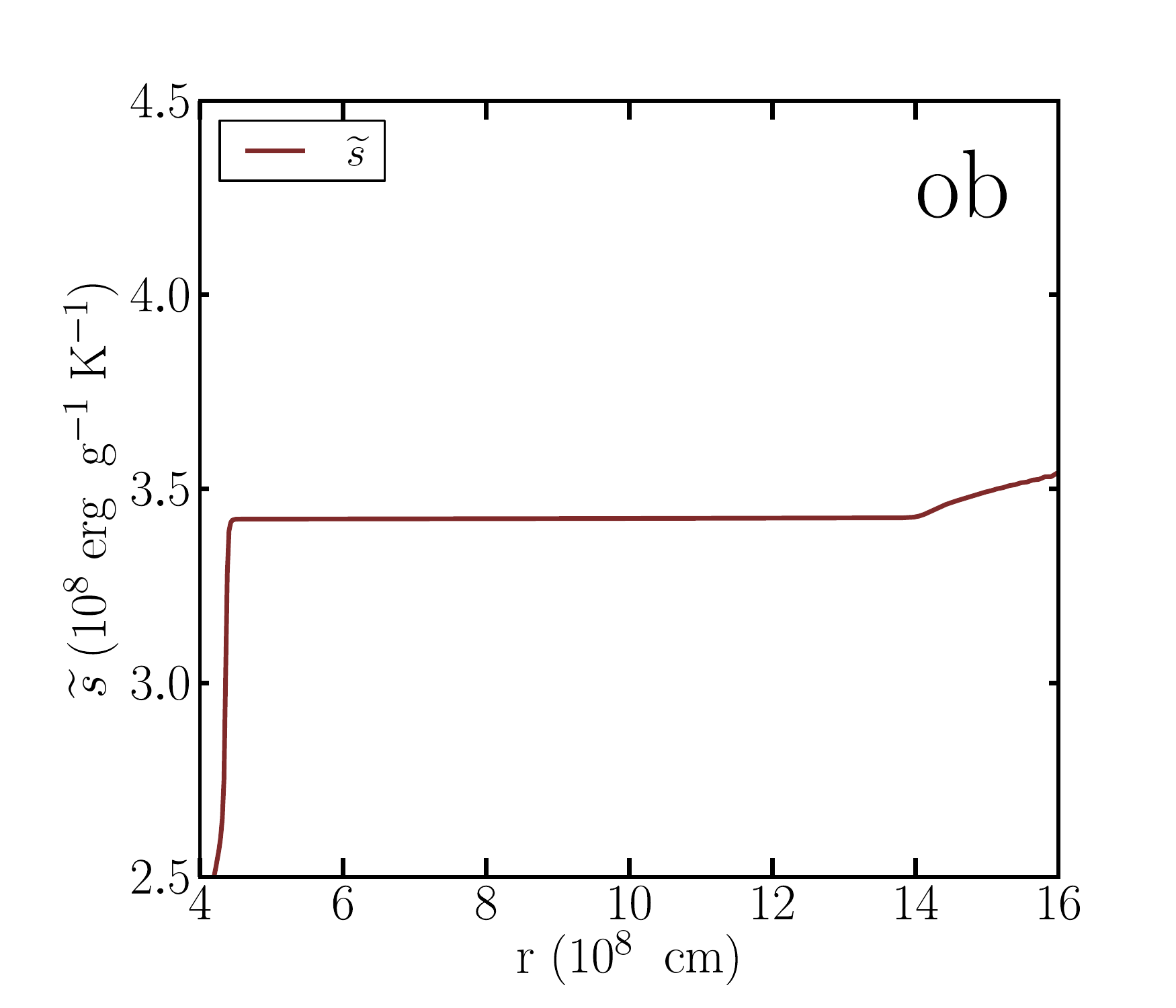}}

\caption{Relative root mean square of fluctuations of density $\rho$, temperature $T$ and pressure $P$ (upper panels) for the {\sf ob.3D.1hp} (left), {\sf ob.3D.2hp} (middle) and {\sf ob.3D.4hp} (right) models togehter with square of the flow's Mach number $M_s^2$. Profiles of mean entropy (lower panels) in {\sf ob.3D.1hp} (left), {\sf ob.3D.2hp} (middle) and {\sf ob.3D.4hp} (right). \label{fig:ob-rms-flct}}
\end{figure}

\newpage

\subsubsection{Mean continuity equation and mean radial momentum equation}

\begin{figure}[!h]
\centerline{
\includegraphics[width=6.cm]{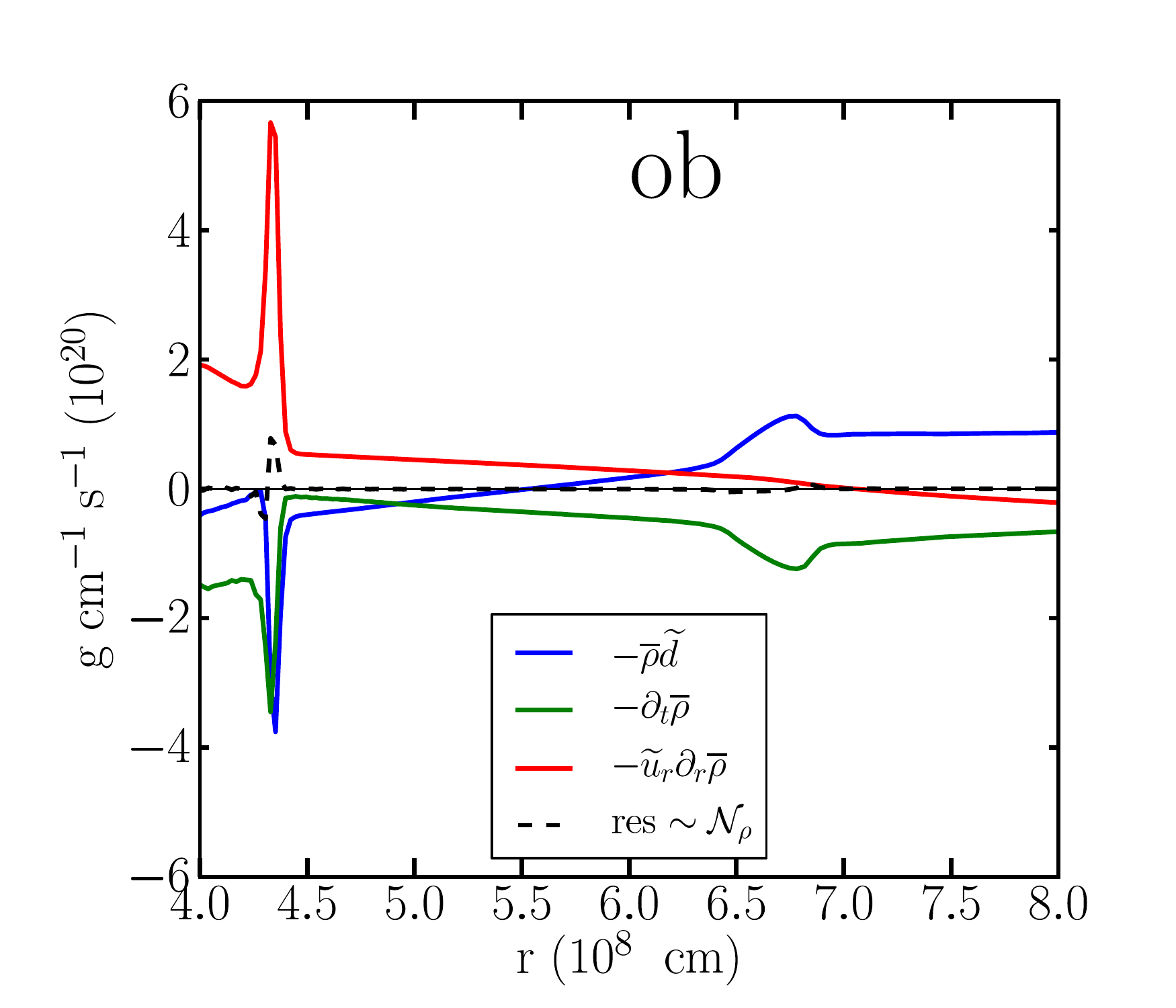}
\includegraphics[width=6.cm]{ob3dB_tavg230_continuity_equation_insf-eps-converted-to.pdf}
\includegraphics[width=6.cm]{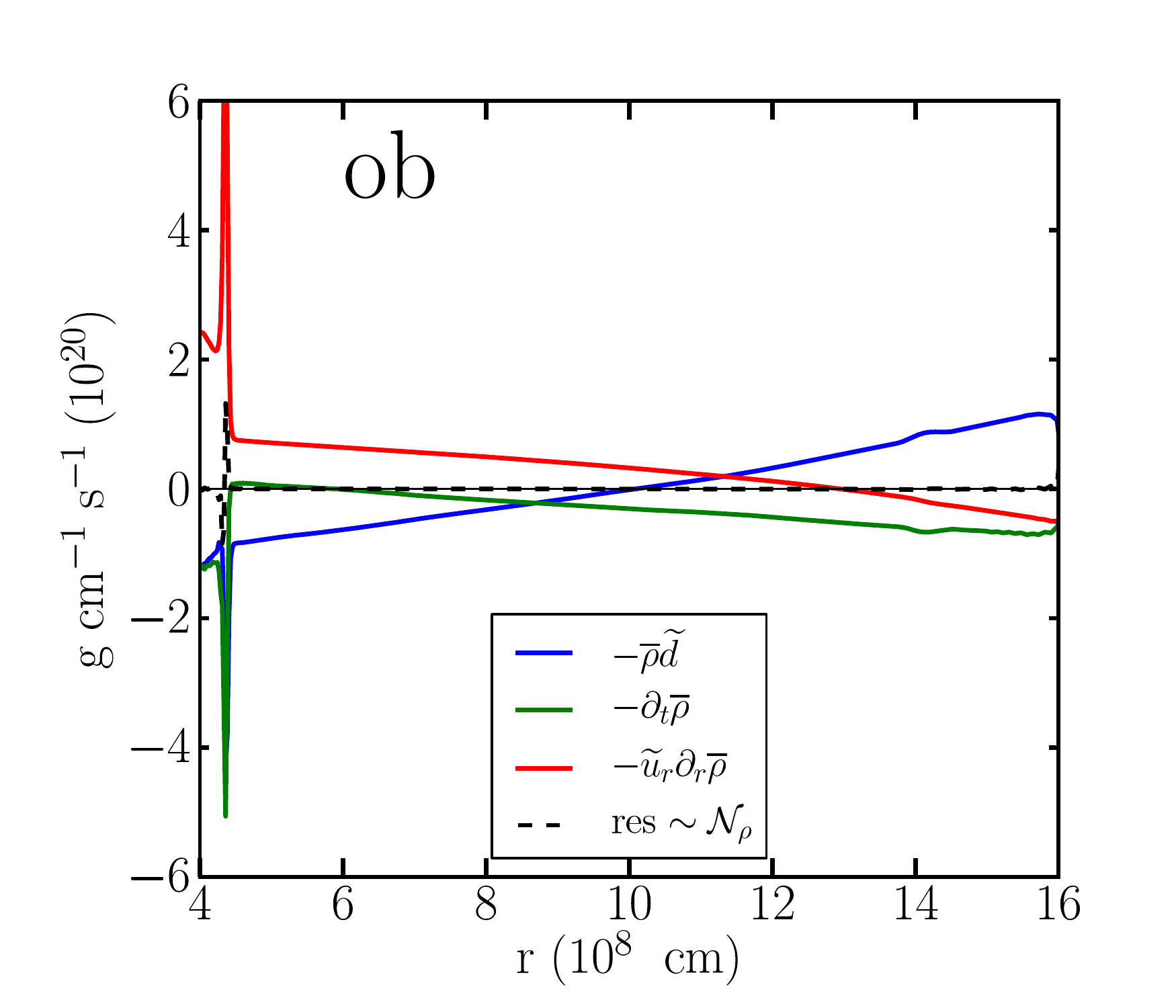}}

\centerline{
\includegraphics[width=6.cm]{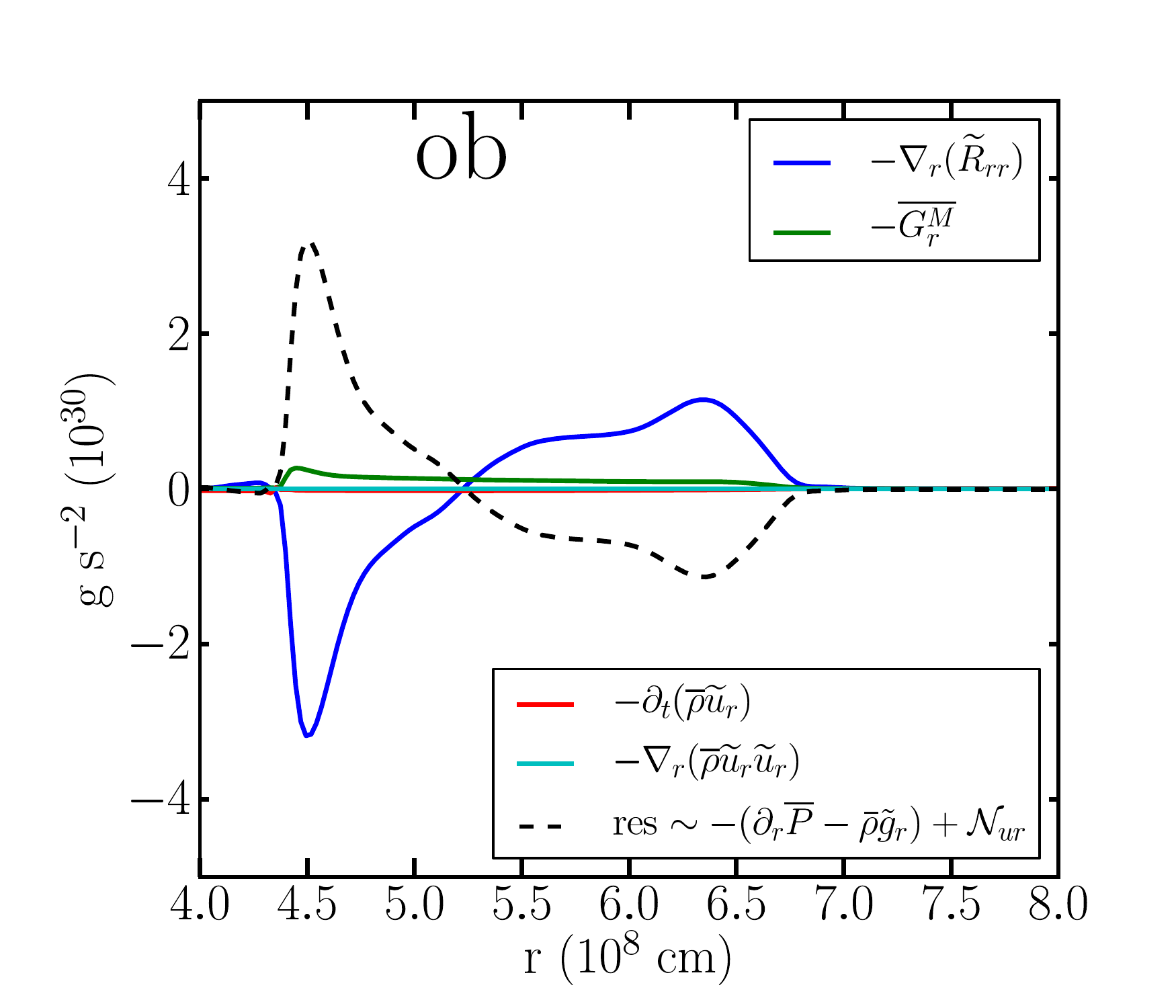}
\includegraphics[width=6.cm]{ob3dB_tavg230_rmomentum_equation_insf-eps-converted-to.pdf}
\includegraphics[width=6.cm]{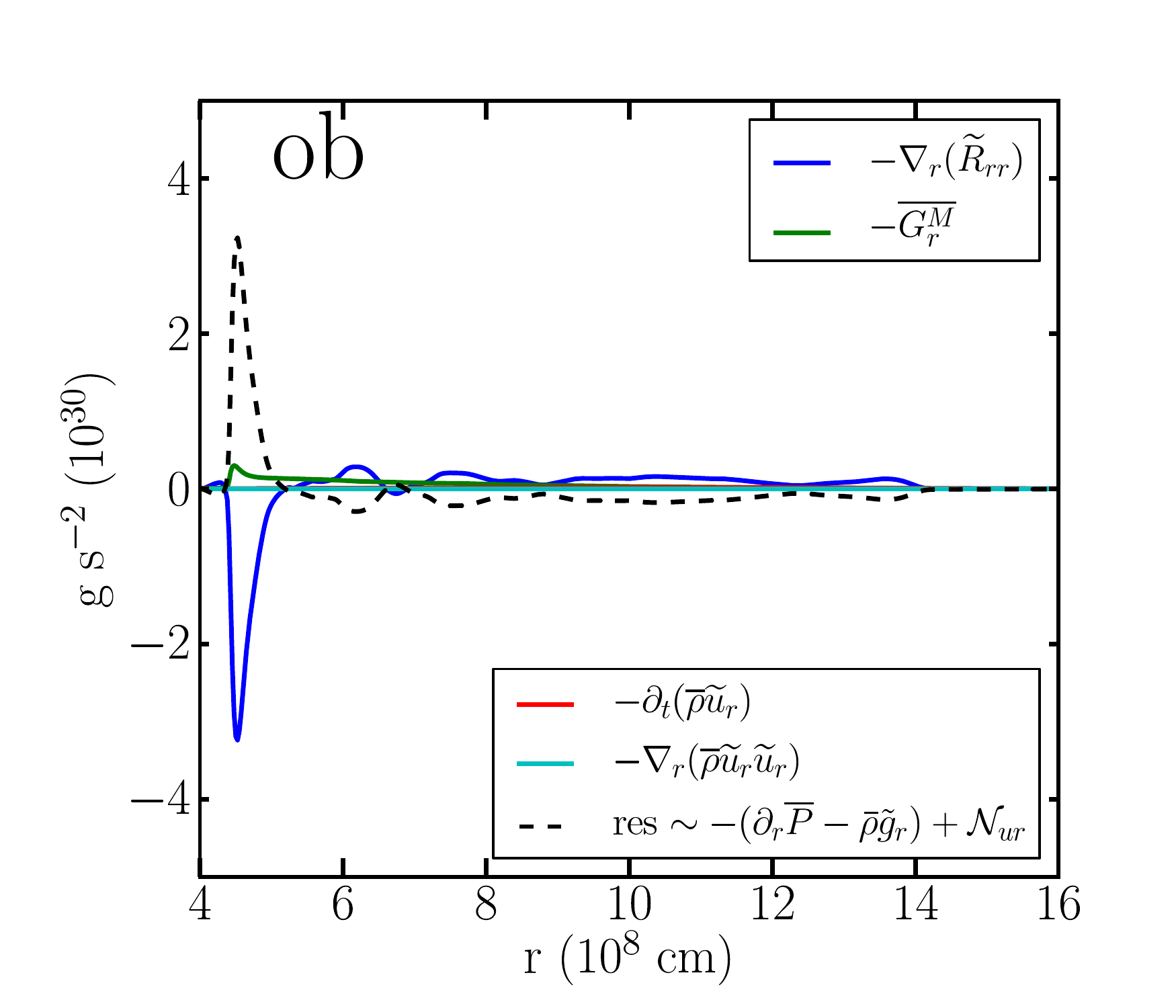}}
\caption{Mean continuity equation (upper panels) and radial momentum equation (lower panels). 1 Hp model {\sf ob.3D.1hp} (left), 2 Hp model {\sf ob.3D.2hp} (middle) and 4 Hp model {\sf ob.3D.4hp} (right)}
\end{figure}

\newpage

\subsubsection{Mean azimuthal and polar momentum equation}

\begin{figure}[!h]
\centerline{
\includegraphics[width=7.0cm]{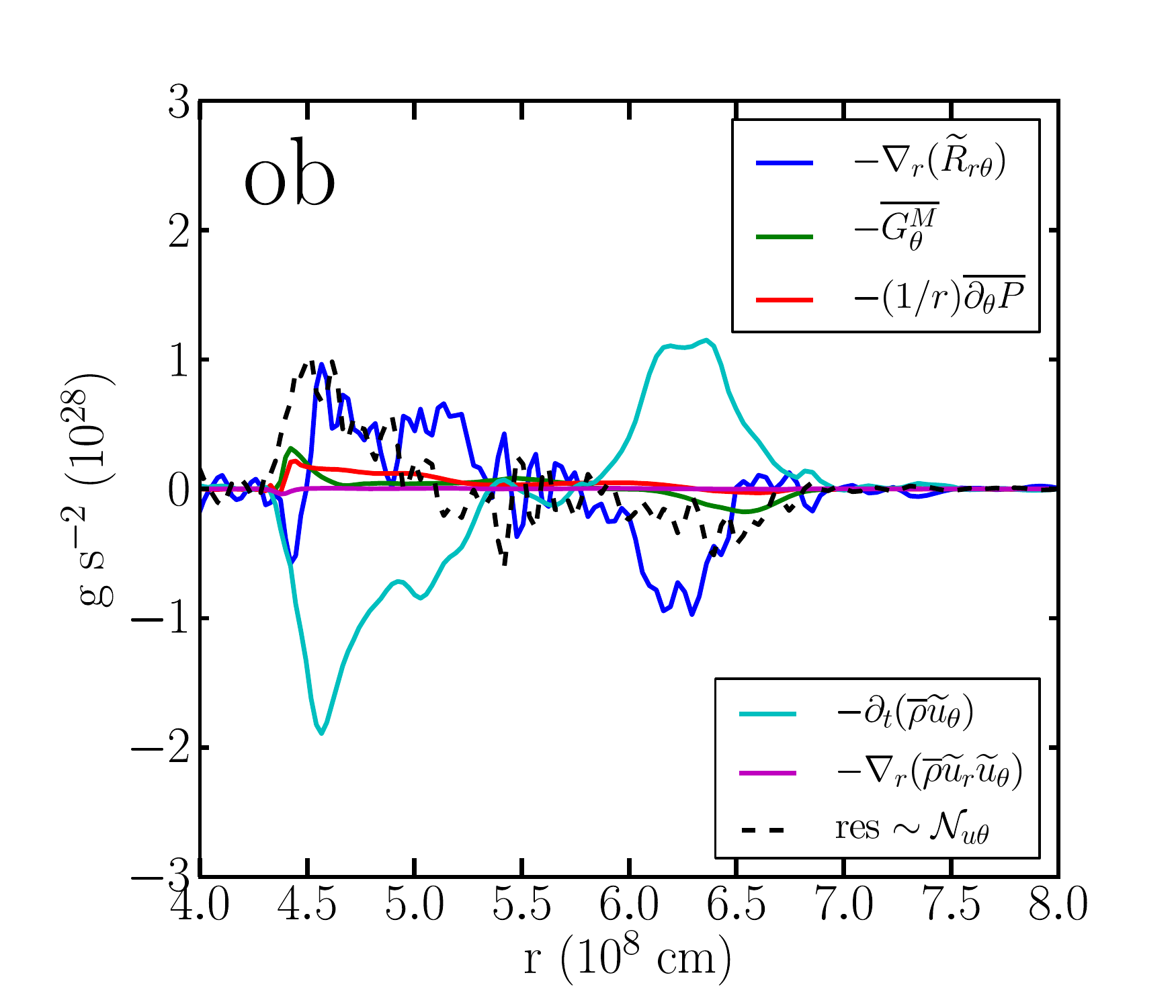}
\includegraphics[width=7.0cm]{ob3dB_tavg230_tmomentum_equation_insf-eps-converted-to.pdf}
\includegraphics[width=7.0cm]{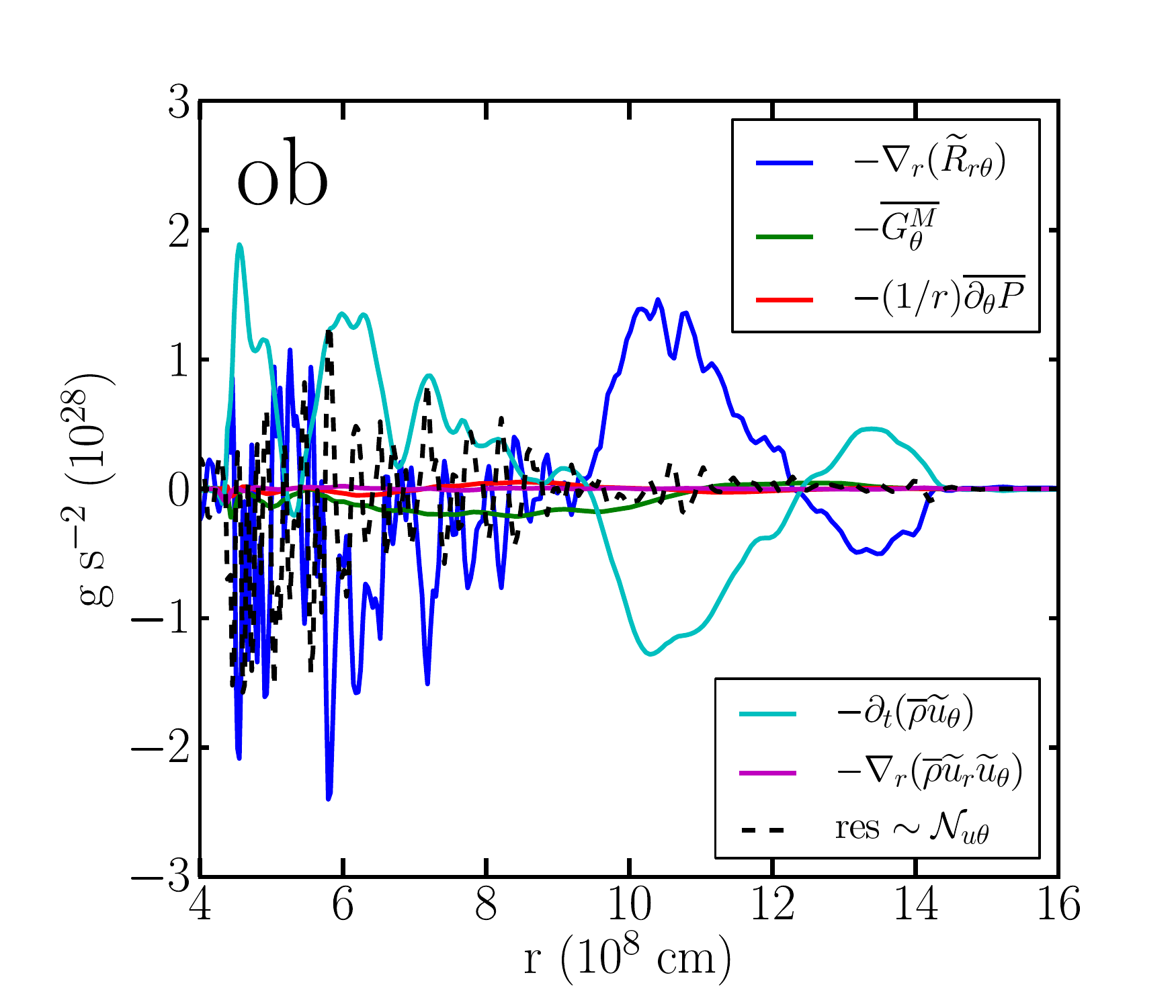}}

\centerline{
\includegraphics[width=7.0cm]{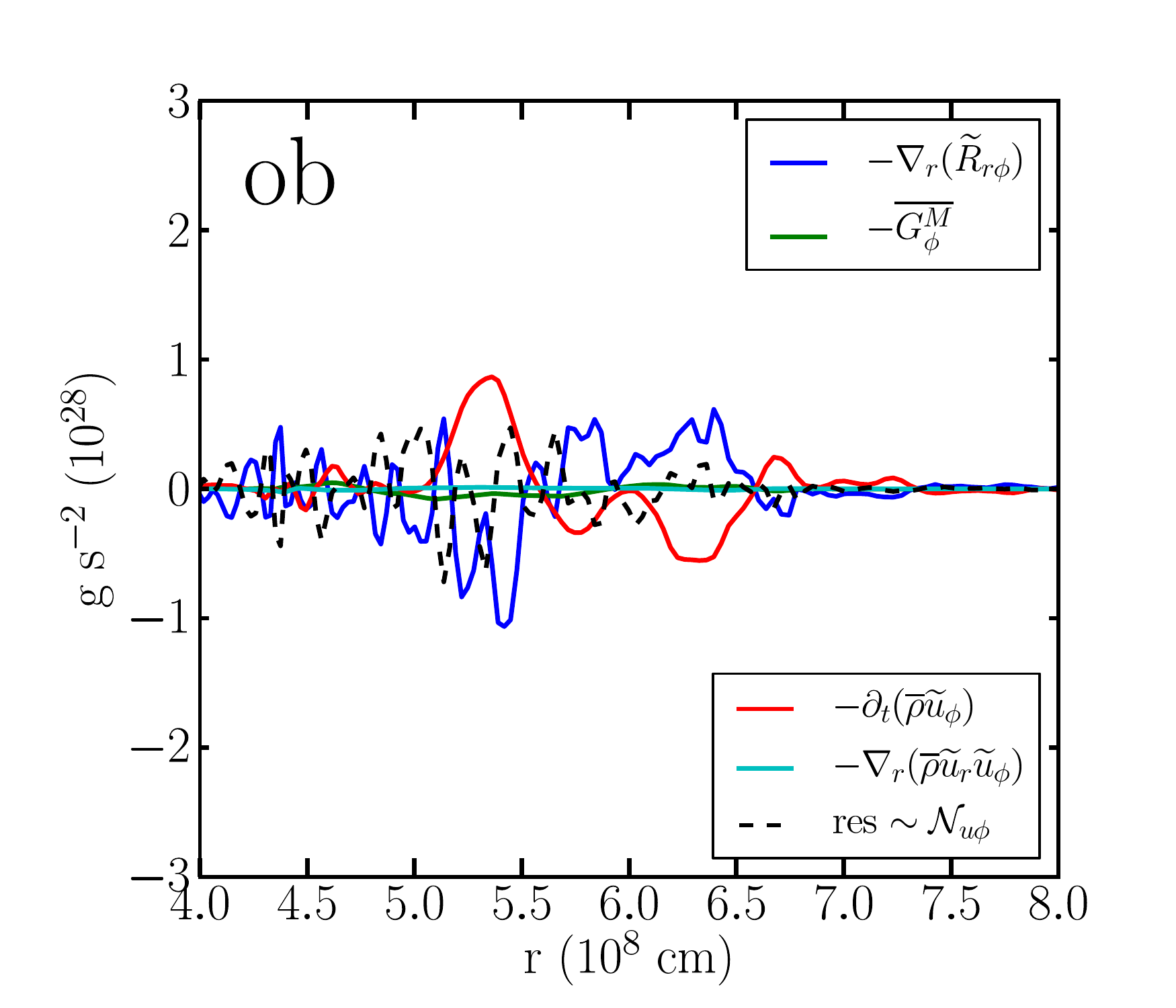}
\includegraphics[width=7.0cm]{ob3dB_tavg230_pmomentum_equation_insf-eps-converted-to.pdf}
\includegraphics[width=7.0cm]{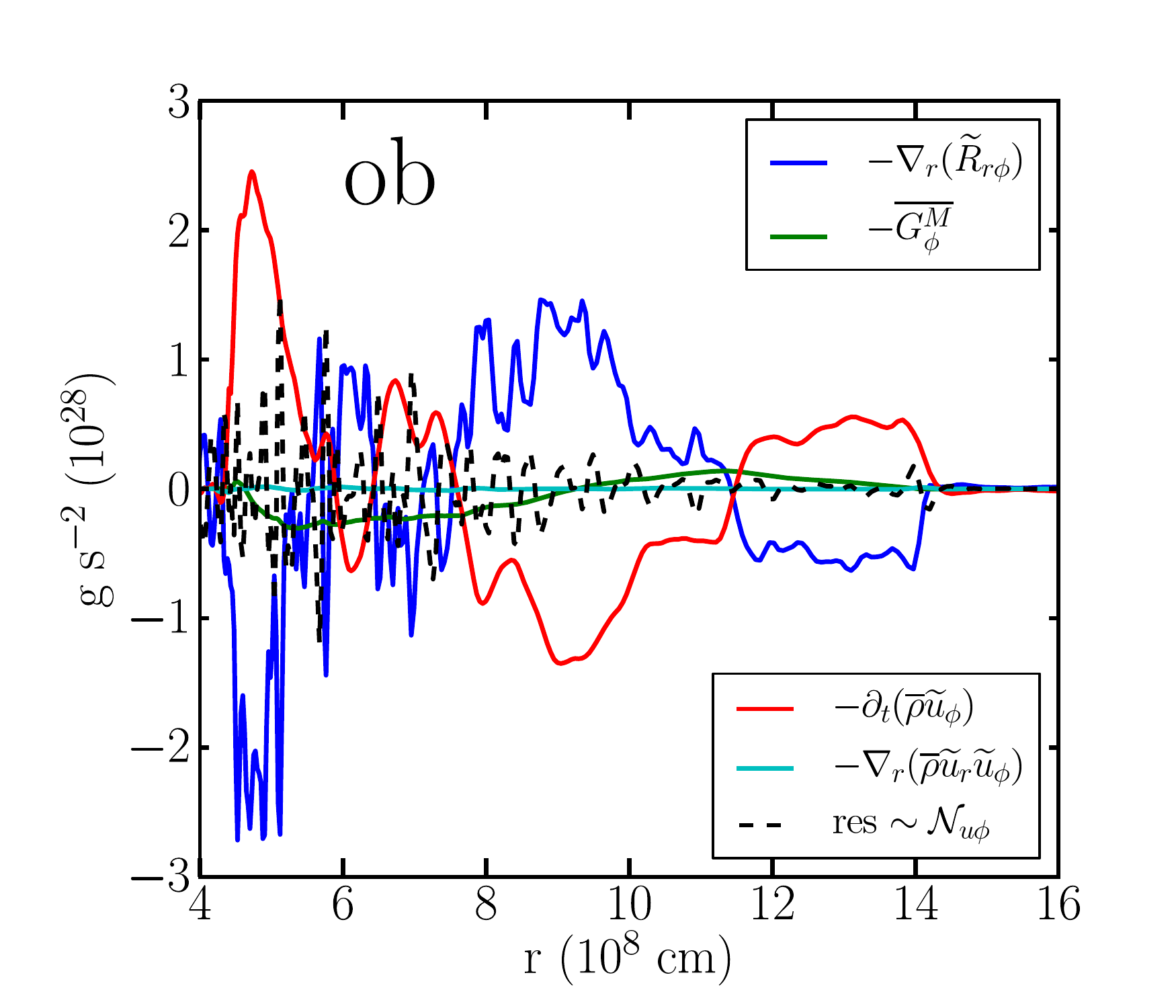}}
\caption{Mean azimuthal equation (upper panels) and mean polar momentum equation (lower panels). 1 Hp model {\sf ob.3D.1hp} (left), 2 Hp model {\sf ob.3D.2hp} (middle) and 4 Hp model {\sf ob.3D.4hp} (right).}
\end{figure}

\newpage

\subsubsection{Mean total energy equation and turbulent kinetic energy equation}

\begin{figure}[!h]
\centerline{
\includegraphics[width=7.cm]{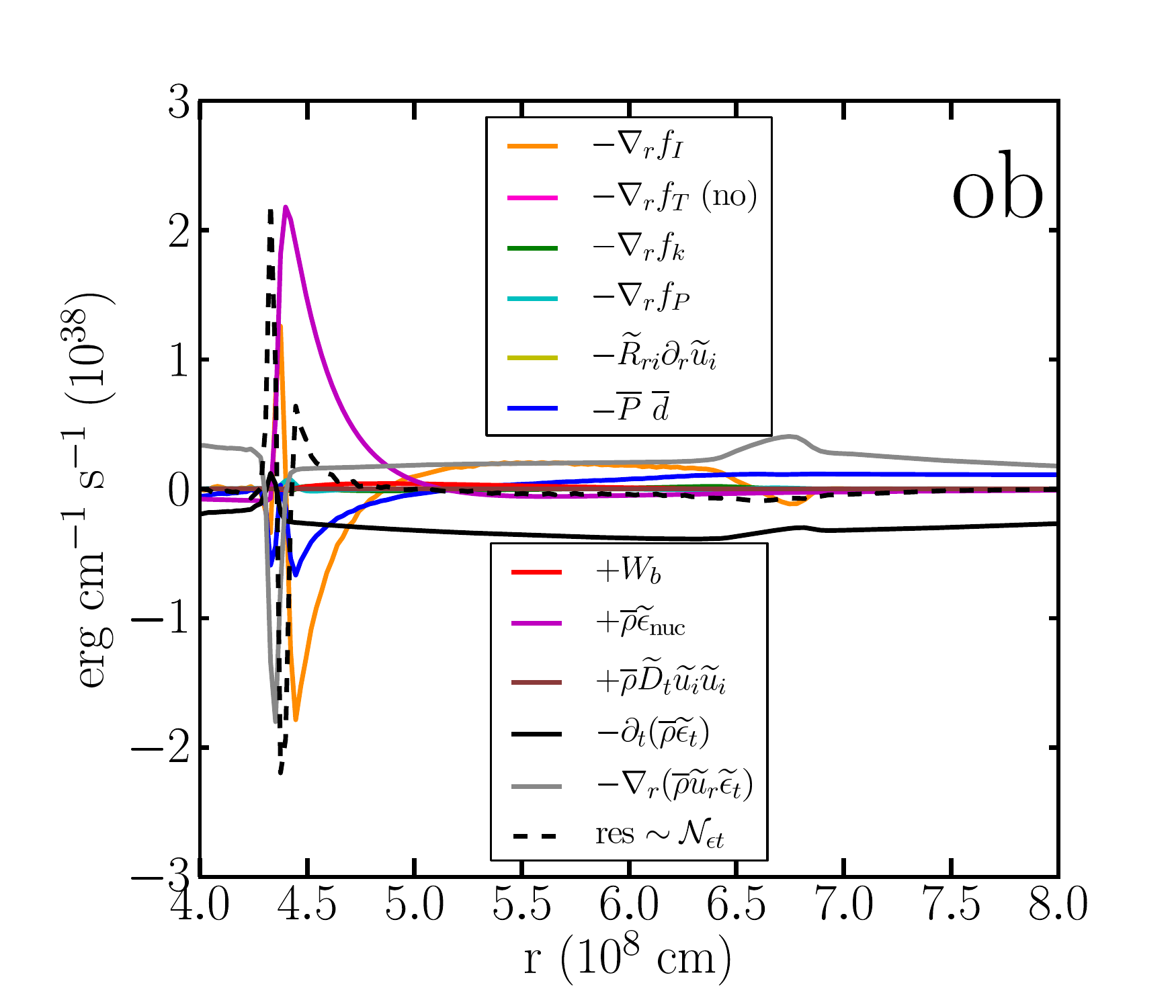}
\includegraphics[width=7.cm]{ob3dB_tavg230_total_energy_equation_insf-eps-converted-to.pdf}
\includegraphics[width=7.cm]{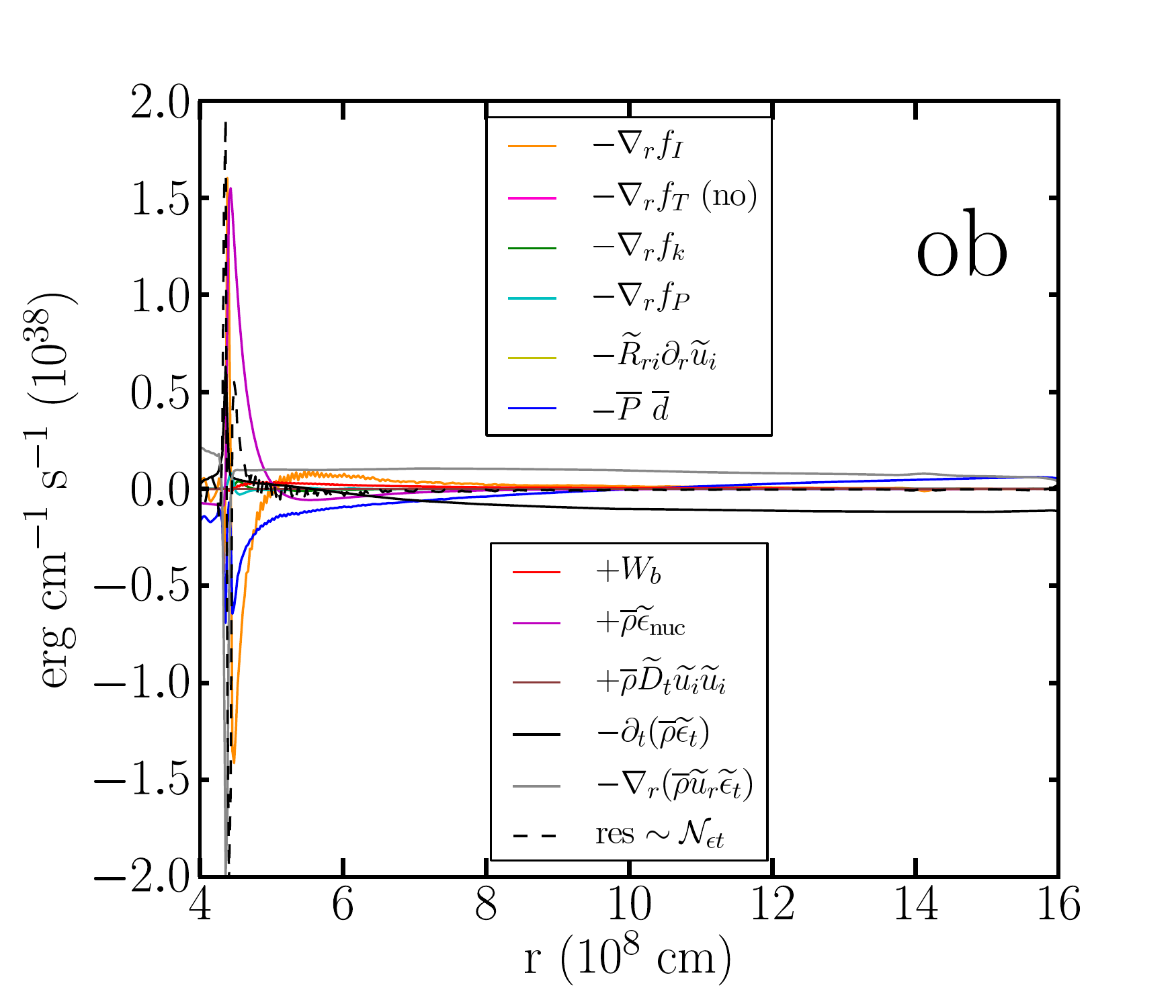}}

\centerline{
\includegraphics[width=7.cm]{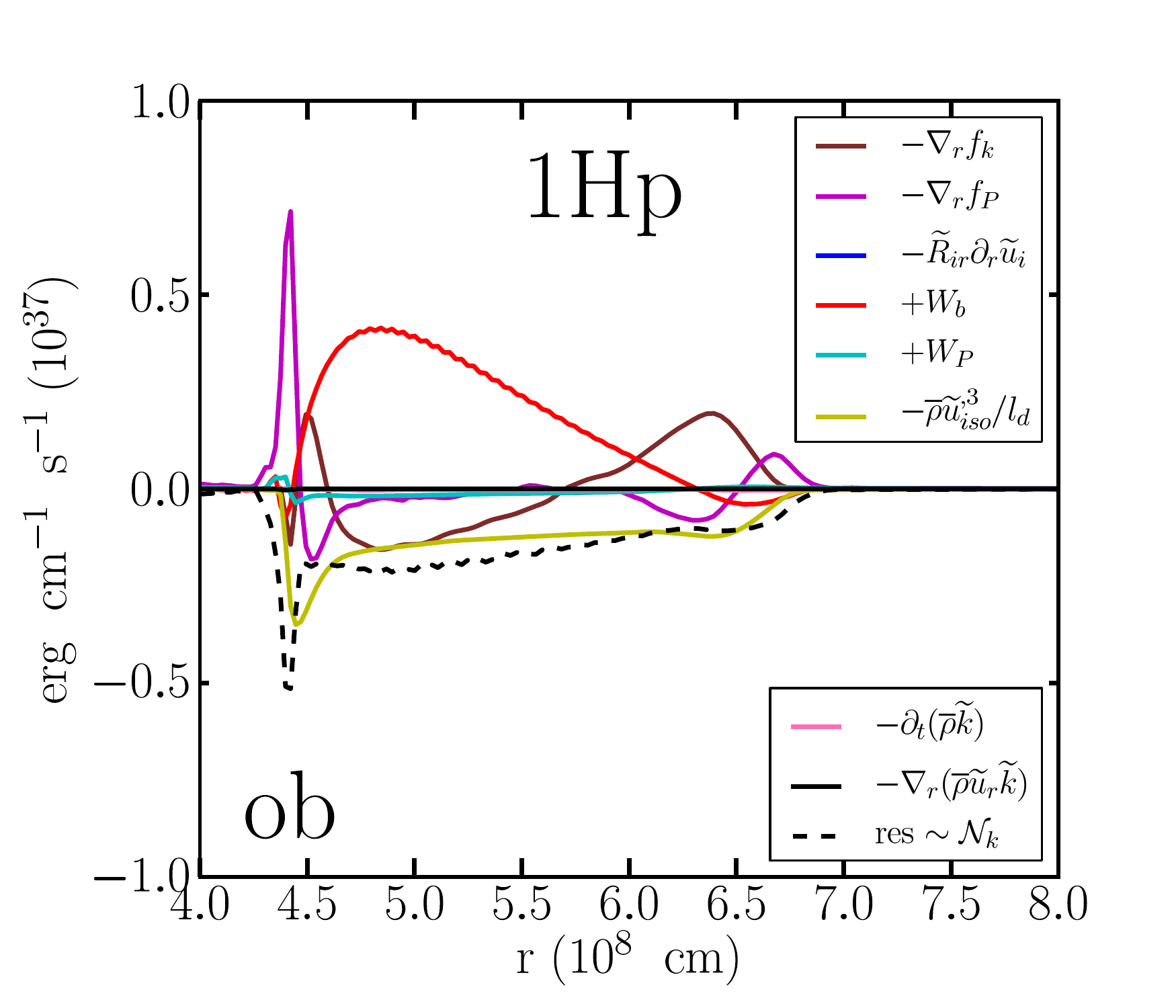}
\includegraphics[width=7.cm]{ob3dB_tavg230_mfields_k_equation_insf-eps-converted-to.pdf}
\includegraphics[width=7.cm]{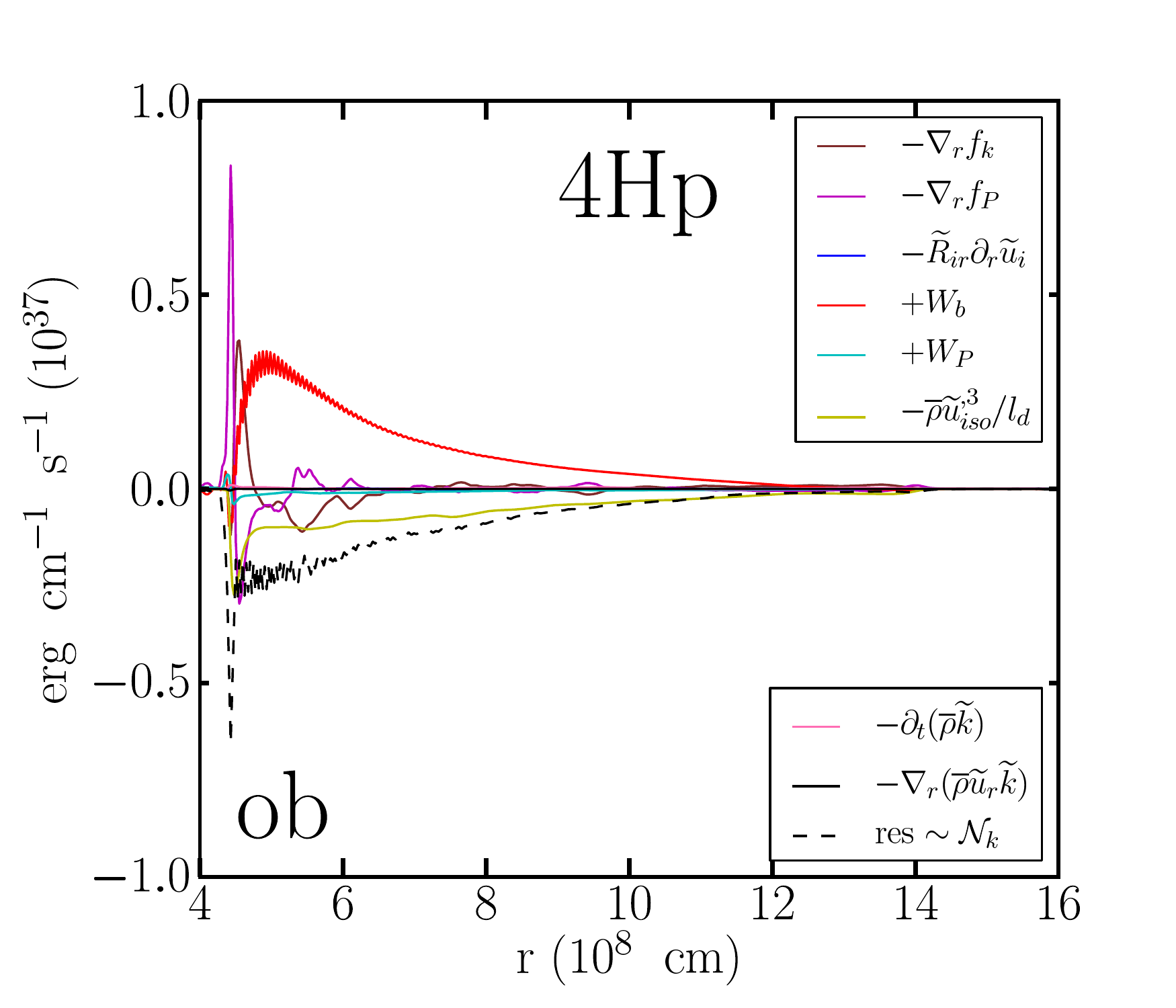}}
\caption{Mean total energy equation (upper panels) and mean turbulent kinetic energy equation (lower panels). 1 Hp model {\sf ob.3D.1hp} (left), 2 Hp model {\sf ob.3D.2hp} (middle) and 4 Hp model {\sf ob.3D.4hp} (right).}
\end{figure}

\newpage

\subsubsection{Mean turbulent kinetic energy equation (radial + horizontal part)}

\begin{figure}[!h]
\centerline{
\includegraphics[width=7.cm]{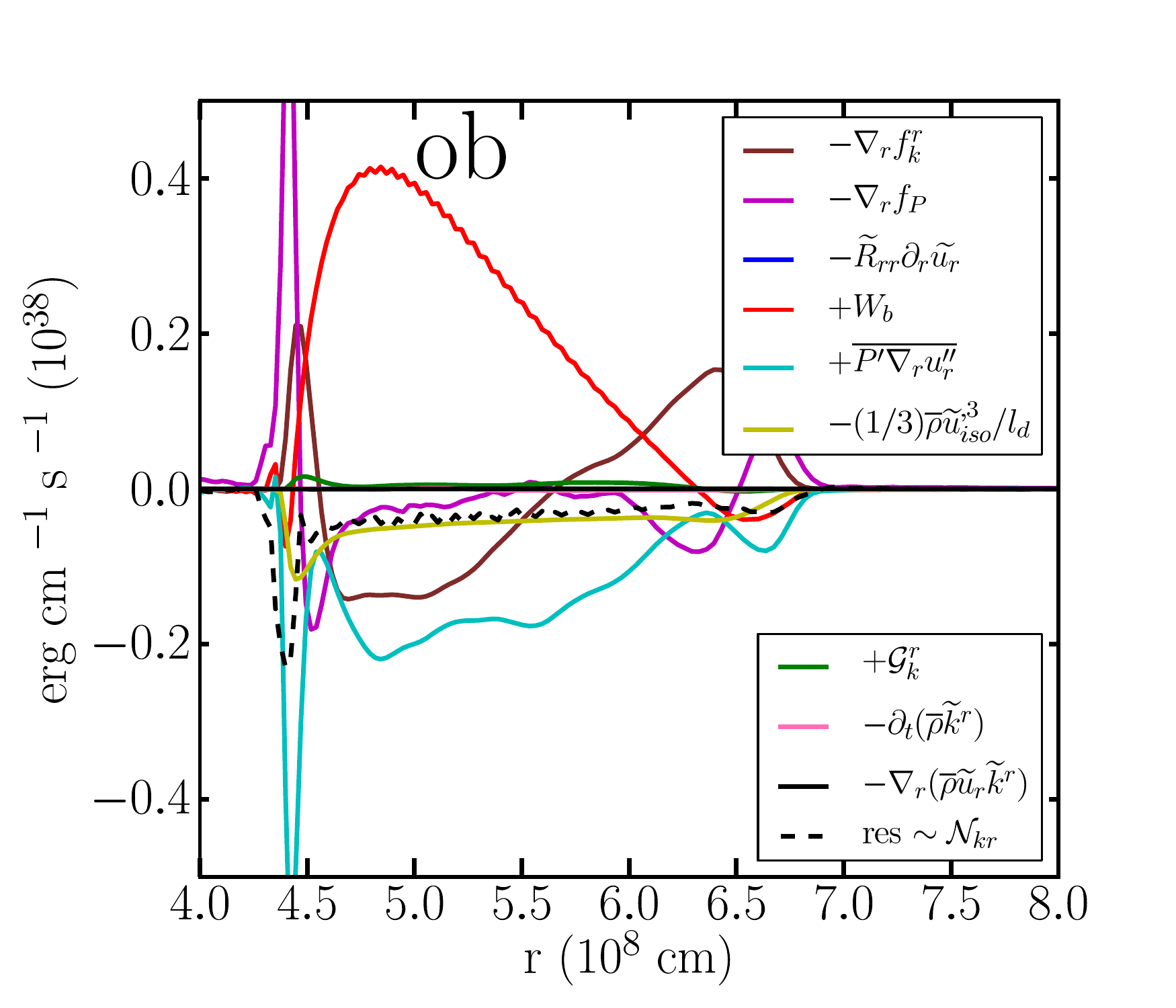}
\includegraphics[width=7.cm]{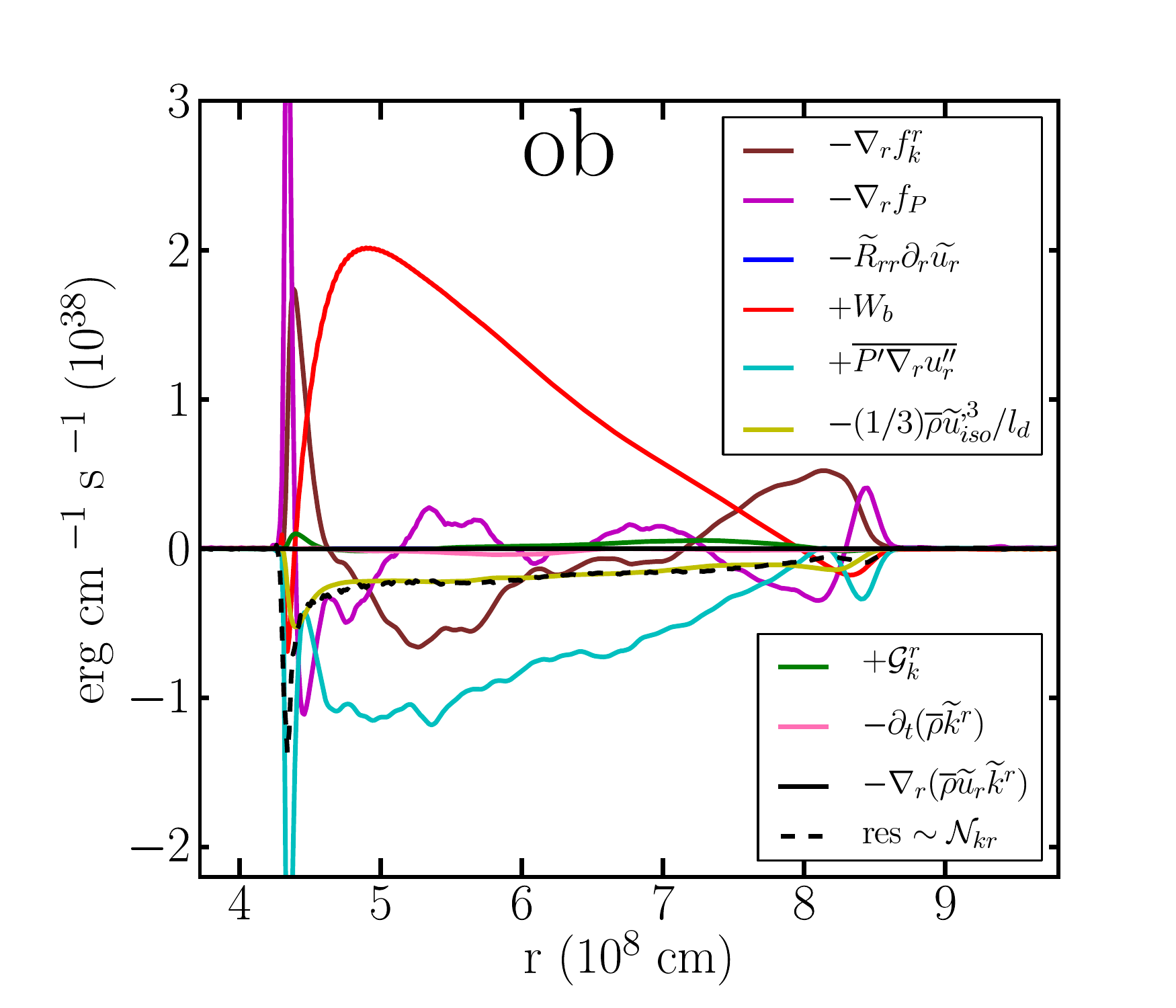}
\includegraphics[width=7.cm]{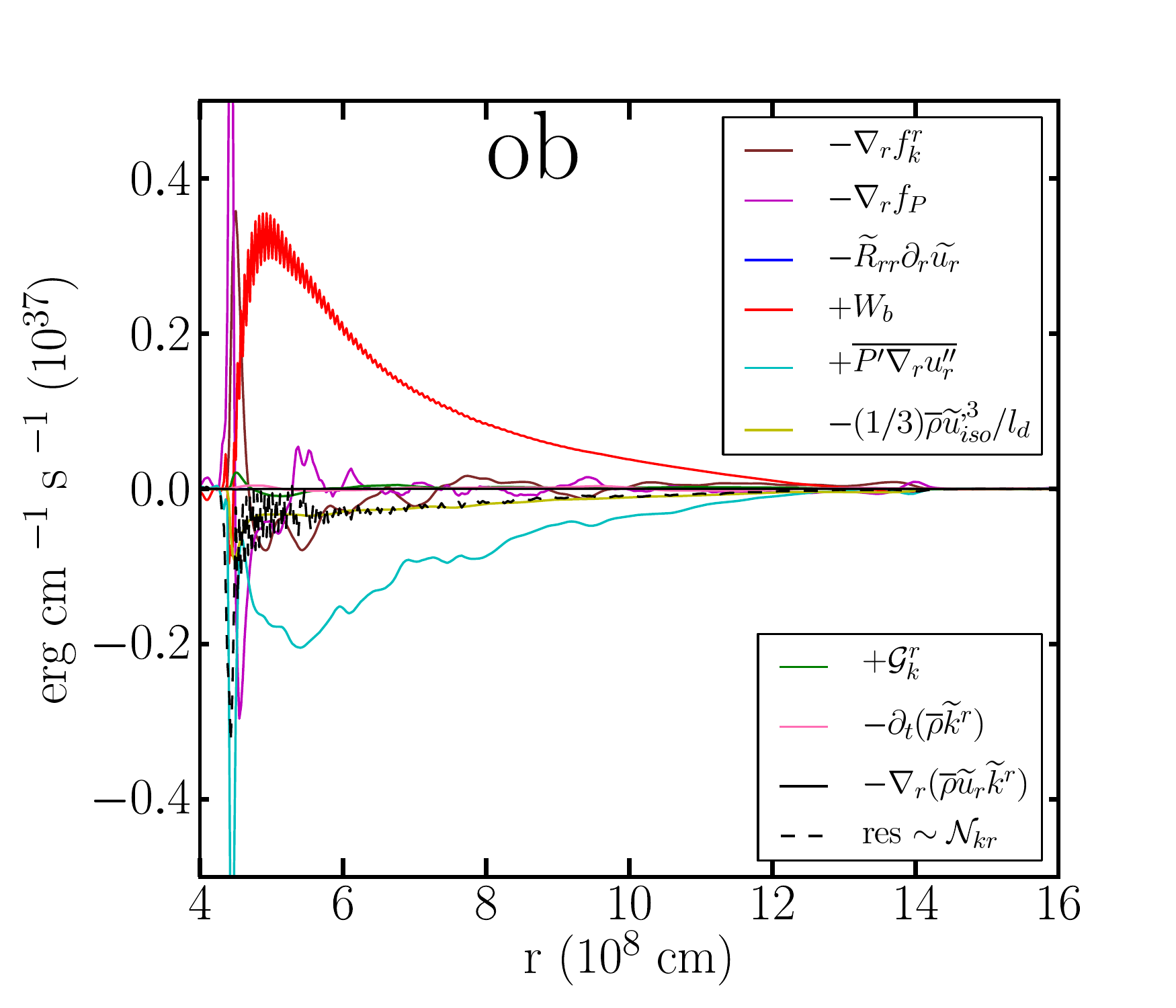}}

\centerline{
\includegraphics[width=7.cm]{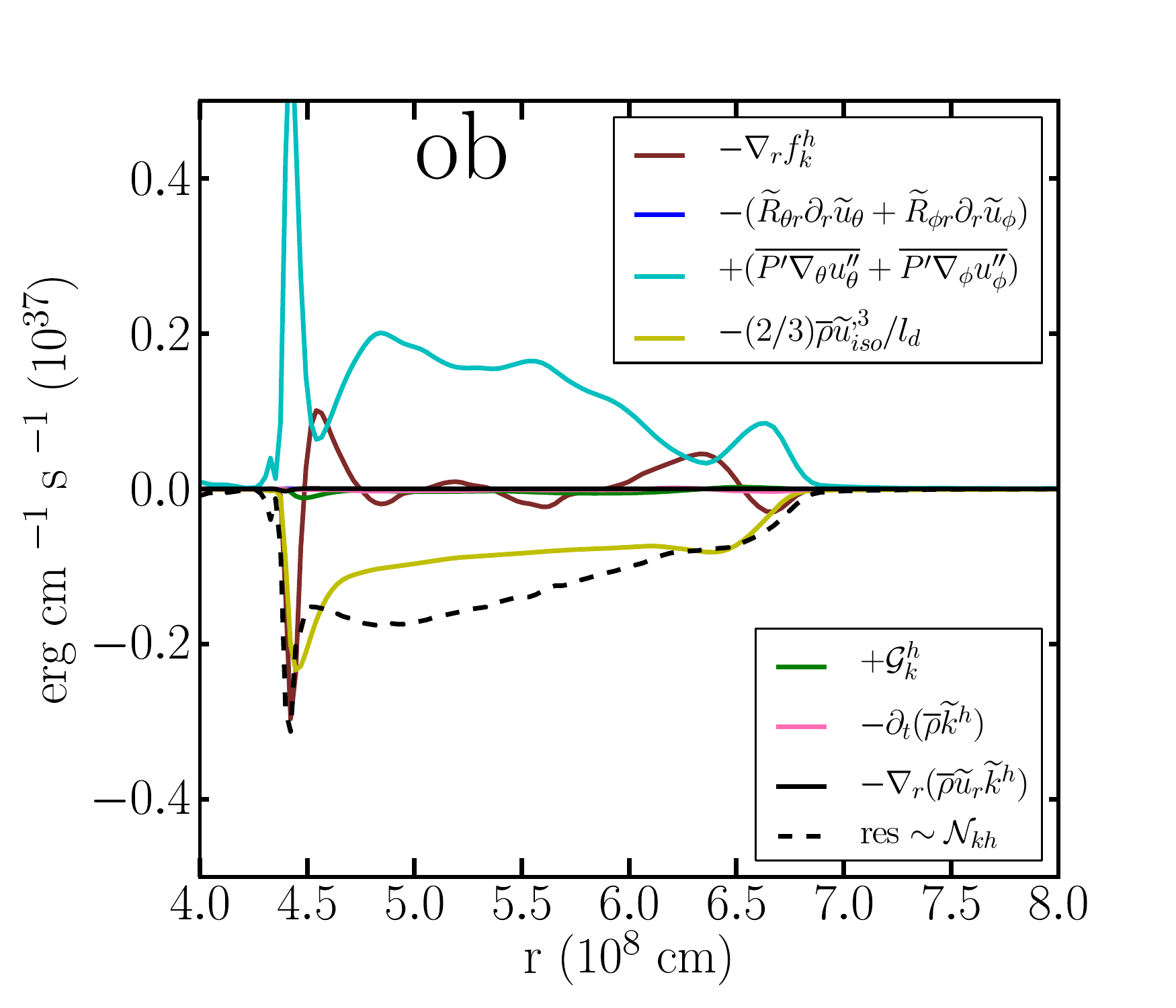}
\includegraphics[width=7.cm]{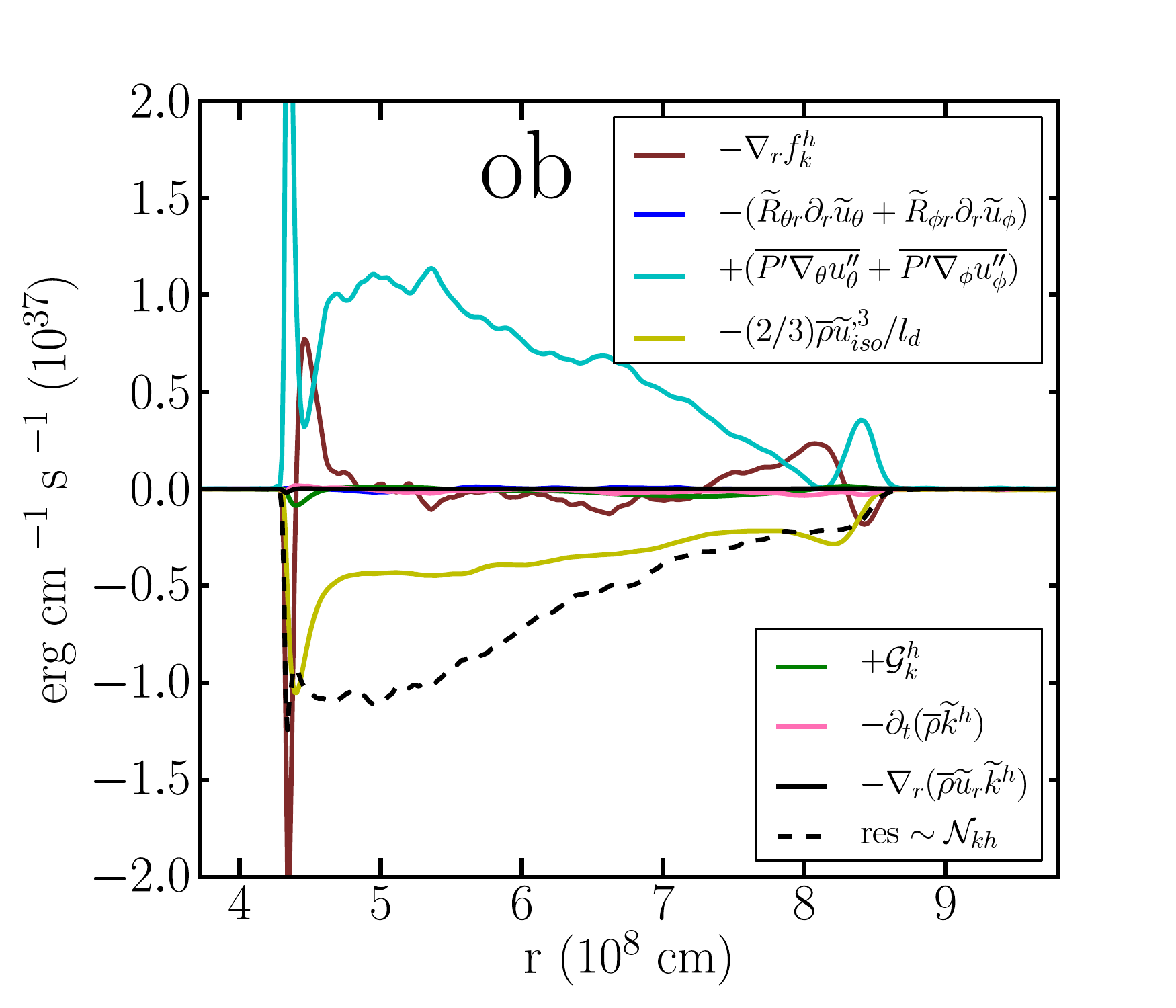}
\includegraphics[width=7.cm]{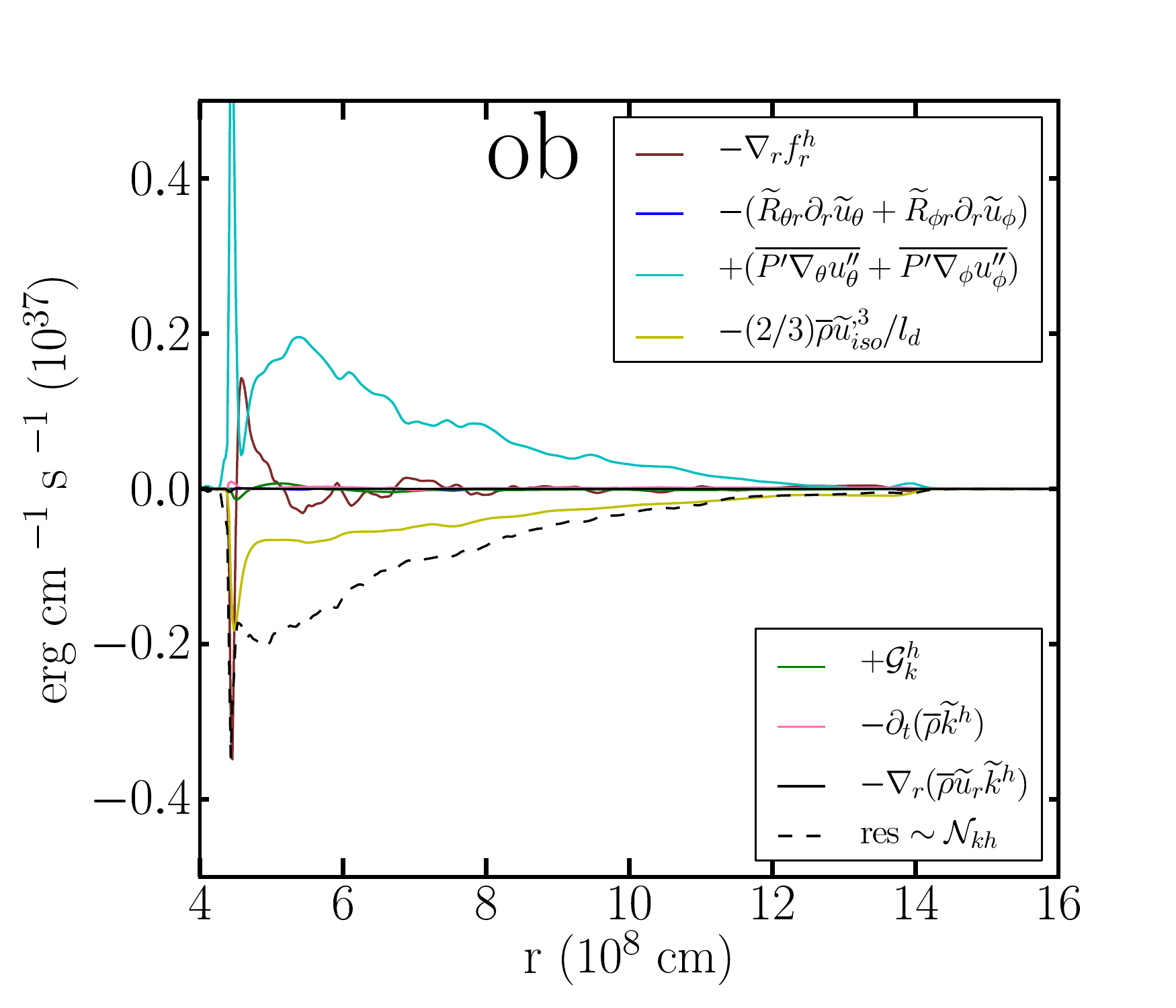}}
\caption{Radial (upper panels) and horizontal (lower panels) part of the mean turbulent kinetic energy equation. 1 Hp model {\sf ob.3D.1hp} (left), 2 Hp model {\sf ob.3D.2hp} (middle) and 4 Hp model {\sf ob.3D.4hp} (right).}
\end{figure}

\newpage

\subsubsection{Mean turbulent mass flux equation and mean density-specific volume covariance equation}

\begin{figure}[!h]
\centerline{
\includegraphics[width=7.cm]{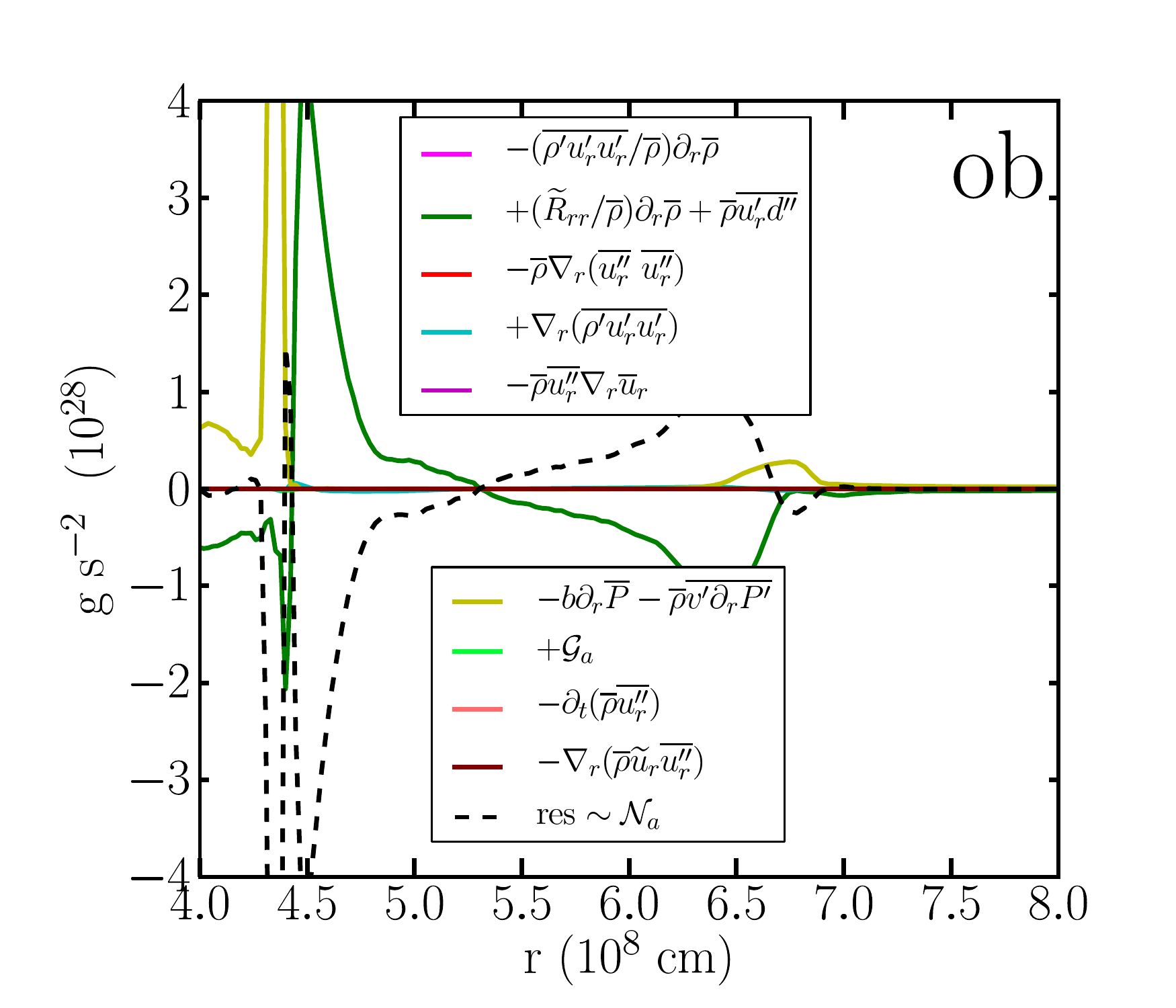}
\includegraphics[width=7.cm]{ob3dB_tavg230_mfields_a_equation_insf-eps-converted-to.pdf}
\includegraphics[width=7.cm]{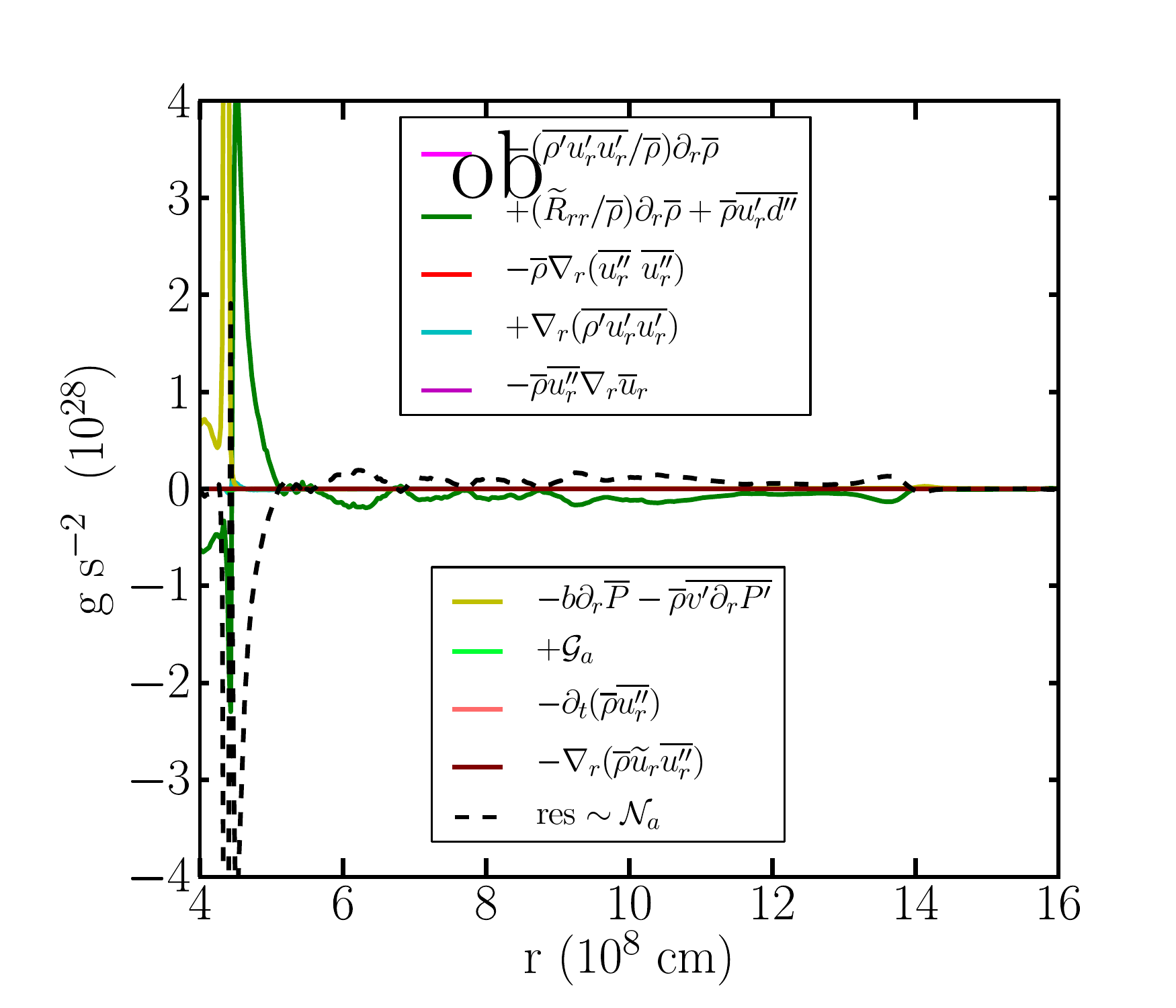}}

\centerline{
\includegraphics[width=7.cm]{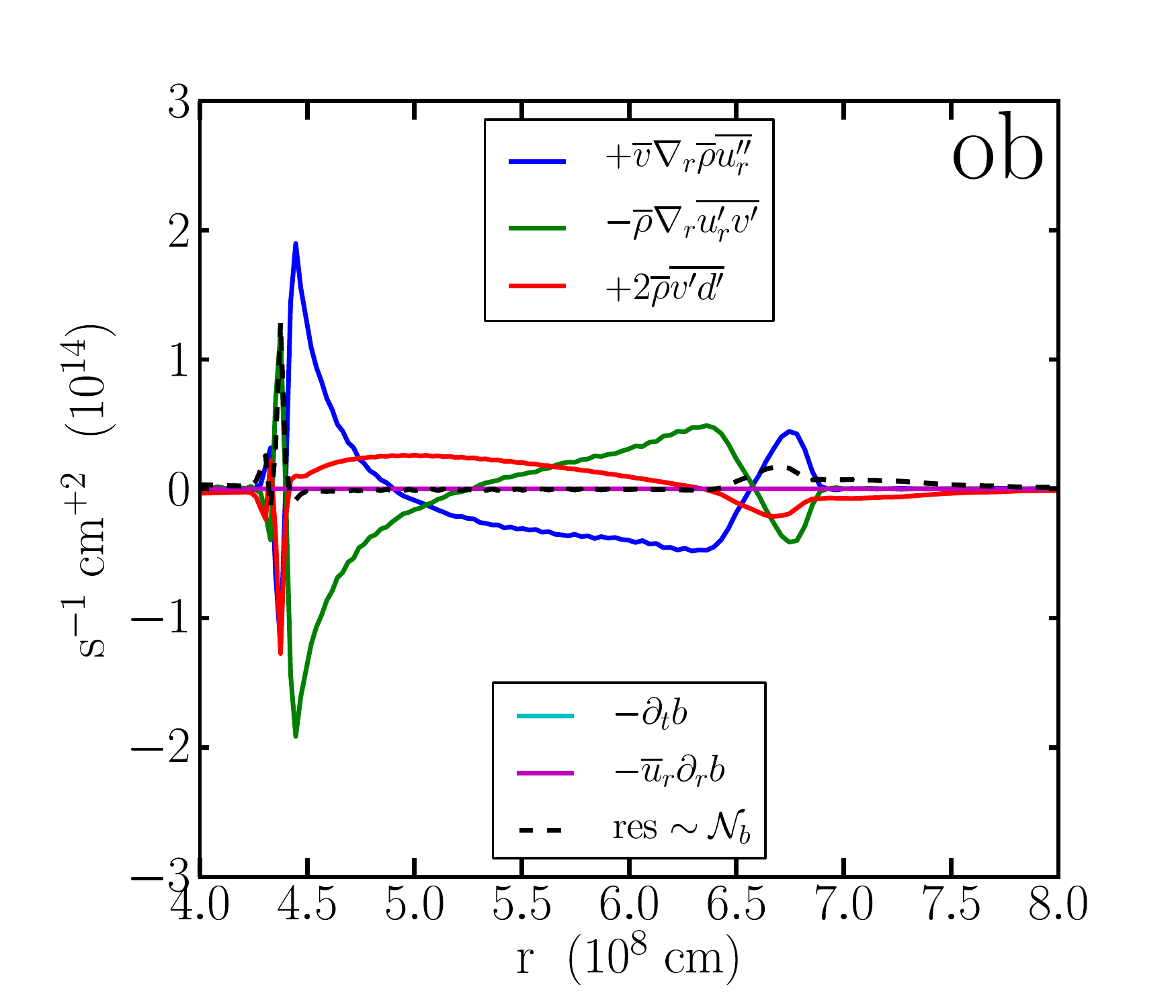}
\includegraphics[width=7.cm]{ob3dB_tavg230_mfields_b_equation_insf-eps-converted-to.pdf}
\includegraphics[width=7.cm]{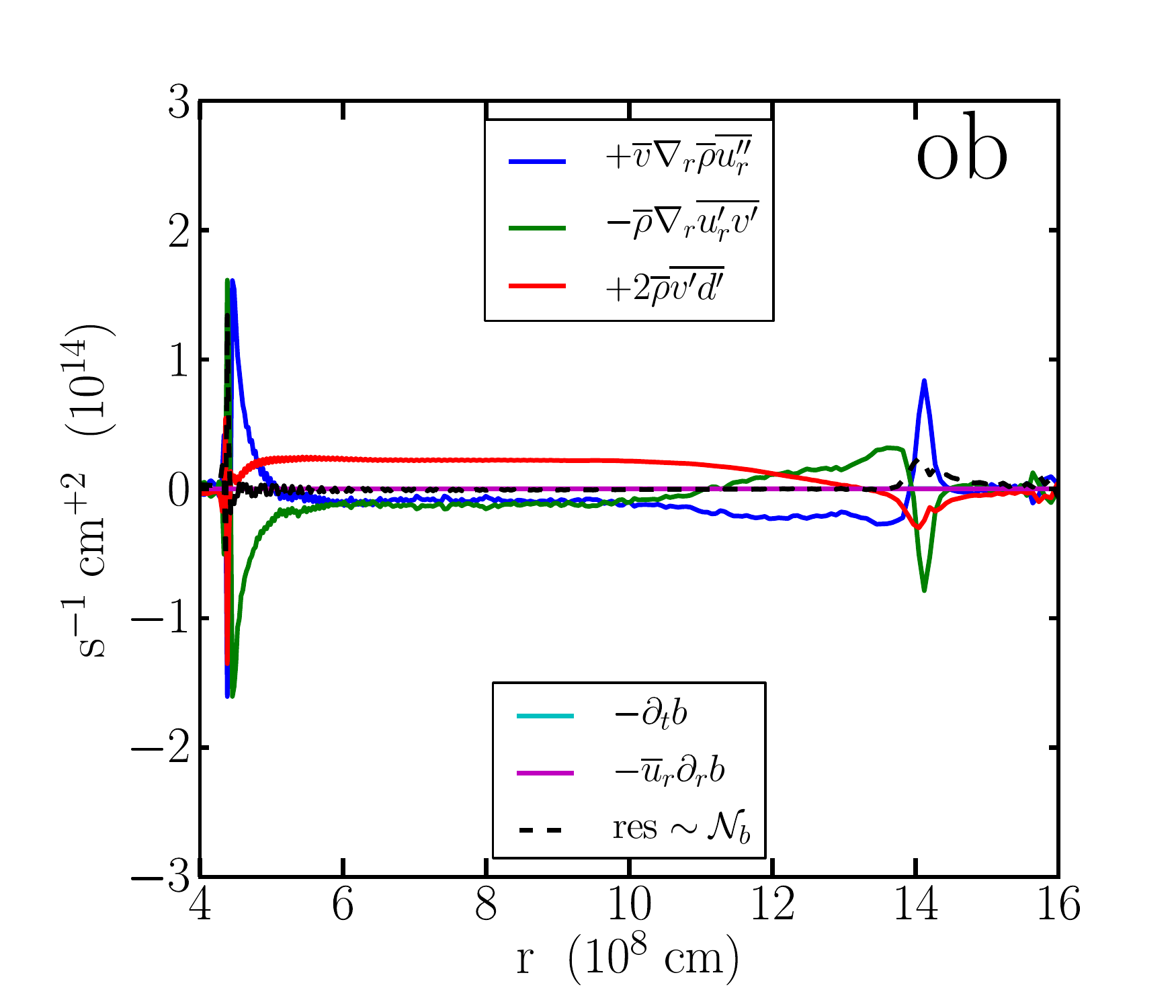}}
\caption{Mean turbulent mass flux equation (upper panels) and density-specific volume covariance equation (lower panels). 1 Hp model {\sf ob.3D.1hp} (left), 2 Hp model {\sf ob.3D.2hp} (middle) and 4 Hp model {\sf ob.3D.4hp} (right).}
\end{figure}
 
\newpage

\subsubsection{Mean specific angular momentum equation and internal energy flux equation}

\begin{figure}[!h]
\centerline{
\includegraphics[width=7.cm]{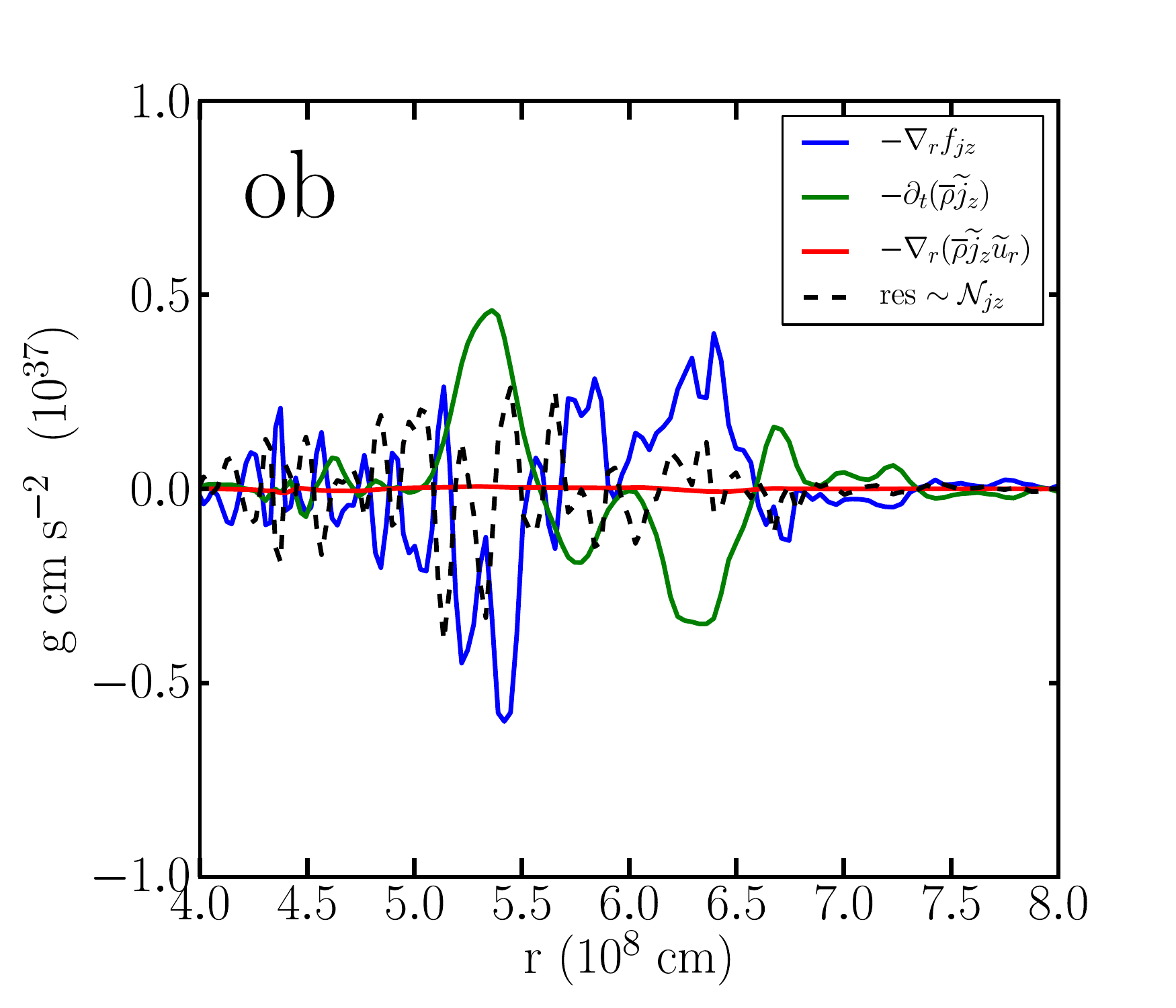}
\includegraphics[width=7.cm]{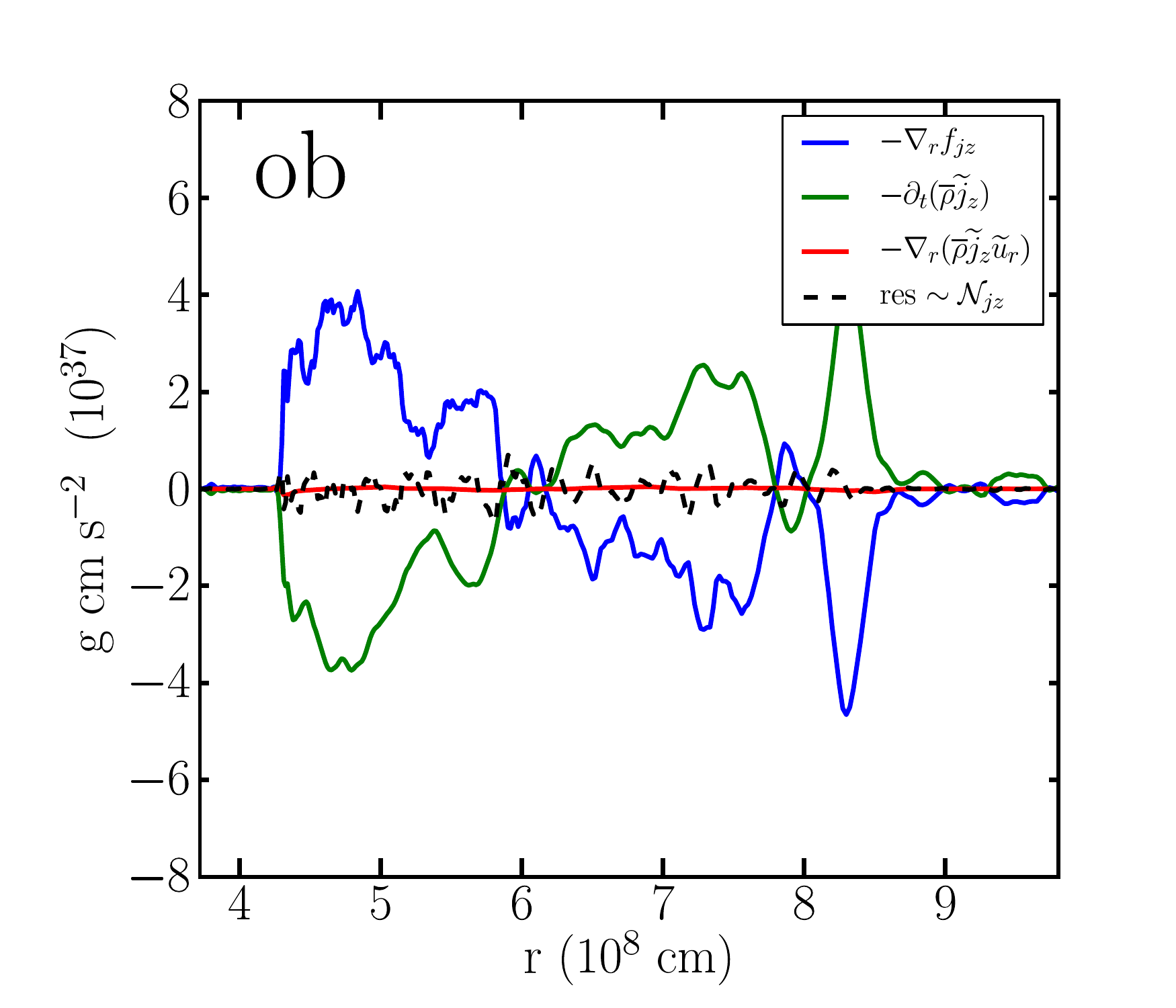}
\includegraphics[width=7.cm]{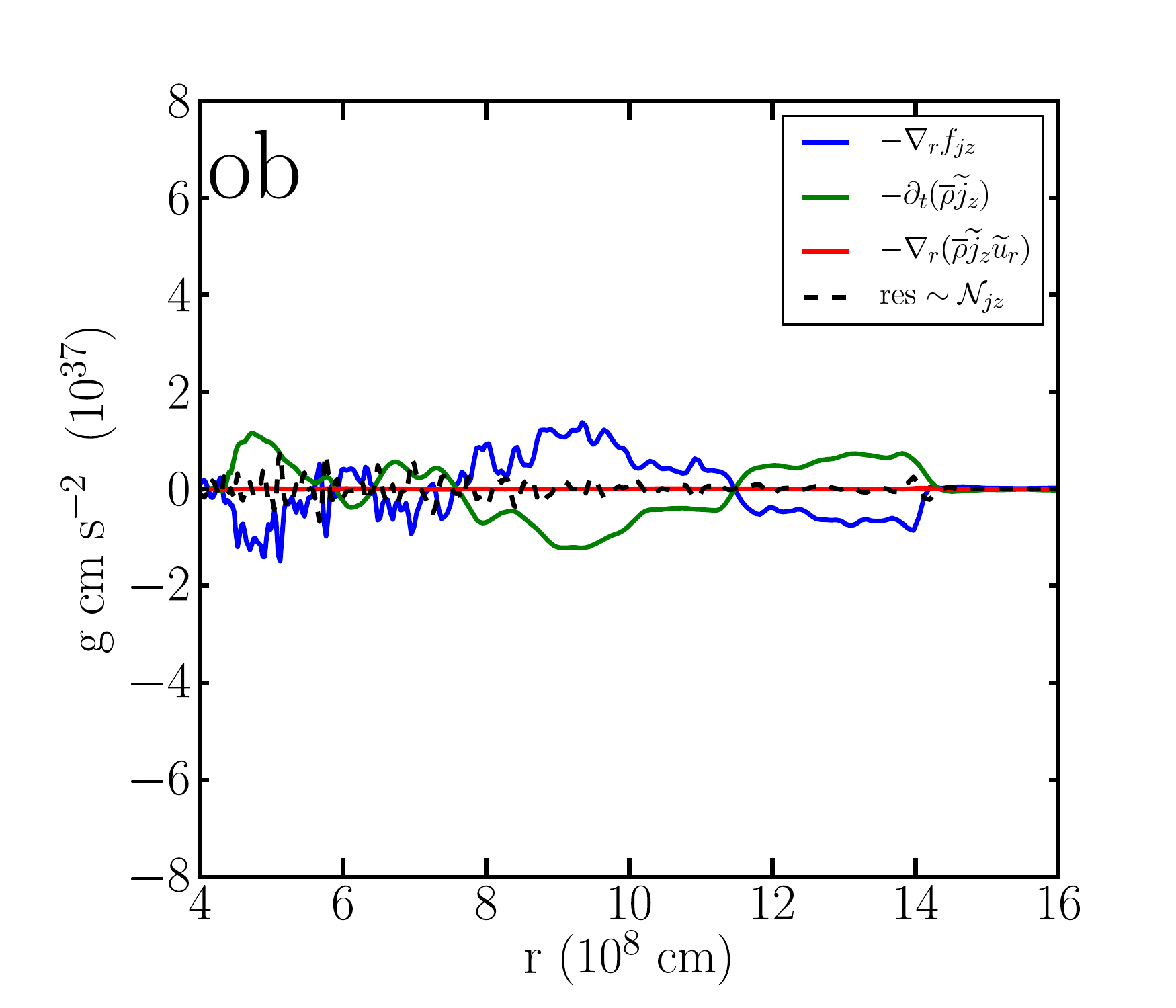}}

\centerline{
\includegraphics[width=7.cm]{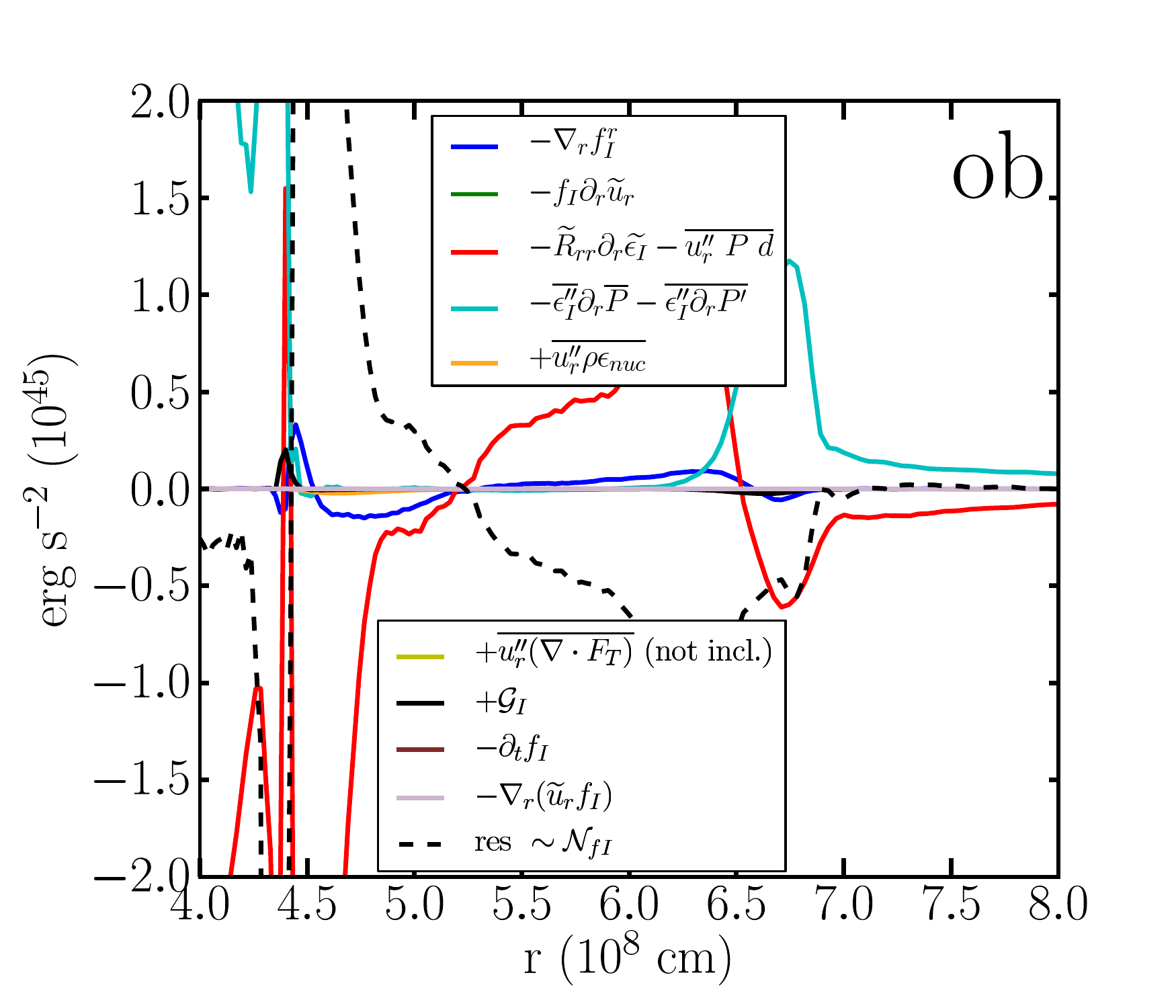}
\includegraphics[width=7.cm]{ob3dB_tavg230_mfields_i_equation_insf-eps-converted-to.pdf}
\includegraphics[width=7.cm]{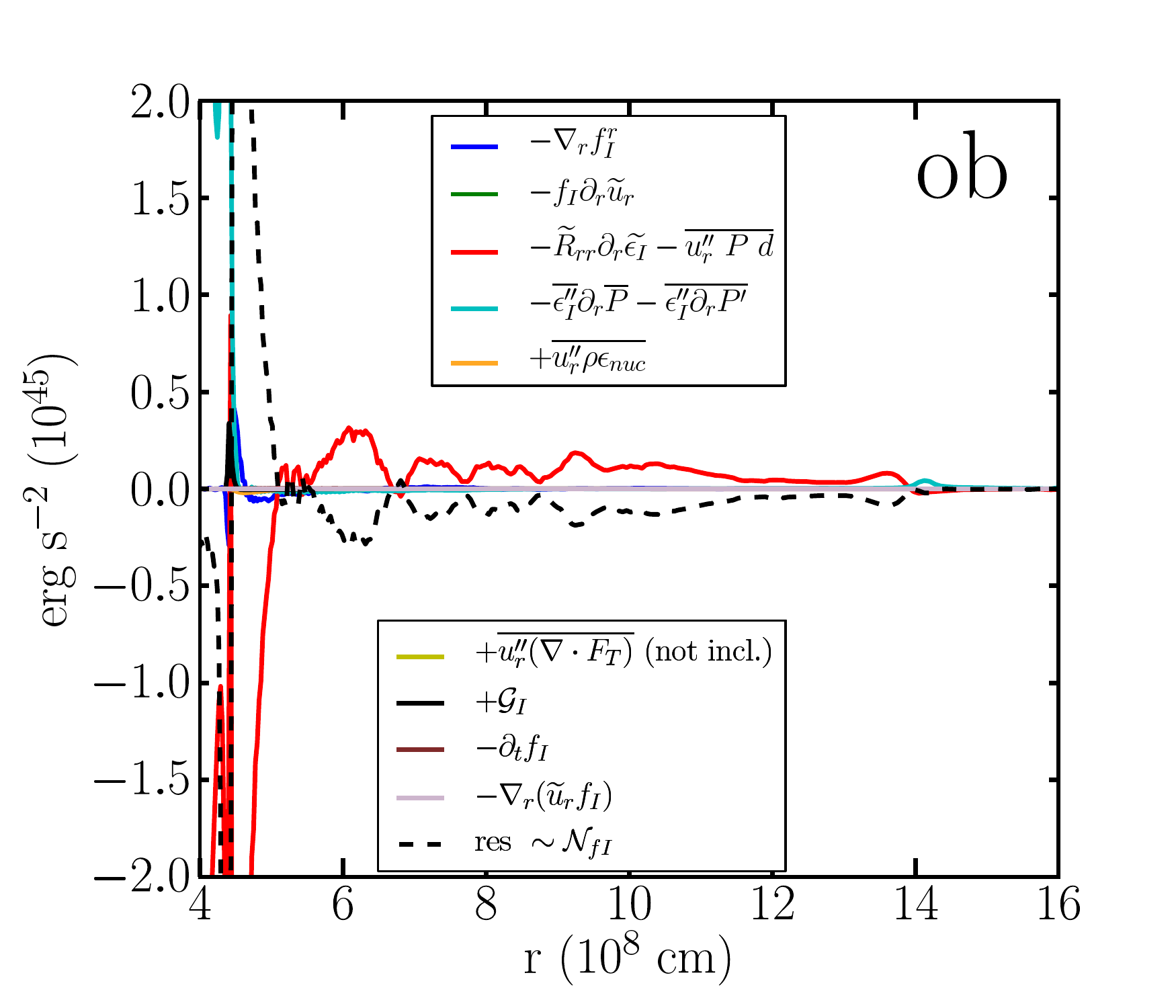}}
\caption{Mean specific angular momentum equation (upper panels) and mean turbulent internal energy flux equation (lower panels). 1 Hp model {\sf ob.3D.1hp} (left), 2 Hp model {\sf ob.3D.2hp} (middle) and 4 Hp model {\sf ob.3D.4hp} (right).}
\end{figure}

\newpage

\subsubsection{Mean entropy equation and mean entropy flux equation}

\begin{figure}[!h]
\centerline{
\includegraphics[width=7.cm]{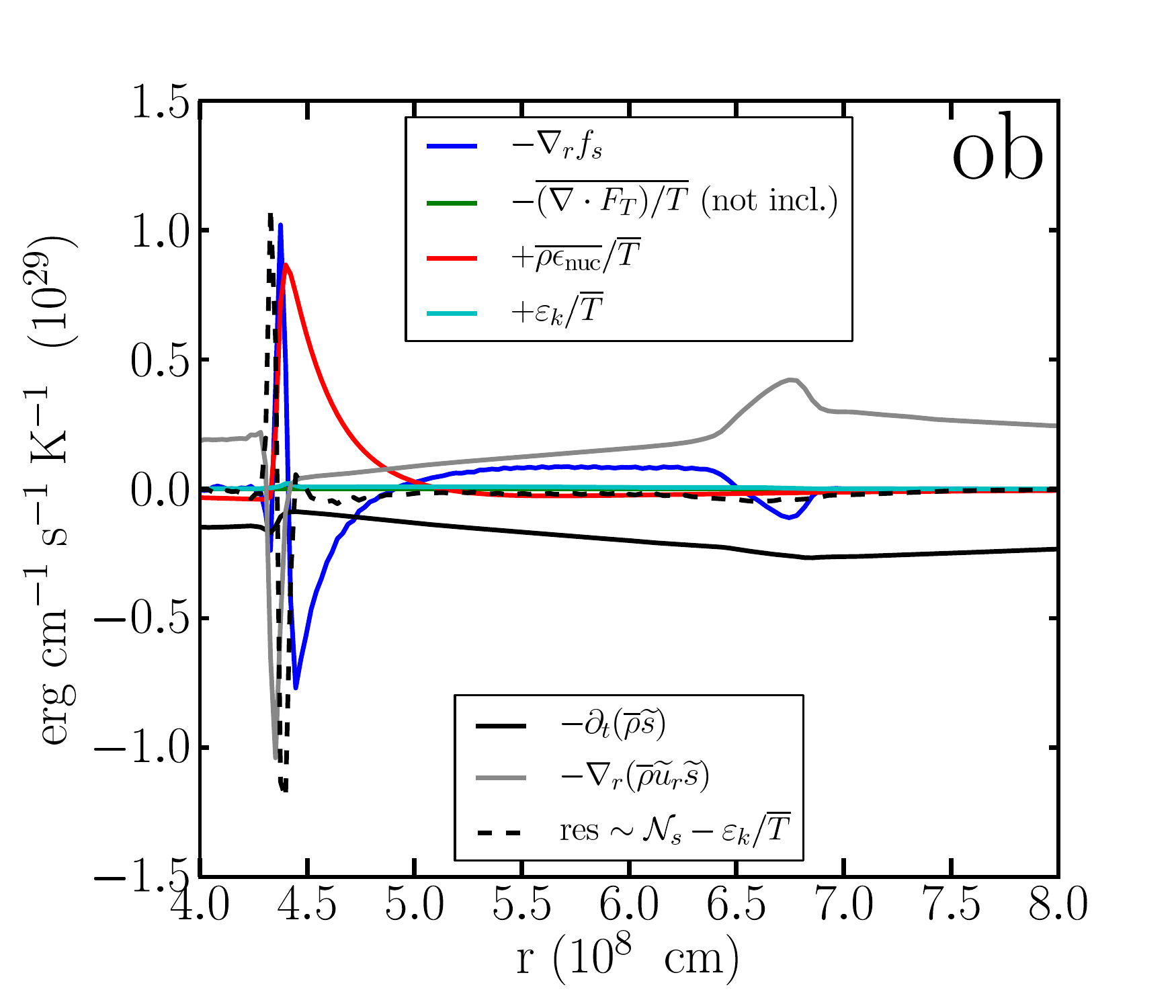}
\includegraphics[width=7.cm]{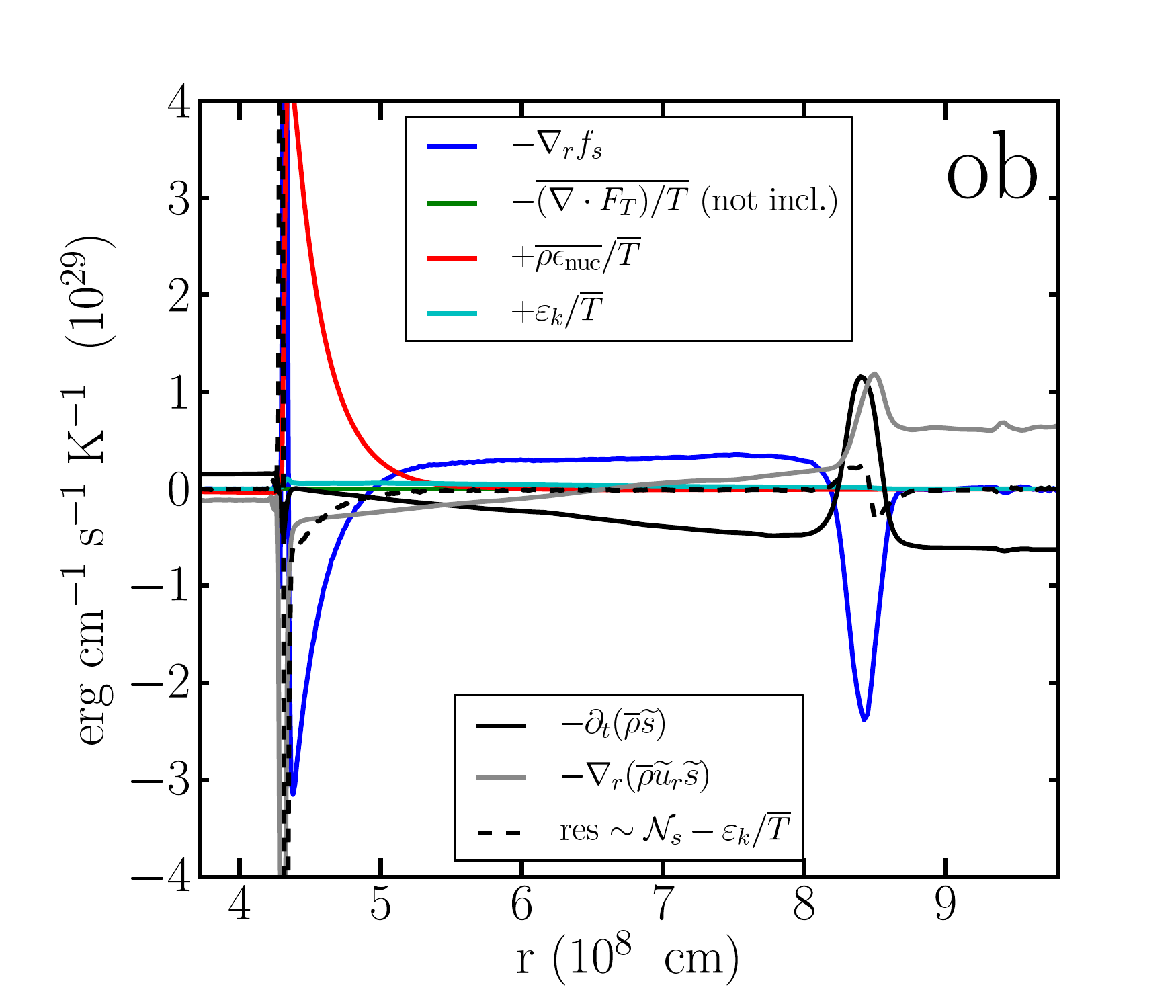}
\includegraphics[width=7.cm]{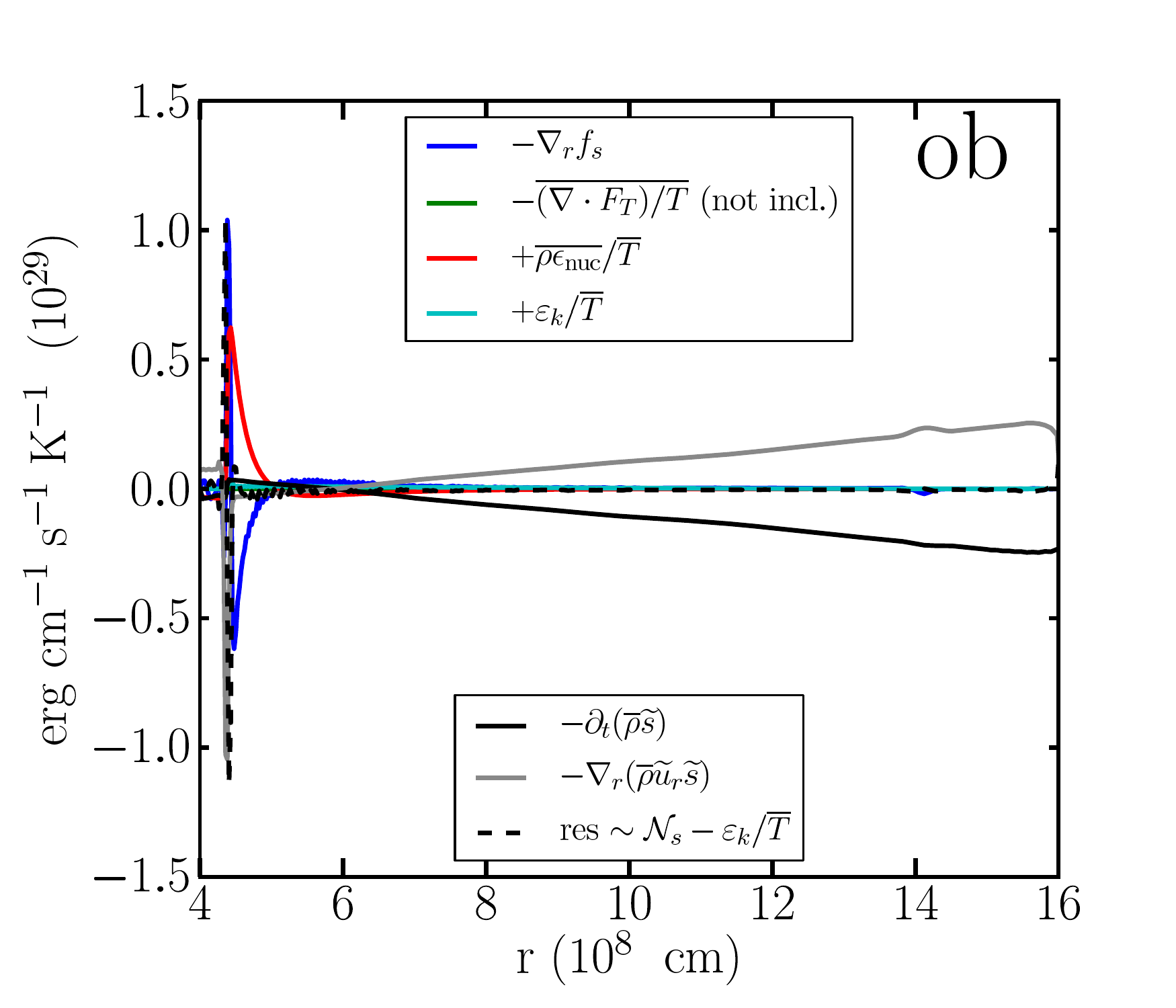}}

\centerline{
\includegraphics[width=7.cm]{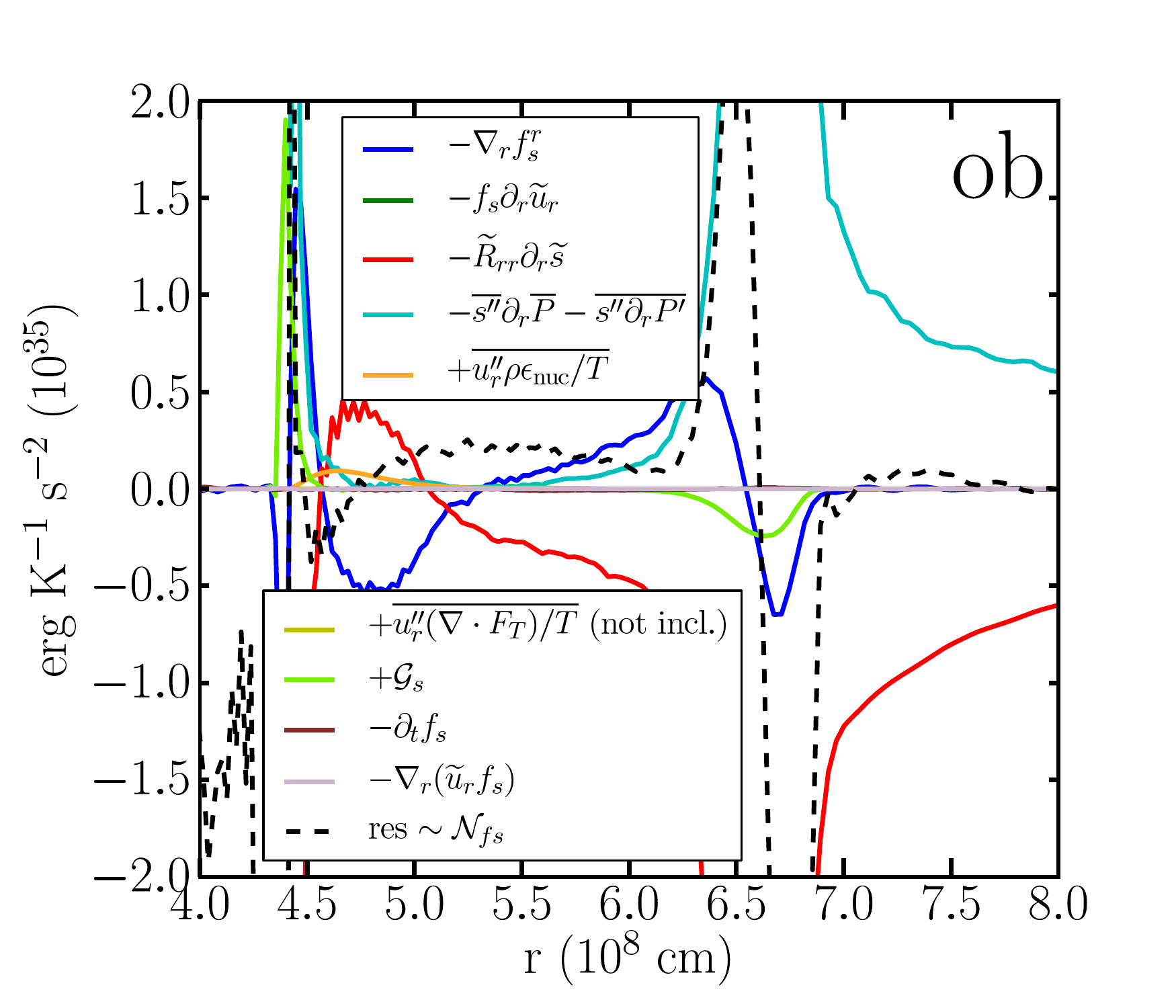}
\includegraphics[width=7.cm]{ob3dB_tavg230_mfields_s_equation_insf-eps-converted-to.pdf}
\includegraphics[width=7.cm]{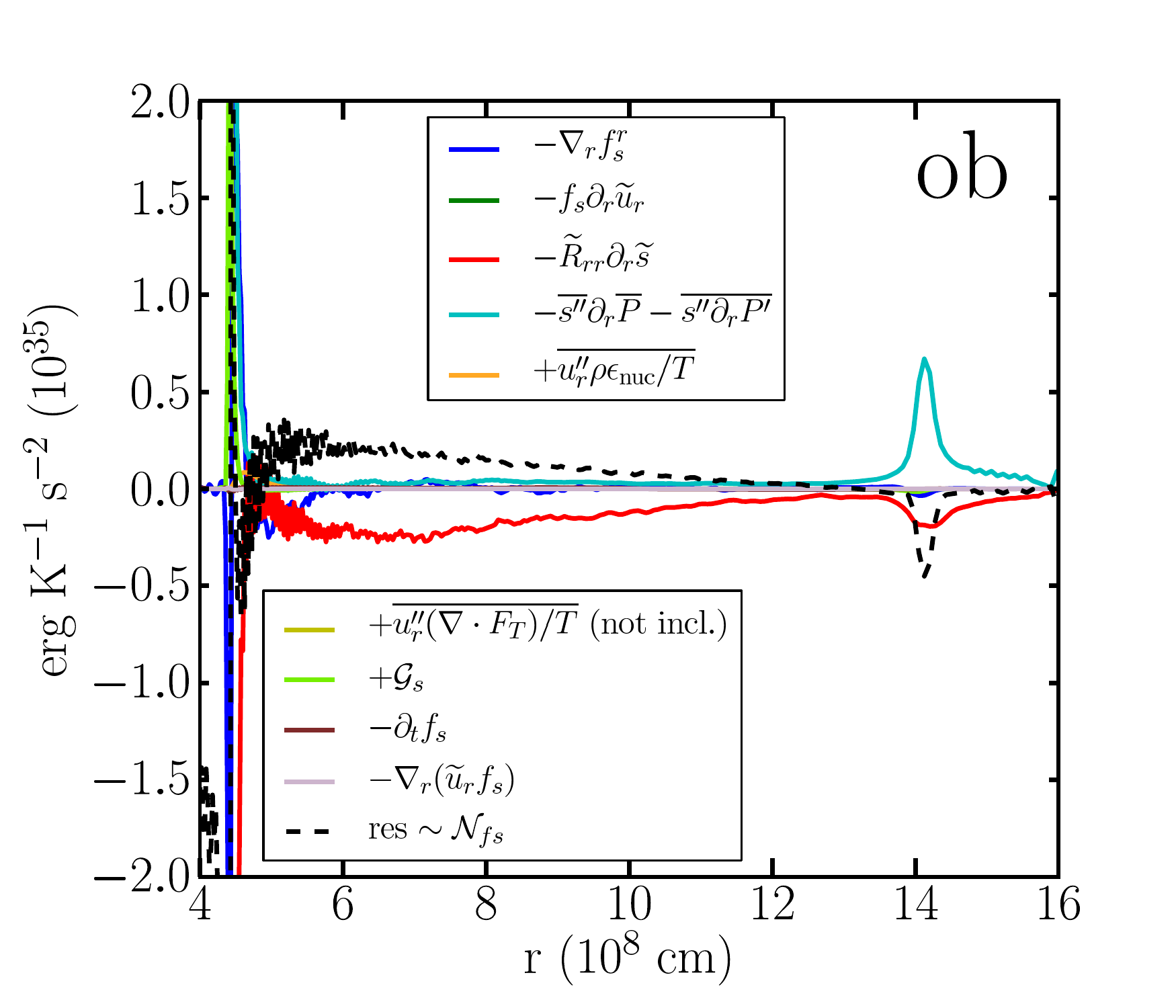}}
\caption{Mean entropy equation (upper panels) and mean entropy flux equation (lower panels). 1 Hp model {\sf ob.3D.1hp} (left), 2 Hp model {\sf ob.3D.2hp} (middle) and 4 Hp model {\sf ob.3D.4hp} (right).}
\end{figure}

\newpage

\subsubsection{Mean turbulent kinetic energy and mean velocities}

\begin{figure}[!h]
\centerline{
\includegraphics[width=7.cm]{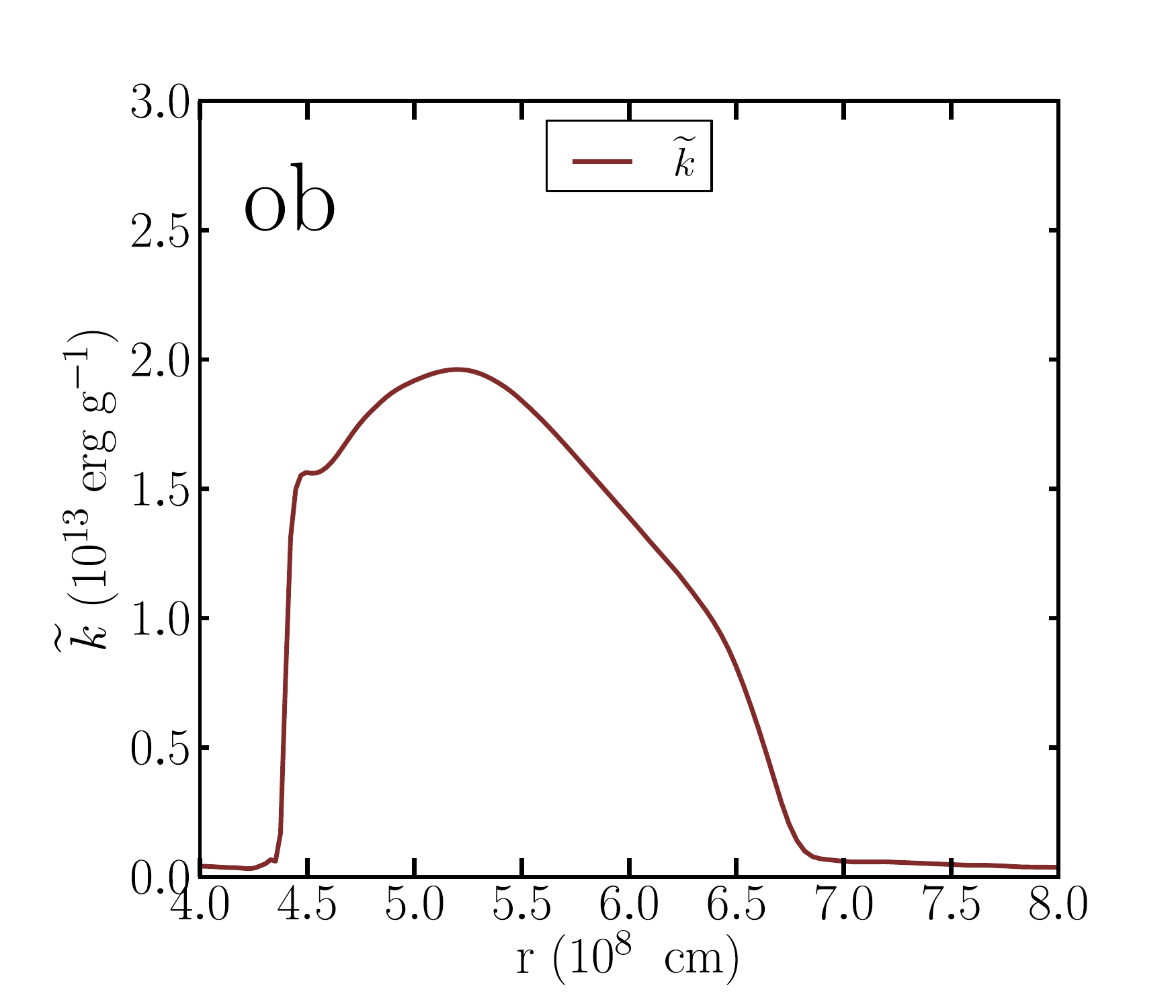}
\includegraphics[width=7.cm]{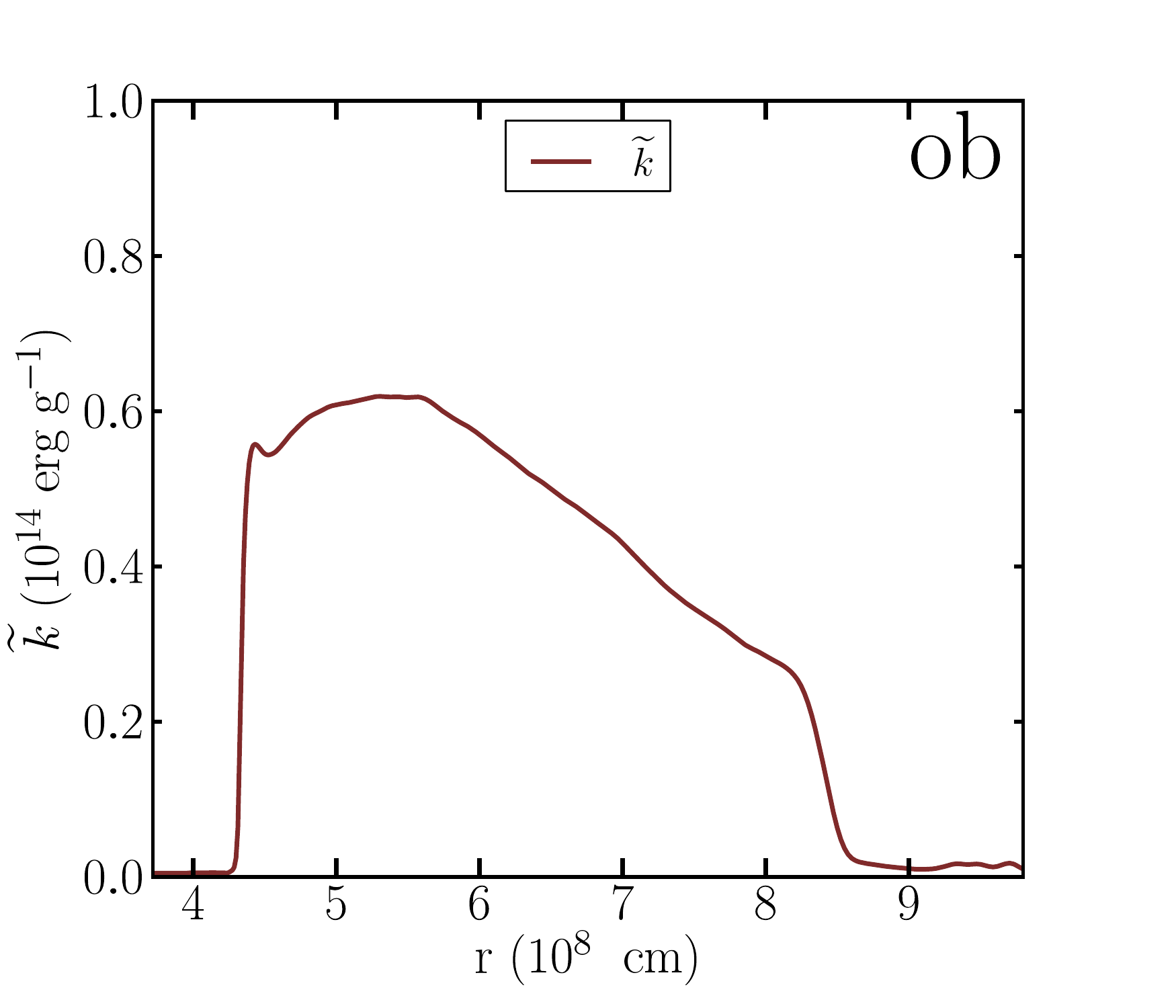}
\includegraphics[width=7.cm]{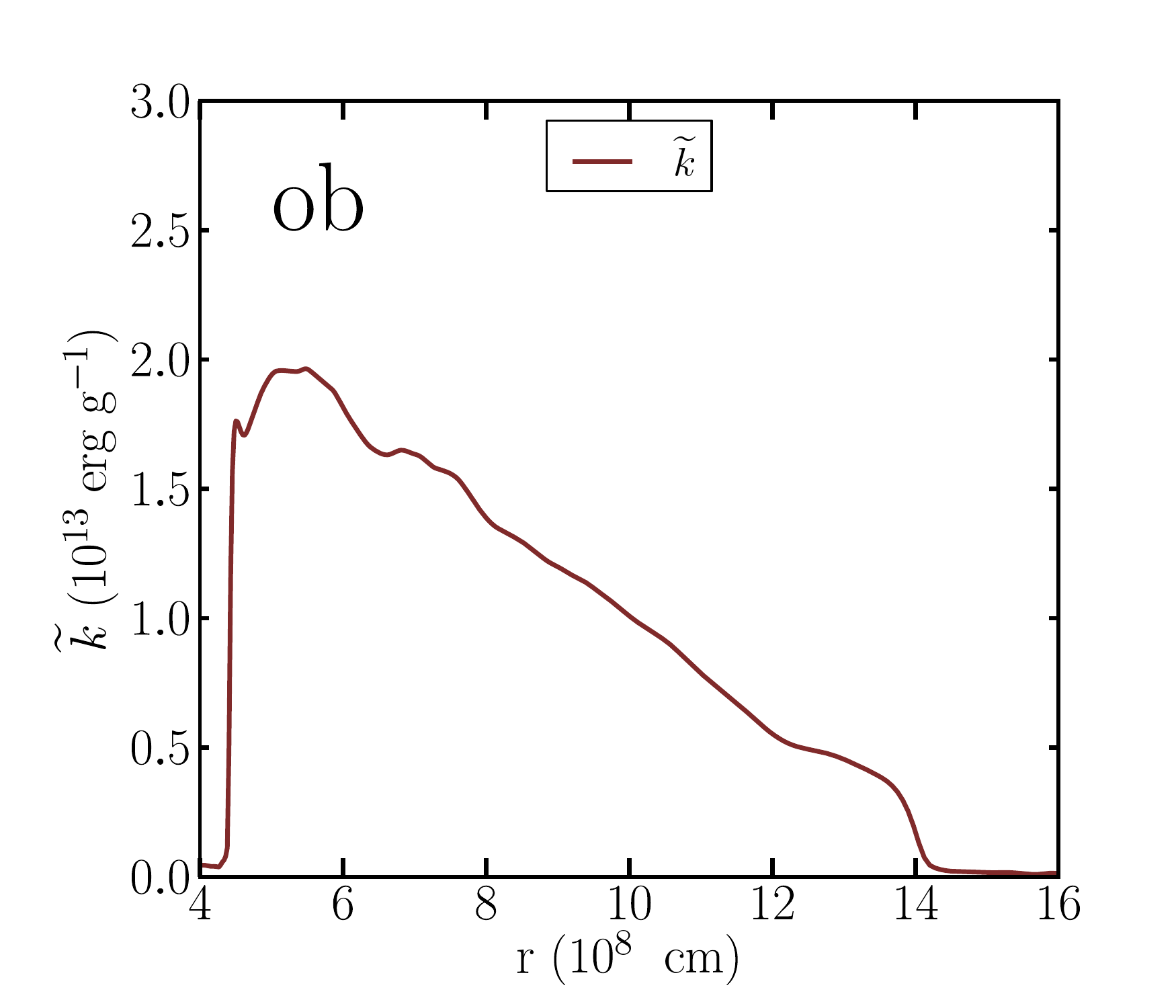}}

\centerline{
\includegraphics[width=7.cm]{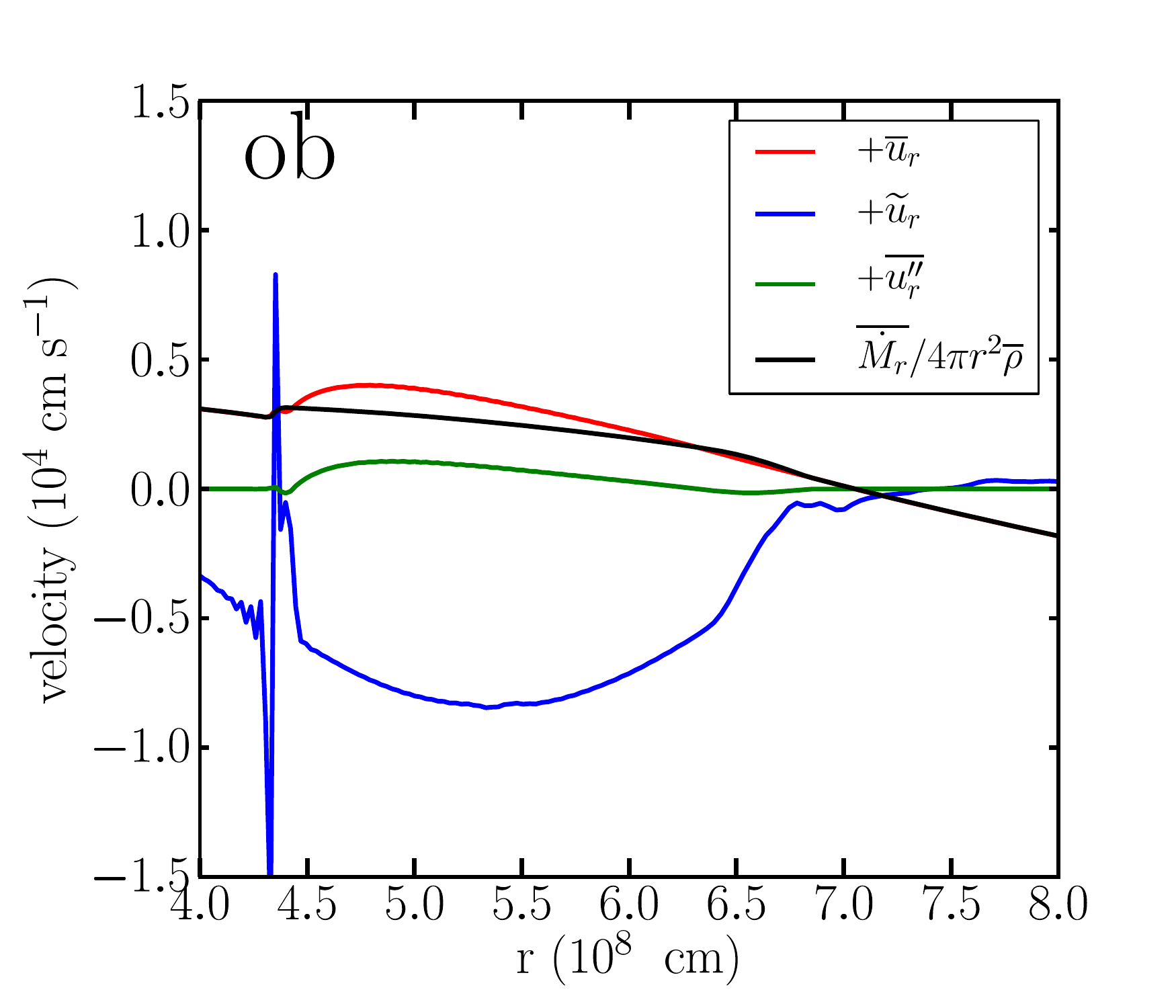}
\includegraphics[width=7.cm]{ob3dB_tavg230_mean_velocities_insf-eps-converted-to.pdf}
\includegraphics[width=7.cm]{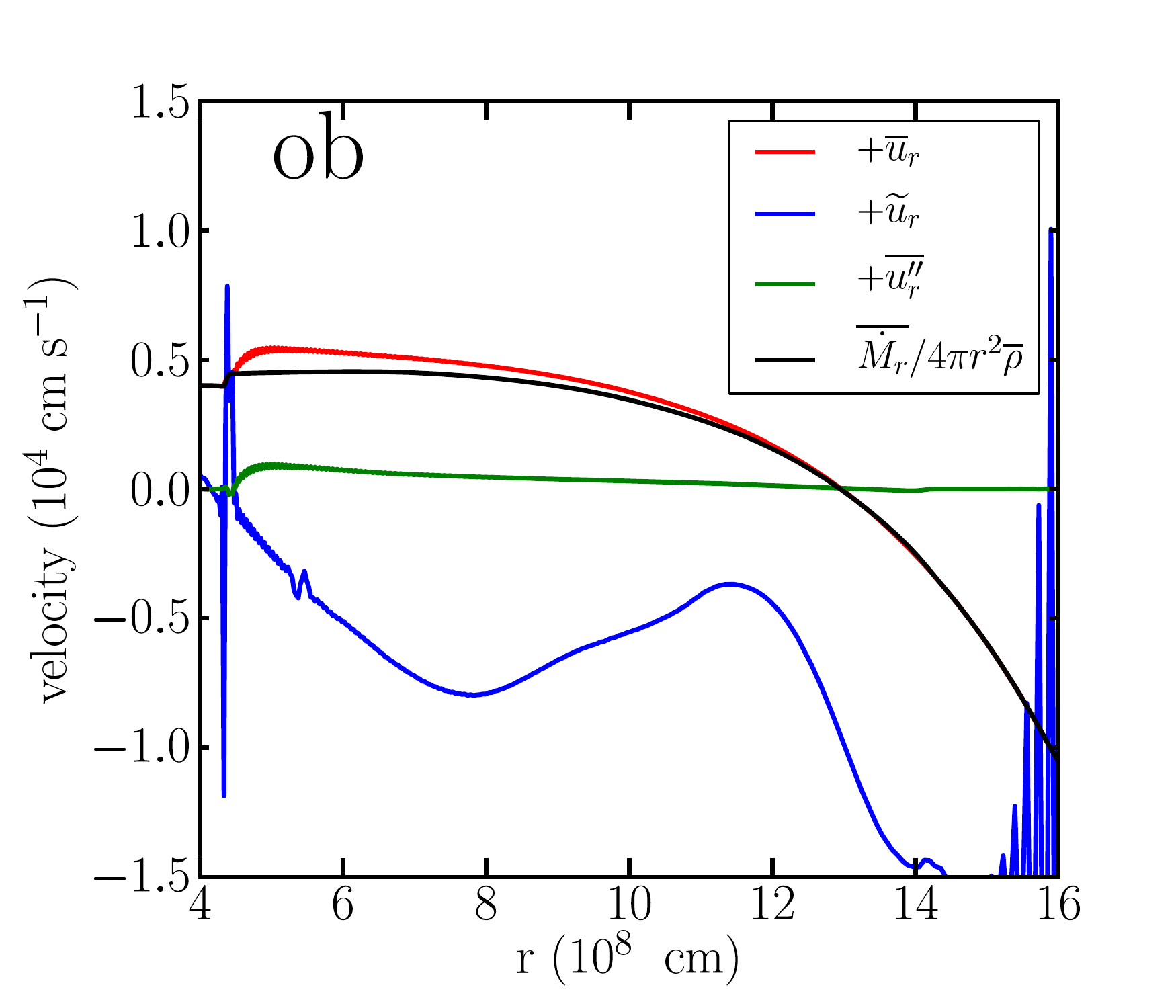}}
\caption{Mean turbulent kinetic energy (upper panels) and mean velocities (lower panels). 1 Hp model {\sf ob.3D.1hp} (left), 2 Hp model {\sf ob.3D.2hp} (middle) and 4 Hp model {\sf ob.3D.4hp} (right).}
\end{figure}

\newpage

\subsection{Red giant convection envelope model}

\subsubsection{Mean continuity equation and mean radial momentum equation}

\begin{figure}[!h]
\centerline{
\includegraphics[width=6.5cm]{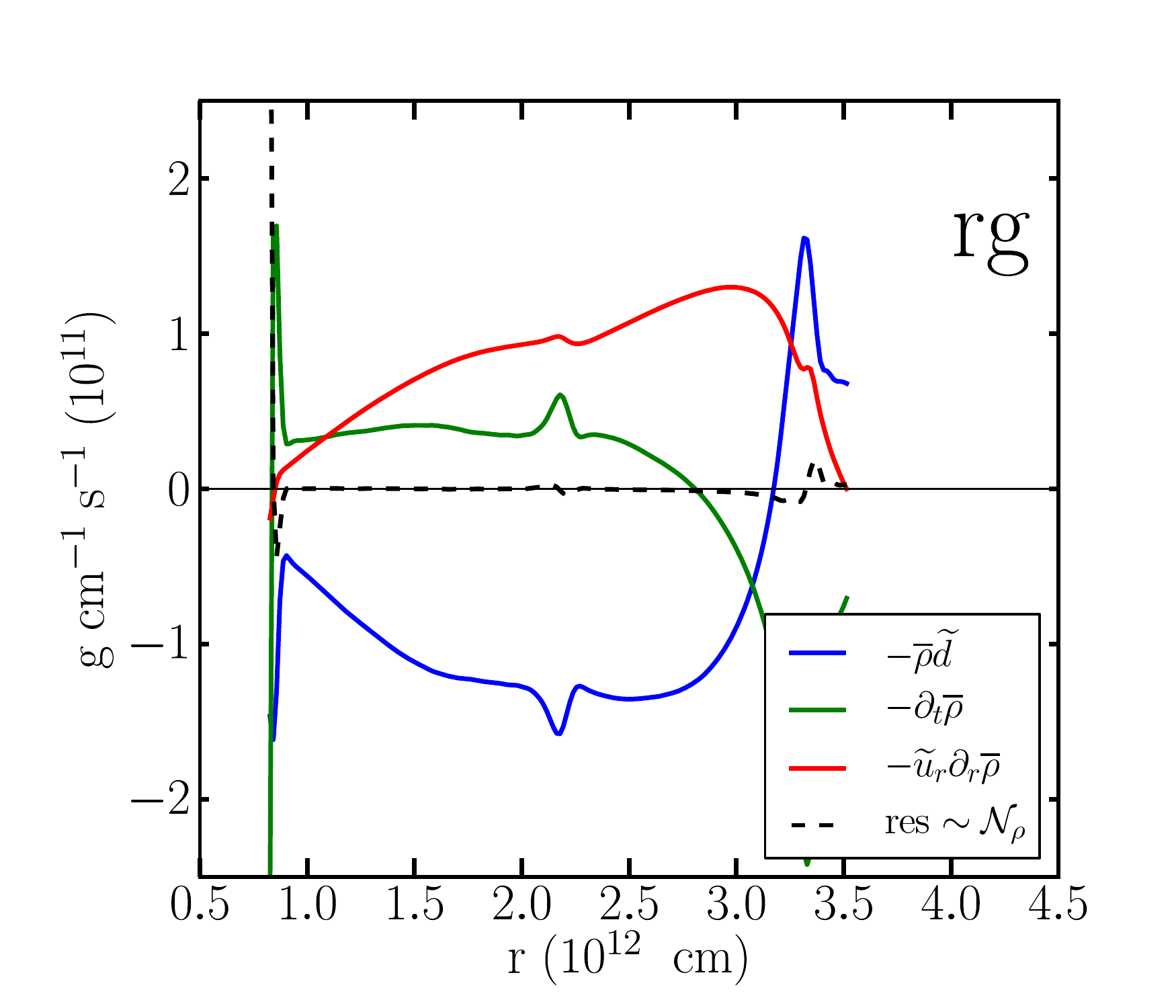}
\includegraphics[width=6.5cm]{rgmrez_tavg800_continuity_equation_insf-eps-converted-to.pdf}}

\centerline{
\includegraphics[width=6.5cm]{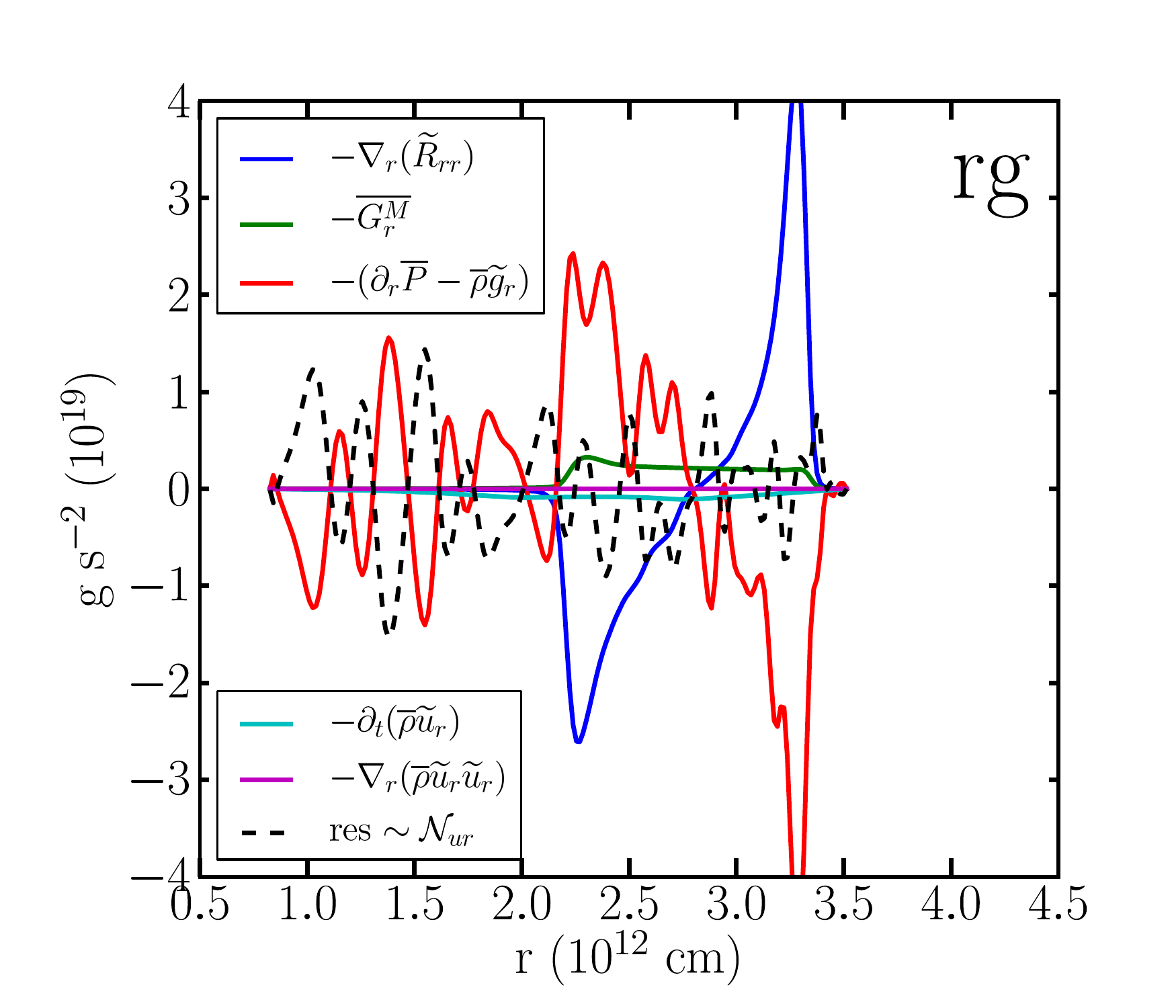}
\includegraphics[width=6.5cm]{rgmrez_tavg800_rmomentum_equation_insf-eps-converted-to.pdf}}
\caption{Mean continuity equation (upper panels) and radial momentum equation (lower panels). 4 Hp model {\sf rg.3D.4hp} (left) and 7 Hp model {\sf rg.3D.mrez} (right).}
\end{figure}
 
\newpage

\subsubsection{Mean azimuthal and polar momentum equation}

\begin{figure}[!h]
\centerline{
\includegraphics[width=7.cm]{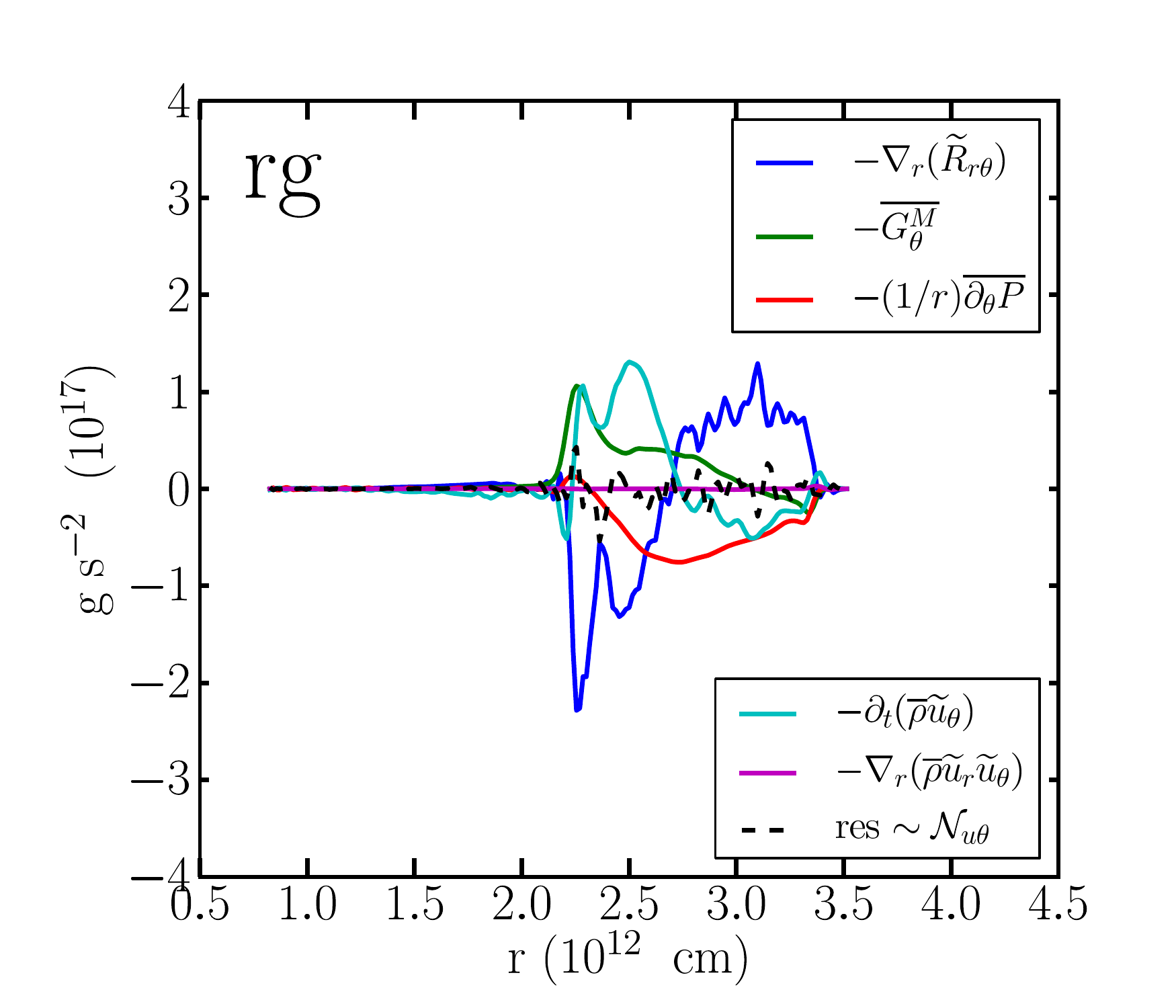}
\includegraphics[width=7.cm]{rgmrez_tavg800_tmomentum_equation_insf-eps-converted-to.pdf}}

\centerline{
\includegraphics[width=7.cm]{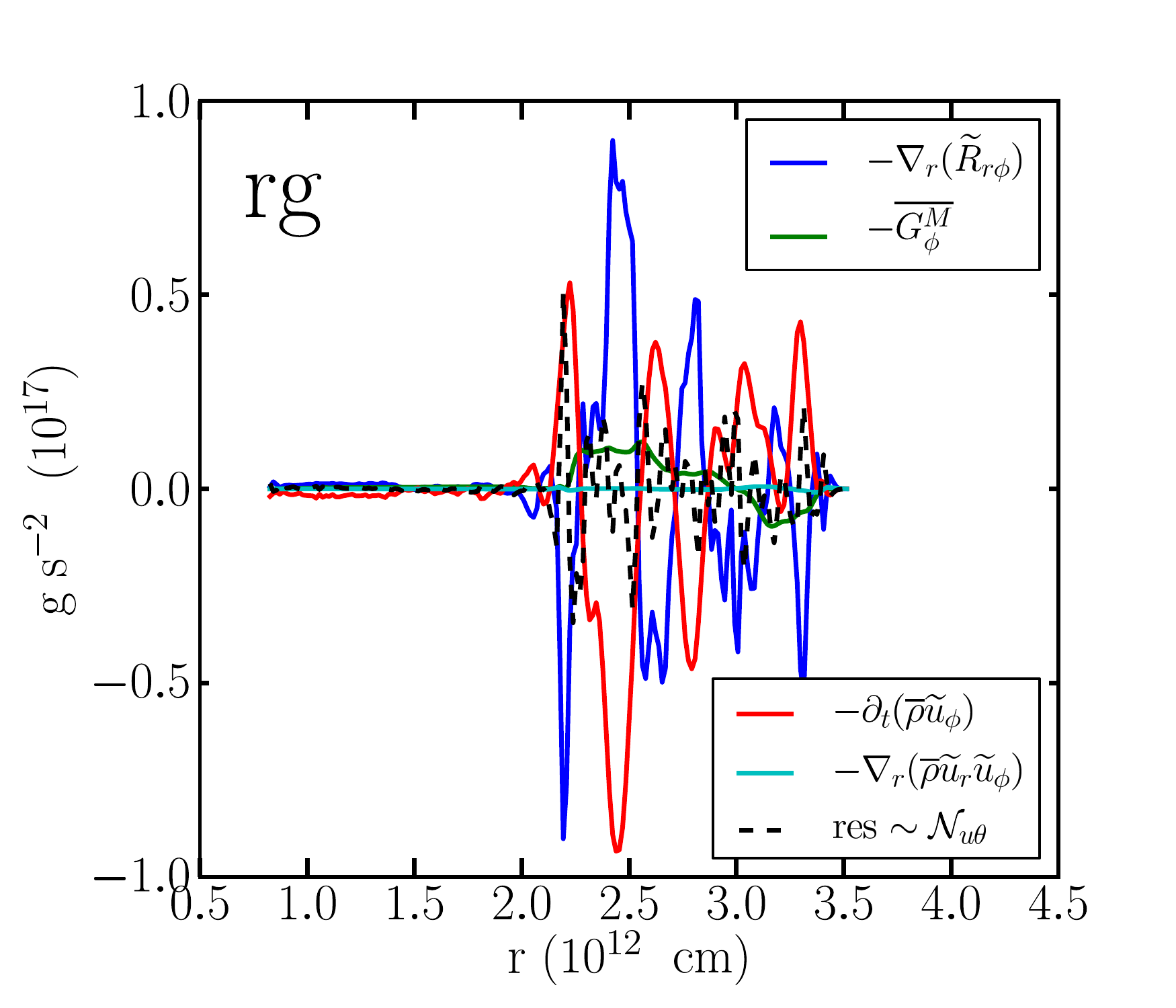}
\includegraphics[width=7.cm]{rgmrez_tavg800_pmomentum_equation_insf-eps-converted-to.pdf}}
\caption{Mean azimuthal equation (upper panels) and mean polar momentum equation (lower panels). 4 Hp model {\sf rg.3D.4hp} (left) and 7 Hp model {\sf rg.3D.mrez} (right).}
\end{figure}

\newpage

\subsubsection{Mean total energy equation and mean turbulent kinetic energy equation}

\begin{figure}[!h]
\centerline{
\includegraphics[width=6.5cm]{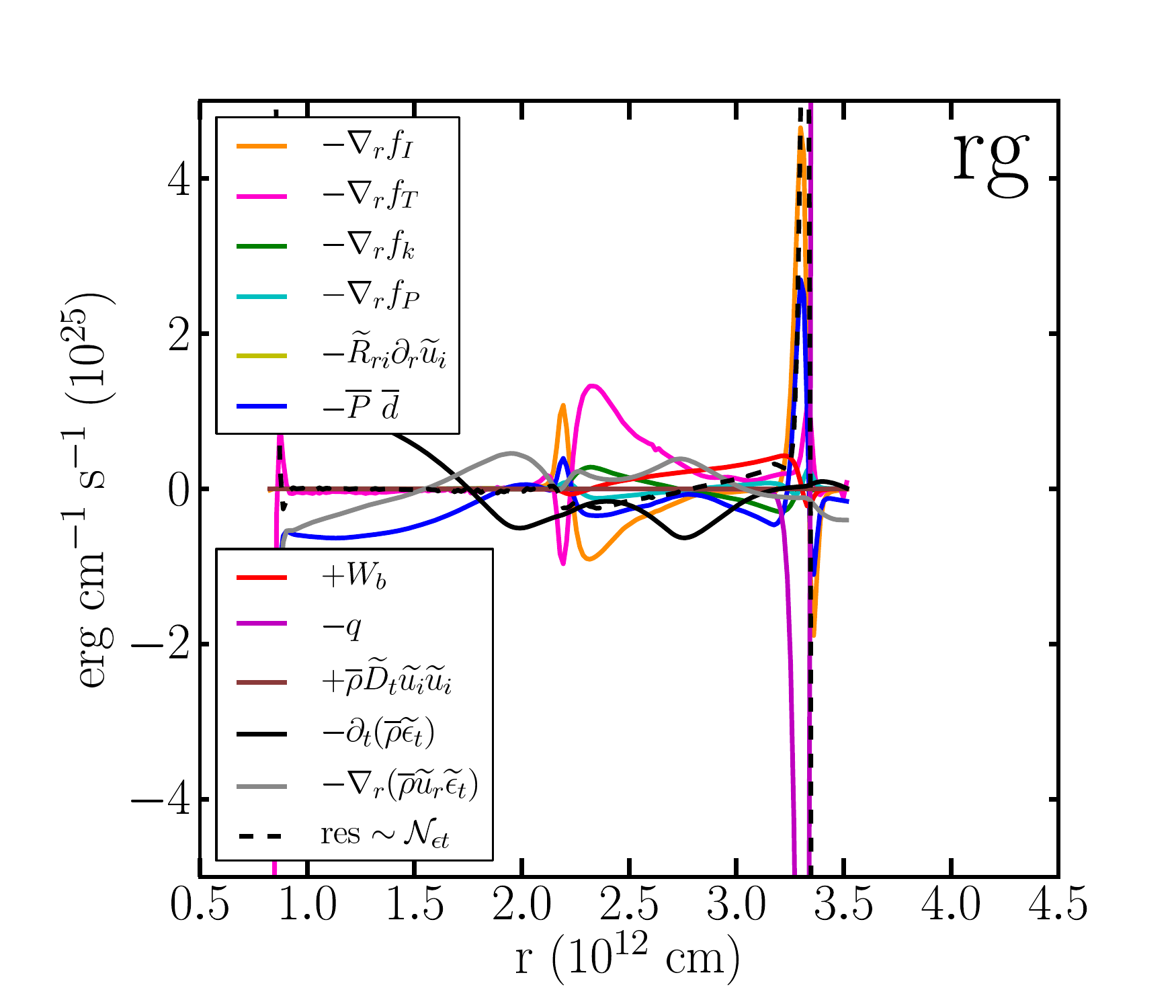}
\includegraphics[width=6.5cm]{rgmrez_tavg800_total_energy_equation_insf-eps-converted-to.pdf}}

\centerline{
\includegraphics[width=6.5cm]{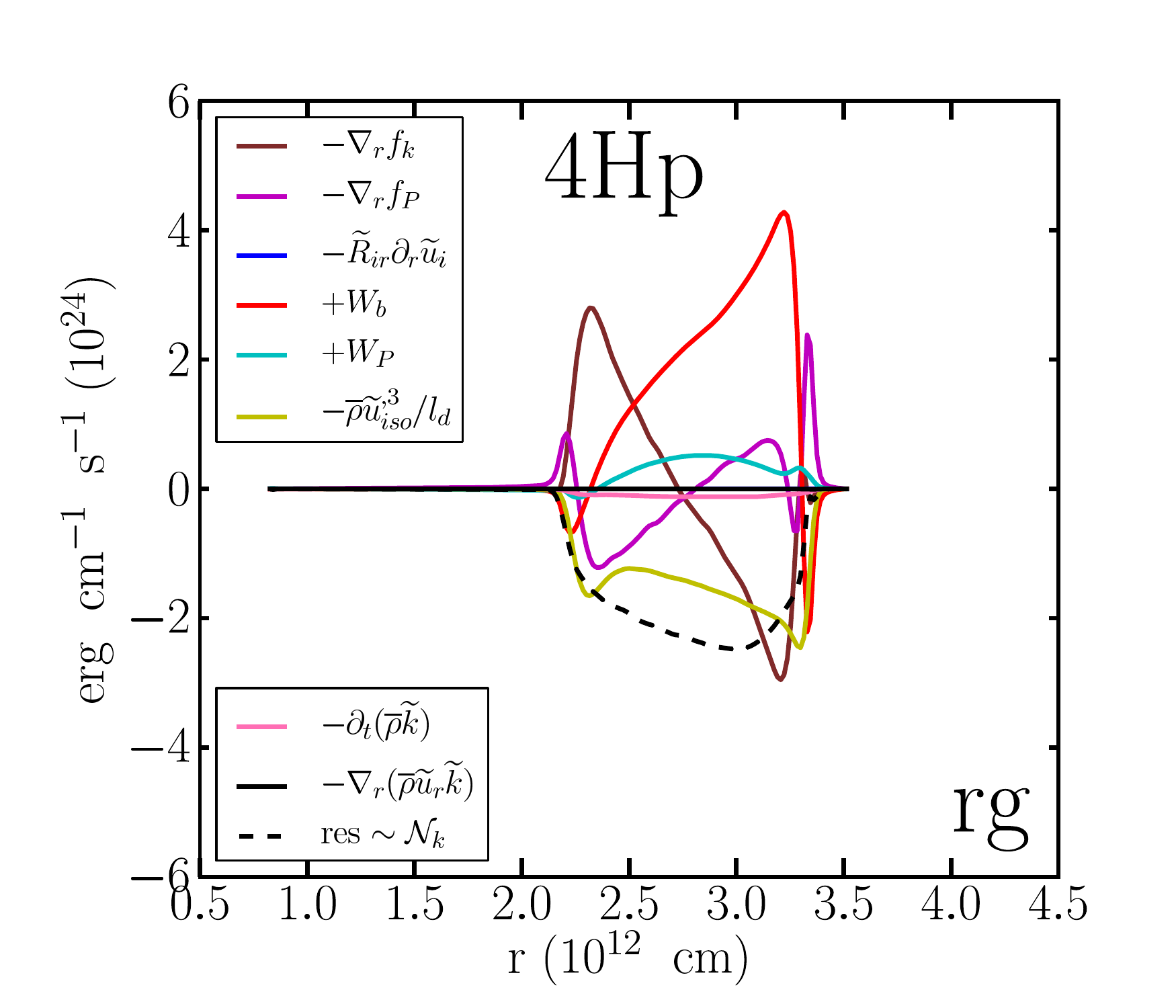}
\includegraphics[width=6.5cm]{rgmrez_tavg800_mfields_k_equation_insf-eps-converted-to.pdf}}
\caption{Mean total energy equation (upper panels) and mean turbulent kinetic energy equation (lower panels). 4 Hp model {\sf rg.3D.4hp} (left) and 7 Hp model {\sf rg.3D.mrez} (right).}
\end{figure}

\newpage

\subsubsection{Mean turbulent kinetic energy equation (radial + horizontal part)}

\begin{figure}[!h]
\centerline{
\includegraphics[width=6.5cm]{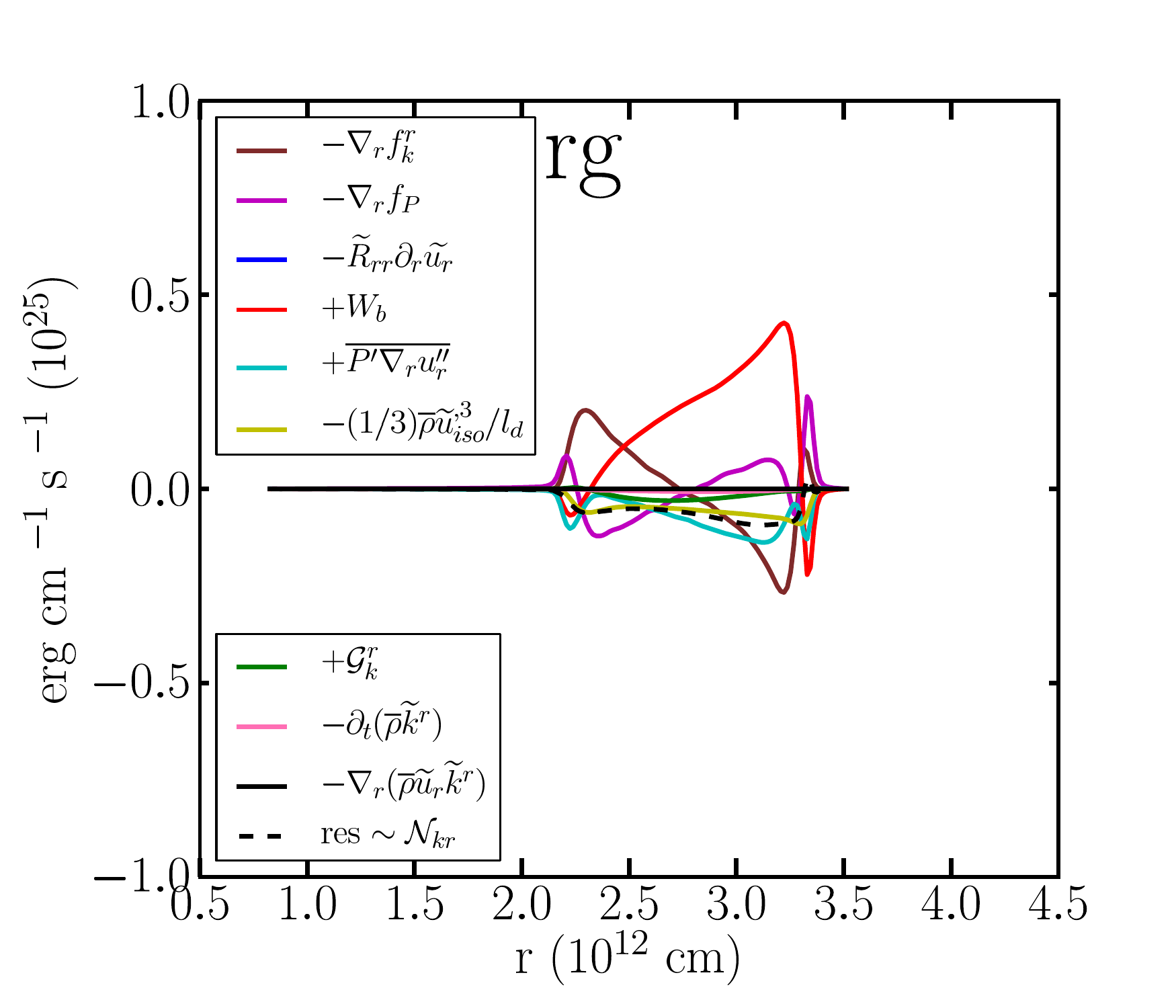}
\includegraphics[width=6.5cm]{rgmrez_tavg800_mfields_k_equation_rad_insf-eps-converted-to.pdf}}

\centerline{
\includegraphics[width=6.5cm]{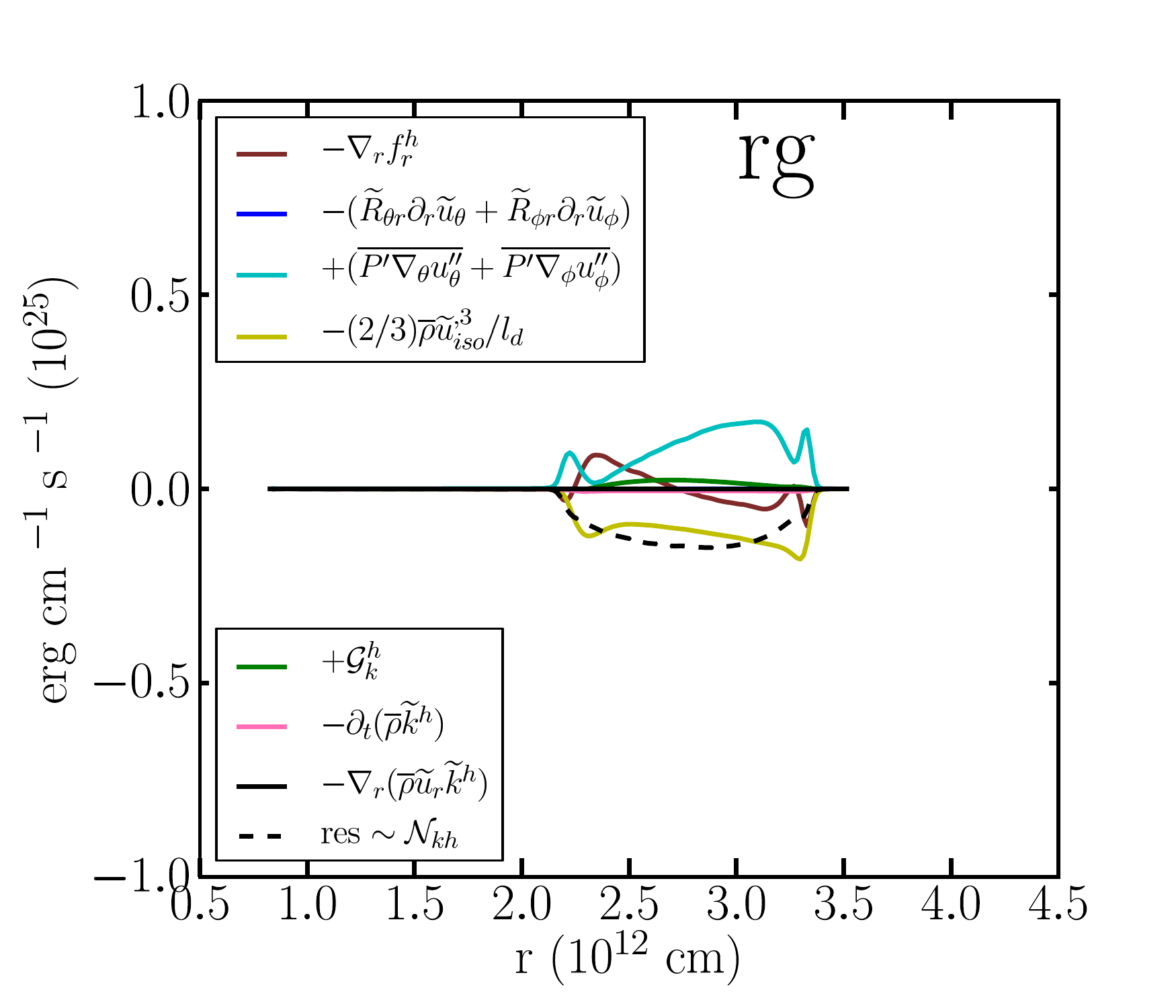}
\includegraphics[width=6.5cm]{rgmrez_tavg800_mfields_k_equation_hor_insf-eps-converted-to.pdf}}
\caption{Radial (upper panels) and horizontal (lower panels) part of the mean turbulent kinetic energy equation. 4 Hp model {\sf rg.3D.4hp} (left) and 7 Hp model {\sf rg.3D.mrez} (right).}
\end{figure}

\newpage

\subsubsection{Mean turbulent mass flux and mean density-specific volume covariance equation}

\begin{figure}[!h]
\centerline{
\includegraphics[width=6.5cm]{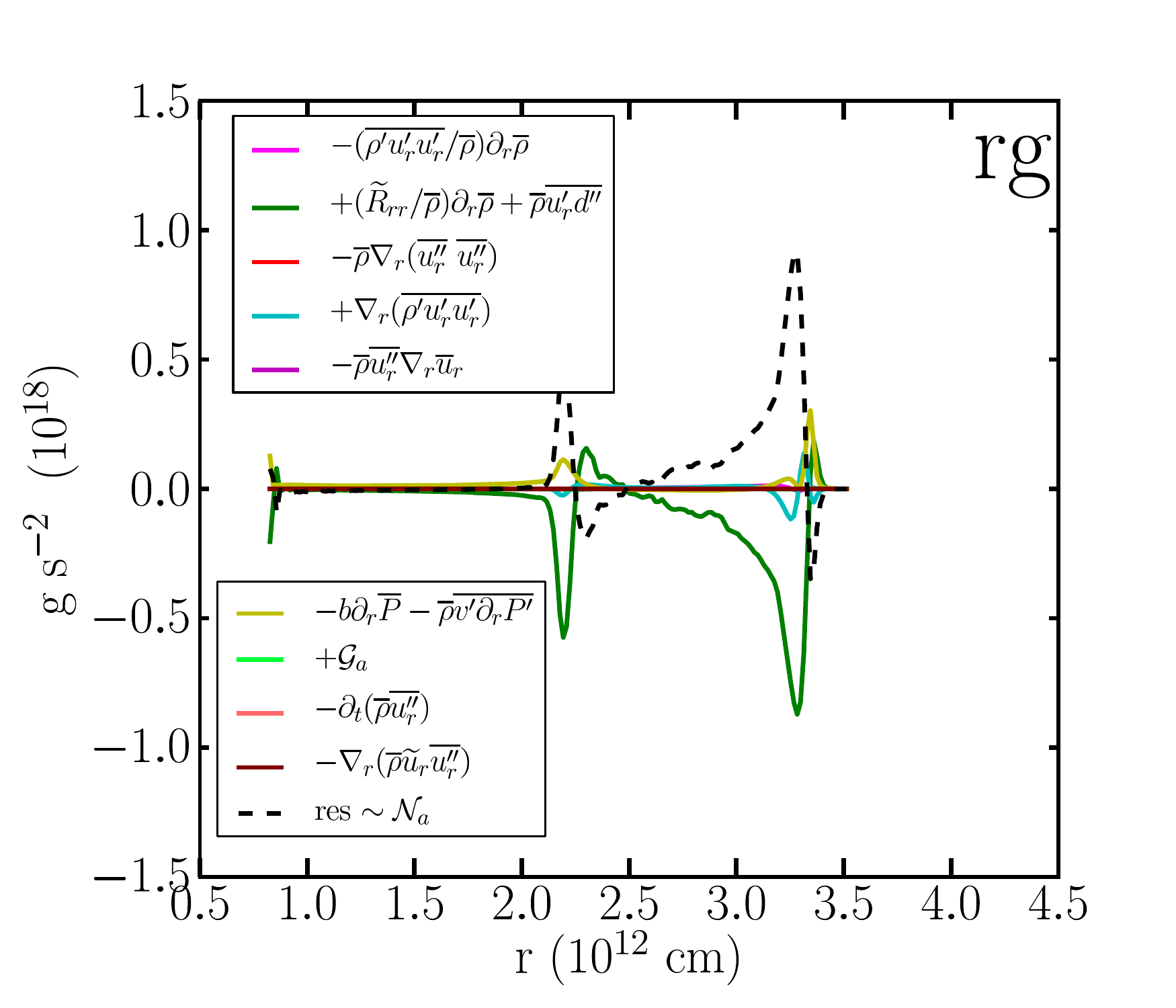}
\includegraphics[width=6.5cm]{rgmrez_tavg800_mfields_a_equation_insf-eps-converted-to.pdf}}

\centerline{
\includegraphics[width=6.5cm]{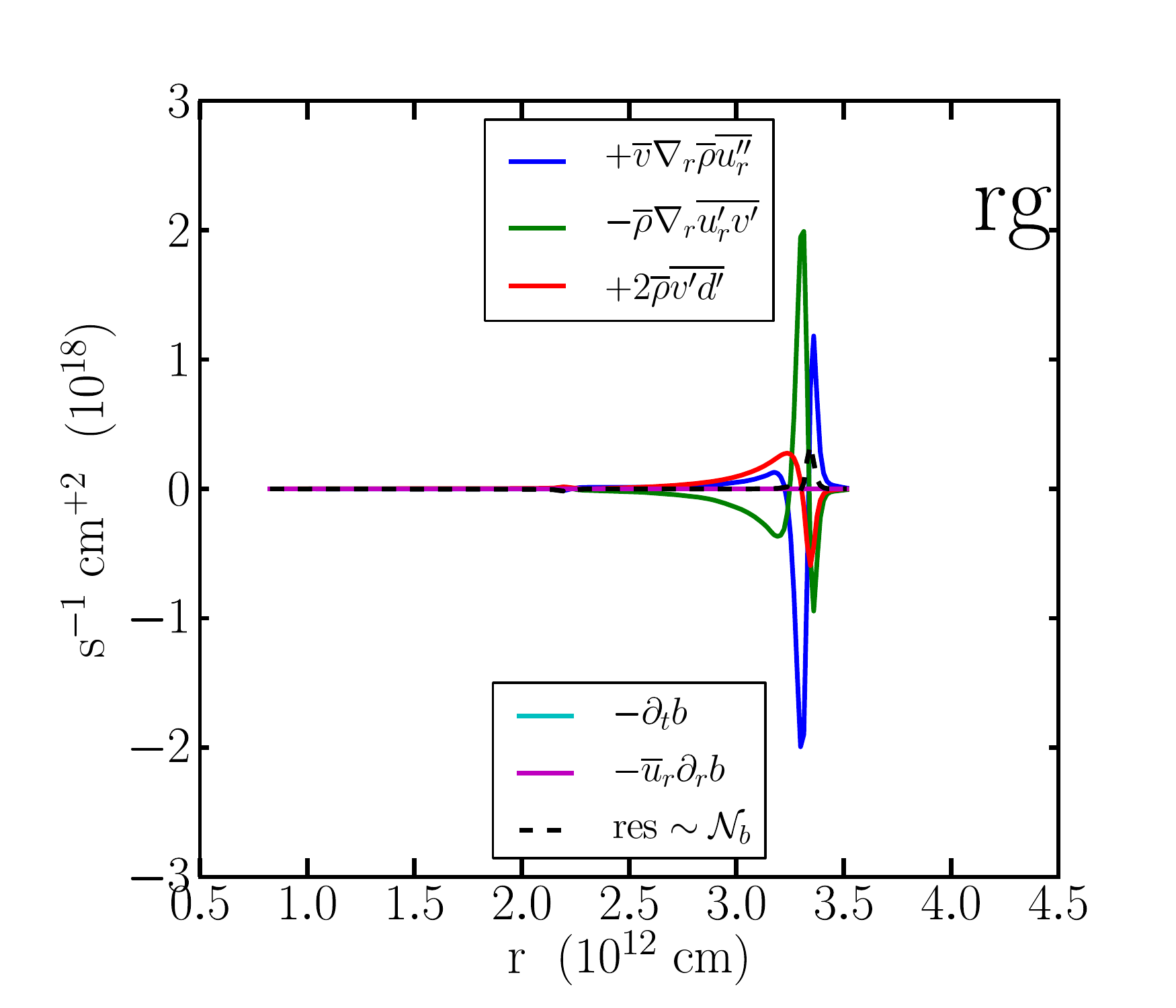}
\includegraphics[width=6.5cm]{rgmrez_tavg800_mfields_b_equation_insf-eps-converted-to.pdf}}
\caption{Mean turbulent mass flux equation (upper panels) and density-specific volume covariance equation (lower panels). 4 Hp model {\sf rg.3D.4hp} (left) and 7 Hp model {\sf rg.3D.mrez} (right).}
\end{figure}

\newpage

\subsubsection{Mean specific angular momentum equation and internal energy flux equation}

\begin{figure}[!h]
\centerline{
\includegraphics[width=6.5cm]{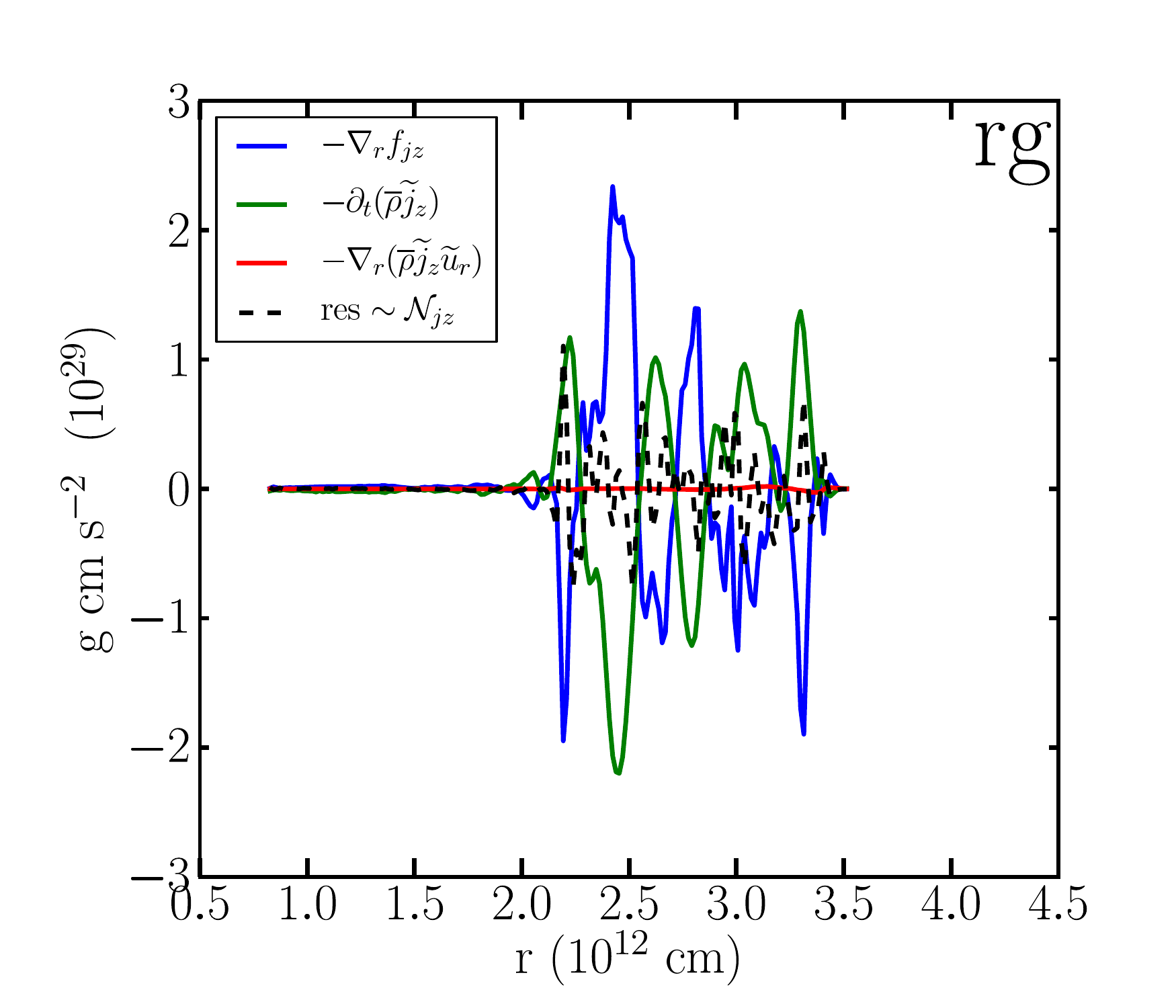}
\includegraphics[width=6.5cm]{rgmrez_tavg800_jz_equation_insf-eps-converted-to.pdf}}

\centerline{
\includegraphics[width=6.5cm]{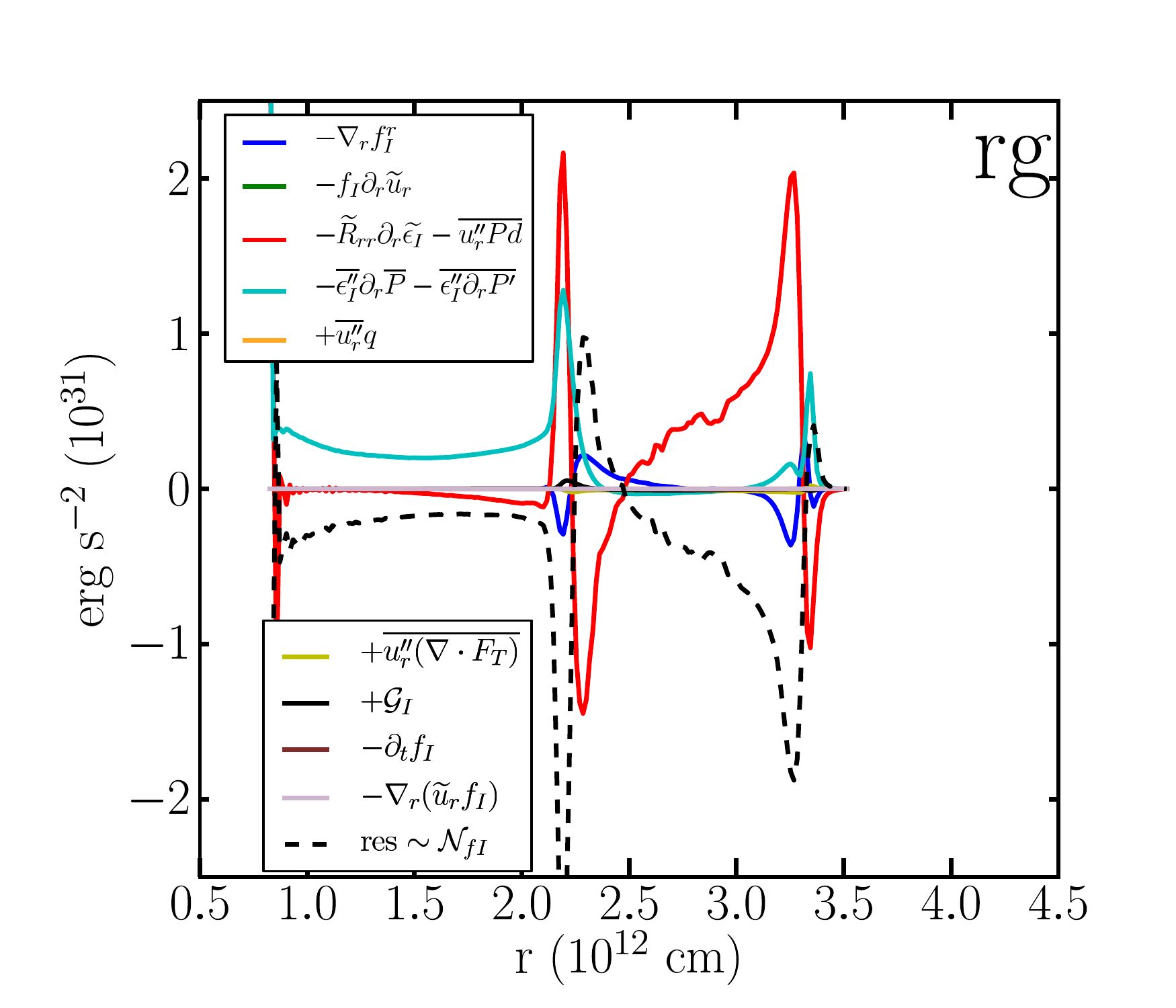}
\includegraphics[width=6.5cm]{rgmrez_tavg800_mfields_i_equation_insf-eps-converted-to.pdf}}
\caption{Mean specific angular momentum equation (upper panels) and mean turbulent internal energy flux equation (lower panels). 4 Hp model {\sf rg.3D.4hp} (left) and 7 Hp model {\sf rg.3D.mrez} (right).}
\end{figure}

\newpage

\subsubsection{Mean entropy equation and mean entropy flux equation}

\begin{figure}[!h]
\centerline{
\includegraphics[width=6.5cm]{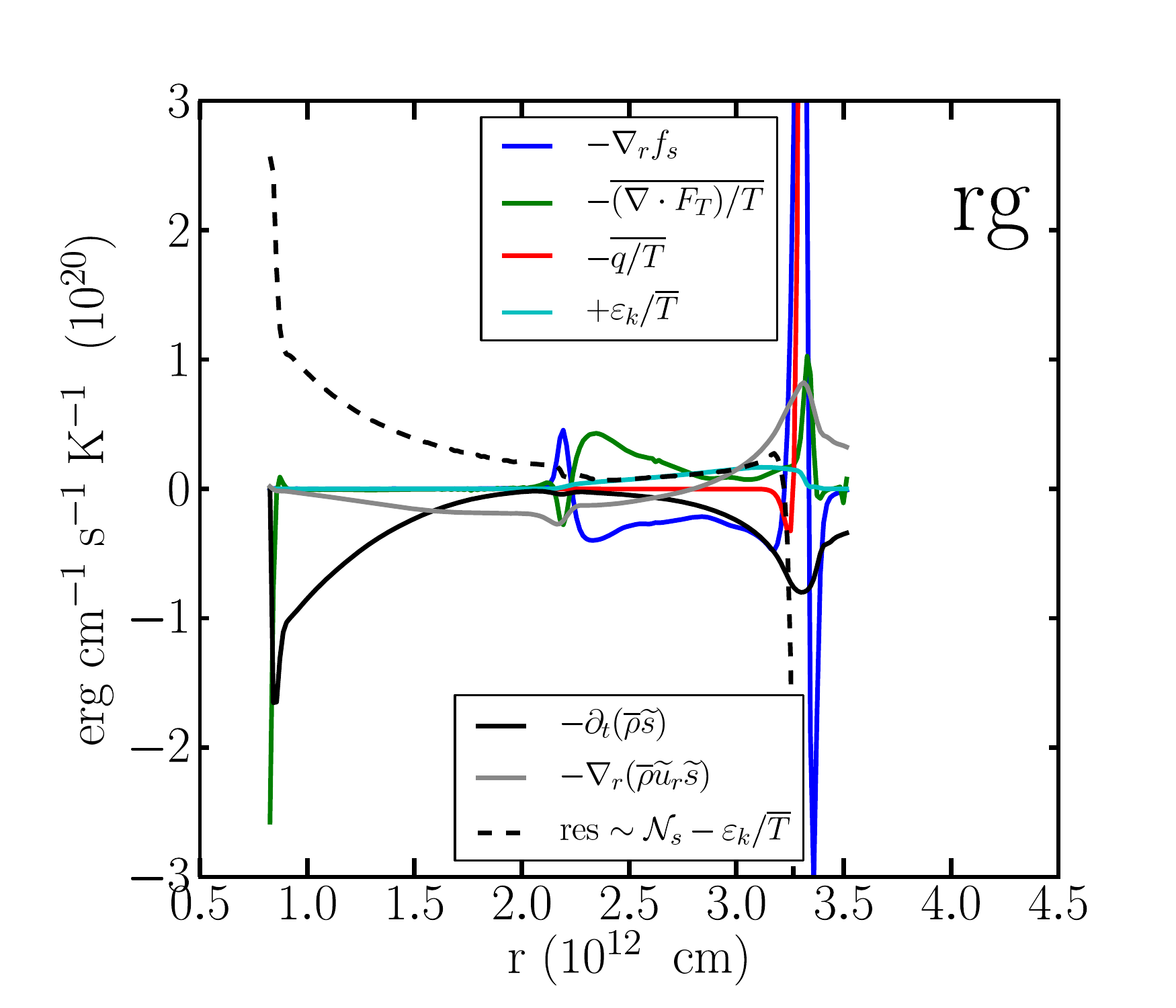}
\includegraphics[width=6.5cm]{rgmrez_tavg800_entropy_equation_insf-eps-converted-to.pdf}}

\centerline{
\includegraphics[width=6.5cm]{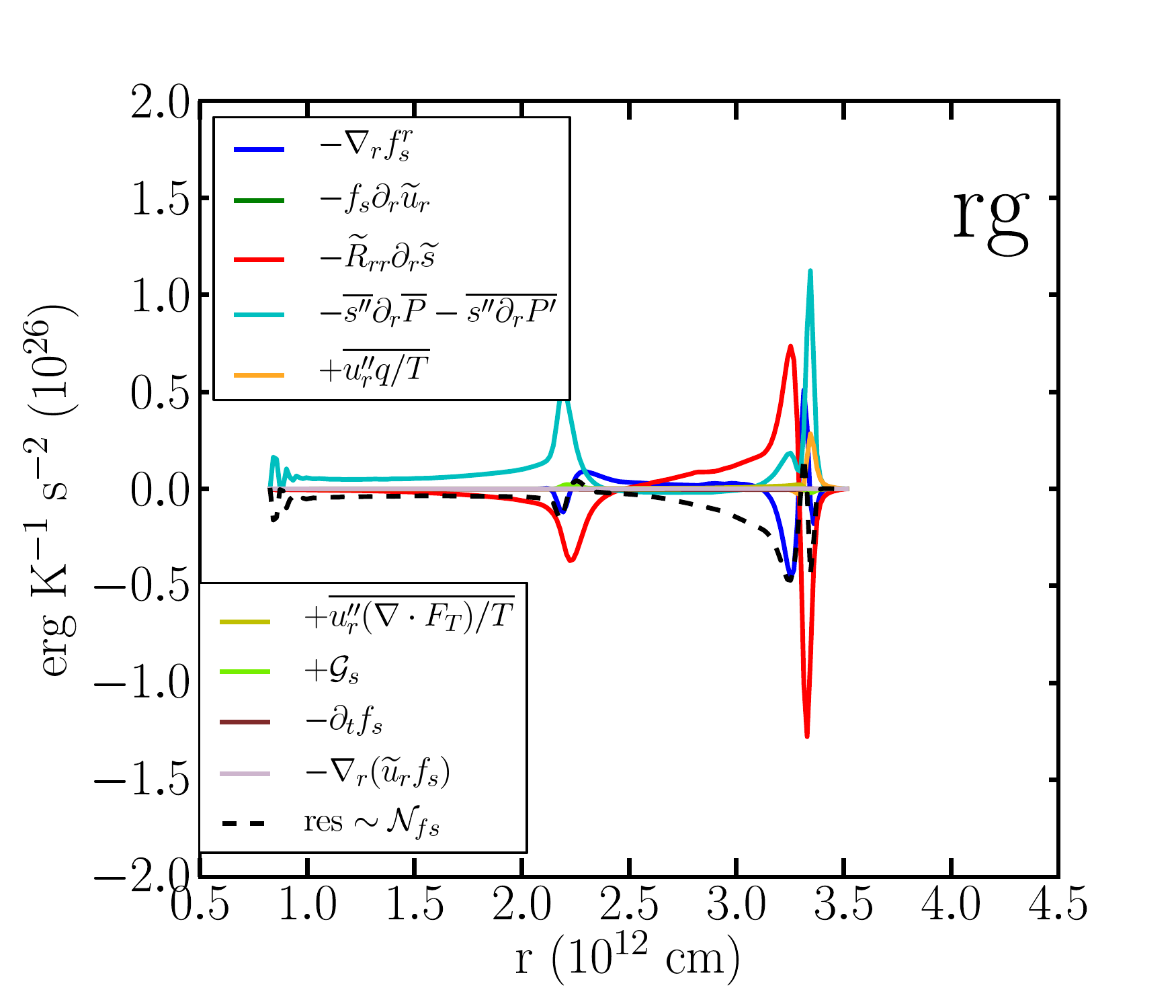}
\includegraphics[width=6.5cm]{rgmrez_tavg800_mfields_s_equation_insf-eps-converted-to.pdf}}
\caption{Mean entropy equation (upper panels) and mean entropy flux equation (lower panels). 4 Hp model {\sf rg.3D.4hp} (left) and 7 Hp model {\sf rg.3D.mrez} (right).}
\end{figure}

\newpage

\subsubsection{Mean turbulent kinetic energy and mean velocities}

\begin{figure}[!h]
\centerline{
\includegraphics[width=6.5cm]{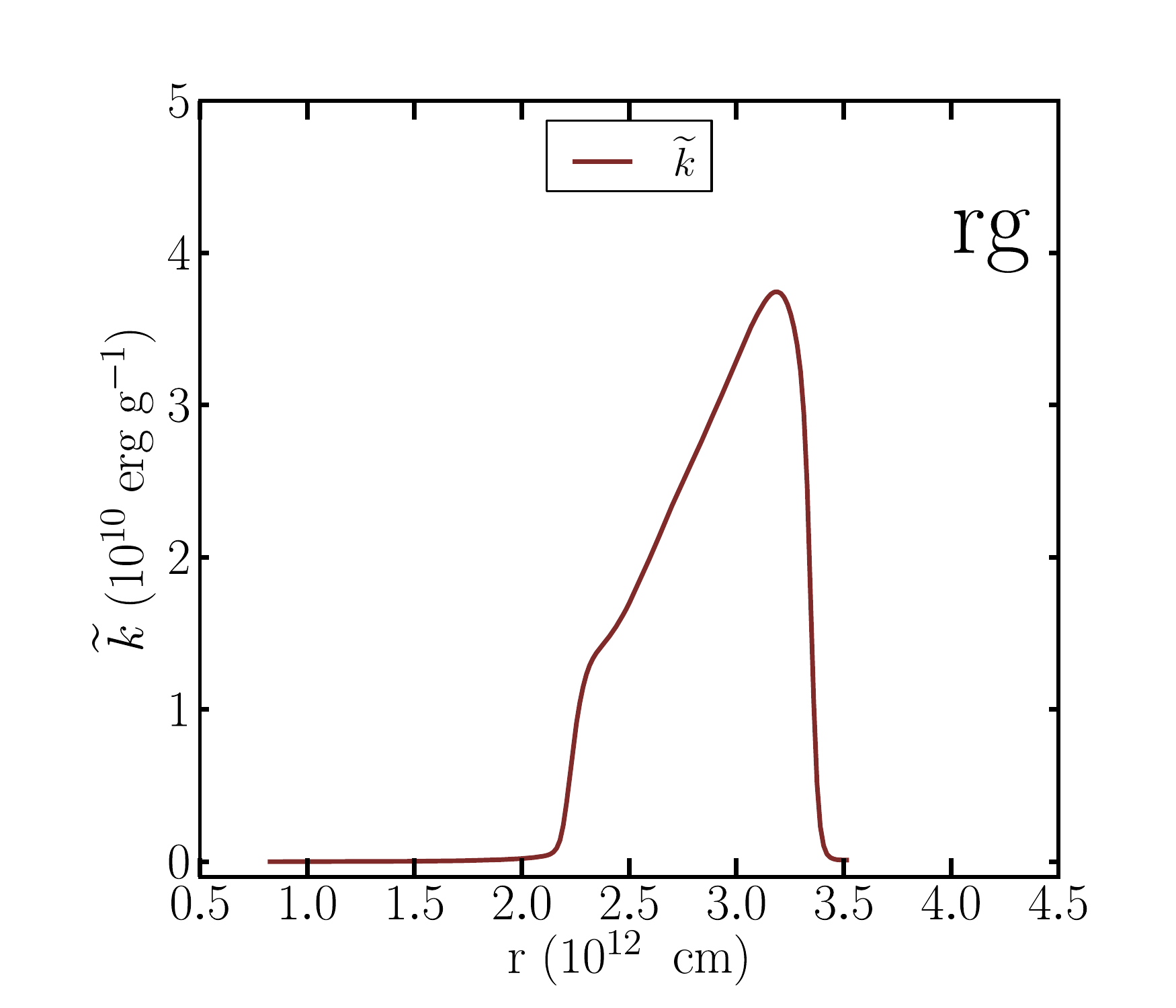}
\includegraphics[width=6.5cm]{rgmrez_tavg800_mean_k_insf-eps-converted-to.pdf}}

\centerline{
\includegraphics[width=6.5cm]{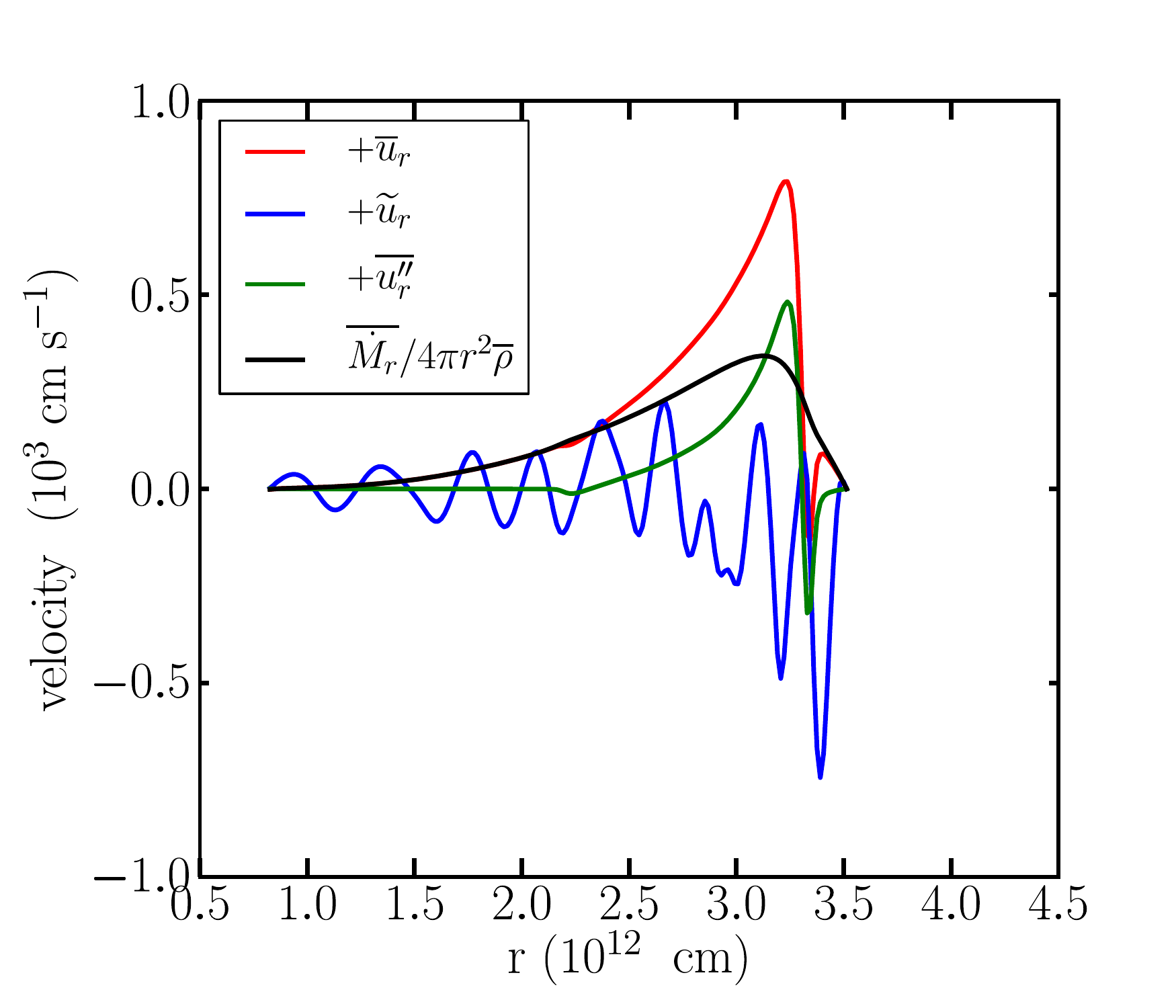}
\includegraphics[width=6.5cm]{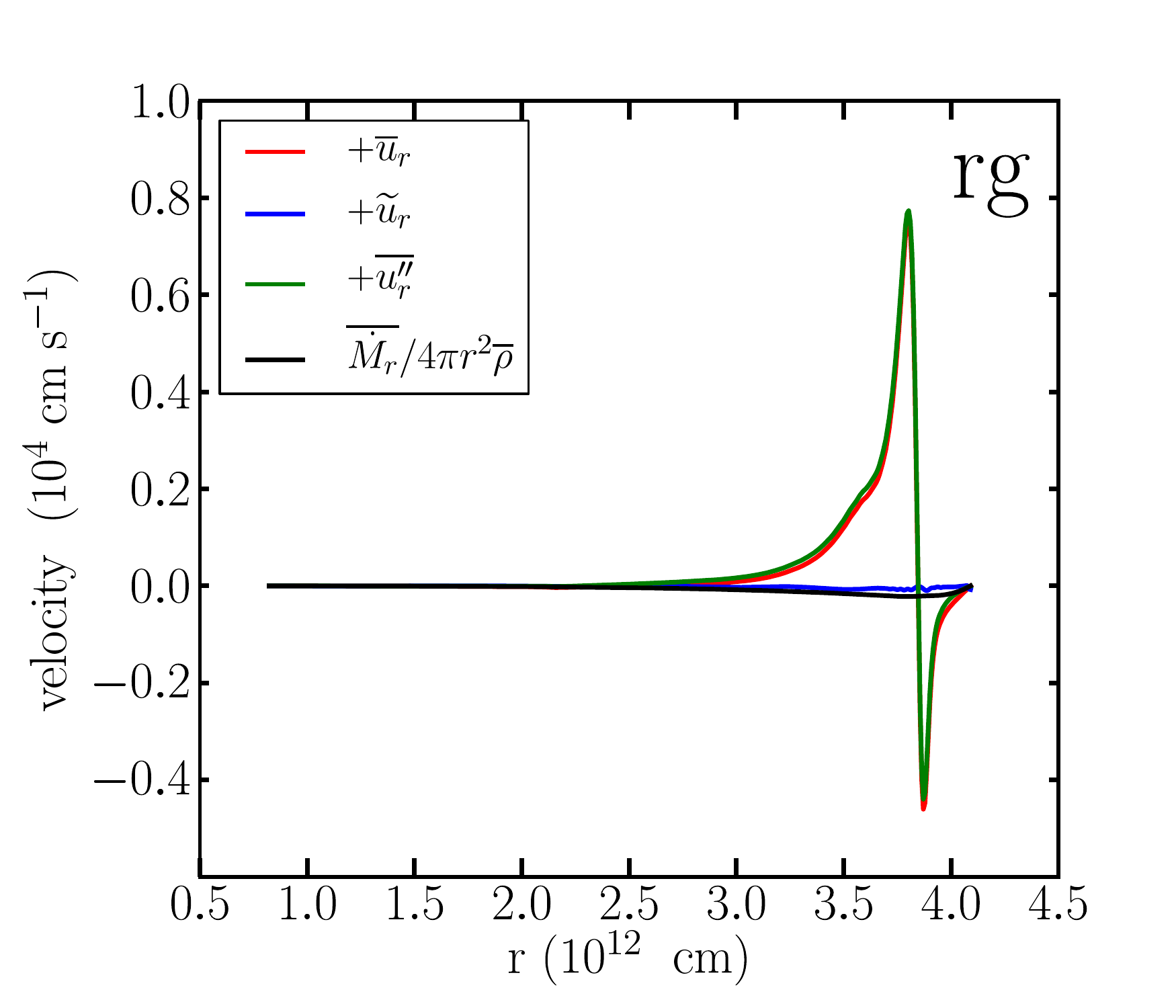}}
\caption{Mean turbulent kinetic energy (upper panels) and mean velocities (lower panels). 4 Hp model {\sf rg.3D.4hp} (left) and 7 Hp model {\sf rg.3D.mrez} (right).}
\end{figure}

\newpage

\section{Dependence on Convection Zone Driving Source Profile (Heating and Cooling)}

\subsection{Oxygen burning shell models}

\subsubsection{Mean continuity equation and mean radial momentum equation}

\begin{figure}[!h]
\centerline{
\includegraphics[width=5.6cm]{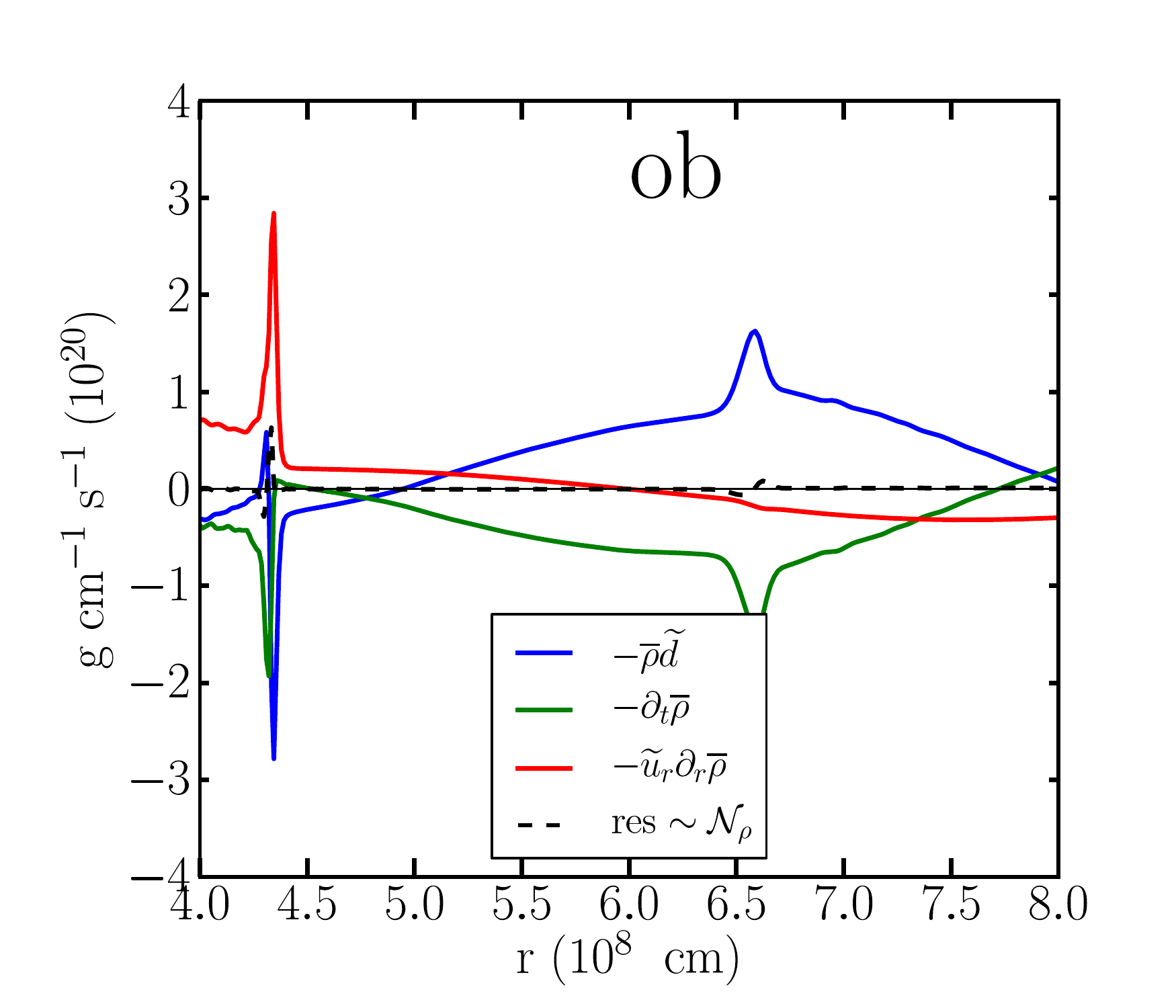}
\includegraphics[width=5.6cm]{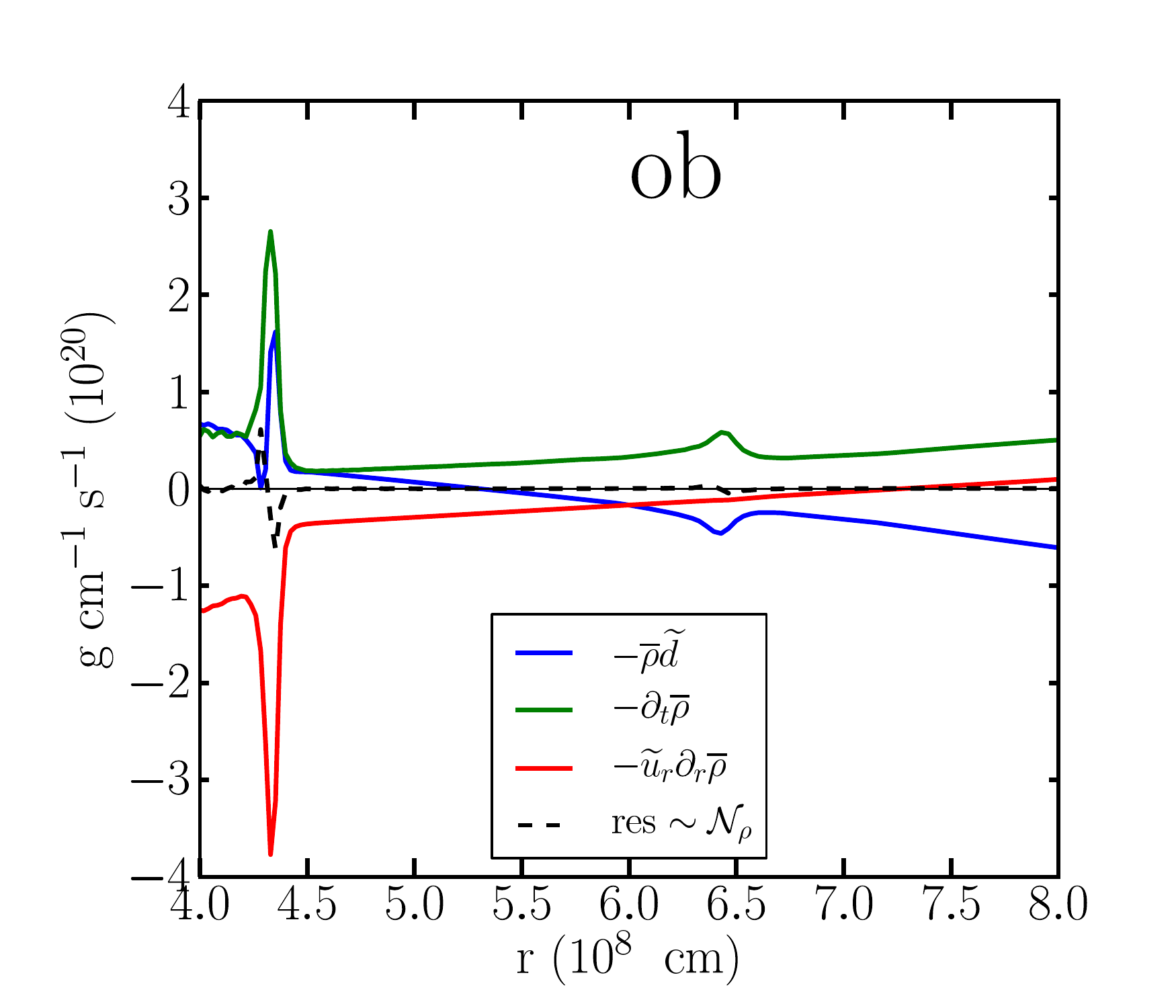}}

\centerline{
\includegraphics[width=5.6cm]{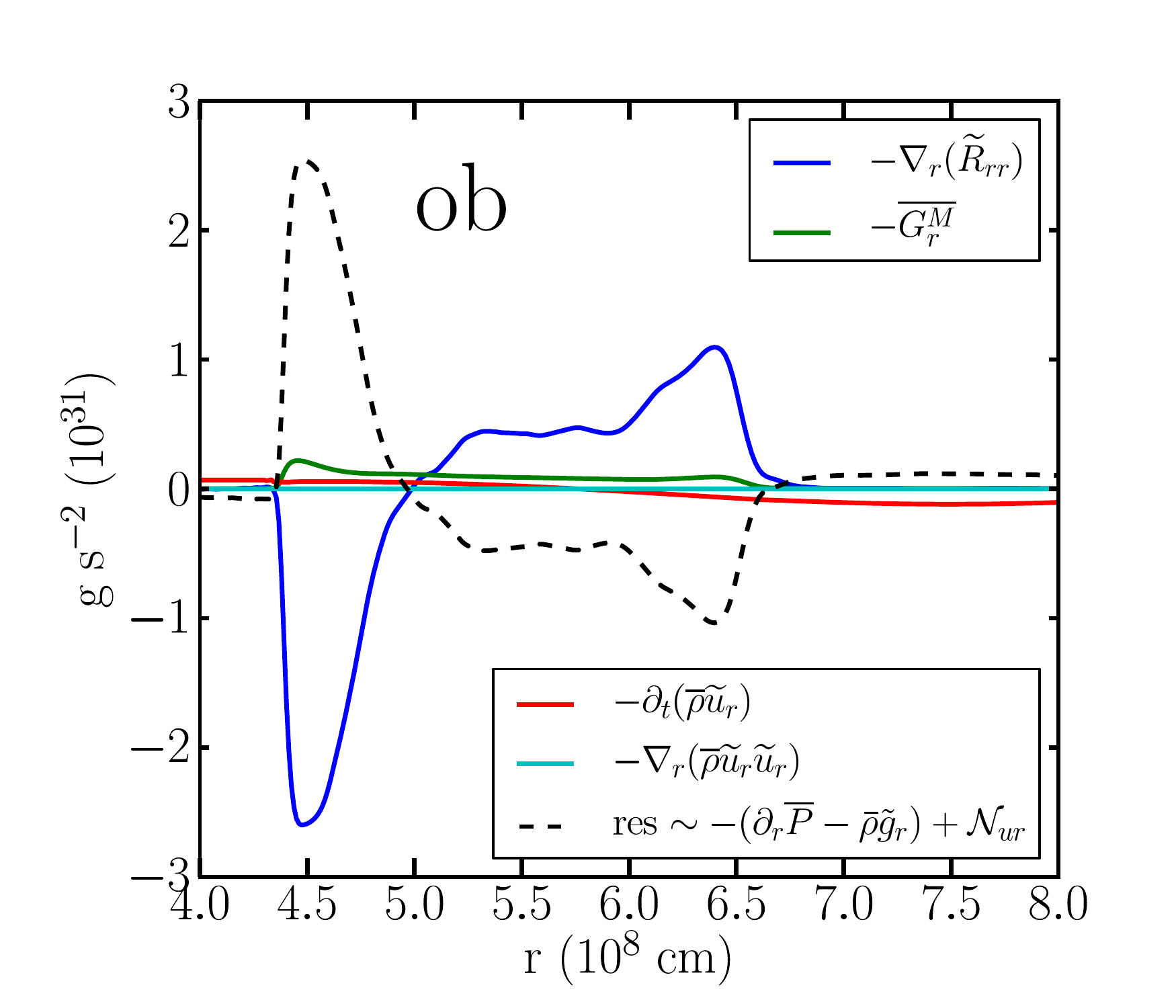}
\includegraphics[width=5.6cm]{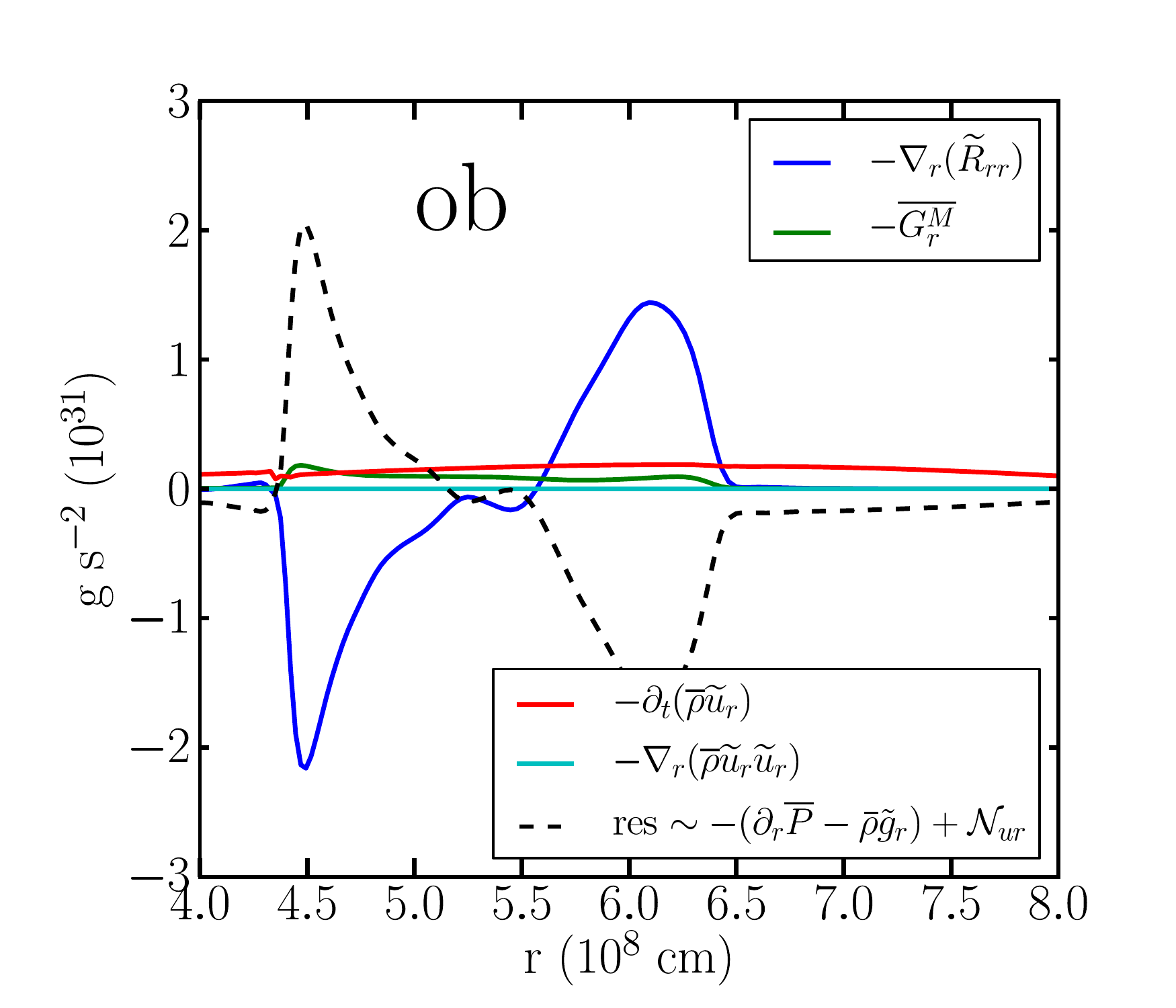}}
\caption{Mean continuity equation (upper panels) and radial momentum equation (lower panels). Model with volumetric heating at the bottom of convection zone {\sf ob.3D.1hp.vh} (left) and model with volumetric cooling at the top of convection zone {\sf ob.3D.1hp.vc} (right).}
\end{figure}

\newpage

\subsubsection{Mean azimuthal and polar momentum equation}

\begin{figure}[!h]
\centerline{
\includegraphics[width=6.5cm]{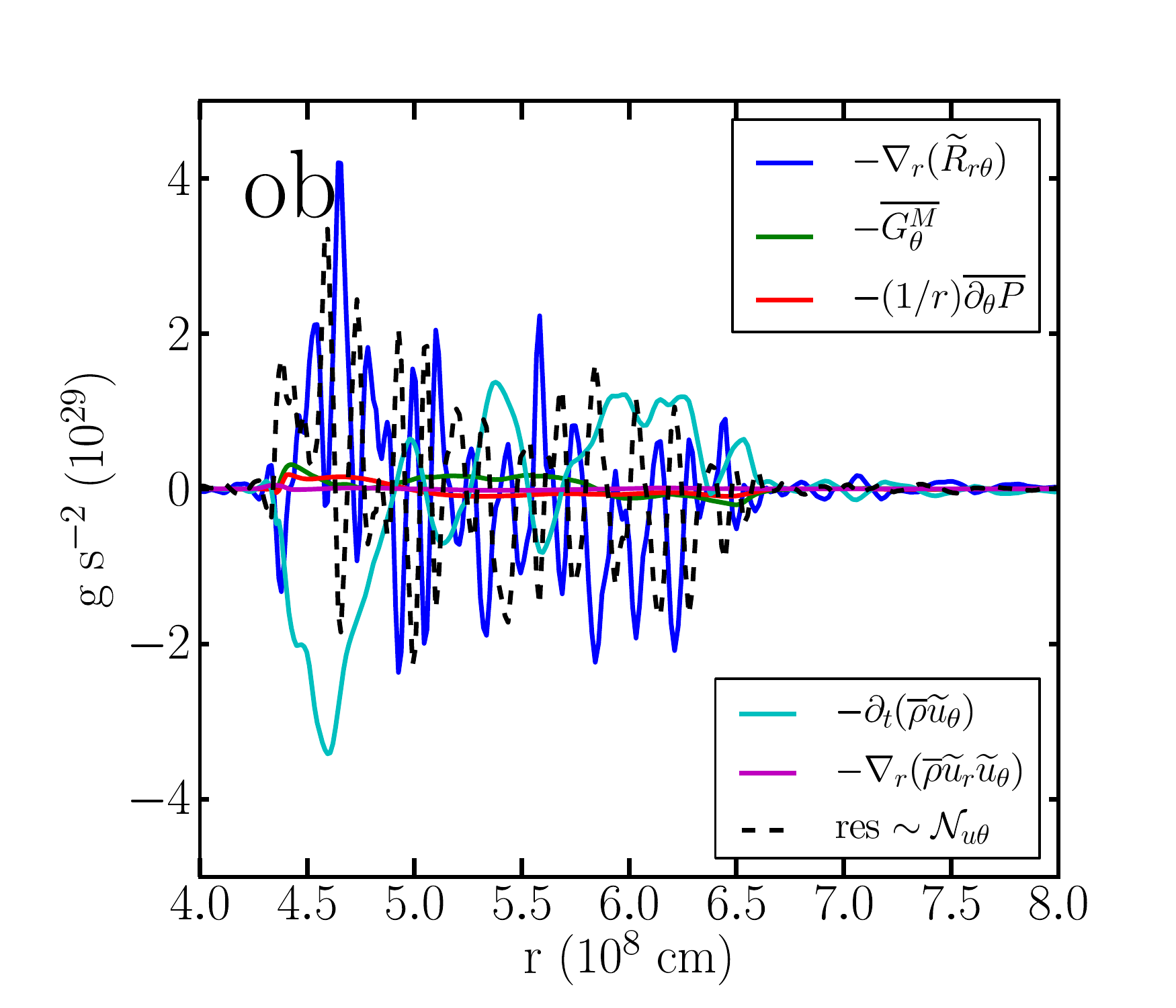}
\includegraphics[width=6.5cm]{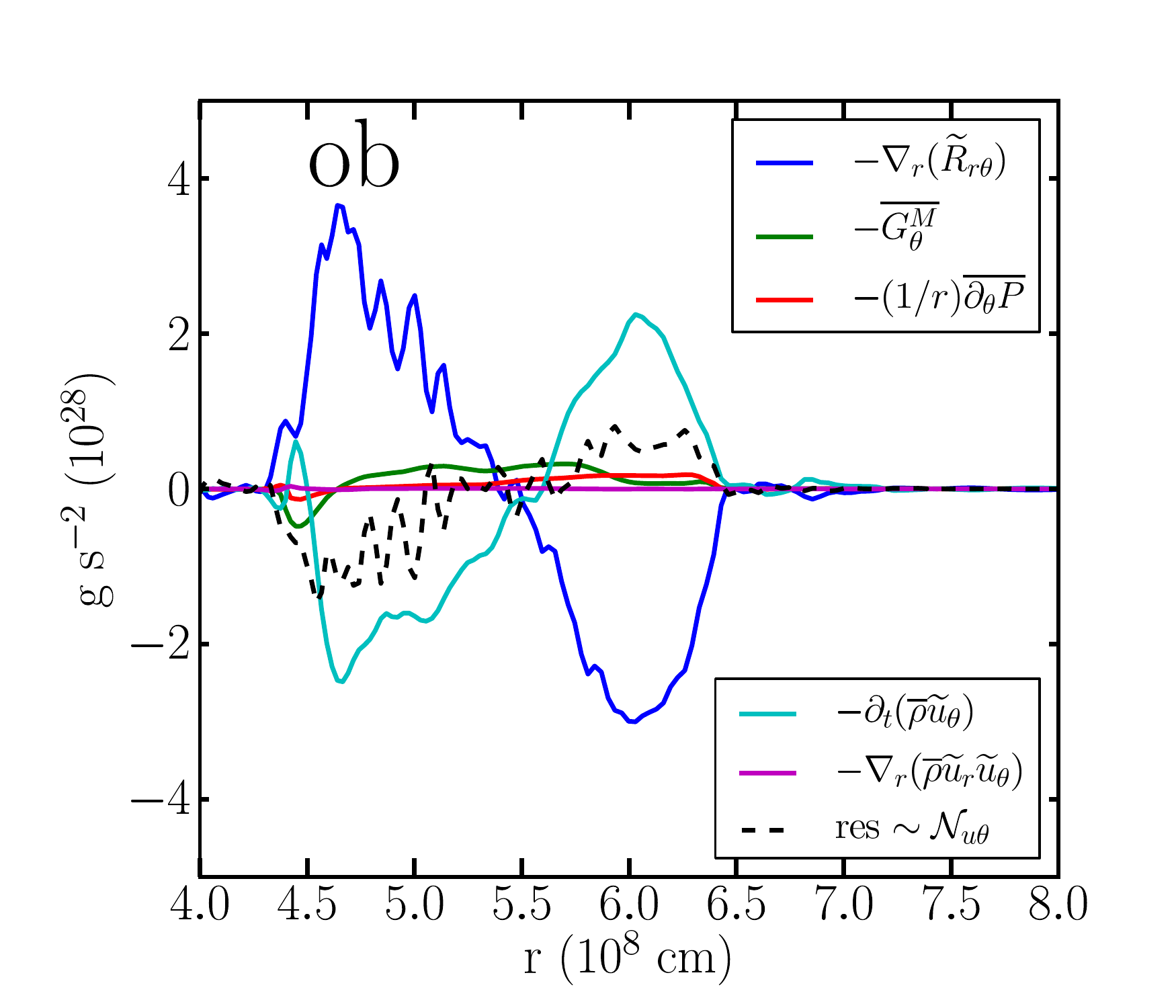}}

\centerline{
\includegraphics[width=6.5cm]{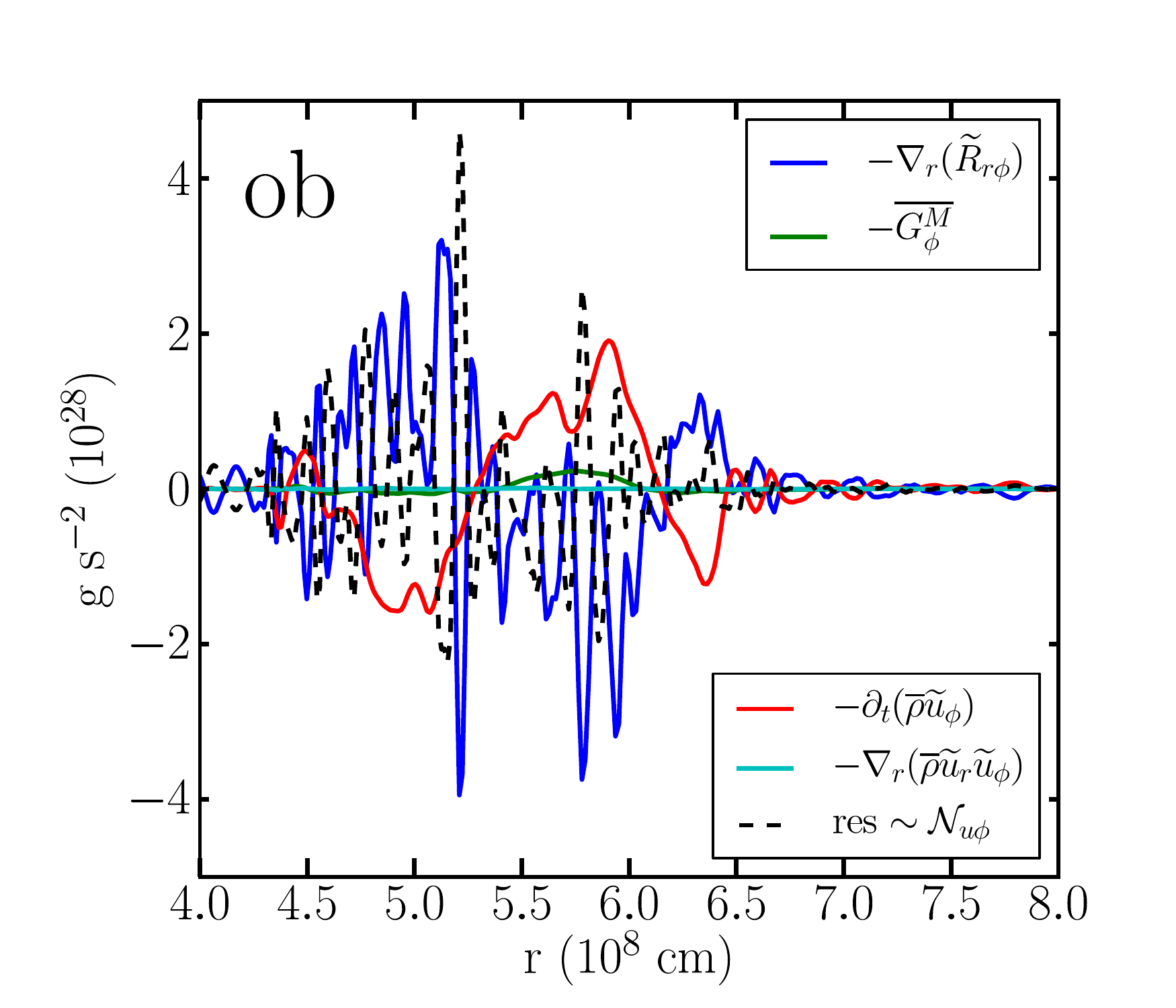}
\includegraphics[width=6.5cm]{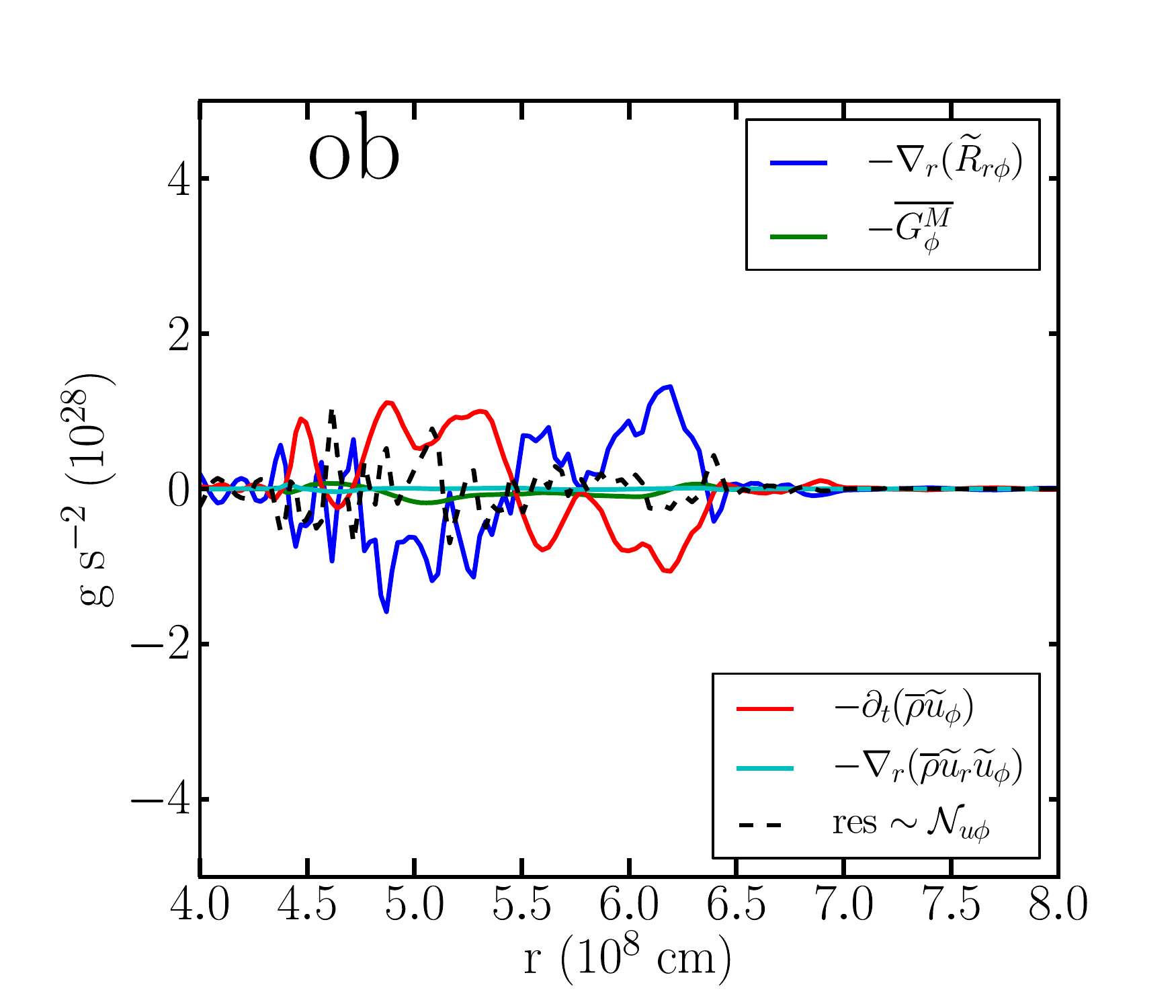}}
\caption{Mean azimuthal equation (upper panels) and mean polar momentum equation (lower panels). Model with volumetric heating at the bottom of convection zone {\sf ob.3D.1hp.vh} (left) and model with volumetric cooling at the top of convection zone {\sf ob.3D.1hp.vc} (right).}
\end{figure}

\newpage

\subsubsection{Mean total energy equation and mean turbulent kinetic energy equation}

\begin{figure}[!h]
\centerline{
\includegraphics[width=6.6cm]{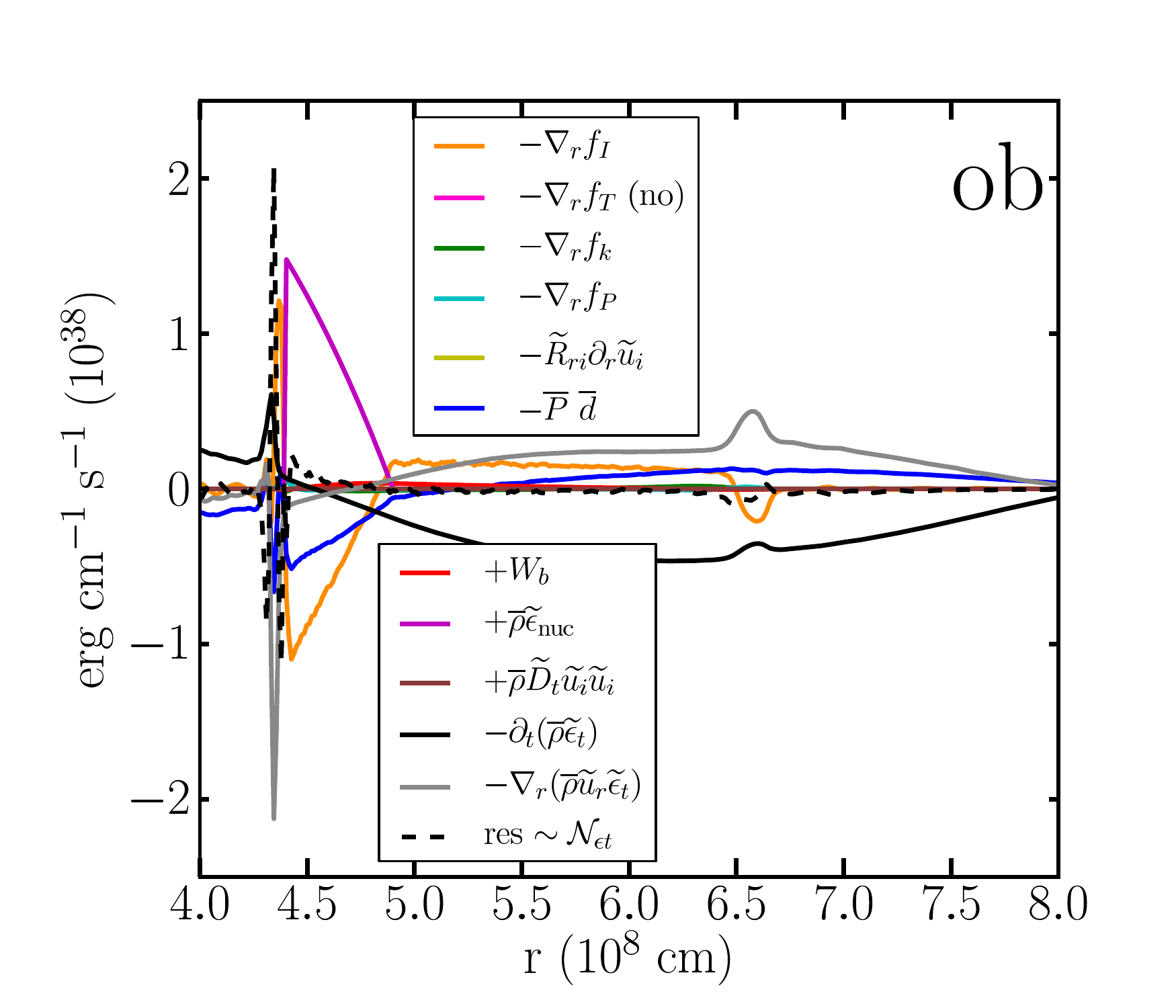}
\includegraphics[width=6.6cm]{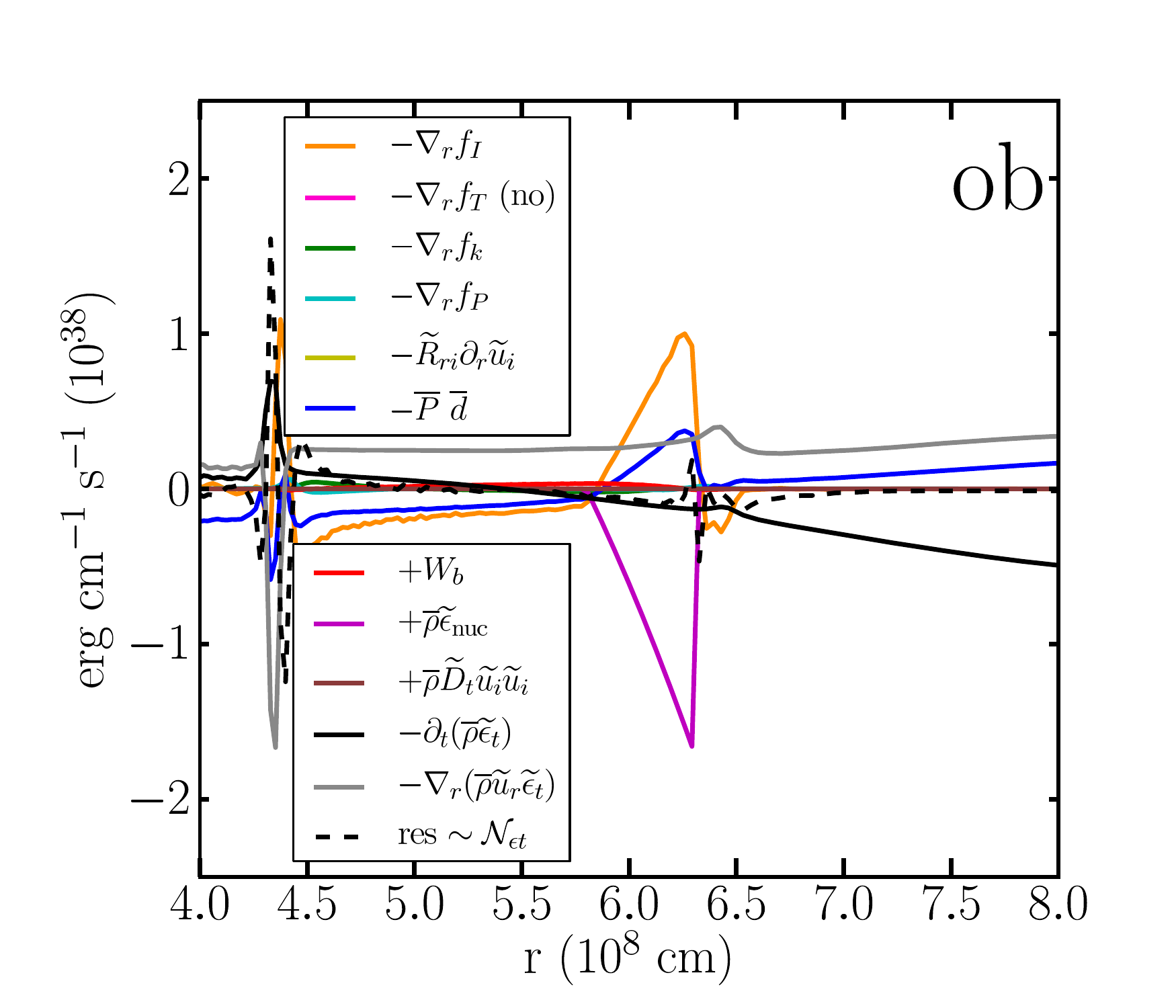}}

\centerline{
\includegraphics[width=6.6cm]{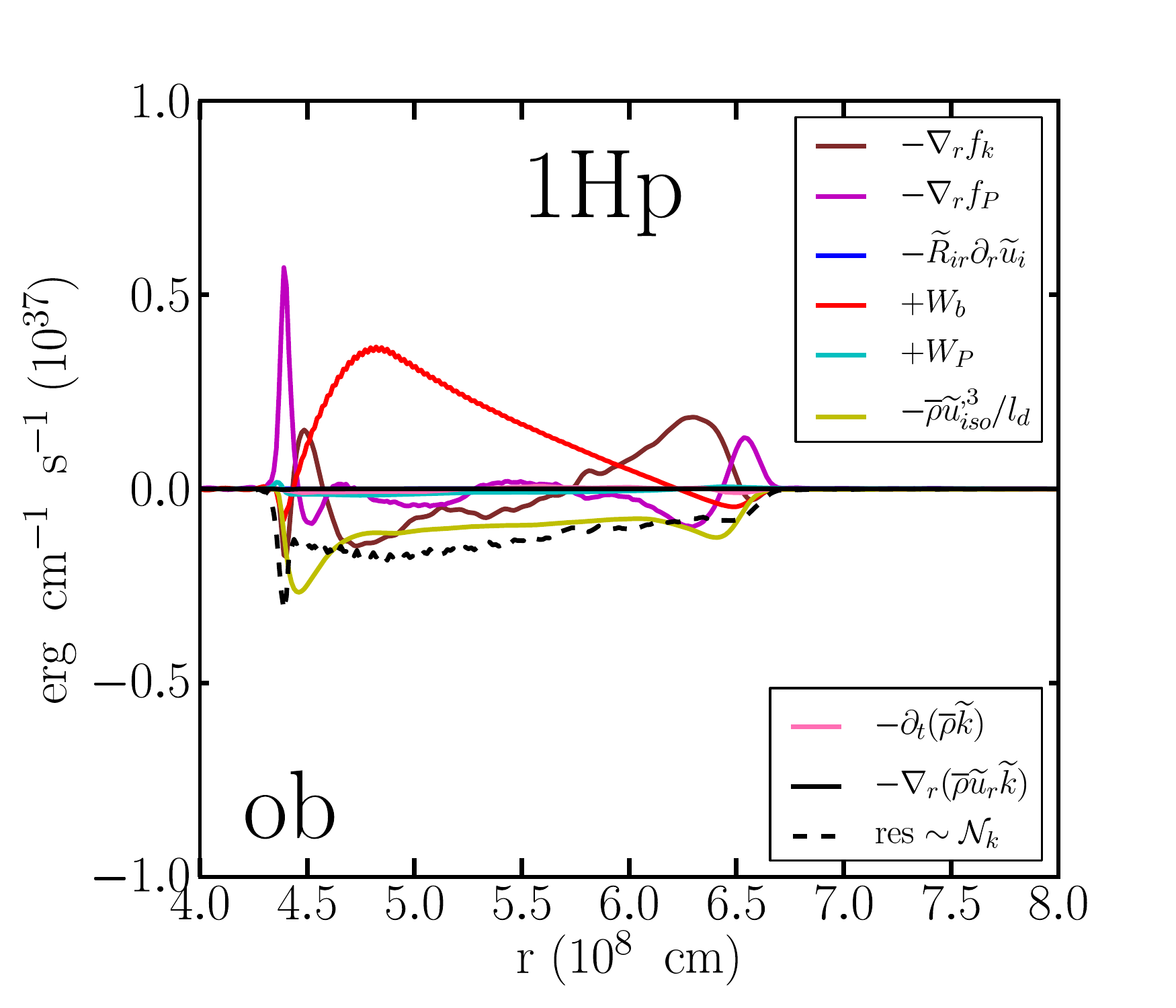}
\includegraphics[width=6.6cm]{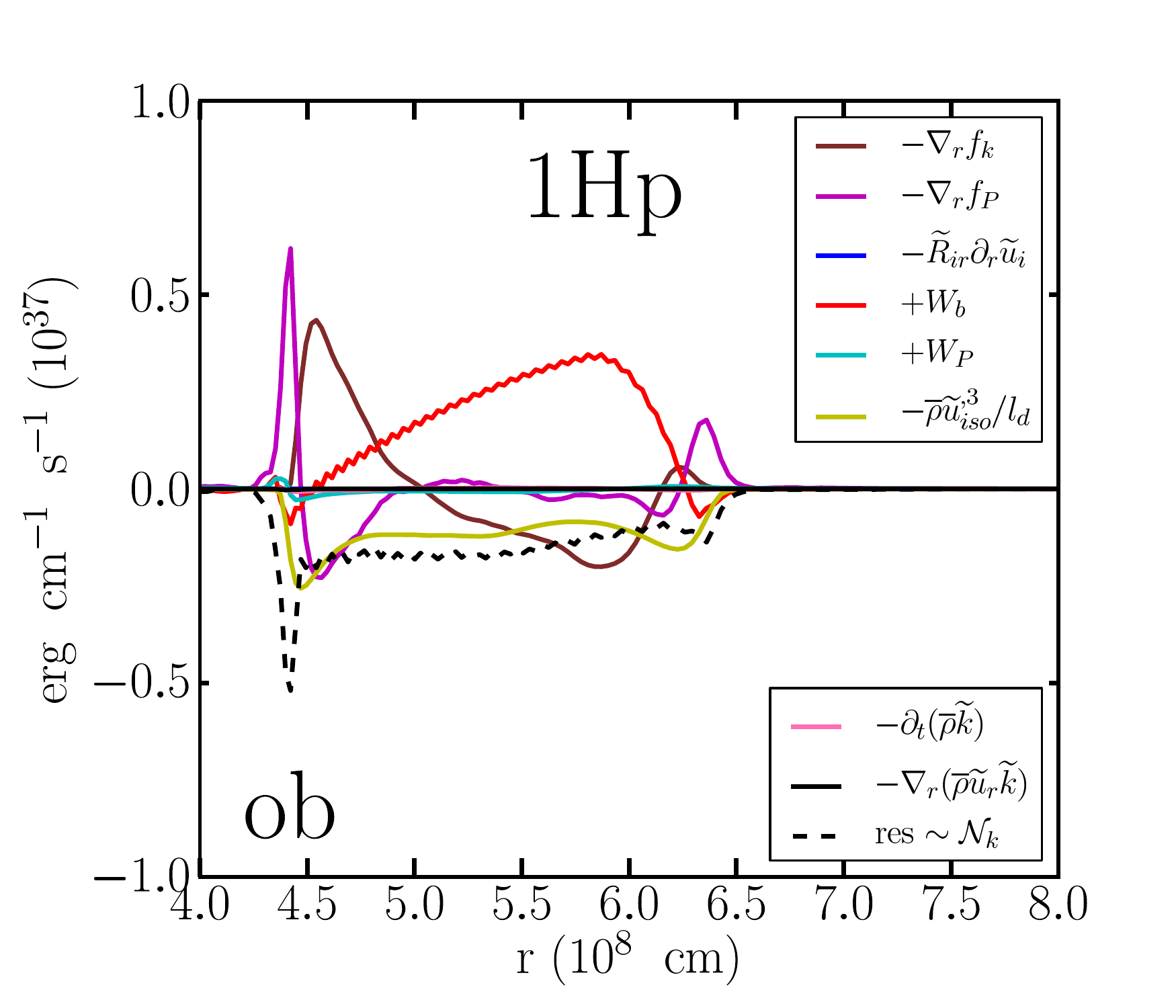}}
\caption{Mean total energy equation (upper panels) and mean turbulent kinetic energy equation (lower panels). Model with volumetric heating at the bottom of convection zone {\sf ob.3D.1hp.vh} (left) and model with volumetric cooling at the top of convection zone {\sf ob.3D.1hp.vc} (right).}
\end{figure}

\newpage

\subsubsection{Mean turbulent kinetic energy equations (radial + horizontal part)}

\begin{figure}[!h]
\centerline{
\includegraphics[width=6.6cm]{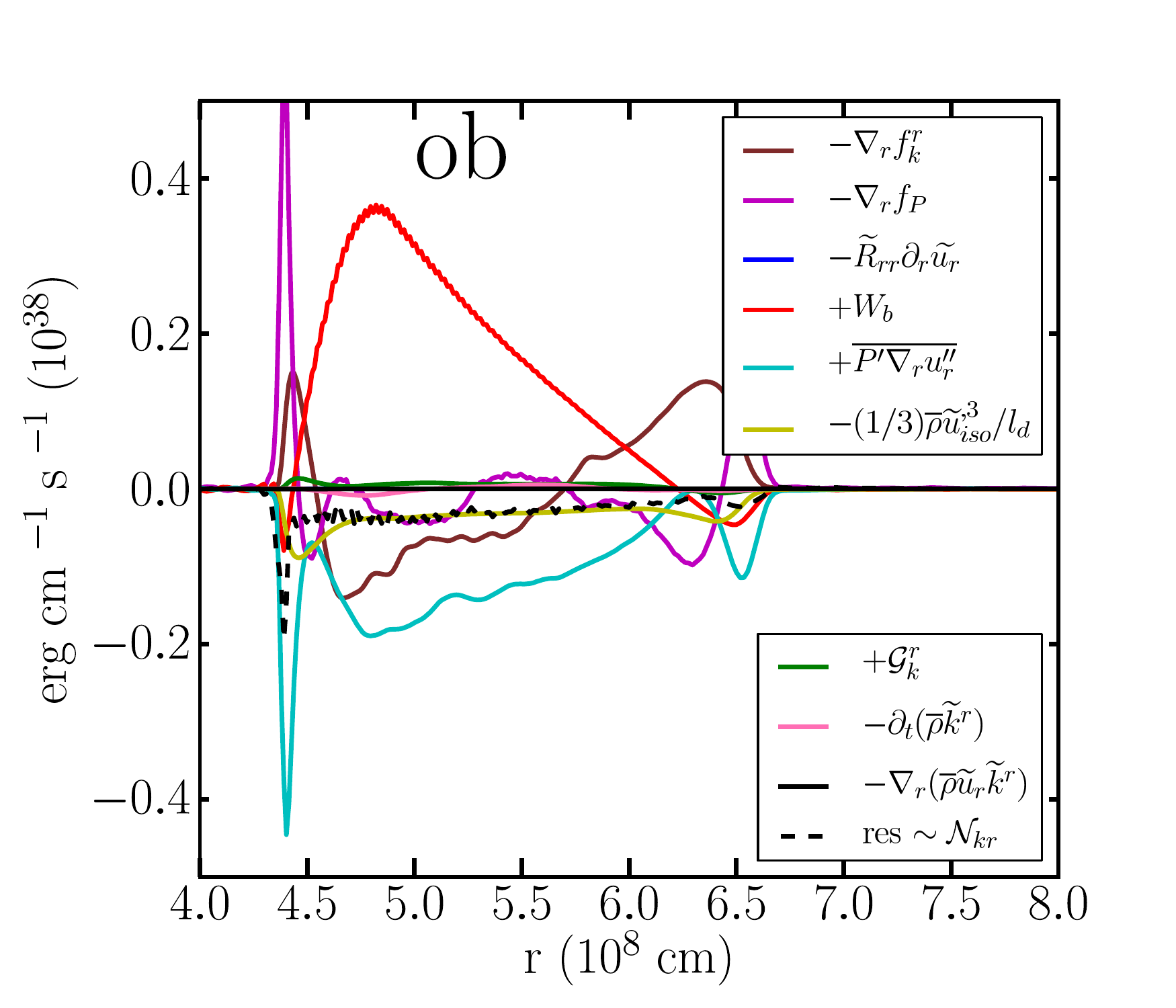}
\includegraphics[width=6.6cm]{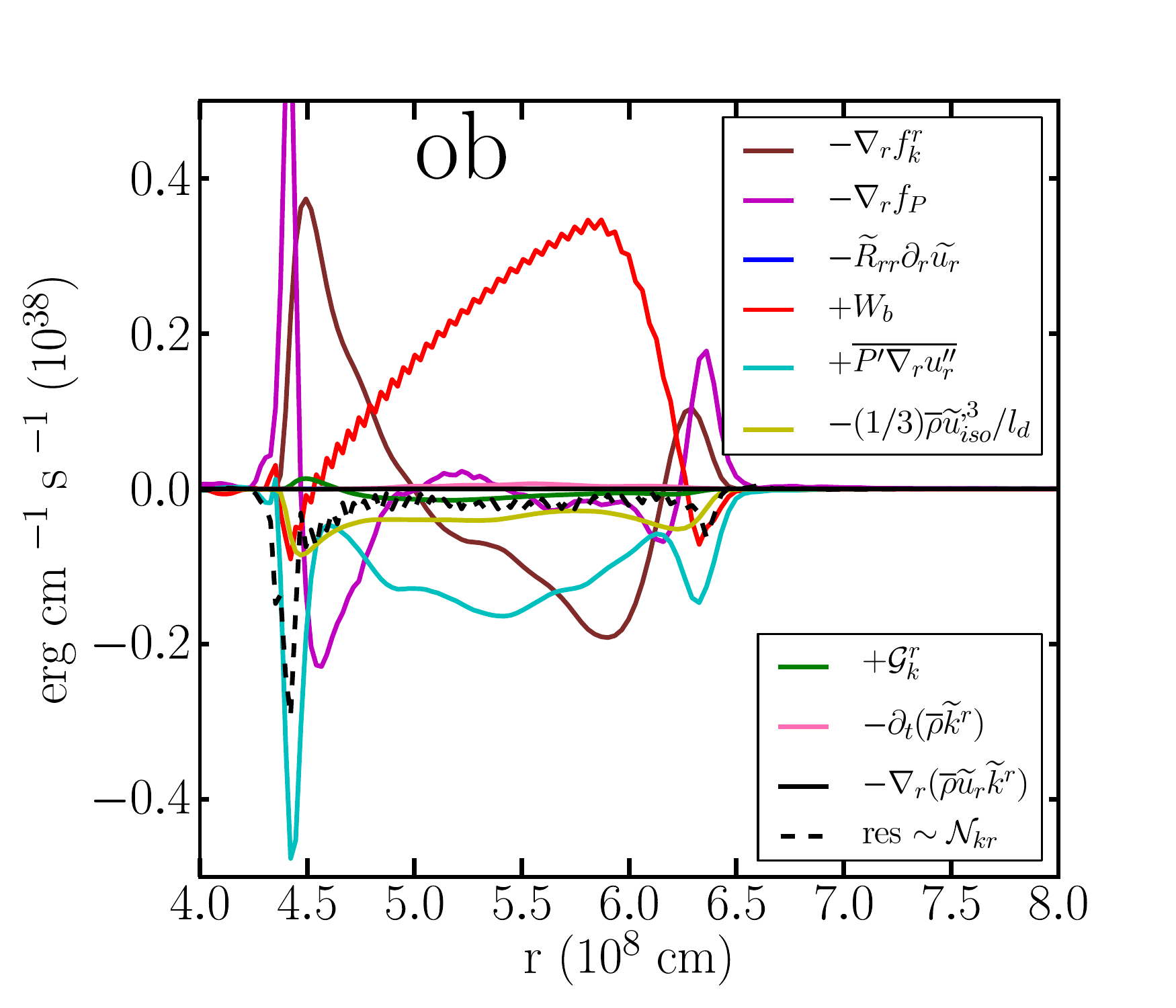}}

\centerline{
\includegraphics[width=6.6cm]{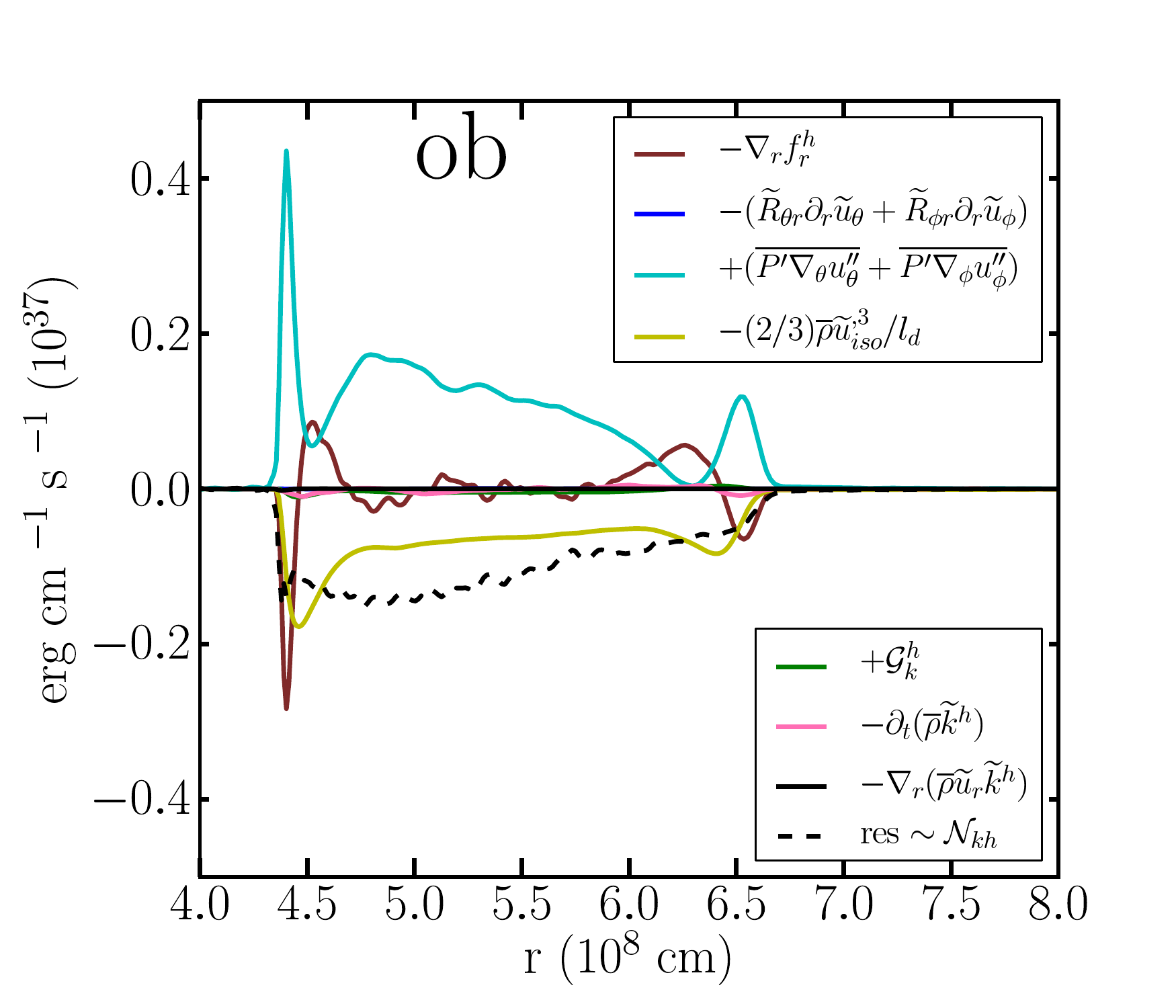}
\includegraphics[width=6.6cm]{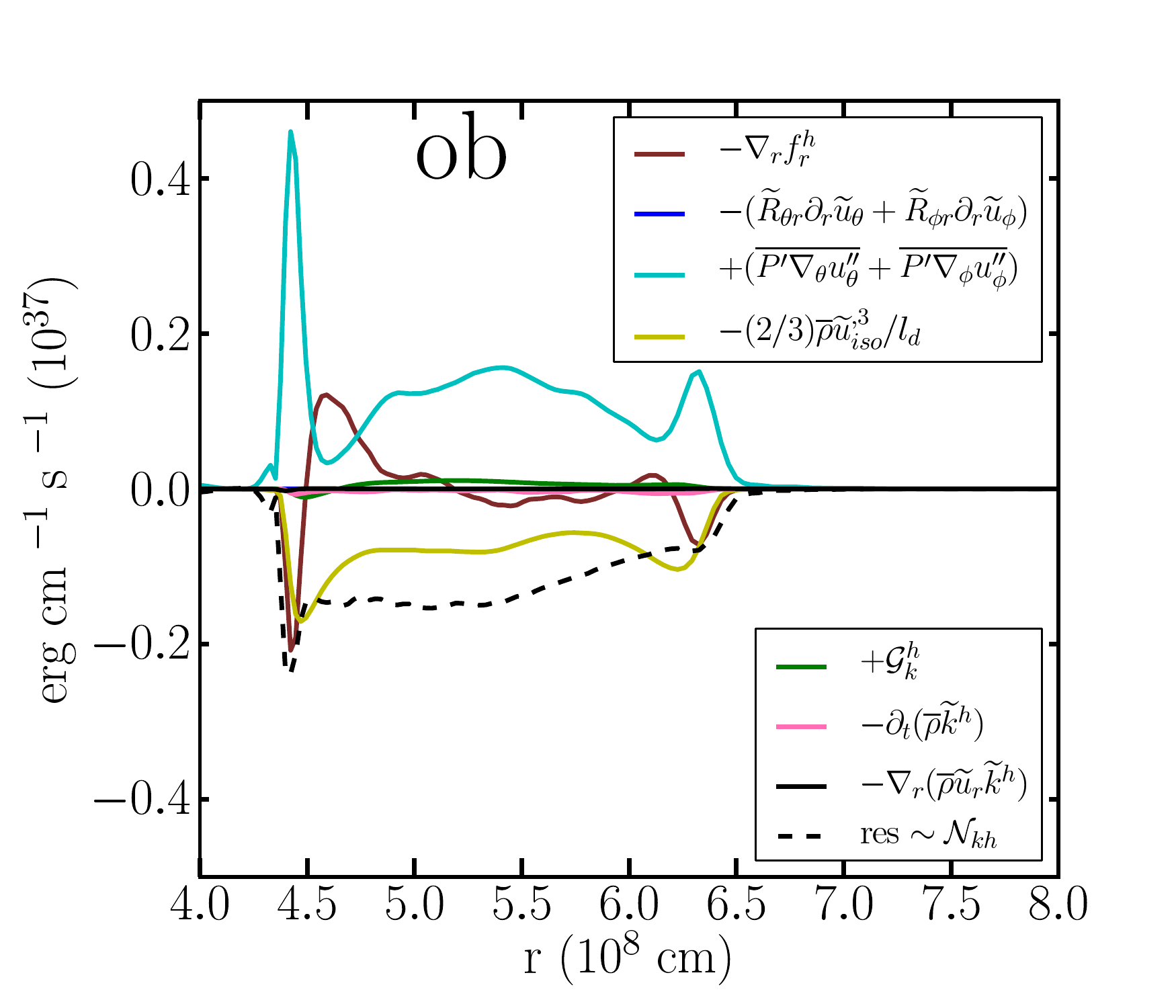}}
\caption{Radial (upper panels) and horizontal (lower panels) part of the mean turbulent kinetic energy equation. Model with volumetric heating at the bottom of convection zone {\sf ob.3D.1hp.vh} (left) and model with volumetric cooling at the top of convection zone {\sf ob.3D.1hp.vc} (right).}
\end{figure}

\newpage

\subsubsection{Mean turbulent mass flux and mean density-specific volume covariance equations}

\begin{figure}[!h]
\centerline{
\includegraphics[width=6.6cm]{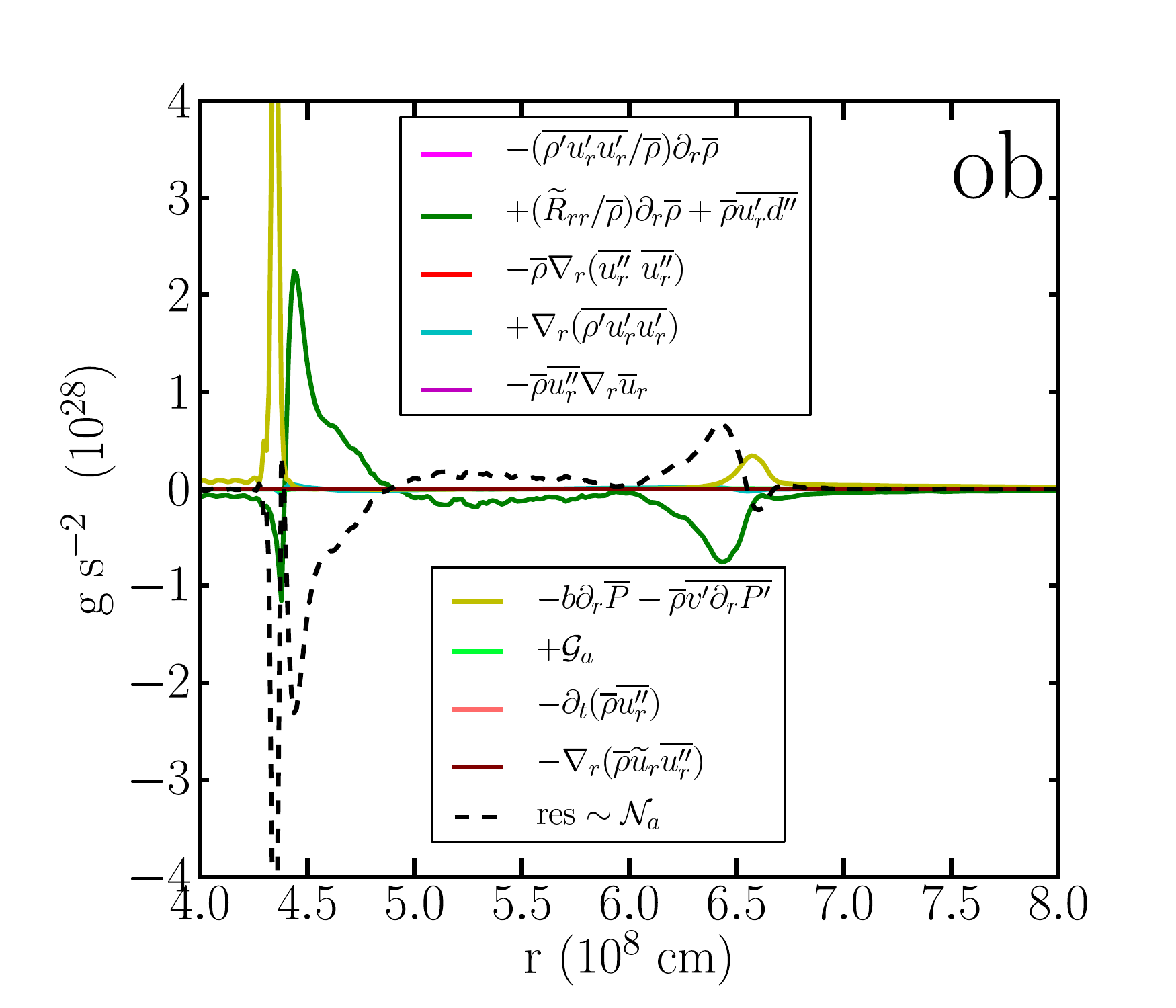}
\includegraphics[width=6.6cm]{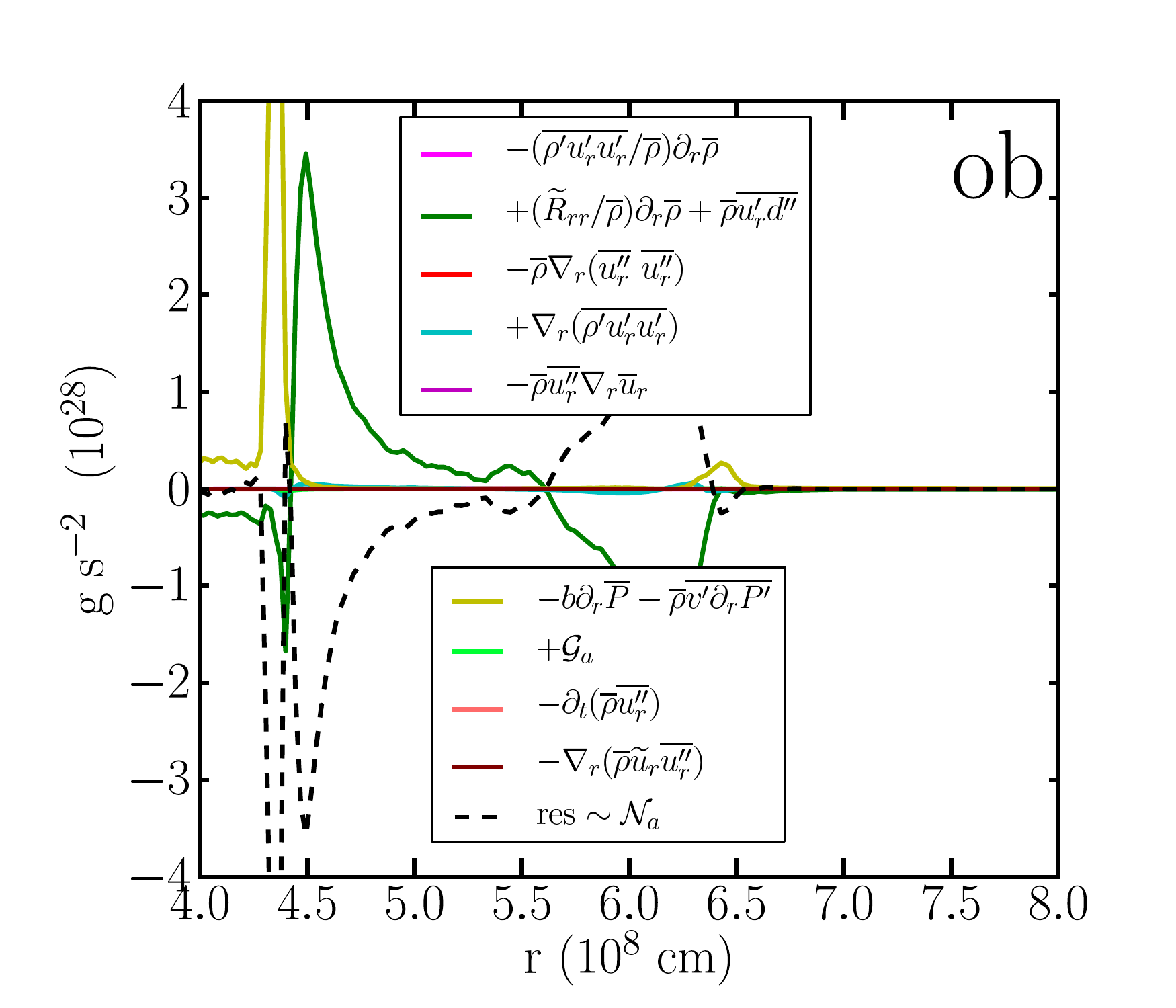}}

\centerline{
\includegraphics[width=6.6cm]{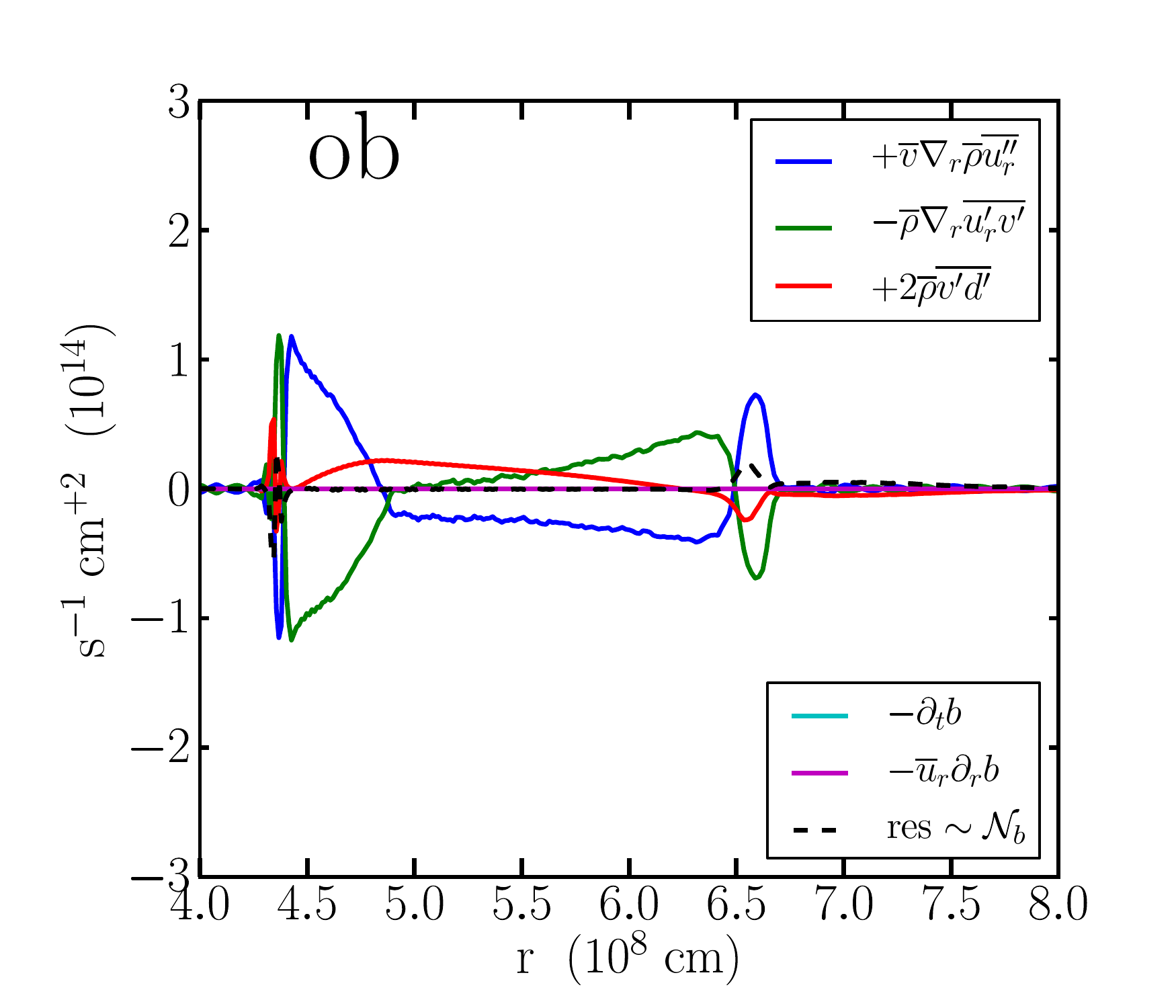}
\includegraphics[width=6.6cm]{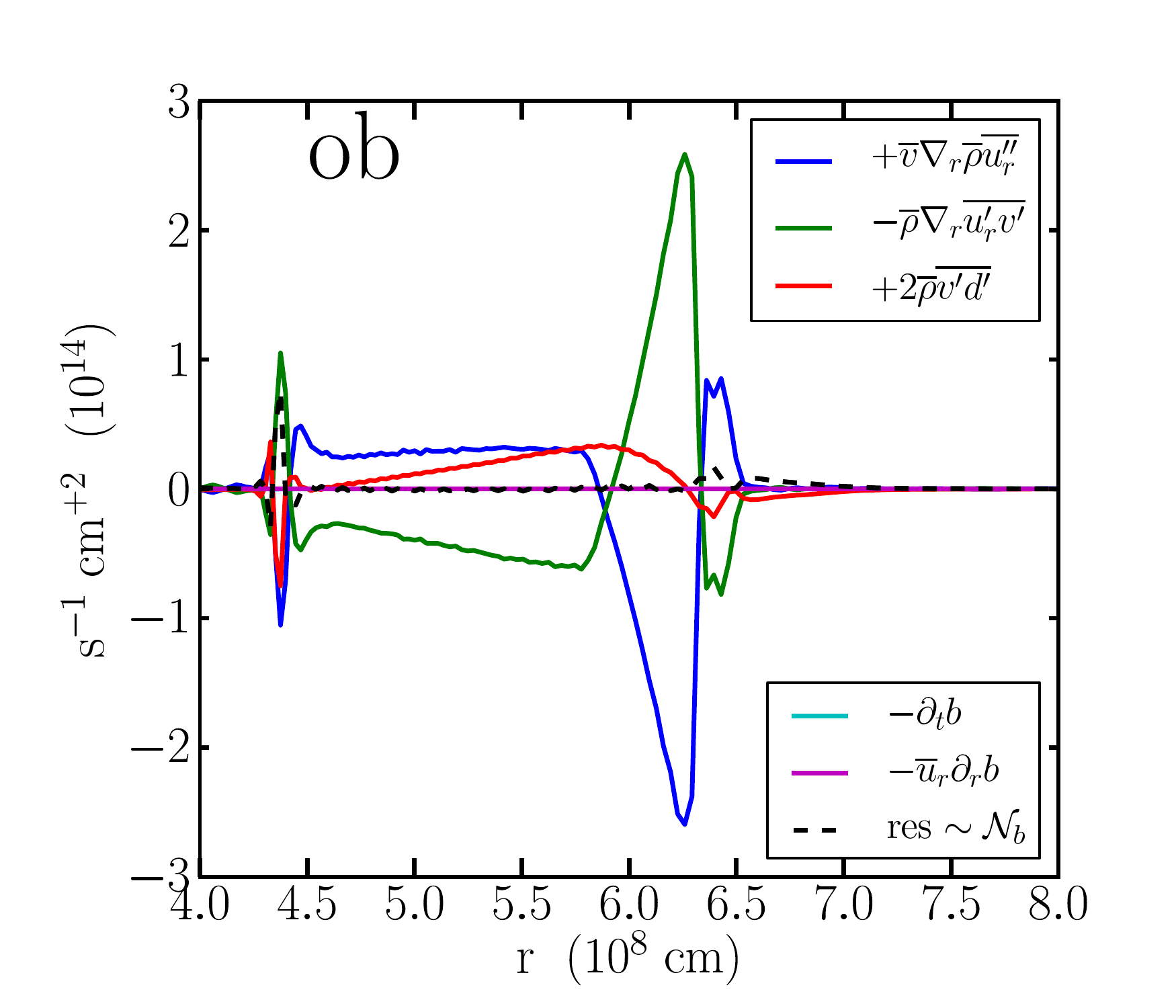}}
\caption{Mean turbulent mass flux equation (upper panels) and density-specific volume covariance equation (lower panels). Model with volumetric heating at the bottom of convection zone {\sf ob.3D.1hp.vh} (left) and model with volumetric cooling at the top of convection zone {\sf ob.3D.1hp.vc} (right).}
\end{figure}

\newpage

\subsubsection{Mean specific angular momentum equation and mean internal energy flux equation}

\begin{figure}[!h]
\centerline{
\includegraphics[width=6.6cm]{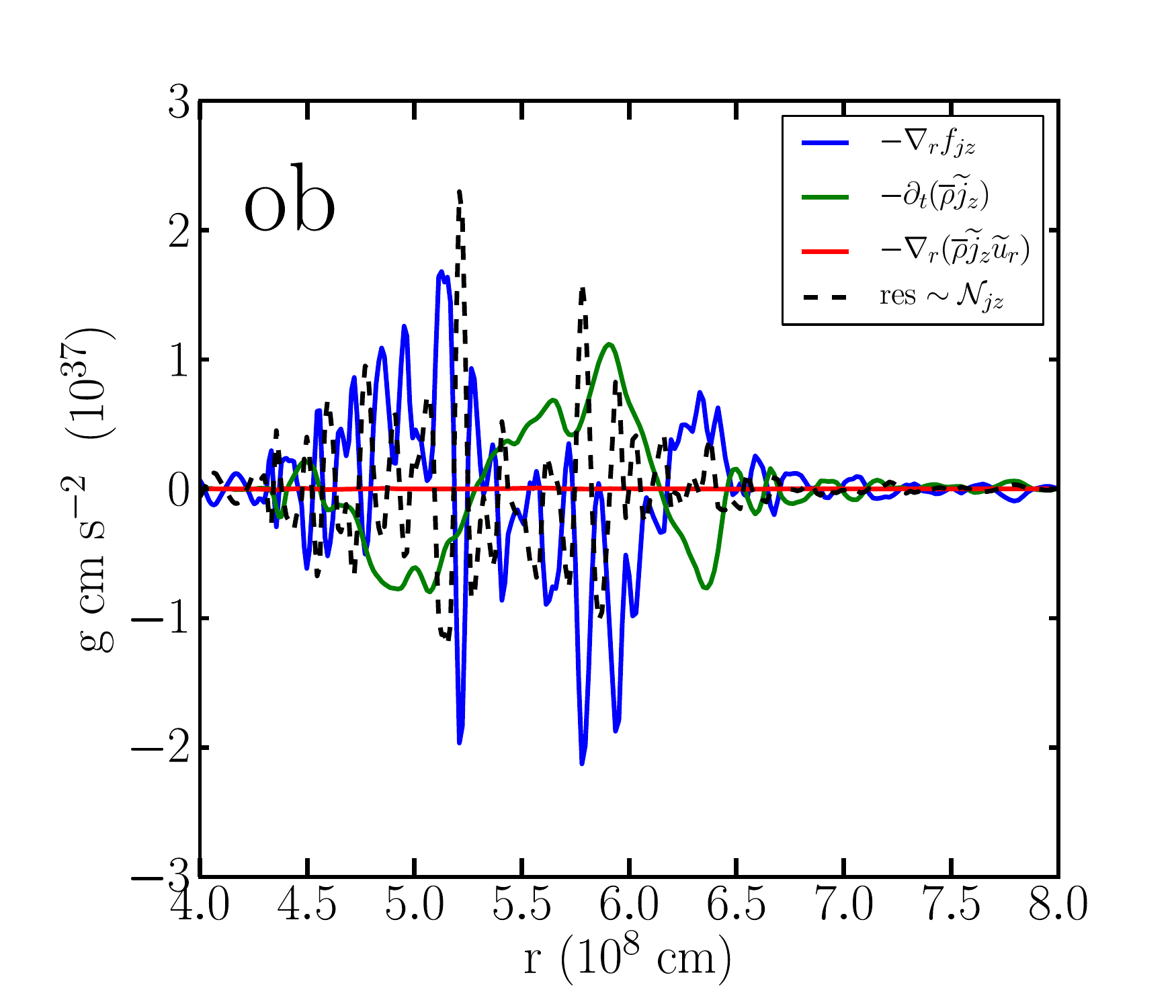}
\includegraphics[width=6.6cm]{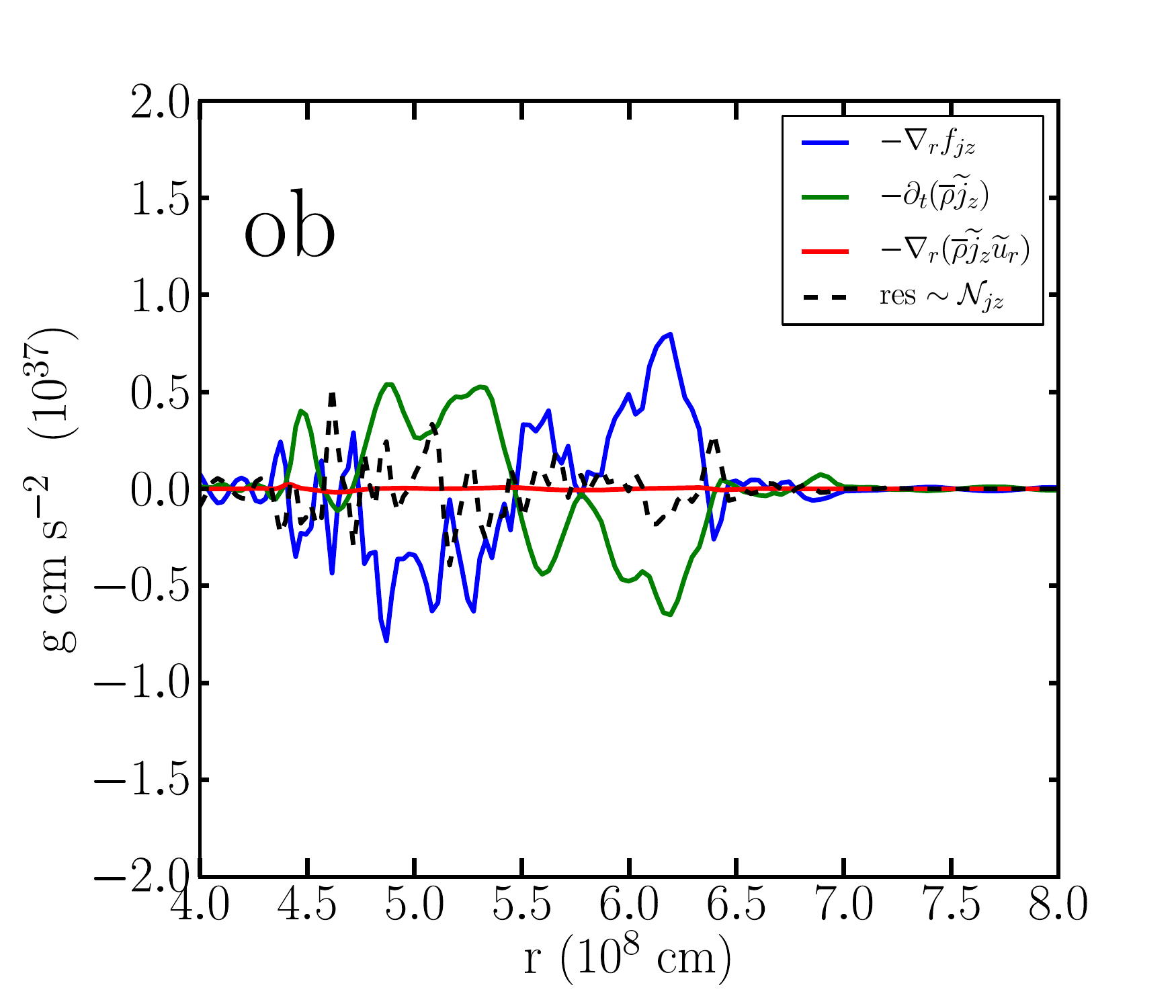}}

\centerline{
\includegraphics[width=6.6cm]{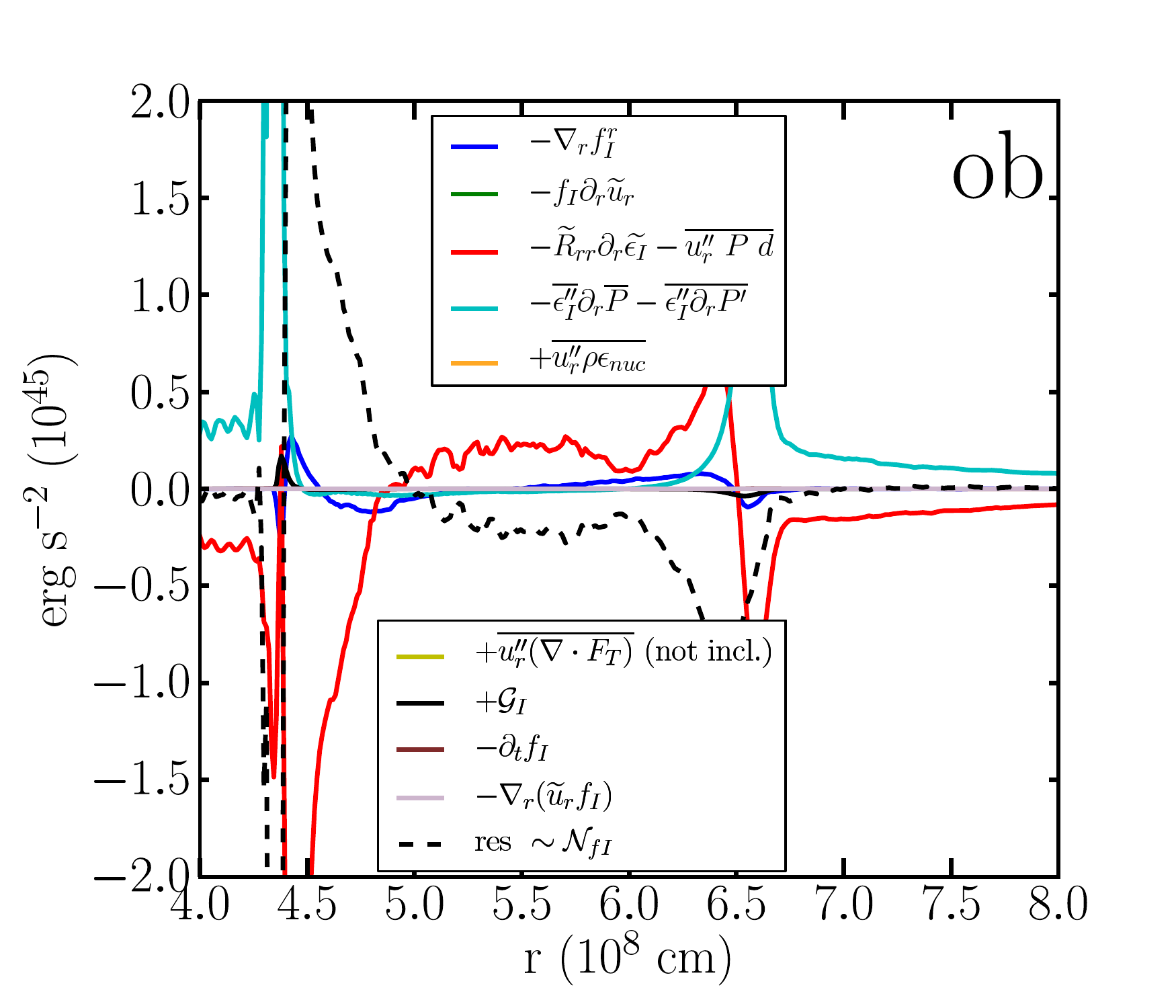}
\includegraphics[width=6.6cm]{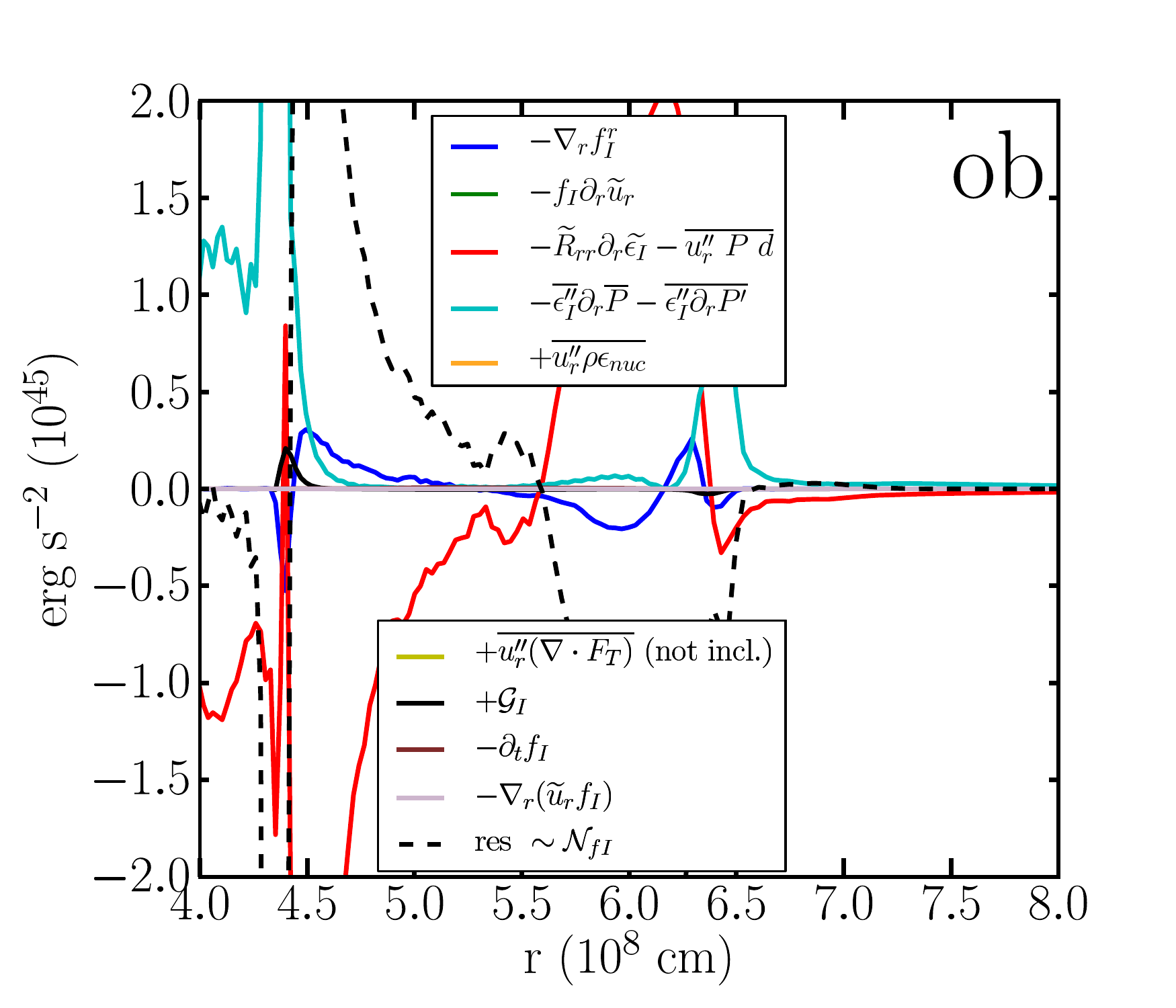}}
\caption{Mean specific angular momentum equation (upper panels) and mean turbulent internal energy flux equation (lower panels). Model with volumetric heating at the bottom of convection zone {\sf ob.3D.1hp.vh} (left) and model with volumetric cooling at the top of convection zone {\sf ob.3D.1hp.vc} (right).}
\end{figure}

\newpage

\subsubsection{Mean turbulent kinetic energy and mean velocities}

\begin{figure}[!h]
\centerline{
\includegraphics[width=6.6cm]{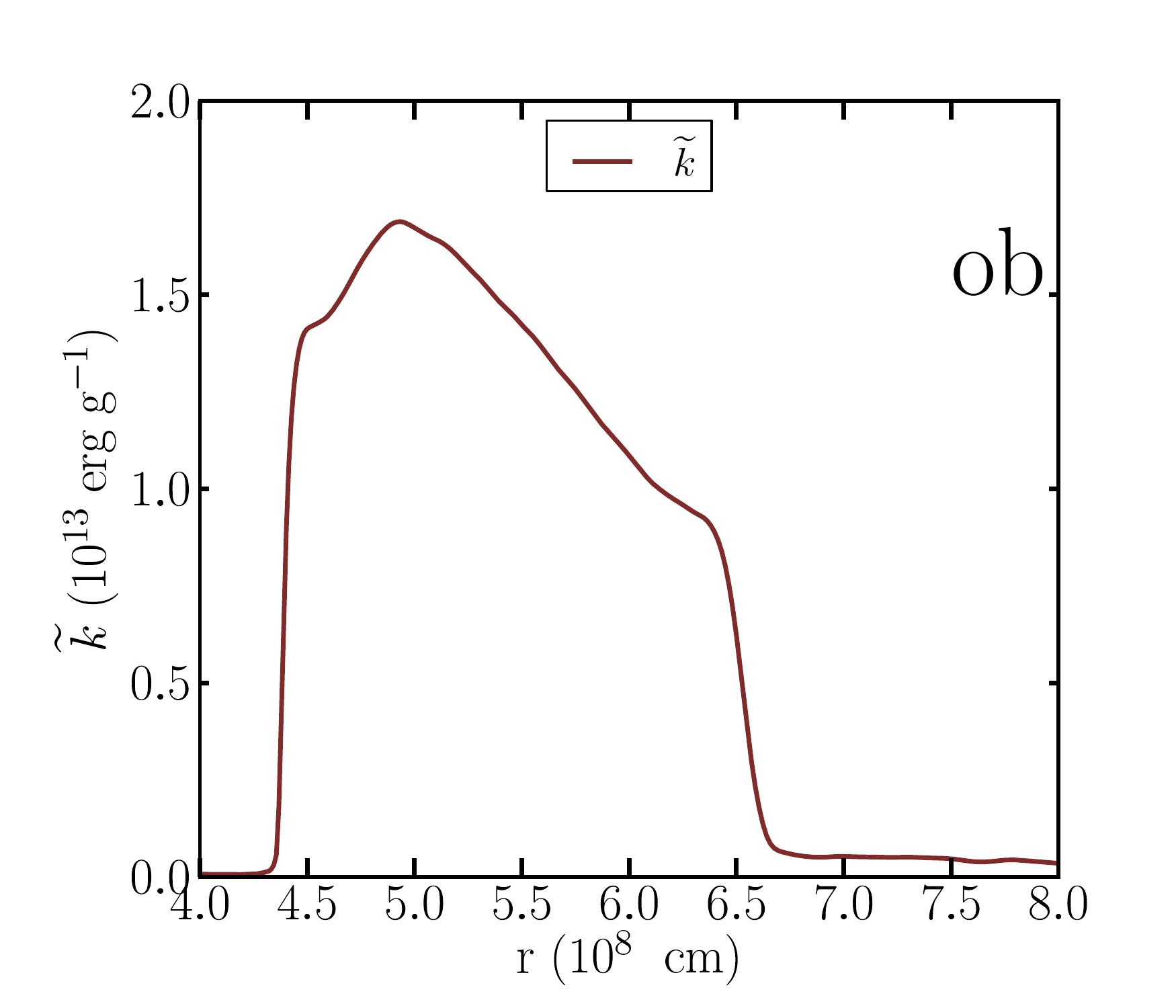}
\includegraphics[width=6.6cm]{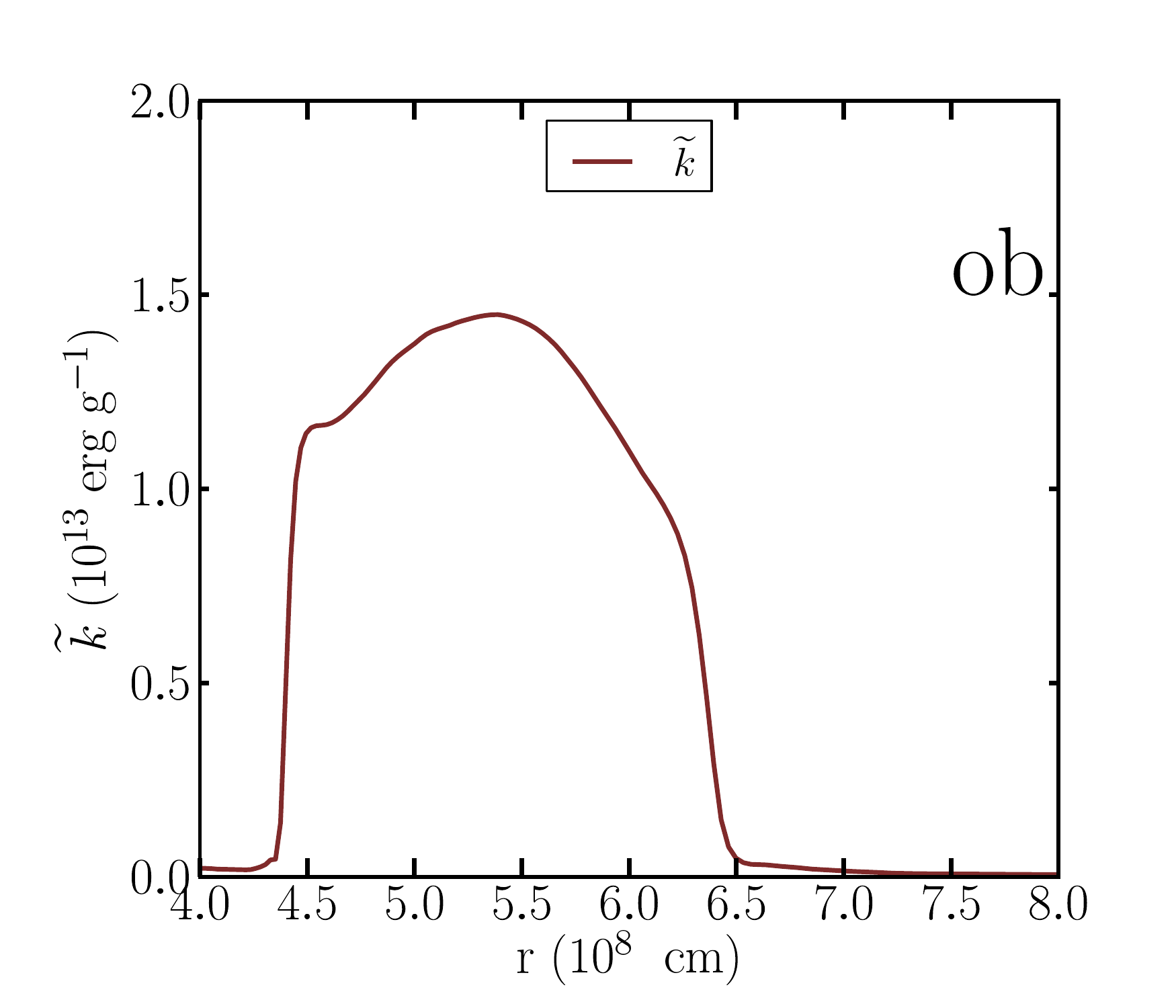}}

\centerline{
\includegraphics[width=6.6cm]{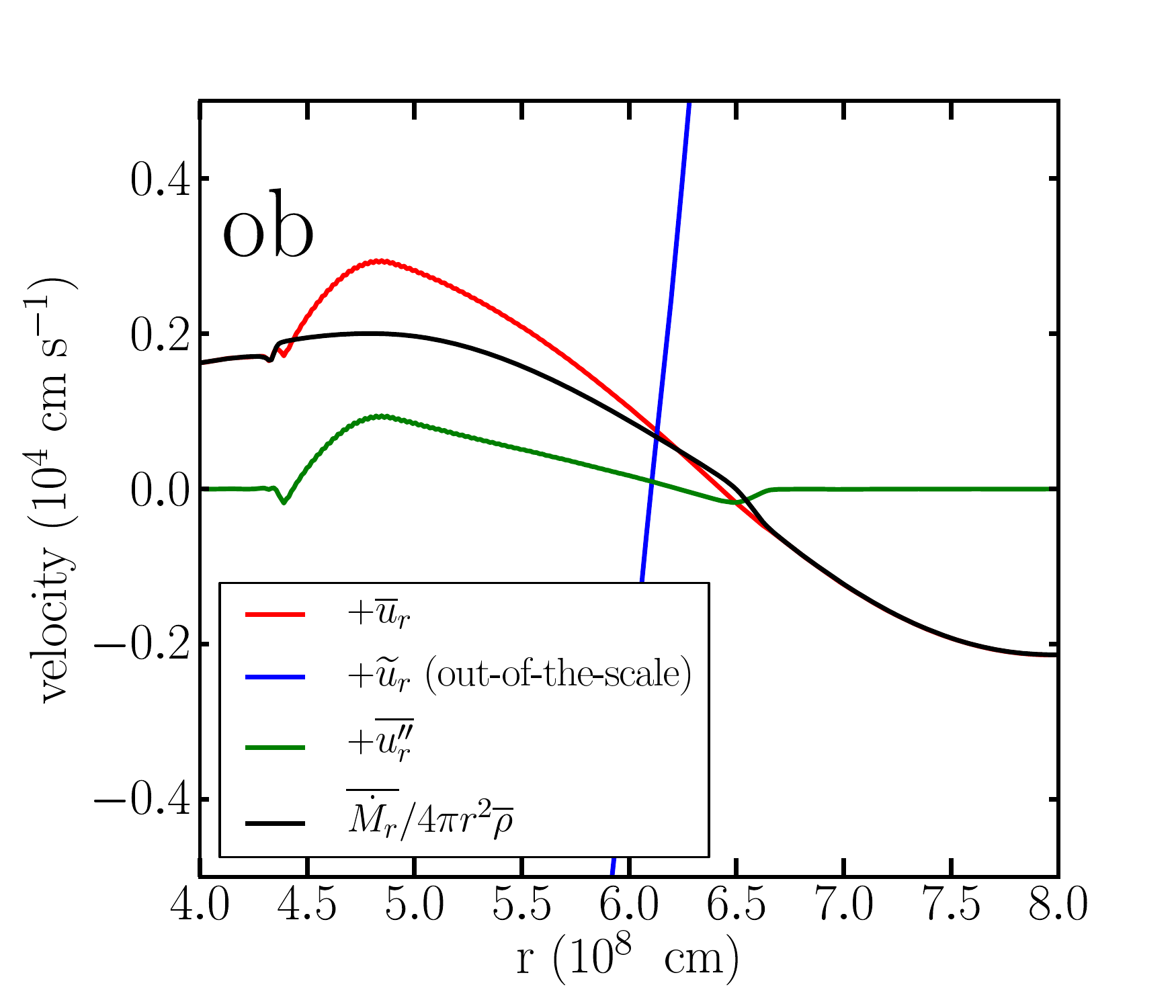}
\includegraphics[width=6.6cm]{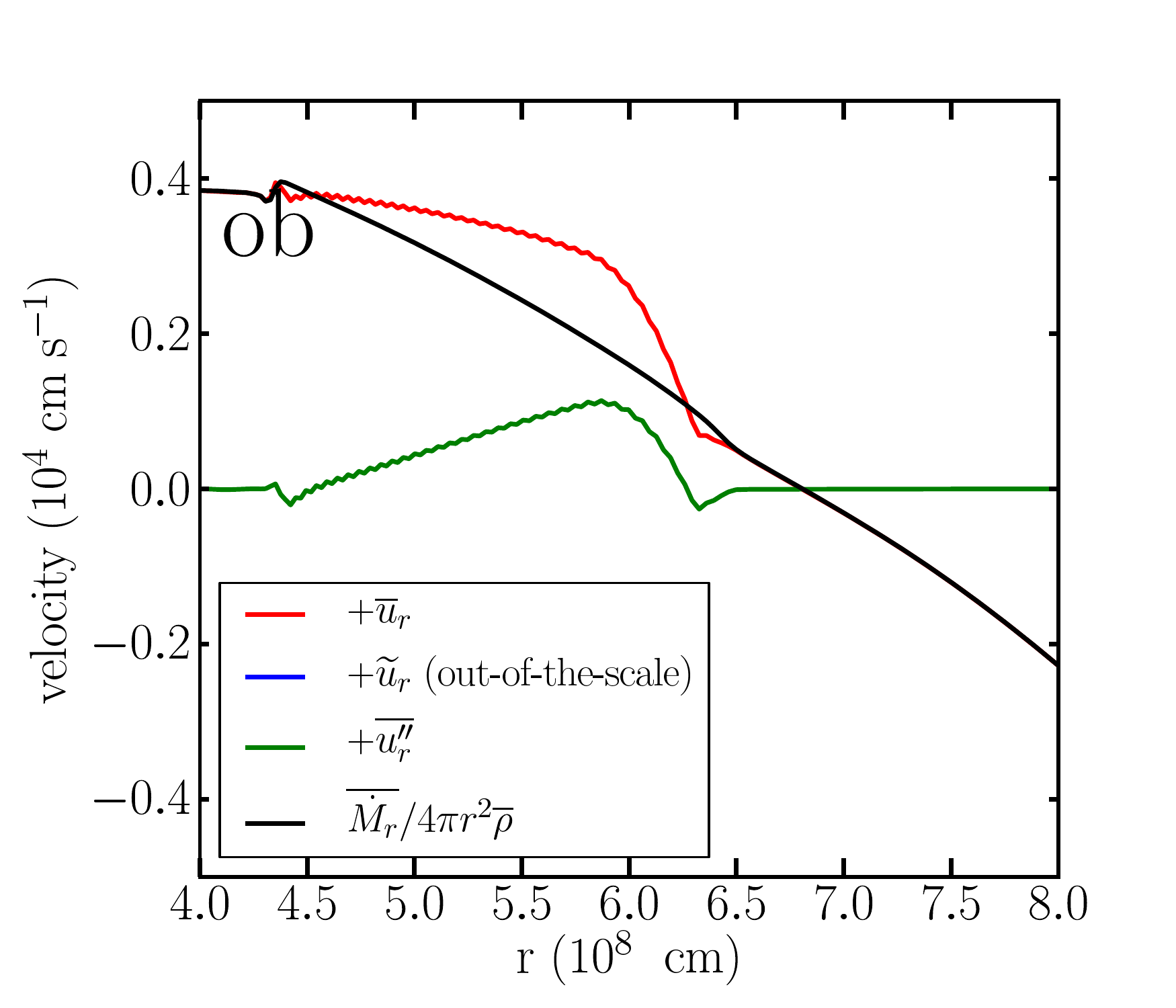}}
\caption{Mean turbulent kinetic energy (upper panels) and mean background velocities (lower panels). Model with volumetric heating at the bottom of convection zone {\sf ob.3D.1hp.vh} (left) and model with volumetric cooling at the top of convection zone {\sf ob.3D.1hp.vc} (right).}
\end{figure}

\newpage

\section{Assessment of Some One-Point Turbulence Closure Model Assumptions}
\label{sec:model-comparison}

\subsection{Downgradient approximations}

\begin{align}
f_k = (C \ \rho \ \sqrt{\fht{k}} \ l_d) \ \partial_r \fht{k} \ \ \ \ \ \ \ \ \ \ \ f_I = (C \ \rho \ \sqrt{\fht{k}} \ l_d) \ \partial_r \fht{\epsilon}_I \ \ \ \ \ \ \ \ \ \ \ f_I = (C \ \rho \ \fht{k}^2 / \varepsilon_k) \ \partial_r \fht{\epsilon}_I
\end{align}

\begin{figure}[!h]
\centerline{
\includegraphics[width=5.2cm]{obmrez_tavg230_mean_k_insf-eps-converted-to.pdf}
\includegraphics[width=5.2cm]{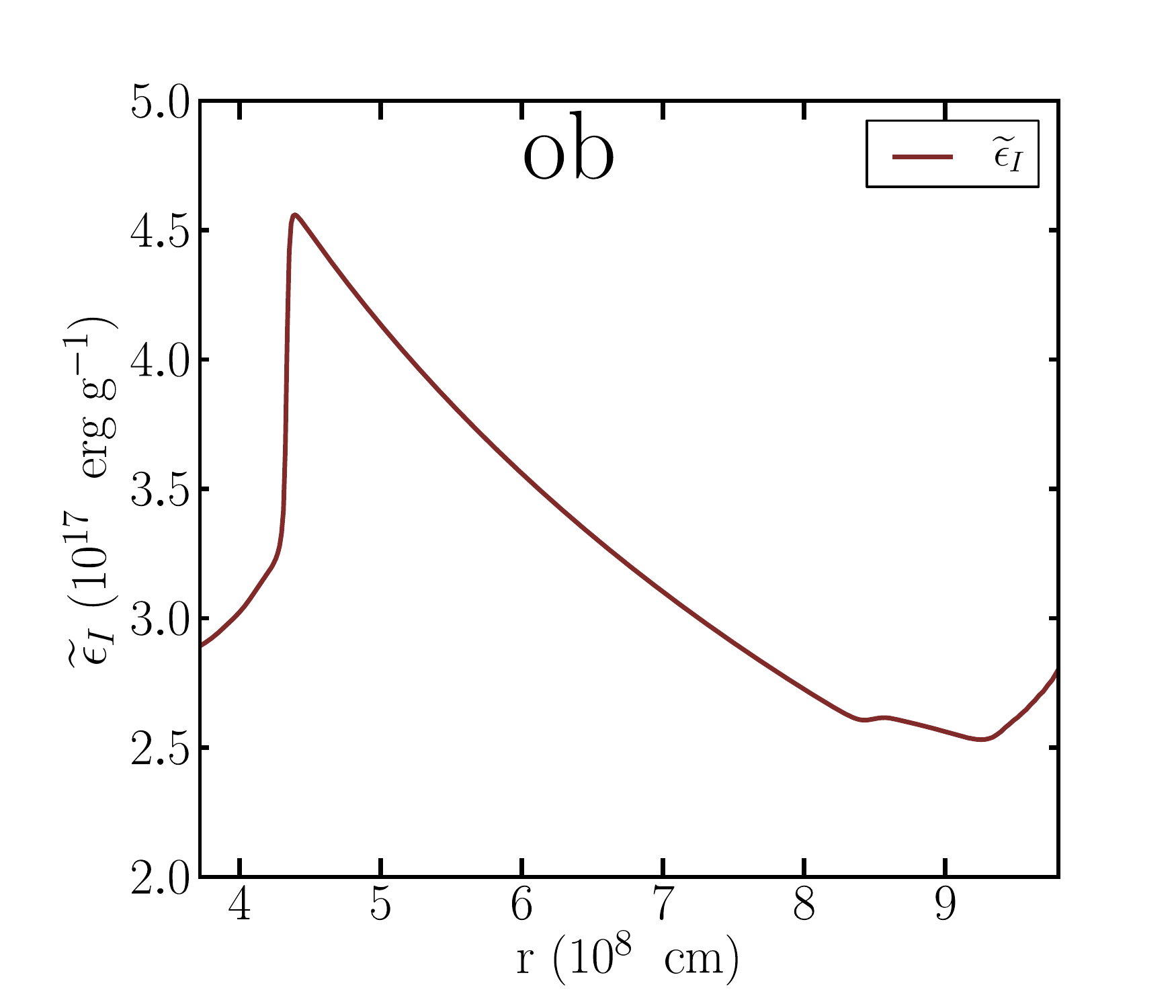}
\includegraphics[width=5.2cm]{rgmrez_tavg800_mean_k_insf-eps-converted-to.pdf}
\includegraphics[width=5.2cm]{rgmrez_tavg800_mean_ei_insf-eps-converted-to.pdf}}

\centerline{
\includegraphics[width=5.2cm]{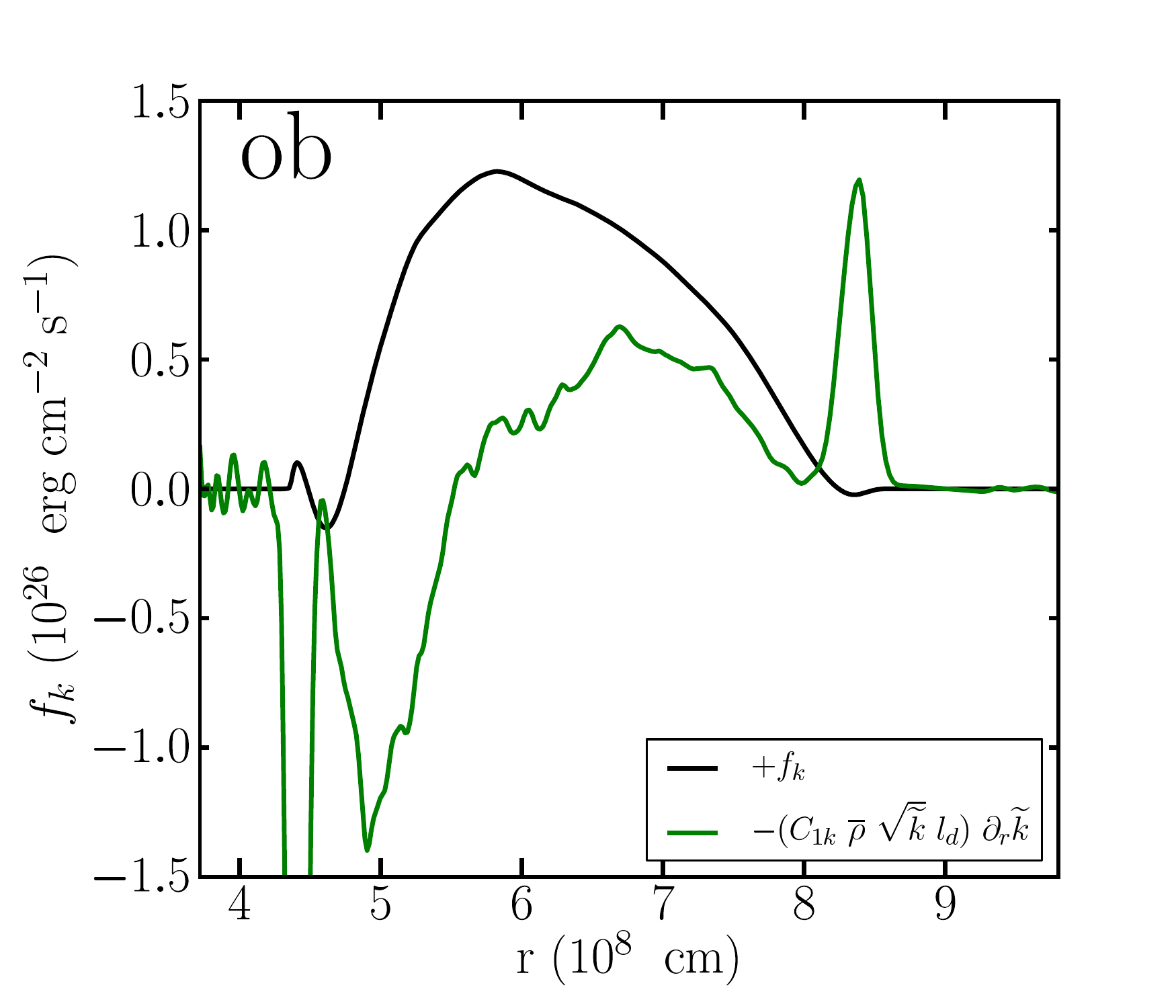}
\includegraphics[width=5.2cm]{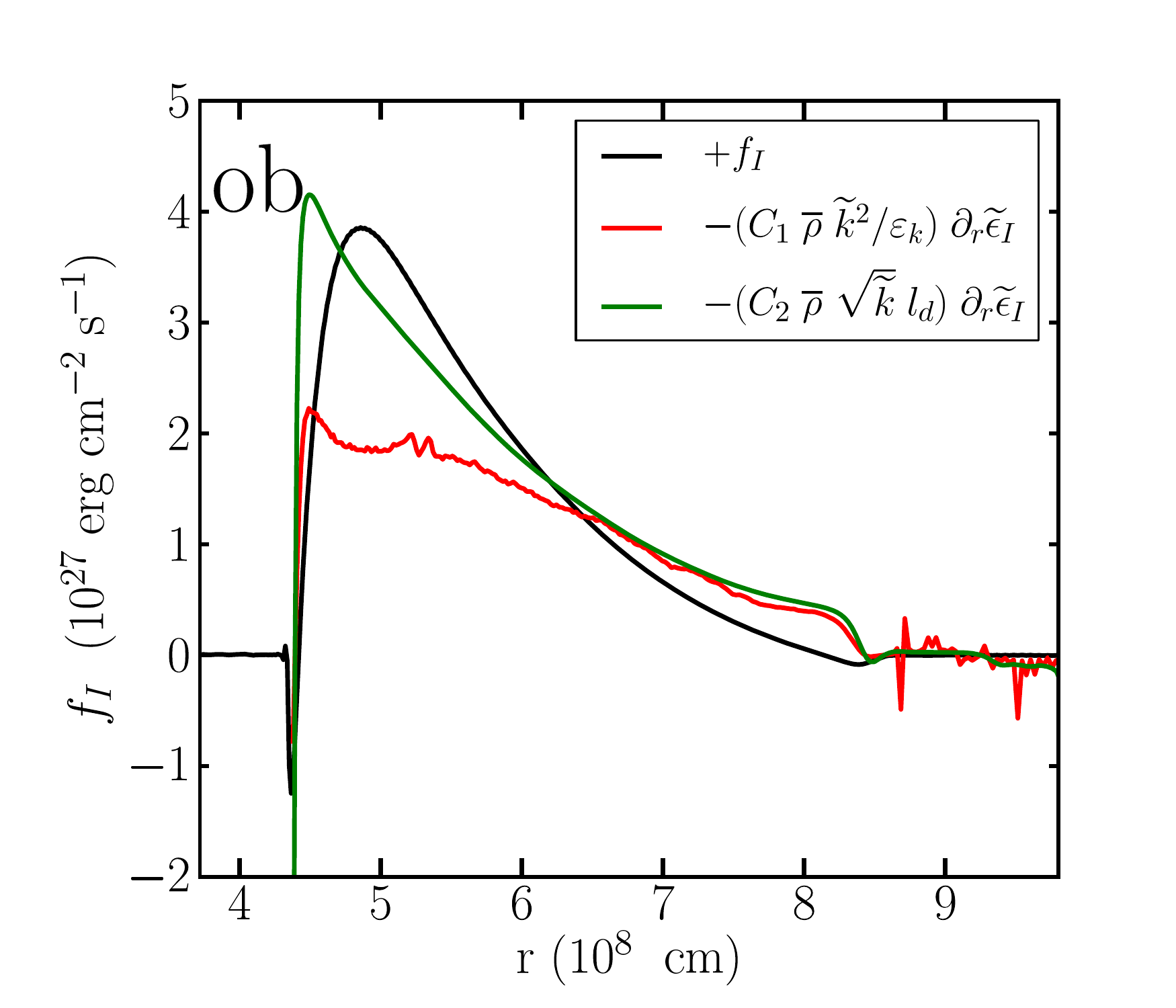}
\includegraphics[width=5.2cm]{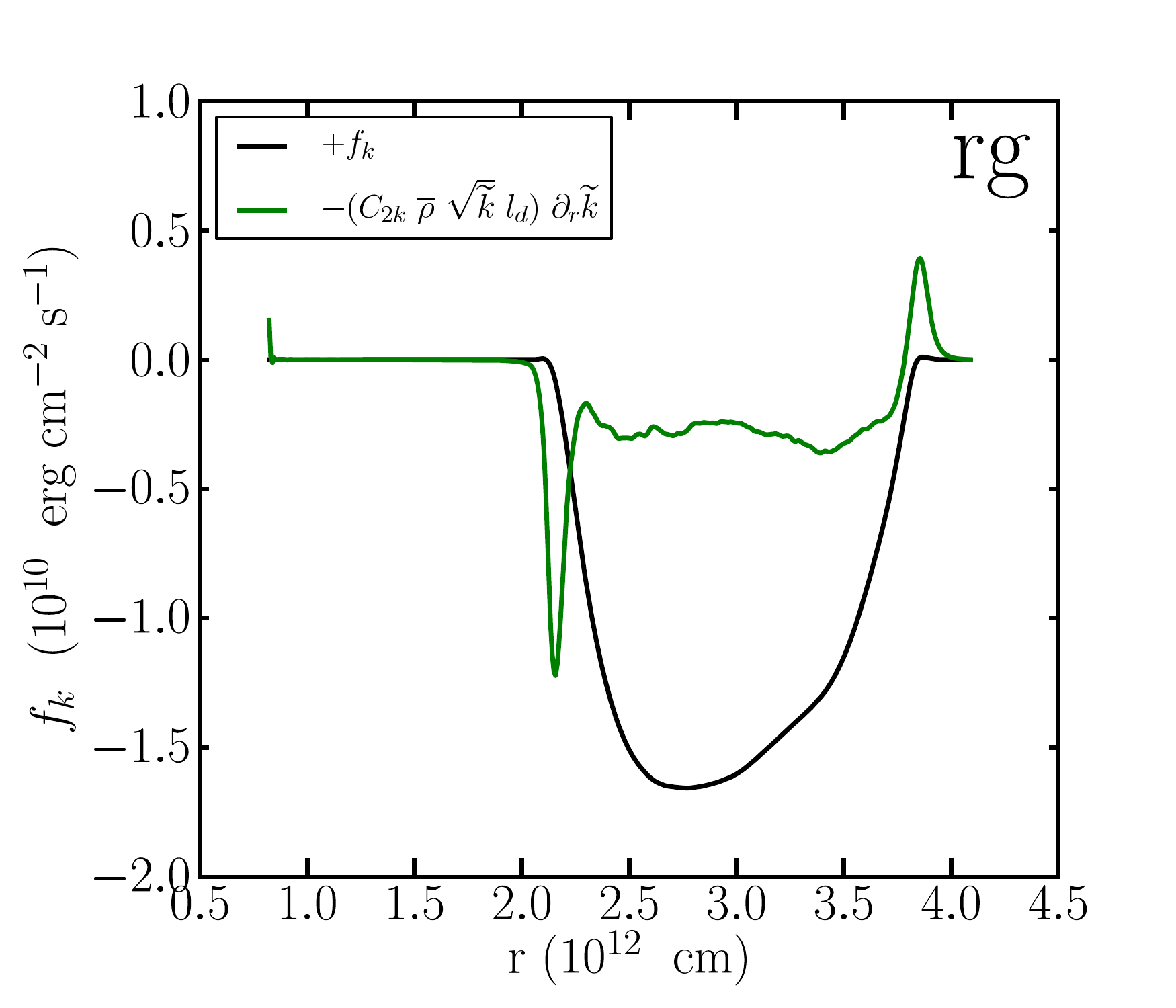}
\includegraphics[width=5.2cm]{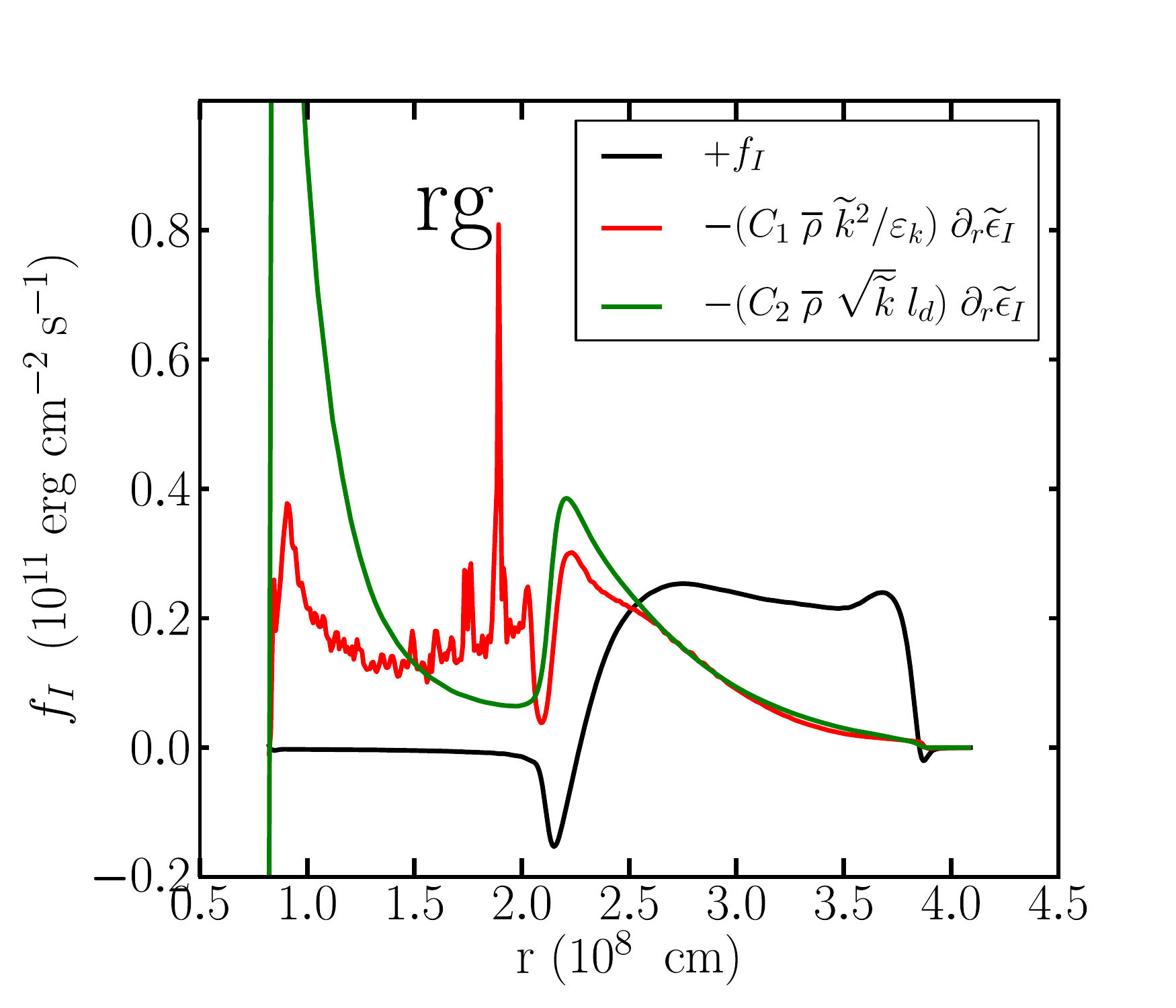}}
\caption{Profiles of mean turbulent kinetic energy and mean internal energy (upper panels) and various downgradient approximations to turbulent kinetic energy flux and internal energy flux (lower panels) for the oxygen burning shell model {\sf ob.3D.mr} (ob) and red giant envelope convection {\sf rg.3D.mr} (rg). $l_d$ is dissipation length scale and $C$ is model constant. \label{fig:downgrad_approx}}
\end{figure}

\newpage

\subsection{Various approximations from Besnard-Harlow-Rauenzahn (BHR)}

\begin{align}
\eht{\rho'u'_ru'_r} = C_{1a} \frac{l_d}{\sqrt{\fht{k}}} \fht{R}_{rr} \partial_r \overline{u''_r} \ \ \ \ \ \ \ \ \ \ \eht{v' \partial_r P'} = -C_{2a} \frac{\sqrt{k}}{l_d} \eht{u''_r}  \ \ \ \ \ \ \ \ \ \ \eht{v'u'_r} = -C_{1b} \frac{l_d}{\sqrt{k}} \frac{\fht{R}_{rr}}{\eht{\rho}}\frac{\partial}{\partial r} \left( \frac{1+b}{\eht{\rho}} \right) \ \ \ \ \ \ \ \ \ \ \eht{v'd'} =  -C_{2b} \frac{\sqrt{k}}{l_d} \frac{b}{\eht{\rho}} \nonumber
\end{align}

\begin{figure}[!h]
\centerline{
\includegraphics[width=5.3cm]{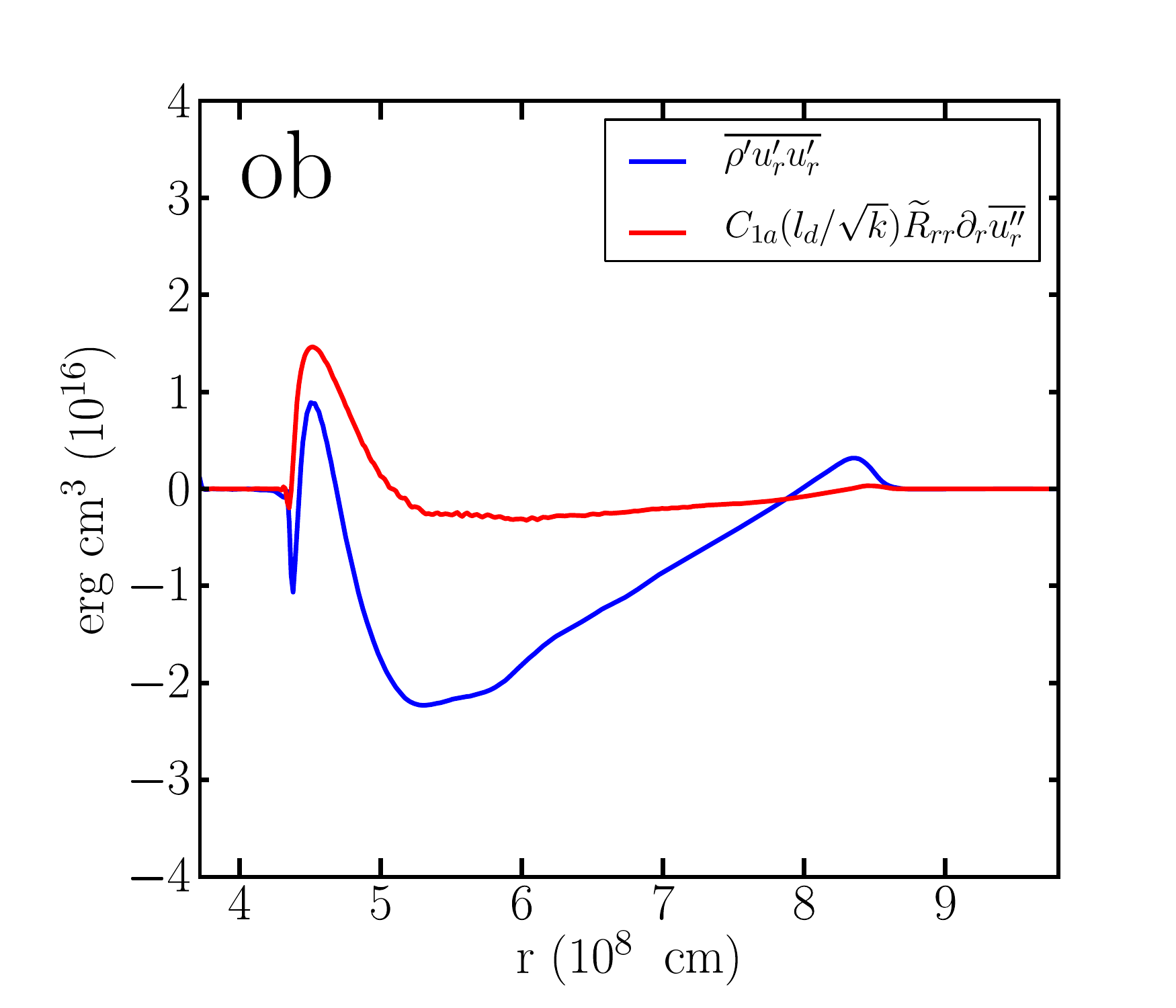}
\includegraphics[width=5.3cm]{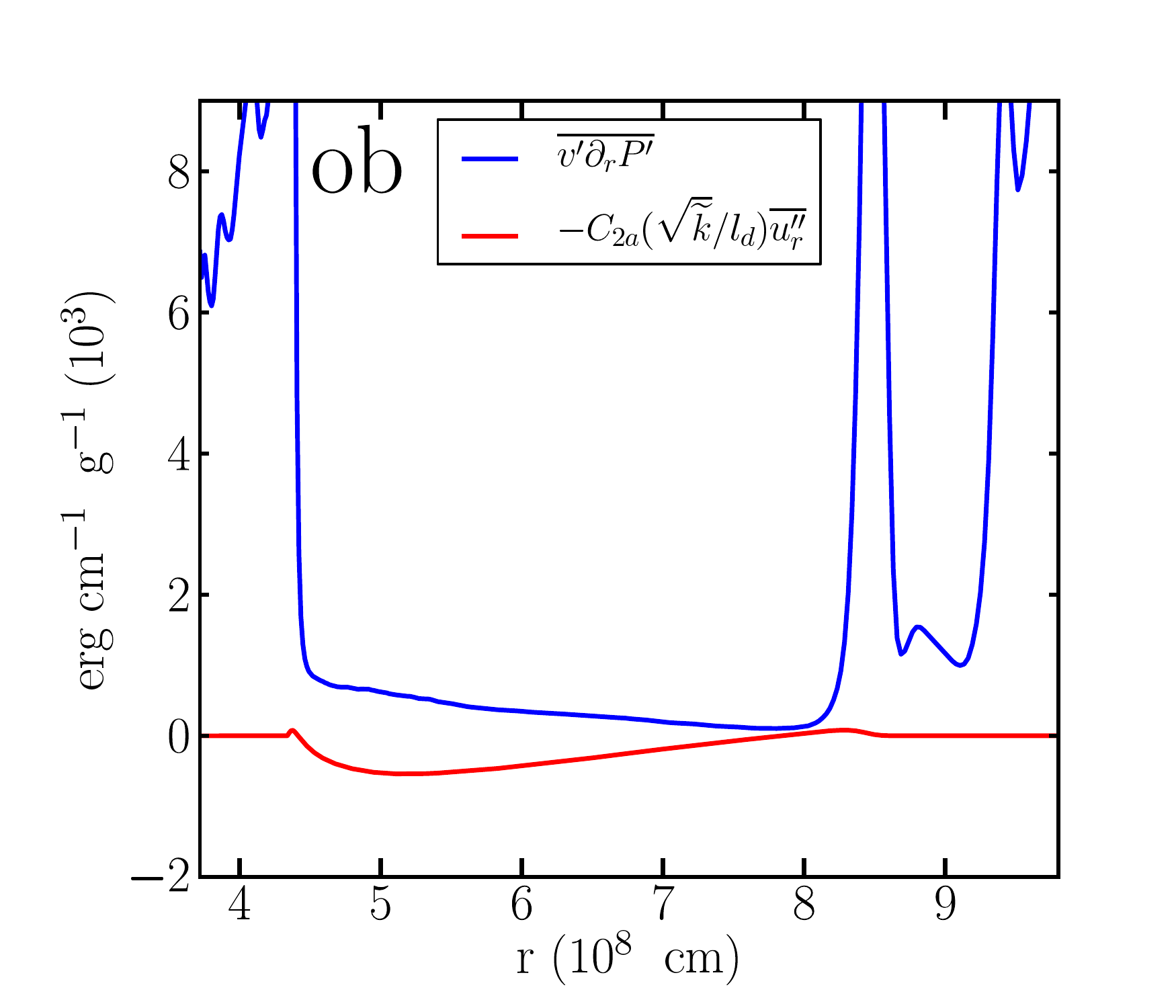}
\includegraphics[width=5.3cm]{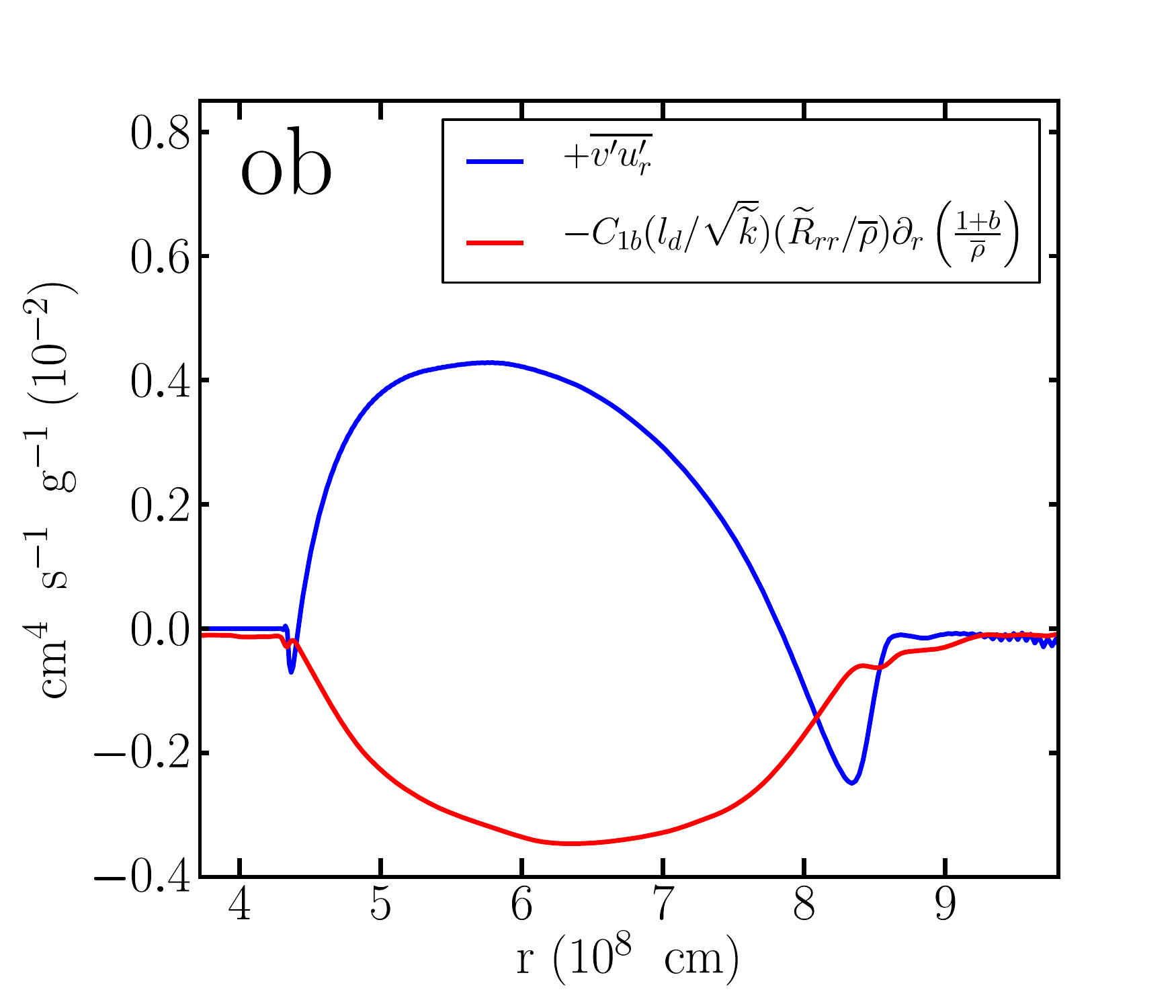}
\includegraphics[width=5.3cm]{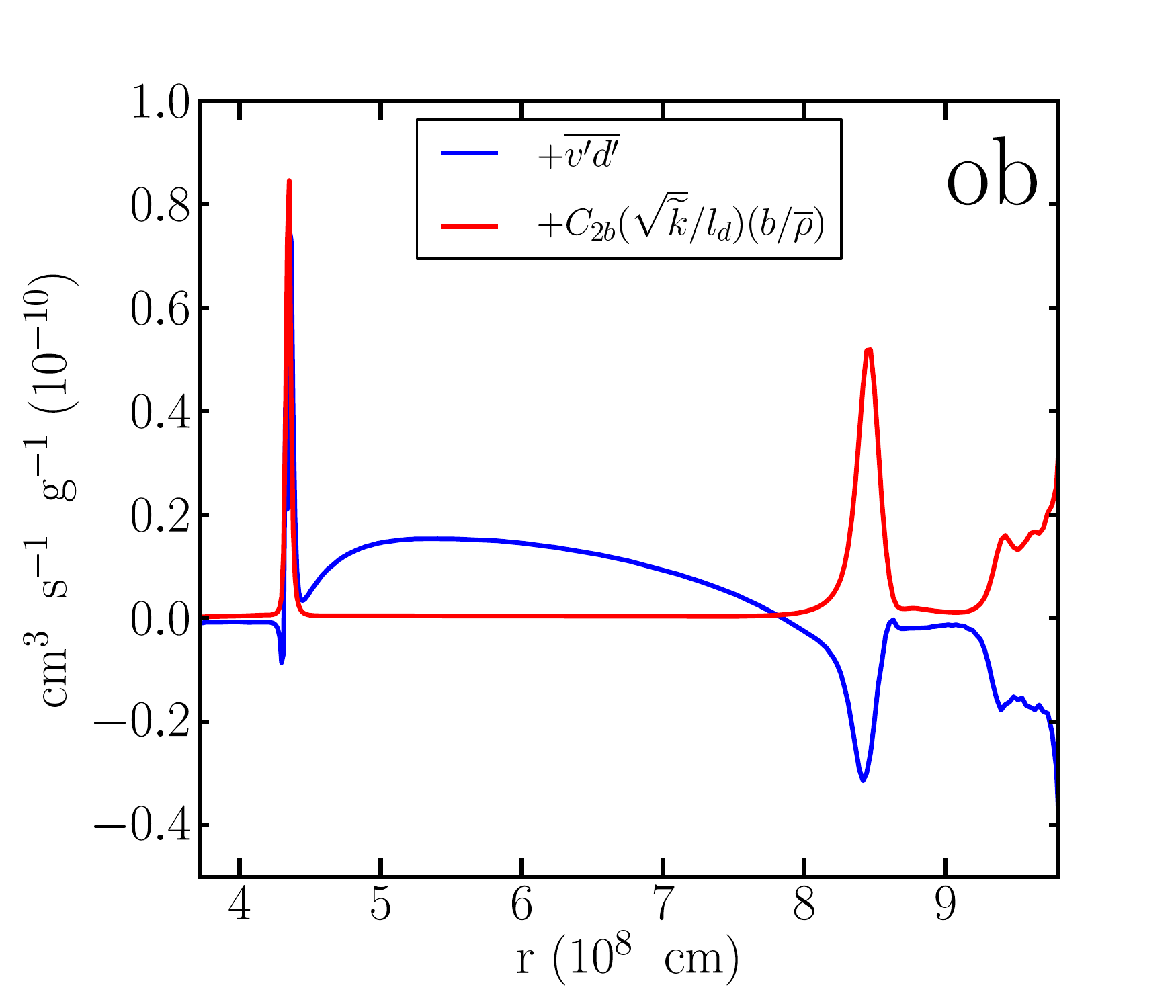}}

\centerline{
\includegraphics[width=5.3cm]{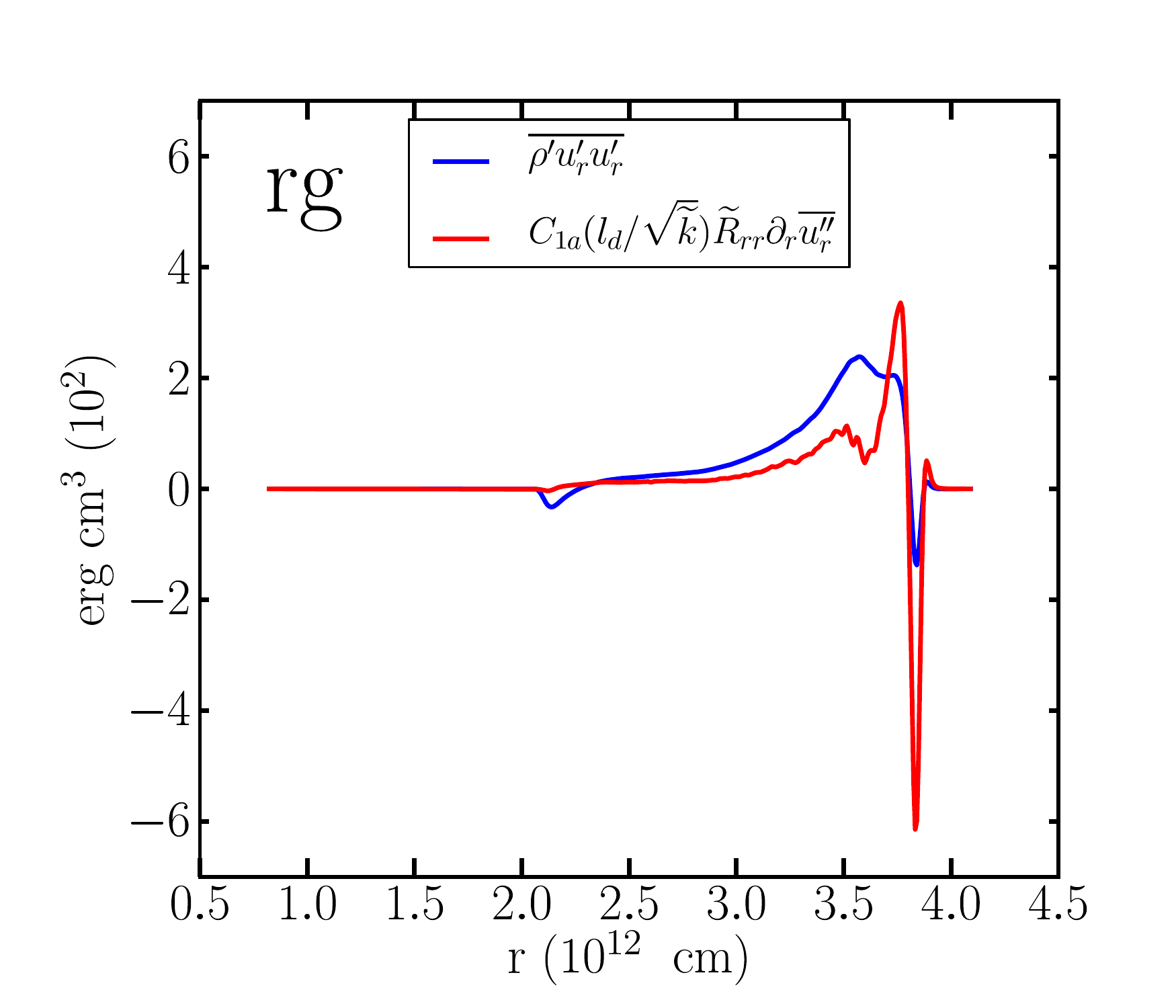}
\includegraphics[width=5.3cm]{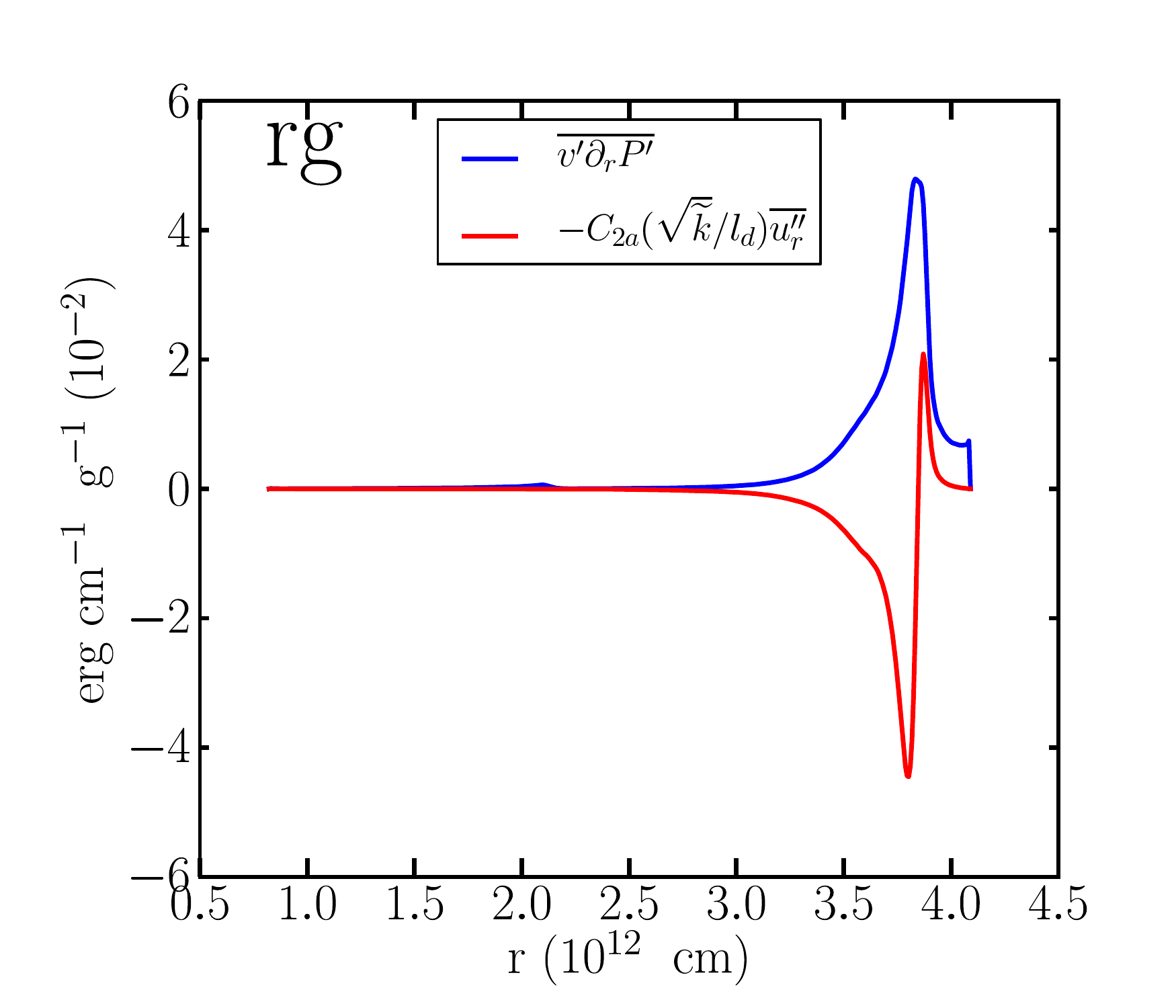}
\includegraphics[width=5.3cm]{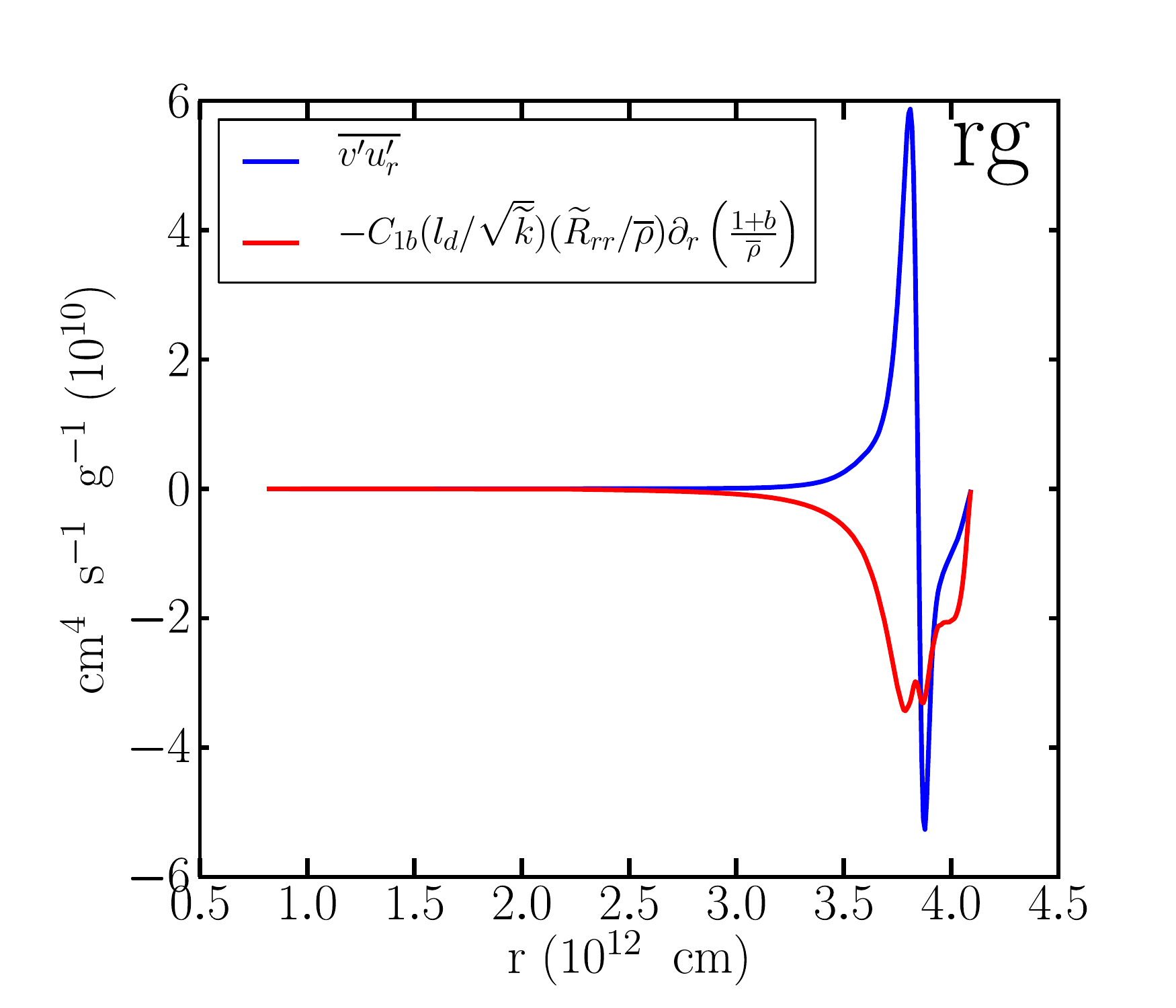}
\includegraphics[width=5.3cm]{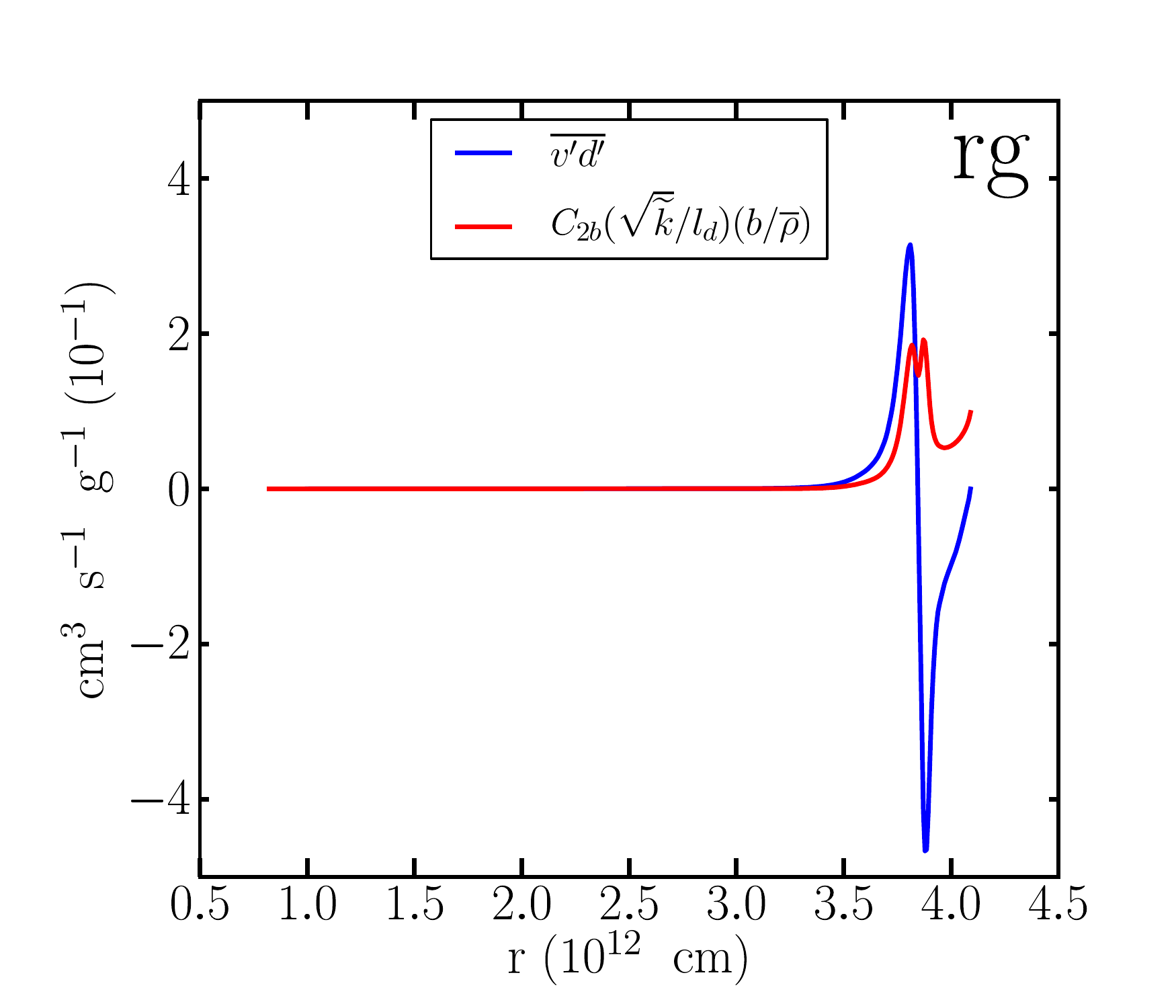}}
\caption{Various approximations taken from Besnard-Harlow-Rauenzahn (BHR) model for the  oxygen burning shell model {\sf ob.3D.mr} (ob) and red giant envelope convection {\sf rg.3D.mr} (rg). $l_d$ is dissipation length scale and $C$ is model constant. \label{fig:bhr-models}}
\end{figure}

\newpage

\subsection{Quasi-normal approximation and decay-rate assumption model}

\begin{align}
\eht{a'b'c'd'} = & \ \overline{a'b'} \ \overline{c'd'} + \overline{a'c'} \ \overline{b'd'} + \overline{a'd'} \ \overline{b'c'} \hspace{2.cm} & \overline{a'b'} = & \ (l_d^2 / \Delta) \overline{\partial_k a' \partial_k b'} & \ \ \ \mbox{(original formulations)} \\
\fht{a''b''c''d''} = & \ \fht{a''b''} \ \fht{c''d''} + \fht{a''c''} \ \fht{b''d''} + \fht{a''d''} \ \fht{b''c''} \hspace{2.cm} & \fht{a''b''} = & \ (l_d^2 / \Delta) \overline{\partial_k a'' \partial_k b''} & \ \ \ \mbox{(assumed Favre equivalents)}
\end{align}

\begin{figure}[!h]
\centerline{
\includegraphics[width=5.9cm]{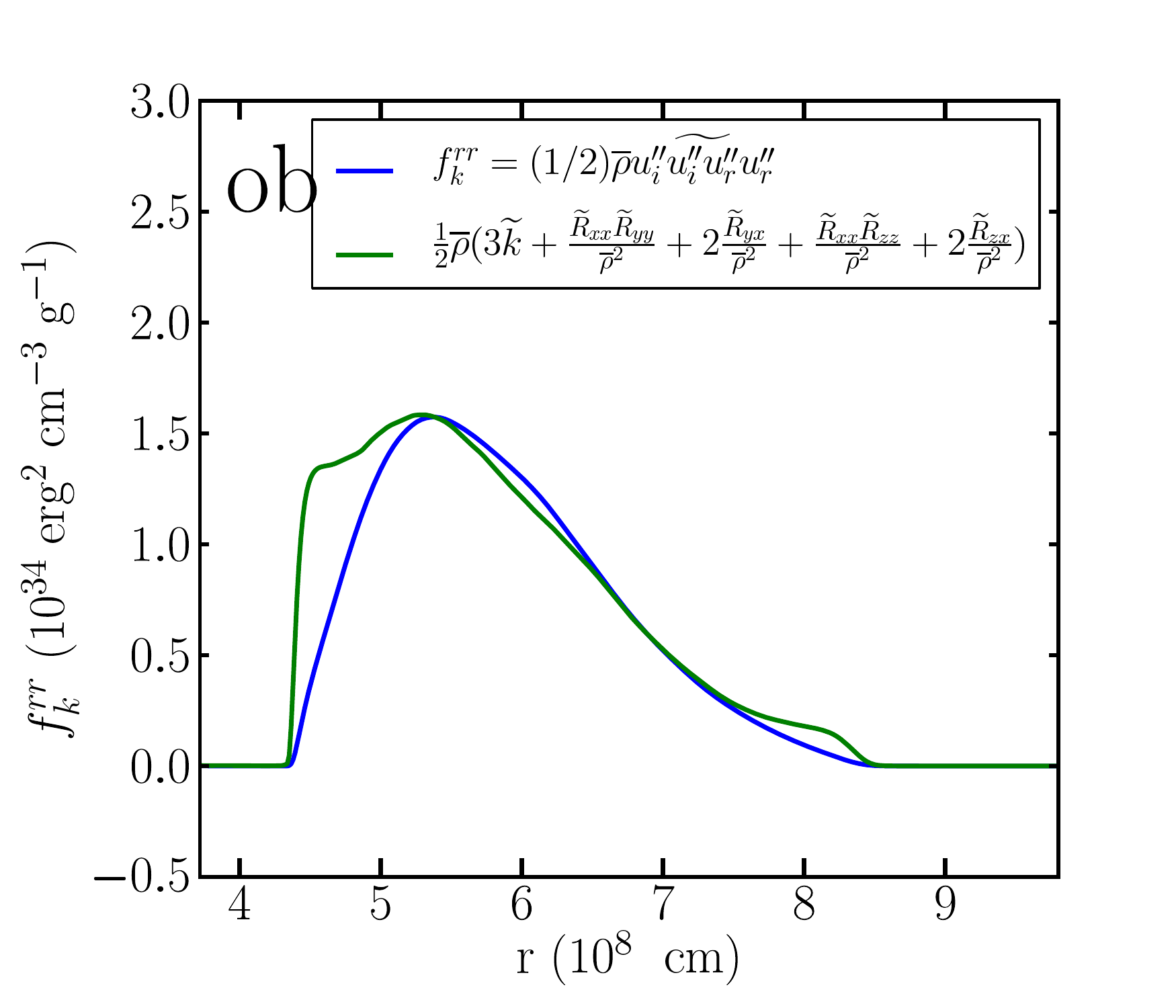}
\includegraphics[width=5.9cm]{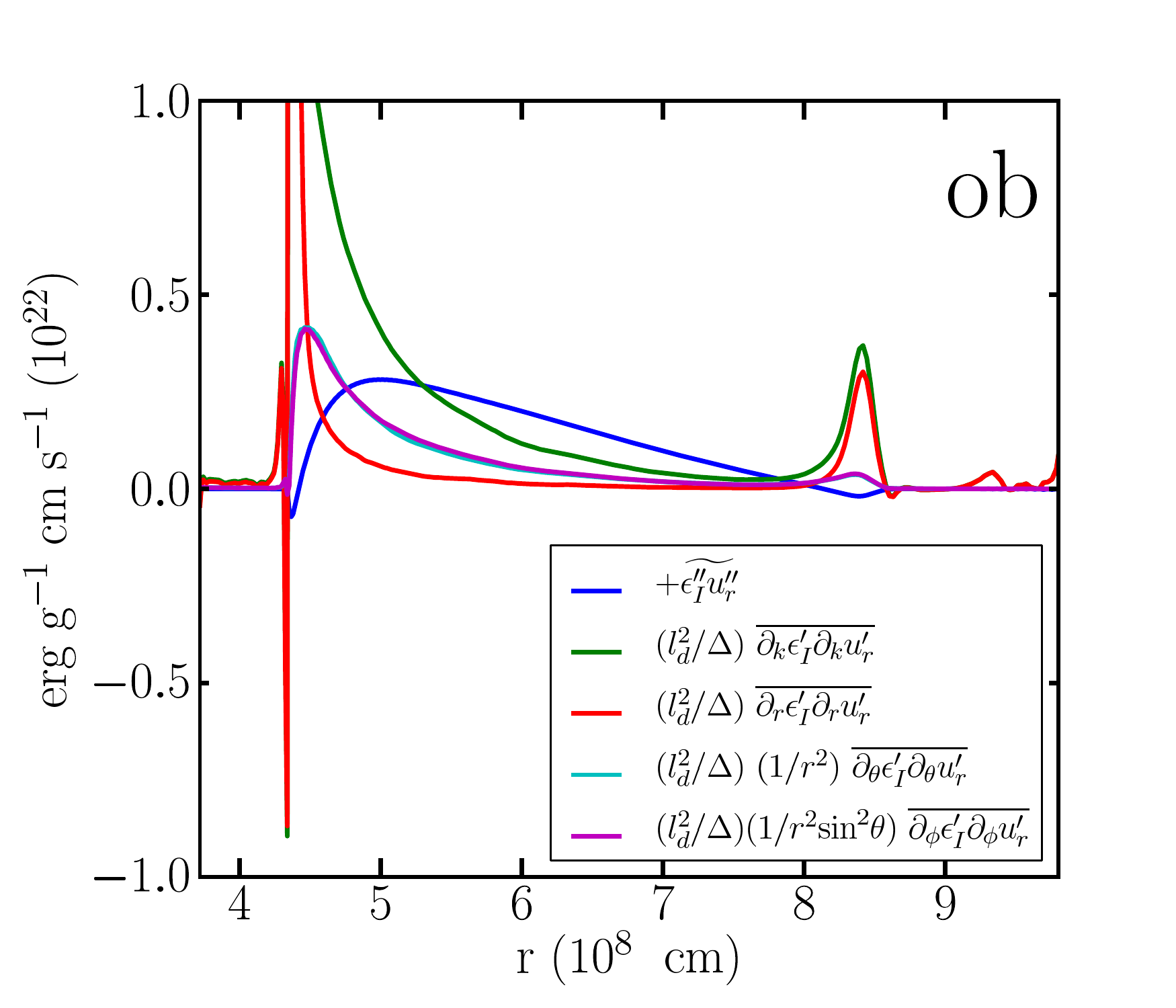}
\includegraphics[width=5.9cm]{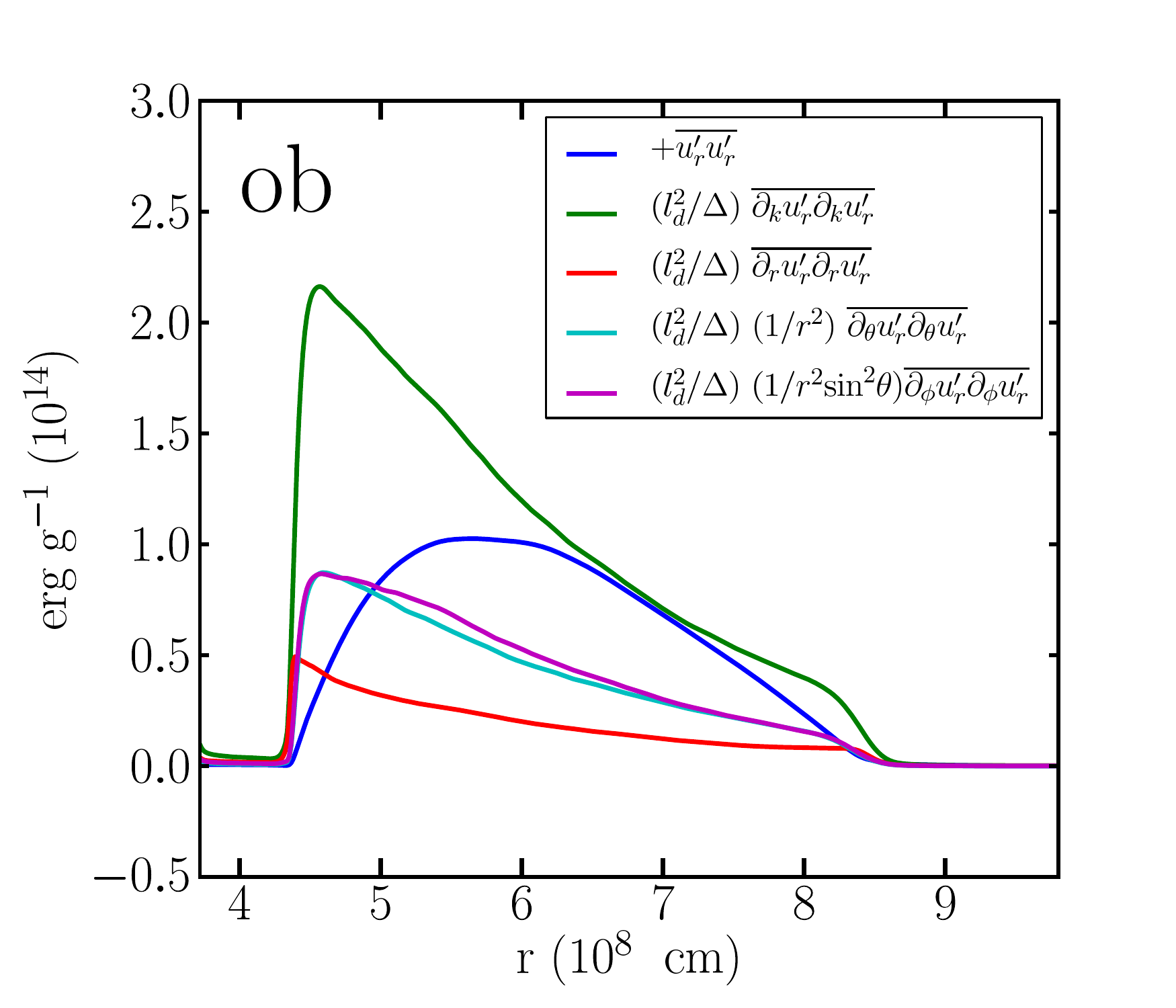}}

\centerline{
\includegraphics[width=5.9cm]{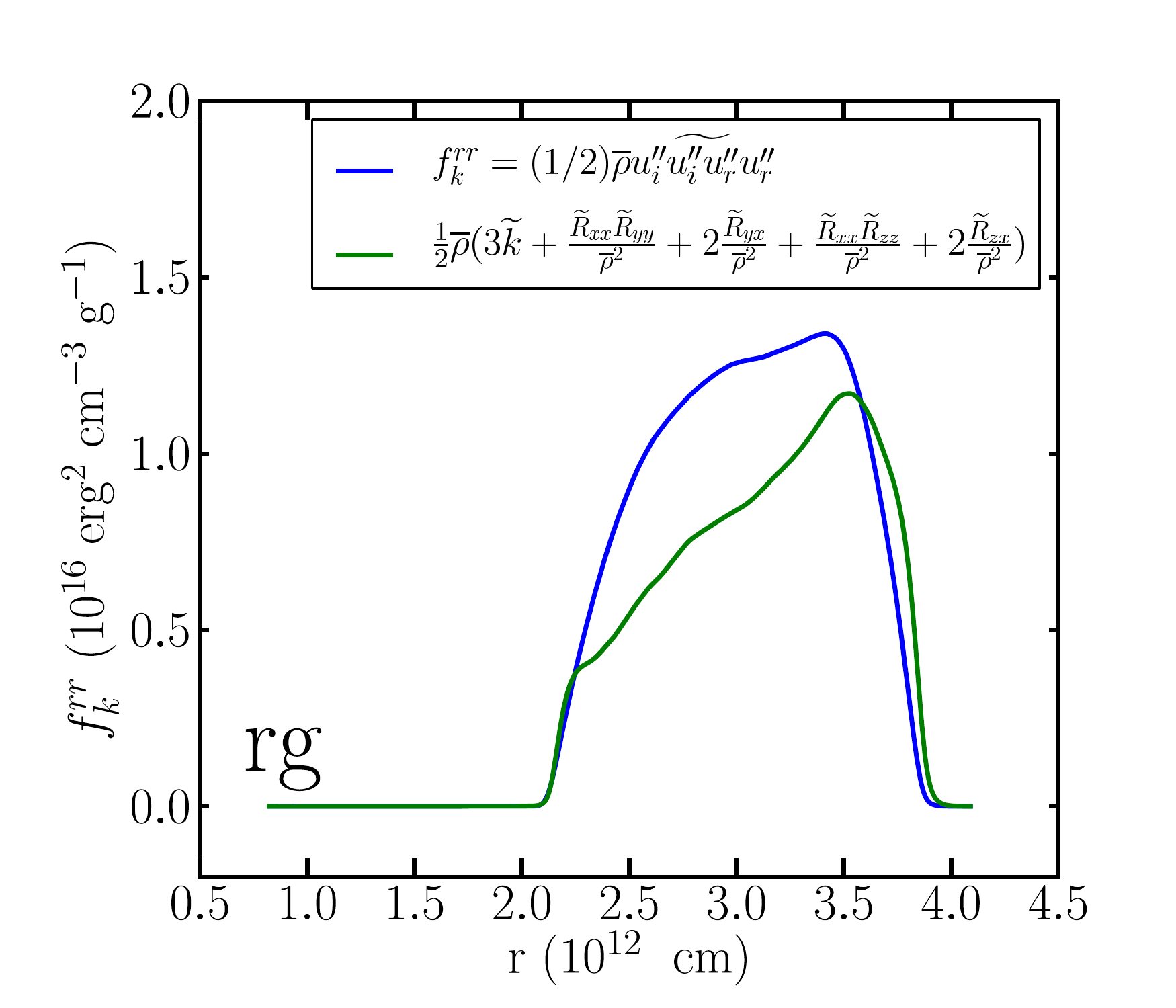}
\includegraphics[width=5.9cm]{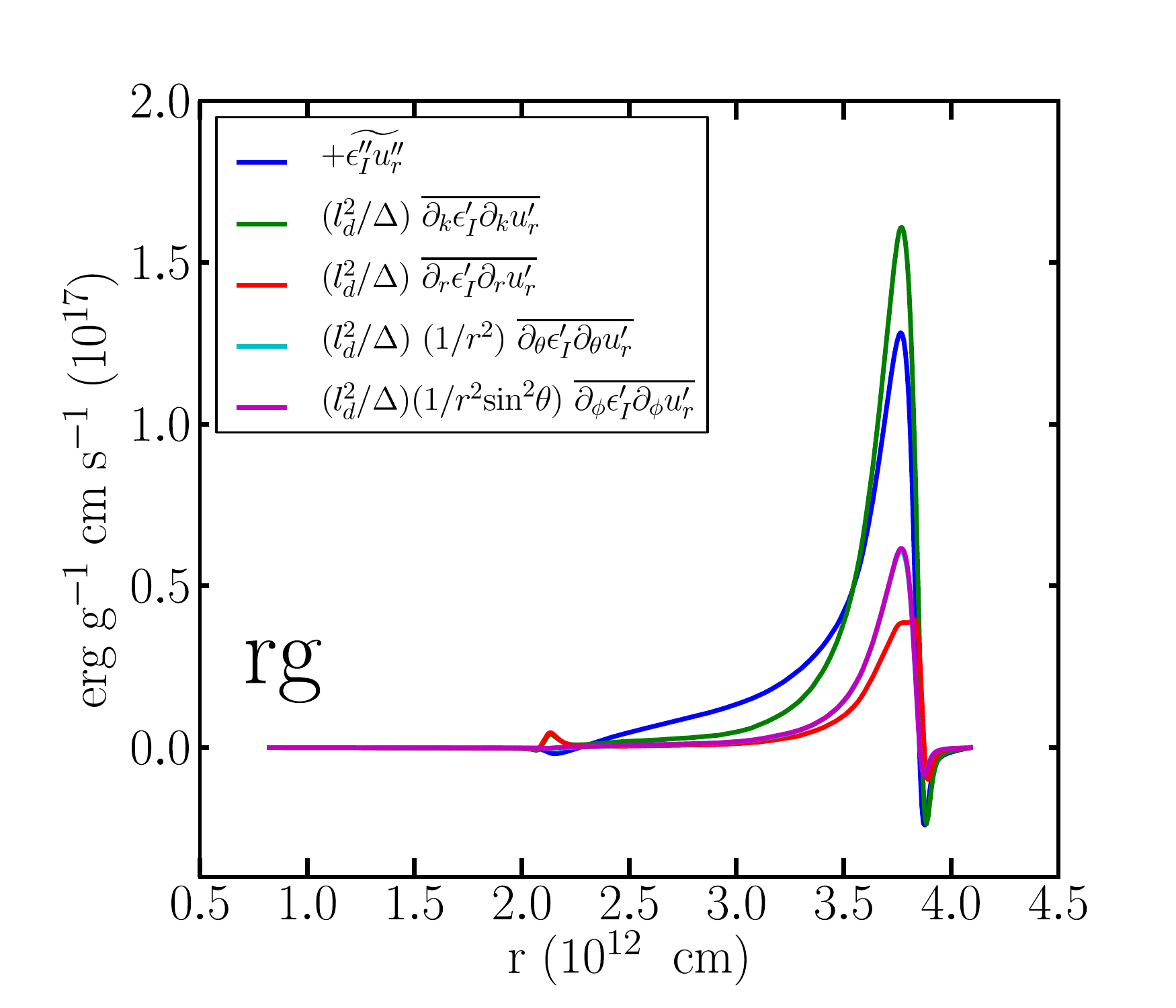}
\includegraphics[width=5.9cm]{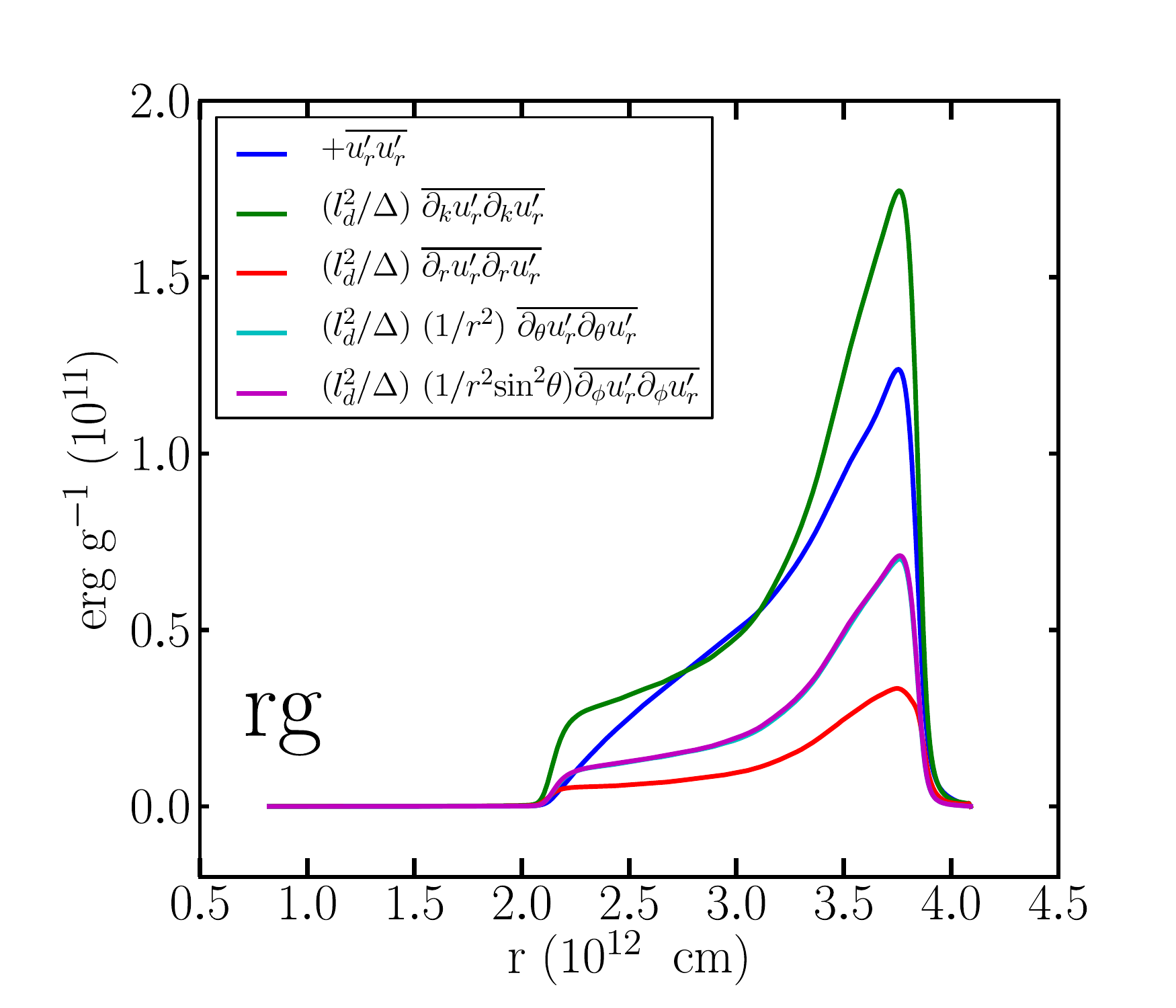}}
\caption{Quasi-normal approximations (left panels) and decay-rate assumption models (middle, right panels) for the  oxygen burning shell model {\sf ob.3D.mr} (ob) and red giant envelope convection {\sf rg.3D.mr} (rg). \label{fig:qn-dc-models}}
\end{figure}

\newpage






\section{Fourier scale analysis}

\par We present some preliminary scale analysis in this section including cumulative spectra for a variety of covariances at a location within the convection zone in \S\ref{sec:scale-cumulative} as well as radial profiles for mean-field data which has been separated into large and small components in \S\S\ref{sec:scale-profiles-1} to \ref{sec:scale-profiles-3}.

\subsection{Cumulative Fourier spectra for Covariances}\label{sec:scale-cumulative}

\par The mean "energy" spectrum $\eht{E_{ab}}(r,k,t_c)$ for a given covariance is found by taking the space-time average of the product of the fluctuations after writing each fluctuation as a Fourier series

\begin{align}
\eht{\ef{a}\ef{b}}(r,t_c) =  & \frac{1}{\Delta \Omega T}\int \ef{a}(r,\theta,\phi,t) \ef{b}(r,\theta,\phi,t) d\Omega  dt = \frac{1}{\Delta \Omega  T}\int \sum_{\vec{l}} \sum_{\vec{m}} \widehat{\ef{a}}(r,\vec{l},t)\widehat{\ef{b}}(r,\vec{m},t)e^{i(\vec{l}+\vec{m})\cdot\vec{x}} d\Omega  dt  \nonumber \\
  = & \frac{1}{T}\int \sum_{\vec{l}} \widehat{\ef{a}}(r,\vec{l},t)\widehat{\ef{b}}^*(r,\vec{l},t) dt   = \sum_{\vec{l}} \overline{\widehat{\ef{a}}(r,\vec{l},t)\widehat{\ef{b}}^*(r,\vec{l},t)}  \nonumber \\
  = &   \sum_{k = |\vec{l}|} \eht{E_{ab}}(r,k,t_c)
\end{align}

\noindent where $k = |\vec{l}| = ( { l_{\phi}^2 + l_{\theta}^2})^{1/2}$ is the horizontal wavenumber magnitude and $\vec{x} = (\theta,\phi)$ is the angular position vector and $^*$ indicates complex conjugation. By this definition, $E_{ab}(r,k,t)$ involves a sum of the terms $ \overline{\widehat{\ef{a}}\widehat{\ef{b}}^*}$ over annuli of unit width in the plane of the wave vector $\vec{l}$ components $l_{\theta}$ and $l_{\phi}$ centered on $k$. Here, $d\Omega = d\theta d\phi$ rather than the solid angle increment.  This choice greatly simplifies the algebra and results in a negligible difference compared to solid angle averages for the small wedges studied here. In the remainder of this section, our data is presented in terms of a normalized wavenumber $\hat{k} = k/ k_0$ where $k_0 \equiv 2\pi/{\mathcal L}$ is the lowest wavenumber in the domain of width ${\mathcal L}$ so that an integer wavenumber $\hat{k}$ represents a Fourier mode of wavelength ${\mathcal L}/\hat{k}$. We drop the hat over the wavenumber in what follows.


\par Finally, the cumulative spectrum $E_c(k)$ is defined by the normalized partial sum

\begin{align}
E_{c}(\overline{\ef{a}\ef{b}})(r,k,t) = \frac{\sum_{1 \le l \le k} \eht{E}_{ab}(r,l,t)}{\eht{\ef{a}\ef{b}}}.
\end{align}

\newpage

\begin{figure}[!h]
\centerline{
\includegraphics[width=7.cm]{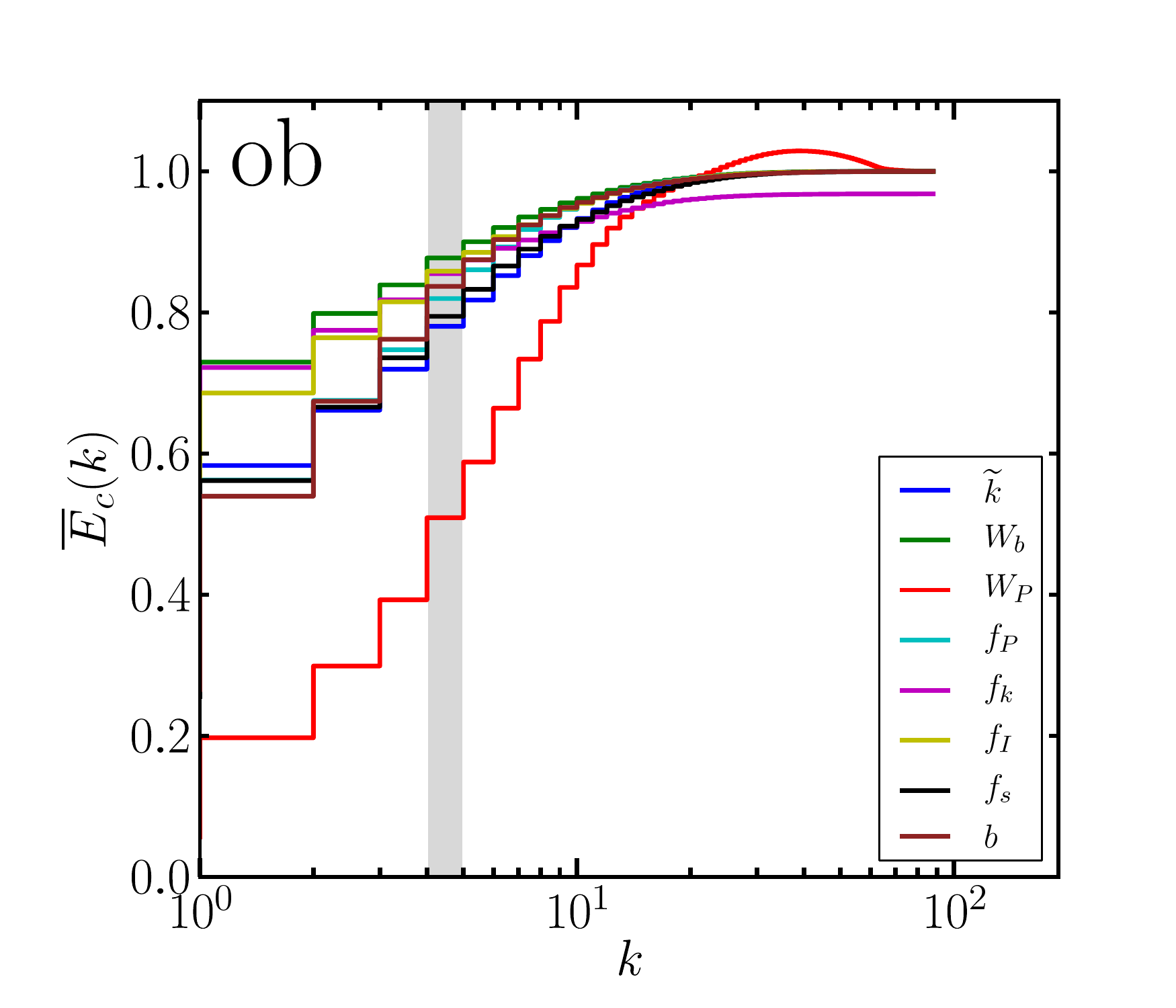}
\includegraphics[width=7.cm]{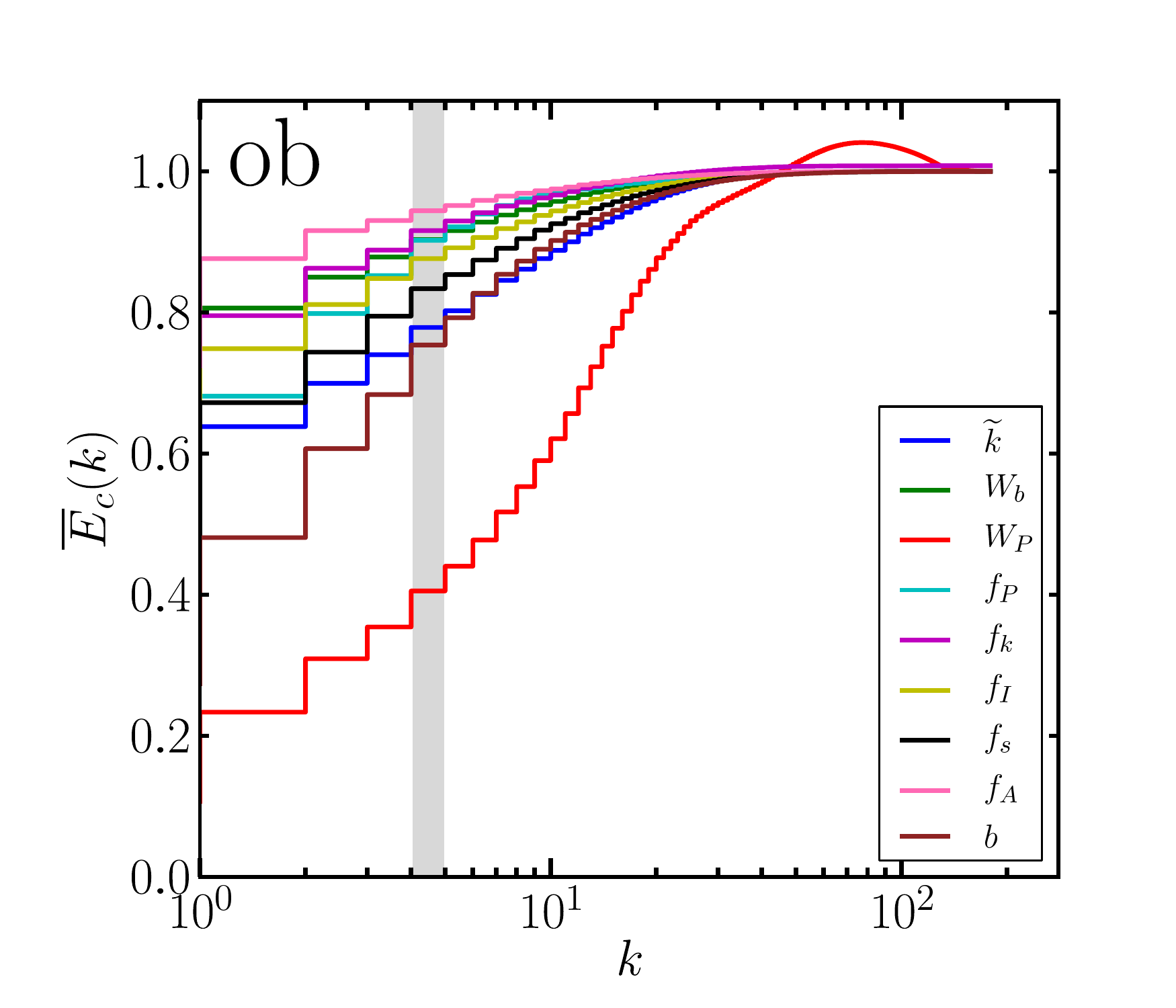}
\includegraphics[width=7.cm]{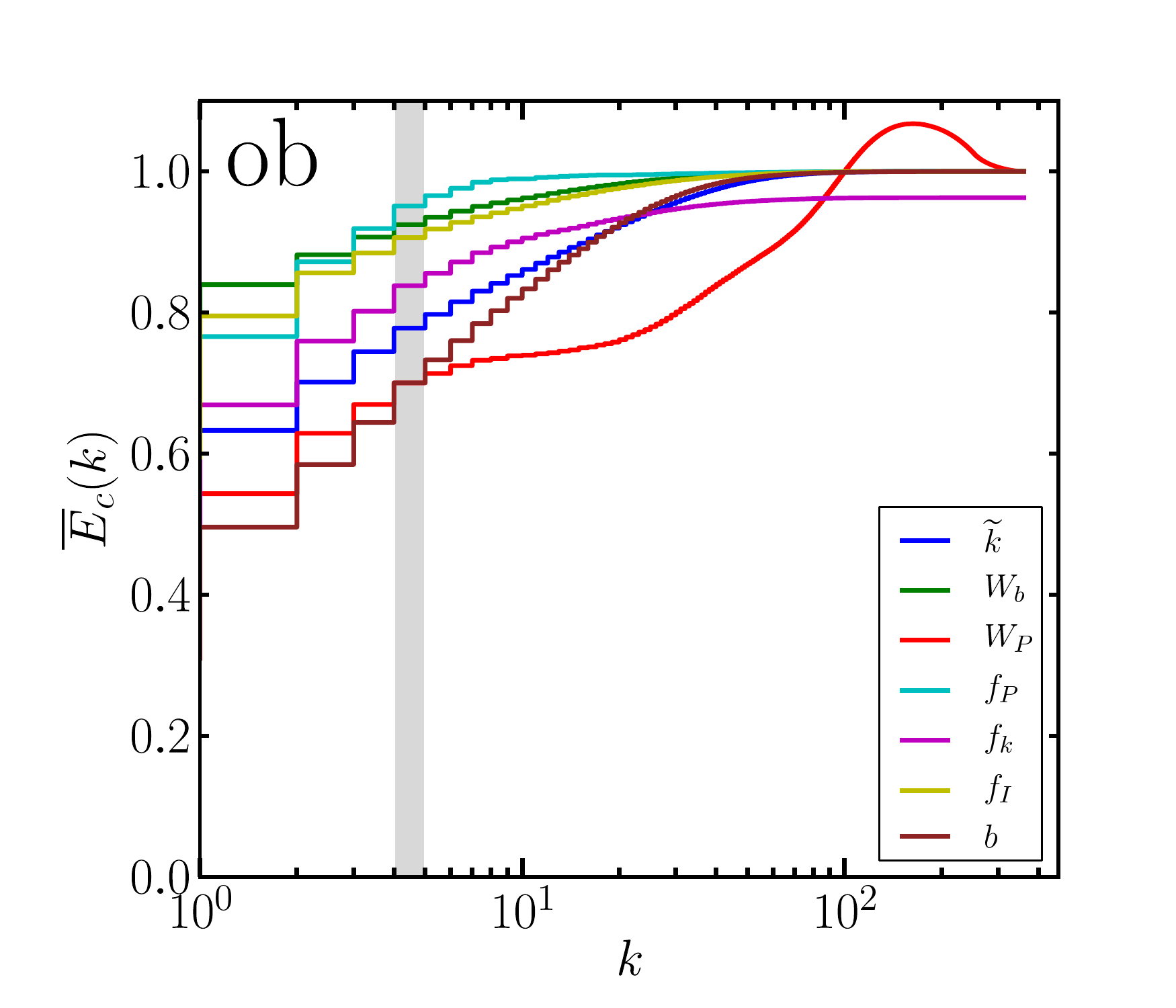}}

\centerline{
\includegraphics[width=7.cm]{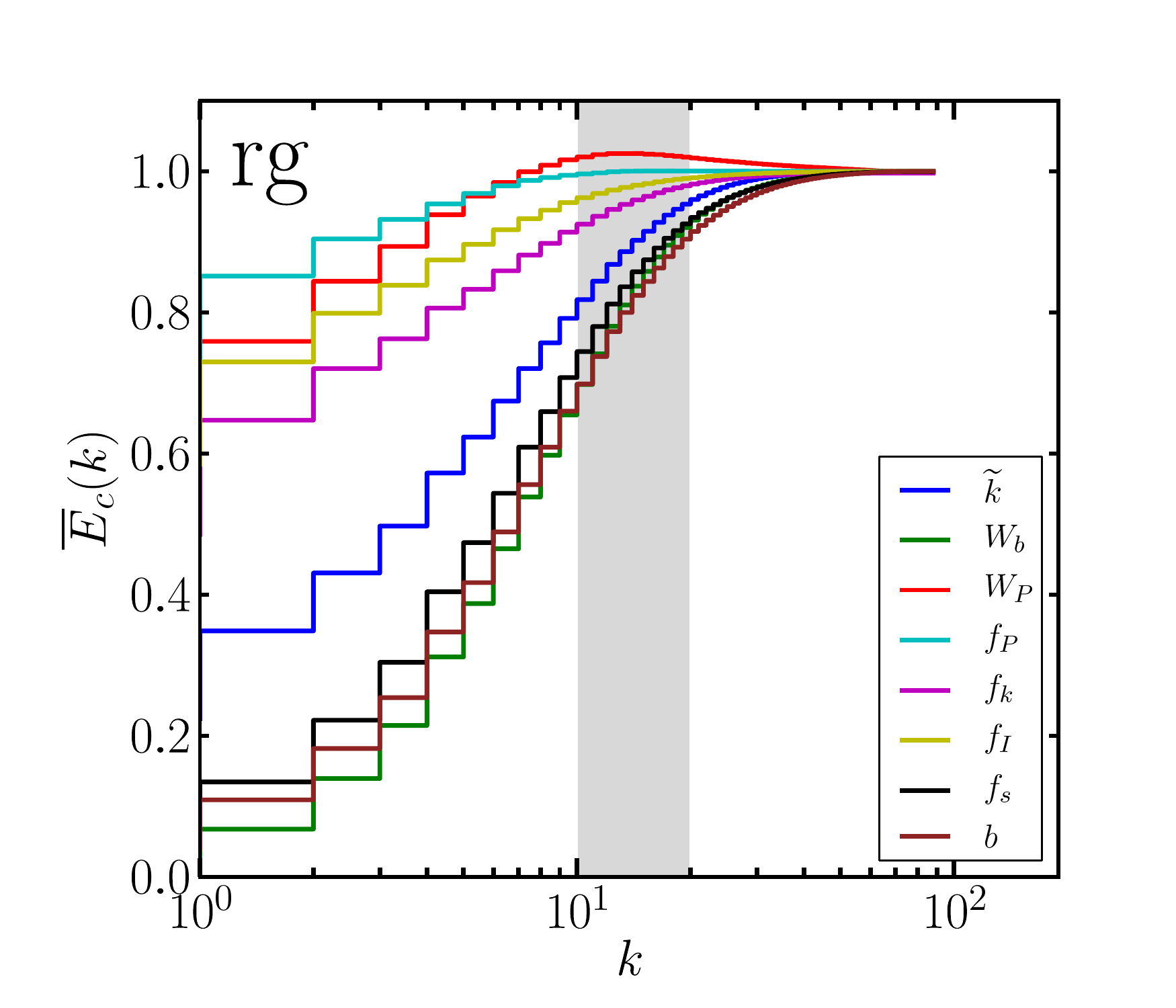}}

\caption{Cumulative Fourier spectra of relevant mean fields for the oxygen burning shell models (upper panels) {\sf ob.3D.lr} (left), {\sf ob.3D.mr} (middle) and {\sf ob.3D.hr} (right) derived at radius where corresponding mean field has a maximum value. The same is done for red giant envelope convection model (lower panel) {\sf rg.3D.lr}. The shaded vertical lines separates cumulative contributions from large and small scales.}
\end{figure}

\newpage


\clearpage

\subsection{Fourier scale decomposition of turbulent kinetic energy equation (ob.3D.mr)}\label{sec:scale-profiles-1}

\begin{figure}[!h]
\centerline{
\includegraphics[width=6.3cm]{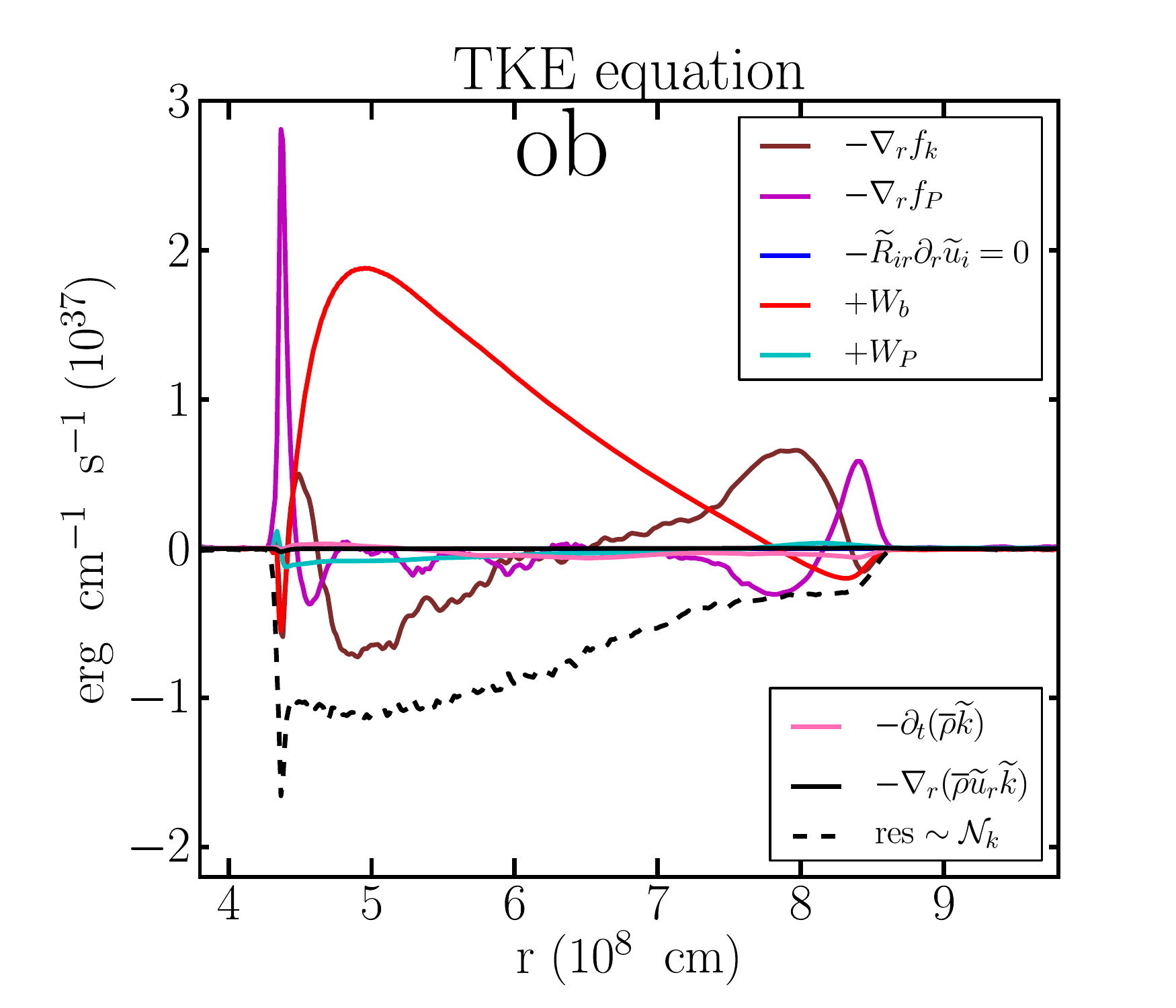}
\includegraphics[width=6.3cm]{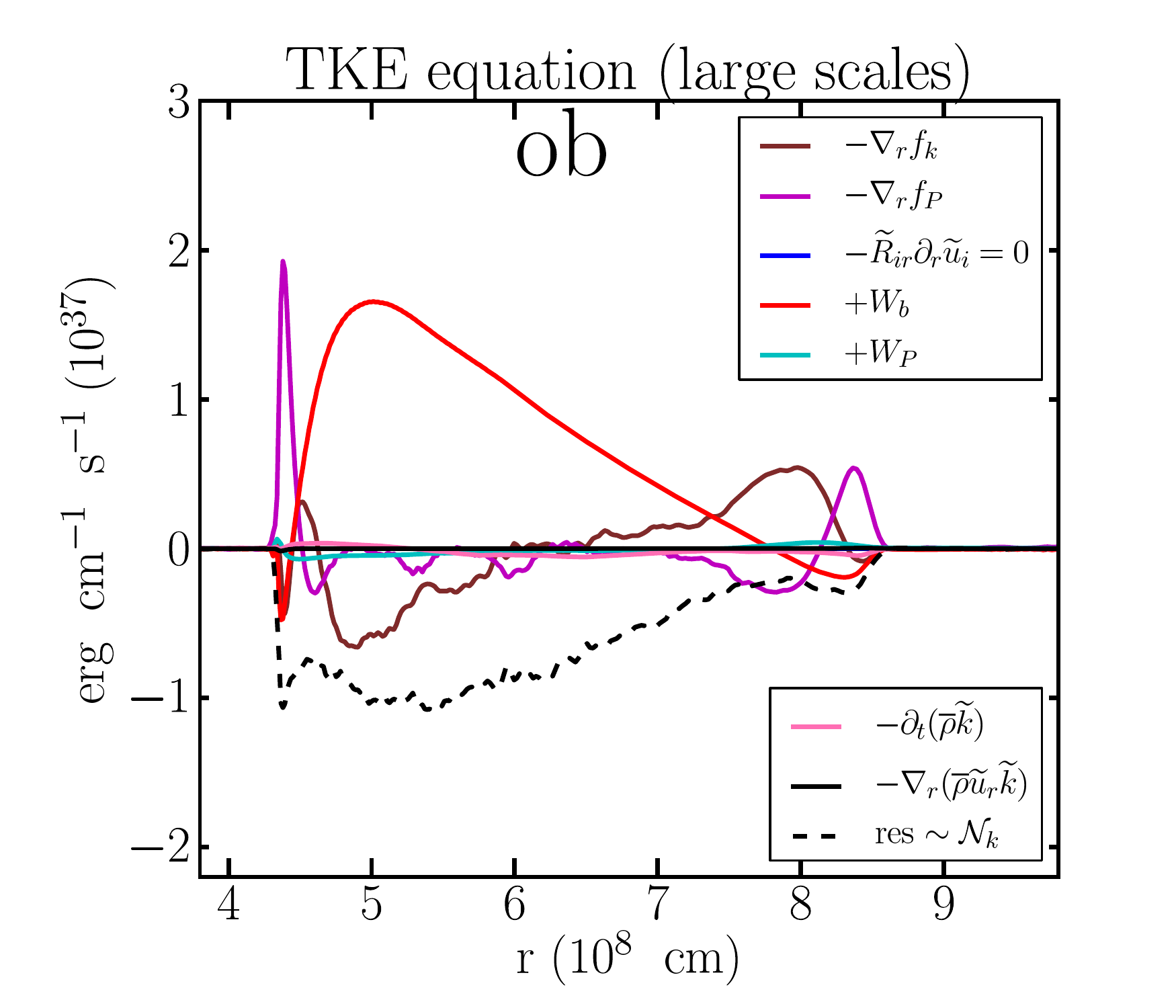}
\includegraphics[width=6.3cm]{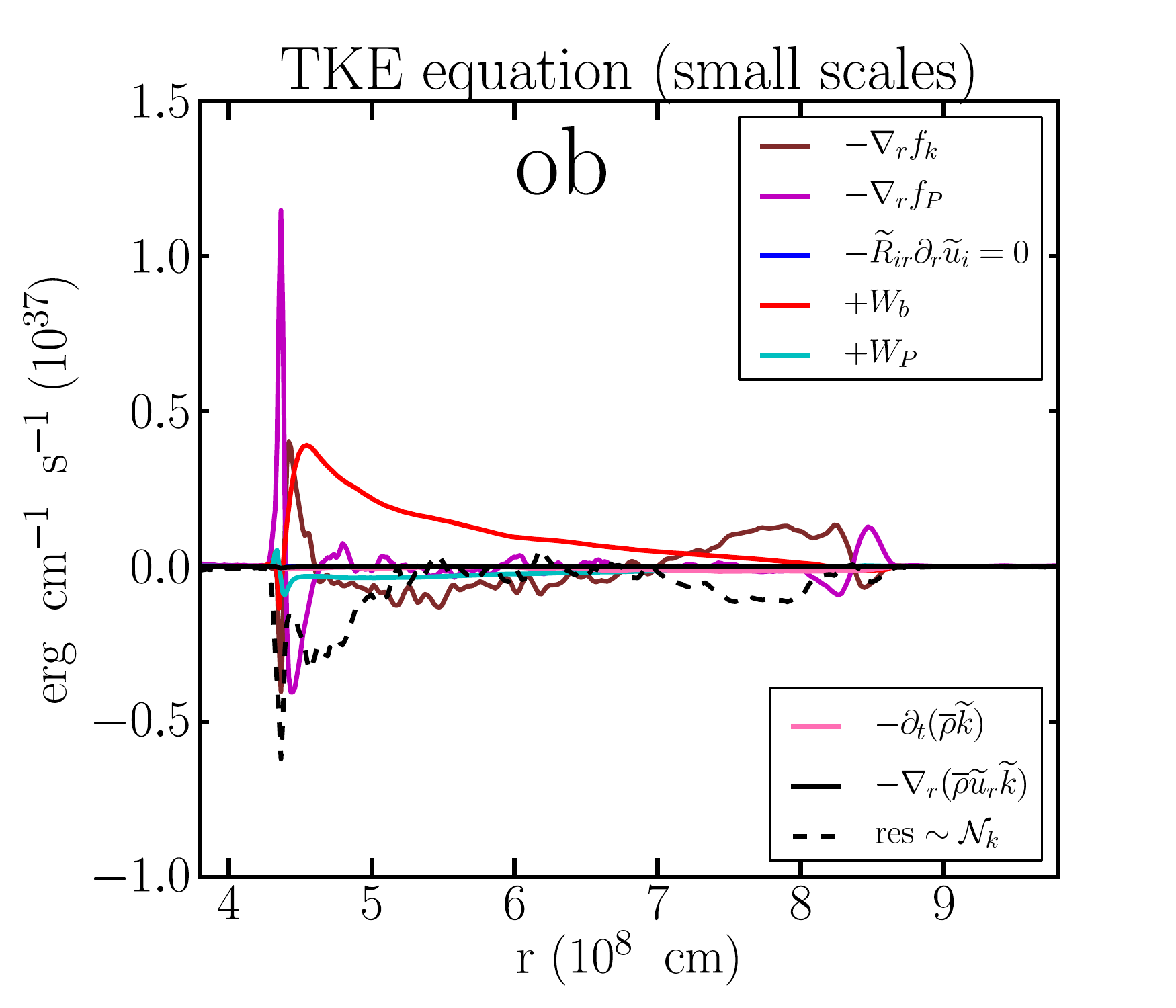}}

\centerline{
\includegraphics[width=6.3cm]{obmrez_mfields_k_equation_insf-eps-converted-to.pdf}
\includegraphics[width=6.3cm]{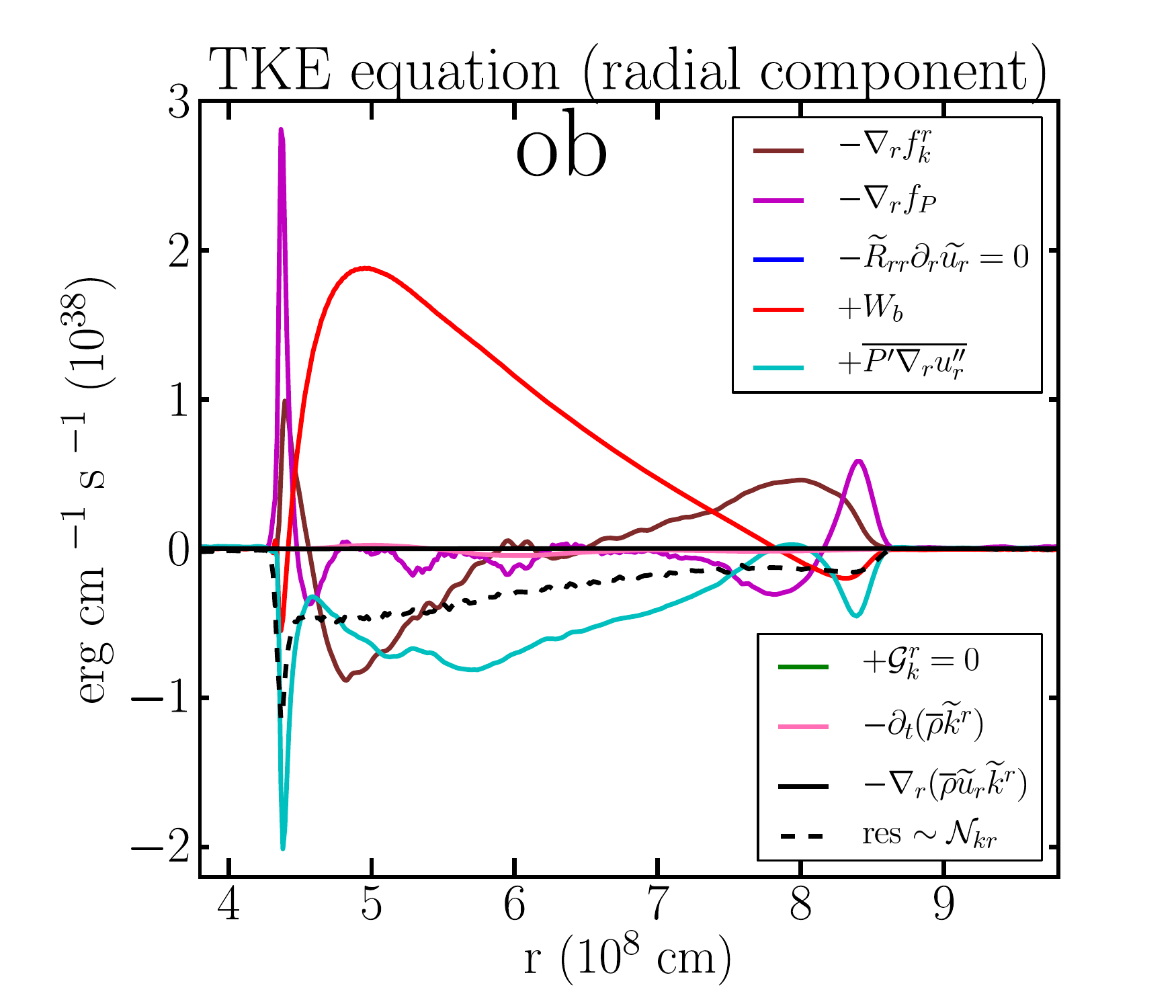}
\includegraphics[width=6.3cm]{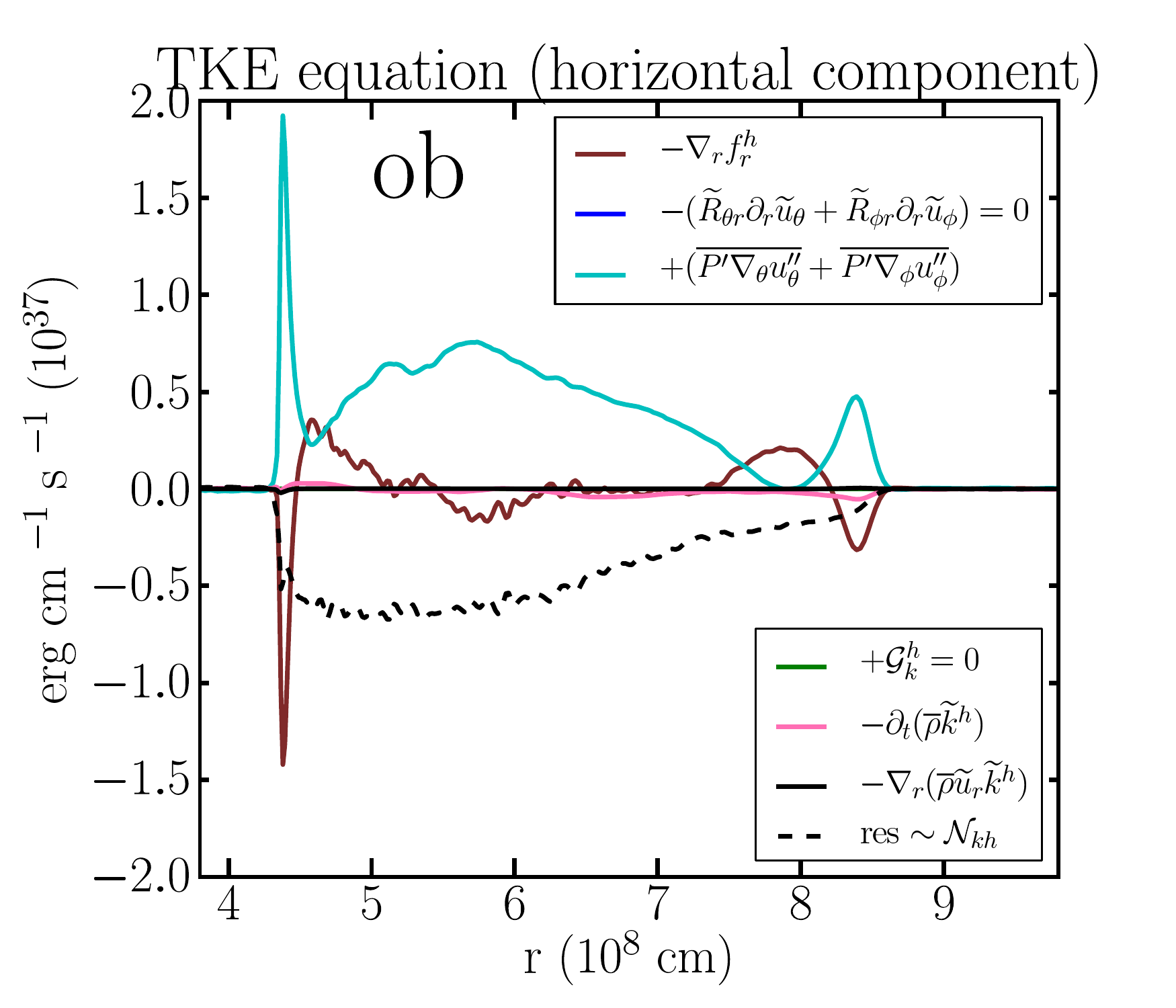}}
\caption{Upper panels: Fourier scale decomposition of mean fields in turbulent kinetic energy equation (left) into large (middle) and small (right) scale component. Scale separation wave number was taken to be 4. Averaging was performed over 230 s at around central time 505 s. Lower panels:  Decomposition of mean fields in turbulent kinetic energy equation (left) into radial (middle) and horizontal components (right).}
\end{figure}

\newpage

\subsection{Fourier scale decomposition of radial and horizontal part of turbulent kinetic energy equation (ob.3D.mr)}\label{sec:scale-profiles-2}

\begin{figure}[!h]
\centerline{
\includegraphics[width=6.2cm]{obmrez_mfields_k_equation_rad_insf-eps-converted-to.pdf}
\includegraphics[width=6.2cm]{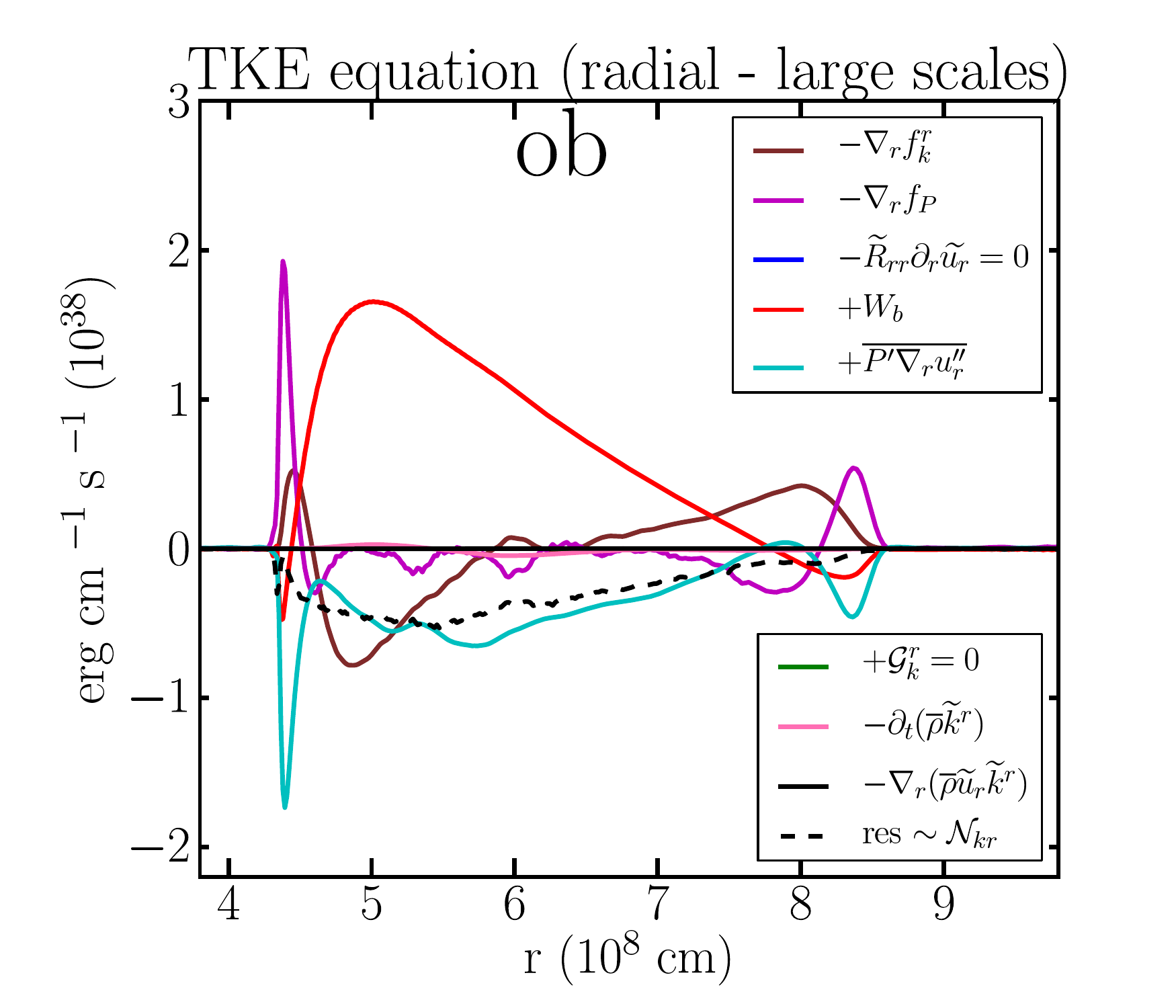}
\includegraphics[width=6.2cm]{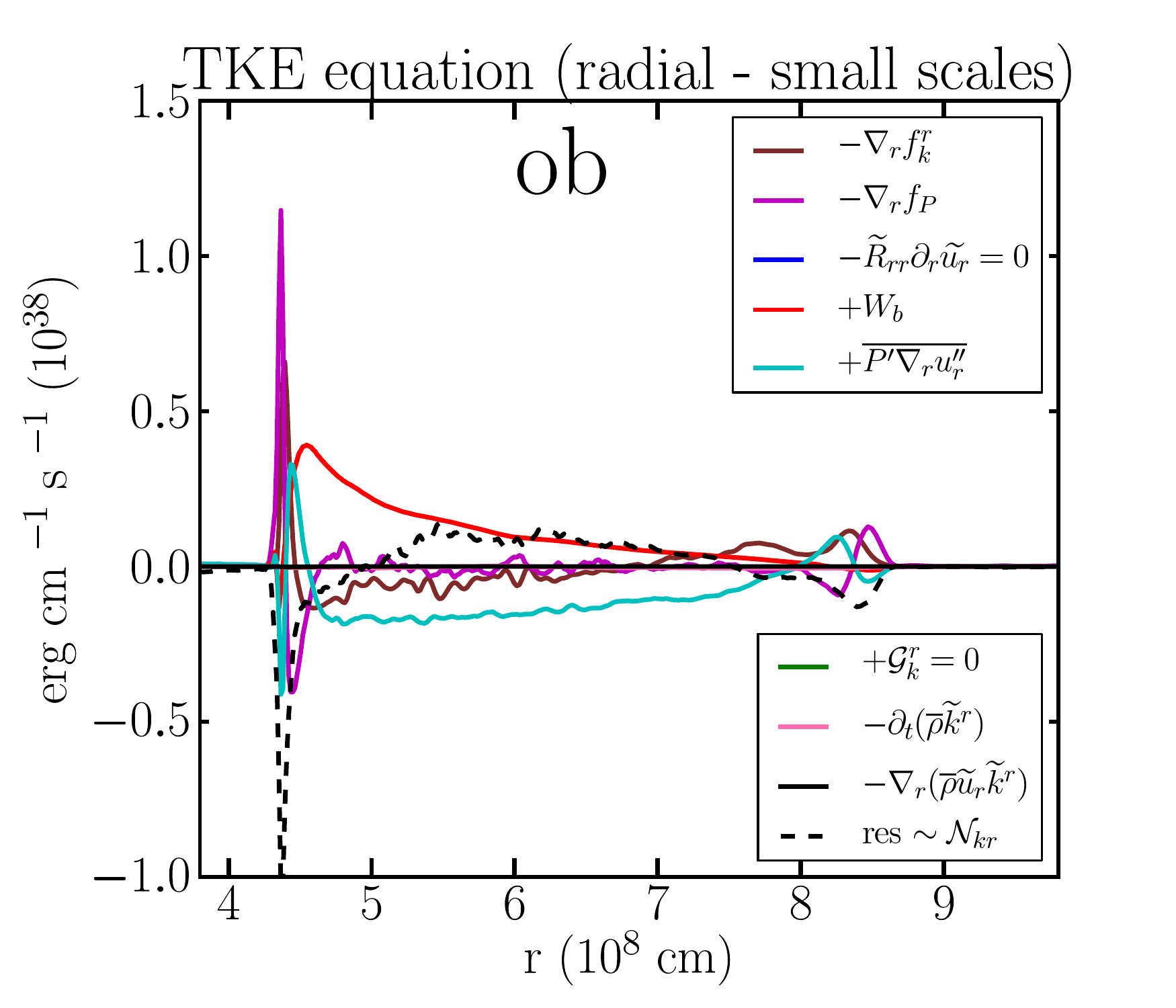}}

\centerline{
\includegraphics[width=6.2cm]{obmrez_mfields_k_equation_hor_insf-eps-converted-to.pdf}
\includegraphics[width=6.2cm]{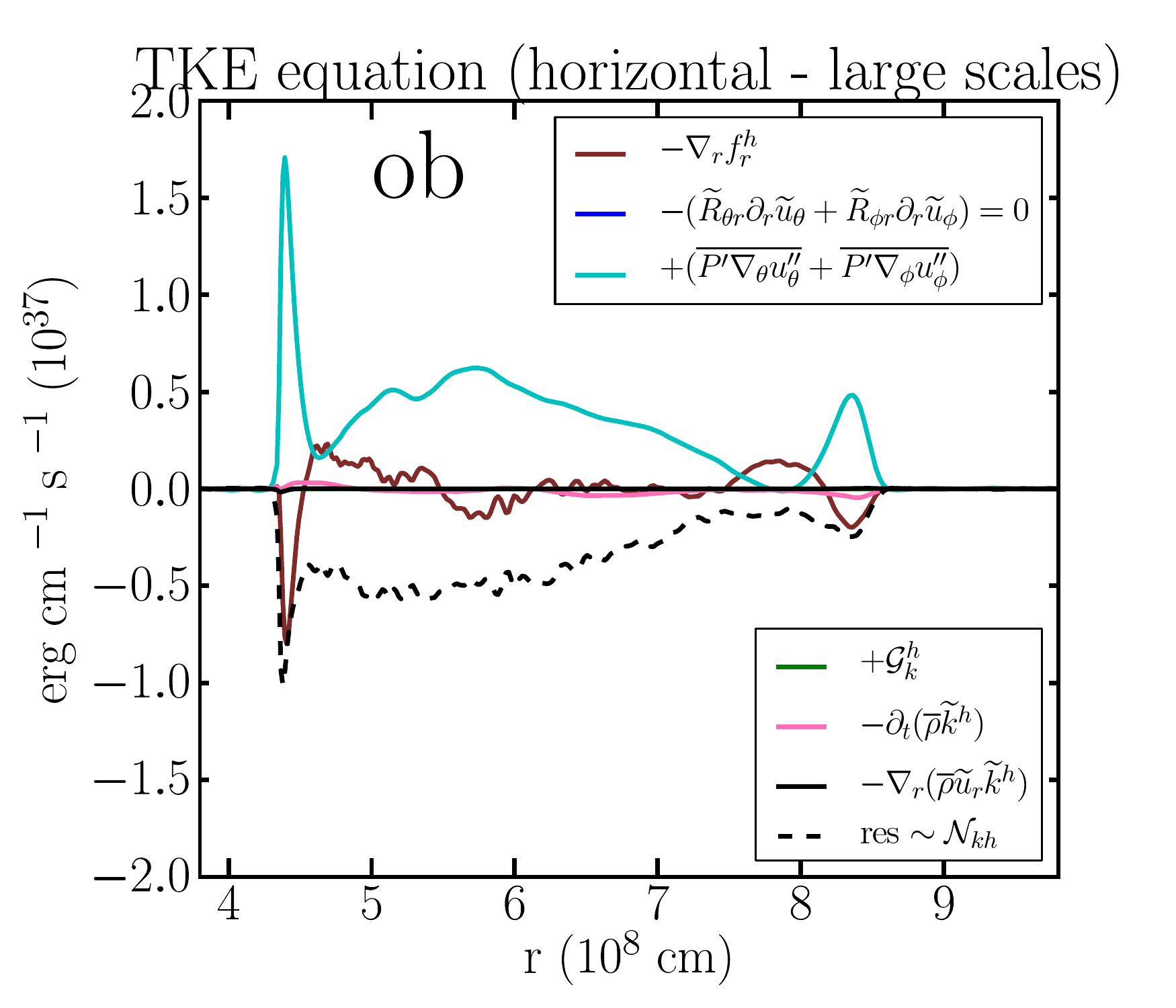}
\includegraphics[width=6.2cm]{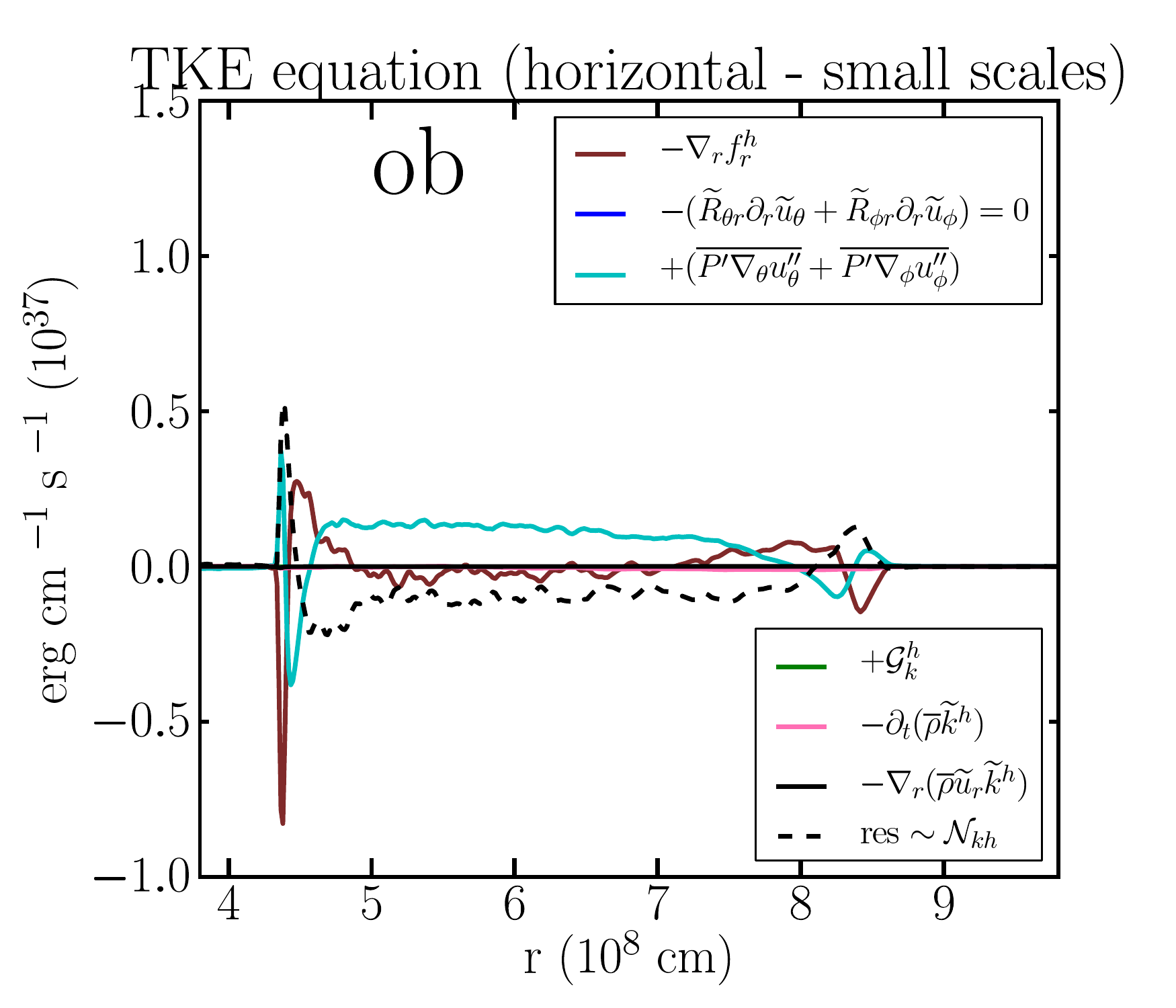}}
\caption{Upper panels: Fourier scale decomposition of radial part of mean fields in turbulent kinetic energy equation (left) into large (middle) and small (right) scale component. Lower panels: The same for the horizontal part of turbulent kinetic energy equation.  Scale separation wave number was taken to be 4. Averaging was performed over 230 s at around central time 505 s.}
\end{figure}

\newpage

\subsection{Fourier scale decomposition of individual mean fields in turbulent kinetic energy equation (ob.3D.mr)}\label{sec:scale-profiles-3}

\begin{figure}[!h]
\centerline{
\includegraphics[width=6.3cm]{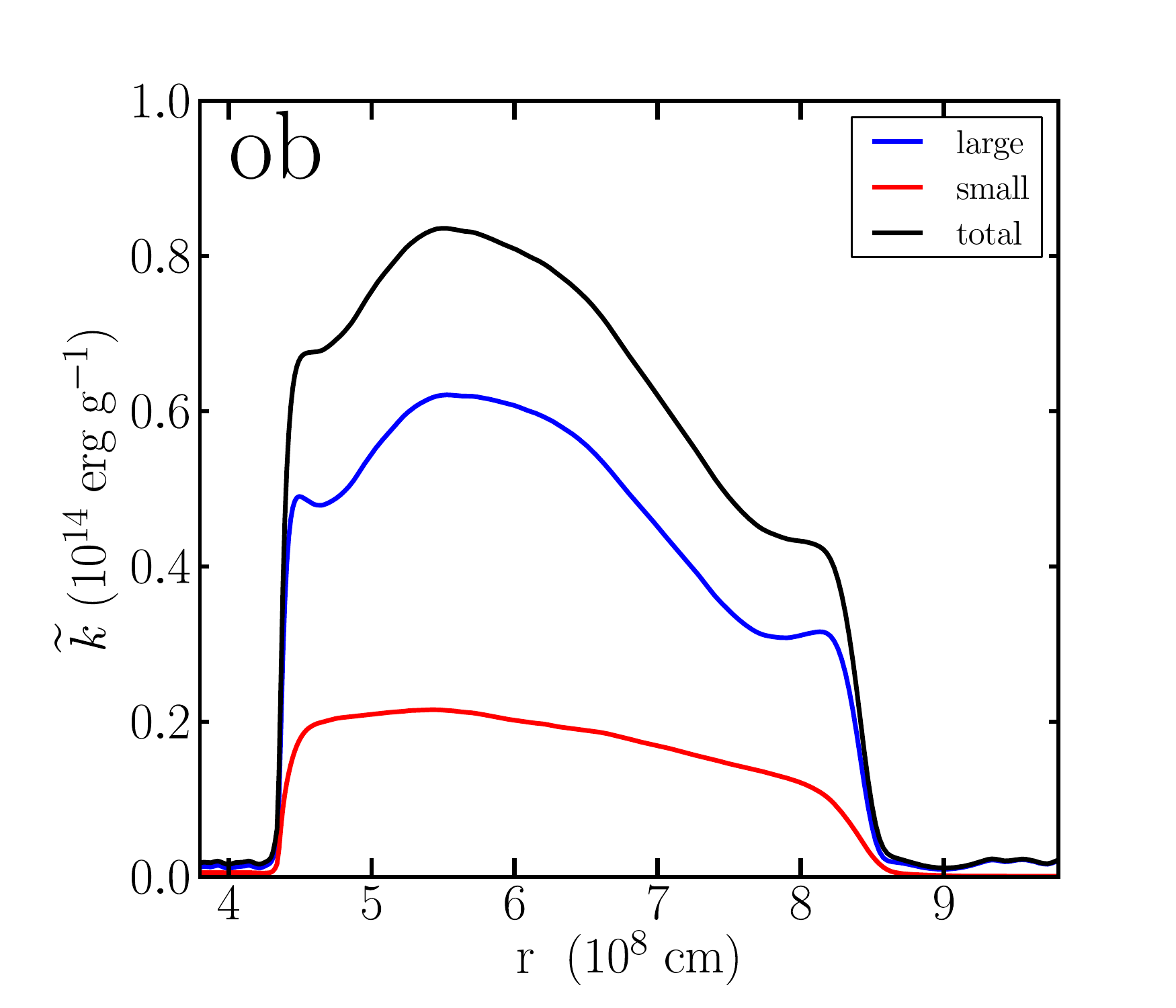}
\includegraphics[width=6.3cm]{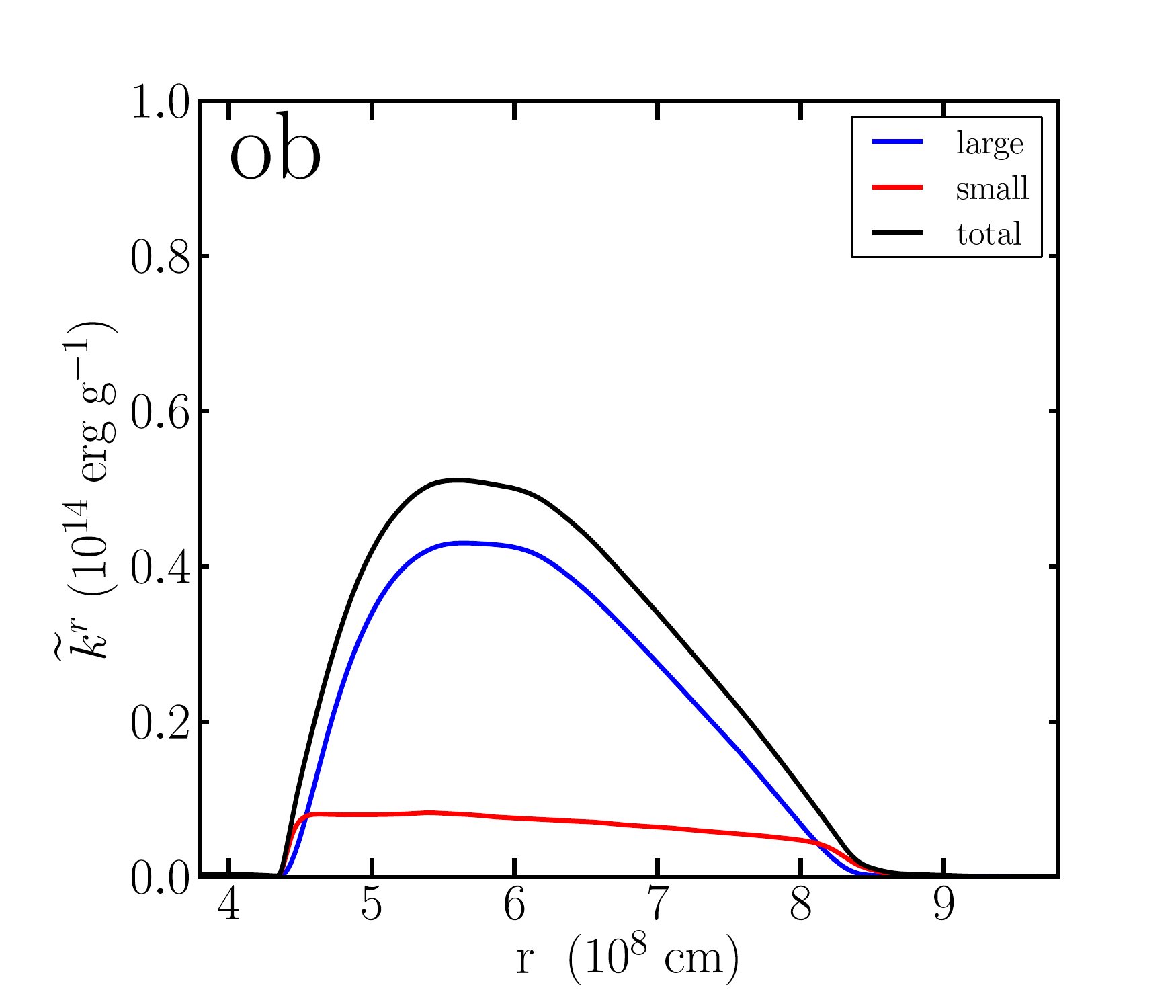}
\includegraphics[width=6.3cm]{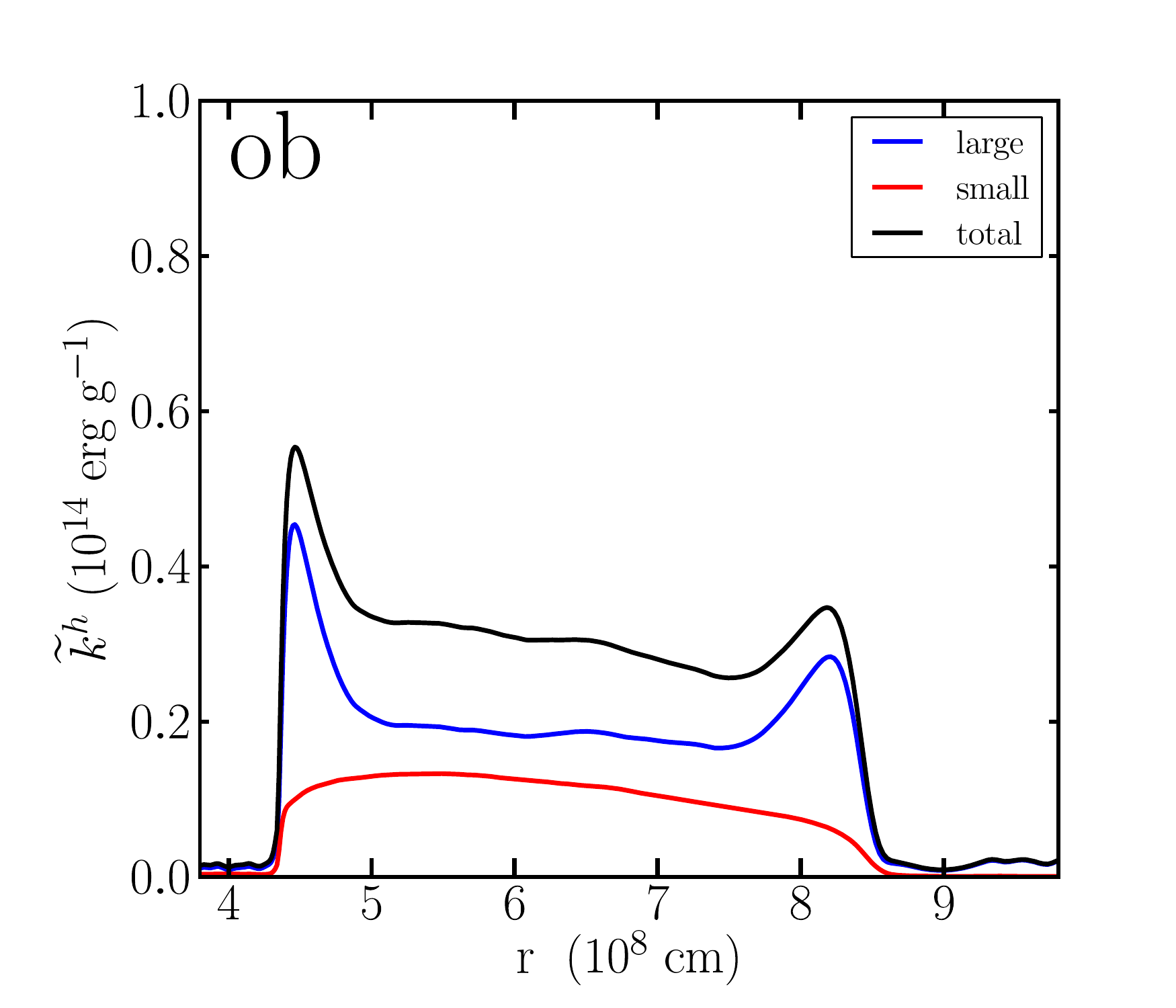}}

\centerline{
\includegraphics[width=6.3cm]{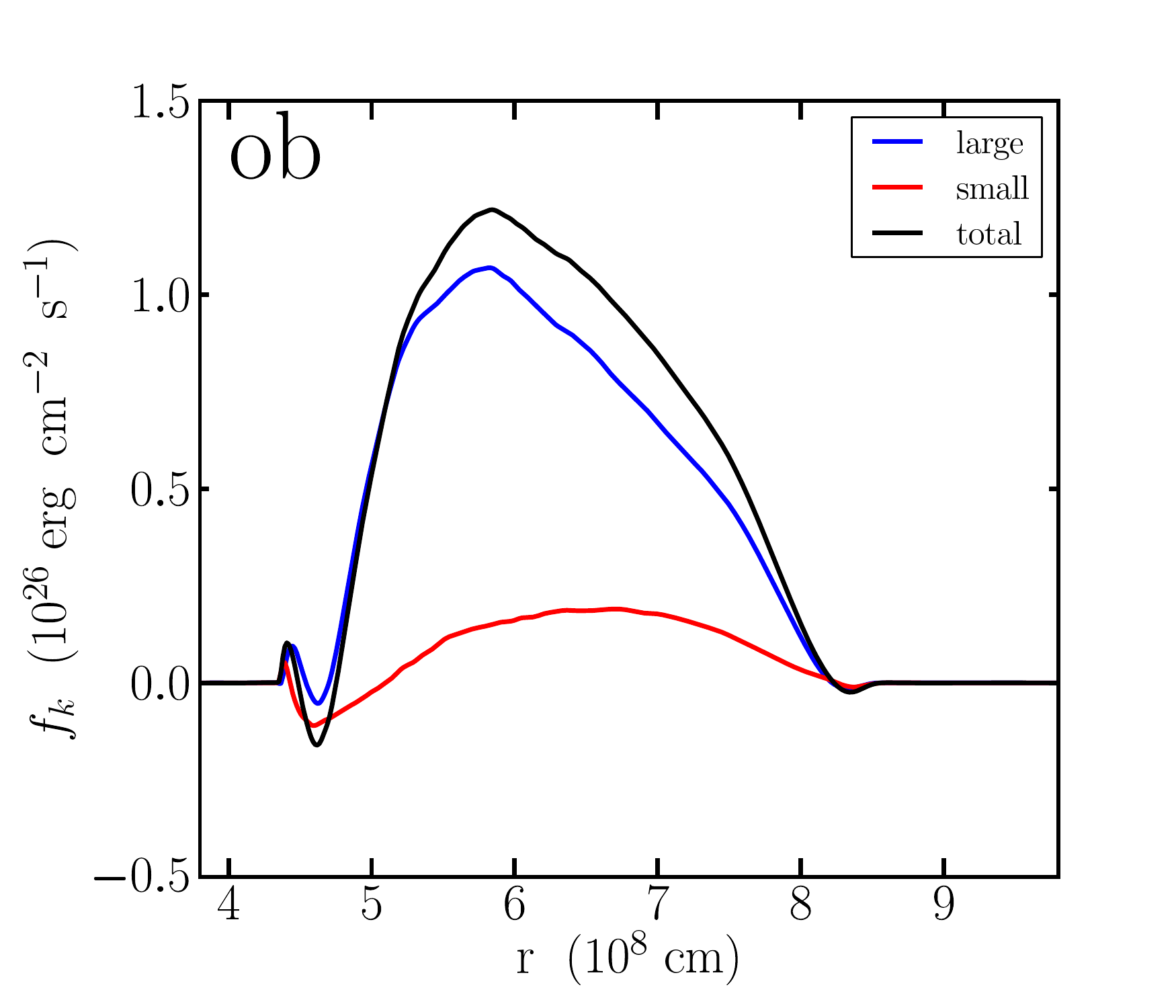}
\includegraphics[width=6.3cm]{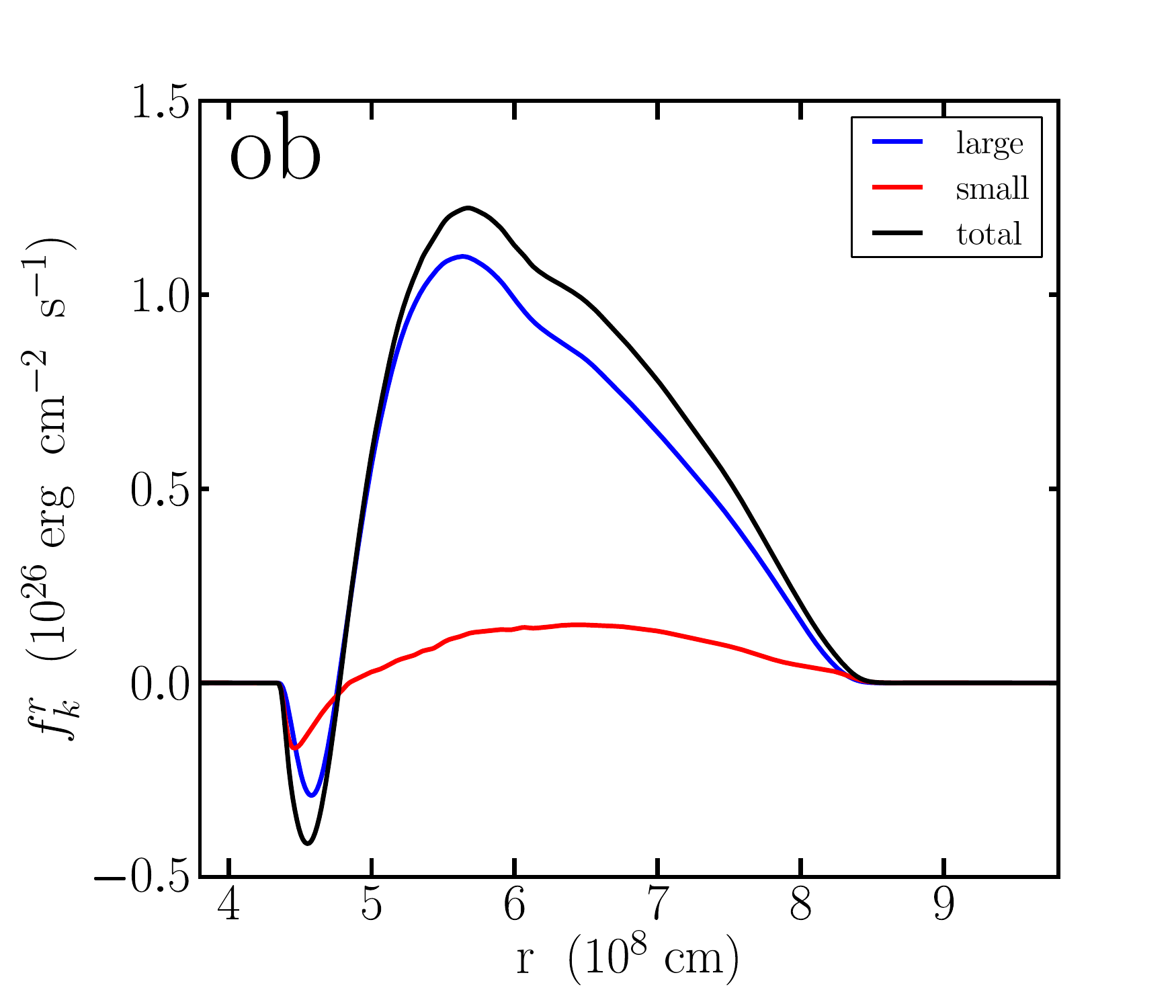}
\includegraphics[width=6.3cm]{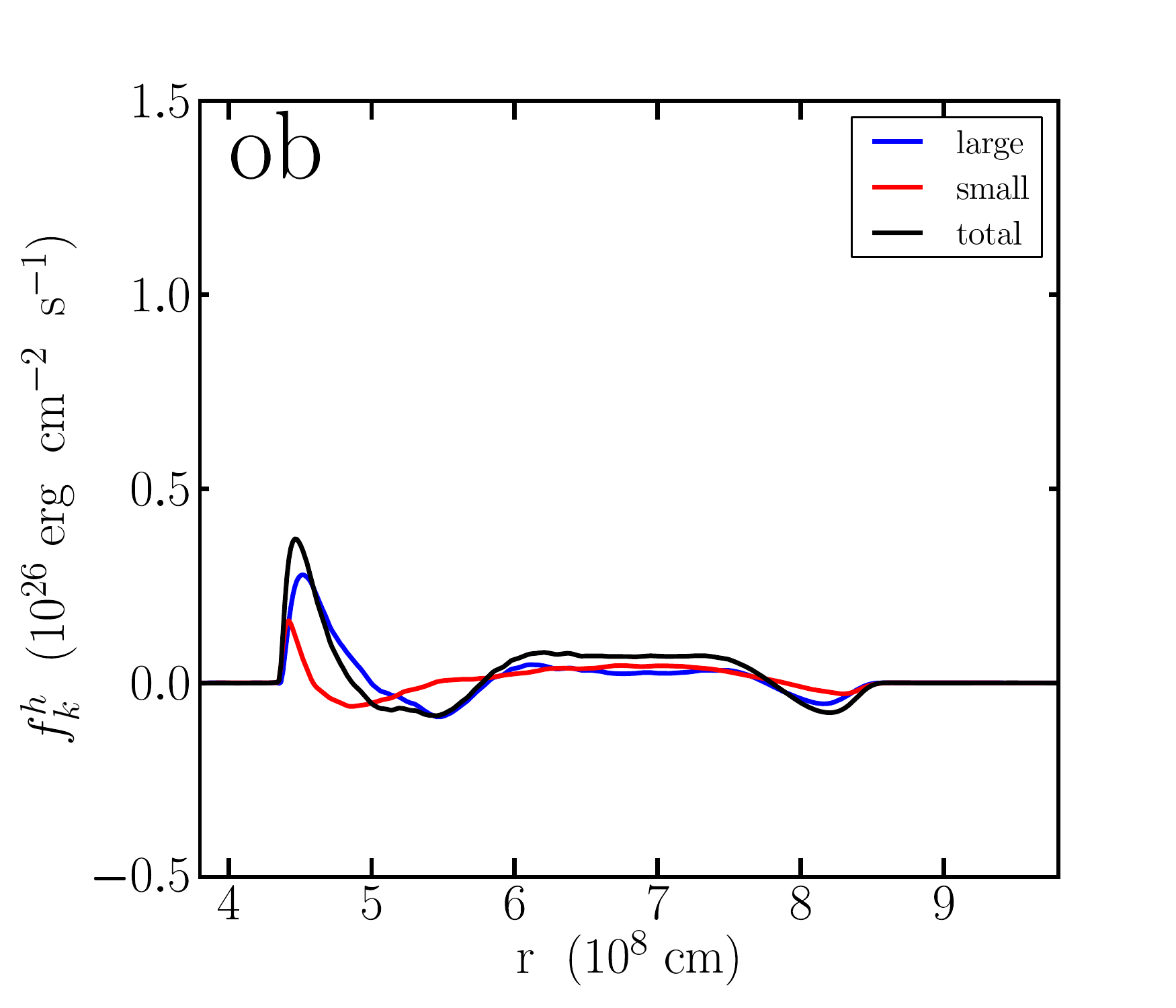}}
\caption{Upper panels: Fourier scale decoposition of turbulent kinetic energy (left), radial part of turbulent kinetic energy (middle) and horizontal part of turbulent kinetic energy (right). Lower panels: The same is done for the turbulent kinetic energy flux. Separation wave number was taken to be 4. Averaging was performed over 230 s at around central time 505 s.}
\end{figure}

\newpage

\begin{figure}[!h]
\centerline{
\includegraphics[width=6.3cm]{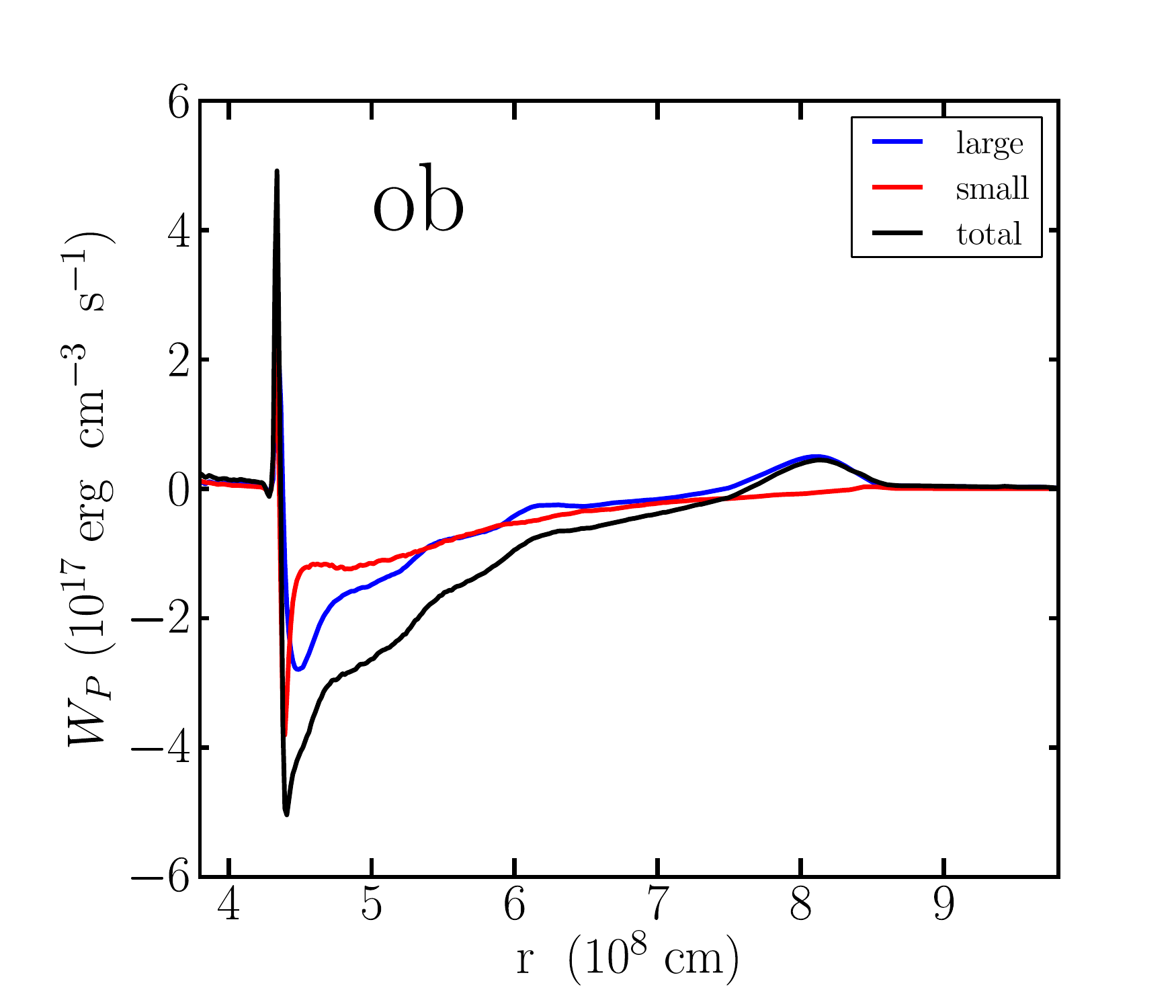}
\includegraphics[width=6.3cm]{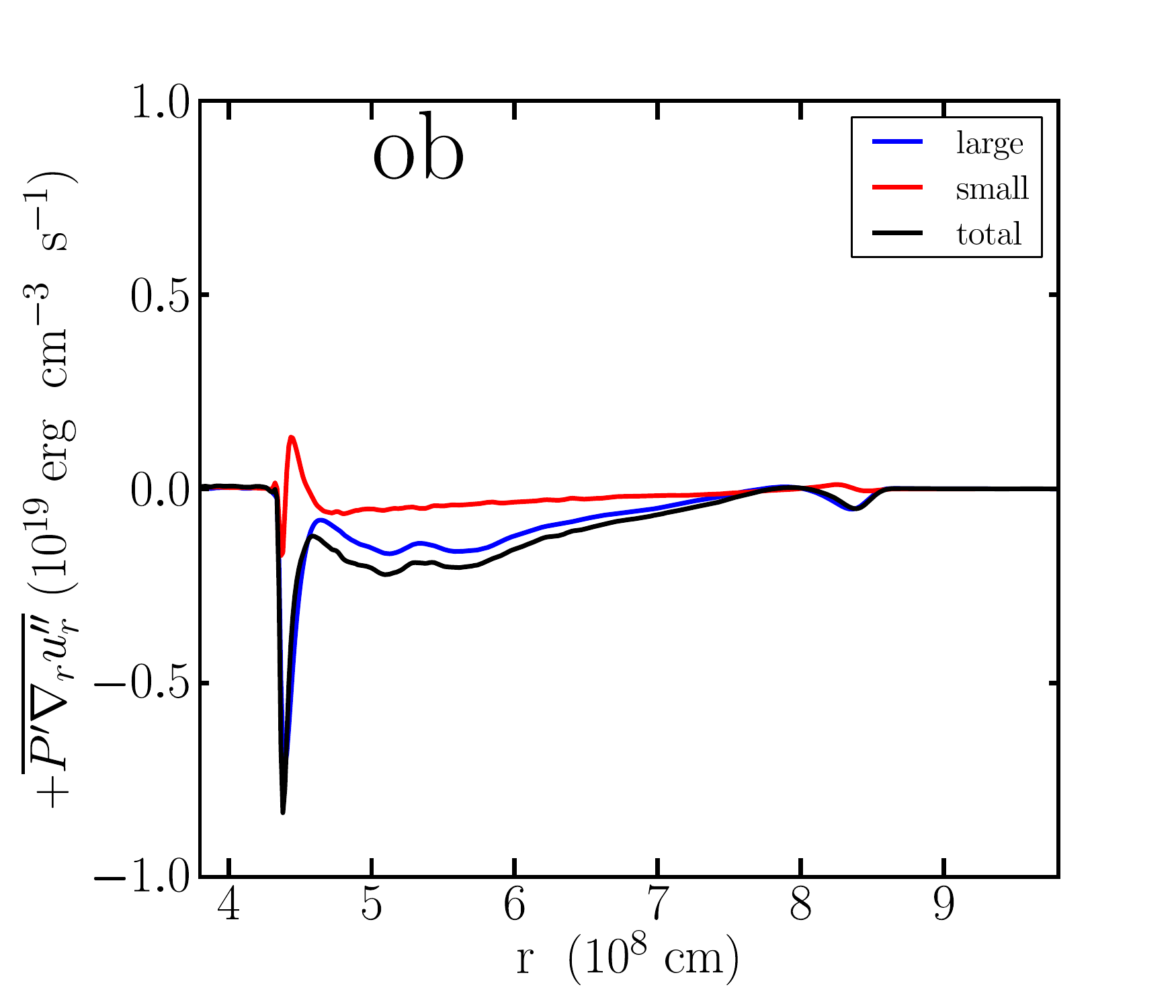}
\includegraphics[width=6.3cm]{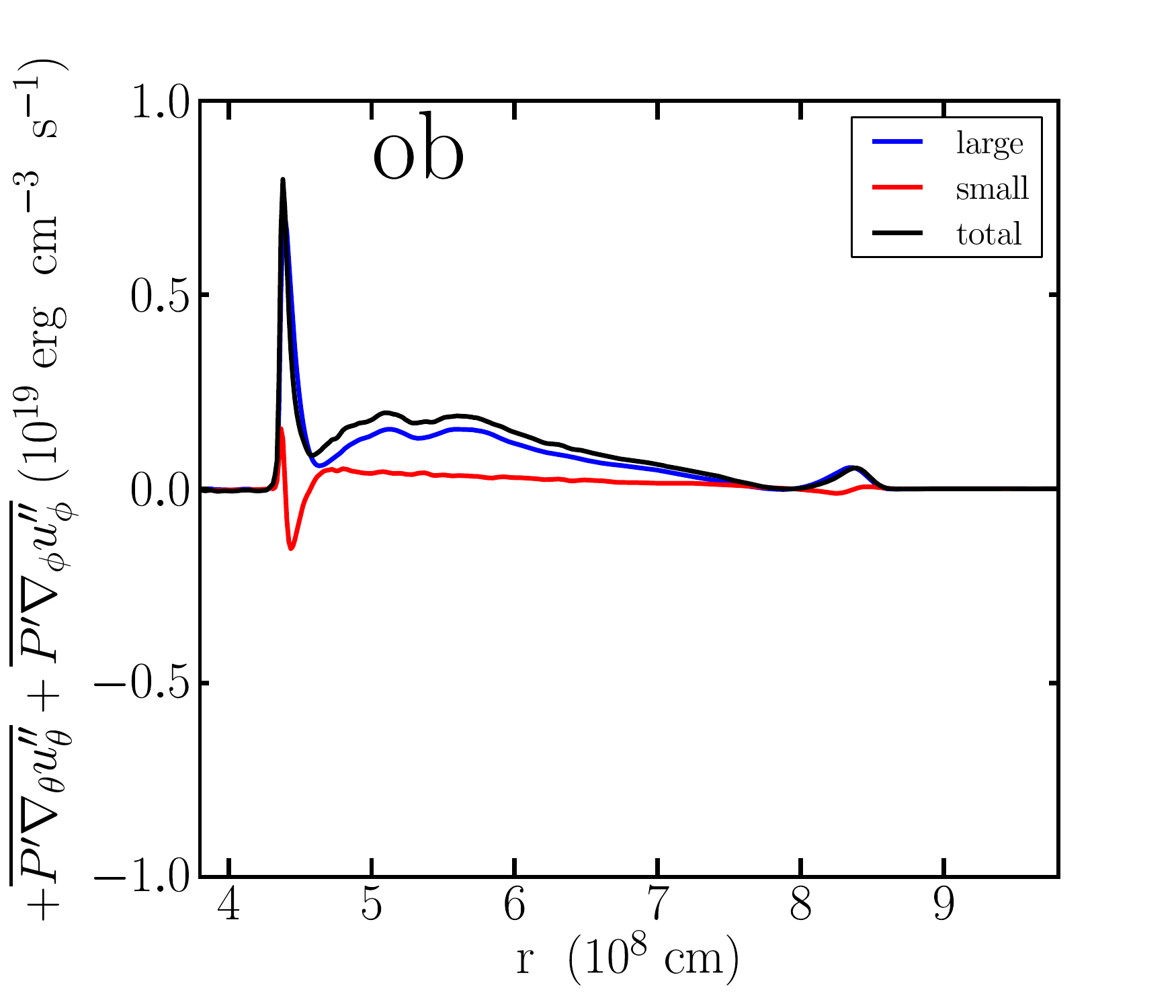}}

\centerline{
\includegraphics[width=6.3cm]{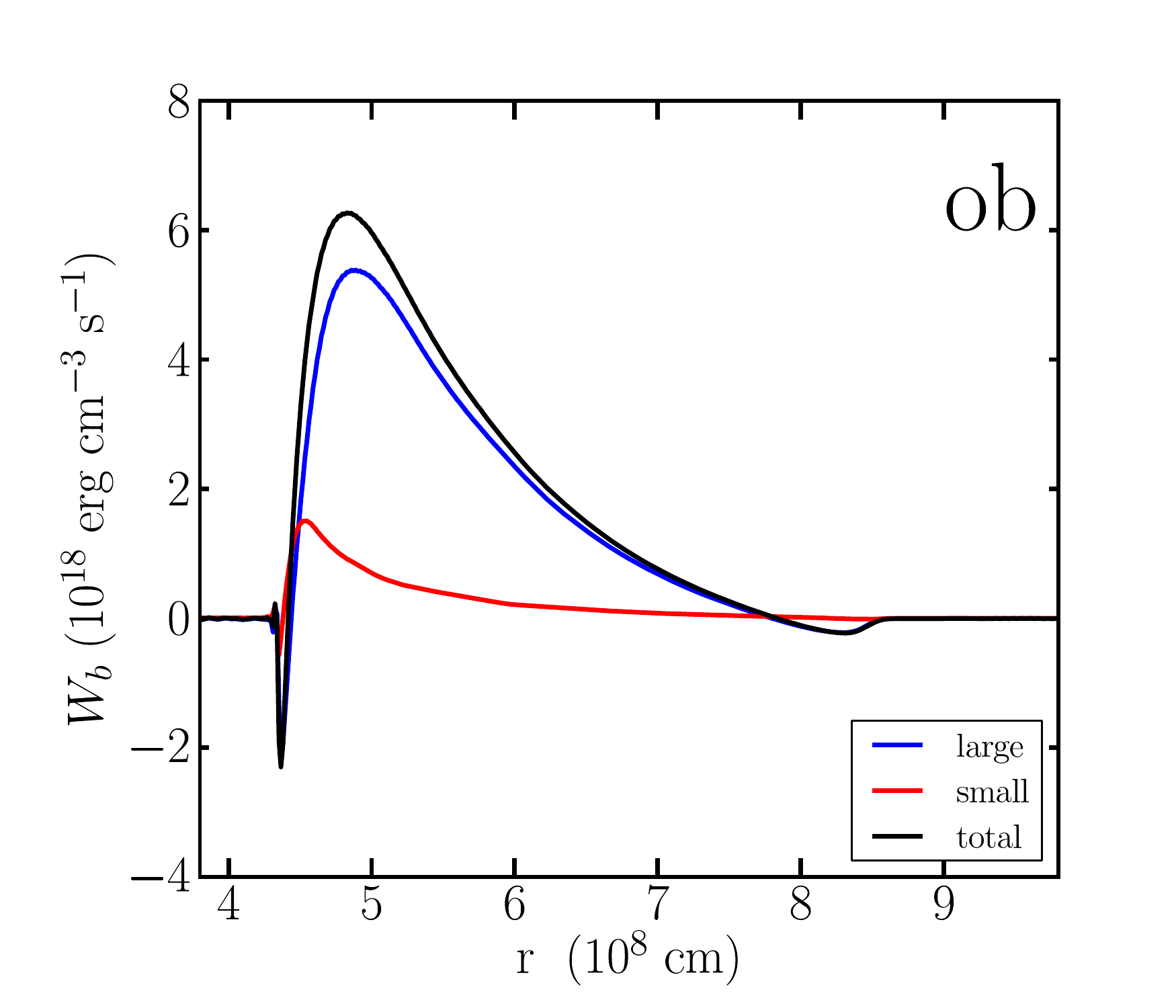}
\includegraphics[width=6.3cm]{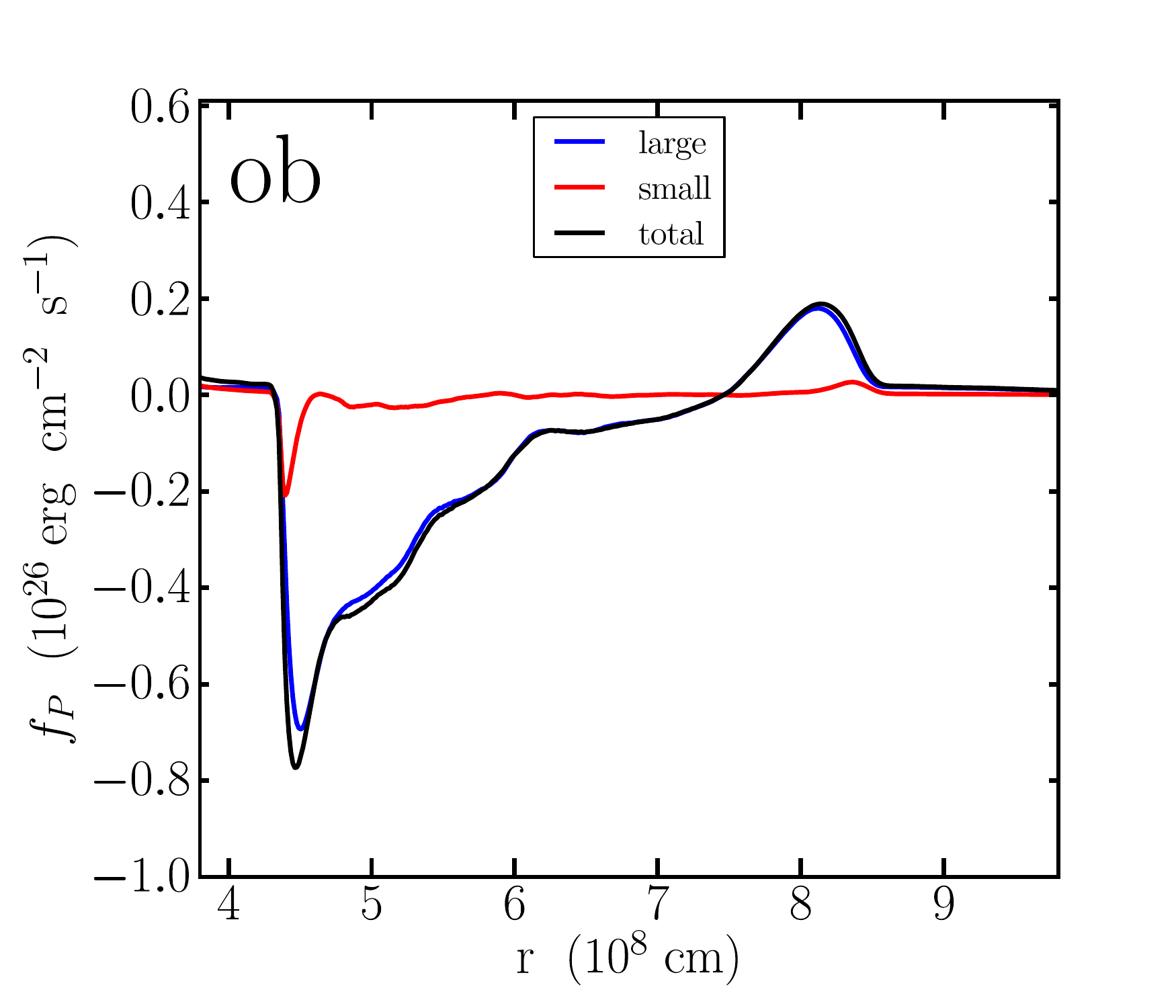}}
\caption{Upper panels: Fourier scale decomposition of the turbulent pressure dilataion (left), radial pressure strain (middle) and horizontal pressure strain (right). Lower panels: Fourier scale decomposition of the buoyancy work $W_b$ and acoustic flux $f_P$. Separation wave number was taken to be 4. Averaging was performed over 230 s at around central time 505 s.}
\end{figure}

\newpage

\section{Properties of hydrogen injection flash and core helium flash simulation data}

\subsection{Summary of the hydrogen injection flash and core helium flash simulations and their properties}


\begin{table}[!h]
\centerline{
\begin{tabular}{|l c c c|}
\hline
{\bf Parameter}  & {\sf hif.3D} & {\sf hif.3D} & {\sf chf.3D} \\
Grid zoning & 375$\times 45^2$ & 375$\times 45^2$ & 270$\times 30^2$ \\
$r_\mathrm{in}$, $r_\mathrm{out}$ ($10^{8}$ cm) & 2., 16. & 2., 16. & 2., 12. \\
$\Delta \theta$, $\Delta \phi$ & 45$^\circ$ & & 30$^\circ$ \\
& & & \\
{\bf Convection zone} & He-burning (CVZ-1) & CNO-burning (CVZ-2) & He-burning (CVZ)\\
$r^c_\mathrm{in}$, $r^c_\mathrm{out}$ ($10^{8}$ cm) & 4.7, 9.55 & 9.55, 11.3 & 4.7, 9.4\\
CZ stratification ($H_p$)         & 2.4   & 1.3 & 2.3 \\
$\Delta t_\mathrm{av}$ (s)          & $4000$ & $4000$ & $18000$ \\
$v_\mathrm{rms}$ ($10^5$ cm/s)      & 8.2 & 11.8 & 8.5 \\
$\tau_\mathrm{conv}$ (s)     & 1180  & 290 & 1100 \\
$P_{turb}/P_{gas} (10^{-5})$          & 2.6 & 3.7  & 3.3 \\ 
$L_\mathrm{heat}$ ($10^{43}$ erg/s)  & 1.2  & 1.8 & 1.2 \\
$L_\mathrm{d}$ ($10^{41}$ erg/s)     & 5.3  & 6.1 & 6. \\
$l_\mathrm{d}$ ($10^{8}$ cm)        & 5.6  & 2.9  & 4.8 \\
$\tau_\mathrm{d}$ (s)            & 337.4   & 123.6 & 281.9\\
$\tau_\mathrm{dr}$ (s)           & 719.7   & 175.8 & 659.7 \\
$\tau_\mathrm{dh}$ (s)           & 211.2   & 116.1 & 142.5 \\
\hline
\end{tabular}}
\caption{boundaries of computational domain $r_\mathrm{in}$, $r_\mathrm{out}$; boundaries of convection zone at bottom and top $r_\mathrm{b}^c$, $r_\mathrm{t}^c$; angular size of computational domain $\Delta \theta$, $\Delta \phi$ ; depth of convection zone ``CZ stratification'' in pressure scale height $H_P$; averaging timescale of mean fields analysis $\Delta t_\mathrm{av}$; global rms velocity $v_\mathrm{rms}$; convective turnover timescale $\tau_\mathrm{conv}$; average ratio of turbulent ram pressure and gas pressure $p_{turb}/p_{gas}$; total luminosity of the hydrodynamic model $L$; total rate of kinetic energy dissipation $L_d$; dissipation length-scale  $l_d$; turbulent kinetic energy dissipation time-scale $\tau_d$; radial turbulent kinetic energy dissipation time-scale $\tau_{dr}$; horizontal turbulent kinetic energy dissipation time-scale $\tau_{dh}$. The numerical values may vary in time up to 20$\%$ due to limited amount of data for averaging out the time dependence.\label{tab:chf-hif-models} }
\end{table}

\newpage

\subsection{Snapshots of turbulent kinetic energy in a meridional plane}

\vspace{1.cm}

\begin{figure}[!h]
\centerline{
\includegraphics[width=11.8cm]{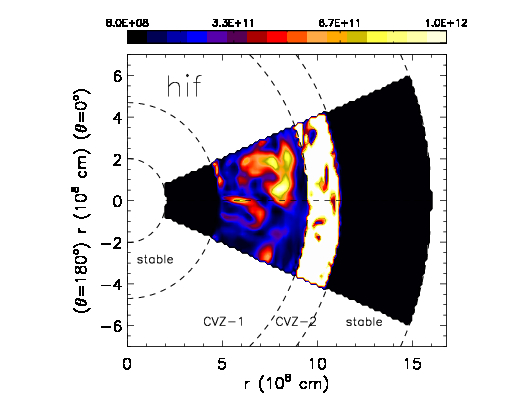}
\includegraphics[width=11.8cm]{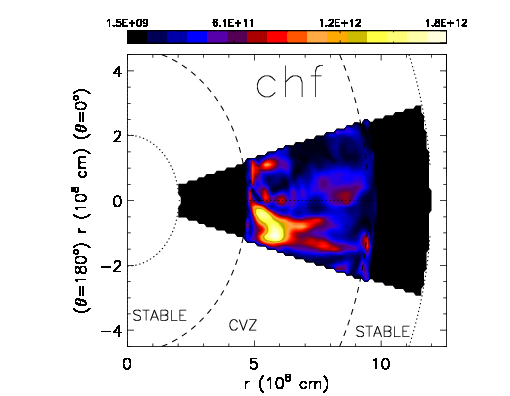}}
\caption{Snapshots of turbulent kinetic energy (in erg g$^{-1}$) in a meridional plane.}
\label{fig:ob-rg-tke-cuts}
\end{figure}

\newpage

\subsection{Background structure of our models}

\begin{figure}[!h]
\centerline{
\includegraphics[width=7.cm]{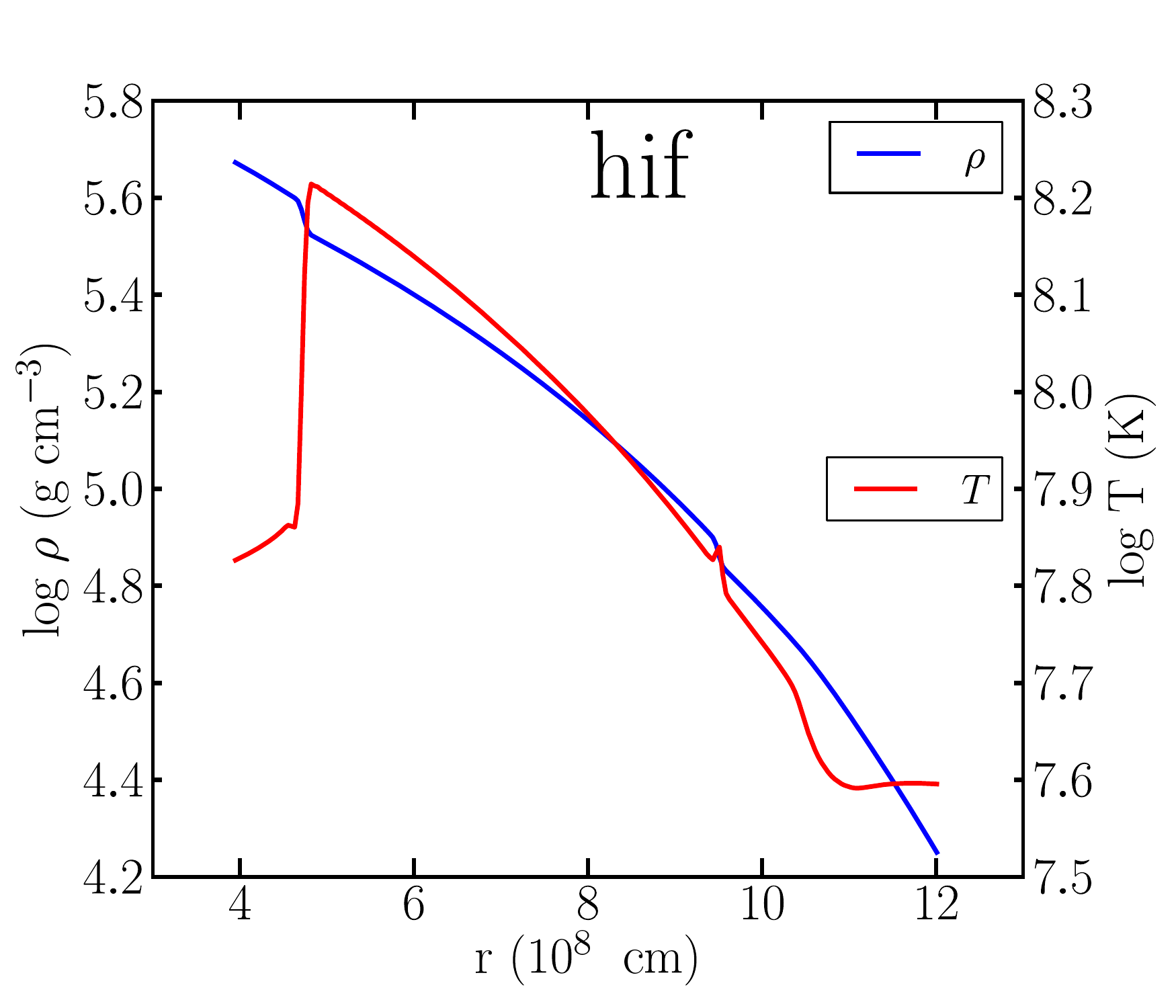}
\includegraphics[width=7.cm]{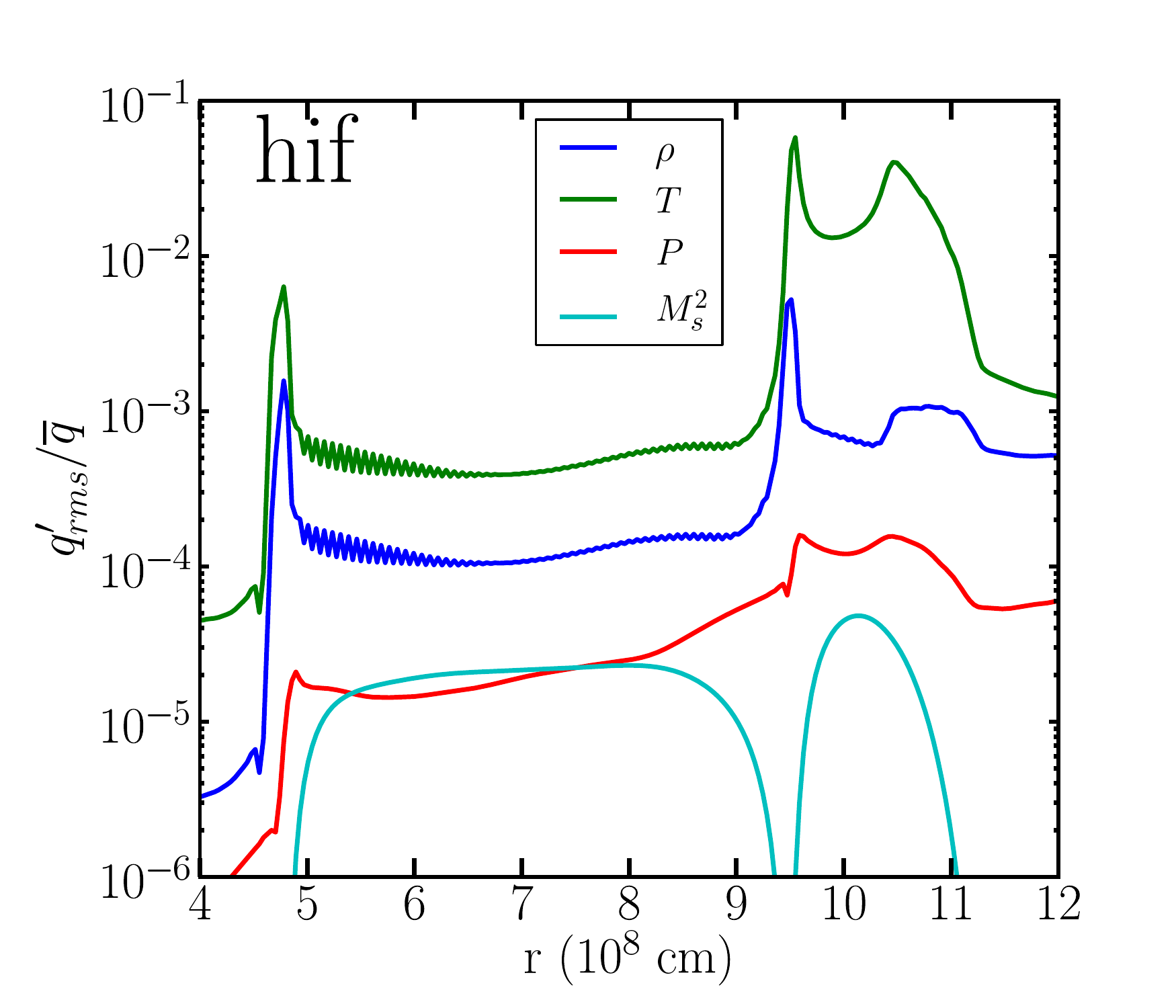}
\includegraphics[width=7.cm]{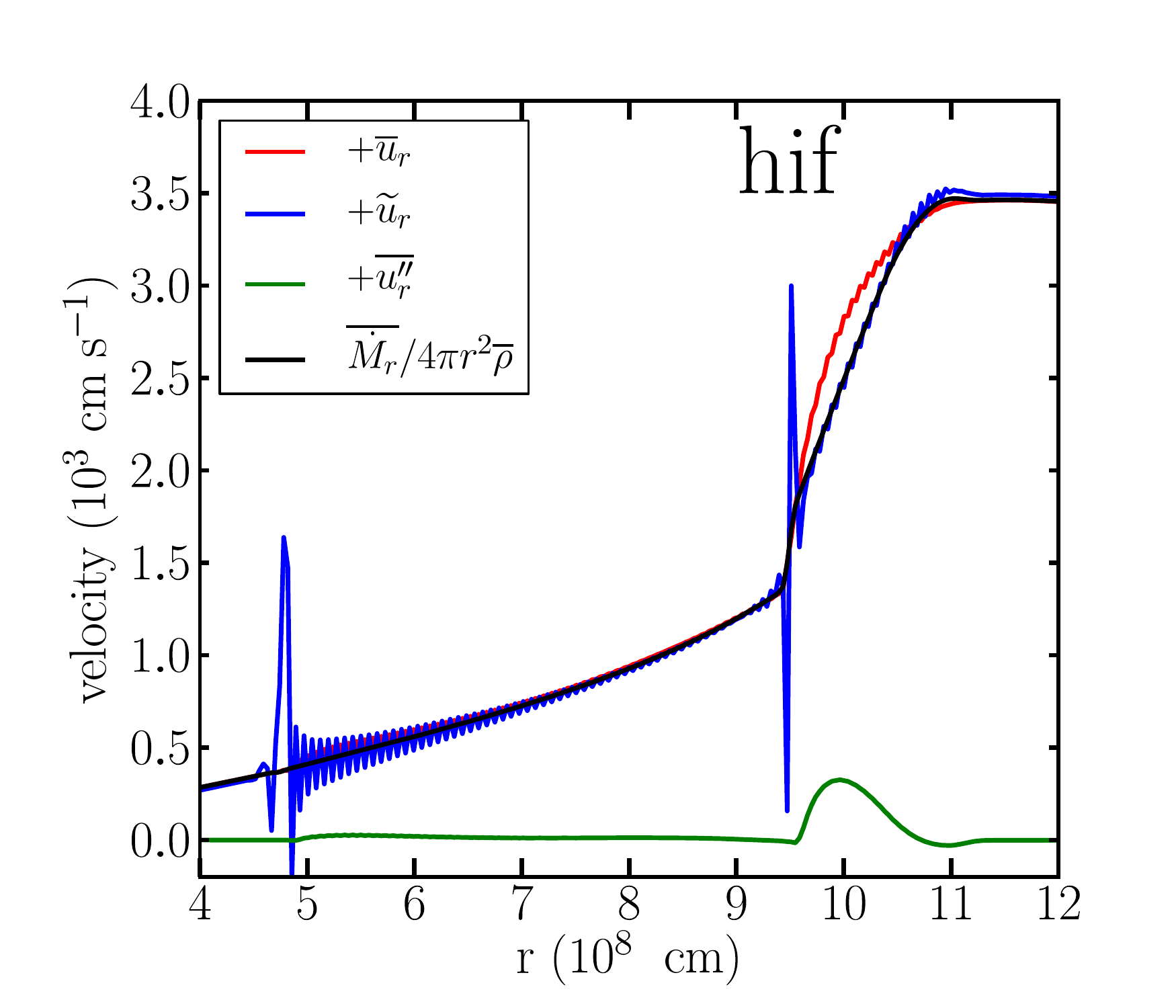}}

\centerline{
\includegraphics[width=7.cm]{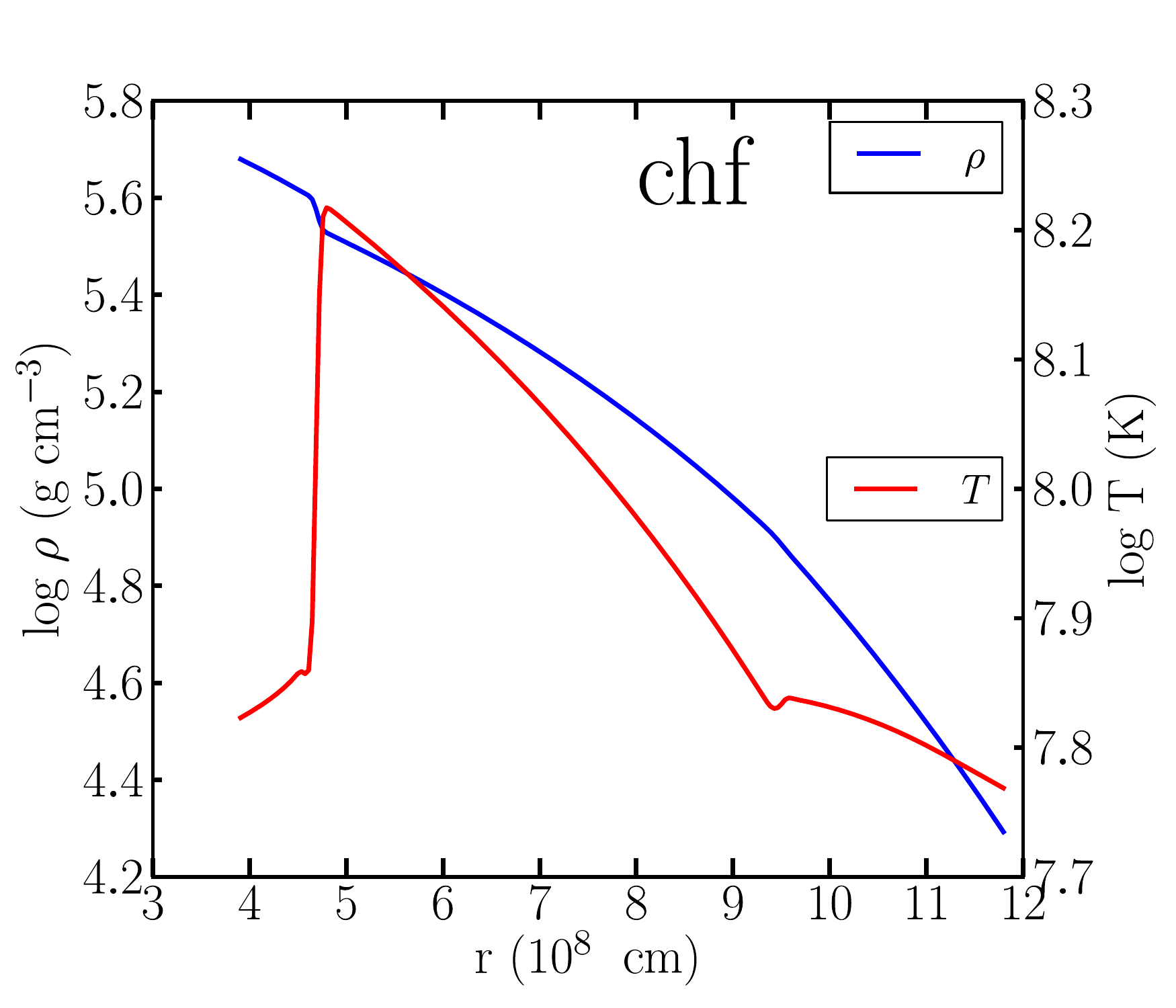}
\includegraphics[width=7.cm]{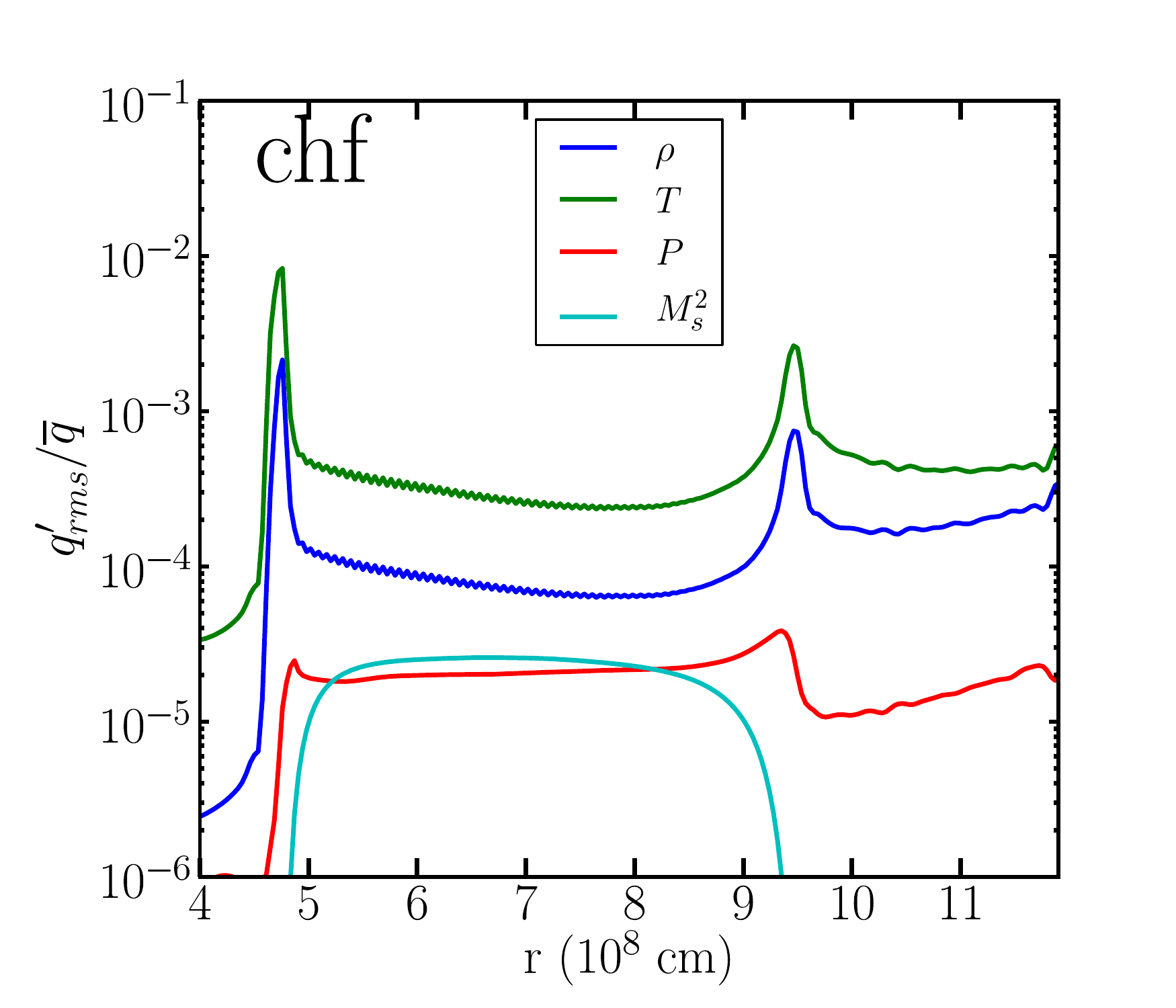}
\includegraphics[width=7.cm]{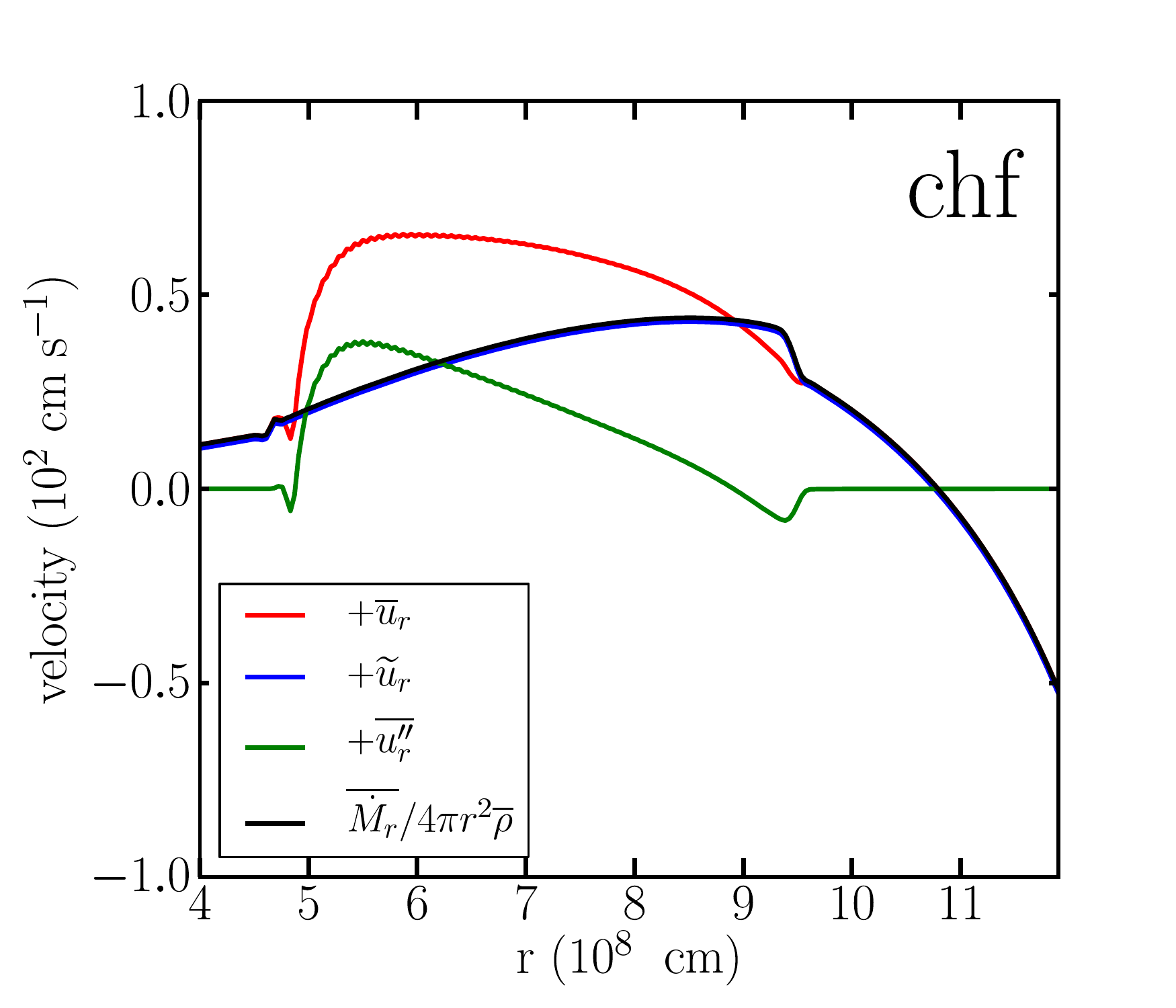}}
\caption{Properties of our data. Model {\sf hif.3D} (upper panels) and model {\sf chf.3D} (lower panels). \label{fig:data}}
\end{figure}

\newpage

\section{Profiles and integral budgets of mean fields}

\subsection{Mean continuity equation}

\begin{align}
\fav{D}_t \av{\rho} =& -\av{\rho} \fav{d} + {\mathcal N_\rho}  
\end{align}

\begin{figure}[!h]
\centerline{
\includegraphics[width=6.cm]{hif3d_tavg4000_initial_model_rho_t-eps-converted-to.pdf}
\includegraphics[width=6.cm]{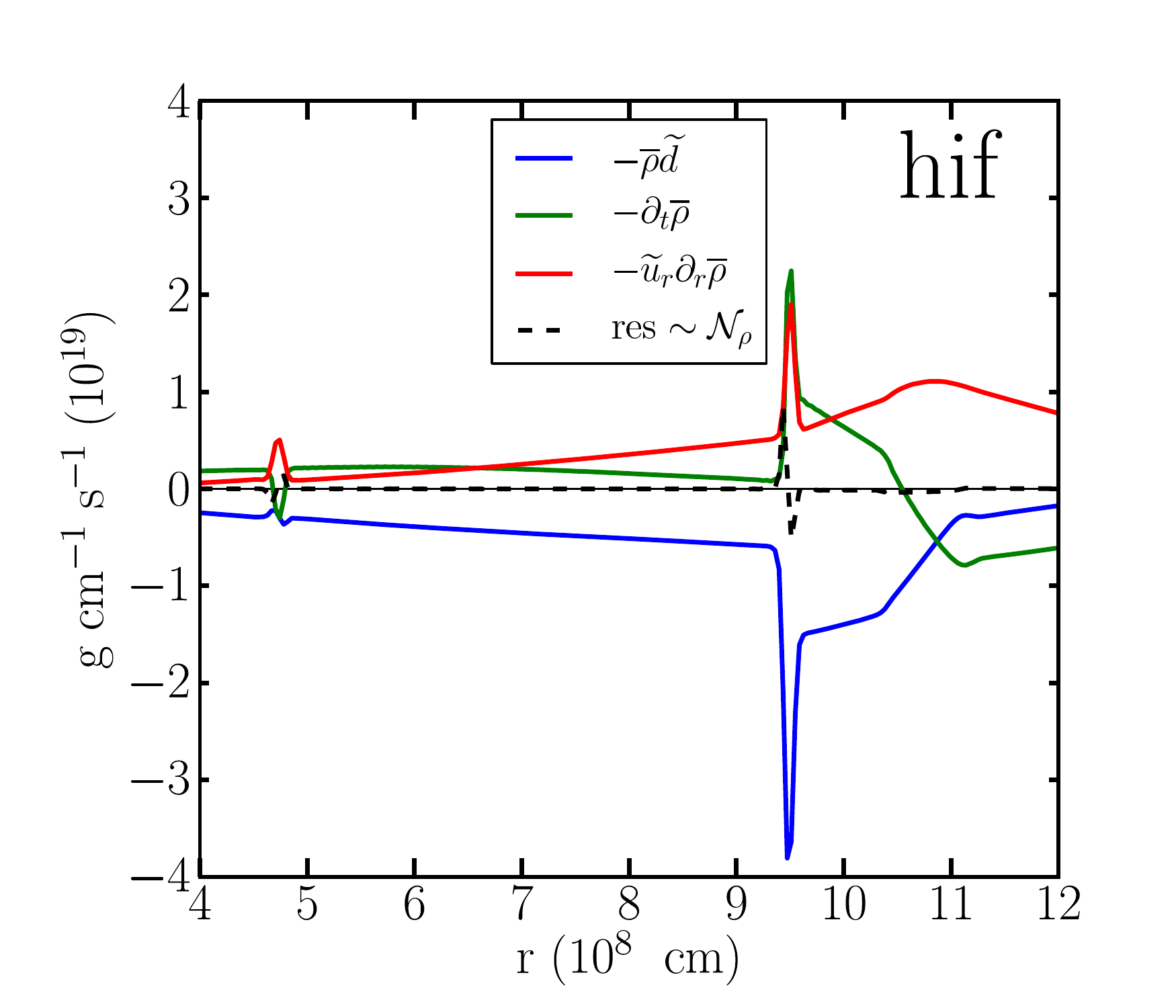}
\includegraphics[width=6.cm]{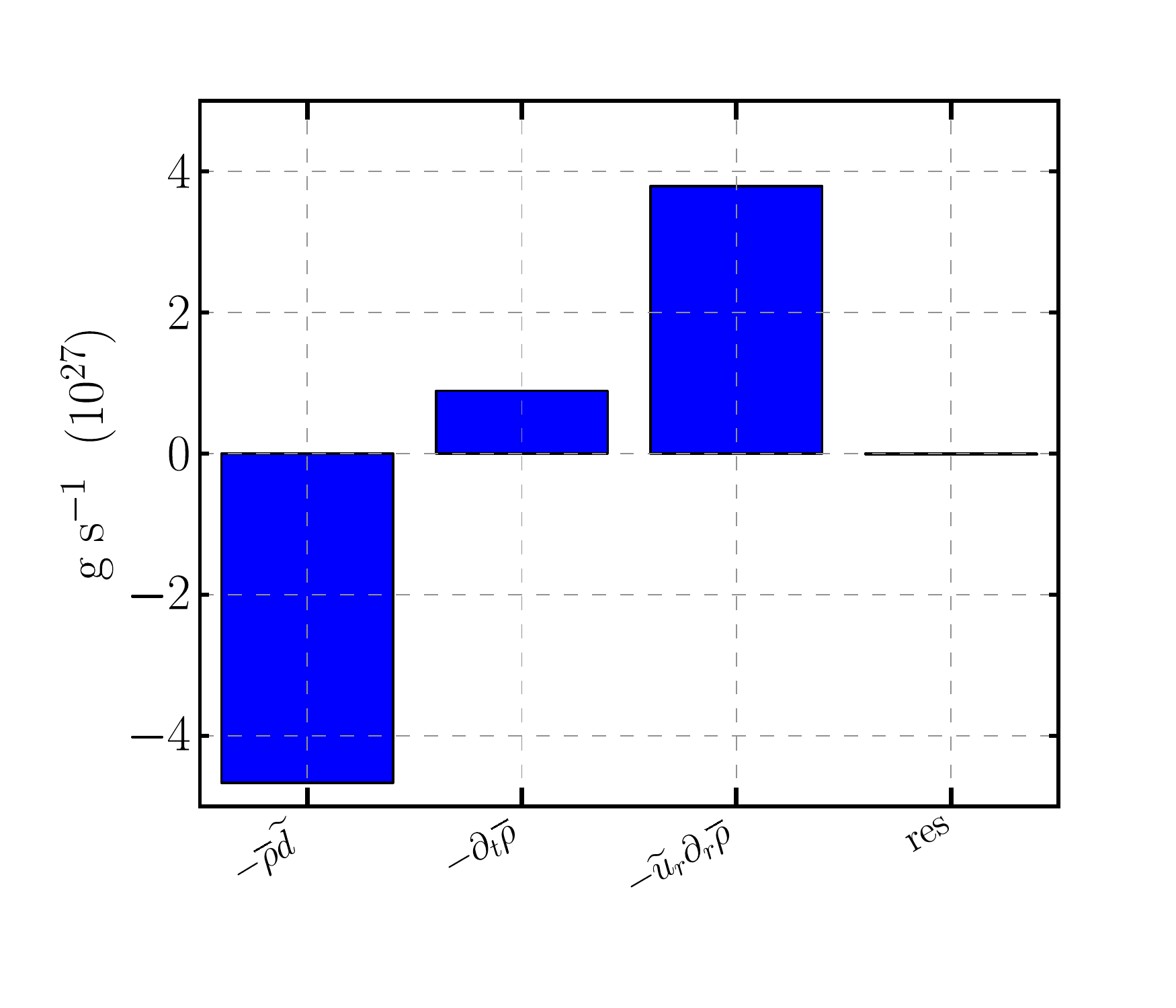}}

\centerline{
\includegraphics[width=6.cm]{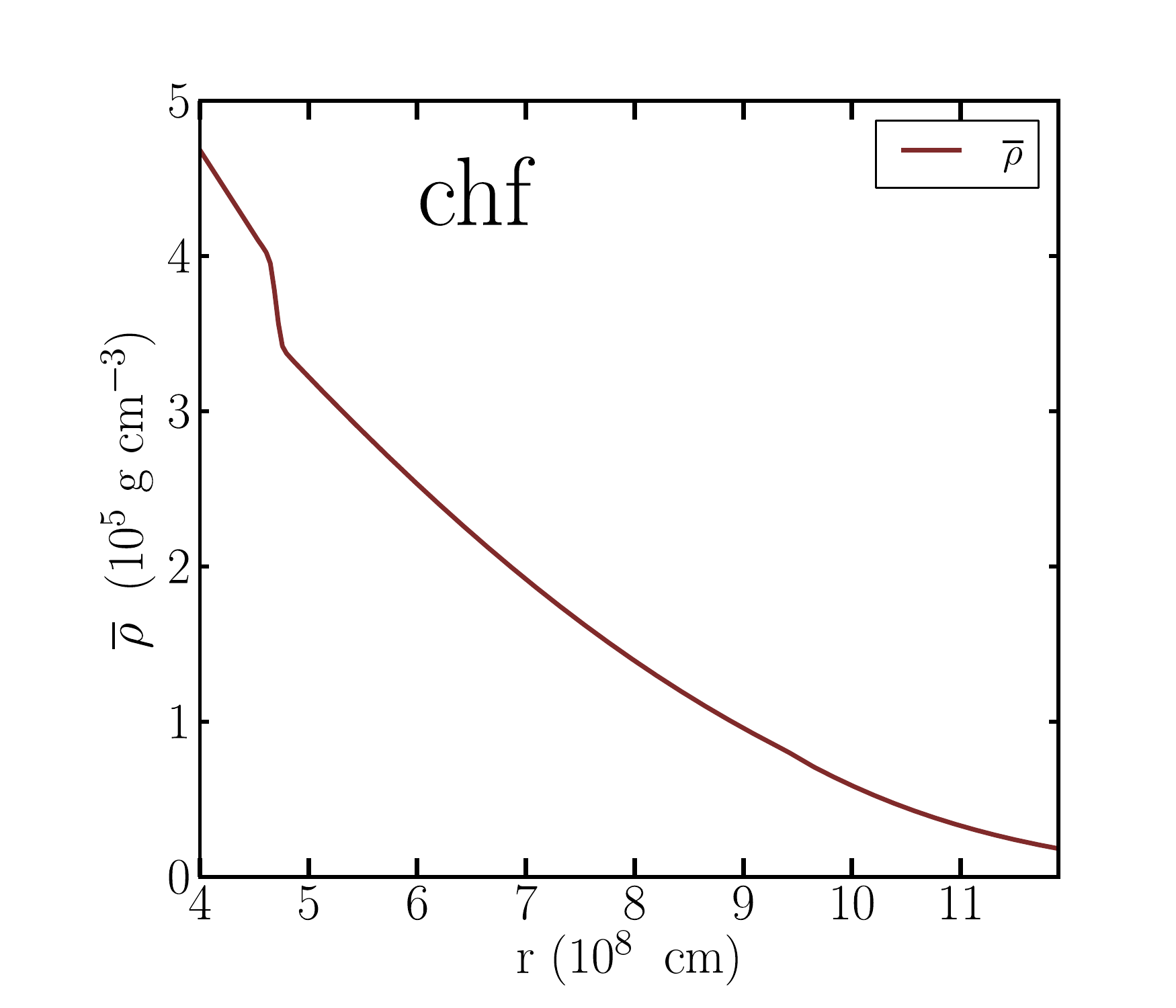}
\includegraphics[width=6.cm]{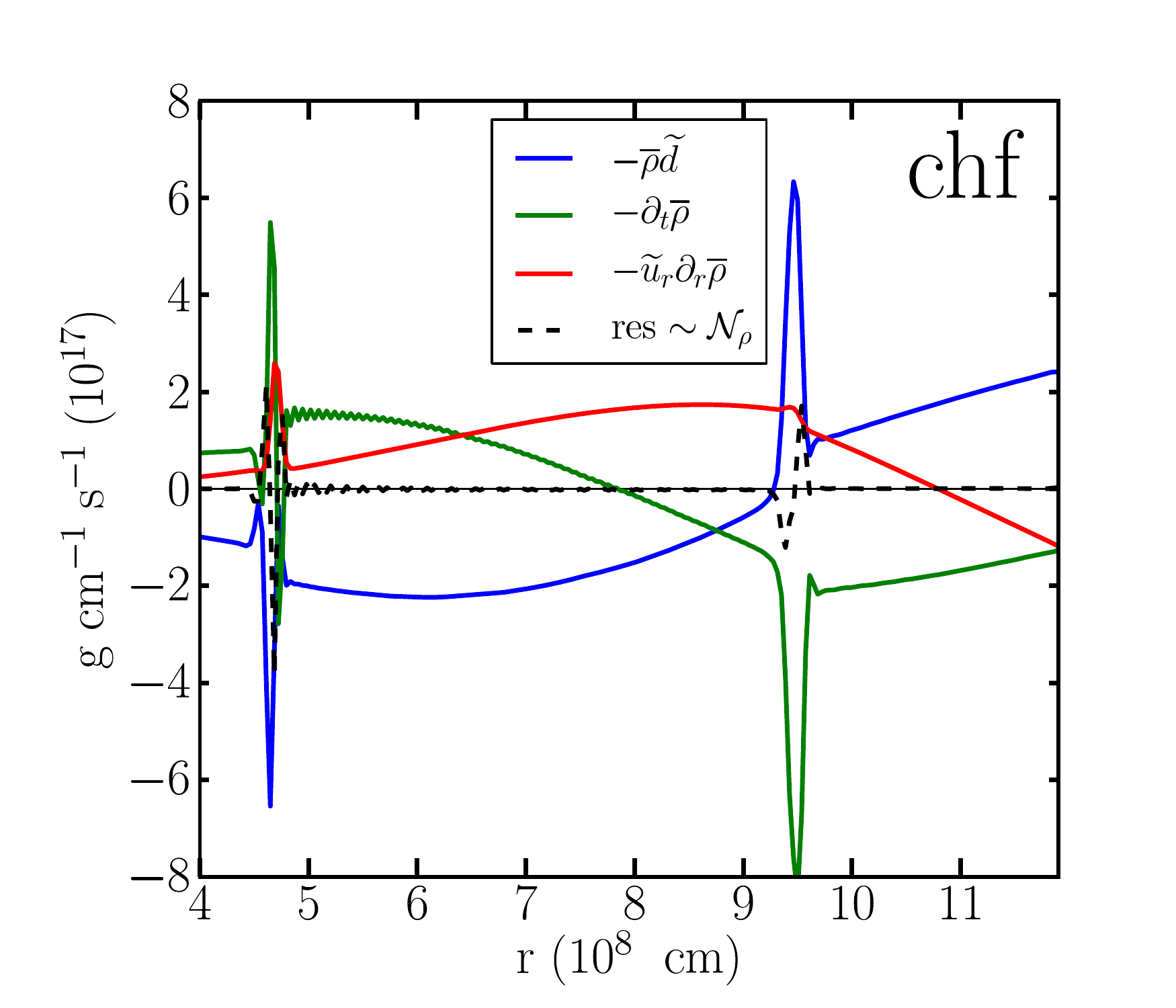}
\includegraphics[width=6.cm]{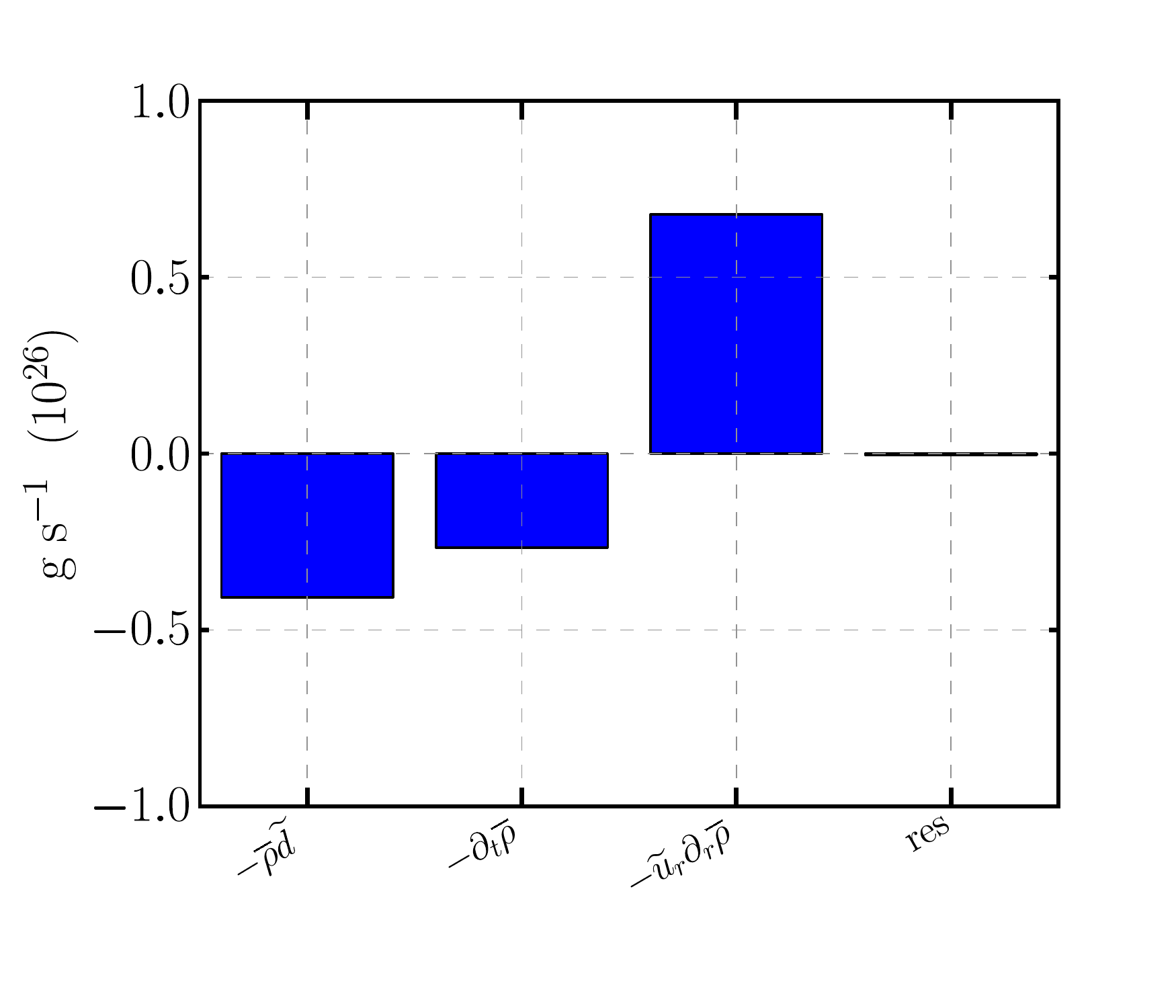}}
\caption{Mean continuity equation. Model {\sf hif.3D} (upper panels) and model {\sf chf.3D} (lower panels)}
\end{figure}

\newpage

\subsection{Mean radial momentum equation}

\begin{align}
\av{\rho}\fav{D}_t\fav{u}_r = & -\nabla_r \fav{R}_{rr} -\av{G^{M}_r} - \partial_r \av{P} + \av{\rho}\fav{g_r} + {\mathcal N_{ur}}
\end{align}

\begin{figure}[!h]
\centerline{
\includegraphics[width=6.5cm]{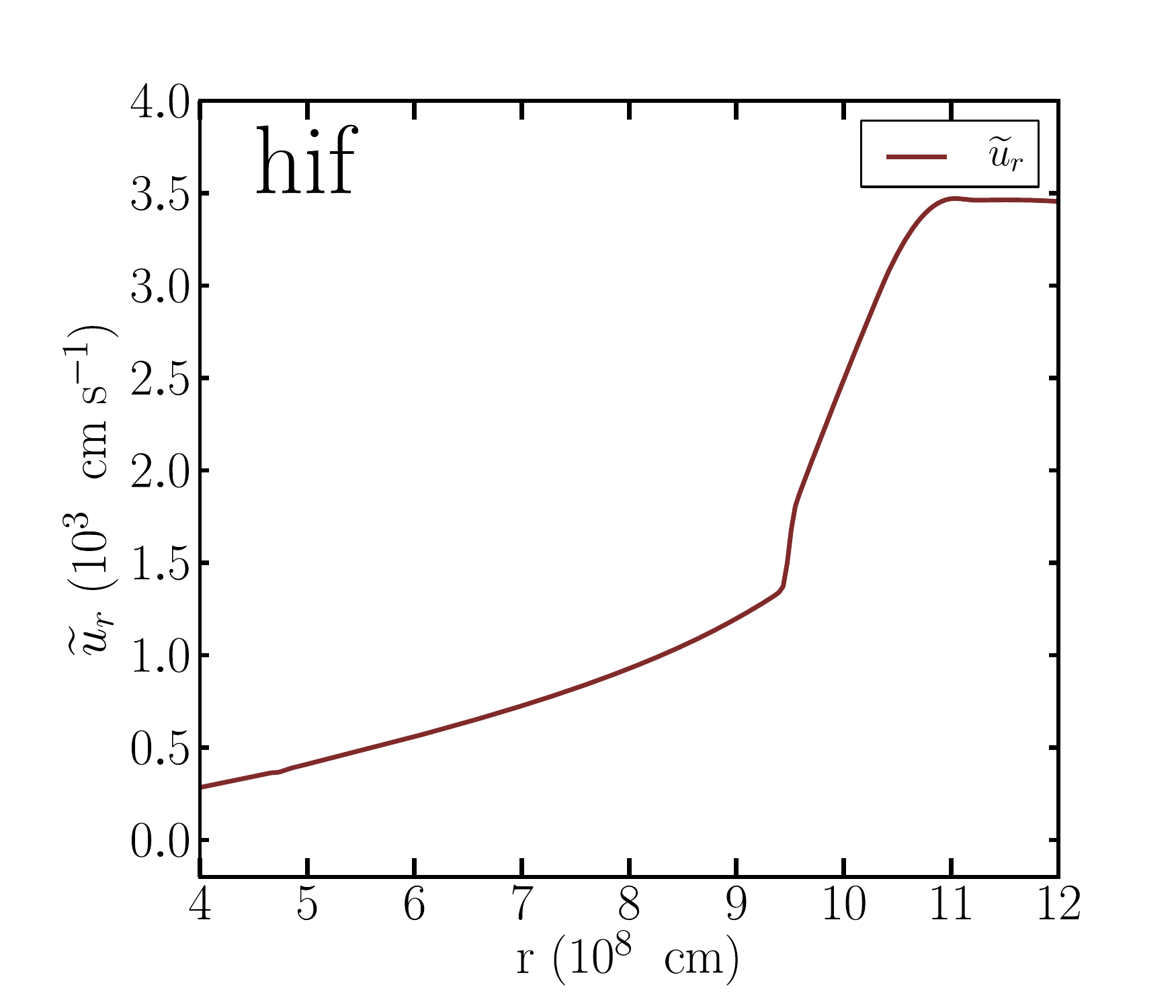}
\includegraphics[width=6.5cm]{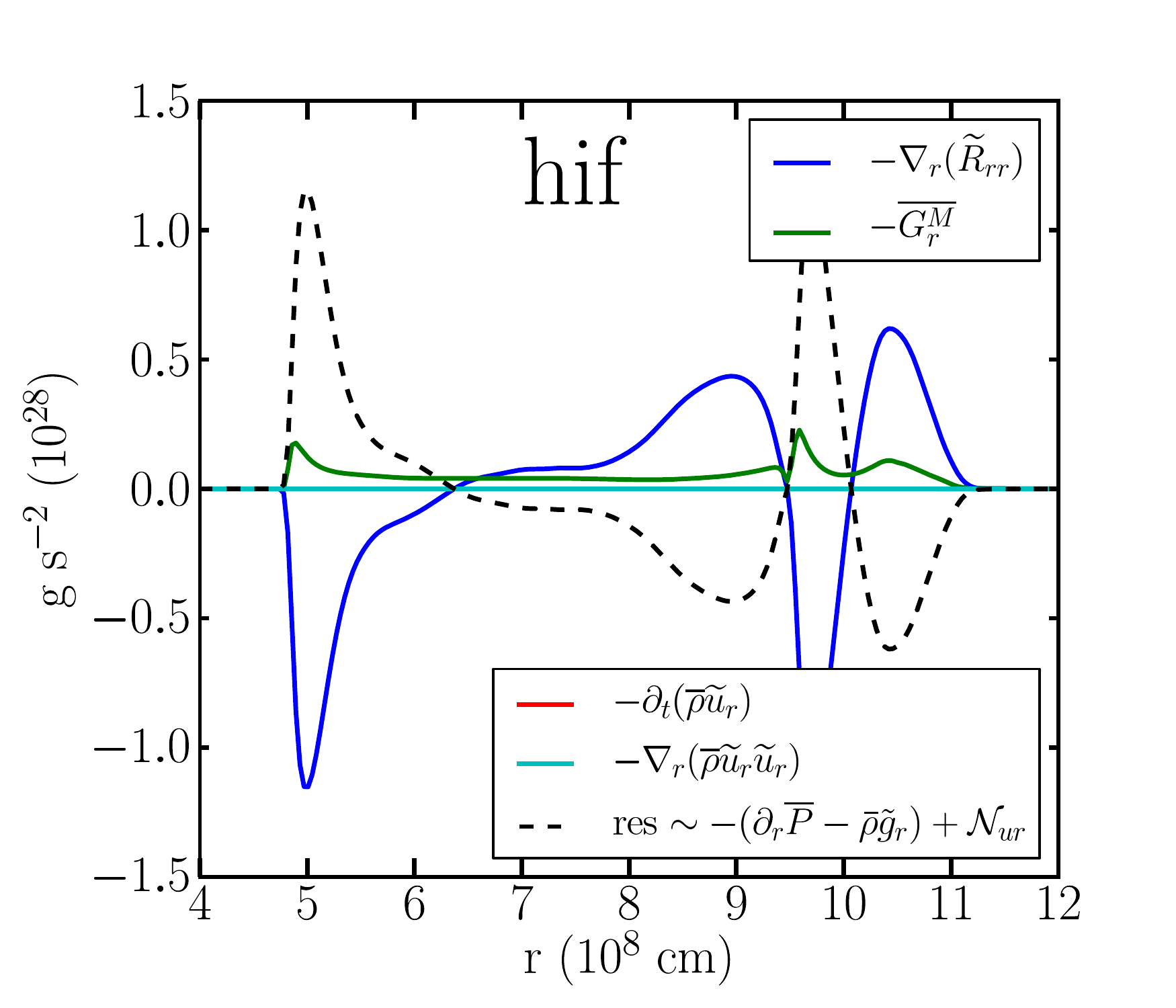}
\includegraphics[width=6.5cm]{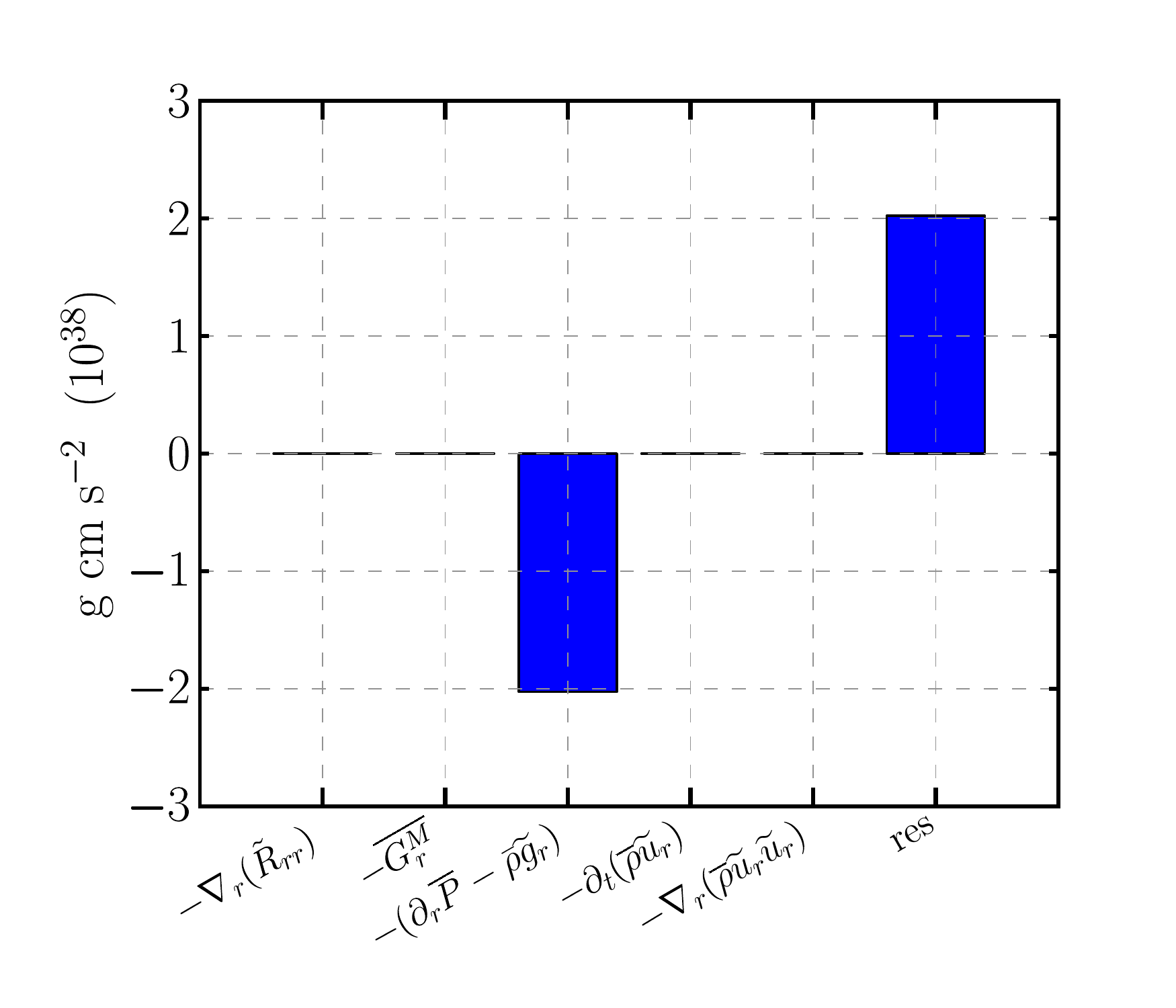}}

\centerline{
\includegraphics[width=6.5cm]{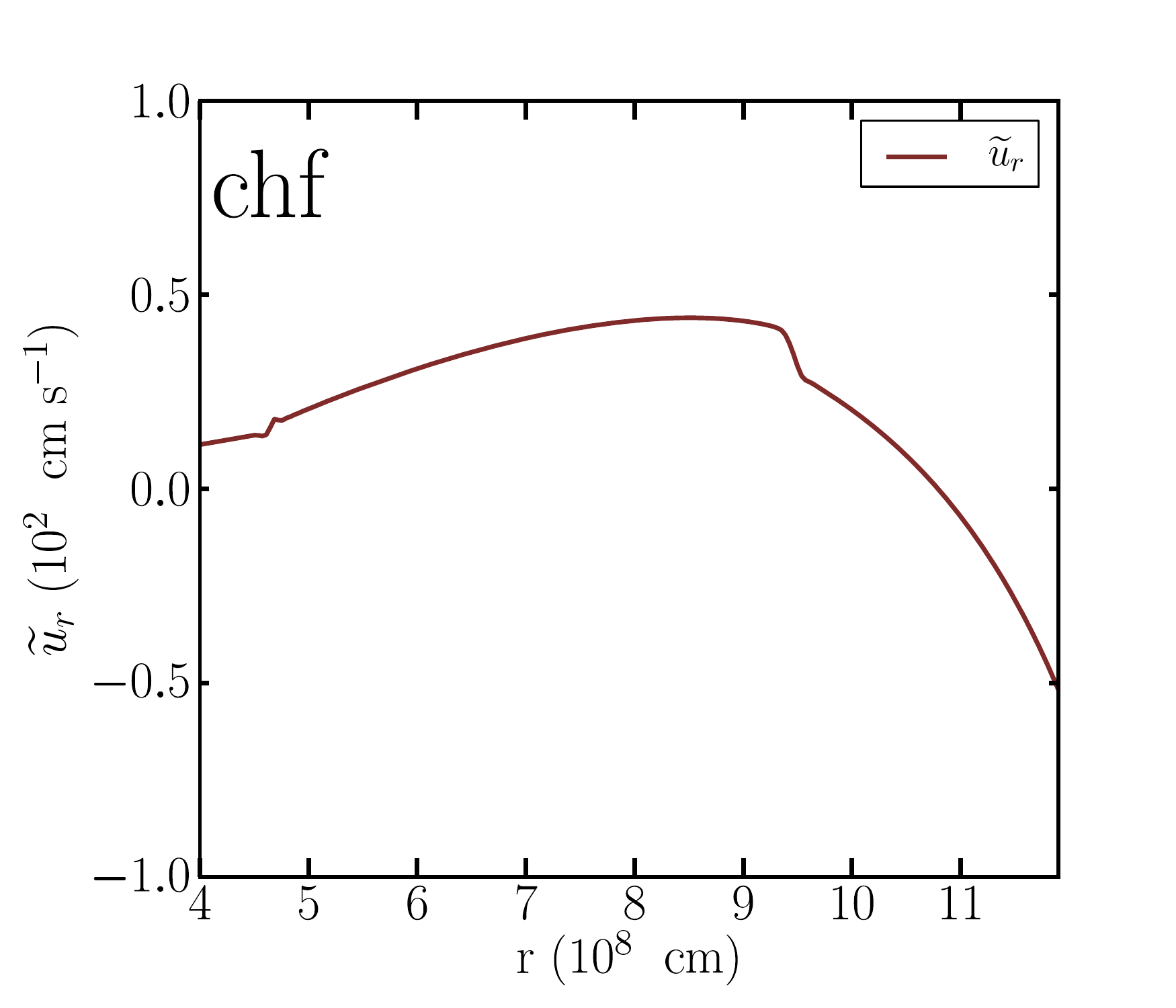}
\includegraphics[width=6.5cm]{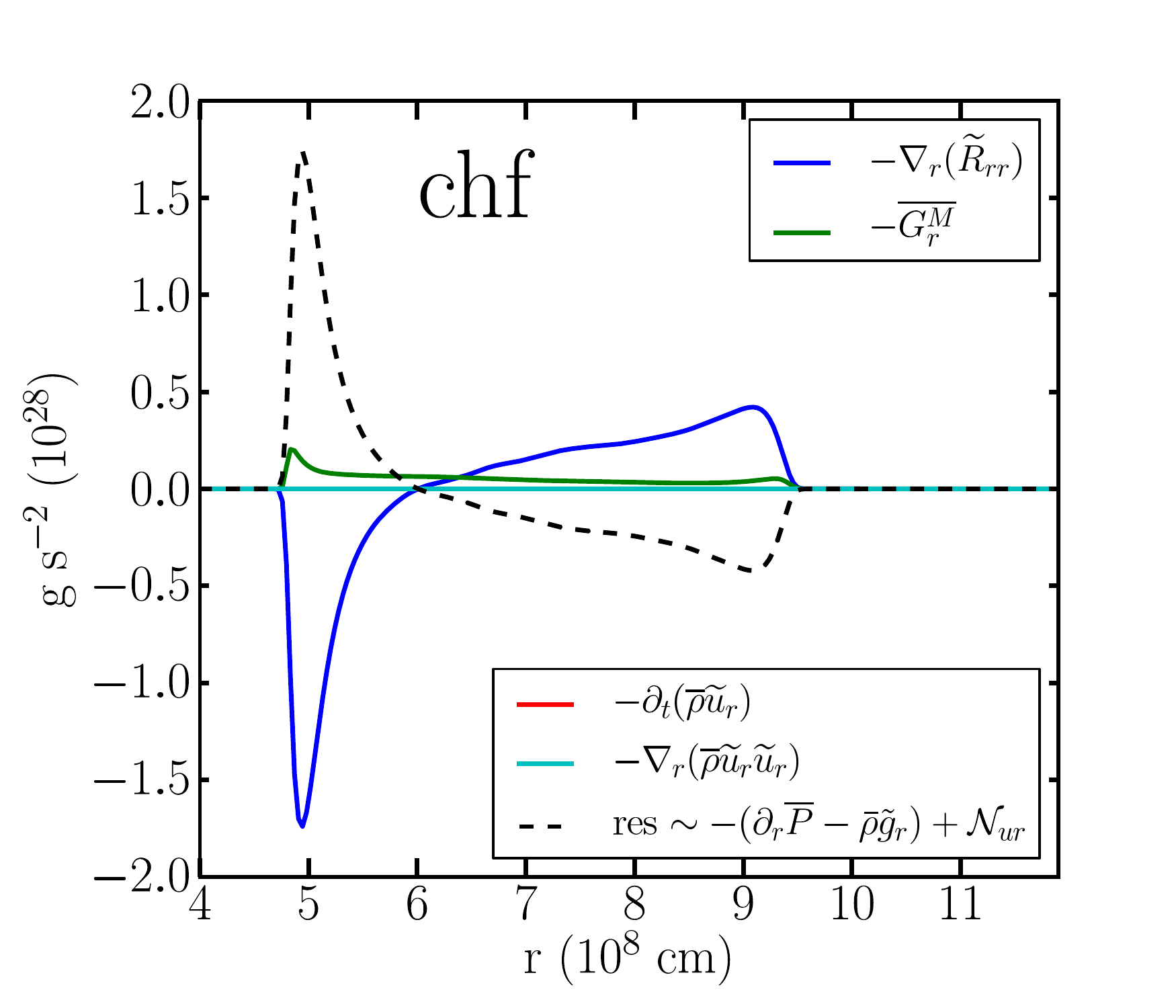}
\includegraphics[width=6.5cm]{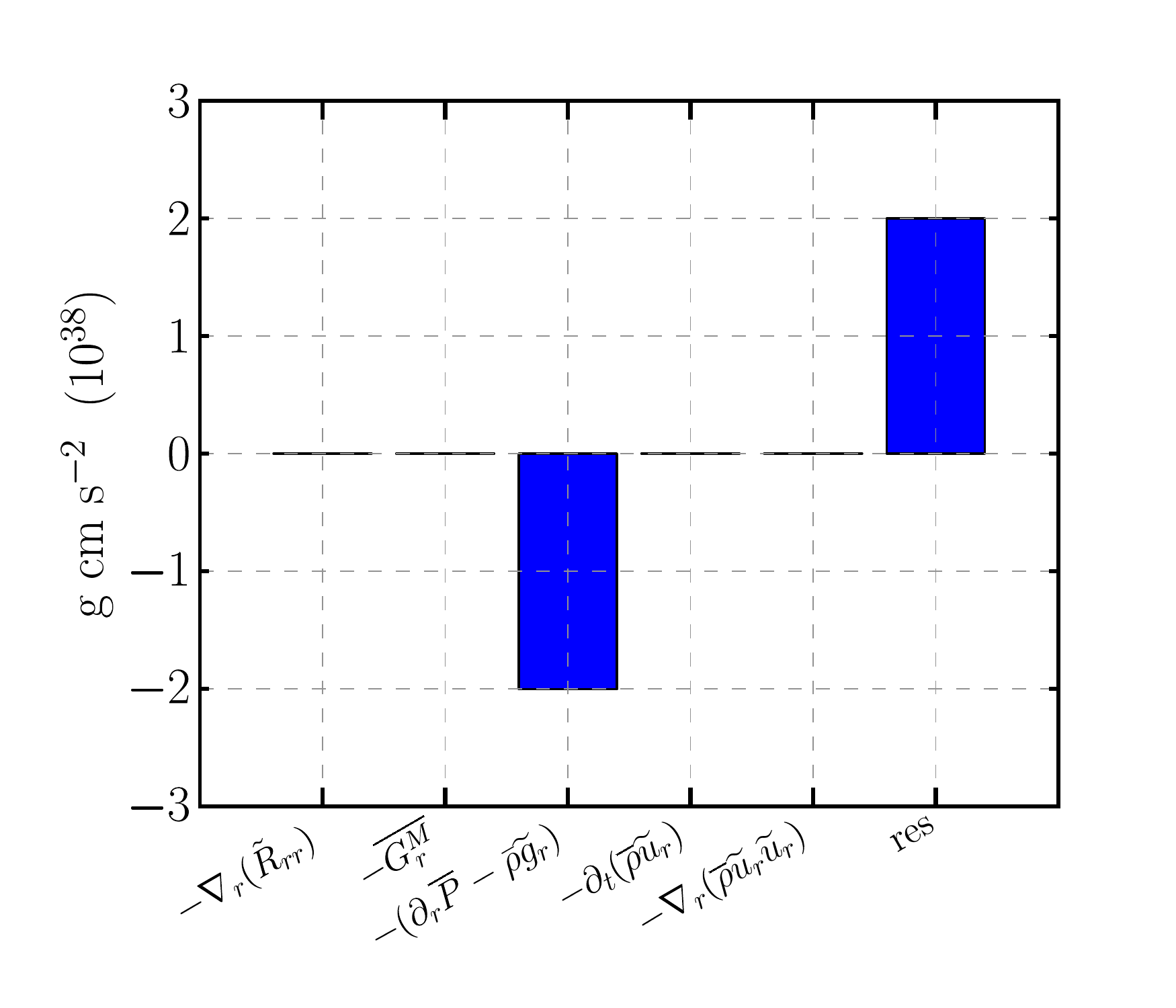}}
\caption{Mean radial momentum equation. Model {\sf hif.3D} (upper panels) and model {\sf chf.3D} (lower panels)}
\end{figure}

\newpage

\subsection{Mean azimuthal momentum equation}

\begin{align}
\av{\rho}\fav{D}_t\fav{u}_\theta = & -\nabla_r \fav{R}_{\theta r} -\av{G^{M}_\theta} - (1/r)\overline{\partial_\theta P} + {\mathcal N_{u \theta}} 
\end{align}

\begin{figure}[!h]
\centerline{
\includegraphics[width=6.5cm]{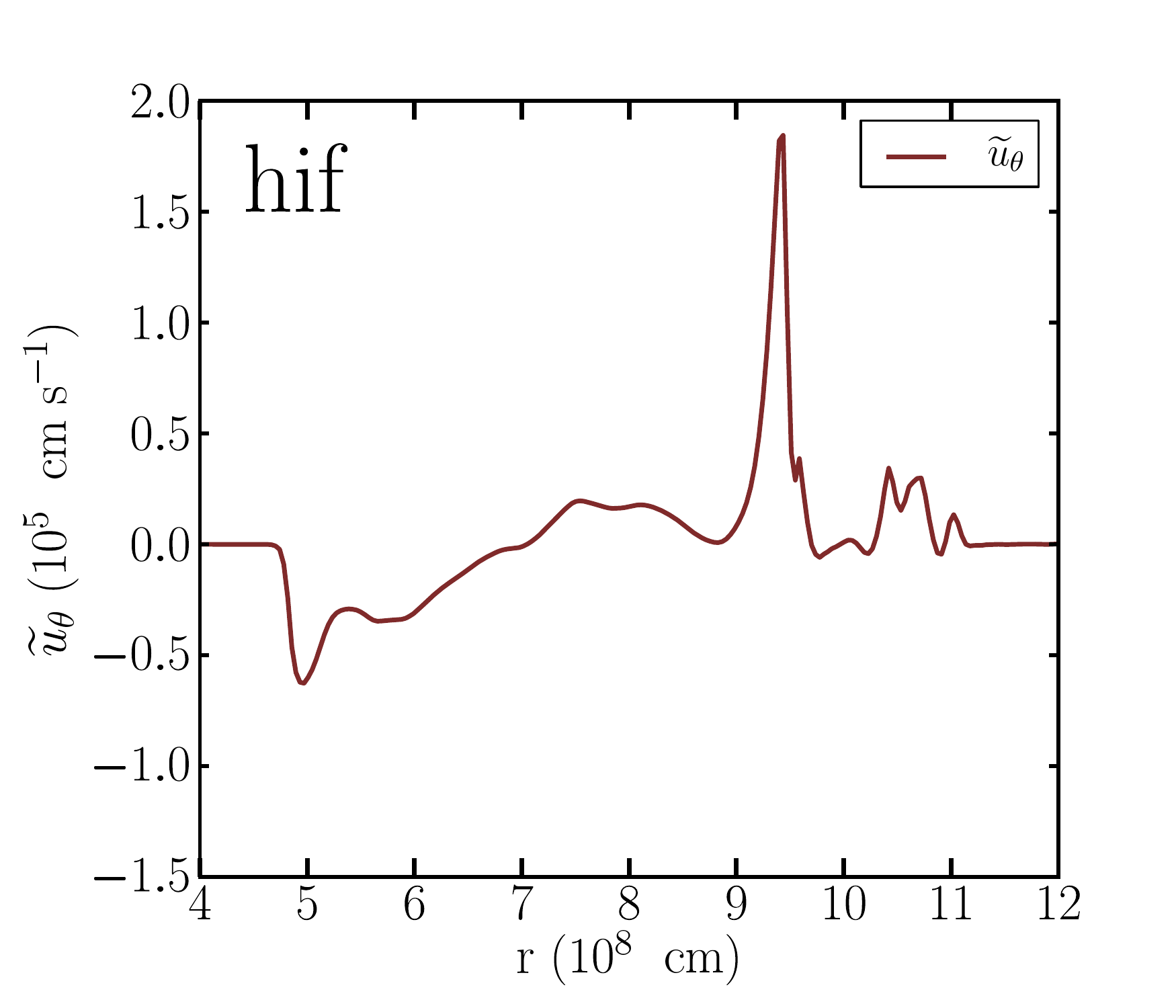}
\includegraphics[width=6.5cm]{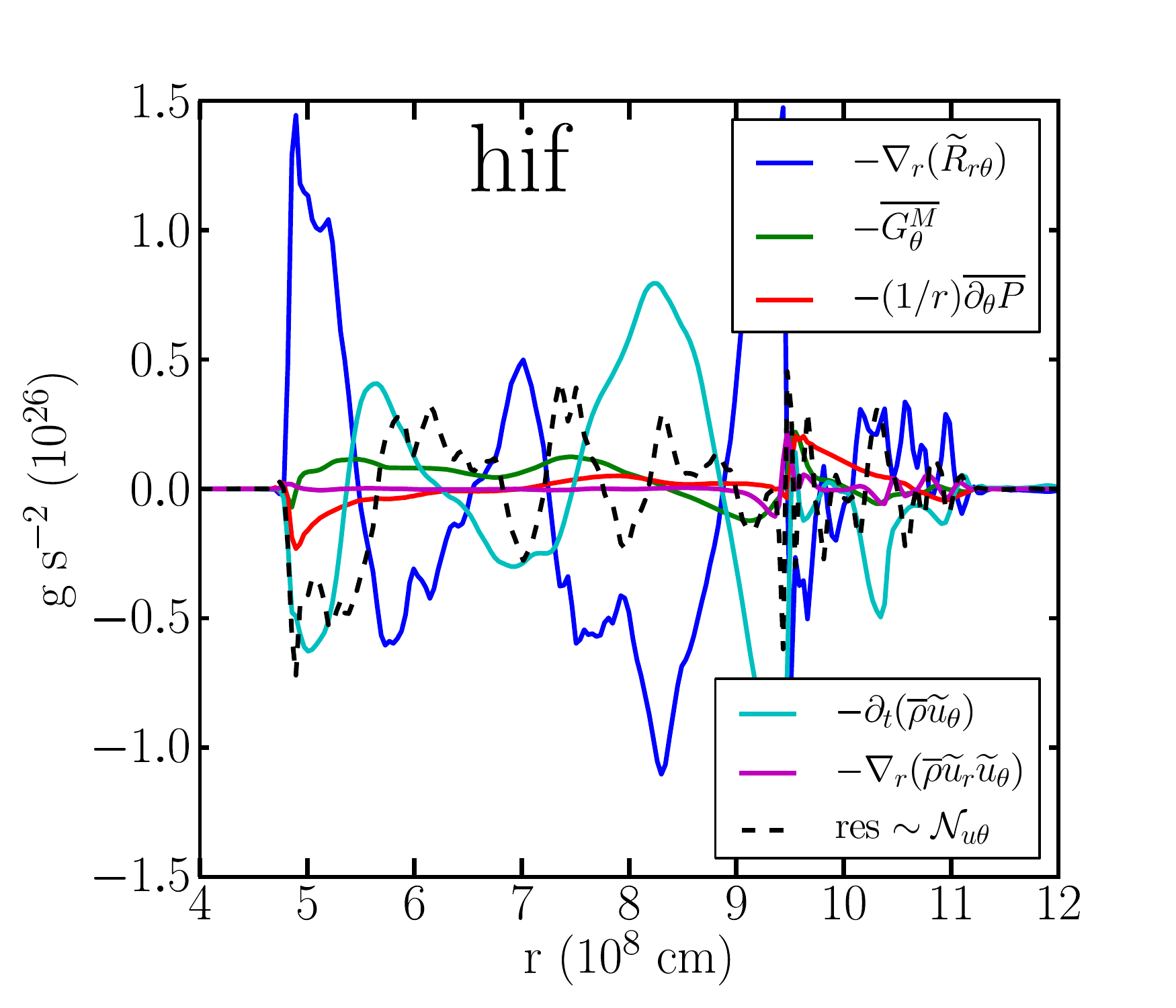}
\includegraphics[width=6.5cm]{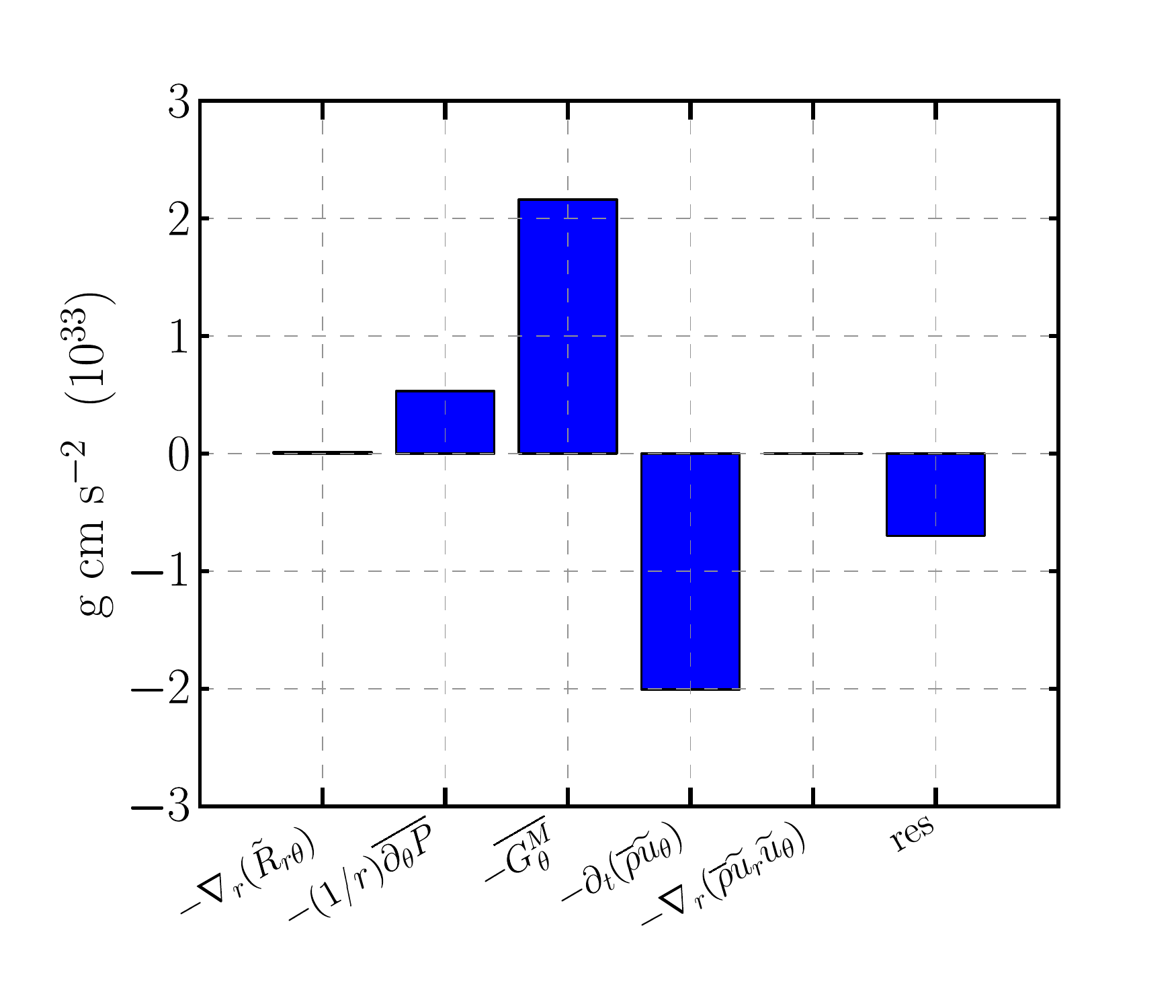}}

\centerline{
\includegraphics[width=6.5cm]{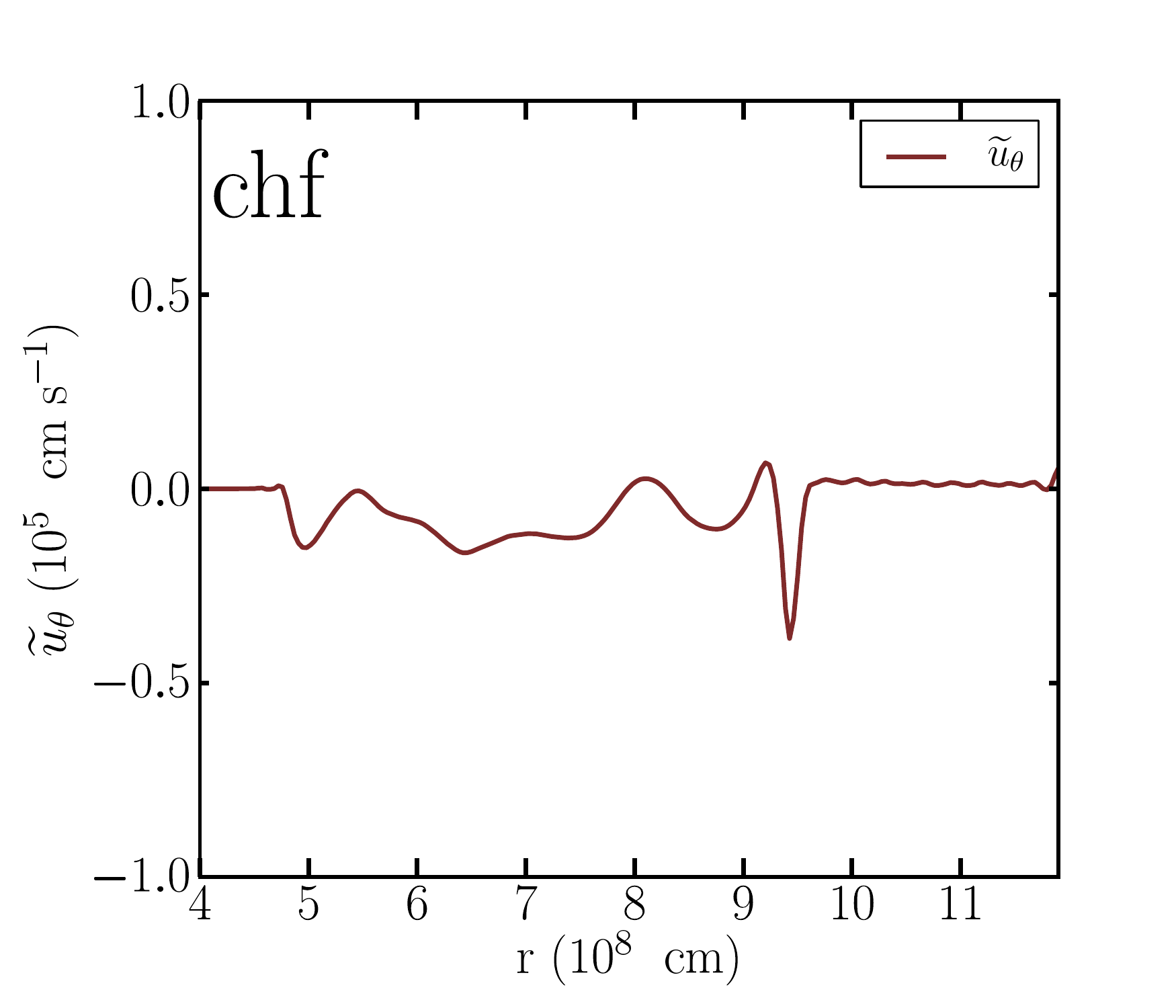}
\includegraphics[width=6.5cm]{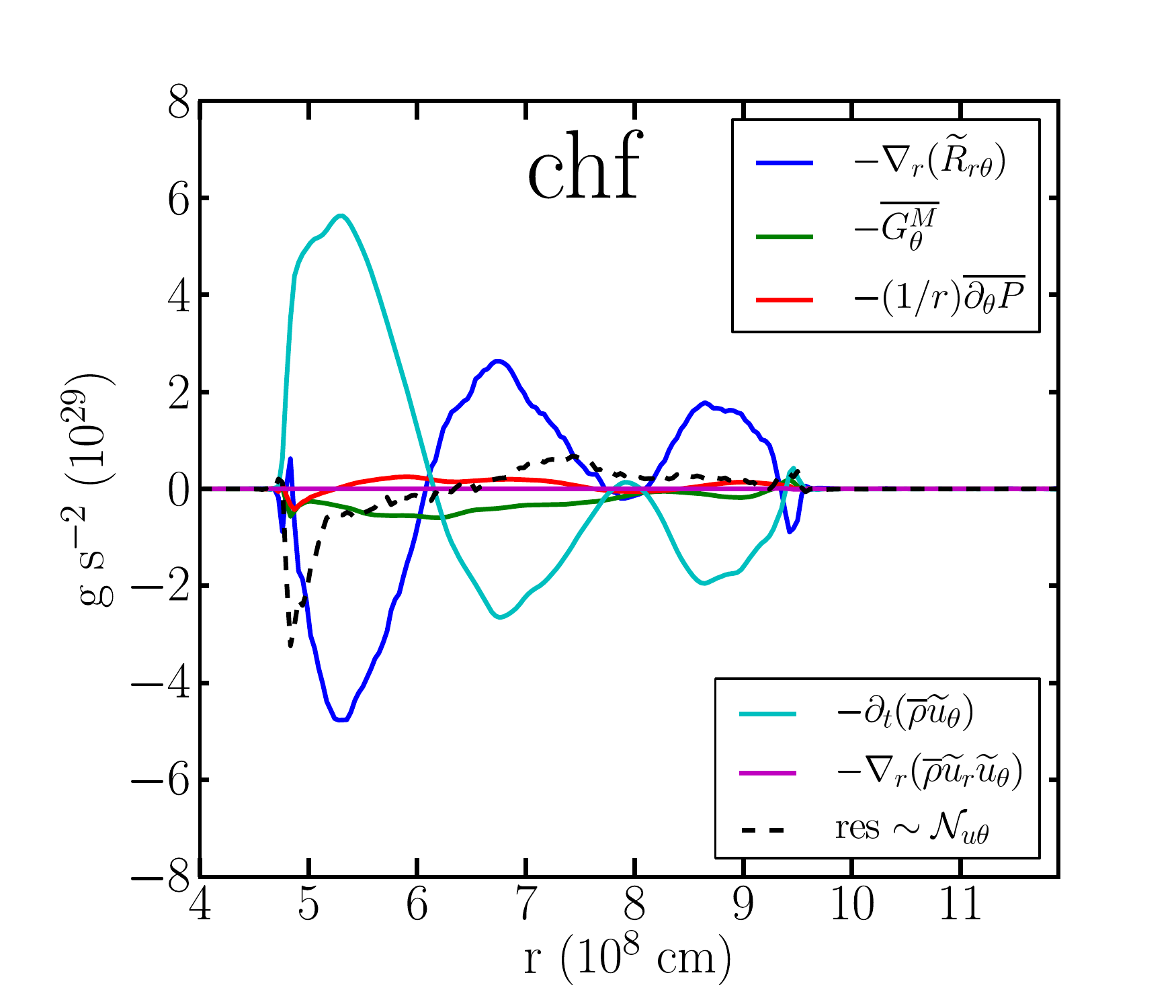}
\includegraphics[width=6.5cm]{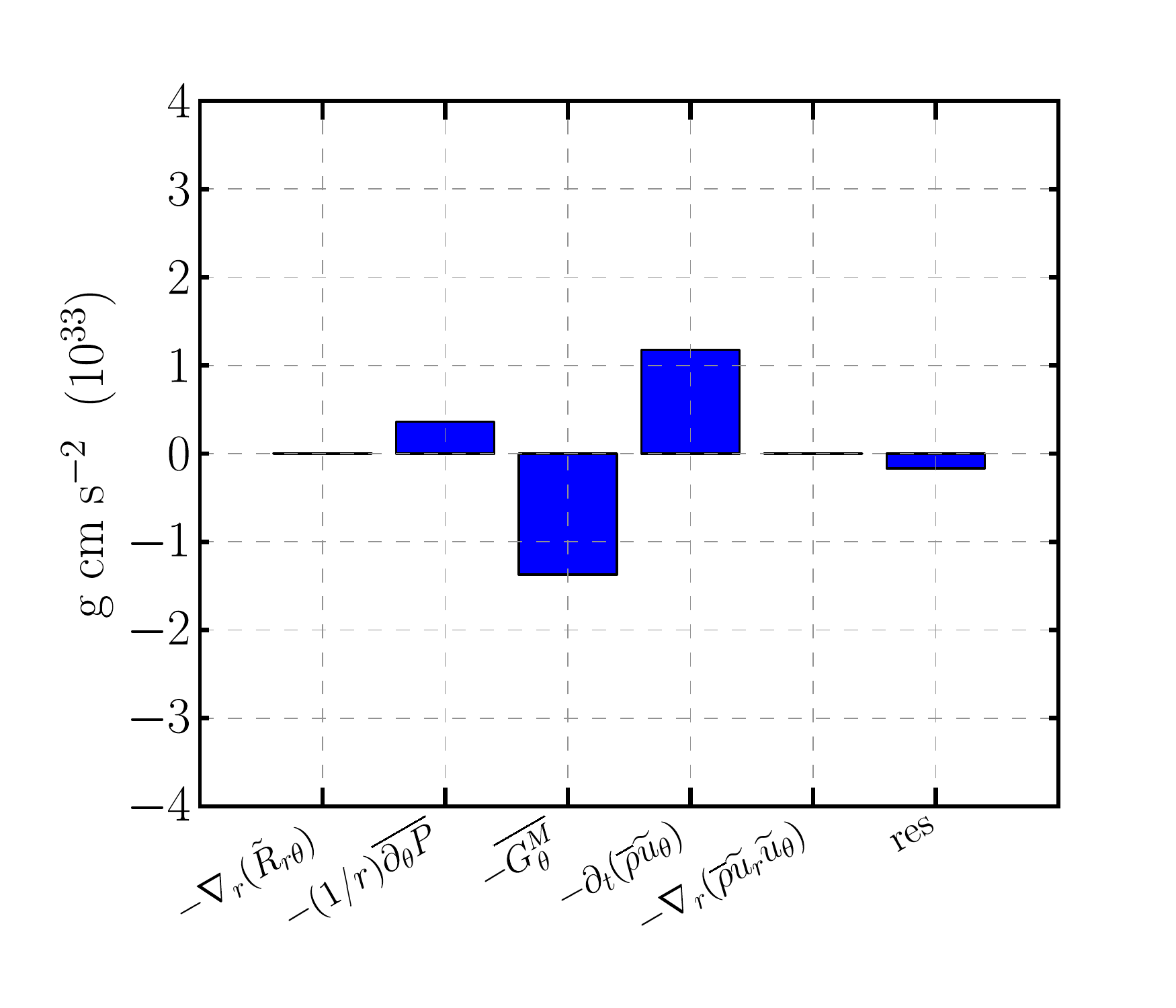}}

\caption{Mean azimuthal momentum equation. Model {\sf hif.3D} (upper panels) and model {\sf chf.3D} (lower panels)}
\end{figure}

\newpage

\subsection{Mean polar momentum equation}

\begin{align}
\av{\rho}\fav{D}_t\fav{u}_\phi = & -\nabla_r \fav{R}_{\phi r} -\av{G^{M}_\phi} + {\mathcal N_{u \phi}}
\end{align}

\begin{figure}[!h]
\centerline{
\includegraphics[width=6.5cm]{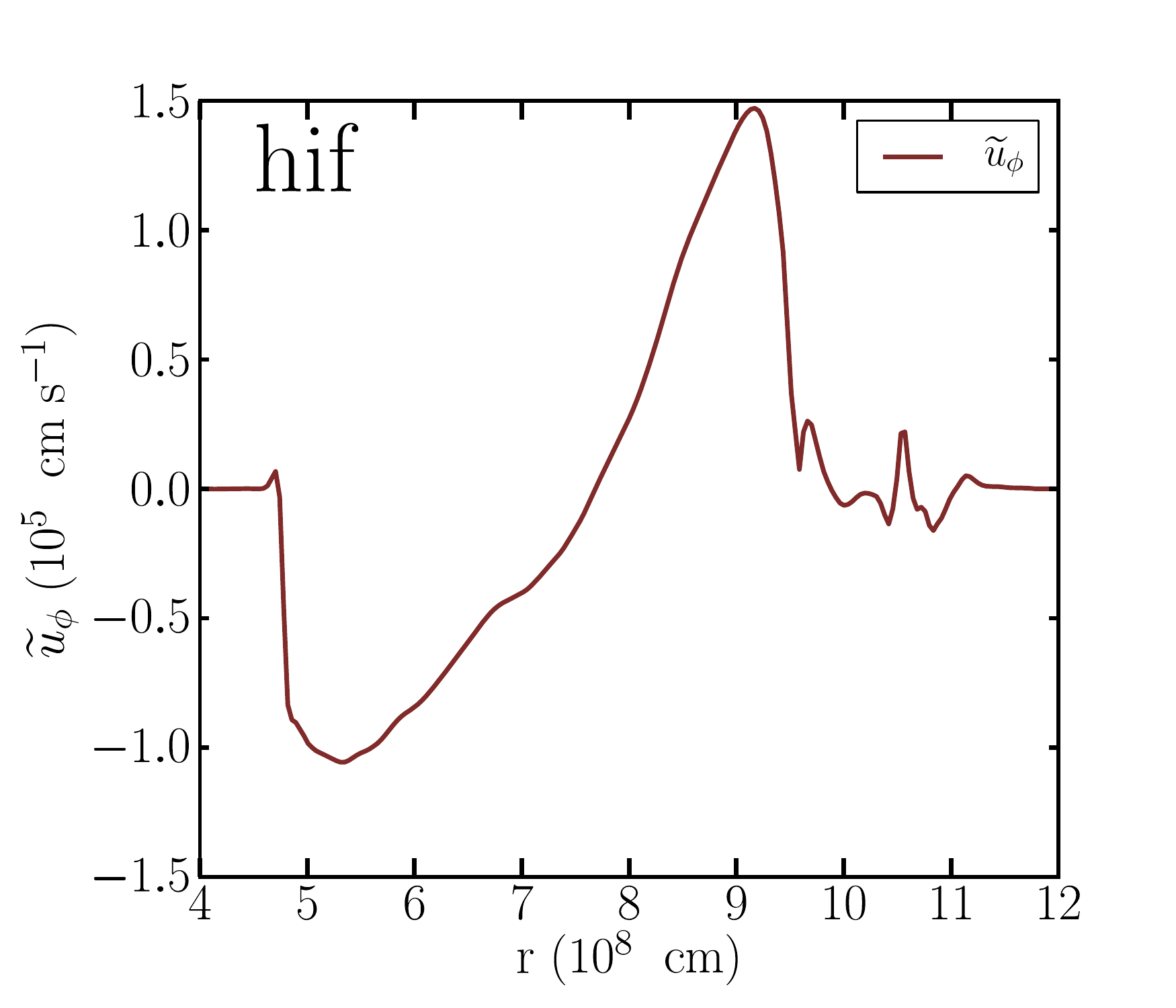}
\includegraphics[width=6.5cm]{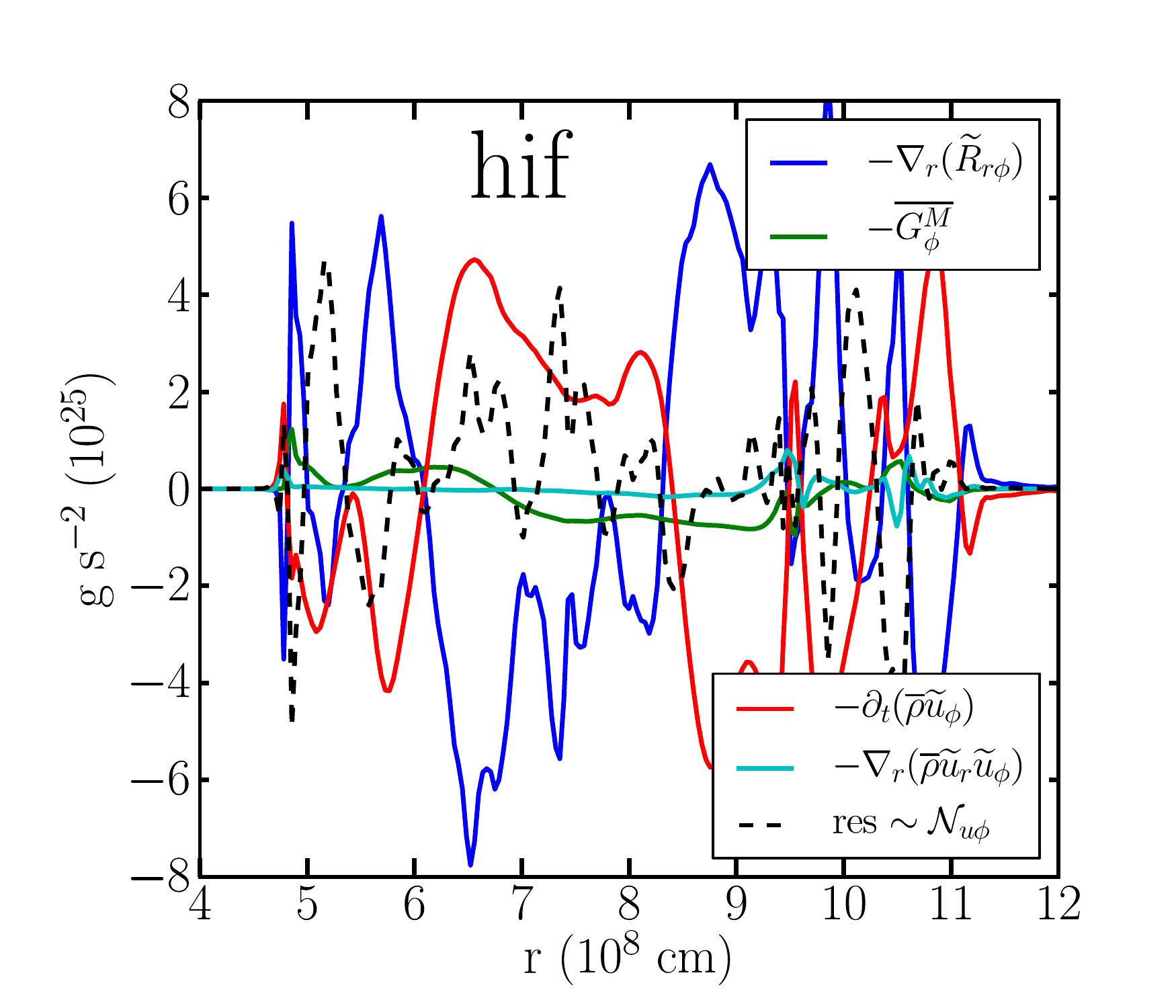}
\includegraphics[width=6.5cm]{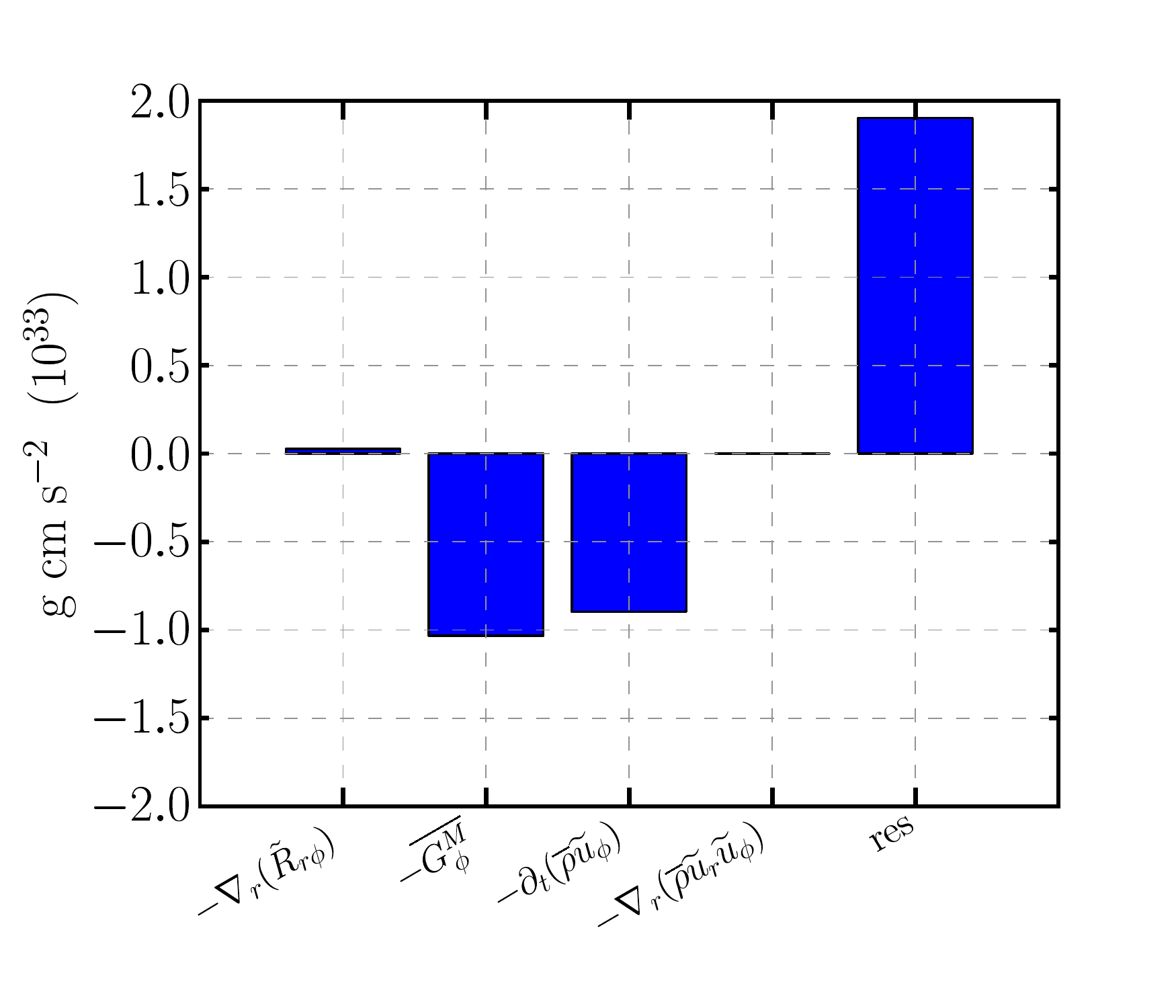}}

\centerline{
\includegraphics[width=6.5cm]{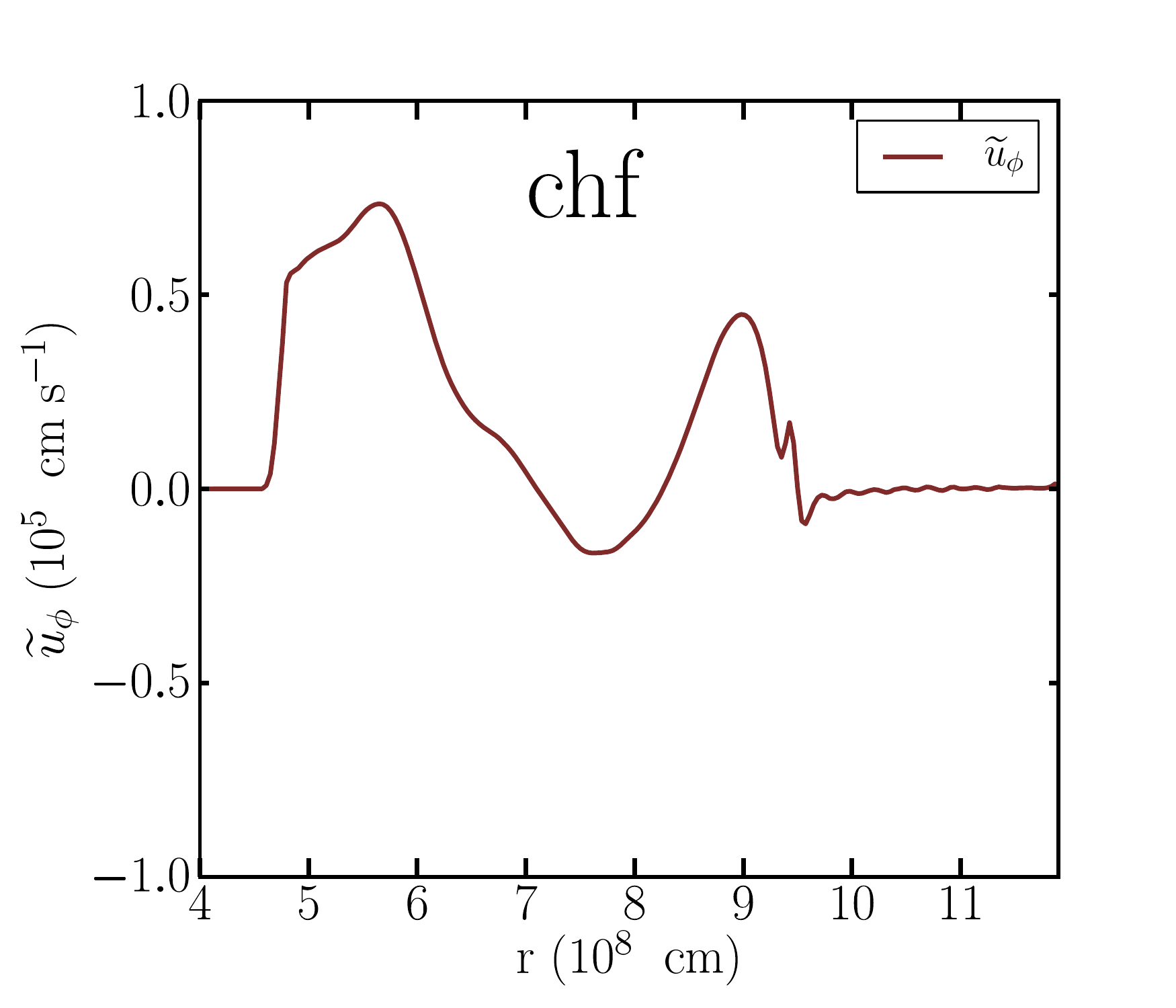}
\includegraphics[width=6.5cm]{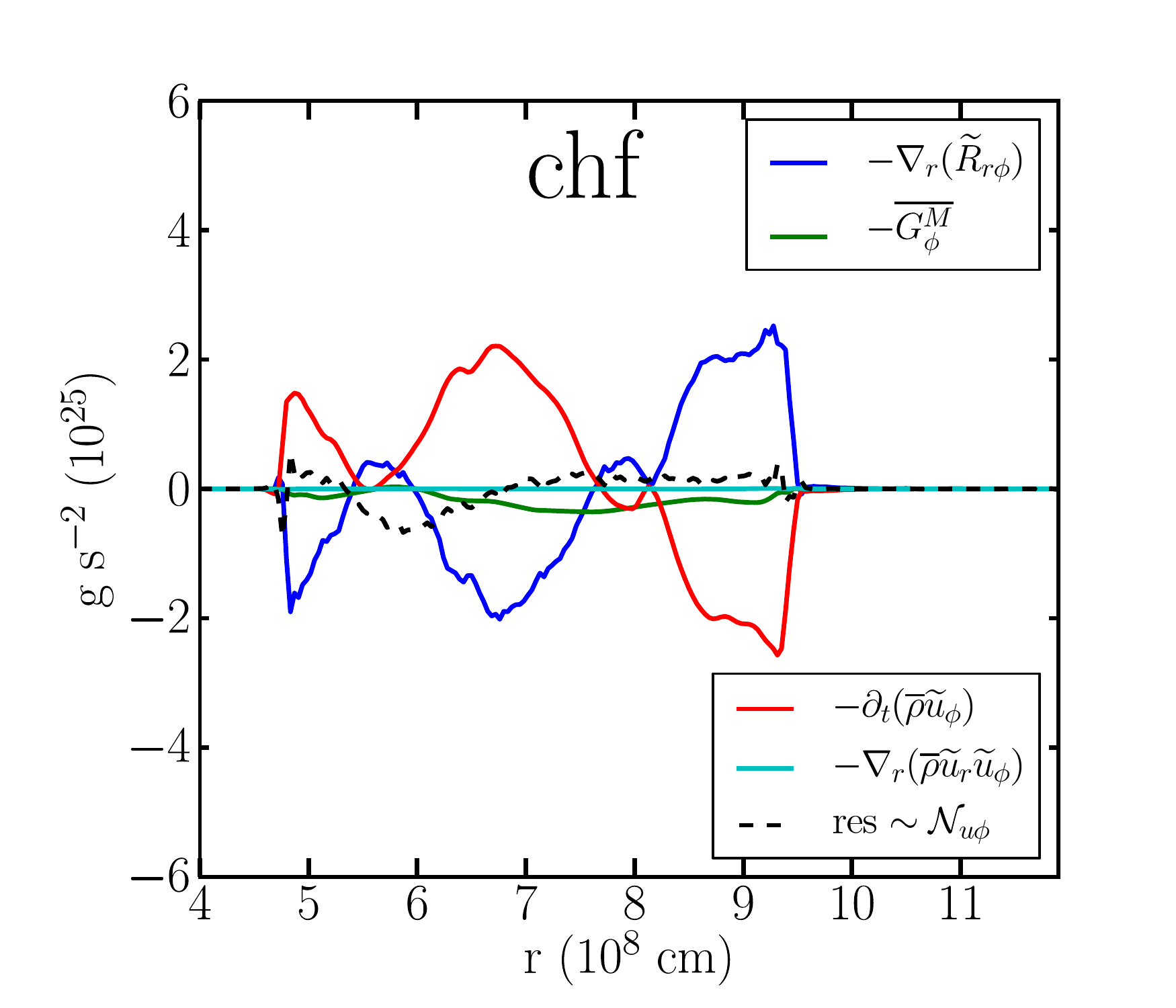}
\includegraphics[width=6.5cm]{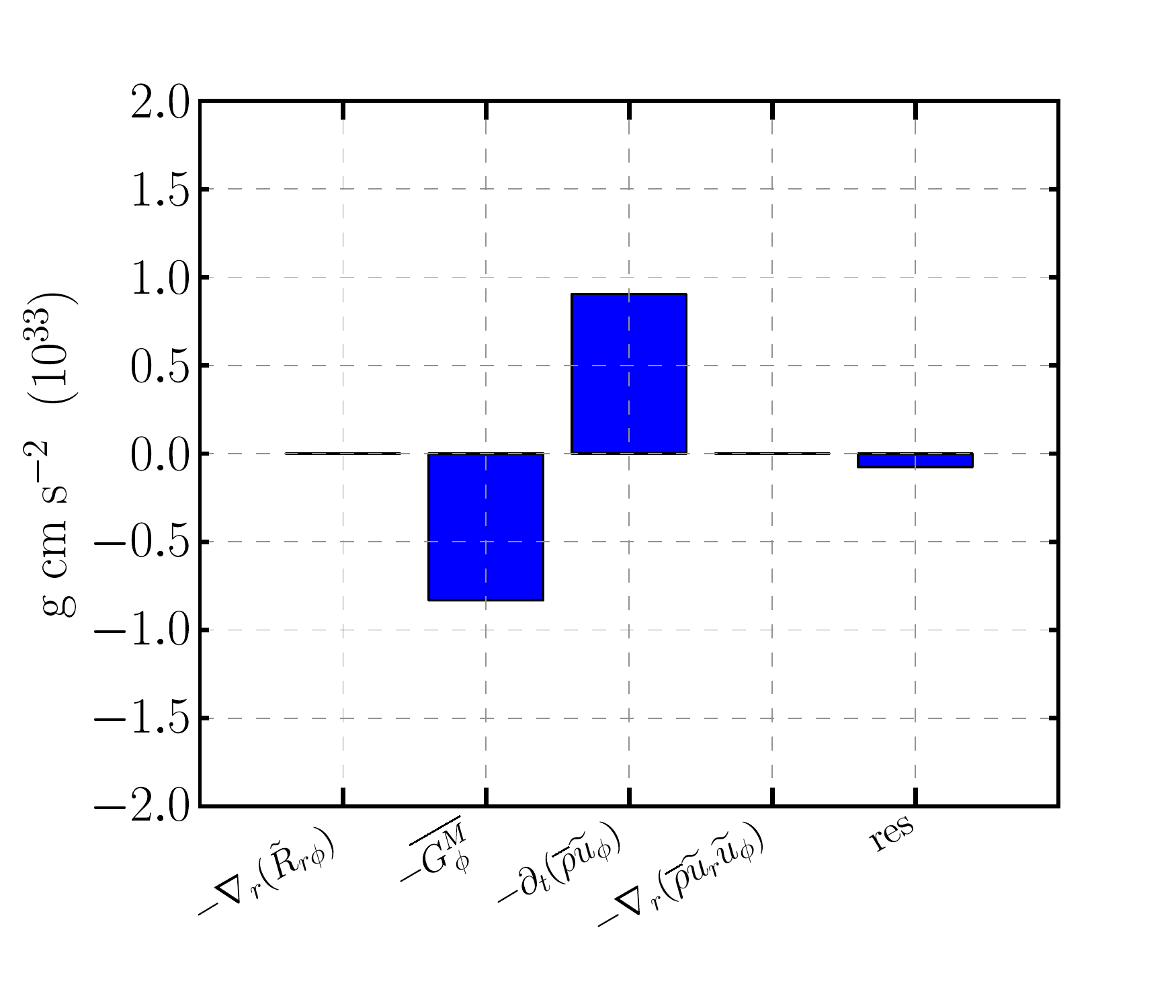}}

\caption{Mean polar momentum equation. Model {\sf hif.3D} (upper panels) and model {\sf chf.3D} (lower panels)}
\end{figure}

\newpage
 
\subsection{Mean internal energy equation}

\begin{align}
\av{\rho} \fav{D}_t \fav{\epsilon}_I = & - \nabla_r  ( f_I + f_T ) - \av{P} \ \av{d} - W_P  + {\mathcal S} + {\mathcal N_{\epsilon I}} 
\end{align}

\begin{figure}[!h]
\centerline{
\includegraphics[width=6.5cm]{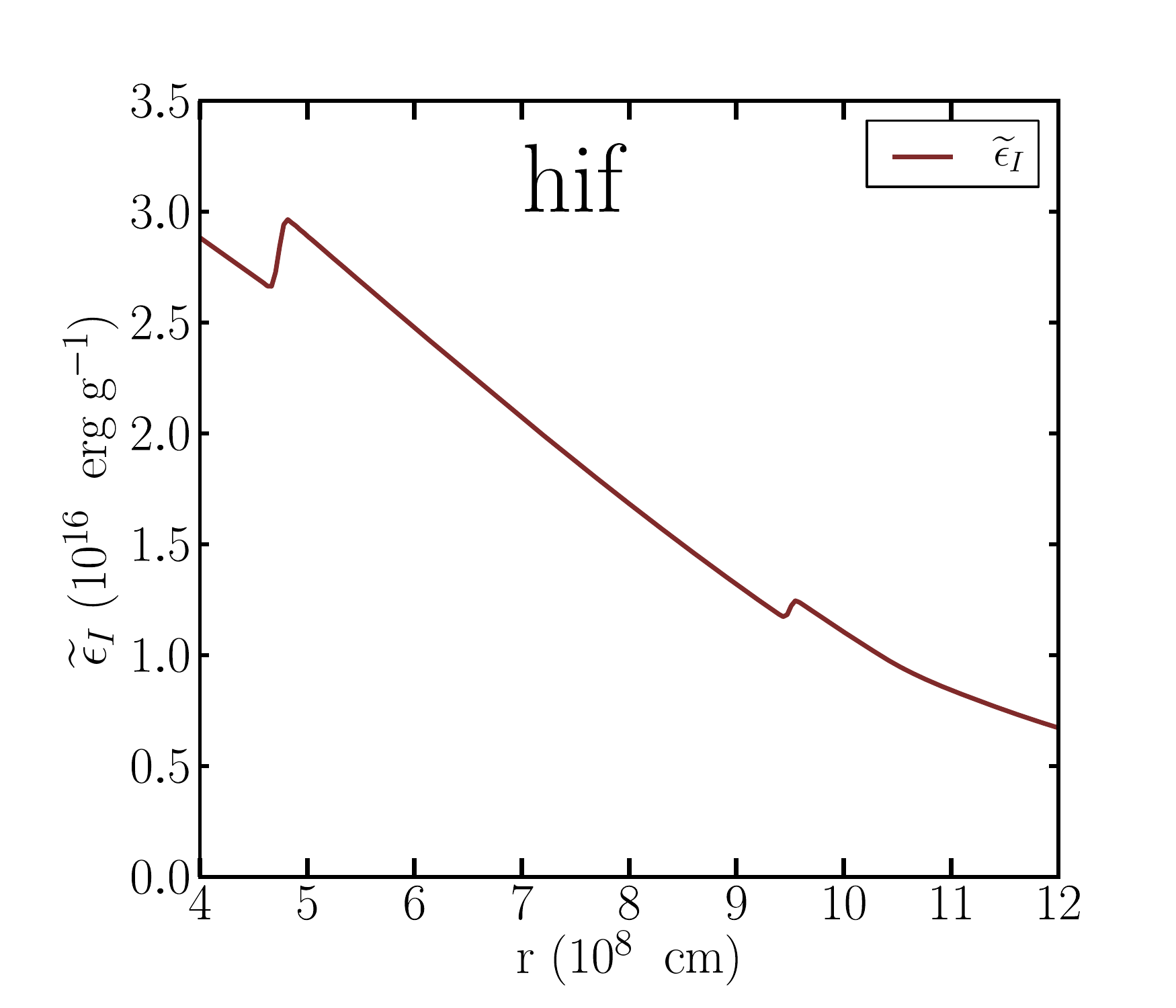}
\includegraphics[width=6.5cm]{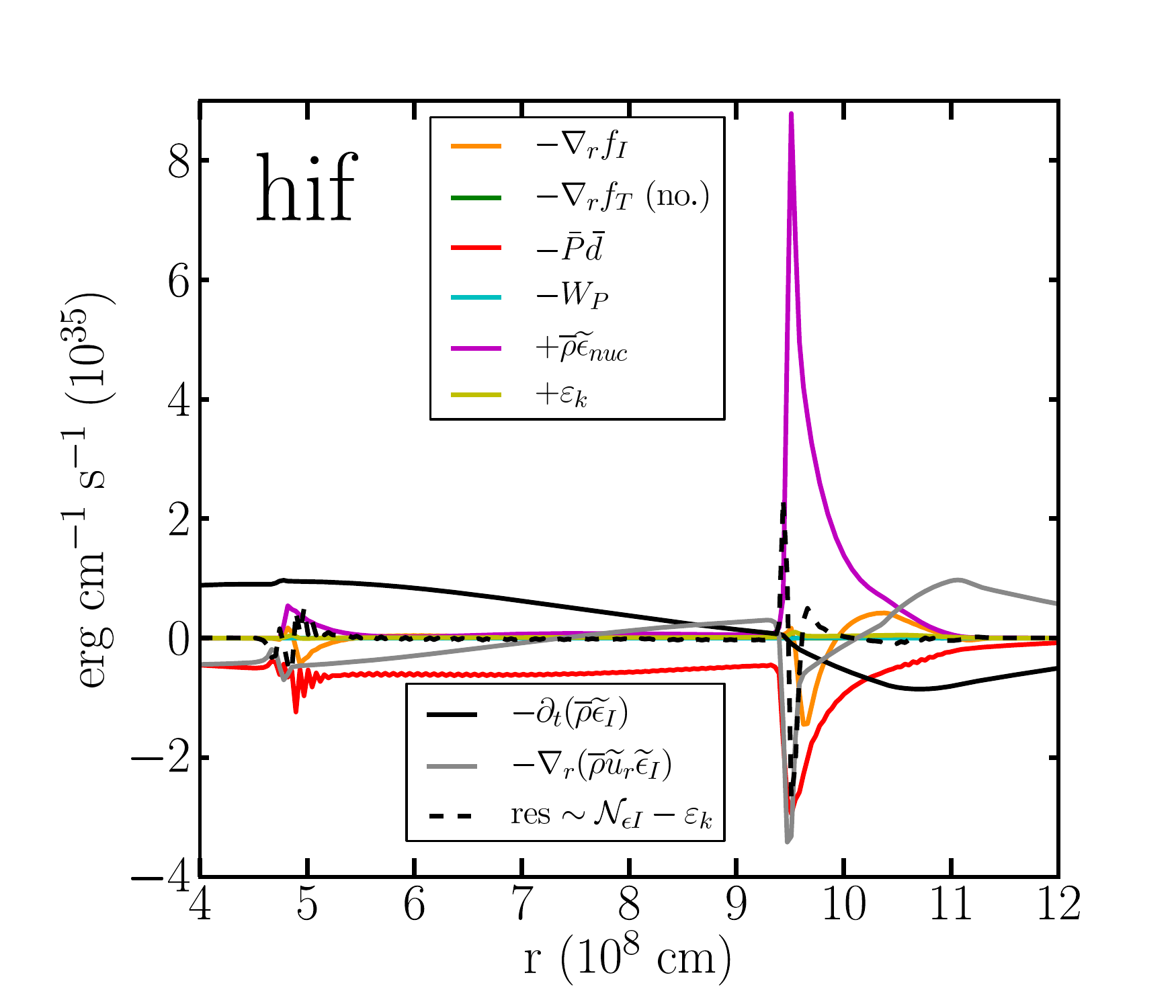}
\includegraphics[width=6.5cm]{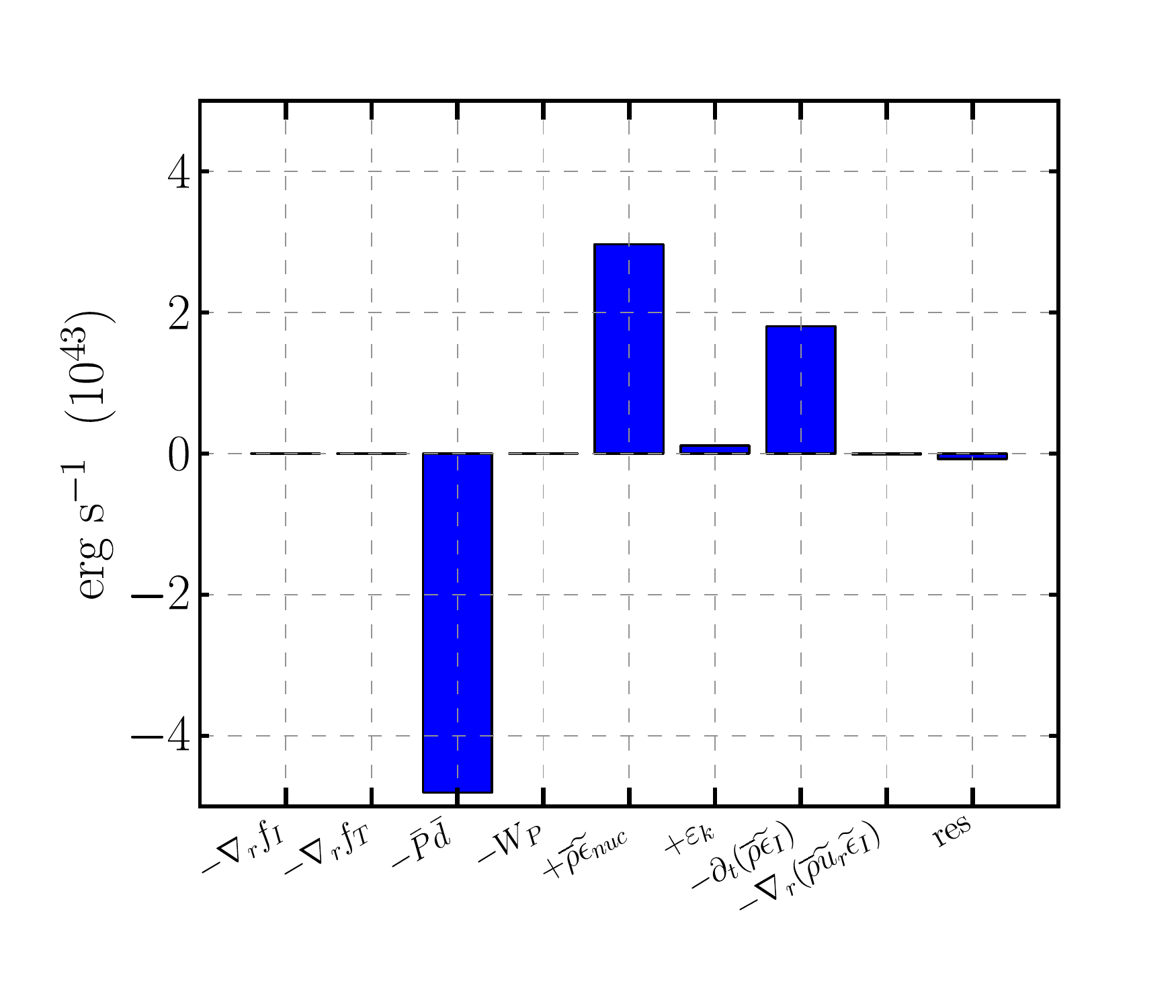}}

\centerline{
\includegraphics[width=6.5cm]{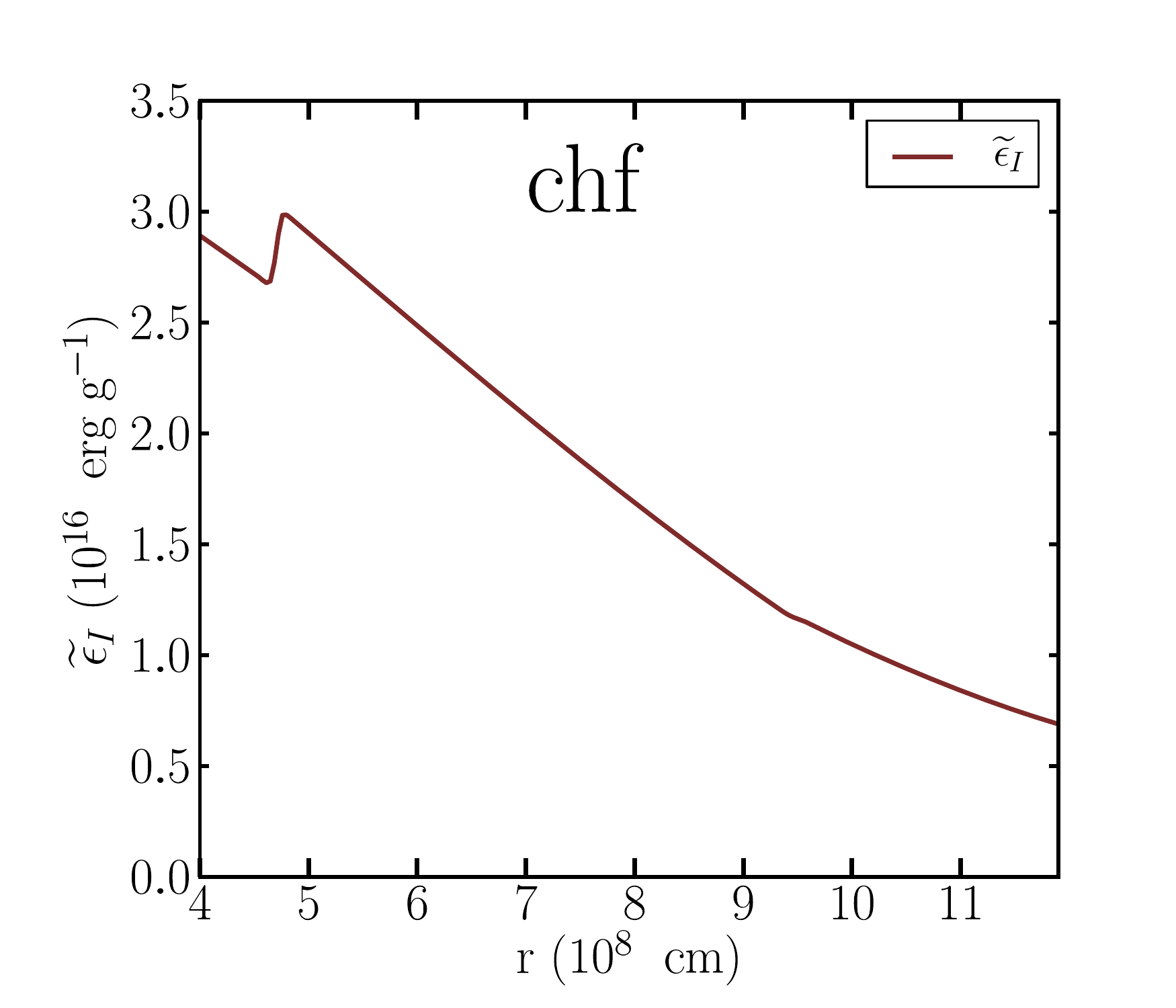}
\includegraphics[width=6.5cm]{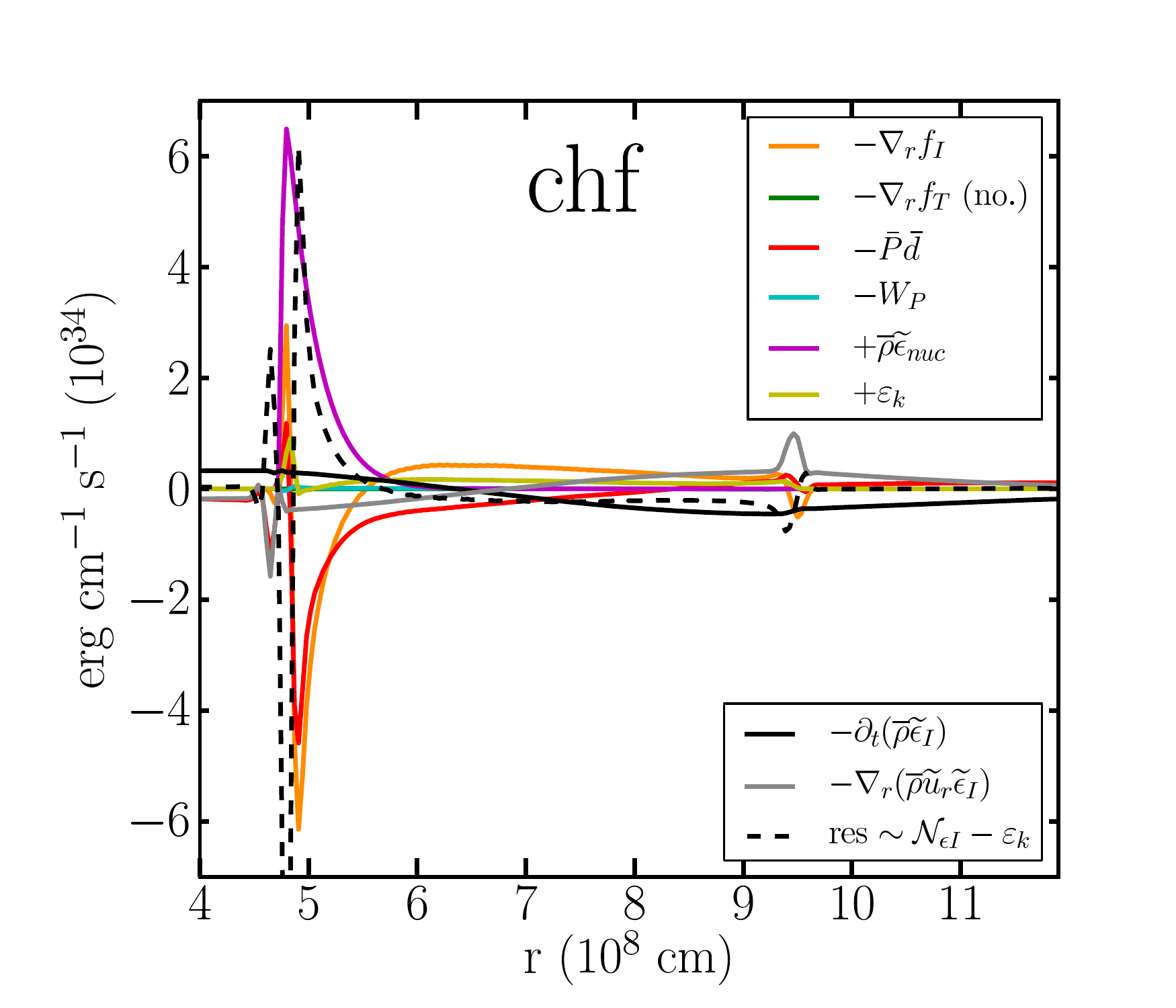}
\includegraphics[width=6.5cm]{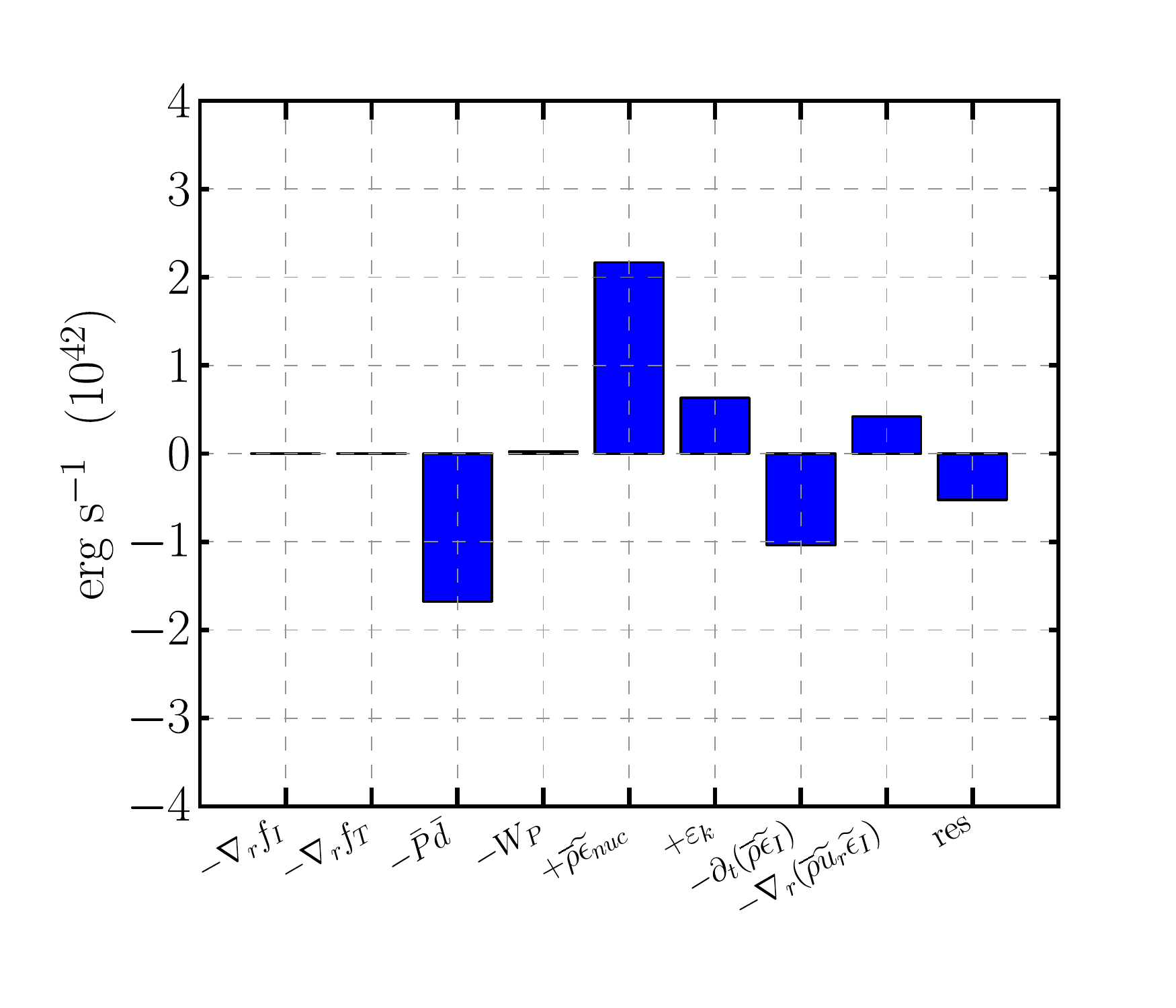}}

\caption{Mean internal energy equation. Model {\sf hif.3D} (upper panels) and model {\sf chf.3D} (lower panels)}
\end{figure}

\newpage
 
\subsection{Mean kinetic energy equation}

\begin{align}
\av{\rho} \fav{D}_t \fav{\epsilon}_k = &  -\nabla_r  ( f_k +  f_P ) - \fht{R}_{ir}\partial_r \fht{u}_i + W_b + W_P +\av{\rho}\fav{D}_t (\fav{u}_i \fav{u}_i / 2) + {\mathcal N_{\epsilon k}} \label{eq:rans_mke} 
\end{align}

\begin{figure}[!h]
\centerline{
\includegraphics[width=6.5cm]{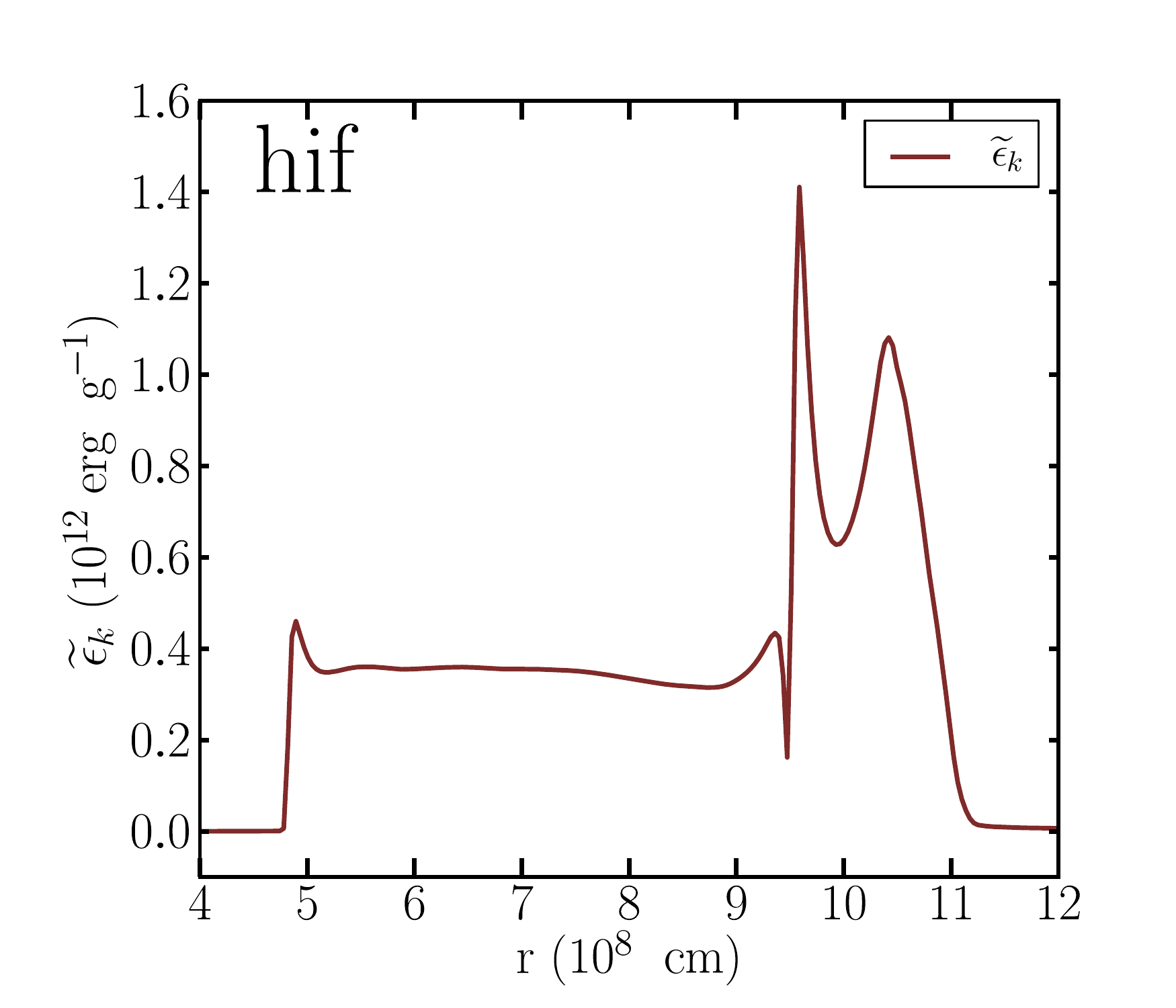}
\includegraphics[width=6.5cm]{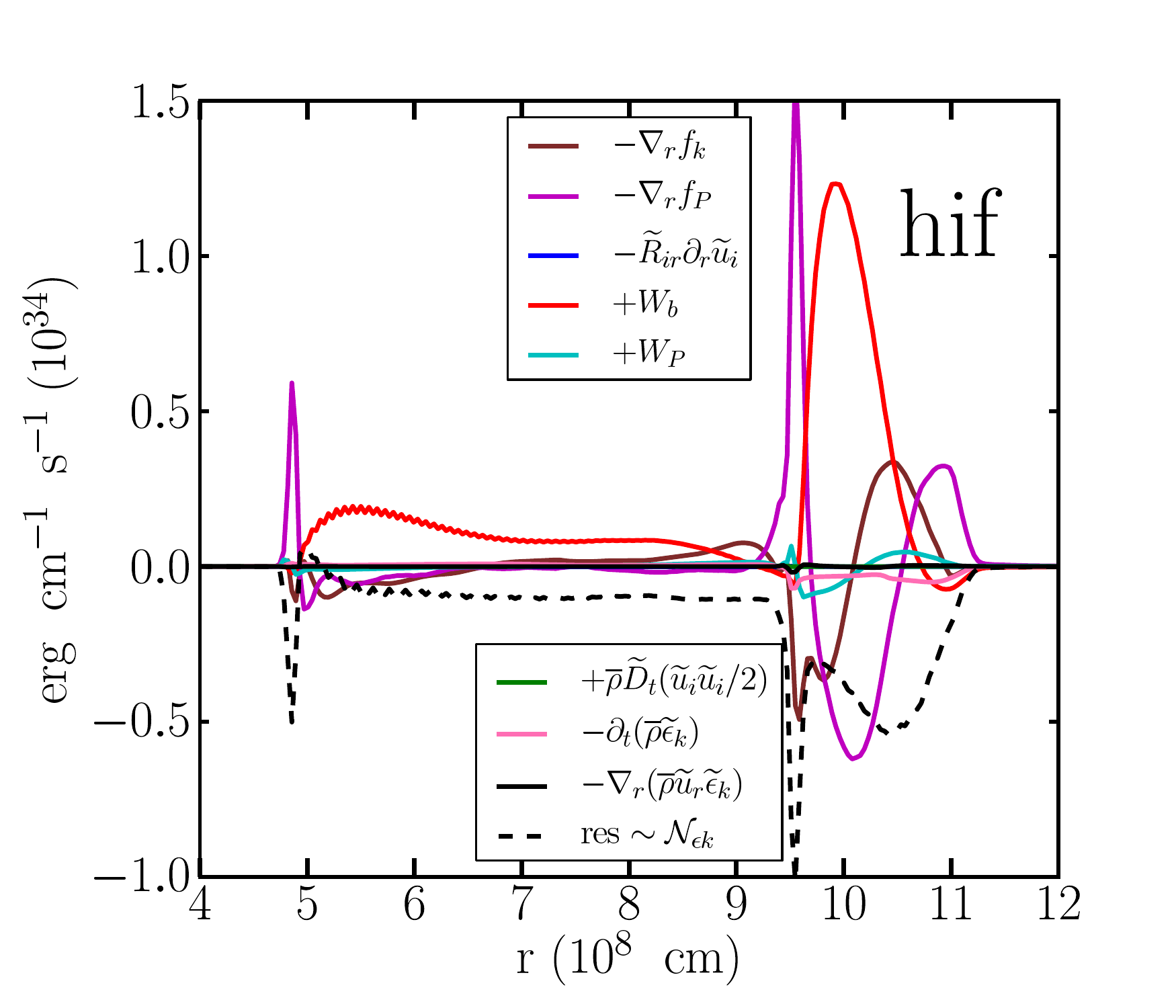}
\includegraphics[width=6.5cm]{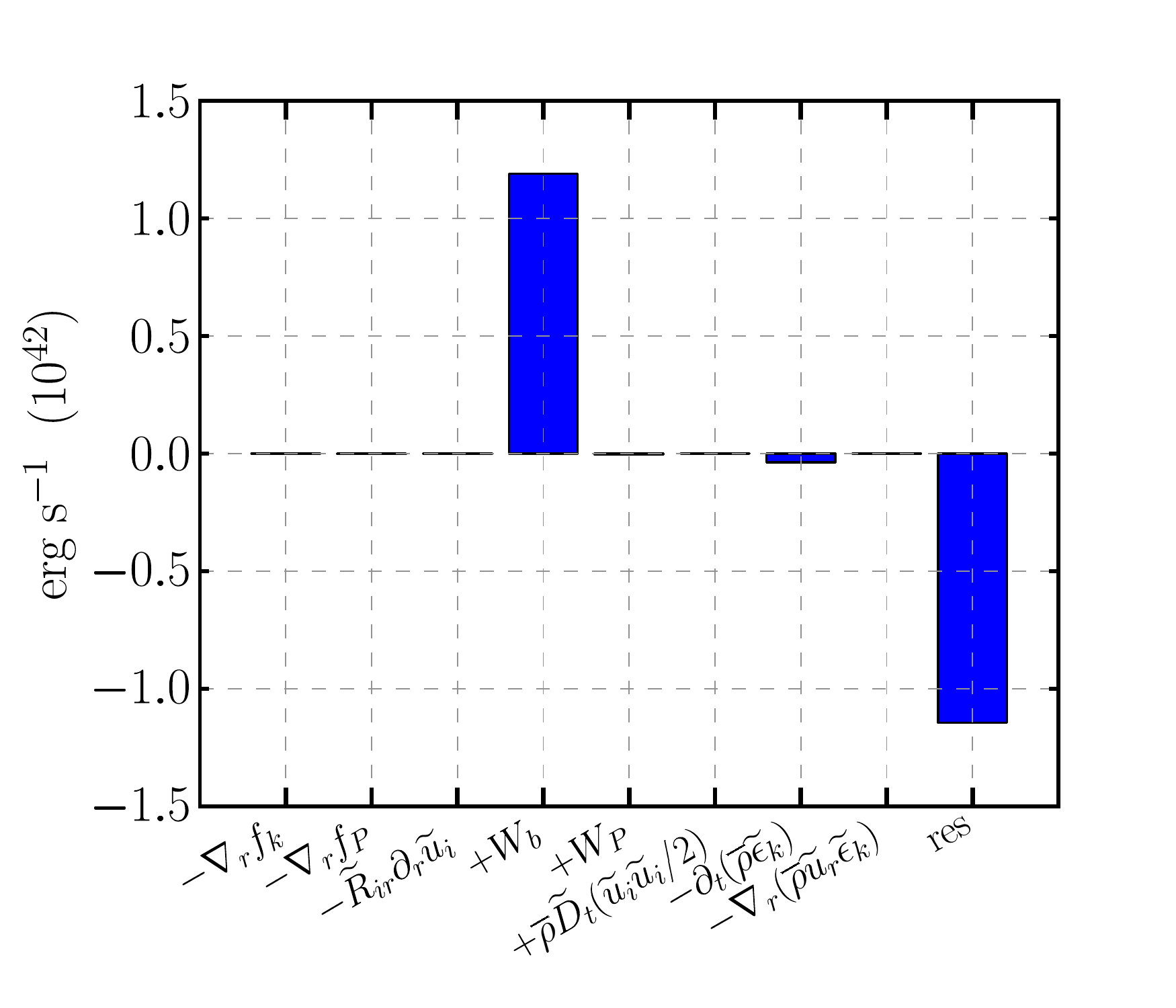}}

\centerline{
\includegraphics[width=6.5cm]{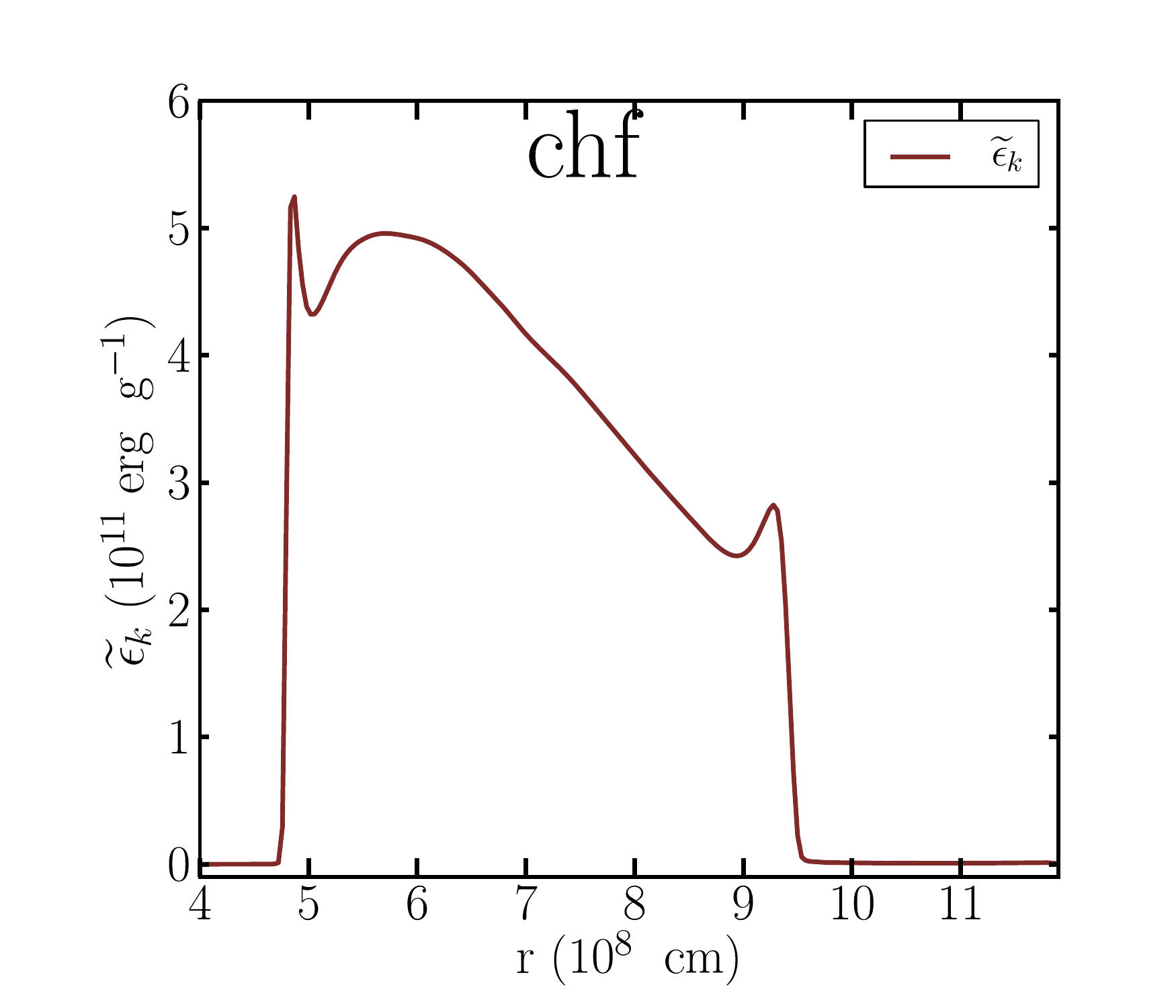}
\includegraphics[width=6.5cm]{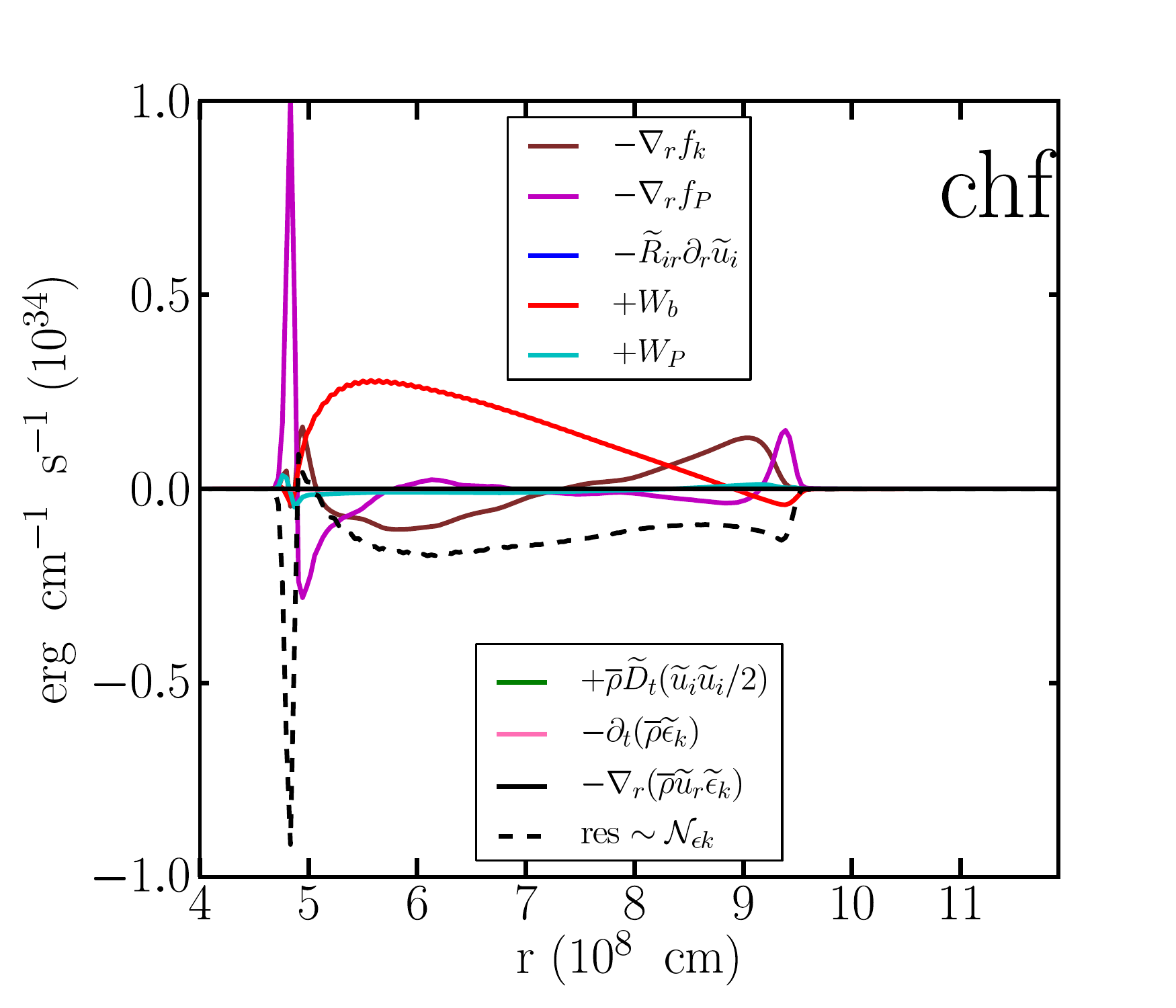}
\includegraphics[width=6.5cm]{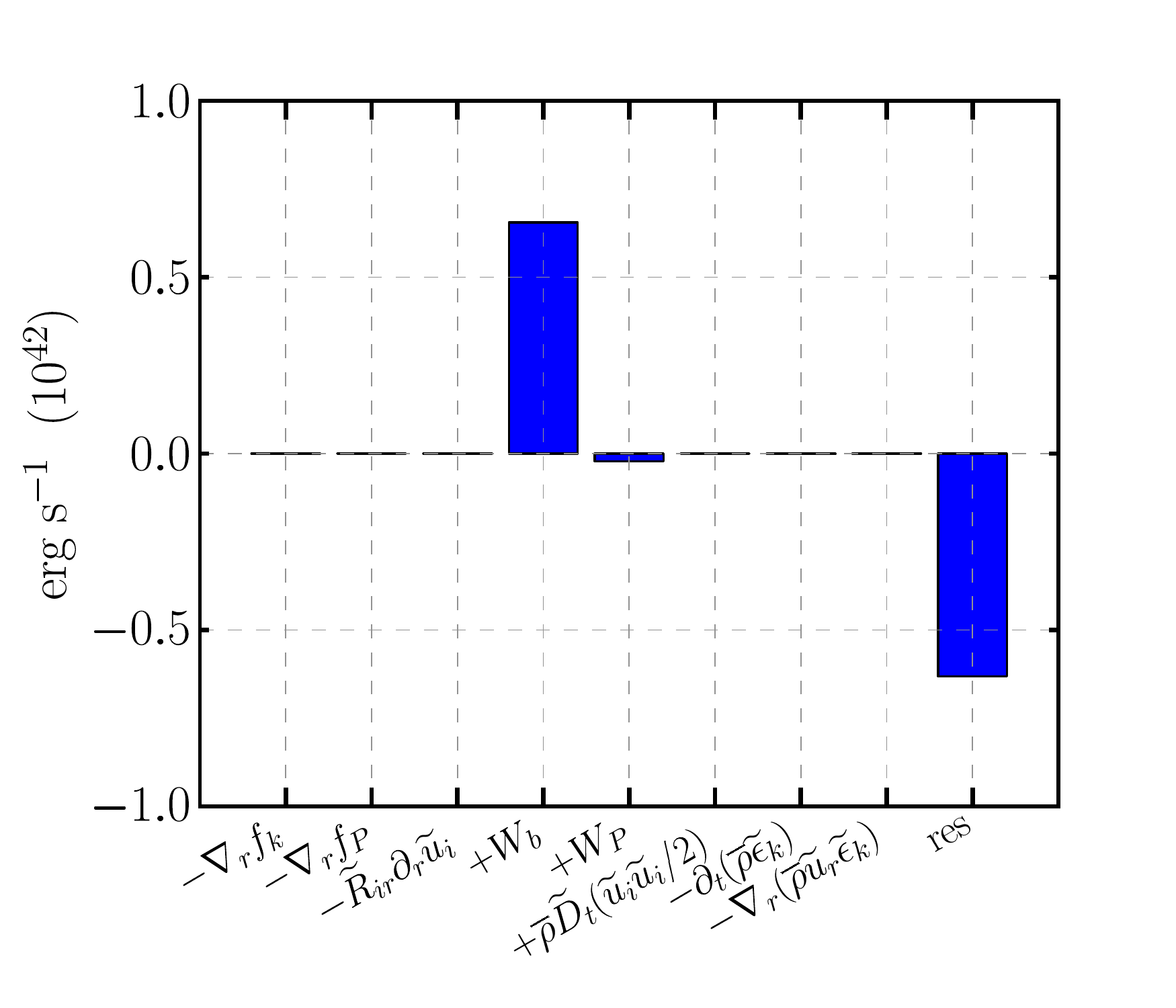}}

\caption{Mean kinetic energy equation. Model {\sf hif.3D} (upper panels) and model {\sf chf.3D} (lower panels)}
\end{figure}

\newpage

\subsection{Mean total energy equation}

\begin{align}
\av{\rho} \fav{D}_t \fav{\epsilon}_t = &  - \nabla_r ( f_I + f_T + f_k + f_P ) - \fht{R}_{ir}\partial_r \fht{u}_i - \av{P} \ \av{d} + W_b + {\mathcal S} + \av{\rho}\fav{D}_t (\fav{u}_i \fav{u}_i / 2) + {\mathcal N_{\epsilon t}} 
\end{align}

\begin{figure}[!h]
\centerline{
\includegraphics[width=6.5cm]{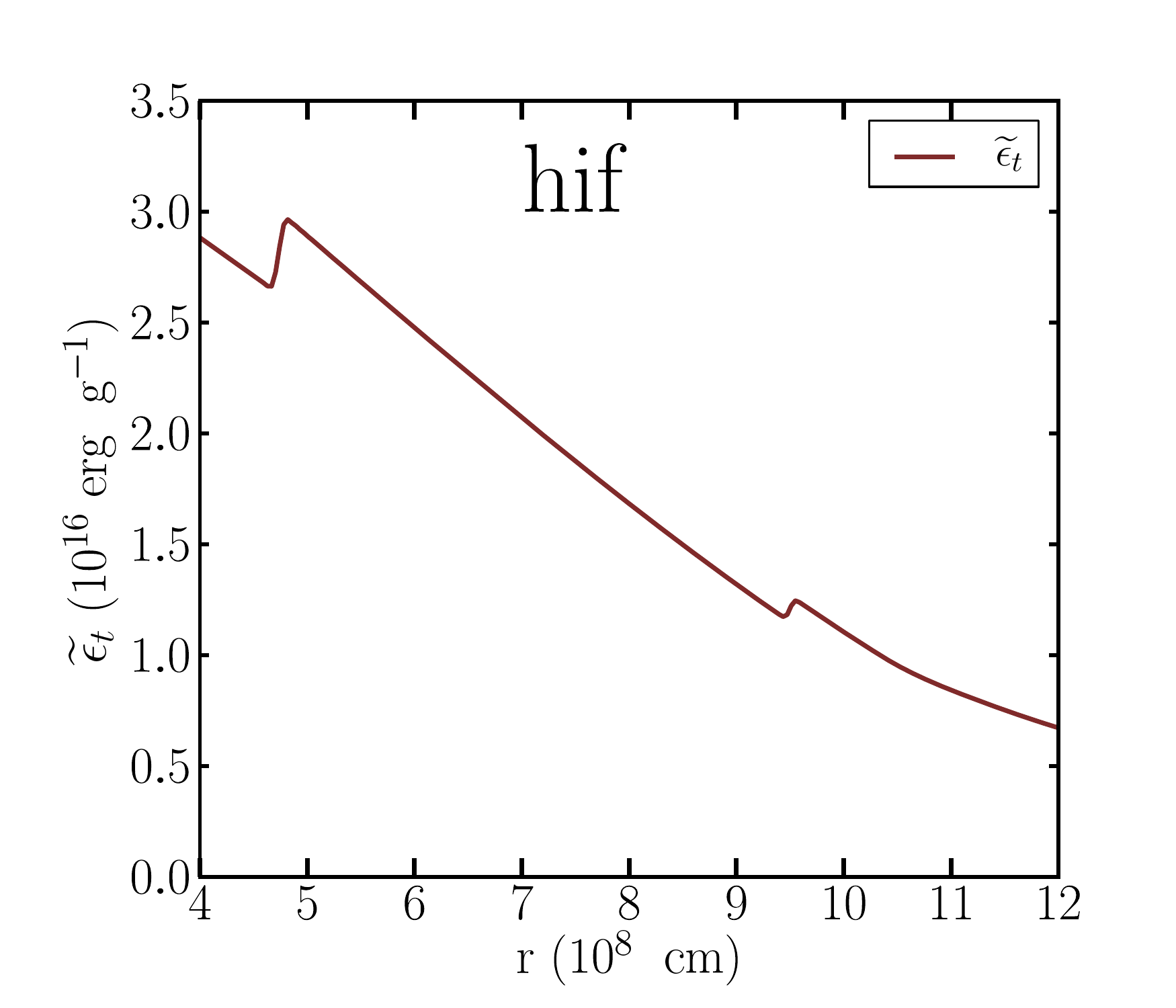}
\includegraphics[width=6.5cm]{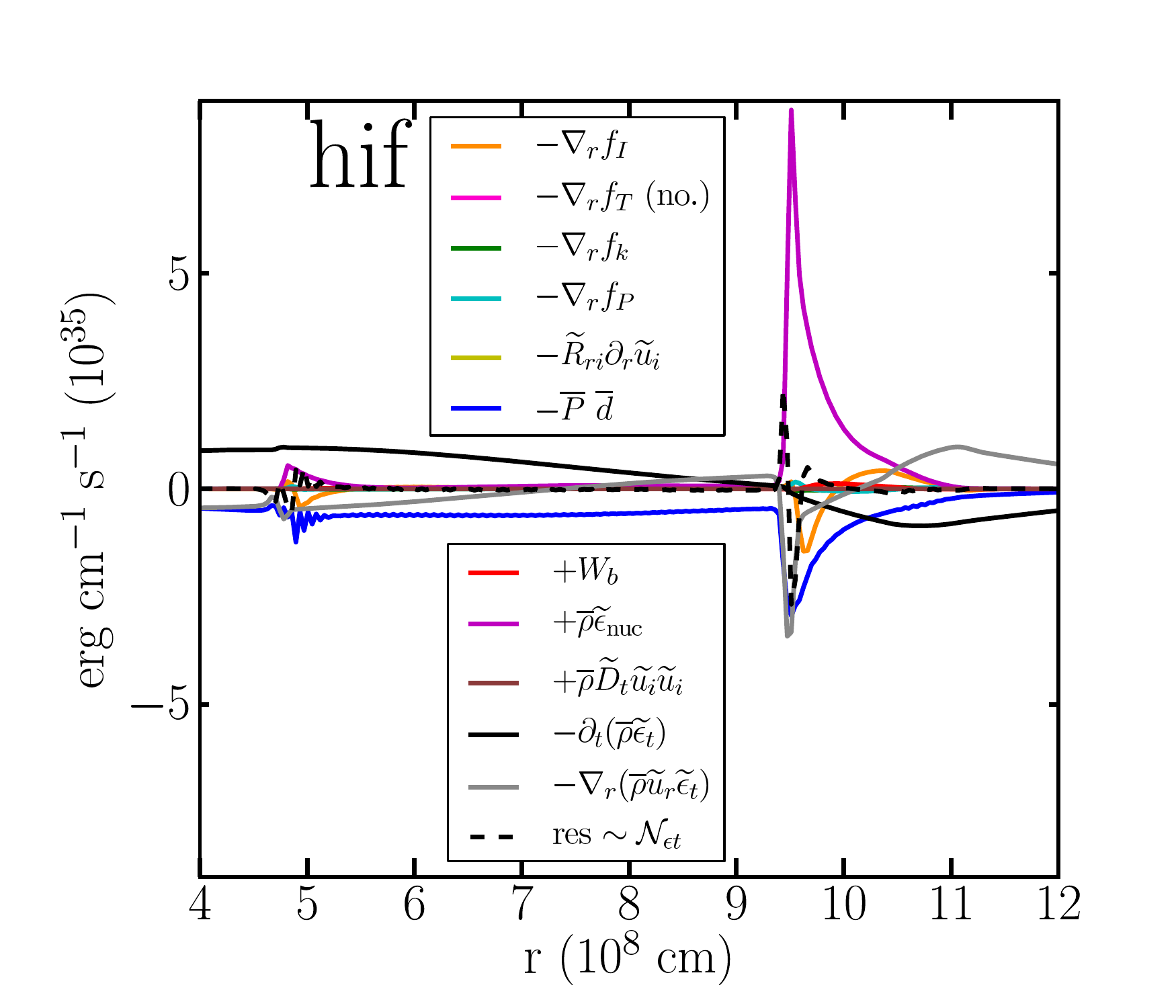}
\includegraphics[width=6.5cm]{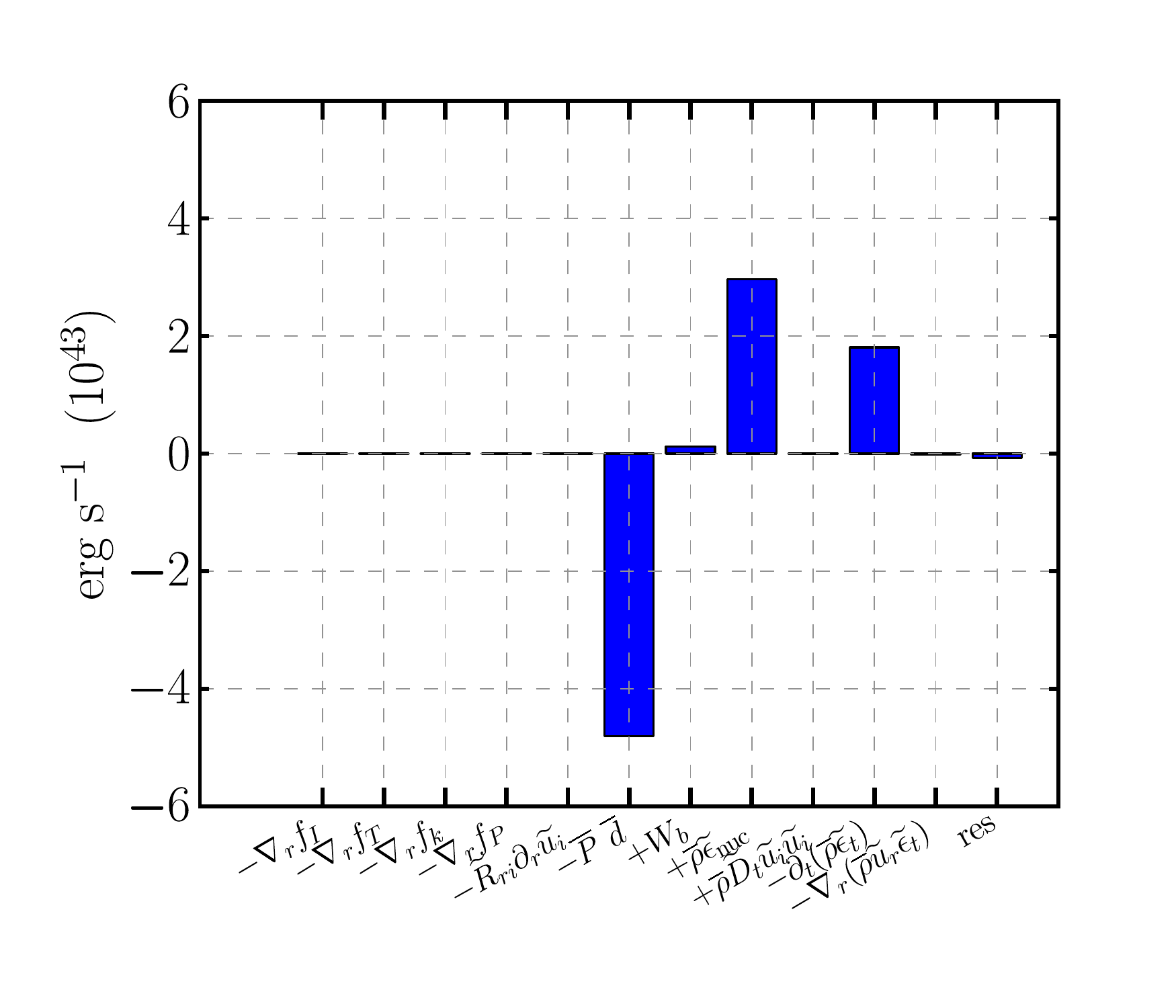}}

\centerline{
\includegraphics[width=6.5cm]{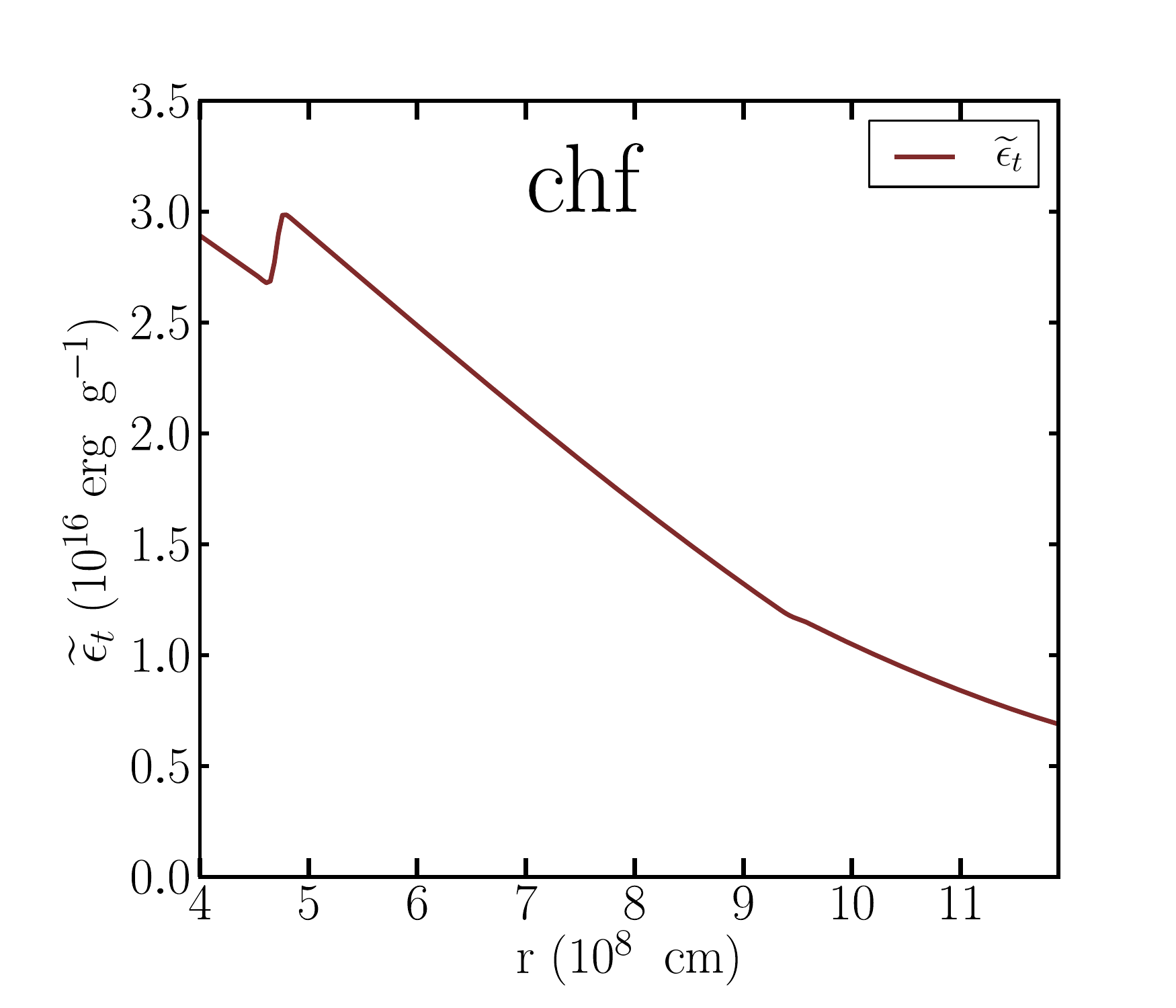}
\includegraphics[width=6.5cm]{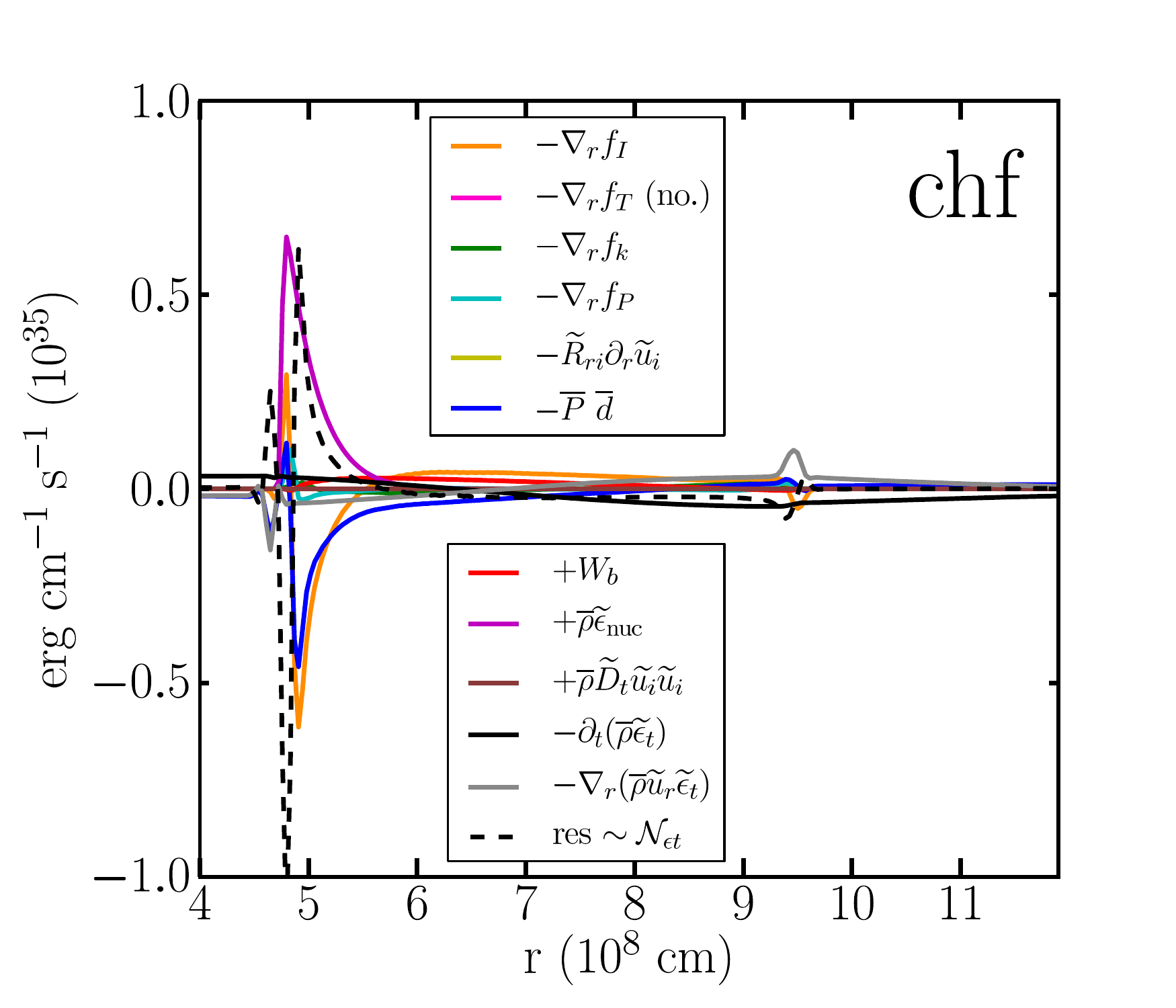}
\includegraphics[width=6.5cm]{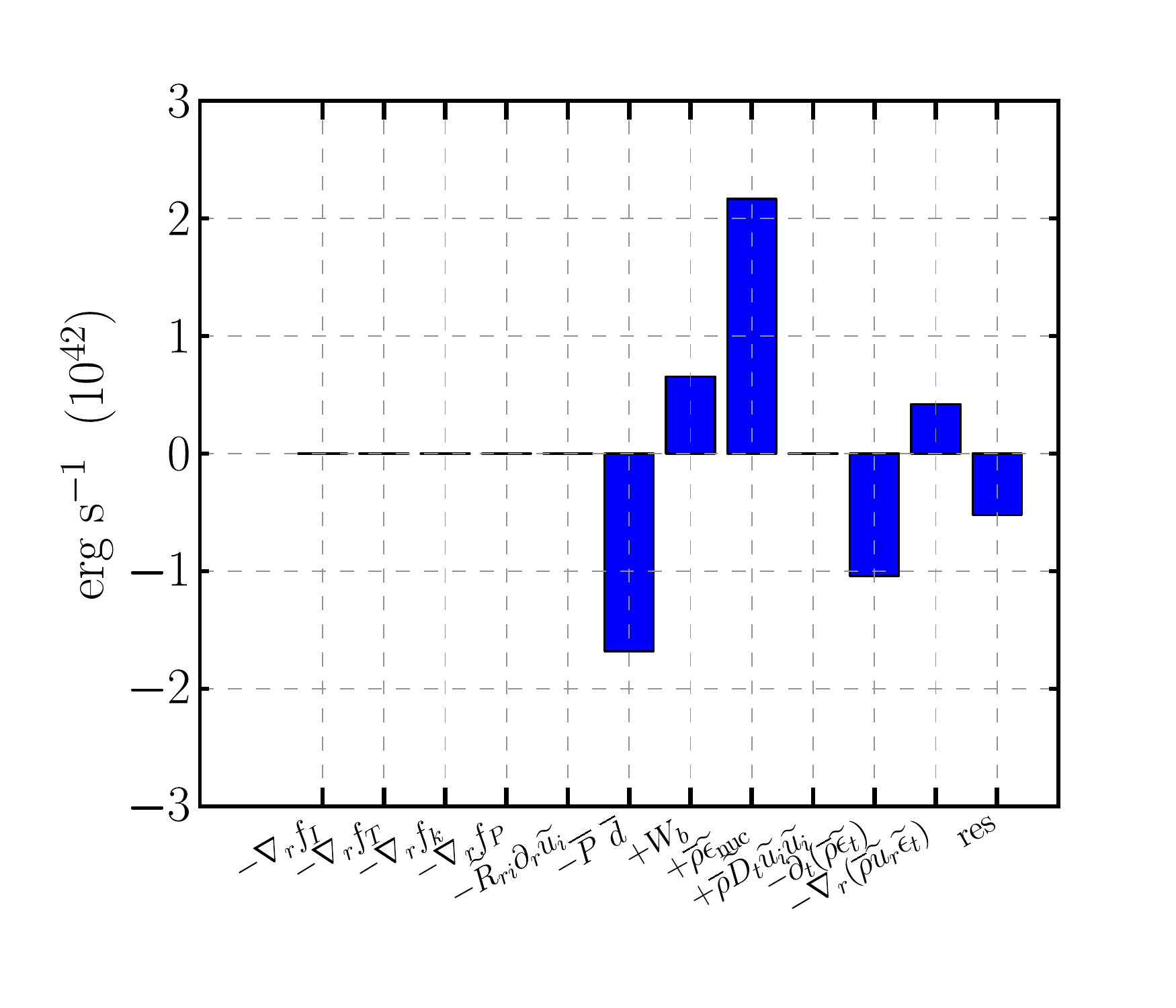}}

\caption{Mean total energy equation. Model {\sf hif.3D} (upper panels) and model {\sf chf.3D} (lower panels)}
\end{figure}

\newpage

\subsection{Mean entropy equation}

\begin{align}
\av{\rho} \fav{D}_t \fav{s} = & - \nabla_r  f_s    - \av{(\nabla \cdot F_T)/T}+ \av{{\mathcal S}/T} + {\mathcal N_s}   
\end{align}

\begin{figure}[!h]
\centerline{
\includegraphics[width=6.5cm]{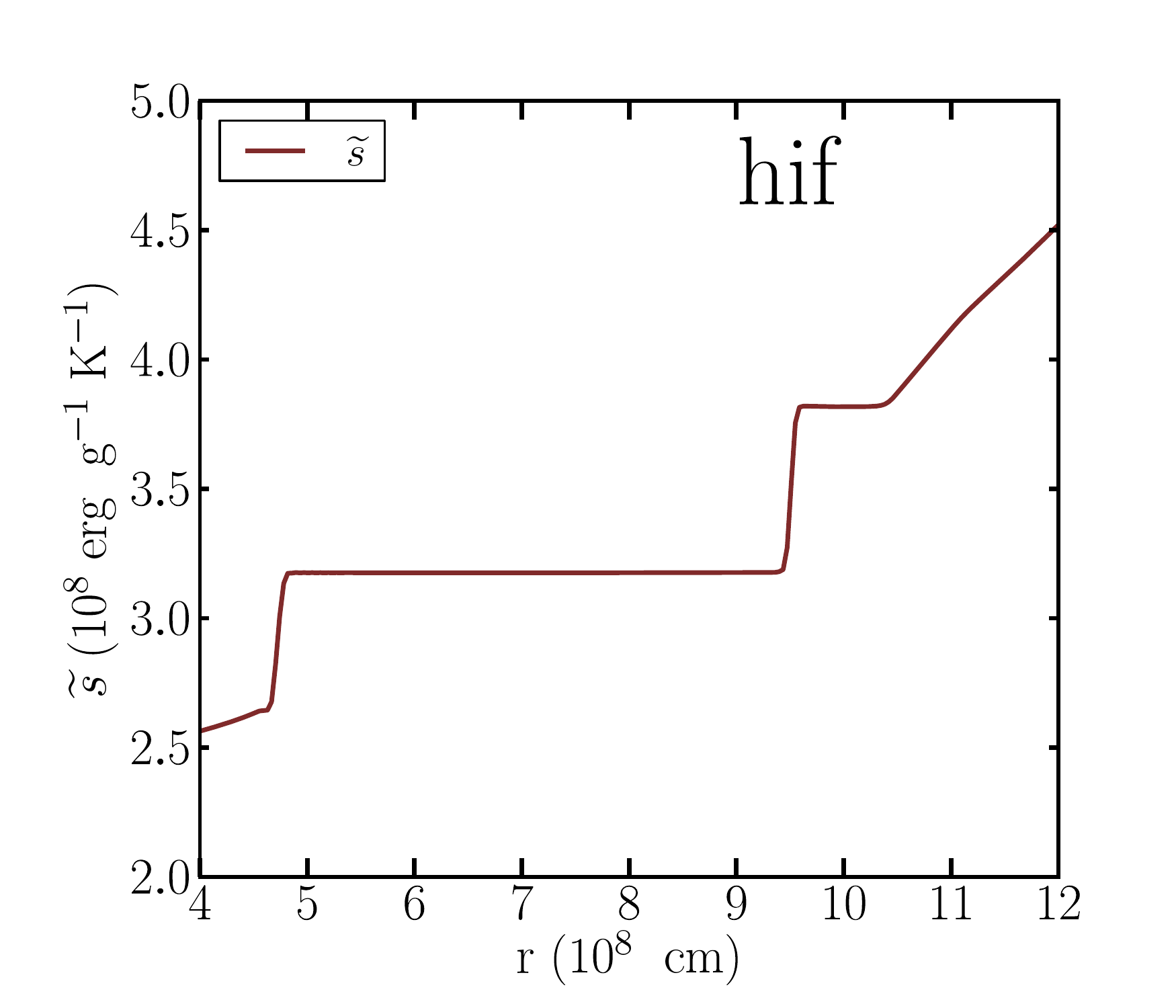}
\includegraphics[width=6.5cm]{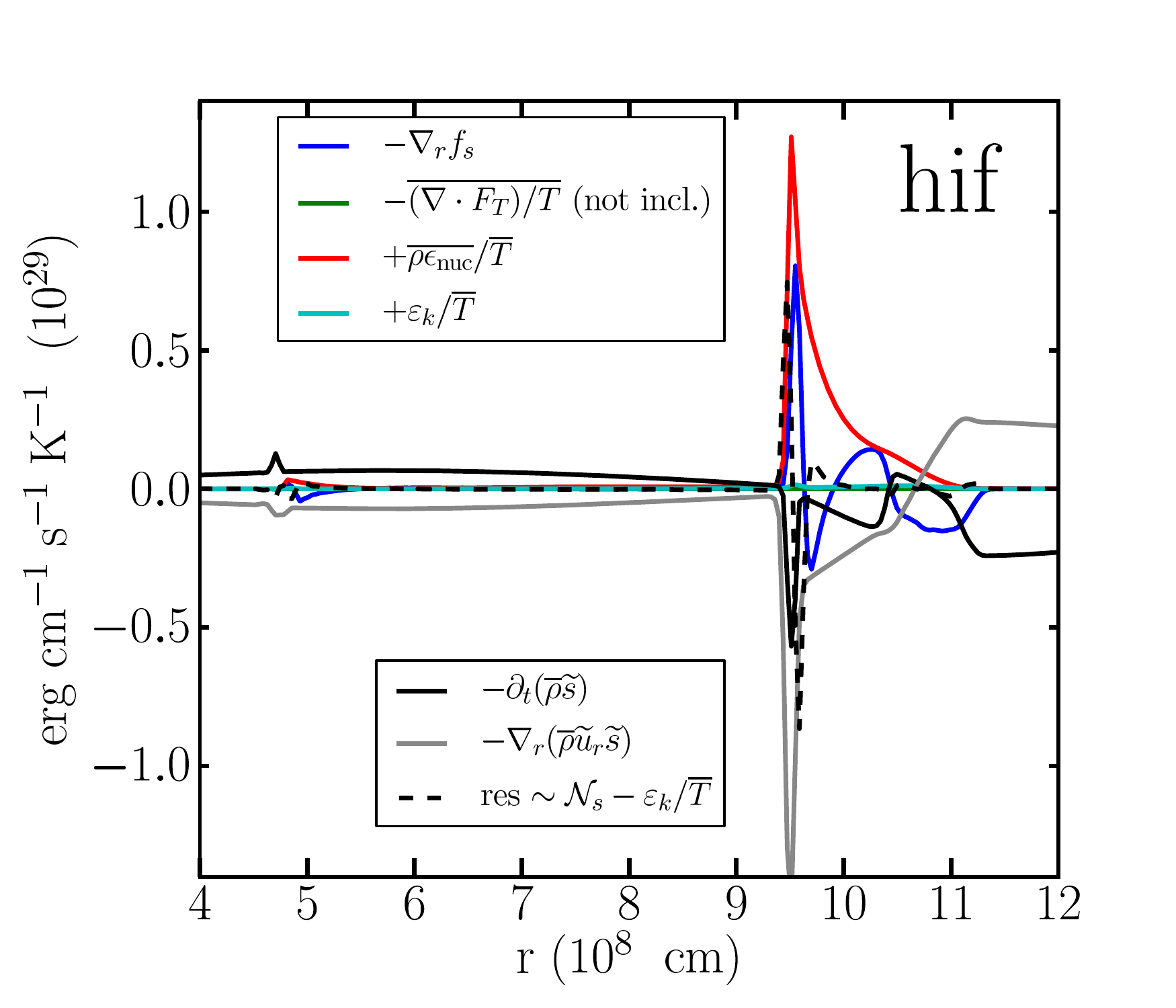}
\includegraphics[width=6.5cm]{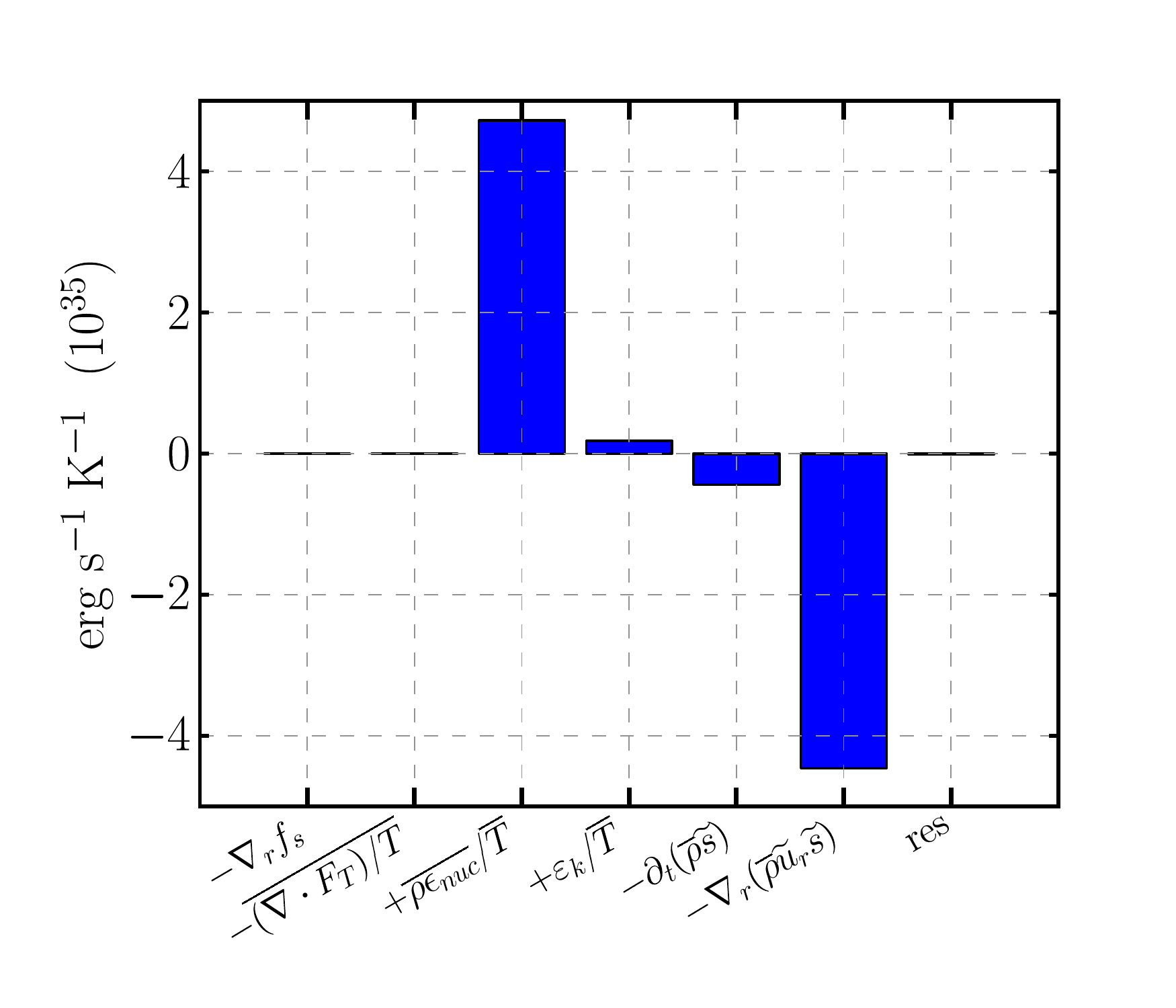}}

\centerline{
\includegraphics[width=6.5cm]{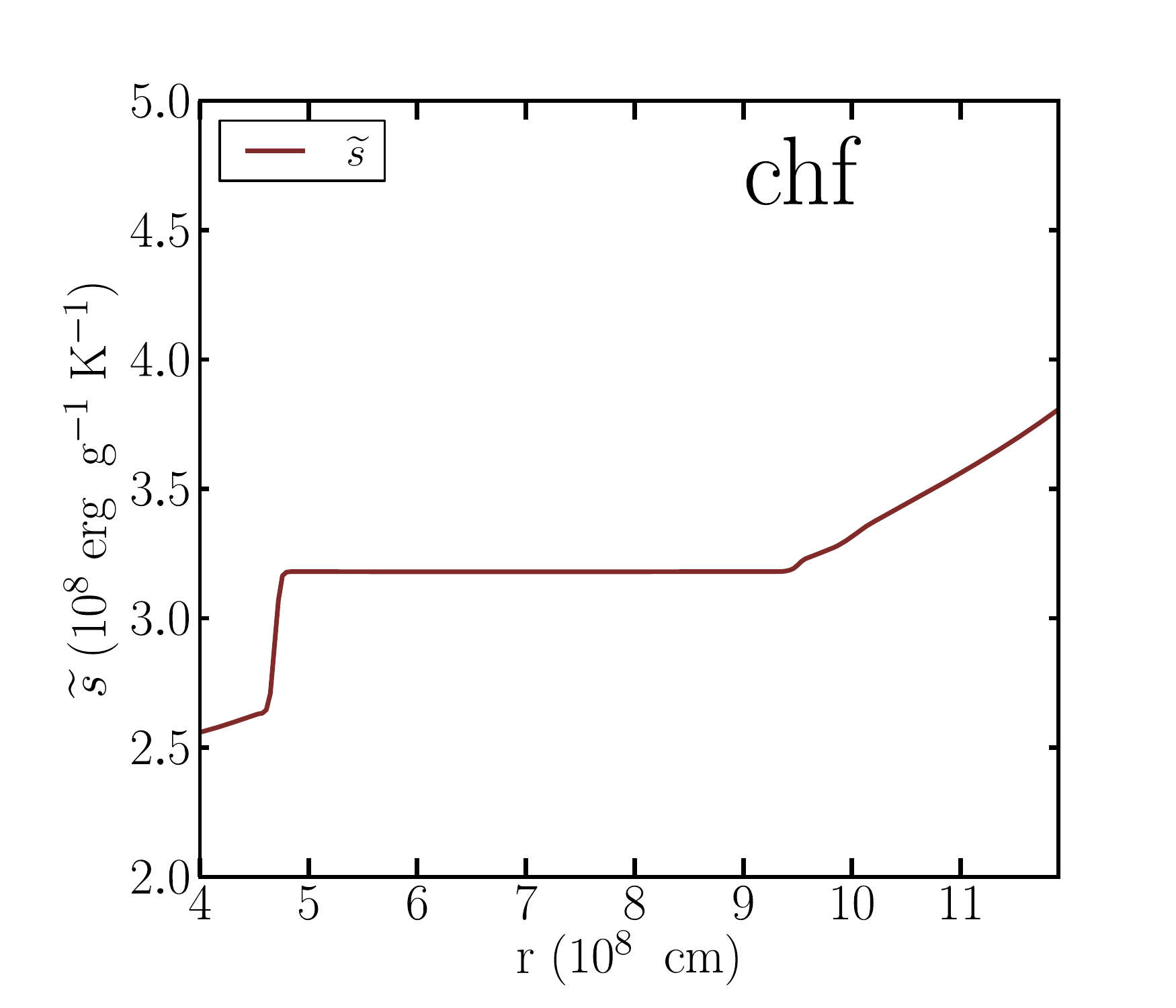}
\includegraphics[width=6.5cm]{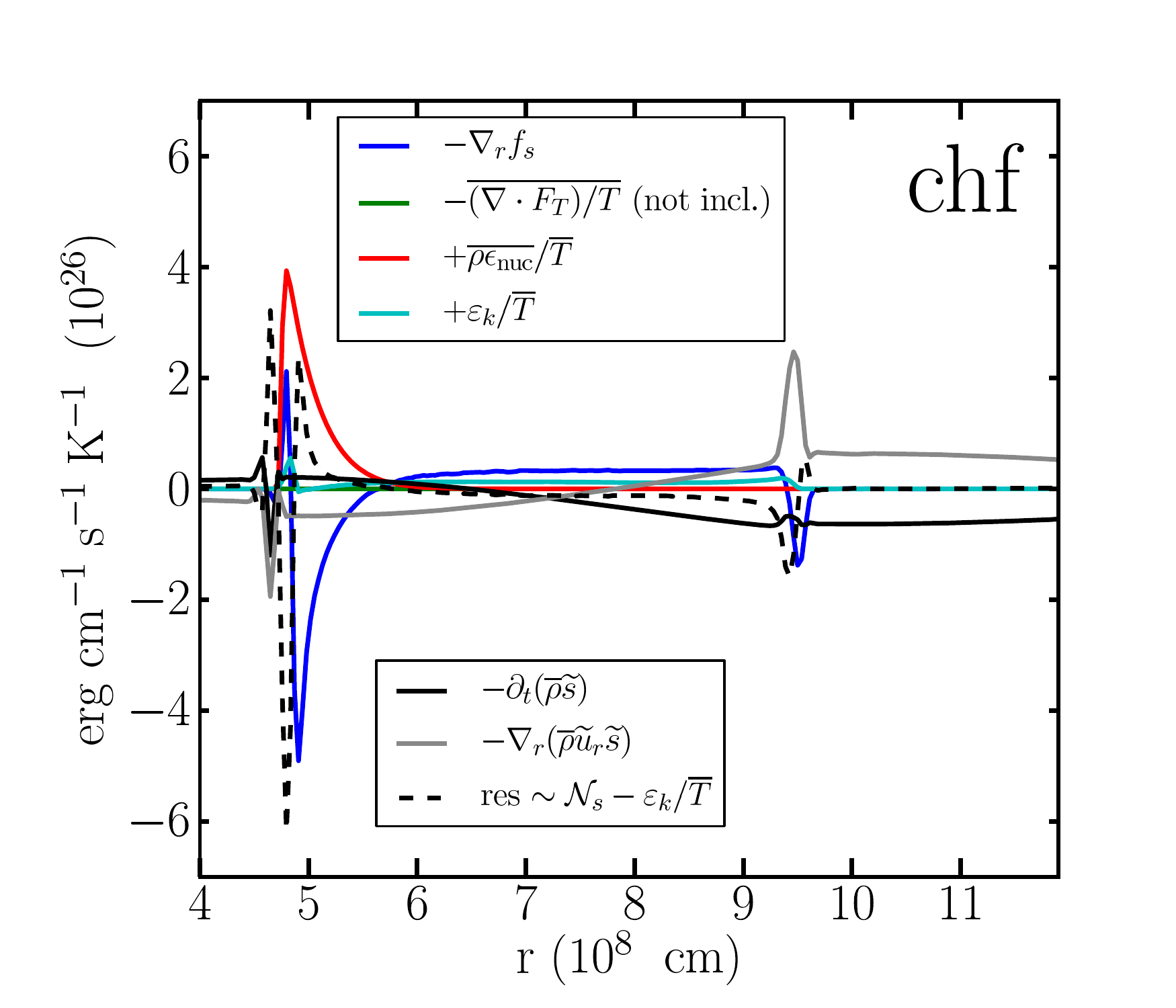}
\includegraphics[width=6.5cm]{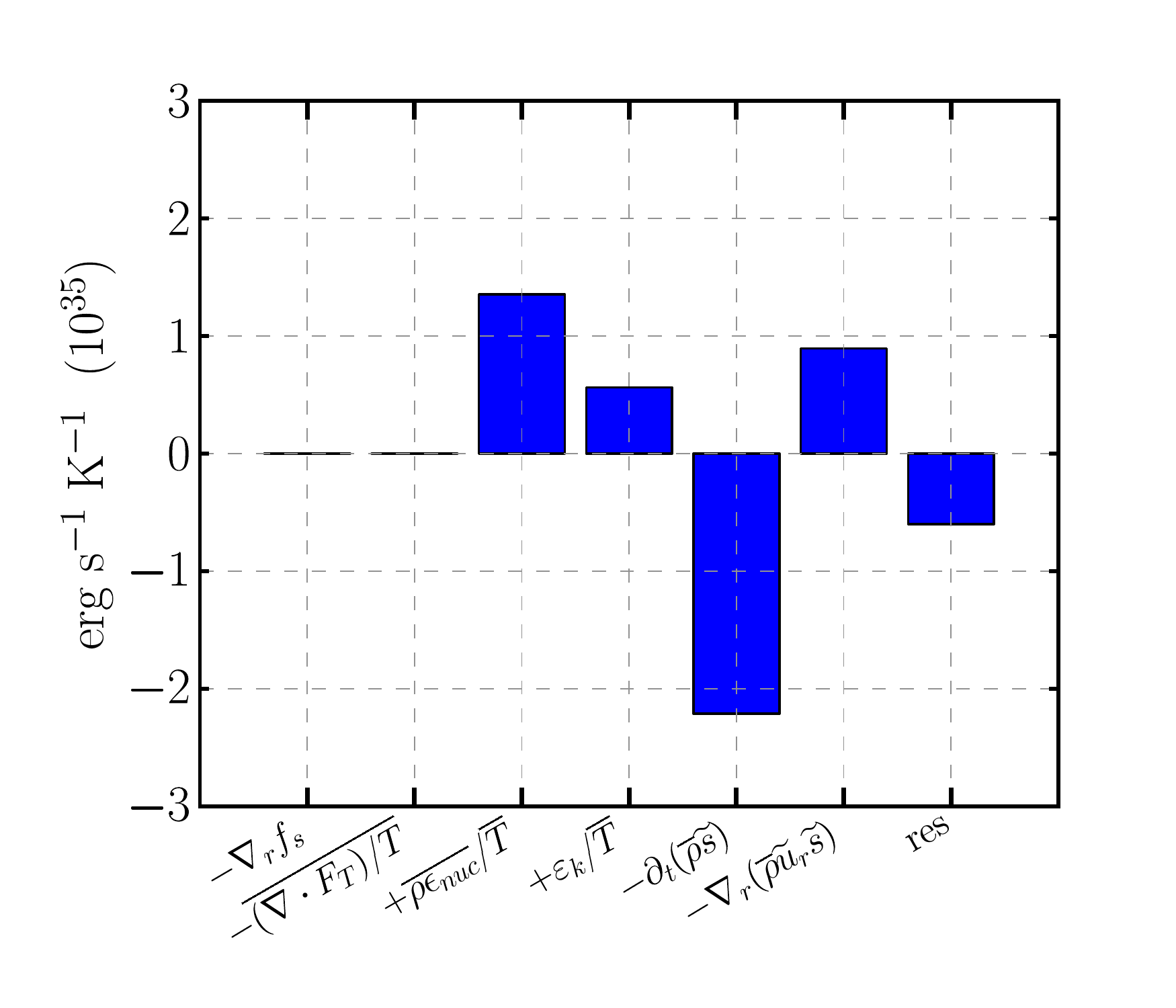}}

\caption{Mean entropy equation. Model {\sf hif.3D} (upper panels) and model {\sf chf.3D} (lower panels)}
\end{figure}

\newpage

\subsection{Mean pressure equation }

\begin{align}
\av{D}_t \av{P} = & -\nabla_r f_P - \Gamma_1 \eht{P} \ \eht{d} + (1 -\Gamma_1) W_P + (\Gamma_3 -1){\mathcal S} + (\Gamma_3 - 1)\nabla_r f_T + {\mathcal N_P} 
\end{align}

\begin{figure}[!h]
\centerline{
\includegraphics[width=6.5cm]{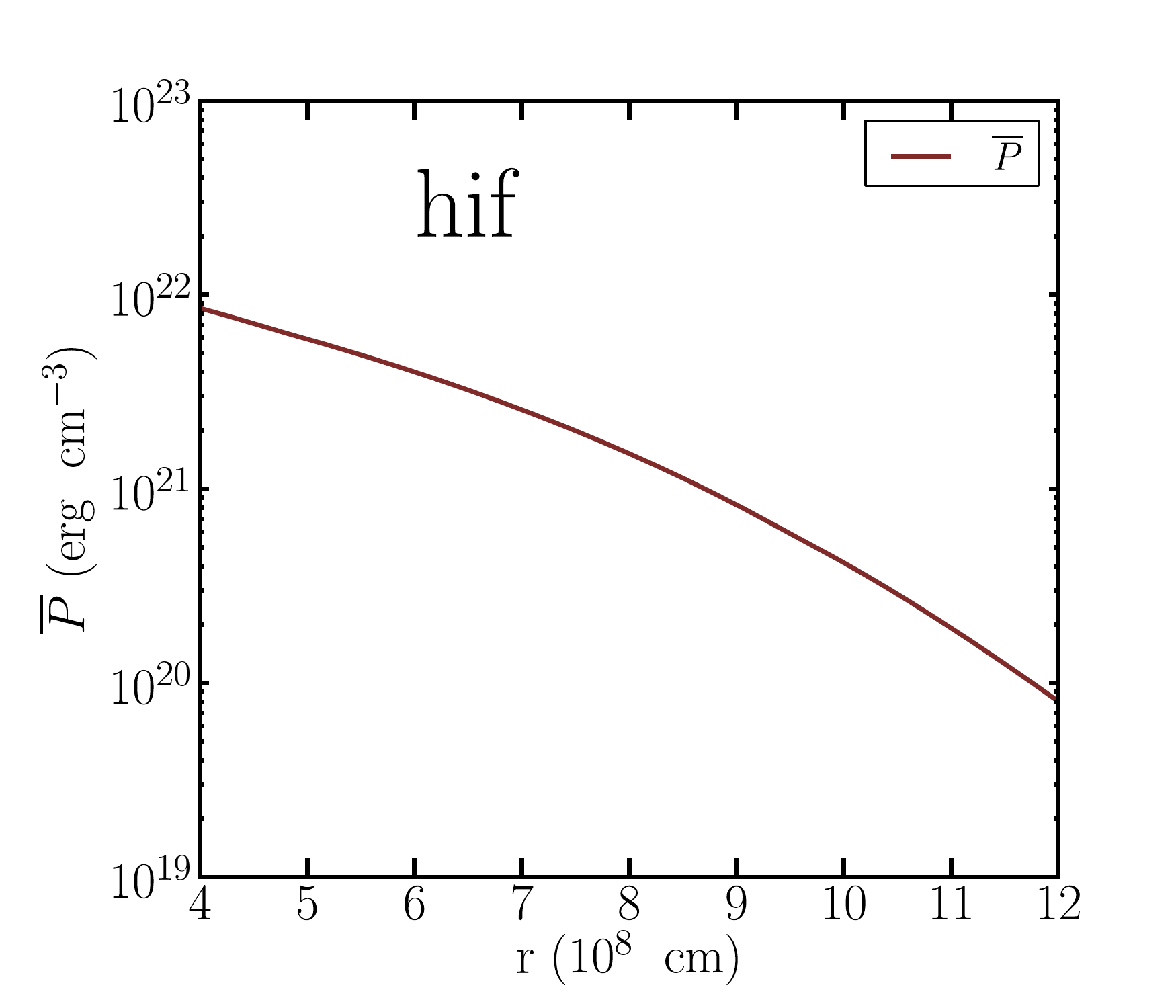}
\includegraphics[width=6.5cm]{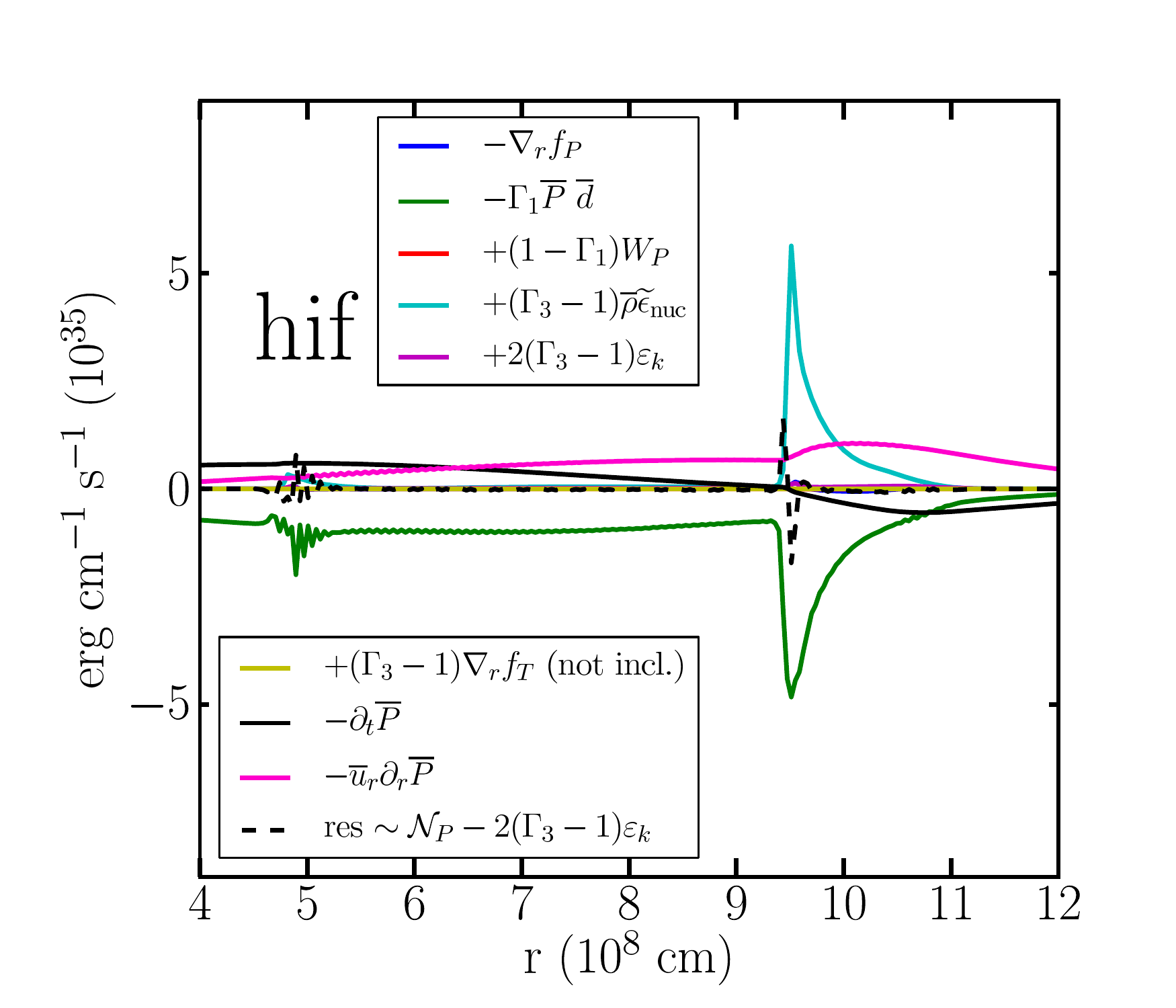}
\includegraphics[width=6.5cm]{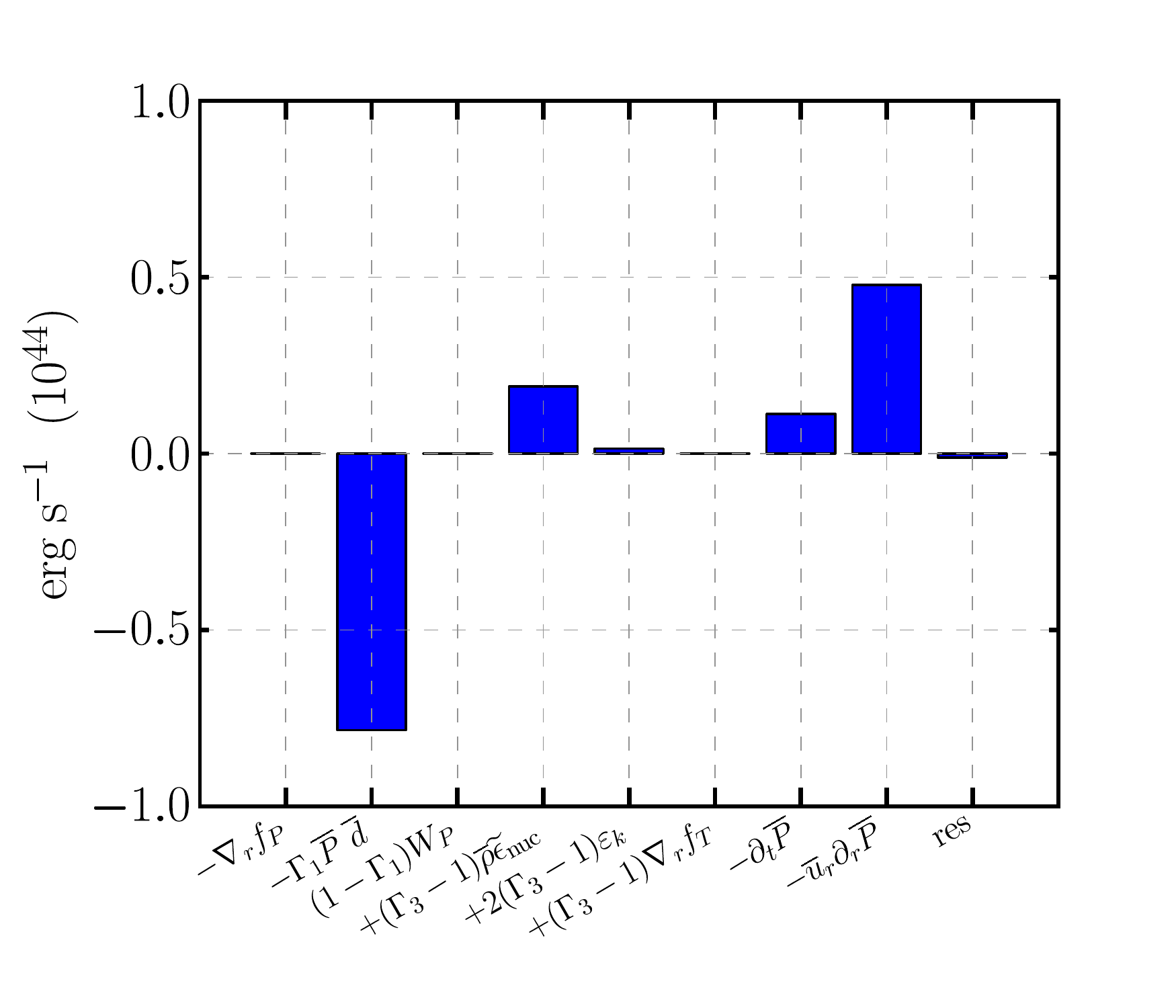}}

\centerline{
\includegraphics[width=6.5cm]{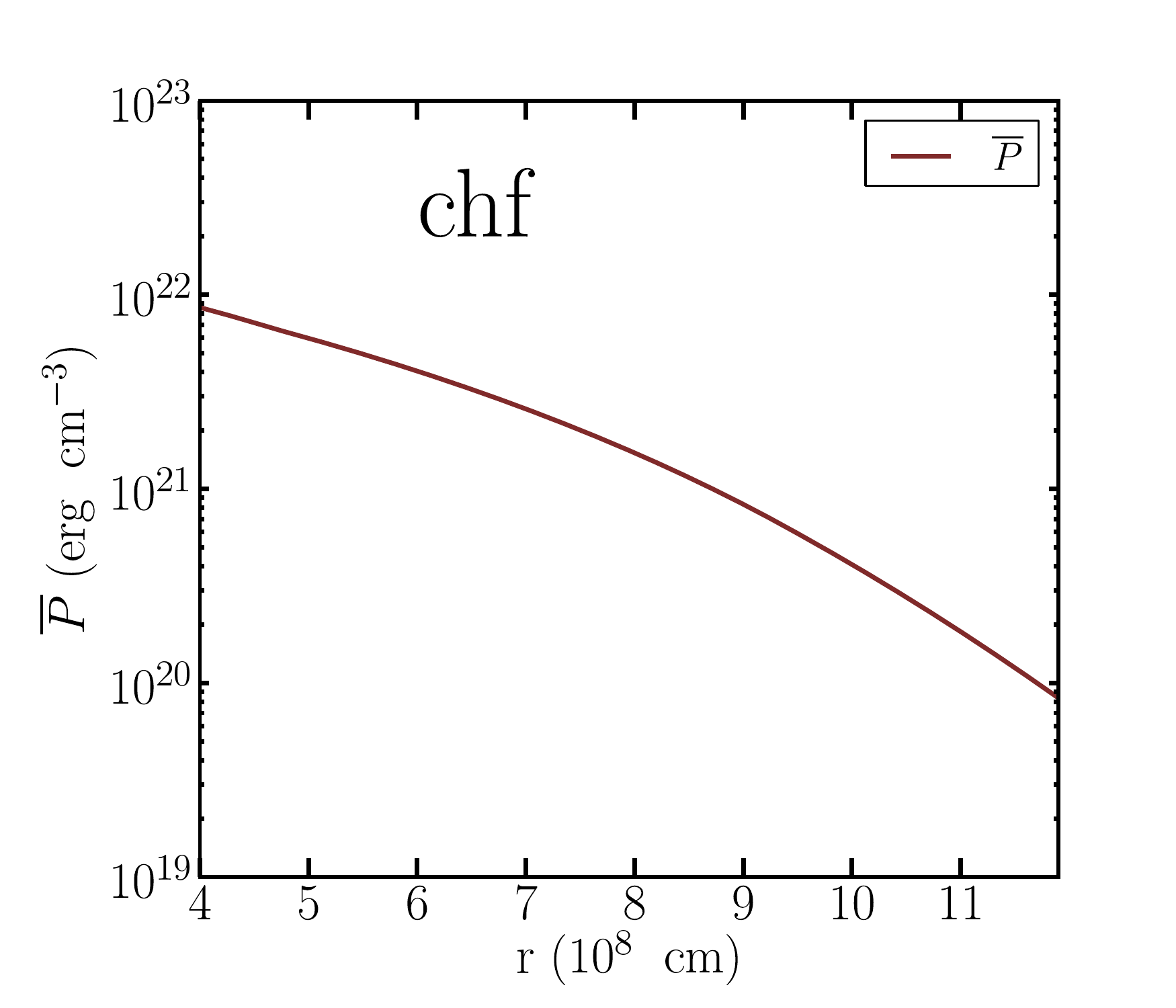}
\includegraphics[width=6.5cm]{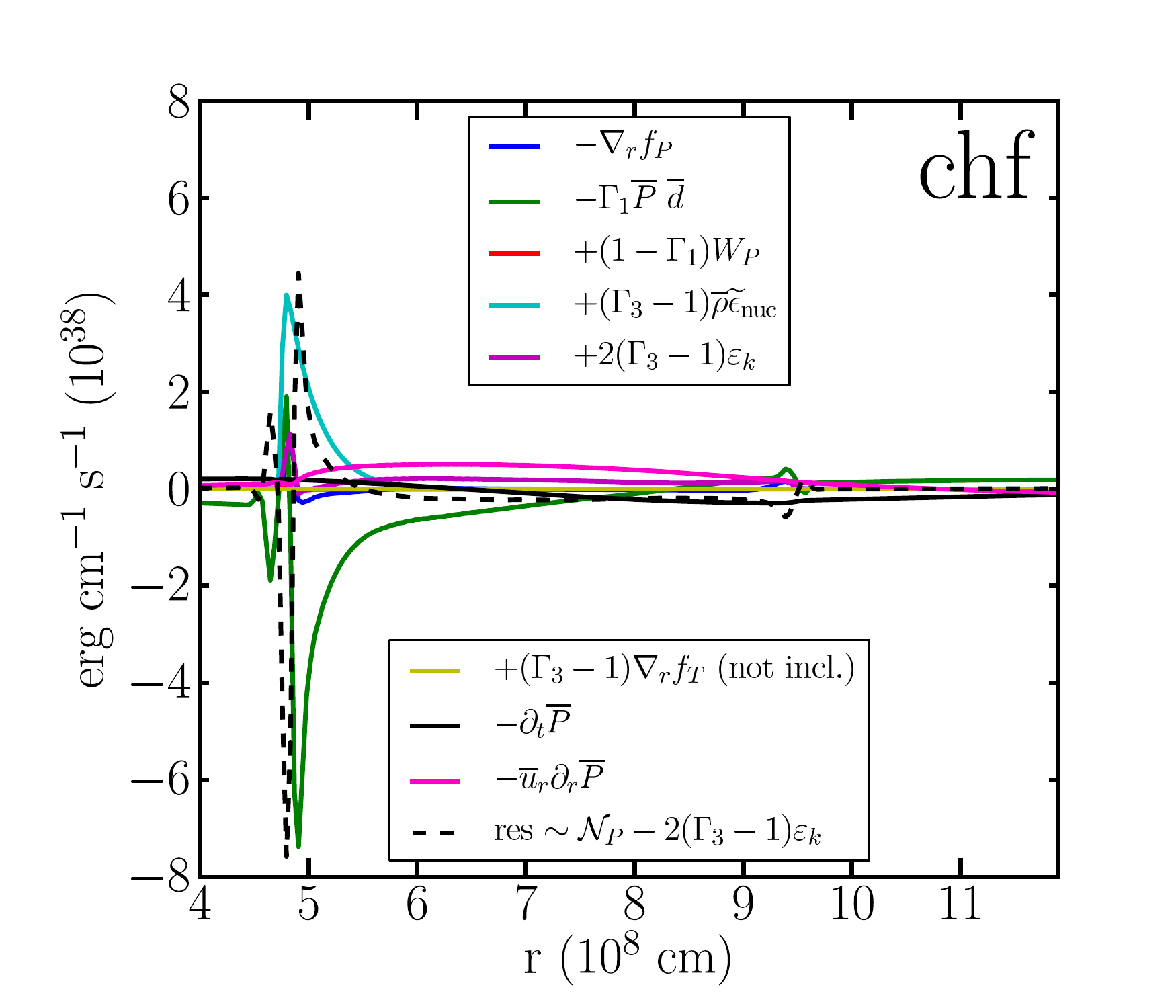}
\includegraphics[width=6.5cm]{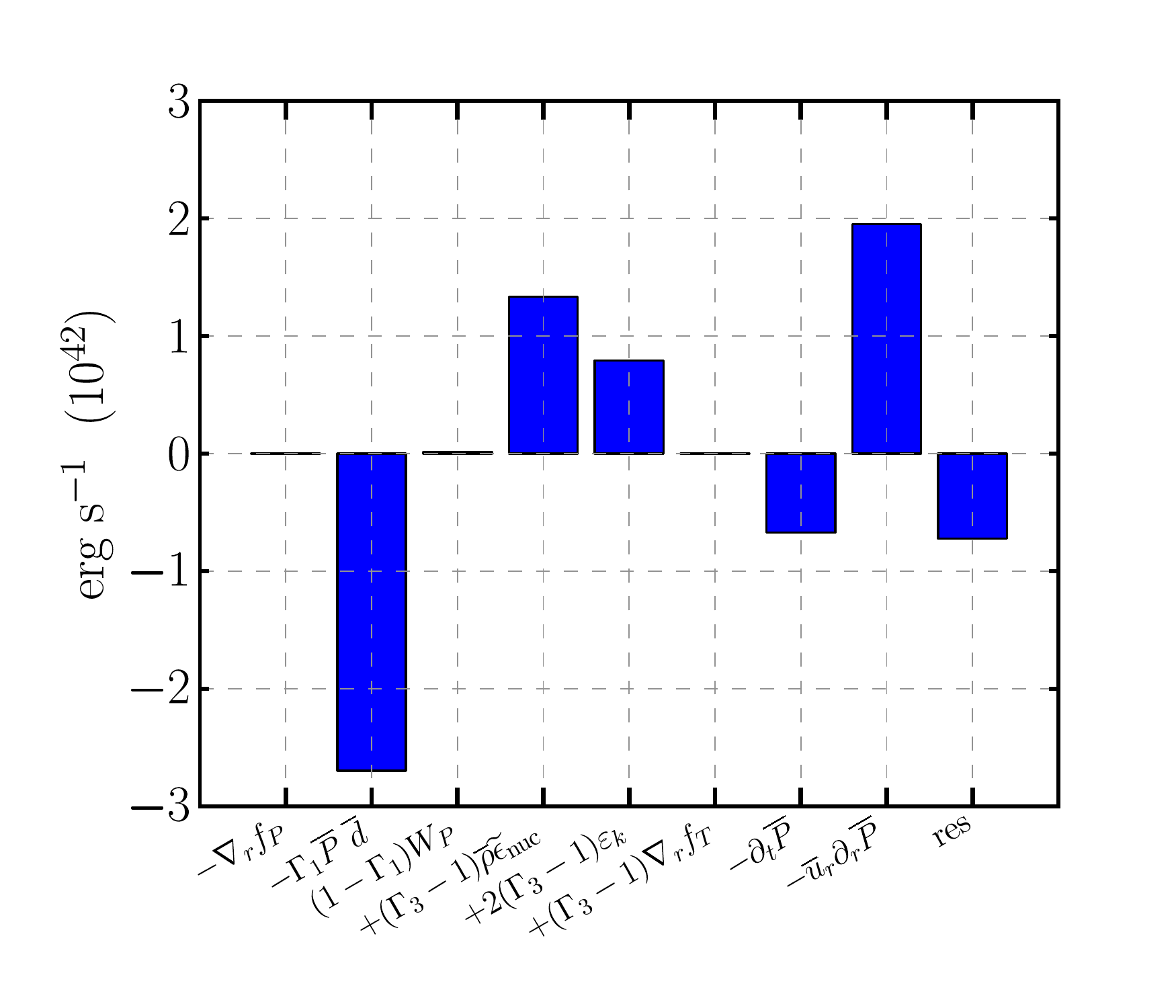}}
\caption{Mean pressure equation. Model {\sf hif.3D} (upper panels) and model {\sf chf.3D} (lower panels)}
\end{figure}

\newpage

\subsection{Mean enthalpy equation}

\begin{align}
\erho\fav{D}_t \fav{h} = & -\nabla_r f_h - \Gamma_1\eht{P} \ \eht{d} - \Gamma_1 W_P + \Gamma_3 {\mathcal S} + \Gamma_3 \nabla_r f_T +  {\mathcal N_h} \label{eq:rans_h}
\end{align}

\begin{figure}[!h]
\centerline{
\includegraphics[width=6.5cm]{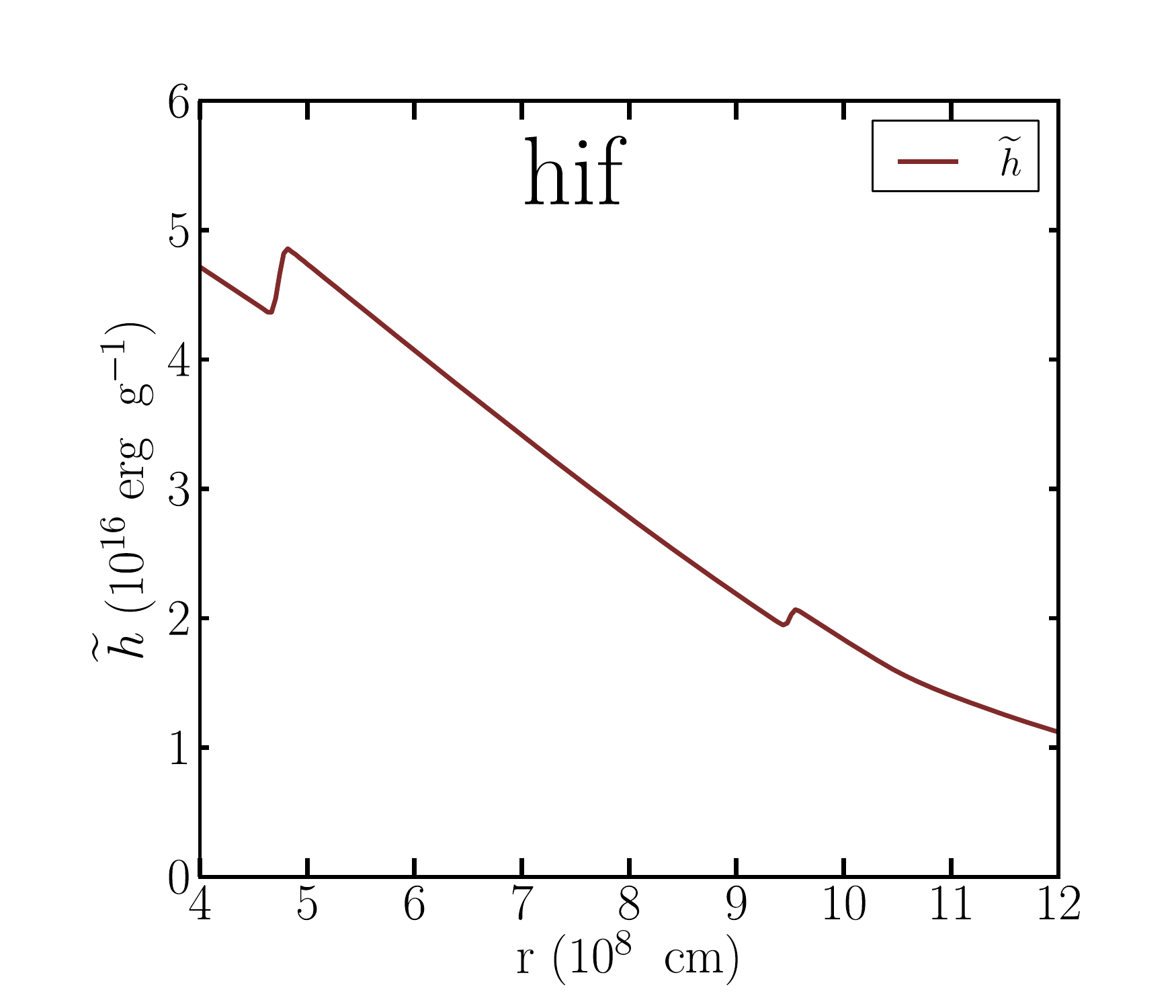}
\includegraphics[width=6.5cm]{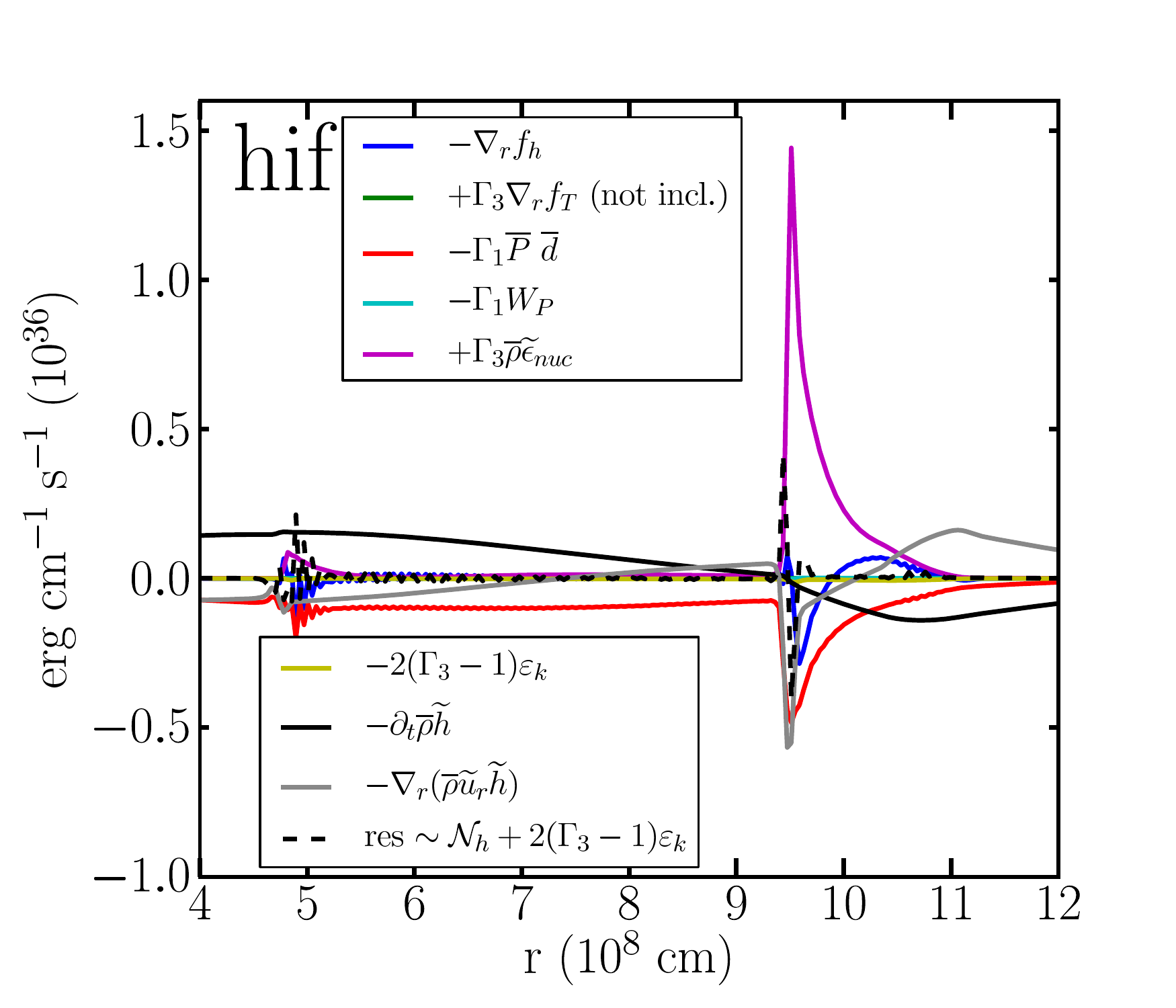}
\includegraphics[width=6.5cm]{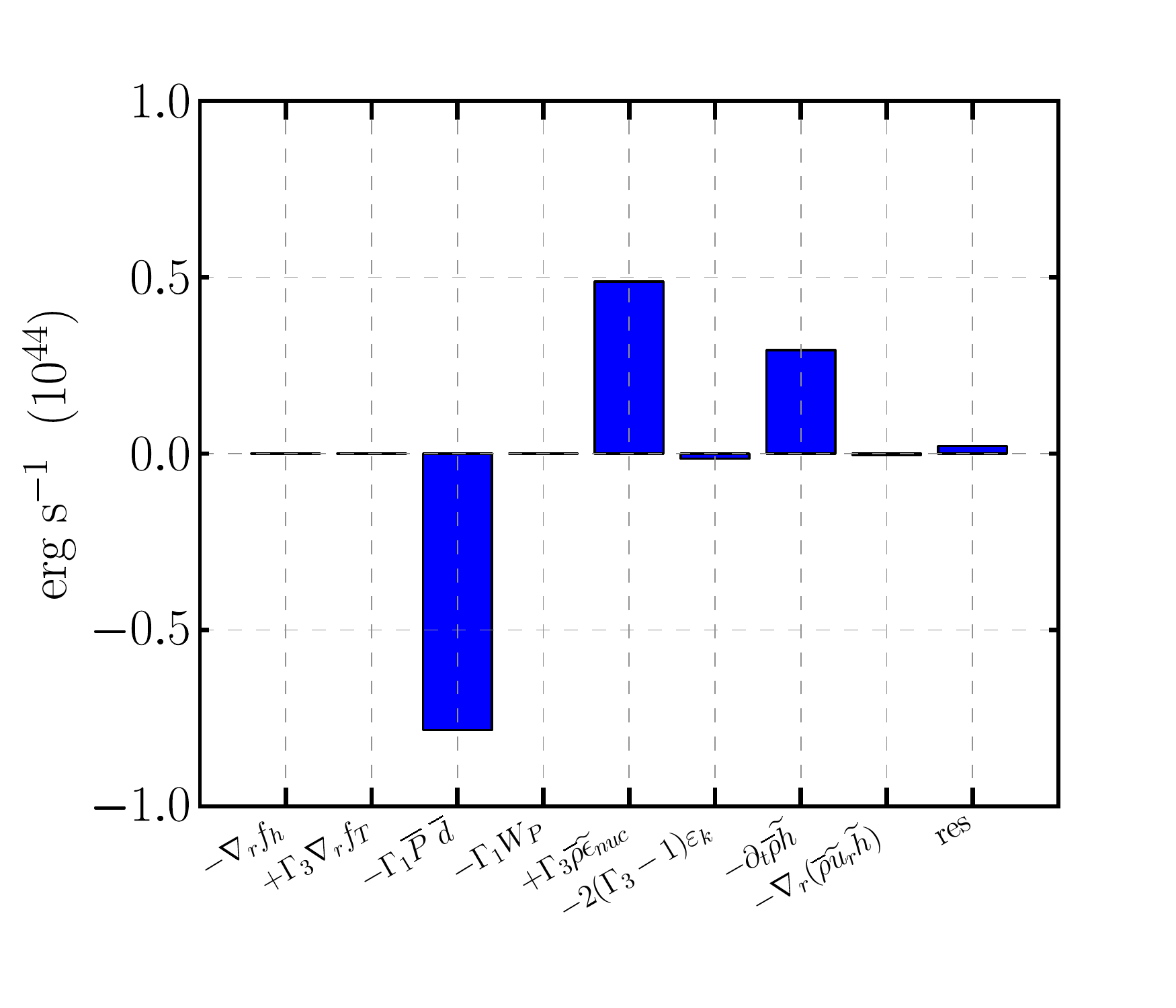}}

\centerline{
\includegraphics[width=6.5cm]{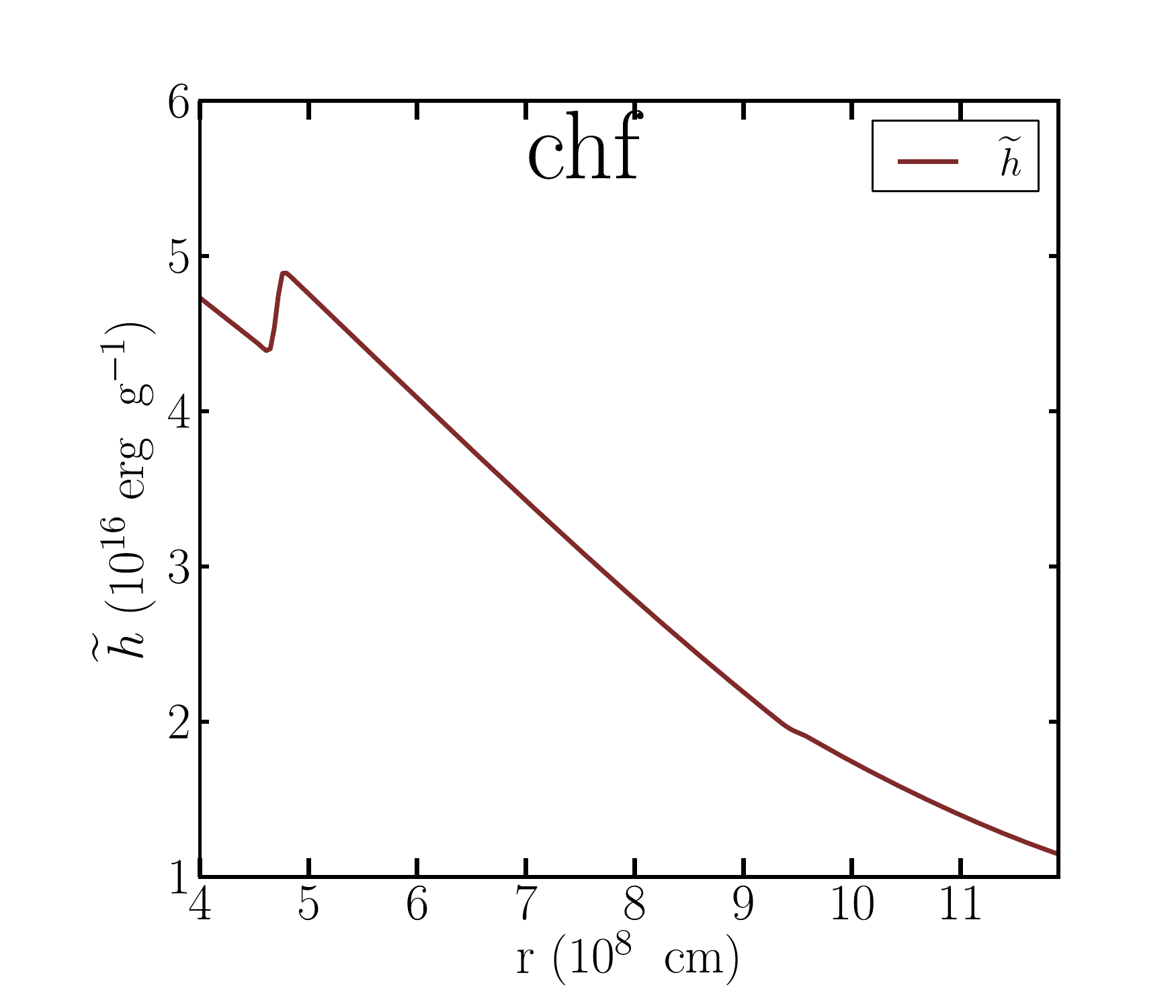}
\includegraphics[width=6.5cm]{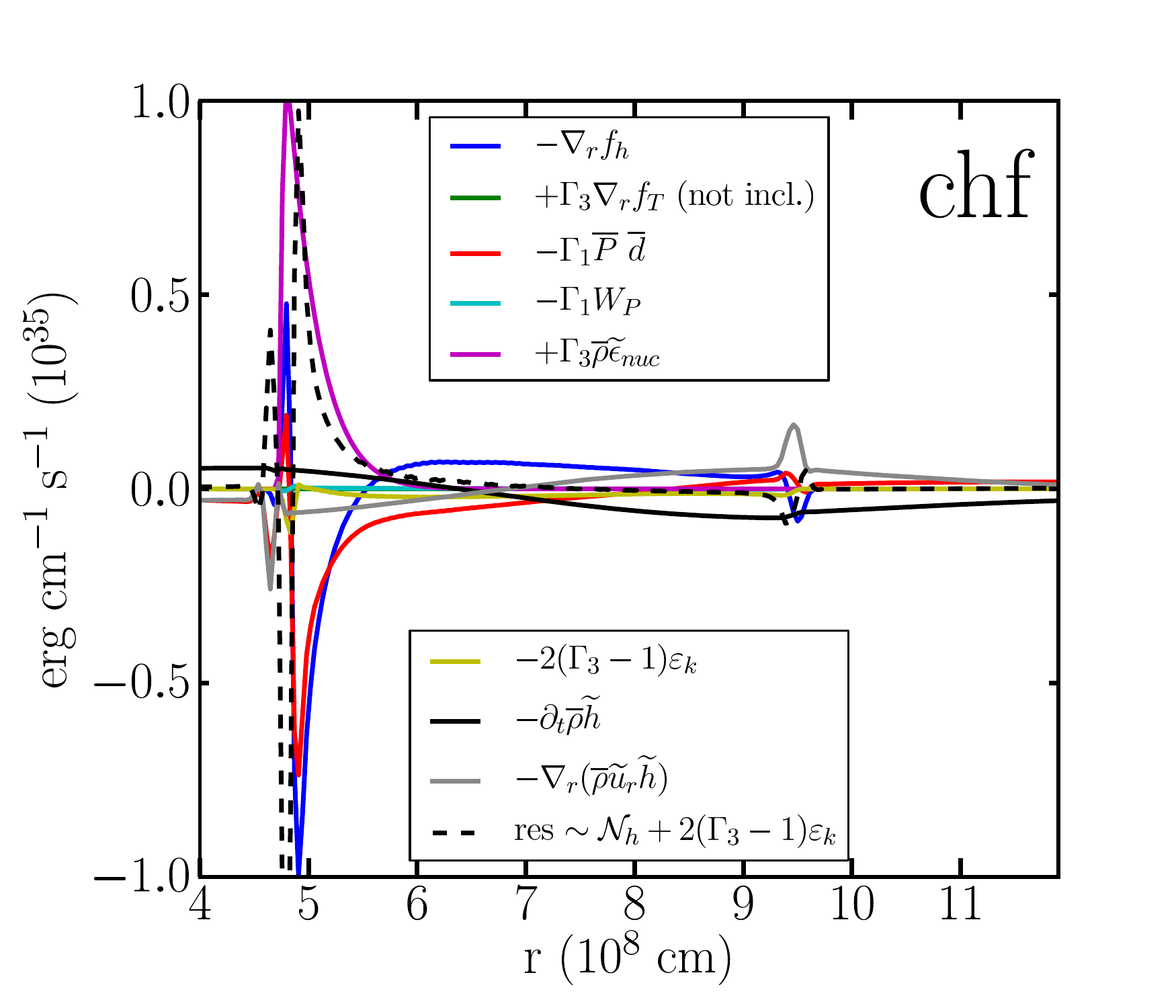}
\includegraphics[width=6.5cm]{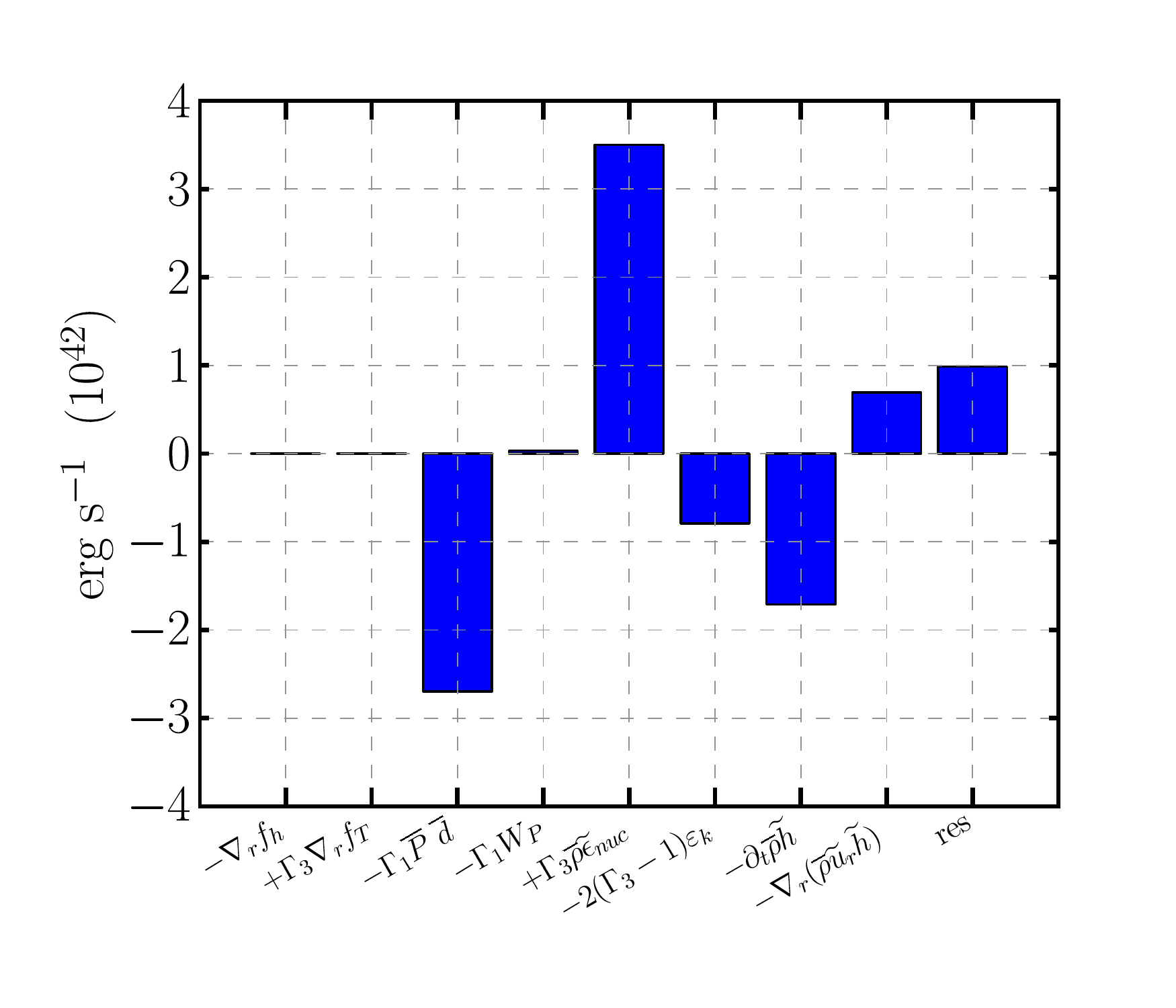}}

\caption{Mean enthalpy equation. Model {\sf hif.3D} (upper panels) and model {\sf chf.3D} (lower panels)}
\end{figure}

\newpage

\subsection{Mean angular momentum equation (z-component)}

\begin{align}
\erho\fav{D}_t \fav{j}_z = & -\nabla_r f_{jz} + {\mathcal N_{jz}} 
\end{align}

\begin{figure}[!h]
\centerline{
\includegraphics[width=6.5cm]{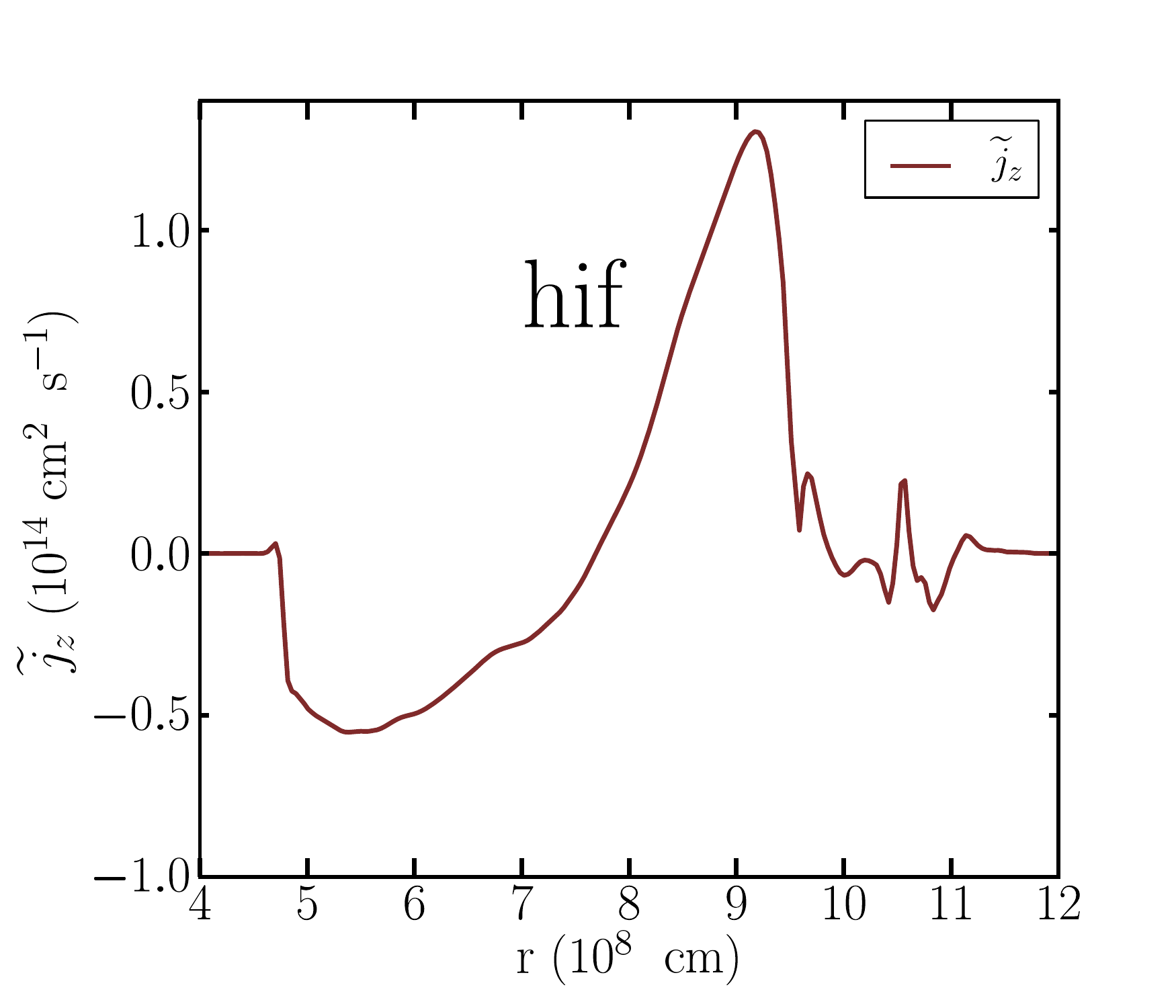}
\includegraphics[width=6.5cm]{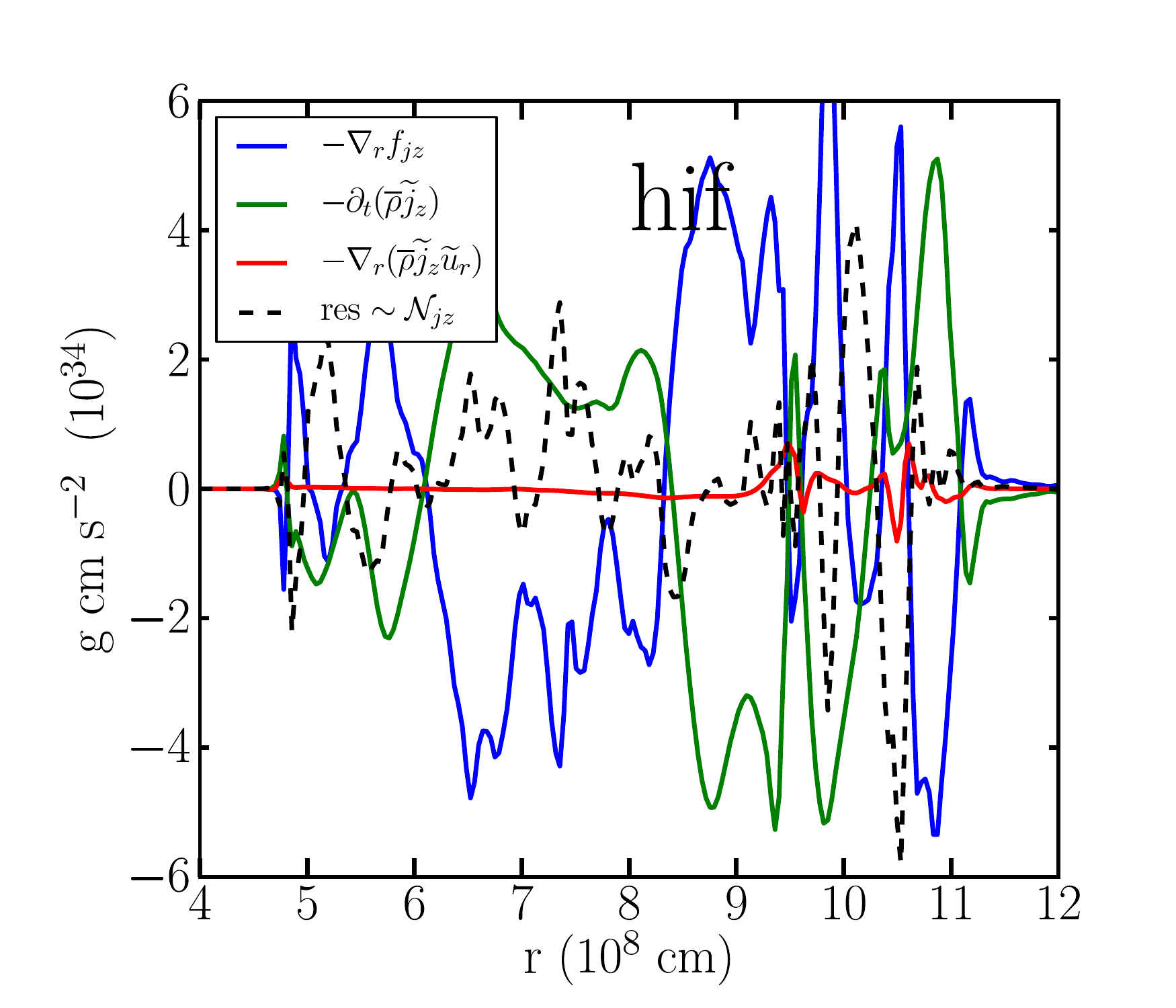}
\includegraphics[width=6.5cm]{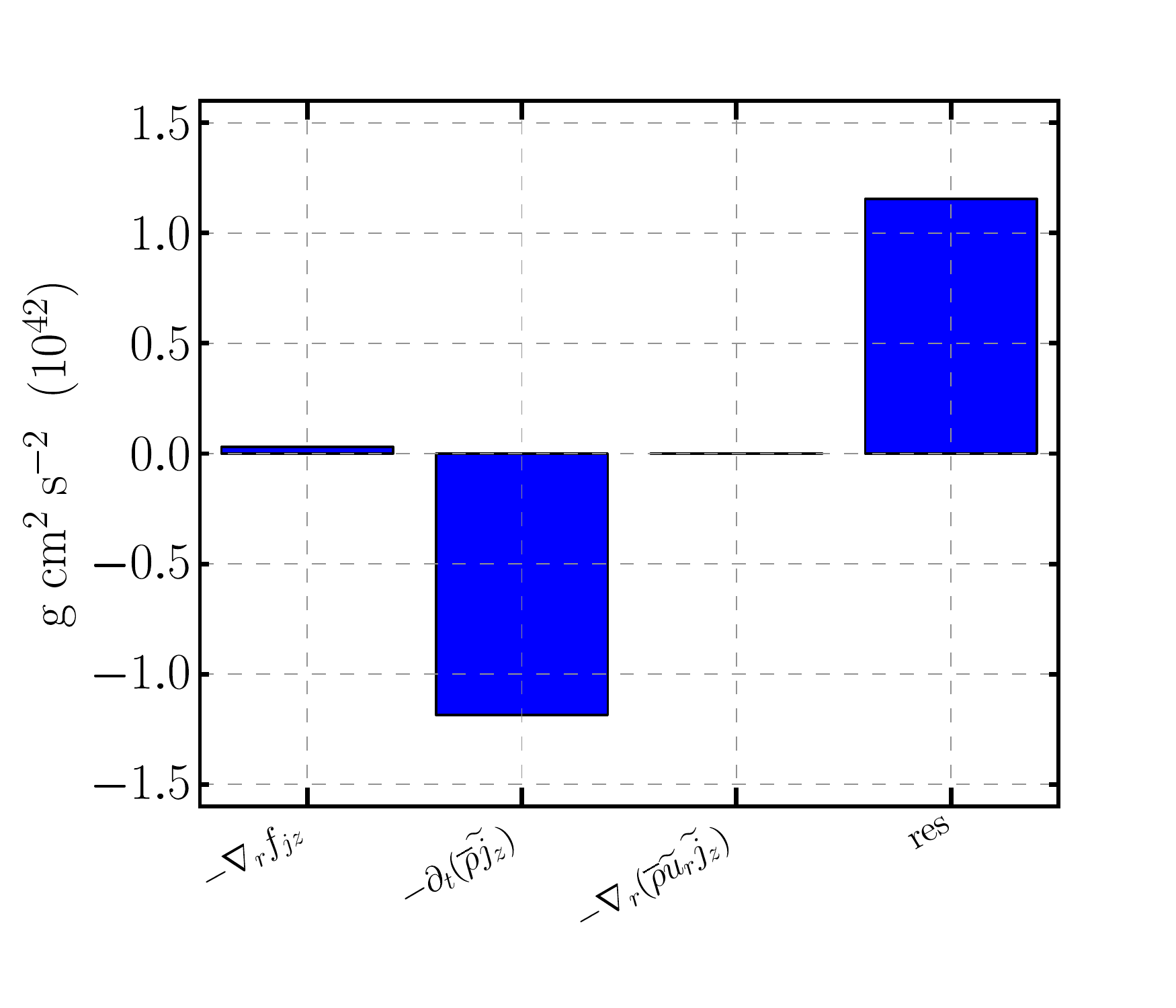}}

\centerline{
\includegraphics[width=6.5cm]{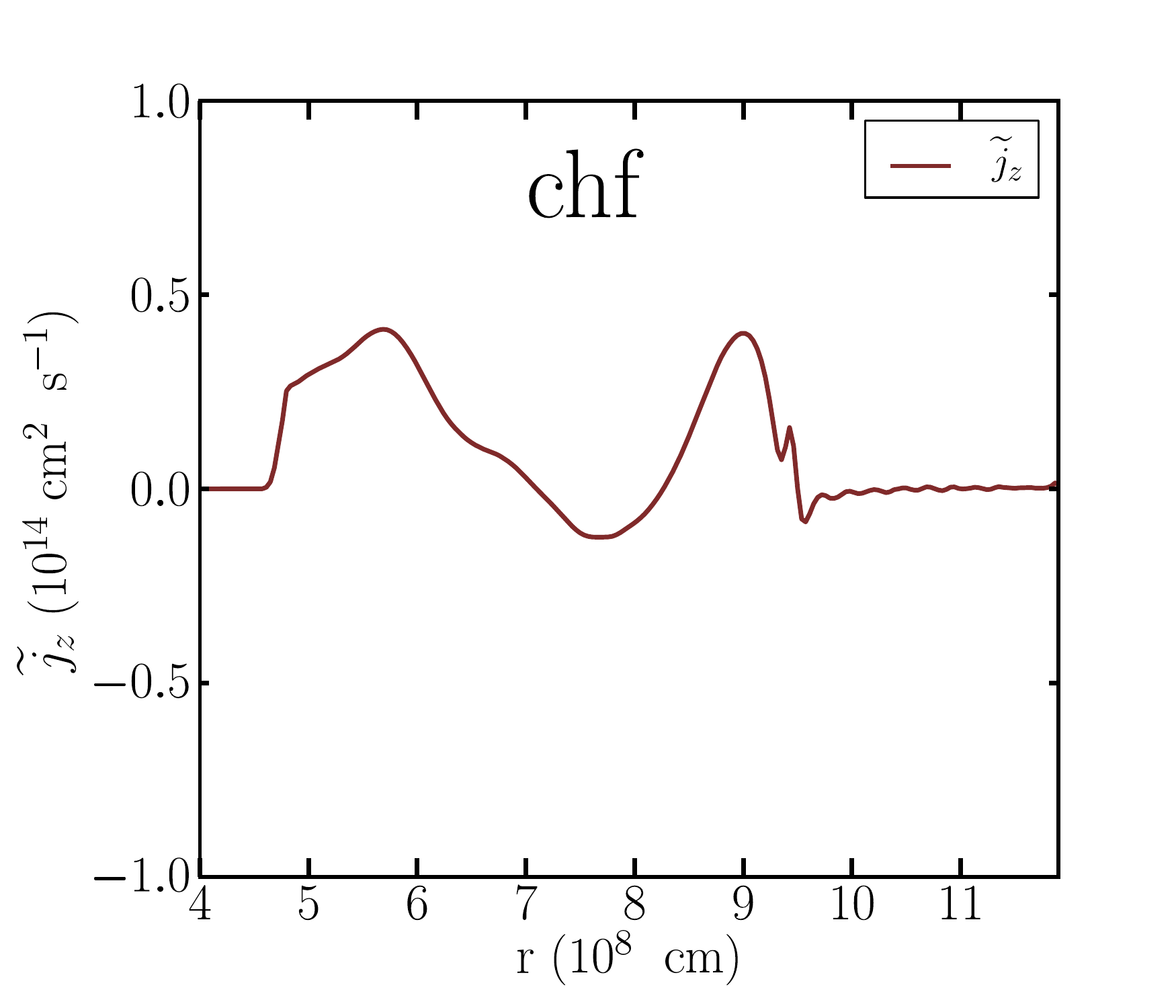}
\includegraphics[width=6.5cm]{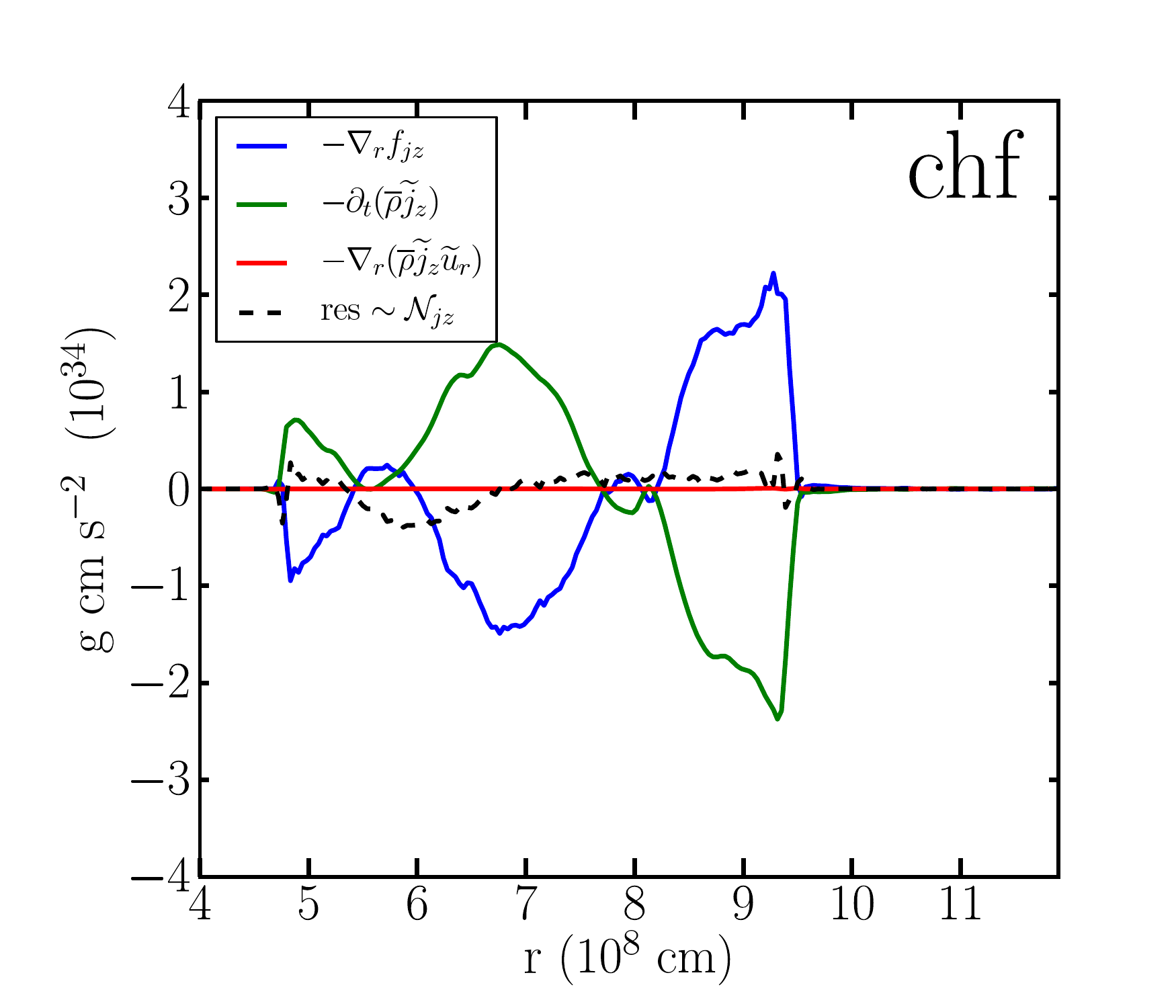}
\includegraphics[width=6.5cm]{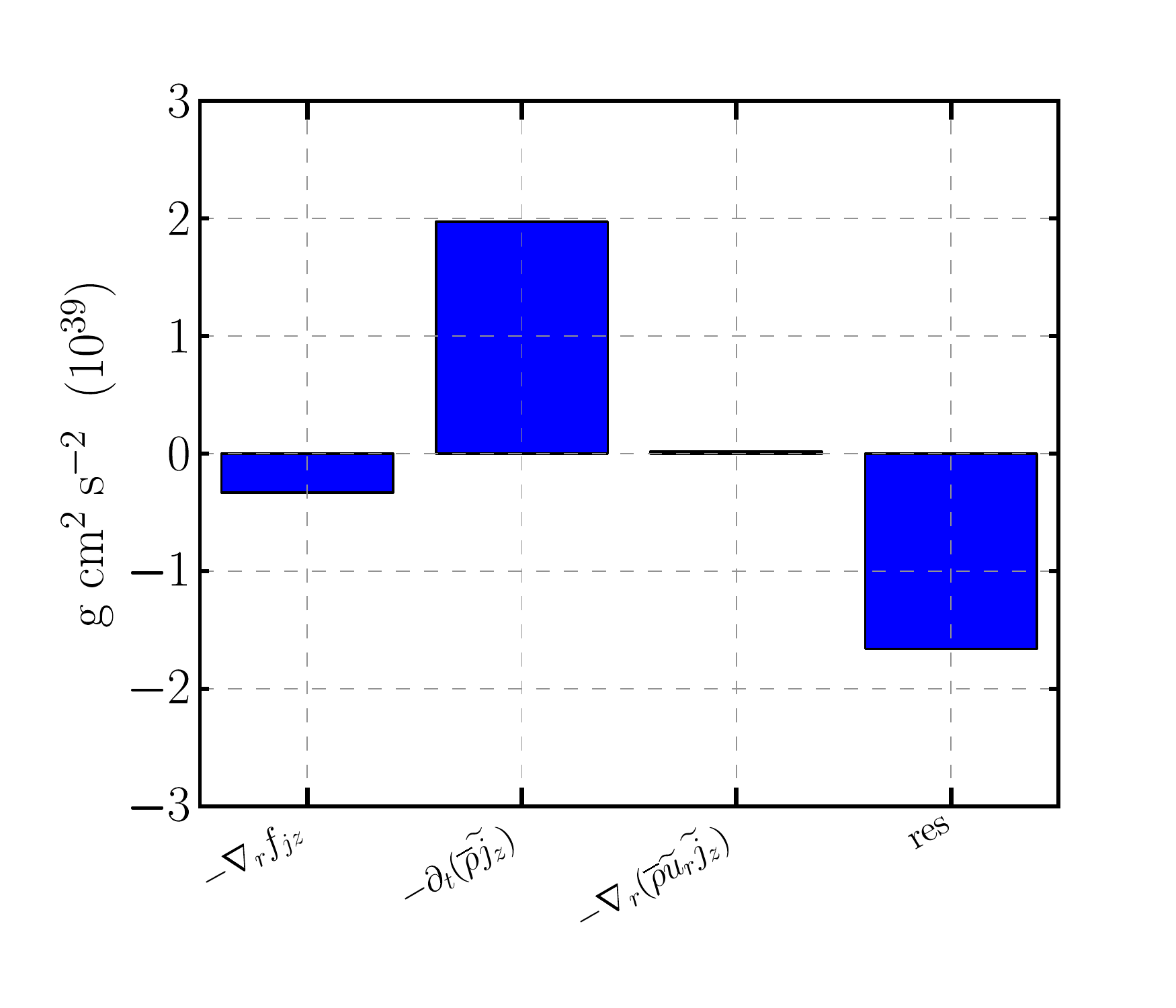}}

\caption{Mean angular momentum equation (z-component). Model {\sf hif.3D} (upper panels) and model {\sf chf.3D} (lower panels)}
\end{figure}

\newpage

\subsection{Mean composition equations}

\begin{align}
\erho\fav{D}_t \fav{X}_\alpha = & -\nabla_r f_\alpha + \av{\rho}\fav{\dot{X}}_\alpha^{\rm nuc} + {\mathcal N_\alpha} 
\end{align}

\begin{figure}[!h]
\centerline{
\includegraphics[width=6.5cm]{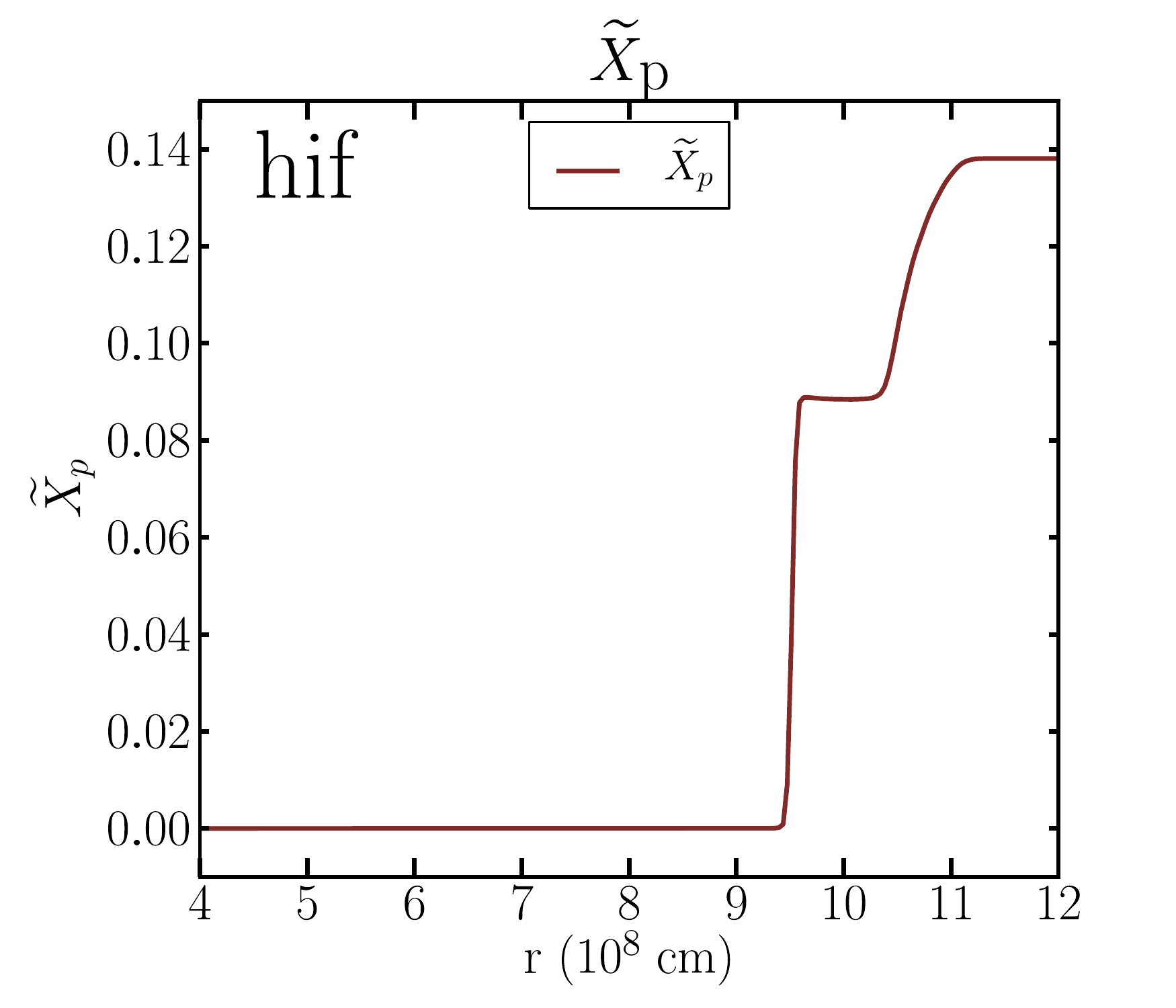}
\includegraphics[width=6.5cm]{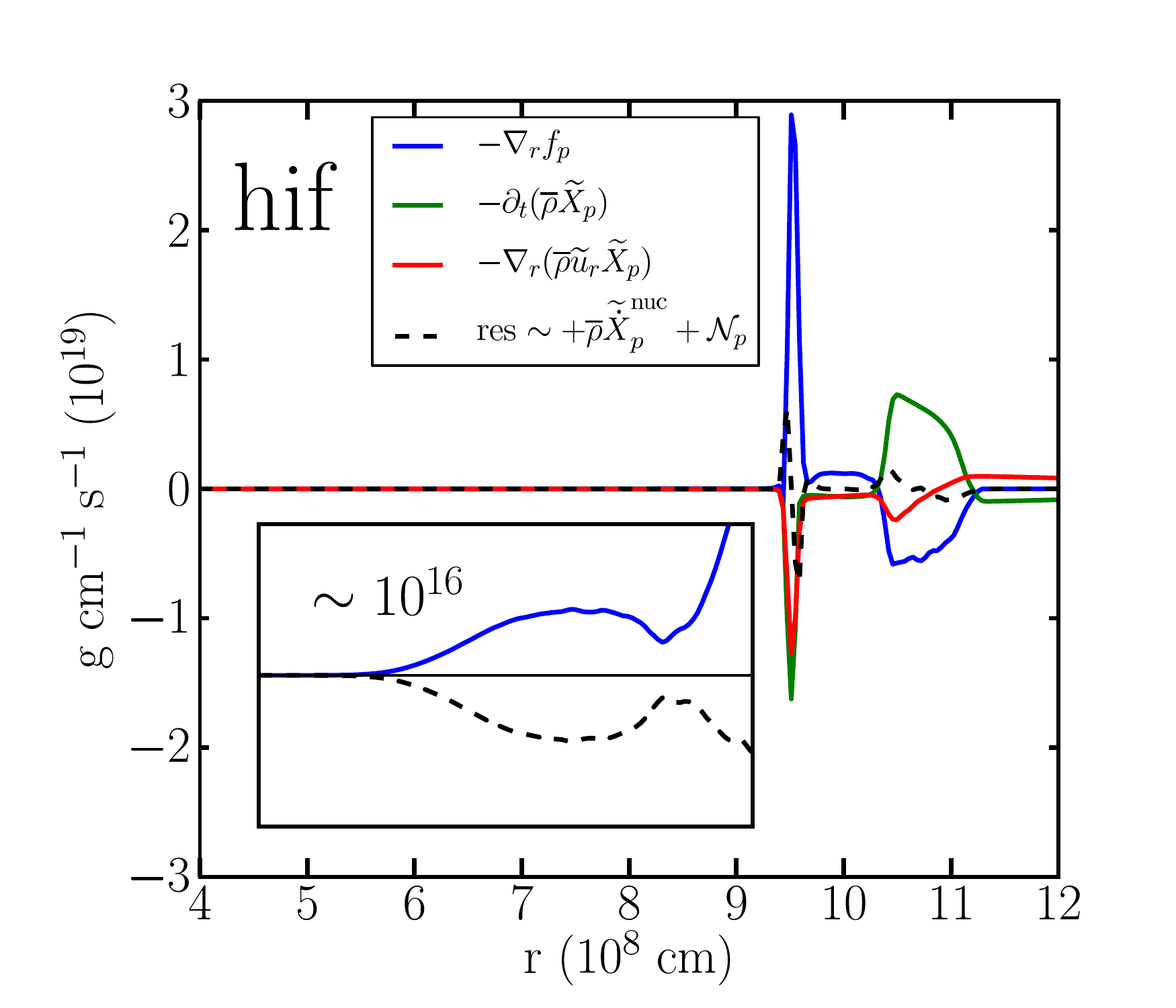}
\includegraphics[width=6.5cm]{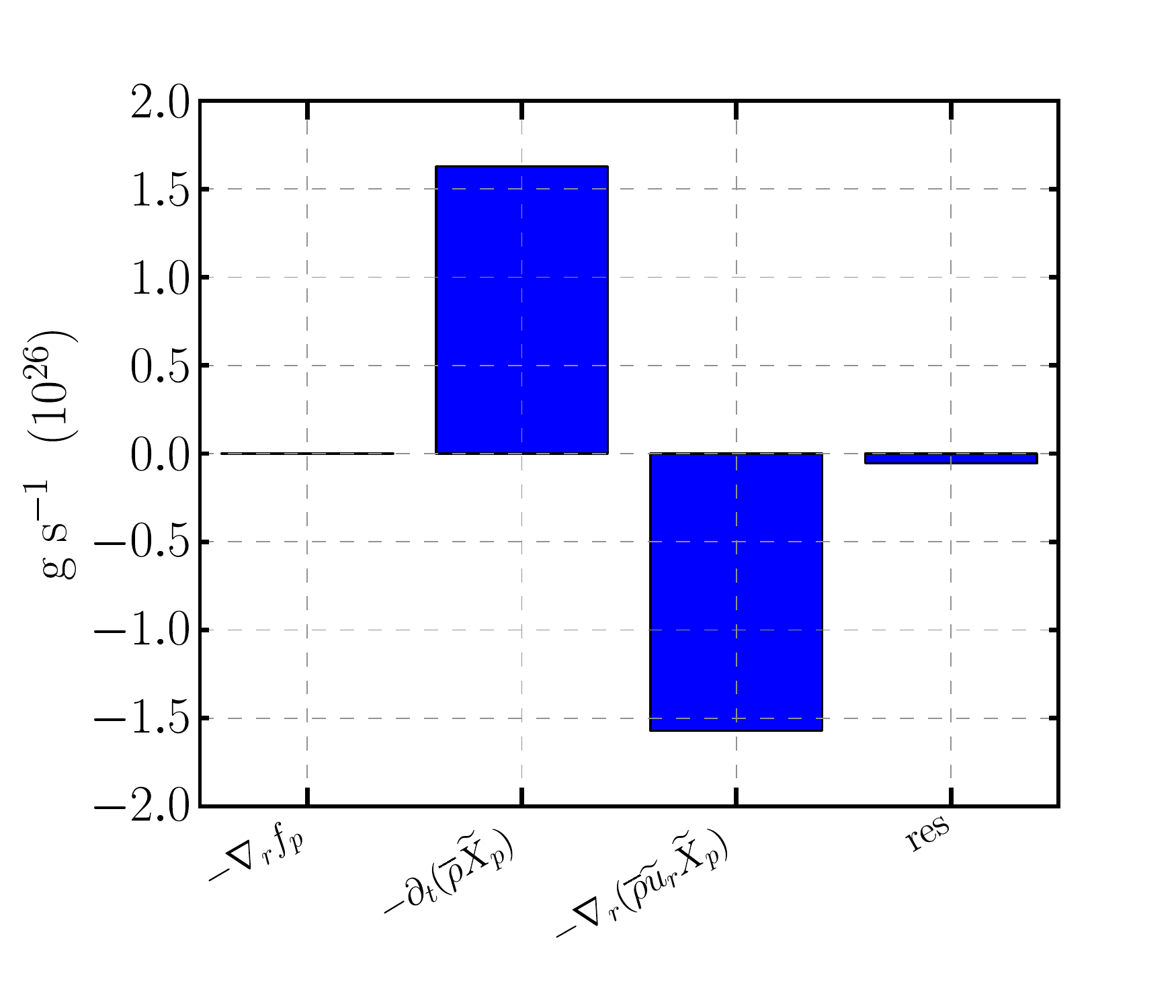}}

\centerline{
\includegraphics[width=6.5cm]{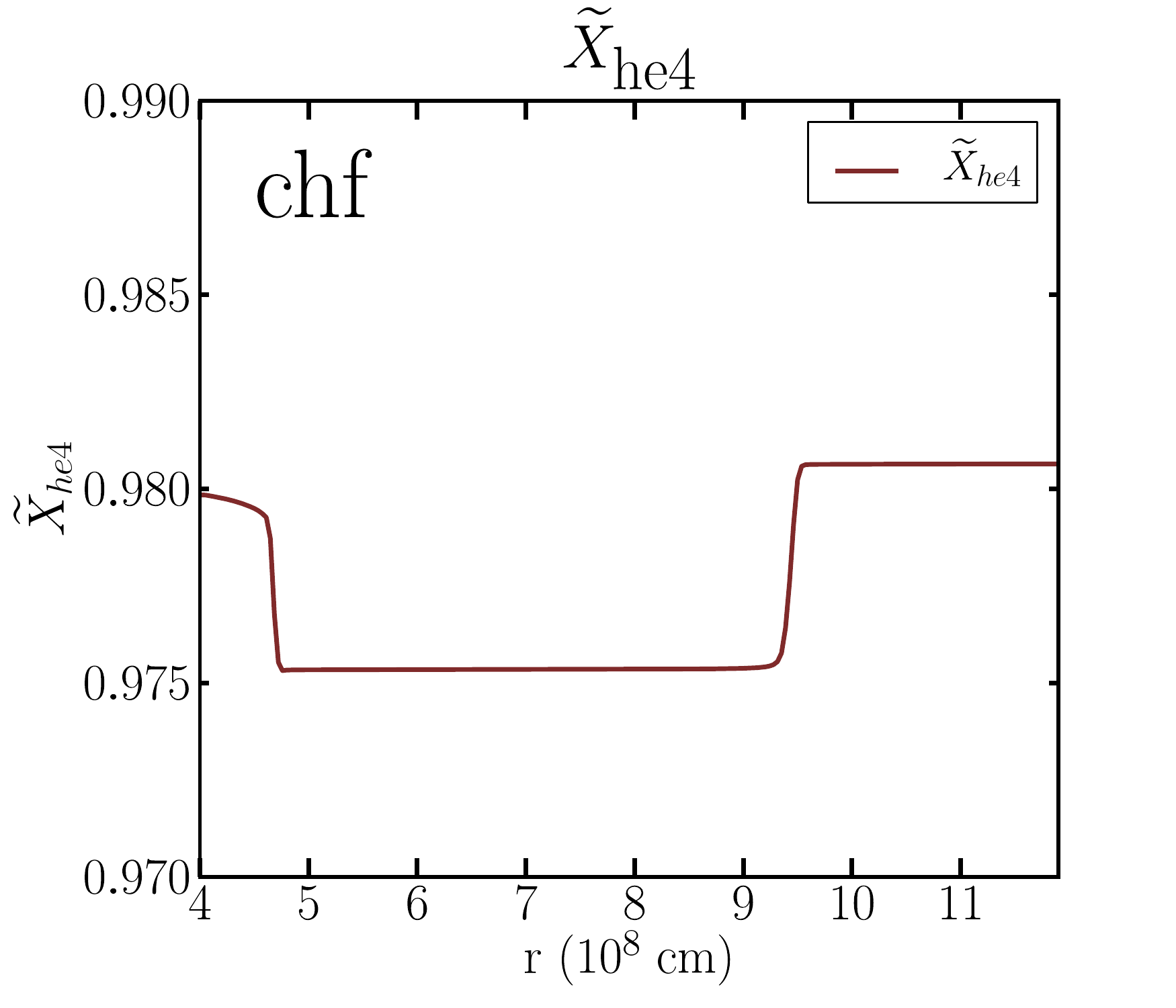}
\includegraphics[width=6.5cm]{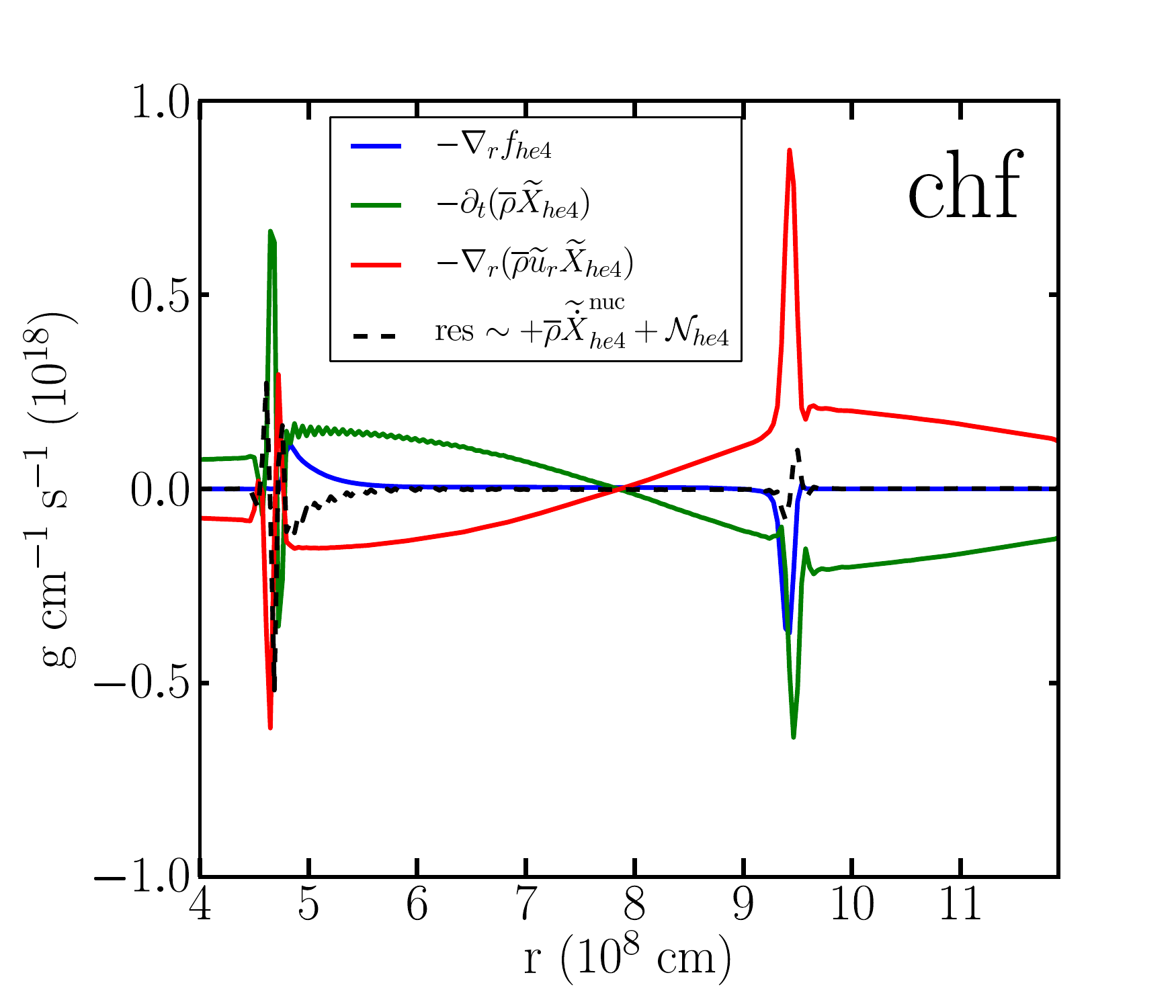}
\includegraphics[width=6.5cm]{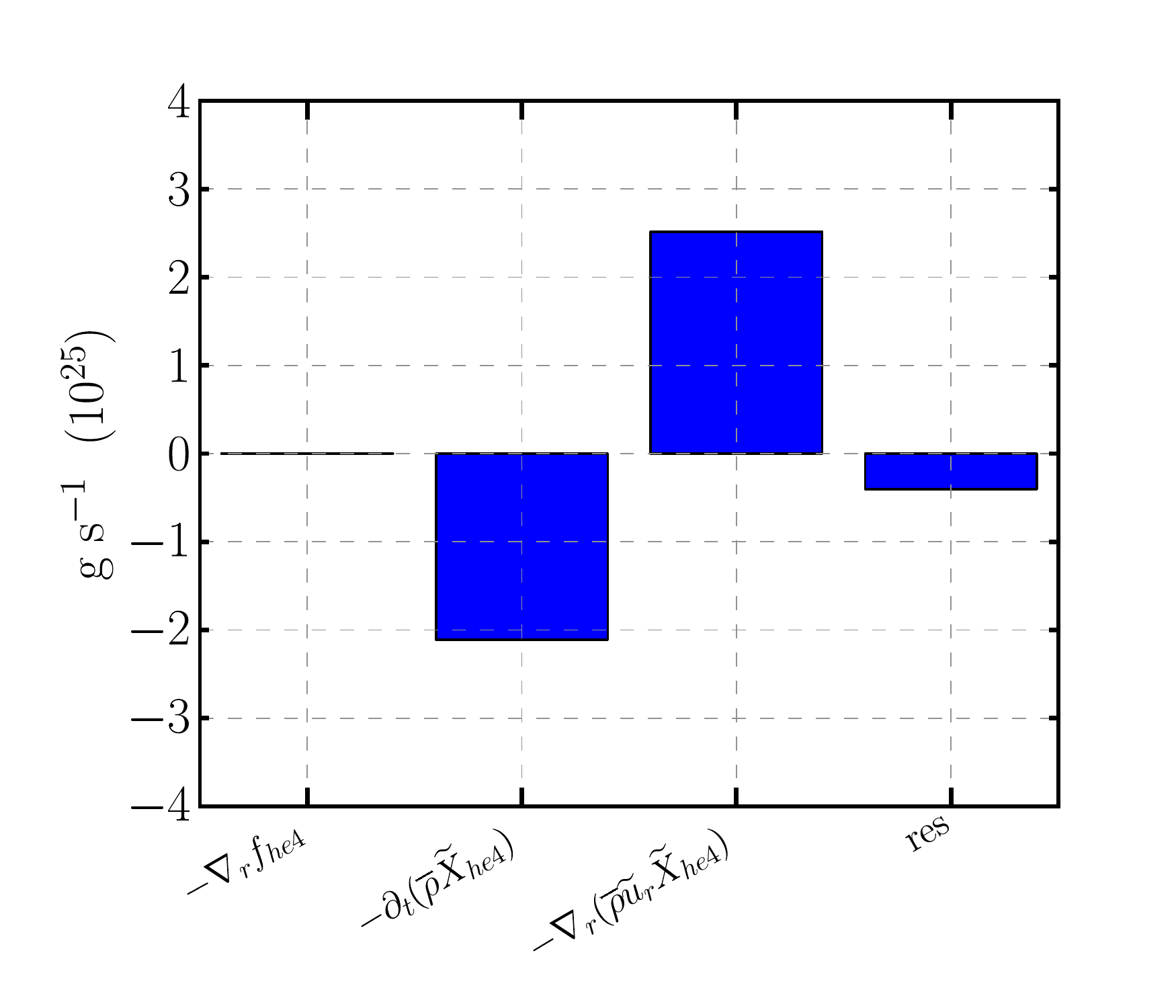}}
\caption{Mean composition equations. Model {\sf hif.3D} (upper panels) and model {\sf chf.3D} (lower panels)}
\end{figure}

\newpage

\subsection{Mean turbulent kinetic energy equation}

\begin{align}
\av{\rho} \fav{D}_t \fav{k}^{ } = & -\nabla_r ( f_k +  f_P ) - \fht{R}_{ir}\partial_r \fht{u}_i + W_b + W_P + {\mathcal N_k}   
\end{align}

\begin{figure}[!h]
\centerline{
\includegraphics[width=6.5cm]{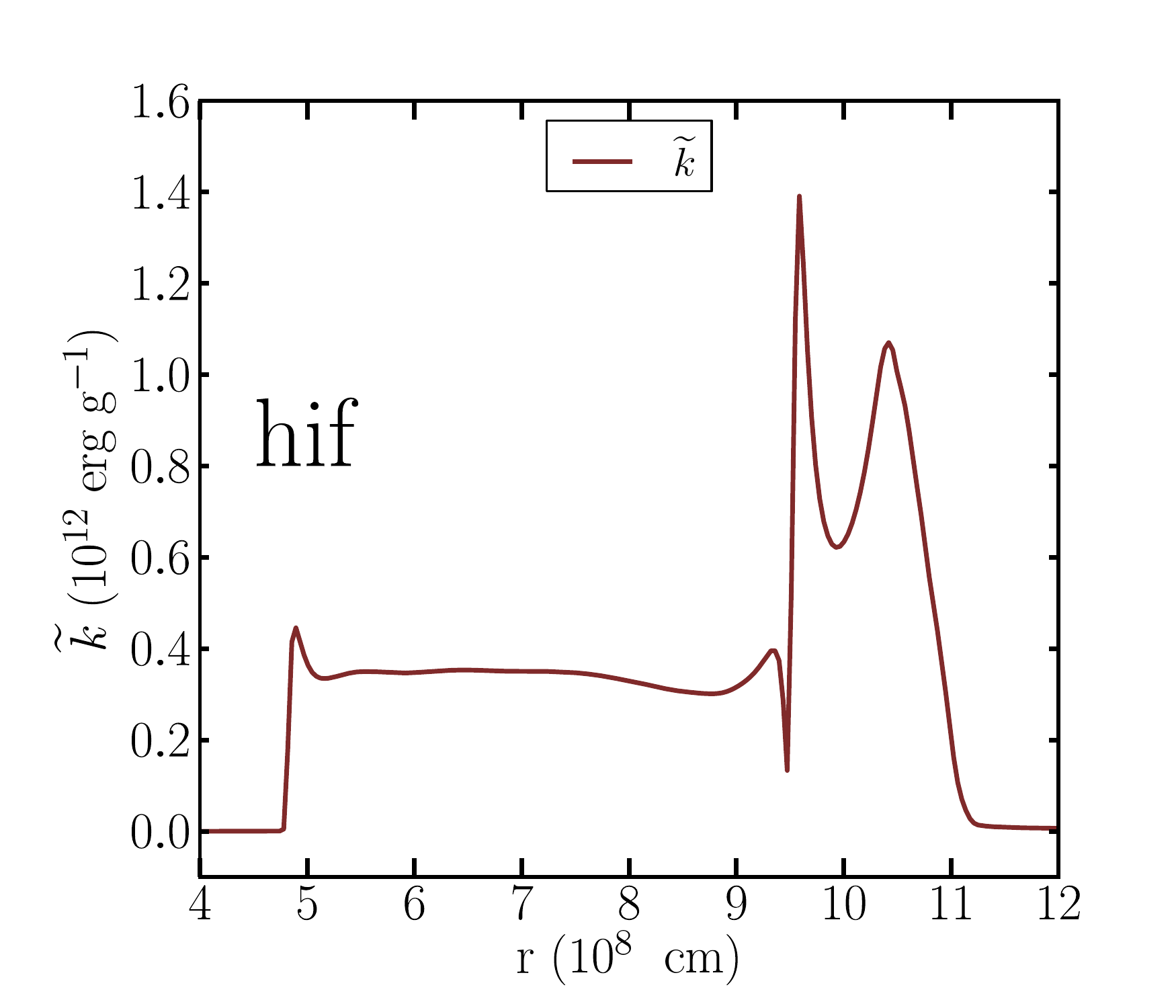}
\includegraphics[width=6.5cm]{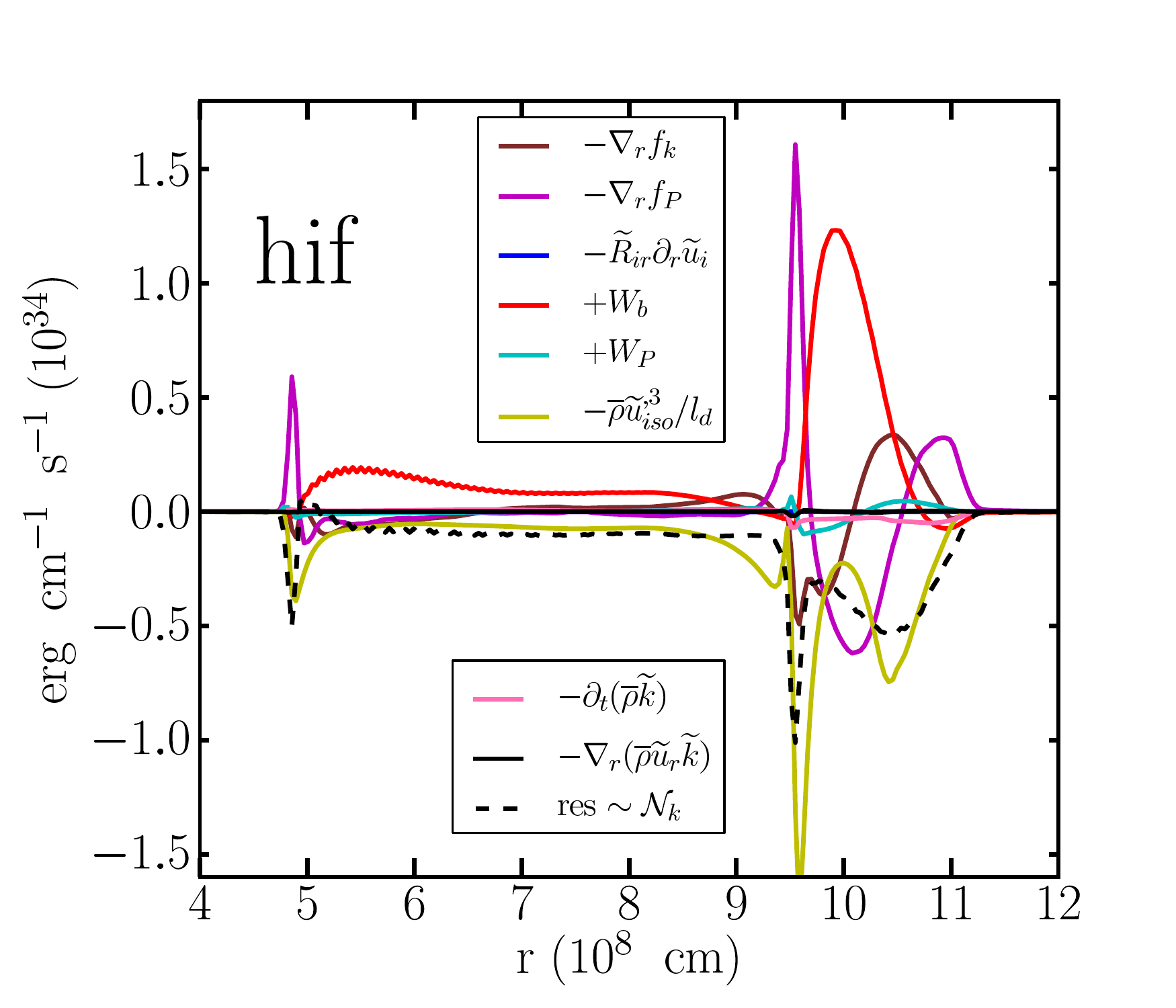}
\includegraphics[width=6.5cm]{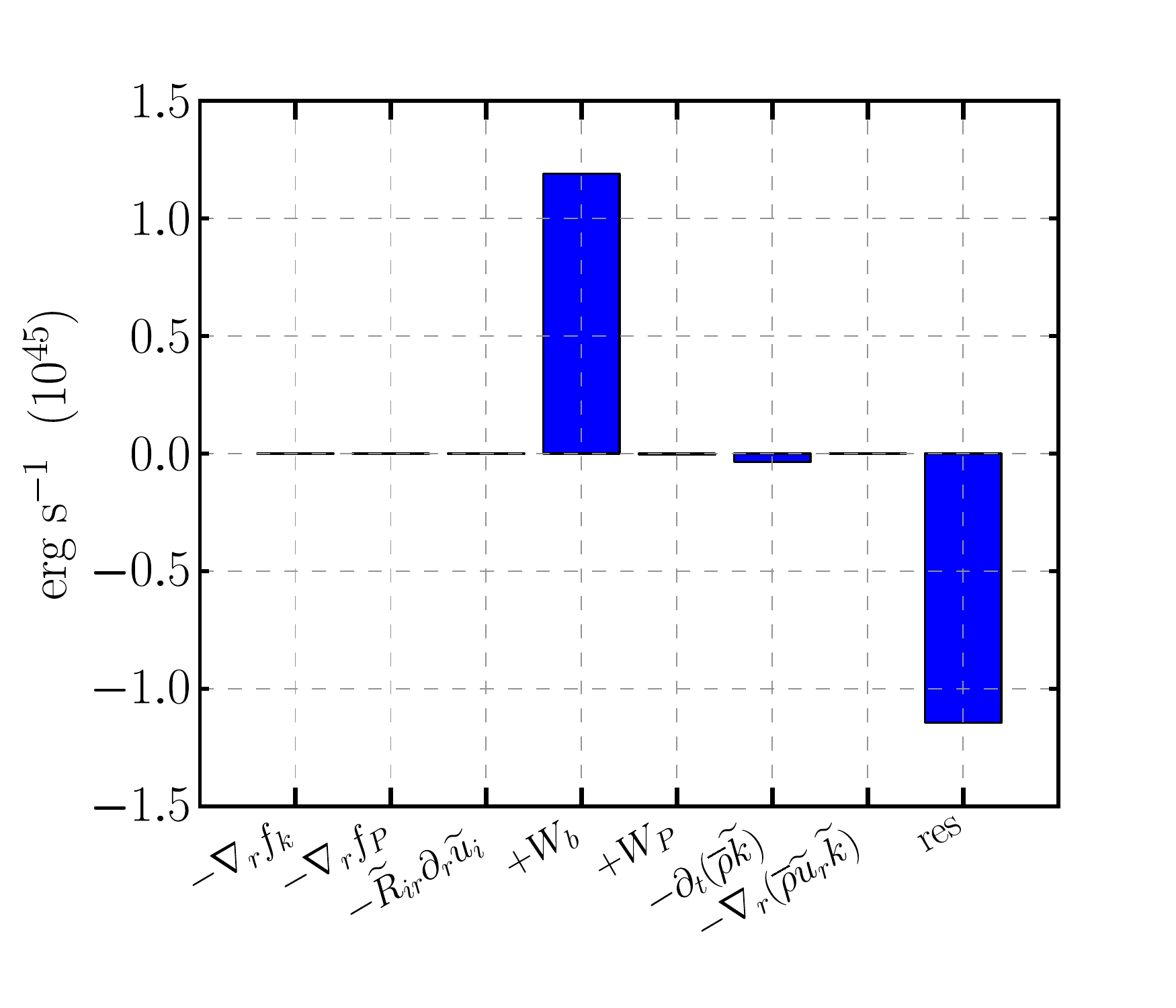}}

\centerline{
\includegraphics[width=6.5cm]{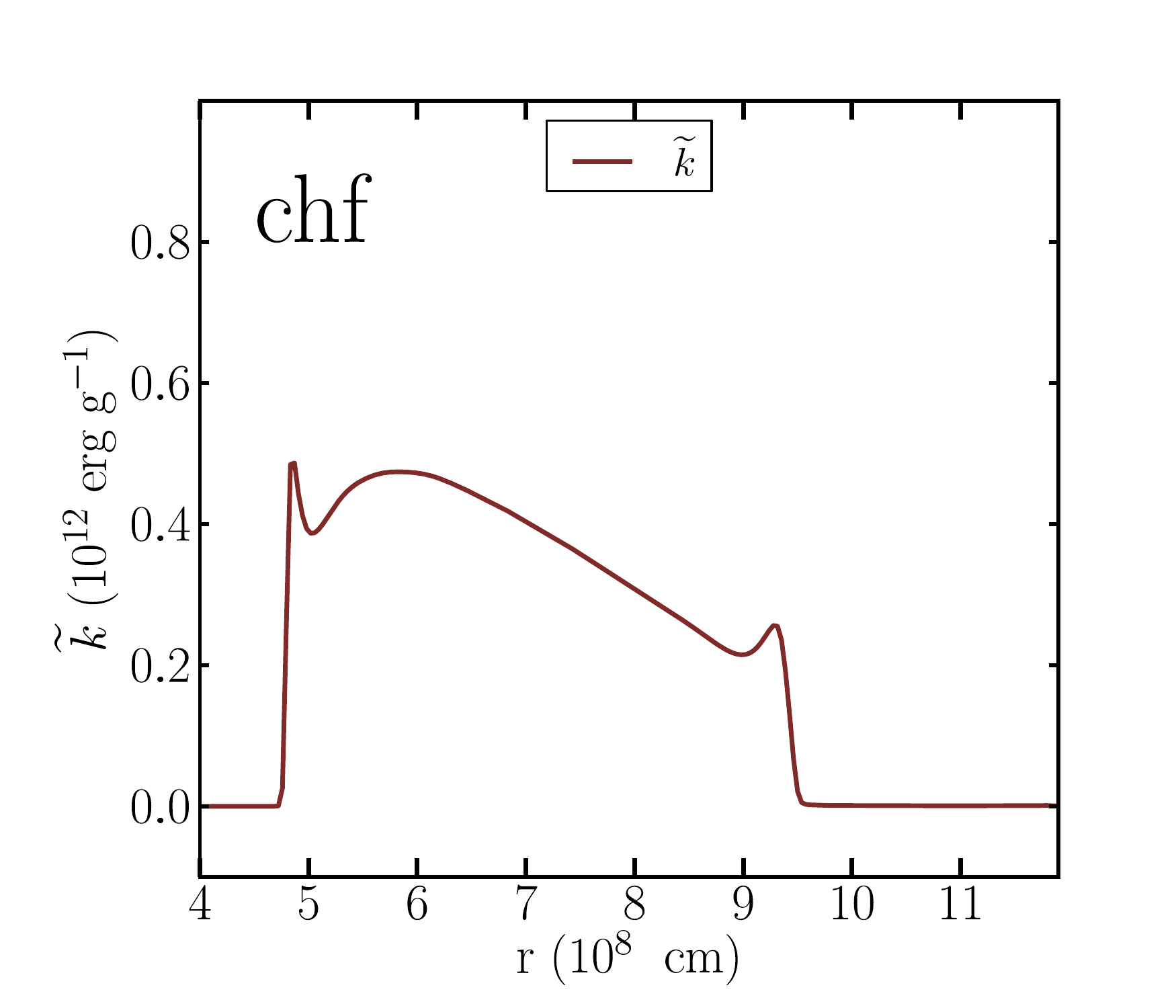}
\includegraphics[width=6.5cm]{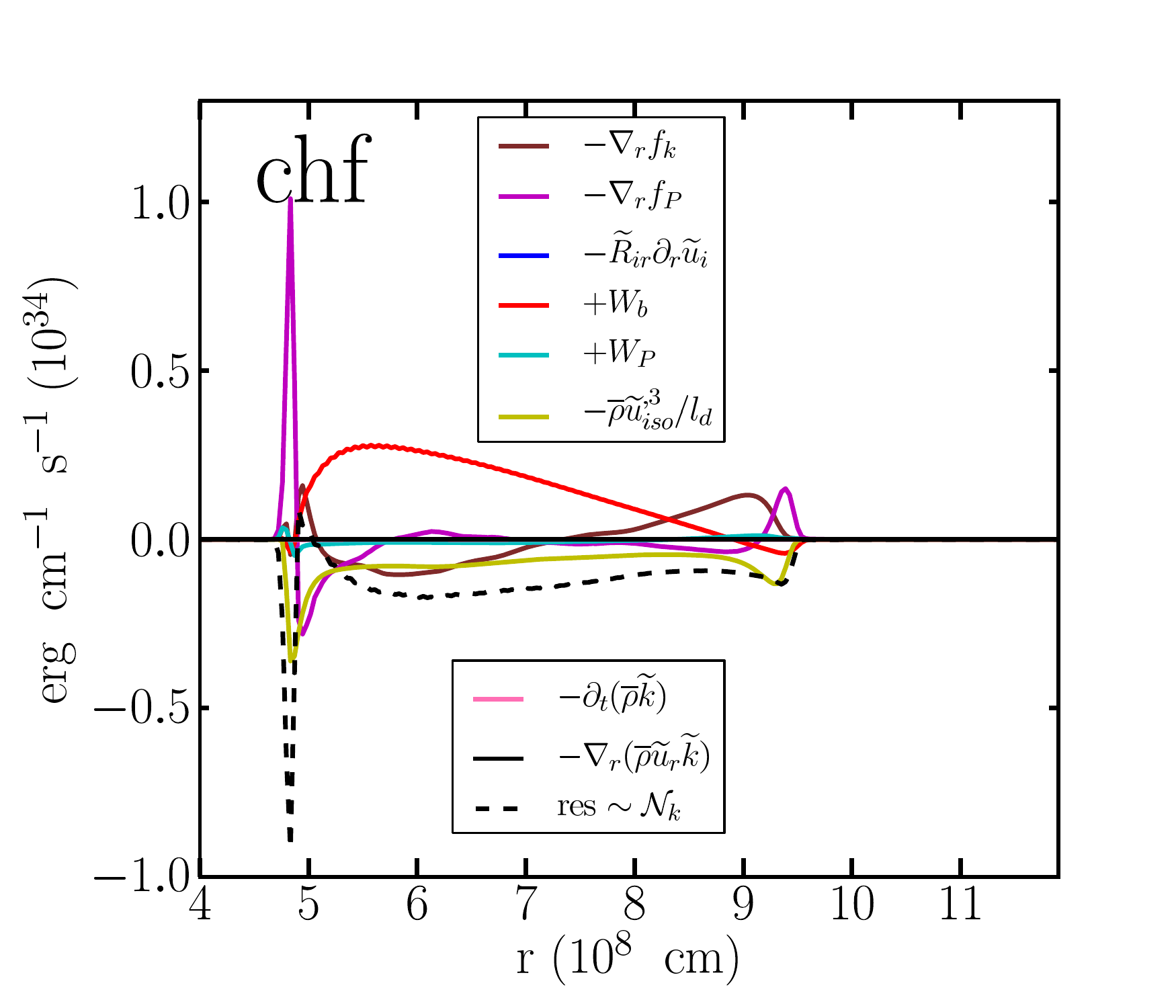}
\includegraphics[width=6.5cm]{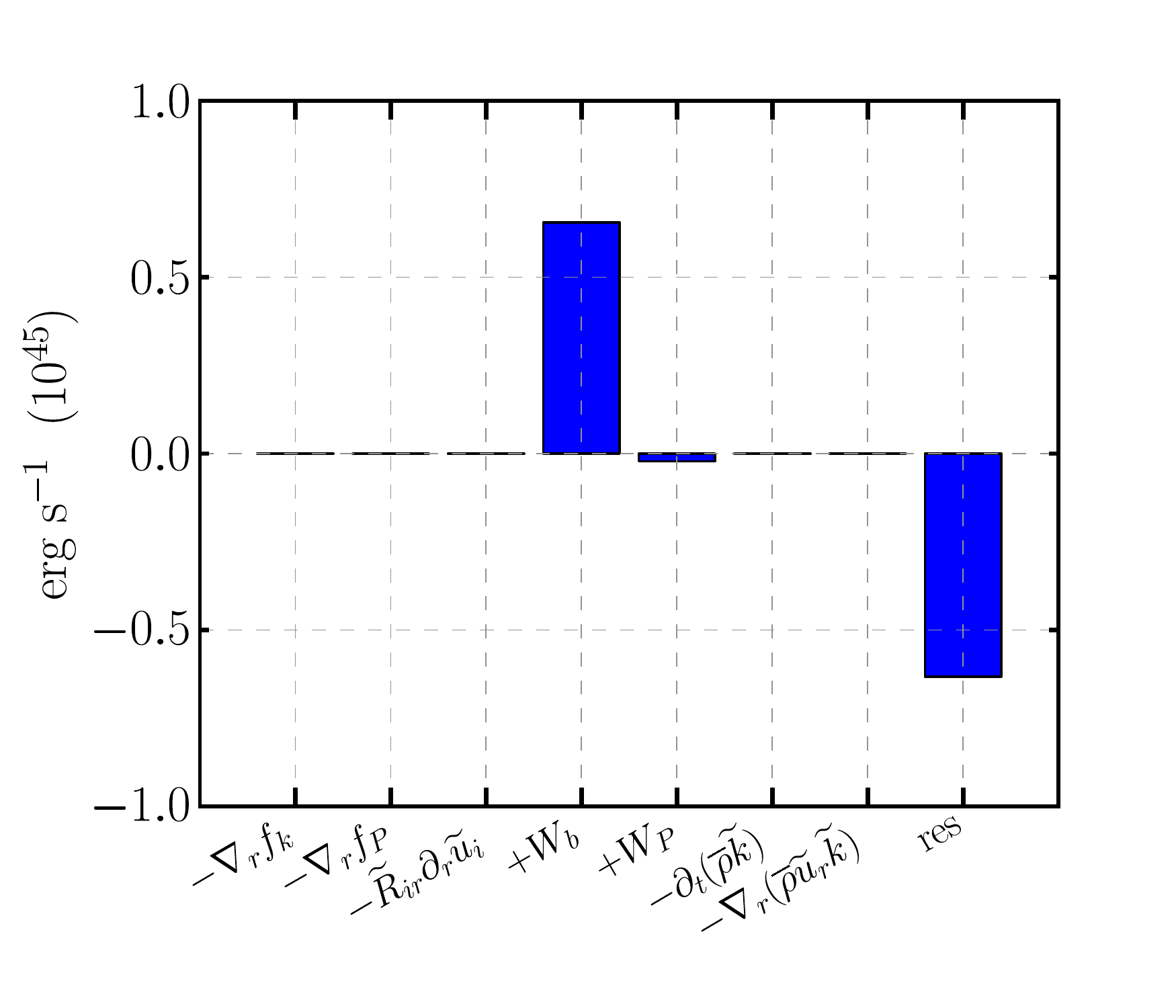}}

\caption{Mean turbulent kinetic energy equation. Model {\sf hif.3D} (upper panels) and model {\sf chf.3D} (lower panels)}
\end{figure}

\newpage

\subsection{Radial part of mean turbulent kinetic energy equation}

\begin{align}
\av{\rho} \fav{D}_t \fav{k}^r =  &  -\nabla_r  ( f_k^r + f_P )  - \fht{R}_{rr}\partial_r \fht{u}_r + W_b  + \eht{P'\nabla_r u''_r} + {\mathcal G_k^r} + {\mathcal N_{kr}} 
\end{align}

\begin{figure}[!h]

\centerline{
\includegraphics[width=6.5cm]{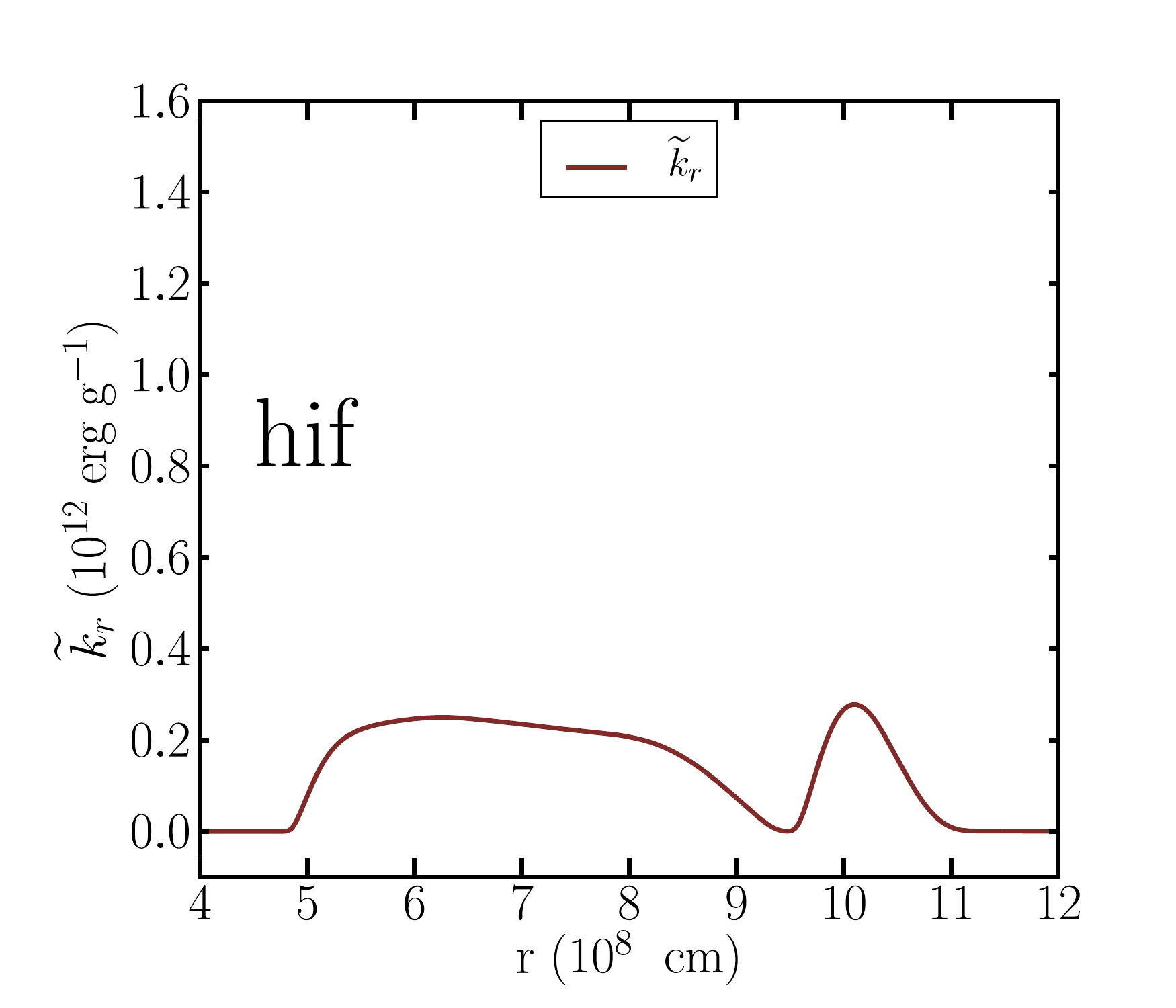}
\includegraphics[width=6.5cm]{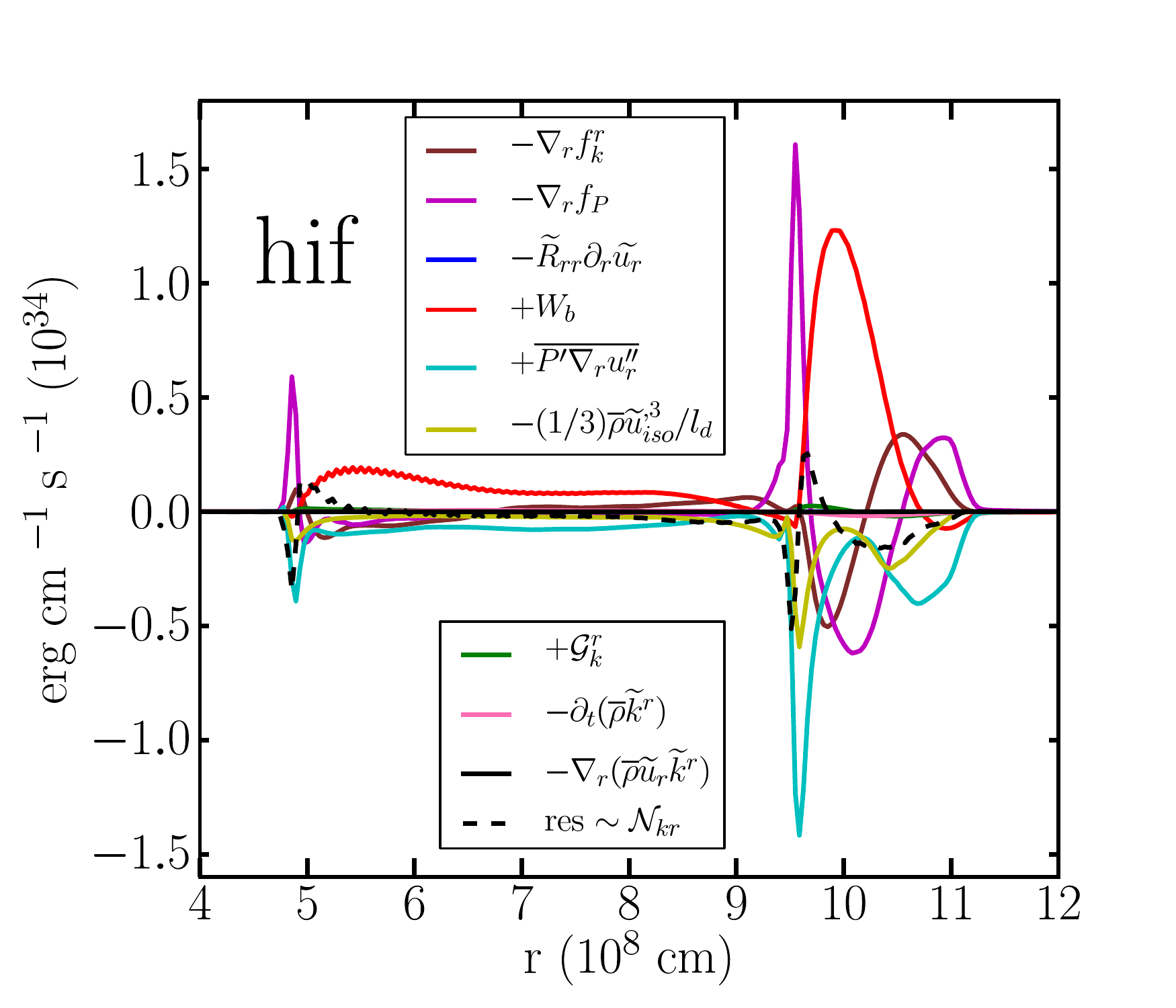}
\includegraphics[width=6.5cm]{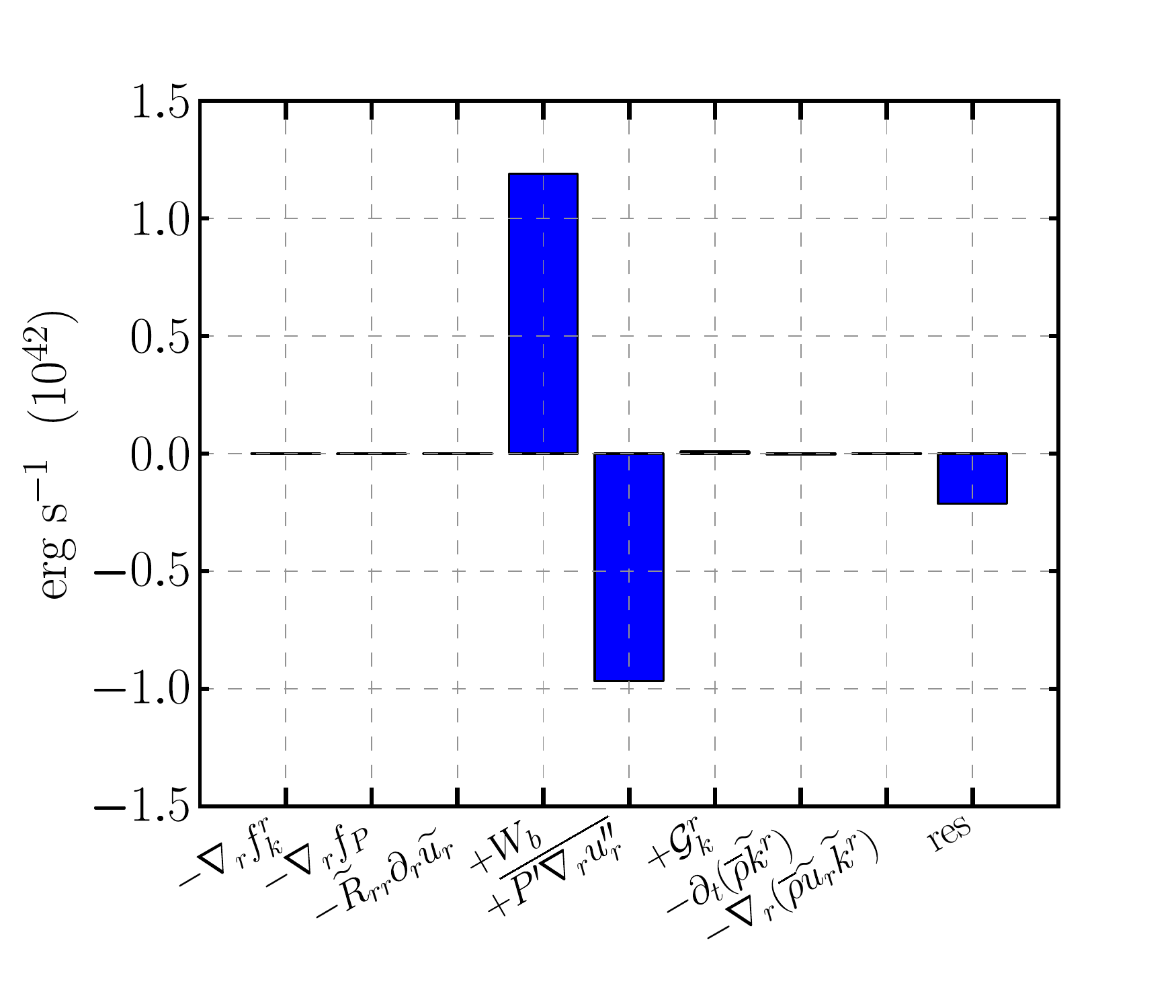}}

\centerline{
\includegraphics[width=6.5cm]{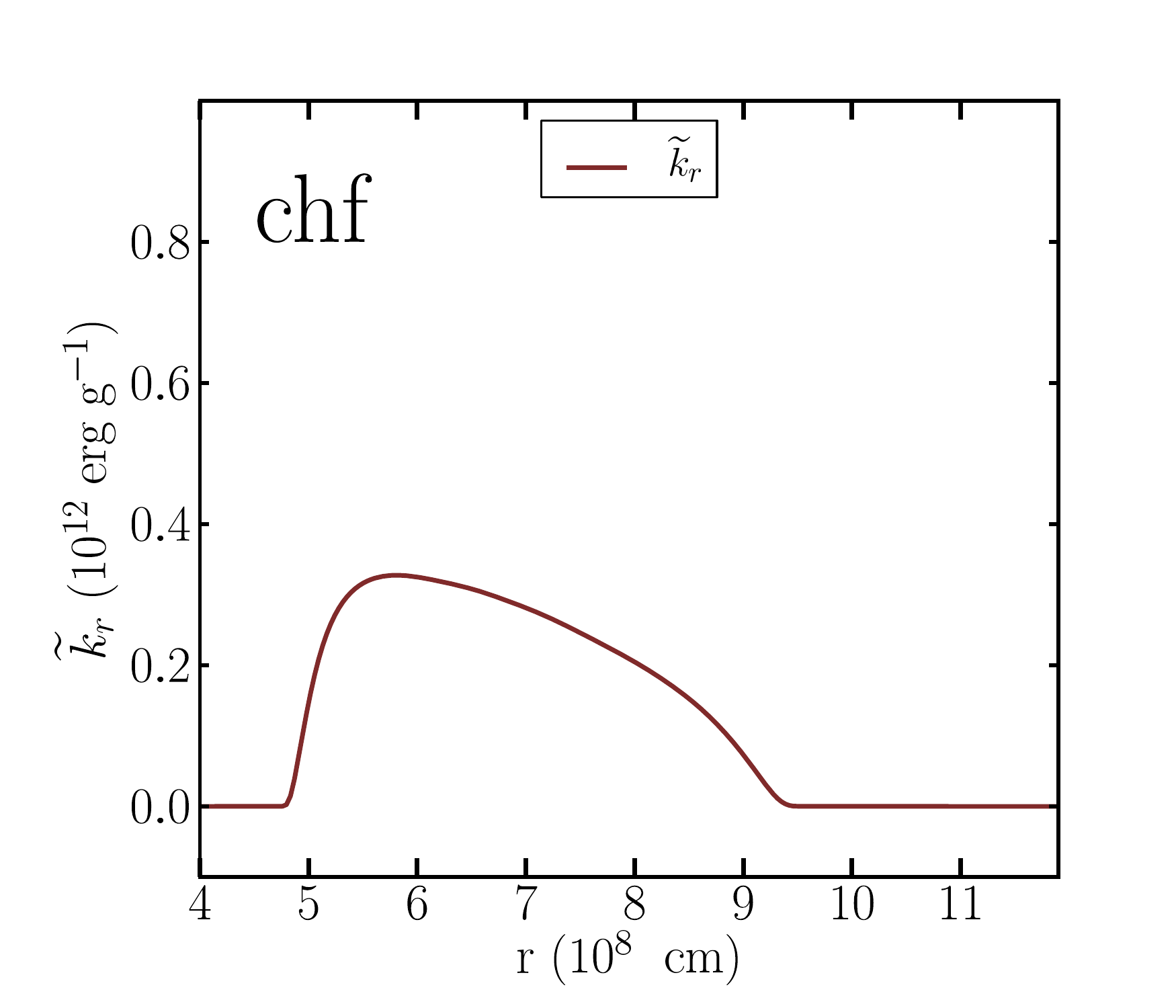}
\includegraphics[width=6.5cm]{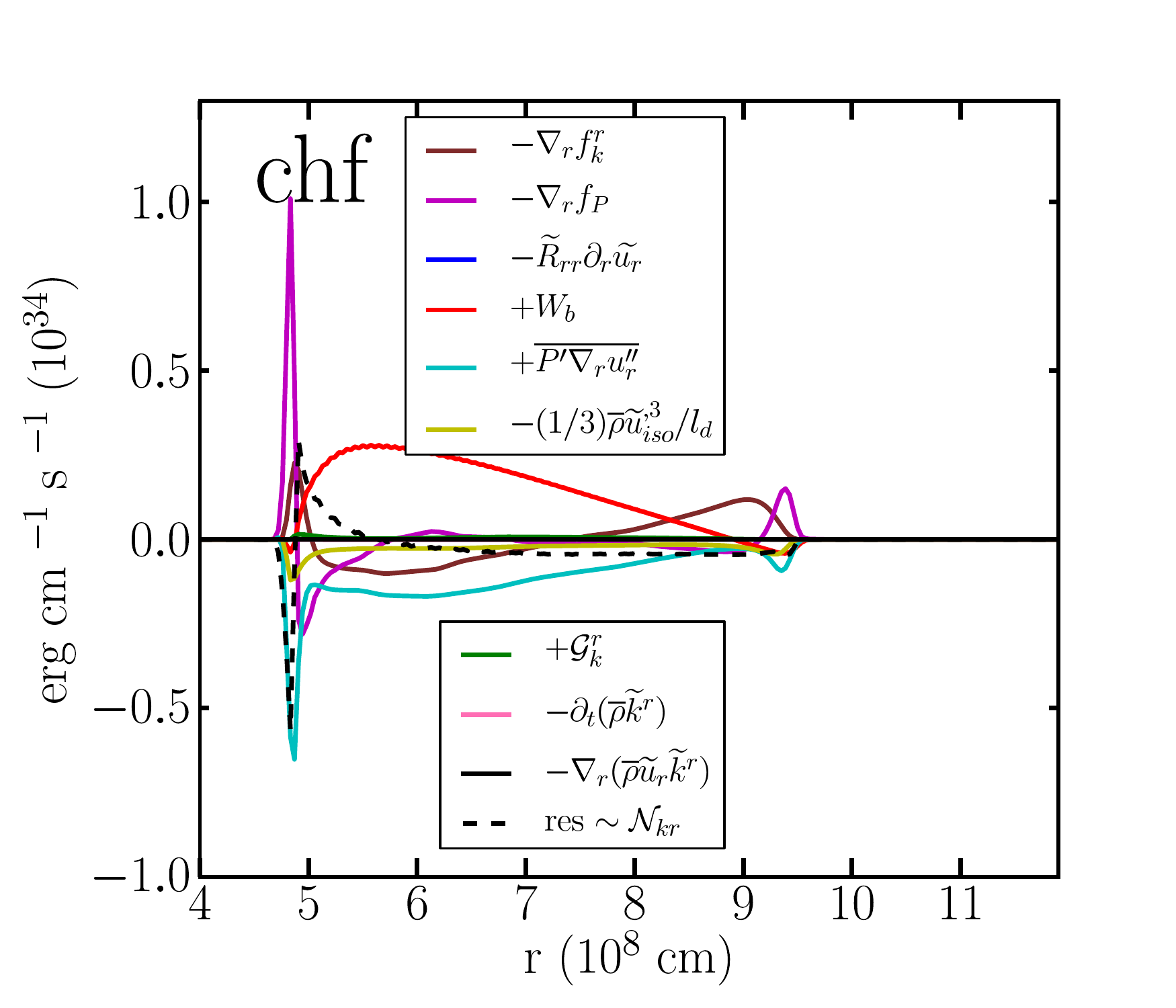}
\includegraphics[width=6.5cm]{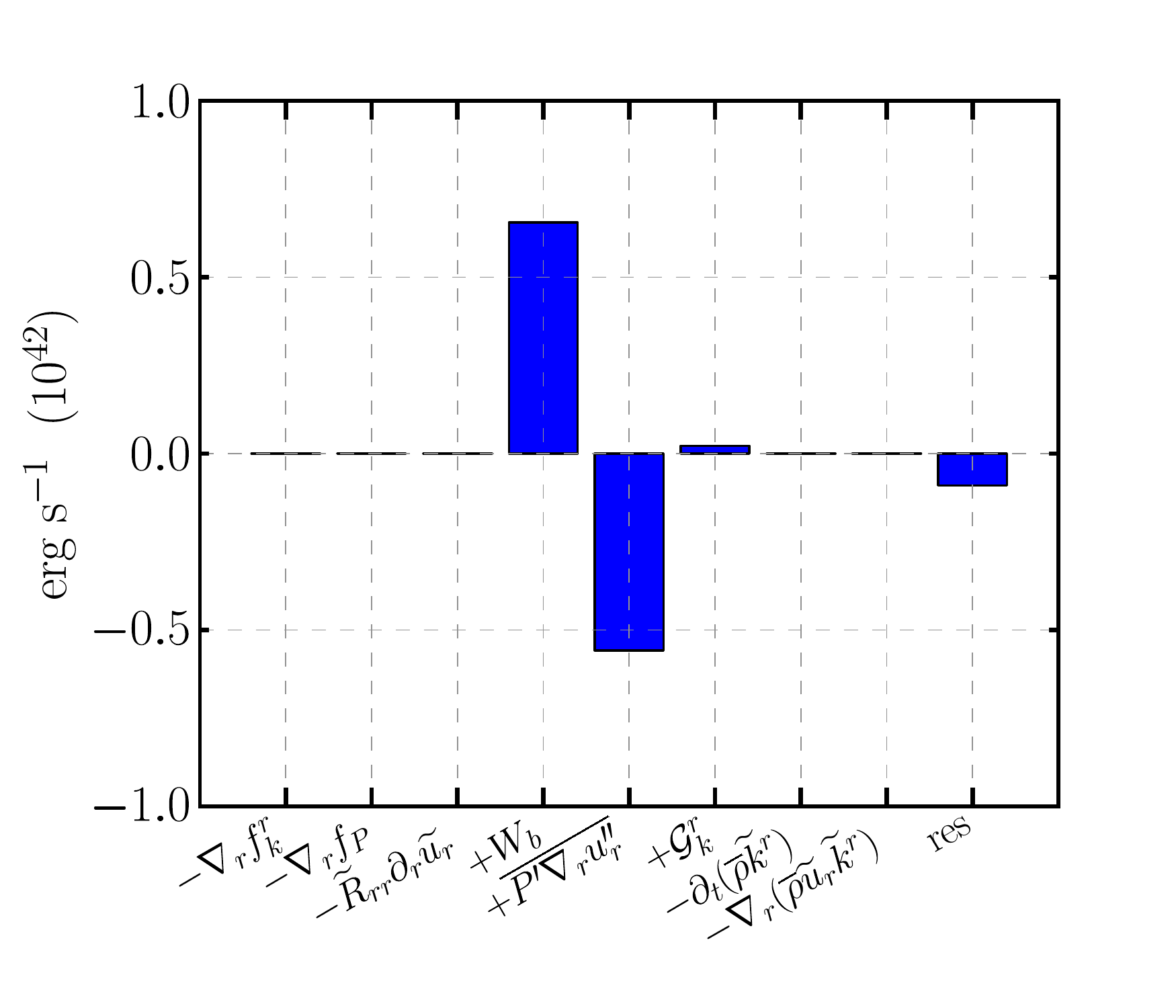}}

\caption{Radial turbulent kinetic energy equation. Model {\sf hif.3D} (upper panels) and model {\sf chf.3D} (lower panels)}
\end{figure}

\newpage

\subsection{Horizontal part of mean turbulent kinetic energy equation}

\begin{align}
\av{\rho} \fav{D}_t \fav{k}^h =  &  -\nabla_r f_k^h - (\fht{R}_{\theta r}\partial_r \fht{u}_\theta + \fht{R}_{\phi r}\partial_r \fht{u}_\phi) + (\eht{P' \nabla_\theta u''_\theta} + \eht{P' \nabla_\phi u''_\phi}) + {\mathcal G_k^h} + {\mathcal N_{kh}} 
\end{align}

\begin{figure}[!h]
\centerline{
\includegraphics[width=6.5cm]{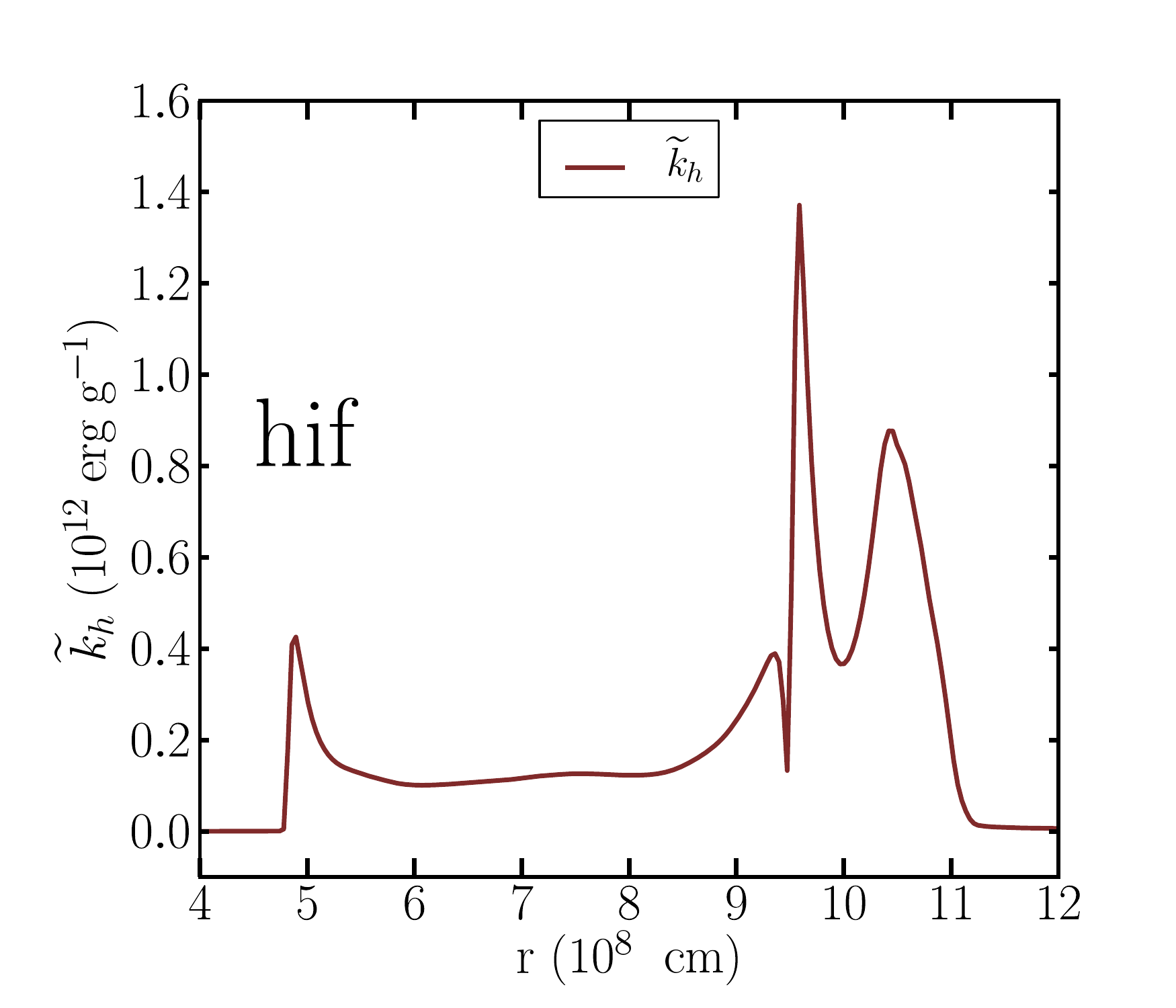}
\includegraphics[width=6.5cm]{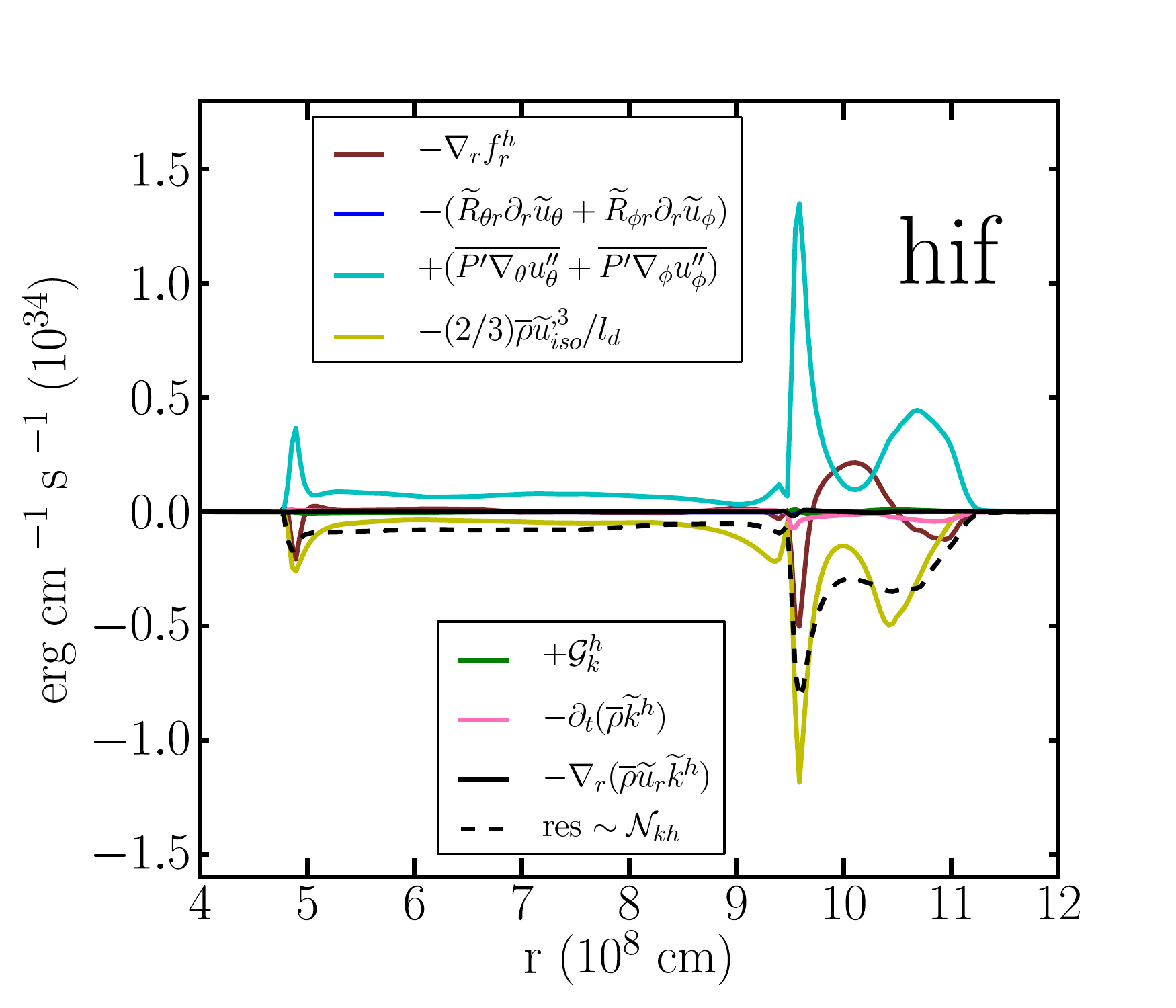}
\includegraphics[width=6.5cm]{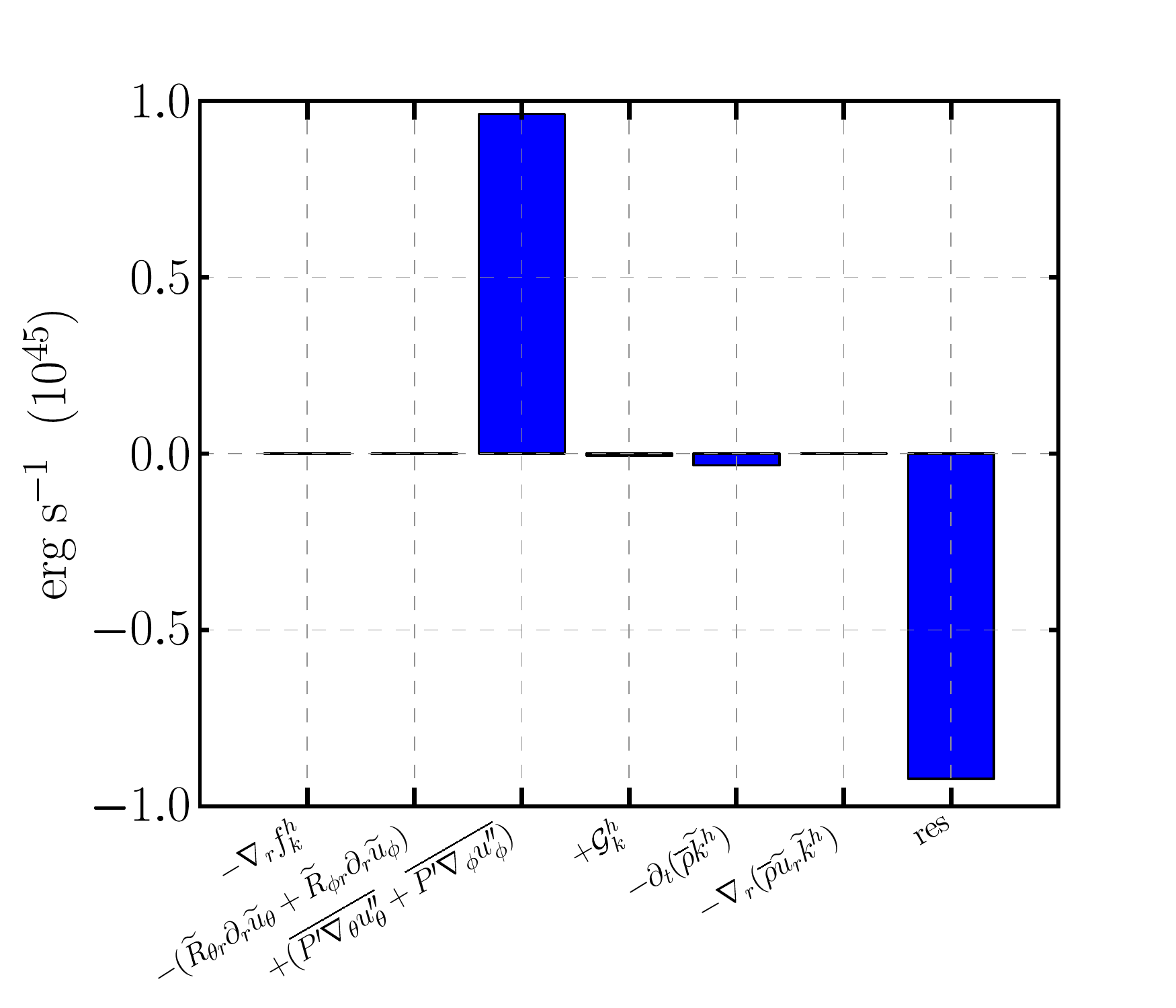}}

\centerline{
\includegraphics[width=6.5cm]{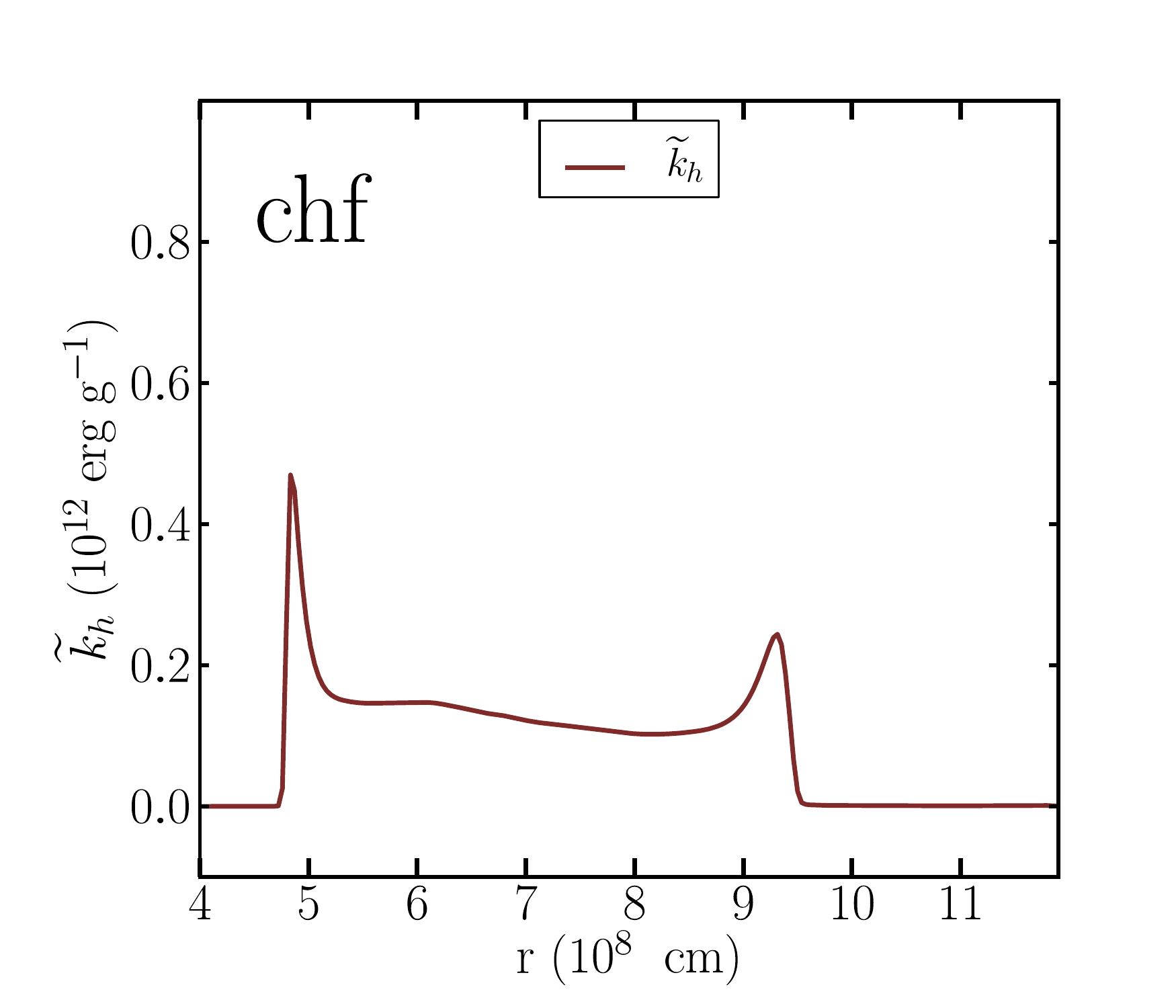}
\includegraphics[width=6.5cm]{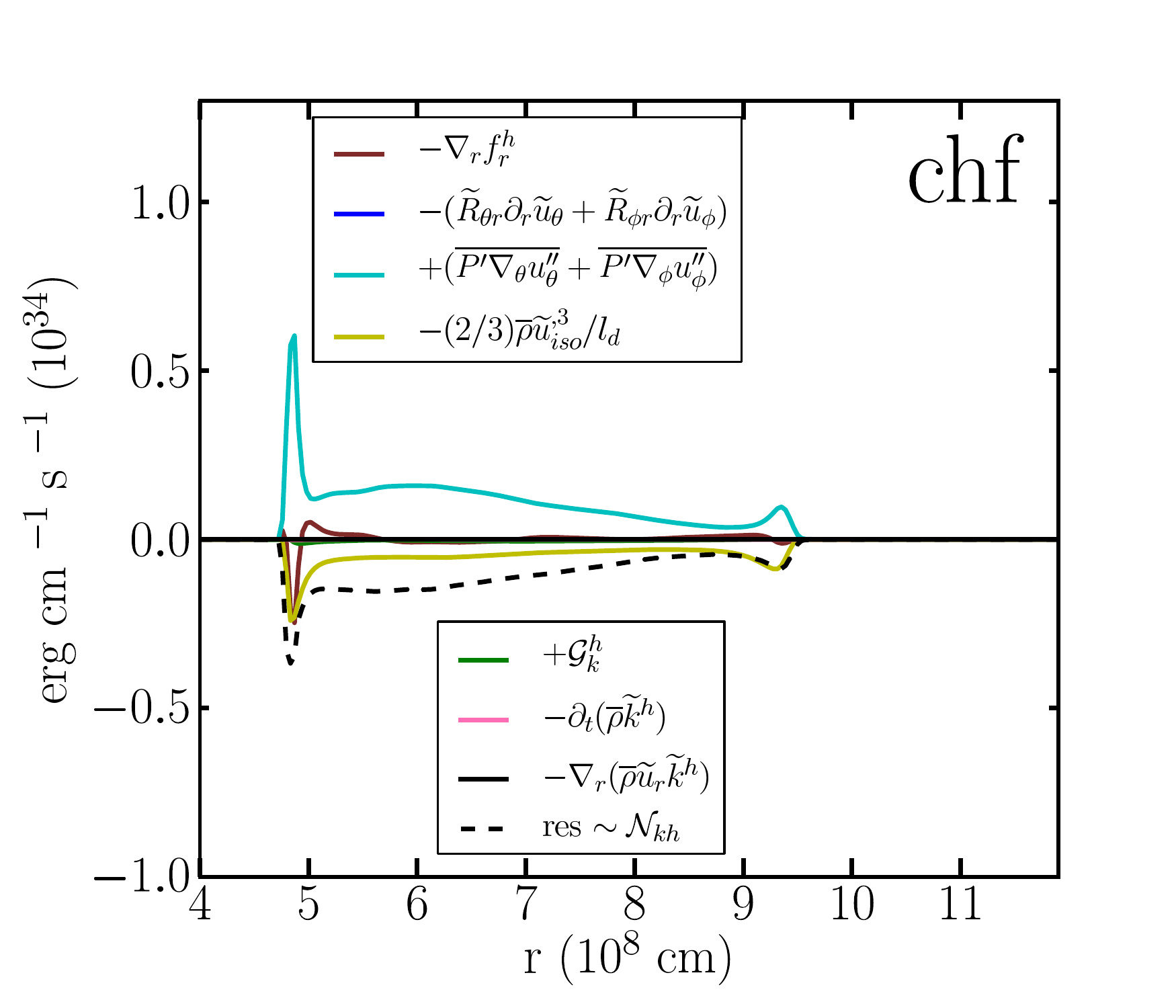}
\includegraphics[width=6.5cm]{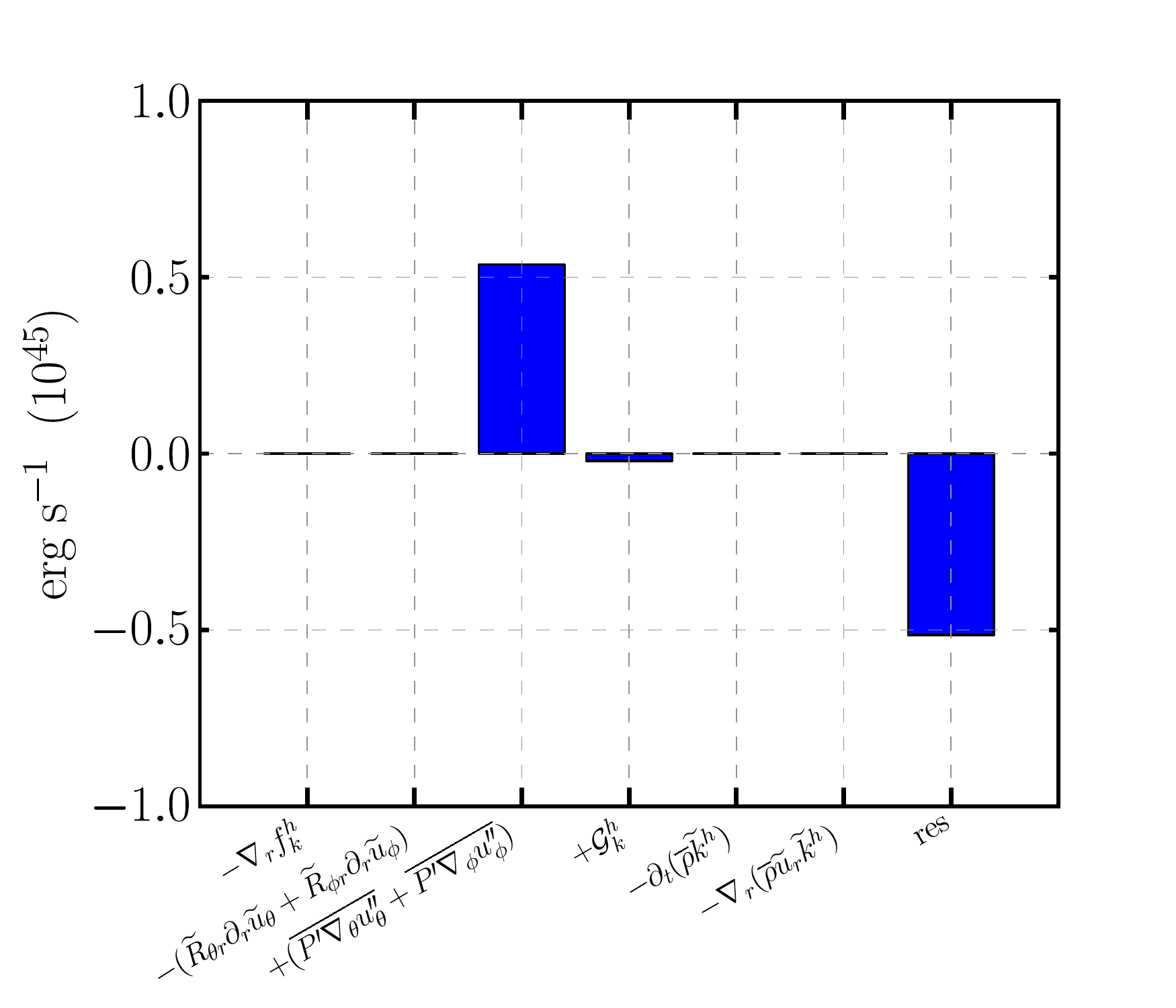}}

\caption{Horizontal turbulent kinetic energy equation. Model {\sf hif.3D} (upper panels) and model {\sf chf.3D} (lower panels)}
\end{figure}

\newpage

\section{Mean field composition data for the  hydrogen injection flash model}

\subsection{Mean H$^1$ and He$^3$ equation}

\begin{figure}[!h]
\centerline{
\includegraphics[width=6.7cm]{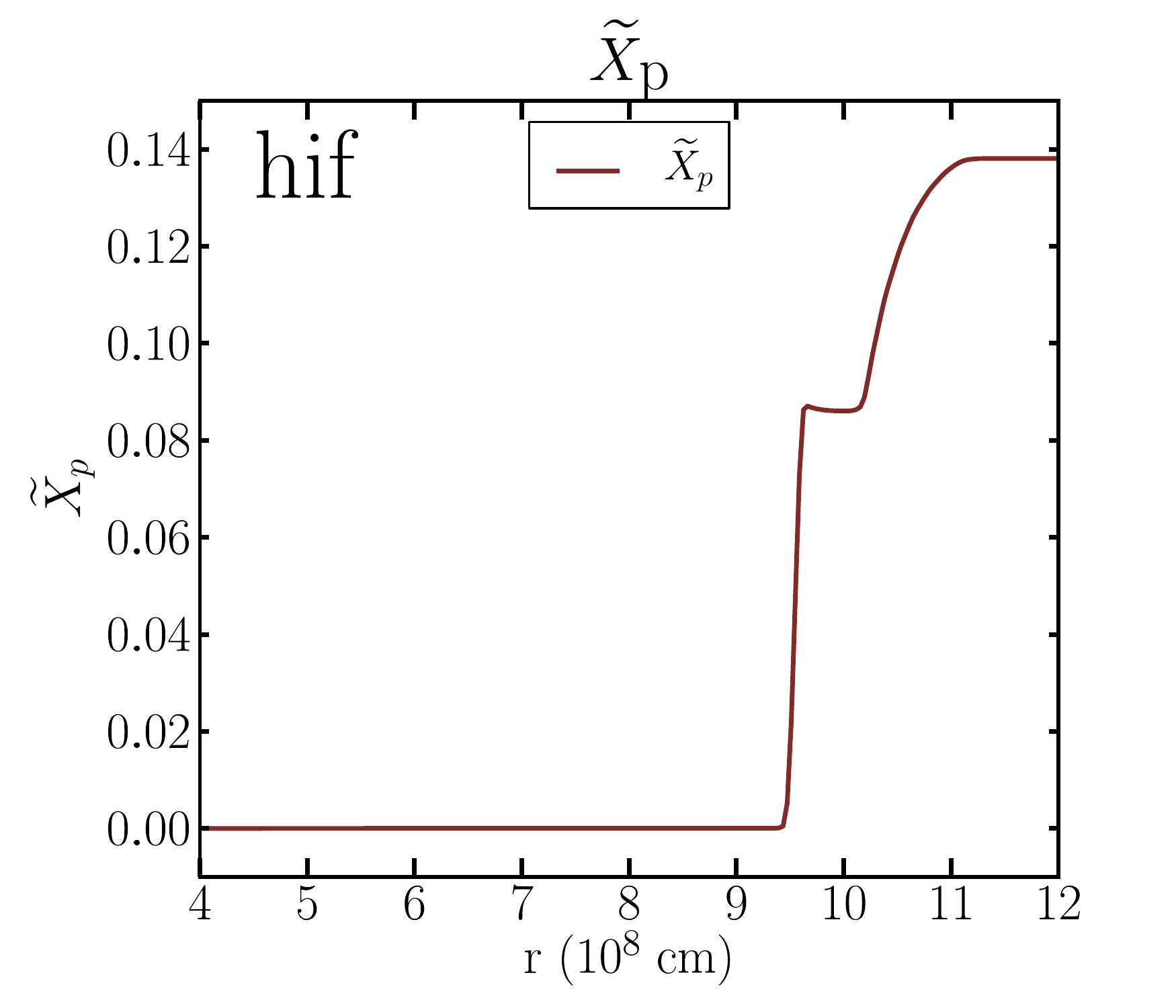}
\includegraphics[width=6.7cm]{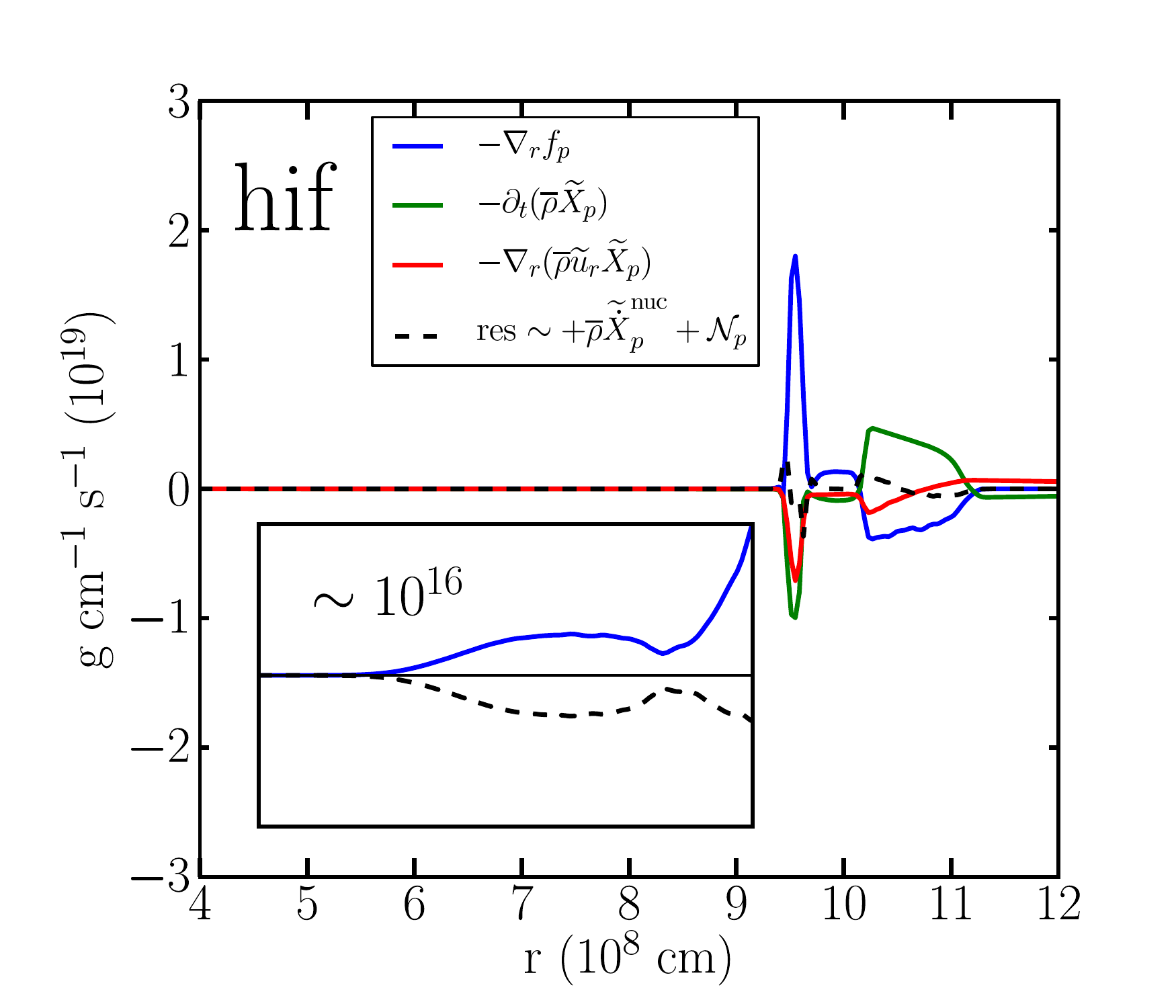}
\includegraphics[width=6.7cm]{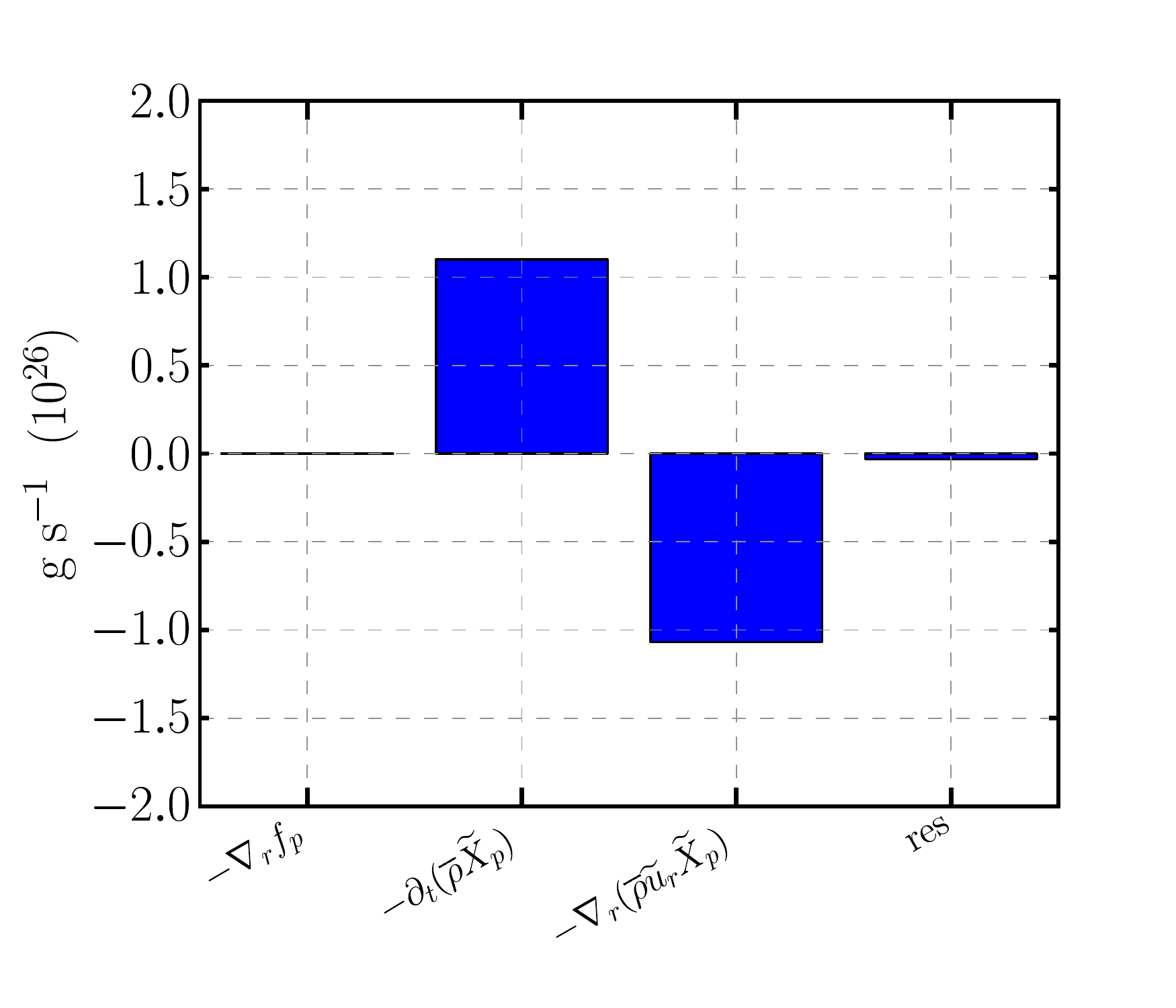}}

\centerline{
\includegraphics[width=6.7cm]{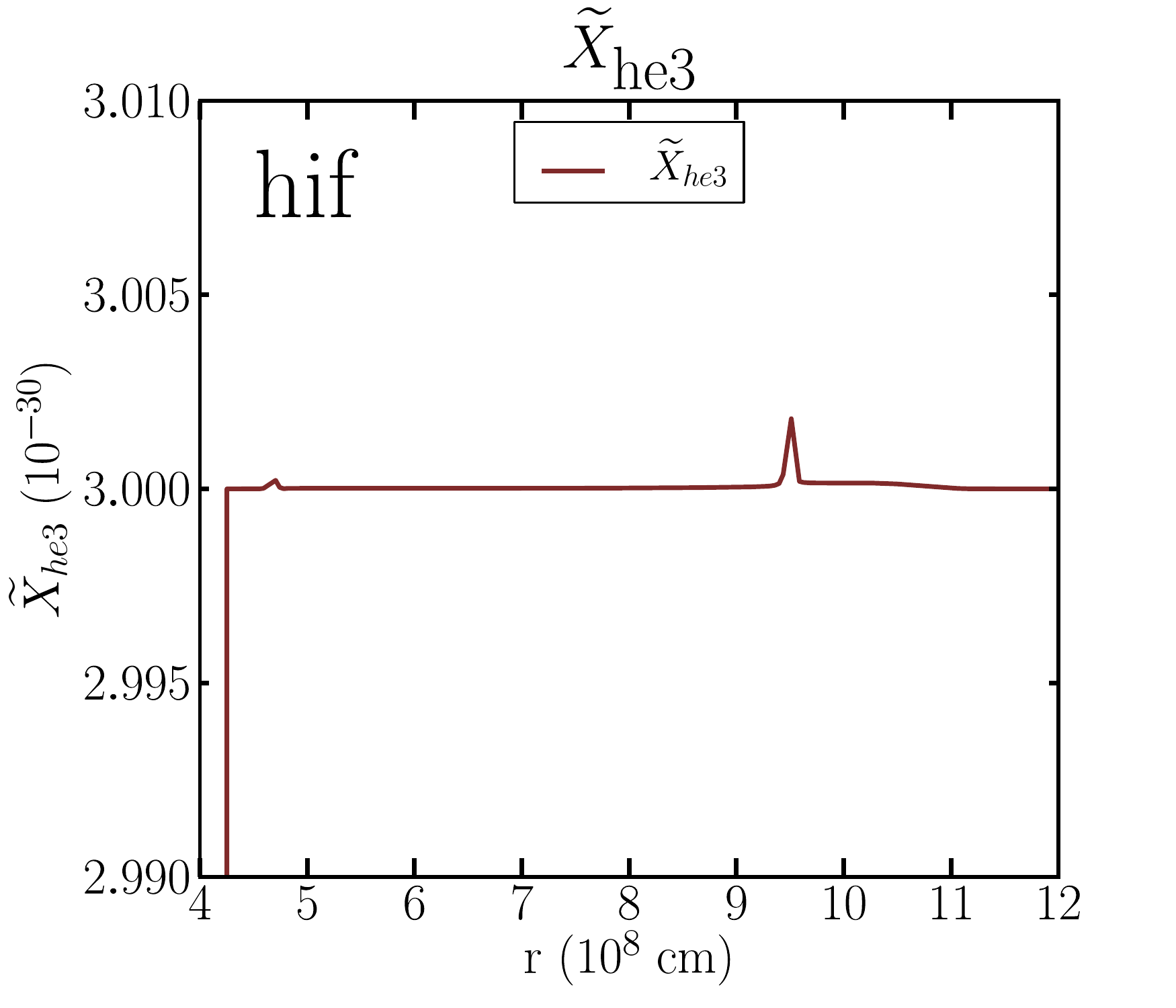}
\includegraphics[width=6.7cm]{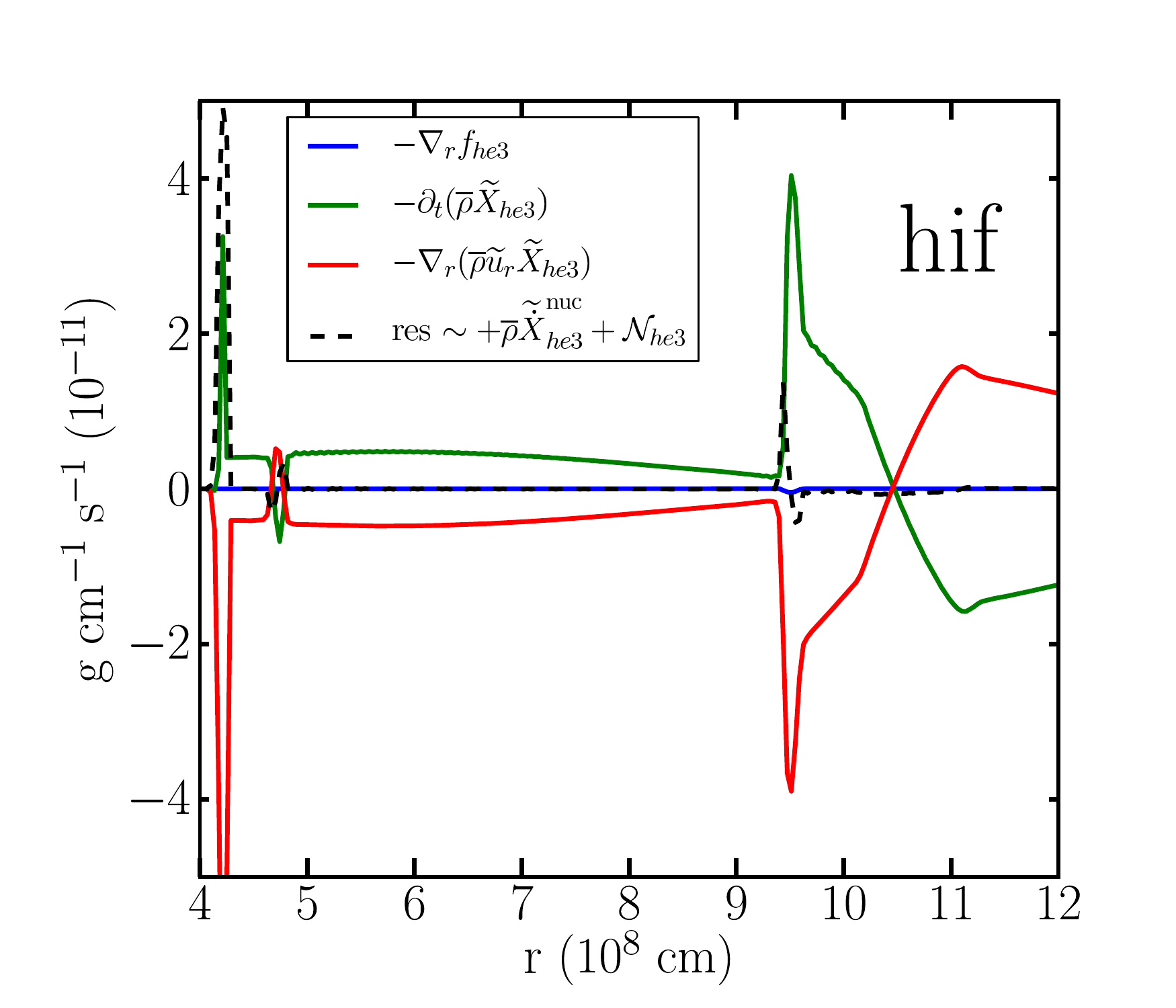}
\includegraphics[width=6.7cm]{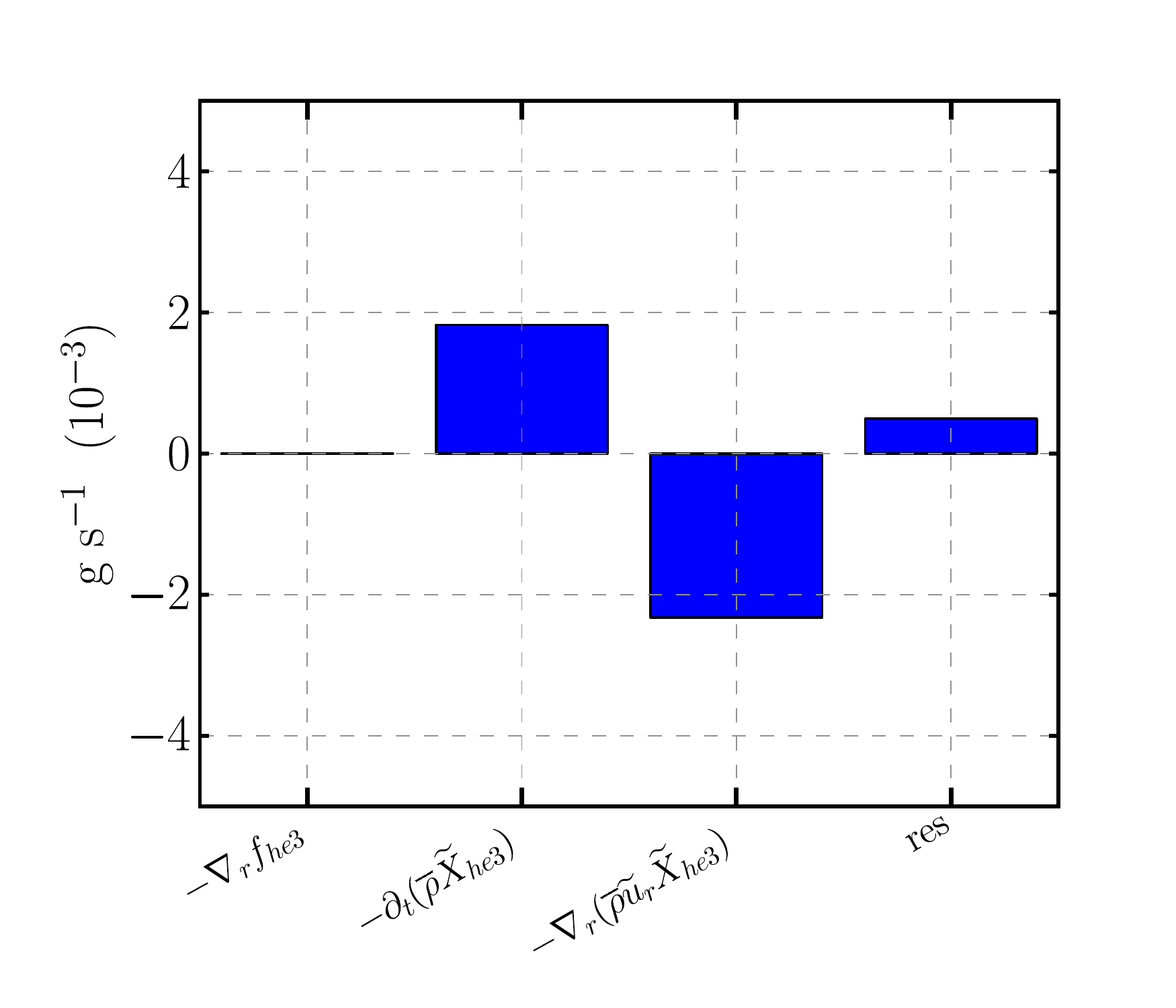}}
\caption{Mean composition equations for {\sf hif.3D}. \label{fig:x-equations}}
\end{figure}

\newpage

\subsection{Mean He$^4$ and C$^{12}$ equation}

\begin{figure}[!h]
\centerline{
\includegraphics[width=6.7cm]{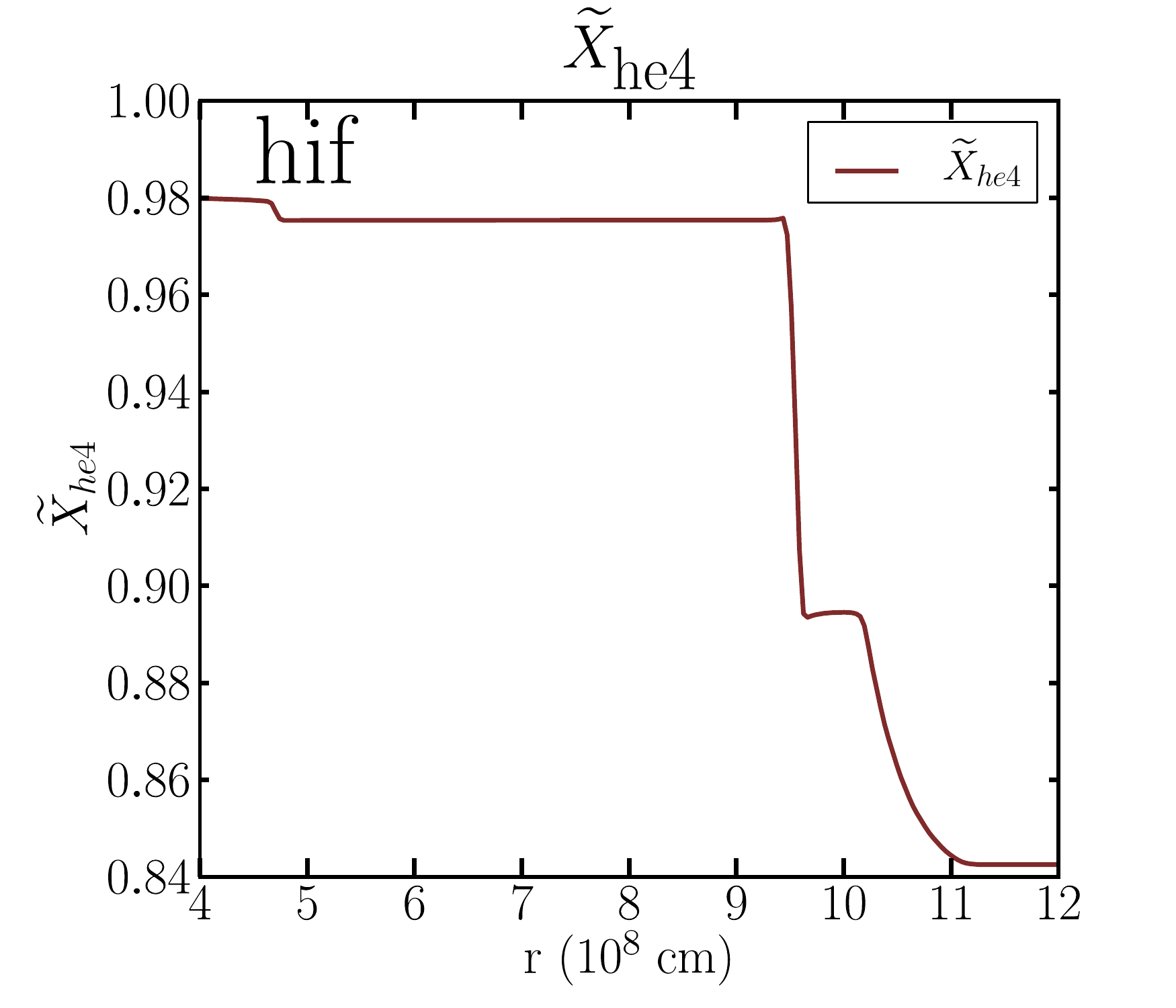}
\includegraphics[width=6.7cm]{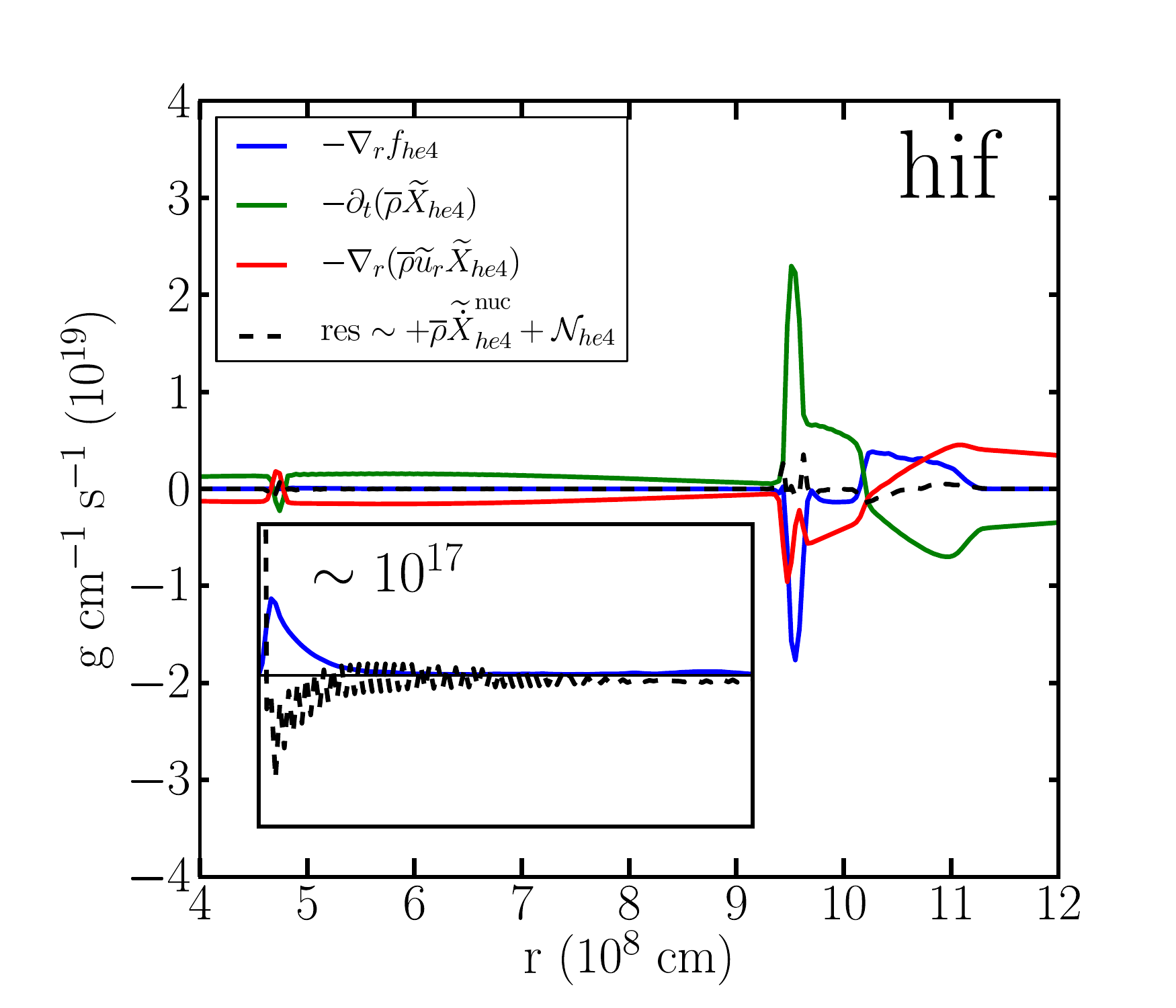}
\includegraphics[width=6.7cm]{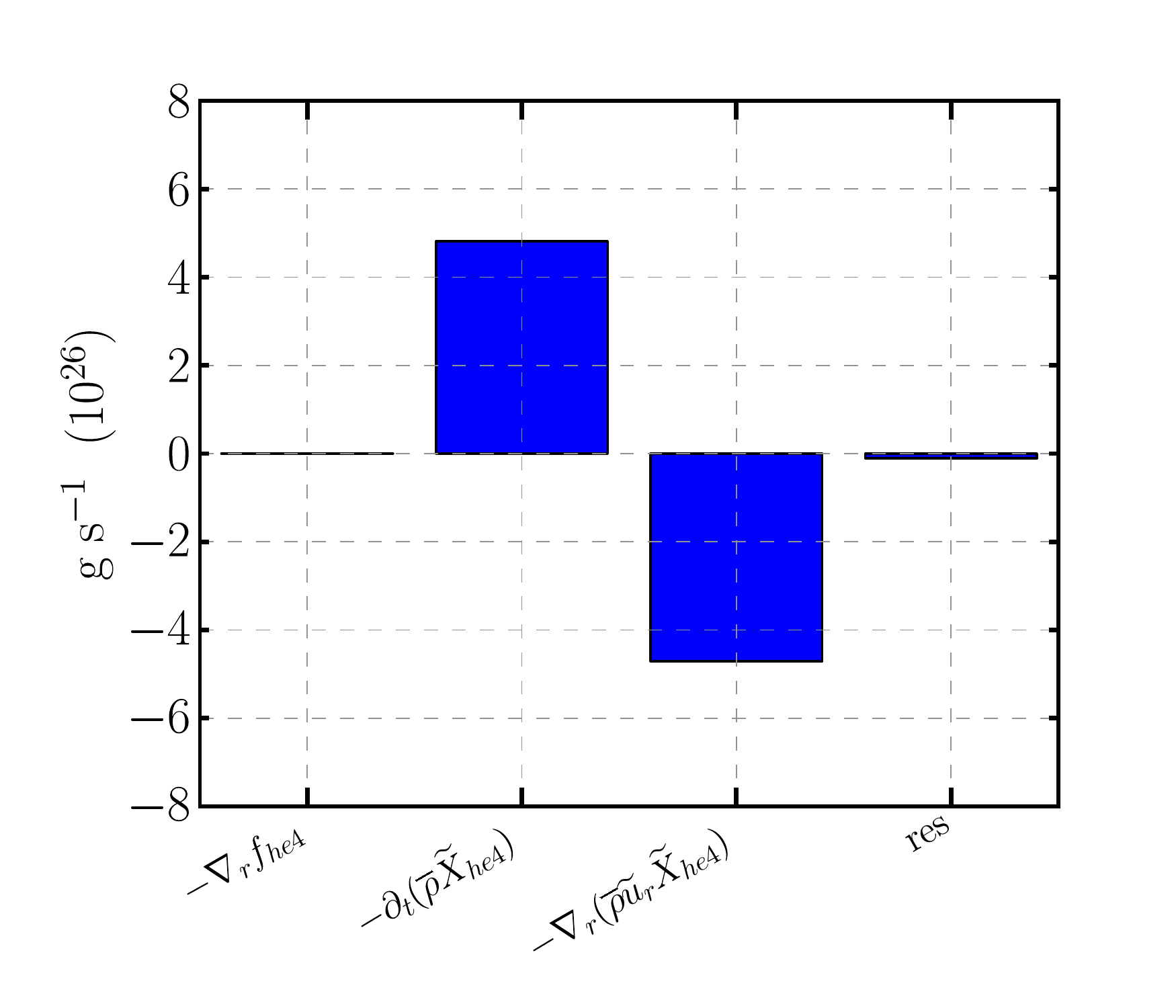}}

\centerline{
\includegraphics[width=6.7cm]{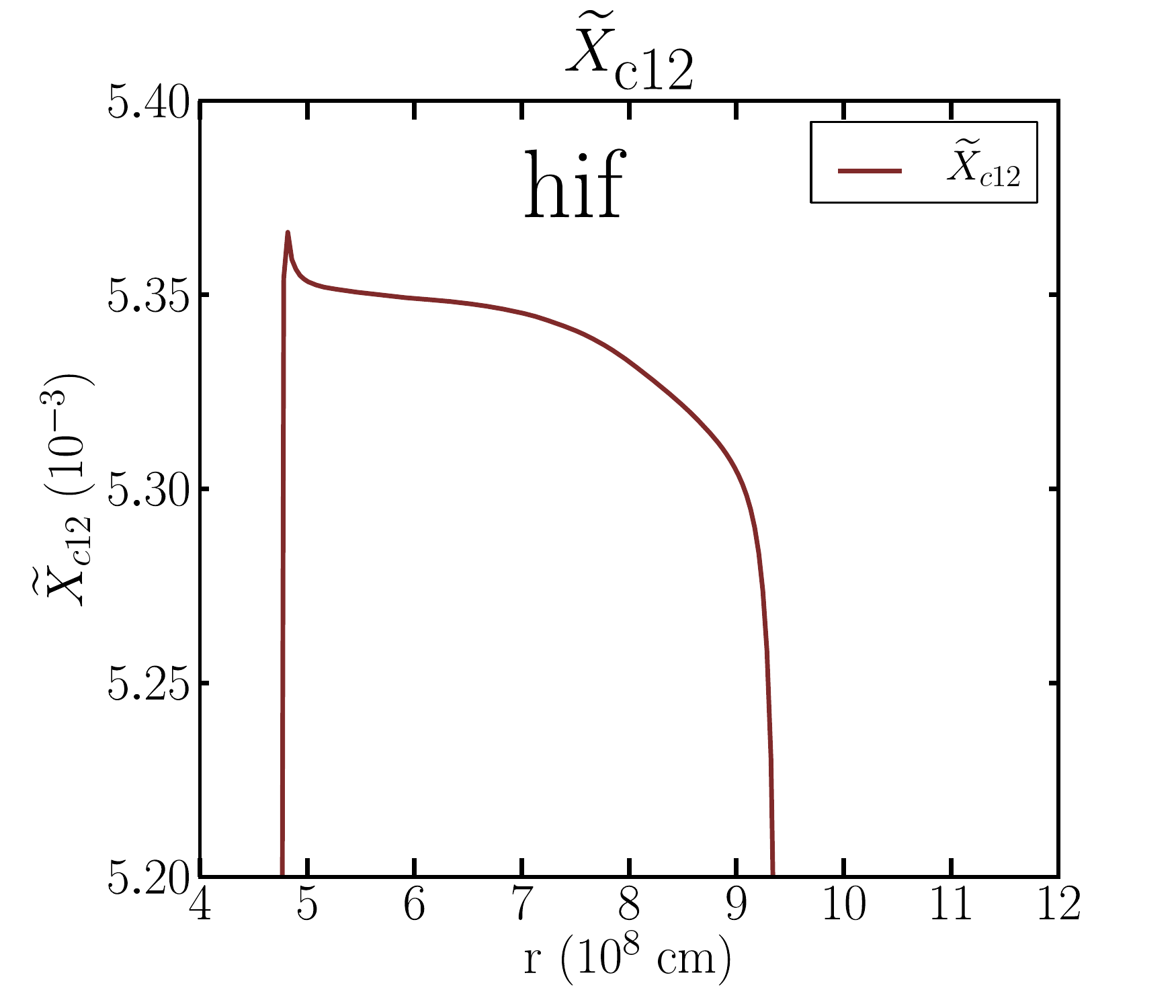}
\includegraphics[width=6.7cm]{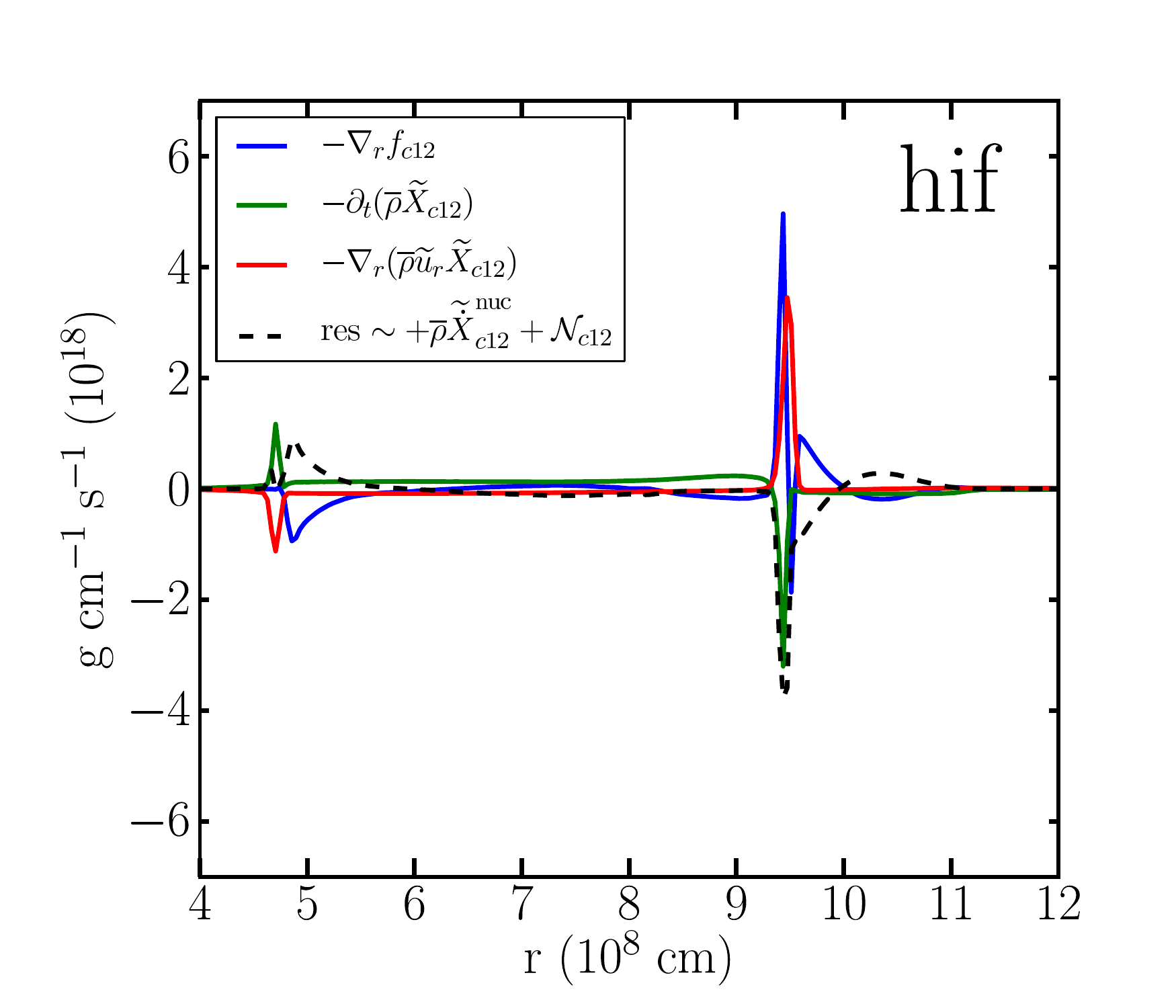}
\includegraphics[width=6.7cm]{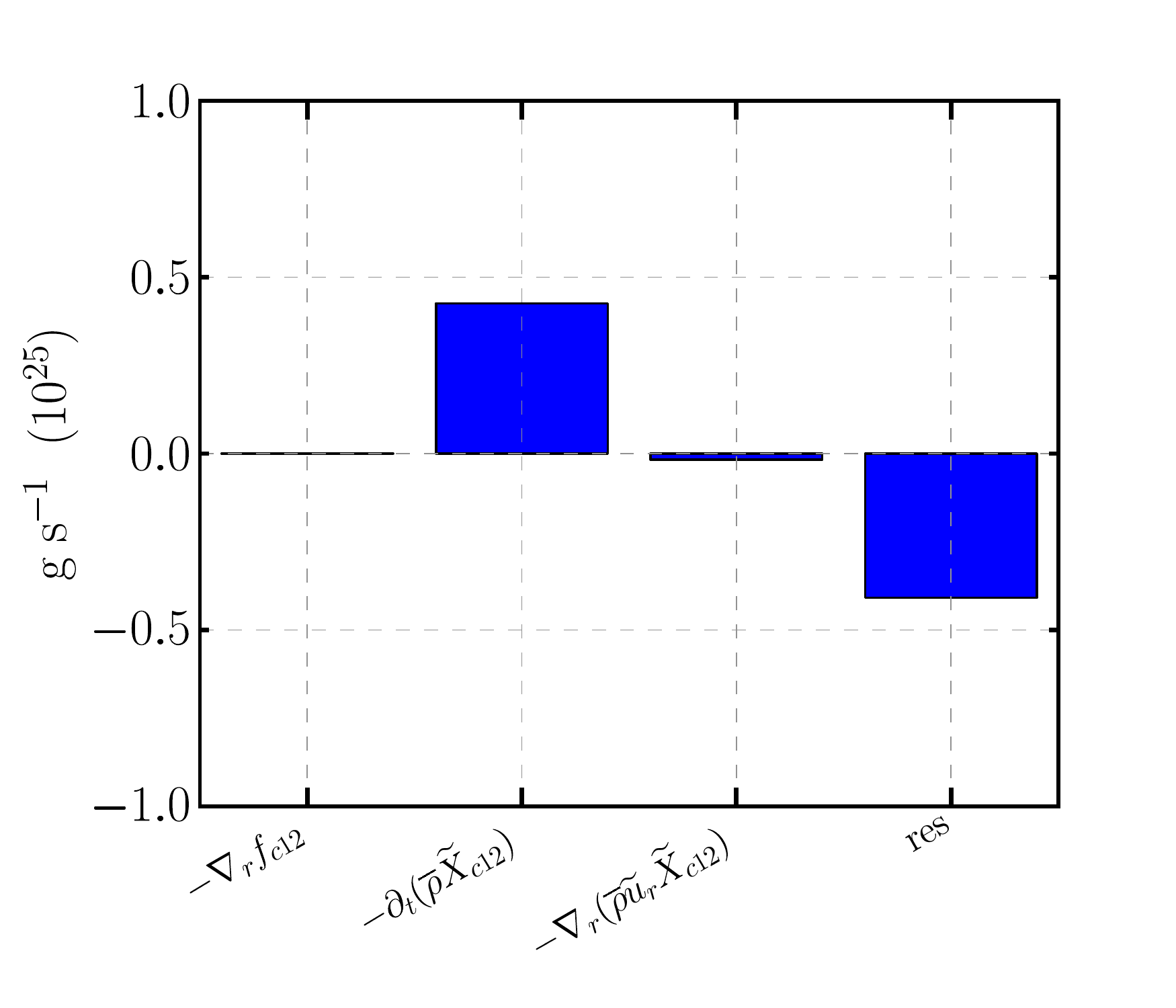}}
\caption{Mean composition equations for {\sf hif.3D}. \label{fig:x-equations}}
\end{figure}

\newpage

\subsection{Mean C$^{13}$ and N$^{13}$ equation}

\begin{figure}[!h]
\centerline{
\includegraphics[width=6.7cm]{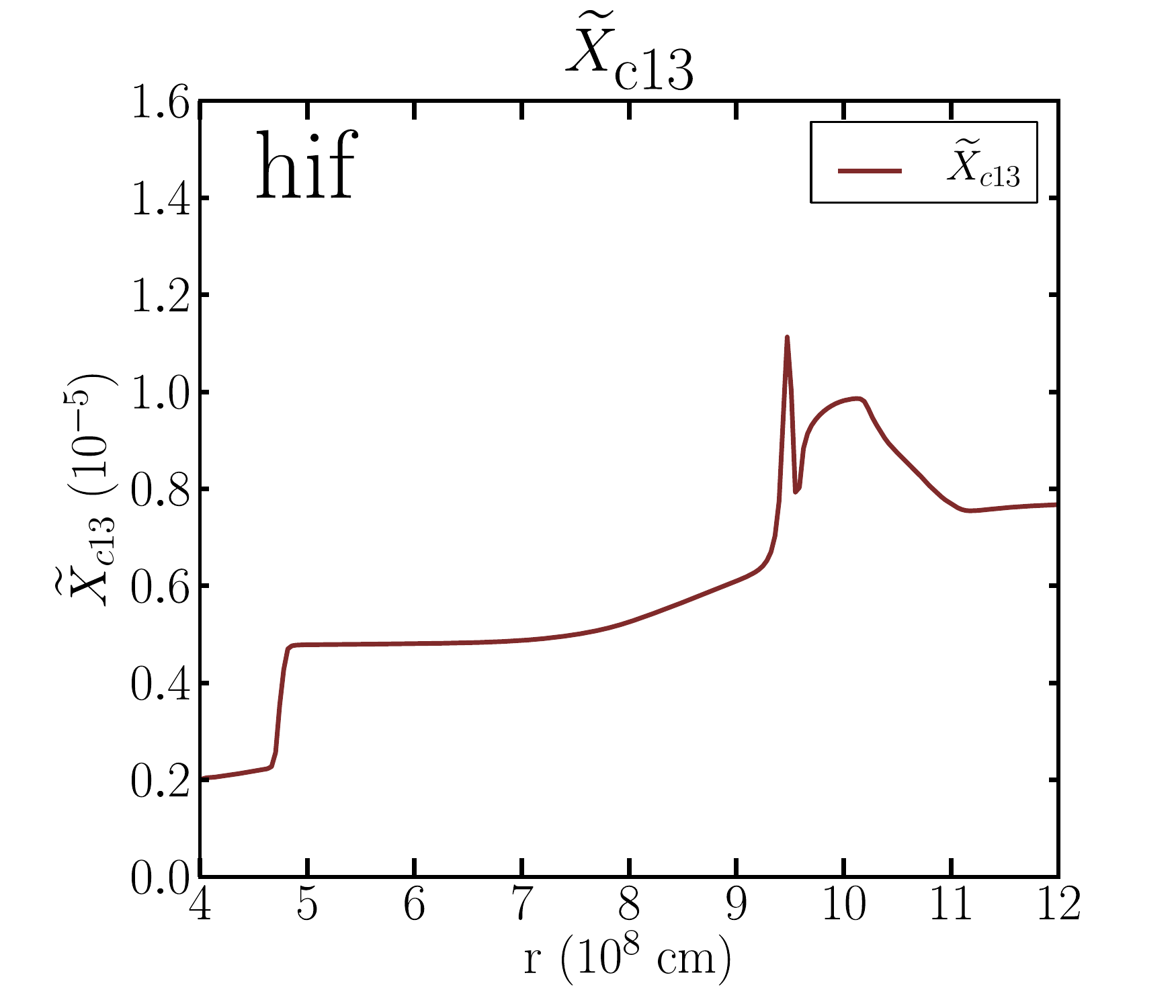}
\includegraphics[width=6.7cm]{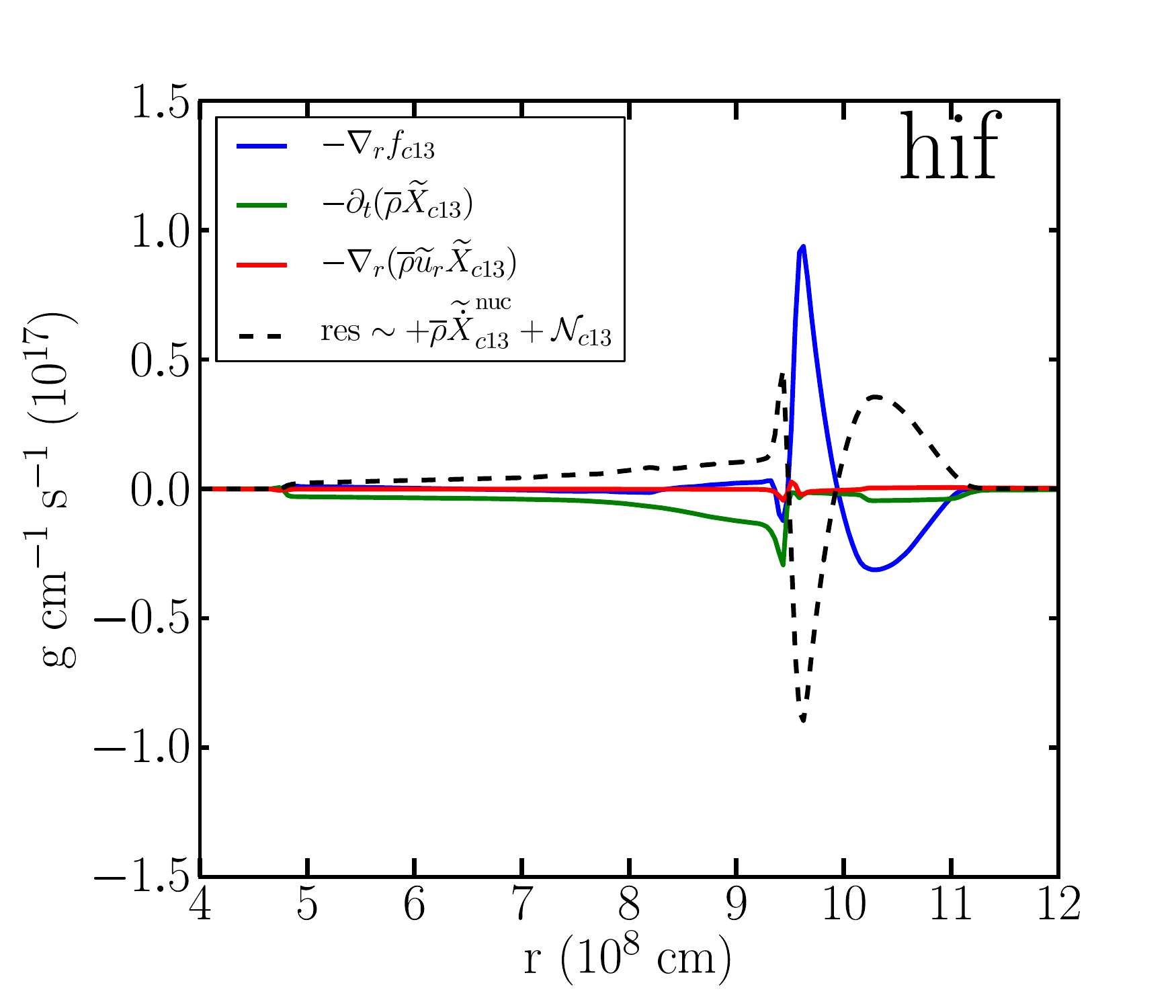}
\includegraphics[width=6.7cm]{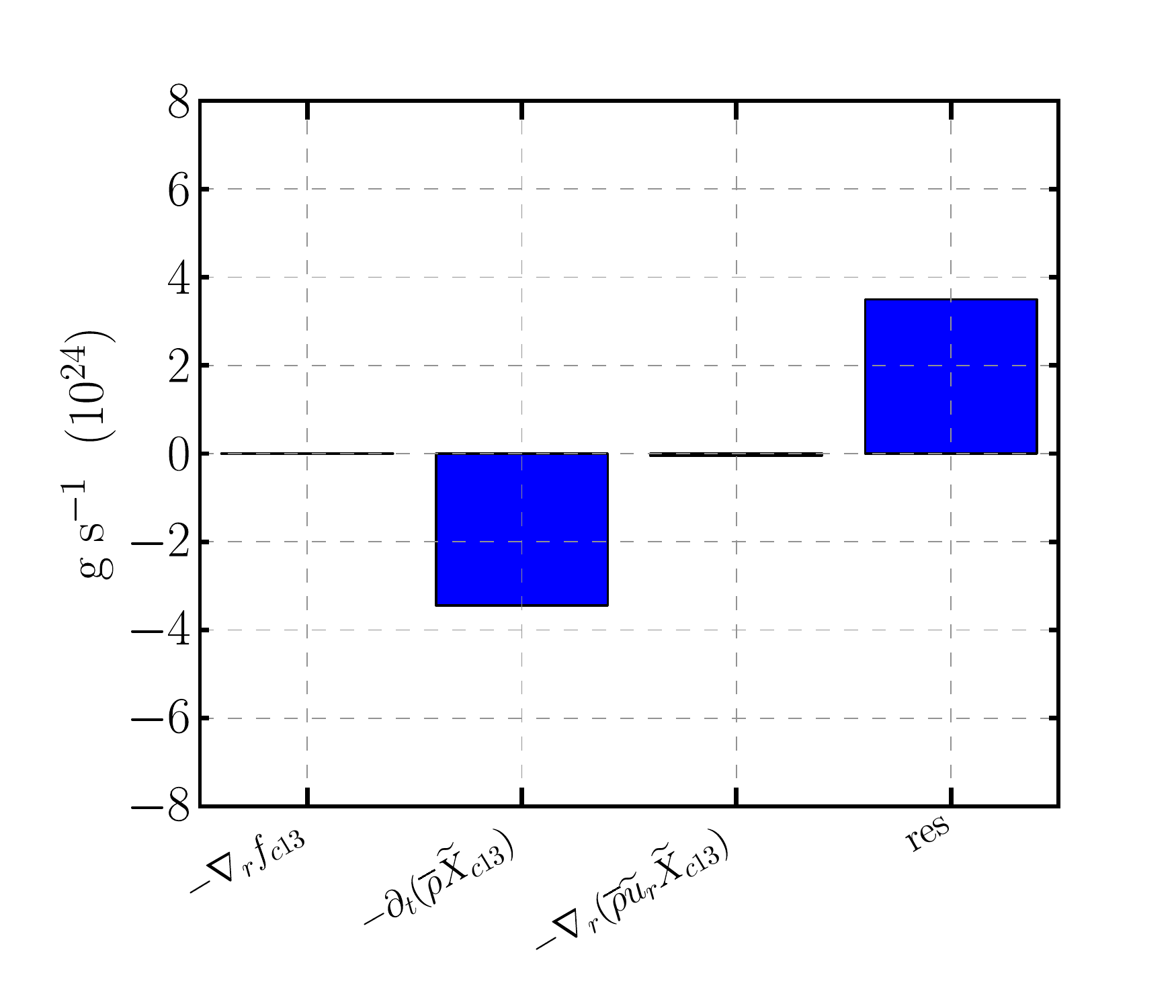}}

\centerline{
\includegraphics[width=6.7cm]{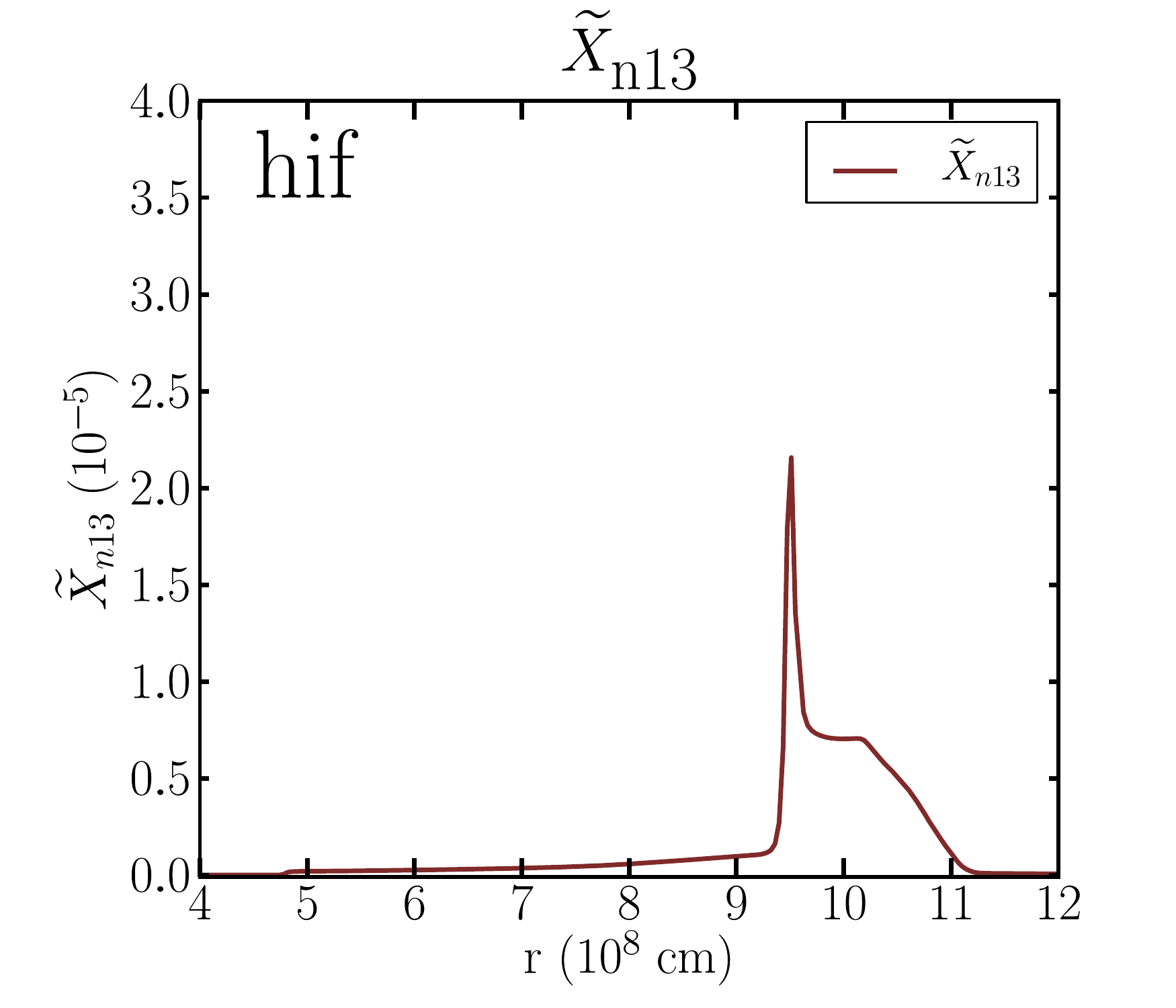}
\includegraphics[width=6.7cm]{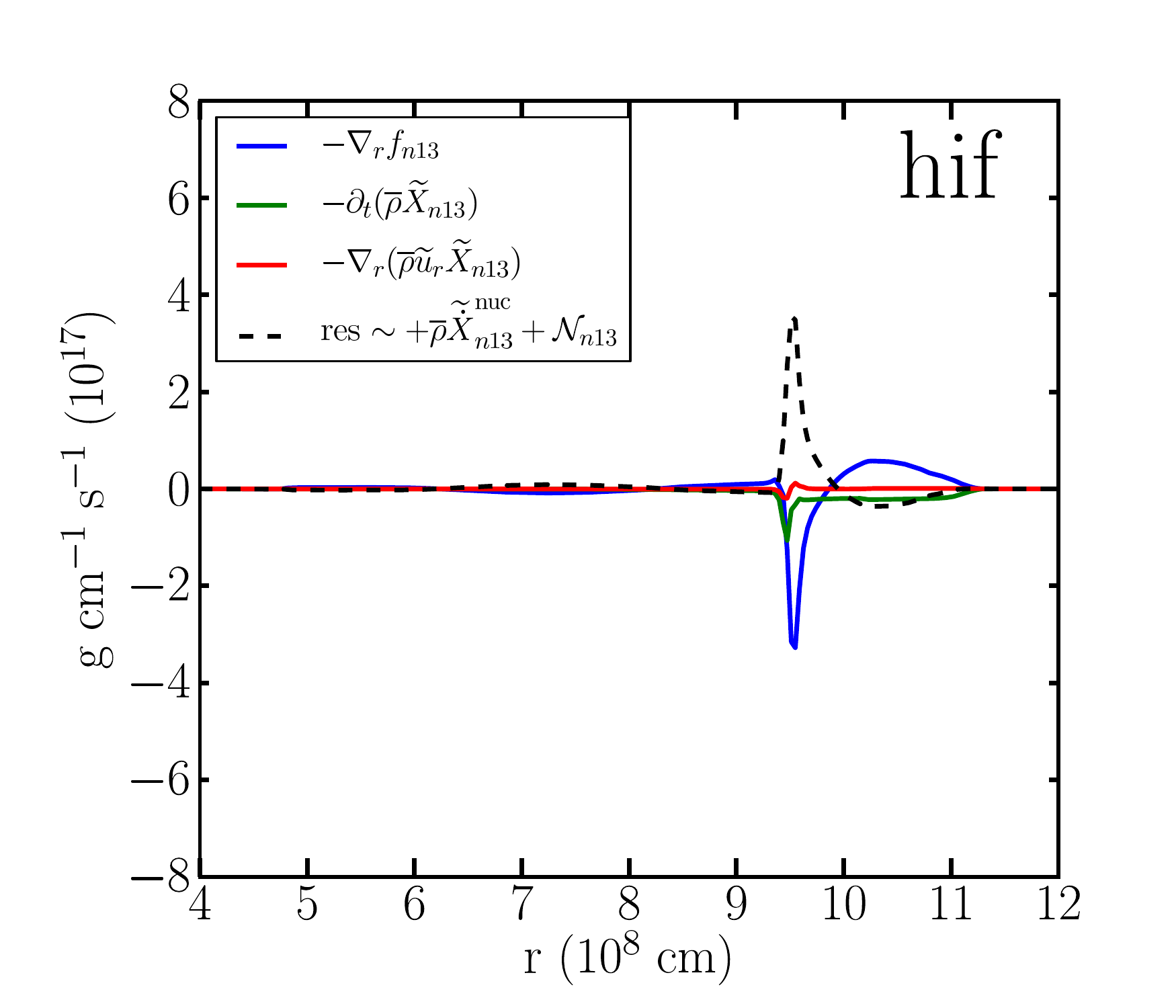}
\includegraphics[width=6.7cm]{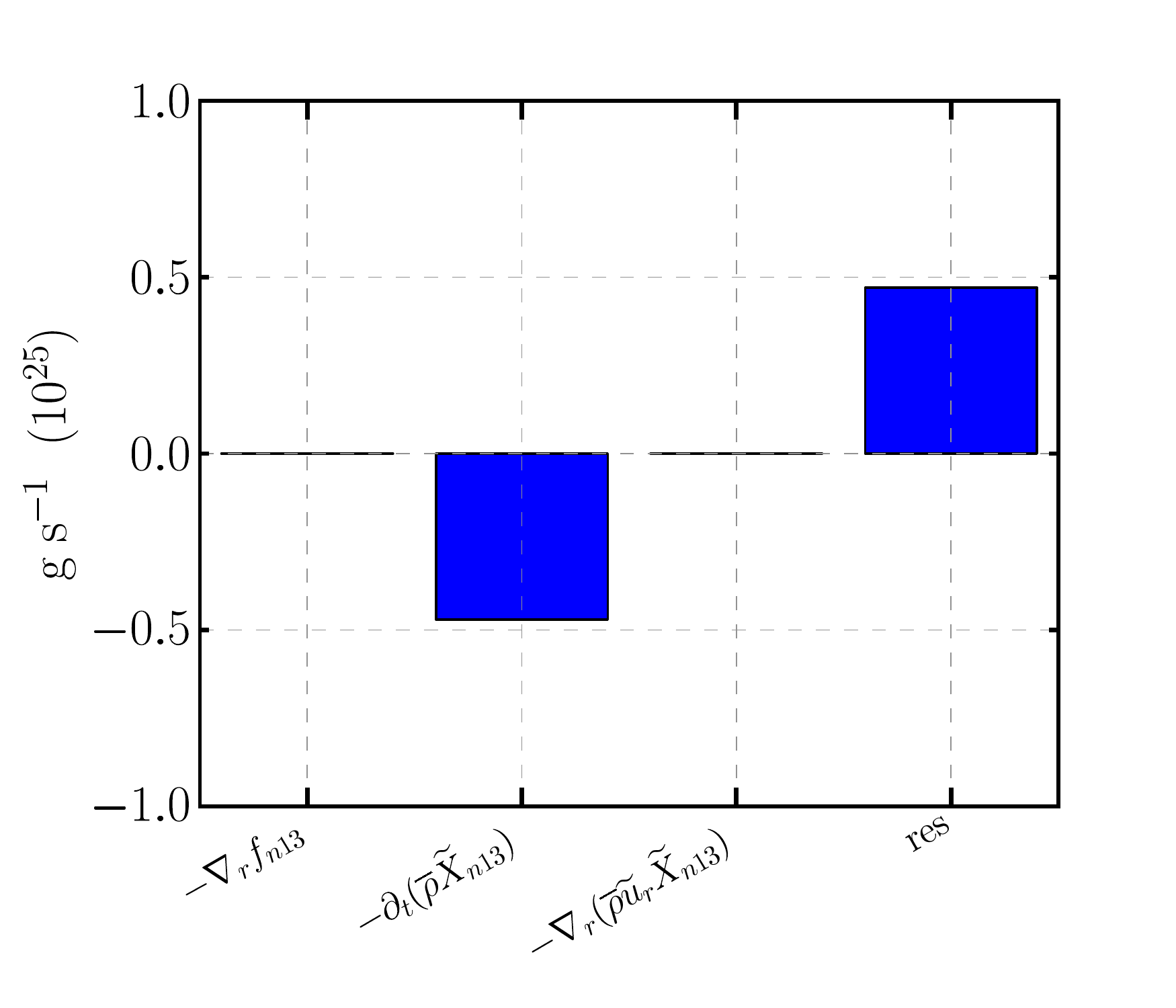}}
\caption{Mean composition equations for {\sf hif.3D}. \label{fig:x-equations}}
\end{figure}

\newpage

\subsection{Mean N$^{14}$ and N$^{15}$ equation}

\begin{figure}[!h]
\centerline{
\includegraphics[width=6.7cm]{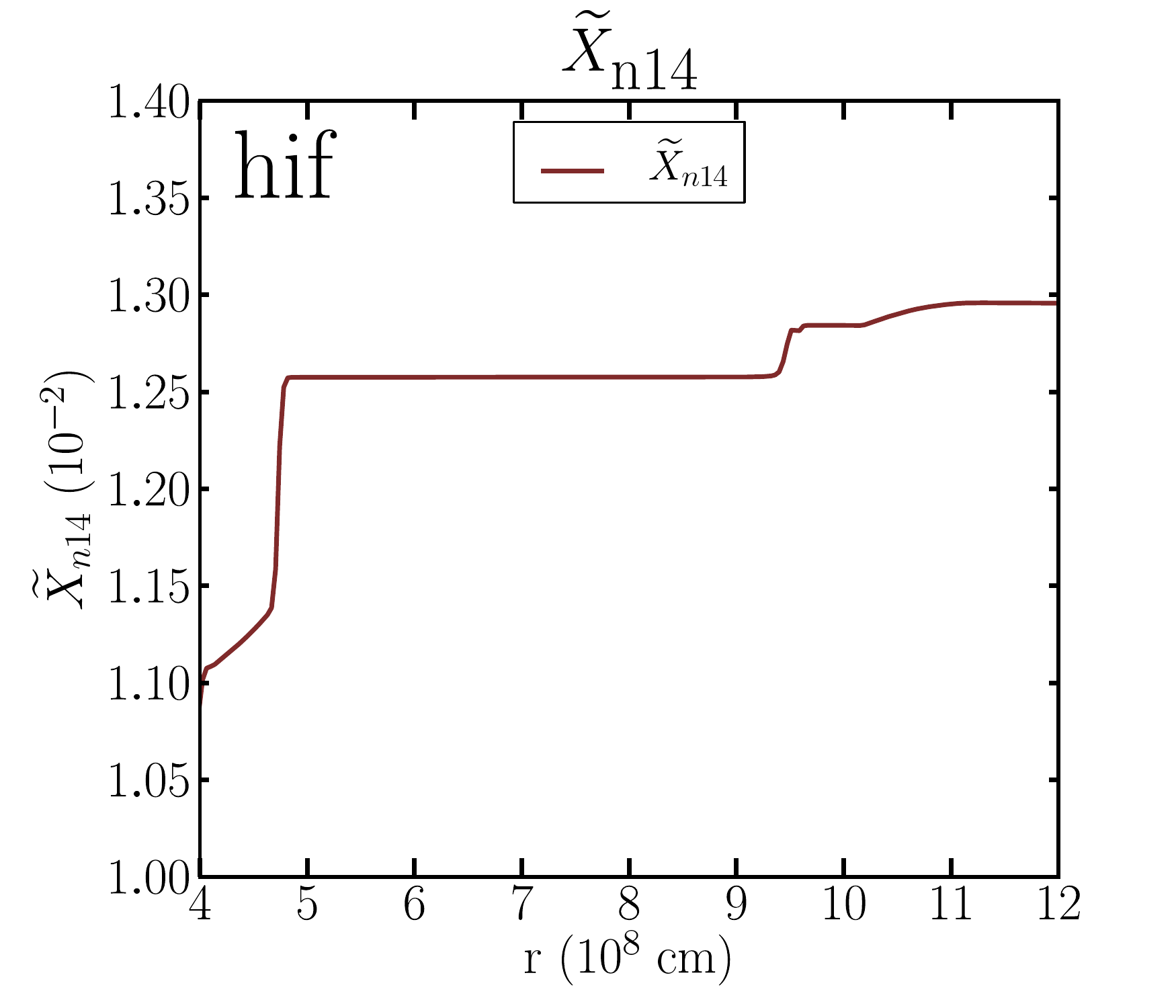}
\includegraphics[width=6.7cm]{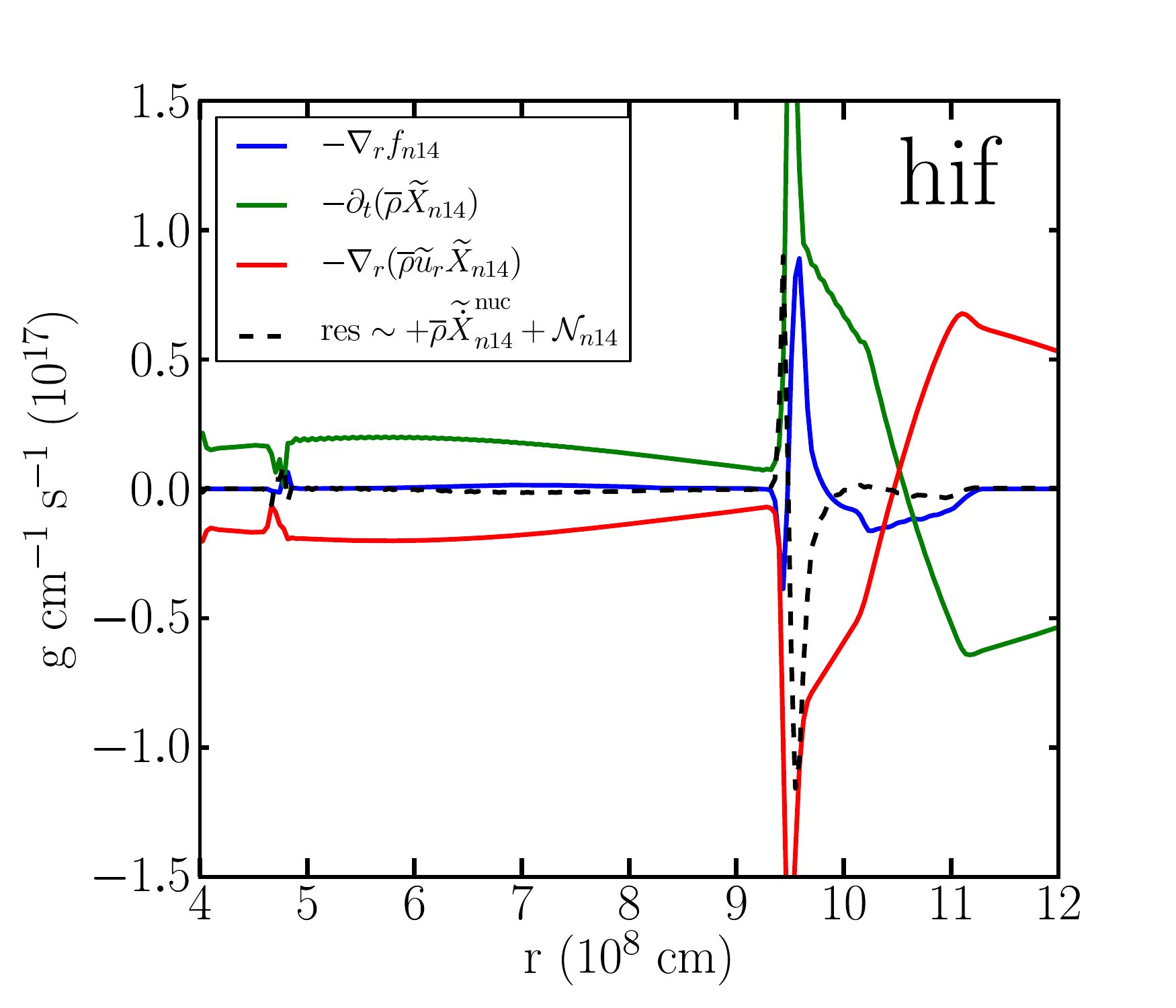}
\includegraphics[width=6.7cm]{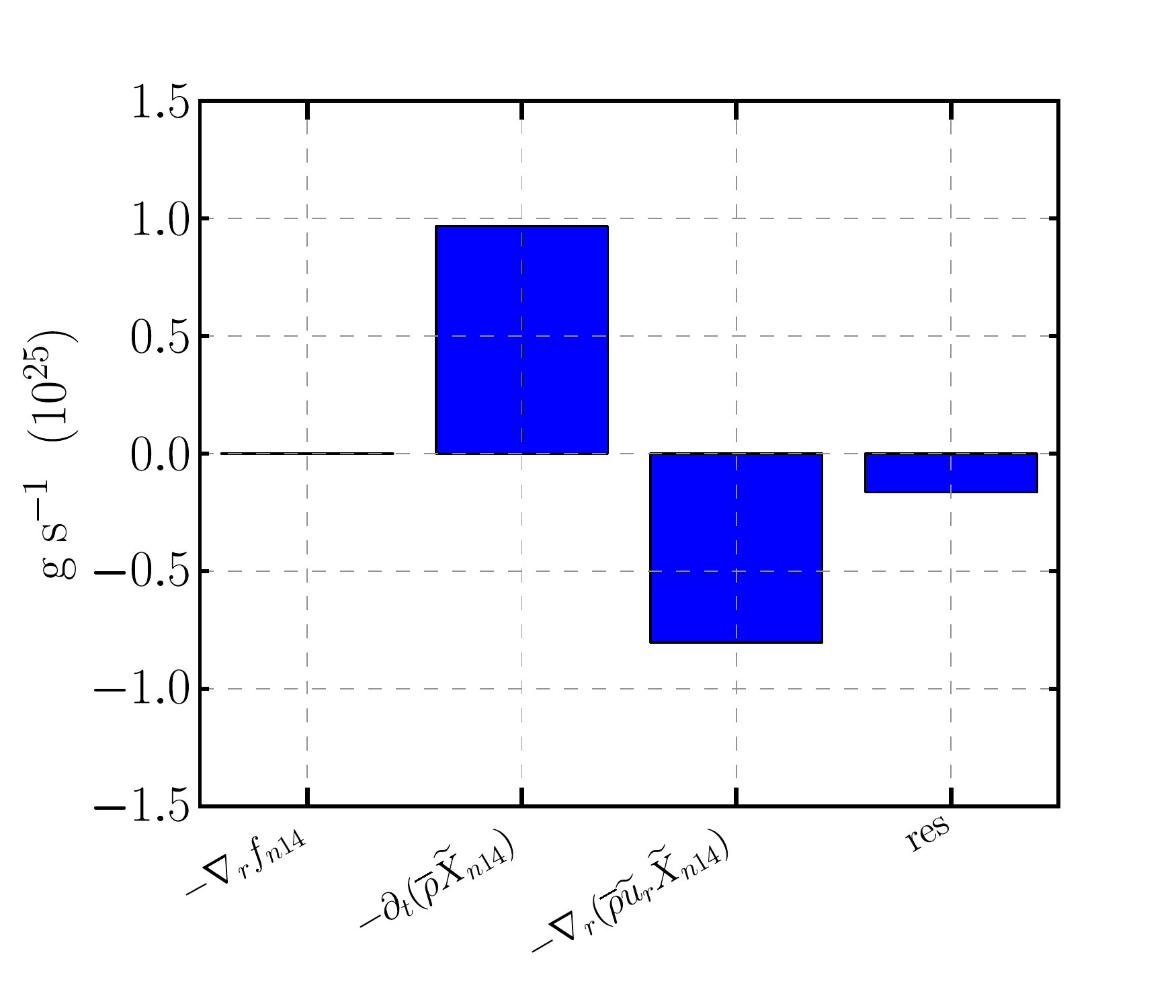}}

\centerline{
\includegraphics[width=6.7cm]{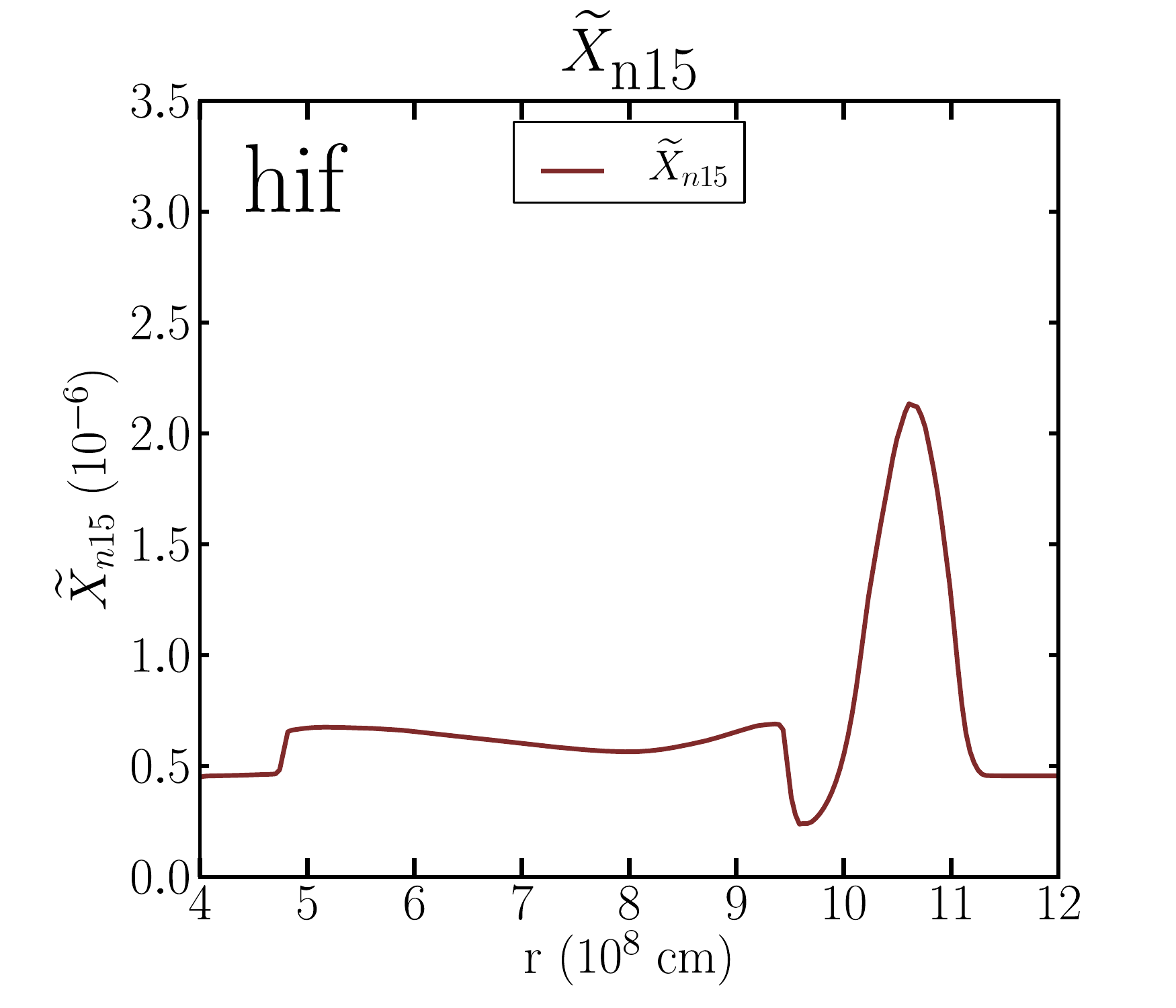}
\includegraphics[width=6.7cm]{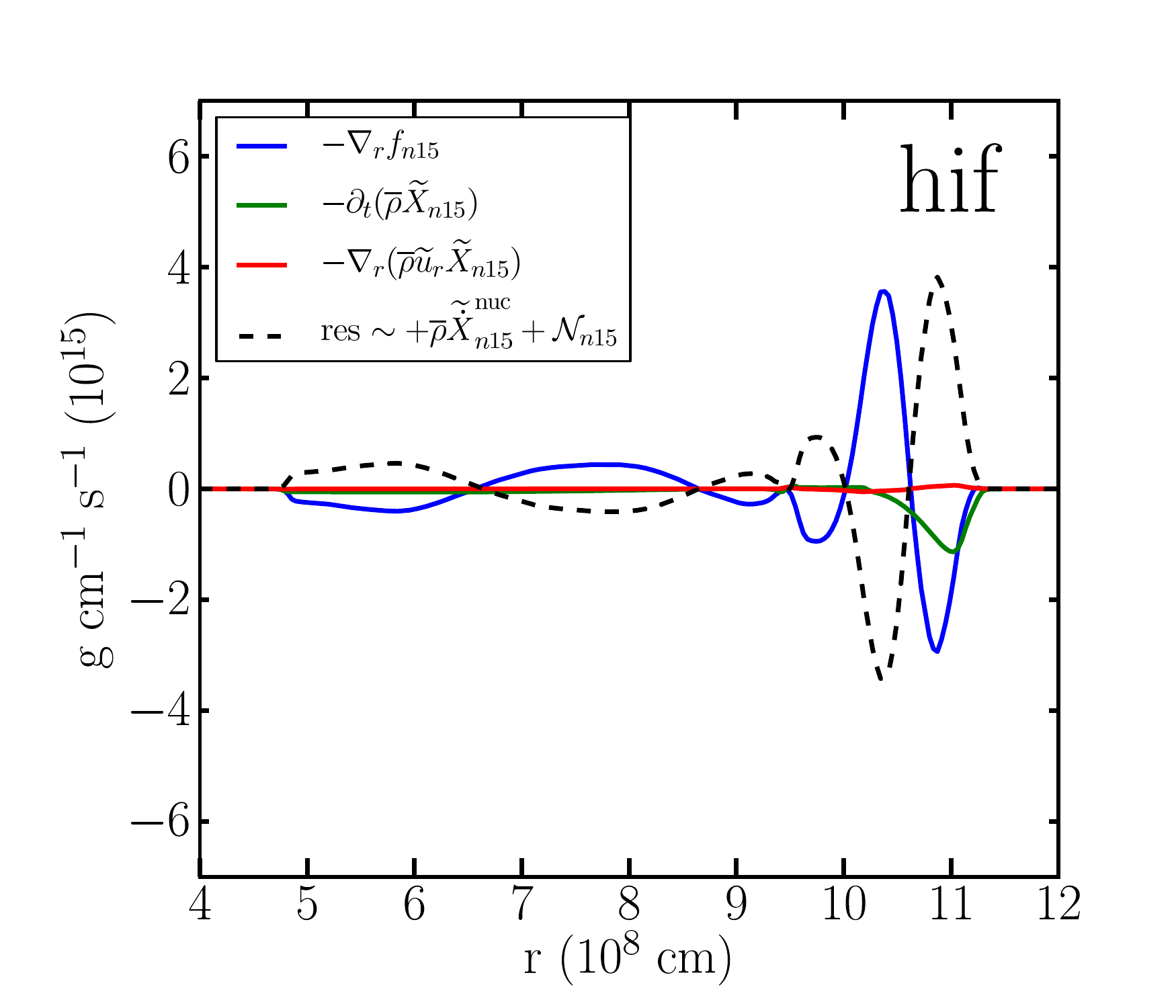}
\includegraphics[width=6.7cm]{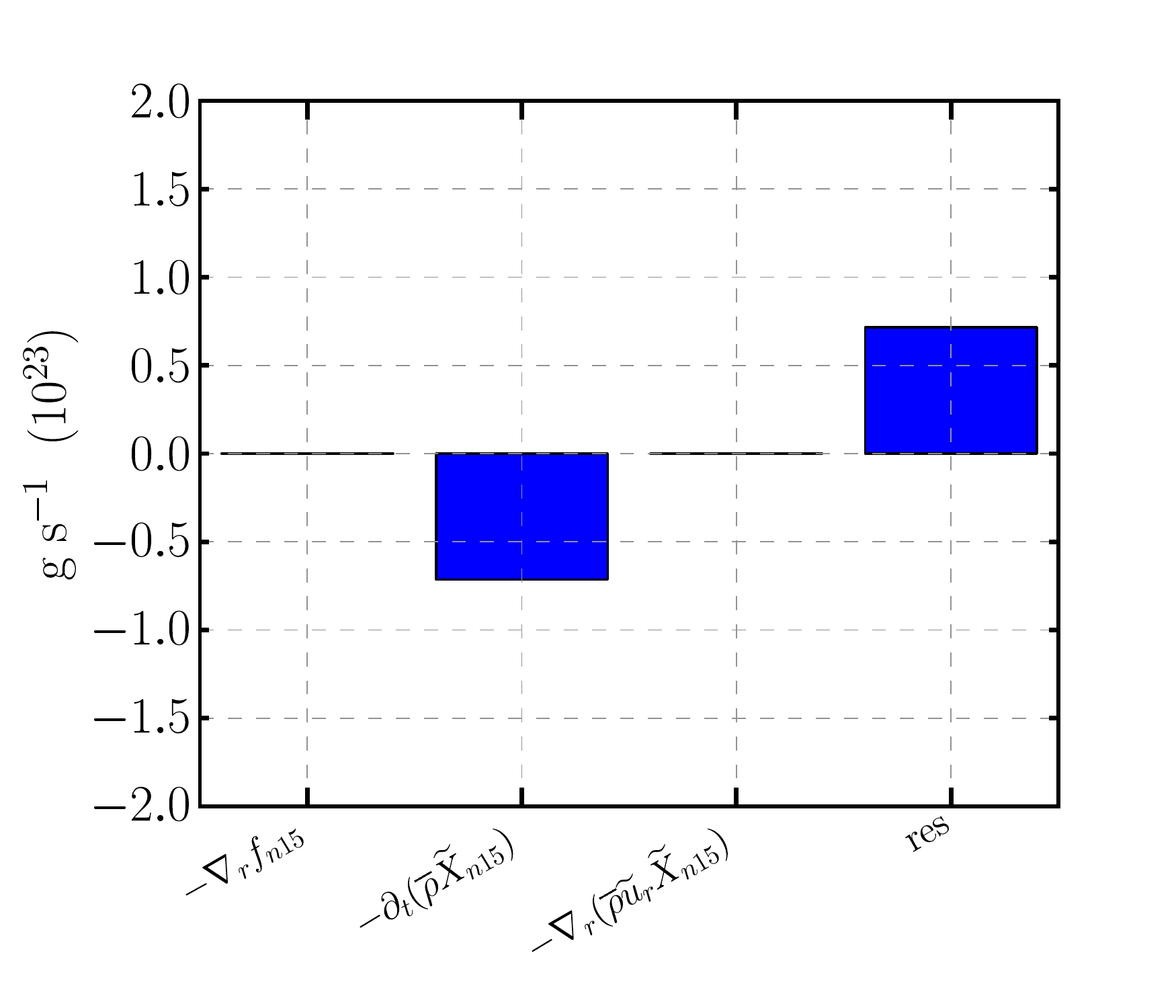}}
\caption{Mean composition equations for {\sf hif.3D}. \label{fig:x-equations}}
\end{figure}

\newpage

\subsection{Mean O$^{15}$ and O$^{16}$ equation}

\begin{figure}[!h]
\centerline{
\includegraphics[width=6.7cm]{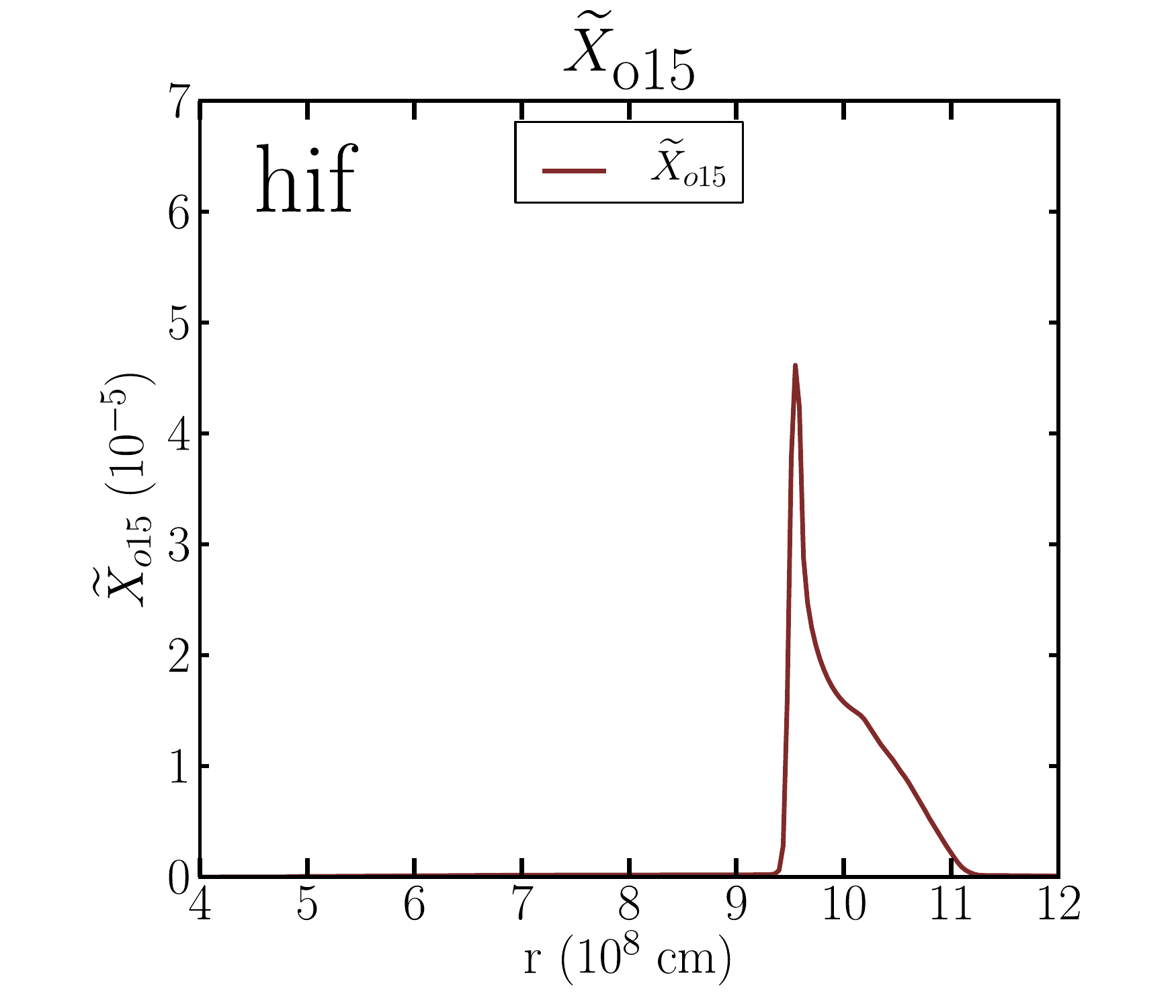}
\includegraphics[width=6.7cm]{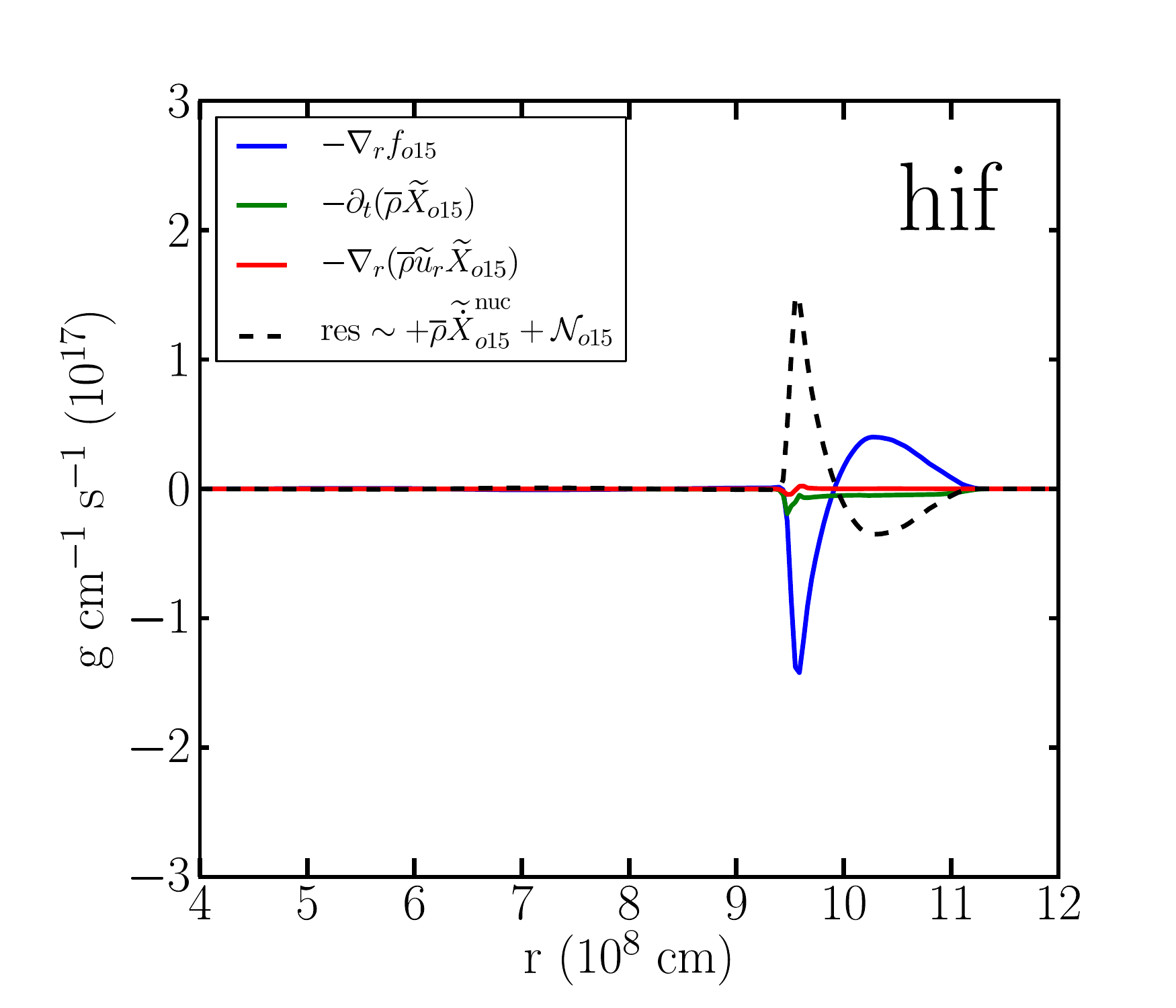}
\includegraphics[width=6.7cm]{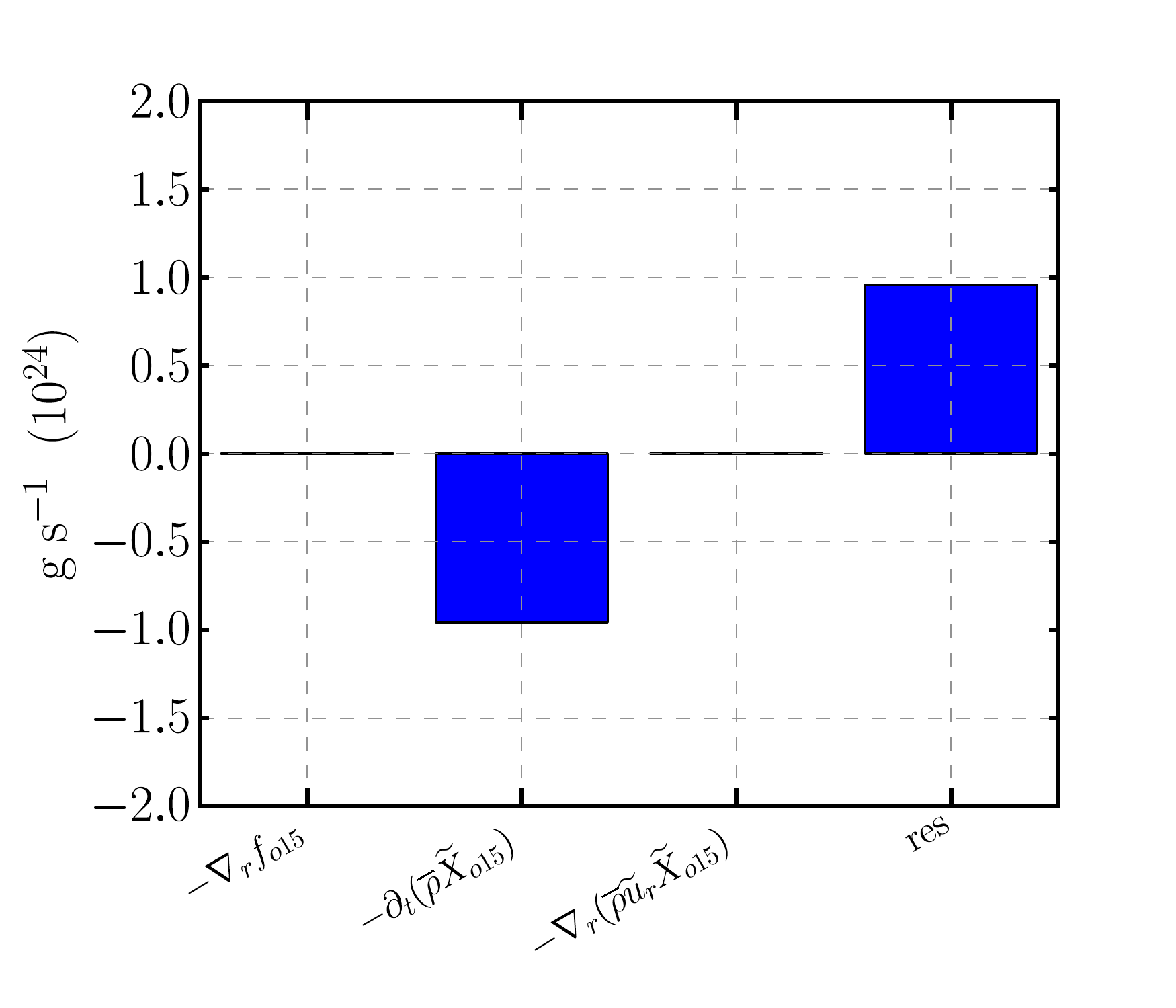}}

\centerline{
\includegraphics[width=6.7cm]{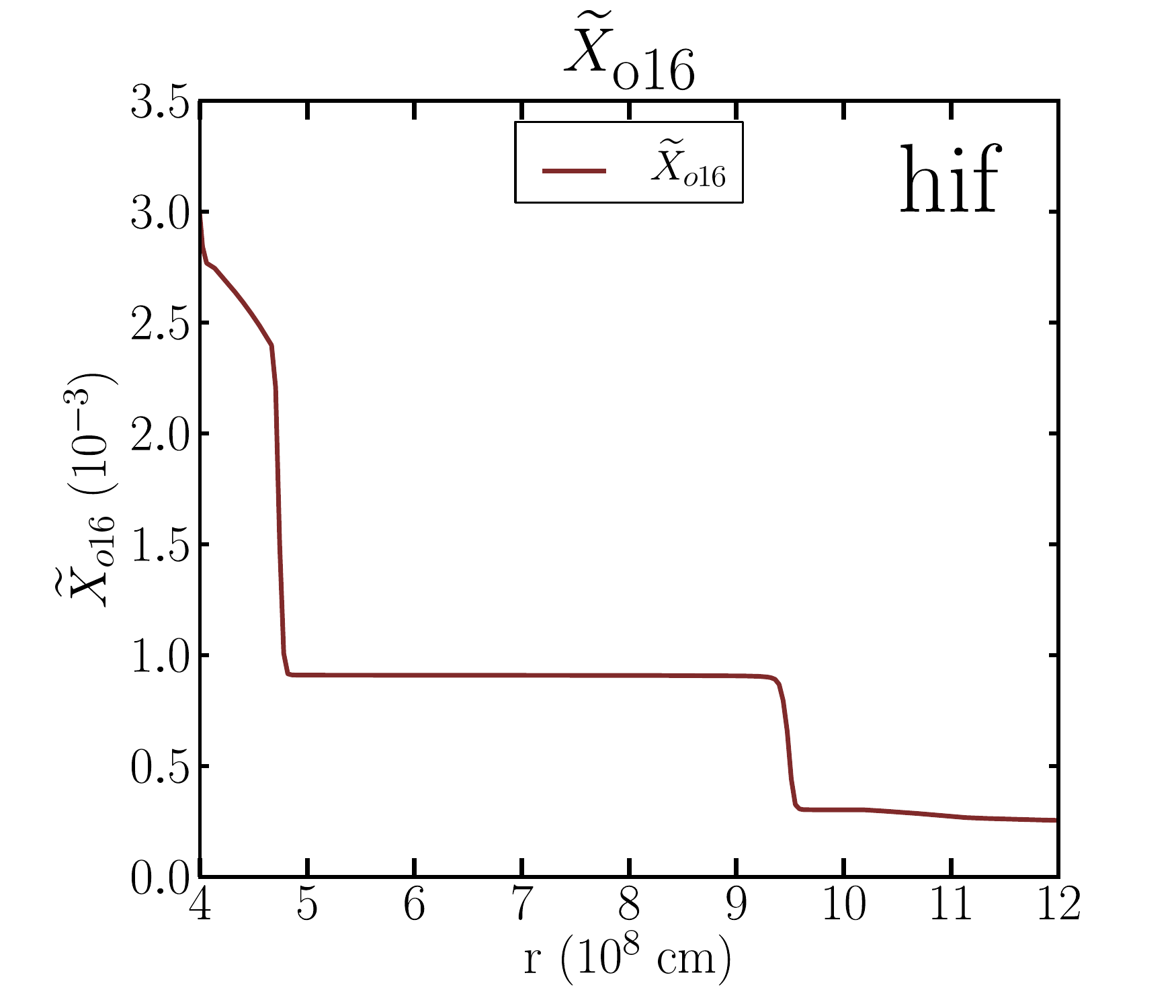}
\includegraphics[width=6.7cm]{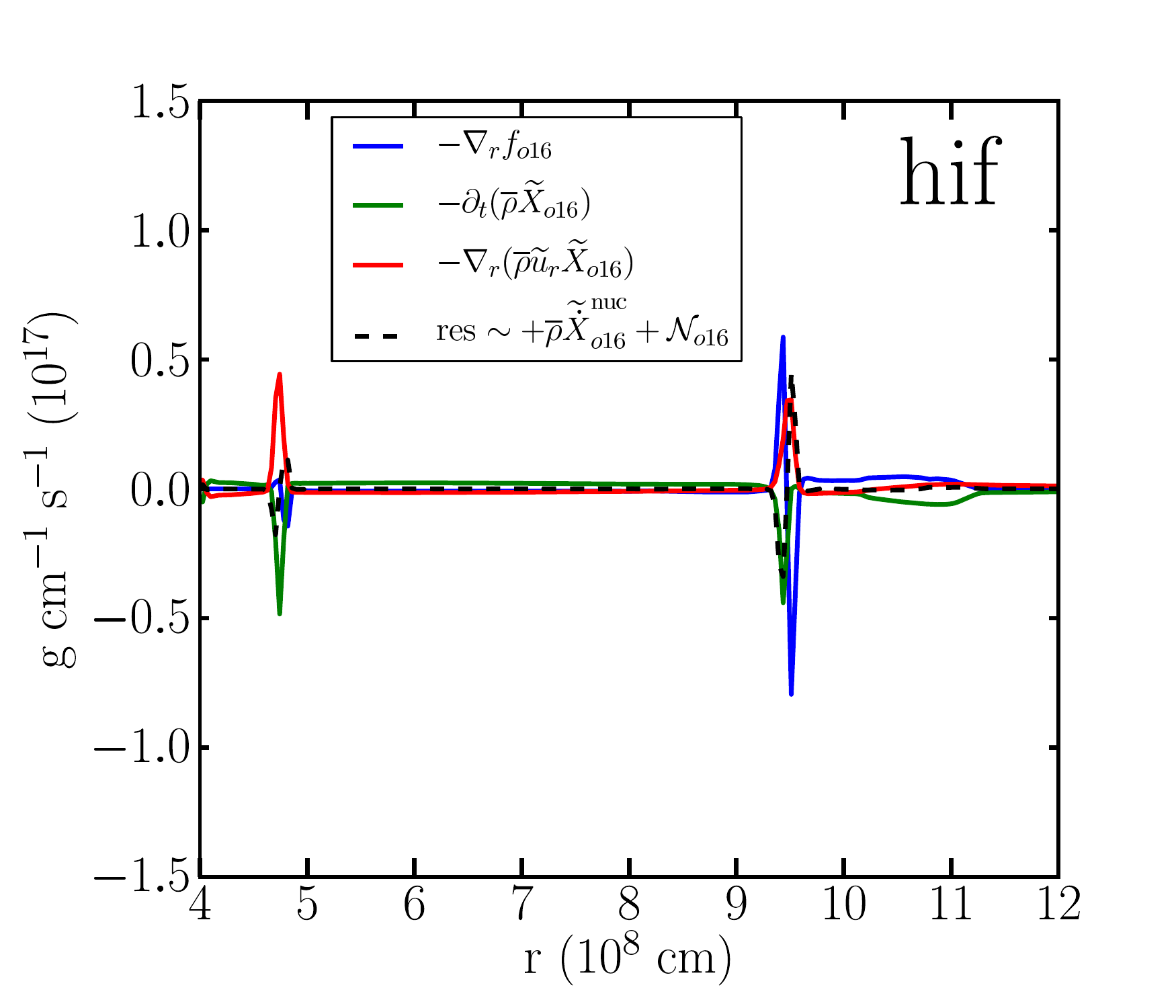}
\includegraphics[width=6.7cm]{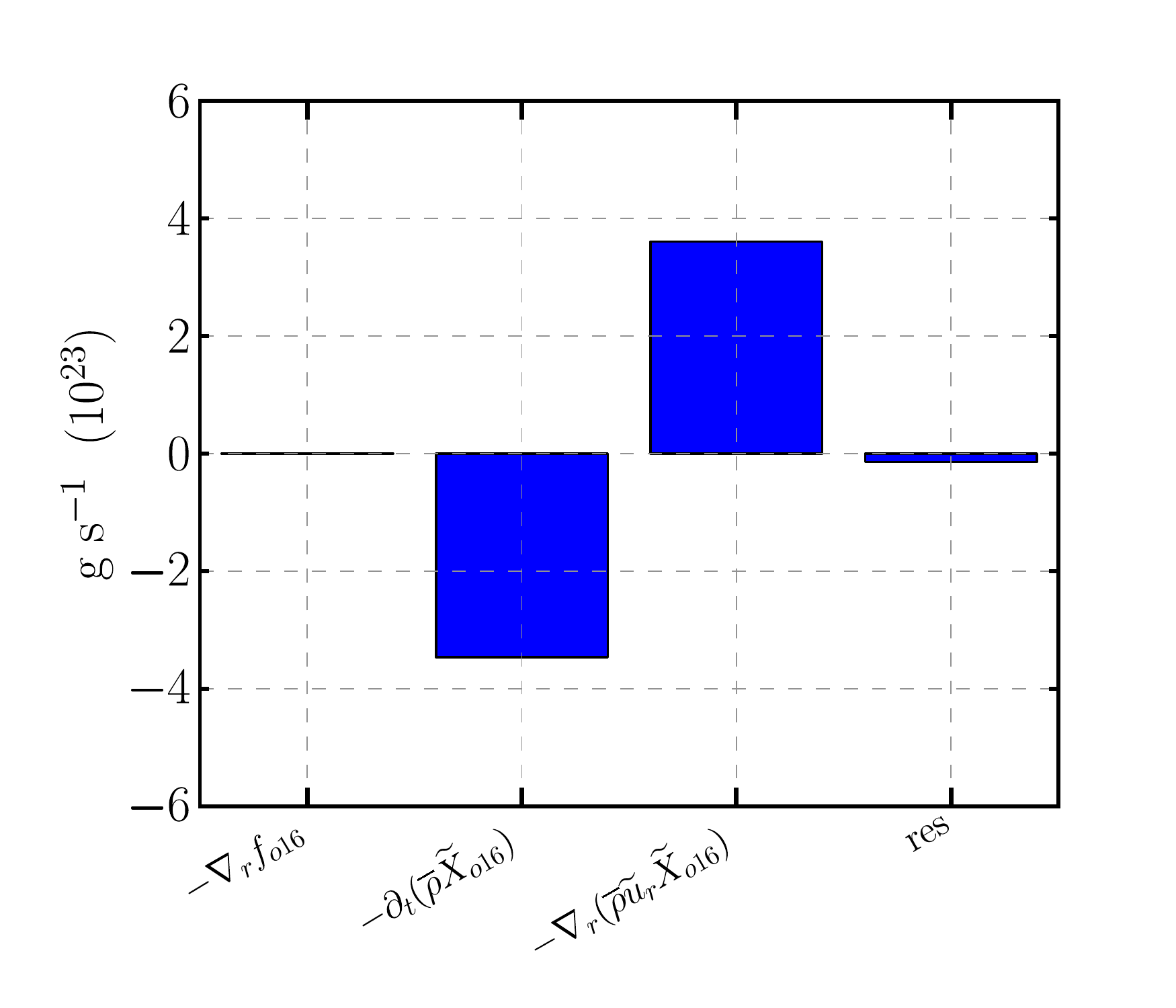}}
\caption{Mean composition equations for {\sf hif.3D}. \label{fig:x-equations}}
\end{figure}

\newpage

\subsection{Mean O$^{17}$ and Ne$^{20}$ equation}

\begin{figure}[!h]
\centerline{
\includegraphics[width=6.7cm]{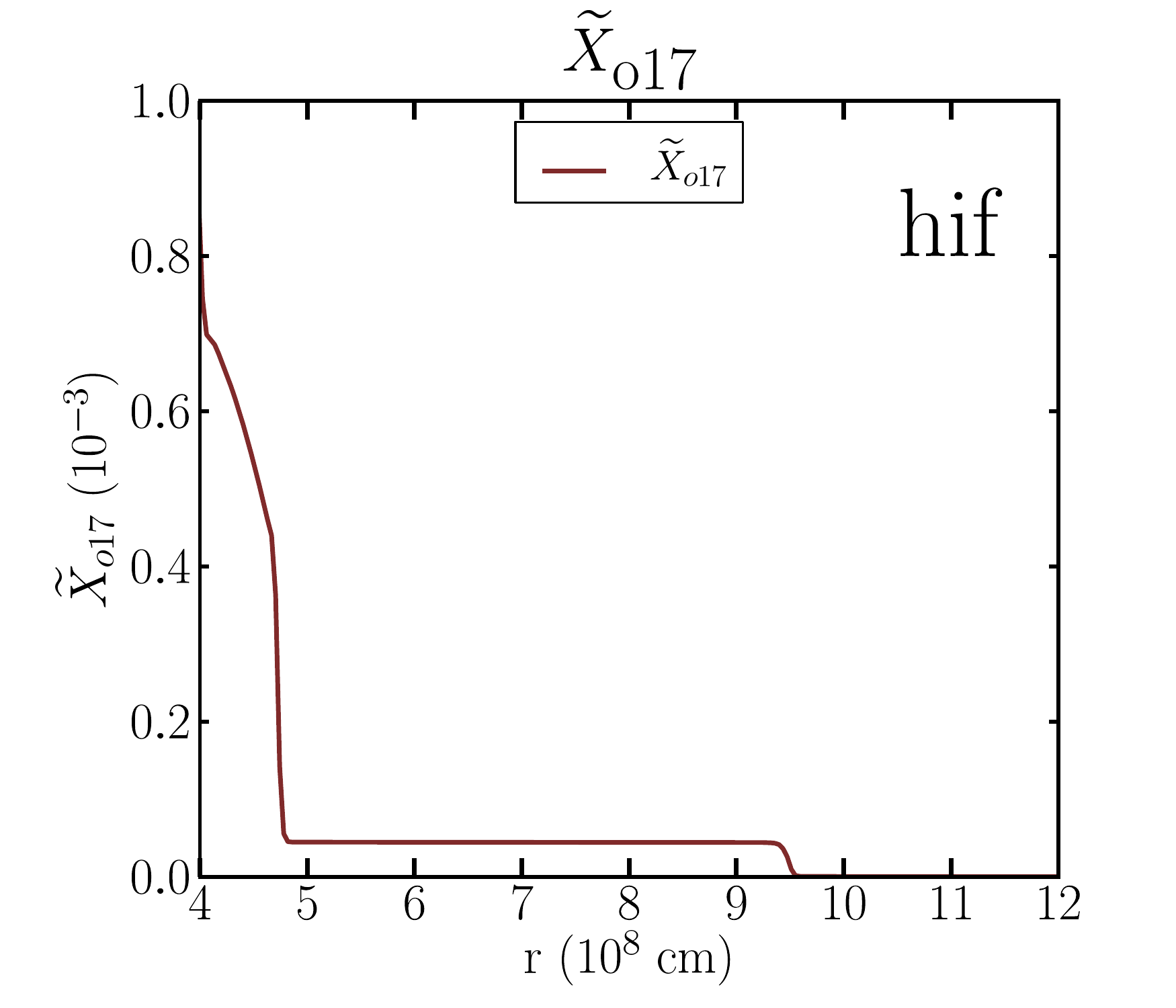}
\includegraphics[width=6.7cm]{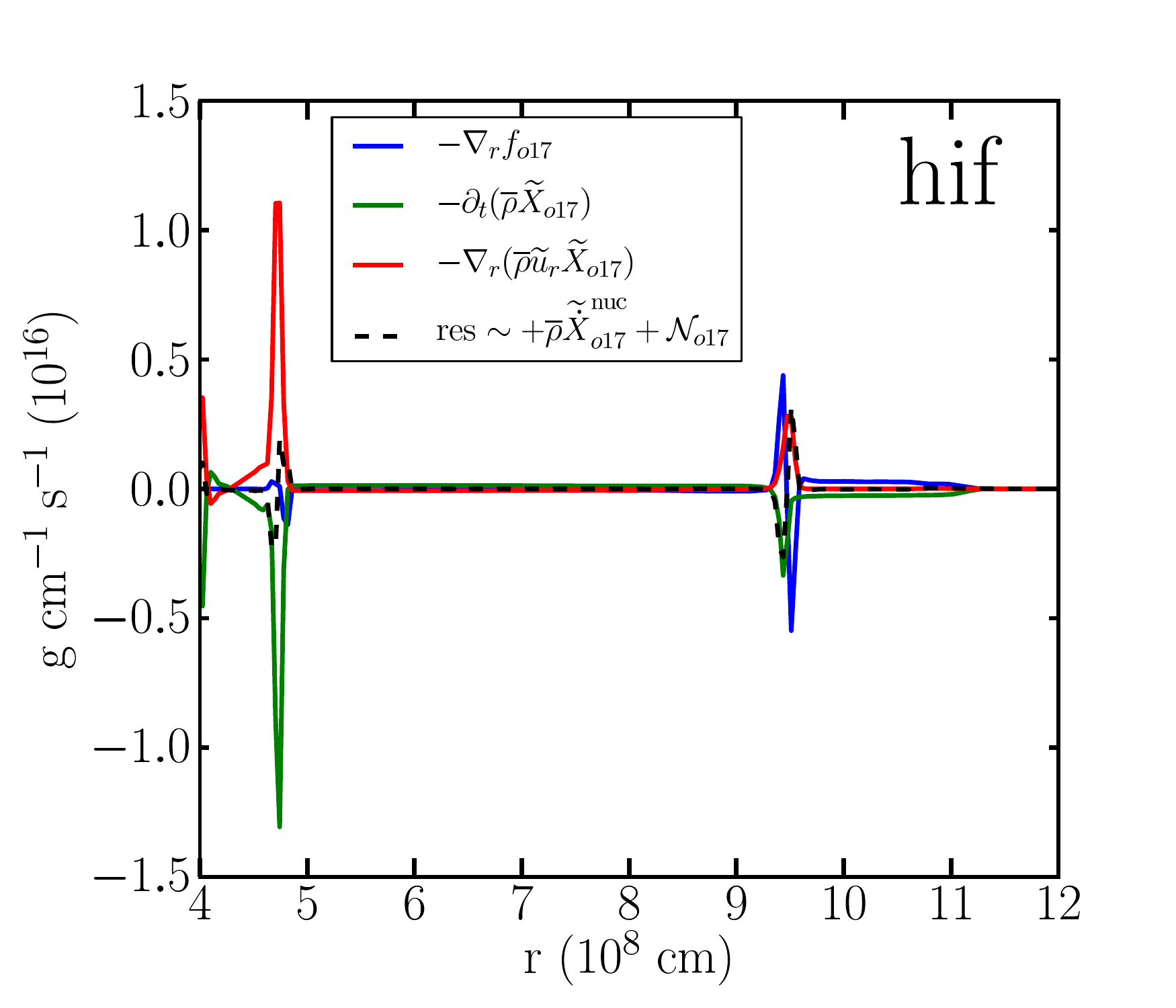}
\includegraphics[width=6.7cm]{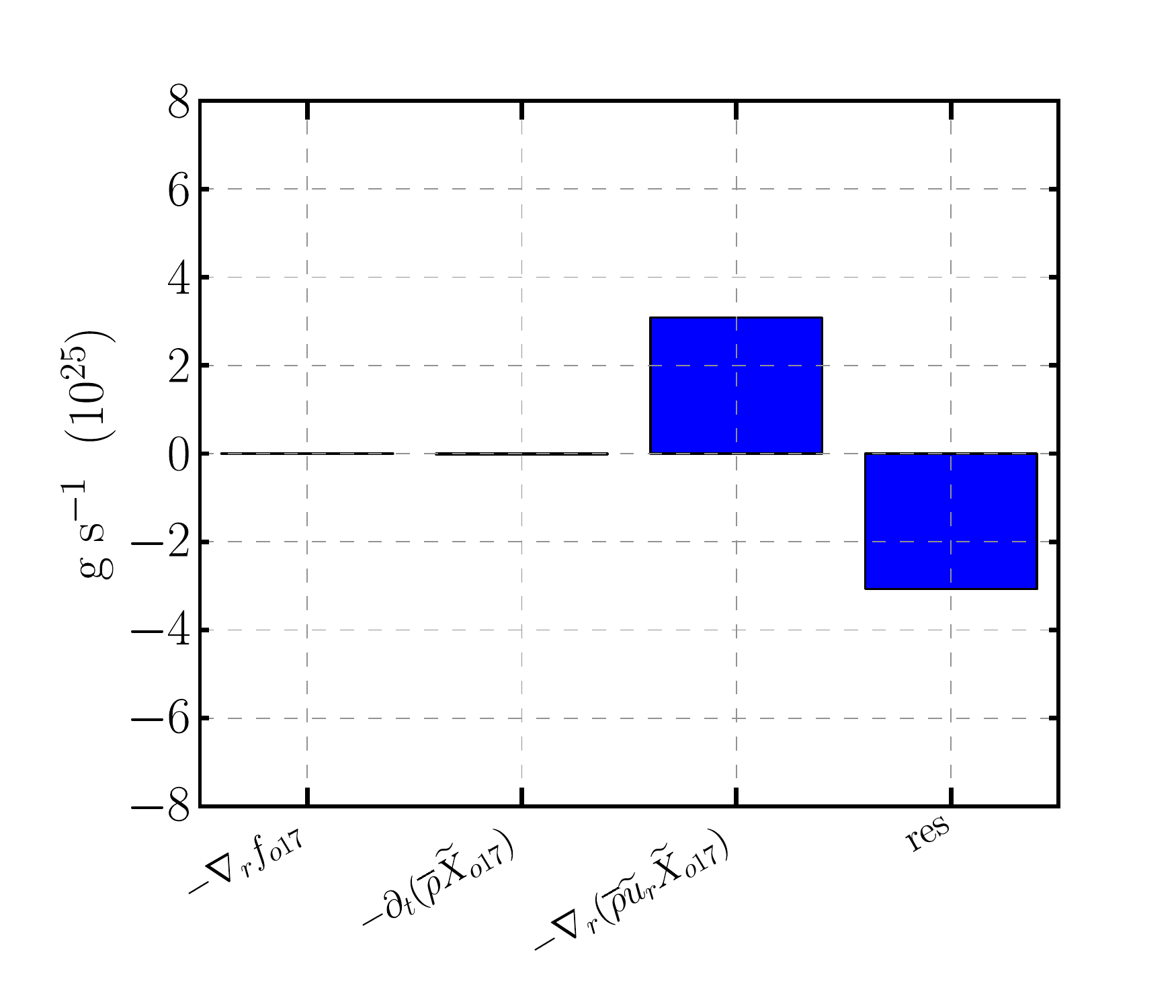}}

\centerline{
\includegraphics[width=6.7cm]{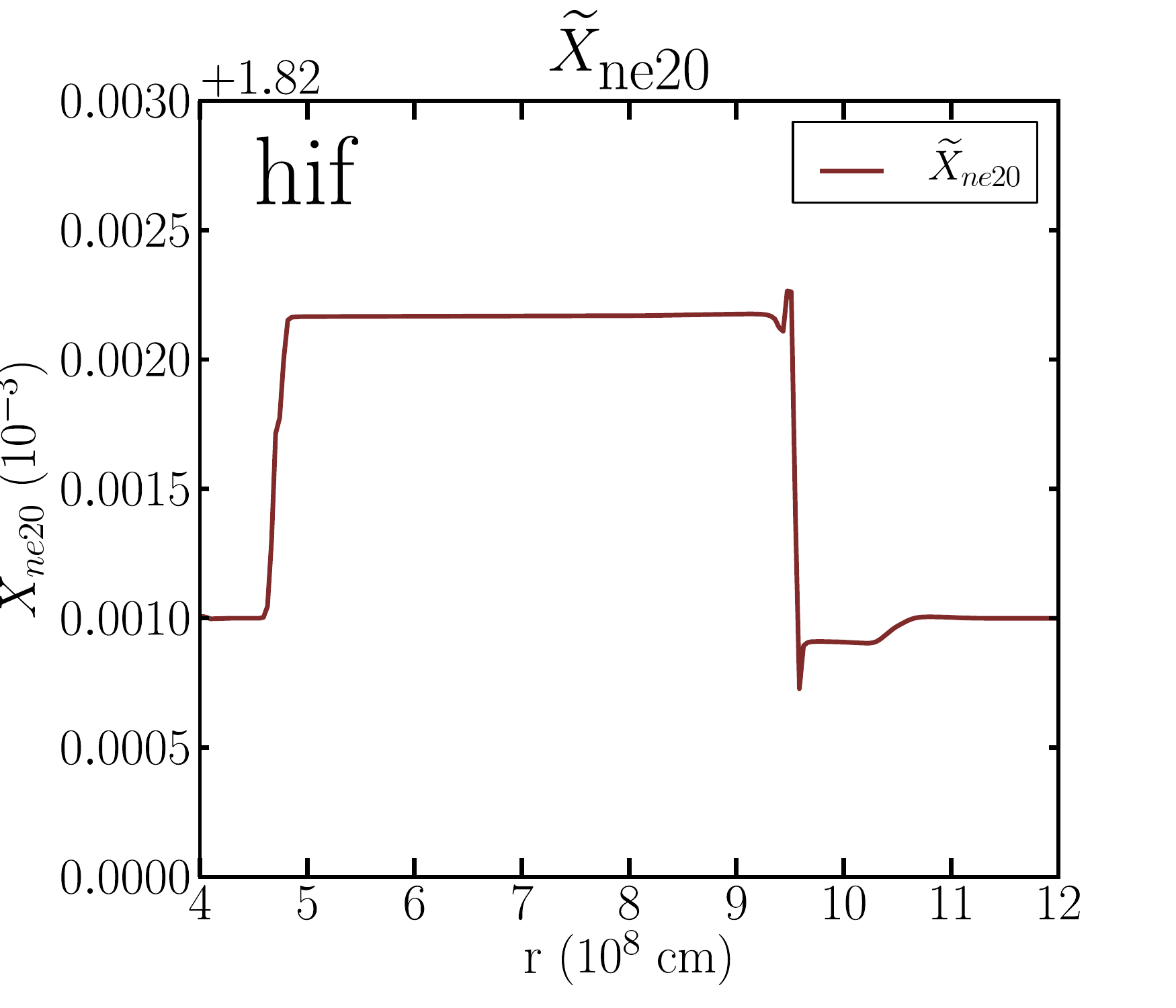}
\includegraphics[width=6.7cm]{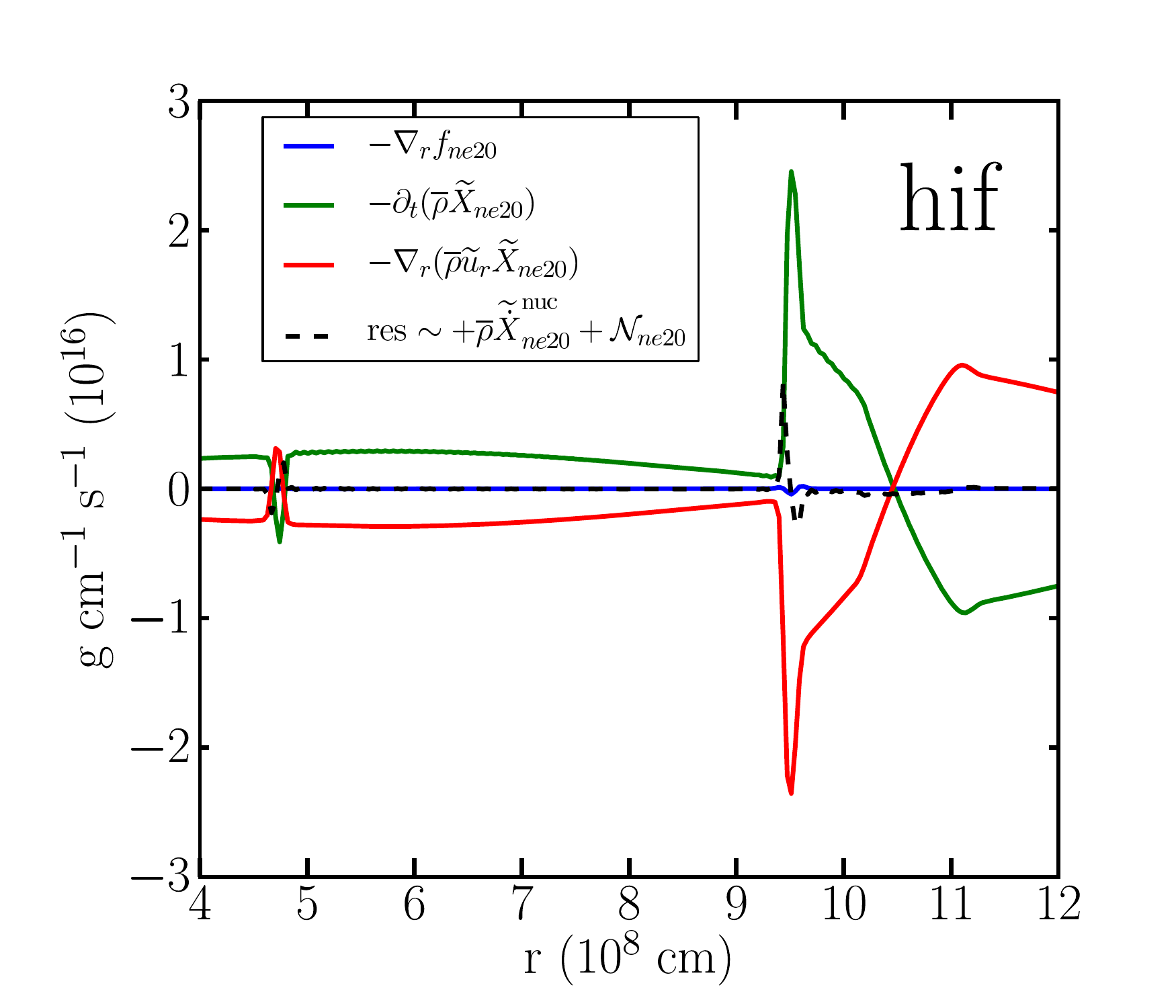}
\includegraphics[width=6.7cm]{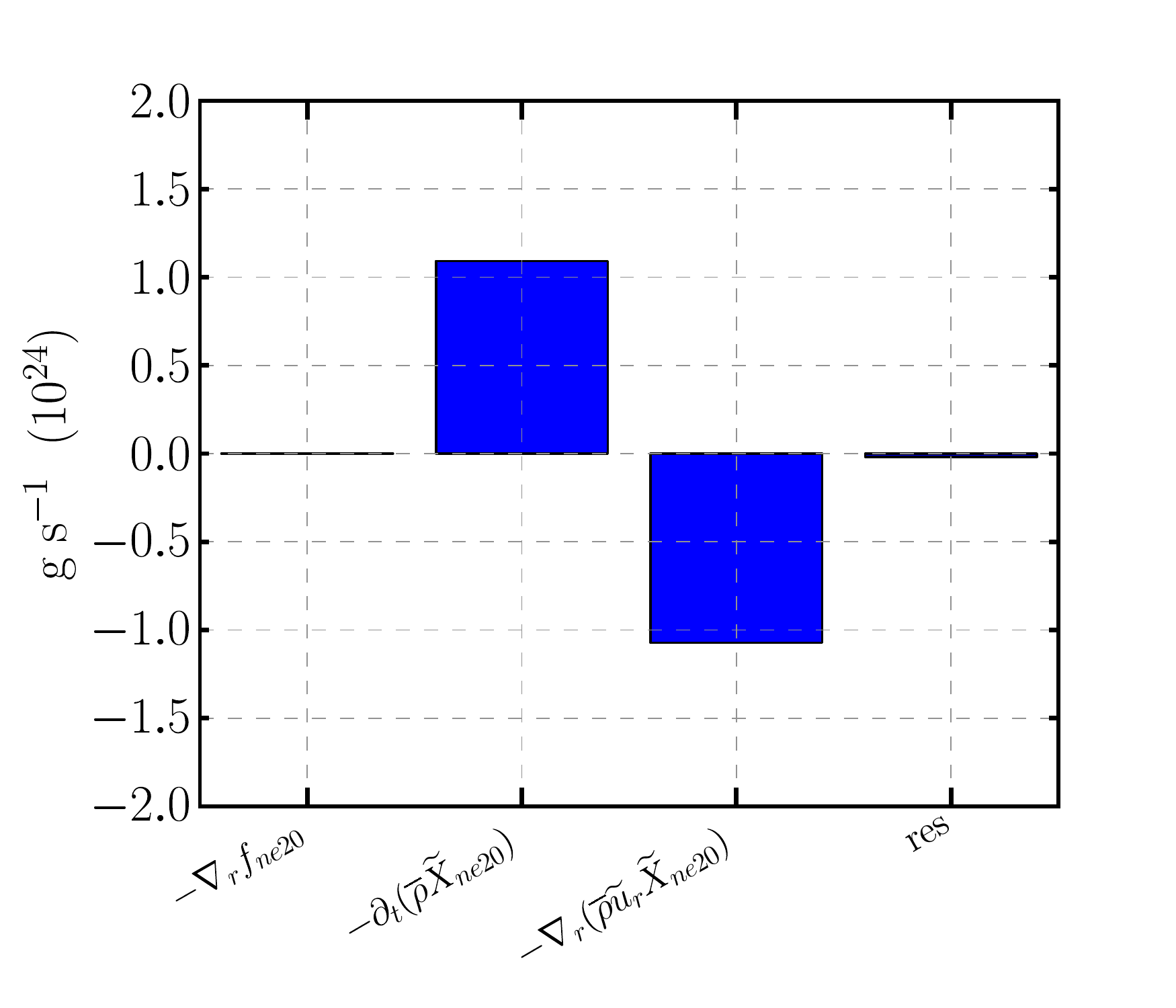}}
\caption{Mean composition equations for {\sf hif.3D}. \label{fig:x-equations}}
\end{figure}

\newpage

\subsection{Mean Mg$^{24}$ and Si$^{28}$ equation}

\begin{figure}[!h]
\centerline{
\includegraphics[width=6.7cm]{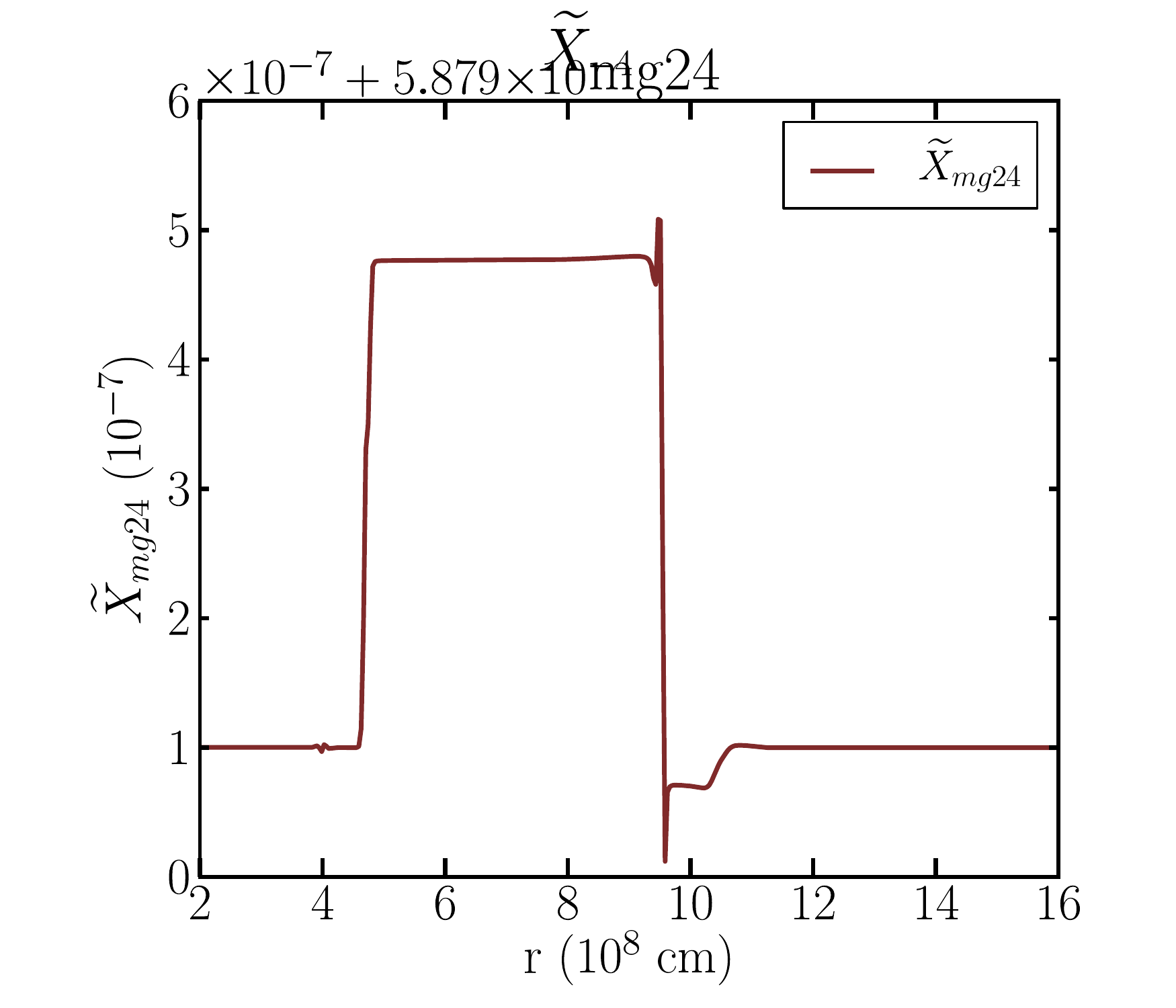}
\includegraphics[width=6.7cm]{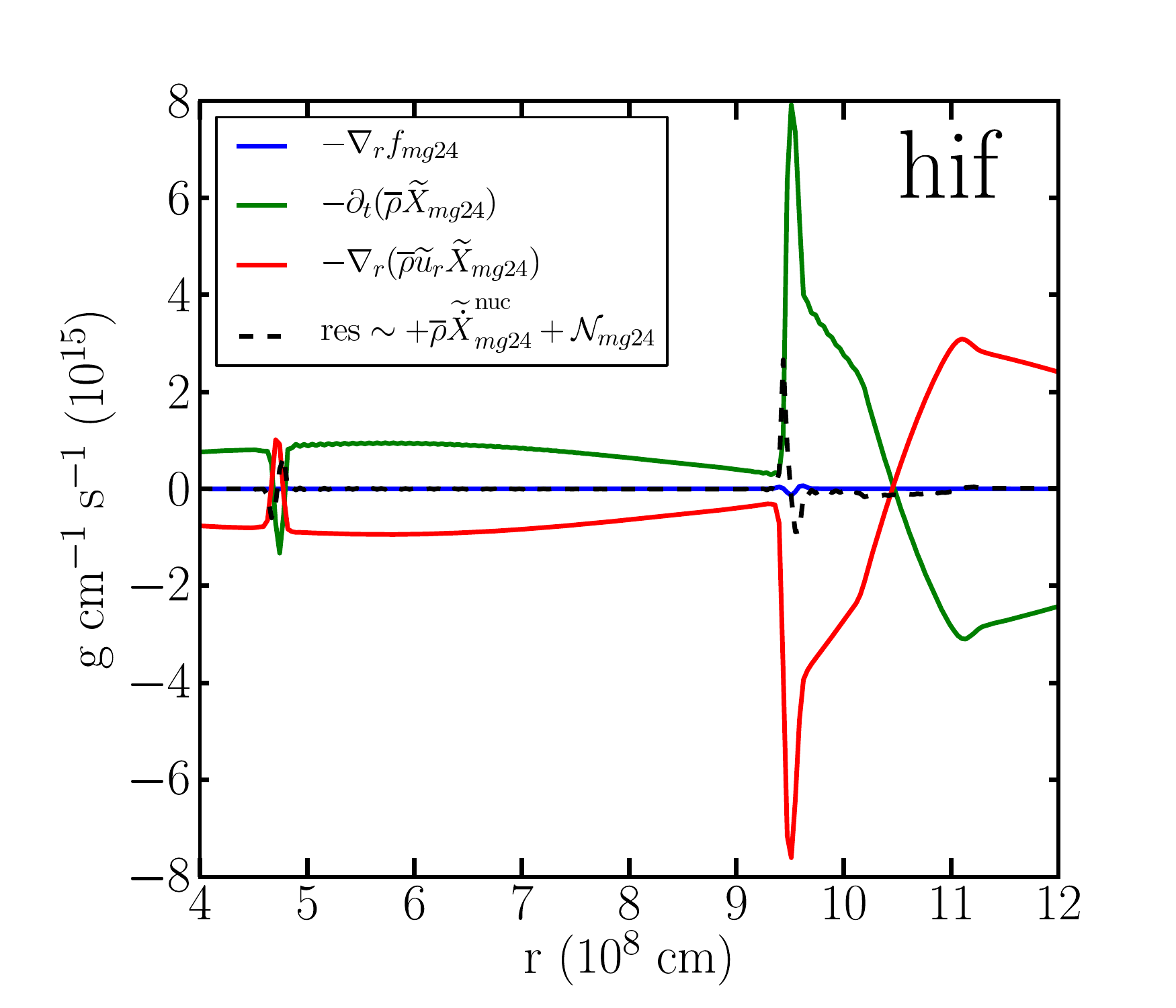}
\includegraphics[width=6.7cm]{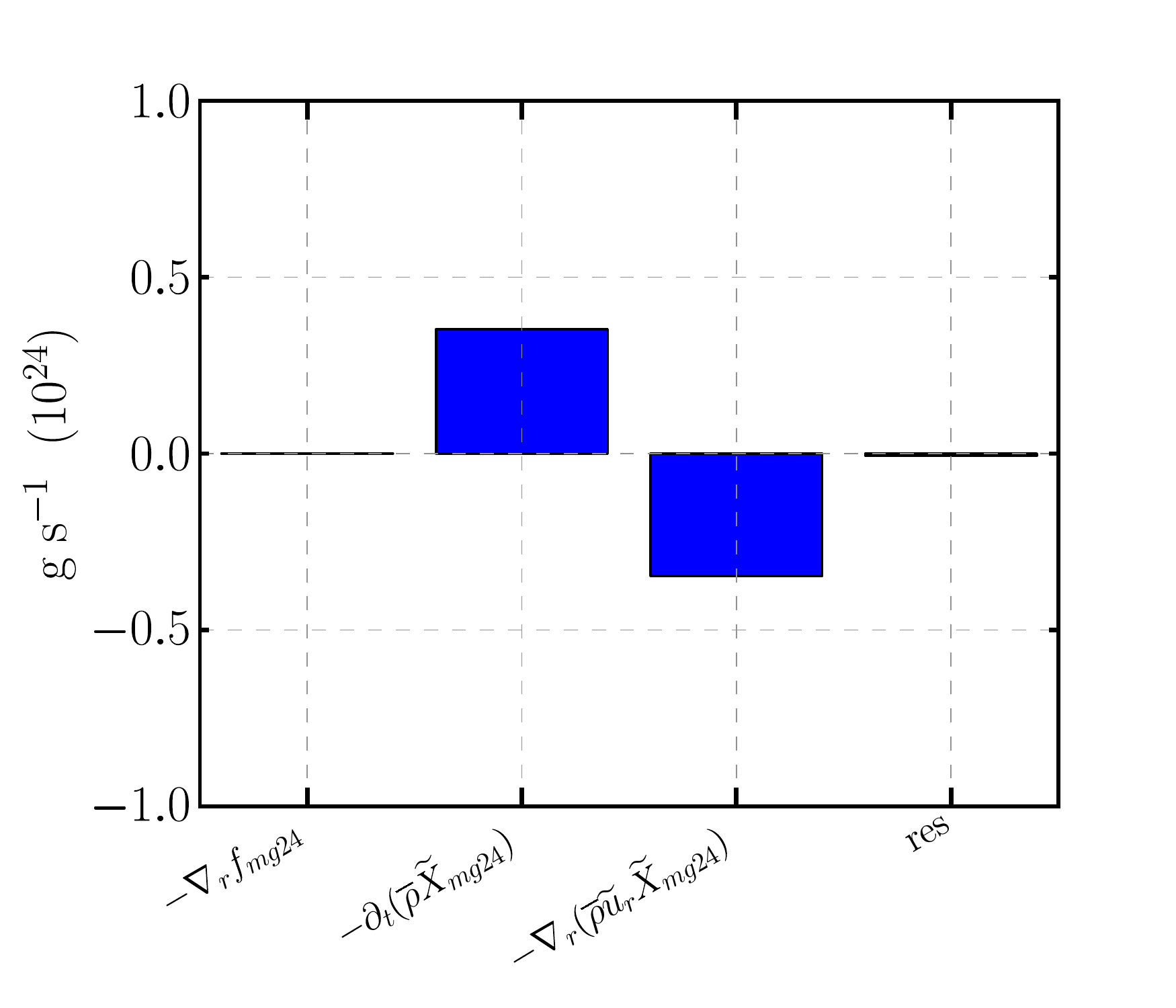}}

\centerline{
\includegraphics[width=6.7cm]{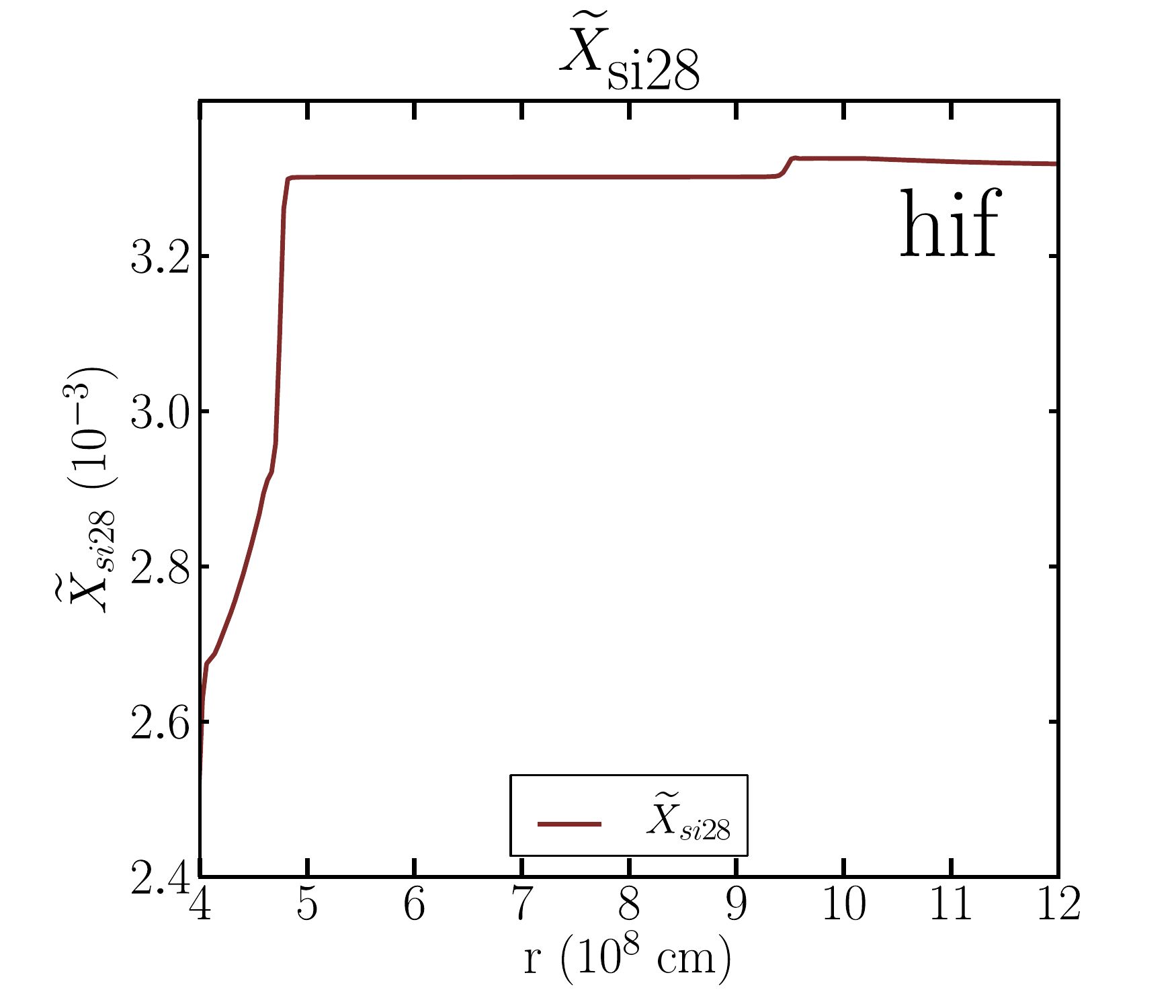}
\includegraphics[width=6.7cm]{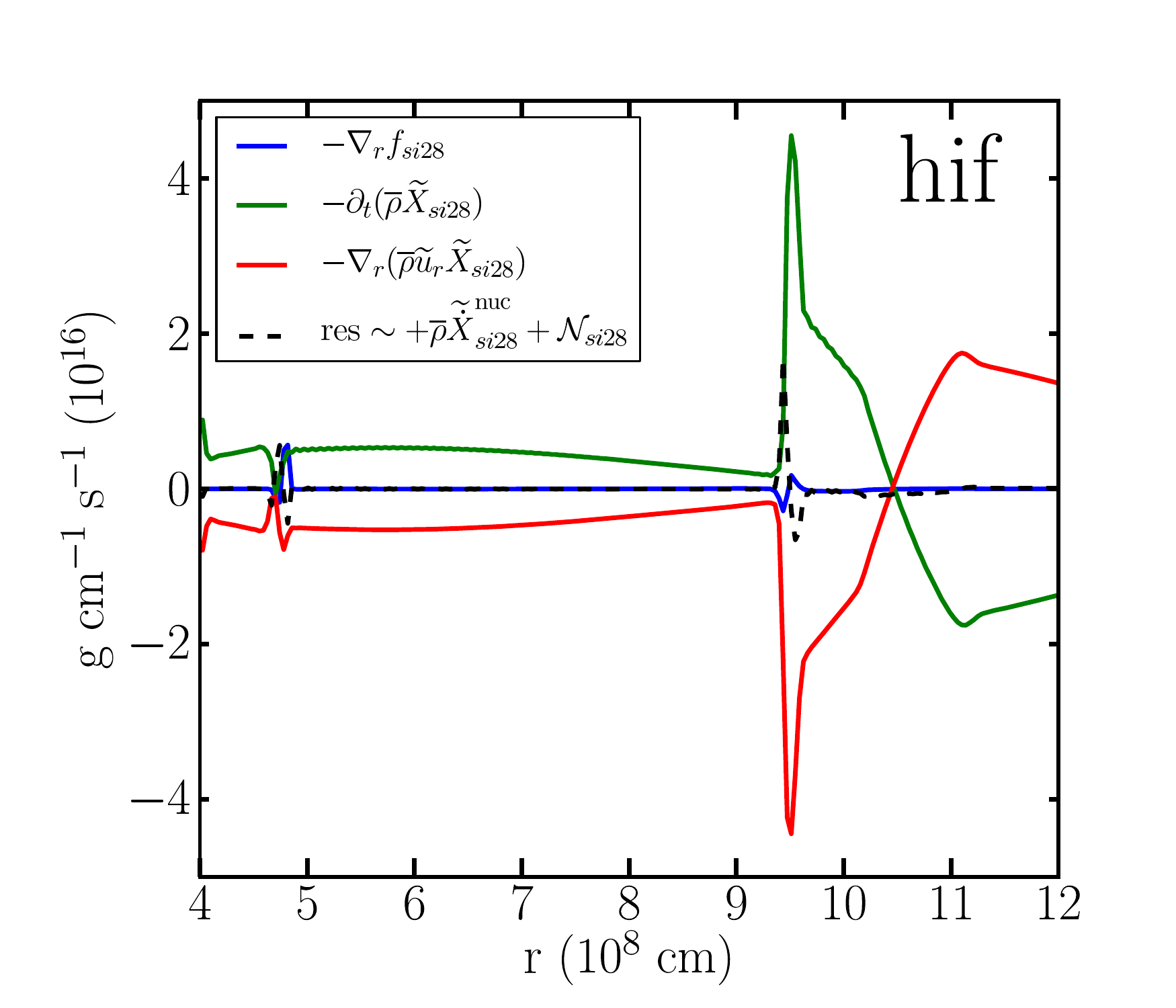}
\includegraphics[width=6.7cm]{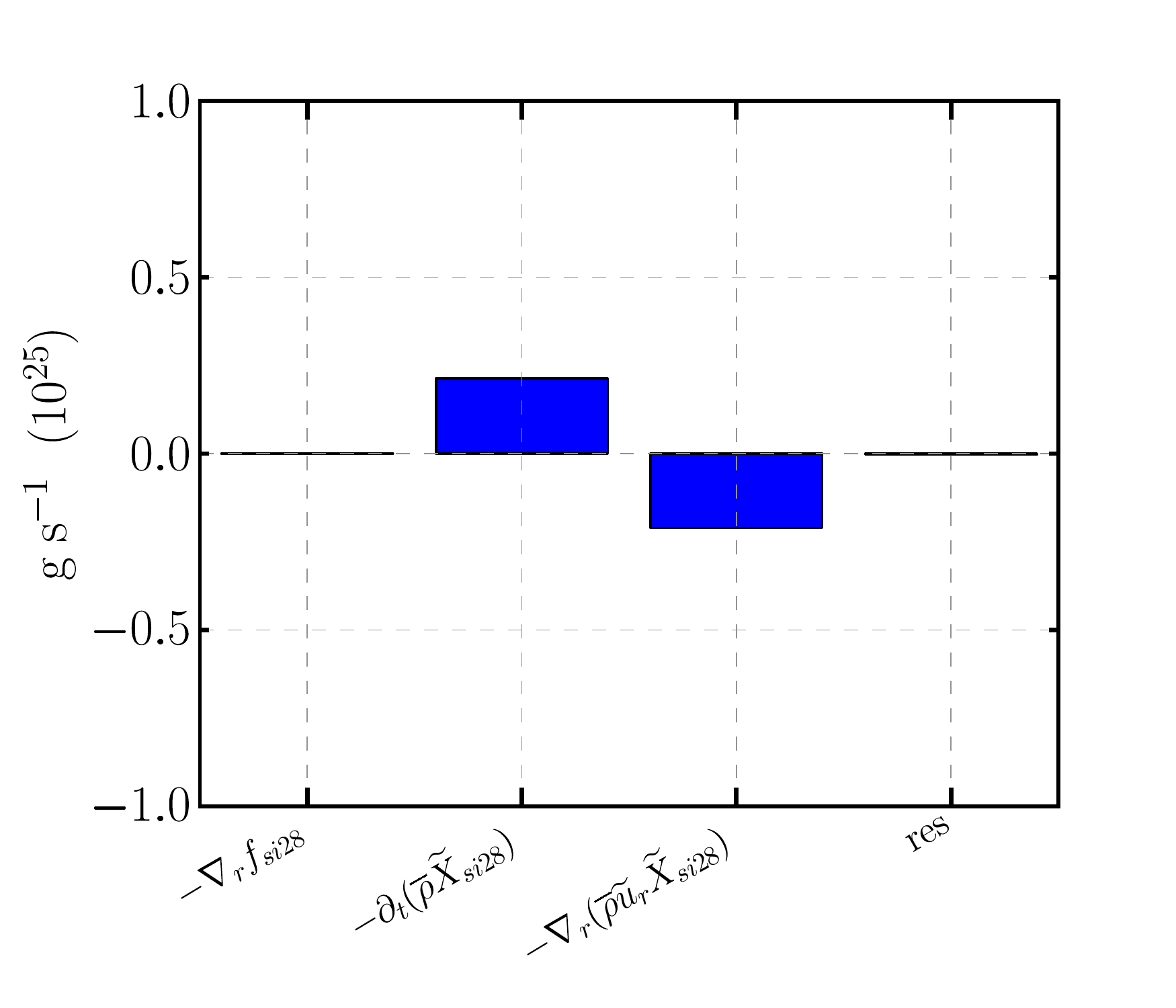}}
\caption{Mean composition equations for {\sf hif.3D} \label{fig:x-equations}}
\end{figure}

\newpage

\section{Mean field composition data for the core helium flash model}

\subsection{Mean He$^{4}$ and C$^{12}$ equation}

\begin{figure}[!h]
\centerline{
\includegraphics[width=6.7cm]{chf3d_tavg12000_mean_xhe4_insf-eps-converted-to.pdf}
\includegraphics[width=6.7cm]{chf3d_tavg12000_xhe4_equation_insf-eps-converted-to.pdf}
\includegraphics[width=6.7cm]{chf3d_tavg12000_xhe4_equation_insf_bar-eps-converted-to.pdf}}

\centerline{
\includegraphics[width=6.7cm]{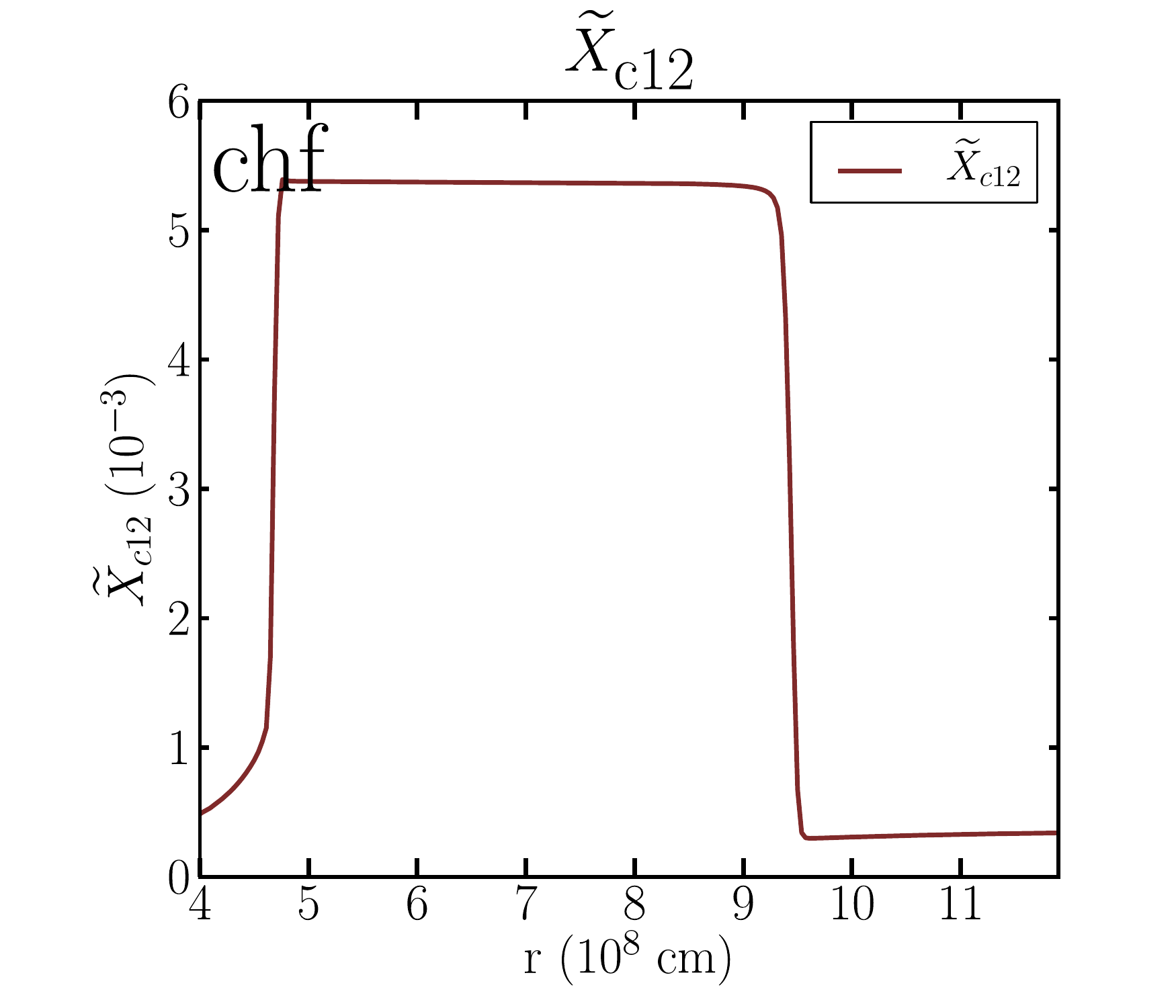}
\includegraphics[width=6.7cm]{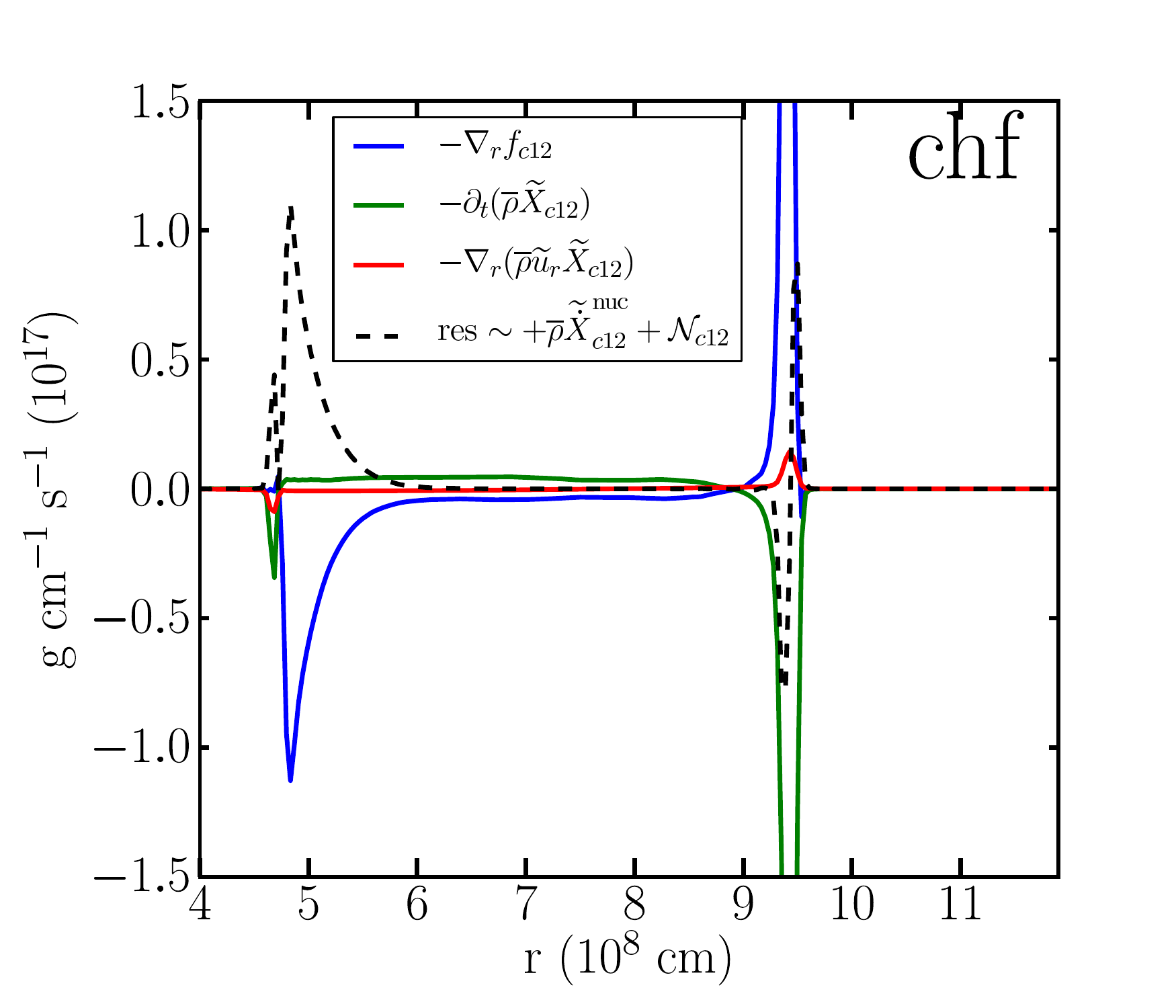}
\includegraphics[width=6.7cm]{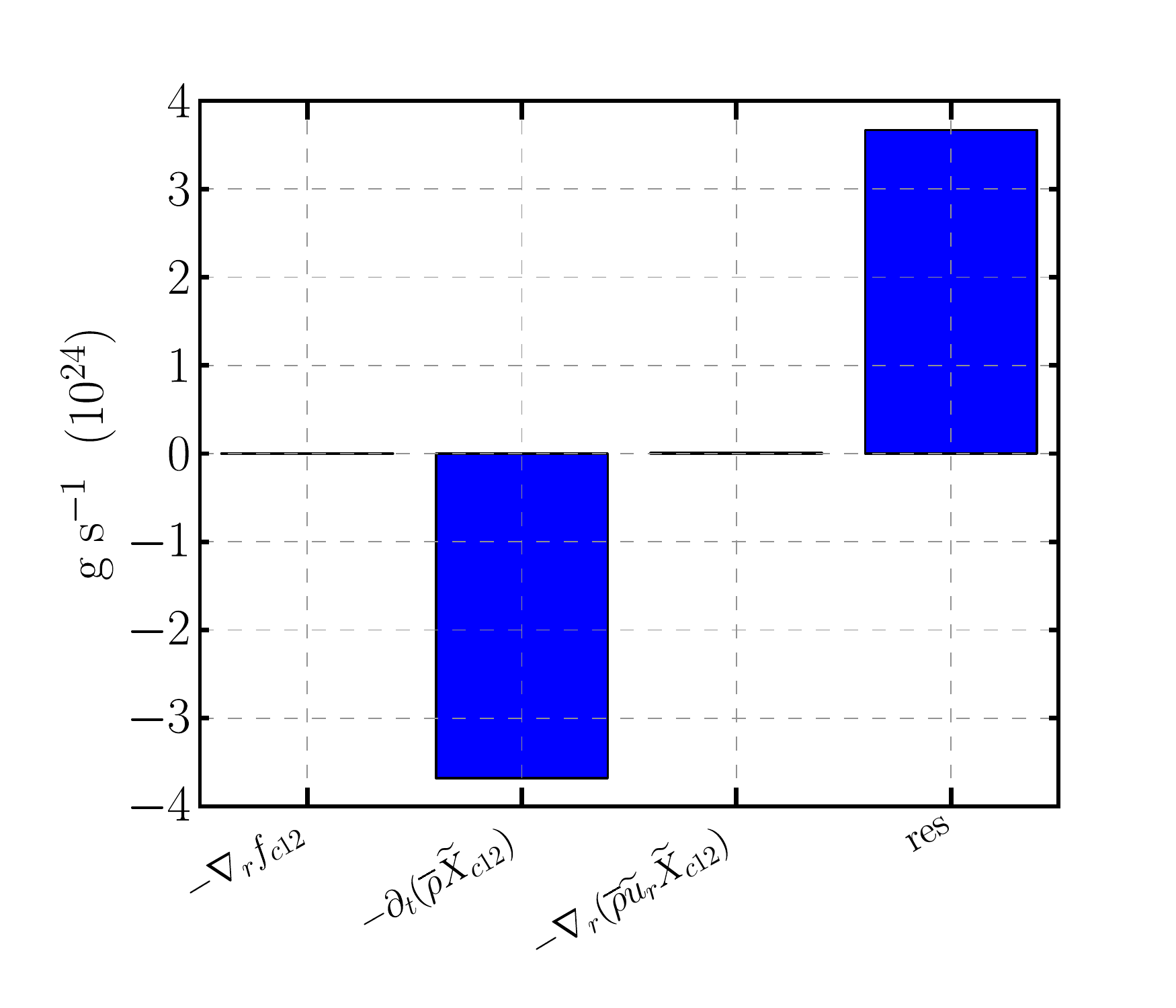}}
\caption{Mean composition equations for {\sf chf.3D}. \label{fig:x-equations}}
\end{figure}

\newpage

\subsection{Mean O$^{16}$ and Ne$^{20}$ equation}

\begin{figure}[!h]
\centerline{
\includegraphics[width=6.7cm]{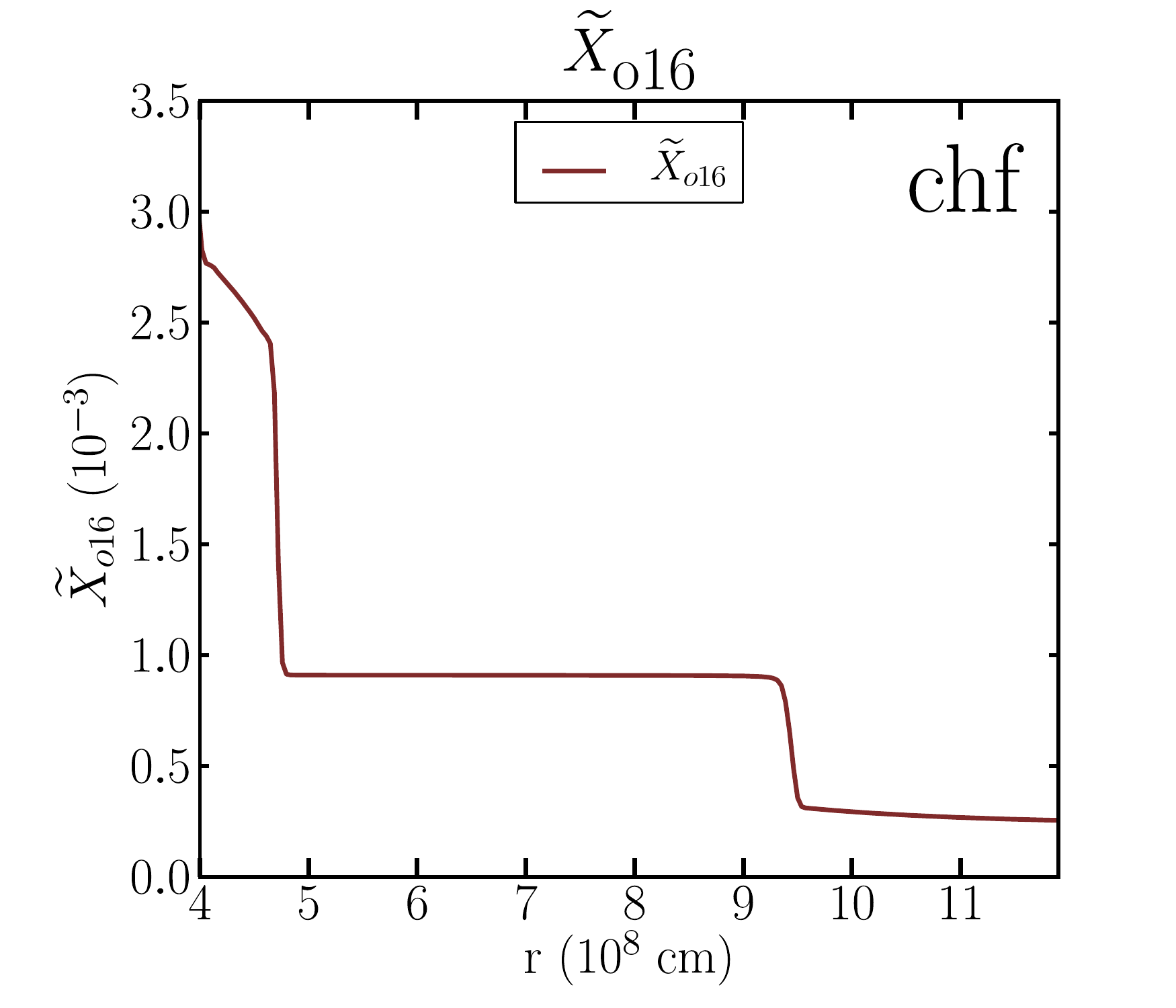}
\includegraphics[width=6.7cm]{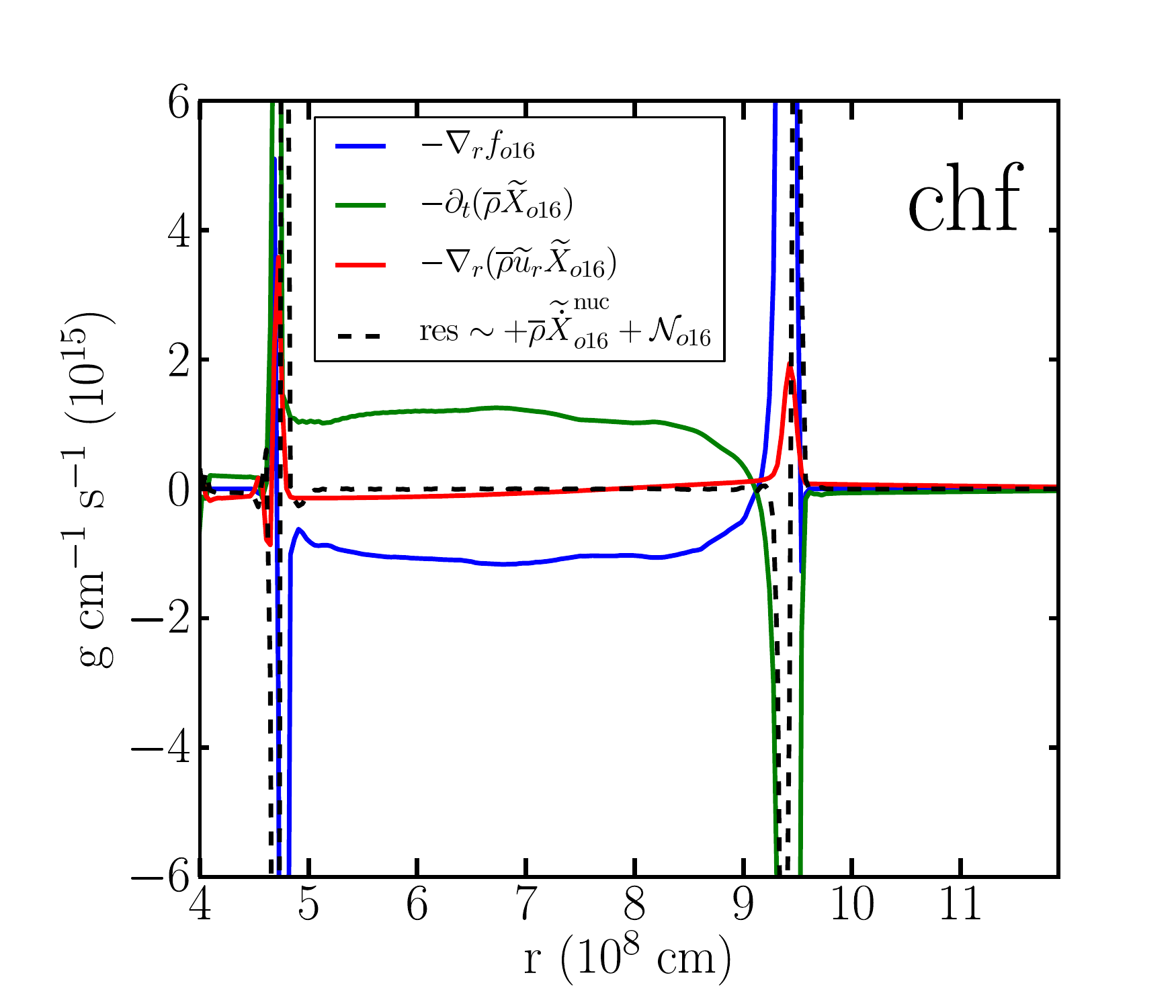}
\includegraphics[width=6.7cm]{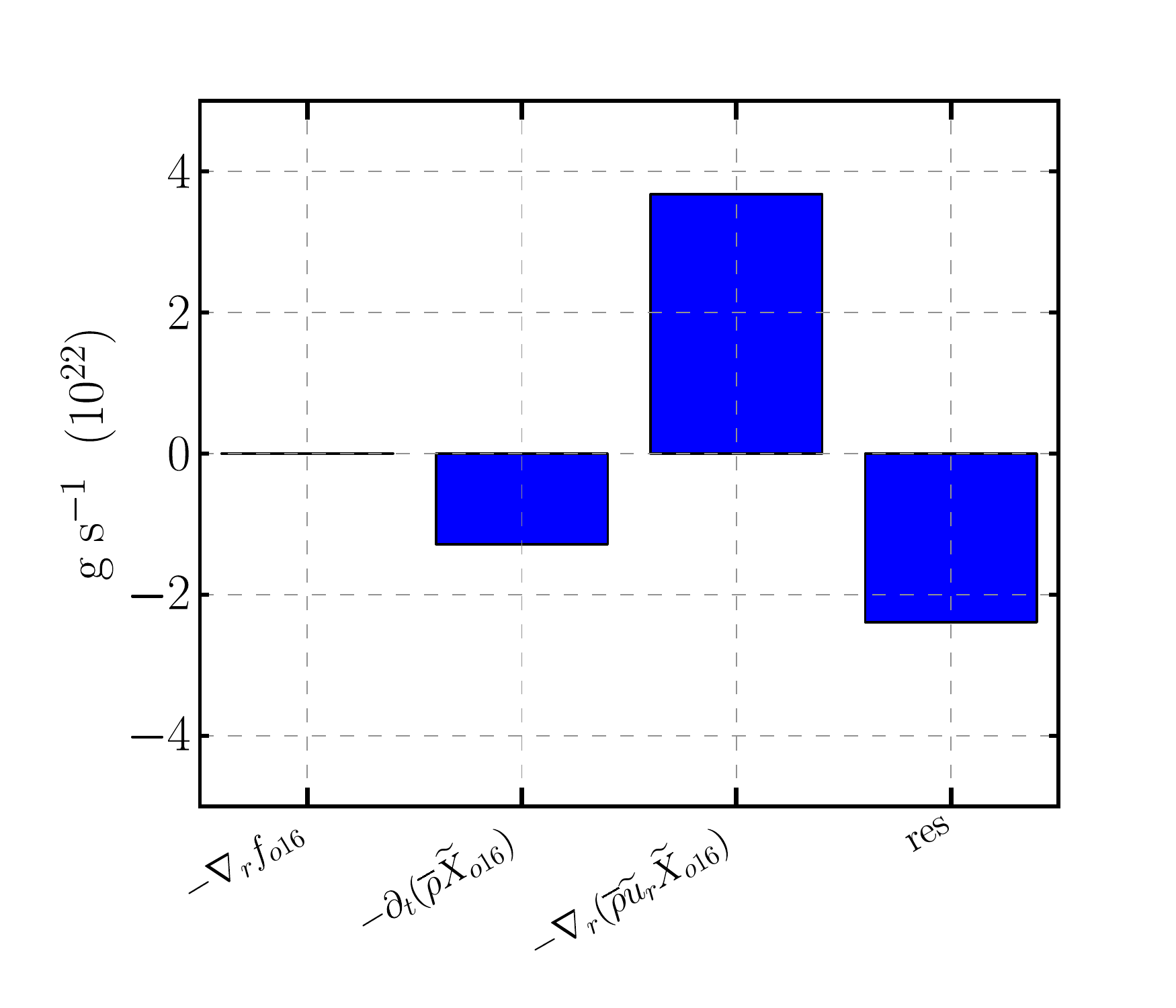}}

\centerline{
\includegraphics[width=6.7cm]{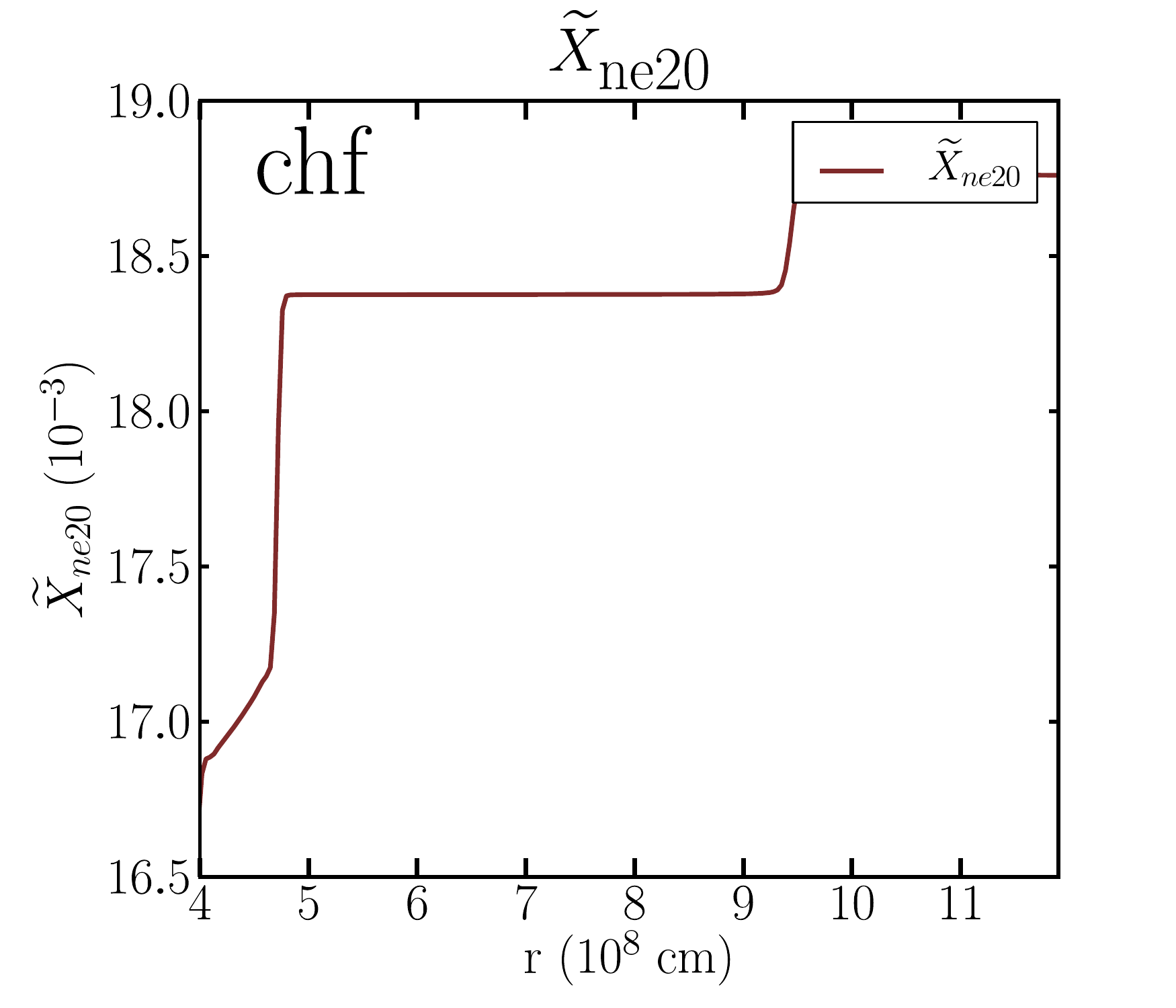}
\includegraphics[width=6.7cm]{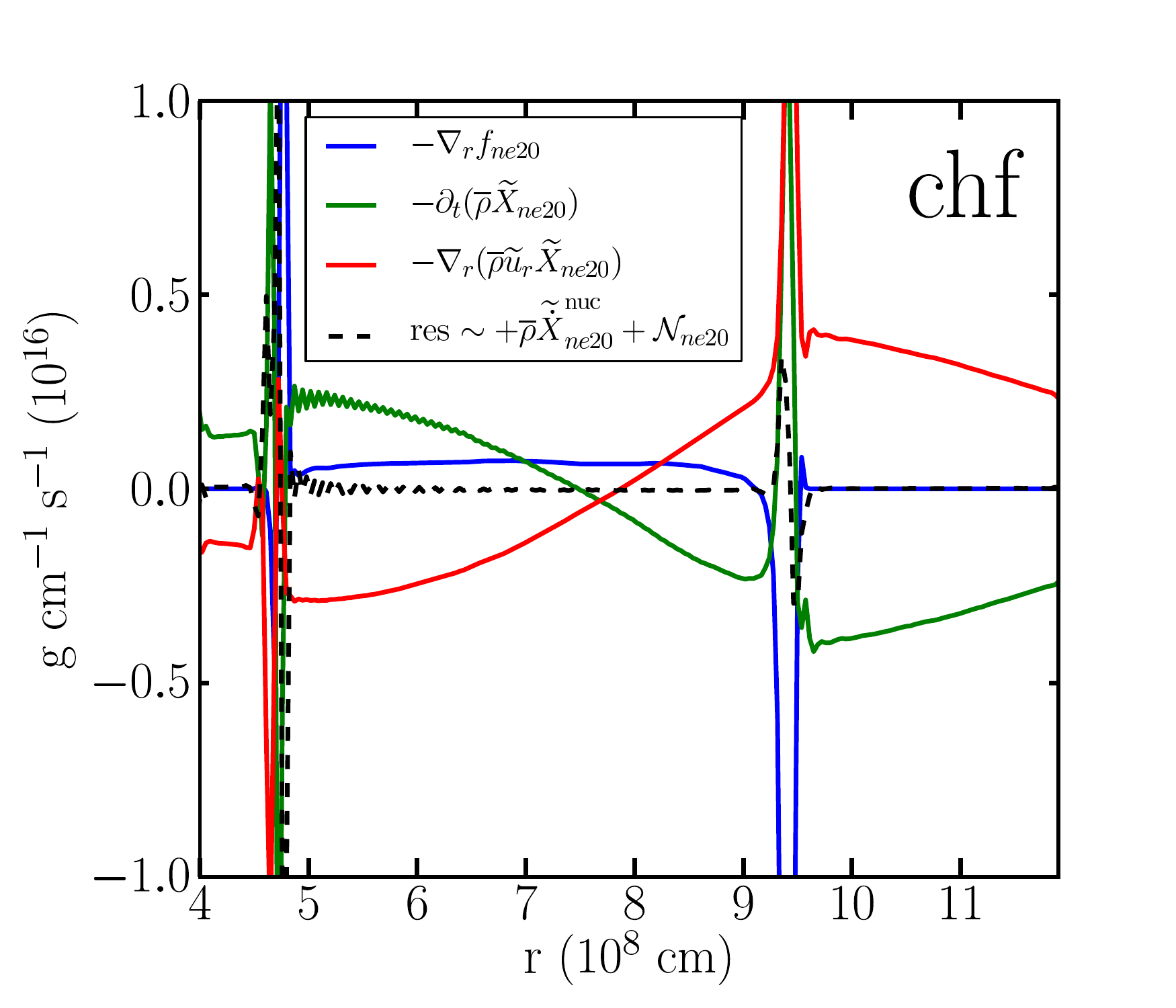}
\includegraphics[width=6.7cm]{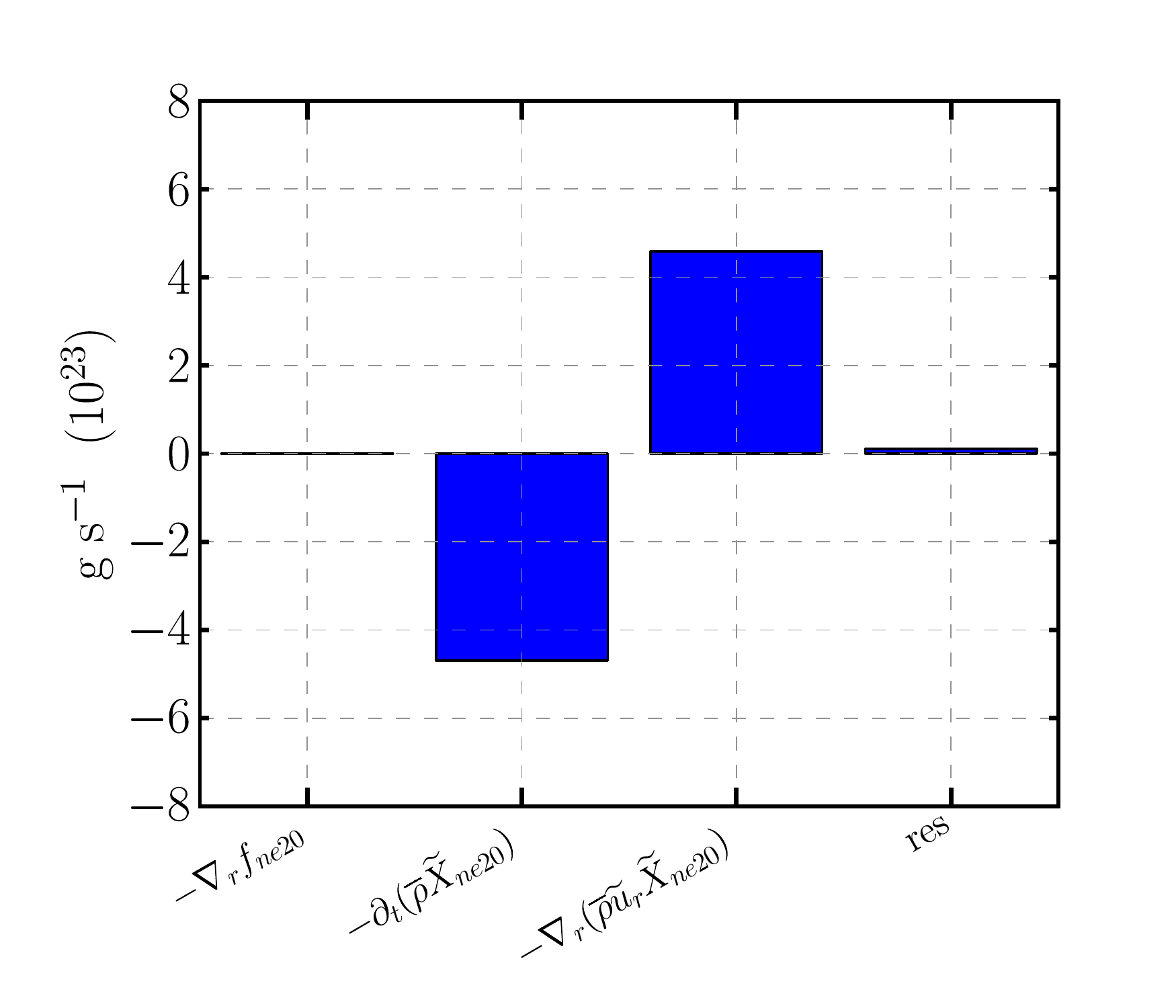}}
\caption{Mean composition equations for {\sf chf.3D}.\label{fig:x-equations}}
\end{figure}

\clearpage

\section{Instantaneous hydrodynamic equations in spherical coordinates (Eulerian form)}
\label{sect:mean-field-derivation}

The hydrodynamic equations of a viscous multi-component reactive gas subject to gravity and thermal transport in spherical coordinates ($r, \theta, \phi$):
%

\fontsize{9pt}{20pt}

\begin{align}
\partial_{t} \big(\rho\big) \ \ \ = & - \left( \frac{1}{r^{2}}\partial_{r}\big(r^{2}[\rho u_{r}~]\big) + \frac{1}{r\sin{\theta}}\partial_{\theta} \big(\sin{\theta} [\rho u_{\theta}]\big) + \frac{1}{r\sin{\theta}}\partial_{\phi} [\rho u_{\phi}] \right) \\ 
\partial_{t} \big(\rho u_{r}\big) = & -\left( \frac{1}{r^{2}} \partial_{r} \big( r^{2} [\rho u_{r}^{2} - \tau_{rr}]\big) + \frac{1}{r\sin{\theta}}\partial_{\theta}(\sin{\theta}[\rho u_{r} u_{\theta} - \tau_{r\theta}]\big) + \frac{1}{r\sin{\theta}}\partial_{\phi}\big([\rho u_{r} u_{\phi} - \tau_{r\phi}]\big) + G_r^M + \partial_{r} P \right) - \rho \partial_{r} \Phi \\
\partial_{t} \big(\rho u_{\theta}\big) = & -\left( \frac{1}{r^{2}} \partial_{r} \big(r^{2} [\rho u_{\theta} u_{r} - \tau_{\theta r}]\big) + \frac{1}{r\sin{\theta}}\partial_{\theta}\big(\sin{\theta}[\rho u_{\theta}^{2} - \tau_{\theta \theta}]\big) + \frac{1}{r\sin{\theta}}\partial_{\phi}[\rho u_{\theta} u_{\phi} - \tau_{\theta \phi}]\big) + G_\theta^M + \frac{1}{r} \partial_{\theta} P \right) - \rho \frac{1}{r} \partial_{\theta} \Phi \\
\partial_{t} \big(\rho u_{\phi}\big) = & -\left( \frac{1}{r^{2}} \partial_{r} \big(r^{2} [\rho u_{\phi} u_r - \tau_{\phi r}]\big) + \frac{1}{r\sin{\theta}}\partial_{\theta}\big(\sin{\theta}[\rho u_{\theta}u_\phi - \tau_{\theta \phi}]\big) + \frac{1}{r\sin{\theta}}\partial_{\phi}\big([\rho u_{\phi}^2 - \tau_{\phi \phi}]\big) + G_\phi^M + \frac{1}{r\sin{\theta}} \partial_{\phi} P \right) - \rho \frac{1}{r \sin{\theta}} \partial_{\phi} \Phi \\
\partial_{t} \big(\rho \epsilon_T\big)  = & -\left( \frac{1}{r^{2}} \partial_{r} (r^{2} [u_{r} \big(\rho \epsilon_T + P\big) - K \partial_{r} T ]\big) + \frac{1}{r\sin{\theta}} \partial_{\theta} \big(\sin{\theta} [u_{\theta}\big(\rho \epsilon_T + P\big) - K\frac{1}{r} \partial_{\theta} T ]\big) + \frac{1}{r \sin{\theta}} \partial_{\phi} [u_{\phi}\big(\rho \epsilon_T + P\big) - K \frac{1}{r \sin{\theta}} \partial_{\phi} T] \right) + \nonumber \\ 
& + \left( \frac{1}{r^2}\partial_r \big( r^2 [u_j \tau_{jr}] \big) + \frac{1}{r\sin{\theta}}\partial_\theta \big( \sin{\theta} [u_j\tau_{j\theta}] \big) + \frac{1}{r\sin{\theta}}\partial_\phi [u_j\tau_{j\phi}] \right) -\rho \big(u_{r}\partial_{r} \Phi + u_{\theta}\frac{1}{r} \partial_{\theta} \Phi + u_{\phi}\frac{1}{r \sin{\theta}}\partial_{\phi} \Phi) + \rho \epsilon_{nuc} \\
\partial_{t} \big(\rho \epsilon_I\big)  = &  - \left( \frac{1}{r^{2}}\partial_{r}\big(r^{2}[\rho u_{r} \epsilon_I]\big) + \frac{1}{r\sin{\theta}}\partial_{\theta} \big(\sin{\theta} [\rho u_{\theta} \epsilon_I]\big) + \frac{1}{r\sin{\theta}}\partial_{\phi} [\rho u_{\phi} \epsilon_I] \right) - P \left( \frac{1}{r^{2}}\partial_{r}\big(r^{2}[u_{r}~]\big) + \frac{1}{r\sin{\theta}}\partial_{\theta} \big(\sin{\theta} [u_{\theta}]\big) + \frac{1}{r\sin{\theta}}\partial_{\phi} [u_{\phi}]  \right) + \nonumber \\
& + \left( \frac{1}{r^{2}}\partial_{r}\big(r^{2}[K\partial_r T]\big) + \frac{1}{r\sin{\theta}}\partial_{\theta} \big(\sin{\theta} [K\frac{1}{r}\partial_\theta T]\big) + \frac{1}{r\sin{\theta}}\partial_{\phi} [K\frac{1}{r\sin{\theta}}\partial_\phi T] \right) + \left( \tau_{jr} \partial_r u_j + \tau_{j\theta} \frac{1}{r}\partial_\theta u_j + \tau_{j\phi} \frac{1}{r\sin{\theta}}\partial_\phi u_j  \right) + \rho \epsilon_{nuc} \\
\partial_{t} \big(\rho \epsilon_K\big)  = & -\left( \frac{1}{r^2}\partial_r [r^2 \big(\rho u_r \epsilon_K - u_j\tau_{jr}\big)] + \frac{1}{r\sin{\theta}} \partial_\theta [\sin{\theta} \big(\rho u_\theta \epsilon_K - u_j\tau_{j\theta} \big)] + \frac{1}{r\sin{\theta}} \partial_\phi [\rho u_\phi \epsilon_K - u_j\tau_{j\phi}] \right) - \nonumber \\
& -\left( \frac{1}{r^2} \partial_r \big( r^2 [P u_r] \big) + \frac{1}{r \sin{\theta}} \partial_\theta \big(\sin{\theta}[P u_\theta] \big) + \frac{1}{r\sin{\theta}} \partial_\phi [ P u_\phi ] \right) + P \left( \frac{1}{r^{2}}\partial_{r}\big(r^{2}[u_{r}]\big) + \frac{1}{r\sin{\theta}}\partial_{\theta} \big(\sin{\theta} [u_{\theta}]\big) + \frac{1}{r\sin{\theta}}\partial_{\phi} [u_{\phi}]  \right) - \nonumber \\
& -  \left( \tau_{jr} \partial_r u_j + \tau_{j\theta} \frac{1}{r}\partial_\theta u_j + \tau_{j\phi} \frac{1}{r\sin{\theta}}\partial_\phi u_j  \right)  - \rho \big(u_{r}\partial_{r} \Phi + u_{\theta}\frac{1}{r} \partial_{\theta} \Phi + u_{\phi}\frac{1}{r \sin{\theta}}\partial_{\phi} \Phi) \\   
\partial_{t} \big(\rho X_{k}\big) = & -\left( \frac{1}{r^{2}} \partial_{r}(r^{2} [\rho u_{r} X_{k} ] \big) + \frac{1}{r\sin{\theta}}\partial_{\theta}\big(\sin{\theta} [\rho u_{\theta} X_{k} ]\big) + \frac{1}{r \sin{\theta}} \partial_{\phi} [\rho u_{\phi} X_{k}] \right) + \rho \dot{X}_{k}^{n} \ \ \ \ \ \ \ \ \ \ \ \  k = 1 ... N_{nuc} 
\end{align}

\begin{align}
G_r^M = -\frac{(\rho u_{\theta}^{2} - \tau_{\theta\theta})}{r} - \frac{(\rho u_{\phi}^{2} - \tau_{\phi\phi})}{r} \ \ \ G_\theta^M = + \frac{(\rho u_{\theta} u_{r} - \tau_{\theta r})}{r} -  \frac{(\rho u_{\phi}^{2} - \tau_{\phi\phi}) \cos{\theta}}{r \sin{\theta}} \ \ \ G_\phi^M = + \frac{(\rho u_{\phi} u_{r} - \tau_{\phi r})}{r} +  \frac{(\rho u_{\phi} u_{\theta} - \tau_{\phi \theta})\cos{\theta}}{r \sin{\theta}}
\end{align}

\newpage 

\section{Instantaneous hydrodynamic equations in spherical coordinates (Lagrangian form)}

The hydrodynamic equations of a viscous multi-component reactive gas subject to gravity and thermal transport in spherical coordinates ($r, \theta, \phi$) are ($D_t (.) = \partial_t (.) + u_n \partial_n (.)$ is advective derivative):

\fontsize{9pt}{20pt}

\begin{align}
D_{t} \big(\rho\big) \ \ \ = & - \rho \left( \frac{1}{r^{2}}\partial_{r}\big(r^{2}[u_{r}~]\big) + \frac{1}{r\sin{\theta}}\partial_{\theta} \big(\sin{\theta} [u_{\theta}]\big) + \frac{1}{r\sin{\theta}}\partial_{\phi} [u_{\phi}] \right) \\ 
\rho D_{t} \big(u_{r}\big) = & +\left( \frac{1}{r^{2}} \partial_{r} \big( r^{2} [\tau_{rr}]\big) + \frac{1}{r\sin{\theta}}\partial_{\theta}(\sin{\theta}[\tau_{r\theta}]\big) + \frac{1}{r\sin{\theta}}\partial_{\phi}\big([\tau_{r\phi}]\big) - G_r^M - \partial_{r} P \right) + \rho g_r \\
\rho D_{t} \big(u_{\theta}\big) = & +\left( \frac{1}{r^{2}} \partial_{r} \big(r^{2} [\tau_{\theta r}]\big) + \frac{1}{r\sin{\theta}}\partial_{\theta}\big(\sin{\theta}[\tau_{\theta \theta}]\big) + \frac{1}{r\sin{\theta}}\partial_{\phi}[\tau_{\theta \phi}]\big) - G_\theta^M - \frac{1}{r} \partial_{\theta} P \right) + \rho g_\theta \\
\rho D_{t} \big(u_{\phi}\big) = & +\left( \frac{1}{r^{2}} \partial_{r} \big(r^{2} [\tau_{\phi r}]\big) + \frac{1}{r\sin{\theta}}\partial_{\theta}\big(\sin{\theta}[\tau_{\phi \theta}]\big) + \frac{1}{r\sin{\theta}}\partial_{\phi}\big([\tau_{\phi \phi}]\big) - G_\phi^M - \frac{1}{r\sin{\theta}} \partial_{\phi} P \right) + \rho g_\phi \\
\rho D_{t} \big(\epsilon_T\big) = & -\left( \frac{1}{r^{2}} \partial_{r} \big( r^{2} [u_{r} P - K \partial_{r} T ]\big) + \frac{1}{r\sin{\theta}} \partial_{\theta} \big(\sin{\theta} [u_{\theta} P - K\frac{1}{r} \partial_{\theta} T ]\big) + \frac{1}{r \sin{\theta}} \partial_{\phi} [u_{\phi}P - K \frac{1}{r \sin{\theta}} \partial_{\phi} T] \right) - \nonumber \\
& +\left( \frac{1}{r^2}\partial_r [r^2 \big(u_j\tau_{jr} \big)] + \frac{1}{r\sin{\theta}} \partial_\theta [\sin{\theta} \big(u_j\tau_{j\theta} \big)] + \frac{1}{r\sin{\theta}} \partial_\phi [u_j\tau_{j\phi}] \right) +\rho \big(u_{r} g_r + u_{\theta} g_\theta + u_{\phi} g_\phi) + \rho \epsilon_{nuc} \\
\rho D_{t} \big(\epsilon_I\big) = & -P \left( \frac{1}{r^{2}}\partial_{r}\big(r^{2}[u_{r}]\big) + \frac{1}{r\sin{\theta}}\partial_{\theta} \big(\sin{\theta} [u_{\theta}]\big) + \frac{1}{r\sin{\theta}}\partial_{\phi} [u_{\phi}]  \right) + \left( \frac{1}{r^{2}}\partial_{r}\big(r^{2}[K\partial_r T]\big) + \frac{1}{r\sin{\theta}}\partial_{\theta} \big(\sin{\theta} [K\frac{1}{r}\partial_\theta T]\big) + \frac{1}{r\sin{\theta}}\partial_{\phi} [K\frac{1}{r\sin{\theta}}\partial_\phi T] \right) + \nonumber \\
& + \left( \tau_{jr} \partial_r u_j + \tau_{j\theta} \frac{1}{r}\partial_\theta u_j + \tau_{j\phi} \frac{1}{r\sin{\theta}}\partial_\phi u_j  \right) + \rho \epsilon_{nuc} \\
\rho D_{t} \big(\epsilon_K\big) = & +\left( \frac{1}{r^2}\partial_r [r^2 \big(u_j\tau_{jr} \big)] + \frac{1}{r\sin{\theta}} \partial_\theta [\sin{\theta} \big(u_j\tau_{j\theta} \big)] + \frac{1}{r\sin{\theta}} \partial_\phi [u_j\tau_{j\phi}] \right) -\left( \frac{1}{r^2} \partial_r \big( r^2 [P u_r] \big) + \frac{1}{r \sin{\theta}} \partial_\theta \big(\sin{\theta}[P u_\theta] \big) + \frac{1}{r\sin{\theta}} \partial_\phi [ P u_\phi ] \right) + \nonumber \\
& + P \left( \frac{1}{r^{2}}\partial_{r}\big(r^{2}[u_{r}~]\big) + \frac{1}{r\sin{\theta}}\partial_{\theta} \big(\sin{\theta} [u_{\theta}]\big) + \frac{1}{r\sin{\theta}}\partial_{\phi} [u_{\phi}]  \right) - \left( \tau_{jr} \partial_r u_j + \tau_{j\theta} \frac{1}{r}\partial_\theta u_j + \tau_{j\phi} \frac{1}{r\sin{\theta}}\partial_\phi u_j  \right) +\rho \big(u_{r} g_r + u_{\theta} g_\theta + u_{\phi} g_\phi) \\
\rho D_{t} \big(X_{k}\big) = & + \rho \dot{X}_{k}^{n} \ \ \ \ \ \ \ \ \ \ \ \  k = 1 ... N_{nuc}
\end{align}

\begin{align}
G_r^M = -\frac{(\rho u_{\theta}^{2} - \tau_{\theta\theta})}{r} - \frac{(\rho u_{\phi}^{2} - \tau_{\phi\phi})}{r} \ \ \ G_\theta^M = + \frac{(\rho u_{\theta} u_{r} - \tau_{\theta r})}{r} -  \frac{(\rho u_{\phi}^{2} - \tau_{\phi\phi}) \cos{\theta}}{r \sin{\theta}} \ \ \ G_\phi^M = + \frac{(\rho u_{\phi} u_{r} - \tau_{\phi r})}{r} +  \frac{(\rho u_{\phi} u_{\theta} - \tau_{\phi \theta})\cos{\theta}}{r \sin{\theta}}
\end{align}

\fontsize{12pt}{20pt}

\noindent
where $\rho$, $u_r$, $u_{\theta}$, $u_{\phi}$, $P$, $\epsilon_T$, $\epsilon_I$, $\epsilon_K$, $T$, $\epsilon_{nuc}$, $X_k$, and $\dot{X}_k^n$ are the density, the radial velocity, the $\theta$-velocity, the rotation velocity, the pressure, the total specific energy, the specific internal energy, the specific kinetic energy, the temperature, the energy generation rate per mass due to reactions, the mass fraction of species $k$, and the change of this mass fraction due to reactions, respectively.  $N_{nuc}$ is the number of species the gas is composed. $\tau_{ij} = 2 \mu S_{ij} - 2/3 \mu \nabla \cdot {\bf u} \delta_{ij}$ is the viscous stress, where $S_{ij} = \frac{1}{2} (\partial_i u_j + \partial_j u_i)$ is strain-rate, $\mu = \rho \nu$ is dynamic viscosity and $\nu$ is the kinematic viscosity. K is thermal conductivity. $g_i$ is gravitational acceleration in $r, \theta, \phi$ and $\Phi$ is gravitational potential. $G$ are geometric terms. 

\section{Reynolds decomposition}\label{sec:reynolds-decomposition}

\noindent 
Reynolds decomposition:

\begin{equation}
A(r,\theta,\phi) = \eht{A}(r) + A'(r,\theta,\phi)
\end{equation}

\noindent
Definition of the averaging space-time operator:

\begin{equation}
\eht{A(r)} = \frac{1}{\Delta T \Delta \Omega} \int_{\Delta T}\int_{\Delta \Omega} A(r,
\theta,\phi) \ dt \ d\Omega
\end{equation}

\noindent
Some properties of the operator:

\begin{equation}
\eht{A'} = 0 
\end{equation}

\begin{equation}
\eht{\eht{A}}=\eht{A}
\end{equation}

\begin{equation}
(A')'=A'
\end{equation}

\begin{equation}
\eht{\eht{A}B} = \eht{A} \ \eht{B}
\end{equation}

\begin{equation}
\eht{AB} = \eht{A} \ \eht{B} + \eht{A'B'}
\end{equation}

\begin{equation}
\eht{A'B'} = \eht{A'B}
\end{equation}

\section{Favre decomposition}\label{sec:favre-decomposition}

\noindent 
Favre decomposition:

\begin{equation}
F = \fht{F}(r) + F''(r,\theta,\phi)
\end{equation}

\noindent
Definition of the averaging operator:

\begin{equation}
\fht{F} = \frac{\eht{\rho F}}{\eht{\rho}}
\end{equation}

\noindent
Some properties of the operator: 

\begin{equation}
\eht{\rho F''} = \fht{F''} = 0
\end{equation}

\begin{equation}
\fht{F} = \eht{F} + \frac{\eht{\rho F'}}{\eht{\rho}} 
\end{equation}

\begin{equation}
F'' = F' - \frac{\eht{\rho F'}}{\eht{\rho}} \rightarrow \eht{F''} = -\frac{\eht{\rho F'}}{\eht{\rho}}
\end{equation}

\begin{equation}
\eht{\rho F''G''} = \eht{\rho}\fht{F''G''}
\end{equation}

\newpage

\section{Derivation of first order moments}

\subsection{Mean continuity equation}

We begin by instantaneous 3D continuity equation and apply "ensemble'' (space-time) averaging (Sect.\ref{sect:intro}):

\fontsize{9pt}{20pt}

\begin{align}
\partial_{t} \rho = & - \left( \frac{1}{r^{2}}\partial_{r}\big(r^{2}[\rho u_{r}~]\big) + \frac{1}{r\sin{\theta}}\partial_{\theta} \big(\sin{\theta} [\rho u_{\theta}]\big) + \frac{1}{r\sin{\theta}}\partial_{\phi} [\rho u_{\phi}] \right) \\
\partial_{t} \eht{\rho} = & -\left( \eht{\frac{1}{r^{2}}\partial_{r}\big(r^{2}[\rho u_{r}~]\big)} + \cancelto{0}{\eht{\frac{1}{r\sin{\theta}}\partial_{\theta} \big(\sin{\theta} [\rho u_{\theta}]\big)}} + \cancelto{0}{\eht{\frac{1}{r\sin{\theta}}\partial_{\phi} [\rho u_{\phi}]}} \right) \ \ \ \mbox{``ensemble'' (space-time) averaging} \\
\partial_t \eht{\rho} = & -\frac{1}{r^{2}}\partial_{r}\big(r^{2}[\eht{\rho} \fht{u}_{r}~]\big) \\
\partial_t \eht{\rho} = & -\fht{u}_r \partial_r \eht{\rho} -\eht{\rho} \frac{1}{r^{2}}\partial_{r} (r^2 \fht{u}_r) \\
\partial_t \eht{\rho} + \fht{u}_r \partial_r \eht{\rho} = & -\eht{\rho}\frac{1}{r^{2}}\partial_{r} (r^2 \fht{u}_r) \\
{\color{red} \fht{D}_t \eht{\rho} =} & {\color{red} -\eht{\rho} \fht{d}}
\end{align}

\fontsize{12pt}{20pt}

\subsection{Mean radial momentum equation}

We begin by instantaneous 3D radial momentum equation and apply "ensemble'' (space-time) averaging (Sect.\ref{sect:intro}):

\fontsize{9pt}{20pt}

\begin{align}
\partial_{t} \rho u_{r} = & -\left( \frac{1}{r^{2}} \partial_{r} \big( r^{2} [\rho u_{r}^{2} - \tau_{rr}]\big) + \frac{1}{r\sin{\theta}}\partial_{\theta}(\sin{\theta}[\rho u_{r} u_{\theta} - \tau_{r\theta}]\big) + \frac{1}{r\sin{\theta}}\partial_{\phi}\big([\rho u_{r} u_{\phi} - \tau_{r\phi}]\big) + G_r^M + \partial_{r} P \right) + \rho g_r \\
\partial_{t} \eht{\rho u_{r}}= & -\left( \eht{\frac{1}{r^{2}} \partial_{r} \big( r^{2} [\rho u_{r}^{2} - \tau_{rr}]\big)} + \cancelto{0}{\eht{\frac{1}{r\sin{\theta}}\partial_{\theta}(\sin{\theta}[\rho u_{r} u_{\theta} - \tau_{r\theta}]\big)}} + \cancelto{0}{\eht{\frac{1}{r\sin{\theta}}\partial_{\phi}\big([\rho u_{r} u_{\phi} - \tau_{r\phi}]\big)}} + \eht{G_r^M} + \partial_{r} \eht{P} \right) + \eht{\rho g_r} \\
\partial_{t} \eht{\rho {u}_{r}}= & -\frac{1}{r^{2}} \partial_{r} \big( r^{2} [\eht{\rho u_{r} u_{r}}] \big) - \cancelto{0}{\frac{1}{r^{2}} \partial_{r} \big(r^2 \eht{\tau}_{rr}\big)} - \eht{G_r^M} - \partial_{r} \eht{P} + \eht{\rho g_r} \ \ \ \mbox{we neglect mean viscosity $\eht{\tau}$} 
\end{align}

\newpage

\fontsize{9pt}{20pt}

\begin{align}
\partial_{t} \eht{\rho} \fht{u_{r}} = & -\frac{1}{r^{2}} \partial_{r} \big( r^{2} [\eht{\rho} \fht{u_{r} u_{r}}] \big) - \eht{G_r^M} - \partial_{r} \eht{P} + \eht{\rho} \fht{g}_r \\
\partial_{t} \eht{\rho} \fht{u_{r}} = & -\frac{1}{r^{2}} \partial_{r} \big( r^{2} [\eht{\rho} \fht{u_{r}} \fht{u_{r}} + \eht{\rho}\fht{u''_r u''_r}] \big) - \eht{G_r^M} - \partial_{r} \eht{P} + \eht{\rho} \fht{g}_r \\
\partial_{t} \eht{\rho} \fht{u_{r}} + \frac{1}{r^{2}} \partial_{r} \big( r^{2} [\eht{\rho} \fht{u_{r}} \fht{u_{r}}] \big) = & -\frac{1}{r^{2}} \partial_{r} \big( \eht{\rho}\fht{u''_r u''_r} \big) - \eht{G_r^M} - \partial_{r} \eht{P} + \eht{\rho} \fht{g}_r \\
{\color{red} \eht{\rho} \fht{D}_t \fht{u}_r = } & {\color{red}-\nabla_r \fht{R}_{rr}  - \eht{G_r^M} - \partial_{r} \eht{P} + \eht{\rho} \fht{g}_r}
\end{align}

\fontsize{12pt}{20pt}

\subsection{Mean polar momentum equation}

We begin by instantaneous 3D polar momentum equation and apply "ensemble'' (space-time) averaging (Sect.\ref{sect:intro}):

\fontsize{9pt}{20pt}

\begin{align}
\partial_{t} \rho u_{\theta} = & -\left( \frac{1}{r^{2}} \partial_{r} \big(r^{2} [\rho u_{\theta} u_{r} - \tau_{\theta r}]\big) + \frac{1}{r\sin{\theta}}\partial_{\theta}\big(\sin{\theta}[\rho u_{\theta}^{2} - \tau_{\theta \theta}]\big) + \frac{1}{r\sin{\theta}}\partial_{\phi}[\rho u_{\theta} u_{\phi} - \tau_{\theta \phi}]\big) + G_\theta^M + \frac{1}{r} \partial_{\theta} P \right) - \rho \frac{1}{r} \partial_{\theta} \Phi \\
\partial_{t} \eht{\rho u_{\theta}} = & -\left( \eht{\frac{1}{r^{2}} \partial_{r} \big(r^{2} [\rho u_{\theta} u_{r} - \tau_{\theta r}]\big)} + \cancelto{0}{\eht{\frac{1}{r\sin{\theta}}\partial_{\theta}\big(\sin{\theta}[\rho u_{\theta}^{2} - \tau_{\theta \theta}]\big)}} + \cancelto{0}{\eht{\frac{1}{r\sin{\theta}}\partial_{\phi}[\rho u_{\theta} u_{\phi} - \tau_{\theta \phi}]\big)}} + \eht{G_\theta^M} + \frac{1}{r} \eht{\partial_{\theta} P} + \eht{\rho g_\theta}\right) \\
\partial_{t} \eht{\rho u_{\theta}} = & -\frac{1}{r^{2}} \partial_{r} \big( r^{2} [\eht{\rho u_{\theta} u_{r}}] \big) + \cancelto{0}{\frac{1}{r^{2}} \partial_{r} \big(r^{2} \eht{\tau_{\theta r}} \big)} - \eht{G_\theta^M} - \frac{1}{r} \eht{\partial_{\theta} P} + \cancelto{0}{\eht{\rho g_\theta}} \ \ \ \mbox{we neglect mean viscosity $\eht{\tau}$ and gravity in $\theta$} \\
\partial_{t} \eht{\rho} \fht{u}_{\theta} = & -\frac{1}{r^{2}} \partial_{r} \big( r^{2} [\eht{\rho} \fht{u_{\theta} u_{r}}] \big) - \eht{G_\theta^M} - \frac{1}{r} \eht{\partial_{\theta} P}\\
\partial_{t} \eht{\rho} \fht{u}_{\theta} = & -\frac{1}{r^{2}} \partial_{r} \big( r^{2} [\eht{\rho} \fht{u}_{\theta} \fht{u}_{r} + \eht{\rho}\fht{u''_\theta u''_r} \big) - \eht{G_\theta^M} - \frac{1}{r} \eht{\partial_{\theta} P}\\ 
\partial_{t} \eht{\rho} \fht{u}_{\theta} + \frac{1}{r^{2}} \partial_{r} \big( r^{2} [\eht{\rho} \fht{u}_{\theta} \fht{u}_{r}] \big) = & -\frac{1}{r^{2}} \partial_{r} \big( r^{2} \eht{\rho}\fht{u''_\theta u''_r} \big) - \eht{G_\theta^M} - \frac{1}{r} \eht{\partial_{\theta} P}\\
{\color{red} \eht{\rho}\fht{D}_t \fht{u}_\theta =} & {\color{red} -\nabla_r \fht{R}_{\theta r} - \eht{G_\theta^M} - (1/r) \eht{\partial_{\theta} P}}
\end{align}

\fontsize{12pt}{20pt}

\subsection{Mean azimutal momentum equation}

We begin by instantaneous 3D azimutal momentum equation and apply "ensemble'' (space-time) averaging (Sect.\ref{sect:intro}):

\fontsize{9pt}{20pt}

\begin{align}
\partial_{t} \big(\rho u_{\phi}\big) = & -\left( \frac{1}{r^{2}} \partial_{r} \big(r^{2} [\rho u_{\phi} u_r - \tau_{\phi r}]\big) + \frac{1}{r\sin{\theta}}\partial_{\theta}\big(\sin{\theta}[\rho u_{\theta}u_\phi - \tau_{\theta \phi}]\big) + \frac{1}{r\sin{\theta}}\partial_{\phi}\big([\rho u_{\phi}^2 - \tau_{\phi \phi}]\big) + G_\phi^M + \frac{1}{r\sin{\theta}} \partial_{\phi} P \right) - \\
& - \rho \frac{1}{r \sin{\theta}} \partial_{\phi} \Phi \nonumber \\
\partial_{t} \eht{\rho u_{\phi}} = & -\left( \eht{\frac{1}{r^{2}} \partial_{r} \big(r^{2} [\rho u_{\phi} u_r - \tau_{\phi r}]\big)} + \cancelto{0}{\eht{\frac{1}{r\sin{\theta}}\partial_{\theta}\big(\sin{\theta}[\rho u_{\theta}u_\phi - \tau_{\theta \phi}]\big)}} + \cancelto{0}{\eht{\frac{1}{r\sin{\theta}}\partial_{\phi}\big([\rho u_{\phi}^2 - \tau_{\phi \phi}]\big)}} + \eht{G_\phi^M} + \cancelto{0}{\eht{\frac{1}{r\sin{\theta}} \partial_{\phi} P}} \right) + \nonumber \\ 
+ \eht{\rho g_\phi} \\
\partial_{t} \eht{\rho u_{\phi}} = & -\frac{1}{r^{2}} \partial_{r} \big(r^{2} [\eht{\rho u_{\phi} u_r}] \big) - \cancelto{0}{\frac{1}{r^{2}} \partial_{r} \big(r^{2} \eht{\tau_{\phi r}} \big)} + \eht{G_\phi^M} - \cancelto{0}{\eht{\rho g_\phi}} \ \ \ \mbox{we neglect mean viscosity $\tau$ and gravity in $\phi$} \\
\partial_{t} \eht{\rho} \fht{u}_{\phi} = & -\frac{1}{r^{2}} \partial_{r} \big(r^{2} [\eht{\rho} \fht{u_{\phi} u_r}] \big) + \eht{G_\phi^M} \\
\partial_{t} \eht{\rho} \fht{u}_{\phi} = & -\frac{1}{r^{2}} \partial_{r} \big(r^{2} [\eht{\rho} \fht{u}_{\phi} \fht{u}_r - \eht{\rho} \fht{u''_\phi u''_r}] \big) - \eht{G_\phi^M} \\
\partial_{t} \eht{\rho} \fht{u}_{\phi} + \frac{1}{r^{2}} \partial_{r} \big(r^{2} [\eht{\rho} \fht{u}_{\phi} \fht{u}_r]\big) = & -\frac{1}{r^{2}} \partial_{r} \big(r^{2} \eht{\rho} \fht{u''_\phi u''_r} \big) - \eht{G_\phi^M} \\
{\color{red} \eht{\rho}\fht{D}_t \fht{u}_\phi =} & {\color{red} -\nabla_r \fht{R}_{\phi r} - \eht{G_\phi^M}}
\end{align}

\fontsize{12pt}{20pt}

\subsection{Mean internal energy equation}

We begin by instantaneous 3D internal energy equation and apply "ensemble'' (space-time) averaging (Sect.\ref{sect:intro}):

\fontsize{9pt}{20pt}

\begin{align}
\partial_{t} \big(\rho \epsilon_I\big)  = &  - \left( \frac{1}{r^{2}}\partial_{r}\big(r^{2}[\rho u_{r} \epsilon_I]\big) + \frac{1}{r\sin{\theta}}\partial_{\theta} \big(\sin{\theta} [\rho u_{\theta} \epsilon_I]\big) + \frac{1}{r\sin{\theta}}\partial_{\phi} [\rho u_{\phi} \epsilon_I] \right) - P \left( \frac{1}{r^{2}}\partial_{r}\big(r^{2}[u_{r}~]\big) + \frac{1}{r\sin{\theta}}\partial_{\theta} \big(\sin{\theta} [u_{\theta}]\big) + \frac{1}{r\sin{\theta}}\partial_{\phi} [u_{\phi}]  \right) + \nonumber \\
& + \left( \frac{1}{r^{2}}\partial_{r}\big(r^{2}[K\partial_r T]\big) + \frac{1}{r\sin{\theta}}\partial_{\theta} \big(\sin{\theta} [K\frac{1}{r}\partial_\theta T]\big) + \frac{1}{r\sin{\theta}}\partial_{\phi} [K\frac{1}{r\sin{\theta}}\partial_\phi T] \right) + \left( \tau_{jr} \partial_r u_j + \tau_{j\theta} \frac{1}{r}\partial_\theta u_j + \tau_{j\phi} \frac{1}{r\sin{\theta}}\partial_\phi u_j  \right) + \rho \epsilon_{\rm nuc} 
\end{align}

\begin{align}
\partial_{t} \eht{\rho \epsilon_I}  = &  - \left( \eht{\frac{1}{r^{2}}\partial_{r}\big(r^{2}[\rho u_{r} \epsilon_I]\big)} + \cancelto{0}{\eht{\frac{1}{r\sin{\theta}}\partial_{\theta} \big(\sin{\theta} [\rho u_{\theta} \epsilon_I]\big)}} + \cancelto{0}{\eht{\frac{1}{r\sin{\theta}}\partial_{\phi} [\rho u_{\phi} \epsilon_I]}} \right) - \eht{P \left( \frac{1}{r^{2}}\partial_{r}\big(r^{2}[u_{r}~]\big) + \frac{1}{r\sin{\theta}}\partial_{\theta} \big(\sin{\theta} [u_{\theta}]\big) + \frac{1}{r\sin{\theta}}\partial_{\phi} [u_{\phi}]  \right)} + \nonumber \\
& + \left( \eht{\frac{1}{r^{2}}\partial_{r}\big(r^{2}[K\partial_r T]\big)} + \cancelto{0}{\eht{\frac{1}{r\sin{\theta}}\partial_{\theta} \big(\sin{\theta} [K\frac{1}{r}\partial_\theta T]\big)}} + \cancelto{0}{\eht{\frac{1}{r\sin{\theta}}\partial_{\phi} [K\frac{1}{r\sin{\theta}}\partial_\phi T]}} \right) + \left( \eht{\tau_{jr} \partial_r u_j + \tau_{j\theta} \frac{1}{r}\partial_\theta u_j + \tau_{j\phi} \frac{1}{r\sin{\theta}}\partial_\phi u_j}  \right) + \eht{\rho \epsilon_{\rm nuc}} \\
\partial_{t} \eht{\rho \epsilon_I}  = &  - \frac{1}{r^{2}}\partial_{r}\big(r^{2}[\eht{\rho u_{r} \epsilon_I}]\big) - \eht{P d} + \frac{1}{r^{2}}\partial_{r}\big(r^{2}[\eht{\chi \partial_r T}]\big) + \eht{\tau_{ji} \partial_i u_j} + \eht{\rho \epsilon_{\rm nuc}} \\
\partial_{t} \eht{\rho} \fht{\epsilon_I}  = &  - \frac{1}{r^{2}}\partial_{r}\big(r^{2}[\eht{\rho} \fht{u_{r} \epsilon_I}]\big) - \eht{P d} + \frac{1}{r^{2}}\partial_{r}\big(r^{2}[\eht{\chi \partial_r T}]\big) + \eht{\tau_{ji} \partial_i u_j} + \eht{\rho \epsilon_{\rm nuc}} \\
\partial_{t} \eht{\rho} \fht{\epsilon_I}  = &  - \frac{1}{r^{2}}\partial_{r}\big(r^{2}[\eht{\rho} \fht{u}_{r} \fht{\epsilon}_I]\big) -\frac{1}{r^{2}}\partial_{r}\big(r^{2}[\eht{\rho} \fht{u''_{r} \epsilon''_I}]\big)  - \eht{P d} + \frac{1}{r^{2}}\partial_{r}\big(r^{2}[\eht{\chi \partial_r T}]\big) + \cancelto{0}{\eht{\tau_{ji}} \partial_i \eht{u_j}} + \eht{\tau'_{ji}\partial_i u'_j} + \eht{\rho} \fht{\epsilon}_{\rm nuc} \ \ \ \mbox{we neglect $\eht{\tau}$} \\
\eht{\rho}\fht{D}_t \fht{\epsilon}_I = & -\nabla_r f_I - \eht{P d} + \nabla_r f_T + \eht{\tau'_{ji}\partial_i u'_j} + \eht{\rho} \fht{\epsilon}_{\rm nuc} \\
\eht{\rho}\fht{D}_t \fht{\epsilon}_I = & -\nabla_r (f_I + f_T) - \eht{P} \ \eht{d} - \eht{P'd'} + \varepsilon_k + \eht{\rho} \fht{\epsilon}_{\rm nuc} \\
{\color{red} \eht{\rho}\fht{D}_t \fht{\epsilon}_I =} & {\color{red}-\nabla_r (f_I + f_T) - \eht{P} \ \eht{d} - W_P + \varepsilon_k + \eht{\rho} \fht{\epsilon}_{\rm nuc}}
\end{align}

\fontsize{12pt}{20pt}

\subsection{Mean kinetic energy equation}

We begin by instantaneous 3D kinetic energy equation and apply "ensemble'' (space-time) averaging (Sect.\ref{sect:intro}):

\fontsize{9pt}{20pt}

\begin{align}
\partial_{t} \big(\rho \epsilon_K\big)  = & -\left( \frac{1}{r^2}\partial_r [r^2 \big(\rho u_r \epsilon_K - u_j\tau_{jr}\big)] + \frac{1}{r\sin{\theta}} \partial_\theta [\sin{\theta} \big(\rho u_\theta \epsilon_K - u_j\tau_{j\theta} \big)] + \frac{1}{r\sin{\theta}} \partial_\phi [\rho u_\phi \epsilon_K - u_j\tau_{j\phi}] \right) - \nonumber \\
& -\left( \frac{1}{r^2} \partial_r \big( r^2 [P u_r] \big) + \frac{1}{r \sin{\theta}} \partial_\theta \big(\sin{\theta}[P u_\theta] \big) + \frac{1}{r\sin{\theta}} \partial_\phi [ P u_\phi ] \right) + P \left( \frac{1}{r^{2}}\partial_{r}\big(r^{2}[u_{r}]\big) + \frac{1}{r\sin{\theta}}\partial_{\theta} \big(\sin{\theta} [u_{\theta}]\big) + \frac{1}{r\sin{\theta}}\partial_{\phi} [u_{\phi}]  \right) - \nonumber \\
& -  \left( \tau_{jr} \partial_r u_j + \tau_{j\theta} \frac{1}{r}\partial_\theta u_j + \tau_{j\phi} \frac{1}{r\sin{\theta}}\partial_\phi u_j  \right)  - \rho \big(u_{r}\partial_{r} \Phi + u_{\theta}\frac{1}{r} \partial_{\theta} \Phi + u_{\phi}\frac{1}{r \sin{\theta}}\partial_{\phi} \Phi) 
\end{align}

\begin{align}
\partial_{t} \eht{\rho \epsilon_K} = & -\left( \eht{\frac{1}{r^2}\partial_r [r^2 \big(\rho u_r \epsilon_K - u_j\tau_{jr}\big)]} + \cancelto{0}{\eht{\frac{1}{r\sin{\theta}} \partial_\theta [\sin{\theta} \big(\rho u_\theta \epsilon_K - u_j\tau_{j\theta} \big)]}} + \cancelto{0}{\eht{\frac{1}{r\sin{\theta}} \partial_\phi [\rho u_\phi \epsilon_K - u_j\tau_{j\phi}]}} \right) - \nonumber \\
& -\left( \eht{\frac{1}{r^2} \partial_r \big( r^2 [P u_r] \big)} + \cancelto{0}{\eht{\frac{1}{r \sin{\theta}} \partial_\theta \big(\sin{\theta}[P u_\theta] \big)}} + \cancelto{0}{\eht{\frac{1}{r\sin{\theta}} \partial_\phi [ P u_\phi ]}} \right) + \eht{P \left( \frac{1}{r^{2}}\partial_{r}\big(r^{2}[u_{r}]\big) + \frac{1}{r\sin{\theta}}\partial_{\theta} \big(\sin{\theta} [u_{\theta}]\big) + \frac{1}{r\sin{\theta}}\partial_{\phi} [u_{\phi}]  \right)} - \nonumber \\
& -  \left( \eht{\tau_{jr} \partial_r u_j + \tau_{j\theta} \frac{1}{r}\partial_\theta u_j + \tau_{j\phi} \frac{1}{r\sin{\theta}}\partial_\phi u_j}  \right)  + \eht{\rho \big( u_{r} g_r + \cancelto{0}{u_{\theta} g_\theta} + \cancelto{0}{u_{\phi} g_\phi \big)}} \\
\partial_{t} \eht{\rho \epsilon_K} = & -\frac{1}{r^2}\partial_r [r^2 \big(\eht{\rho u_r \epsilon_K}\big)] -\frac{1}{r^2}\partial_r [r^2 \eht{(u_j\tau_{jr})}] -\frac{1}{r^2} \partial_r \big( r^2 [\eht{P u_r}] \big) + \eht{P d} -\eht{\tau_{ji} \partial_i u_j}  + \eht{\rho u_{r} g_r} \\
\eht{\rho}\fht{D}_t \fht{\epsilon}_K = & -\frac{1}{r^2}\partial_r [r^2 \big(\eht{\rho u''_r \epsilon''_K}\big)] -\cancelto{0}{\frac{1}{r^2}\partial_r [r^2 (\eht{u_j} \ \eht{\tau_{jr}})]} - \frac{1}{r^2}\partial_r [r^2 \eht{(u'_j \tau'_{jr})}] - \frac{1}{r^2} \partial_r \big( r^2 [\eht{P} \eht{u_r}] \big) - \frac{1}{r^2} \partial_r \big( r^2 [\eht{P' u'_r}] \big) + \eht{P} \ \eht{d} + \eht{P'd'} -\varepsilon_k  - \eht{\rho} \eht{u''_{r}} \fht{g}_r + \eht{\rho} \ \eht{u}_r \fht{g}_r \\ 
\eht{\rho}\fht{D}_t \fht{\epsilon}_K = & -\frac{1}{r^2}\partial_r [r^2 \big(\eht{\rho u''_r \epsilon''_K}\big)] - \frac{1}{r^2}\partial_r [r^2 \eht{(u'_j \tau'_{jr})}] - \frac{1}{r^2} \partial_r \big( r^2 [\eht{P} \eht{u_r}] \big) - \frac{1}{r^2} \partial_r \big( r^2 [\eht{P' u'_r}] \big) + \eht{P} \ \eht{d} + W_P -\varepsilon_k  + W_b + \eht{\rho} \ \eht{u}_r \fht{g}_r \\
\eht{\rho}\fht{D}_t \fht{\epsilon}_K = & -\nabla_r \eht{\rho u''_r \epsilon''_K} - \nabla_r f_\tau - (\eht{P}\nabla_r \eht{u}_r + \eht{u}_r \partial_r \eht{P}) - \nabla_r f_P + \eht{P} \ \eht{d} + W_P -\varepsilon_k  + W_b + \eht{\rho} \ \eht{u}_r \fht{g}_r \\
\eht{\rho}\fht{D}_t \fht{\epsilon}_K = & -\nabla_r \eht{\rho u''_r \epsilon''_K} - \nabla_r f_\tau - \eht{P}\ \eht{d} - \eht{\rho} \ \eht{u}_r \fht{g}_r - \nabla_r f_P + \eht{P} \ \eht{d} + W_P -\varepsilon_k  + W_b + \eht{\rho} \ \eht{u}_r \fht{g}_r \\
{\color{red} \eht{\rho}\fht{D}_t \fht{\epsilon}_K = } & {\color{red} -\nabla_r \eht{\rho u''_r \epsilon''_K} - \nabla_r (f_\tau + f_P ) + W_P -\varepsilon_k  + W_b }
\end{align}

Second way:

\begin{align}
\eht{\rho}\fht{D}_t \fht{\epsilon}_K = & +\eht{\rho}\fht{D}_t \fht{u}_i \fht{u}_i + \eht{\rho}\fht{D}_t \fht{u''_i u''_i} \\
\eht{\rho}\fht{D}_t \fht{\epsilon}_K = & +\eht{\rho}\fht{D}_t \fht{u}_i \fht{u}_i + \eht{\rho}\fht{D}_t \fht{k} \\
{\color{red} \eht{\rho}\fht{D}_t \fht{\epsilon}_K = }& {\color{red} -\nabla_r  ( f_k +  f_P + f_\tau) - \fht{R}_{ir}\partial_r \fht{u}_i + W_b + W_P -\varepsilon_k +\av{\rho}\fav{D}_t (\fav{u}_i \fav{u}_i / 2) }
\end{align}

where equation for the $\fht{k}$ is derived later.

\fontsize{12pt}{20pt}

\subsection{Mean total energy equation}

\fontsize{9pt}{20pt}

\begin{align}
\eht{\rho}\fht{D}_t \fht{\epsilon}_t = & + \eht{\rho}\fht{D}_t \fht{\epsilon}_I + \eht{\rho}\fht{D}_t \fht{\epsilon}_K \\
{\color{red} \av{\rho} \fav{D}_t \fav{\epsilon}_t =} & {\color{red} -\nabla_r ( f_I + f_T + f_k + f_P + f_\tau)  - \av{P} \ \av{d} - \fht{R}_{ir}\partial_r \fht{u}_i + W_b + {\mathcal S} + \av{\rho}\fav{D}_t (\fav{u}_i \fav{u}_i / 2)} 
\end{align}

\fontsize{12pt}{20pt}

\subsection{Mean pressure equation}

We begin by deriving 3D instantaneous pressure equation and then apply "ensemble'' (space-time) averaging (Sect.\ref{sect:intro}):

\fontsize{9pt}{20pt}

\begin{align}
dP = & \frac{\partial P}{\partial \rho}\big|_{\epsilon_I} d\rho + \frac{\partial P}{\partial \epsilon_I}\big|_\rho d\epsilon_I = \frac{P}{\rho}(1-\Gamma_3+\Gamma_1)d\rho + \rho(\Gamma_3 -1)d\epsilon_I \\
D_t P = & + \frac{P}{\rho}(1-\Gamma_3+\Gamma_1) D_t \rho + (\Gamma_3 -1)\rho D_t\epsilon_I \\
D_t P = & -(1-\Gamma_3+\Gamma_1)Pd + (\Gamma_3 -1)(-Pd + {\mathcal S} + \nabla \cdot F_T + \tau_{ij}\partial_i u_j) \\
\partial_t P =  & - u_n \partial_n P -(1-\Gamma_3+\Gamma_1)Pd + (\Gamma_3 -1)(-Pd + {\mathcal S} + \nabla \cdot F_T + \tau_{ij}\partial_i u_j) \\
\partial_t P = & -\left( \frac{1}{r^{2}}\partial_{r}\big(r^{2}[P u_{r}~]\big) + \frac{1}{r\sin{\theta}}\partial_{\theta} \big(\sin{\theta} [P u_{\theta}]\big) + \frac{1}{r\sin{\theta}}\partial_{\phi} [P u_{\phi}] \right) + (1-\Gamma_1)Pd + (\Gamma_3-1)({\mathcal S} + \nabla \cdot F_T + \tau_{ij}\partial_i u_j) \\
\partial_t \eht{P} = & -\left( \eht{\frac{1}{r^{2}}\partial_{r}\big(r^{2}[P u_{r}~]\big)} + \cancelto{0}{\eht{\frac{1}{r\sin{\theta}}\partial_{\theta} \big(\sin{\theta} [P u_{\theta}]\big)}} + \cancelto{0}{\eht{\frac{1}{r\sin{\theta}}\partial_{\phi} [P u_{\phi}]}} \right) + (1-\Gamma_1)\eht{Pd} + (\Gamma_3-1)({\mathcal S} + \eht{\nabla \cdot F_T} + \eht{\tau_{ij}\partial_i u_j}) \\
\partial_t \eht{P} = & -\frac{1}{r^{2}}\partial_{r}\big(r^{2}[\eht{P} \eht{u}_{r}]\big) - \frac{1}{r^{2}}\partial_{r}\big(r^{2}[\eht{P'u'_r}]\big) + (1-\Gamma_1)\eht{P} \ \eht{d} + (1-\Gamma_1)\eht{P'd'} + (\Gamma_3-1)({\mathcal S} + \frac{1}{r^{2}}\partial_{r}\big(r^{2} \eht{\chi \partial_r T} \big) + \cancelto{0}{\eht{\tau_{ij}} \partial_i \eht{{u'_j}}} + \eht{\tau'_{ij}\partial_i u'_j}) \\
\partial_t \eht{P} = & -\nabla_r \eht{P} \eht{u}_{r} - \nabla_r f_P + (1-\Gamma_1)\eht{P} \ \eht{d} + (1-\Gamma_1)W_P + (\Gamma_3-1)({\mathcal S} + \nabla_r f_T + \varepsilon_k) \\
\partial_t \eht{P} = & -\eht{u}_r \partial_r \eht{P}- \eht{P} \ \eht{d} - \nabla_r f_P + (1-\Gamma_1)\eht{P} \ \eht{d} + (1-\Gamma_1)W_P + (\Gamma_3-1)(\eht{{\mathcal S}} + \nabla_r f_T + \varepsilon_k) \\
\partial_t \eht{P} + \eht{u}_r \partial_r \eht{P} = & -\nabla_r f_P  -\Gamma_1 \eht{P} \ \eht{d} + (1-\Gamma_1)W_P + (\Gamma_3-1)({\mathcal S} + \nabla_r f_T + \varepsilon_k) \\
{\color{red} \eht{D}_t \eht{P} = }& {\color{red} -\nabla_r f_P  -\Gamma_1 \eht{P} \ \eht{d} + (1-\Gamma_1)W_P + (\Gamma_3-1)({\mathcal S} + \nabla_r f_T + \varepsilon_k)}
\end{align}

\fontsize{12pt}{20pt}

\subsection{Mean enthalpy equation}

We start from the total energy equation, where we can subsitute $\rho \epsilon_t = \rho h + \rho \epsilon_k - P$ (for clarity reasons we use here more compact vector notation):

\fontsize{9pt}{20pt}

\begin{align}
\partial_t \epsilon_t + \vec{\nabla} \cdot \big( (\rho \epsilon_t + P) \vec{u} \big) = \rho \vec{u} \cdot \vec{g} + \vec{\nabla} \cdot F_T + {\mathcal S}  &&  \\
\partial_t (\rho h + \rho \epsilon_k - P) + \vec{\nabla} \cdot (\rho h \vec{u} + \rho \epsilon_k \vec{u} ) = \rho \vec{u} \cdot \vec{g} + \vec{\nabla} \cdot F_T + {\mathcal S} & & \\ 
\partial_t \rho h + {\color{brown}\partial_t \rho \epsilon_k} - {\color{blue}\partial_t P} = - \vec{\nabla} \cdot (\rho h \vec{u} + \rho \epsilon_k \vec{u}) + \rho \vec{u} \cdot \vec{g} + \vec{\nabla} \cdot F_T + {\mathcal S} & & \\
\partial_t \rho h + {\color{brown}\left[-\vec{\nabla} \cdot (\rho \epsilon_k \vec{u}) - \vec{\nabla} \cdot (P \vec{u}) + P (\vec{\nabla} \cdot \vec{u}) + \rho \vec{u} \cdot \vec{g} + \nabla_i u_j \tau_{ji} \right]} - {\color{blue}\left[-\vec{\nabla} \cdot (P \vec{u}) + (1-\Gamma_1)P \vec{\nabla} \cdot \vec{u} + (\Gamma_3 - 1)({\mathcal S} + \vec{\nabla} \cdot F_T + \tau_{ij}\partial_j u_i) \right]} = && \\
= - \vec{\nabla} \cdot (\rho h \vec{u} + \rho \epsilon_k \vec{u}) + \rho \vec{u} \cdot \vec{g} + \vec{\nabla} \cdot F_T + {\mathcal S} &&   \\
\partial_t \rho h + {\color{brown}\left[-\cancel{\vec{\nabla} \cdot (\rho \epsilon_k \vec{u})} - \cancel{\vec{\nabla} \cdot (P \vec{u})} + \cancel{P (\vec{\nabla} \cdot \vec{u})} + \cancel{\rho \vec{u} \cdot \vec{g}} + \nabla_i u_j \tau_{ji} \right]} + {\color{blue}\cancel{\vec{\nabla} \cdot (P \vec{u})} - \cancel{P\vec{\nabla} \cdot \vec{u}} +\Gamma_1 P \vec{\nabla} \cdot \vec{u} - \Gamma_3 ({\mathcal S} + \vec{\nabla} \cdot F_T) + (\cancel{{\mathcal S}} + \cancel{\vec{\nabla} \cdot F_T})} - = && \\
{\color{blue} - (\Gamma_3 -1) \tau_{ij}\partial_j u_i} = - \vec{\nabla} \cdot (\rho h \vec{u} + \cancel{\rho \epsilon_k \vec{u}}) + \cancel{\rho \vec{u} \cdot \vec{g}} + \cancel{\vec{\nabla} \cdot F_T} + \cancel{{\mathcal S}} &&   
\end{align}

So, from the above we have:

\begin{align}
\partial_t \rho h + & \nabla_i u_j \tau_{ji} + \Gamma_1 P \vec{\nabla} \cdot \vec{u} - \Gamma_3 {\mathcal S} - \Gamma_3 \vec{\nabla} \cdot F_T - (\Gamma_3 -1) \tau_{ij}\partial_j u_i = -\vec{\nabla} \cdot (\rho h \vec{u}) & \\
\rho D_t h = & -\Gamma_1 P \vec{\nabla} \cdot \vec{u} + \Gamma_3 {\mathcal S} + \Gamma_3 \vec{\nabla} \cdot F_T - \nabla_i u_j \tau_{ji} +  (\Gamma_3 -1) \tau_{ij}\partial_j u_i \\
\eht{\rho}\fht{D}_t \fht{h} = & -\nabla_r f_h - \Gamma_1\eht{P} \ \eht{d} - \Gamma_1 W_P + \Gamma_3 {\mathcal S} + \Gamma_3 \nabla_r f_T - \nabla_r \eht{u_j \tau_{jr}} +  (\Gamma_3 -1) \eht{\tau_{ij}\partial_j u_i} \\
{\color{red} \eht{\rho}\fht{D}_t \fht{h} =} & {\color{red}-\nabla_r (f_h + f_\tau) - \Gamma_1\eht{P} \ \eht{d} - \Gamma_1 W_P + \Gamma_3 {\mathcal S} + \Gamma_3 \nabla_r f_T +  (\Gamma_3 -1)\varepsilon_k}  
\end{align}

\fontsize{12pt}{20pt}


\subsection{Mean temperature equation}

We begin by deriving 3D instantaneous temperature equation and then apply "ensemble'' (space-time) averaging (Sect.\ref{sect:intro}):

\fontsize{9pt}{20pt}

\begin{align}
dT = & \frac{\partial T}{\partial \rho}\big|_{\epsilon_I} d\rho + \frac{\partial T}{\partial \epsilon_I}\big|_\rho d\epsilon_I = \left( \frac{T}{\rho}(\Gamma_3 -1) - \frac{P}{\rho^2}\frac{1}{c_v} \right) d\rho + \frac{1}{c_v} d\epsilon_I \\
D_t T = & +\frac{T}{\rho}(\Gamma_3 -1) D_t\rho - \frac{P}{\rho^2}\frac{1}{c_v} D_t \rho + \frac{1}{c_v} D_t \epsilon_I \\
D_t T = & -\frac{T}{\rho}(\Gamma_3 -1)(\rho d) + \frac{P}{\rho^2}\frac{1}{c_v}(\rho d) + \frac{1}{c_v}\left(-\frac{Pd}{\rho} + \frac{\nabla \cdot F_T}{\rho} + \frac{\tau_{ij}\partial_j u_i}{\rho} + \frac{{\mathcal S}}{\rho} \right) \\
D_t T = & -(\Gamma_3 -1)Td + \frac{\nabla \cdot F_T}{c_v \rho} + \frac{\tau_{ij}\partial_j u_i}{c_v \rho} + \frac{{\mathcal S}}{c_v \rho} \\
\partial_t T = & -u_n \partial_n T  -(\Gamma_3 -1)Td + \frac{\nabla \cdot F_T}{c_v \rho} + \frac{\tau_{ij}\partial_j u_i}{c_v \rho} + \frac{{\mathcal S}}{c_v \rho} \\
\partial_t T = & -\left( \frac{1}{r^{2}}\partial_{r}\big(r^{2}[T u_{r}~]\big) + \frac{1}{r\sin{\theta}}\partial_{\theta} \big(\sin{\theta} [T u_{\theta}]\big) + \frac{1}{r\sin{\theta}}\partial_{\phi} [T u_{\phi}] \right) + Td  -(\Gamma_3 -1)Td + \frac{\nabla \cdot F_T}{c_v \rho} + \frac{\tau_{ij}\partial_j u_i}{c_v \rho} + \frac{{\mathcal S}}{c_v \rho} \\
\partial_t \eht{T} = & -\left( \eht{\frac{1}{r^{2}}\partial_{r}\big(r^{2}[T u_{r}~]\big)} + \cancelto{0}{\eht{\frac{1}{r\sin{\theta}}\partial_{\theta} \big(\sin{\theta} [T u_{\theta}]\big)}} + \cancelto{0}{\eht{\frac{1}{r\sin{\theta}}\partial_{\phi} [T u_{\phi}]}} \right) + \eht{Td}  -(\Gamma_3 -1)\eht{Td} + \eht{\frac{\nabla \cdot F_T}{c_v \rho}} + \eht{\frac{\tau_{ij}\partial_j u_i}{c_v \rho}} + \eht{\frac{{\mathcal S}}{c_v \rho}} \\
\partial_t \eht{T} = & -\nabla_r \eht{T}\eht{u}_r -\nabla_r f_T + \eht{Td}  -\Gamma_3 \eht{Td} + \eht{Td} + \eht{(\nabla \cdot F_T)/(c_v \rho)} + \eht{(\tau_{ij}\partial_j u_i)/(c_v \rho)} + \eht{{\mathcal S}/(c_v \rho)} \\
\partial_t \eht{T} + \eht{u}_r \partial_r \eht{T} = & -\eht{T}\ \eht{d} -\nabla_r f_T + \eht{Td}  -\Gamma_3 \eht{Td} + \eht{Td} + \eht{(\nabla \cdot F_T)/(c_v \rho)} + \eht{(\tau_{ij}\partial_j u_i)/(c_v \rho)} + \eht{{\mathcal S}/(c_v \rho)} \\
\eht{D}_t \eht{T} = & -\eht{T}\ \eht{d} -\nabla_r f_T + \eht{Td}  -\Gamma_3 \eht{Td} + \eht{Td} + \eht{(\nabla \cdot F_T)/(c_v \rho)} + \eht{(\tau_{ij}\partial_j u_i)/(c_v \rho)} + \eht{{\mathcal S}/(c_v \rho)} \\
\eht{D}_t \eht{T} = & -\eht{T}\ \eht{d} -\nabla_r f_T + \eht{Td}  -\Gamma_3 \eht{Td} + \eht{T} \ \eht{d} + \eht{T'd'} + \eht{(\nabla \cdot F_T)/(c_v \rho)} + \eht{(\tau_{ij}\partial_j u_i)/(c_v \rho)} + \eht{{\mathcal S}/(c_v \rho)} \\
\eht{D}_t \eht{T} = & -\nabla_r f_T + \eht{Td}  -\Gamma_3 \eht{Td} + \eht{T'd'} + \eht{(\nabla \cdot F_T)/(c_v \rho)} + \eht{(\tau_{ij}\partial_j u_i)/(c_v \rho)} + \eht{{\mathcal S}/(c_v \rho)} \\
\eht{D}_t \eht{T} = & -\nabla_r f_T + \eht{T} \ \eht{d} + \eht{T'd'} -\Gamma_3 (\eht{T} \ \eht{d} + \eht{T'd'}) + \eht{T'd'} + \eht{(\nabla \cdot F_T)/(c_v \rho)} + \eht{(\tau_{ij}\partial_j u_i)/(c_v \rho)} + \eht{{\mathcal S}/(c_v \rho)}\\
{\color{red} \eht{D}_t \eht{T} =} & {\color{red}-\nabla_r f_T + (1-\Gamma_3)\eht{T} \ \eht{d} + (2-\Gamma_3)\eht{T'd'} + \eht{(\nabla \cdot F_T)/(c_v \rho)} + \eht{(\tau_{ij}\partial_j u_i)/(c_v \rho)} + \eht{{\mathcal S}/(c_v \rho)}}
\end{align}

\fontsize{12pt}{20pt}

\subsection{Mean angular momentum equation (z-component)}

Z component of the specific angular momentum is defined as  $j_z = r \sin{\theta} u_\phi$. We begin by multiplying the instantaneous 3D azimutal momentum equation by $r \sin{\theta}$, neglect viscosity and $\phi$ component of gravity and obtain (Sect.\ref{sect:intro}):

\fontsize{9pt}{20pt}

\begin{align}
r\sin{\theta} \partial_{t} \big(\rho u_{\phi}\big) = & -r\sin{\theta}\left( \frac{1}{r^{2}} \partial_{r} \big(r^{2} [\rho u_{\phi} u_r - \cancelto{0}{\tau_{\phi r}}]\big) + \frac{1}{r\sin{\theta}}\partial_{\theta}\big(\sin{\theta}[\rho u_{\theta}u_\phi - \cancelto{0}{\tau_{\theta \phi}}]\big) + \frac{1}{r\sin{\theta}}\partial_{\phi}\big([\rho u_{\phi}^2 - \cancelto{0}{\tau_{\phi \phi}}]\big) + G_\phi^M + \frac{1}{r\sin{\theta}} \partial_{\phi} P \right) - \nonumber \\
- \cancelto{0}{r\sin{\theta} \rho \frac{1}{r \sin{\theta}} \partial_{\phi} \Phi} \\
r\sin{\theta} \partial_{t} \big(\rho u_{\phi}\big) = & -r\sin{\theta}\left( \frac{1}{r^{2}} \partial_{r} \big(r^{2} [\rho u_{\phi} u_r]\big) + \frac{1}{r\sin{\theta}}\partial_{\theta}\big(\sin{\theta}[\rho u_{\theta}u_\phi]\big) + \frac{1}{r\sin{\theta}}\partial_{\phi}\big([\rho u_{\phi}^2]\big) + G_\phi^M + \frac{1}{r\sin{\theta}} \partial_{\phi} P \right) \\
\partial_t \rho j_z = & -\frac{1}{r^2}\partial_r (r^2 \rho j_z u_r) - \frac{1}{r\sin{\theta}}\partial_\theta (\sin{\theta}\rho j_z u_\theta) - \frac{1}{r\sin{\theta}}\partial_\phi (\rho j_z u_\phi) - \partial_\phi P \\
\partial_t \eht{\rho j_z} = & -\eht{\frac{1}{r^2}\partial_r (r^2 \rho j_z u_r)} - \cancelto{0}{\eht{\frac{1}{r\sin{\theta}}\partial_\theta (\sin{\theta}\rho j_z u_\theta)}} - \cancelto{0}{\eht{\frac{1}{r\sin{\theta}}\partial_\phi (\rho j_z u_\phi)}} - \cancelto{0}{\eht{\partial_\phi P}}\\
\partial_t \eht{\rho j_z} = & -\frac{1}{r^2}\partial_r (r^2 \eht{\rho j_z u_r})  \\
\partial_t \eht{\rho} \fht{j_z} = & -\frac{1}{r^2}\partial_r (r^2 \eht{\rho} \fht{j_z} \fht{u_r}) -\frac{1}{r^2}\partial_r (r^2 \eht{\rho} \fht{j''_z u''_r})  \\
\partial_t \eht{\rho} \fht{j_z} + \frac{1}{r^2}\partial_r (r^2 \eht{\rho} \fht{j_z} \fht{u_r}) = & -\frac{1}{r^2}\partial_r (r^2 \eht{\rho} \fht{j''_z u''_r})  \\
{\color{red} \eht{\rho}\fht{D}_t \fht{j_z} =} & {\color{red} -\nabla_r f_{jz}} 
\end{align}

\fontsize{12pt}{20pt}


\subsection{Mean $\alpha$ equation}

We begin by 3D instantaneous composition equation and then apply "ensemble'' (space-time) averaging (Sect.\ref{sect:intro}):

\fontsize{9pt}{20pt}

\begin{align}
\partial_{t} \big(\rho X_{\alpha}\big) = & -\left( \frac{1}{r^{2}} \partial_{r}(r^{2} [\rho u_{r} X_{\alpha} ] \big) + \frac{1}{r\sin{\theta}}\partial_{\theta}\big(\sin{\theta} [\rho u_{\theta} X_{\alpha} ]\big) + \frac{1}{r \sin{\theta}} \partial_{\phi} [\rho u_{\phi} X_{\alpha}] \right) + \rho \dot{X}_{\alpha}^{\rm nuc} \ \ \ \ \ \ \ \ \ \ \ \  \alpha = 1 ... N_{nuc} \\
\partial_{t} \eht{\rho X_{\alpha}} = & -\left( \eht{\frac{1}{r^{2}} \partial_{r}(r^{2} [\rho u_{r} X_{\alpha} ] \big)} + \cancelto{0}{\eht{\frac{1}{r\sin{\theta}}\partial_{\theta}\big(\sin{\theta} [\rho u_{\theta} X_{\alpha} ]\big)}} + \cancelto{0}{\eht{\frac{1}{r \sin{\theta}} \partial_{\phi} [\rho u_{\phi} X_{\alpha}]}} \right) + \eht{\rho \dot{X}_{\alpha}^{\rm nuc}} \\
\partial_{t} \eht{\rho} \fht{X}_{\alpha} = & -\frac{1}{r^{2}} \partial_{r} ( r^{2} [\eht{\rho} \fht{u}_{r} \fht{X}_{\alpha} + \eht{\rho}\fht{u''_r X''_\alpha} ] ) + \eht{\rho} \fht{\dot{X}}_{\alpha}^{\rm nuc} \\
\partial_{t} \eht{\rho} \fht{X}_{\alpha} +\frac{1}{r^{2}} \partial_{r} ( r^{2} [\eht{\rho} \fht{u}_{r} \fht{X}_{\alpha}] ) = & -\frac{1}{r^{2}} \partial_{r} (\eht{\rho}\fht{u''_r X''_\alpha}) + \eht{\rho} \fht{\dot{X}}_{\alpha}^{\rm nuc} \\
{\color{red} \eht{\rho}\fht{D}_t \fht{X}_\alpha = }& {\color{red} -\nabla_r f_\alpha + \eht{\rho} \fht{\dot{X}}_{\alpha}^{\rm nuc}}
\end{align}

\fontsize{12pt}{20pt}

\newpage

\subsection{Mean number of nucleons per isotope ($A$) equation} 

We begin by deriving 3D instantaneous A equation and then apply "ensemble'' (space-time) averaging (Sect.\ref{sect:intro}):

\fontsize{8pt}{20pt}

\begin{align}
A = & +\left(\sum_\alpha \frac{X_\alpha}{A_\alpha}  \right)^{-1} \\
D_t A = & +D_t \left(\sum_\alpha \frac{X_\alpha}{A_\alpha}  \right)^{-1} = +D_t \frac{1}{\sum_\alpha (X_\alpha / A_\alpha)} = -\frac{D_t \sum_\alpha (X_\alpha/A_\alpha)}{[\sum_\alpha (X_\alpha/A_\alpha)]^2} = -A^2 D_t \sum_\alpha \frac{X_\alpha}{A_\alpha} \\
D_t A = & -A^2 D_t \sum_\alpha \frac{X_\alpha}{A_\alpha} = -A^2 \sum_\alpha \frac{A_\alpha D_t X_\alpha - X_\alpha D_t A_\alpha}{A^2_\alpha} = -A^2 \sum_\alpha \frac{A_\alpha D_t X_\alpha}{A^2_\alpha} \\
D_t A = & -A^2 \sum_\alpha \frac{A_\alpha D_t X_\alpha}{A^2_\alpha} = -A^2 \sum_\alpha \frac{A_\alpha \dot{X}_\alpha^{\rm nuc}}{A^2_\alpha} = -A^2 \sum_\alpha \frac{\dot{X}_\alpha^{\rm nuc}}{A_\alpha} \\
\rho D_t A = & -\rho A^2 \sum_\alpha \frac{\dot{X}_\alpha^{\rm nuc}}{A_\alpha} \\
\partial_t \rho A = & -\left( \frac{1}{r^{2}} \partial_{r}(r^{2} [\rho u_{r} A ] \big) + \frac{1}{r\sin{\theta}}\partial_{\theta}\big(\sin{\theta} [\rho u_{\theta} A ]\big) + \frac{1}{r \sin{\theta}} \partial_{\phi} [\rho u_{\phi} A] \right) -\rho A^2 \sum_\alpha \frac{\dot{X}_\alpha^{\rm nuc}}{A_\alpha} \\
\partial_t \overline{\rho A} = & -\left( \eht{\frac{1}{r^{2}} \partial_{r}(r^{2} [\rho u_{r} A ] \big)} + \cancelto{0}{\eht{\frac{1}{r\sin{\theta}}\partial_{\theta}\big(\sin{\theta} [\rho u_{\theta} A ]\big)}} + \cancelto{0}{\eht{\frac{1}{r \sin{\theta}} \partial_{\phi} [\rho u_{\phi} A]}} \right) - \eht{\rho A^2 \sum_\alpha \frac{\dot{X}_\alpha^{\rm nuc}}{A_\alpha}} \\
\partial_t \overline{\rho A} = & -\frac{1}{r^{2}} \partial_{r}(r^{2} [\eht{\rho u_{r} A }] \big) - \eht{\rho A^2 \sum_\alpha \frac{\dot{X}_\alpha^{\rm nuc}}{A_\alpha}} \\
\partial_t \overline{\rho} \fht{A} = & -\frac{1}{r^{2}} \partial_{r}(r^{2} [\eht{\rho} \fht{u_{r} A}] \big) - \eht{\rho A^2 \sum_\alpha \frac{\dot{X}_\alpha^{\rm nuc}}{A_\alpha}} \\
\partial_t \overline{\rho} \fht{A} = & -\frac{1}{r^{2}} \partial_{r}(r^{2} [\eht{\rho} \fht{u}_{r} \fht{A}])  -\frac{1}{r^{2}} \partial_{r}(r^{2} [\eht{\rho} \fht{u''_r A''}]) - \eht{\rho A^2 \sum_\alpha \frac{\dot{X}_\alpha^{\rm nuc}}{A_\alpha}} \\
\partial_t \overline{\rho} \fht{A} + \frac{1}{r^{2}} \partial_{r}(r^{2} [\eht{\rho} \fht{u}_{r} \fht{A}]) = &   -\frac{1}{r^{2}} \partial_{r}(r^{2} [\eht{\rho} \fht{u''_r A''}]) - \eht{\rho A^2 \sum_\alpha \frac{\dot{X}_\alpha^{\rm nuc}}{A_\alpha}} \\
{\color{red} \rho \fht{D}_t\fht{A} =} & {\color{red} -\nabla_r f_A - \eht{\rho A^2 \sum_\alpha \frac{\dot{X}_\alpha^{\rm nuc}}{A_\alpha}}}
\end{align}

\fontsize{12pt}{20pt}

\newpage

\subsection{Mean charge per isotope ($Z$) equation}

We begin by deriving 3D instantaneous Z equation and then apply "ensemble'' (space-time) averaging (Sect.\ref{sect:intro}):

\fontsize{7pt}{20pt}

\begin{align}
Z = & + A \sum_\alpha \frac{Z_\alpha A_\alpha}{A_\alpha} \\
D_t Z = & +D_t \left(A \sum_i \frac{Z_\alpha A_\alpha}{A_\alpha}\right) = \sum_\alpha \frac{Z_\alpha X_\alpha}{A_\alpha} D_t A + A\sum_\alpha D_t \frac{Z_\alpha X_\alpha}{A_\alpha} \\
D_t Z = & -\sum_\alpha \frac{Z_\alpha X_\alpha}{A_\alpha}  A^2 \sum_\alpha \frac{\dot{X}_\alpha^{\rm nuc}}{A_\alpha} + A\sum_\alpha \frac{A_\alpha D_t Z_\alpha X_\alpha - Z_\alpha X_\alpha D_t A_\alpha}{A_\alpha^2} \\
D_t Z = & -Z A \sum_\alpha \frac{\dot{X}_\alpha^{\rm nuc}}{A_\alpha} + A \sum_\alpha \frac{A_\alpha D_t Z_\alpha X_\alpha}{A_\alpha^2} \\
D_t Z = & -Z A \sum_\alpha \frac{\dot{X}_\alpha^{\rm nuc}}{A_\alpha} + A\sum_\alpha \frac{Z_\alpha D_t X_\alpha + X_\alpha D_t Z_\alpha}{A^2_\alpha} \\
D_t Z = & -Z A \sum_\alpha \frac{\dot{X}_\alpha^{\rm nuc}}{A_\alpha} + A\sum_\alpha \frac{Z_\alpha \dot{X}_\alpha^{\rm nuc}}{A_\alpha} \\
\rho D_t Z = & -\rho Z A \sum_\alpha \frac{\dot{X}_\alpha^{\rm nuc}}{A_\alpha} + \rho A \sum_\alpha \frac{Z_\alpha \dot{X}_\alpha^{\rm nuc}}{A_\alpha} \\
\partial_t \overline{\rho Z} = & -\left( \eht{\frac{1}{r^{2}} \partial_{r}(r^{2} [\rho u_{r} Z ] \big)} + \cancelto{0}{\eht{\frac{1}{r\sin{\theta}}\partial_{\theta}\big(\sin{\theta} [\rho u_{\theta} Z ]\big)}} + \cancelto{0}{\eht{\frac{1}{r \sin{\theta}} \partial_{\phi} [\rho u_{\phi} Z]}} \right) -\eht{\rho Z A \sum_\alpha \frac{\dot{X}_\alpha^{\rm nuc}}{A_\alpha}} + \eht{\rho A \sum_\alpha \frac{Z_\alpha \dot{X}_\alpha^{\rm nuc}}{A_\alpha}}\\
\partial_t \overline{\rho Z} = & -\frac{1}{r^{2}} \partial_{r}(r^{2} [\eht{\rho u_{r} Z} ] \big) -\eht{\rho Z A \sum_\alpha \frac{\dot{X}_\alpha^{\rm nuc}}{A_\alpha}} + \eht{\rho A \sum_\alpha \frac{Z_\alpha \dot{X}_\alpha^{\rm nuc}}{A_\alpha}} \\
\partial_t \overline{\rho} \fht{Z} = & -\frac{1}{r^{2}} \partial_{r}(r^{2} [\eht{\rho} \fht{u_{r} Z}]) -\eht{\rho Z A \sum_\alpha \frac{\dot{X}_\alpha^{\rm nuc}}{A_\alpha}} + \eht{\rho A \sum_\alpha \frac{Z_\alpha \dot{X}_\alpha^{\rm nuc}}{A_\alpha}} \\
\partial_t \overline{\rho} \fht{Z} = & -\frac{1}{r^{2}} \partial_{r}(r^{2} [\eht{\rho} \fht{u_{r}} \fht{Z}]) -\frac{1}{r^{2}} \partial_{r}(r^{2} [\eht{\rho} \fht{u''_{r} Z''}]) -\eht{\rho Z A \sum_\alpha \frac{\dot{X}_\alpha^{\rm nuc}}{A_\alpha}} + \eht{\rho A \sum_\alpha \frac{Z_\alpha \dot{X}_\alpha^{\rm nuc}}{A_\alpha}} \\
\partial_t \overline{\rho} \fht{Z}  +\frac{1}{r^{2}} \partial_{r}(r^{2} [\eht{\rho} \fht{u_{r}} \fht{Z}]) = & -\frac{1}{r^{2}} \partial_{r}(r^{2} [\eht{\rho} \fht{u''_{r} Z''}]) -\eht{\rho Z A \sum_\alpha \frac{\dot{X}_\alpha^{\rm nuc}}{A_\alpha}} + \eht{\rho A \sum_\alpha \frac{Z_\alpha \dot{X}_\alpha^{\rm nuc}}{A_\alpha}} \\
{\color{red} \eht{\rho}\fht{D}_t \fht{Z} =} & {\color{red} -\nabla_r f_Z -\eht{\rho Z A \sum_\alpha \frac{\dot{X}_\alpha^{\rm nuc}}{A_\alpha}} + \eht{\rho A \sum_\alpha \frac{Z_\alpha \dot{X}_\alpha^{\rm nuc}}{A_\alpha}}}
\end{align}

\fontsize{12pt}{20pt}

\subsection{Mean entropy equation}

We can derive the mean entropy equation in the following way (Sect.\ref{sect:intro}):

\fontsize{9pt}{20pt}

\begin{align}
\rho D_t s = & +(-\nabla \cdot F_T + {\mathcal S} + \varepsilon_k)/T \\
\partial_t \rho s + \nabla_r (\rho s u_r) =  & +(-\nabla \cdot F_T + {\mathcal S} + \varepsilon_k)/T \\
\partial_t \eht{\rho s} + \nabla_r \eht{\rho u_r s} = & +\eht{(-\nabla \cdot F_T + {\mathcal S} + \varepsilon_k)/T} \\
\partial_t \eht{\rho}\fht{s} + \nabla_r (\eht{\rho} \fht{u}_r \fht{s}) = & -\nabla_r (\eht{\rho} \fht{s''u''_r}) - \eht{\nabla \cdot F_T / T} + \eht{{\mathcal S}/T} + \eht{{\varepsilon_k}/T} \\
{\color{red} \eht{\rho} \fht{D}_t \fht{s} =} & {\color{red}-\nabla_r f_s - \eht{\nabla \cdot F_T / T} + \eht{{\mathcal S}/T} + \eht{{\varepsilon_k}/T} }
\end{align}

\fontsize{12pt}{20pt}

\newpage

\section{General formula for second and third order moments and variances}

\subsection{Second-order moments}

In order to calculate evolution equations for correlations of two arbitrary fluctuations, we can derive the following general formula.

\fontsize{11pt}{20pt}

\begin{align}
\eht{\rho}\fht{D}_t \fht{c''d''} - \eht{\rho D_t c''d''} & = \eht{\rho} \big(\partial_t \fht{c''d''} + \fht{u}_n \partial_n \fht{c''d''} \big) - \eht{\rho \big( \partial_t c''d'' + u_n \partial_n c''d'' \big)} = \eht{\rho} \partial_t \fht{c''d''} + \eht{\rho}\fht{u}_n \partial_n \fht{c''d''} - \eht{\rho \partial_t c''d''} - \eht{\rho u_n \partial_n c''d''} = \nonumber \\
& = \eht{\rho} \partial_t \fht{c''d''} + \eht{\rho} \fht{u}_n \partial_n \fht{c''d''} - \big(\eht{\partial_t \rho c''d''} - \eht{c''d''\partial_t \rho} \big) - \eht{\rho u_n \partial_n c''d''} = \\
& = \eht{\rho} \partial_t \fht{c''d''} + \eht{\rho} \fht{u}_n \partial_n \fht{c''d''} - \partial_t \eht{\rho} \fht{c''d''} - \eht{c''d''\partial_n \rho u_n} - \eht{\rho u_n \partial_n c''d''} = \\
& = \eht{\rho} \partial_t \fht{c''d''} + \eht{\rho} \fht{u}_n \partial_n \fht{c''d''} - \big(\eht{\rho} \partial_t \fht{c''d''} + \fht{c''d''} \partial_t \eht{\rho} \big) - \eht{c''d''\partial_n \rho u_n} - \eht{\rho u_n \partial_n c''d''} = \\
& = \eht{\rho} \partial_t \fht{c''d''} + \eht{\rho} \fht{u}_n \partial_n \fht{c''d''} - \big(\eht{\rho} \partial_t \fht{c''d''} - \fht{c''d''} \partial_n \eht{\rho}\fht{u}_n \big) - \eht{c''d''\partial_n \rho u_n} - \eht{\rho u_n \partial_n c''d''} = \\
& =  \eht{\rho} \partial_t \fht{c''d''} -  \eht{\rho} \partial_t \fht{c''d''} + \eht{\rho} \fht{u}_n \partial_n \fht{c''d''} +  \fht{c''d''} \partial_n \eht{\rho}\fht{u}_n - \eht{\partial_n \rho u_n c''d''} = \\
& = \partial_n \eht{\rho} \fht{u}_n \fht{c''d''} - \big( \eht{\partial_n \rho \fht{u}_n c''d''} + \eht{\partial_n \rho u''_n c''d'' }  \big) = -\eht{\partial_n \rho u''_n c''d''}
\end{align}

\begin{align}
\eht{\rho}\fht{D}_t \fht{c''d''} = \eht{\rho D_t c''d''} -\eht{\partial_n \rho u''_n c''d''} = \eht{c'' \rho D_t d''} + \eht{d'' \rho D_t c''} -\eht{\partial_n \rho u''_n c''d''}
\end{align}

\begin{align}
\rho D_t c'' = \rho D_t c - \rho D_t \fht{c} = \rho D_t c - \rho \fht{D}_t \fht{c} - \rho u''_n \partial_n \fht{c} =  \rho D_t c - \frac{\rho}{\eht{\rho}} \left[ \eht{\rho} \fht{D}_t \fht{c} \right] - \rho u''_n \partial_n \fht{c} \\
\rho D_t d'' = \rho D_t d - \rho D_t \fht{d} = \rho D_t d - \rho \fht{D}_t \fht{d} - \rho u''_n \partial_n \fht{d} =  \rho D_t d - \frac{\rho}{\eht{\rho}} \left[ \eht{\rho} \fht{D}_t \fht{d} \right] - \rho u''_n \partial_n \fht{d}  
\end{align}

\begin{align}
\eht{\rho}\fht{D}_t \fht{c''d''} =  & \ \eht{c''\left(\rho D_t d - \frac{\rho}{\eht{\rho}}\left[\eht{\rho}\fht{D}_t \fht{d} \right] - \rho u''_n\partial_n\fht{d} \right)} + \ \eht{d''\left(\rho D_t c - \frac{\rho}{\eht{\rho}}\left[\eht{\rho}\fht{D}_t \fht{c} \right] - \rho u''_n\partial_n\fht{c} \right)} - \eht{\partial_n {\rho c''d'' u''_n}}
\end{align}

\begin{align}
{\color{red} \eht{\rho}\fht{D}_t \fht{c''d''} = } & {\color{red} +\eht{c''\rho D_t d} - \eht{\rho}\fht{c''u''_n}\partial_n\fht{d} + \eht{d''\rho D_t c} - \eht{\rho}\fht{d''u''_n}\partial_n\fht{c} - \eht{\partial_n{\rho c''d'' u''_n}} }
\label{eq:second-order-moments}
\end{align}

\fontsize{12pt}{20pt}

\newpage

\subsection{Third-order moments}

In order to calculate evolution equations for correlations of three arbitrary fluctuations, we can derive the following general formula.

\fontsize{11pt}{20pt}

\begin{align}
\eht{\rho}\fht{D}_t \fht{c''d''e''} - \eht{\rho D_t c''d''e''} & = \ \eht{\rho}\big(\partial_t \fht{c''d''e''} + \fht{u}_n\partial_n \fht{c''d''e''} \big) - \eht{\rho ( \partial_t c''d''e'' + u_n \partial_n c''d''e'' )} = \\
& = \ \eht{\rho}\partial_t \fht{c''d''e''} + \eht{\rho}\fht{u}_n\partial_n \fht{c''d''e''} - \eht{\rho \partial_t c''d''e''} - \eht{\rho u_n \partial_n c''d''e''} = \\
& = \ \eht{\rho}\partial_t \fht{c''d''e''} + \eht{\rho}\fht{u}_n\partial_n \fht{c''d''e''} - (\eht{\partial_t \rho c''d''e''} - \eht{c''d''e''\partial_t \rho}) - \eht{\rho u_n \partial_n c''d''e''} = \\
& = \ \eht{\rho}\partial_t \fht{c''d''e''} + \eht{\rho}\fht{u}_n\partial_n \fht{c''d''e''} - \partial_t \eht{\rho} \fht{c''d''e''} - (\eht{c''d''e''\partial_n \rho u_n} + \eht{\rho u_n \partial_n c''d''e''}) = \\
& = \ \eht{\rho}\partial_t \fht{c''d''e''} + \eht{\rho}\fht{u}_n\partial_n \fht{c''d''e''} - (\eht{\rho}\partial_t \fht{c''d''e''} + \fht{c''d''e''}\partial_t \eht{\rho} ) - \eht{\partial_n c''d''e'' \rho u_n} =  \\
& = \ \eht{\rho}\fht{u}_n\partial_n \fht{c''d''e''} + \fht{c''d''e''}\partial_n \eht{\rho}\fht{u}_n  - \eht{\partial_n c''d''e'' \rho \fht{u}_n} - \eht{\partial_n c''d''e'' \rho u''_n} = \\
& = \ - \eht{\partial_n c''d''e'' \rho u''_n} \\
\end{align}

\begin{align}
\eht{\rho D_t c''d''e''} = & \ \eht{\rho c''d'' D_t e''} + \eht{\rho c''e'' D_t d''} + \eht{\rho d''e'' D_t c''}
\end{align}

\begin{align}
\eht{c''d''\rho D_t e''} & = \ \eht{c''d''\rho D_t e} - \eht{c''d''\rho D_t \fht{e}} = \eht{c''d''\rho D_t e} - \eht{c''d''\rho (\partial_t \fht{e} + u_n \partial_n \fht{e}) } = \\
& = \ \eht{c''d''\rho D_t e} - \eht{c''d''\rho \partial_t \fht{e}} - \eht{c''d''\rho u_n \partial_n \fht{e}} = \eht{c''d''\rho D_t e} - \eht{\rho} \fht{c''d''}\partial_t \fht{e} - (\eht{c''d''\rho \fht{u}_n \partial_n \fht{e}} + \eht{c''d''\rho u''_n \partial_n \fht{e}} ) =  \\
& = \ \eht{c''d''\rho D_t e} - (\eht{\rho}\fht{c''d''}\partial_t \fht{e} + \eht{\rho}\fht{c''d''}\fht{u}_n \partial_n \fht{e}) - \eht{\rho} \fht{c''d''u''_n}\partial_n\fht{e} = \\
& \ = \eht{c''d''\rho D_t e} - \eht{\rho}\fht{c''d''}\fht{D}_t \fht{e} - \eht{\rho}\fht{c''d''u''_n} \partial_n \fht{e} 
\end{align}

\begin{align}
{\color{red} \eht{\rho}\fht{D}_t \fht{c''d''e''}} & = {\color{red} \ \eht{c''d''\rho D_t e} - \eht{\rho}\fht{c''d''}\fht{D}_t \fht{e} - \eht{\rho}\fht{c''d''u''_n} \partial_n \fht{e} +} \\
& {\color{red}\ + \eht{c''e''\rho D_t d} - \eht{\rho}\fht{c''e''}\fht{D}_t \fht{d} - \eht{\rho}\fht{c''e''u''_n} \partial_n \fht{d} +} \\
& {\color{red} \ + \eht{d''e''\rho D_t c} - \eht{\rho}\fht{d''e''}\fht{D}_t \fht{c} - \eht{\rho}\fht{d''e''u''_n} \partial_n \fht{c} -} \\
& {\color{red} \ - \eht{\partial_n c''d''e'' \rho u''_n}}    
\end{align}

\fontsize{12pt}{20pt}

\subsection{Reynolds and Favrian variance}

The Reynolds variance can be derived in following way:

\fontsize{11pt}{20pt}

\begin{align}
\fht{D}_t \eht{c'd'} - \eht{D_t c'd'} & = +\partial_t \eht{c'd'} + \fht{u}_n \partial_n \eht{c'd'} - (\partial_t \eht{c'd'} + \eht{u_n \partial_n c'd'}) = \\
& = +\partial_t \eht{c'd'} + \fht{u}_n \partial_n \eht{c'd'} - \partial_t \eht{c'd'} - \fht{u}_n \partial_n \eht{c'd'} - \eht{u''_n \partial_n c'd'} = \\
& = - \eht{u''_n \partial_n c'd'}
\end{align}

Next step:

\begin{align}
\fht{D}_t \eht{c'd'} = \eht{D_t c'd'} - \eht{u''_n \partial_n c'd'} = \eht{c' D_t d'} + \eht{d' D_t c'} - \eht{u''_n \partial_n c'd'}
\label{eq:reynolds-variance-general}
\end{align}

Next step:

\begin{align}
D_t c' = & D_t c - D_t \eht{c} = D_t c - \fht{D}_t \eht{c} - u''_n \partial_n \eht{c} \\
D_t d' = & D_t d - D_t \eht{c} = D_t d - \fht{D}_t \eht{d} - u''_n \partial_n \eht{d}
\end{align}

Now let's put these equations in the Equation~\ref{eq:reynolds-variance-general}:

\begin{align}
\fht{D}_t \eht{c'd'} & = +\eht{c' D_t d'} + \eht{d' D_t c'} - \eht{u''_n \partial_n c'd'} = \\
& = +\eht{c' (D_t d - \fht{D}_t \eht{d} - u''_n \partial_n \eht{d})} + \eht{d' (D_t c - \fht{D}_t \eht{c} - u''_n \partial_n \eht{c})} - \eht{u''_n \partial_n c'd'} = \\
& = +\eht{c'D_t d} - \eht{c'u''_n}\partial_n \eht{d} + \eht{d' D_t c} - \eht{d'u''_n}\partial_n \eht{c} - \eht{u''_n \partial_n c'd'} = \\
& = +\eht{c'D_t d} - \eht{c'u''_n}\partial_n \eht{d} + \eht{d' D_t c} - \eht{d'u''_n}\partial_n \eht{c} - \partial_n \eht{u''_n c'd'} + \eht{c'd'\partial_n u''_n}
\end{align}

From this general formula by substituting $d = c$ we get:

\begin{align}
{\color{red} \fht{D}_t \eht{c'c'}} & = + 2 \eht{c' D_t c} - 2 \eht{c'u''_n} \partial_n \eht{c} - \eht{u''_n \partial_n c'c'} = \\
& = {\color{red}+2 \eht{c' D_t c} - 2 \eht{c'u''_n} \partial_n \eht{c} - \partial_n \eht{u''_n c' c'} + \eht{c' c' \partial_n u''_n}}
\end{align}

\noindent
The Favrian variance can be easily derived from general equation for second-order moments Equation~\ref{eq:second-order-moments} ie.

\begin{align}
\eht{\rho}\fht{D}_t \fht{c''d''} = & +\eht{c''\rho D_t d} - \eht{\rho}\fht{c''u''_n}\partial_n\fht{d} + \eht{d''\rho D_t c} - \eht{\rho}\fht{d''u''_n}\partial_n\fht{c} - \eht{\partial_n{\rho c''d'' u''_n}}  
\end{align}

Now, substitute $d = c$ and you'll get the equation for Favrian variance:

\begin{align}
{\color{red}\eht{\rho}\fht{D}_t \fht{c''c''} =} & {\color{red}+2 \eht{c''\rho D_t c} - 2\eht{\rho}\fht{c''u''_n}\partial_n\fht{c} - \eht{\partial_n{\rho c''c'' u''_n}}} 
\end{align}

\fontsize{12pt}{20pt}

\newpage

\section{Derivation of second-order moments equations}

\subsection{Reynolds stress equation}

We can derive the Reynolds stress equation using the general formula for second order moments, where we substitute $c''$ with $u''_i$ and $d''$ with $u''_j$.

\begin{align}
\overline{\rho}\widetilde{D}_t \widetilde{c''d''} = & \ \overline{c'' \rho D_t d} - \overline{\rho} \widetilde{c''u''_n}\partial_n \widetilde{d} + \overline{d'' \rho D_t c} - \overline{\rho} \widetilde{d''u''_n}\partial_n \widetilde{c} - \overline{\partial_n \rho c''d''u''_n} \\
\underbrace{\eht{\rho}\fht{D}_t \fht{u''_i u''_j}}_\text{$\eht{\rho}\fht{D}_t (\fht{R}_{ij} / \eht{\rho})$} = & \ \eht{u''_i \rho D_t u_j} - \underbrace{\eht{\rho} \fht{u''_i u''_n} \partial_n \fht{u}_j}_\text{$\fht{R}_{in}\partial_n \fht{u}_j$} + \eht{u''_j \rho D_t u_i} - \underbrace{\eht{\rho} \fht{u''_j u''_n} \partial_n \fht{u}_i}_\text{$\fht{R}_{jn}\partial_n \fht{u}_i$} - \underbrace{\eht{\partial_n \rho u''_i u''_j u''_n}}_\text{$\eht{\nabla \cdot \rho u''_i u''_j u''_n}$}
\end{align}

\noindent
So, the general formula for Reynolds stress $\fht{R}_{ij}$ is:

\begin{align}
\eht{\rho}\fht{D}_t (\fht{R}_{ij} / \eht{\rho}) = -\left(\fht{R}_{in}\partial_n \fht{u}_j + \fht{R}_{jn}\partial_n \fht{u}_i \right) -\left( \nabla_r \eht{\rho}\fht{u''_i u''_j u''_r} + \eht{G_{rr}^R} \right) + \eht{u''_i \rho D_t u_j} + \eht{u''_j \rho D_t u_i} 
\end{align}

\subsubsection{Mean equation for $\fht{R}_{rr}$}

\begin{align}
\eht{\rho}\fht{D}_t \big( \fht{R}_{rr} / \eht{\rho} \big) = -\left(\fht{R}_{rn}\partial_n \fht{u}_r + \fht{R}_{rn}\partial_n \fht{u}_r \right) -\left( \nabla_r 2 f_k^r + \eht{G_{rr}^R} \right) + 2  \eht{u''_r \rho D_t u_r} 
\end{align}

\noindent
where

\begin{align}
\eht{u''_r \rho D_t u_r} = & \eht{u''_r \left( \frac{1}{r^{2}} \partial_{r} \big( r^{2} [\tau_{rr}]\big) + \frac{1}{r\sin{\theta}}\partial_{\theta}(\sin{\theta}[\tau_{r\theta}]\big) + \frac{1}{r\sin{\theta}}\partial_{\phi}\big([\tau_{r\phi}]\big) - G_r^M - \partial_{r} P + \rho g_r \right) } \\
\eht{u''_r \rho D_t u_r} = & \eht{u''_r \left( \nabla_r \tau_{rr} + \nabla_\theta \tau_{r\theta} + \nabla_\phi \tau_{r\phi}  - G_r^M - \partial_{r} P + \rho g_r \right) }
\end{align}

\noindent
Some terms can be further manipulated in following way:

\begin{align}
+\eht{u''_r  \nabla_r \tau_{rr}} = & \ \eht{u''_r} \nabla_r \eht{\tau_{rr}} + \eht{u''_r \nabla_r \tau'_{rr}} =  \cancelto{0}{\eht{u''_r} \nabla_r \eht{\tau_{rr}}} + \nabla_r (\eht{u''_r \tau'_{rr}}) - \eht{\tau'_{rr}\partial_r u''_r} \\
+\eht{u''_r  \nabla_\theta \tau_{r\theta}} = & \ \cancelto{0}{\eht{u''_r} \nabla_\theta \eht{\tau_{r\theta}}} + \eht{u''_r \nabla_\theta \tau'_{r\theta}} = \cancelto{0}{\eht{\nabla_\theta (u''_r \tau'_{r\theta})}} - \eht{\tau'_{r\theta}\frac{1}{r}\partial_\theta u''_r }  & \\
+\eht{u''_r  \nabla_\phi \tau_{r\phi}} = \ & \cancelto{0}{\eht{u''_r} \nabla_\phi \eht{\tau_{r\phi}}} + \eht{u''_r \nabla_\phi \tau'_{r\phi}} = \cancelto{0}{\eht{\nabla_\phi (u''_r \tau'_{r\phi})}} - \eht{\tau'_{r\phi}\frac{1}{r\sin{\theta}}\partial_\phi u''_r } & \\
-\eht{u''_r G_r^M} = & \ -\eht{u''_r G_r^M} & \\
-\eht{u''_r \partial_r P} = & \ -\eht{u''_r} \partial_r \eht{P} - \eht{u''_r \partial_r P'} = -\eht{u''_r} \partial_r \eht{P} - \nabla_r \eht{(u''_r P')} + \eht{P'\nabla_r u''_r}  & \\ 
+\eht{u''_r \rho g_r} \sim & \ \eht{u''_r \rho} \eht{g}_r = 0 &
\end{align}

\noindent 
Final equation is:

\begin{align}
\eht{\rho}\fht{D}_t \left( \fht{R}_{rr} / \eht{\rho} \right) = & \ -2\fht{R}_{rr}\partial_r \fht{u}_r - \nabla_r \fht{F}_{rrr}^R - \eht{G_{rr}^R}  + 2\nabla_r (\eht{u''_r \tau'_{rr}}) -2\eht{u''_r G_r^M} - 2 \eht{u''_r} \partial_r \eht{P} - 2 \nabla_r \eht{(u''_r P')} + 2 \eht{P'\nabla_r u''_r} - \\
& \ -2 \big( \eht{\tau'_{rr}\partial_r u''_r} + \eht{\tau'_{r\theta}\frac{1}{r}\partial_\theta u''_r } + \eht{\tau'_{r\phi}\frac{1}{r\sin{\theta}}\partial_\phi u''_r }  \big) \nonumber \\
\eht{\rho}\fht{D}_t \left( \fht{R}_{rr} / \eht{\rho} \right) = & \ -\nabla_r ( 2 f_k^r + 2 f_P + 2 f_\tau^r ) + 2 W_b - 2\fht{R}_{rr}\partial_r \fht{u}_r + 2\eht{P'\nabla_r u''_r} - 2\overline{u''_r G^{M}_r} - \eht{G^{R}_{rr}} - 2\varepsilon_k^r
\end{align}

\subsubsection{Mean equation for $\fht{R}_{\theta \theta}$}

\begin{align}
\eht{\rho}\fht{D}_t \big( \fht{R}_{\theta \theta} / \eht{\rho} \big) = -\left(\fht{R}_{\theta n}\partial_n \fht{u}_\theta + \fht{R}_{\theta n}\partial_n \fht{u}_\theta \right) -\left( \nabla_r 2 f_k^\theta - \eht{G_{\theta \theta}^R} \right) + 2  \eht{u''_\theta \rho D_t u_\theta} \\
\end{align}

\noindent
where

\begin{align}
\eht{u''_\theta \rho D_t u_\theta} = & \ \eht{u''_\theta \left( \frac{1}{r^{2}} \partial_{r} \big(r^{2} [\tau_{\theta r}]\big) + \frac{1}{r\sin{\theta}}\partial_{\theta}\big(\sin{\theta}[\tau_{\theta \theta}]\big) + \frac{1}{r\sin{\theta}}\partial_{\phi}[\tau_{\theta \phi}]\big) - G_\theta^M - \frac{1}{r} \partial_{\theta} P + \rho g_\theta \right) } \\
\eht{u''_\theta \rho D_t u_\theta} = & \ \eht{u''_\theta \big( \nabla_r \tau_{\theta r} + \nabla_\theta \tau_{\theta \theta} + \nabla_\phi \tau_{\theta \phi} - G_\theta^M - \frac{1}{r} \partial_{\theta} P + \rho g_\theta \big) }
\end{align}

\noindent
Some terms can be further manipulated in following way:

\begin{align}
+\eht{u''_\theta \nabla_r \tau_{\theta r}} = & \ \cancelto{0}{\eht{u''_\theta}\nabla_r \eht{\tau_{\theta r}}} + \nabla_r (\eht{u''_\theta \tau'_{\theta r}}) - \eht{\tau'_{\theta r}\partial_r u''_\theta}\\
+\eht{u''_\theta \nabla_\theta \tau_{\theta \theta}} = & \ -\eht{\tau'_{\theta \theta} \frac{1}{r}\partial_\theta u''_\theta} \\
+\eht{u''_\theta \nabla_\phi \tau_{\theta \phi}} = & \ -\eht{\tau'_{\theta \phi} \frac{1}{r\sin{\theta}} \partial_\phi u''_\theta} \\
-\eht{u''_\theta G_\theta^M} = & \ -\eht{u''_\theta G_\theta^M} \\
-\eht{u''_\theta \frac{1}{r}\partial_\theta P } = & \ -\eht{u''_\theta}\frac{1}{r}\partial_\theta \eht{P} - \eht{u''_\theta \frac{1}{r}\partial_\theta P'} = -\cancelto{0}{\eht{u''_\theta}\frac{1}{r}\partial_\theta \eht{P}} - \cancelto{0}{\eht{\frac{1}{r\sin{\theta}}\partial_\theta (\sin{\theta} u''_\theta P')}} + \eht{P'\frac{1}{r\sin{\theta}}\partial_\theta (\sin{\theta}u''_\theta}) \\
+\eht{u''_\theta \rho g_\theta} = 0 & \ 
\end{align}

\noindent
Final equation is:

\begin{align}
\eht{\rho}\fht{D}_t \big( \fht{R}_{\theta \theta} / \eht{\rho} \big) = & -2\fht{R}_{\theta r}\partial_r \fht{u}_\theta -\nabla_r 2 f_k^\theta - \eht{G_{\theta \theta}^R} + 2\nabla_r (\eht{u''_\theta \tau'_{\theta r}}) - 2\eht{u''_\theta G_\theta^M} + 2\eht{P'\nabla_\theta u''_\theta} - \nonumber \\ 
& - 2 \left(\eht{\tau'_{\theta r}\partial_r u''_\theta} - \eht{\tau'_{\theta \theta} \frac{1}{r}\partial_\theta u''_\theta} - \eht{\tau'_{\theta \phi} \frac{1}{r\sin{\theta}} \partial_\phi u''_\theta} \right) \\
\eht{\rho}\fht{D}_t \left( \fht{R}_{\theta \theta} / \eht{\rho} \right) = & -\nabla_r ( 2 f_k^\theta + 2 f_\tau^\theta ) - 2\fht{R}_{\theta r}\partial_r \fht{u}_\theta +2\eht{P' \nabla_\theta u''_\theta}  - 2\overline{u''_\theta G^{M}_\theta} - \eht{G^{R}_{\theta \theta}} - 2 \varepsilon_k^\theta 
\end{align}

\subsubsection{Mean equation for $\fht{R}_{\phi \phi}$}

\begin{align}
\eht{\rho}\fht{D}_t \big( \fht{R}_{\phi \phi} / \eht{\rho} \big) = -\left(\fht{R}_{\phi n}\partial_n \fht{u}_\phi + \fht{R}_{\phi n}\partial_n \fht{u}_\phi \right) -\left( \nabla_r 2 f_k^\phi + \eht{G_{\phi \phi}^R} \right) + 2  \eht{u''_\phi \rho D_t u_\phi} 
\end{align}

\noindent
where

\begin{align}
\eht{u''_\phi \rho D_t u_\phi} = & \ \eht{u''_\phi \left( \frac{1}{r^{2}} \partial_{r} \big(r^{2} [\tau_{\phi r}]\big) + \frac{1}{r\sin{\phi}}\partial_{\phi}\big(\sin{\phi}[\tau_{\phi \theta}]\big) + \frac{1}{r\sin{\phi}}\partial_{\phi}[\tau_{\phi \phi}]\big) - G_\phi^M - \frac{1}{r\sin{\theta}} \partial_{\phi} P + \rho g_\phi \right) } \\
\eht{u''_\phi \rho D_t u_\phi} = & \ \eht{u''_\phi \big( \nabla_r \tau_{\phi r} + \nabla_\theta \tau_{\phi \theta} + \nabla_\phi \tau_{\phi \phi} - G_\phi^M - \frac{1}{r\sin{\theta}} \partial_{\phi} P + \rho g_\phi \big) }
\end{align}

\noindent
Some terms can be further manipulated in following way:

\begin{align}
+\eht{u''_\phi \nabla_r \tau_{\phi r}} = & \ \cancelto{0}{\eht{u''_\phi}\nabla_r \eht{\tau_{\phi r}}} + \nabla_r (\eht{u''_\phi \tau'_{\phi r}}) - \eht{\tau'_{\phi r}\partial_r u''_\phi}\\
+\eht{u''_\phi \nabla_\theta \tau_{\phi \theta}} = & \ -\eht{\tau'_{\phi \theta} \frac{1}{r}\partial_\theta u''_\phi} \\
+\eht{u''_\phi \nabla_\phi \tau_{\phi \phi}} = & \ -\eht{\tau'_{\phi \phi} \frac{1}{r\sin{\theta}} \partial_\phi u''_\phi} \\
-\eht{u''_\phi G_\phi^M} = & \ -\eht{u''_\phi G_\phi^M} \\
-\eht{u''_\phi \frac{1}{r\sin{\theta}}\partial_\phi P } = & \ -\eht{u''_\phi}\frac{1}{r\sin{\theta}}\partial_\phi \eht{P} - \eht{u''_\phi \frac{1}{r\sin{\theta}}\partial_\phi P'} = -\cancelto{0}{\eht{u''_\phi}\frac{1}{r\sin{\theta}}\partial_\phi \eht{P}} - \cancelto{0}{\eht{\frac{1}{r\sin{\theta}}\partial_\phi (u''_\phi P')}} + \eht{P'\frac{1}{r\sin{\theta}}\partial_\phi (u''_\phi}) \\
+\eht{u''_\phi \rho g_\phi} = 0 & \ 
\end{align}

\noindent
Final equation is:

\begin{align}
\eht{\rho}\fht{D}_t \big( \fht{R}_{\phi \phi} / \eht{\rho} \big) = & -2\fht{R}_{\phi r}\partial_r \fht{u}_\phi -\nabla_r 2 f_k^\phi - \eht{G_{\phi \phi}^R} + 2\nabla_r (\eht{u''_\phi \tau'_{\phi r}}) - 2\eht{u''_\phi G_\phi^M} + 2\eht{P'\nabla_\phi u''_\phi} - \nonumber \\ 
& - 2 \left(\eht{\tau'_{\phi r}\partial_r u''_\phi} - \eht{\tau'_{\phi \theta} \frac{1}{r}\partial_\theta u''_\phi} - \eht{\tau'_{\phi \phi} \frac{1}{r\sin{\theta}} \partial_\phi u''_\phi} \right) \\
\eht{\rho}\fht{D}_t \left( \fht{R}_{\phi \phi} / \eht{\rho} \right) = & -\nabla_r ( 2 f_k^\phi + 2 f_\tau^\phi) - 2\fht{R}_{\phi r}\partial_r \fht{u}_\phi +2\eht{P' \nabla_\phi u''_\phi}  - 2\overline{u''_\phi G^{M}_\phi} - \eht{G^{R}_{\phi \phi}} - 2\varepsilon_k^\phi 
\end{align}

\newpage

\subsection{Turbulent kinetic energy equations}

\begin{align}
\fht{k} = & +\frac{1}{2}\fht{R}_{ii} / \eht{\rho}  \\
\av{\rho} \fav{D}_t \fav{k}^{ } = & -\nabla_r ( f_k +  f_P ) - \fht{R}_{ir}\partial_r \fht{u}_i + W_b + W_P + {\mathcal N_k}  \label{eq:rans_tke_appendix} \\
\fht{k}^r = & +\frac{1}{2}\fht{R}_{rr} / \eht{\rho}  \\
\av{\rho} \fav{D}_t \fav{k}^r =  &  -\nabla_r  ( f_k^r + f_P )  - \fht{R}_{rr}\partial_r \fht{u}_r + W_b  + \eht{P'\nabla_r u''_r} + {\mathcal G_k^r} + {\mathcal N_{kr}} \label{eq:rans_ekin_r} \\
\fht{k}^h = & +\fht{k}_\theta + \fht{k}_\phi = +\frac{1}{2} \big( \fht{R}_{\theta \theta} + \fht{R}_{\phi \phi} \big) / \eht{\rho} \\
\av{\rho} \fav{D}_t \fav{k}^h =  &  -\nabla_r f_k^h - (\fht{R}_{\theta r}\partial_r \fht{u}_\theta + \fht{R}_{\phi r}\partial_r \fht{u}_\phi) + (\eht{P' \nabla_\theta u''_\theta} + \eht{P' \nabla_\phi u''_\phi}) + {\mathcal G_k^h} + {\mathcal N_{kh}} \label{eq:rans_ekin_h} 
\end{align}

\subsection{Turbulent mass flux equation}

The turbulent mass flux equation can be derived in the following way:

\fontsize{10pt}{20pt}

\begin{align}
\rho D_t \fht{c} - \rho \fht{D}_t \fht{c} = & \ \rho \partial_t \fht{c} + \rho u_n \partial_n \fht{c} - [\rho \partial_t \fht{c} + \rho \fht{u}_n \partial_n \fht{c}_n] = \rho (u_n - \fht{u}_n)\partial_n \fht{c} =  \rho u''_n \partial_n \fht{c} \\
\rho D_t c'' =  & \ \rho D_t c - \rho D_t \fht{c} = \rho D_t c -  \rho \fht{D}_t \fht{c} - \rho u''_n \partial_n \fht{c} = \rho D_t c -  \frac{\rho}{\eht{\rho}} [\eht{\rho}\fht{D}_t \fht{c}] - \rho u''_n \partial_n \fht{c} \\
\rho D_t u_r'' =  & +\left( \frac{1}{r^{2}} \frac{\partial}{\partial r} \big( r^{2} [\tau_{rr}]\big) + \frac{1}{r\sin{\theta}}\frac{\partial}{\partial \theta}(\sin{\theta}[\tau_{r\theta}]\big) + \frac{1}{r\sin{\theta}}\frac{\partial}{\partial \phi}\big([\tau_{r\phi}]\big) - G_r^M - \frac{\partial P}{\partial r} \right) + \rho g_r + \\ 
 & + \frac{\rho}{\eht{\rho}}\left( \frac{1}{r^2}\dr r^2 (\fht{R}_{rr}-\eht{\tau_{rr}}) + \eht{G_r^M} + \dr \eht{P} - \eht{\rho}\fht{g}_r \right) - \rho u''_n \partial_n \fht{u_r}  & \
\end{align}

\begin{align}
\fht{D}_t \eht{u''_i} = & \ \eht{D_t u''_i - u''_n\partial_n u''_i} = \eht{\partial_t u''_i}  + \eht{u_n \partial_n u''_i} - \eht{u''_n \partial_n u''_i} = \eht{\partial_t u''_i}  + \eht{u_n \partial_n u''_i} + \eht{\fht{u_n} \partial_n u''_i} - \eht{u_n \partial_n u''_i} = \eht{\partial_t u''_i} + \eht{\fht{u_n} \partial_n u''_i} = \fht{D}_t \eht{u''_i} \\
\eht{\rho}\fht{D_t}\eht{u''_i} = & \ \eht{\frac{\eht{\rho}}{\rho}[\rho D_t u''_i]} - \eht{\eht{\rho}u''_n\partial_n u''_i} 
\end{align}

\begin{align}
\eht{\rho}\fht{D_t}\eht{u''_r} = & \ \eht{\frac{\eht{\rho}}{\rho}[\rho D_t u''_r]} - \eht{\eht{\rho}u''_n\partial_n u''_r} = \nonumber \\ 
& = \eht{\frac{\eht{\rho}}{\rho} \left[ \frac{1}{r^{2}} \frac{\partial}{\partial r} \big( r^{2} \tau_{rr} \big) - G_r^M - \frac{\partial}{\partial r}P + \rho g_r + \frac{\rho}{\eht{\rho}}\left(\frac{1}{r^2}\dr r^2 (\fht{R}_{rr}-\eht{\tau_{rr}}) + \eht{G_r^M} + \dr \eht{P} - \eht{\rho}\fht{g_r} \right) - \rho u''_n \partial_n \fht{u_r} \right]} - \eht{\eht{\rho}u''_n\partial_n u''_r} = \\
& = \eht{\frac{\eht{\rho}}{\rho} \left[ \frac{1}{r^{2}} \frac{\partial}{\partial r} \big( r^{2} \tau_{rr} \big) \right] -  \left[ \frac{1}{r^{2}} \frac{\partial}{\partial r} \big( r^{2} \tau_{rr} \big) \right]} -\eht{\frac{\eht{\rho}}{\rho}G_r^M + \eht{G_r^M}} - \eht{\frac{\eht{\rho}}{\rho} \left[\frac{\partial}{\partial r}P \right] +  \left[\frac{\partial}{\partial r}P \right]} + \frac{1}{r^2}\dr r^2 (\fht{R}_{rr}) - \eht{\eht{\rho} u''_n \partial_n \fht{u_r}} - \eht{\eht{\rho}u''_n\partial_n u''_r} \\
& = \eht{\left[ \frac{\eht{\rho}}{\rho} - 1 \right] \frac{1}{r^{2}} \frac{\partial}{\partial r} \big( r^{2} \tau_{rr} \big)} - \eht{\left[\frac{\eht{\rho}}{\rho} - 1 \right] G_r^M} - \eht{\left[\frac{\eht{\rho}}{\rho} - 1 \right]\frac{\partial}{\partial r} P} + \frac{1}{r^2}\dr r^2 (\fht{R}_{rr}) - \eht{\rho}\eht{u''_n\partial_n u_r } = \\
& = +\frac{1}{r^2}\dr r^2 (\fht{R}_{rr}) - \eht{\rho}\eht{u''_n\partial_n u_r}  - \eht{\frac{\rho'}{\rho} \frac{1}{r^{2}} \frac{\partial}{\partial r} \big( r^{2} \tau_{rr} \big)} + \eht{\frac{\rho'}{\rho} G_r^M} + \eht{\frac{\rho'}{\rho}\frac{\partial}{\partial r} P} = \\
& = +\frac{1}{r^2}\dr r^2 (\fht{R}_{rr}) - \eht{\rho}\eht{u''_n\partial_n u_r}  - \eht{\rho' v \frac{1}{r^{2}} \frac{\partial}{\partial r} \big( r^{2} (\eht{\tau_{rr}} + \tau'_{rr} \big)} + \eht{\rho' v \frac{\partial}{\partial r} (\eht{P} + P')} + \eht{\rho' v G_r^M} = \\
& = +\frac{1}{r^2}\dr r^2 (\fht{R}_{rr}) - \eht{\rho}\eht{u''_n\partial_n u_r} - \eht{\rho' v} \left(\frac{1}{r^{2}} \frac{\partial}{\partial r} \big( r^{2} \eht{\tau_{rr}} \big) - \frac{\partial}{\partial r} \eht{P} \right) - \eht{\rho' v \left(\frac{1}{r^{2}} \frac{\partial}{\partial r} \big( r^{2} \tau'_{rr} \big) - \frac{\partial}{\partial r} P' \right) }  + \eht{\rho' v G_r^M} = \\
& = +\frac{1}{r^2}\dr r^2 (\fht{R}_{rr}) - \eht{\rho}\eht{u''_n\partial_n u_r} + b\left(\frac{1}{r^{2}} \frac{\partial}{\partial r} \big( r^{2} \eht{\tau_{rr}} \big) - \frac{\partial}{\partial r} \eht{P} \right) - \eht{\rho' v \left(\frac{1}{r^{2}} \frac{\partial}{\partial r} \big( r^{2} \tau'_{rr} \big) - \frac{\partial}{\partial r} P' \right) }  + \eht{\rho' v G_r^M} = \\
& = +\nabla_r ( \fht{R}_{rr} ) - \eht{\rho}\eht{{\bf{u''}} \cdot \nabla u_r} -b \nabla_r \eht{\tau}_{rr} - b\partial_r \eht{P} + \eht{\rho' v \partial_r P'} - \eht{\rho' v \nabla_r  \tau'_{rr} }  + \eht{\rho' v G_r^M} 
\end{align}

\begin{align}
{\color{red} \eht{\rho}\fht{D}_t \eht{u''_r} }& =  +\nabla_r ( \fht{R}_{rr} ) - \eht{\rho}\eht{u''_n \partial_n u_r} -b\nabla_r \eht{\tau}_{rr}- b\partial_r \eht{P} + \eht{\rho' v \partial_r P'} - \eht{\rho' v \nabla_r  \tau'_{rr} }  + \eht{\rho' v G_r^M} = \\
& =  +\nabla_r ( \fht{R}_{rr} ) - \eht{\rho}\eht{u''_n}\partial_n \eht{u}_r - \eht{\rho}\partial_n \eht{u''_n u'_r} + \eht{\rho} \eht{u'_r\partial_n u''_n} - b\partial_r \eht{P} + \eht{\rho' v \partial_r P'} - \eht{\rho' v \nabla_r  \tau'_{rr} }  + \eht{\rho' v G_r^M} = \\
& = (+\nabla_r ( \fht{R}_{rr} ) - \eht{\rho}\nabla_r \eht{u''_r u'_r}) - \eht{\rho}\eht{u''_r} \nabla_r \eht{u}_r + \eht{\rho} \eht{u'_r d''} - b\partial_r \eht{P} + \eht{\rho' v \partial_r P'} - \eht{\rho' v \nabla_r  \tau'_{rr} }  + \eht{\rho' v G_r^M} = \\
& {\color{red} = -(\eht{\rho'u'_ru'_r}/\eht{\rho})\partial_r\eht{\rho} + (\fht{R}_{rr}/\eht{\rho})/\partial_r \eht{\rho} - \eht{\rho} \nabla_r (\eht{u''_r} \ \eht{u''_r}) + \nabla_r \overline{\rho' u'_r u'_r} - \eht{\rho}\eht{u''_r} \nabla_r \eht{u}_r + \eht{\rho} \eht{u'_r d''} - b\partial_r \eht{P} + \eht{\rho' v \partial_r P'} - \eht{\rho' v \nabla_r  \tau'_{rr} }  + \eht{\rho' v G_r^M}} 
\end{align}

\newpage

\fontsize{12pt}{20pt}

\subsection{Density-specific volume covariance equation}

The density-specific volume ($v = 1/\rho$) covariance ($b = -\eht{\rho' v'}$) equation can be derived from the continuity equation in the following way. 

\fontsize{9pt}{20pt}

\begin{align}
\partial_t \rho + \partial_n (\rho u_n) = & \ 0 \\
\partial_t \eht{\rho} + \fht{u}_n \partial_n \eht{\rho} = & \ -\eht{\rho} \partial_n \fht{u}_n \\
\partial_t \eht{\rho} + \eht{u}_n \partial_n \eht{\rho} - \eht{u''_n} \partial_n \eht{\rho} = & \ - \eht{\rho}\partial_n \eht{u}_n + \eht{\rho} \partial_n \eht{u''_n} \\
\eht{D}_t \eht{\rho} = & \ +\overline{u''_n} \partial_n \eht{\rho} - \eht{\rho} \partial_n \eht{u}_n + \eht{\rho} \partial_n \eht{u''_n} \\
\eht{D}_t \eht{\rho} = & \ +( \overline{u''_n} \partial_n \eht{\rho} +  \eht{\rho} \partial_n \eht{u''_n} ) - \eht{\rho} \partial_n \eht{u}_n \\
\eht{D}_t \eht{\rho} = & \ - \eht{\rho} \partial_n \eht{u}_n + \partial_n ( \overline{u''_n} \eht{\rho} )
\end{align}

\begin{align}
\partial_t \rho + \partial_n (\rho u_n) = & \ 0 \\
\partial_t (1/v) + \partial_n (u_n / v) = & \ 0 \\
-\partial_t v + v\partial_n u_n - u_n\partial_n v = & \ 0 \\
\partial_t v - v\partial_n u_n - u_n\partial_n v + u_n\partial_n v + u_n\partial_n v = & \ 0 \\
\partial_t v - \partial_n (v u_n) + 2 u_n \partial_n v = & \ 0 \\
\partial_t \eht{v} - \partial_n (\eht{v u_n}) + 2 \eht{u_n \partial_n v} = & \ 0 \\
\eht{D}_t \eht{v} = & +\eht{v}\partial_n \eht{u}_n - \partial_n \eht{u'_n v'} + 2 \eht{v'\partial_n u'_n}  
\end{align}

\begin{align}
b = -\eht{\rho'v'} = \eht{\rho}\eht{v} - 1 \\
\eht{D}_t b = \eht{\rho}\eht{D}_t \eht{v} + \eht{v} \ \eht{D}_t \eht{\rho} 
\end{align}

Using previously derived equations for $\eht{D}_t \eht{v}$ and $\eht{D}_t \eht{\rho}$ we get:

\begin{align}
\eht{D}_t b = -\eht{\rho}\partial_n \eht{\rho' u'_n} + 2 \eht{\rho}\eht{v'\partial_n u'_n} + \eht{v}\partial_n \eht{\rho} \eht{u''_n}
\end{align}

\subsection{Mean internal energy flux equation}

We can derive the internal energy flux equation using the general formula for second order moments, where we substitute $c$ with  $\epsilon_I$ and $d$ with $u_i$.

\begin{align}
\overline{\rho}\widetilde{D}_t \widetilde{c''d''} = & \overline{c'' \rho D_t d} - \overline{\rho} \widetilde{c''u''_n}\partial_n \widetilde{d} + \overline{d'' \rho D_t c} - \overline{\rho} \widetilde{d''u''_n}\partial_n \widetilde{c} - \overline{\partial_n \rho c''d''u''_n} \\
\erho \fav{D}_t (f_I / \eht{\rho}) = &  {\mathcal N_{fI}} -\nabla_r f_I^r  - f_I \partial_r \fht{u}_r  - \fht{R}_{rr} \partial_r \fht{\epsilon_I} - \eht{\epsilon''_I} \partial_r \eht{P} - \eht{\epsilon''_I \partial_r P'}  - \eht{u''_r \left( P d \right)}  + \overline{u''_r ({\mathcal S} + \nabla \cdot F_T)} + {\mathcal G_I} + {\mathcal N_{fI}}\label{eq:rans_fi}
\end{align}

\subsection{Mean enthalpy flux equation}

We can derive the enthalpy flux equation using the general formula for second order moments, where we substitute $c$ with  $h$ and $d$ with $u_i$.

\begin{align}
\overline{\rho}\widetilde{D}_t \widetilde{c''d''} = & \overline{c'' \rho D_t d} - \overline{\rho} \widetilde{c''u''_n}\partial_n \widetilde{d} + \overline{d'' \rho D_t c} - \overline{\rho} \widetilde{d''u''_n}\partial_n \widetilde{c} - \overline{\partial_n \rho c''d''u''_n} \\
\erho \fav{D}_t (f_h / \eht{\rho}) = &  -\nabla_r f_h^r - f_h \partial_r \fht{u}_r - \fht{R}_{rr} \partial_r \fht{h} -\eht{h''}\partial_r \eht{P} - \eht{h''\partial_r P'} - \Gamma_1\eht{u''_r \left( P d \right) } + \Gamma_3 \overline{u''_r ({\mathcal S} + \nabla \cdot F_T)} + {\mathcal G_h} + {\mathcal N_{h \ }} \label{eq:rans_fh} 
\end{align}

\subsection{Mean entropy flux equation}

We can derive the entropy flux equation using the general formula for second order moments, where we substitute $c$ with $s$ and $d$ with $u_i$.

\begin{align}
\overline{\rho}\widetilde{D}_t \widetilde{c''d''} = & \overline{c'' \rho D_t d} - \overline{\rho} \widetilde{c''u''_n}\partial_n \widetilde{d} + \overline{d'' \rho D_t c} - \overline{\rho} \widetilde{d''u''_n}\partial_n \widetilde{c} - \overline{\partial_n \rho c''d''u''_n} \\
\erho \fav{D}_t (f_s / \eht{\rho}) = &  -\nabla_r f_s^r - f_s \partial_r \fht{u}_r - \fht{R}_{rr} \partial_r \fht{s} -\eht{s''}\partial_r \eht{P} - \eht{s''\partial_r P'} + \eht{u''_r ( {\mathcal S} + \nabla \cdot F_T)  / T} + {\mathcal G_s} + {\mathcal N_{fs}}  \label{eq:rans_fs}
\end{align}

\subsection{Mean composition flux equation}

We can derive the composition flux equation using the general formula for second order moments, where we substitute $c$ with $X_\alpha$ and $d$ with $u_i$.

\begin{align}
\overline{\rho}\widetilde{D}_t \widetilde{c''d''} = & \overline{c'' \rho D_t d} - \overline{\rho} \widetilde{c''u''_n}\partial_n \widetilde{d} + \overline{d'' \rho D_t c} - \overline{\rho} \widetilde{d''u''_n}\partial_n \widetilde{c} - \overline{\partial_n \rho c''d''u''_n} \\
\erho \fav{D}_t (f_\alpha / \eht{\rho}) = &  -\nabla_r f_\alpha^r  - f_\alpha \partial_r \fht{u}_r - \fht{R}_{rr} \partial_r \fht{X}_\alpha -\eht{X''_\alpha} \partial_r \eht{P} - \eht{X''_\alpha \partial_r P'} + \overline{u''_r \rho \dot{X}_\alpha^{\rm nuc}} + {\mathcal G_\alpha} + {\mathcal N_{f\alpha}} \label{eq:rans_falpha} \\
\end{align}

\subsection{Mean A and Z flux equations}

We can derive the composition flux equation using the general formula for second order moments, where we substitute $c$ with $A$ or $Z$ and $d$ with $u_i$.

\begin{align}
\overline{\rho}\widetilde{D}_t \widetilde{c''d''} = & \overline{c'' \rho D_t d} - \overline{\rho} \widetilde{c''u''_n}\partial_n \widetilde{d} + \overline{d'' \rho D_t c} - \overline{\rho} \widetilde{d''u''_n}\partial_n \widetilde{c} - \overline{\partial_n \rho c''d''u''_n} \\
\erho \fav{D}_t (f_A / \eht{\rho}) = &  \ {\mathcal N_{fA}} -\nabla_r f_A^r - f_A \partial_r \fht{u}_r - \fht{R}_{rr} \partial_r \fht{A} -\eht{A''} \partial_r \eht{P} - \eht{A'' \partial_r P'} - \overline{u''_r \rho A^2\Sigma_\alpha \dot{X}_\alpha^{\rm nuc} / A_\alpha} + {\mathcal G_A}                 \label{eq:rans_fabar} \\
\erho \fav{D}_t (f_Z / \eht{\rho}) = &  \ {\mathcal N_{fZ}} -\nabla_r f_Z^r  - f_Z \partial_r \fht{u}_r - \fht{R}_{rr} \partial_r \fht{Z} -\eht{Z''} \partial_r \eht{P} - \eht{Z'' \partial_r P'} - \overline{u''_r \rho Z A \Sigma_\alpha (\dot{X}_\alpha^{\rm nuc}/ A_\alpha)} - \nonumber \\
& - \overline{u''_r \rho A \Sigma_\alpha (Z_\alpha \dot{X}_\alpha^{\rm nuc} / A_\alpha)}  + {\mathcal G_Z}   \label{eq:rans_fzbar} 
\end{align}

\subsection{Mean angular momentum flux equation}

We can derive the angular momentum flux equation using the general formula for second order moments, where we substitute $c$ with $j_z$ and $d$ with $u_i$.

\begin{align}
\overline{\rho}\widetilde{D}_t \widetilde{c''d''} = & \overline{c'' \rho D_t d} - \overline{\rho} \widetilde{c''u''_n}\partial_n \widetilde{d} + \overline{d'' \rho D_t c} - \overline{\rho} \widetilde{d''u''_n}\partial_n \widetilde{c} - \overline{\partial_n \rho c''d''u''_n} \\
\erho \fav{D}_t (f_{jz} / \rho) = & -\nabla_r f_{jz}^r  - f_{jz} \partial_r \fht{u}_r - \fht{R}_{rr} \partial_r \fht{j_z} -\eht{j''_z} \partial_r \eht{P} - \eht{j''_z \partial_r P'} + {\mathcal G_{jz}} + {\mathcal N_{jz}} \label{eq:rans_fjz}
\end{align}

\section{Derivation of Reynolds and Favrian variance equations}

The variance evolution equations can be found by using the general formula for second-order moments and similar algebraic manipulation presented in previous sections.

\newpage

\section{Divergence of tensors in spherical geometry up to third order}

\noindent
{\bf{BACKGROUND READING:}} \\

\vspace{0.3cm}

\noindent
CONTINUUM MECHANICS (Lecture Notes) \\
Zdenek Martinec, Department of Geophysics, Faculty of Mathematics and Physics, Charles University in Prague 

\fontsize{9pt}{20pt}

\begin{align}
\nabla (.) = \sum_n \frac{{\bf{e_n}}}{h_n}\frac{\partial (.)}{\partial x_n} \ \ \ \ \mbox{: nabla operator} & & {\bf{V}} = &  \sum_i V_i {\bf{{e_i}}} & & \mbox{: tensor of first order (vector)}\\
& & {\bf{S}} = &  \sum_{ij} S_{ij} ({\bf{e_i}} \otimes {\bf{e_j}}) & & {\mbox{: tensor of second order}}\\
& & {\bf{T}} = &  \sum_{ijk} T_{ijk} ({\bf{e_i}} \otimes {\bf{e_j}} \otimes {\bf{e_k}}) & & \mbox{: tensor of third order}
\end{align}

\begin{align}
\nabla \cdot {\bf{V}} = & \sum_i \frac{1}{h_i} \left[ \frac{\partial V_i}{\partial x_i} + \sum_m \Gamma_{mi}^i V_m \right] & & \mbox{: div of first order tensor (vector)} \\
\nabla \cdot {\bf{S}} = & \sum_{ij} \frac{1}{h_i} \left[ \frac{\partial S_{ij}}{\partial x_i} + \sum_m \Gamma_{mi}^i S_{mj} + \sum_m \Gamma_{mi}^j S_{im} \right] {\bf{e_j}} & & \mbox{: div of second order tensor} \\
\nabla \cdot {\bf{T}} = & \sum_{ijk} \frac{1}{h_i} \left[ \frac{\partial T_{ijk}}{\partial x_i} + \sum_m \Gamma_{mi}^i T_{mjk} + \sum_m \Gamma_{mi}^j T_{imk} + \sum_m \Gamma_{mi}^k T_{ijm} \right]  ({\bf{e_j}} \otimes {\bf{e_k}}) & & \mbox{: div of third order tensor}
\end{align}

\begin{align}
& x_1 = r & & x_2 = \theta & & x_3 = \phi & & \mbox{(coordinates)} \\
& {\bf{e_1 = e_r}} & &  {\bf{e_2 = e_\theta}} & & {\bf{e_3 = e_\phi}} && \mbox{(unit base vectors)} \\
& h_1 = h_r = 1 & &  h_2 = h_\theta = r & & h_3 = h_\phi = r \sin{\theta} & & \mbox{(scale factors)}
\end{align}

\begin{align}
\begin{pmatrix}
\Gamma_{r\theta}^\theta = \ 1  &  \Gamma_{r\phi}^\phi = \ \sin{\theta} & \Gamma_{\theta \phi}^{\phi} = \ \cos{\theta}   \\
\Gamma_{\theta \theta}^r = -1   &  \Gamma_{\phi \phi}^r = -\sin{\theta}  & \Gamma_{\phi \phi}^{\theta} = -\cos{\theta}  
\end{pmatrix} &&  \mbox{Christoffel symbols}
\end{align}

\newpage

\noindent
{{\bf{Divergence of first order tensor}} $\nabla \cdot {\bf{V}}$}

\begin{align}
\frac{1}{r^2} \frac{\partial (r^2 V_r) }{\partial_r} + \frac{1}{r \sin{\theta}}\frac{\partial}{\partial \theta}(V_\theta \sin{\theta}) + \frac{1}{r \sin{\theta}} \frac{\partial V_\phi}{\partial \phi}
\end{align}

\noindent
{{\bf{Divergence of second order tensor}} $\nabla \cdot {\bf{S}}$}

\begin{align}
S_r ({\bf{e_r}}): & \hspace{1cm} \frac{1}{r^2}\dr (r^2 S_{rr}) + \frac{1}{r\sin{\theta}}\frac{\partial}{\partial \theta} (\sin{\theta} \ S_{\theta r}) + \frac{1}{r\sin{\theta}}\frac{\partial S_{\phi r}}{\partial \phi} - \frac{S_{\theta\theta}}{r} - \frac{S_{\phi\phi}}{r} \\
S_\theta ({\bf{e_\theta}}): & \hspace{1cm} \frac{1}{r^2}\dr (r^2 S_{r \theta}) + \frac{1}{r\sin{\theta}}\frac{\partial}{\partial \theta} (\sin{\theta} \ S_{\theta\theta}) + \frac{1}{r\sin{\theta}}\frac{\partial S_{\phi \theta}}{\partial \phi} + \frac{S_{\theta r}}{r} - \frac{S_{\phi\phi} \cos{\theta}}{r\sin{\theta}} \\
S_\phi ({\bf{e_\phi}}): & \hspace{1cm} \frac{1}{r^2}\dr (r^2 S_{r \phi}) + \frac{1}{r\sin{\theta}}\frac{\partial}{\partial \theta} (\sin{\theta} \ S_{\theta \phi}) + \frac{1}{r\sin{\theta}}\frac{\partial S_{\phi\phi}}{\partial \phi} + \frac{S_{\phi r}}{r} + \frac{S_{\phi \theta} \cos{\theta}}{r\sin{\theta}}
\end{align}

\noindent
{{\bf{Divergence of third order tensor}} $\nabla \cdot {\bf{T}}$}

\begin{align}
T_{rr} \ ({\bf{e_r \otimes e_r}}): & \hspace{1cm} \frac{1}{r^2}\dr (r^2 T_{rrr}) + \frac{1}{r\sin{\theta}}\frac{\partial}{\partial \theta} (\sin{\theta} \ T_{\theta rr}) + \frac{1}{r\sin{\theta}}\frac{\partial T_{\phi rr}}{\partial \phi} - \frac{T_{\theta\theta r}}{r} - \frac{T_{\theta r \theta}}{r} -\frac{T_{\phi\phi r}}{r} - \frac{T_{\phi r \phi}}{r} \\
T_{r\theta} \ ({\bf{e_r \otimes e_\theta}}): & \hspace{1cm} \frac{1}{r^2}\dr (r^2 T_{rr \theta}) + \frac{1}{r\sin{\theta}}\frac{\partial}{\partial \theta} (\sin{\theta} \ T_{\theta r \theta}) + \frac{1}{r\sin{\theta}}\frac{\partial T_{\phi r \theta}}{\partial \phi} - \frac{T_{\theta \theta \theta}}{r} + \frac{T_{\theta rr}}{r} - \frac{T_{\phi \phi \theta}}{r} - \frac{T_{\phi r \phi} \cos{\theta}}{r\sin{\theta}} \\
T_{r \phi} \ ({\bf{e_r \otimes e_\phi}}): & \hspace{1cm} \frac{1}{r^2}\dr (r^2 T_{rr \phi}) + \frac{1}{r\sin{\theta}}\frac{\partial}{\partial \theta} (\sin{\theta} \ T_{\theta r \phi}) + \frac{1}{r\sin{\theta}}\frac{\partial T_{\phi r \phi}}{\partial \phi} + \frac{T_{\theta \theta \phi}}{r} - \frac{T_{\phi \phi \phi}}{r} + \frac{T_{\phi r \phi} \cos{\theta}}{r\sin{\theta}} \\
T_{\theta r} \ ({\bf{e_\theta \otimes e_r}}): & \hspace{1cm} \frac{1}{r^2}\dr (r^2 T_{r \theta r}) + \frac{1}{r\sin{\theta}}\frac{\partial}{\partial \theta} (\sin{\theta} \ T_{\theta \theta r}) + \frac{1}{r\sin{\theta}}\frac{\partial T_{\phi \theta r}}{\partial \phi} + \frac{T_{\theta rr}}{r} - \frac{T_{\theta \theta \theta}}{r} -\frac{T_{\phi \phi r} \cos{\theta}}{r \sin{\theta}} - \frac{T_{\phi \theta \phi}}{r}  \\
T_{\theta \theta} \ ({\bf{e_\theta \otimes e_\theta}}): & \hspace{1cm} \frac{1}{r^2}\dr (r^2 T_{r \theta \theta}) + \frac{1}{r\sin{\theta}}\frac{\partial}{\partial \theta} (\sin{\theta} \ T_{\theta \theta \theta}) + \frac{1}{r\sin{\theta}}\frac{\partial T_{\phi \theta \theta}}{\partial \phi} + \frac{T_{\theta r \theta}}{r} + \frac{T_{\theta \theta r}}{r} - \frac{T_{\phi \phi \theta} \cos{\theta}}{r\sin{\theta}} - \frac{T_{\phi \theta \phi} \cos{\theta}}{r\sin{\theta}} \\
T_{\theta \phi} \ ({\bf{e_\theta \otimes e_\phi}}): & \hspace{1cm} \frac{1}{r^2}\dr (r^2 T_{r \theta \phi}) + \frac{1}{r\sin{\theta}}\frac{\partial}{\partial \theta} (\sin{\theta} \ T_{\theta \theta \phi}) + \frac{1}{r\sin{\theta}}\frac{\partial T_{\phi \theta \phi}}{\partial \phi} + \frac{T_{\theta r \phi}}{r} + \frac{T_{\phi \theta r}}{r} + \frac{T_{\phi \theta \theta} \cos{\theta}}{r\sin{\theta}} \\
T_{\phi r} \ ({\bf{e_\phi \otimes e_r}}): & \hspace{1cm} \frac{1}{r^2}\dr (r^2 T_{r \phi r}) + \frac{1}{r\sin{\theta}}\frac{\partial}{\partial \theta} (\sin{\theta} \ T_{\theta \phi r}) + \frac{1}{r\sin{\theta}}\frac{\partial T_{\phi \phi r}}{\partial \phi} - \frac{T_{\theta \phi \theta}}{r} + \frac{T_{\phi rr}}{r} + \frac{T_{\phi \theta r} \cos{\theta}}{r\sin{\theta}} - \frac{T_{\phi \phi \phi}}{r} \\
T_{\phi \theta} \ ({\bf{e_\phi \otimes e_\theta}}): & \hspace{1cm} \frac{1}{r^2}\dr (r^2 T_{r \phi \theta}) + \frac{1}{r\sin{\theta}}\frac{\partial}{\partial \theta} (\sin{\theta} \ T_{\theta \phi \theta}) + \frac{1}{r\sin{\theta}}\frac{\partial T_{\phi \phi \theta}}{\partial \phi} + \frac{T_{\theta \phi r}}{r} + \frac{T_{\phi r \theta}}{r} + \frac{T_{\phi \theta \theta} \cos{\theta}}{r\sin{\theta}} - \frac{T_{\phi \theta \phi} \cos{\theta}}{r\sin{\theta}}  \\
T_{\phi \phi} \ ({\bf{e_\phi \otimes e_\phi}}): & \hspace{1cm} \frac{1}{r^2}\dr (r^2 T_{r \phi \phi}) + \frac{1}{r\sin{\theta}}\frac{\partial}{\partial \theta} (\sin{\theta} \ T_{\theta \phi \phi}) + \frac{1}{r\sin{\theta}}\frac{\partial T_{\phi \phi \phi}}{\partial \phi} + \frac{T_{\phi r \phi}}{r} + \frac{T_{\phi \theta \phi} \cos{\theta}}{r\sin{\theta}} + \frac{T_{\phi \phi r}}{r} + \frac{T_{\phi \phi \theta} \cos{\theta}}{r\sin{\theta}}
\end{align}

\section{Periodic boundary conditions and properties of divergence angular components}\label{sec:periodic-bc}

All of the simulation data studied in this document use domains that are wedges in a spherical coordinate system and have period boundary conditions in the $\theta$ and $\phi$ directions. Therefore,  the angular components of divergence terms vanish upon averaging.

\begin{align}
\theta \in & [\theta_L,\theta_R] \\ 
\phi \in & [\phi_L,\phi_r ] 
\end{align}

\begin{align} 
\langle \nabla \cdot {\bf q} \ (r,t) \rangle = \frac{1}{\Delta\Omega}\int \nabla \cdot {\bf q} \ (r,\theta,\phi,t')~d\Omega & \ \ \ \ \ \ {\bf q} = q(q_r,q_\theta,q_\phi) 
\end{align}

\noindent
Proof:

\begin{align}
\langle \nabla_\theta q_\theta \rangle = & \int_{\Delta \Omega} \frac{1}{r \sin{\theta}} \partial_\theta (\sin{\theta}[q_\theta]) \ d \Omega =  \int_{\Delta \theta} \int_{\Delta \phi} \frac{1}{r \sin{\theta}} \partial_\theta (\sin{\theta}[q_\theta]) \ \sin{\theta} \ d\theta \ d\phi = \nonumber \\ 
= & \frac{1}{r} \int_{\Delta \theta} \int_{\Delta \phi} \partial_\theta (\sin{\theta}[q_\theta]) \ d\theta \ d\phi = 
\frac{1}{r} \int_{\Delta \theta} \partial_\theta \langle\sin{\theta}[q_\theta]\rangle_\phi \ d\theta = \frac{1}{r}(\langle\sin{\theta}[q_\theta] \rangle_\phi)_{\theta_R}^{\theta_L} = 0 \\
\\
\langle \nabla_\phi q_\phi \rangle = & \int_{\Delta \Omega} \frac{1}{r\sin{\theta}}\partial_\phi [q_\phi] = \int_{\Delta \theta} \int_{\Delta \phi} \frac{1}{r \sin{\theta}} \partial_\theta [q_\phi] \ \sin{\theta} \ d\theta \ d\phi = \nonumber \\
= & \frac{1}{r} \int_{\Delta \theta} \partial_\phi [q_\phi] \ d\theta d\phi = \frac{1}{r} \int_{\Delta \theta} \partial_\phi \langle\rho U_\phi\rangle_\theta \ d\phi = \frac{1}{r} (\langle q_\phi\rangle_\theta)_{\phi_R}^{\phi_L} = 0
\end{align}

\bibliography{referenc}

\end{document}